\documentclass[12pt,a4paper,twoside,english]{book}

  \usepackage{titletoc} 
  \usepackage[toctitles]{titlesec} 

\usepackage[french,english]{babel}

\usepackage{ae,lmodern}
\usepackage{ifluatex}
\ifluatex 
    \usepackage{fontspec}
    \defaultfontfeatures{Ligatures={TeX}}
\else
    \usepackage[utf8]{inputenc}
    \usepackage[T2A,T1]{fontenc}
    \usepackage{textcomp}
    \usepackage{bookman}
    \usepackage{palatino}
    \usepackage{kerkis}
\fi
\usepackage{amsmath}
\usepackage{amssymb}
\usepackage[nointegrals]{wasysym}
\usepackage{bm}
\usepackage{eurosym}
\usepackage{ulem}
\usepackage{anyfontsize}

\renewcommand{\emph}[1]{\textit{#1}}

\usepackage[table]{xcolor}
\definecolor{coltit}{rgb}{0.4,0.4,0.4}  
\definecolor{coltoc}{rgb}{1,1,0.8}      
\definecolor{colurl}{rgb}{0.,0.6,0.}    
\definecolor{colcite}{rgb}{0.,0.6,0.8}  
\definecolor{collink}{rgb}{0.,0.,0.9}   
\definecolor{colsec}{rgb}{0.7,0.1,0.}   
\definecolor{colhead}{rgb}{0.6,0.4,0.6} 
\definecolor{colcap}{rgb}{0.,0.6,0.4}   
\definecolor{coltabhead}{rgb}{0.,0.9,0.9}  
\definecolor{coltabcell}{rgb}{0.4,1.,0.8}  
\definecolor{coltabsep}{rgb}{0.7,1.,0.8}  
\definecolor{coleqnum}{rgb}{0.4,0.,0.4}  
\definecolor{coltip}{rgb}{0.4,0.4,0.}   
\usepackage[colorlinks=true,
            breaklinks=true,
            pdfstartview=FitV, 
            linkcolor=collink, 
            citecolor=colcite, 
            urlcolor=colurl, 
            backref=true]{hyperref}

\usepackage{tabularx}
\usepackage{multirow}
\usepackage{graphicx}
\usepackage{wrapfig}
\usepackage{longtable}
\usepackage{rotating}
\usepackage{arydshln}

\newcolumntype{L}[1]%
              {>{\raggedright\let\newline\\\arraybackslash\hspace{0pt}}m{#1}}
\newcolumntype{C}[1]%
              {>{\centering\let\newline\\\arraybackslash\hspace{0pt}}m{#1}}
\newcolumntype{R}[1]%
              {>{\raggedleft\let\newline\\\arraybackslash\hspace{0pt}}m{#1}}

\usepackage{multicol}
\usepackage[most]{tcolorbox}
\usepackage{etoolbox}
\usepackage{ifthen}
\usepackage{calc}

\usepackage{tikz}
\usetikzlibrary{arrows,shapes}

\graphicspath{{Figures/}{./Figures/}{../Figures/}}
\DeclareGraphicsExtensions{.png,.pdf,.jpg}

\usepackage{geometry}
\newcommand{\eg}{\textit{e.g.}}
\newcommand{\ie}{\textit{i.e.}}
\newcommand{\cf}{\textit{cf.}}

\newcommand{\etc}{\textit{etc.}}
\newcommand{\etal}{\textit{et al.}}

\newcommand{\ncode}[1]{\texttt{#1}}

\newcommand{\implic}{$\Rightarrow$}

\newcommand{\affilCEA}{Universit\'e Paris-Saclay, Universit\'e de Paris, CEA, CNRS, Astrophysique, Instrumentation, Modélisation Paris-Saclay, 91191 Gif-sur-Yvette, France}
\newcommand{\myemail}{frederic.galliano@cea.fr}

\newcommand{\seppar}{\vspace*{10pt}}

\newcommand{\citengl}[1]{``\textit{#1}''}
\newcommand{\citext}[1]{\textit{#1}}
\newcommand{\expression}[1]{\textit{#1}}
\newcommand{\familyname}[1]{\textsc{#1}}

\newcommand{\sms}[1]{{\mbox{{\scriptsize #1}}}}

\newcommand{\dd}{\textnormal{d}}
\newcommand{\ddiff}{\,\textnormal{d}}

\newcommand{\E}[1]{\times10^{#1}}

\newcommand{\lvec}{\overrightarrow}
\newcommand{\vect}[1]{\protect\overrightarrow{#1}}
\newcommand{\mat}[1]{\protect\overleftrightarrow{#1}}

\providecommand{\refeq}[1]{\hyperref[#1]{Eq.\ (\ref{#1})}}
\newcommand{\refeqp}[1]{\hyperref[#1]{(Eq.\ \ref{#1})}}
\newcommand{\refeqnp}[1]{\hyperref[#1]{Eq.\ \ref{#1}}}
\newcommand{\refeqs}[2]{\ifthenelse{\equal{#2}{}}%
                      {\hyperref[#1]{Eqs.\ (\ref{#1})}}%
                      {\hyperref[#1]{Eqs.\ (\ref{#1})}~--~\hyperref[#2]{(\ref{#2})}}}
\newcommand{\refeqsnp}[2]{\ifthenelse{\equal{#2}{}}%
                        {\hyperref[#1]{Eqs.\ \ref{#1}}}%
                        {\hyperref[#1]{Eqs.\ \ref{#1}}~--~\hyperref[#2]{\ref{#2}}}}

\newcommand{\reftab}[1]{\hyperref[#1]{Table \ref{#1}}}
\newcommand{\reftabs}[2]{\ifthenelse{\equal{#2}{}}%
                       {\hyperref[#1]{Tables \ref{#1}}}%
                       {\hyperref[#1]{Tables \ref{#1}}~--~\hyperref[#2]{\ref{#2}}}}

\newcommand{\reffig}[1]{\hyperref[#1]{Fig.\ \ref{#1}}}
\newcommand{\reffigs}[2]{\ifthenelse{\equal{#2}{}}%
                       {\hyperref[#1]{Figs.\ \ref{#1}}}%
                       {\hyperref[#1]{Figs.\ \ref{#1}}~--~\hyperref[#2]{\ref{#2}}}}
\newcommand{\refsubfig}[2]{\hyperref[#1]{Fig.\ \ref{#1}.#2}}

\newcommand{\refchap}[1]{\hyperref[#1]{Chap.\ \ref{#1}}}
\newcommand{\refchaps}[2]{\ifthenelse{\equal{#2}{}}%
                       {\hyperref[#1]{Chaps.\ \ref{#1}}}%
                       {\hyperref[#1]{Chaps.\ \ref{#1}}~--~\hyperref[#2]{\ref{#2}}}}

\newcommand{\refsec}[1]{\hyperref[#1]{Sect.\ \ref{#1}}}
\newcommand{\refsecs}[2]{\ifthenelse{\equal{#2}{}}%
                       {\hyperref[#1]{Sects.\ \ref{#1}}}%
                       {\hyperref[#1]{Sects.\ \ref{#1}}~-- \hyperref[#2]{\ref{#2}}}}
\newcommand{\refS}[1]{\hyperref[#1]{\S\ref{#1}}}

\newcommand{\refapp}[1]{\hyperref[#1]{Appendix \ref{#1}}}
\newcommand{\refapps}[2]{\ifthenelse{\equal{#2}{}}%
                       {\hyperref[#1]{Appendices \ref{#1}}}%
                       {\hyperref[#1]{Appendices \ref{#1}}~--~\hyperref[#2]{\ref{#2}}}}

\newcounter{textlistctr}

\newcommand{\squishlist}{
   \begin{list}{$\bullet$}
    { \setlength{\itemsep}{0pt}      \setlength{\parsep}{3pt}
      \setlength{\topsep}{3pt}       \setlength{\partopsep}{0pt}
      \setlength{\leftmargin}{1.5em} \setlength{\labelwidth}{1em}
      \setlength{\labelsep}{0.5em} } }
\newcommand{\squishlisttwo}{
   \begin{list}{$\bullet$}
    { \setlength{\itemsep}{0pt}    \setlength{\parsep}{0pt}
      \setlength{\topsep}{0pt}     \setlength{\partopsep}{0pt}
      \setlength{\leftmargin}{2em} \setlength{\labelwidth}{1.5em}
      \setlength{\labelsep}{0.5em} } }
\newcommand{\squishend}{
    \end{list}  }

\newcounter{obsrefctr}
\newcommand{\sectnn}[1]{\section*{#1}%
                      \addcontentsline{toc}{starsection}{#1}}

\newcommand{\spitz}{\textit{Spitzer}}

\newcommand{\hersc}{\textit{Herschel}}
\newcommand{\planck}{\textit{Planck}}

\newcommand{\IRACiv}{IRAC$_\sms{8$\mu m$}$}

\newcommand{\WISEiii}{WISE$_\sms{12$\mu m$}$}

\newcommand{\MIPSi}{MIPS$_\sms{24$\mu m$}$}

\newcommand{\SPIREii}{SPIRE$_\sms{350$\mu m$}$}
\newcommand{\SPIREiii}{SPIRE$_\sms{500$\mu m$}$}

\newcommand{\citeprep}[1]%
           {\citeauthor{#1}, \textcolor{colcite}{\textit{in prep.}}}
\newcommand{\citetprep}[1]%
           {\citeauthor{#1} \textcolor{colcite}{(\textit{in prep.})}}
\newcommand{\citepprep}[1]%
            {(\citeauthor{#1}, \textcolor{colcite}{\textit{in prep.}})}
\newcommand{\citesubm}[1]%
           {\citeauthor{#1}, \textcolor{colcite}{\textit{submitted}}}
\newcommand{\citetsubm}[1]%
           {\citeauthor{#1} \textcolor{colcite}{(\textit{submitted})}}
\newcommand{\citepsubm}[1]%
           {(\citeauthor{#1}, \textcolor{colcite}{\textit{submitted}})}
\newcommand{\citepress}[1]%
           {\citeauthor{#1}, \textcolor{colcite}{\textit{in press}}}
\newcommand{\citetpress}[1]%
           {\citeauthor{#1} \textcolor{colcite}{(\textit{in press})}}
\newcommand{\citeppress}[1]%
           {(\citeauthor{#1}, \textcolor{colcite}{\textit{in press}})}

\newcommand{\M}[1]{M$\;$#1}

\newcommand{\ngc}[1]{NGC$\;$#1}

\newcommand{\um}[1]{UM$\;$#1}

\newcommand{\rcw}[1]{RCW$\;$#1}

\newcommand{\xxxdor}{30$\;$Doradus}

\newcommand{\hen}{He$\;$2-10}

\newcommand{\izw}{I$\;$Zw$\;$18}
\newcommand{\iizw}{II$\;$Zw$\;$40}

\newcommand{\snia}{SN$\,$\textsc{i}a}
\newcommand{\snii}{SN$\,$\textsc{ii}}

\newcommand{\mic}{\mu\textnormal{m}}

\newcommand{\Lsun}{\textnormal{L}_\odot}
\newcommand{\Msun}{\textnormal{M}_\odot}
\newcommand{\Zsun}{\textnormal{Z}_\odot}

\newcommand{\emic}{\;\mic}

\newcommand{\eLsun}{\;\Lsun}
\newcommand{\eMsun}{\;\Msun}
\newcommand{\eZsun}{\;\Zsun}

\newcommand{\tmic}{$\mic$}

\newcommand{\tMsun}{$\Msun$}

\newcommand{\ariii}{Ar$\,$\textsc{iii}}
\newcommand{\arii}{Ar$\,$\textsc{ii}}
\newcommand{\ci}{C$\,$\textsc{i}}
\newcommand{\cii}{C$\,$\textsc{ii}}
\newcommand{\feii}{Fe$\,$\textsc{ii}}
\newcommand{\ha}{H$\alpha$}

\newcommand{\hi}{H$\,$\textsc{i}}
\newcommand{\hii}{H$\,$\textsc{ii}}
\newcommand{\lya}{Ly$\alpha$}

\newcommand{\neiii}{Ne$\,$\textsc{iii}}
\newcommand{\neii}{Ne$\,$\textsc{ii}}

\newcommand{\oi}{O$\,$\textsc{i}}
\newcommand{\oiv}{O$\,$\textsc{iv}}
\newcommand{\siii}{S$\,$\textsc{iii}}
\newcommand{\siv}{S$\,$\textsc{iv}}
\newcommand{\siII}{Si$\,$\textsc{ii}}

\newcommand{\hmol}{H$_\textnormal{2}$}

\newcommand{\hiline}{[\hi]$_{21\,\textnormal{cm}}$}

\newcommand{\ciiline}{[\cii]$_{158\mu\textnormal{m}}$}
\newcommand{\lyaline}{\lya$_{121.6\textnormal{nm}}$}
\newcommand{\haline}{\ha$_{656.3\textnormal{nm}}$}

\newcommand{\neiiiline}{[\neiii]$_{15.56\mu\textnormal{m}}$}
\newcommand{\neiiline}{[\neii]$_{12.81\mu\textnormal{m}}$}

\newcommand{\niiline}{[N$\,$\textsc{ii}]$_{122\mu\textnormal{m}}$}

\newcommand{\oiline}{[\oi]$_{63\mu\textnormal{m}}$}

\newcommand{\sivline}{[\siv]$_{10.51\mu\textnormal{m}}$}

\newcommand{\COio}{$^{12}$CO(J$=$1$\rightarrow$0)$_{2.6\textnormal{mm}}$}
\newcommand{\COiitoi}{$^{12}$CO(J$=$2$\rightarrow$1)$_{1.3\textnormal{mm}}$}
\newcommand{\COiiitoii}{$^{12}$CO(J$=$3$\rightarrow$2)$_{867\mu\textnormal{m}}$}

\newcommand{\hmoloo}{\hmol$\,$0-0}
\newcommand{\hmolooo}{\hmoloo$\,\textnormal{S(0)}_{28.3\mu\textnormal{m}}$}
\newcommand{\hmolooi}{\hmoloo$\,\textnormal{S(1)}_{17.0\mu\textnormal{m}}$}

\newcommand{\doctype}{\textsc{F.\ Galliano's Research Note Series}}

\usepackage{sectsty}
\usepackage{titling}
\usepackage{authblk}
\usepackage{natbibtoc}
\usepackage{emptypage}
\usepackage{fancyhdr}
\usepackage{lscape}
\usepackage{lastpage}
\usepackage{enumitem}
\usepackage{xpatch}

\pretitle{\begin{center}\bfseries\LARGE\color{coltit}}
\posttitle{\par\end{center}\vskip 0.5em}
\chapterfont{\color{colsec}}
\sectionfont{\color{colsec}}
\subsectionfont{\color{colsec}}
\subsubsectionfont{\color{colsec}}
\paragraphfont{\color{colsec}}

\definecolor{colitem}{rgb}{0.4,0,0.7}

\usetikzlibrary{shadows}
\newlength{\tmpShadow}
\newcommand{\ourShadow}[2]{%
  \settowidth{\tmpShadow}{#1}
  \addtolength{\tmpShadow}{.1em}
  \raisebox{-0.25ex}{\textcolor{gray!70}{#1}}%
  \kern-\tmpShadow%
  \textcolor{#2}{#1}}
\newcommand*{\MyTriangle}{\ourShadow{$\blacktriangleright$}{colitem}}
\newcommand*{\MyBall}{\tikz\draw[baseline,ball color=colitem,draw=colitem] circle (2.5pt);}

\setlist[itemize]{label=\MyTriangle}
\setlist[itemize,1]{label=\MyBall}
\setlist[enumerate]{label=\textcolor{colitem}{\textbf{\textit{(\roman*)}}}}
\setlist[enumerate,1]{label=\textcolor{colitem}{\textbf{\textit{\arabic*.}}}}
\setlist[enumerate,2]{label=\textcolor{colitem}{\textbf{\textit{\alph*)}}}}
\setlist[description,1]{font=\bfseries\color{colitem}}
\setlist[description,2]{font=\itshape\color{colitem}}
\newlist{inlinelist}{enumerate*}{1}
\setlist*[inlinelist,1]{label=\textcolor{colitem}{\textit{(\roman*)}}}
\newlist{inlinelistalph}{enumerate*}{1}
\setlist*[inlinelistalph,1]{label=\textcolor{colitem}{\textit{(\alph*)}}}

\xpretocmd\headrule{\color{colhead}}{}{\PatchFailed}
\xpretocmd\footrule{\color{colhead}}{}{\PatchFailed}
\author{\href{mailto:\myemail}{Fr\'ed\'eric \textsc{Galliano}}}
\affil{{\small\href{http://irfu.cea.fr/dap/index.php}{\affilCEA}}}

  \usepackage{colortbl}
  \usepackage{makecell}
  \usepackage[upright]{fourier} 
  \usepackage{modiagram}
  \usepackage{pdfcomment}

  \usepackage{amsfonts}
  \definecolor{Prune}{RGB}{99,0,60}
  \usepackage{mdframed}
  \usepackage{scrextend}
  \usepackage[absolute]{textpos} 
  \usepackage{array}

  \renewcommand{\hrulefill}{\leavevmode\leaders\hrule height 1pt\hfill}  
  \renewcommand{\headrule}{\vspace{-6pt}%
              \textcolor{colhead}{\hrulefill\raisebox{-2.1pt}%
                {\quad\decofourleft\decosix\decofourright\quad}%
                 \hrulefill}}
  \renewcommand{\footrule}{%
              \textcolor{colhead}{\hrulefill\raisebox{-2.1pt}%
                {\quad\decofourleft\leafNE\decofourright\quad}%
                 \hrulefill}\par}

  \renewcommand{\thechapter}{\Roman{chapter}}
  \newcommand{\newchapter}[1]{\chapter{#1}
    \addcontentsline{lof}{chapter}{\numberline{\thechapter}#1}
    \addcontentsline{lot}{chapter}{\numberline{\thechapter}#1}}

  \newcommand{\CClicence}{Licensed under    
              \href{https://creativecommons.org/licenses/by-sa/4.0/}{CC BY-SA 4.0}.}
  \newcommand{\cclicence}{licensed under    
              \href{https://creativecommons.org/licenses/by-sa/4.0/}{CC BY-SA 4.0}}
  \usepackage[font={color=colcap},figurename=Figure,labelfont=sc]{caption}
  \newcommand{\newcap}[2]{\caption[#1]{\textsl{#1.} #2}}

  \makeatletter
  \patchcmd{\listoffigures}
    {\@mkboth{\MakeUppercase\listfigurename}{\MakeUppercase\listfigurename}}
  {\addcontentsline{toc}{chapter}{\listfigurename}\markboth{\listfigurename}{}}
    {}{}
  \patchcmd{\listoftables}
    {\@mkboth{\MakeUppercase\listtablename}{\MakeUppercase\listtablename}}
    {\addcontentsline{toc}{chapter}{\listtablename}\markboth{\listtablename}{}}
    {}{}
  \patchcmd{\tableofcontents}
    {\@mkboth{\MakeUppercase\contentsname}{\MakeUppercase\contentsname}}
    {\addcontentsline{toc}{chapter}{\contentsname}\markboth{\contentsname}{}}
    {}{}
  \makeatother

  \usepackage{tocbasic}
  \DeclareTOCStyleEntry[dynnumwidth]{tocline}{figure}
  \DeclareTOCStyleEntry[dynnumwidth]{tocline}{table}

  \DeclareMathOperator\erf{erf}

  \newcommand{\mytip}[2]{\pdftooltip{\textcolor{coltip}{#1}}{#2}}

  \newcommand{\hthreeD}{\mytip{3D}{Three Dimensional}}
  \newcommand{\hHAC}{\mytip{a-C(:H)}{Partially hydrogenated amorphous 
                                   carbon grains}}
  \newcommand{\hACF}{\mytip{ACF}{AutoCorrelation Function}}
  \newcommand{\hAGB}{\mytip{AGB}{Asymptotic Giant Branch stars}}
  \newcommand{\hAGN}{\mytip{AGN}{Active Galactic Nucleus}}
  \newcommand{\hAME}{\mytip{AME}{Anomalous Microwave Emission}}
  \newcommand{\hBCD}{\mytip{BCD}{Blue Compact Dwarf galaxies}}
  \newcommand{\hBEMBB}{\mytip{BEMBB}{Broken-Emissivity Modified Black 
                                    Body}}
  \newcommand{\hBG}{\mytip{BG}{Big Grains}}
  \newcommand{\hBH}{\mytip{BH}{Black Hole}}
  \newcommand{\hCGS}{\mytip{CGS}{Centimetre-Gram-Second}}
  \newcommand{\hCCD}{\mytip{CCD}{Charge-Coupled Device}}
  \newcommand{\hCDF}{\mytip{CDF}{Cumulative Distribution Function}}
  \newcommand{\hCNP}{\mytip{CNP}{Cosmic NanoParticles (a.k.a. dust)}}
  \newcommand{\hCIB}{\mytip{CIB}{Cosmic Infrared Background}}
  \newcommand{\hCMB}{\mytip{CMB}{Cosmic Microwave Background}}
  \newcommand{\hCNM}{\mytip{CNM}{Cold Neutral Medium}}
  \newcommand{\hCSD}{\mytip{CSD}{CircumStellar Dust}}
  \newcommand{\hDCD}{\mytip{DCD}{Disordered Charge Distribution}}
  \newcommand{\hDDA}{\mytip{DDA}{Discrete Dipole Approximation}}
  \newcommand{\hDGL}{\mytip{DGL}{Diffuse Galactic Light}}
  \newcommand{\hDIB}{\mytip{DIB}{Diffuse Interstellar Bands}}
  \newcommand{\hDLA}{\mytip{DLA}{Damped Lyman-Alpha systems}}
  \newcommand{\hEMT}{\mytip{EMT}{Effective Medium Theory}}
  \newcommand{\hERE}{\mytip{ERE}{Extended Red Emission}}
  \newcommand{\hETG}{\mytip{ETG}{Early-Type Galaxy (elliptical)}}
  \newcommand{\heVSG}{\mytip{eVSG}{evaporating Very Small Grains}}
  \newcommand{\hFIR}{\mytip{FIR}{Far-InfraRed (40-200 microns)}}
  \newcommand{\hFUV}{\mytip{FUV}{Far-UltraViolet (120-200 nm)}}
  \newcommand{\hFWHM}{\mytip{FWHM}{Full Width at Half Maximum}}
  \newcommand{\hGEMS}{\mytip{GEMS}%
            {Glass with Embedded Metals and Sulfides}}
  \newcommand{\hGRB}{\mytip{GRB}{Gamma-Ray Burst}}
  \newcommand{\hHB}{\mytip{HB}{Hierarchical Bayesian}}
  \newcommand{\hHDR}{\mytip{HDR}{Habilitation à Diriger des Recherches}}
  \newcommand{\hHIM}{\mytip{HIM}{Hot Ionized Medium}}
  \newcommand{\hICM}{\mytip{ICM}{InterClump Medium}}
  \newcommand{\hIDP}{\mytip{IDP}{Interplanetary Dust Particles}}
  \newcommand{\hiid}{\mytip{iid}{independent, identically distributed}}
  \newcommand{\hIMF}{\mytip{IMF}{Initial Mass Function}}
  \newcommand{\hIR}{\mytip{IR}{InfraRed (0.8-200 microns)}}
  \newcommand{\hISD}{\mytip{ISD}{InterStellar Dust}}
  \newcommand{\hISM}{\mytip{ISM}{InterStellar Medium}}
  \newcommand{\hISRF}{\mytip{ISRF}{InterStellar Radiation Field}}
  \newcommand{\hISS}{\mytip{ISS}{International Space Station}}
  \newcommand{\hLIRG}{\mytip{LIRG}{Luminous InfraRed Galaxies}}
  \newcommand{\hLMC}{\mytip{LMC}{Large Magellanic Cloud}}
  \newcommand{\hLIMS}{\mytip{LIMS}{Low- and Intermediate-Mass Stars}}
  \newcommand{\hLTG}{\mytip{LTG}{Late-Type Galaxy (spiral)}}
  \newcommand{\hMBB}{\mytip{MBB}{Modified Black Body}}
  \newcommand{\hMCRT}{\mytip{MCRT}{Monte-Carlo Radiative Transfer}}
  \newcommand{\hMCMC}{\mytip{MCMC}{Markov Chain Monte-Carlo}}
  \newcommand{\hMIR}{\mytip{MIR}{Mid-InfraRed (5-40 microns)}}
  \newcommand{\hMKS}{\mytip{MKS}{Meter-Kilogram-Second}}
  \newcommand{\hMKSA}{\mytip{MKSA}{Meter-Kilogram-Second-Ampere}}
  \newcommand{\hML}{\mytip{ML}{Machine-Learning}}
  \newcommand{\hMLE}{\mytip{MLE}{Maximum-Likelihood Estimation}}
  \newcommand{\hMS}{\mytip{MS}{Main Sequence star}}
  \newcommand{\hMW}{\mytip{MW}{Milky Way}}
  \newcommand{\hNHST}{\mytip{NHST}{Null Hypothesis Significance Test}}
  \newcommand{\hNIR}{\mytip{NIR}{Near-InfraRed (0.8-5 microns)}}
  \newcommand{\hNLP}{\mytip{NLP}{Natural Language Processing}}
  \newcommand{\hNS}{\mytip{NS}{Neutron Star}}
  \newcommand{\hNUV}{\mytip{NUV}{Near-UltraViolet (300-380 nm)}}
  \newcommand{\hOOP}{\mytip{OOP}{Out-Of-Plane bending mode of PAHs}}
  \newcommand{\hPAH}{\mytip{PAH}{Polycyclic Aromatic Hydrocarbon}}
  \newcommand{\hPCA}{\mytip{PCA}{Principal Component Analysis}}
  \newcommand{\hPDF}{\mytip{PDF}{Probability Density Function}}
  \newcommand{\hPDR}{\mytip{PDR}{PhotoDissociation Regions}}
  \newcommand{\hPN}{\mytip{PN}{Planetary Nebulae}}
  \newcommand{\hppb}{\mytip{ppb}{part per billion}}
  \newcommand{\hppp}{\mytip{ppp}{posterior predictive p-value}}
  \newcommand{\hQSO}{\mytip{QSO}{Quasi-Stellar Object}}
  \newcommand{\hRAT}{\mytip{RAT}{Radiative Alignment Torques}}
  \newcommand{\hSSC}{\mytip{SSC}{Super Star Cluster}}
  \newcommand{\hSED}{\mytip{SED}{Spectral Energy Distribution}}
  \newcommand{\hSF}{\mytip{SF}{Star Formation}}
  \newcommand{\hSFH}{\mytip{SFH}{Star Formation History}}
  \newcommand{\hSFR}{\mytip{SFR}{Star Formation Rate}}
  \newcommand{\hSI}{\mytip{SI}{International System of units}}
  \newcommand{\hSLED}{\mytip{SLED}{Spectral Line Energy Distribution}}
  \newcommand{\hSMC}{\mytip{SMC}{Small Magellanic Cloud}}
  \newcommand{\hSN}{\mytip{SN}{SuperNova}}
  \newcommand{\hSNIa}{\mytip{\snia}{Type Ia SuperNova}}
  \newcommand{\hSNII}{\mytip{\snii}{Type II SuperNova}}
  \newcommand{\hSNR}{\mytip{SNR}{SuperNova Remnant}}
  \newcommand{\hsSFR}{\mytip{sSFR}{specific Star Formation Rate 
            (SFR/stellar mass)}}
  \newcommand{\hSUE}{\mytip{SUE}{Skewed Uncertainty Ellipse}}
  \newcommand{\hTLS}{\mytip{TLS}{Two-Level System}}
  \newcommand{\hTPAGB}{\mytip{TPAGB}{Thermally-Pulsing Asymptotic Giant 
                                   Branch star}}
  \newcommand{\hUIB}{\mytip{UIB}{Unidentified Infrared Bands}}
  \newcommand{\hULIRG}{\mytip{ULIRG}{UltraLuminous InfraRed Galaxies}}
  \newcommand{\hUV}{\mytip{UV}{UltraViolet (10-380 nm)}}
  \newcommand{\hVCD}{\mytip{VCD}{Very Cold Dust}}
  \newcommand{\hVSG}{\mytip{VSG}{Very Small Grain}}
  \newcommand{\hYSO}{\mytip{YSO}{Young Stellar Object}}
  \newcommand{\hWD}{\mytip{WD}{White Dwarf}}
  \newcommand{\hWIM}{\mytip{WIM}{Warm Ionized Medium}}
  \newcommand{\hWNM}{\mytip{WNM}{Warm Neutral Medium}}
  
  \newcommand{\hZAMS}{\mytip{ZAMS}{Zero-Age Main Sequence}}

  \newcommand{\hAKARI}{\mytip{AKARI}{IR space telescope (2005-2007)}}
  \newcommand{\hALMA}{\mytip{ALMA}{Atacama Large Millimeter/submillimeter 
                     Array (2011)}}
  \newcommand{\hAPEX}{\mytip{APEX}{Atacama Pathfinder Experiment 
                    (submm telescope; 2004-)}}
  \newcommand{\hATHENA}{\mytip{ATHENA}{Advanced Telescope for High ENergy 
                                     Astrophysics (~2030)}}
  \newcommand{\hBLAST}{\mytip{BLAST}{Balloon-borne Large Aperture 
                      Submillimeter Telescope (1997-2010)}}
  \newcommand{\hCOBE}{\mytip{COBE}{COsmic Background Explorer (MIR-to-cm 
                    telescope; 1989-1993)}}
  \newcommand{\hCSO}{\mytip{CSO}{Caltech Submilleter Observatory
                   (1986-2015)}}
  \newcommand{\hDIRBE}{\mytip{DIRBE}{Diffuse Infrared Background 
                     Experiment (IR instrument onboard COBE)}}
  \newcommand{\hDMR}{\mytip{DMR}{Differential Microwave Radiometer (cm 
                   instrument onboard COBE)}}
  \newcommand{\hFIRAS}{\mytip{FIRAS}{Far-InfraRed Absolute 
                     Spectrophotometer (instrument onboard COBE)}}

  \newcommand{\hFTS}{\mytip{FTS}{Fourier Transform Spectrometer}}
  \newcommand{\hGaia}{\mytip{Gaia}{Visible space telescope (2013-2020)}}
  \newcommand{\hhersc}{\mytip{\hersc}{FIR-submm space telescope 
                     (2009-2013)}}

  \newcommand{\hHST}{\mytip{HST}{Hubble Space Telescope (1990-)}}
  \newcommand{\hIRAC}{\mytip{IRAC}{InfraRed Array Camera (MIR instrument 
                    onboard Spitzer)}}
  \newcommand{\hIRAS}{\mytip{IRAS}{InfraRed Astronomical Satellite
                    (1983)}}

  \newcommand{\hIRS}{\mytip{IRS}{InfraRed Spectrograph (instrument 
                   onboard Spitzer)}}
  
  \newcommand{\hISO}{\mytip{ISO}{Infrared Space Observatory (1995-1998)}}
  \newcommand{\hIUE}{\mytip{IUE}{International Ultraviolet Explorer 
                   (1978-1996)}}
  \newcommand{\hJCMT}{\mytip{JCMT}{James Clerk Maxwell Telescope (submm 
                    observatory; 1987-)}}
  \newcommand{\hJWST}{\mytip{JWST}{James Webb Space Telescope (IR space 
                    telescope; 2021-)}}
  \newcommand{\hKAO}{\mytip{KAO}{Kuiper Airborne Observatory (IR 
                   telescope; 1974-1995)}}

  \newcommand{\hMUSE}{\mytip{MUSE}{Multi Unit Spectroscopic Explorer}}
  \newcommand{\hNCT}{\mytip{NCT}{Nuclear Compton Telescope}}
  \newcommand{\hNIKA}{\mytip{NI\-KA2}{New IRAM Kids Arrays (millimeter 
                                  camera)}}

  \newcommand{\hOAO}{\mytip{OAO}{Orbiting Astronomical Observatory 
                   (1968)}}
  \newcommand{\hPACS}{\mytip{PACS}{Photodetector Array Camera and 
                    Spectrometer (instrument onboard Herschel)}}
  \newcommand{\hPILOT}{\mytip{PILOT}{Polarized Instrument for the 
                     Long-wavelength Observation of the Tenuous ISM 
                     (balloon; 2015-2017)}}
  \newcommand{\hplanck}{\mytip{\planck}{submm-mm space telescope 
                                     (2009-2013)}}
  \newcommand{\hPRONAOS}{\mytip{PRONAOS}{PROjet National pour 
                       l'Observation Submillimétrique (balloon; 
                       1994-1999)}}
  \newcommand{\hSOFIA}{\mytip{SOFIA}{Stratospheric Observatory for 
                     Infrared Astronomy (2010-)}}
  \newcommand{\hSPICA}{\mytip{SPICA}{SPace Infrared telescope for 
                            Cosmology and Astrophysics 
                            (canceled 2030 mission)}}
  \newcommand{\hSPIRE}{\mytip{SPIRE}{Spectral and Photometric Imaging 
                     REceiver (instrument onboard Herschel)}}
  \newcommand{\hspitz}{\mytip{\spitz}{IR space telescope (2003-2009)}}
  \newcommand{\hTIR}{\mytip{TIR}{Total InfraRed}}
  
  \newcommand{\hVLT}{\mytip{VLT}{Very Large Telescope}}
  
  \newcommand{\hWISE}{\mytip{WISE}{Wide-field Infrared survey Explorer 
                    (2009-2011)}}
  \newcommand{\hWMAP}{\mytip{WMAP}{Wilkinson Microwave Anisotropy Probe
                                 (2001-2010)}}

  \newcommand{\hdustiness}{\mytip{dustiness}{Dust-to-gas mass ratio}}
  \newcommand{\hDustiness}{\mytip{Dustiness}{Dust-to-gas mass ratio}}

  \newcommand{\hDGS}{\mytip{DGS}{Dwarf Galaxy Sample; Herschel program}}
  \newcommand{\hDustPedia}{\mytip{DustPedia}{European collaboration aimed
            at providing a census of dust properties in nearby 
            galaxies}}
  \newcommand{\hSAGE}{\mytip{SAGE/HERITAGE}{Spitzer and Herschel surveys 
                                          of the Magellanic clouds}}

  \newcommand{\minitoc}%
    {\begin{tcolorbox}[breakable,colback=coltoc,colframe=white,center,%
                      width=0.9\textwidth,before 
                      upper={\let\clearpage\relax}]%
      \textbf{\textcolor{colsec}{\Large Contents}}
      \stopcontents\startcontents\printcontents{}{1}{}%
    \end{tcolorbox}}

  \renewcommand{\sectnn}[1]{\setcounter{secnumdepth}{0}\section{#1}%
            \setcounter{secnumdepth}{3}}

  \newcommand{\proba}[1]{p\left(#1\right)}
  \newcommand{\tproba}[1]{$\proba{#1}$}
  \newcommand{\pcond}[2]{p\left(#1\middle|#2\right)}
  \newcommand{\tpcond}[2]{$\pcond{#1}{#2}$}

  \newcommand{\Pcond}[2]{P\left(#1\middle|#2\right)}
  \newcommand{\tPcond}[2]{$\Pcond{#1}{#2}$}
  \newcommand{\vecprob}[1]{\vec{#1}}
  \newcommand{\CR}[1]{CR_{95\,\%}\left(#1\right)}
  \newcommand{\tCR}[1]{$\tCR{#1}$}
  \newcommand{\CI}[1]{CI_{95\,\%}\left(#1\right)}
  \newcommand{\tCI}[1]{$\tCI{#1}$}

  \defcitealias{desert90}{\tt DBP90}
  \defcitealias{draine09}{D09}
  \defcitealias{galliano21}{G21}
  \defcitealias{galliano18a}{\tt HerBIE}
  \defcitealias{jones17}{\tt THEMIS}
  \defcitealias{mathis77}{MRN}
  \defcitealias{varosi99}{VD99}
  \defcitealias{zubko04}{ZDA04}
  \makeatletter
    \newcommand*{\hyperlinkcite}[1]{\hyper@link{cite}{cite.#1}}
  \makeatother

  \newcommand{\citesmart}[2]{\begin{flushright}%
               \begin{minipage}{0.7\textwidth}%
                 \begin{flushright}%
                   \textcolor[rgb]{0.4,0.4,0.}{\textit{#1} \\ ~ \\ #2}
                 \end{flushright}%
               \end{minipage}%
             \end{flushright}\seppar}

  \newcommand{\takeaway}[1]%
    {\begin{quote}
      \textcolor{colitem}{\lefthand}~#1
    \end{quote}}

  \def\dout{\bgroup\markoverwith{\lower-0.2ex\hbox
            {\kern-.03em\vbox{\hrule width.2em\kern0.45ex\hrule}\kern-.03em}}%
             \ULon}\MakeRobust\dout

  \newlength\myheight
  \newlength\mydepth
  \settototalheight\myheight{Xygp}
  \newcommand*\inlinegraphics[1]{%
    \settototalheight\myheight{Xygp}%
    \settodepth\mydepth{Xygp}%
    \raisebox{-\mydepth}{\includegraphics[height=\myheight]{#1}}}

  \newcommand{\abstractEN}{Interstellar dust is a key physical ingredient 
    of galaxies, obscuring star formation, regulating the heating and 
    cooling of the gas, and building-up chemical complexity.
    In this manuscript, I give a wide review of interstellar dust 
    properties and some of the modern techniques used to study it.
    I start with a general introduction presenting the main concepts, 
    in molecular and solid-state physics, required to understand the 
    contemporary literature on the subject.
    I then review the empirical evidence we currently use to constrain 
    state-of-the-art dust models.
    Follows a long discussion about our current understanding of the 
    grain properties of nearby galaxies, with an emphasis on the 
    results from spectral energy distribution modeling.
    The following chapter presents dust evolution at all scales.
    I review the different microphysical evolution processes, and the 
    way they are accounted for in cosmic dust evolution models.
    I give my take on the origin of interstellar dust in galaxies of 
    different metallicities.
    The last chapter focusses on methodology.
    I give an introduction to the Bayesian method and compare it to 
    frequentist techniques.
    I discuss the epistemological consequences of the two approaches, 
    and show why the field  of interstellar dust requires a 
    probabilistic viewpoint.
    I end the manuscript with a summary of the major breakthroughs 
    achieved in the past decade, and delineate a few prospectives for 
    the next decade.}

\begin{document}
  \selectlanguage{english}%
  \startcontents%
  \frontmatter%

  \begin{titlepage}
    \thispagestyle{empty}
    \newgeometry{left=7.5cm,bottom=2cm, top=1cm, right=1cm}
    \tikz[remember picture,overlay] \node[opacity=1,inner sep=0pt] at   
         (-28mm,-135mm){\includegraphics{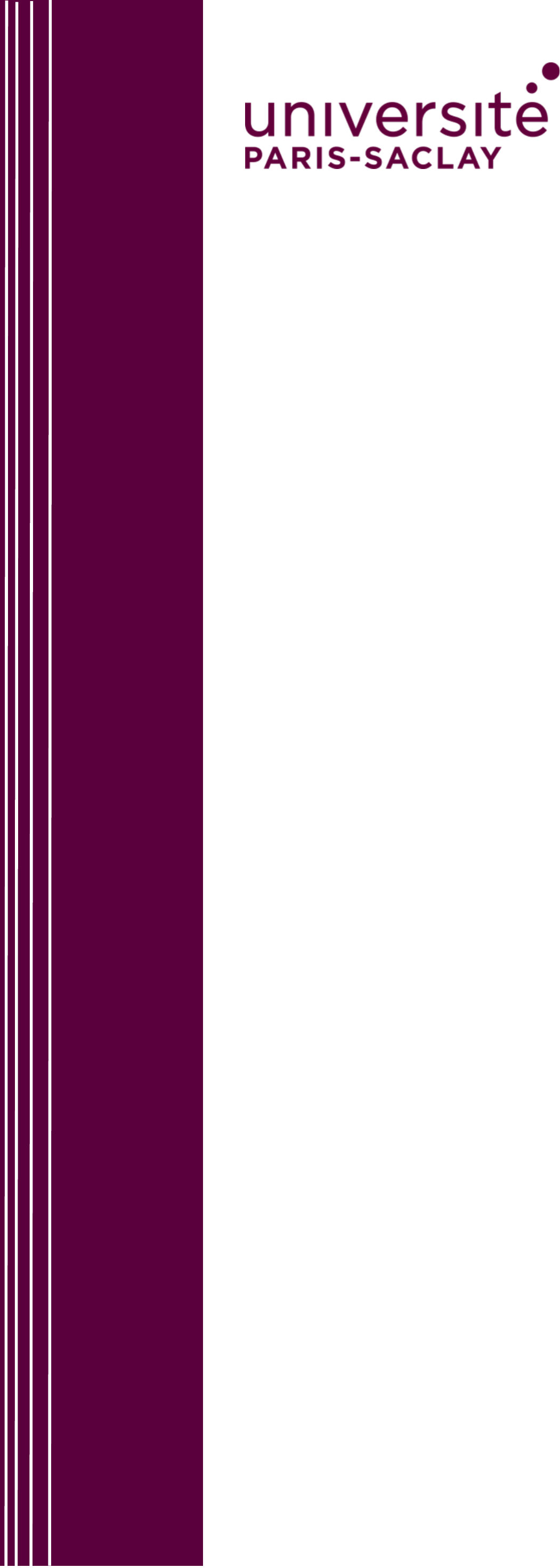}};

    \fontfamily{fvs}\fontseries{m}\selectfont

    \color{white}
    \begin{picture}(0,0)
      \put(-180,-735){\rotatebox{90}{\Huge\textbf{Habilitation à diriger des  
                                                  recherches}}}
    \end{picture}

    \flushright
    \vspace{50mm}
    \color{Prune}
    \fontfamily{fvs}\fontseries{m}\fontsize{22}{26}\selectfont
    A Nearby Galaxy Perspective on Interstellar Dust Properties and their   
    Evolution

    \fontfamily{fvs}\fontseries{m}\fontsize{8}{12}\selectfont
    \normalsize
    \vspace{1.5cm}
    \color{black}
    \textbf{Habilitation à diriger des recherches de l'Université Paris-Saclay}

    \vspace{30mm}
    \textbf{Habilitation présentée et soutenue à Gif-sur-Yvette, 
            le vendredi 14 janvier 2022, par}\\
    \bigskip
    {\Large {\color{Prune} \textbf{Frédéric GALLIANO}}} \\
    Département d'Astrophysique, CEA Paris-Saclay

    \vspace{\fill}

    \flushleft \small \textbf{Composition du jury:}
    \bigskip
    \scriptsize
    \begin{tabular}{|p{8cm}l}
      \arrayrulecolor{Prune}
      \textbf{Véronique BUAT} & Rapportrice \\ 
      Professeure, Laboratoire d'Astrophysique de Marseille &
      \\
      \textbf{Stéphane CHARLOT} & Examinateur \\ 
      Directeur de recherche, Institut d'Astrophysique de Paris &
      \\
      \textbf{Vassilis CHARMANDARIS} & Rapporteur \\ 
      Professeur, Université de Crète, Grèce &   
      \\
      \textbf{François-Xavier DÉSERT} & Examinateur \\ 
      Astronome, Institut de Planétologie et d'Astrophysique de Grenoble &
      \\
      \textbf{Thomas HENNING} & Rapporteur \\ 
      Professeur, Institut Max Planck d'Astronomie, Heidelberg, Allemagne &
      \\
      \textbf{Laurent VERSTRAETE} & Président \\ 
      Professeur, Institut d'Astrophysique Spatiale, Orsay &
      \\
    \end{tabular}
  \end{titlepage}
  \thispagestyle{empty}~\clearpage

  \thispagestyle{empty}
  \begin{tcolorbox}[breakable,colback=coltoc,colframe=white,center,%
                     width=0.9\textwidth,before 
                     upper={\let\clearpage\relax}]%
    \tableofcontents%
  \end{tcolorbox}
  \clearpage
  \begin{tcolorbox}[breakable,colback=coltoc,colframe=white,center,%
                     width=0.9\textwidth,before 
                     upper={\let\clearpage\relax}]%
    \listoffigures%
  \end{tcolorbox}
  \clearpage
  \begin{tcolorbox}[breakable,colback=coltoc,colframe=white,center,%
                     width=0.9\textwidth,before 
                     upper={\let\clearpage\relax}]%
    \listoftables%
  \end{tcolorbox}
  \setcounter{tocdepth}{4}


\chapter{Abstract}
\markboth{Abstract}{}
\citesmart{The truth is the whole.}{\citep[George Wilhelm Friedrich 
                                           \familyname{Hegel};][]{hegel1807}}
\abstractEN


\chapter{Introduction}
\label{chap:intro}
\markboth{Introduction}{}
\citesmart{The unity of all science consists alone in its method, not in its 
           material.}{\citep[Karl \familyname{Pearson};][]{pearson1892}}

\sectnn{Interstellar Dust: A Key To Understanding Galaxy Evolution}

Understanding galaxy evolution is one of the main objectives of observational cosmology, as it allows mapping the history of the Universe, from the dark ages to the present times \citep[\eg][]{madau14,buat15}.
At the center of this evolution lies the \expression{InterStellar Medium} (\hISM).
This complex intertwining of ionized, atomic and molecular gas phases mixed with dust grains, fills the volume of a galaxy, ultimately leading to star formation (\hSF), by gravitational collapse \citep[\eg][]{klessen16}.
Although accounting for only $\simeq\textnormal{1}\,\%$ of its mass, dust is an essential component of the \hISM.
It consists of solid particles ($\textnormal{0.3}\;\textnormal{nm}\lesssim \textnormal{radius}\lesssim\textnormal{0.3}\emic$) made out of the available heavy elements, predominantly arranged in silicate and carbonaceous compounds
\citep[\eg][]{draine03c}.
These grains, absorbing and scattering starlight, have a radical impact on a galaxy.
\begin{itemize}
  \item 
    They have a nefarious role of obscuring our direct view of star formation.
    They normally re-radiate $\simeq\textnormal{25}\,\%$ of the stellar power 
    in the \expression{InfraRed} (\hIR; \cf~\reftab{tab:spectralrange}), and up 
    to $\simeq\textnormal{99}\,\%$ in ultraluminous \hIR\ galaxies 
    \citep[\eg][]{bianchi18,clements96}.
  \item 
    Dust is an essential ingredient of star formation, as it contributes to
    \citep[\eg][]{li03b}:
    \begin{itemize}
      \item 
        radiatively evacuating the gravitational energy of collapsing 
        molecular clouds;
      \item 
        shielding the molecules from starlight, which protect them from
        destruction and reduces their ionization fraction, allowing the 
        formation of protostellar cores.
    \end{itemize}
  \item 
    In addition, grains are responsible for the heating of the gas in 
    \expression{PhotoDissociation Regions} (\hPDR), by the photoelectric effect 
    \citep{draine78,kimura16}.
  \item 
    They are also catalysts of numerous chemical reactions, including the 
    formation of the most abundant molecule in the Universe, \hmol\ 
    \citep{gould63,bron14}.
  \item 
    Elongated grains tend to align with the magnetic field.
    Polarized extinction (in the visible) and emission (in the \hIR) by grains 
    are therefore one of the most popular tools to study the magnetic field in 
    the \hISM\ 
    \citep[\eg][]{planck-collaboration16g,planck-collaboration16f,guillet18}.
\end{itemize}
\takeaway{A detailed knowledge of the dust properties and their evolution is therefore imperative in order to both interpret observations of galaxies and model their \hISM.}

Dust physics is characterized by the great complexity of its make-up, as the number of ways to combine elements to build interstellar solids is virtually limitless.
Most of the progress in this field thus relies on empirical constraints:
observations and laboratory experiments on cosmic dust analogs.
Our current knowledge of \expression{interstellar dust} (\hISD) properties is however hampered by several factors \citep[][for a review]{galliano18}.
First, observations of interstellar regions are always the superimposition, along the line of sight and within the telescope beam, of a range of physical conditions: 
\begin{inlinelist}
  \item 
    intensity and hardness of the \expression{InterStellar Radiation Field} 
    (\hISRF);
  \item 
    gas density; and 
  \item 
    presence of shocks.
\end{inlinelist}
Consequently, since we can never accurately recover the \expression{3-Dimensional} (\hthreeD) structure of a region, several degeneracies between the grain constitution and their excitation prevent a unique solution.
Second, the grain constitution\footnote{Stoichiometry, chemical composition, solid-state structure, size distribution and abundance relative to the gas.} is known to evolve under the effects of \hISRF\ and gas density \citep[\eg][]{draine09,jones13,ysard15}.
It is thus likely that, in addition to variations of excitation conditions, \hISD\ observables are coming from a combination of altered grain mixtures.
Finally, the derivation of precise dust properties, even from observations towards a uniform, uncontaminated region, is limited by an incomplete spectral coverage and by instrumental uncertainties.
\takeaway{It follows that a rigorous attempt at quantifying grain parameters and their evolution must account for these factors in both the choices of astrophysical targets and modeling approach.}

\sectnn{The Relevance of Nearby Galaxies}

Due to their proximity, \hISM\ regions of our own galaxy, the \expression{Milky Way} (\hMW), can be observed with the finest linear resolution.
\hMW\ studies have consequently laid the ground for the development of physical dust models \citep[][for a review]{draine03c}.
They are however limited by the small range of environmental conditions they span.
\begin{itemize}
  \item 
    There are no really massive star-forming regions in the \hMW.
    The only \expression{Super Star Cluster} 
    \citep[\hSSC; $M_\star\gtrsim10^5\;\Msun$; \eg][]{johnson01,dowell08}
    is Westerlund~2 in \rcw{49} \citep{moffat91}.
  \item
    As in most spiral galaxies, there is a radial gradient of 
    metallicity \citep[the mass fraction of elements heavier than He, noted 
                       $Z$; \eg][]{asplund09}.
    It however ranges narrowly 
    \citep[$0.7\;\Zsun\lesssim Z\lesssim 2\;\Zsun$;][]{henry99}.
  \item 
    The massive central black hole, Sgr~A$^\star$ of the Galactic center is 
    relatively passive \citep{mezger96} compared to 
    \expression{Active Galactic Nuclei} (\hAGN), such as \ngc{1068} 
    \citep[\eg][]{le-floch01}.
\end{itemize}
\hMW\ studies are also limited by the confusion along the sightline, as we are seeing the projected material of the entire disk.
Finally, distances of interstellar clouds can be difficult to estimate for the same reason, although \hthreeD\ maps of the \hMW\ are becoming more precise \citep[\eg][]{lallement18}.

In contrast, nearby galaxies \citep[closer than $\simeq\textnormal{100}$~Mpc;][]{galliano18} represent an under-tapped population with several potentials.
First, they harbor a wider range of environmental parameters, allowing us, in particular, to probe dust in extreme conditions.
\begin{itemize}
  \item 
    Nearby galaxies, especially \expression{Blue Compact Dwarfs} (\hBCD), 
    contain numerous \hSSC s 
    \citep[\eg][]{oconnell94,johnson00,martin-hernandez05}. 
    They are ideal laboratories to understand the impact of massive star 
    formation on the \hISM.
  \item 
    Extremely low-metallicity objects, such as \izw\ 
    \citep[$Z\simeq1/35\;\Zsun$;][]{izotov99}, allow us to study dust in 
    environments where the chemical enrichment resembles primordial galaxies.
  \item
    \hAGN s have a radical impact on the \hISM\ of their host galaxies, heavily 
    processing the grains \citep[\eg\ their crystalline fraction;][]{spoon06}.
    In addition, bright \hAGN s can be used to study grains in absorption
    \citep[\eg][]{spoon02,mason07}.
\end{itemize}
Second, face-on galaxies, observed at high Galactic latitude, provide clearer sightlines than in the \hMW.
Finally, the lower linear resolution we can reach in nearby objects ($\simeq\textnormal{100}$~pc to 1~kpc in the \hIR) is the ideal length scale to adopt a statistical description of the distribution of clouds and stars\footnote{The average distance between stars is $\simeq\textnormal{1}$~pc, and the typical size of molecular clouds ranges from $\simeq\textnormal{1}$ to $\simeq100$~pc \citep[\eg][]{solomon87}.}, whereas detailed, parsec-scale \hMW\ studies are left to the nearly impossible task of inferring the precise geometry of each single cloud and the position in space of the surrounding stars.

\sectnn{Scope of the Manuscript}

Most of my career until now has been devoted to studying the dust properties and their evolution in nearby galaxies.
I have chosen to focus the present manuscript on the following directions.
\begin{itemize}
  \item 
    It is primarily a synthesis of my achievements in this field.
    The goal is not to repeat the content of my publications, but to put them
    in perspective with other studies.
    For that reason, the presentation of my results occupy only a small fraction
    of the manuscript.
  \item 
    It also provides a review of our current understanding of \hISD,
    and outlines the most promising directions this field should explore during
    the next decade.
  \item 
    Finally, I have tried to compile introductory material, concepts and figures
    that could be useful to students starting in this field, but that are 
    otherwise scattered across the literature.
    Unless otherwise noted in the caption, the figures presented here are original and are 
    \cclicence\footnote{This means you can freely reproduce or modify a figure without 
                        my permission, as long as you credit my name, by citing this HDR, 
                        and give the link to the license.}.
\end{itemize}
It is divided as follows.
\begin{description}
  \item[\refchap{chap:propaedeutics}] 
    provides a reminder of the main concepts at the foundation of \hISD\
    physics.
  \item[\refchap{chap:dustmodels}] 
    gives a general introduction about the most reliable observational 
    evidences we have about \hISD, and the current models
    attempting at synthesizing them.
  \item[\refchap{chap:dustprop}] 
    reviews dust properties in the \hMW\ and nearby galaxies, and the way 
    they are constrained.
  \item[\refchap{chap:dustevol}] 
    reviews evidences of dust evolution and models accounting for their 
    formation and destruction at the scale of a galaxy.
  \item[\refchap{chap:method}] 
    is a more original take on my methodological approach, motivated by some
    epistemological concepts.
  \item[\refchap{chap:prosp}] 
    is a summary of what we have learned about \hISD\ in the past 
    decade and what we should do during the next one.
\end{description}


\mainmatter
\newchapter{Propaedeutics in Dust Physics}
\label{chap:propaedeutics}
\citesmart{As soon as we thought something, look in what sense the opposite is 
           true.}{\citep[Simone \familyname{Weil};][]{weil47}}
\minitoc
Most reviews about \hISD\ \citep[{\eg}][]{draine03c,whittet03,tielens05,draine11b,jones16a,jones16b,jones16c,galliano18} assume that the reader has a good knowledge of solid-state physics\footnote{The book by \citet{kruegel03} is a notable exception. It starts from elementary electrodynamics and atomic physics.}.
It is however not always the case, especially among students.
The present chapter is intended to synthesize the basic knowledge necessary to understand contemporary publications in the field.
We have opted for a simple presentation of the most important concepts, accompanied with a few original figures.
We present the most important formulae and refer the reader to reliable textbooks for their proofs. 
We have tried to answer everything students always wanted to know about dust, but were afraid to ask.

\section{The Make-Up of Solids}

Atoms can be combined to form molecules or solids.
The properties of these compounds depend greatly on the way their constitutive atoms are bonded together, by their electrons.
Electrons being fermions (their spin is 1/2), the \textit{Pauli exclusion principle} implies that each of them must occupy a different state in a system, characterized by its wave function \citep[\eg\ Tome II, Chapter XIV of][]{cohen-tannoudji96}.
The electronic shells of atoms, the bonds of molecules and the band structure of solids all follow from this principle.

  \subsection{Atomic Structure}
  \label{sec:atom}

    \subsubsection{The Hydrogen Atom}

\begin{figure}[htbp]
  \includegraphics[width=\textwidth]{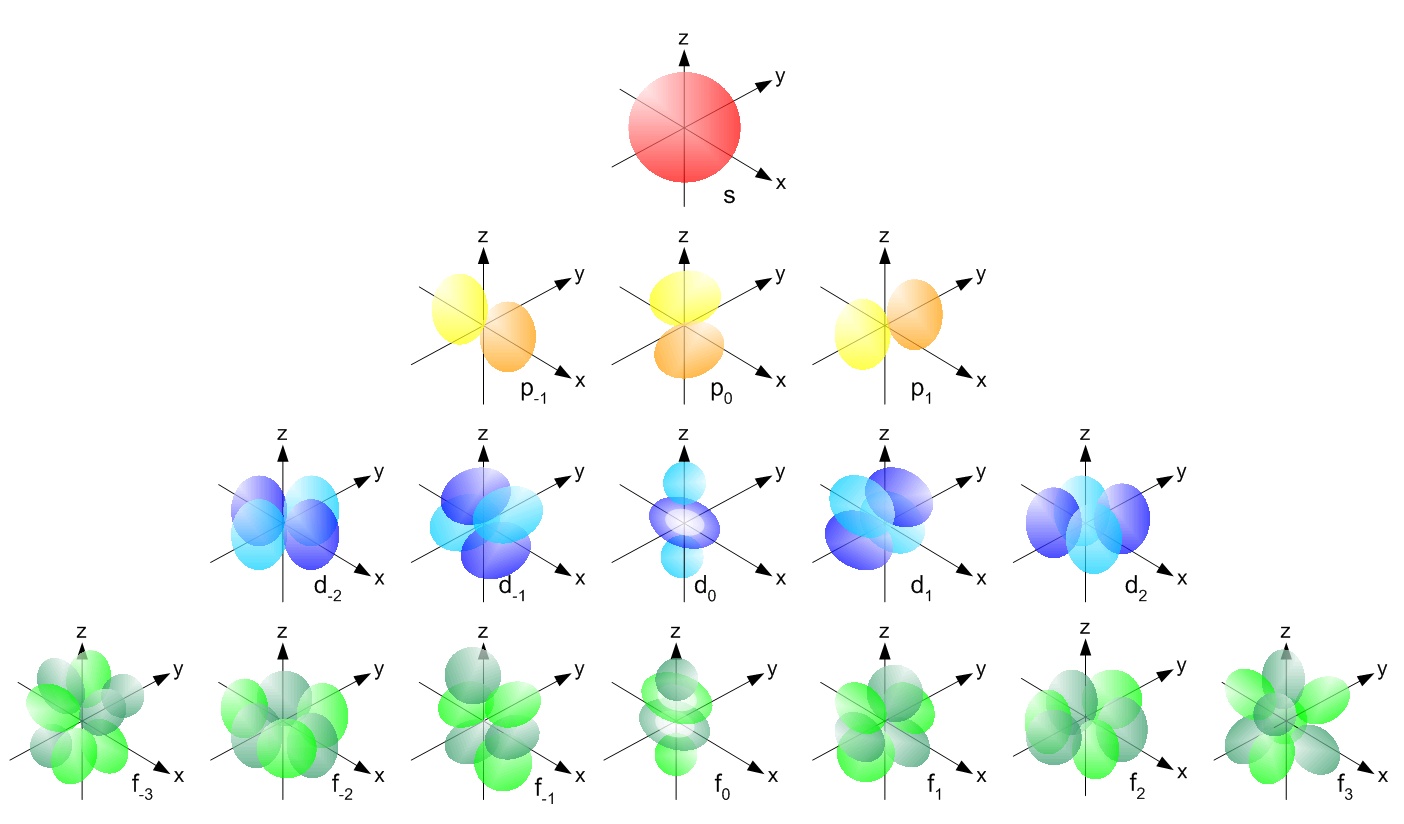}
  \newcap{Shape of s, p, d, f electron orbitals}{These shapes represent the 
     surfaces encompassing the region where the electron has a $90\,\%$     
     probability of presence.
     The orbitals are identified by the letter corresponding to the value of $l$
     (\cf\ \reftab{tab:orbitals}), with the value of $m_l$ as an index.
     \uline{Credit:} \href{https://files.mtstatic.com/site_4334/8855/0?Expires=1608204440&Signature=ozFk6-bH5qAkf3Q6wCt7xw2Yyb-JZlFbob-XCIaFdea71~wQ-80Hr0JHav2ozh2VnKprKQB4AYPfgKBLtNUgipG2-nicXDL2OCGCmVtnCnYjMQiTiN6QEiEKq~VV2cupIQF0bjy8GsNE1gzRh2wTfLZ~o~NUfAKCHfYsEHNia34_&Key-Pair-Id=APKAJ5Y6AV4GI7A555NA}{UCDavis Chemwiki}, licensed under
          \href{https://creativecommons.org/licenses/by-nc-sa/3.0/us/}{CC BY-NC-SA 3.0 US}.}
  \label{fig:orbitals}
\end{figure}

The electrons of an atom each have a distinct wave function, $\Psi$, called \expression{orbital}.
The probability density function of presence of the electron is proportional to $|\Psi|^2$.
These orbitals are solutions to the Schrödinger equation, in the electrostatic well created by the nucleus, assumed to be infinitely heavy\footnote{The electron-to-proton mass ratio is $m_e/m_p\simeq5\E{-4}$ (\cf\ \reftab{tab:constants}).}.
These solutions are relatively simple in the case of the hydrogen atom \citep[\cf\ \eg\ Chap.~3 of][]{bransden83}. 
Its single electron can occupy different shells, characterized by their \expression{principal quantum number}, $n$.
This quantum number determines the energy level of the orbitals ($E\propto1/n^2$) as well as their size  (\cf\ \reftab{tab:qnum}).
Each individual shell is divided into subshells, characterized by the \expression{azimuthal number}, $l$, quantifying the angular momentum of the electron ($L\propto\sqrt{l(l+1)}$), which can be 0, for s subshells (\cf\ \reftab{tab:orbitals}).
The orbitals in each subshell are combinations of spherical harmonic functions.
These functions are anisotropic. 
The \expression{magnetic quantum number}, $m_l$, quantifies their orientation in space, as displayed in \reffig{fig:orbitals}.
Finally, the \expression{spin quantum number}, $m_s$, quantifies the direction of the electronic spin (up or down).
\begin{table}[htbp]
  \centering
  \setlength\arrayrulewidth{2pt}
  \arrayrulecolor{white}
  \begin{tabularx}{\linewidth}%
    {|>{\columncolor{coltabcell}}l%
     |>{\columncolor{coltabcell}}l%
     |>{\columncolor{coltabcell}}l%
     |>{\columncolor{coltabcell}}X|}
    \hline
      \rowcolor{coltabhead}
      \textbf{Name} & \textbf{Symbol} & \textbf{Values} 
      & \textbf{Signification} \\
    \hline
      \cellcolor{coltabhead}
      Principal & $n$ & $1, 2, \ldots, \infty$ 
      & Energy ($E\propto1/n^2$ for H) or size of the shell \\
    \hline
      \cellcolor{coltabhead}
      Azimuthal & $l$ & $0, 1, \ldots, n-1$ 
      & Angular momentum ($L\propto\sqrt{l(l+1)}$) \\
    \hline
      \cellcolor{coltabhead}
      Magnetic & $m_l$ & $l, l-1, \ldots, -l$ 
      & Orientation (spherical harmonic combination) \\
    \hline
      \cellcolor{coltabhead}
      Spin & $m_s$ & $+1/2, -1/2$ & Magnetic moment (spin direction) \\
    \hline
  \end{tabularx}
  \newcap{Atomic quantum numbers}{Adapted from Table 7.2 of \citet{atkins92}.}
  \label{tab:qnum}
\end{table}
\begin{table}[htbp]
  \centering
  \setlength\arrayrulewidth{2pt}
  \arrayrulecolor{white}
  \begin{tabularx}{\linewidth}%
    {|>{\columncolor{coltabcell}}l%
     |>{\columncolor{coltabcell}}l%
     |>{\columncolor{coltabcell}}l%
     |>{\columncolor{coltabcell}}l%
     |>{\columncolor{coltabcell}}l%
     |>{\columncolor{coltabcell}}l%
     |>{\columncolor{coltabcell}}X|}
    \hline
      \rowcolor{coltabhead}
      \textbf{Principal}
      & \textbf{Bohr} 
      & \multicolumn{4}{l|}{\textbf{Azimuthal quantum number}} 
      & \cellcolor{white} \\
      \rowcolor{coltabhead}
      \textbf{quantum} 
      & \textbf{shells} 
      & \cellcolor{coltabsep}s & \cellcolor{coltabsep}p 
      & \cellcolor{coltabsep}d & \cellcolor{coltabsep}f 
      & \cellcolor{coltabhead}\textbf{Subshell letter} \\
      \rowcolor{coltabhead}
      \textbf{number} & & \cellcolor{coltabsep}2 & \cellcolor{coltabsep}6 
      & \cellcolor{coltabsep}10 & \cellcolor{coltabsep}14 
      & \textbf{Number of electrons per subshell} \\
    \hline
      \cellcolor{coltabhead}
      $n=1$ & K & $l=0$ & & & & \cellcolor{white} \\
    \hline
      \cellcolor{coltabhead}
      $n=2$ & L & $l=0$ & $l=1$ & & & \cellcolor{white} \\
    \hline
      \cellcolor{coltabhead}
      $n=3$ & M & $l=0$ & $l=1$ & $l=2$ & & \cellcolor{white} \\
    \hline
      \cellcolor{coltabhead}
      $n=4$ & N & $l=0$ & $l=1$ & $l=2$ & $l=3$ & \cellcolor{white} \\
    \hline
  \end{tabularx}
  \newcap{Atomic shell structure}{We have highlighted the correspondence with 
          the shells of Bohr's pre-quantum atomic model.
          The letters s, p, d, f are used to label orbitals with different
          values of $l$.}
  \label{tab:orbitals}
\end{table}

    \subsubsection{Polyelectronic Atoms}

\begin{figure}[htbp]
  \includegraphics[width=\textwidth]{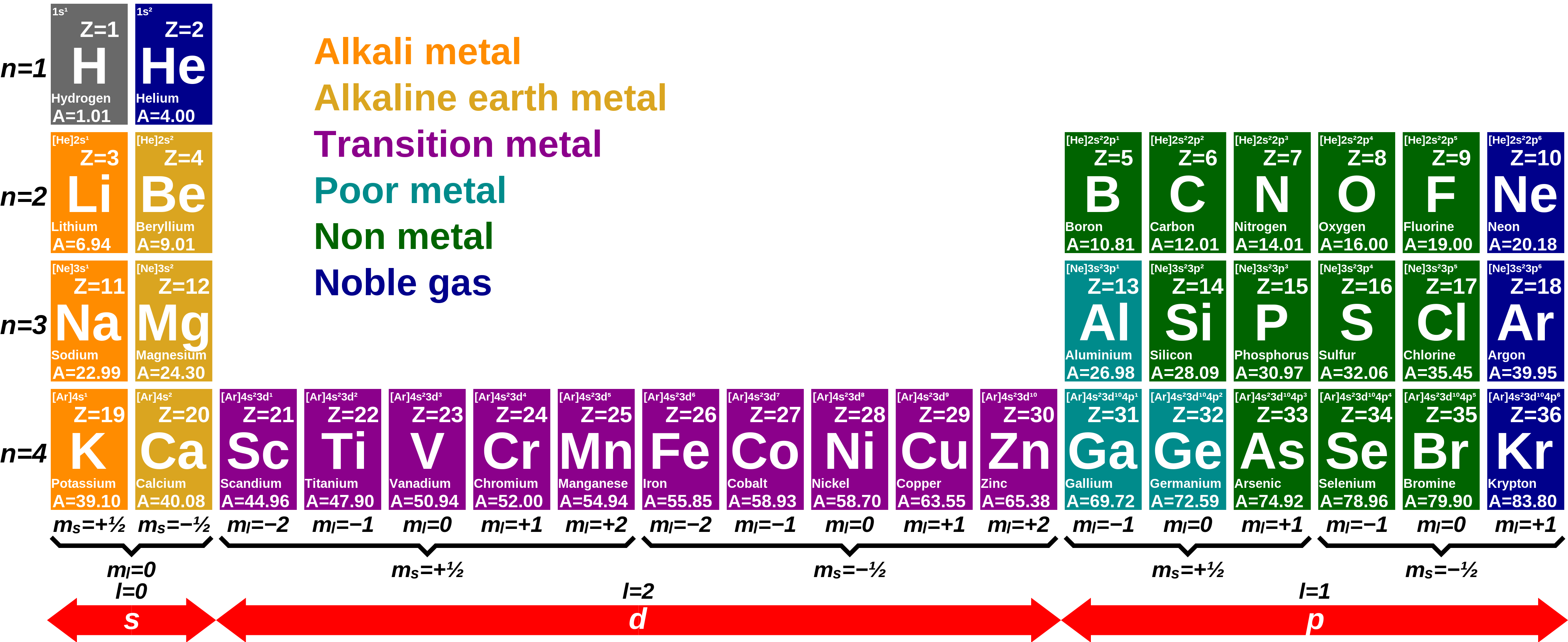}
  \newcap{Periodic table of elements}{We show only the first four rows,
          as they contain all the elements relevant to \hISM\ physics.
          Z is the charge and A, the atomic weight.
          In each cell, we show the electronic configuration in the fundamental
          state.
          We do not repeat the part of the configuration corresponding to the 
          noble gas of the previous row ([He], [Ne], [Ar]).
          We have annotated the table with the four quantum numbers of the
          last electron.
          \CClicence}
  \label{fig:periodic}
\end{figure}

The different energy levels of an atom can be populated by excitation (collision or photon absorption).
In the fundamental state of the H atom, the electron occupies the $n=1$ level.
The electrons of a polyelectronic atom in its fundamental state occupy the lowest energy orbitals available.
Each electron must have a unique set of the four quantum numbers (\reftab{tab:qnum}).
The nucleus charge, $Z$, is higher, which causes the inner shells to be closer to the nucleus.
The electric field seen by the outer shells is thus partially screened by the inner shells.
A fundamental difference between polyelectronic atoms and hydrogen is however the mutual repulsion of the electrons \citep[\cf\ \eg\ Chap.~7 of][]{bransden83}.
The different subshells (s, p, d, f) of a given shell having different geometries, the mutual repulsion depends on $l$.
The subshells of a given $n$ thus have different energies.
The electronic configuration of atoms in their fundamental state results from the ranking in energy of these levels.
The possible number of electrons per subshell is given in \reftab{tab:orbitals}.
We have represented the electronic configuration of atoms in the periodic table (\reffig{fig:periodic}).

    \subsubsection{The Valence Shell}

The outer shell is called the \textit{valence} shell. 
It contains the electrons responsible for molecular bonds (\cf\ \refsec{sec:bond}) and shaping the optical properties of solids (\cf\ \refsec{sec:lightsolid}).
We will see that the nature of the chemical bond depends on the tendency of its atoms:
\begin{inlinelist}
 \item to share electrons;
 \item to form \expression{cations}, by losing one or several electrons; or
 \item to form \expression{anions}, by gaining one or several electrons.
\end{inlinelist}

\begin{figure}[htbp]
  \includegraphics[width=\textwidth]{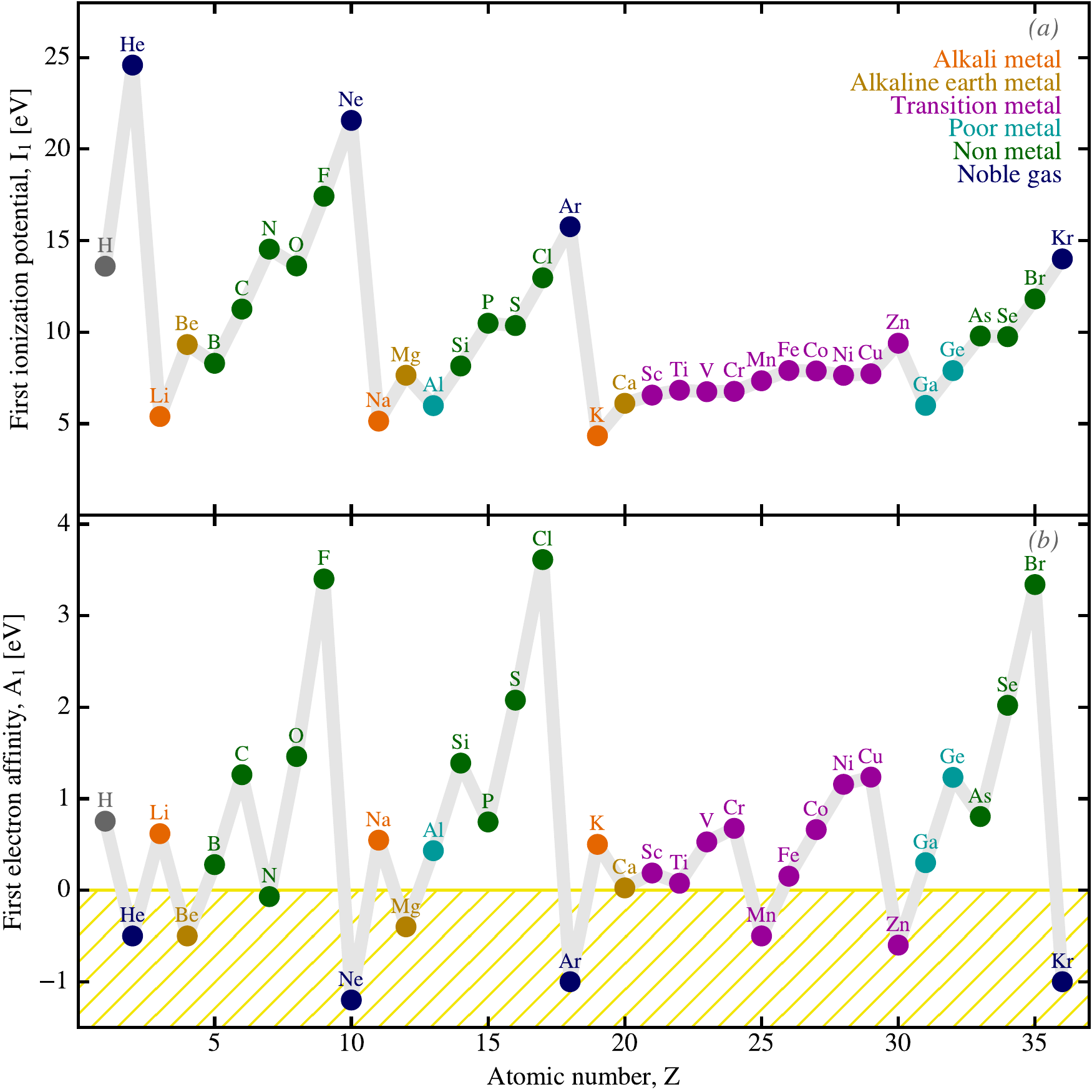}
  \newcap{First ionization potentials and electron affinities}%
         {In panel~\textit{(a)}, we display the first ionization potential.
          It is the minimum energy required to expel the most loosely bound 
          electron, in the fundamental state.
          It is the energy lost during the gas phase reaction: 
          $X\longrightarrow X^++e^-$.
          In panel~\textit{(b)}, we display the first electron affinity, which 
          is the energy gained by adding an electron to the atom.
          It is the energy released during the gas phase reaction:
          $X+e^-\longrightarrow X^-$.
          Elements with a negative electron affinity (yellow hatched area) 
          do not form stable anions.
          We display the different groups with the same color code as in 
          \reffig{fig:periodic}. 
          \CClicence}
  \label{fig:Eion1}
\end{figure}

\paragraph{Ionization potential.}
Panel~\textit{(a)} of \reffig{fig:Eion1} shows the \expression{first ionization potential}, $I_1$, of the elements in \reffig{fig:periodic}.
Atoms with a low $I_1$ tend to form stable cations.
Noble gases have the largest $I_1$ of their row, because their last shell is full.
More generally, \reffig{fig:Eion1} shows that $I_1$ increases from the left to the right of the periodic table (\reffig{fig:periodic}), as the valence shell gets fuller.
This is because, at a given $n$, moving to the right of the table increases the effective nucleus charge $Z-Z_\sms{subshells}$, making the valence shell more tightly bound.
It also decreases from the top to the bottom, as the energy level of the outer shell decreases with $n$ as $1/n^2$.
\takeaway{Metals tend to form stable cations.}

\paragraph{Electron affinity.}
Panel~\textit{(b)} of \reffig{fig:Eion1} shows the \expression{first electron affinity}, $A_1$, of the elements in \reffig{fig:periodic}.
Atoms with a high electron affinity tend to form stable anions.
$A_1$ follows roughly the same trend as $I_1$, with some exceptions.
Noble gases have their last shell full, they therefore tend to remain neutral and have a negative $A_1$.
Alkaline earth metals also have negative $A_1$, because of the energy difference between their full $n$s and empty $n$p shells.
\takeaway{Non metals tend to form stable anions.}
 
\paragraph{Electronegativity.}
The balance between $I_1$ and $A_1$ gives an idea of the tendency of an atom to attract electrons in a bond.
Mulliken's \expression{electronegativity}, $\chi$, is defined as \citep[\eg][]{huheey93}:
\begin{equation}
  \chi = 0.187\times\left(\frac{I_1}{\textnormal{1 eV}}
                         +\frac{A_1}{\textnormal{1 eV}}\right)+0.17.
\end{equation}
If we except noble gases, we see that $\chi$ will be higher to the right of the periodic table, and will decrease from the top to the bottom.

    \subsubsection{Orbital Hybridisation}

\begin{figure}[htbp]
  \begin{tabular}{ccc}
    \includegraphics[width=0.3\textwidth]{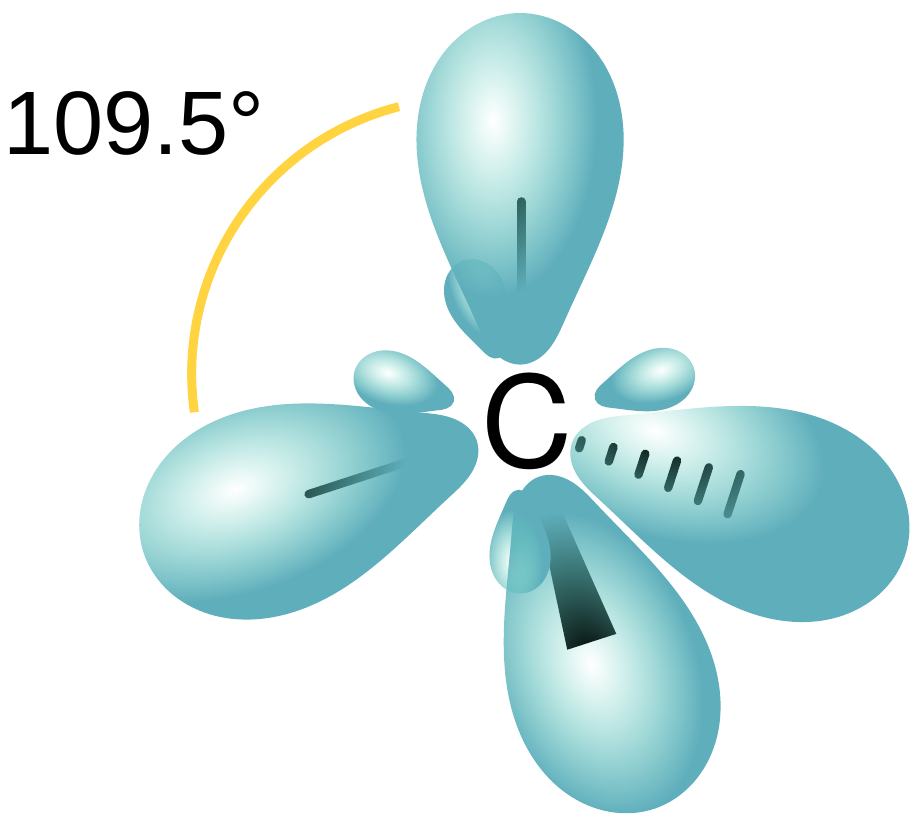} &
    \includegraphics[width=0.3\textwidth]{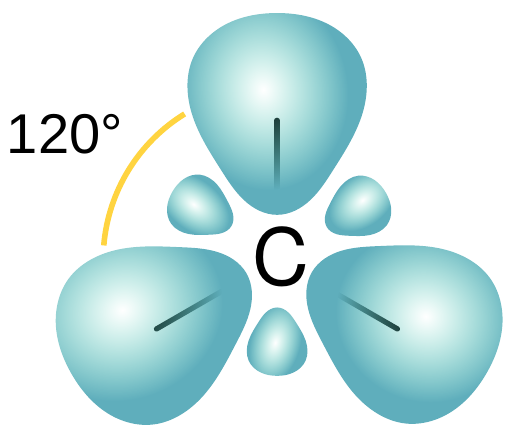} &
    \includegraphics[width=0.3\textwidth]{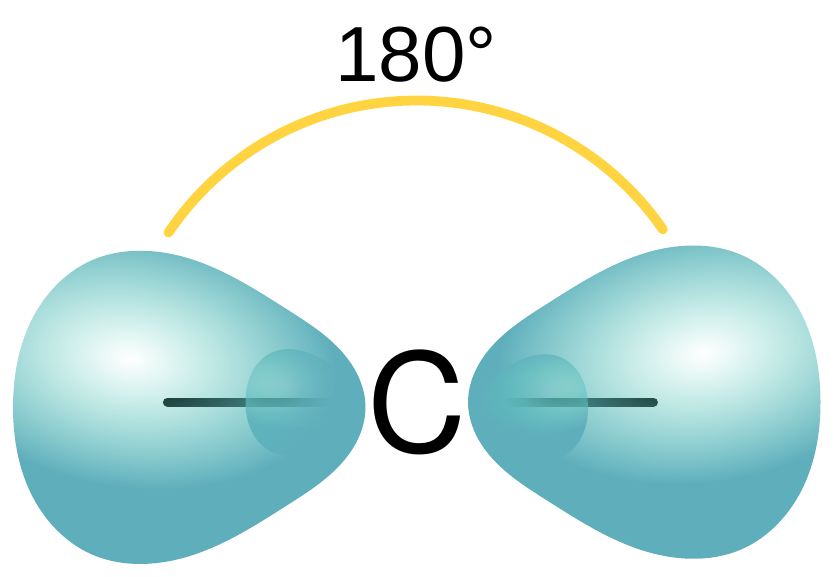} \\
    \textit{(a)} sp$^3$ hybrid & \textit{(b)} sp$^2$ hybrid 
    & \textit{(c)} sp hybrid \\
  \end{tabular}
  \newcap{Hybrid orbitals of carbon}%
         {The left image shows the shape of the four sp$^3$ orbitals of a C 
          atom.
          They are arranged in a tetrahedral geometry.
          The center image shows the shape of the three sp$^2$ orbitals of a C 
          atom.
          They are arranged in a trigonal geometry.
          We have not displayed the remaining p orbital of this C atom.
          The right image shows the shape of the two sp orbitals of a C atom.
          They are arranged along a line.
          We have not displayed the remaining two p orbitals of this C atom.
          Each bond is represented with a black line indicating its direction:
          \begin{inlinelist}
            \item thin solid lines show bonds within the plane of the image;
            \item the thick solid line shows a bond pointing toward the reader;
            \item the dashed line shows a bond pointing in the opposite 
              direction.
          \end{inlinelist}
          \uline{Credit:} adapted from 
          \href{https://en.wikipedia.org/wiki/Orbital_hybridisation}%
               {Wikipedia's Orbital hybridisation article} (changed the central letter), 
               licensed under 
                \href{https://creativecommons.org/licenses/by-sa/3.0/deed.en}%
                     {CC BY-SA 3.0}.}
  \label{fig:hybrid}
\end{figure}

To explain the shape of molecules, the concept of \expression{hybrid orbitals} was introduced in the 1930's by Linus \familyname{Pauling}.
The principle is that orbitals with similar energies can be linearly combined together to form new orbitals with different shapes\footnote{The Schrödinger equation is linear. Any combination of solutions is a solution.}.
The most relevant example to \hISD\ is the hybridisation of carbon.
A 2s electron can be promoted to 2p, resulting in the following configuration.
\begin{center}
  \begin{tabular}{cp{1cm}c}
    C\par
    \begin{MOdiagram}[style=fancy]
      \AO{s}[label={1s}]{0}
      \AO(30pt){s}[label={2s}]{0}
      \AO(60pt){p}[label[y]={2p}]{0;up,up}
    \end{MOdiagram}
    &&
    C$^*$\par
    \begin{MOdiagram}[style=fancy]
      \AO{s}[label={1s}]{0}
      \AO(30pt){s}[label={2s}]{0;up}
      \AO(60pt){p}[label[y]={2p}]{0;up,up,up}
    \end{MOdiagram}
  \end{tabular} 
\end{center}
The carbon is now in an excited state, noted C$^*$.
The promotion requires 4.2~eV.
From this new state, the following combinations of orbitals are possible.
\begin{description}
  \item[sp$^3$ hybrids] are 1/4 s and 3/4 p.
    This hybridisation results in four sp$^3$ orbitals arranged in a 
    tetrahedron, shown in \refsubfig{fig:hybrid}{a}.
    For instance, C in methane (CH$_4$) is sp$^3$ hybridized. The electronic 
    configuration of the $n=2$ shell becomes the following.
    \begin{center}
      \begin{MOdiagram}[style=fancy]
        \AO{s}[label={sp$^3$}]{0;up}
        \AO(20pt){p}[label={sp$^3$}]{0;up,up,up}
      \end{MOdiagram}
    \end{center}
  \item[sp$^2$ hybrids] are 1/3 s and 2/3 p.
    This hybridisation results in one standard p orbital and three sp$^2$ 
    orbitals trigonally-arranged, shown in \refsubfig{fig:hybrid}{b}.
    For instance, C in benzene (C$_6$H$_6$) is sp$^2$ hybridized. The electronic 
    configuration of the $n=2$ shell becomes the following.
    \begin{center}
      \begin{MOdiagram}[style=fancy]
        \AO{s}[label={sp$^2$}]{0;up}
      \AO(20pt){p}[label[x]={sp$^2$},label[y]={sp$^2$},label[z]={p}]{0;up,up,up}
      \end{MOdiagram}
    \end{center}
  \item[sp hybrid] is 1/2 s and 1/2 p. 
    This hybridisation results in two standard p orbitals and two sp orbitals
    linearly-arranged, shown in \refsubfig{fig:hybrid}{c}.
    For instance, C in acetylene (C$_2$H$_2$) is sp hybridized. The electronic 
    configuration of the $n=2$ shell becomes the following.
    \begin{center}
      \begin{MOdiagram}[style=fancy]
        \AO{s}[label={sp}]{0;up}
        \AO(20pt){p}[label[x]={sp},label[y]={p},label[z]={p}]{0;up,up,up}
      \end{MOdiagram}
    \end{center}
\end{description}

    \subsection{Molecular Bonding}
    \label{sec:bond}

Chemical bonds are the result of the overlap between the outer orbitals of two atoms whose valence shell is not full \citep[\cf\ \eg\ Chap.~8 of][]{atkins92}.
Despite their mutual repulsion, sharing electrons leads to a lower energy state, in which a stable bonded molecule is formed.
Covalent, ionic and metallic bonds, that we will define below, typically have dissociation energies of a few eV.
The atomic spacing in a molecule or a solid is of the order of a few $\r{A}$.
\begin{figure}[htbp]
  \includegraphics[width=\textwidth]{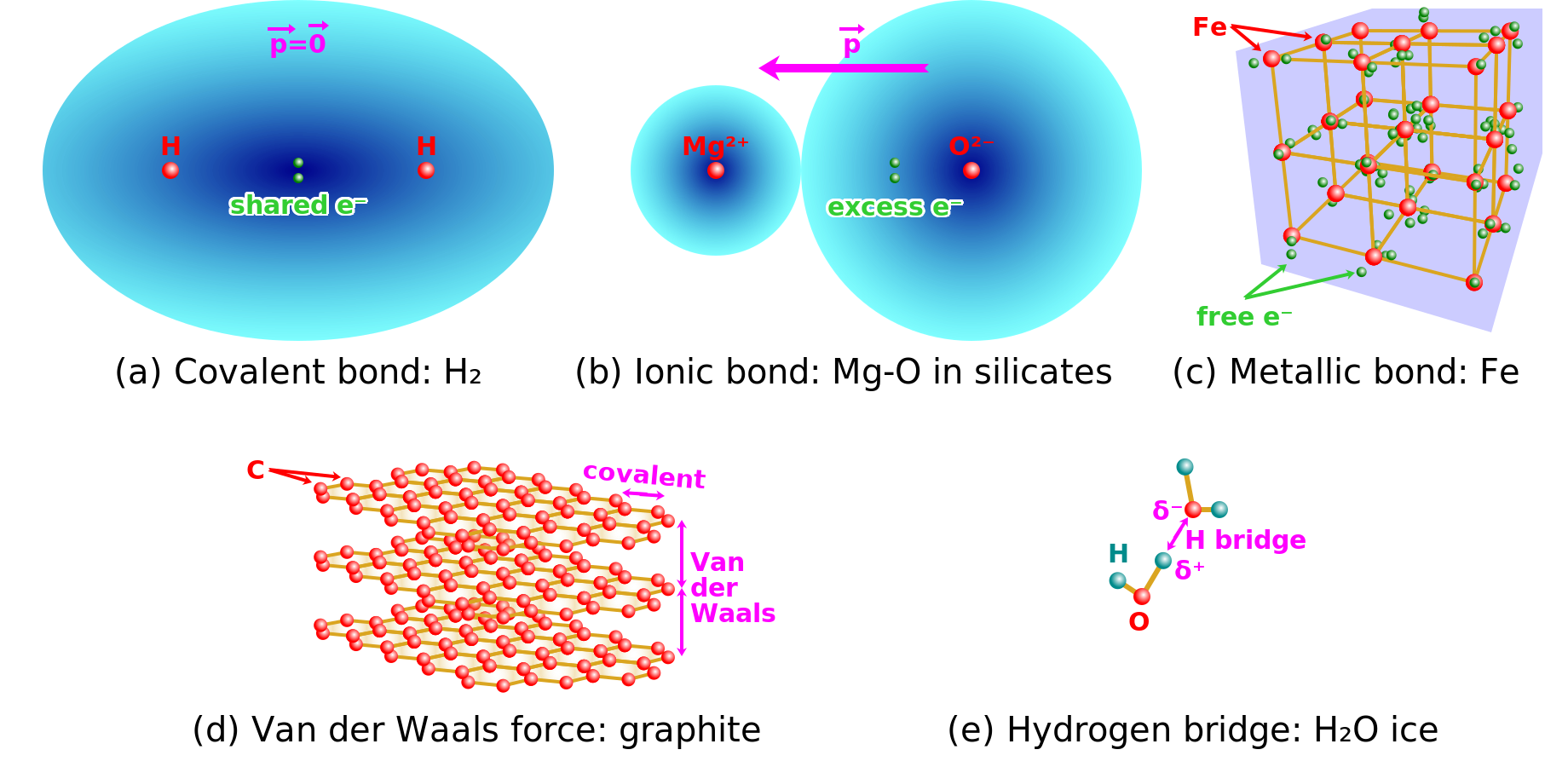}
  \newcap{Different types of molecular bonds}%
         {We have noted $\vec{p}$, the dipole moment created by the bond.
          In general, $\vec{p}=\iiint_V\vec{r}.\rho(\vec{r})\ddiff V$,
          where $\rho(\vec{r})$ is the charge number density.
          In the case of MgO, it reduces to $e.\vect{\textnormal{O.Mg}}$,
          where $\vect{\textnormal{O.Mg}}$ is the distance vector between the 
          two ions.
          The electrons are noted $e^-$, and $\delta^+$ and $\delta^-$ 
          are the excess dipolar charges of both signs. 
          \CClicence}
  \label{fig:bondtypes}
\end{figure}

    \subsubsection{Covalent Bonds}

\begin{figure}[htbp]
  \begin{tabular}{cc}
    \includegraphics[width=0.35\textwidth]{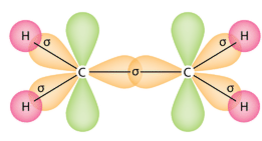} &
    \includegraphics[width=0.62\textwidth]{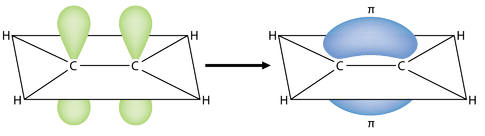} \\
    \textit{(a)} $\sigma$ bonds &
    \textit{(b)} overlap of two p orbitals to form one $\pi$ bond
  \end{tabular}
  \newcap{An example of $\sigma$ and $\pi$ bonds}%
          {We demonstrate the two types of bonds on ethylene (C$_2$H$_4$).
           On the left image: the s orbital of each H atom is shown in red;
           the p orbital of each C atom is shown in green; 
           and the three sp$^2$ hybridized orbitals of each C atom are in 
           orange.
           The frontal overlap of the sp$^2$ hybridized orbitals of two C atoms,
           as well as the overlap of one sp$^2$ hybridized C orbital with the s 
           orbital of an H atom, both form a $\sigma$ bond.
           On the middle image, we show the plane of the molecule with the
           p orbital of each C atom in green.
           The right image shows the overlap of these two p orbitals to form 
           one single $\pi$ bond, in blue, on both sides of the plane.
           \uline{Credit:} adapted from the  
  \href{https://chem.libretexts.org/Bookshelves/Introductory_Chemistry/Book\%3A_Introductory_Chemistry_(CK-12)/09\%3A_Covalent_Bonding/9.18\%3A_Sigma_and_Pi_Bonds}{Chemistry Library},
  licensed under \href{https://creativecommons.org/licenses/by-nc/4.0/}{CC BY-NC 4.0}.}
  \label{fig:sigmapi}
\end{figure}

A \expression{covalent bond} is formed when a pair of electrons with anti-parallel spin is shared between two atoms of similar electronegativity.
The more the orbitals overlap, the stronger the bond is. 
The electron density is the highest between the two atoms, resulting in a directional bond.
For instance, the \hmol\ (\refsubfig{fig:bondtypes}{a}) and CO molecule bonds, as well as the C-C and C-H bonds in hydrocarbons, are all covalent.
Covalent bonds are weakly polar.
Symmetric molecules such as \hmol\ are non-polar, whereas asymmetric molecules, such as CO are polar, because of the difference in electronegativity of C and O.
\takeaway{Covalent bonds are preferentially formed between non metals.}
Covalent bonds are of one of the two following types.
\begin{description}
  \item[$\bm{\sigma}$ bonds]
    result from the frontal overlap of two s, p or sp$^n$ ($n=1,2,3$) orbitals.
    These bonds have a rotational symmetry around their axis
    (\refsubfig{fig:sigmapi}{a}).
    The electron density is maximum between the two atoms.
    It is the strongest covalent bond.
    The C--C bond of ethane (C$_2$H$_6$) is a $\sigma$ bond.
  \item[$\bm{\pi}$ bonds]
    result from the side-by-side overlap of the two lobes of two p orbitals 
    (\refsubfig{fig:sigmapi}{b}).
    The electron density is maximum above and below the plane of the molecule 
    and zero between them.
    These bonds are weaker than $\sigma$ bonds.
    In the double C$=$C bond of ethylene (C$_2$H$_4$), there are one $\sigma$ 
    and one $\pi$ bonds (\reffig{fig:sigmapi}).
    In the triple C$\equiv$C bond of acetylene (C$_2$H$_2$), there are one 
    $\sigma$ bond and two $\pi$ bonds.
\end{description}
Finally, some transitions in interstellar solids involve \expression{antibonding} orbitals.
When a bond forms, both a bonding and an antibonding molecular orbitals with different energy levels become available.
This is demonstrated for the H$_2^+$ molecule, with the Schödinger equation, in Chap.~9 of \citet{bransden83}.
Bonding orbitals have a lower energy level than the dissociated atoms, thus favoring a stable molecule.
On the contrary, the population of an antibonding orbital makes the molecule unstable.
For instance, the splitting of molecular orbitals for both $\sigma$ and $\pi$ bonds of ethylene (\reffig{fig:sigmapi}) are the following.
\begin{center}
  \begin{tabular}{cc}
    \begin{MOdiagram}[names,style=fancy]
      \atom[\textcolor{blue}{C}]{left}{ 1s = {0; up} }
      \atom[\textcolor{blue}{C}]{right}{ 1s = {0; up } }
      \molecule[\textcolor{blue}{C--C}]{ 1sMO = {.75;pair},
                 label={1sigma={\chemsigma$_{\textnormal{sp}^2-\textnormal{sp}^2}$},
                       1sigma*={\chemsigma$^*_{\textnormal{sp}^2-\textnormal{sp}^2}$}} }
      \EnergyAxis[title]
      \node[below,yshift=-5] at (1sleft) {2sp$^2$};
      \node[below,yshift=-5] at (1sright) {2sp$^2$};
      \node[right] at (1sigma.-45) {\footnotesize\textcolor{red}%
                                                           {(bonding)}};
      \node[right,yshift=12] at (1sigma*.-45){\footnotesize\textcolor{red}%
                                                               {(antibonding)}};
    \end{MOdiagram}
    &
    \begin{MOdiagram}[names,style=fancy]
      \atom[\textcolor{blue}{C}]{left}{ 1s = {  0; up } }
      \atom[\textcolor{blue}{C}]{right}{ 1s = {  0; up } }
      \molecule[\textcolor{blue}{C--C}]{ 1sMO = {.75; pair }, 
                       label={1sigma={\chempi$_{\textnormal{p}-\textnormal{p}}$},
                       1sigma*={\chempi$^*_{\textnormal{p}-\textnormal{p}}$}}  }
      \EnergyAxis[title]
      \node[below,yshift=-5] at (1sleft) {2p};
      \node[below,yshift=-5] at (1sright) {2p};
      \node[right] at (1sigma.-45) {\footnotesize\textcolor{red}%
                                                           {(bonding)}};
      \node[right,yshift=12] at (1sigma*.-45){\footnotesize\textcolor{red}%
                                                               {(antibonding)}};
    \end{MOdiagram}
  \end{tabular}
\end{center}
We emphasize those are electronic levels of molecules. 
Molecules also have rotational and vibrational modes that will be discussed in \refsec{sec:IRfeatures}.

    \subsubsection{Ionic Bonds}

An \expression{ionic bond} is formed between two atoms of significantly different electronegativities.
The electron is transferred from the cation to the anion, resulting in a polar bond.
The adhesion is due to long-range Coulomb forces ($\propto1/r^2$) between the two ions. 
Ionic bonding is non-directional as the electron cloud stays centered around the atoms.
The most relevant example to \hISD\ is the O$^{2-}$--$\,$Mg$^{2+}$ bond in silicates (\refsubfig{fig:bondtypes}{b}).
\takeaway{Ionic bonds are preferentially formed between a metal and a non metal.}
Covalent and ionic bonds are two extreme cases. 
Most bonds involving at least one non metal are intermediate between both.

    \subsubsection{Metallic Bonds}
Metals can easily be ionized.
Bonding several metal atoms therefore results in a lattice of cations bathed in a sea of free valence electrons.
The electrons are not attached to a particular atom and can be found anywhere in the solid.
This explains the electric and thermal conductivities of metals.
\refsubfig{fig:bondtypes}{c} represents solid iron.
\takeaway{Metallic bonds are formed between a large number of metal atoms.}

    \subsubsection{Intermolecular Attraction}

Weaker forms of attraction between molecules exist. 
Their dissociation energy is typically of the order of $\simeq0.1$~eV.
They are relevant to \hISD\ studies.
\begin{description}
  \item[Van der Waals bonds] are due to the induced dipole attraction of 
    neutral atoms and molecules.
    Their potential energy drops as $1/r^6$. 
    They are in particular responsible for binding graphene sheets together 
    in graphite (\refsubfig{fig:bondtypes}{d}).
  \item[Hydrogen bridges] are formed when the induced dipole of the
    H atom of a molecule is attracted by the induced dipole of a strongly 
    electronegative atom in another molecule.
    For instance, hydrogen bridges tie the H$_2$O molecules together in water  
    ice (\refsubfig{fig:bondtypes}{e}).
\end{description}

  \subsection{The Solid State}
  \label{sec:solid}

    \subsubsection{The Different Types of Solids}
    \label{sec:metalvsdieletric}

There are two main types of solids: \expression{insulators} and \expression{conductors}.
Their properties are radically different.
Their difference originates in the type of chemical bond making up their crystal lattice.
\begin{description}
  \item[Insulators,] also called \expression{dielectrics}, are solids, whose 
    atoms are tied together with covalent or ionic bonds.
    The valence electrons are therefore located around their specific atoms and 
    can not move freely through the lattice.
    Consequently, when an electric field is applied, it induces a polarization 
    of the bonds, distorting them, but no current is flowing.
    For instance, silicates are dielectric materials.
  \item[Conductors] are solids, whose atoms are tied together with metallic  
    bonds.
    Their valence electrons are therefore free to move through the solid.
    Consequently, when an electric field is applied, a current is flowing.
    For instance, iron is a metallic conductor.
    There are a few non-metal conductors, such as graphite, which is classified 
    as a \expression{semimetal}.
    This peculiar property is due to the delocalized electrons within the 
    aromatic cycles constituting graphite (\cf\ \refsec{sec:dustanalog}).
\end{description}
There is a third, intermediate type of solids, called \expression{semiconductors}.
Semiconductors are insulators at $T=0$~K and conductors at ambient temperatures.
Several cosmic dust candidates belong to this category.
We will define it more precisely in \refsec{sec:fermi}.

    \subsubsection{The Band Structure of Solids}

\begin{figure}[htbp]
  \includegraphics[width=\textwidth]{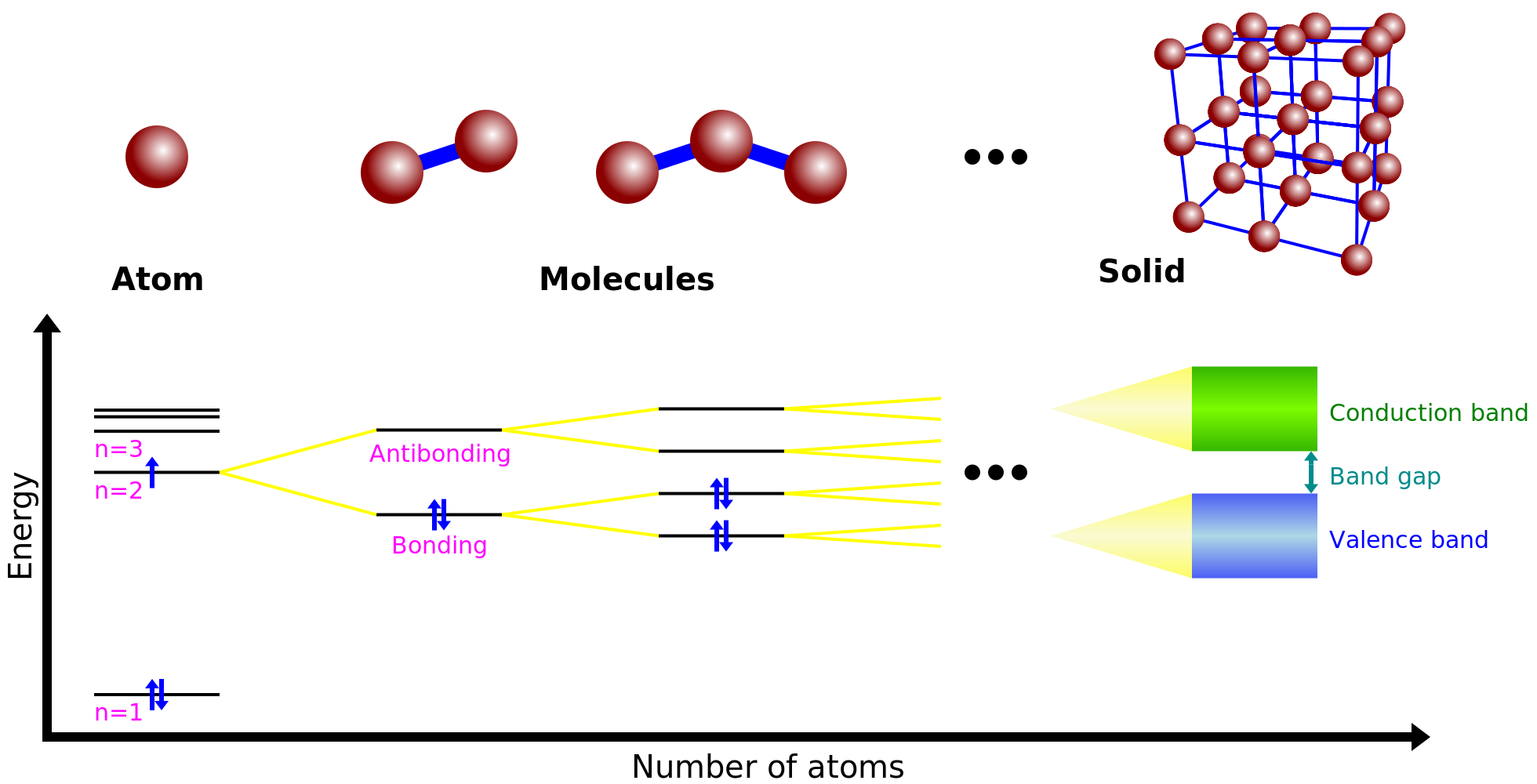}
  \newcap{Origin of the band structure of a solid}%
         {From the left to the right, we represent:
          \begin{inlinelist}
            \item typical discrete atomic levels,
            \item the successive splitting of molecular orbitals,
            \item resulting in the quasi continuous distribution of levels
              in bands.
          \end{inlinelist}
          Electrons are represented with a vertical blue arrow (up or down),
          corresponding to their spin.
          \CClicence}
  \label{fig:bandstruct}
\end{figure}

A solid can be idealized as a periodic lattice of atoms bonded to each other.
The permitted energy levels of a single valence electron, in the periodic electrostatic potential created by this lattice, are a series of continuous functions, also called \expression{bands} \citep[\eg\ Chap.~8 of][for a derivation from the Schrödinger equation]{ashcroft76}.
This can be viewed as a generalization of the molecular level splitting  (\reffig{fig:bandstruct}).
The spacing between a large number of levels is so small that it appears continuous.
At $T=0$~K, the lowest energy bands are filled in priority.
Two of these bands are particularly important.
\begin{description}
  \item[The valence band] is the highest energy band populated by valence 
    electrons, at $T=0$~K.
  \item[The conduction band] is the lowest energy band where electrons can move 
    freely through the solid.
    It is the band immediately superior to the valence band.
\end{description}
The energy difference between the top of the valence band and the bottom of the conduction band is called the \expression{band gap}, noted $E_g$ (\reffig{fig:bandstruct}).

    \subsubsection{The Fermi Level}
    \label{sec:fermi}

\begin{figure}[htbp]
  \includegraphics[width=\textwidth]{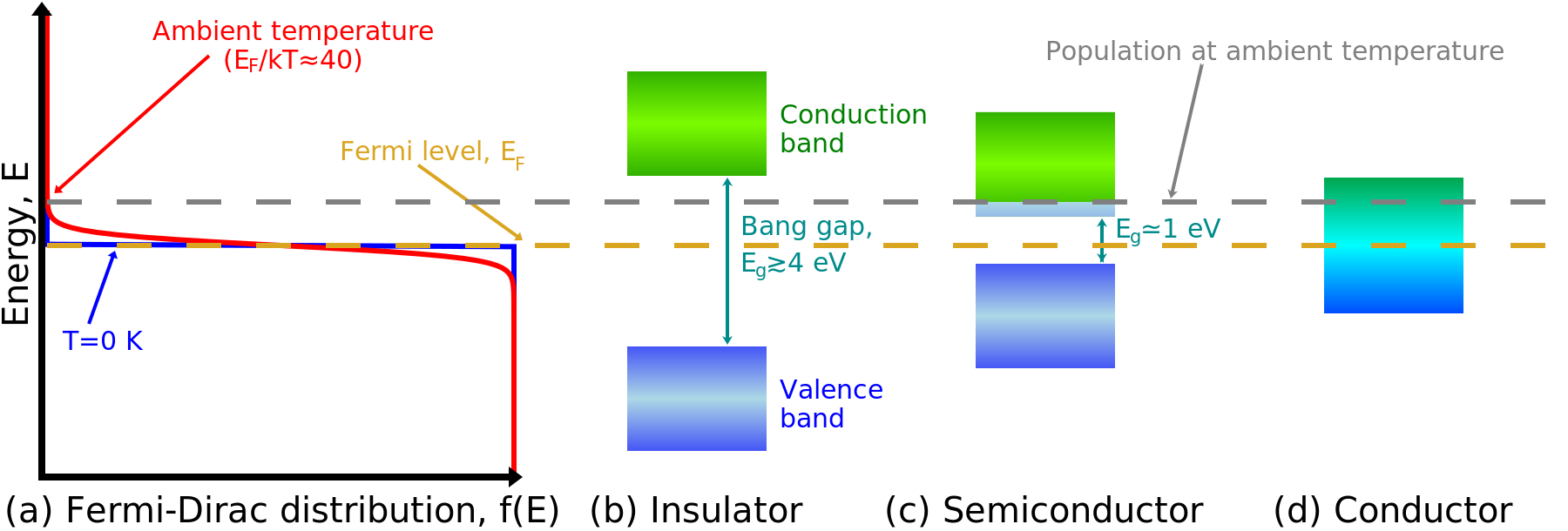}
  \newcap{The Fermi level and the different types of solid}%
         {The left plot shows the rotated Fermi-Dirac distribution
          \refeqp{eq:FermiDirac}, for two values of the temperature, $T=0$~K
          and $T\simeq300$~K.
          The three diagrams on the right show the valence and conduction
          bands relative to the Fermi level, $E_F$, for insulators, 
          semiconductors and conductors.
          For conductors, the valence band is also the conduction band.
          \CClicence}
  \label{fig:fermilevel}
\end{figure}

The probability distribution of identical fermions, such as electrons in a solid, over the energy states of a system at temperature $T$, is given by the \expression{Fermi-Dirac distribution}:
\begin{equation}
  f(E) = \frac{1}{\displaystyle\exp\left(\frac{E-E_F}{kT}\right)+1},
  \label{eq:FermiDirac}
\end{equation}
where $k$ is the \expression{Boltzmann constant} (\cf\ \reftab{tab:constants}), $E$ denotes the different energy levels and $E_F$ is the \expression{Fermi level}\footnote{In the general Fermi-Dirac distribution, the Fermi level, which is proper to solids, is replaced by the \expression{chemical potential} of the system, $\mu$. In our case, the Fermi level is the chemical potential of an electron.}.
This distribution is displayed in \refsubfig{fig:fermilevel}{a}.
The Fermi level is an intrinsic quantity characterizing a solid.
It is the energy required to add an electron to the system.
It also corresponds to the maximum energy an electron can have at $T=0$~K.
The latter interpretation of $E_F$ can be seen in \refsubfig{fig:fermilevel}{a}.
The blue curve shows \refeq{eq:FermiDirac} at $T=0$~K:
\begin{inlinelist}
  \item it gives equal probability to electrons to occupy energy levels $E\le E_F$;
  \item it gives zero probability to energy levels $E>E_F$.
\end{inlinelist}
The actual number density of electrons, $n_e$, is:
\begin{equation}
  n_e = \int_{-\infty}^\infty g(E)f(E)\ddiff E,
\end{equation}
where $g(E)$ is the density of states per infinitesimal energy bin. 
This density of states corresponds to the band structure. 
It is 0 between the bands.
We emphasize that $E_F$ can fall between two bands.
It does not necessarily correspond to an actual allowed level.
This is demonstrated in \reffig{fig:fermilevel}.
\begin{description}
  \item[Insulators] have their valence and conduction bands widely spread apart
    (\refsubfig{fig:fermilevel}{b}).
    At ambient temperature, no electron will populate the conduction band.
    It is another way to see that their valence electrons are localized 
    around their cations.
  \item[Semiconductors] have their valence and conduction bands close to each 
    other (\refsubfig{fig:fermilevel}{c}).
    They are insulators at $T=0$~K, but their conduction band can be populated
    at ambient temperature ($E_g$ gets closer to $kT$).
  \item[Conductors] are solids for which the valence and the conduction bands 
    are the same (\refsubfig{fig:fermilevel}{d}).
    The Fermi level is within the band.
    It is another way to see that the valence electrons are free to move 
    through the lattice at any temperature.
\end{description}

  \subsection{Interstellar Dust Candidates}
  \label{sec:dustanalog}

We briefly review here the constitution of the most likely \hISD\ grain candidates.
Some general properties are given in \reftab{tab:minerals}.
Their optical properties are extensively discussed in \refsec{sec:Qabs}.

    \subsubsection{Silicates}
    \label{sec:silicates}

The different types of silicates are built around silica tetrahedra (SiO$_4^{4-}$), paired with various cations to produce a neutral compound \citep[\cf\ \eg][for a review]{henning10}.
The silica tetrahedra have a central Si$^{4+}$ cation tied to four O$^{2-}$ anions with covalent/ionic bonds.
In the \hISM, the most widely available divalent cations that can be paired with silica tetrahedra are Mg$^{2+}$ and Fe$^{2+}$ (\cf\ \refsec{sec:depletions}).
Silicates have two strong features at 9.7~\tmic\ (Si--O stretching) and 18~\tmic\ (O--Si--O bending).
They are ubiquitous:
\begin{inlinelist}
  \item they are the main constituent of Earth's crust;
  \item they are also found in Solar system and \expression{CircumStellar Dust}
    (\hCSD);
  \item they account for probably 2/3 of interstellar grain mass 
    \citep[\eg][]{draine03c};
  \item their features are observed in distant galaxies, in absorption 
    \citep[\eg][]{marcillac06} and in emission \citep[\eg][]{hony11}.
\end{inlinelist}
Interstellar silicates are widely amorphous \citep[\eg][]{kemper04}.
Crystalline silicates have additional distinctive narrow features, due to SiO$_4$ as well as (Fe,Mg)--O vibrations, in the 9.0--12.5~\tmic\ and 14-22~\tmic\ ranges, with a few bands above 33~\tmic.
The following two types of silicates are the most relevant to \hISD\
(\cf\ \reftab{tab:minerals}).

\begin{figure}[htbp]
  \includegraphics[width=\textwidth]{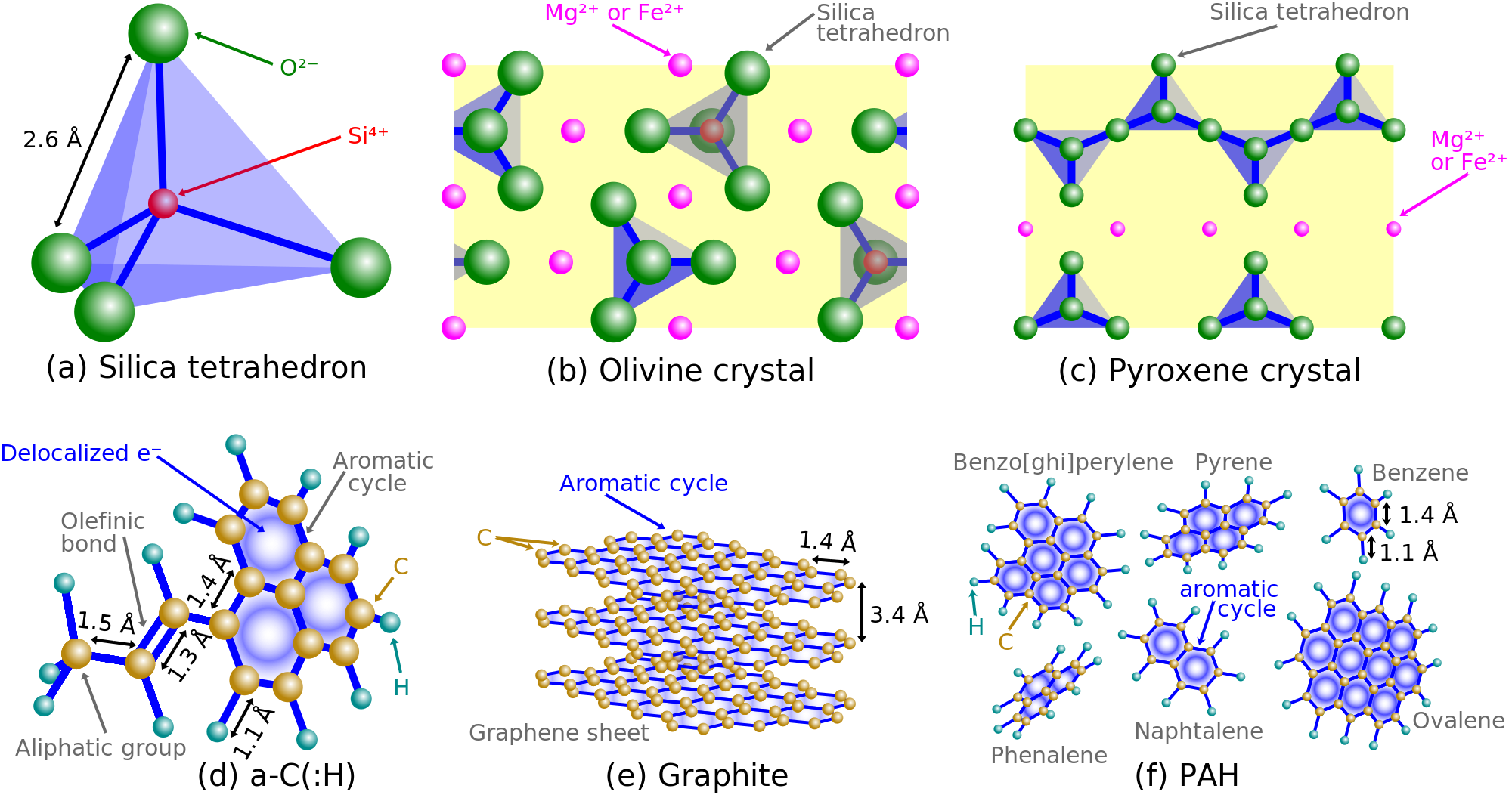}
  \newcap{Structure of interstellar dust candidates}%
         {The (a) image shows the silica tetrahedron, which is the building 
          block of silicates.
          The (b) image shows the structure of crystalline olivine.
          Some silica tetrahedra are pointing toward the reader above the 
          yellow plane, and some are pointing backward. 
          The latter appear dimmer and the Si atom are visible in red.
          The (c) image shows the structure of crystalline pyroxene.
          Its most important feature is the chain of alternate silica 
          tetrahedra.
          They share one O atom, explaining why the stoichiometry of pyroxene 
          is different from olivine.
          The (d) image shows the diversity of carbon pairing in an \hHAC.
          Every C atom is sp$^2$ hybridized, except the aliphatic C, which is 
          sp$^3$ hybridized.
          Every bond is a $\sigma$ bond except:
          \begin{inlinelist}
            \item the 6 delocalized electrons within each aromatic cycles are
              $\pi$ bond pairing of p orbitals;
            \item one of the olefinic bonds is a $\pi$ bond.
          \end{inlinelist}
          The example we have shown corresponds to a very small grain. 
          In a larger \hHAC, most of these bonds would be linked into the rest 
          of a contiguous \hthreeD\ network \citep[\eg][for more realistic 
          structures]{micelotta12}.
          The (e) image shows that graphite is the stack of graphene sheets.
          Each graphene sheet is made exclusively of aromatic cycles.
          The (f) image displays a few different \hPAH s.
          \CClicence}
  \label{fig:crystal}
\end{figure}

\begin{figure}[htbp]
  \begin{tabular}{ccccc}
    \includegraphics[width=0.17\textwidth]{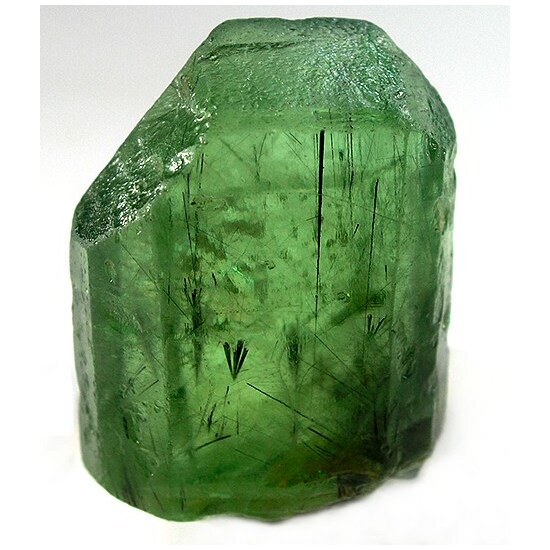} &
    \includegraphics[width=0.17\textwidth]{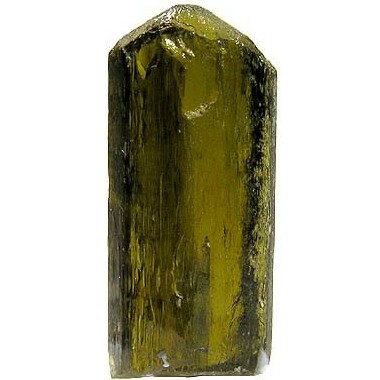} &
    \includegraphics[width=0.17\textwidth]{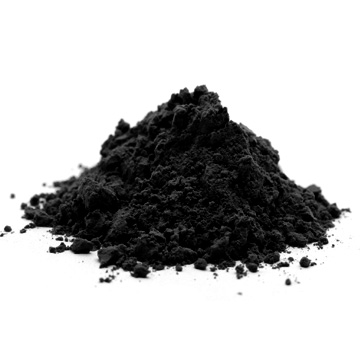} &
    \includegraphics[width=0.17\textwidth]{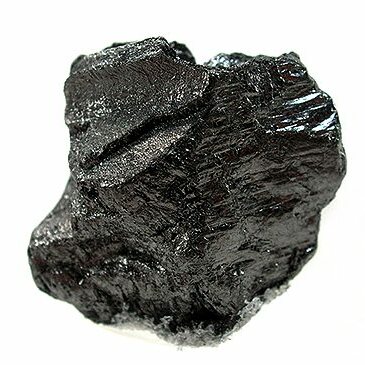} &
    \includegraphics[width=0.17\textwidth]{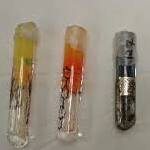} \\
    \textit{(a)} Forsterite & \textit{(b)} Enstatite 
    & \textit{(c)} Soot $\simeq$ a-C(:H) & \textit{(d)} Graphite 
    & \textit{(e)} \hPAH s \\
  \end{tabular}
  \newcap{Appearance of various minerals}%
         {The \textit{(a)} and \textit{(b)} images show crystalline silicates.
          The soot powder \textit{(c)} is an approximate analog of 
          \hHAC, on the a-C end, except that soot can contain radicals with O, 
          N, \etc\ 
          These images are useful to visualize the differences in optical
          properties of these compounds.
          They however show macroscopic samples.
          Even the powder in image \textit{(c)} is 
          made of particles much larger than 1~\tmic, which is the maximum size 
          of interstellar grains.
          \uline{Credit:} 
          \begin{inlinelistalph}
            \item forsterite from
            \href{https://commons.wikimedia.org/wiki/File:Forsterite-Ludwigite-34581.jpg}%
                 {Rob \familyname{Lavinsky}}, licensed under 
            \href{https://creativecommons.org/licenses/by-sa/3.0/deed.en}{CC BY-SA 3.0};
            \item enstatite from
            \href{https://commons.wikimedia.org/wiki/File:Enstatite-83152.jpg}%
                 {Rob \familyname{Lavinsky}}, licensed under 
            \href{https://creativecommons.org/licenses/by-sa/3.0/deed.en}{CC BY-SA 3.0};
            \item soot from  
           \href{https://en.wikipedia.org/wiki/Carbon_black\#/media/File:Carbon_black.jpg}%
            {Wikipedia},
            not licensed;
            \item graphite from
            \href{https://commons.wikimedia.org/wiki/File:Graphite-tn19a.jpg}%
              {Rob \familyname{Lavinsky}}, licensed under
            \href{https://creativecommons.org/licenses/by-sa/3.0/deed.en}{CC BY-SA 3.0};
            \item \hPAH s from the Astrochemistry Lab, NASA Ames Research 
          Center, with permission from Lou \familyname{Allamandola}.
          \end{inlinelistalph}}
  \label{fig:minerals}
\end{figure}

\paragraph{Olivine.} 
Olivine have the general formula (Mg,Fe)$_2$SiO$_4$, with different proportions of Mg and Fe.
Its crystalline structure is represented on \refsubfig{fig:crystal}{b}.
The two following compounds are the extreme cases of the Fe-to-Mg ratio.
\begin{description}
  \item[Forsterite] is the Mg-end of the series, with formula Mg$_2$SiO$_4$.
  \item[Fayalite] is the Fe-end of the series, with formula Fe$_2$SiO$_4$.
\end{description}
Olivine have an olive green color (\refsubfig{fig:minerals}{a}).

\paragraph{Pyroxene.} 
Pyroxene have the general formula (Mg,Fe)SiO$_3$, with different proportions of Mg and Fe.
They are constituted of silica tetrahedron chains, sharing one O atom (\refsubfig{fig:crystal}{c}), which explains their stoichiometry.
The two following compounds are the extreme cases of the Fe-to-Mg ratio.
\begin{description}
  \item[Enstatite] is the Mg-end of the series, with formula MgSiO$_3$.
  \item[Ferrosilite] is the Fe-end of the series, with formula FeSiO$_3$.
\end{description}
Pyroxene can be darker than olivine (\refsubfig{fig:minerals}{b}).
\takeaway{In general, silicates are translucent minerals.
They are gemstones, used in jewelry.}

    \subsubsection{Hydrogenated Amorphous Carbon}

This is a broad class of solids, noted \hHAC\ \citep[a notation introduced by][]{jones12b}. 
Carbon atoms can be paired in the following ways (\refsubfig{fig:crystal}{d}).
\begin{description}
  \item[Aromatic cycles] are hexagonal rings made of six sp$^2$ hybridized C 
    atoms. 
    Two of the three available sp$^2$ orbitals of each C atom make $\sigma$ 
    bonds, tying the cycle together.
    The last sp$^2$ orbital can be used to make a $\sigma$ bond with another C 
    atom, extending the compound, or with an H atom, ending the solid in this 
    direction.
    The six remaining p orbitals of the cycle make a sort of ring-shaped $\pi$ 
    bond.
    The electrons of these bonds are delocalized, they do not belong to a 
    specific C atom, but they are confined to the cycle.
    This is why compounds with aromatic cycles have some properties of 
    metals: electric conductivity and shiny appearance.
  \item[Aliphatic groups] are centered around a sp$^3$ hybridized C atom.
    Its four sp$^3$ orbitals can be paired to other C atoms or to H atoms, 
    forming $\sigma$ bonds.
  \item[Olefinic bonds] are alkene-type double bonds between two sp$^2$ 
    hybridized C atoms (\reffig{fig:sigmapi}).
    There is one $\sigma$ bond bridging two sp$^2$ orbitals, and one $\pi$
    bond linking the p orbital of each C atom.
\end{description}
The hydrogenation of \hHAC\ influences directly their band gap \citep{jones12b}.
Generally, H-poor \hHAC, which can be noted a-C, are sp$^2$ dominated (aromatic/olefinic), and have a low band gap ($E_g\simeq0.4-0.7$~eV).
On the contrary, H-rich \hHAC, which can be noted a-C:H, are mostly aliphatic (sp$^3$), and have a larger band gap ($E_g\simeq1.2-2.5$~eV).
The aromatic domains are responsible for bright features at 3.3, 6.2, 7.7, 8.6 and 11.3~\tmic, that will be extensively discussed in \refsec{sec:PAH}, whereas the main aliphatic feature is at 3.4~\tmic.
An important feature at 2175~$\r{A}$ (\refsec{sec:extinction}) is thought to originate in the transition between the $\pi$ and $\pi^*$ bands of sp$^2$ domains \citep[\eg][]{draine01}.
\takeaway{\hHAC\ tend to be more opaque than silicates
(\refsubfig{fig:minerals}{c}).}

    \subsubsection{Graphite}

Graphite is a mineral made of the stacking of graphene sheets, bonded by Van Der Waals interactions (\refsec{sec:bond}).
Graphene sheets are planar compounds exclusively constituted of aromatic cycles (\refsubfig{fig:crystal}{e}).
Pure graphite is solely made of sp$^2$ carbon.
Its aromaticity explains its shiny silver metallic appearance (\refsubfig{fig:minerals}{d}).
It has a strong $\pi\rightarrow\pi^*$ transition around 2175~$\r{A}$. 
The exact central wavelength however depends on the size and shape of the particles, and pure graphite seems too wide to account for the interstellar feature \citep[\eg][]{draine93,voshchinnikov04,papoular09}.
Graphite also has a broad band at 30~\tmic, seen parallel to the sheets, which corresponds to the oscillation frequency of the delocalized $\pi$ electrons \citep[\eg][]{venghaus77,draine07}.

    \subsubsection{Polycyclic Aromatic Hydrocarbons (PAH)}

\hPAH s are a class of molecules made of aromatic cycles, with peripheral H atoms (\refsubfig{fig:crystal}{f}).
They have the aromatic features of a-C, as well as the $\pi\rightarrow\pi^*$ transition around 2175~$\r{A}$ \citep[\eg][]{joblin92}.
Similarly to graphite, the exact central wavelength depends on the particle size and shape \citep[\eg][]{duley98}.
They can be colorful (\cf\ \refsubfig{fig:minerals}{e}).
They are highly flammable and carcinogenic.
\begin{table}[htbp]
  \centering
  \setlength\arrayrulewidth{2pt}
  \arrayrulecolor{white}
  \begin{tabularx}{\linewidth}%
    {|>{\columncolor{coltabcell}}l%
     |>{\columncolor{coltabcell}}l%
     |>{\columncolor{coltabcell}}l%
     |>{\columncolor{coltabcell}}l%
     |>{\columncolor{coltabcell}}X|}
    \hline
      \rowcolor{coltabhead}
      \textbf{Name}
      & \textbf{Stoichiometry} 
      & \textbf{Density}
      & \textbf{Melting}
      & \textbf{Main spectroscopic features} \\
    \hline
      \rowcolor{coltabsep}
      \multicolumn{5}{|c|}{\textsc{Silicates}} \\
    \hline
      \cellcolor{coltabhead}
      Forsterite & Mg$_2$SiO$_4$ & 3.3 g/cm$^3$ & 2200 K & 9.7, 18 \tmic \\
      \cellcolor{coltabhead}
      Fayalite   & Fe$_2$SiO$_4$ & 4.4 g/cm$^3$ & 1500 K & 9.7, 18 \tmic \\
      \cellcolor{coltabhead}
      Enstatite   & MgSiO$_3$ & 3.2 g/cm$^3$ & 2100 K & 9.7, 18 \tmic \\
      \cellcolor{coltabhead}
      Ferrosilite & FeSiO$_3$ & 4.0 g/cm$^3$ & 1200 K & 9.7, 18 \tmic \\
    \hline
      \rowcolor{coltabsep}
      \multicolumn{5}{|c|}{\textsc{Carbonaceous}} \\
    \hline
      \cellcolor{coltabhead}
      \hHAC\  & C$_n$H$_m$ &  1.8--2.1 g/cm$^3$ & N/A 
         & 2175~$\r{A}$, 3.3, 3.4, 6.2, 7.7, 8.6, 11.3 \tmic \\
      \cellcolor{coltabhead}
      Graphite & C$_n$      &  2.3 g/cm$^3$ & 3600 K & 2175~$\r{A}$, 30 \tmic \\
      \cellcolor{coltabhead}
      \hPAH\   & C$_n$H$_m$ &  2.2 g/cm$^3$ & N/A 
        & 2175~$\r{A}$, 3.3, 6.2, 7.7, 8.6, 11.3 \tmic \\
    \hline
  \end{tabularx}
  \newcap{General properties of interstellar dust candidates}%
         {The values of the density and melting temperature are 
          approximate.
          They vary between samples and experimental conditions.}
  \label{tab:minerals}
\end{table}

\section{The Interaction of Light with Solids}
\label{sec:lightsolid}

\begin{figure}[htbp]
  \includegraphics[width=\textwidth]{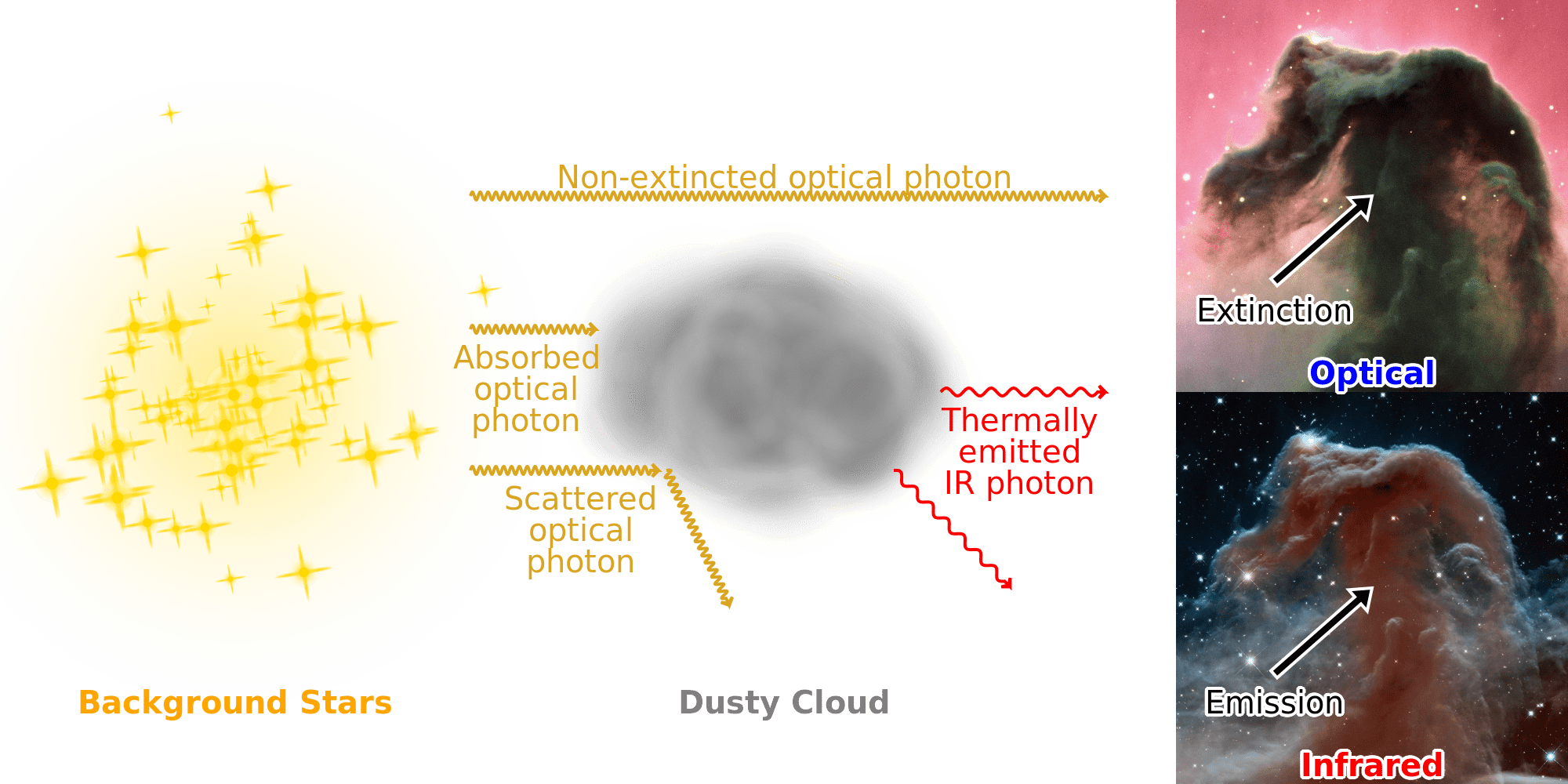}
  \newcap{Absorption, scattering and emission}%
         {The images on the right are the Horsehead nebula.
          \CClicence\
          \uline{Credit:} Horsehead nebula images from 
          \href{https://esahubble.org/images/heic1307b/}{NASA, ESA, and the Hubble 
                Heritage Team (AURA/STScI); ESO}, licensed under 
          \href{https://creativecommons.org/licenses/by/4.0/}{CC BY 4.0}.}
  \label{fig:absem}
\end{figure}

\begin{figure}[htbp]
  \includegraphics[width=\textwidth]{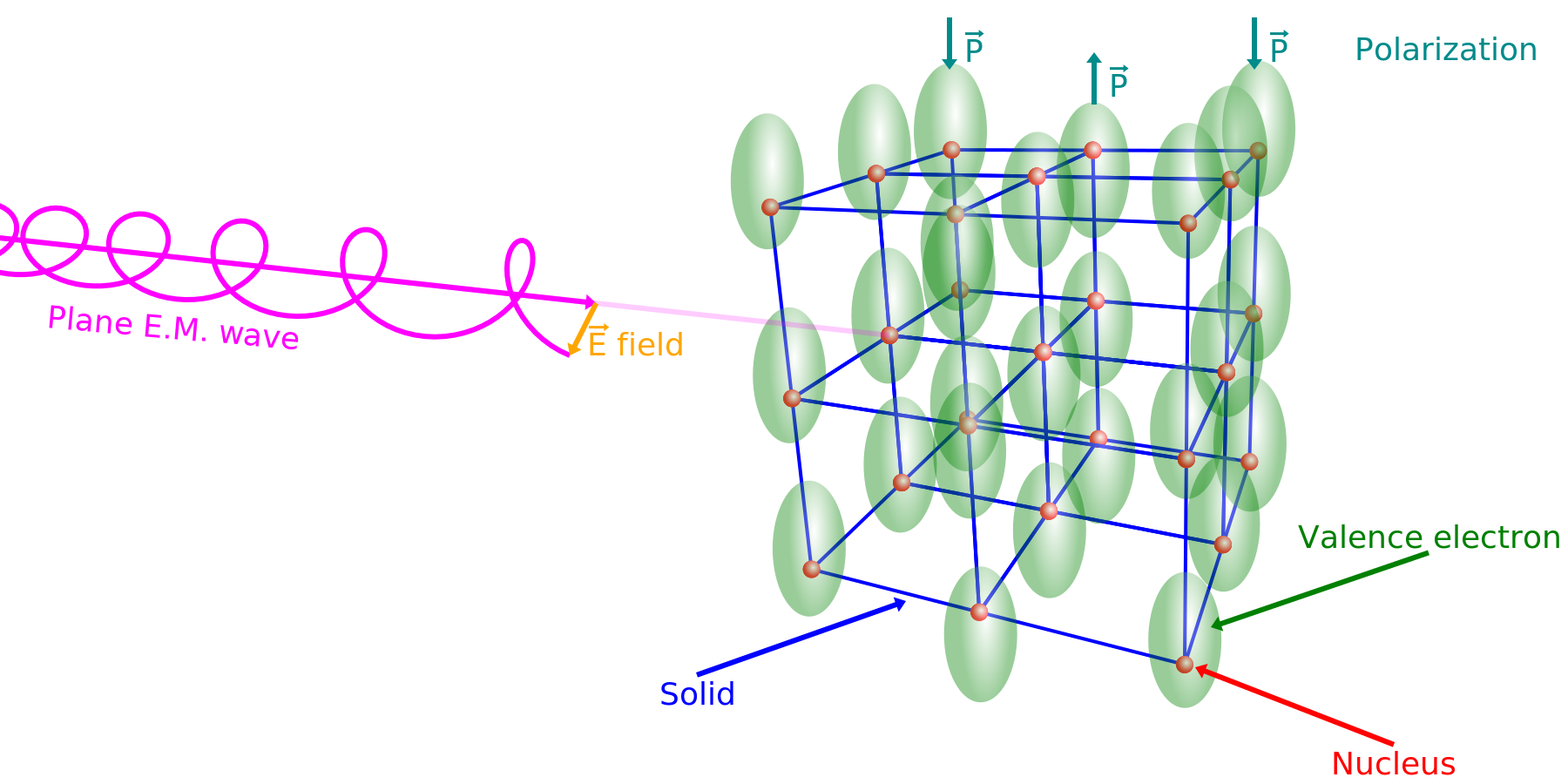}
  \newcap{Effect of an electromagnetic wave on a dielectric}%
         {An incoming, circularly polarized, electromagnetic wave is figured 
          in magenta. 
          The cube on the right represents a solid.
          The nuclei, assumed to be fixed, are the red spheres.
          The valence electrons are the green ellipsoids.
          They are displaced out of their equilibrium positions by the 
          electromagnetic wave, inducing a time-dependent polarization.
          \CClicence}
  \label{fig:light_solid}
\end{figure}

The interaction of an electromagnetic wave with dust grains results in the three following phenomena (\cf\ \reffig{fig:absem}).
\begin{description}
  \item[Absorption:] a fraction of the electromagnetic energy is stored into 
    the grain;
  \item[Scattering:] the wave vector of the fraction that is not absorbed 
    changes direction, its field polarization changes, but its frequency is not 
    affected;
  \item[Emission:] the energy stored in the grain is ulteriorly re-emitted in 
    the \hIR.
\end{description}
The sum of absorption and scattering is called \expression{extinction}.
These three phenomena can be modeled, assuming valence electrons are harmonic oscillators.
This way, the response to an electromagnetic wave of bonds in a dielectric or free electrons in a metal can be quantified.
This is illustrated in \reffig{fig:light_solid}.
Throughout this manuscript, we use $\lambda$, $\nu=c/\lambda$ and $\omega=2\pi\nu$ to respectively denote the \expression{wavelength}, \expression{frequency} and \expression{angular frequency} of an electromagnetic wave, $c$ being the \expression{speed of light} (\cf\ \reftab{tab:constants}).

  \subsection{Bonds as Harmonic Oscillators}
  \label{sec:HObond}

The harmonic oscillator model is particularly useful to describe the way bonds react to electromagnetic waves.

    \subsubsection{The Harmonic Oscillator Amplitude}

The position along the $x$-axis of a unidimensional harmonic oscillator of mass, $m$, as a function of time, $t$, follows the equation:
\begin{equation}
  \underbrace{m\frac{\dd^2 x(t)}{\dd t^2}}_\sms{inertia} 
   + \underbrace{b\frac{\dd x(t)}{\dd t}}_\sms{friction} 
   + \underbrace{k_ex(t)}_\sms{restoring force} 
  = \underbrace{F(t)}_\sms{external force},
  \label{eq:HO0}
\end{equation}
where $F$ is the external force applied to the oscillator, $k_e$ is the strength of the restoring force, and $b$ is a dissipation constant.
This equation is simply the Newton law ($\mathbb{F}=m\ddot{x}$), where the force, $\mathbb{F}$, has three components:
\begin{inlinelist}
  \item the external force, $F$, displacing the oscillator out of its 
    equilibrium position ($x=0$);
  \item the restoring force, $-k_ex$, which is proportional to the 
    distance, meaning it is stronger when the oscillator is farther away from 
    its equilibrium position;
  \item the friction force, $-b\dot{x}$, proportional to the velocity, having 
    the effect of slowing down the oscillator.
\end{inlinelist}

\begin{figure}[htbp]
  \begin{tabular}{cc}
    \includegraphics[width=0.48\textwidth]{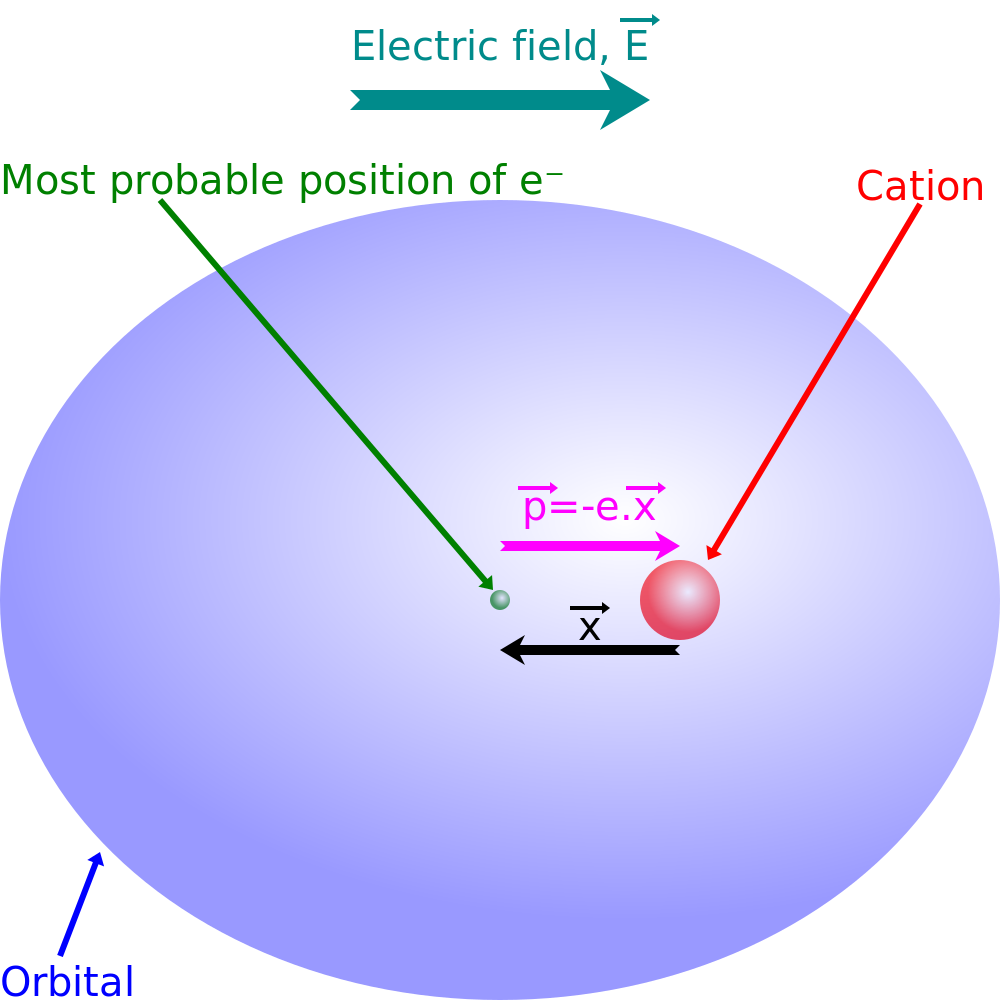}
    & \includegraphics[width=0.48\textwidth]{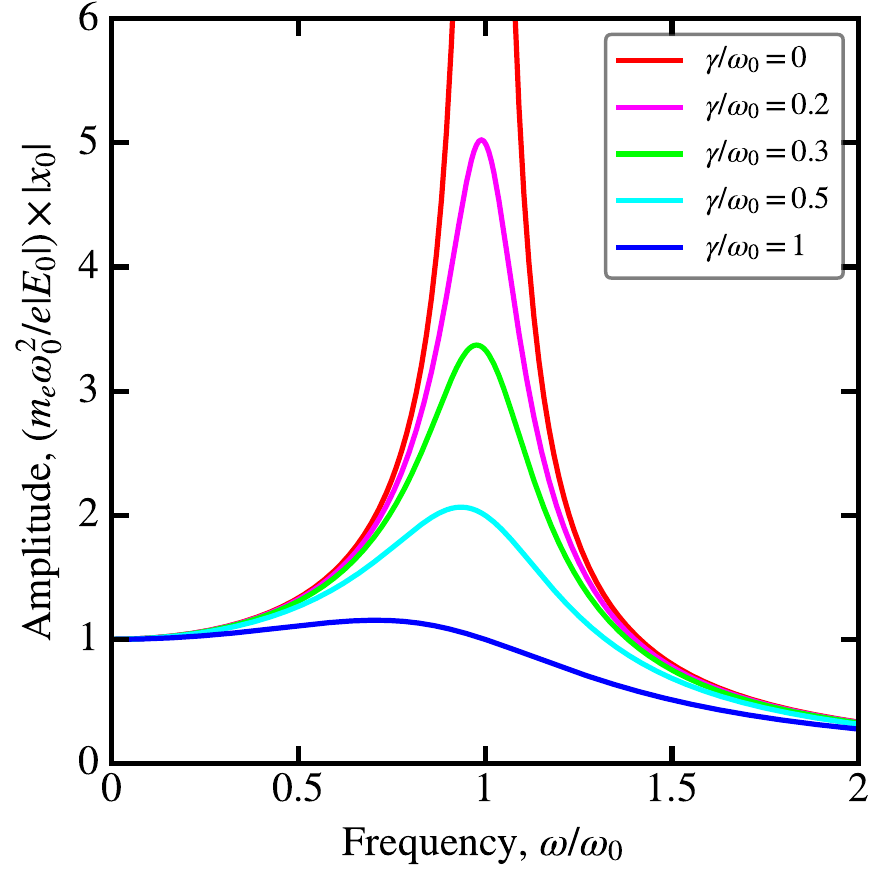} \\
  \end{tabular}
  \newcap{Amplitude of a forced harmonic oscillator}%
         {The left image represents the effect of an electromagnetic
          field, $\vect{E}$, on an orbital. 
          The displacement between the nucleus and the most probable position 
          of the electron induces a permanent dipole, $\vect{p}=-e.\vect{x}$.
          In this simple picture, the nucleus (cation) is assumed fixed and the
          electron oscillates around it.
          The right panel shows the amplitude of the motion of the electron 
          \refeqp{eq:HOx0mod} as a function of frequency, for different values 
          of the friction, $\gamma$.
          We have normalized both axes in order to display dimensionless 
          quantities.
          \CClicence}
  \label{fig:HOx0}
\end{figure}

In the case of the motion of an electron, excited by the Lorentz force of a complex, harmonic plane electromagnetic wave with angular frequency $\omega$, $F(t)=e\,E_0\exp(-i\omega t)$, \refeq{eq:HO0} can be rewritten:
\begin{equation}
  \frac{\dd^2x(t)}{\dd t^2} +\gamma\frac{\dd x(t)}{\dd t} + \omega_0^2x(t)
  = \frac{e\,E_0\exp(-i\omega t)}{m_e},
  \label{eq:HO}
\end{equation}
where $m_e$ is the electron mass.
We have also introduced the \expression{natural frequency}, $\omega_0$, and the \expression{damping constant}, $\gamma$:
\begin{eqnarray}
  \omega_0\equiv\sqrt{\frac{k_e}{m_e}} & \mbox{ and } 
    & \gamma\equiv\frac{b}{m_e}.
  \label{eq:HOfrequency}
\end{eqnarray}
In this case, the restoring force is created by the atom's electrostatic potential well, and the friction can be interpreted as collisions of the electron with the lattice.
The solution to \refeq{eq:HO} has the form $x(t)=x_0\exp(-i\omega t)$, with complex amplitude \citep[\eg][]{levi16}:
\begin{equation}
  x_0 = \frac{eE_0}{m_e(\omega_0^2-\omega^2-i\omega\gamma)}.
  \label{eq:HOx0}
\end{equation}
It is important to consider both $\vect{x}$ and $\vect{E}$ as complex quantities, since there is a phase shift induced by the dissipation term.
The module of $x_0$, giving the physical value of the amplitude, is:
\begin{equation}
  |x_0| = \frac{e|E_0|}{m_e\sqrt{\left(\omega^2-\omega_0^2\right)^2 
          +\gamma^2\omega^2}}.
  \label{eq:HOx0mod}
\end{equation}
This is the classic harmonic oscillator solution.
It is represented in \reffig{fig:HOx0}.
The amplitude is maximum at the \expression{resonant frequency}, $\omega_r=\sqrt{\omega_0^2-\gamma^2/2}$.
If there is no dissipation ($\gamma=0$), the amplitude becomes infinite, and the electron escapes.

    \subsubsection{The Plasma Frequency}
Free electrons oscillate around heavy cations at the \expression{plasma frequency}, $\omega_p$.
Its formula is \citep[\cf\ \eg\ Chap.~1 of][for a simple proof]{kruegel03}:
\begin{equation}
  \omega_p \equiv \sqrt{\frac{n_ee^2}{m_e\epsilon_{0}}},
  \label{eq:plasma}
\end{equation}
where $n_e$ is the number density of free electrons.
It applies to metals, as well as actual gaseous plasmas.
These media absorb and scatter electromagnetic waves with frequencies lower than $\omega_p$.
For instance, in the case of the Earth's ionosphere, $n_e\simeq10^{12}$~m$^{-3}$, thus $\omega_p\simeq60$~MHz.
This explains why amateur radio operators communicate over long distances at frequencies lower than this value, to benefit from the reflection of their transmission on the ionosphere \citep[\eg][]{perry18}.
In the case of a metal, with density $n_e\simeq10^{29}$~m$^{-3}$, $\omega_p\simeq20$~PHz, corresponding to a wavelength $\lambda_p\simeq0.1\emic$, in the \expression{UltraViolet} (\hUV; \cf~\reftab{tab:spectralrange}).
This explains the shiny appearance of metals, as they are able to reflect the visible light, which has a lower frequency than \hUV\ photons.
It happens that the expression of $\omega_p$ also appears in the optical properties of dielectrics, and we will use it extensively.

    \subsubsection{The Dielectric Function}

\begin{figure}[htbp]
  \includegraphics[width=\textwidth]{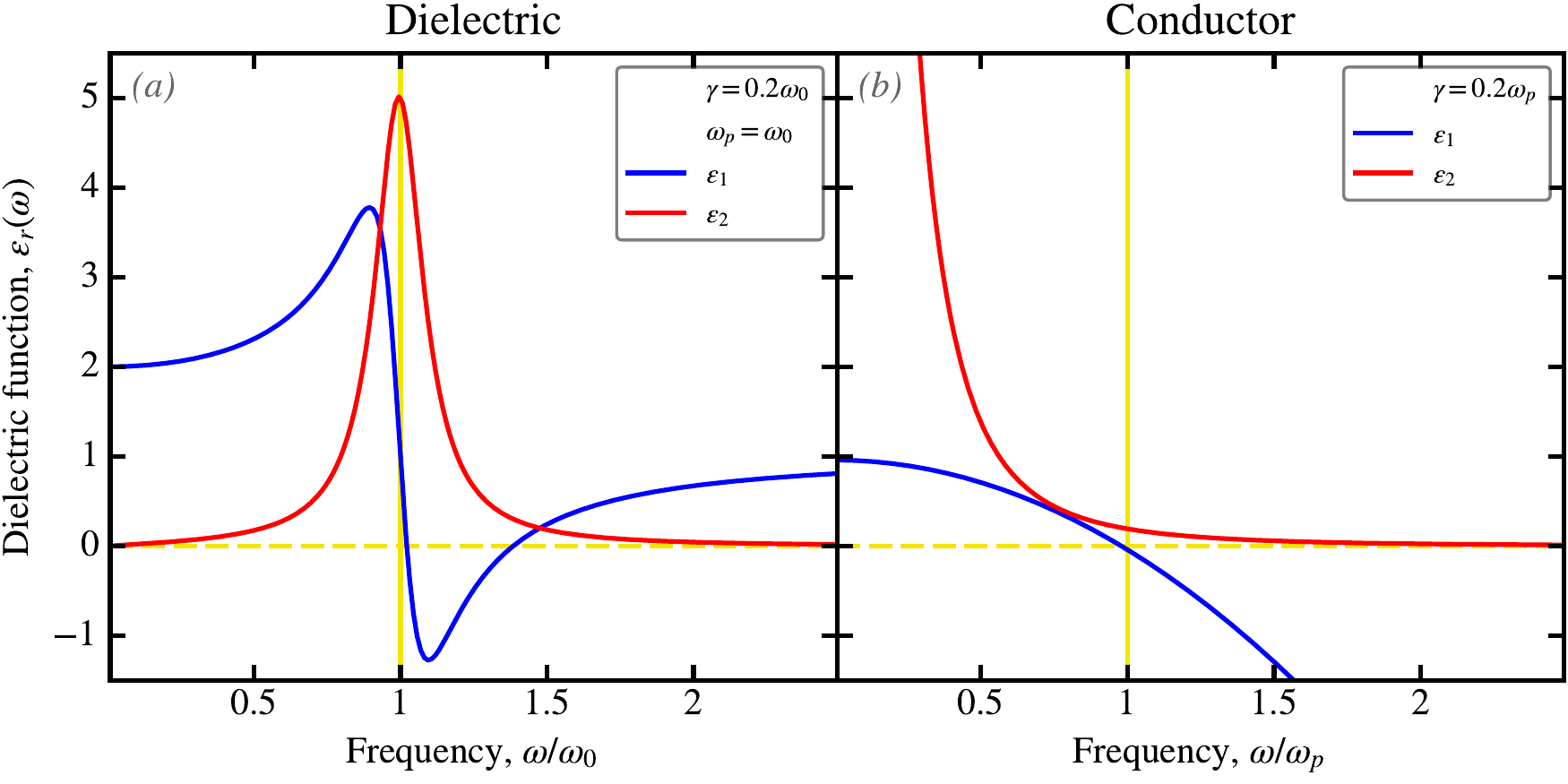}
  \newcap{Idealized optical constants}%
         {Both panels show dielectric functions where the bonds 
         (in the dielectric case) or the free electrons 
         (in the case of a conductor) are treated as harmonic oscillators.
         Panel~\textit{(a)} displays \refeq{eq:eps}.
         We see that the imaginary part, $\epsilon_2$, which
         corresponds to the absorption, peaks around the resonant frequency and 
         drops rapidly to zero on both sides.
         Panel~\textit{(b)} displays \refeq{eq:epsmet}.
         We see that the absorption by a conductor rises rapidly below the 
         plasma frequency.
         It is another way to witness the fact that metals can absorb any light 
         below $\omega_p$.
         At high frequency, $\epsilon_2\rightarrow0$.
         \CClicence}
  \label{fig:eps}
\end{figure}

In a dielectric, an electromagnetic wave polarizes the bonds.
If we consider each bond as a dipole with moment $\vect{p}$, the \expression{polarization density} is defined as:
\begin{equation}
  \vect{P} = n_e\vect{p},
  \label{eq:P}
\end{equation}
where $n_e$ is the number density of valence electrons.
The induced polarization density is directly related to the electric field:
\begin{equation}
  \vect{P} = \epsilon_0\chi\vect{E},
  \label{eq:suscept}
\end{equation}
where $\chi$ is the \expression{electric susceptibility}.
The \expression{electric displacement field}, $\vect{D}$, which accounts for the charge displacement induced by an electric field $\vect{E}$ is defined as:
\begin{equation}
  \vect{D}=\epsilon\vect{E},
  \label{eq:D}
\end{equation}
where $\epsilon$ is the \expression{electric permittivity} of the medium.
The \expression{relative electric permittivity}, $\epsilon_r$ is defined such that: $\epsilon=\epsilon_r\epsilon_0$.
It is a macroscopic quantity, as no medium is truly continuous.
At atomic scales, \refeq{eq:D} can be broken into two terms:
\begin{equation}
  \vect{D} = \underbrace{\epsilon_0\vect{E}}_\sms{vacuum between atoms} 
           + \underbrace{\vect{P}}_\sms{induced dipoles}
           = \epsilon_0\underbrace{\left(1+\chi\right)}_{\epsilon_r}\vect{E}.
\end{equation}
The second equality derives from \refeq{eq:suscept}.
It also implies that $\epsilon_r=1+\chi$.
The induced polarization is, using \refeq{eq:HOx0} and \refeq{eq:P}:
\begin{equation}
  \vect{P} = n_e e \vect{x} 
    = \underbrace{\frac{n_e e^2}{m_e}
      \frac{1}{\omega_0^2-\omega^2-i\omega\gamma}}_{\epsilon_0\chi}\vect{E}.
\end{equation}
Finally, a bit of algebra and introducing the plasma frequency \refeqp{eq:plasma}, gives:
\begin{equation}
  \epsilon_r(\omega) = 
  \underbrace{1+\frac{\omega_p^2(\omega_0^2-\omega^2)}%
              {(\omega_0^2-\omega^2)^2+\gamma^2\omega^2}}_{\epsilon_1(\omega)}
  +i\underbrace{\frac{\omega_p^2\gamma\omega}%
               {(\omega_0^2-\omega^2)^2+\gamma^2\omega^2}}_{\epsilon_2(\omega)}.
  \label{eq:eps}
\end{equation}
This dispersion relation is called the \expression{dielectric function}.
It is displayed in \refsubfig{fig:eps}{a}.
It can be related to the refractive index of the material, $m$.
Indeed, plane electromagnetic waves propagating in a dielectric have a phase velocity $v_\sms{ph}=1/\sqrt{\epsilon\mu}$, where $\mu$ is the \expression{magnetic susceptibility} \citep[\cf\ \eg\ Chap.~7 of][for a derivation from Maxwell's equations]{jackson99}.
This phase velocity can also be expressed as a function of the speed of light, $c$: $v_\sms{ph}=c/m$, where $m$ is the refractive index.
Since we can decompose $\mu=\mu_r\mu_0$, and because $\epsilon_0\mu_0=1/c^2$, we have:
\begin{equation}
  m(\omega)\equiv\sqrt{\epsilon_r(\omega)\mu_r(\omega)}=n(\omega)+ik(\omega).
  \label{eq:nandk}
\end{equation}
The refractive index is sometimes referred to as the \expression{optical constants}, or simply the \expression{``$n$ and $k$''}.
In a nonmagnetic medium, $\mu_r=1$, thus:
\begin{equation}
  \left\{
  \begin{array}{rcl}
    \epsilon_1(\omega) & = & n^2(\omega)-k^2(\omega) \\
    \epsilon_2(\omega) & = & 2n(\omega)k(\omega).
  \end{array}
  \right.
\end{equation}
The two complex quantities $\epsilon(\omega)$ and $m(\omega)$ contain the same information.
\refeq{eq:eps} corresponds to one single type of harmonic oscillator, that is to one mode of one type of bond.
An actual dielectric is usually the linear combination of several resonances, such as \refeq{eq:eps}, with different sets of $\omega_p$, $\omega_0$, and $\gamma$.

    \subsubsection{Harmonic Oscillator Cross-Section}

The flux carried by a plane electromagnetic wave is given by the time average of the \expression{Poynting vector}:
\begin{equation}
  \langle|\vect{\mathcal{P}}|\rangle=\frac{\epsilon_0c}{2}|\vect{E_0}|^2.
\end{equation}
The power radiated by a dipole, harmonically oscillating, is \citep[\cf\ \eg\ Chap.~9 of][]{jackson99}:
\begin{equation}
  W_\sms{rad} 
  =\frac{2}{3}\frac{\langle\ddot{|\vect{p}|}^2\rangle}{4\pi\epsilon_0c^3}
  = \frac{\omega^4e^2|x_0|^2}{12\pi\epsilon_0c^3}.
\end{equation}
This power is the radiation in response to the excitation by the incident wave.
This is the scattering contribution.
The \expression{scattering cross-section} of this harmonic oscillator is thus simply:
$C_\sms{sca} = W_\sms{rad}/\langle|\vect{\mathcal{P}}|\rangle.$
Replacing $|x_0|$ by \refeq{eq:HOx0mod}, we obtain:
\begin{equation}
  C_\sms{sca}(\omega)
    = C_\sms{T}\frac{\omega^4}{(\omega^2-\omega_0^2)^2+\gamma^2\omega^2},
  \label{eq:HOCsca}
\end{equation}
where we have introduced the \expression{Thomson cross-section}:
\begin{equation}
  C_\sms{T}
    \equiv\frac{8\pi}{3}\left(\frac{e^2}{4\pi\epsilon_0m_ec^2}\right)^2
                \simeq 6.66\E{-29}\;\textnormal{m}^2.
  \label{eq:sigmaT}
\end{equation}
Now, the absorbed power comes from the dissipation into the solid.
The dissipation force in \refeq{eq:HO} is $\vect{F}_\sms{dis}=-m_e\gamma\dot{\vect{x}}$. 
The dissipated power is thus the work of this force:
\begin{equation}
  W_\sms{dis} = m_e\gamma|\dot{\vect{x}}|^2 = \frac{1}{2}m_e\gamma\omega^2|x_0|^2.
\end{equation}
The absorption cross-section of the harmonic oscillator is therefore:
$C_\sms{abs} = W_\sms{dis}/\langle|\vect{\mathcal{P}}|\rangle$.
Using \refeq{eq:HOx0mod} and \refeq{eq:sigmaT}, we obtain:
\begin{equation}
  C_\sms{abs}(\omega) = \frac{e^2}{m_e\epsilon_0c}
      \frac{\gamma\omega^2}{(\omega^2+\omega_0^2)^2+\gamma^2\omega^2}
  \label{eq:HOCabs}
\end{equation}
It is interesting to look at the limiting behavior of both $C_\sms{sca}$ and $C_\sms{abs}$ \citep[see also Chap.~1 of][]{kruegel03}.
\begin{description}
  \item[At high frequency] ($\omega\gg\omega_0$; short wavelength), we have:
\begin{equation}
  \left\{
  \begin{array}{rcl}
    C_\sms{sca}(\omega)
      &\simeq&C_\sms{T}
    \\
    C_\sms{abs}(\omega)
      &\simeq&\displaystyle
      \frac{e^2}{4m_e\epsilon_0c}\frac{\gamma}{\omega^2}.
  \end{array}
  \right.
\end{equation}
  \item[Around the resonant frequency] ($\omega\simeq\omega_0$), we have:
\begin{equation}
  \left\{
  \begin{array}{rcl}
    C_\sms{sca}(\omega)
      &\simeq&\displaystyle
      \frac{C_\sms{T}}{4}\frac{\omega_0^2}{(\omega-\omega_0)^2+(\gamma/2)^2}
    \\
    C_\sms{abs}(\omega)
      &\simeq&\displaystyle
      \frac{e^2}{4m_e\epsilon_0c}\frac{\gamma}{(\omega-\omega_0)^2+(\gamma/2)^2}.
  \end{array}
  \right.
\end{equation}
It shows that around the resonant frequency, both $C_\sms{sca}$ and $C_\sms{abs}$ have \expression{Lorentz profiles} centered at $\omega_0$, with \expression{Full Width at Half Maximum} (\hFWHM), $\gamma$.
  \item[At short frequency] ($\omega\ll\omega_0$; long wavelength):
\begin{equation}
  \left\{
  \begin{array}{rcl}
    C_\sms{sca}(\omega) 
      &\simeq& 
      \displaystyle
      C_\sms{T}\left(\frac{\omega}{\omega_0}\right)^4
  \\
  C_\sms{abs}(\omega) 
     &\simeq& 
    \displaystyle
     \frac{e^2\gamma}{m_e\epsilon_0c\omega_0^2}
     \left(\frac{\omega}{\omega_0}\right)^2.
  \end{array}
  \right.
\end{equation}
Those approximations are particularly useful.
\takeaway{For dielectrics, $C_\sms{sca}\propto1/\lambda^4$ and $C_\sms{abs}\propto1/\lambda^2$ at long wavelength.}
\end{description}

    \subsubsection{Optical Constants of Conductors}

\refeq{eq:eps} is valid for a dielectric, as it assumes the medium is only constituted of dipoles.
This is not the case in a conductor where there are also free charges.
An electromagnetic wave induces a current, $\vect{j}$, related to the electric field, $\vect{E}$, by the \expression{conductivity}, $\sigma$:
\begin{equation}
  \vect{j}=\sigma\vect{E}=n_ee\vect{v},
  \label{eq:vfree}
\end{equation}
where the second equality relates the current to its microscopic origin, the velocity of free electrons, $\vect{v}$.
Maxwell's equations for plane waves give \citep[\eg\ Chap.~7 of][]{jackson99}:
\begin{equation}
  \epsilon_r(\omega)=\underbrace{\epsilon_\sms{d}(\omega)}_\sms{bound electrons}
                    +\underbrace{i\frac{\sigma(\omega)}{\epsilon_0\omega}}_\sms{free electrons},
  \label{eq:epsmet}
\end{equation}
where $\epsilon_\sms{d}(\omega)$ is the leftover dielectric term.
We can apply the harmonic oscillator model again to these free electrons.
The difference is that there is no restoring force, $\omega_0=0$.
From \refeq{eq:HOx0}, the velocity of free electrons becomes:
\begin{equation}
  |\vect{v}| = \frac{e|\vect{E}|}{m_e(\gamma-i\omega)}.
\end{equation}
Injecting this quantity in \refeq{eq:vfree}, we obtain:
\begin{equation}
  \sigma(\omega) = \epsilon_0\omega_p^2\frac{\gamma}{\gamma^2+\omega^2}
                 + i\epsilon_0\omega_p^2\frac{\omega}{\gamma^2+\omega^2}.
\end{equation}
Focussing on the free electron term in \refeq{eq:epsmet}, \ie\ assuming $\epsilon_\sms{d}=0$, we get:
\begin{equation}
  \epsilon_r(\omega) 
  = \underbrace{1 - \frac{\omega_p^2}{\gamma^2+\omega^2}}_{\epsilon_1(\omega)}
  +i \underbrace{\frac{\gamma}{\omega}\frac{\omega_p^2}{\gamma^2+\omega^2}}_{\epsilon_2(\omega)}.
\end{equation}
It is also known as the \expression{Drude model}.
This equation is displayed in \refsubfig{eq:eps}{b}.
Interestingly enough, the cross-sections of \refeq{eq:HOCsca} and \refeq{eq:HOCabs} apply also to conductors, with $\omega_0=0$.
We can easily derive their limiting behavior.
\takeaway{Conductors have the same behavior than dielectrics at long wavelength: $C_\sms{sca}\propto1/\lambda^4$ and $C_\sms{abs}\propto1/\lambda^2$.}

    \subsubsection{The Kramers-Kronig Relations}
    \label{sec:KK}

The residue theorem implies that, if $f(x)$ is a complex function of the complex variable $x$, analytical over $\Im(x)\ge 0$, and dropping faster than $1/|x|$, we have the relation:
\begin{equation}
  f(\omega) = \frac{1}{i\pi}\:P\int_{-\infty}^\infty\frac{f(x)}{x-\omega}\ddiff x,
\end{equation}
with $\omega$ real and positive.
We have used the \expression{Cauchy principal value}:
\begin{equation}
  P\int_{-\infty}^\infty\frac{f(x)}{x-\omega}\ddiff x
  = \lim_{\delta\rightarrow0}\left(\int_{-\infty}^{\omega-\delta}
                              \frac{f(x)}{x-\omega}\ddiff x
                              +
                              \int_{\omega+\delta}^\infty
                              \frac{f(x)}{x-\omega}\ddiff x\right),
\end{equation}
which is simply an integral avoiding the singularity in $x=\omega$.
Decomposing $f(x)=f_1(x)+if_2(x)$, we obtain cross-relations between $f_1(x)$ and $f_2(x)$.
These general mathematical relations are usually applied to the susceptibility, from which we derive the dielectric function \citep[\eg\ Chap.~21 of ][]{draine11b}:
\begin{equation}
  \left\{
  \begin{array}{rcl}
    \epsilon_1(\omega) -1& = &\displaystyle
      \frac{2}{\pi}\:P\int_0^\infty\frac{x\epsilon_2(x)}{x^2-\omega^2}\ddiff x
      \\
    \epsilon_2(\omega) & = &\displaystyle
      -\frac{2}{\pi}\omega\:
      P\int_0^\infty\frac{\epsilon_1(x)-1}{x^2-\omega^2}\ddiff x.
  \end{array}
  \right.
  \label{eq:KK}
\end{equation}
These relations are known as the \expression{Kramers-Kronig relations}.

\paragraph{Implications.}
We only need to specify $\epsilon_1$ or $\epsilon_2$ at all frequencies, and can use \refeq{eq:KK} to derive the other one.
These relations are used to check laboratory data consistency \citep[\eg][]{zubko96}.

\paragraph{Interpretation.}
They are a consequence of the causality requirement for a linear system (here, we have $\vect{P}=\epsilon_0\chi\vect{E}$).
In our case, they impose that the response of the polarization does not precede the effect of the electric field.
Sect.~62 of \citet{landau60}, Chap.~21 of \citet{draine11b}, and Chap.~2 of \citet{kruegel03} discuss these relations more extensively.

\paragraph{Constraint on the Cross-Section.}
They give some constraints on the long wavelength behavior of the dielectric function.
Let's assume that $\epsilon_2(\omega)\propto\omega^{\beta-1}$ for $\omega<\delta$, for an arbitrary $\delta$.
The first relation tells us that:
\begin{equation}
  \epsilon_1(0)=1+\frac{2}{\pi}\int_0^\delta x^{\beta-2}\ddiff x
               +\int_\delta^\infty\frac{\epsilon_2(x)}{x}\ddiff x.
\end{equation}
The second integral is finite by requirement.
For the first integral to be finite, we need to have $\beta>1$.
At long wavelength, \refeq{eq:eps} tells us that, for a dielectric:
\begin{equation}
  \left\{
  \begin{array}{ccccc} 
    \epsilon_1(\omega)&\underset{\omega\to0}{\to}&
      \displaystyle1+\left(\frac{\omega_p}{\omega_0}\right)^2&=&\mbox{const} \\
    \epsilon_2(\omega)&\underset{\omega\to0}{\to}&
      \displaystyle\frac{\omega_p^2\gamma}{\omega_0^4}\omega
      &\ll&\epsilon_1(\omega).
  \end{array}
  \right.
  \label{eq:epslim}
\end{equation}
We will see in \refeq{eq:Rayleigh} that, at long wavelength, $C_\sms{abs}\propto\epsilon_2/[(\epsilon_1+2)^2+\epsilon_2^2]/\lambda$.
Using \refeq{eq:epslim}, we get $C_\sms{abs}(\lambda)\propto\epsilon_2(\lambda)/\lambda\propto\lambda^{-\beta}$.
We will see in \refsec{sec:mixT} that this $\beta$ parameter is sometimes referred to as the \expression{emissivity index}. 
\takeaway{Assuming that $C_\sms{abs}(\lambda)\propto\lambda^{-\beta}$ at long wavelengths, we need to have $\beta>1$.}

  \subsection{Grain Optical Properties}
  \label{sec:Qabs}

    \subsubsection{Why Are Most Dust Features in the MIR?}
    \label{sec:IRfeatures}

\begin{figure}[htbp]
  \includegraphics[width=\textwidth]{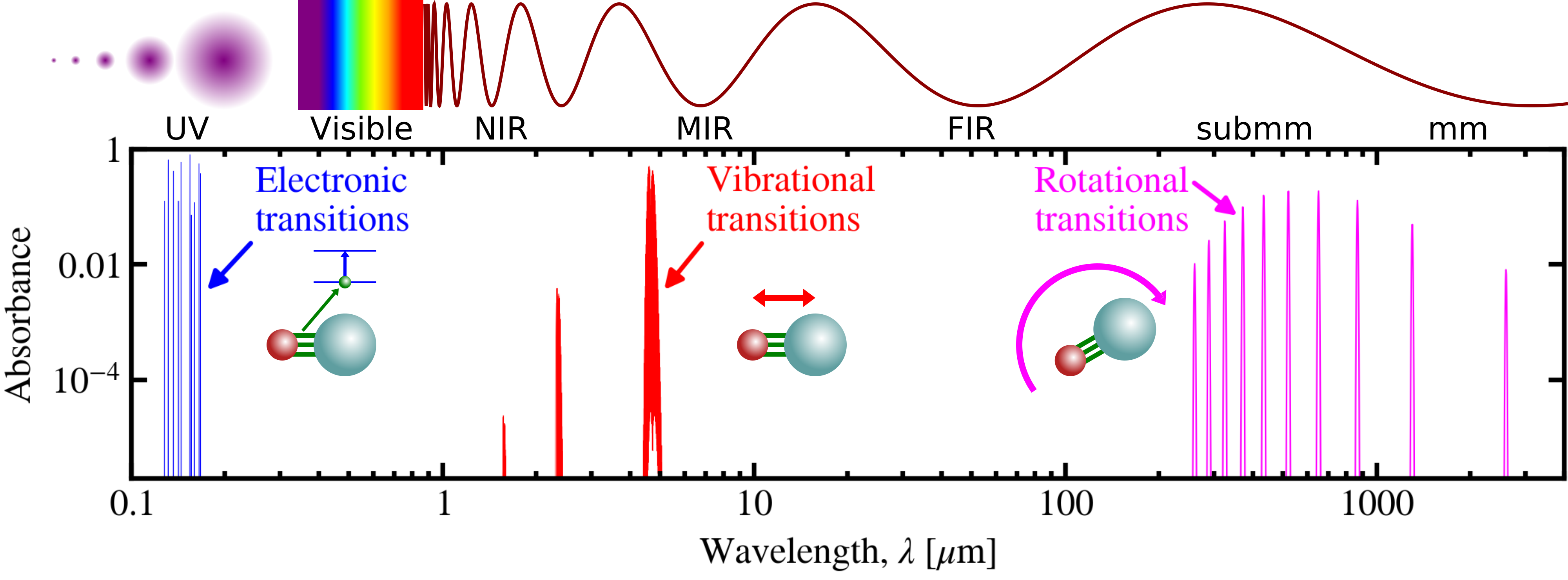}
  \newcap{Molecular transitions}{The different types of transitions are 
          illustrated with the CO molecule.
          \CClicence}
  \label{fig:elrotvib}
\end{figure}
All but one spectral features of the interstellar grain candidates we have listed in \reftab{tab:minerals} are in the \hMIR.
This is a general trend \citep[\eg\ Table 1 of][]{vandertak18}.
It can be understood by making an analogy with the different types of molecular transitions \citep[\cf\ \eg\ Chap.~2 of][for a review]{tielens05}.
Those are illustrated in \reffig{fig:elrotvib}.
\begin{description}
  \item[Electronic transitions] are transitions between the quantum
    harmonic oscillator levels of the bonding electron.
    In the case of solids, they are transitions between bands, such as
    the $\pi\rightarrow\pi^\star$ transition of aromatic carbon, 
    at 2175~$\r{A}$ (\cf\ \refsec{sec:dustanalog}).
    The natural frequency of these resonances is given by 
    \refeq{eq:HOfrequency}: $\omega_0=\sqrt{k_e/m_e}$.
    The energy of these resonances is comparable to the binding energy, or the 
    band gap.
    They typically range between $\simeq4$ and $\simeq20$~eV.
    They are thus in the \hUV\ domain ($\lambda\simeq0.06-0.30\emic$).
  \item[Vibrational transitions] are associated with the stretching 
    or bending of a bond.
    These modes involve the motion of the nuclei, which are much heavier than
    the electrons.
    Their frequency is $\omega_v=\sqrt{k_v/\mu_{1,2}}$, where 
    $\mu_{1,2}=m_1m_2/(m_1+m_2)$ is the reduced mass of the two atoms, $m_1$ and 
    $m_2$.
    Typically, $\mu_{1,2}=0.9, 6, 10\times m_p$ for C--H, C--C and Si--O bonds, 
    respectively ($m_p$ is the proton mass; \cf\ \reftab{tab:constants}).
    At first order, the new force constant is similar to previously, 
    $k_e\simeq k_v$.
    The frequency is now reduced by a factor 
    $\simeq\sqrt{m_e/\mu_{1,2}}\simeq0.007-0.02$.
    These transitions are thus in the \hMIR\ ($\lambda\simeq2-40\emic$).
    They are the most relevant transitions for \hISD.
  \item[Rotational transitions] are associated with the rotation of the 
    molecule.
    Their energy depends on the centrifugal force, which reduces the frequency
    by a factor $\simeq m_e/m_p$.
    These transitions are thus in the millimeter regime.
    Most dust grains do not have detectable rotational transitions, because of 
    their inertia.
    Only the smallest, charged grains have a non-negligible rotational emission
    that will be discussed in \refsec{sec:AMEobs}.
\end{description}

    \subsubsection{Dielectric Functions of Realistic Materials}

\begin{figure}[htbp]
  \includegraphics[width=\textwidth]{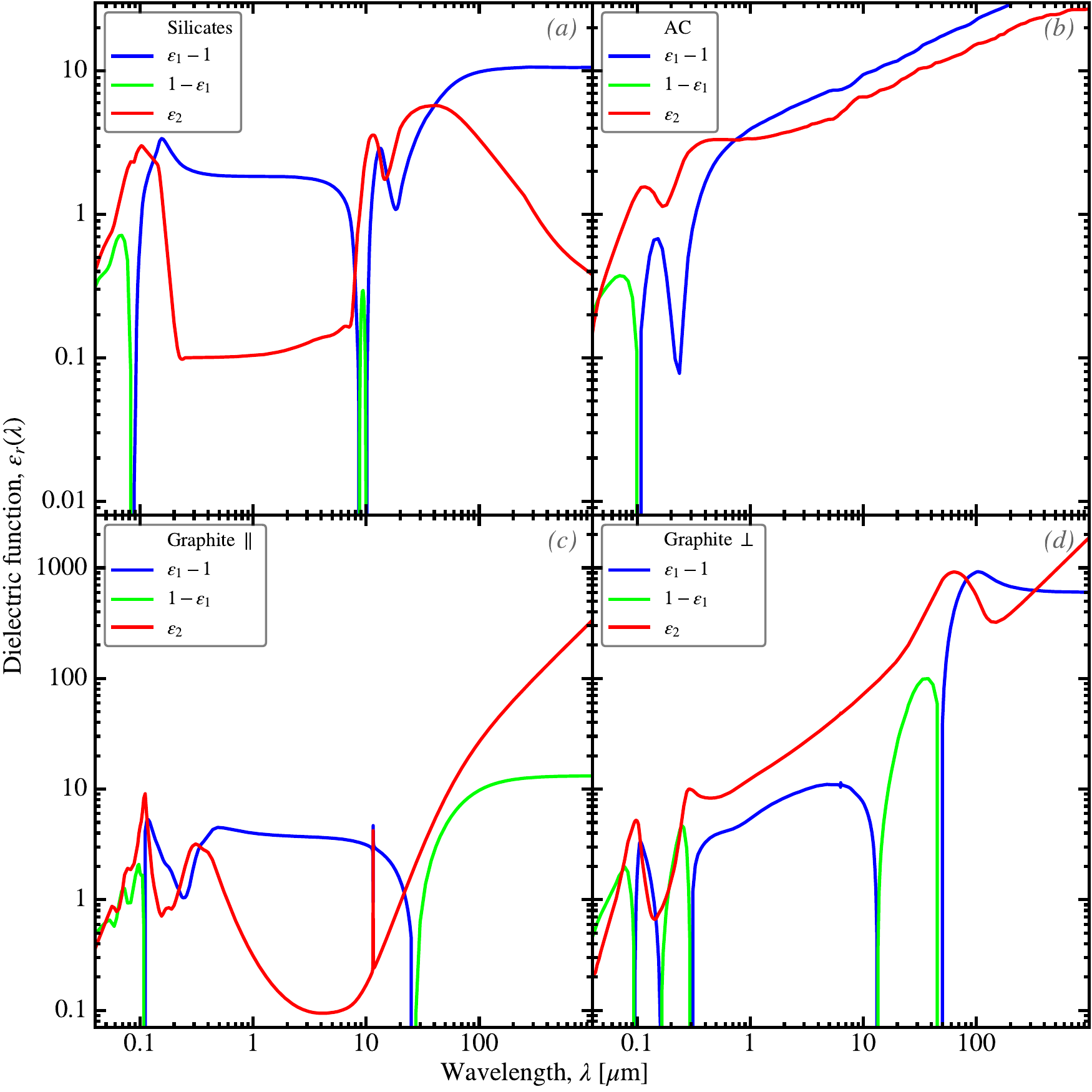}
  \newcap{Dielectric functions of different materials}%
         {In each panel, we show the real part of the dielectric function as
          $\epsilon_1-1$ (blue) and $1-\epsilon_1$ (green), as this quantity
          can be negative.
          The imaginary part, $\epsilon_2$, is in red.
          Panel~\textit{(a)} shows the astronomical silicates by
          \citet{draine03,draine03b}.
          Panel~\textit{(b)} shows the ACAR type of amorphous carbon by 
          \citet{zubko96}. 
          These are produced by arc discharges between amorphous carbon  
          electrodes in a 10~mbar Ar atmosphere.
          Panel~\textit{(c)} and \textit{(d)} show the graphite by
          \citet{draine03,draine03b}.
          Graphite is anisotropic, as we have seen in 
          \refsubfig{fig:crystal}{e}.
          It is usually approximated by mixing 2/3 of the optical properties
          parallel to the graphene sheets \textit{(c)}, and 1/3 perpendicular
          \textit{(d)}.
          \CClicence}
  \label{fig:epsgrain}
\end{figure}

The dielectric functions of \refeq{eq:eps} and \refeq{eq:epsmet} correspond to simple cases where there is only one type of oscillator.
Realistic materials have more complex structures, with several modes per bond.
Deriving dielectric functions of potential interstellar grain analogs is the subject of a rich literature.
There are three types of approaches to determine the dispersion relation of a medium.
\begin{description}
  \item[The theoretical knowledge] of the microscopic structure of the crystal 
    can be used to determine the different resonances that we have demonstrated 
    in \refsec{sec:HObond}.
    The resonant or plasma frequency, as well as the collisional rates 
    ($\gamma$) have to be known.
    The Kramers-Kronig relations (\refsec{sec:KK}) can be used to obtain 
    complementary constraints.
  \item[Laboratory data] about the spectral profile of some features, or the 
    density of the material, or any relevant characteristics can also be used.
    The \hISM\ community is active in this field, as our results largely depend 
    on the accurate values of atomic and molecular data.
    Several teams focus their research effort on laboratory measures of dust 
    analogs.
  \item[Astronomical observations] can be used to constrain some features.
    The most famous example is the use of observations of the 9.7 and 18~\tmic\
    band profiles to build the optical properties of \expression{astronomical 
    silicates}, by \citet{draine84}.
    Indeed, we have seen in \refsec{sec:dustanalog} that there is a diversity
    of silicate composition, and we do not know which one is relevant to 
    \hISD\ (it is likely a mixture).
\end{description}
These approaches are not exclusive and are usually combined as observations and laboratory data are always partial.
The work by \citet{draine84} was the first study to use these principles to derive the \hUV-to-mm dielectric functions of graphite and astronomical silicates.
We now have a better knowledge of the dispersion relations of several important materials:
\begin{inlinelist}
  \item silicates \citep[\eg][]{laor93,weingartner01,draine03,draine03b};
  \item amorphous carbon 
    \citep[\eg][]{rouleau91,zubko96,jones12a,jones12b,jones12c};
  \item graphite \citep[\eg][]{laor93,draine03,draine03b,draine16};
  \item \hPAH s \citep[\eg][]{li01,draine07}; and
  \item composite grains \citep[\eg][]{kohler14,kohler15}.
\end{inlinelist}
\reffig{fig:epsgrain} shows a few examples.

    \subsubsection{Computing Grain Cross-Sections}
    \label{sec:calcQabs}

The dielectric functions are intensive quantities characterizing the bulk optical properties of solids, but independent of their size and shape.
To compute usable absorption and scattering cross-sections, there is one last step to do.
Let's assume our grains are spheres of radius $a$.
\begin{itemize}
  \item
The extinction cross-section of the grain can be written:
\begin{equation}
  C_\sms{ext}(\lambda,a) = \underbrace{\pi a^2}_\sms{geometric cross-section}
                          \times
                          \underbrace{Q_\sms{ext}(\lambda,a)}_\sms{efficiency}.
  \label{eq:Qext}
\end{equation}
This expression is simply the geometric cross-section, $\pi a^2$, times an \expression{extinction efficiency}, $Q_\sms{ext}$, which is a dimensionless quantity.
We can express the scattering and absorption cross-section the same way:
\begin{eqnarray}
  C_\sms{sca}(\lambda,a) & = & \pi a^2 Q_\sms{sca}(\lambda,a) \\
  \label{eq:Qsca}  
  C_\sms{abs}(\lambda,a) & = & \pi a^2 Q_\sms{abs}(\lambda,a).
  \label{eq:Qabs}  
\end{eqnarray}
Out of these three efficiencies, only two are independent, as we have $Q_\sms{ext}=Q_\sms{sca}+Q_\sms{abs}$.
  \item
Another useful quantity that can be derived from the efficiencies is the \expression{albedo}, quantifying the fraction of the incident light that is scattered by the grain:
\begin{equation}
  \tilde\omega(\lambda,a) \equiv \frac{C_\sms{sca}(\lambda,a)}{C_\sms{ext}(\lambda,a)}
                    = \frac{Q_\sms{sca}(\lambda,a)}{Q_\sms{ext}(\lambda,a)}.
  \label{eq:albedo}
\end{equation}
  \item
Let's call $\theta$ the angle between the directions of the incident and scattered light.
The probability distribution of scattering angles is called the \expression{scattering phase function}:
\begin{equation}
  \Phi(\cos\theta,\lambda,a) \equiv \frac{1}{C_\sms{sca}(\lambda,a)}
    \frac{\dd \mathcal{C}_\sms{sca}(\cos\theta,\lambda,a)}{\dd\Omega},
  \label{eq:phasefunction}
\end{equation}
where $\mathcal{C}_\sms{sca}(\cos\theta,\lambda,a)$ is the \expression{differential scattering cross-section} (\ie\ the cross-section for scattering in a given direction), and $\dd\Omega=\dd\cos\theta\ddiff\phi$ is the solid angle element.
It is normalized over all directions, such that:
\begin{equation}
  \iint_\Omega\Phi(\cos\theta,\lambda,a)\ddiff\Omega
  =2\pi\int_{-1}^1\Phi(\cos\theta,\lambda,a)\ddiff\cos\theta=1.
\end{equation}
For isotropic scattering, we have $\Phi(\cos\theta,\lambda,a)=1/4\pi$.
The first moment of this distribution is called the \expression{asymmetry parameter}, defined as:
\begin{equation}
  g(\lambda,a) = \langle\cos\theta\rangle 
    = 2\pi\int_{-1}^1\Phi(\cos\theta,\lambda,a)\cos\theta\ddiff\cos\theta.
  \label{eq:costheta}
\end{equation}
This parameter is a direct product of Mie theory.
Forward and backward scattering correspond to $\langle\cos\theta\rangle\simeq1$ 
and $\langle\cos\theta\rangle\simeq-1$, respectively, whereas isotropic scatterers have $\langle\cos\theta\rangle\simeq0$.
There are approximate analytical phase functions.
The most famous is from \citet{henyey41}:
\begin{equation}
  \Phi(\cos\theta,\lambda,a) = \frac{1}{4\pi}
    \frac{1-g^2(\lambda,a)}{(1+g^2(\lambda,a)-2g(\lambda,a)\cos\theta)^{3/2}}.
  \label{eq:HG41}
\end{equation}
Other distributions have been proposed \citep[\eg][]{draine03}.
\end{itemize}
The treatment to compute $Q_\sms{sca}$, $Q_\sms{abs}$ and $g$ depends on the value of the \expression{size parameter}:
\begin{equation}
  x = \frac{2\pi a}{\lambda}.
\end{equation}
As long as we do not zoom in scales where the hypothesis of a continuous medium breaks down, that is scales of a few $\r{A}$ (\ie\ grains made of a few atoms, or hard X-ray photons), the estimation of the efficiencies of a grain only depends on $x$ and $m$.
The \expression{Mie theory} \citep[\cf\ \eg\ Chap.~4 of][]{bohren83}\footnote{See also B.~Draine's public code, \href{https://www.astro.princeton.edu/~draine/scattering.html}{bhmie.f}, implementing the algorithm in Appendix~A of \citet{bohren83}.} is the central tool to compute grain cross-sections.
It is a numerical method, exactly solving Maxwell's equations for the scattering of a plane electromagnetic wave by a homogeneous sphere of known refractive index, $m(\lambda)$.
Several regimes can be identified, depending on the value of the size parameter.
They are illustrated in \reffig{fig:mie_regimes}.
In \reffig{fig:cross_sections}, we show the actual cross-sections of silicate and graphite grains of different sizes.

\begin{figure}[htbp]
  \includegraphics[width=\textwidth]{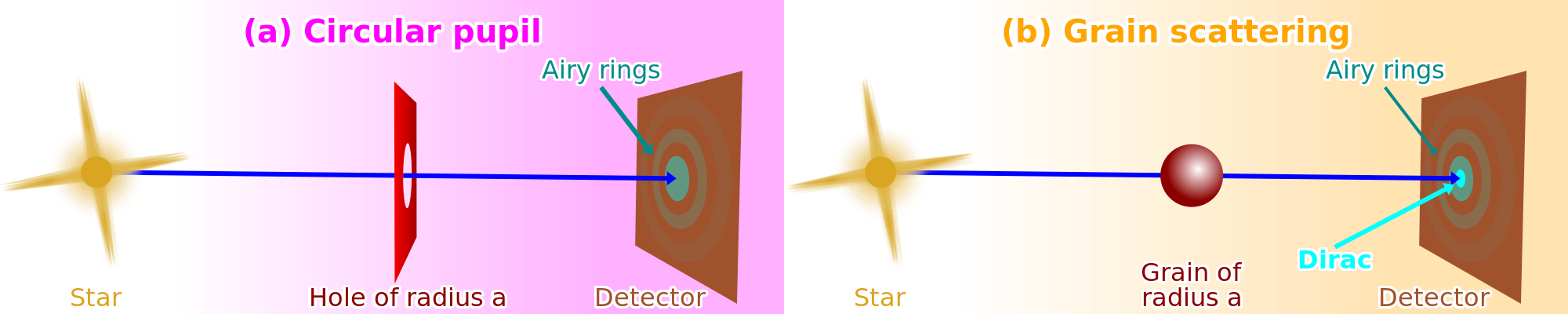}
  \newcap{Illustration of the extinction paradox}%
         {We have represented two situations.
          In panel~\textit{(a)}, we show the radiation from a distant star
          passing through a circular pupil of radius $a$, and projected onto
          the plane of a detector.
          We assume that:
          \begin{inlinelist}
            \item the star is very distant, so that the incident wave 
              can be considered planar;
            \item the detector is at a distant from the pupil much larger
              than $a$.
          \end{inlinelist}
          The observed pattern is a series of Airy rings.
          In panel~\textit{(b)}, we show the same stellar radiation 
          absorbed and scattered by a dust grain of the same radius $a$ as 
          the pupil, at the same distance from a detector.
          The observed pattern is made of the same Airy rings, but there is 
          a bright spot in the center.
          This property is a result of \expression{Babinet's theorem}.
          This theorem is based on the fact that the wave function of a 
          plane wave, $\psi(x,y)$, diffracted by a pupil of transmission 
          $T(x,y)$, is proportional to the Fourier transform of the pupil, 
          $\mathcal{F}(T)$.
          Here, the direction represented by the blue arrow is the $z$ axis.
          The $x$ and $y$ axes are the coordinates in the plane of the 
          pupil and of the detector, both being parallel.
          If we have two pupils, $T_{(a)}$ and $T_{(b)}$, that are 
          complementary, that is $T_{(a)}(x,y)+T_{(b)}(x,y)=1$, 
          $\forall(x,y)$, we have, by linearity of the Fourier transform 
          $\psi_{(a)}(x,y)+\psi_{(b)}(x,y)\propto
           \mathcal{F}(T_{(a)})+\mathcal{F}(T_{(b)})=\delta(x,y)$, 
          where $\delta$ is the Dirac distribution centered in the middle 
          of the screen.
          In our case, the two complementary pupils are the circular hole 
          in panel \textit{(a)}, and the grain, appearing as a circular 
          screen to the stellar radiation, in panel \textit{(b)}.
          The intensity is the squared module of the wave function, 
          $I(x,y)=|\psi(x,y)|^2$.
          We see that $\psi_{(b)}(x,y)=\psi_0\delta(x,y)-\psi_{(a)}(x,y)$, 
          thus $I_{(b)}(x,y)=I_\sms{spot}\delta(x,y)+I_{(a)}(x,y)$,
          confirming that the pattern in panel~\textit{(b)} is the pattern 
          in panel~\textit{(a)} plus a bright spot.
          In panel~\textit{(b)}, the bright spot is just the incident wave
          (rays parallel to the $z$ axis).
          We note that $I_{(a)}(x,y)=I_{(b)}(x,y)$, except in $(0,0)$.
          The incident light of intensity $I_0$ on the grain surface 
          ($\pi a^2$) is absorbed.
          Thus, the absorbed power is $P_\sms{abs}=\pi a^2I_0$.
          Now, the Airy pattern is the fraction diffracted by the grain 
          contour, that is the scattered light.
          Babinet's theorem tells us that it is identical in both cases
          we have represented.
          We clearly see in panel \textit{(a)} that the scattered power is
          also $P_\sms{sca}=\pi a^2 I_0$.
          The grain thus extinct a power $P_\sms{ext}=2\pi a^2I_0$.
          This is the extinction paradox:
          \begin{inlinelist}
            \item
              the flux incident on the grain surface is fully absorbed;
            \item
              the contour of the grain scatters the same power at 
              small angles, corresponding to the Airy rings in both panels.
          \end{inlinelist}
          \CClicence}
  \label{fig:baboulinet}
\end{figure}

\begin{figure}[htbp]
  \includegraphics[width=\textwidth]{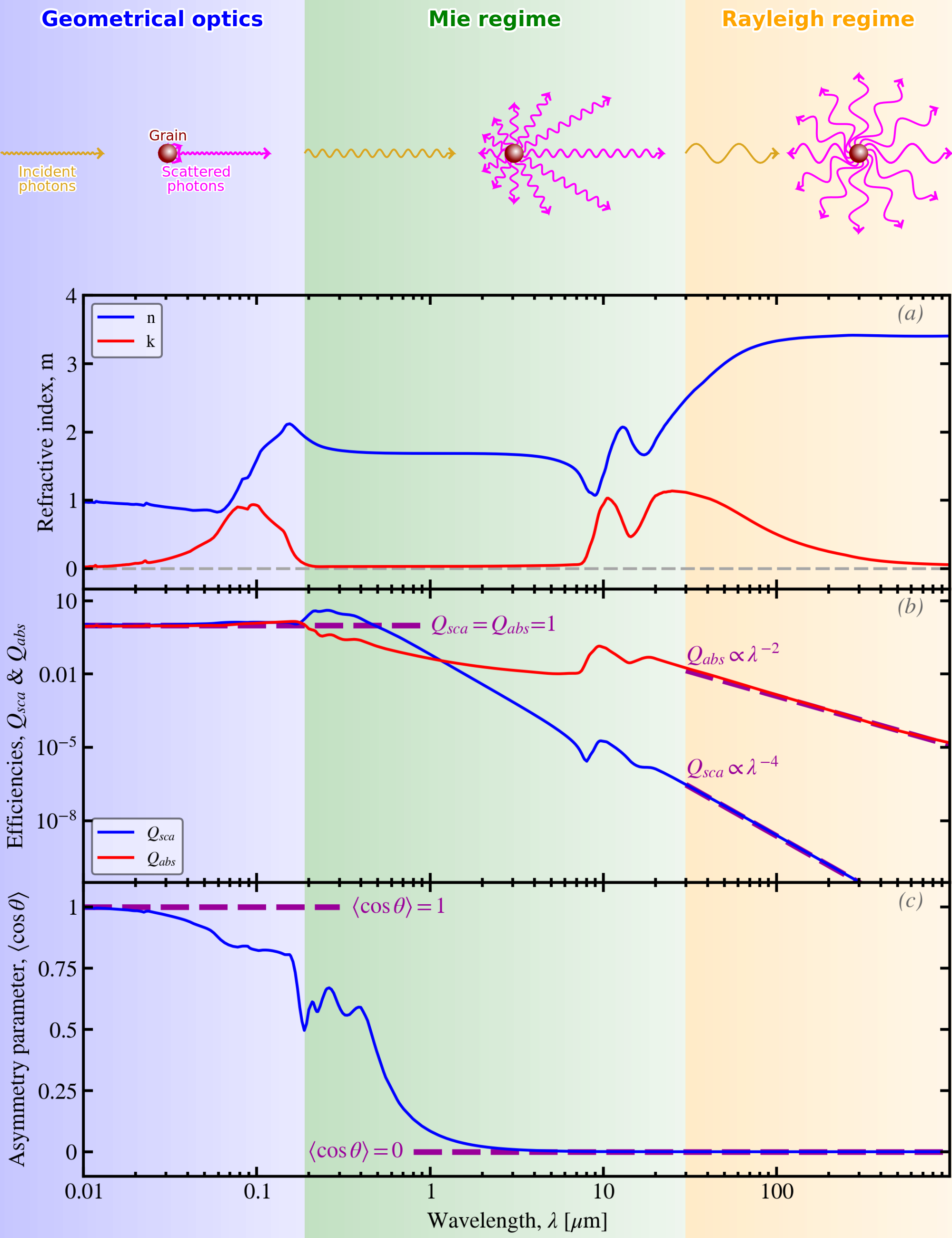}
  \newcap{Mie, Rayleigh and geometric optics regime}%
         {The top pictures represent the scattering pattern (in magenta) 
          of a spherical grain (red sphere), in the three regimes.
          The direction of the incident photon is shown in yellow.
          The scattering pattern has been calculated using the 
          \citet{henyey41} phase function.
          The length of each wiggly arrow is proportional to its scattering 
          probability.
          The three panels plot below show the optical properties of a 
          $a=0.1$~\tmic\ radius silicate, by \citet{draine03,draine03b}.
          \CClicence}
  \label{fig:mie_regimes}
\end{figure}

\begin{figure}[htbp]
  \includegraphics[width=\textwidth]{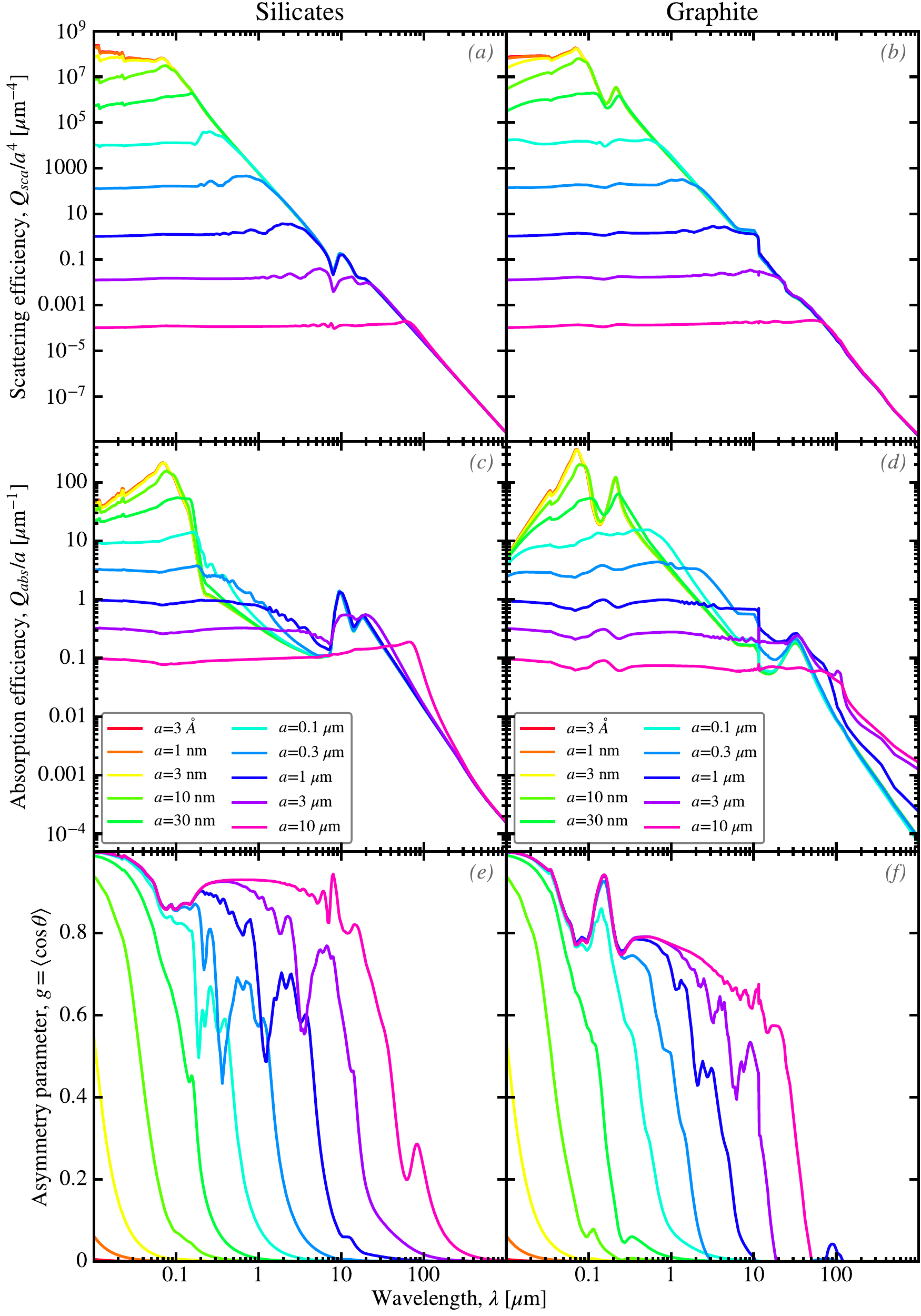}
  \newcap{Cross-sections of silicate and graphite grains}%
         {The color lines show the optical properties of 
          grains with different radius, $a$, from 
          \citet{draine03,draine03b}.
          \CClicence}
  \label{fig:cross_sections}
\end{figure}

\paragraph{Geometrical optics.}
It is the regime for which $x\gg1$ \citep[\cf\ \eg\ Chap.~7 of][]{bohren83}.
For interstellar grains, which have submicronic sizes, it corresponds to \hUV\ wavelengths and shorter.
This regime is more relevant to circumstellar dust, where grains can be significantly larger.
In geometrical optics, the undulatory nature of light is put aside.
Instead, light is modeled as rays, using the formalism of Fresnel.
Mie theory is valid in this regime, but numerical problems start arising.
\begin{itemize}
  \item
    The important feature of this regime is that, 
    $Q_\sms{sca}\simeq Q_\sms{abs}\simeq1$ (\cf\ 
    \refsubfig{fig:mie_regimes}{b}).
    This can also be seen in panels \textit{(a)} to \textit{(d)} of 
    \reffig{fig:cross_sections}.
    When the size of the grain increases, the range where both $Q_\sms{sca}$ 
    and $Q_\sms{abs}$ are flat extends to shorter wavelengths.
    The cross-sections are independent of wavelength, as the grains 
    are effectively behaving as opaque circular screens.
    It also means that the cross-section is proportional to the area of the 
    grain, but independent of its volume.
  \item
    $Q_\sms{sca}\simeq Q_\sms{abs}\simeq1$ implies that $Q_\sms{ext}\simeq2$, 
    meaning that the extinction cross-section is twice the geometric 
    cross-section.
    This counter-intuitive result is called the \expression{extinction 
    paradox} \citep[Chap.~4 of][for an extensive discussion]{bohren83}.
    It is a real diffraction effect, that is not actually predicted by the 
    methods of geometrical optics, for which $Q_\sms{ext}\simeq1$.
    Its resolution lies in accounting for the diffraction around the grain.
    It is illustrated in \reffig{fig:baboulinet}.
    The exact interpretation of this paradox is still debated nowadays 
    \citep[\eg][]{berg11}.
  \item
    Another feature of this regime is that grains are efficient forward 
    scatterers ($\langle\cos\theta\rangle\simeq1$; 
    \refsubfig{fig:mie_regimes}{c}).
\end{itemize}

\paragraph{The Mie regime.} 
It corresponds to grain sizes comparable to the wavelength of the incident light ($x\simeq1$).
Mie theory is valid outside this regime, but this is the regime where none of the other approximations are valid.
\begin{itemize}
  \item
    The optical properties have non-trivial features, depending on $x$ and 
    on the actual resonances of $m$.
    For instance, panels~\textit{(a)} and \textit{(c)} of 
    \reffig{fig:cross_sections} show the 9.7 and 18~\tmic\ features of 
    silicates.
    These features disappear only for radii $a\gtrsim10$~\tmic, as the 
    geometrical optics regime extends in the mid-\hIR\ (\hMIR; 
    \cf~\reftab{tab:spectralrange}), in this case.
  \item
    The scattering pattern is non-trivial, 
    $\langle\cos\theta\rangle$ varying between 0 and 1.
    It is schematically represented in the central cartoon at the top of 
    \reffig{fig:mie_regimes}, using the \refeq{eq:HG41}.
    It shows that backward scattering is likely, although forward 
    scattering is more probable.
  \item Finally, we notice small oscillations in the visible and near-\hIR\
    (\hNIR; \cf~\reftab{tab:spectralrange}) 
    ranges, mostly visible in the $Q_\sms{abs}$ of micronic size grains
    (panels \textit{(c)} and \textit{(d)} of \reffig{fig:cross_sections}).
    These patterns are called \expression{interference structure}.
    They are due to the interference between the incident and scattered
    waves.
    They are more prominent for weakly absorbent material, such as silicate,
    which is more transparent than graphite (\cf\ \refsec{sec:dustanalog}).
\end{itemize}

\paragraph{The Rayleigh regime.} 
It corresponds to grain sizes significantly smaller than the wavelength of the incident light, $x\ll1$ \citep[\cf\ \eg\ Chap.~5 of][]{bohren83}.
In the case of interstellar grains, it applies essentially to the \hNIR\ regime and longward.
The refraction index needs to be small, too: $|m|x\ll1$.
The solutions are analytic \citep[\eg][]{li08}:
\begin{equation}
  \left\{
  \begin{array}{rcl}
    C_\sms{sca}(\lambda,a) & = & \displaystyle\frac{128\pi^5}{3}
    \left|\frac{\epsilon_r(\lambda)-1}{\epsilon_r(\lambda)+2}\right|^2
    \frac{a^6}{\lambda^4}
    \\
    C_\sms{abs}(\lambda,a) & = & \displaystyle
      24\pi^2\frac{\epsilon_2(\lambda)}%
                 {(\epsilon_1(\lambda)+2)^2+\epsilon_2^2(\lambda)}
      \frac{a^3}{\lambda}. 
  \end{array}
  \right.
  \label{eq:Rayleigh}
\end{equation}
\begin{itemize}
  \item 
    \refeq{eq:Rayleigh} and panels \textit{(a)} to \textit{(d)} of 
    \reffig{fig:cross_sections} show that $Q_\sms{sca}/a^4$ and 
    $Q_\sms{abs}/a$ are independent of radius in this regime.
    This explains why small grains have a negligible albedo, whereas
    large grains are efficient scatterers.
  \item 
    The fact that $Q_\sms{abs}/a$ is independent of radius
    implies that the absorption cross-section is proportional 
    to the grain volume: $C_\sms{abs}\propto a^3$.
    The dust mass can thus be probed by measures in absorption or 
    in emission.
    An interpretation of this property is that, the wavelength being 
    comparable to the size of the grain, each bond can 
    interact with the electromagnetic field, independently of its location 
    within the solid.
    The interaction of the incident light with the grain is thus 
    proportional
    to the total number of oscillators, which is proportional to the mass.
  \item
    It explains why scattering appears negligible in this regime:
    $(C_\sms{sca}\propto a^6)\ll(C_\sms{abs}\propto a^3)$.
    Finally, we can see that scattering is isotropic
    ($\langle\cos\theta\rangle\simeq0$).
\end{itemize}
\takeaway{The absorption cross-section of most interstellar grains, in the 
          \hNIR-to-mm window, is proportional to their volume.
          The dust mass can thus be probed by absorption or emission 
          measures.}

    \subsubsection{Beyond Homogeneous Spheres}
    \label{sec:QabsnonMie}

Mie theory is restricted to homogeneous spheres.
However, there are several observational indications that this hypothesis is not fully accurate (\cf\ \refsec{sec:mantles}).
\begin{itemize}
  \item Grains are rapidly formed and destroyed in the \hISM, 
    implying that they likely are composites of several materials with 
    different dielectric functions, and voids.
  \item The polarization of starlight and of \hISM\ emission in the far-\hIR\ 
    (\hFIR; \cf\ \reftab{tab:spectralrange})
    indicates that at least some of the grains are elongated.
\end{itemize}
There are several methods to estimate cross-sections of grains beyond the hypothesis of homogeneous spheres.

\paragraph{Effective medium theory (EMT).} 
\hEMT\ is a class of methods to replace the individual dielectric functions of a composite material by an average, $\epsilon_\sms{av}(\lambda)$.
Cross-sections can then be estimated using Mie theory or any other approximation.
\hEMT\ assumes that the different domains are smaller than the wavelength and well-mixed in the grain.
There are different \expression{mixing rules} \citep[\cf\ \eg\ Chap.~8 of][]{bohren83}.
It seems that their accuracy depends on the type of sample they are applied to, as independent studies find better agreements with one or the other \citep[\eg][]{abeles76,perrin90}.
\begin{description}
  \item[Maxwell Garnett's rule] assumes that the medium is constituted of 
    a matrix with dielectric function $\epsilon_m(\lambda)$, and some
    inclusions.
    The inclusions can be of $N$ different types, with  dielectric 
    functions $\epsilon_i(\lambda)$ ($i=1\ldots N$), and volume filling 
    factors, $\phi_i$.
    It is implicit that $\phi_i\ll1$.
    \begin{equation}
       \epsilon_\sms{av}(\lambda) 
       = \frac{\epsilon_m(\lambda)+\sum_{i=1}^N\phi_ic_i\epsilon_i(\lambda)}%
              {1+\sum_{i=1}^N\phi_ic_i},
    \end{equation}
    with:
    \begin{equation}
      c_i=\frac{3\epsilon_m(\lambda)}%
               {\epsilon_i(\lambda)+2\epsilon_m(\lambda)}.
    \end{equation}
  \item[Bruggeman's rule] does not put a hierarchy between a predominant 
    matrix and a few inclusions.
    The dielectric function is the solution of:
    \begin{equation}
      \sum_{i=1}^N\phi_i\frac{\epsilon_i(\lambda)-\epsilon_\sms{av}(\lambda)}%
                           {\epsilon_i(\lambda)+2\epsilon_\sms{av}(\lambda)}
      =0.
    \end{equation}
\end{description}

\begin{figure}[htbp]
  \begin{tabular}{ccc}
    \includegraphics[width=0.31\textwidth]{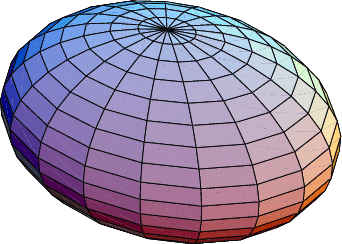} &
    \includegraphics[height=0.31\textwidth,angle=90]{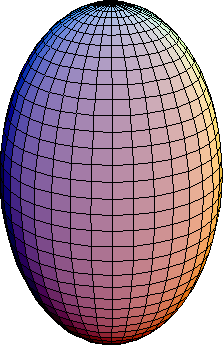}
    & \includegraphics[width=0.31\textwidth]{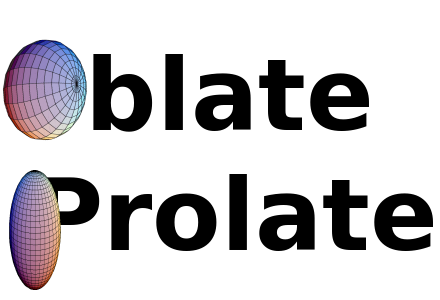}
    \\
    \textit{(a)} Oblate & \textit{(b)} Prolate 
    & \textit{(c)} Mnemotechnics\\
  \end{tabular}
  \newcap{Oblate and prolate spheroids}%
         {\uline{Credit:} volumes produced with 
          \href{https://mathworld.wolfram.com/Ellipsoid.html}{Mathematica}.}
  \label{fig:spheroids}
\end{figure}

\paragraph{Ellipsoids in the Rayleigh regime.}
They have general analytical solutions \citep[\eg\ Chap.~5 of][]{bohren83}.
When the three axes of the ellipsoid are aligned on the coordinates $x,y,z$ and if the the electric field is along $x$, we have:
\begin{equation}
  \left\{
  \begin{array}{rcl}
    C_\sms{sca}(\lambda,a) & = & \displaystyle
      \frac{128\pi^5}{27}
      \left|\frac{\epsilon_r(\lambda)-1}%
                 {1+(\epsilon_r(\lambda)-1)L_x}\right|^2
      \frac{a^6}{\lambda^4} \\
    C_\sms{abs}(\lambda,a) & = & \displaystyle
       \frac{8\pi^2}{3}\Im\left[\frac{\epsilon_r(\lambda)-1}%
                                       {1+(\epsilon_r(\lambda)-1)L_x}\right]
       \frac{a^3}{\lambda},
  \end{array}
  \right.
\end{equation}
where $L_x$ is the \expression{shape factor}. 
Noting $l_a>l_b$ the two lengths, oblate spheroids have dimensions along the three axes $(l_a,l_a,l_b)$, whereas prolate spheroids have dimensions $(l_a,l_b,l_b)$ (\cf\ \reffig{fig:spheroids}).
With these notations, the shape factor is \citep[\eg\ Chap.~22 of][]{draine11b}:
\begin{equation}
  \left\{
  \begin{array}{rcll}
    L_x & = & \displaystyle
     \frac{1+\xi^2}{\xi^2}\left[1-\frac{\arctan\xi}{\xi}\right]
     & \mbox{for oblate spheroids} \\
    L_x & = & \displaystyle
     \frac{1-\xi^2}{\xi^2}\left[\frac{1}{2\xi}
     \ln\left(\frac{1+\xi}{1-\xi}\right)-1\right]
     & \mbox{for prolate spheroids},
  \end{array}
  \right.
\end{equation}
with $\xi^2=|1-(l_b/l_a)^2|$.
In case of randomly oriented ellipsoids, we simply need to take the arithmetic mean of the cross-sections along the three axes.

\begin{figure}[htbp]
  \begin{tabular}{cc}
    \includegraphics[width=0.48\textwidth]{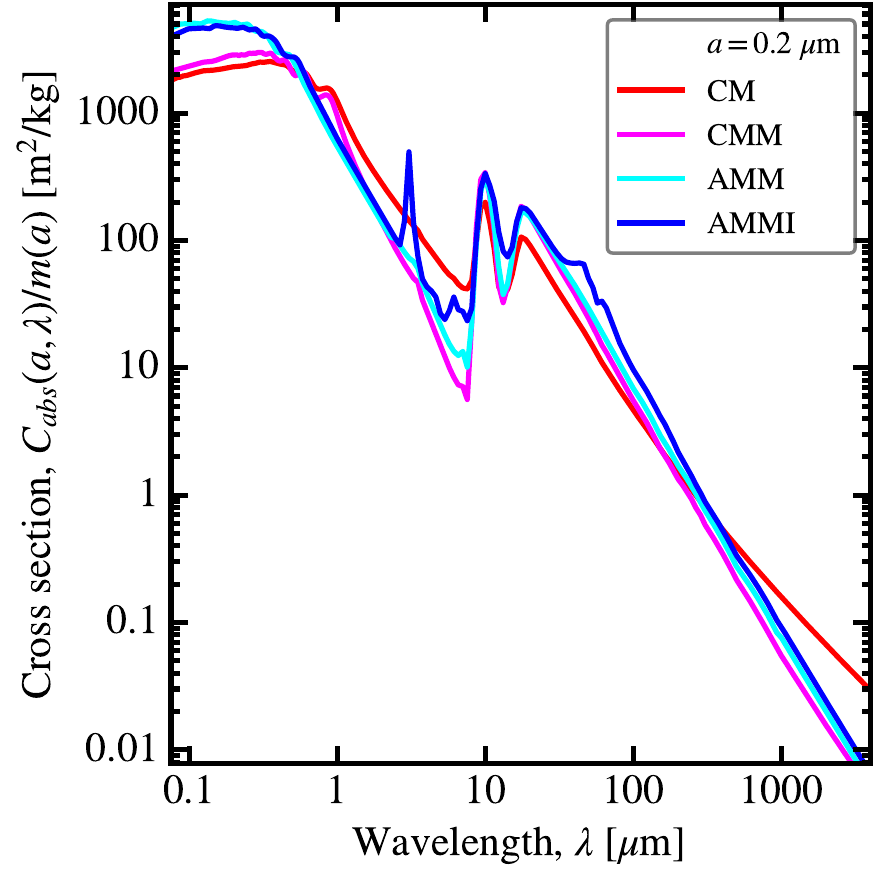} & 
    \includegraphics[width=0.48\textwidth]{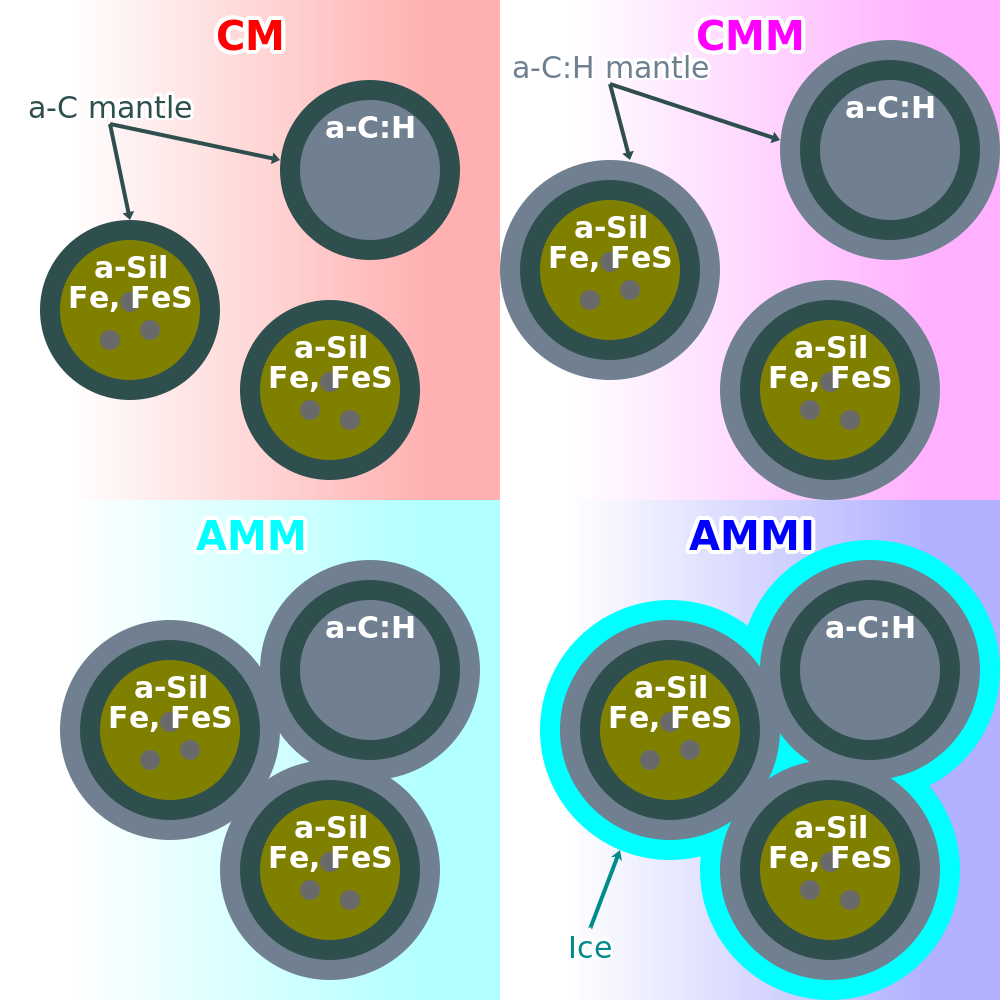} \\
  \end{tabular}
  \newcap{Absorption cross-sections of grain aggregates computed with DDA}%
         {The curves in the left panel are the absorption cross-sections of
          grains of radius $a=0.2$~\tmic:
          $C_\sms{abs}(a,\lambda)/m(a)=3/(4\rho)\times 
           Q_\sms{abs}(a,\lambda)/a$, where $\rho$ is the density.
          The four curves correspond to the four main mixtures of the 
          THEMIS model \citep{jones13,jones17}, represented in the right
          panel:
          \begin{inlinelist}
            \item the \expression{Core-Mantle} (CM) mixture is made
              amorphous Forsterite and Enstatite with Fe and FeS (troilite)
              inclusions, and aliphatic amorphous carbon (a-C:H);
              both grains are coated with an aromatic mantle (a-C);
            \item the \expression{Core-Mantle-Mantle} (CMM) is the CM
              mixture with additional coating by aliphatic material;
            \item the \expression{Aggregate-Mantle-Mantle} (AMM) mixture
              is constituted of aggregated CMM grains;
            \item the \expression{Aggregate-Mantle-Mantle-Ice} (AMMI)
              mixture is the AMM aggregates coated with water ice.
          \end{inlinelist}
          The CM optical properties are from \citet{jones13}.
          The CMM, AMM and AMMI optical properties have estimated by 
          \citet{kohler15} using \hDDA.
          Notice the apparition of the 3~\tmic\ water ice band 
          (\reffig{fig:MIRext}) in the AMMI model.
          \CClicence}
  \label{fig:aggregates}
\end{figure}

\paragraph{Discrete Dipole Approximation (DDA).}
\hDDA\ \citep[][]{purcell73}\footnote{See the public code \href{http://ddscat.wikidot.com/}{DDSCAT} by \citet{draine94}.} allows the user to model complex composite grains as arbitrary arrays of independent domains.
These domains are approximated by a series of discrete dipoles, which must be much smaller than the incoming wavelength.
This method is computer intensive, but very flexible \citep[\cf\ \eg\ the results of][]{kohler15,ysard18}.
We show some of the results of \citet{kohler15} in \reffig{fig:aggregates}.
This figure exhibits an important result we will discuss later:
\takeaway{the addition of mantles and the aggregation of grains tend to increase the absorptivity per unit mass in the \hFIR\ window.}

    \subsubsection{Polarization}
    \label{sec:intropola}

\begin{figure}[htbp]
  \includegraphics[width=\textwidth]{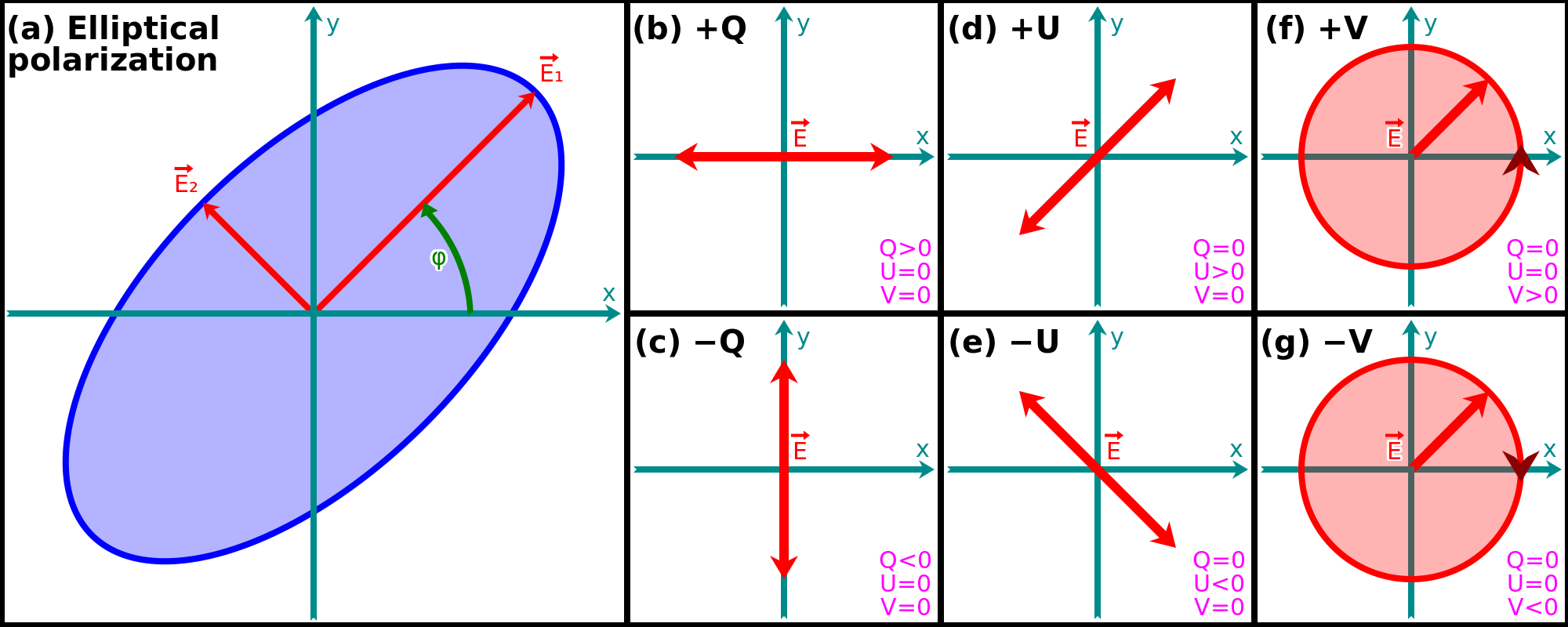}
  \newcap{Stokes parameters}%
         {Panel~\textit{(a)} displays the notations we have used to parametrize
          the elliptical polarization.
          Panels~\textit{(b)} to \textit{(g)} demonstrate the type of 
          polarization each parameter represents (linear or circular, in 
          different directions), when the rest is zero.
          \CClicence}
  \label{fig:stokes}
\end{figure}

The real part of the electric field of a monochromatic, plane electromagnetic wave, propagating along the $z$-axis, can be written at time $t$ and at $z=0$:
\begin{equation}
  \Re({\vect{E}}) = \vect{E_1}\cos\omega t + \vect{E_2}\sin\omega t,
\end{equation}
where $\vect{E_1}\perp\vect{E_2}$ \citep[\cf\ \eg\ Chap.~2 of][]{bohren83}.
This is the parametric equation of an ellipse in the $(x,y)$ plane
(\cf\ \refsubfig{fig:stokes}{a}).
It is the most general case, called \expression{elliptical polarization}.
It is fully characterized by the modules of $\vect{E_1}$ and $\vect{E_2}$, the angle $\varphi$ and the rotation direction $\eta=\pm1$ (-1: clockwise; +1:counterclockwise).
Particular cases are:
\begin{inlinelist}
  \item \expression{linear polarization}, if $|\vect{E_1}|=0$ or 
    $|\vect{E_2}|=0$; and
  \item \expression{circular polarization}, if $|\vect{E_1}|=|\vect{E_2}|$.
\end{inlinelist}

\paragraph{The Stokes parameters.} 
They constitute a four element vector easier to observationally measure than the ellipse parameters.
They are noted $(I,Q,U,V)=\vect{S}$.
$I$ is the total intensity of the beam, that can be partially polarized.
The three others can be expressed as:
\begin{equation}
  \left\{
  \begin{array}{rcl}
    Q & = & \left(|\vect{E_1}|^2 - |\vect{E_2}|^2\right)\cos(2\varphi) \\
    U & = & \left(|\vect{E_1}|^2 - |\vect{E_2}|^2\right)\sin(2\varphi) \\
    V & = & 2\eta|\vect{E_1}||\vect{E_2}|.
  \end{array}
  \right.
\end{equation}
The interpretation of these parameters is illustrated in \reffig{fig:stokes}.
They allow us to decompose the light into its linear and circular polarization components.
The intensity of the polarized light is $I_p=\sqrt{Q^2+U^2+V^2}\le I$.
It is also frequent to quote the \expression{linearly polarized intensity}:
\begin{equation}
  P \equiv \sqrt{Q^2+U^2},
  \label{eq:pol}
\end{equation}
and the \expression{linear polarization fraction}, $P/I$.

\begin{figure}[!htbp]
  \includegraphics[width=\textwidth]{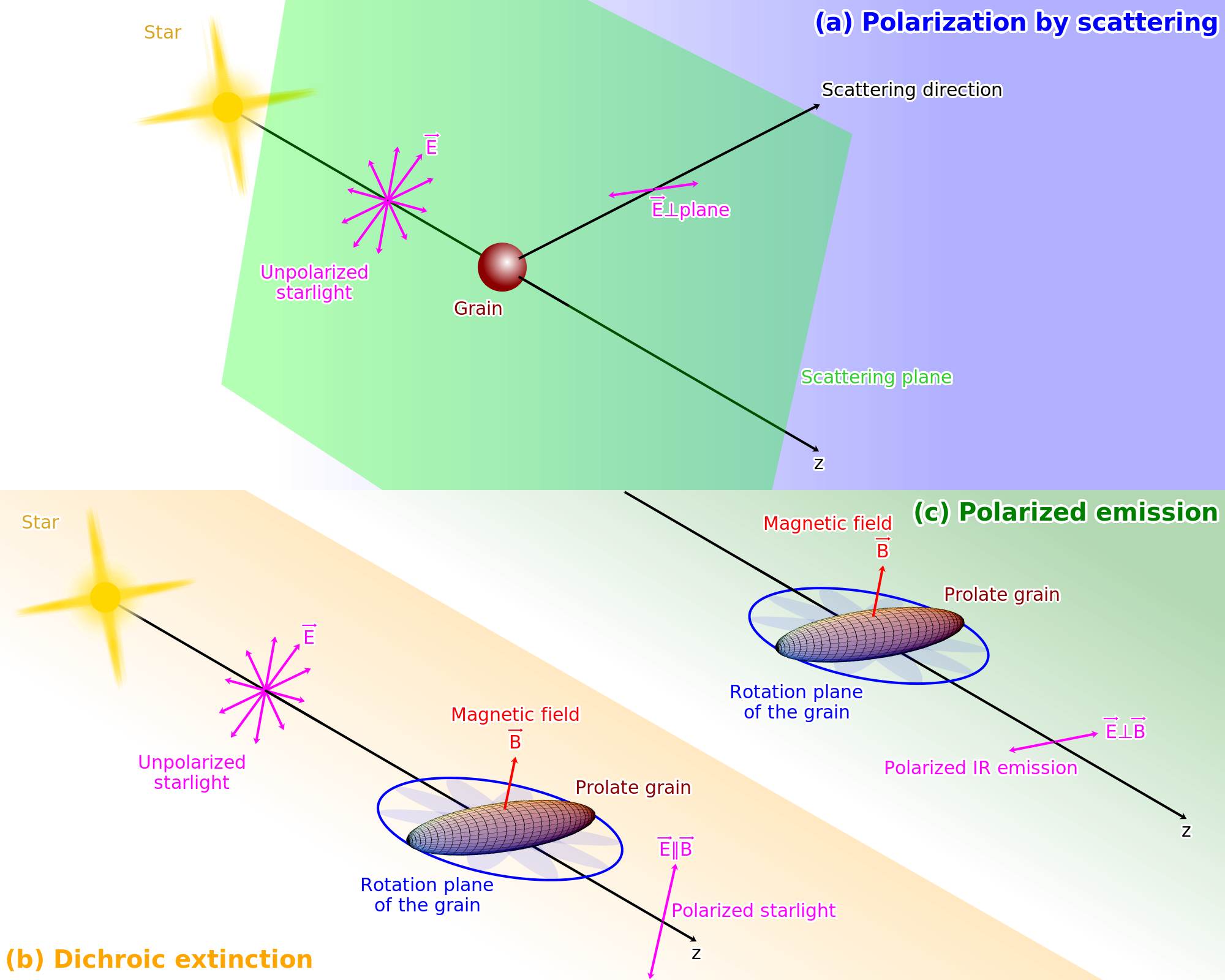}
  \newcap{The three types of grain-induced polarization of light}%
         {Panel~\textit{(a)} represents the unpolarized starlight radiation
          being scattered by a grain.
          The electric field of the scattered light oscillates predominantly 
          perpendicularly to the scattering plane.
          The larger the scattering angle is, the higher the linear polarization
          fraction is.
          Panels~\textit{(b)} and \textit{(c)} represent a single prolate grain 
          whose rotation axis is aligned with the magnetic field, $\vect{B}$.
          The case of oblate grains is more trivial as their rotation axis is 
          their symmetry axis.
          Panel~\textit{(b)} illustrates dichroic extinction of starlight in the
          visible/near-\hIR.
          The grain extinct preferentially the component of the electric field
          oscillating in its rotation plane.
          The extincted starlight is thus polarized parallel to $\vect{B}$.
          Panel~\textit{(c)} shows the \hIR\ emission of the same prolate grain 
          is polarized along its major axis, perpendicular to $\vect{B}$.
          In both panels \textit{(b)} and \textit{(c)}, we have represented the 
          optimal case, when the magnetic field is perpendicular to the 
          sightline.
          In the more general case, the polarization will be relative to the
          projection of $\vect{B}$ on the plane of the sky.
          \CClicence}
  \label{fig:polarization}
\end{figure}

\paragraph{Grain alignment.}
Asymmetric grains tend to be aligned with the magnetic field.
If we consider the simple case of spheroidal grains (\cf\ \reffig{fig:spheroids}): 
\begin{inlinelist}
  \item the rotation axis of oblate grains is along their symmetry axis;
  \item the rotation axis of prolate grains is perpendicular to their long
        axis, their cross-section thus needs to be integrated over their 
        spinning dynamics \citep[\eg][]{guillet18}.
\end{inlinelist}
The rotation axis of the grains tends to align with the magnetic field, $\vect{B}$.
This is represented in panels~\textit{(b)} and \textit{(c)} of \reffig{fig:polarization}.
Several mechanisms have been proposed to explain this alignment \citep[{\cf}][for a review]{andersson15}.
Nowadays, \expression{Radiative Alignment Torques} \citep[\hRAT;][]{dolginov76} are favored, because they provide the best account of the observational constraints.
This complex scenario is based on the fact that irregular grains have different cross-sections for clockwise and counterclockwise circularly polarized light.
Light scattering on such grains therefore provides a torque that increases the angular momentum of the grains.
If these grains have paramagnetic inclusions (such as iron), the grain rotation precesses and aligns with $\vect{B}$.
Although the alignment is caused by the radiation field, it is independent of its direction.
This mechanism becomes inefficient at high optical depth, which is consistent with observations.

\paragraph{Polarization by scattering.} 
It is represented on \refsubfig{fig:polarization}{a}.
When an incident beam is scattered by a grain (this grain does not have to be asymmetric), the electric field component in the scattering plane is diminished, inducing a polarization perpendicular to the plane.
The larger the scattering angle is, the larger the polarization.
The polarization of an incident Stokes vector, $\vect{S_i}$, resulting in a scattered beam, $\vect{S_s}$, is described as: $\vect{S_s}=\mat{M}\vect{S_i}$, where $\mat{M}$ is the $4\times4$ \expression{Müller matrix} \citep[\cf\ Chap.~3 of][for different examples of Müller matrices]{bohren83}.
This polarization process is not related to grain alignment with the magnetic field, but it depends on the distribution of stars and dust clouds \citep[\eg][]{wood97a}.

\paragraph{Dichroic extinction.} 
It is the selective extinction of the electric field oscillating along the major axis of an elongated grains.
It is represented in \refsubfig{fig:polarization}{b}.
Since grains in the diffuse \hISM\ tend to align their rotation axis with the magnetic field, they polarize starlight parallel to $\vect{B}$.
For this reason, the polarization of starlight has historically been used to map the magnetic field of the \hMW\ \citep{mathewson70}.

\paragraph{Polarized emission.} 
The polarization of the emission by elongated grains has been predicted by \citet{stein66}.
Such grains emit \hIR\ light preferentially polarized along the direction of their major axis.
Since their major axis is perpendicular to the magnetic field, their \hIR\ emission is perpendicular to $\vect{B}$.
It is represented in \refsubfig{fig:polarization}{c}.
The \hFIR\ polarized emission has been extensively used to map the magnetic field in the \hMW, with the \hplanck\ satellite \citep[\eg][]{planck-collaboration16f}.
\takeaway{Grain polarization is parallel to the magnetic field in the visible and perpendicular in the \hIR: $\vect{P_\sms{vis}}\parallel\vect{B}$ and $\vect{P_\sms{IR}}\perp\vect{B}$.}

  \subsection{Heat Capacities}
  \label{sec:heatcap}

In \refsec{sec:HObond}, presenting the harmonic oscillations of valence electrons, we argued that the energy dissipation was due to collisions with the lattice.
The energy absorbed by the grain is thus redistributed throughout the lattice and stored in the harmonic oscillations of its atoms.

    \subsubsection{Distribution of Harmonic Oscillators}

The energy levels of a one-dimensional quantum harmonic oscillator, of natural frequency $\nu$, are \citep[\eg\ Chap.~2 of][]{atkins05}:
\begin{equation}
  E_n = h\nu\left(n+\frac{1}{2}\right) \;\;\;\mbox{ with }\;\;\; n=0,1,\ldots,
  \label{eq:quantHO}
\end{equation}
where $h$ is the \expression{Planck constant} (\cf\ \reftab{tab:constants}).
At thermal equilibrium, the probability distribution of a large ensemble of such harmonic oscillators, at temperature $T$, is the \expression{Boltzmann distribution}:
\begin{equation}
  p_n(T) = \frac{\exp\left(\displaystyle\frac{-E_n}{kT}\right)}{Z(T)},
  \label{eq:boltzHO}
\end{equation}
where we have introduced the \expression{partition function}:
\begin{equation}
  Z(T) = \sum_{n=0}^\infty\exp\left(-\frac{E_n}{kT}\right)
      = \frac{\displaystyle\exp\left(-\frac{h\nu}{kT}\right)}%
             {\displaystyle1-\exp\left(-\frac{h\nu}{kT}\right)}.
\end{equation}
The second equality comes from injecting \refeq{eq:quantHO} into the first equality.
The mean energy of this ensemble of oscillators is the first moment of the distribution in \refeq{eq:boltzHO}:
\begin{equation}
  \langle E\rangle = \frac{h\nu}{2}
     + \frac{h\nu}{\displaystyle\exp\left(\frac{h\nu}{kT}\right)-1}.
  \label{eq:Eeinstein}
\end{equation}
Since each one of the $N$ atoms of the lattice has three degrees of freedom (along the three Cartesian axes $x,y,z$), the number of oscillators is $3N$\footnote{This is for $N\gg1$. The actual number of degrees of freedom is 3N-6, subtracting the three translatory and three rotational possible motions of the grain as a whole.}.
The \expression{internal energy} of the grain, which is the energy stored in all its oscillators, is thus:
\begin{equation}
  U(T) = 3N\langle E\rangle = 3Nh\nu\left(\frac{1}{2}
     + \frac{1}{\displaystyle\exp\left(\frac{h\nu}{kT}\right)-1}\right).
  \label{eq:Ueinstein}
\end{equation}

    \subsubsection{Debye's Model}
    \label{sec:debye}

\begin{figure}[htbp]
  \includegraphics[width=\textwidth]{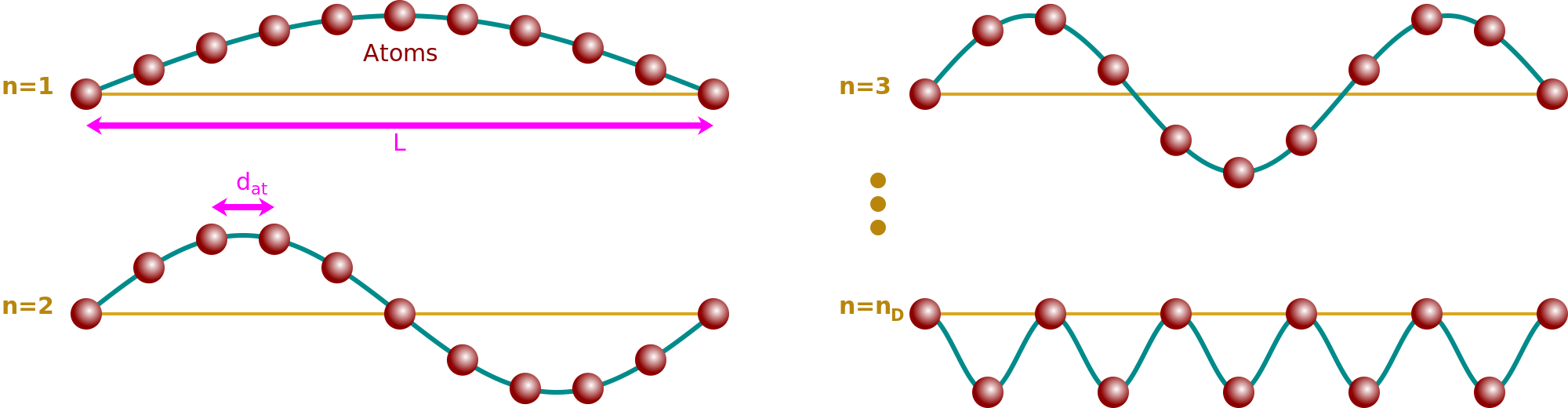}
  \newcap{Phonon modes}%
         {We represent the simplest case of a string of atoms (red spheres).
          The total length of the solid is materialized by the yellow 
          horizontal line.
          The two atoms at each end of this line are fixed.
          The modes are thus quantified.
          The shortest possible wavelength is $2d_\sms{at}$, corresponding to the
          $n=n_\sms{D}$ mode.
          \CClicence}
  \label{fig:phonons}
\end{figure}

\begin{figure}[htbp]
  \includegraphics[width=\textwidth]{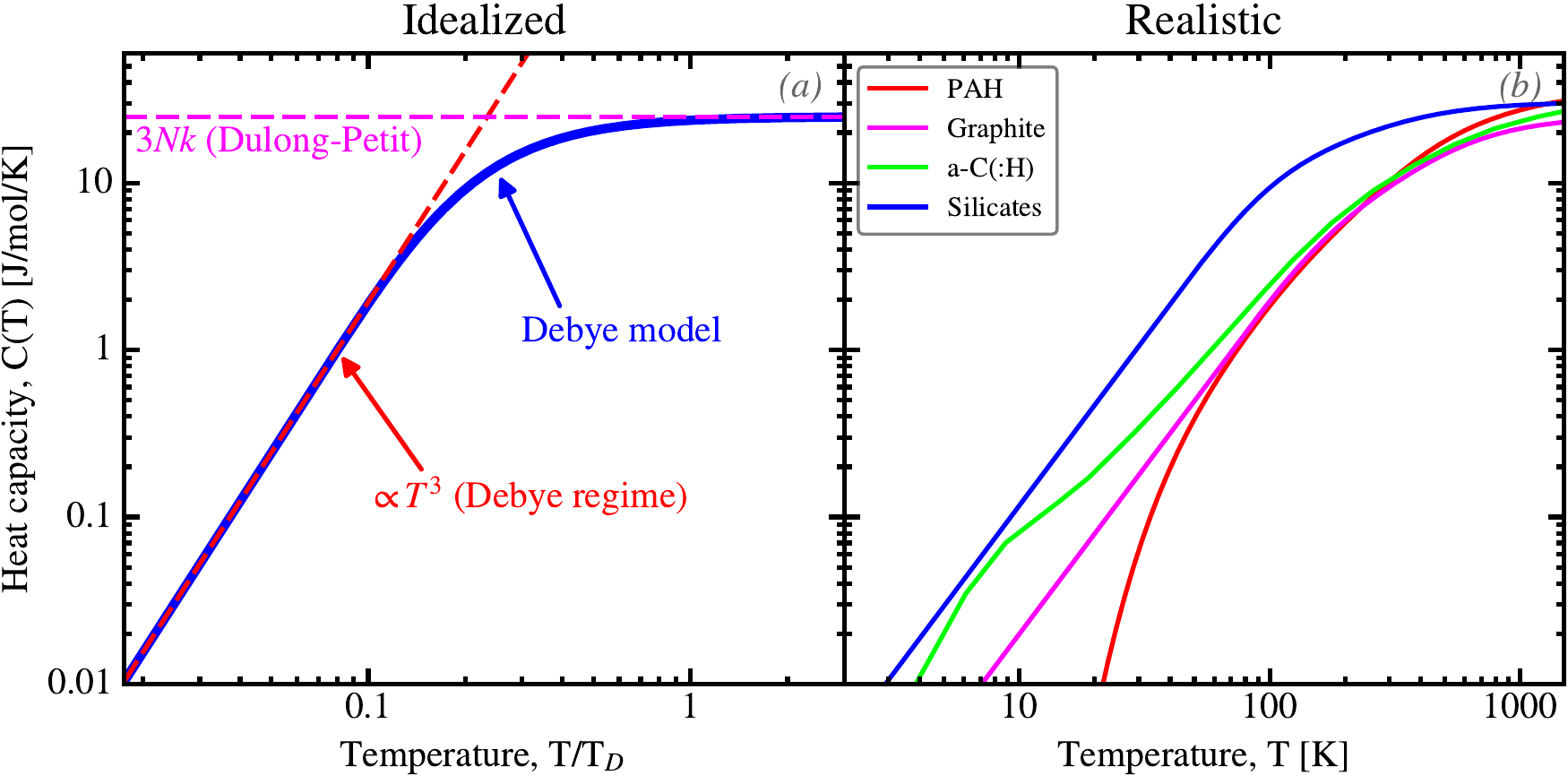}
  \newcap{Heat capacities.}%
         {Panel~\textit{(a)} shows the Debye model (\refeq{eq:Cdebye}; blue),
         with the two limiting regimes:
         \begin{inlinelist}
           \item the classical Dulong-Petit regime (magenta dashed line); and
           \item the quantum Debye regime (red dashed line).
         \end{inlinelist}
         Panel~\textit{(b)} shows the heat capacities of realistic materials:
         \hPAH, graphite and silicate from \citet{draine01} and \hHAC\ from 
         \citet{jones13}.
         \CClicence}
  \label{fig:heatcapacities}
\end{figure}

\refeq{eq:Ueinstein} considers that each atom oscillates independently of its neighbors. 
In reality, there are collective vibrational modes in a crystal lattice.
These modes actually are sound waves, propagating at the sound speed of the material, $c_s$.
Because the size of a grain is finite, the number of these possible modes is quantified.
Indeed, if $L$ is the size of the grain along one dimension, the wavelength of the modes along this dimension is $\lambda_n=2L/n$, with $n=1,2,\ldots$
(\cf\ \reffig{fig:phonons}).
The shortest possible wavelength corresponds to oscillations of adjacent atoms in opposition of phase: $\lambda_\sms{D}=2d_\sms{at}$, where $d_\sms{at}$ is the interatomic distance.
These quantified modes can be treated as quasi-particles, called \expression{phonons}.
These phonons have energies $hc_s/\lambda_n$, thus:
\begin{equation}
  E_n = \frac{nhc_s}{2L} \;\;\;\mbox{ for }\;\;\; n=1,\ldots,n_\sms{D}.
\end{equation}
We now need to integrate \refeq{eq:Eeinstein} over the different modes:
\begin{equation}
  U(T) = U_0 + \int_0^{\nu_\sms{D}}\frac{h\nu}%
        {\displaystyle\exp\left(\frac{h\nu}{kT}\right)-1}g(\nu)\ddiff\nu,
  \label{eq:Udebye0}
\end{equation}
where $g(\nu)$ is the density of modes with frequency $\nu$, and $U_0$ is a constant coming from the 1/2 term in \refeq{eq:Eeinstein}.
It can be shown that \citep[\cf\ Chap.~III.E of][]{diu97}:
\begin{equation}
  g(\nu) = \frac{9N\nu^2}{\nu_\sms{D}^3},
\end{equation}
where the Debye frequency can be explicited as a function of the density of atoms, $n_\sms{at}$:
\begin{equation}
  \nu_\sms{D} = c_s\sqrt[3]{\frac{9n_\sms{at}}{4\pi}}.
\end{equation}
From $\nu_\sms{D}$, we can also define the \expression{Debye temperature}, $T_\sms{D}\equiv h\nu_\sms{D}/k$.
\refeq{eq:Udebye0} thus becomes:
\begin{equation}
  U(T) = U_0 + \frac{9N}{\nu_\sms{D}^3}\int_0^{\nu_\sms{D}}
    \frac{h\nu^3}{\displaystyle\exp\left(\frac{h\nu}{kT}\right)-1}\ddiff\nu,
\end{equation}
and the \expression{Debye heat capacity} can be derived:
\begin{equation}
  C(T) = \frac{\partial U}{\partial T} 
       = 9kN\left(\frac{T}{T_\sms{D}}\right)^3\int_0^{T_\sms{D}/T}
         \frac{x^4e^x}{\left(e^x-1\right)^2}\ddiff x.
  \label{eq:Cdebye}
\end{equation}
It is represented in \refsubfig{fig:heatcapacities}{a}.
\begin{description}
  \item[The Dulong-Petit regime]
    is the limiting behavior of \refeq{eq:Cdebye} at high temperature, namely:
    \begin{equation}
      C(T) \simeq 3Nk \;\;\;\mbox{ for }\;\;\; T\gg T_\sms{D}.
    \end{equation}
    It is the classical expression of heat capacity.
    It is constant becomes it assumes that energy can be indifferently stored 
    in oscillators, ignoring their limited number.
  \item[The Debye regime] 
    is the low-temperature limit of \refeq{eq:Cdebye}, namely:
    \begin{equation}
      C(T) \simeq \frac{12\pi^4}{5}Nk\left(\frac{T}{T_\sms{D}}\right)^3
       \;\;\;\mbox{ for }\;\;\; T\ll T_\sms{D}.
      \label{eq:debyeprox}
    \end{equation}
    It accounts for the quantification of the modes.
    It provides a correct agreement with laboratory measurements.
\end{description}

    \subsubsection{Heat Capacities of Realistic Materials}

The Debye model is an idealization providing a good approximation.
It has however several limitations.
\begin{description}
  \item[Conduction electrons] contribute to the heat capacity of metals, and 
    dominate at low temperatures.
    Their contribution to the heat capacity is 
    \citep[\cf\ \eg\ Chap.~2 of][]{ashcroft76}:
    \begin{equation}
      C_\sms{cond}(T) = \frac{\pi^2Nk}{2}\frac{T}{T_\sms{F}},
    \end{equation}
    where $T_{F}=E_{F}/k$ is the Fermi temperature \refeqp{eq:FermiDirac}.
  \item[Laboratory data] can be used to determine the Debye and Fermi 
    temperatures of the compound.
    If the structure of the grain is too complex, the heat capacity can be
    fitted on experimental measurements \citep[\eg][]{draine01}.
    We show the heat capacity of various interstellar grain candidates in
    \refsubfig{fig:heatcapacities}{b}.
\end{description}

  \subsection{Heating and Cooling}
  \label{sec:heatcool}

    \subsubsection{Kirchhoff's Law}
    \label{sec:kirchhoff}

Let's consider a grain at thermal equilibrium with a radiation source, such as the light from a star.
The \expression{specific intensity} received by the grain, $I_\nu(\lambda,\Omega)$, is the electromagnetic power per unit frequency, area (A) and solid angle ($\Omega$)\footnote{Throughout this manuscript, we use the subscript $\nu$ to exclusively denote \expression{spectral densities}, that is quantities per unit frequency, $f_\nu$.
Such a quantity can also be expressed per unit wavelength: $f_\lambda=f_\nu\ddiff\nu/\dd\lambda=f_\nu c/\lambda^2$.
Quantities depending on the frequency, but not per unit frequency, should not be written with subscript $\nu$: \dout{$\kappa_\nu$}$\longrightarrow\kappa(\nu)$.}:
$\dd E_\star=I_\nu\ddiff t\ddiff\nu\ddiff A\ddiff\Omega$ (we discuss this quantity in more details in \refsec{sec:RT}).
The \expression{absorption coefficient} of this grain, $\alpha(\lambda)$, is the fraction of this specific intensity it absorbs per unit length, $l$:
$\dd I_\nu=-\alpha I_\nu\ddiff l$.
The \expression{emission coefficient} of this grain, $j_\nu(\lambda)$, is the power it emits per unit frequency, volume (V) and solid angle (it is isotropic):
$\dd E_\sms{em}=j_\nu\ddiff t\ddiff\nu\ddiff V\ddiff\Omega$.
\citet{kirchhoff60}'s law states that the ratio $j_\nu(\lambda)/\alpha(\lambda)=f_\nu(T,\lambda)$ is a universal function depending only on $T$ and $\lambda$ \citep[\eg][for a review]{robitaille09}.
\citet{planck1900} later gave an analytical expression of this empirical function, assuming the energy levels were discrete, providing a quantum formulation of the \expression{black body} radiation.
It became the \expression{Planck function}, $f_\nu(T,\lambda)=B_\nu(T,\lambda)$, where:
\begin{equation}
  B_\nu(T,\lambda) = \frac{2hc}{\lambda^3}
    \frac{1}{\displaystyle\exp\left(\frac{hc}{\lambda kT}\right)-1}.
  \label{eq:BB}
\end{equation}

\paragraph{Opacity.} 
The \expression{mass absorption coefficient} of a grain, $\kappa_\sms{abs}(\lambda)$, is its absorption cross-section per unit mass: $\kappa_\sms{abs}(\lambda)=\alpha(\lambda)/\rho$.
For a single spherical grain, it is, using \refeq{eq:Qabs}:
\begin{equation}
  \kappa_\sms{abs}(a,\lambda) = 
     \frac{\overbrace{C_\sms{abs}(a,\lambda)}^\sms{cross-section}}%
          {\underbrace{\displaystyle\frac{4\pi}{3}a^3\rho}_\sms{mass}}
          = \frac{3}{4\rho}\frac{Q_\sms{abs}(a,\lambda)}{a}.
  \label{eq:kappa}
\end{equation}
This quantity is often referred to as the \expression{opacity}.
In this manuscript, we however extend this term to its scattering component, too.
We will therefore call $\kappa\equiv\kappa_\sms{ext}=\kappa_\sms{sca}+\kappa_\sms{abs}$, the \textit{opacity}, $\kappa_\sms{abs}$ and $\kappa_\sms{sca}$ being called the \expression{absorption} and \expression{scattering opacities}, respectively.
We have seen in \refsec{sec:Qabs} that $Q_\sms{abs}/a$ is practically independent of radius for most interstellar grains in the \hNIR\ regime and longward, thus:
\takeaway{the \hNIR-to-mm opacity of interstellar grains having the same homogeneous composition is independent of their radius.}

\begin{figure}[htbp]
  \includegraphics[width=\textwidth]{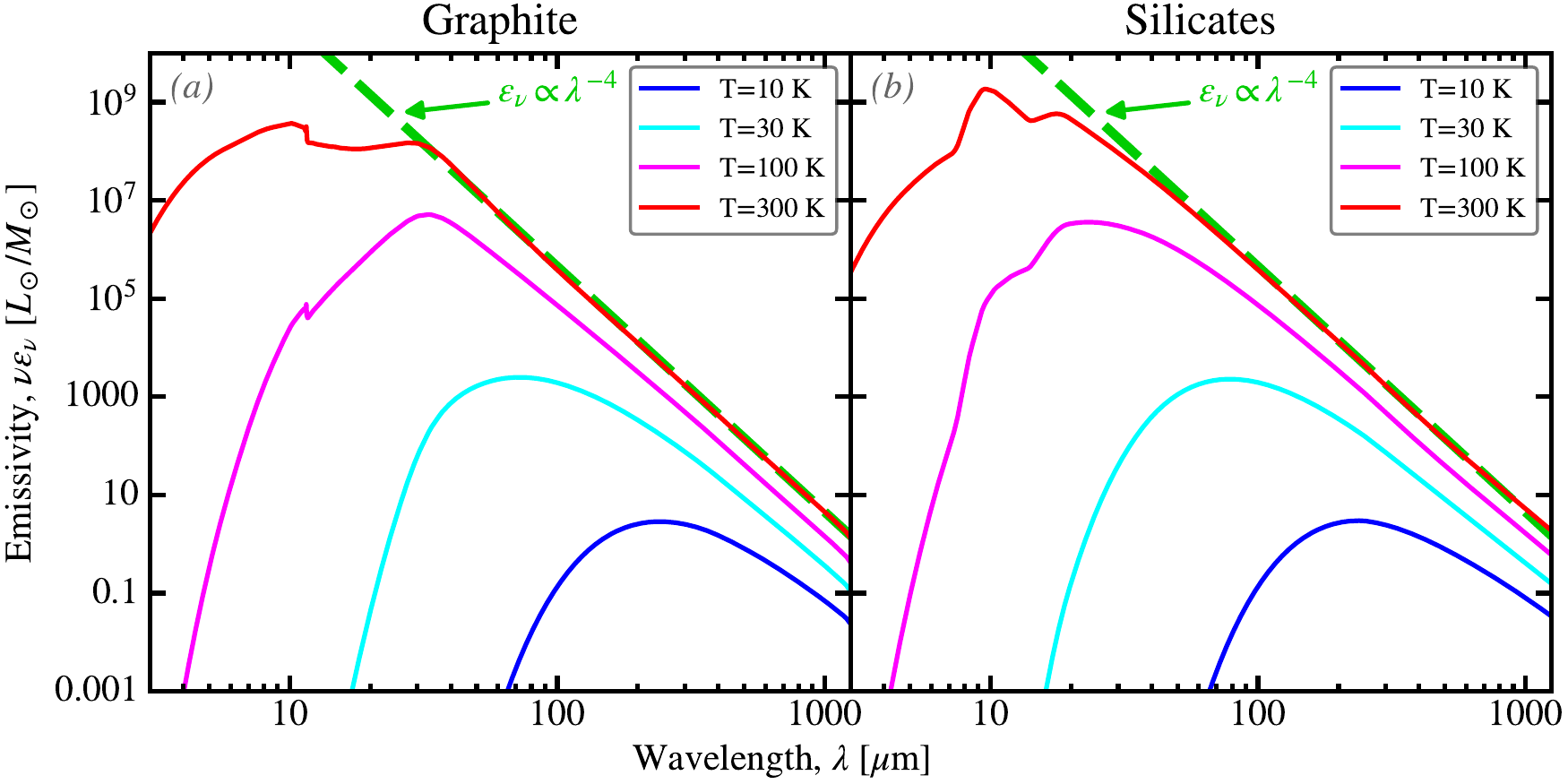}
  \newcap{Emissivity of grains at thermal equilibrium with the radiation field}%
         {We show the emission spectrum of spherical graphite \textit{(a)} and 
          silicates \textit{(b)} from \citet{draine03,draine03b}, at different 
          equilibrium temperatures \refeqp{eq:BBQ}.
          We have overlaid in green the Rayleigh-Jeans approximation 
          ($\beta=2$ for both compounds).
          \CClicence}
  \label{fig:BBQ}
\end{figure}

\paragraph{Emissivity.} 
The \expression{emissivity} of a grain, $\epsilon_\nu(\lambda)$, is the power it emits per unit frequency and mass ($\dd m=\rho\dd V$): $\dd E_\sms{em}=\epsilon_\nu(\lambda)\rho\ddiff t\ddiff\nu\ddiff V\ddiff\Omega/4\pi$.
The last differential element simply denotes an average over solid angles.
We thus see that $\epsilon_\nu=4\pi j_\nu/\rho$.
Kirchhoff's law then becomes:
\begin{equation}
  \epsilon_\nu(\lambda)=4\pi\kappa_\sms{abs}(\lambda)\times B_\nu(T,\lambda).
  \label{eq:BBQ}
\end{equation}
\refeq{eq:BBQ} is the emission spectrum of grains at thermal equilibrium with the radiation field.
We show a few examples in \reffig{fig:BBQ}.
The limiting behavior of \refeq{eq:BBQ} when $\lambda\gg hc/kT$ is given by the \expression{Rayleigh-Jeans law}:
\begin{equation}
  \epsilon_\nu(\lambda)\simeq8\pi kT\frac{\kappa_\sms{abs}(\lambda)}{\lambda^2}.
  \label{eq:RJBBQ}
\end{equation}

\paragraph{Modified Black Body (MBB).} 
A \hMBB\ is an idealized body, at thermal equilibrium with the radiation field, that does not perfectly absorb all frequencies.
In other words, it has a non zero albedo.
It is an imperfect black body, sometimes also called \expression{grey body}.
\refeq{eq:BBQ} is a \hMBB.
In the \hISM\ literature, \hMBB\ usually refers to the case where we approximate the opacity as a power-law:
\begin{equation}
  \kappa_\sms{abs}(\lambda) \simeq 
    \kappa_0\left(\frac{\lambda_0}{\lambda}\right)^\beta,
  \label{eq:kappaMBB}
\end{equation}
where $\lambda_0$ is a reference wavelength, and $\kappa_0$, the opacity at $\lambda_0$.
This approximation was popularized by \citet{hildebrand83}.
We have seen in \refsec{sec:HObond} that $\beta\simeq2$ for typical grains, and that we must have $\beta>1$ in order to satisfy the Kramers-Kronig relations (\cf\ \refsec{sec:KK}).
\refeq{eq:RJBBQ} implies that $\epsilon_\nu(\lambda)\propto\lambda^{2+\beta}$, in the Rayleigh-Jeans regime.
We have shown this relation in \reffig{fig:BBQ}.
\begin{description}
  \item[The net flux] radiated by a black body is:
  \begin{equation}
    F_\sms{BB}(T) = \int_0^\infty\pi B_\nu(T,\nu)\ddiff\nu=\sigma T^4,
  \end{equation}
  where $\sigma$ is the \expression{Stefan-Boltzmann constant} 
  (\cf\ \reftab{tab:constants}).
  In the case of a \hMBB, the emitted power per unit mass is:
  \begin{equation}
    \frac{P_\sms{MBB}(\beta,T)}{M_\sms{MBB}} 
     = \int_0^\infty4\pi\kappa_0\left(\frac{\lambda_0}{\lambda}\right)^\beta 
       B_\lambda(T,\lambda)\ddiff\lambda
     = \frac{8\pi\kappa_0\lambda_0^\beta(kT)^{4+\beta}}{c^{2+\beta}h^{3+\beta}}
       \Gamma(4+\beta)\zeta(4+\beta),
  \end{equation}
  where $\Gamma$ is the \expression{gamma function}, and $\zeta$ is the
  \expression{Riemann zeta function}.
  \item[Wien's law] states that the emission peak of $B_\nu(T,\lambda)$ is
    located at $\lambda_\sms{max}(T)=5.0996\E{3}/T\emic$.
    For a \hMBB, the wavelength peak of 
    $\nu\epsilon_\nu(\beta,T,\lambda)$\footnote{In general, we prefer 
    displaying $\nu f_\nu$ quantities than simply $f_\nu$ or $f_\lambda$,
    as it represents better the energy balance.} is located at:
    \begin{equation}
      \lambda_\sms{max}(\beta,T) = \frac{hc}{kT}
    \frac{1}{(4+\beta)+W\left(-(4+\beta)\exp\left[-(4+\beta)\right]\right)},
    \end{equation}
    where $W$ is the \expression{Lambert W function}.
\end{description}

    \subsubsection{Equilibrium Heating}
    \label{sec:Teq}

\refeq{eq:BBQ} expresses the general emissivity of a grain at thermal equilibrium with the radiation field. 
To use this formula, we need to determine the equilibrium or \expression{steady-state} temperature of the grain.
This is simply performed by equating the absorbed and emitted powers.
It is convenient to define the \expression{mean intensity} of the \hISRF:
\begin{equation}
  J_\nu(\lambda) 
    = \frac{1}{4\pi}\iint_\sms{sphere}I_\nu(\lambda,\Omega)\ddiff\Omega,
\end{equation}
which is simply the specific intensity from the stars, averaged over solid angle.
The power absorbed by the grain is:
\begin{equation}
  P_\sms{abs}(a) = \iint_\sms{sphere}\int_0^\infty J_\nu(\nu)
                \times\pi a^2Q_\sms{abs}(a,\nu)\ddiff\nu\ddiff\Omega
  = \int_0^\infty4\pi J_\nu(\nu)\times\pi a^2Q_\sms{abs}(a,\nu)\ddiff\nu.
  \label{eq:PabsBBQ}
\end{equation}
Similarly, the emitted power is:
\begin{equation}
  P_\sms{em}(a,T) = \iint_\sms{sphere}\int_0^\infty B_\nu(T,\nu)
                \times\pi a^2Q_\sms{abs}(a,\nu)\ddiff\nu\ddiff\Omega
  = \int_0^\infty4\pi B_\nu(T,\nu)\times\pi a^2Q_\sms{abs}(a,\nu)\ddiff\nu.
  \label{eq:PemBBQ}
\end{equation}
The \expression{equilibrium temperature}, $T_\sms{eq}$, is then simply the numerical solution to $P_\sms{abs}(a)=P_\sms{em}(a,T_\sms{eq})$.
Several quantities can be precomputed to simplify this task.

\begin{figure}[htbp]
  \includegraphics[width=\textwidth]{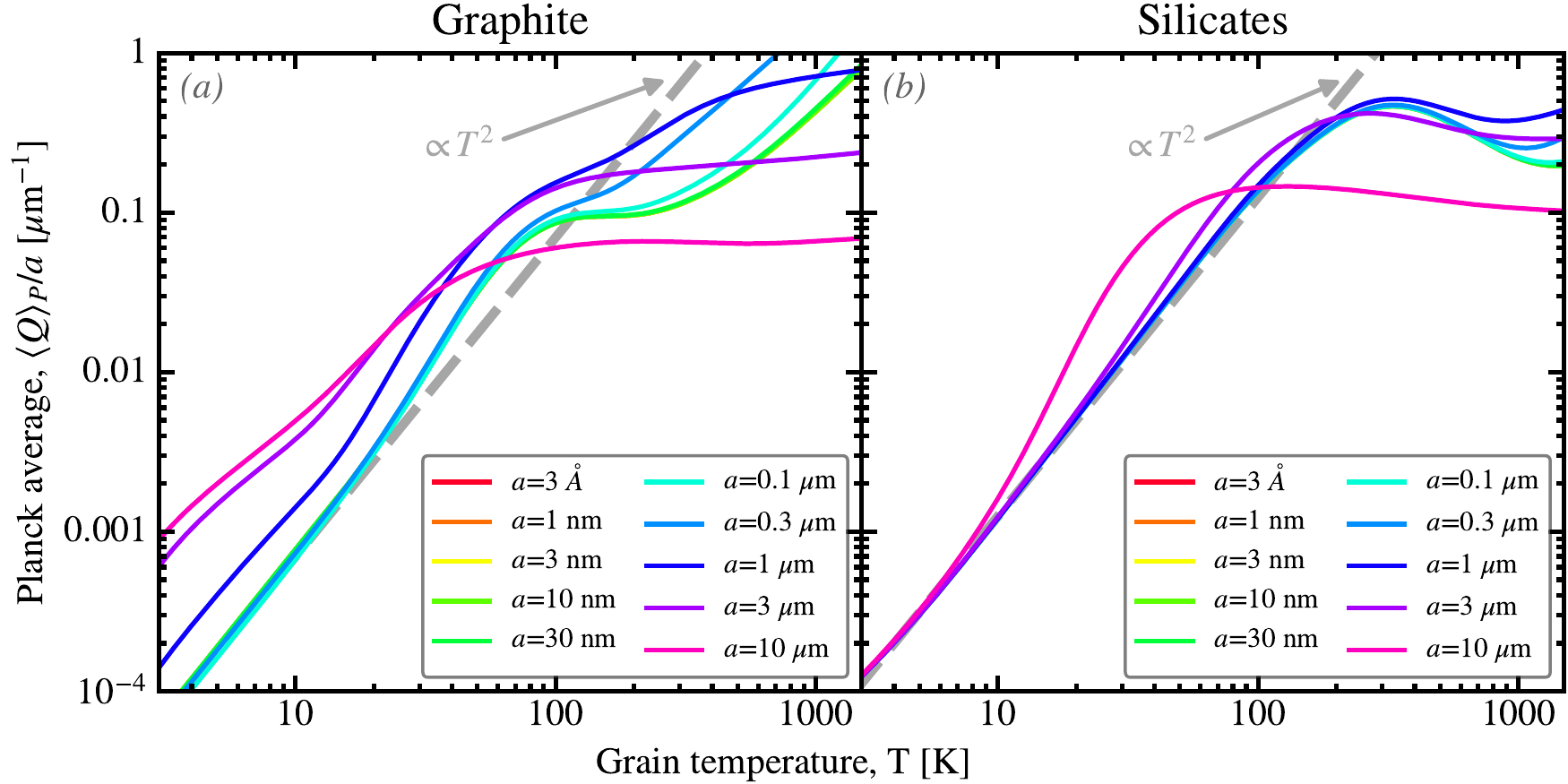}
  \newcap{Planck averages of graphite and silicates}%
         {We show the result of \refeq{eq:planckav} applied to the
          graphite \textit{(a)} and silicates \textit{(b)} of 
          \citet{draine03,draine03b}, for several radii, $a$.
          We overplot the approximation of \refeq{eq:planckavMBB} in grey.
          \CClicence}
  \label{fig:planckav}
\end{figure}

\paragraph{Planck average.} 
The \expression{Planck average} of a grain is defined as:
\begin{equation}
  \langle Q\rangle_P(a,T) \equiv\frac{\pi}{\sigma T^4}
    \int_0^\infty Q_\sms{abs}(a,\nu)B_\nu(T,\nu)\ddiff\nu.
  \label{eq:planckav}
\end{equation}
This quantity needs to be computed only once for a range of temperatures.
\refeq{eq:PemBBQ} then simplifies: $P_\sms{em}(a,T)=4\pi a^2\sigma T^4\langle Q\rangle_P(a,T)$.
Since interstellar grains emit predominantly in the \hIR, where the approximation of \refeq{eq:kappaMBB} is valid, we can derive an analytical expression of \refeq{eq:planckav}, knowing that $\sigma\equiv2\pi^5k^4/15h^3c^2$:
\begin{equation}
  \langle Q\rangle_P(a,T) = \frac{15}{\pi^4}Q_\sms{abs}(a,\lambda_0)
    \Gamma(4+\beta)\zeta(4+\beta)\left(\frac{\lambda_0kT}{hc}\right)^\beta.
  \label{eq:planckavMBB}
\end{equation}
We show the Planck average of typical grains in \reffig{fig:planckav}.
We can see that $\langle Q\rangle_P$ is almost independent of radius for grains smaller than $0.1\emic$.

\paragraph{ISRF-averaged efficiency.} 
The \hISRF-averaged efficiency is the equivalent of the Planck average for the absorbed power:
\begin{equation}
  \langle Q\rangle_\star(a)\equiv\frac{\pi}{J_\star}
     \int_0^\infty Q_\sms{abs}(a,\nu)J_\nu(\nu)\ddiff\nu,
  \label{eq:ISRFav}
\end{equation}
where $J_\star=\pi\displaystyle\int_0^\infty J_\nu(\nu)\ddiff\nu$.
This quantity is less general than \refeq{eq:planckavMBB} as it needs to be evaluated for each particular shape of the \hISRF.
\refeq{eq:PabsBBQ} then simplifies: $P_\sms{abs}(a)=4\pi a^2J_\star\langle Q\rangle_\star(a)$.
The equilibrium temperature is thus the solution to: $\sigma T^4\langle Q\rangle_P(a,T)=J_\star\langle Q\rangle_\star(a)$.

\begin{figure}[htbp]
  \includegraphics[width=\textwidth]{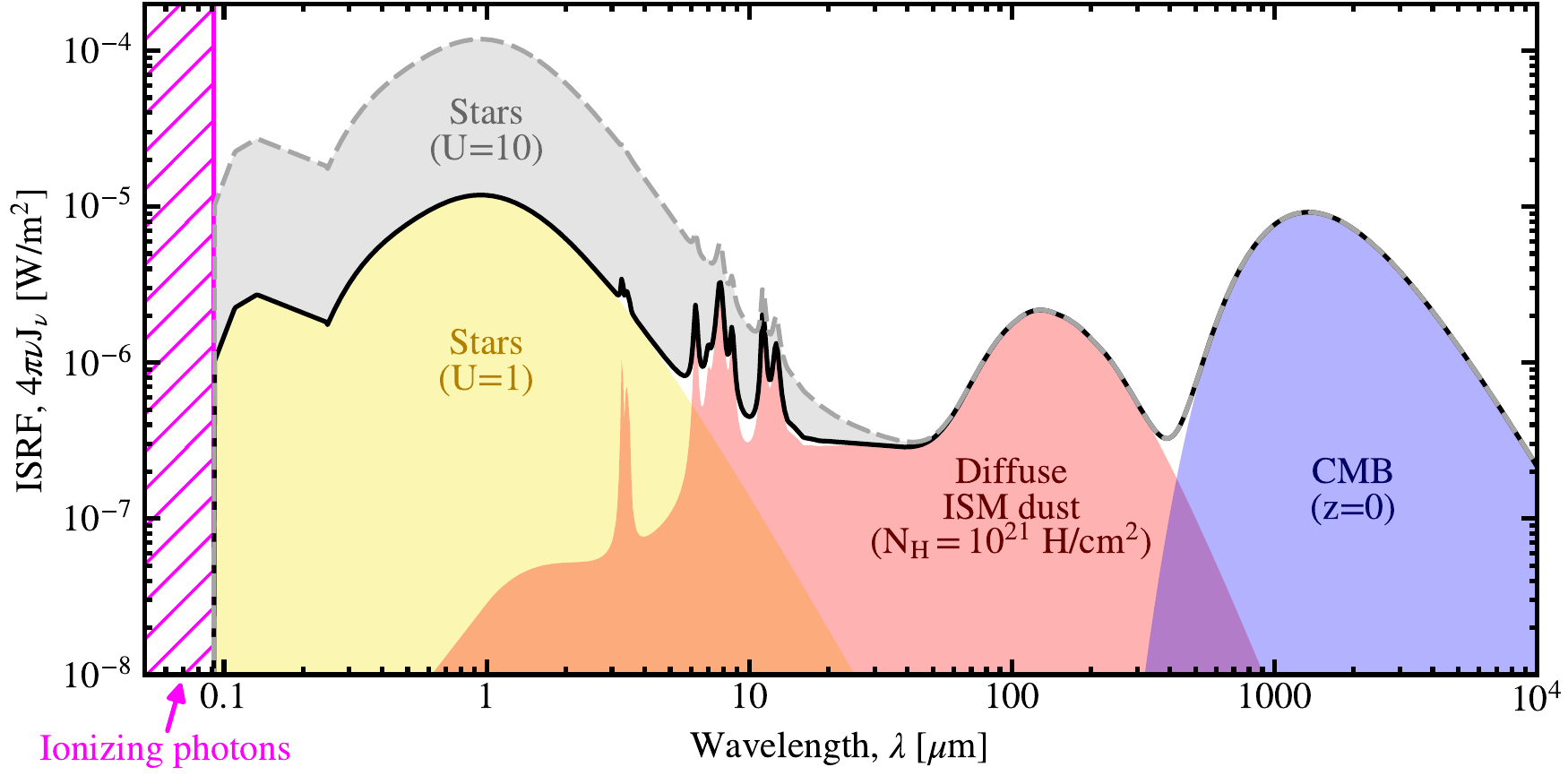}
  \newcap{Diffuse Galactic ISRF}%
         {This black line represents the average \hISRF\ experienced in the 
          diffuse \hISM\ of the \hMW.
          The stellar component (yellow) has been modeled by \citet{mathis83}.
          The \hUV\ and \hNIR\ bumps are the contributions by young and old 
          stars, respectively.
          This \hISRF\ represents the neutral \hISM, there is therefore no 
          emission shortward the Lyman break ($\lambda_\sms{Lyc}=0.0912\emic$).
          This stellar \hISRF\ can be scaled by the factor $U$.
          We have represented the $U=10$ case in grey.
          The diffuse dust emission is shown in red.
          The original \citet{mathis83} work did not have the constraints we 
          have today on this component.
          The emission represented here is the \citet{jones17} model, for 
          a typical hydrogen column density
          $N_\sms{H}=10^{21}\;\textnormal{cm}^{-2}$.
          The blue component is the \expression{Cosmic Microwave Background}
          (\hCMB), which is a perfect black body at $T_\sms{CMB}(z=0)=2.73$~K 
          \citep{mather94}.
          At higher redshift, $z$, the temperature of this component is 
          $T_\sms{CMB}(z)=(1+z)\times2.73$~K.
          The \expression{Cosmic Infrared Background} 
          \citep[\hCIB; \eg][]{dole06} is not represented here as its 
          \expression{Spectral Energy Distribution} (\hSED) is 
          similar to the dust emission, and is slightly lower.
          \CClicence}
  \label{fig:ISRF}
\end{figure}

\paragraph{The diffuse ISRF.} 
The diffuse \hISRF\ of the \hMW\ has been modeled by \citet{mathis83}.
It is represented in \reffig{fig:ISRF}.
This \hISRF\ is commonly used to describe grain heating (\ie\ how much power a grain absorbs) in the \hMW, and also in nearby galaxies.
Most of the heating is provided by the stellar component, as the integrand in \refeq{eq:PabsBBQ} is $J_\nu Q_\sms{abs}\propto J_\nu/\lambda^2$.
The long wavelengths have a negligible weight.
This stellar \hISRF\ can be scaled by a dimensionless factor $U$, to account for variations of the stellar density.
This scaling factor is not totally realistic, as regions with high radiation densities ($U\gtrsim10^3$) usually are star-forming regions, containing young star associations.
The \hUV\ bump of the stellar \hISRF, corresponding to these young stars in \reffig{fig:ISRF}, would be dominating the emission, while the \hNIR\ bump, corresponding to older stellar populations, would be, at first order constant.
This is however not very important for grains at thermal equilibrium, as their spectrum depends only on the total absorbed power, given by $\langle Q\rangle_\star$.

\begin{figure}[htbp]
  \includegraphics[width=\textwidth]{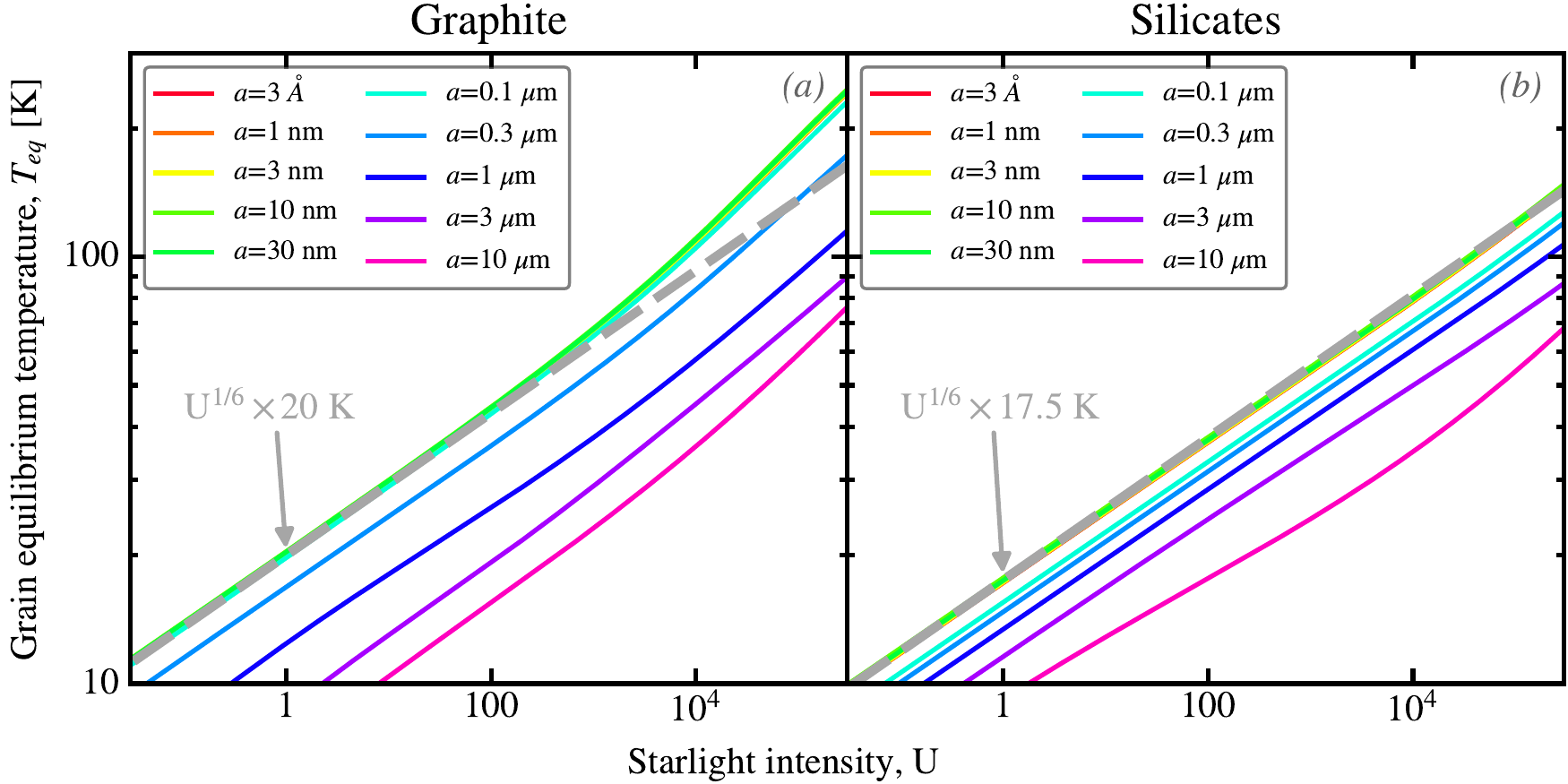}
  \newcap{Grain equilibrium temperatures}%
         {We show the equilibrium temperatures derived by equating
          \refeq{eq:PabsBBQ} and \refeq{eq:PemBBQ} for the graphite of 
          \citet{laor93} and the silicates of \citet{draine03,draine03b}.
          For grains smaller than $\simeq20$~nm, this temperature does not
          have reality, as these grains are out of equilibrium with the \hISRF.
          The grains are heated by the stellar \hISRF\ of \citet{mathis83}, 
          scaled by the factor $U$.
          We see that grains smaller than $\simeq0.1\emic$ roughly have the
          same temperature.
          This is because these grains are essentially in the Rayleigh regime
          over most of the visible-to-\hNIR\ range.
          In this regime, the dashed grey line is a good approximation.
          \CClicence}
  \label{fig:Teq}
\end{figure}

\paragraph{Equilibrium temperatures.} 
The equilibrium temperature of typical grains is shown in \reffig{fig:Teq}, as a function of $U$.
Assuming the heating is solely provided by the stellar component in \reffig{fig:ISRF}, $J_\star(U)=U\times J_\star(1)$ and $\langle Q\rangle_\star(U,a)=\langle Q\rangle_\star(1,a)$.
In this case, the integrated mean intensity is:
\begin{equation}
  \int_0^\infty4\pi J_\nu(U=1,\nu)\ddiff\nu
   =4\times J_\star(1)=2.2\E{-5}\;\textnormal{W/m}^2.
  \label{eq:U}
\end{equation}
The equilibrium temperature is thus:
\begin{equation}
  T_\sms{eq}(U,a) = {\underbrace{\left(\frac{\pi^4J_\star(1)
                                \langle Q\rangle_\star(a)}%
                      {\sigma Q_\sms{abs}(a,\lambda_0)
                       \Gamma(4+\beta)
           \zeta(4+\beta)}\right)}_\sms{weakly dependent on $a$}}^{1/(4+\beta)}
           \left(\frac{hc}{k\lambda_0}\right)^{\beta/(4+\beta)}
           U^{1/(4+\beta)}.
  \label{eq:U2T}
\end{equation}
For grains smaller than $a\simeq0.1\emic$, we thus have $T_\sms{eq}(U)\propto U^{1/(4+\beta)}$.
We show the equilibrium temperature of graphite and silicates, varying $U$ in \reffig{fig:Teq}.
We see that:
\begin{itemize}
  \item for graphite, $T_\sms{eq}^\sms{gra}(U)\simeq U^{1/6}\times20$~K;
  \item for silicates, $T_\sms{eq}^\sms{sil}(U)\simeq U^{1/6}\times17.5$~K.
\end{itemize}
\takeaway{Interstellar grains of a given homogeneous composition, at thermal equilibrium with the radiation field, mostly have the same temperature.}

    \subsubsection{Stochastic Heating}
    \label{sec:stochastheat}

\paragraph{Absorption and cooling times.}
Not all grains are at thermal equilibrium with the \hISRF.
To determine if this is the case, we need to estimate the \expression{photon absorption rate} of the grain:
\begin{equation}
  \frac{1}{\tau_\sms{abs}(U,a)} 
   = \int_0^\infty \pi a^2Q_\sms{abs}(a,\nu)\frac{4\pi J_\nu(\nu)}{h\nu}\ddiff\nu
  \propto a^3 U,
  \label{eq:tauabs}
\end{equation}
where the proportionality has been derived using \refeq{eq:Rayleigh}.
The grain absorption timescale, $\tau_\sms{abs}$, gives the average time between two photon absorptions.
We also need to estimate the \expression{cooling rate} of the grain:
\begin{equation}
  \frac{1}{\tau_\sms{cool}(a,T)} = \frac{P_\sms{em}(a,T)}{H(T)}
   \propto T^{3-n+\beta},
  \label{eq:taucool}
\end{equation}
where $H(T)$ is the \expression{enthalpy} of the grain at temperature $T$:
\begin{equation}
  H(T) = \int_0^T C(T^\prime)\ddiff T^\prime.
  \label{eq:enthalpy}
\end{equation}
It is the vibrational energy content of the grain.
The proportionality of \refeq{eq:taucool} has been derived from \refeq{eq:planckavMBB} for $P_\sms{em}$, and from the low-temperature behaviour of $H(a,T)\propto a^3 T^{n+1}$ (\refeqnp{eq:debyeprox}; $n=3$ corresponding to the three-dimensional Debye model).
The cooling time, $\tau_\sms{cool}$, is independent of the grain size.

\begin{figure}[htbp]
  \includegraphics[width=\textwidth]{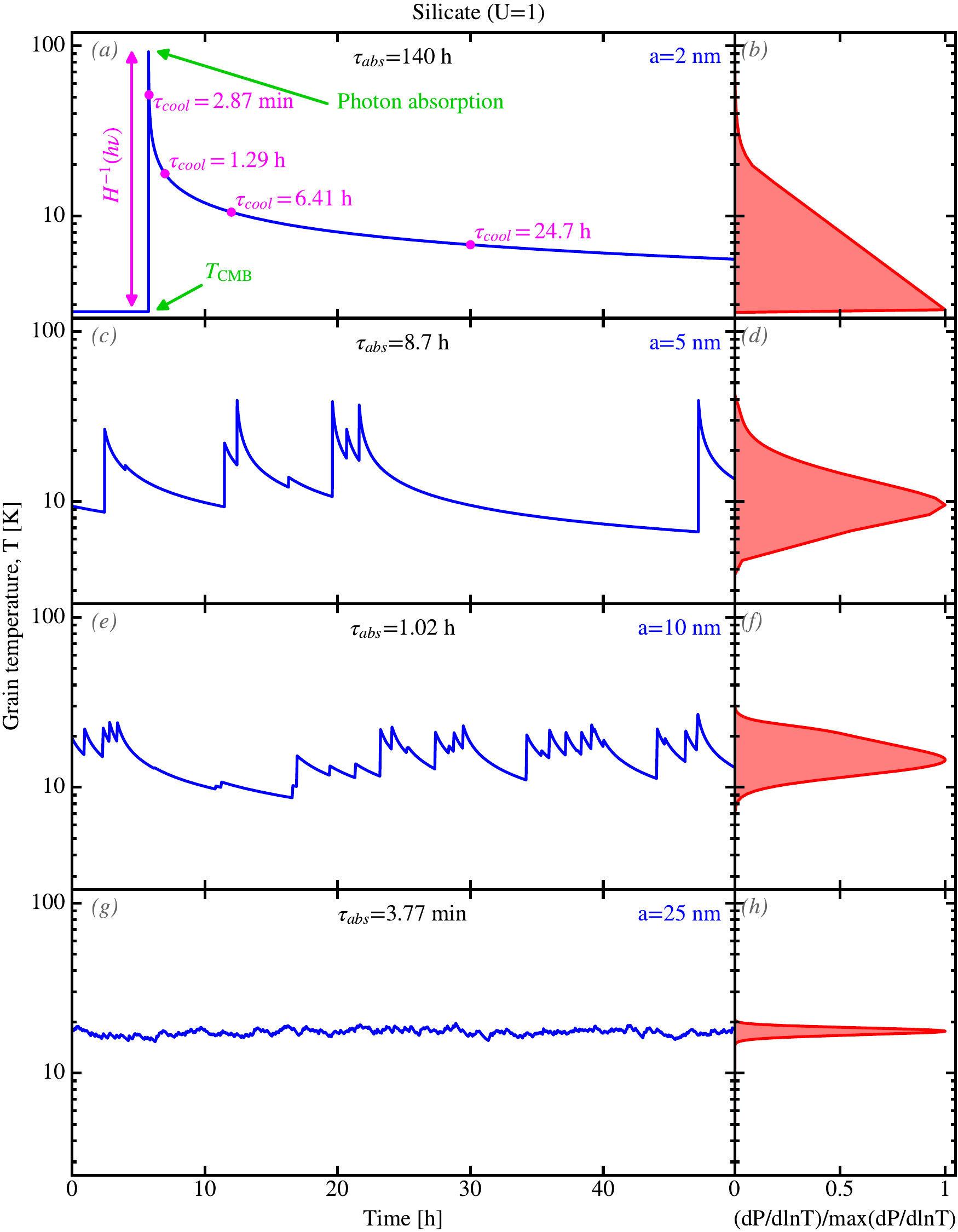}
  \newcap{Temperature fluctuations of grains with different radii}%
         {The left panels show the time variation of the temperature of 
          silicate grains \citep{draine03,draine03b}, exposed to the 
          \citet{mathis83} \hISRF\ with $U=1$.
          The radius of the grain, $a$, increases downward.
          The right panels show the corresponding probability distribution
          of the temperature.
          These simulations were performed using the \citet{draine85} 
          Monte-Carlo method.
          See \citet{draine03c} for a similar simulation with graphite.
          \CClicence}
  \label{fig:flucT_a}
\end{figure}

\begin{figure}[htbp]
  \includegraphics[width=\textwidth]{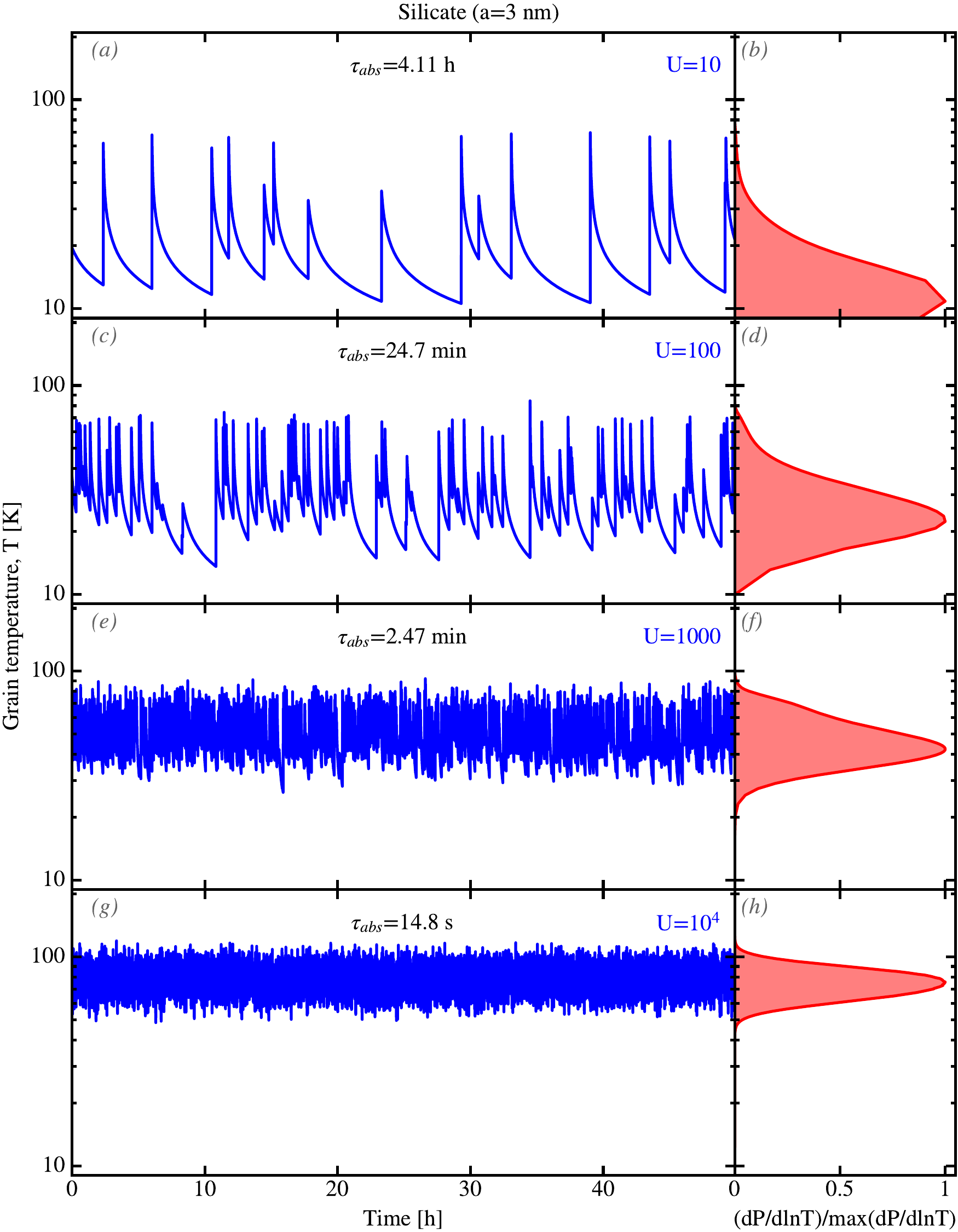}
  \newcap{Temperature fluctuations of grains with different starlight 
          intensities}%
         {This is a variation on \reffig{fig:flucT_a}.
          The grain radius is constant ($a=3$~nm), and the starlight intensity,
          $U$, increases downward.
          \CClicence}
  \label{fig:flucT_U}
\end{figure}

\paragraph{The temperature fluctuations.}
If $\tau_\sms{abs}\gtrsim\tau_\sms{cool}$, the grain has the time to significantly cool down between two photon absorptions.
Its temperature is thus changing with time.
This is represented on the simulation in \refsubfig{fig:flucT_a}{a}, as follows.
\begin{enumerate}
  \item A grain, with $a=2$~nm, starts at $T=T_\sms{CMB}$.
  \item It then receives a photon after $\simeq6$~h, which causes its 
    temperature to spike to $T_\sms{s}\simeq100$~K.
    The value of $T_\sms{s}$ is such that 
    $H(T_\sms{s})-H(T_\sms{CMB})=h\nu_\sms{s}$, where $h\nu_\sms{s}$ is the energy 
    of the incident photon.
    It is the only photon it receives within the 50~h displayed here.
    Its absorption time is indeed $\tau_\sms{abs}\simeq140$~h.
  \item The grain then cools down by radiating.
    The heat capacity we have used here \citep{draine01} has $n\simeq2$.
    The cooling time is thus $\tau_\sms{cool}\propto T^{-3}$.
    The cooling time is short when the grain is hot, but decreases with the 
    temperature.
    Such a grain spends most of its time at low temperatures.
\end{enumerate}
In practice, we do not see the emission of the grain varying with time, as observations encompass a statistical number of grains, with different time histories.
What we observe is an \expression{ergodic}\footnote{The ergodicity is a principle stating that the steady-state statistical distribution of the properties of a large number of identical particles is the average of their properties over time.} average: small grains appear to have a temperature distribution, represented in \refsubfig{fig:flucT_a}{b} for $a=2$~nm.
When the radius of the grain increases, the absorption time decreases as $\tau_\sms{abs}\propto a^{-3}$, as shown in the remaining panels of \reffig{fig:flucT_a}.
The number of temperature spikes increases, as the grain being larger, it intercepts more photons.
The temperature of the spikes also decreases with $a$, as the grain stores the energy of a single photon in a larger number of phonon modes.
For large enough grains (\refsubfig{fig:flucT_a}{g}), the temperature fluctuations become negligible. 
The grain appears to have a single temperature; its temperature distribution tends toward a Dirac distribution (\refsubfig{fig:flucT_a}{h}):
it has reached thermal equilibrium.
\reffig{fig:flucT_U} shows the same type of simulations, fixing the radius of the grain, and varying the starlight intensity.
In this case, the heating rate increases linearly with $U$ (downward in \reffig{fig:flucT_U}).

\begin{figure}[htbp]
  \begin{tabular}{cc}
    \includegraphics[width=0.48\textwidth]{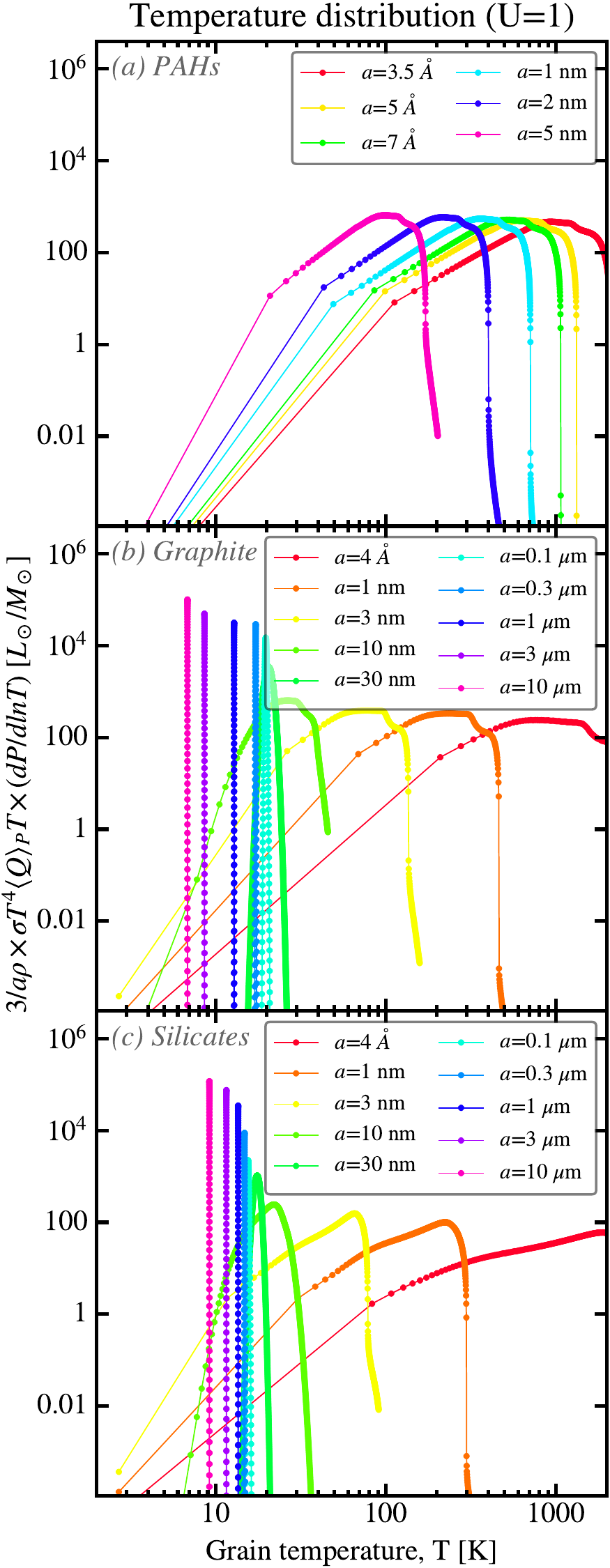} &
    \includegraphics[width=0.48\textwidth]{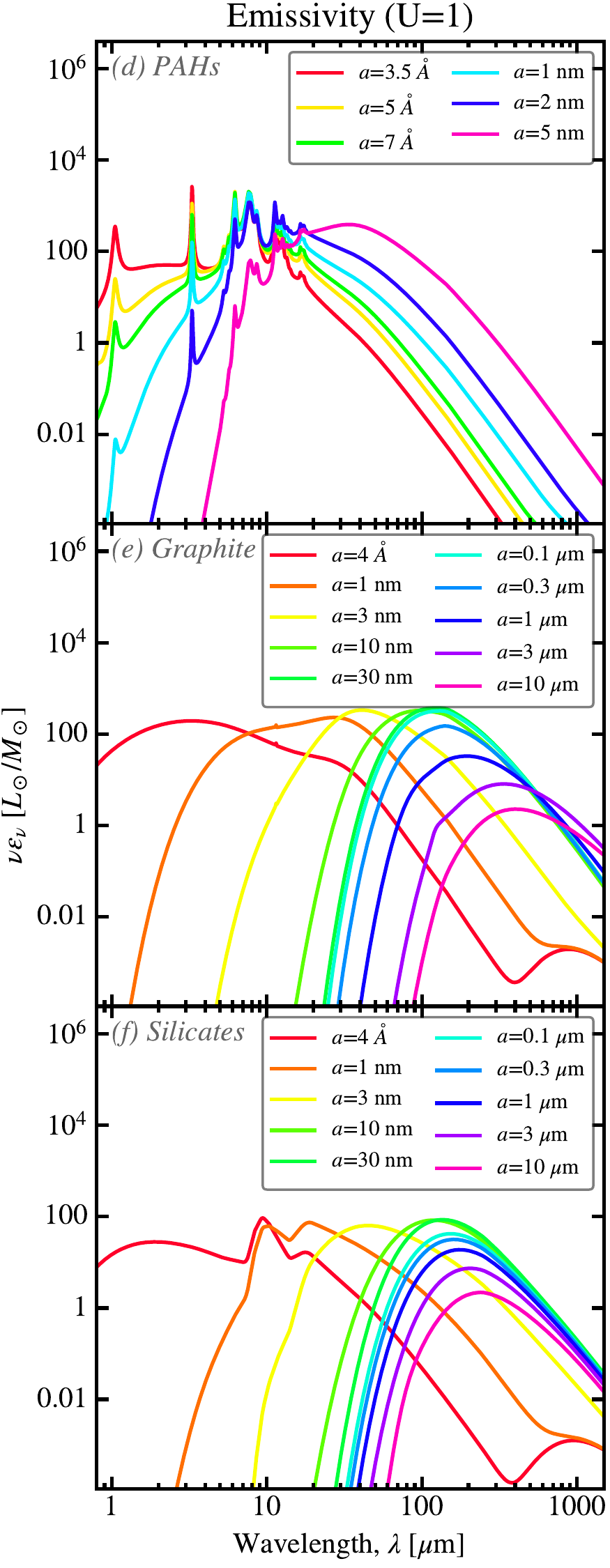} \\
  \end{tabular}
  \newcap{Stochastically heated grains}%
         {The left panels show the temperature distribution of grains with
          different radii.
          The dots represent the sampling of our adaptative energy grid.
          The \hPAH s are a mixture of neutral ($50\,\%$) and ionized ($50\,\%$)
          molecules \citep{draine07}, the graphite and silicates are from
          \citet{draine03,draine03b}.
          The right panels show the corresponding emissivities.
          \CClicence}
  \label{fig:stochastheat}
\end{figure}

\paragraph{Out-of-equilibrium emission.}
At each time, in the simulations of \reffigs{fig:flucT_a}{fig:flucT_U}, the emission of the grain is: 
$\epsilon_\nu(a,\nu,t)=3\pi/\rho\times Q_\sms{abs}(a,\nu)/a\times B_\nu(T(t),\nu)$.
To average over time, we simply need to integrate over the temperature distribution:
\begin{equation}
  \epsilon_\nu(a,\nu)=\frac{3\pi}{\rho}\frac{Q_\sms{abs}(a,\nu)}{a}
     \int_0^\infty B_\nu(T,\nu)\frac{\dd P(T,a)}{\dd T}\ddiff T.
  \label{eq:dPdT}
\end{equation}
The left panels of \reffig{fig:stochastheat} show the temperature distributions of \hPAH s, graphite and silicates of different sizes.
The corresponding emission spectra, computed using \refeq{eq:dPdT}, are shown in the right panels.
We see that the smallest grains fluctuate to the highest temperatures.
Their emission is thus the broadest, and extends to the shortest wavelengths.
An important difference between equilibrium and stochastic heating is that the emission spectrum of small grains depends not only on the intensity of the \hISRF, but also on its hardness.
The latter can be quantified with the mean energy of stellar photons: 
\begin{equation}
  \langle h\nu\rangle\equiv\frac{\displaystyle\int_0^\infty J_\nu(\nu)\ddiff\nu}%
                              {\displaystyle\int_0^\infty
                               \frac{J_\nu(\nu)}{h\nu}\ddiff\nu}.
\end{equation}
This parameter determines the average temperature spikes.
This is the reason why stochastic heating is sometimes referred to as \expression{single photon heating}, \expression{transient heating} or \expression{quantum heating}.
The \hISRF\ intensity roughly scales with the emissivity, but does not affect its spectral shape.

\paragraph{Numerical methods.}
To compute the emission spectrum of an out-of-equilibrium grain, we need to calculate its temperature distribution, $\dd P/\dd T$ \refeqp{eq:dPdT}.
This is done numerically.
Several methods can be found in the literature.
\begin{description}
  \item[The Monte-Carlo method] \citep{draine85} consists in simulating the 
    temporal evolution of the grain, by drawing random stellar photons.
    This is the method we have used in \reffigs{fig:flucT_a}{fig:flucT_U}.
    This method is easy to implement, but it is not the most numerically
    efficient.
  \item[Solving the integral equation] governing $\dd P/\dd T$ \citep{desert86}.
    This method is efficient and is used by the \ncode{DustEM} code 
    \citep{compiegne11}.
  \item[The transition matrix method] \citep{guhathakurta89} consists in 
    building a squared matrix whose elements are the probability per unit time
    that a grain will transit from a state to another.
    The row and columns of this matrix correspond to the final and initial 
    energy bins.
    This matrix can then be diagonalized to compute $\dd P/\dd T$.
    This is the method we have implemented in our code.
    It has been used in \reffig{fig:stochastheat}.
    We have implemented an adaptative grid in energy, represented by the dots 
    in the left panels of \reffig{fig:stochastheat}.
    The grid is refined to ensure the accuracy of the emission spectrum.
    We can see that more points are needed at high temperatures.
  \item[ISRF moment approximations] \citep{natale15} consists in 
    interpolating
    a precomputed grid, characterizing the \hISRF\ by its first two moments: 
    its intensity, $U$, and the mean energy of the photons, 
    $\langle h\nu\rangle$.
    This method is not exact, but it is fast enough to be implemented in 
    radiative transfer simulations.
\end{description}

\begin{figure}[htbp]
  \includegraphics[width=\textwidth]{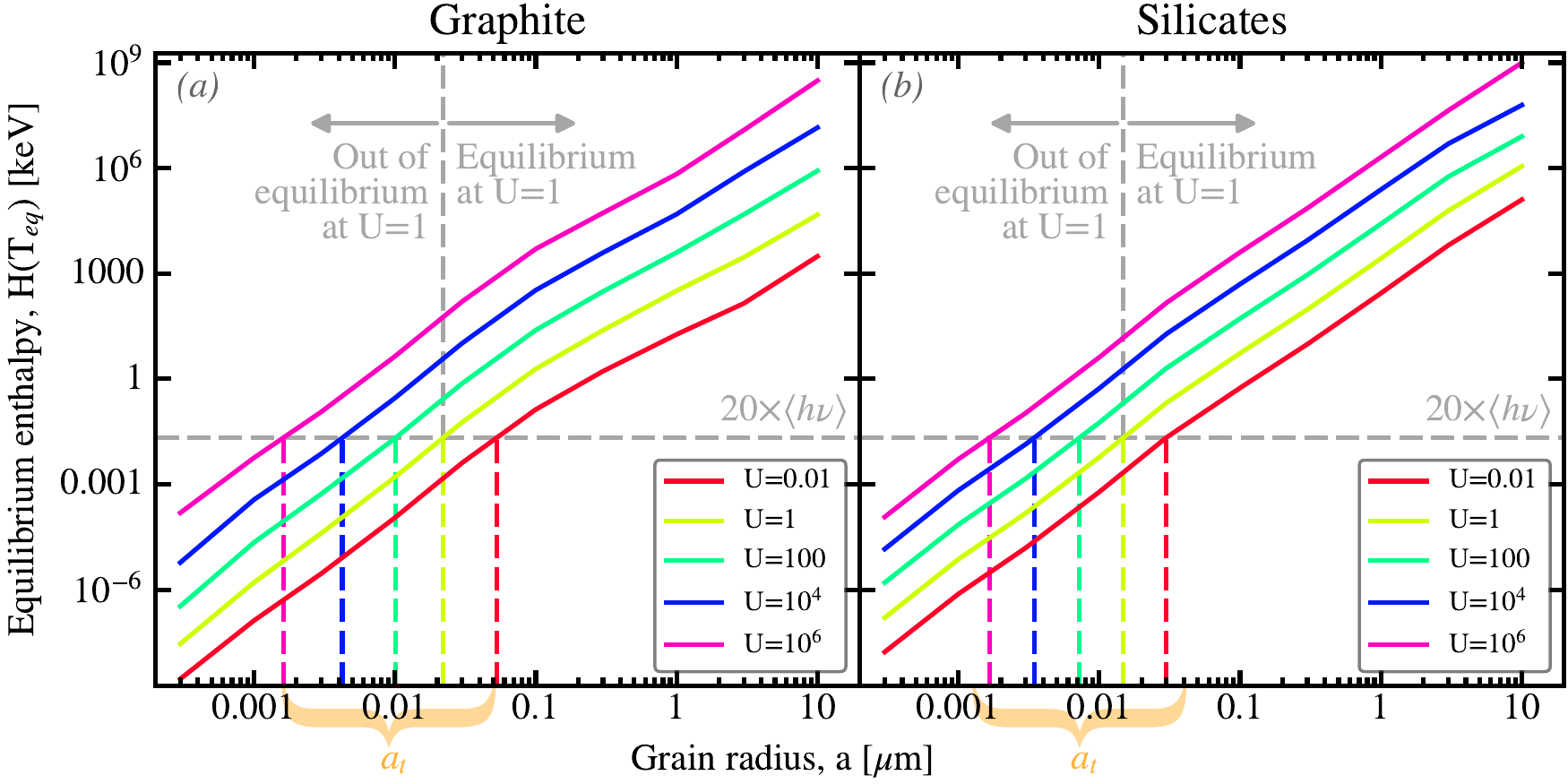}
  \newcap{Transition radius for stochastically heated grains}%
         {These two panels show the enthalpy at the hypothetical equilibrium
          temperature, for the graphite of \citet{laor93} and the silicates
          of \citet{draine03,draine03b}.
          The horizontal dashed grey line represents the
          $20\times\langle h\nu\rangle$ level, for the \citet{mathis83} \hISRF.
          The enthalpies above this line correspond to 
          $H(T_\sms{eq})\gg\langle h\nu\rangle$, that is to equilibrium grains.
          The vertical dashed color lines, show the transition radius for each
          $U$. 
          Left of this line, grains are stochastically heated.
          \CClicence}
  \label{fig:transition}
\end{figure}

\paragraph{Equilibrium criterion.}
A simple criterion to determine if a grain is at thermal equilibrium with the \hISRF\ consists in comparing its equilibrium enthalpy with the average energy of a stellar photon.
\reffig{fig:transition} compares these two quantities for graphite and silicates, varying $a$ and $U$.
\takeaway{A grain is at thermal equilibrium if $H(a,T_\sms{eq})\gg\langle h\nu\rangle$.}
\begin{description}
  \item[The transition radius] between stochastically heated and equilibrium
    grains can therefore be estimated as the radius for which 
    $H(a_t,T_\sms{eq})\simeq20\times\langle h\nu\rangle$.
    This value is indicative as the transition to equilibrium is a smooth, 
    continuous process.
    These transition radii are the vertical dashed lines in 
    \reffig{fig:transition}.
    The mean stellar photon energy of the \citet{mathis83} \hISRF\ is a 
    constant:
    $\langle h\nu\rangle\simeq1.05$~eV.
    The equilibrium enthalpy behaves as 
    $H(a_t,T_\sms{eq})\propto a_t^3T_\sms{eq}^{n+1}$.
    Since $T_\sms{eq}\propto U^{1/(4+\beta)}$, we have
    $a_t\propto U^{-(n+1)/3(4+\beta)}\simeq U^{-1/6}$ ($n=2$ and $\beta=2$ for the 
    grains in \reffig{fig:transition}).
    The following approximations provide good fits:
    \begin{itemize}
      \item for graphite, 
        $a_t^\sms{gra}(U)\simeq22\;\textnormal{nm}\times U^{-1/6}$;
      \item for silicates, 
        $a_t^\sms{sil}(U)\simeq15\;\textnormal{nm}\times U^{-1/6}$.
    \end{itemize}
  \item[Role of the \hCMB:]
note that, contrary to a common misconception, the smallest grains are not in thermal equilibrium with the \hCMB. 
For instance, a 3~$\r{A}$ silicate has an enthalpy at 2.7~K of $H(T_\sms{CMB})\simeq0.5\;\mu$eV, while the average photon energy received from the \hCMB\ is $\langle h\nu_\sms{CMB}\rangle\simeq1$~meV. 
The effect of considering photons with $\lambda>1000\emic$ as a source of \expression{continuous heating} \citep[following][]{guhathakurta89}, creates an
artificial minimum temperature close to the actual \hCMB\ temperature. 
This is the bump peaking at $\lambda\simeq1$~mm for the smallest grains in panels~\textit{(e)} and \textit{(f)} of \reffig{fig:stochastheat}.
This is an artefact due to this approximation.
Fortunately, this artefact does not affect the emitted spectrum, integrated over the size distribution, in any detectable way.
\end{description}

    \subsubsection{Collisionnal Heating}
    \label{sec:coll}

In addition to photon absorption, collisions with gas particles can, in specific conditions, contribute to dust heating.
Obviously, this will happen when the temperature of the gas is the hottest, such as in a plasma.
The collision rate of gas particles, following a Maxwell-Boltzmann distribution, with a dust grain is:
\begin{equation}
  \frac{1}{\tau_\sms{coll}}\simeq
    \underbrace{\sqrt{\frac{8kT}{\pi m}}}_\sms{mean velocity}
    \underbrace{\frac{1}{n\pi a^2}}_\sms{mean free path},
\end{equation}
where $n$ is the density of the gas, and $m$, the mass of the particles.
Assuming that the protons and the electrons are thermalized, the ratio of their collision rates is $\tau_\sms{coll}(e^-)/\tau_\sms{coll}(\textnormal{H}^+)=\sqrt{m_\sms{e}/m_\sms{p}}\simeq0.02$ (\cf\ \reftab{tab:constants}).
The collisions with the protons can thus be neglected.

\begin{figure}[htbp]
  \includegraphics[width=\textwidth]{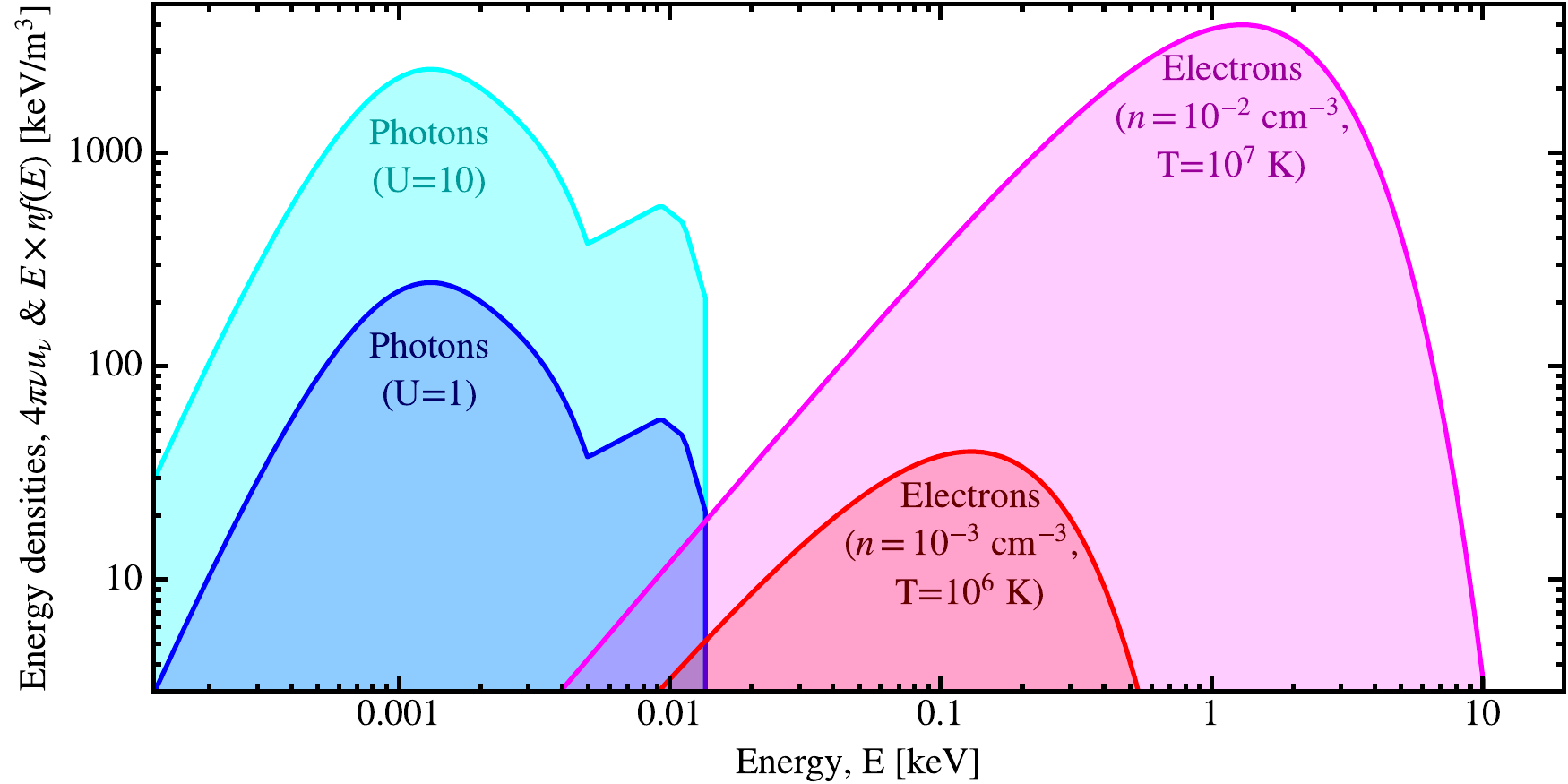}
  \newcap{Photon and electron energy densities}%
         {The blue and cyan curves show the energy per unit volume 
          ($u_\nu=4\pi cJ_\nu$) of the stellar photons, for $U=1$ and $U=10$.
          The electron energy density is the Maxwell-Boltzmann distribution
          \refeqp{eq:MBel}: $n\times f(E)$.
          We have plotted it for densities and temperatures typical of 
          coronal plasmas: $n=10^{-3}-10^{-2}$~cm$^{-3}$ and $T=10^6-10^7$~K.
          \CClicence}
  \label{fig:Edens}
\end{figure}

\paragraph{Electronic heating rate.}
The Maxwell-Boltzmann distribution of the electrons of energy, $E$, can be written:
\begin{equation}
   f(E,T) = 
    \frac{2}{\sqrt{\pi}}\sqrt{\frac{E}{(kT)^3}}\exp\left(-\frac{E}{kT}\right).
  \label{eq:MBel}
\end{equation}
This distribution is normalized: $\int_0^\infty f(E,T)\ddiff E=1$.
It is displayed in \reffig{fig:Edens} and compared to the stellar radiation field.
The collisional cross-section is very poorly constrained.
At low energies, it should be close to the geometric cross-section, $\pi a^2$.
However, at high energies, electrons can pass thought the grain.
\citet{dwek86} proposed the following cross-section:
\begin{equation}
  \sigma_\sms{coll}(a,E) = \pi a^2\times
  \left\{\begin{array}{ll}
    1 & \mbox{for } E < E_\star(a) \\
    \displaystyle
    1 - \left[1-\left(\frac{E_\star(a)}{E}\right)^{3/2}\right]^{2/3}
    & \mbox{for } E\ge E_\star(a),
  \end{array}\right.
\end{equation}
where $E_\star(a)$ is the threshold energy.
According to the fit of experimental data shown in Fig.~1 of \citet{dwek87}, this threshold energy is \citep[\cf\ the discussion in][]{bocchio13}:
\begin{equation}
  \left\{\begin{array}{rcll}
    E_\star^\sms{car}(a) & \simeq 
      & \displaystyle10\;\textnormal{keV}
        \times\left(\frac{a}{1\emic}\right)^{2/3}
      & \mbox{for carbonaceous} \\ 
    E_\star^\sms{sil}(a) & \simeq
      & \displaystyle14\;\textnormal{keV}
        \times\left(\frac{a}{1\emic}\right)^{2/3}
      & \mbox{for silicates.} \\ 
  \end{array}\right.
\end{equation}
Assuming the grains are at rest, the collision rate for a given energy, $E$, is now:
\begin{equation}
  \frac{1}{\tau_\sms{coll}(a,E)} = n\times\sigma_\sms{coll}(a,E)\times v(E),
\end{equation}
where the velocity of an electron with energy $E$ is:
\begin{equation}
  v(E) = \sqrt{\frac{2E}{m}}.
\end{equation}
The collisional power received by the grain is finally the integral of the energy deposit per unit time, $E/\tau_\sms{coll}$, over the whole Maxwell-Boltzmann distribution:
\begin{equation}
  P_\sms{coll}(a,n,T) = \int_0^\infty\frac{E}{\tau_\sms{coll}(a,E)}f(E,T)\ddiff E.
  \label{eq:Pcoll}
\end{equation}

\begin{figure}[htbp]
  \includegraphics[width=\textwidth]{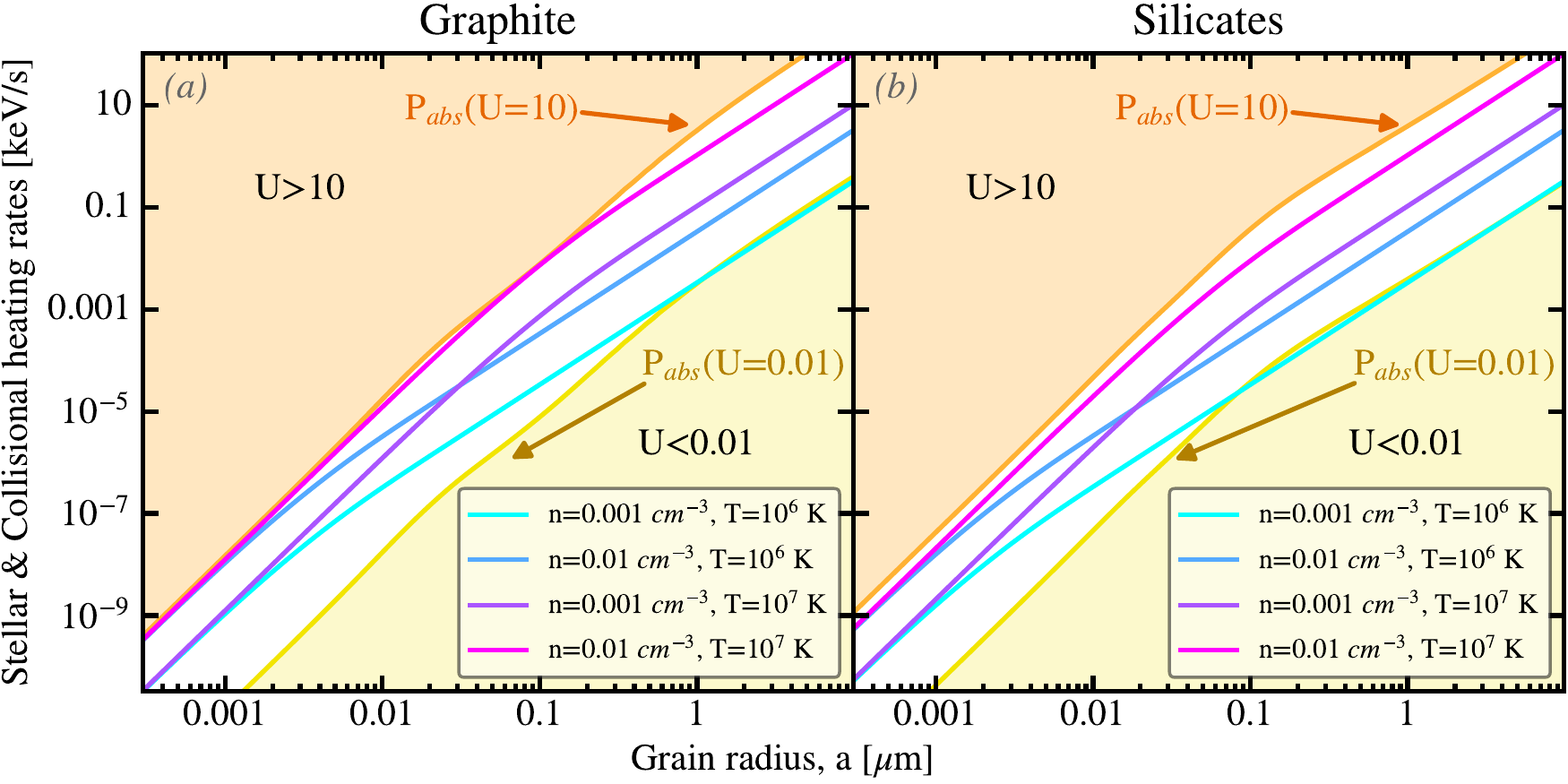}
  \newcap{Collisional heating rate}%
         {In each panel, the cyan, blue, purple and magenta lines show the 
          collisional heating rates from \refeq{eq:Pcoll}, for different values
          of $n$ and $T$.
          We compare the stellar photon heating rate in yellow ($U=0.01$) and
          orange ($U=10$).
          The area colored in yellow corresponds to $U<0.01$, where photon 
          heating is negligible.
          The area colored in orange corresponds to $U>10$, where photon
          heating dominates.
          \CClicence}
  \label{fig:Pcoll}
\end{figure}

\paragraph{Cases where collisions dominate the heating.}
Coronal plasmas, that can be found in the \expression{Hot Ionized Medium} (\hHIM; \cf\ \reftab{tab:ISMism}) of the \hMW\ or the halo of elliptical galaxies, have typical temperatures of $T\simeq10^6-10^7$~K.
This gas has been heated by successive \expression{SuperNova} (\hSN) blast waves.
It has a low density ($n\simeq10^{-3}-10^{-2}$~cm$^{-3}$), but can have a large filling factor (it fills $50\,\%$ of the volume of the \hMW).
At these temperatures, it is responsible for a diffuse X-ray emission.
When dust grains are present in such a gas, collisions can dominate the heating, depending on the balance between the photon and electron energy densities (\cf\ \reffig{fig:Edens}).
\begin{description}
  \item[The collisional heating rate] is displayed as a function of the grain
    radius in \reffig{fig:Pcoll}.
    This figure basically shows that, if a grain is in a coronal plasma, 
    collisional heating dominates for $U<0.01$, and photon heating dominates
    if $U>10$. 
    In between, both can play a role, depending on $n$ and $T$.
  \item[The transition radius] between stochastically-heated and 
    equilibrium grains can be computed similarly to \refeq{fig:transition}, 
    using the heating rate of \reffig{fig:Pcoll}.
    This time, the single heating events are due to electron collisions, with
    average energies: $\langle E\rangle=3/2kT\simeq0.13$~keV$\times T/10^6$~K.
    They are much higher than the typical \hISRF\ photon energy, 
    $\langle h\nu\rangle\simeq1$~eV.
    The stochastic heating will be the most efficient for the lowest densities
    and highest temperatures.
    For the extreme case, $n=10^{-3}$~cm$^{-3}$ and $T=10^7$~K, the transition
    radii are: $a_t^\sms{gra}\simeq58$~nm and $a_t^\sms{sil}\simeq35$~nm.
\end{description}


\newchapter{Dust Observables and Models}
\label{chap:dustmodels}
\citesmart{All models are wrong, but some are 
           useful.}{\citep[George E.\ P.\ \familyname{Box};][]{box79}}
\minitoc
The present chapter is intended to provide a comprehensive overview of the current state of the field.
It discusses both the observables that can be used to constrain dust properties, and the current state-of-the-art models.
It bridges the basic physical knowledge of \refchap{chap:propaedeutics} with their current application.

\section{A Brief History of Interstellar Dust Studies}
\label{sec:history}

\begin{figure}[htbp]
  \begin{tabular}{ccc}
    \includegraphics[width=0.45\textwidth]{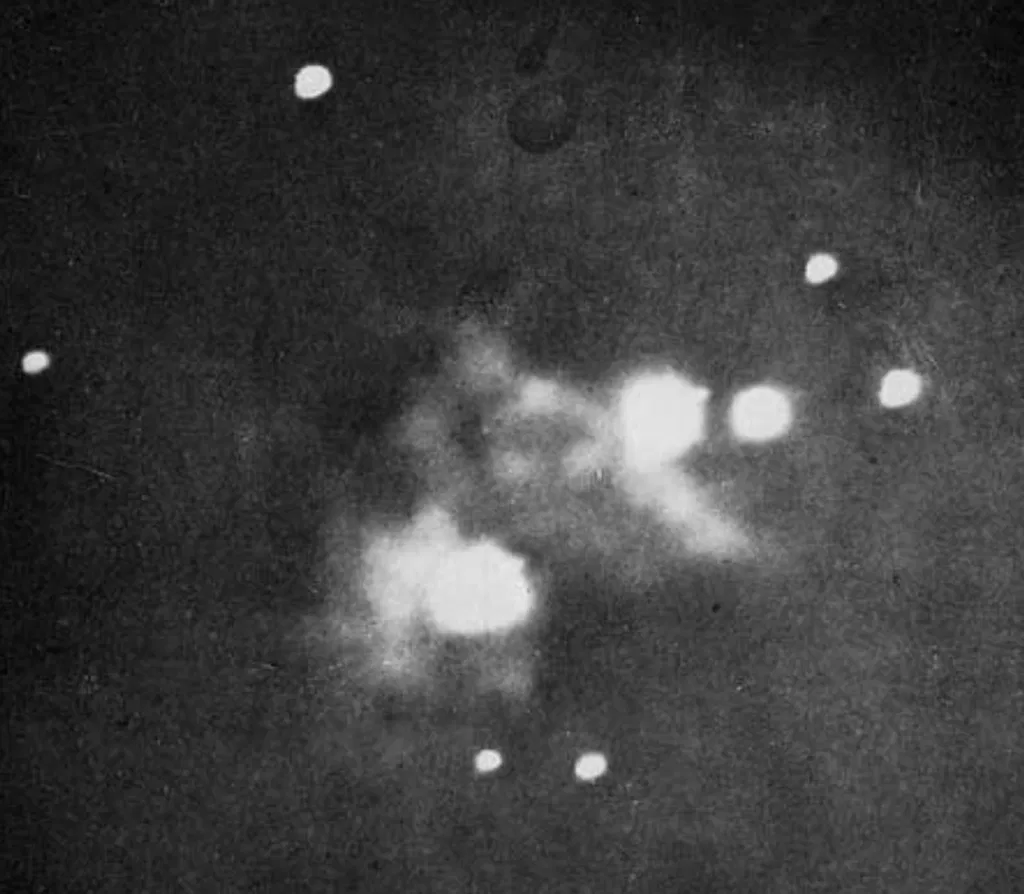} &
    \includegraphics[width=0.31\textwidth]{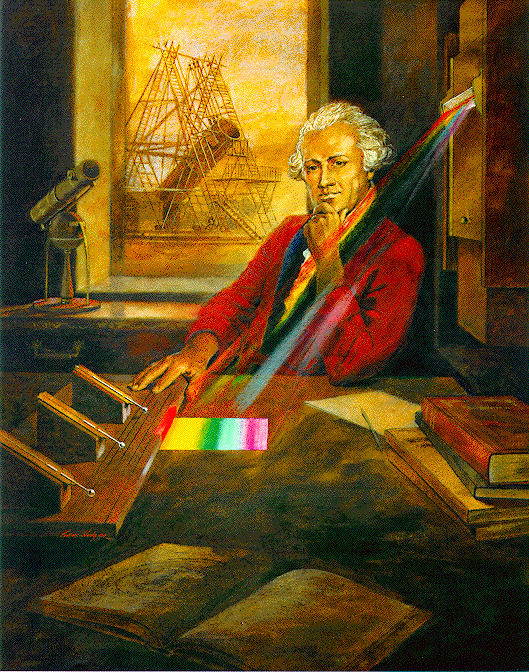} &
    \includegraphics[width=0.177\textwidth]{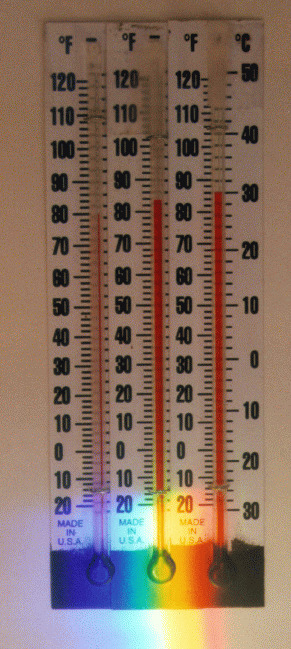} \\
    \textit{(a)}~First photograph of Orion (1880) &
    \textit{(b)}~Sir William \familyname{Herschel} &
    \textit{(c)}~Reenactment \\
  \end{tabular}
  \newcap{Early ISM and IR astronomy}%
         {Panel~\textit{(a)} shows the 1880 picture of the Orion nebula taken 
          at the Lick observatory, by Henry \familyname{Draper}.
          Panel~\textit{(b)} represents Sir William \familyname{Herschel}
          measuring the \hSED\ of the Sun.
          We can see in the background his 1.2~m speculum mirror telescope.
          Panel~\textit{(c)} shows a simple recreation of his experiment with 
          commercial thermometers.
          We see that the temperature is the highest on the last thermometer, 
          after the red.
          \uline{Credit:}
          \begin{inlinelistalph}
            \item \href{https://commons.wikimedia.org/wiki/File:Orion_1880.jpg}{Henry 
              \textsc{Draper}}, public domain;
            \item 
       \href{http://elsofista.blogspot.com/2010/10/el-descubrimiento-del-infrarrojo.html}%
             {El Sofista}, not licensed;
         \item \href{https://coolcosmos.ipac.caltech.edu/page/lesson_herschel_experiment}%
           {courtesy of NASA/JPL-Caltech}.
         \end{inlinelistalph}}
  \label{fig:draper}
\end{figure}
\noindent
The History of our progressive understanding of \hISD\ is driven by the successive technological innovations that allowed us to shed light on its nature.
\begin{description}
  \item[Telescopes] allowed \expression{deep-sky} 
    observations already through the XVIII$^\textnormal{th}$ century 
    \citep[\eg][]{wilson07}.
    Charles \familyname{Messier}'s catalog of \citengl{nebulae} was first 
    established in 1774.
    The quality and size of mirrors increased through the 
    XIX$^\textnormal{th}$ century, and at the beginning of the XX$^\textnormal{th}$ 
    century, the first meter-class telescopes, with silvered-glass 
    mirrors\footnote{Before that, telescope mirrors were made of the lower 
    optical quality speculum metal alloy.
    The process of silvering glass mirrors was invented by Léon 
    \familyname{Foucault} in the 1860's.}, were built.
    The Mount Wilson Observatory was founded in 1904, and its 1.5-m Hale 
    telescope was commissioned in 1908.
  \item[Photographic plates] turned astronomical observations into a 
    reproducible empirical science. 
    They required long exposures and reliable tracking of the sky's apparent 
    rotation.
    Photography was invented by Nicéphore \familyname{Niépce} in the 1820's.
    His associate, Louis \familyname{Daguerre}, perfected the technique and
    commercialized the first \expression{daguerreotype}\footnote{A 
    daguerreotype is the capture of an image directly on a chemically-treated 
    metal plate, without the recourse to a negative.} in 1839.
    The first attempts to capture images of the sky (the Moon and the Sun), were
    performed using this invention, in the 1840's.
    Emulsion plates were substituted to daguerreotypes, and the first 
    deep-sky picture (the Orion nebula; \refsubfig{fig:draper}{a}) was taken by 
    Henry \familyname{Draper} in 1880 \citep{barker88}.
  \item[Solid-state physics] was given a strong impulse, after World War II, 
    by the prospective development of electronics \citep{martin13}.
    The transistor was invented in 1947, at Bell laboratories in New Jersey.
    This impulse led to the development of technical and conceptual tools
    for the optical and electronic properties of solids that our field would
    benefit from.
  \item[The introduction of computers] revolutionized all scientific fields.
    The principle of the computer was laid out in Alan \familyname{Turing}'s
    seminal 1937 article \citep{turing37}.
    The first computers were developed during World War II to break the German
    encryption codes \citep[\eg][]{mcgrayne11}.
    They started to be used in astrophysics during the 1950's, to compute
    stellar structures.
    Before that, some calculations were impossible.
    For instance, the first dust model of the 1930's was made of small iron 
    particles 
    \citep[\eg][]{schoenberg37,greenstein38}, in part because Mie computations 
    for small metal spheres were easier on paper than for large dielectrics 
    \citep{van-de-hulst86}. 
    The first dust radiative transfer numerical computations were performed
    in the early 1970's, using iterative methods \citep{mathis70} and
    Monte-Carlo methods \citep{mattila70}.
  \item[The development of modern detectors] solved the issues of photography: 
    \begin{inlinelist}
      \item non-linear response;
      \item restricted dynamic range;
      \item low detection efficiency;
      \item reciprocity failure; and
      \item adjacency effects
    \end{inlinelist}
    \citep{boksenberg82}.
    The first \expression{Charge-Coupled Device} (\hCCD) was invented in 1969 at
    Bell laboratories \citep{amelio70}.
    In the \hIR, photomultipliers and bolometers were developed in the 1930's, 
    and found important military applications during World War II and later: 
    night vision and guiding rockets \citep[][for an extensive 
    review]{rogalsky12}.
    The first \hIR\ thermal detector had in fact been built 150 years earlier 
    by Sir William \familyname{Herschel}, in 1800.
    He used a prism to split the Sun light over several thermometers 
    \citep[\cf\ \refsubfig{fig:draper}{b};][]{rogalsky12}.
    He found that the highest temperature was beyond the red, that is in the 
    infrared\footnote{The photosphere of the Sun is 
    indeed a $T\simeq5800$~K black body, peaking
    at $\lambda_\sms{max}\simeq0.88\emic$.
    This is the first \hSED\ in history.} (\refsubfig{fig:draper}{c}).
  \item[The possibility to send airborne and space observatories] opened the 
    spectral windows where the atmosphere is opaque (\cf\ 
    \refsec{sec:atmosphere}).
    The interest to send telescopes in space was first advocated for by
    Lyman \familyname{Spitzer} Jr., in 1946 \citep{spitzer46}.
    The first successful space telescope was launched in 1968.
    It was the \expression{Orbiting Astronomical Observatory} (\hOAO), operating
    in the \hUV\ \citep{code70}.
    The first airborne observatory was the balloon experiment 
    \expression{Stratoscope I}, operating in the \hIR, in 1958.
\end{description}
The next technological innovation that will revolutionize our field is difficult to predict. 
We can however note that quantum computers should permit \expression{ab initio} calculations of the properties and evolution of complex molecules and solids, in the near future.
If this is true, it should allow us to precisely characterize the dust composition in different environments.

  \subsection{The Challenges of Observing Interstellar Regions}

    \subsubsection{Limitations Due to the Atmosphere}
    \label{sec:atmosphere}

\begin{figure}[!htbp]
  \includegraphics[width=\textwidth]{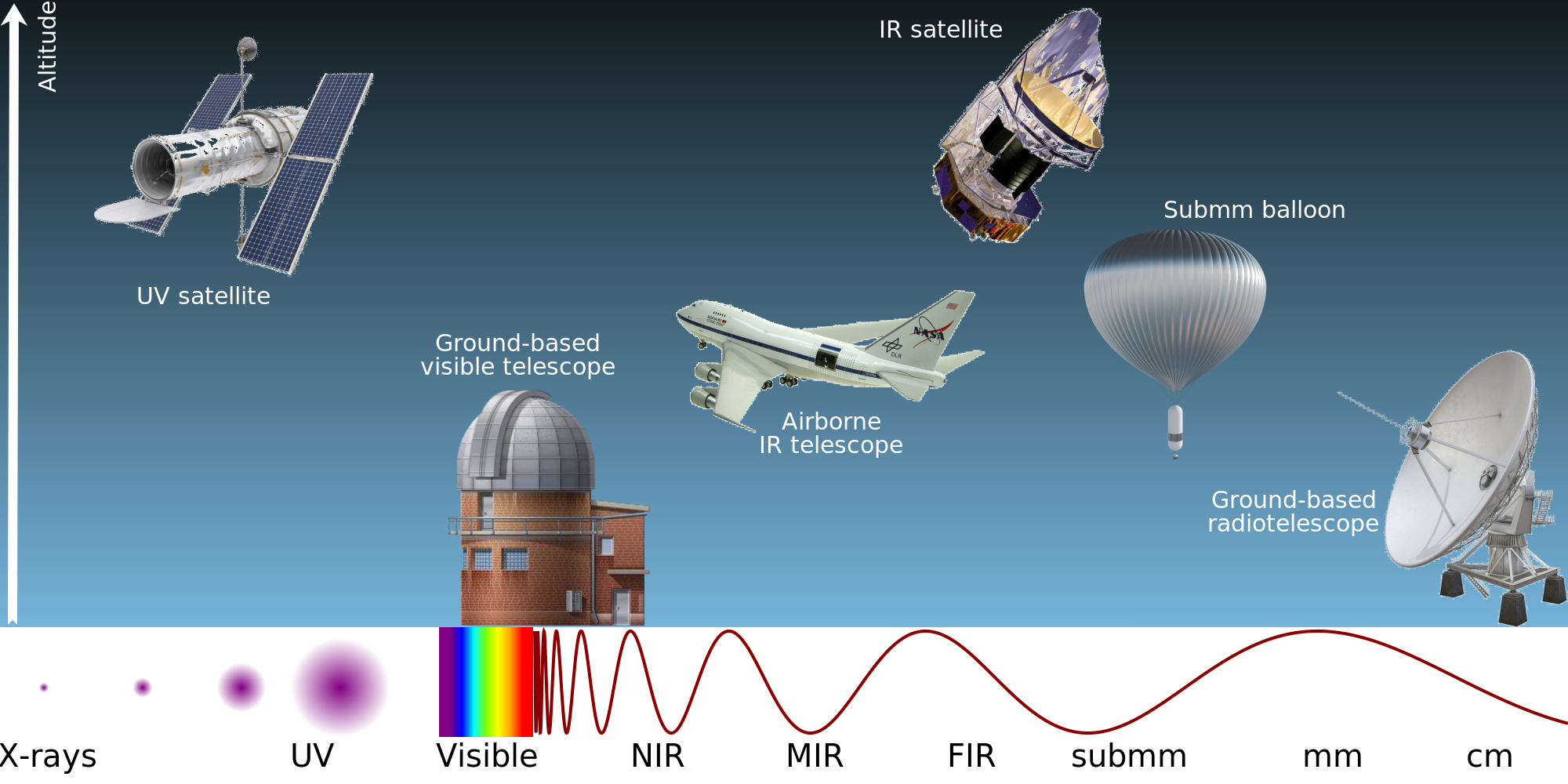}
  \includegraphics[width=\textwidth]{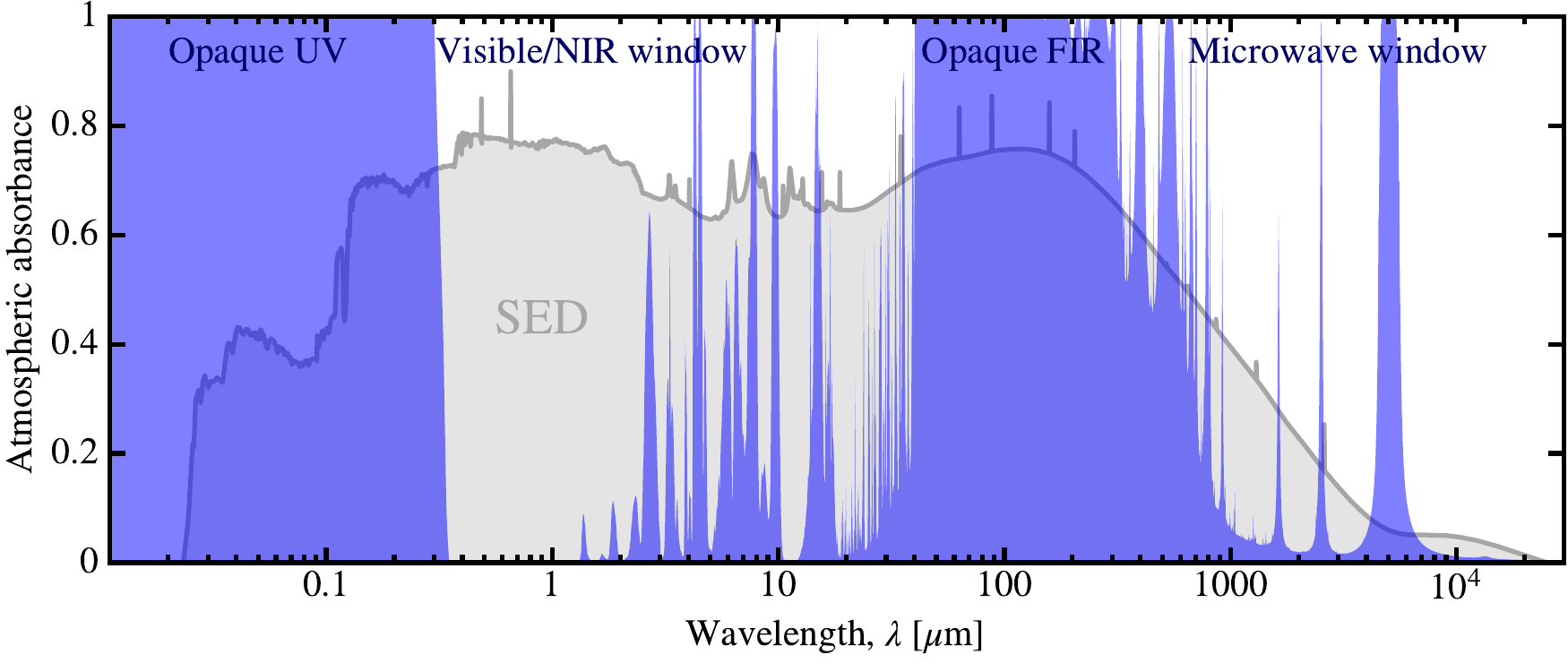}
  \newcap{Absorbance of Earth's atmosphere}%
         {The bottom panel shows the absorbance of the atmosphere, in blue.
          Most of the absorption is due to H$_2$O, with some contribution by
          N$_2$, O$_2$, O$_3$, N$_2$O, CH$_4$ and CO$_2$.
          We have displayed in grey, in the background, the typical \hSED\ of a 
          galaxy, for reference.
          We see that the \hUV\ and \hFIR\ windows are totally opaque from the 
          ground (\cf\ \reftab{tab:spectralrange} for the denomination of the 
          different spectral windows).
          Consequently, observations in these spectral windows can only be
          achieved from space, or above the troposphere (stratospheric airplane 
          or balloon).
          \CClicence}
  \label{fig:atmosphere}
\end{figure}

The atmosphere of Earth is transparent in only a few spectral windows.
The bottom panel of \reffig{fig:atmosphere} shows its absorbance as a function of wavelength.
We see in particular that the \hUV\ and \hFIR\ ranges are completely opaque, and thus inaccessible from the ground.
This is the reason why the first evidences for \hISD\ came from the extinction of starlight in the visible range.
The last forty years have seen the development of space and airborne observatories in the \hUV\ and \hIR\ windows, providing us with a panchromatic view of \hISD\ properties (top panel of \reffig{fig:atmosphere}).

    \subsubsection{Historical Ground-Based Observatories}

\paragraph{The meter-class visible telescopes.}
The first discoveries about \hISD\ were performed in the visible domain, by understanding the extinction toward stars.
Robert \familyname{Trumpler}'s seminal paper \citep{trumpler30} was based on observations made at the Lick observatory, near San Jose, as well as Edward \familyname{Barnard}'s most famous observations \citep{barnard1899,barnard19}.
Such telescopes, all across the world, were the main source of empirical constraints on dust, until the end of the 1970's.

\paragraph{The first MIR telescopes.}
Ground-based observations, at high altitude, in dry regions of the globe such as the Mauna Kea, are possible in the \hMIR\ (\cf\ \reffig{fig:atmosphere}).
\hIR\ astronomy really started in the 1960's \citep[\cf][for an historical review]{walker00}.
The first \hIR\ surveys of the northern and southern skies were performed by \citet{neugebauer68} and \citet{price68}, at 2~\tmic.
A generation of 2-to-3-m \hMIR\ telescopes were commissioned at the end of the 1970's, such as the \expression{Wyoming InfraRed Observatory} (WIRO; 1977; $\diameter=2.3$~m), the \expression{United Kingdom InfraRed Telescope} (UKIRT; 1978; $\diameter=3.8$~m) and the \expression{InfraRed Telescope Facility} (IRTF; 1979; $\diameter=3$~m).
Current large telescopes such as Subaru (Mauna Kea; $\diameter=8.2$~m) or the \expression{Very Large Telescope} (\hVLT; Paranal; $\diameter=8.2$~m) operate in the visible-to-\hMIR\ range.

\paragraph{The submillimeter observatories.}
On the other side of the \hFIR\ atmospheric absorption, the submillimeter domain can be observed from the ground in dry conditions.
The first ground-based submillimeter observatories appeared in 1990's.
Among them are: the \expression{Caltech Submilleter Observatory} (\hCSO; Mauna Kea; $\diameter=10$~m; 1986-2015), the \expression{James Clerk Maxwell Telescope} (\hJCMT; Mauna Kea; $\diameter=15$~m; 1987), the \expression{Atacama Pathfinder Experiment} (\hAPEX; Atacama desert; $\diameter=12$~m; 2004) and the \expression{Atacama Large Millimeter/submillimeter Array} (\hALMA; Atacama desert; interferometer; 2011).

    \subsubsection{Airborne Observatories}

\begin{figure}[htbp]
  \begin{tabular}{ccc}
    \includegraphics[width=0.345\textwidth]{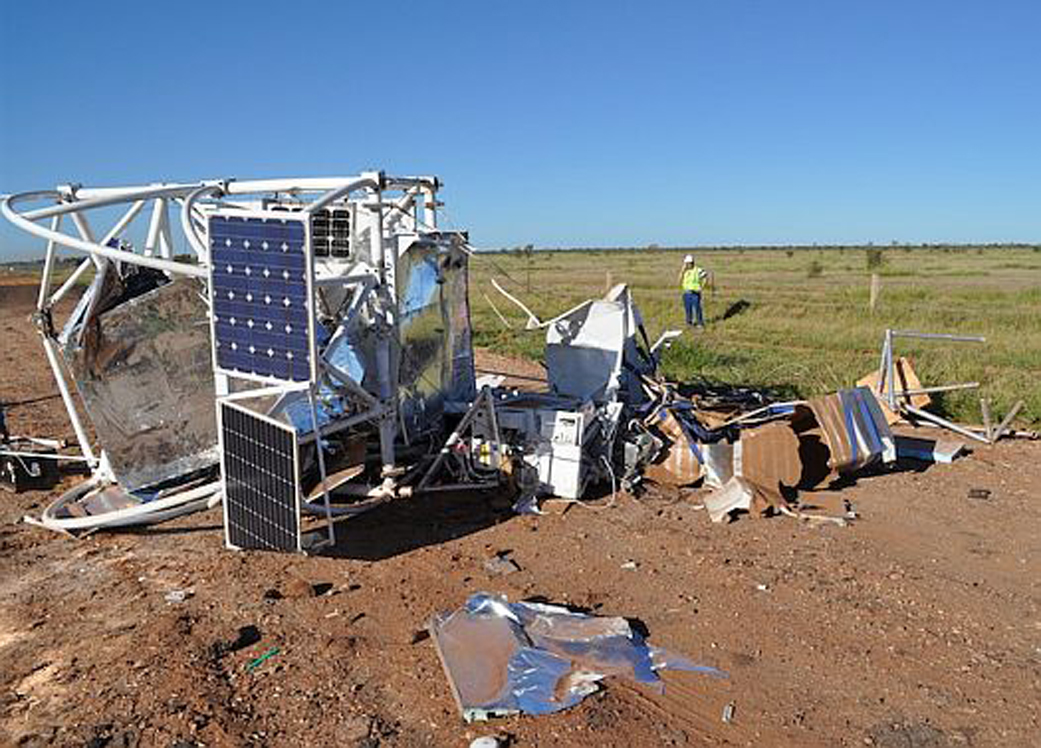} &
    \includegraphics[width=0.275\textwidth]{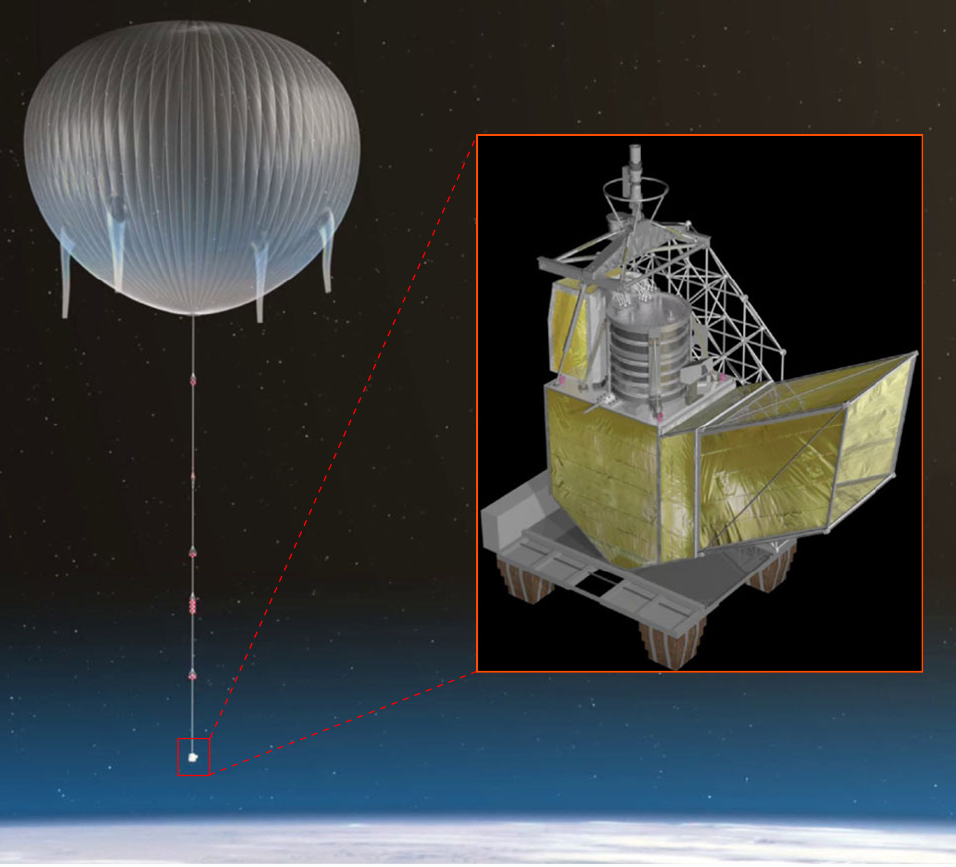} &
    \includegraphics[width=0.31\textwidth]{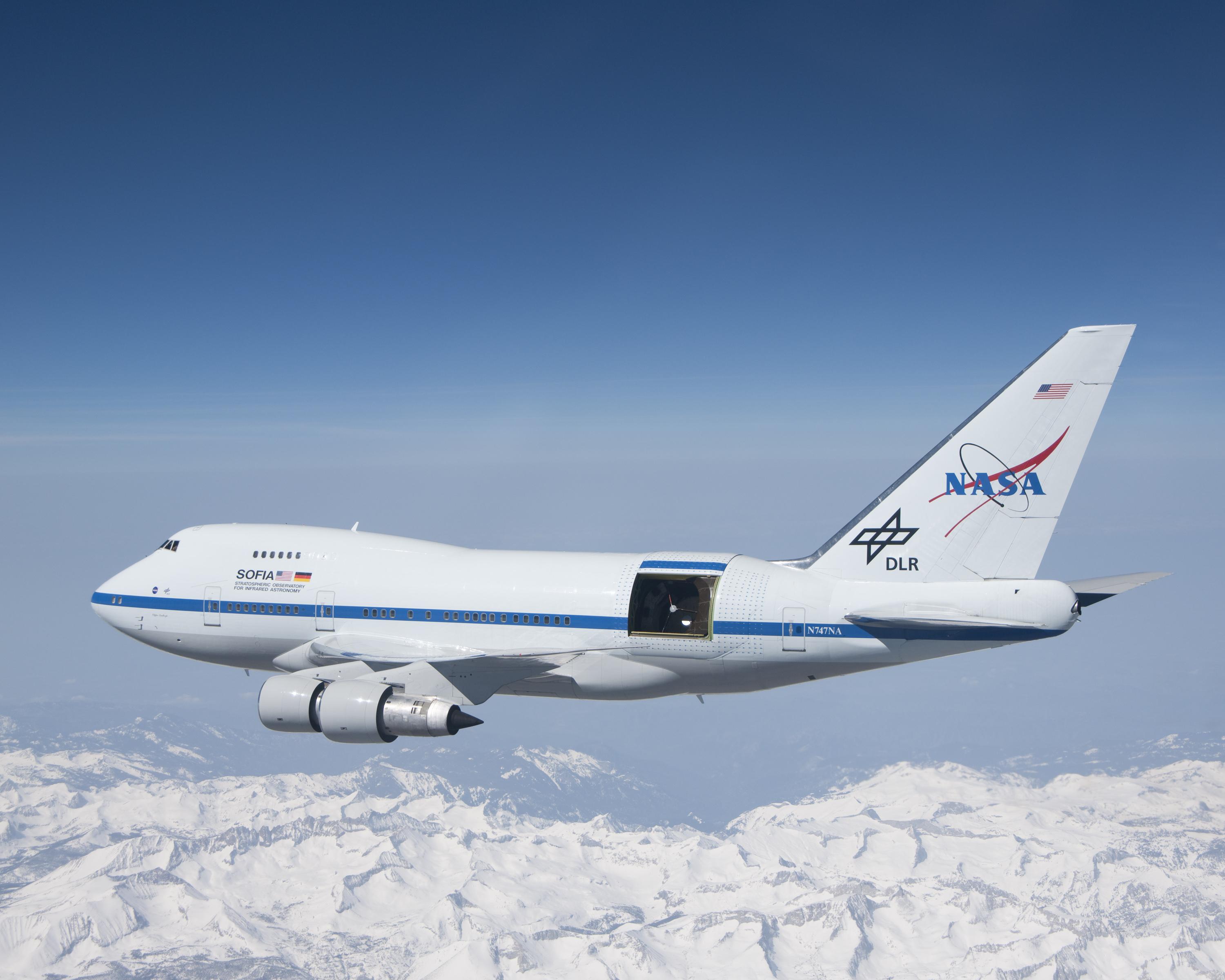} \\
    \textit{(a)} Balloon crash &
    \textit{(b)} \hPILOT &
    \textit{(c)} \hSOFIA \\
  \end{tabular}
  \newcap{Airborne observatories}%
         {Panel~\textit{(a)} shows the 2010 crash site of the 
          \expression{Nuclear Compton Telescope} (\hNCT) in order to illustrate
          the challenges of such observation campaigns.
          Panel~\textit{(b)} shows an artist rendering of the \hPILOT\ balloon
          with a zoom on the gondola where the telescope is \citep{bernard16}.
          Panel~\textit{(c)} shows a picture of \hSOFIA\ in flight, with the 
          telescope door open.
          \uline{Credit:}
          \begin{inlinelistalph}
     \item \href{https://www.nasa.gov/home/hqnews/2010/oct/HQ_10-269_Balloon_Mishap.html}%
            {courtesy of NASA}; 
     \item \citet{bernard16}, with permission from Jean-Philippe \familyname{Bernard}; 
     \item \href{https://www.nasa.gov/mission_pages/SOFIA/overview/index.html}%
            {courtesy of NASA}.
          \end{inlinelistalph}}
  \label{fig:airborne}
\end{figure}
\paragraph{Balloons.}
Stratospheric balloons can reach altitudes of $\simeq40$~km, well above most water vapor absorption.
They can observe for several days continuously, but landing is hazardous (\refsubfig{fig:airborne}{a}).
The first \hIR\ balloon was launched from Johns Hopkins in 1959, and a balloon sent by the Goddard Institute of Space Sciences mapped the sky at 100~\tmic\ in 1966 \citep{walker00}.
During the past two decades, several balloons provided observations of the dust continuum intensity and polarization in different regions of the sky, including:
the \expression{Balloon-borne Large Aperture Submillimeter Telescope} \citep[\hBLAST; 1997-2010; $\diameter=2$~m;][]{pascale08}; the \expression{PROjet National pour l'Observation Submillimétrique} \citep[\hPRONAOS; 1994-1999; $\diameter=2$~m;][]{serra02}; the \expression{Polarized Instrument for the Long-wavelength Observation of the Tenuous \hISM} \citep[\hPILOT; 2015-2017; $\diameter=0.73$~m;][\refsubfig{fig:airborne}{b}]{bernard16}.

\paragraph{Airplanes.}
Airplanes can fly up to $\simeq15$~km altitude and operate during $\simeq10$~h.
Numerous flights can be scheduled, contrary to balloons, which usually do only a few flights in their whole lifetime.
The telescope motion being limited by its orientation perpendicular to the plane (\refsubfig{fig:airborne}{c}), the flight path has to be adapted to the observed source.
The \expression{Kuiper Airborne Observatory} \citep[\hKAO; 1974-1995; $\diameter=0.9$~m;][]{erickson13} was a transport jet plane converted into an observatory.
The \expression{Stratospheric Observatory for Infrared Astronomy} \citep[\hSOFIA; 2010-; $\diameter=2.5$~m;][\refsubfig{fig:airborne}{c}]{young12} is its current successor.
It is a retired Boeing 747, modified to host the telescope.

    \subsubsection{Space Telescopes}

\paragraph{IRAS.}
The \expression{InfraRed Astronomical Satellite} \citep[\hIRAS; 1983; $\diameter=0.57$~m;][]{neugebauer84} was the first observatory to perform an all-sky survey at \hIR\ wavelengths.
It mapped the sky in four broadbands centered at $\lambda=12$, 25, 60 and 100~\tmic, with angular resolutions of $0.5^\prime-2^\prime$.
It opened the \hIR\ window, which was largely unexplored at the time.
It discovered more than $300\,000$ point sources, many of them being starburst galaxies.
These new objects, with deeply embedded star formation at the scale of the whole galaxy, emitting more than $95\,\%$ of their luminosity in the \hIR, were unexpected \citep[\eg][for a review]{soifer87}.
The new categories of \expression{Luminous InfraRed Galaxies} (\hLIRG) and \expression{UltraLuminous InfraRed Galaxies} (\hULIRG) were created to describe what had been observed.
Dusty disks around stars were also discovered \citep{beichman87}.
By accessing the cold grain emission, the first reliable dust masses of galaxies and Galactic clouds could be estimated.
The \hIR\ emission provided a new constraint that shaped modern dust models \citep{desert90}.
\hIRAS\ data are still used nowadays \citep[\eg][, hereafter \citetalias{galliano21}]{galliano21}.

\paragraph{COBE.}
The \expression{COsmic Background Explorer} \citep[\hCOBE; 1989-1993; $\diameter=0.2$~m;][]{boggess92} was aimed at mapping the \hCMB, as its name indicates.
However, two of its three instruments were used to map the whole sky in the \hMIR\ and \hFIR, providing the main constraints on the emission of dust models until the \hplanck\ mission \citep{sodroski94,dwek97}.
The third instrument, covering the microwave range was also instrumental in providing the first evidence of spinning grains.
\begin{description}
  \item[DIRBE] (\expression{Diffuse Infrared Background Experiment}) was an 
    instrument observing through ten broadbands at: $\lambda=1.25$, 2.2, 3.5, 
    4.9, 12, 25, 60, 100, 140 and 240~\tmic\ \citep{hauser98}.
  \item[FIRAS] (\expression{Far-InfraRed Absolute Spectrophotometer}) was a 
    low spectral resolution spectro-imager, observing between 
    $\lambda=100$~\tmic\ and 10~mm \citep{mather99}.
    At long wavelengths, the angular resolution was only $7^\circ$.
  \item[DMR] (\expression{Differential Microwave Radiometer}) was mapping the
    sky in three broadbands centered at $\lambda=3.3$, 5.6 and 9.5~cm
    \citep{smoot94}.
\end{description}

\paragraph{ISO.}
The \expression{Infrared Space Observatory} \citep[\hISO; 1995-1998; $\diameter=0.6$~m;][]{kessler96} was the first mission to extensively perform spectroscopy over the whole \hIR\ range.
For that reason, it provided a wealth of data about all spectral features: silicates \citep{molster05}, \hPAH s \citep{abergel05,sauvage05}, ices \citep{dartois05}.
Studies of \hIR\ gas lines also took off: molecular \citep{habart05} and ionized \citep{peeters05}.
Finally, it refined our knowledge, through dust tracers, of star formation at all scales \citep{nisini05,verma05,elbaz05}.
There were four instruments onboard.
\begin{description}
  \item[ISOCAM] \citep{cesarsky96cam} was a low spectral resolution \hMIR\ 
    spectro-imaging camera, in the $\lambda=2.5-17$~\tmic\ range.
    It also had twenty broad and narrow bands in the same range.
  \item[SWS] \citep{de-graauw96} was a medium to high spectral resolution 
    ($R\equiv\lambda/\Delta\lambda=1\,000-35\,000$) 
    \hMIR\ spectrometer in the $\lambda=2.4-45$~\tmic\ range.
  \item[LWS] \citep{clegg96} was a low to medium spectral resolution 
    ($R\simeq150-9700$) \hFIR\ spectrometer, in the $\lambda=43-198$~\tmic\ 
    range.
    Combined together, SWS and LWS provided several continuous spectra over the
    whole \hIR\ range \citep[$\lambda=2.4-198$~\tmic; \eg][]{peeters02}, that 
    have never been equaled.
  \item[ISOPHOT] \citep{lemke96} was a photometer observing through several 
    broad and narrow band filters, in the  $\lambda=2.5-240$~\tmic.
\end{description}

\paragraph{Spitzer.} 
The \expression{Spitzer space telescope} \citep[cryogenic operation: 2003-2009; $\diameter=0.85$~m;][]{werner04} was the successor of \hISO.
Its larger mirror size and more modern detectors allowed it to refine our understanding of what \hISO\ discovered, and observe a significantly larger number of targets.
Its angular resolution was $40^{\prime\prime}$ at 160~\tmic.
It had three instruments onboard.
\begin{description}
  \item[IRAC] \citep[\expression{InfraRed Array Camera};][]{fazio04} was 
    performing photometry through four broadbands, centered at $\lambda=3.6$, 
    4.5, 5.8 and 8.0~\tmic.
  \item[IRS] \citep[\expression{InfraRed Spectrograph};][]{houck04irs} was a
    medium and high spectral resolution ($R=90-600$) spectro-imager, observing 
    in the $\lambda=5.3-38$~\tmic\ range.
  \item[MIPS] \citep[\expression{Multiband Imaging Photometer for 
    Spitzer};][]{rieke04} was a photometer observing through three broadbands
    centered at $\lambda=24$, 70 and 160~\tmic.
\end{description}

\paragraph{AKARI.}
The \expression{AKARI space telescope} \citep[cryogenic phase: 2006-2008; $\diameter=0.69$~m;][]{murakami07} was comparable to \hspitz.
One of its advantages was its ability to record spectra down to 2~\tmic, while \hspitz/\hIRS\ was limited to 5~\tmic.
\hAKARI\ performed an all-sky survey in several \hMIR\ to \hFIR\ bands.
It had two instruments onboard.
\begin{description}
  \item[IRC] \citep[\expression{InfraRed Camera};][]{onaka07} was a \hMIR\ 
    camera with numerous broad and narrow bands, as well as a low/mid-spectral 
    resolution spectrometer, observing in the $\lambda=1.8-26.5$~\tmic\ range.
  \item[FIS] \citep[\expression{Far-Infrared Surveyor};][]{kawada07} was a 
    \hFIR\ photometer observing through four broadbands centered at 
    $\lambda=65$, 90, 140 and 160~\tmic.
    It also had a \expression{Fourier Transform Spectrometer} (\hFTS) over the
    same range.
\end{description}

\paragraph{WISE.}
The \expression{Wide-field Infrared Survey Explorer} \citep[\hWISE; 2009-2011; $\diameter=0.4$~m;][]{wright10} was a \hMIR\ all sky surveyor.
It mapped the sky through four broad photometric bands centered at $\lambda=3.4$, 4.6, 12 and 22~\tmic.

\paragraph{Herschel.}
The \expression{Herschel space observatory} \citep[2009-2013; $\diameter=3.5$~m;][]{pilbratt10} was a \hFIR-submm mission.
Its large mirror allowed it to reach subarcminute angular resolution at long wavelength ($36^{\prime\prime}$ at $\lambda=500$~\tmic).
Combined with \hspitz\ data at shorter wavelengths, it gives access to the full dust emission and provides the most reliable dust property estimates of galaxies and Galactic regions.
\hhersc\ data allowed us to build large databases of galaxy dust properties \citep[\eg][]{davies17}.
It also allowed us to better constrain the submillimeter grain opacity \citep{meixner10,galliano11}.
Among its discoveries, it demonstrated the filamentary nature of star-forming regions \citep{andre10}.
It had three instruments onboard.
\begin{description}
  \item[PACS] \citep[\expression{Photodetector Array Camera and 
    Spectrometer};][]{poglitsch10} was an imager observing through three 
    broadbands centered at $\lambda=70$, 100 and 160~\tmic.
    It also had a spectrometer that could target specific \hFIR\ lines.
  \item[SPIRE] \citep[\expression{Spectral and Photometric Imaging 
    REceiver};][]{griffin10} was a photometer observing through three broadbands
    centered at $\lambda=250$, 350 and 500~\tmic.
    It also had a \hFTS\ providing a continuous, medium spectral resolution 
    spectrum over the $\lambda=194-671$~\tmic\ spectral range.
  \item[HIFI] \citep[\expression{Heterodyne Instrument for the 
    Far-Infrared};][]{de-graauw10} was a very high spectral resolution 
    ($R\simeq10^7$) spectrometer covering the $\lambda=157-625$~\tmic\ spectral
    range.
    It was designed to accurately measure gas line intensities.
\end{description}

\begin{figure}[!htbp]
  \includegraphics[width=\textwidth]{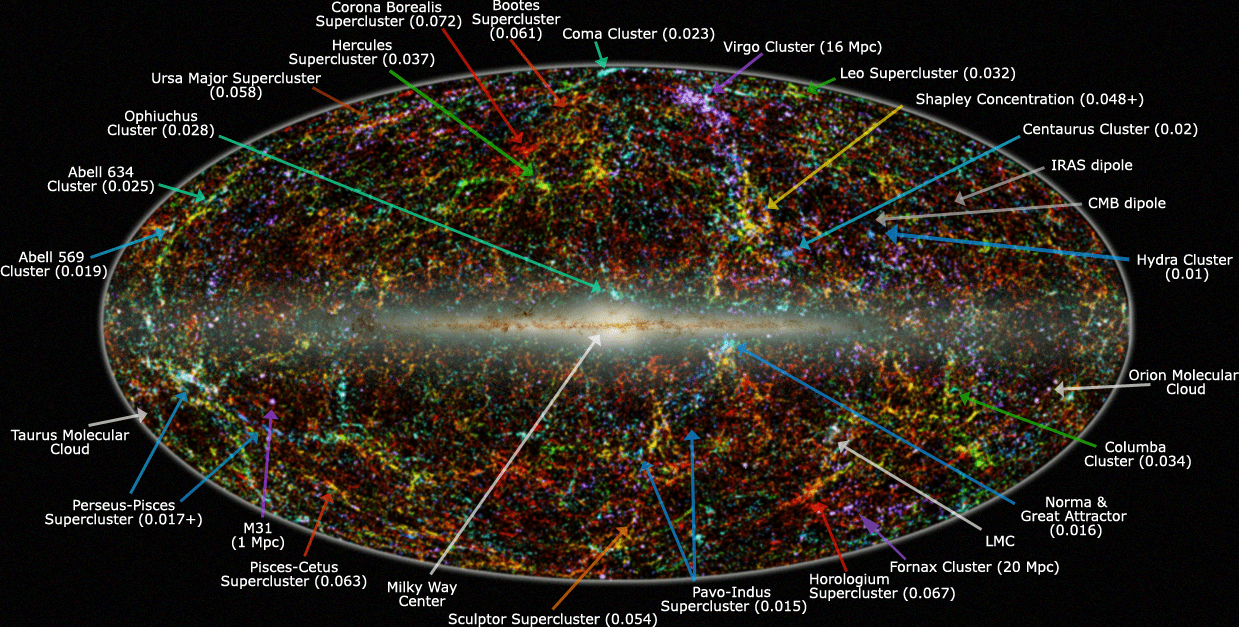}
  \begin{center}
    \textit{(a)} Uranography of all sky map projections
  \end{center}
  \includegraphics[width=\textwidth]{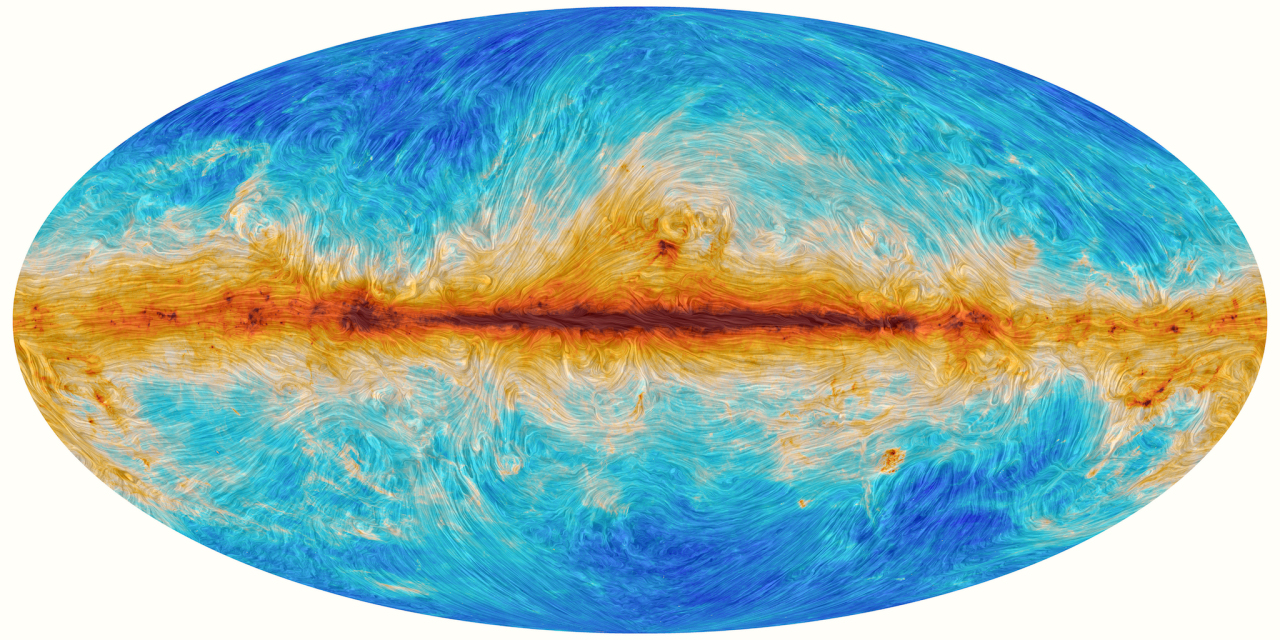}
  \begin{center}
    \textit{(b)} \hplanck\ all sky map of the dust polarized emission
  \end{center}
  \newcap{All sky maps}%
         {Panel~\textit{(a)} describes the main features of the usual all sky
          projections. 
          It shows both Galactic and extragalactic structures.
          Panel~\textit{(b)} shows the all sky map of the dust polarized 
          emission.
          It is a \expression{Van Gogh representation}, where the brush lines
          show the orientation of the projected magnetic field.
          \uline{Credit:} 
          \begin{inlinelistalph}
            \item \citet{jarrett04}, with permission from Tom \familyname{Jarrett};
            \item \href{https://www.notion.so/The-interstellar-magnetic-field-revealed-by-Planck-59e5c044cdde44019c6084fcc0f5ce4c}{copyright ESA/\planck\ collaboration, credit Marc-Antoine 
               \familyname{Miville-Deschênes}}, with his permission.
          \end{inlinelistalph}}
  \label{fig:allsky}
\end{figure}
\paragraph{Planck.}
The \expression{Planck space observatory} \citep[2009-2013; $\diameter=1.5$~m;][]{tauber10} was a \hFIR-to-microwave satellite designed to study the cosmological background.
It is a successor to \hCOBE.
It was launched in the same rocket as \hhersc.
It performed an all sky survey in all its bands (\reffig{fig:allsky}).
\hplanck\ had a larger beam than \hhersc\ ($5^\prime$ at $\lambda=1$~mm), but had an accurate absolute calibration.
Its measure of the emission of the diffuse \hISM\ of the \hMW\ is now the main constraint on dust models \citep[\eg][]{compiegne11}.
\hplanck\ could also measure the linear polarization in all its bands.
It thus provided unique constraints on the grain properties \citep[\eg][]{guillet18} and maps of the Galactic magnetic field \citep[\eg][]{planck-collaboration16g}.
It had two instruments onboard.
\begin{description}
  \item[HFI] \citep[\expression{High Frequency Instrument};][]{lamarre10} was
    a photometer/polarizer observing through six broadbands centered at 
    $\lambda=350$, 550, 850, 1380, 2096 and 2997~\tmic.
  \item[LFI] \citep[\expression{Low Frequency Instrument};][]{mandolesi10} was
    a photometer/polarizer observing through three broadbands centered at 
    $\lambda=4.3$, 6.8 and 10~cm.
\end{description}

\paragraph{The JWST.}
The \expression{James Webb Space Telescope} 
\citep[\hJWST; 2021-; $\diameter=6.5$~m;][]{mcelwain20} should be launched a few months after the time this manuscript is being written.
Its large segmented mirror, that will unfold in space, will allow us to access sub-arcsec resolution in the \hMIR.
It will have four instruments onboard.
\begin{description}
  \item[MIRI] \citep[\expression{Mid-InfraRed Instrument}; 
    $\lambda=5-27$~\tmic;][]{rieke15} contains an camera and an imaging 
    spectrometer.
    It will be the most relevant instrument to \hISD\ studies.
  \item[NIRspec] \citep[\expression{Near-InfraRed Spectrograph}; 
    $\lambda=0.6-5$~\tmic;][]{birkman16} is a \hNIR\ spectrometer.
  \item[NIRcam] \citep[\expression{Near-InfraRed Camera}; 
    $\lambda=0.6-5$~\tmic;][]{beichman12} is a \hNIR\ camera.
  \item[NIRISS] \citep[\expression{Near-InfraRed Imager and Slitless 
    Spectrograph}; $\lambda=0.8-5$\tmic;][]{doyon12} will perform \hNIR\ 
    imaging and spectroscopy.
\end{description}

\paragraph{UV and X-ray Satellites.}
The \hUV\ spectral shape of the extinction curve is an important constraint on dust models.
\hUV\ satellite have one advantage over \hIR\ instruments: they do not need to be cooled down.
\hIR\ instruments indeed need a cryostat to limit their proper emission.
The lifetime of \hIR\ missions is thus the lifetime of their helium supply, typically only a few years, whereas \hUV\ telescopes can operate during several decades.
The most important \hUV\ missions are the following.
\begin{description}
  \item[IUE] \citep[\expression{International Ultraviolet Explorer}; 1978-1996; 
    $\diameter=0.45$~m;][]{boggess78} was the first important \hUV\ mission, 
    expanding our view on dust extinction at $\lambda=115-320$~nm.
  \item[HST] \citep[\expression{Hubble Space Telescope}; 1990-; 
    $\diameter=2.4$~m;][]{burrows91} can take spectra in the near-\hUV\ range.
  \item[FUSE] \citep[\expression{Far Ultraviolet Spectroscopic Explorer}; 
    1999-2007; $\diameter=1.5$~m;][]{moos00} extended the spectral coverage of 
    \hIUE\ at $\lambda=90.5-110.5$~nm.
\end{description}
As we will see in \refsec{sec:Xrays}, the X-ray regime can provide interesting constraints on the dust properties.
The most important missions are: 
ROSAT \citep[1990-1999; $\diameter=0.84$~m; $\lambda=0.06-30$~nm;][]{aschenbach91}, 
XMM-Newton \citep[1999-; $\diameter=0.7$~m; $\lambda=0.1-12$~nm;][]{jansen01} and Chandra \citep[1999-; $\diameter=1.2$~m; $\lambda=0.1-12$~nm;][]{weisskopf02}.
The \expression{Advanced Telescope for High ENergy Astrophysics} \citep[\hATHENA; $\simeq2030$;][]{wilms14} will revolutionize the field.

    \subsubsection{Grain-Collecting Spacecrafts}
    \label{sec:graincollect}

\begin{figure}[htbp]
  \begin{tabular}{ccc}
    \includegraphics[width=0.35\textwidth]{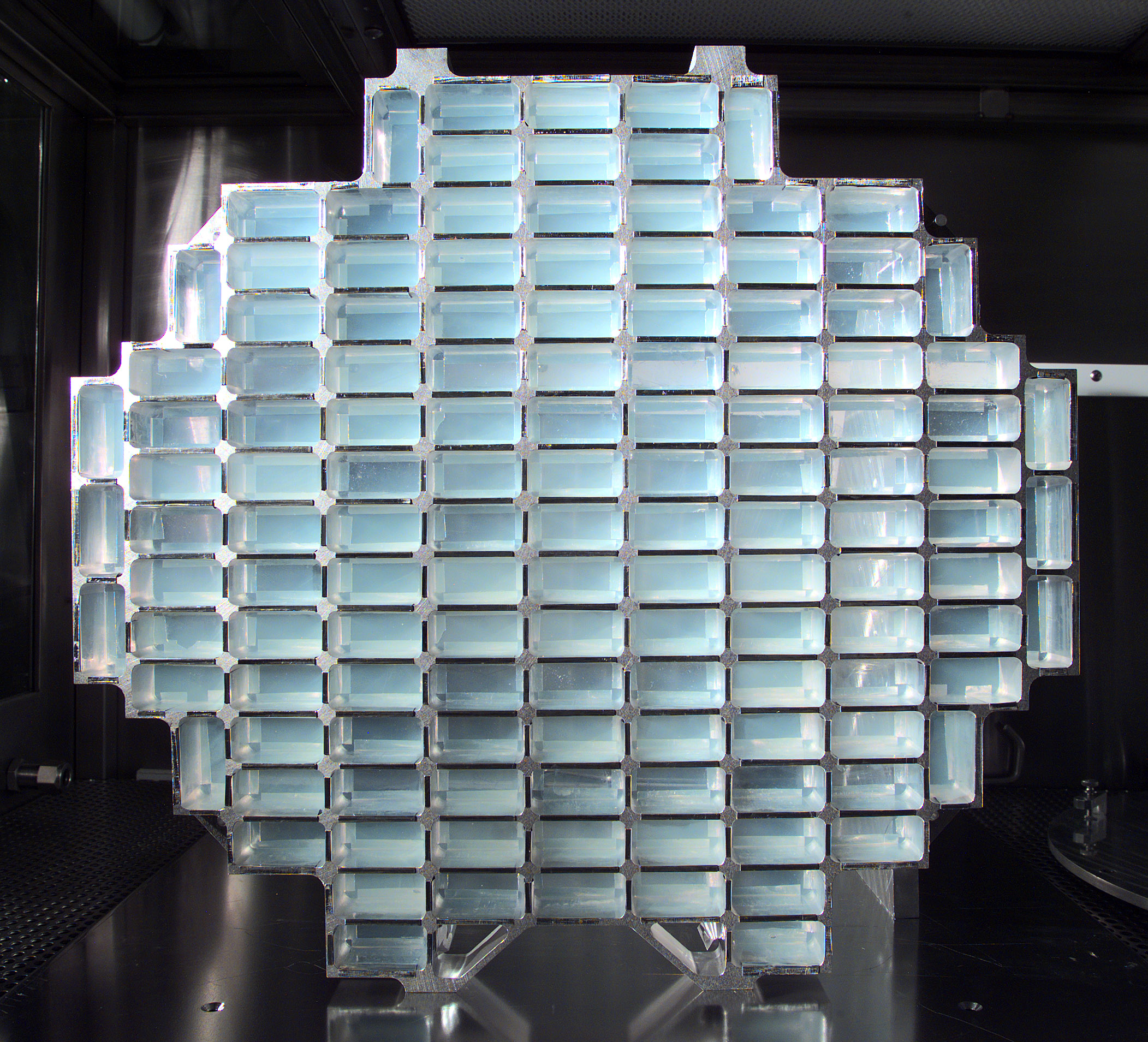} &
    \includegraphics[width=0.25\textwidth]{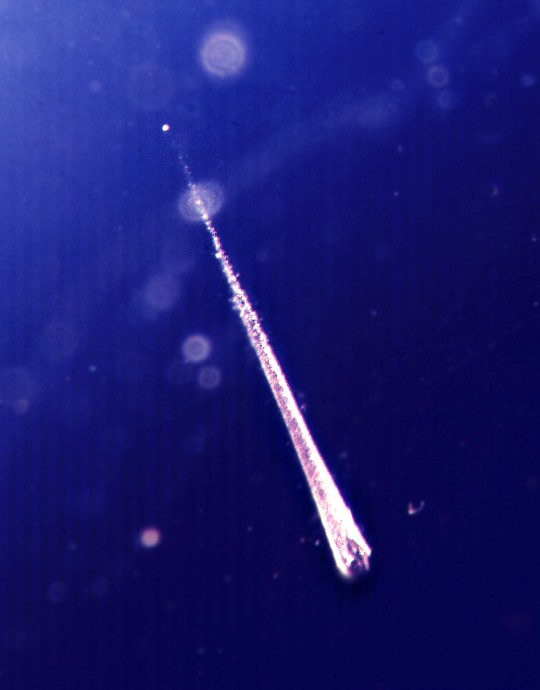} &
    \includegraphics[width=0.335\textwidth]{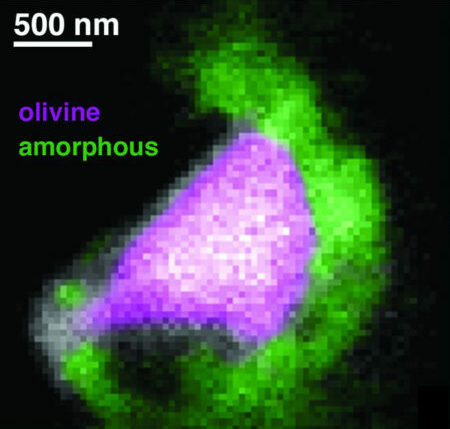} \\
    \textit{(a)} Aerogel honeycomb matrix &
    \textit{(b)} Aerogel dust track &
    \textit{(c)} X-ray image of a grain \\
  \end{tabular}
  \newcap{Analysis of the Stardust mission}%
         {Panel~\textit{(a)} shows the honeycomb matrix of Stardust.
          Each array is filled with an ultralight silica aerogel.
          It is $\simeq1000$ times less dense than glass.
          Dust grains arriving at several km/s are slowed down without being
          pulverized.
          Panel~\textit{(b)} shows the cone-shaped track of one of the grains
          captured in the aerogel.
          Panel~\textit{(c)} shows the X-ray image of one of 
          the grains.
          The magenta part corresponds to olivine crystals, surrounded by 
          non-crystalline magnesium silicate in green.
          \uline{Credit:}
          \begin{inlinelistalph}
            \item \href{https://stardust.jpl.nasa.gov/photo/spacecraft.html}%
                       {courtesy of NASA/JPL}; 
            \item \href{https://stardust.jpl.nasa.gov/photo/aerogel1.html}%
                       {courtesy of NASA/JPL};
            \item \href{https://als.lbl.gov/space-dust-analysis-could-provide-clues-to-solar-system-origins/}{Anna Butterworth/UC Berkeley from STXM data, courtesy of Berkeley Lab.}
          \end{inlinelistalph}}
  \label{fig:stardust}
\end{figure}
\noindent
Electromagnetic waves are not the only vectors of information about \hISD\ we can get.
The motion of the heliosphere relative to the local interstellar cloud creates an inflow of \hISD\ through the Solar system \citep[at 26~km/s; \eg][for a review]{kruger19}.
Contrary to interplanetary grains, this interstellar flow is important at high ecliptic latitude, allowing us to discriminate grains from extrasolar origins.
Several spacecrafts have collected actual interstellar grains, and analyzed them \textit{in situ} or returned them to Earth.
\begin{description}
  \item[Ulysses] \citep[1990-2009;][]{bame92} was a spacecraft designed to 
    analyze the Solar wind. 
    It was the first mission to capture dust grains from interstellar origin.
  \item[Galileo] \citep[1989-2003;][]{johnson92} was a spacecraft sent to study
    Jupiter and its satellites.
    It detected interstellar grains on its way.
  \item[Cassini] \citep[1997-2017;][]{matson02} was a spacecraft sent to study 
    Saturn and its satellites.
    It embarked an instrument called the \expression{Cosmic Dust Analyzer}
    (CDA), recording the size, speed, direction and chemical composition of 
    interstellar grains.
    It identified thirty six of them, smaller than 200~nm \citep{atobelli16}.
    They appeared to be essentially Mg-rich silicates with iron inclusions.
  \item[Stardust] \citep[1999-2006;][]{brownlee03} was a spacecraft that 
    captured grains in a low-density aerogel, and returned them to Earth for 
    laboratory analysis (\reffig{fig:stardust}).
    It identified seven interstellar grains \citep{westphal14}.
    Those were Mg-rich silicates, with sizes $\gtrsim1$~\tmic.
    Two of them had crystalline structures.
\end{description}

  \subsection{Chronology of the Main Breakthroughs}
  \label{sec:chronology}

In what follows, we present the main discoveries about \hISD.
We order the discussion by themes.
\reftab{tab:chronology} puts all these breakthroughs in chronological order.
This is a partial and incomplete review.
We refer the reader to \citet{van-de-hulst86}, \citet{dorschner03}, \citet{whittet03}, \citet{li03b} and \citet{li05}, for more complete historical reviews.

    \subsubsection{Obscuration and Dimming of Starlight}
    \label{sec:histodimming}

\begin{figure}[htbp]
  \begin{tabular}{cc}
    \includegraphics[width=0.39\textwidth]{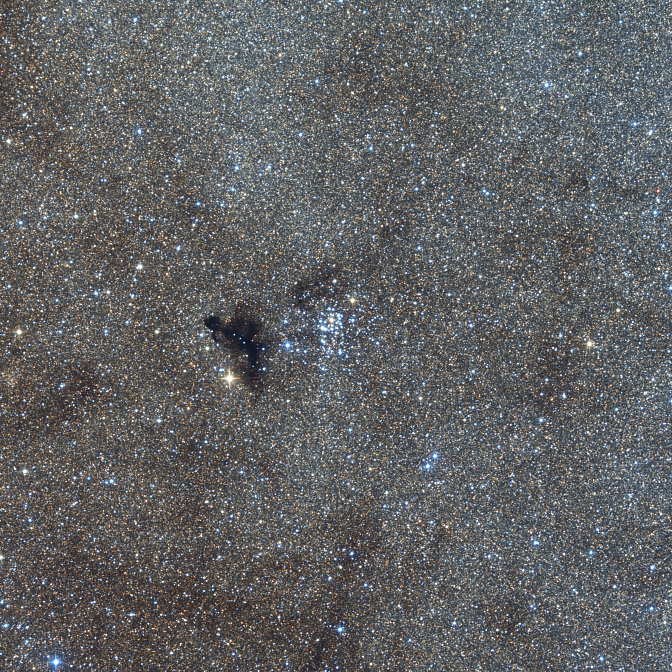} &
    \includegraphics[width=0.57\textwidth]{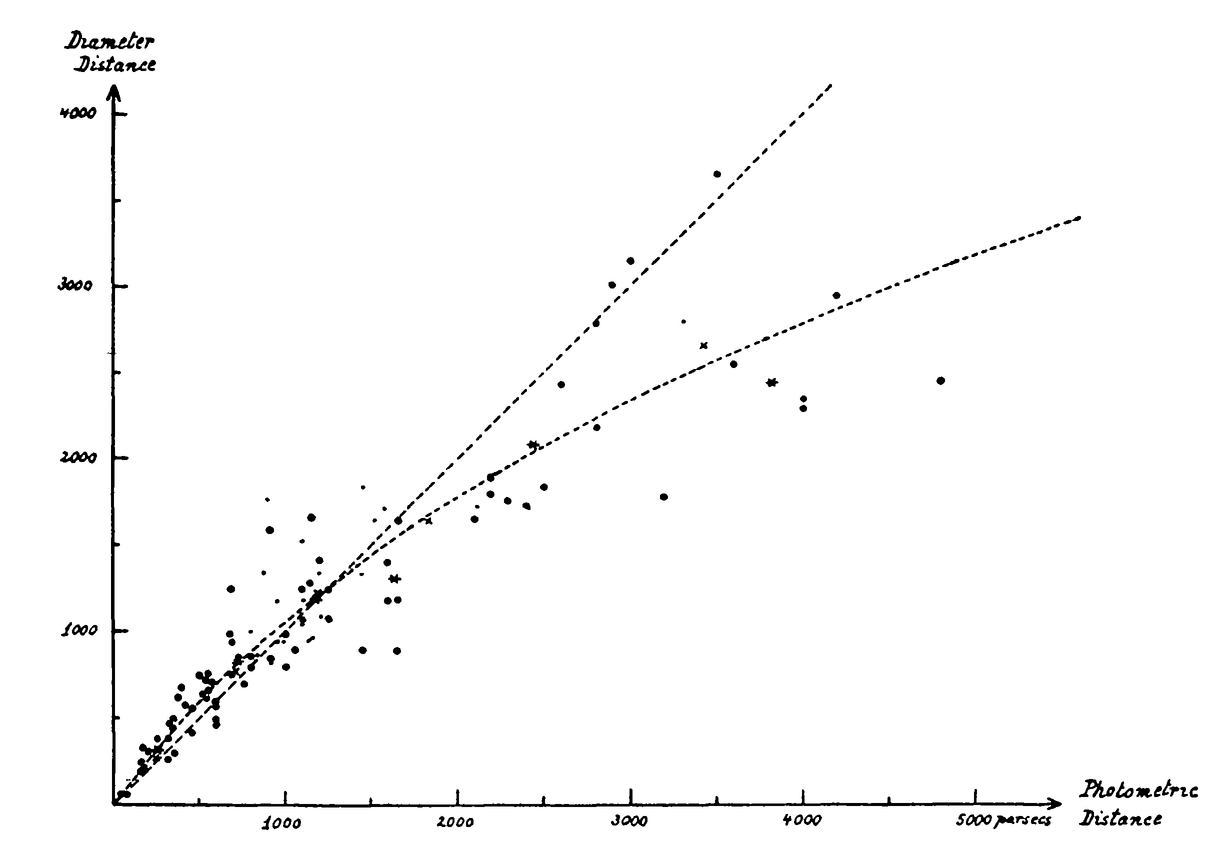} \\
    \textit{(a)}~The Ink Spot nebula &
    \textit{(b)}~\citet{trumpler30}'s relation \\
  \end{tabular}
  \newcap{First evidences of interstellar dust}%
         {Panel~\textit{(a)} shows the \expression{Ink spot} nebula (B$\,$86),
          that was originally mistaken by Herschel as a \citengl{hole in 
          the heavens}.
          Panel~\textit{(b)} shows the relation between the photometric and 
          diameter distances of 100 open clusters \citep[Fig.~1 
          of][]{trumpler30}.
          The non-linearity of this relation provided the first unambiguous 
          evidence for \hISD.
          Notice the absence of error bars:
          this plot is from a long gone epoch, when major scientific
          discoveries could be accompanied by a feeling of airiness and 
          eyeball statistics.
          \uline{Credit:} 
          \begin{inlinelistalph}
            \item \href{http://astro.i-net.hu/node/186}{Gábor Tóth Astrophotography}, 
              licensed under 
              \href{https://creativecommons.org/licenses/by-nc-nd/2.0/}{CC BY-NC-ND};
            \item \citet{trumpler30}.
          \end{inlinelistalph}}
  \label{fig:1stevidence}
\end{figure}

\begin{figure}[htbp]
  \begin{tabular}{ccc}
    \includegraphics[width=0.31\textwidth]{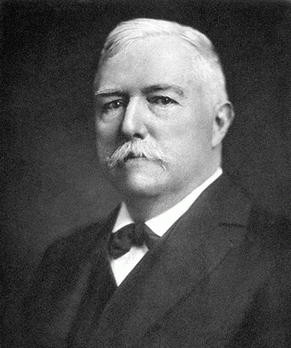} &
    \includegraphics[width=0.31\textwidth]{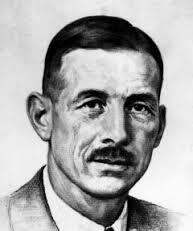} &
    \includegraphics[width=0.31\textwidth]{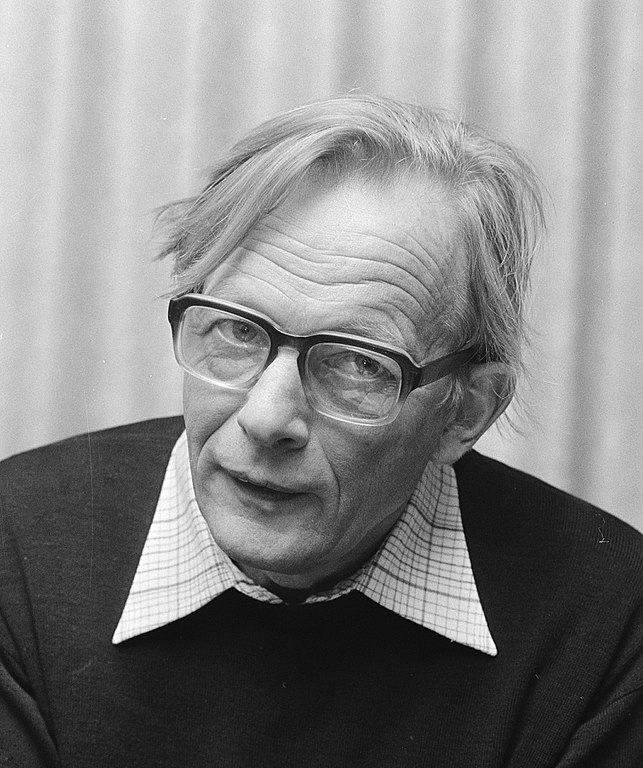} \\
    \textit{(a)} Edward E.\ \familyname{Barnard} &
    \textit{(b)} Robert J.\ \familyname{Trumpler} &
    \textit{(c)} Hendrik C.\ \familyname{van de Hulst} \\
    (1857--1923) & (1886--1956) & (1918--2000) \\
  \end{tabular}
  \newcap{The pioneers}{\uline{Credit:}
    \begin{inlinelistalph}
      \item  
       \href{https://commons.wikimedia.org/wiki/File:EdwardEmersonBarnard.jpg}{Wikipedia}, 
       public domain;
      \item \citet{weaver57};
      \item   
        \href{https://commons.wikimedia.org/wiki/File:Hendrik_C._van_de_Hulst_1977.jpg}%
        {Rob \familyname{Bogaerts}}, licensed under 
        \href{https://creativecommons.org/licenses/by-sa/3.0/nl/}{CC BY-SA 3.0 NL}.
    \end{inlinelistalph}}
  \label{fig:pioneers}
\end{figure}

\paragraph{The discovery of dark nebulae.}
The first evidence of \hISD\ came through the obscuration of visible starlight.
There was a debate during the whole XIX$^\sms{th}$ century about the reality of this obscuration.
\begin{itemize}
  \item Sir William \familyname{Herschel}, in his treaty on 
    \expression{The Construction of the Heavens} \citep{herschel1785}, 
    reported\citengl{an opening or a hole} in the Scorpius constellation.
    This hole was later identified as the \expression{Ink Spot} nebula, B$\,$86 
    \citep[\refsubfig{fig:1stevidence}{a}; \cf][for the historical branching 
    out of this discovery]{steinicke16}.
  \item Edward \familyname{Barnard}'s photographs of Ophiucus showed dark 
    lanes through the nebula \citep{barnard1899}.
    They were at the time interpreted by Agnes \familyname{Clerke}, as 
    \citengl{glades and clearing} in the stellar distribution \citep{clerke03}.
  \item Twenty years later, Edward \familyname{Barnard} 
    (\refsubfig{fig:pioneers}{a}) realized these black patches were actually 
    \citengl{real, obscuring masses, most probably dark nebulae} 
    \citep{barnard19}.
\end{itemize}

\paragraph{The reddening of starlight.}
The selective extinction of starlight provided the first consensual evidence of \hISD.
This was realized at the beginning of the 1930's.
\begin{itemize}  
  \item Already in the middle of the XIX$^\sms{th}$ century, Friedrich Georg 
    Wilhelm \familyname{Struve} found that the stellar volume density is
    decreasing with distance from the Sun \citep{struve1847}.
    This decrease could be explained by a $\simeq1$~mag/kpc absorption by 
    interstellar material.
    This is a factor of $\simeq2$ from the actual value, which is 
    wavelength-dependent.
  \item Dark nebulae and the extinction of starlight were both understood by
    Henry N.~\familyname{Russell} \citep{russell22}.
    In advance on his time, he noticed that \citengl{in certain instances stars 
    embedded in dense luminous nebulosity are abnormally red}.
    He finally deduced that this \citengl{obscuration of light in space, 
    therefore, whether general or selective with respect to wave-length, will 
    be produced mainly by dust particles a few millionths of an inch in 
    diameter} (\ie\ $\simeq0.03-0.1$~\tmic).
  \item The seminal study of Robert J.\ \familyname{Trumpler} 
    \citep[\refsubfig{fig:pioneers}{b};][]{trumpler30} provided the first solid 
    evidence of \hISD.
    His study was based on the observation, at the Lick observatory, of 100 
    open clusters.
    He found that the \expression{diameter distances} (assuming all clusters 
    have the same diameter) are always smaller than the \expression{photometric 
    distances} (based on the stellar spectral types), and this discrepancy 
    increases with distance (\refsubfig{fig:1stevidence}{b}).
    This allowed him to interpret stellar color excesses as selective extinction
    by fine dust particles.
    He inferred particles of average mass $10^{-19}$~g or 
    larger\footnote{Assuming $\rho\simeq3\;\textnormal{g/cm}^3$, the radius 
    of these particles would be $a\simeq2$~nm or larger.}.
    At the same time, \citet{schalen29,schalen31} independently arrived at a 
    similar conclusions, at the Uppsala observatory.
\end{itemize}

  \subsubsection{The Dust Continuum}
  \label{sec:histodustcont}

\paragraph{The Shape of the extinction curve.}
The investigation of the spectral shape of the extinction curve started right after Trumpler's study.
\begin{itemize}
  \item A series of papers \citep{rudnick36,hall37,greenstein38,stebbins39} 
    concluded that $\kappa(\lambda)$ was roughly proportional to $1/\lambda$ in 
    the $\lambda=0.3-1\emic$ range.
    By the end of the 1930's, it was thus clear that the grains responsible for 
    the visible light extinction were not in the Rayleigh regime 
    ($\kappa_\sms{sca}(\lambda)\propto1/\lambda^4$; \refsec{sec:calcQabs}).
  \item \citet{stebbins43} extended the study of extinction curves to the
    $0.35\emic<\lambda<1.03\emic$ range, and found deviations to the
    $1/\lambda$ behavior, due to the now well-known far-\hUV\ rise and \hNIR\ 
    knee (\refsec{sec:extinction}).
  \item During the 1950's and 1960's followed a discussion on the universality 
    of the shape of the extinction curves \citep[\cf\ references in][]{li05}.
  \item Starting in the 1960's, the first airborne and space observatories 
    opened the \hUV\ window, up to the Lyman break \citep{york73}.
    The 2175~$\r{A}$ bump was discovered by \citet[][\cf\ 
    \refsec{sec:histofeat}]{stecher65}.
  \item The first \hUV\ extinction curves in the \expression{Large Magellanic 
    Cloud} (\hLMC) were measured by \citet{borgman75}, and by 
    \citet{koornneef78} in \xxxdor.
    In the \expression{Small Magellanic Cloud} (\hSMC), it was first measured 
    by \citet{rocca-volmerange81}.
    In both cases, the steepness of the curve and the weakness of the bump
    were noted.
  \item \citet*{cardelli89} performed a mathematical fit to a collection of 
    extinction curves toward different sightlines in the \hMW.
    They found that, over the $0.125\emic<\lambda<3.5\emic$ range, the shape is 
    universal and controlled by a single parameter, 
    $R(V)=A(V)/(A(B)-A(V))$
    (where $A(\lambda)$ is the magnitude extinction; \cf\ 
     \refsec{sec:extinction}).
  \item \hISO\ observations of the Galactic center exhibited a flatter \hMIR\ 
    curve \citep{lutz96}, confirmed by \citet{indebetouw05} with \hspitz\ data.
\end{itemize}

\paragraph{Polarization by dichroic extinction.} \citet{hall49} and \citet{hiltner49} found that starlight was linearly polarized. 
\begin{itemize}
  \item \citet{davis51} proposed that this polarization is due to dichroic 
    extinction by non-spherical particles aligned on the magnetic field
    (\cf\ \refsec{sec:intropola}).
  \item \citet{serkowski73} determined the wavelength dependence of the  
    polarization fraction, known as the \expression{Serkowski curve} (\cf\ 
    \refsec{sec:polavis}).
\end{itemize}

\paragraph{Dust emission.}
The thermal emission of heated grains started to be observed in the 1960's.
The presence of very small grains or large molecules with $a\lesssim1$~nm was speculated by \citet[][they are known as \expression{Platt particles}]{platt56}.
\begin{itemize}
  \item \citet{greenberg68} first realized that small grains must be 
    stochastically heated.
  \item \citet{andriesse78} reported the first observational evidence of such
    transiently heated Platt particles in \M{17}.
    The spectral shape of the \hMIR\ spectrum was indeed constant over the 
    region, and much wider than a single grey body emission, consistent with 
    temperatures up to 150~K.
  \item Similarly, the \hNIR\ continuum and 3.3~\tmic\ feature emission of 
    several reflection nebulae was shown by \citet{sellgren83} to be consistent
    with small grains fluctuating up to 1000~K. 
  \item The presence of small grains in the diffuse \hISM\ was clearly 
    evidenced by the 12 and 25~\tmic\ \hIRAS\ emission \citep{boulanger88}.
    It was observed by numerous studies afterward in all types of interstellar 
    regions and galaxies.
\end{itemize}

  \subsubsection{Identification of Dust Features}
  \label{sec:histofeat}

The confirmation of the presence of various solid-state and molecular features was important to better constrain the dust composition.

\paragraph{Silicates.}
The first identification of silicates was reported by \citet{woolf69}, in absorption toward M giant and supergiant stars.
\citet{kemper04} provided a $2\,\%$ upper limit on the crystalline silicate fraction, based on \hISO\ observations toward the Galactic center.
The \hMIR\ features, proper to crystalline silicates were indeed not detected.

\begin{figure}[htbp]
  \begin{tabular}{cc}
    \includegraphics[width=0.52\textwidth]{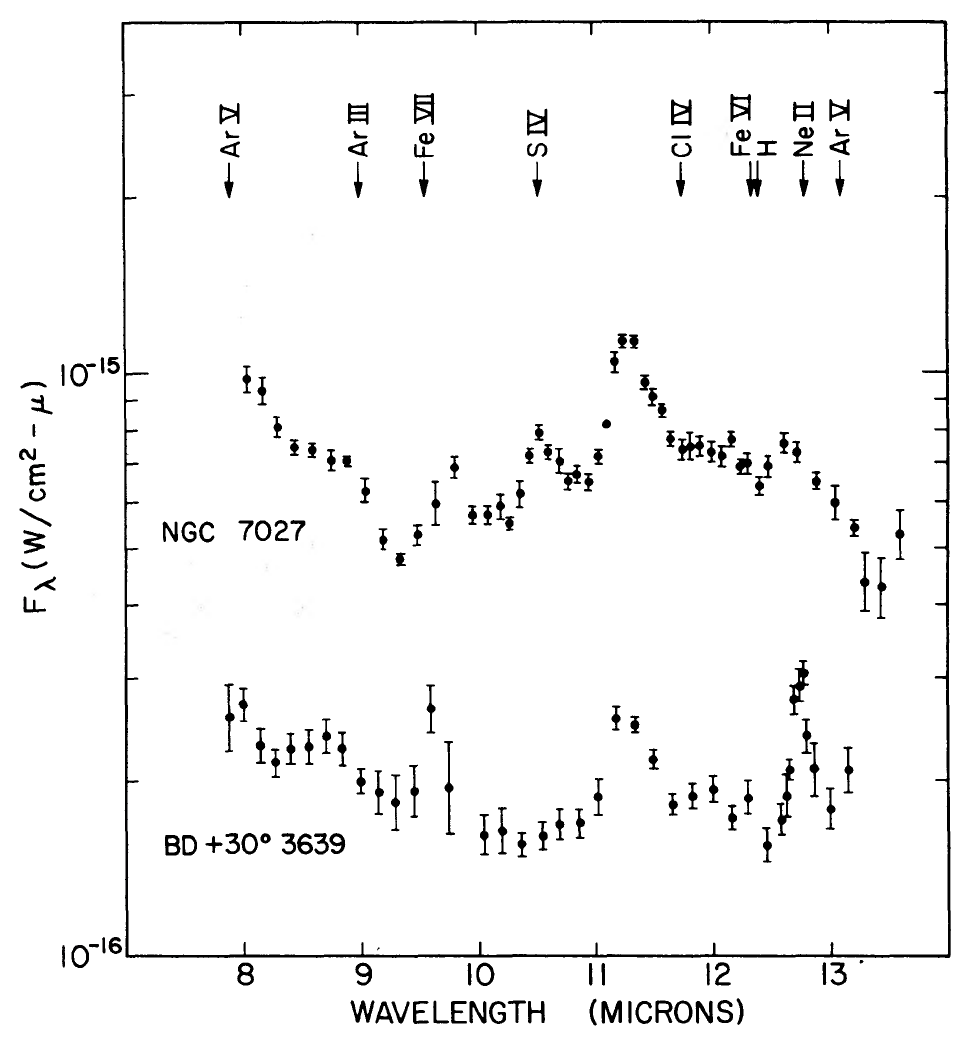} &
    \includegraphics[width=0.44\textwidth]{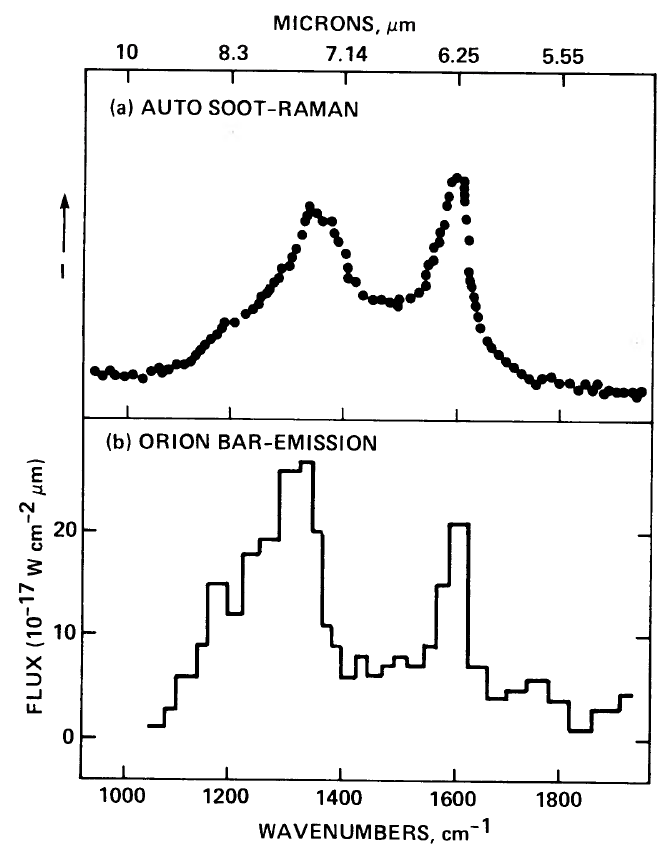} \\
    \textit{(a)} First detection of \hUIB s &
    \textit{(b)} Attribution of \hUIB s to \hPAH s \\
  \end{tabular}
  \newcap{First detection of UIBs}%
         {Panel~\textit{(a)} shows the first detection of the 8.6 and 
          11.3~\tmic\ aromatic features by \citet{gillett73}, in the planetary
          nebulae \ngc{7027} and BD$\,$+30$^\circ$3639.
          The red wing of the 7.7~\tmic\ is also visible.
          Panel~\textit{(b)} shows the qualitative comparison between a 
          laboratory spectrum of soot and the \hUIB s in the Orion bar, by
          \citet{allamandola85}.
          \uline{Credit:}
          \begin{inlinelistalph}
            \item Fig.~1 of \citet{gillett73};
            \item Fig.~1 of \citet{allamandola85}.
          \end{inlinelistalph}}
  \label{fig:1stPAH}
\end{figure}

\paragraph{Carbonaceous grains.}
\hMIR\ aromatic emission features were first detected in the \expression{Planetary Nebula} (\hPN) \ngc{7027} by \citet[][\cf\ \refsubfig{fig:1stPAH}{a}]{gillett73}.
They were called at the time \expression{Unidentified Infrared Bands} (\hUIB s).
They were attributed to the bending and stretching modes of \hPAH s ten years later \citep[][\cf\ \refsubfig{fig:1stPAH}{b}]{duley81,leger84,allamandola85}.
The 3.4~\tmic\ aliphatic feature in absorption was first detected toward the Galactic center by \citet{willner79}.

\paragraph{Ices.}
Ice absorption features have been searched for since the 1940's and the \expression{dirty ice model} (\cf\ \refsec{sec:histomodel}).
\begin{itemize}
  \item The 3.1~\tmic\ H$_2$O ice band was finally  detected by 
    \citet{gillett73b}.
  \item \citet{lacy84} provided the first observational identification of CO 
    ice absorption.
  \item CO$_2$ ice absorption was discovered in \hISO/SWS spectra 
    \citep{de-graauw96,dhendecourt96,guertler96}.
\end{itemize}

\paragraph{X-ray edges.}
The absorption of X-ray photons by inner electronic shells can provide information on the crystalline configuration of solids \citep[\eg][for the theoretical predictions]{forrey98,draine03b}.
The first X-ray absorption edges were detected in Chandra and XMM-Newton data \citep{paerels01}, but their interpretation remained problematic.
More recent studies have been able to constrain grain structures using these features \citep[\eg][]{lee09b}.

  \subsubsection{Dusty Epiphenomena}
  \label{sec:epiph}

\paragraph{Diffuse Interstellar Bands.}
There are unidentified, ubiquitous absorption features in the $\lambda\simeq0.4-2$~\tmic\ range, called \expression{Diffuse Interstellar Bands} (\hDIB s; \cf\ \refsec{sec:DIBs}).
\begin{itemize}
  \item They were first reported by \citet{heger22,heger22b}.
  \item It was only \citet{merrill34} who showed their interstellar nature.
  \item More than 400 of them have been identified toward a diversity of
    sightlines, even in external galaxies \citep{hobbs09}.
    They remain largely unidentified, although four of them have been 
    attributed to C$_{60}^+$ \citep{campbell15,walker15}.
\end{itemize}

\paragraph{Extended Red Emission.}
The \expression{Extended Red Emission} (\hERE) is a broad emission band, found in the $\lambda\simeq0.6-0.9$~\tmic\ range of a diversity of Galactic environments.
It is attributed to dust photoluminescence \citep[\eg][]{witt04}, but the nature of its carriers is still debated.
Photoluminescence is a non-thermal emission process in which, subsequently to 
the absorption of a \hUV\ photon, a grain is brought to an excited electronic 
state.
After partial internal relaxation, a redder photon is emitted, bringing the 
electron back to its fundamental state.
The \hERE\ was first reported in the \expression{Red Rectangle} reflection nebula \citep{schmidt80}.

\paragraph{Spinning Grains.}
The radio emission of fastly spinning dust grains was predicted by several authors \citep{erickson57,hoyle70,ferrara94}.
\begin{itemize}
  \item \citet{kogut96b} detected an \expression{Anomalous Microwave Emission}
    (\hAME) in the \hCOBE/\hDMR\ data.
    This excess could not be explained by thermal dust, free-free and 
    synchrotron emissions.
  \item \citet{draine98,draine98b} showed the \hAME\ could be explained by 
    the rotational emission of small, charged, fastly rotating grains.
\end{itemize}

  \subsubsection{Dust Models}
  \label{sec:histomodel}

\begin{figure}[htbp]
  \begin{tabular}{cc}
    \includegraphics[width=0.48\textwidth]{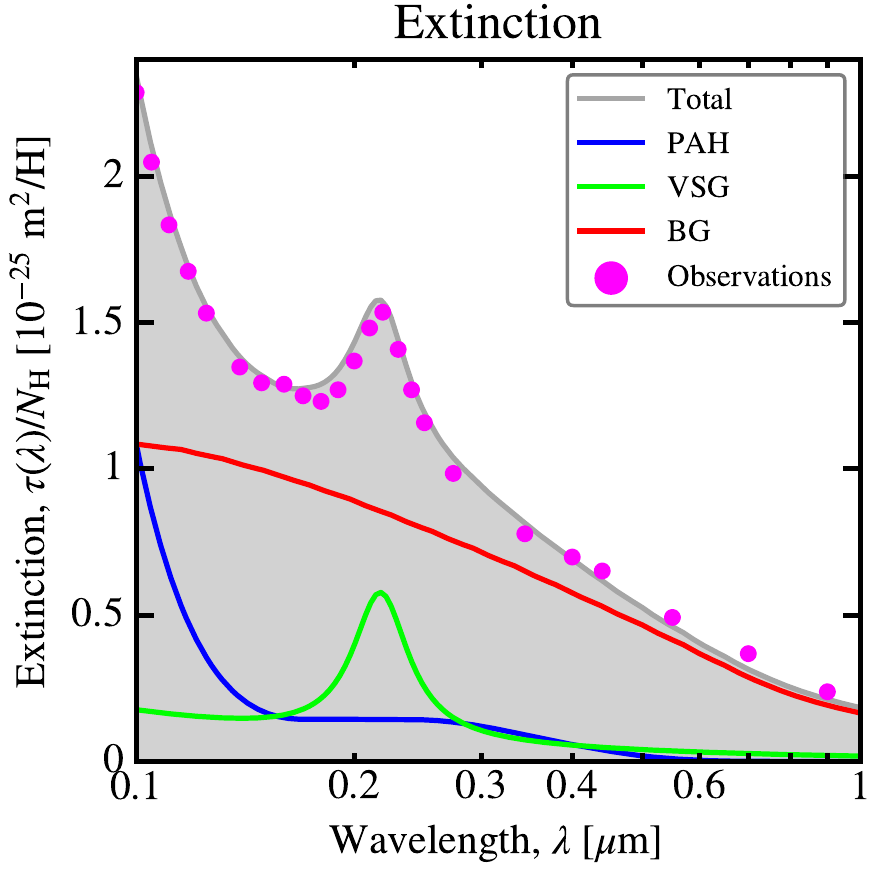} &
    \includegraphics[width=0.48\textwidth]{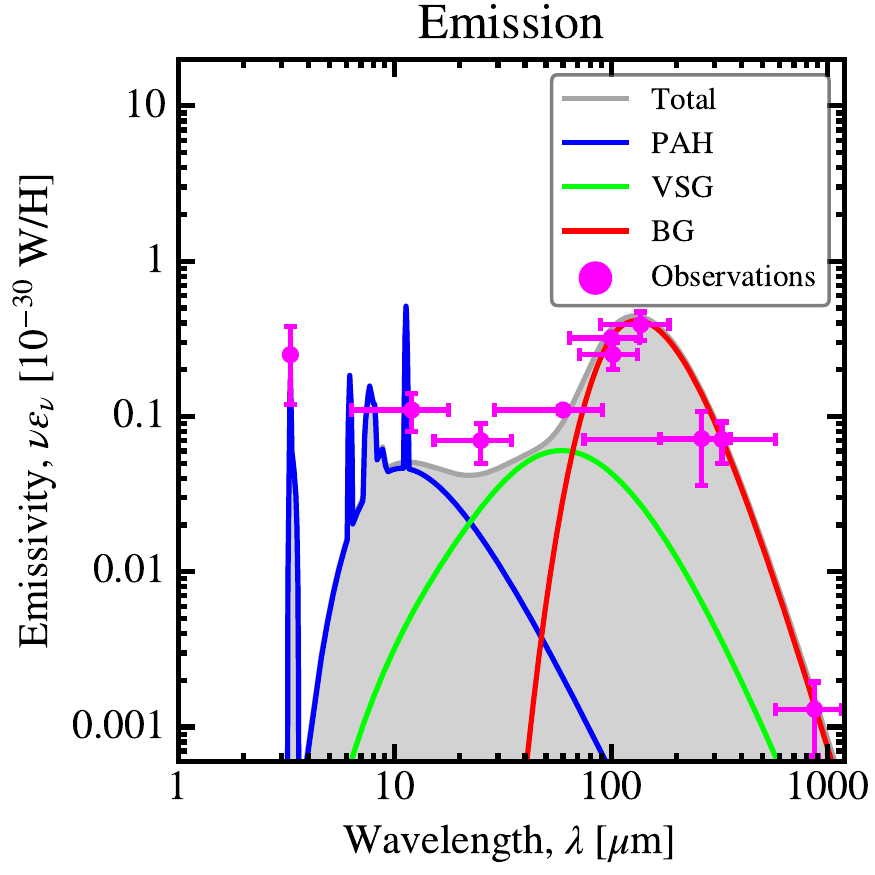} \\
  \end{tabular}
  \newcap{\citet{desert90} dust model}%
         {These two panels show the \citetalias{desert90} model extinction and 
          emission, and how it fits the observations of the diffuse Galactic 
          \hISM.
          We have shown its three components: \hPAH, \hVSG\ and \hBG.
          Notice the vintage-looking square shaped \hPAH\ features of this 
          pre-\hISO\ model.
          \CClicence}
  \label{fig:DBP90}
\end{figure}
\paragraph{First models.}
The first dust models of the 1930's, following Trumpler's study, were mainly speculative.
\begin{itemize}
  \item The first assumptions about the dust composition were made by analogy 
    with micrometeorites.
    \citet{schalen36} and \citet{greenstein38} assumed grains were made of 
    $a=0.01$~\tmic\ iron particles.
    \citet{lindblad35} had assumed dust could form by condensation in space.
  \item The \expression{dirty ice model} \citep{oort46,van-de-hulst49} was the 
    first attempt to base the dust composition on the abundant atoms: H, C, N, 
    O. 
    These atoms would form various ices (H$_2$O, CH$_4$, NH$_3$) nucleating on 
    grain seeds.
    The predicted dust temperature of this model was $\simeq15$~K, close to the
    actual value ($\simeq18$~K).
  \item Several studies hypothesized that graphite could be a major dust 
    constituent, explaining the polarization of starlight, because of its 
    anisotropic crystalline structure \citep{cayrel54,hoyle62}.
    Graphite was supported by the discovery of the 2175~$\r{A}$ bump 
    \citep{stecher65,stecher65b}. 
  \item \citet{donn68} hypothesized that \hPAH s could be responsible for the
    2175~$\r{A}$ bump, which was supported by the laboratory measurements of
    \citet{joblin92}.
  \item The \expression{\citetalias{mathis77} model} \citep*{mathis77} was the 
    first attempt at fitting the average extinction curve of the \hMW\ with 
    a mixture of graphite and silicate grains, with a power-law size 
    distribution:
    \begin{equation}
      f(a)\equiv \frac{\dd N}{\dd a} \propto a^{-3.5} 
      \;\;\;\mbox{ for }\;\;\; a_-<a<a_+,
      \label{eq:MRN}
    \end{equation}
    where $[a_-,a_+]=[0.005,1]$~\tmic\ for graphite and 
    $[a_-,a_+]=[0.025,0.25]$~\tmic\ for silicates.
\end{itemize}

\paragraph{Calculation of the optical properties.}
Dust models rely on the computation of optical properties.
The techniques have improved with time.
The laboratory measurements on the most relevant compounds also expanded.
\begin{itemize}  
  \item Mie theory (\cf\ \refsec{sec:calcQabs}) was independently developed by 
    Gustav \familyname{Mie} and Peter \familyname{Debye} \citep{mie08,debye09}.
  \item The \hDDA\ method (\cf\ \refsec{sec:QabsnonMie}) was developed by 
    \citet{purcell73}.
  \item \citet{draine84} presented the first \hUV-to-mm optical properties of 
    astronomical silicate and graphite.
    These properties have been refined by numerous studies afterward.
  \item Similarly, synthetic optical properties of a mixture of astronomical
    \hPAH s were presented by \citet{li01} and updated by \citet{draine07}.
  \item The properties of various amorphous carbon compounds were       
    inferred from laboratory data by \citet{rouleau91} and \citet{zubko96}.
    \citet{jones12a,jones12b,jones12c} presented a physical parametrization of 
    the optical properties of \hHAC.
\end{itemize}

\paragraph{Elemental depletions.}
Elemental depletions (\cf\ \refsec{sec:depletions}) are an important set of constraints on the dust mass and on the stoichiometry of the dominant grain compounds.
\citet{greenberg74} laid the ground for such studies.
\citet{savage79b} showed that the depletion strength correlates well with the average density of the gas.
Several studies have refined this approach.
\citet{jenkins09} presented a unified view, showing depletions were controlled by a single parameter, correlated with the column density.

\paragraph{Modern panchromatic models.}
With the \hCOBE\ and \hIRAS\ data, dust models started to have the possibility to be constrained by both the emission and the extinction.
These simultaneous constraints are important to break the degeneracy between the composition and the size distribution.
\begin{itemize}
  \item The \citet*{desert90} model (hereafter \citetalias{desert90}) was 
    the first silicate-carbon-\hPAH\ model,
    fitting the extinction and emission of the Galactic diffuse \hISM\ 
    (\reffig{fig:DBP90}).
    This model had three components: the \hPAH s, the \expression{Very Small 
    Grains} (\hVSG; small carbon grains) and the \expression{Big Grains} (\hBG; 
    large carbon-coated silicates).
  \item \citet{dwek97} built the first model constrained by the fine spectral 
    sampling of the \hCOBE\ data (\hDIRBE\ broadbands and \hFIRAS\ spectrum).
  \item \citet*{zubko04} added the elemental depletions to the emission
    and extinction, to constrain more accurately the grain composition.
  \item \citet{jones13,jones17} developed the \expression{The Heterogeneous 
    Evolution Model for Interstellar Solids} (\citetalias{jones17}).
    This model is physically parametrized to account for the known evolution
    of the dust mixture as a function of the physical conditions.
  \item \citet{guillet18} presented a dust model accounting for the polarized
    emission observed by \hplanck.
\end{itemize}

\newcommand{\mission}[1]{\inlinegraphics{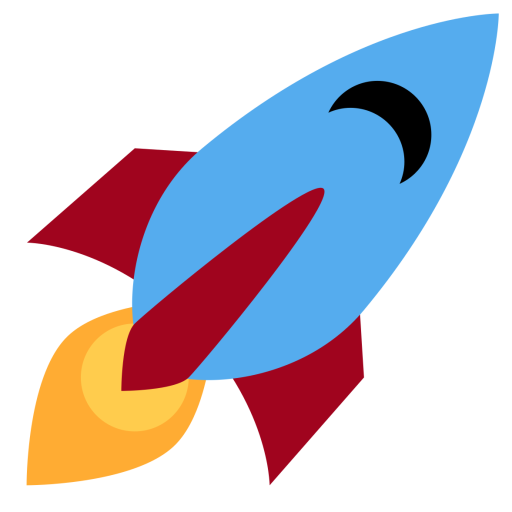}
                       \textcolor[rgb]{0.8,0,0}{#1}}
\newcommand{\observ}[1]{\inlinegraphics{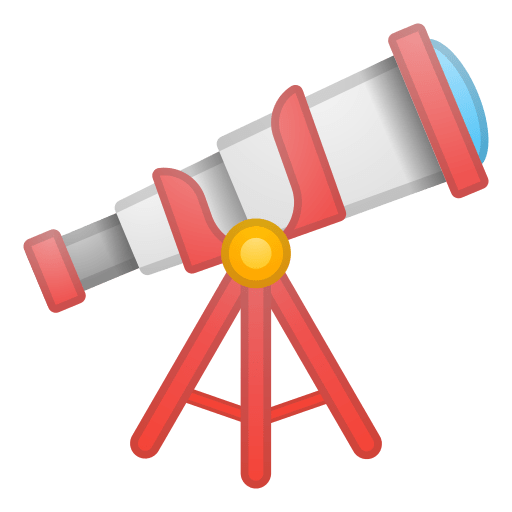}
                       \textcolor[rgb]{0,0,0.8}{#1}}
\newcommand{\concept}[1]{\inlinegraphics{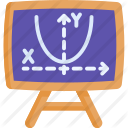}
                       \textcolor[rgb]{0.35,0.35,0.35}{#1}}
\begin{table}[htbp]
  \centering
  \setlength\arrayrulewidth{2pt}
  \arrayrulecolor{white}
  \begin{tabularx}{\linewidth}%
    {|>{\columncolor{coltabcell}}l
     |>{\columncolor{coltabcell}}X|}
    \hline
      \rowcolor{coltabsep}
      \multicolumn{2}{|c|}{\textsc{The Prehistory}} \\
    \hline
      \rowcolor{coltabcell}
      \cellcolor{coltabhead}1785 
        & \observ{Herschel's \expression{Construction of Heavens}} \\
      \cellcolor{coltabhead}1800 
        & \observ{Herschel's discovery of infrared radiation} \\
      \cellcolor{coltabhead}1847 & \observ{Struve's dimming of starlight} \\
      \cellcolor{coltabhead}1880 
        & \observ{First deep-sky photograph by Henry \familyname{Draper}} \\
      \cellcolor{coltabhead}1900 & \concept{Planck's black body radiation} \\
      \cellcolor{coltabhead}1908 & \concept{Mie theory} \\
      \cellcolor{coltabhead}1919 & \observ{Barnard's obscuration} \\
      \cellcolor{coltabhead}1922 
        & \observ{Heger's first observation of \hDIB s} \\
    \hline
      \rowcolor{coltabsep}
      \multicolumn{2}{|c|}{\textsc{The Classical Era}} \\
    \hline
      \rowcolor{coltabcell}
      \cellcolor{coltabhead}1930 & \observ{Trumpler's color excess study} \\
      \cellcolor{coltabhead}1934 & \observ{Interstellar nature of \hDIB s} \\
      \cellcolor{coltabhead}1936 & \concept{Small metallic particle model} \\
      \cellcolor{coltabhead}1949 & \concept{Dirty ice model} \\
      \cellcolor{coltabhead}1949 & \observ{Polarization of starlight} \\
      \cellcolor{coltabhead}1965 
        & \observ{Discovery of the 2175~$\r{A}$ bump} \\
      \cellcolor{coltabhead}1969 
        & \observ{First observation of silicate features} \\
      \cellcolor{coltabhead}1970 
        & \concept{First dust radiative transfer models} \\
      \cellcolor{coltabhead}1973
        & \observ{First detection of the \hUIB s} \\
      \cellcolor{coltabhead}1973
        & \observ{Serkowski curve} \\
      \cellcolor{coltabhead}1977 & \concept{MRN model} \\
      \cellcolor{coltabhead}1978 & \observ{First evidence of small, 
                                            stochastically heated grains} \\
      \cellcolor{coltabhead}1979 
        & \observ{First detection of the 3.4~\tmic\ feature} \\
      \cellcolor{coltabhead}1980 
        & \observ{First detection of \hERE} \\
      \cellcolor{coltabhead}1983 & \concept{\hISRF\ of Mathis, Mezger \&\ 
                                             Panagia} \\
    \hline
      \rowcolor{coltabsep}
      \multicolumn{2}{|c|}{\textsc{The Space Age}} \\
    \hline
      \rowcolor{coltabcell}
      \cellcolor{coltabhead}1983 & \mission{Launch of \hIRAS} \\
      \cellcolor{coltabhead}1984 & \concept{Draine \&\ Lee optical 
                                             properties} \\
      \cellcolor{coltabhead}1984 & \concept{\hPAH s proposed to explain the 
                                             \hUIB s} \\
      \cellcolor{coltabhead}1989 & \mission{Launch of \hCOBE} \\
      \cellcolor{coltabhead}1989 & \observ{Parametrization 
         of the Galactic extinction curve by Cardelli, Clayton \&\ Mathis} \\
      \cellcolor{coltabhead}1990 
        & \concept{Désert, Boulanger \&\ Puget model} \\
      \cellcolor{coltabhead}1995 & \mission{Launch of \hISO} \\
      \cellcolor{coltabhead}1996 & \observ{First detection of \hAME} \\
      \cellcolor{coltabhead}2003 & \mission{Launch of \hspitz} \\
      \cellcolor{coltabhead}2004 & \concept{Zubko, Dwek \&\ Arendt model} \\
      \cellcolor{coltabhead}2009 & \mission{Launch of \hhersc\ \&\ \hplanck}\\
      \cellcolor{coltabhead}2011 & \observ{Revision of dust opacities} \\
      \cellcolor{coltabhead}2013 & \concept{THEMIS model} \\
      \cellcolor{coltabhead}2015 & \observ{Whole-sky maps of the polarized 
        dust emission} \\
      \cellcolor{coltabhead}2015 & \concept{Identification of two \hDIB s} \\
      \cellcolor{coltabhead}2018 & \concept{Polarized dust emission model} \\
      \cellcolor{coltabhead}2021 & \mission{Launch of \hJWST?} \\
      \rowcolor{coltabcell}
      \cellcolor{coltabhead} & \ldots \\
    \hline
      \rowcolor{coltabsep}
      \multicolumn{2}{|c|}{\textsc{The Quantum Age?}} \\
    \hline
  \end{tabularx}
  \newcap{Chronology of the main ISD breakthroughs}%
         {If this chronology happens to be representative in any way, it shows 
          that the recent progress relies more on conceptual breakthroughs and
          space missions, whereas the progress in the early days was mainly 
          observational.}
  \label{tab:chronology}
\end{table}

\section{The Current Empirical Constraints}
\label{sec:dustobs}

We now review the current empirical constraints that are used to build dust models.
These models are calibrated on observations of the diffuse Galactic \hISM.
This medium indeed presents several advantages.
\begin{itemize}
  \item It is optically thin. 
    No radiative transfer is thus needed.
  \item It appears to be rather uniformly illuminated.
    The mixing of heterogeneous physical conditions along the sightline is 
    probably limited.
  \item Dust properties of the diffuse \hISM\ appear rather uniform.
  \item It benefits from a wealth of observations.
\end{itemize}
Observations of the diffuse Galactic \hISM\ are thus the most important ones.
Extragalactic constraints are the focus of \refchap{chap:dustprop}.
It is currently impossible to gather the same type of data set, at the same level of accuracy, in external galaxies.

\begin{figure}[htbp]
  \includegraphics[width=\textwidth]{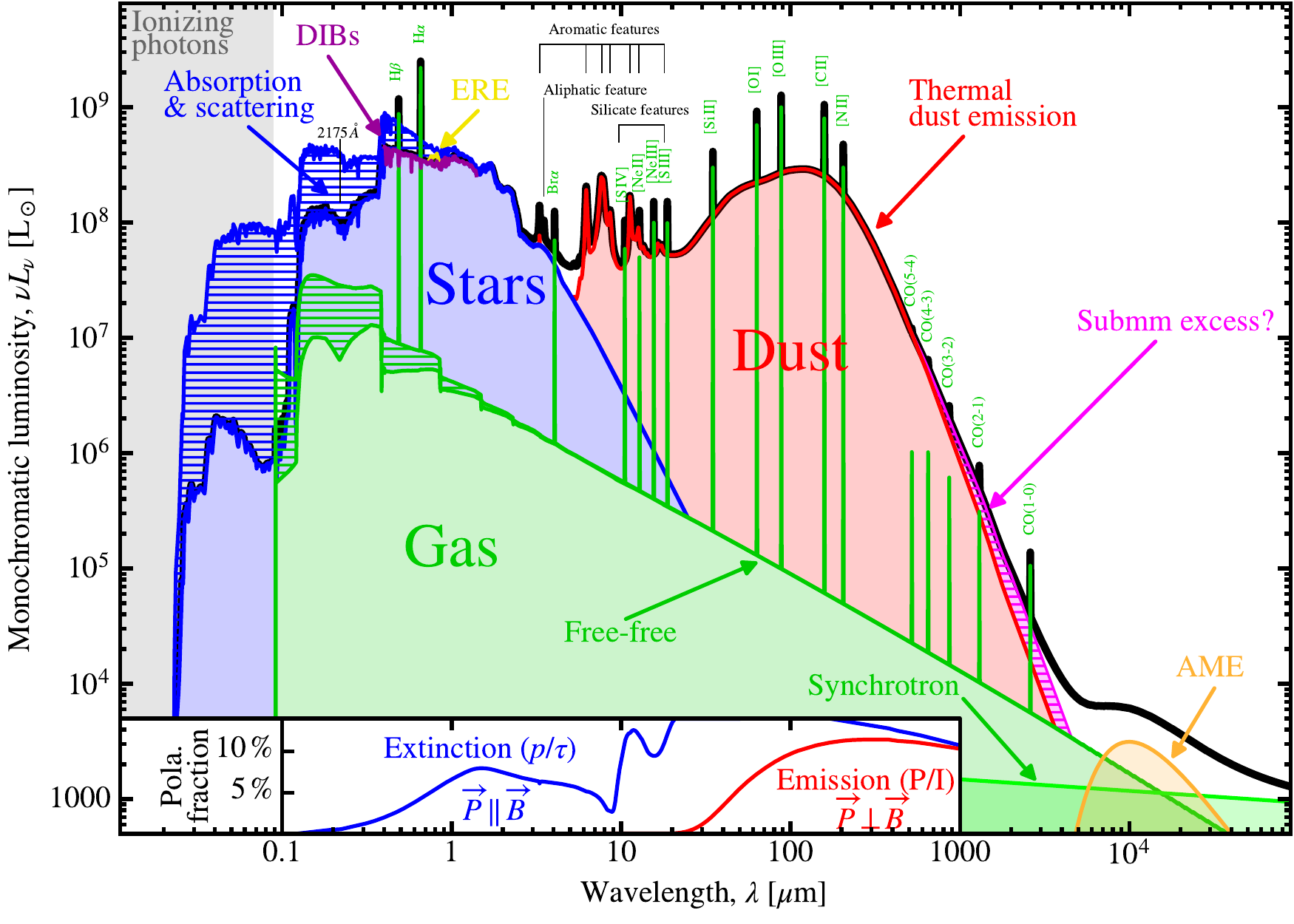}
  \newcap{Panchromatic dust observables}%
         {Spectral energy distribution of a typical late-type galaxy 
          \citep{galliano18}.
          The blue hatched area shows the power absorbed by dust. 
          Typical \hDIB, \hERE\ and \hAME\ spectra are shown, together 
          with the most relevant gas lines. 
          The free–free continuum is emitted by the deceleration of free 
          electrons scattering off ions in ionized regions.
          The synchrotron continuum is emitted by electrons spiraling through 
          the magnetic field. 
          Here, $L_\nu$ is the electromagnetic power emitted per unit frequency. 
          Inset: Model D of \citet{guillet18}, with $G_0=100$; $\tau$ is the 
          optical depth, $p$ is the starlight polarization degree, P is the 
          polarization intensity, I is the total intensity, $\vect{P}$ is the 
          light polarization vector, and $\vect{B}$ is the magnetic field 
          vector.
          \CClicence}
  \label{fig:dustobs}
\end{figure}

Several reviews discuss the available dust observables \citep[\eg][]{draine03c,dwek05,draine09,galliano18,hensley21}.
We have represented on \reffig{fig:dustobs} most of these observables on top of the typical \hSED\ of a gas-rich galaxy.
\takeaway{A fundamental local quantity of the \hISM\ is the \expression{dust-to-gas mass ratio} or \expression{dustiness}\footnote{We are trying to promote the term \expression{dustiness}, introduced by \citetalias{galliano21}, as it is much more concise than \expression{dust-to-gas mass ratio}.}: 
\begin{equation}
  Z_\sms{dust}\equiv \frac{M_\sms{dust}}{M_\sms{gas}}.
  \label{eq:dustiness}
\end{equation}}

  \subsection{Extinction}
  \label{sec:extinction}

Dust extincts light from the X-rays to the \hMIR.
The effect of dust extinction on a background source is sometimes referred to as \expression{reddening}.
It is indeed more important on the blue side of the visible window.

    \subsubsection{UV-to-NIR Extinction}
    \label{sec:extinctionUV}

\paragraph{The extinction magnitude.}
We saw in \refsec{sec:histodimming} that the first dust studies were performed in extinction, in the visible range.
Consequently, extinction properties were characterized by quantities depending on the magnitude system.
The magnitude, $m(\lambda_0)$, of a star of flux $F_\nu(\lambda_0)$, observed through a photometric filter centered at wavelength $\lambda_0$, is:
\begin{equation}
  m(\lambda_0) \equiv 
    -2.5\log\left(\frac{F_\nu(\lambda_0)}{F_\nu^0(\lambda_0)}\right),
\end{equation}
where $F_\nu^0(\lambda_0)$ is the reference flux or \expression{zero-point} of the photometric filter.
The zero-point is a calibration quantity, independent of the observed source.
Two important bands for characterizing extinction are the B and V bands, centered respectively at $\lambda_\sms{B}=0.44\emic$ and $\lambda_\sms{V}=0.55\emic$ (\reftab{tab:magnitude}).
The \expression{total extinction} or \expression{extinction in magnitude} is defined as:
\begin{equation}
  A(\lambda_0)\equiv m_\sms{obs}(\lambda_0)-m_\sms{int}(\lambda_0)
  = 2.5\log\left(\frac{F_\nu^\sms{int}(\lambda_0)}{F_\nu^\sms{obs}(\lambda_0)}\right),
  \label{eq:Alambda}
\end{equation}
where the index \citengl{obs} refers to the \expression{observed} quantity, and \citengl{int} refers to the \expression{intrinsic} quantity, that is the quantity not affected by dust extinction.
In the \hMW, the average V-band extinction over the distance to the star, $d$, is $A(V)/d\simeq1.8\;\textnormal{kpc}^{-1}$ \citep[\eg][]{whittet03}.
$A(\lambda_0)$ can be linked to a more physical quantity, the \expression{optical depth}:
\begin{equation}
  \tau(\lambda) 
    = \underbrace{\kappa(\lambda)}_\sms{dust opacity}
    \times\underbrace{\rho}_\sms{ISM density}
    \times\underbrace{L}_\sms{length of the sightline}
    = \kappa(\lambda)\times \underbrace{\frac{Z_\sms{dust}}{1-Y_\odot-Z_\odot}}_\sms{H mass fraction}
      \times\underbrace{m_\sms{H}}_\sms{H atom mass}
      \times\underbrace{N_\sms{H}}_\sms{column density}.
  \label{eq:tau}
\end{equation}
The expression above has been derived assuming homogeneous properties along the sightline (\cf\ \refsec{sec:RT} for a more rigorous definition of $\tau$).
The observed flux can conveniently be expressed as a function of the optical depth:
\begin{equation}
  F_\nu^\sms{obs}(\lambda) = F_\nu^\sms{int}(\lambda)
    \times\exp\left[-\tau(\lambda)\right].
\end{equation}
\takeaway{\refeq{eq:Alambda} therefore implies that:
$A(\lambda) = 1.086\times\tau(\lambda)$.}

\begin{figure}[htbp]
  \begin{tabular}{cc}
    \includegraphics[width=0.48\textwidth]{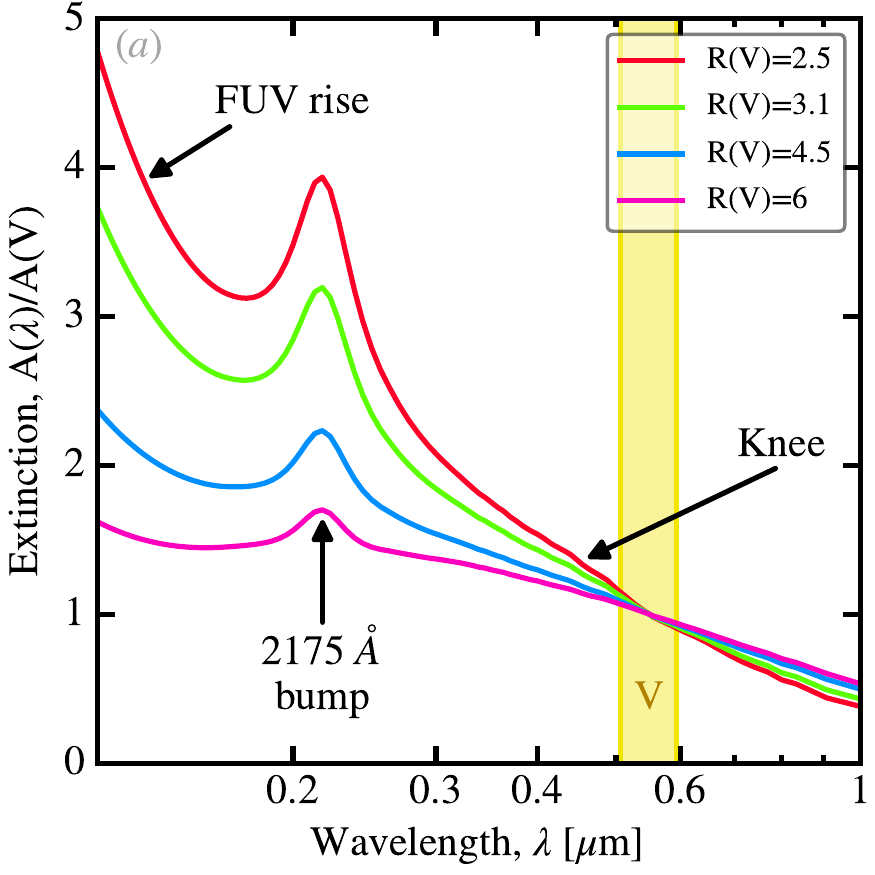} &
    \includegraphics[width=0.48\textwidth]{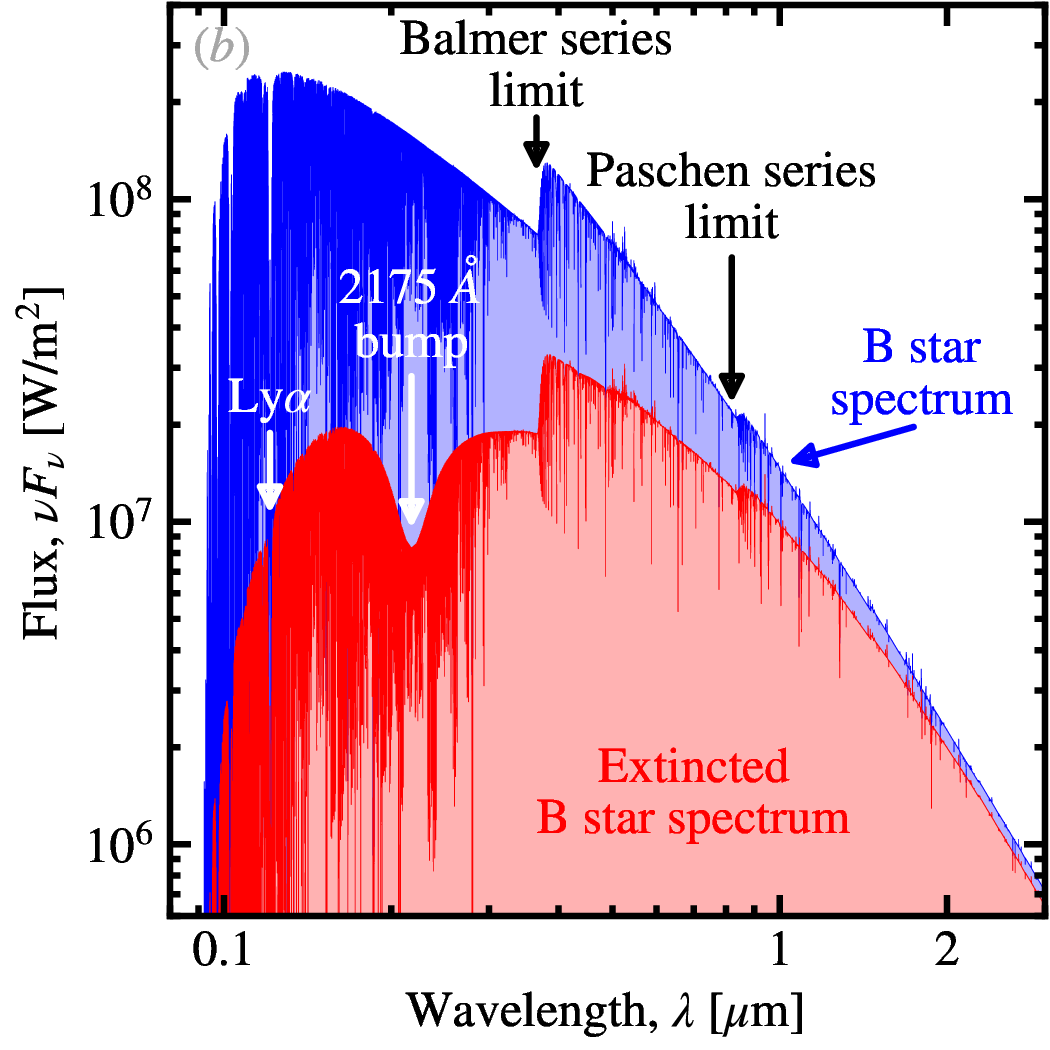} \\
  \end{tabular}
  \newcap{Galactic extinction curves}%
         {Panel~\textit{(a)} represents the average extinction curves from the
          spectroscopic sample of \citet{fitzpatrick19}, for different values
          of R(V).
          Panel~\textit{(b)}: the blue line represents the synthetic, intrinsic
          \hUV-visible \hSED\ of a B star from the \citet{lanz07} library;
          the red line is the extincted \hSED\ with $A(V)=1$ and $R(V)=3.1$.
          \CClicence}
  \label{fig:extinction}
\end{figure}

\paragraph{The selective extinction.}
The spectral shape of the extinction curve varies among sightlines.
It can be quantified by the \expression{selective extinction}, $E(\lambda-\lambda_0)\equiv A(\lambda)-A(\lambda_0)$.
In the \hMW, \citet*{cardelli89} showed that the \hUV-to-\hNIR\ extinction curves follow a universal law, parametrized by the sole visual-to-selective extinction ratio:
\begin{equation}
  R(V)\equiv\frac{A(V)}{A(B)-A(V)}.
\end{equation}
This parametrization is demonstrated in \refsubfig{fig:extinction}{a}.
We see that $A(V)$ is a scaling parameter quantifying the amount of extinction along the sightline.
According to \refeq{eq:tau}, $A(V)\propto Z_\sms{dust}\times N_\sms{H}$.
In the \hMW, there are no drastic \hdustiness\ variations, thus $A(V)\propto N_\sms{H}$.
On the other hand, $R(V)$ is a shape parameter. 
It typically varies between $R(V)\simeq2$ and $R(V)\simeq5$.
On average, $R(V)\simeq3.1$ in the \hMW.
Curves with $R(V)\gtrsim3.1$ tend to be flatter.
\takeaway{The amount of extinction in the \hMW\ is
$N_\sms{H}/E(B-V)\simeq 8.8\E{25}\;\textnormal{m}^{-2}\,\textnormal{mag}^{-1}$ \citep{lenz17}, or, for $R(V)=3.1$,
$N_\sms{H}/A(V)\simeq2.8\E{25}\;\textnormal{m}^{-2}\,\textnormal{mag}^{-1}$.}
The most notable features of the \hUV-to-\hNIR\ extinction curves are the following (\cf\ \refsubfig{fig:extinction}{a}).
\begin{description}
  \item[The Far-\hUV\ (\hFUV) rise] is mainly due to the absorption by small 
    grains, in the Rayleigh regime ($A(\lambda)\propto1/\lambda$; \cf\ 
    \refsec{sec:calcQabs}).
  \item[The 2175~$\r{A}$ bump] is attributed to small sp$^2$-hybridized
    C bonds (\cf\ \refsec{sec:dustanalog}).
  \item[The optical knee] is mainly due to scattering by larger grains.
  \item[The \hNIR\ extinction] can be approximated by a power-law: 
    $A(\lambda)\propto\lambda^{-\alpha}$, with $\alpha\simeq 2.27$ 
    \citep{maiz-apellaniz20}.
\end{description}

\paragraph{Measuring extinction.}
The original method to measure the wavelength dependence of the extinction curve is the \expression{pair method} \citep{stecher65}:
two stars of the same spectral type are observed, one with a low, and one with a high foreground extinction.
The extinction curve is directly derived from the differential \hSED\ or spectrum, assuming the dust properties are uniform along both sightlines.
An alternative to this method consists in replacing the reference star by a synthetic spectrum, knowing the precise spectral type of the star \citep[\eg][]{fitzpatrick05}.
This is demonstrated in \refsubfig{fig:extinction}{b}.

\paragraph{UV-visible scattering.}
Observations of starlight scattering by \hISD\ can constrain the average albedo, $\tilde{\omega}$, and asymmetry parameter, $\langle\cos\theta\rangle$, of the grains (\cf\ \refsec{sec:calcQabs}).
Two types of observations are usually favored to that purpose.
\begin{description}
  \item[The \expression{Diffuse Galactic Light}] (\hDGL) is the scattering of 
    the general \hISRF\ by dust.
    It is the diffuse visible light seen in \hISM\ regions without associated 
    stars.
    The \hDGL\ was first detected by \citet{elvey37}.
    \citet{henyey41} built their scattering phase function (\cf\ 
    \refsec{sec:calcQabs}) to explain this phenomenon.
  \item[Reflection nebulae] are obvious objects to measure $\tilde{\omega}$ and
    $\langle\cos\theta\rangle$, as the visible light comes from a nearby 
    star or cluster, and is scattered on the surface of a cloud facing us.
\end{description}
Both methods converge toward qualitatively consistent results:
\takeaway{\hISD\ has a \hUV-visible albedo around $\tilde{\omega}\simeq0.5$, and is rather forward scattering ($\langle\cos\theta\rangle\gtrsim0.5$), meaning grains are not in the Rayleigh regime (\cf\ \refsec{sec:calcQabs}).}

    \subsubsection{MIR Extinction}
    \label{sec:extinctionMIR}

\begin{figure}[htbp]
  \begin{tabular}{cc}
    \includegraphics[width=0.48\textwidth]{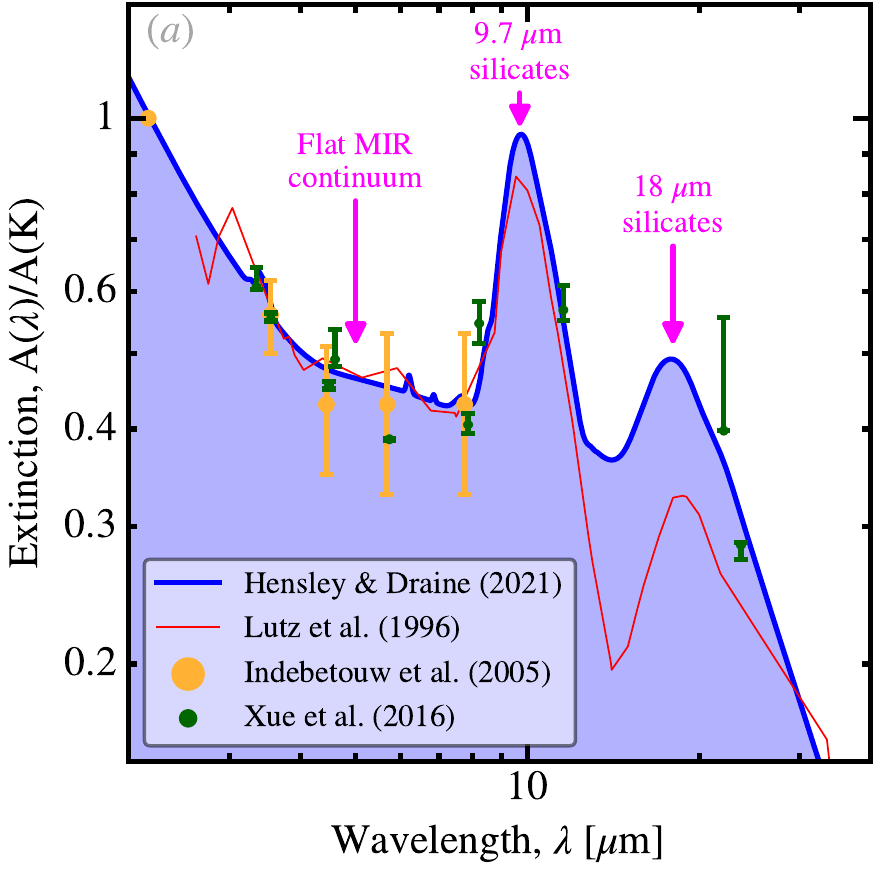} & 
    \includegraphics[width=0.48\textwidth]{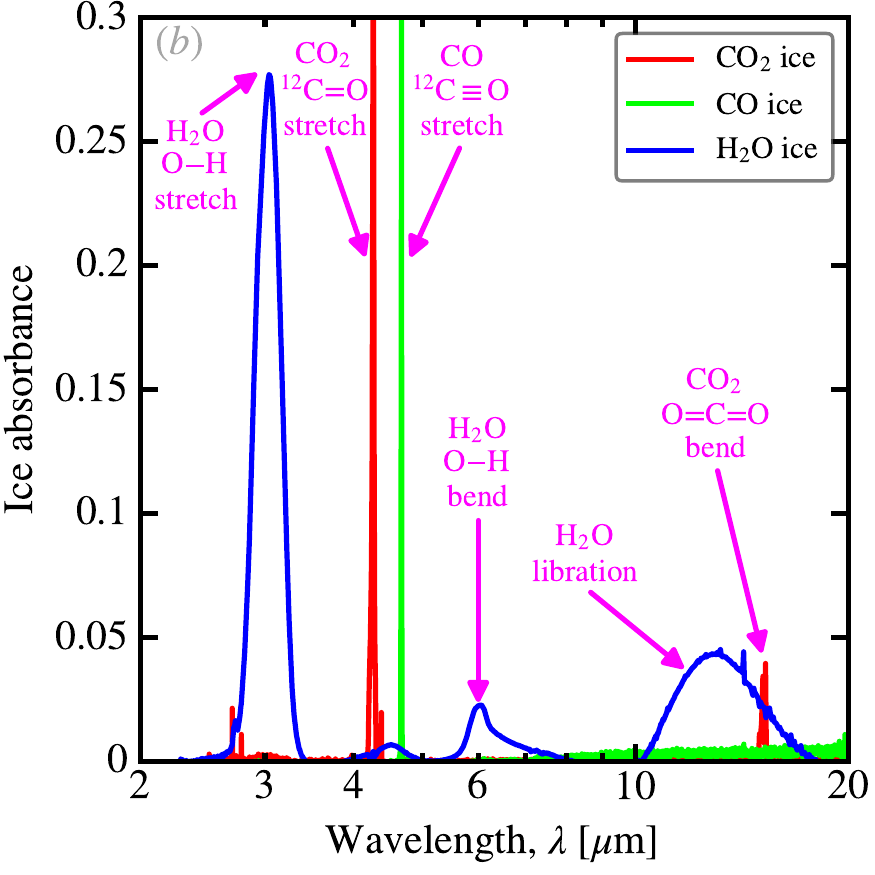} \\
  \end{tabular}
  \newcap{MIR extinction}%
         {Panel~\textit{(a)} shows the synthetic \hMIR\ extinction curve of
          \citet{hensley21} (in blue).
          The \hISO\ observations of the Galactic center by \citet{lutz96} 
          are overlaid in red.
          We also show the \hspitz, \hWISE\ and \hAKARI\ observations 
          \citep{indebetouw05,xue16}.
          Panel~\textit{(b)} shows the absorbance of the most abundant ices,
          taken from the \expression{Leiden Database for Ice}.
          Water ice data are from \citet{oberg07}, CO, from \citet{fraser04},  
          and CO$_2$, from \citet{bisschop07}.
          All ices have been measured at 15~K.
          \CClicence}
  \label{fig:MIRext}
\end{figure}

\paragraph{The MIR continuum.}
The spectral shape of the \hMIR\ extinction is harder to constrain than its \hUV-visible counterpart.
The \hMIR\ range is indeed the domain where the stellar and dust \hSED s intersect (\cf\ \reffig{fig:dustobs}).
It is therefore difficult to precisely model the background sources toward which the extinction is measured.
\begin{itemize}
  \item The Rayleigh-Jeans continuum of old stars, peaking in the \hNIR, such 
    as G and K-type stars, can be used \citep[\eg][]{xue16}.
  \item Otherwise, H recombination lines \citep[\eg][]{lutz96} or H$_2$ 
    rovibrational lines \citep[\eg][]{bertoldi99} provide an alternative.
    The ratio of a series of these lines can indeed be reasonably well modeled.
    Differences between the theoretical and observed ratios result from the 
    extinction.
\end{itemize}
In the \hISO\ days, there was a controversy about the 4-to-8~\tmic\ continuum, which seemed to be following the extrapolation of the \hNIR\ power-law trend \citep[\eg][]{bertoldi99}.
However, \citet{lutz96} showed this continuum toward the Galactic center was relatively flat (\cf\ \refsubfig{fig:MIRext}{a}).
This has been confirmed by ulterior observations with \hspitz, \hWISE\ and \hAKARI\ \citep[\eg][]{indebetouw05,xue16,gordon21}.
It seems to be a general feature of a wide variety of sightlines.
The new synthetic extinction of \citet{hensley21} reproduces this flat continuum (\cf\ \refsubfig{fig:MIRext}{a}).

\paragraph{Silicate features.}
The observed profiles of the Si--O stretching and O--Si--O bending silicate features, at 9.7 and 18~\tmic\ (\cf\ \refsec{sec:dustanalog}), are not perfectly matching those of olivine and pyroxene.
This is the reason of the introduction of the concept of \expression{astrosilicates} by \citet{draine84}.
\begin{itemize}
  \item Interstellar silicates are indeed a mixture of different varieties of 
    compounds, with different stoichiometry, and probably metallic inclusions.
  \item Interstellar silicates are predominantly amorphous.
    The interstellar crystalline silicate fraction seems to be $\lesssim2\,\%$ 
    \citep{demyk99,kemper04,do-duy20}.
\end{itemize}
There are uncertainties about the profile of the features and its potential variations between sightlines.
\begin{description}
  \item[The 9.7~$\bm{\mu}$m feature] has on average a \hFWHM\ of 
    $\simeq2.2$~\tmic.
    $A(V)/\tau(9.7)\simeq9\pm1$ toward the GC \citep{kemper04}, but is higher 
    when averaged over sightlines: $A(V)/\tau(9.7)\simeq19$ 
    \citep{roche84,mathis98}.
    The synthetic extinction of \citet{hensley21} has $A(V)/\tau(9.7)=20$.
  \item[The 18~$\bm{\mu}$m feature] is weaker than the 9.7~\tmic, making its
    characterization more uncertain.
    The depth ratio of the two features is $\tau(9.7)/\tau(18)\simeq2$ 
    \citep{chiar06}.
\end{description}

\paragraph{Ices.}
In regions shielded from the stellar radiation, some molecules can freeze out to form icy mantles onto grains \citep[\cf\ \eg][for a review]{boogert15}.
The dominant species are H$_2$O, CO and CO$_2$.
They are responsible for several \hMIR\ absorption bands, shown in \refsubfig{fig:MIRext}{b}.
These ice features are not observed in the diffuse \hISM.
They start appearing at higher values of $A(V)$, different compositions having different melting points.
They are observed in dense regions, toward molecular clouds, \expression{Young Stellar Objects} (\hYSO) or \hAGN s.

    \subsubsection{X-Rays}
    \label{sec:Xrays}

\paragraph{X-ray halos.}
Although the opacity of typical interstellar grains peaks in the \hUV\ (\cf\ \eg\ \reffig{fig:cross_sections}), grains extinct significantly X-rays.
In this regime, photons have wavelengths approaching the size of the atoms in the grain.
Dust grains, when present along the sightline of an X-ray point source (such as a low-mass X-ray binary), scatter the radiation at small angles, creating an \expression{X-ray halo} \citep{overbeck65,smith98}.
The properties of this halo are complex, as they depend on the grain properties: composition, porosity and maximum size \citep[\eg][]{smith08}.
Such studies are limited by the uncertainty on the distance of the intervening dust and the background source.
They however confirm the low abundance of grains larger than $\simeq0.1$~\tmic\ \citep[\eg][]{valencic15}.
\begin{figure}[htbp]
  \includegraphics[width=\textwidth]{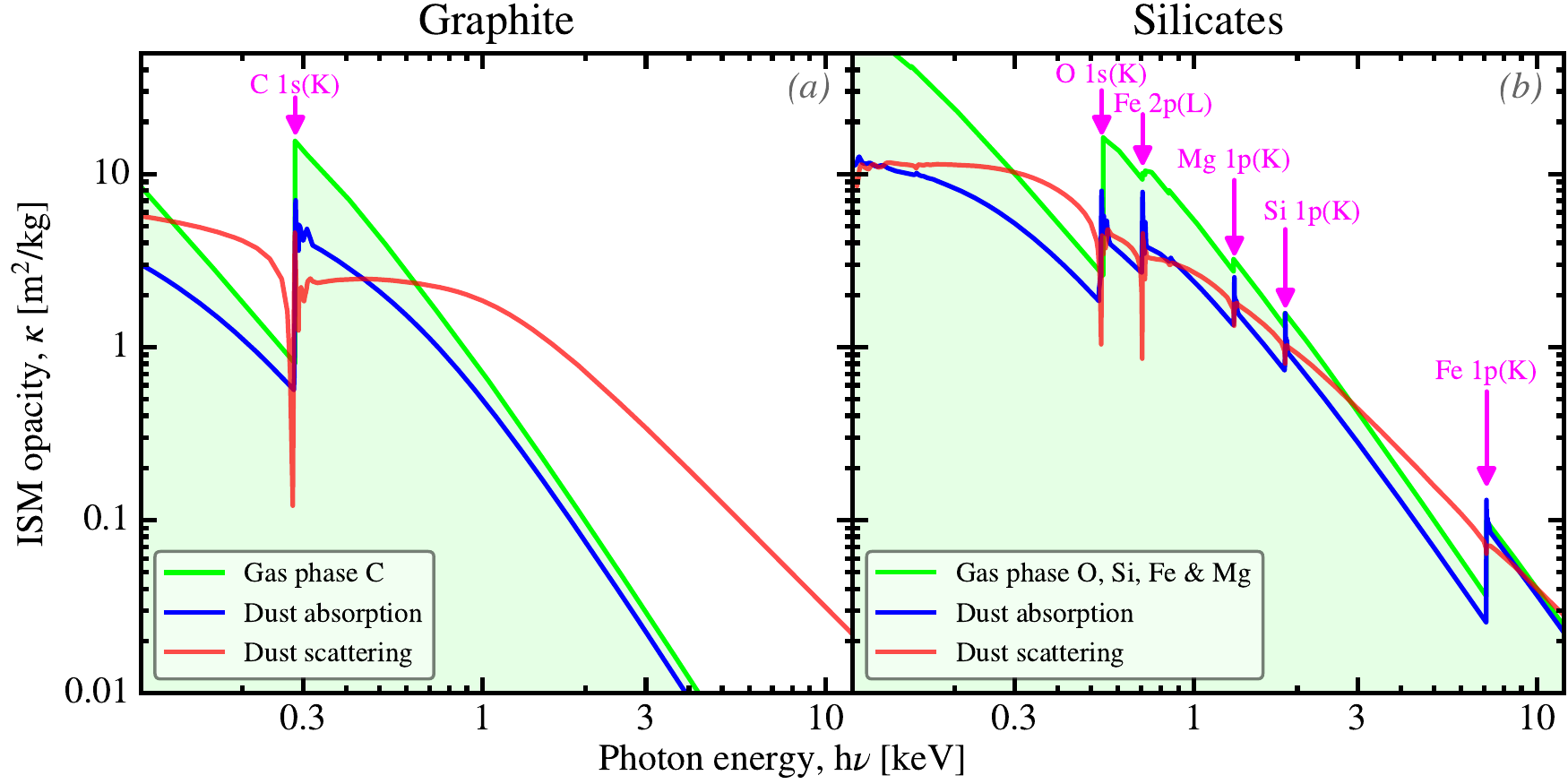}
  \newcap{X-ray edges}%
         {The two panels show the absorption and scattering dust opacities
          (blue and red), and the gas opacity (green).
          The opacities are expressed per unit \hISM\ mass.
          The gas opacity includes only the elements of the grains they are 
          compared to (\ie\ C for graphite, and O, Si, Fe and Mg for silicates).
          The dust cross-sections are from \citet{draine03b}.
          We have assumed an \citetalias{mathis77} size distribution and a 
          Galactic \hdustiness.
          The gas cross-section has been computed with the \ncode{python}
          interface of 
          \href{https://github.com/xraypy/XrayDB}{\ncode{X-ray DB}}.
          We have assumed the Solar abundances of \citet{asplund09}.
          \CClicence}
  \label{fig:Xray}
\end{figure}

\paragraph{X-ray absorption edges.}
Atoms, whether in the gas phase, or locked-up in grains, exhibit X-ray absorption features at specific wavelengths, called \expression{X-ray photoelectric edges} (\cf\ \reffig{fig:Xray}).
These edges correspond to the binding energies of the inner electrons, the letter (K or L in our case) corresponding to Bohr's orbitals (\cf\ \reftab{tab:orbitals}).
The important point is that the energy and the spectral shape of these edges depend on the way the atom is paired \citep[\eg][]{draine03b}.
It is thus possible, using X-ray spectroscopy, to differentiate atoms in the gas and dust phase, but also the crystalline structure of the grains \citep[\eg][]{lee09}.
For instance, \citet{zeegers17} studied the Si K edge along the line of sight of a Galactic X-ray binary.
They were able to constrain the column density and the chemical composition of the silicate grains.
This method was used to show that interstellar silicates are essentially Mg-rich, whereas the iron content is in metallic form \citep{costantini12,rogantini19,westphal19}.
Finally, the crystalline fraction of silicates has been estimated to be in the range $\simeq11-15\,\%$, using X-ray spectra \citep{rogantini19,rogantini20}.
This is significantly higher than the $\simeq2\,\%$ upper limit derived from \hMIR\ spectroscopy (\cf\ \refsec{sec:extinctionMIR}).
This discrepancy might originate in the challenges of X-ray spectroscopy, which requires both high spectral resolution and high signal-to-noise ratios.

    \subsubsection{Dichroic Extinction}
    \label{sec:polavis}

\begin{figure}[htbp]
  \begin{tabular}{cc}
    \includegraphics[width=0.48\textwidth]{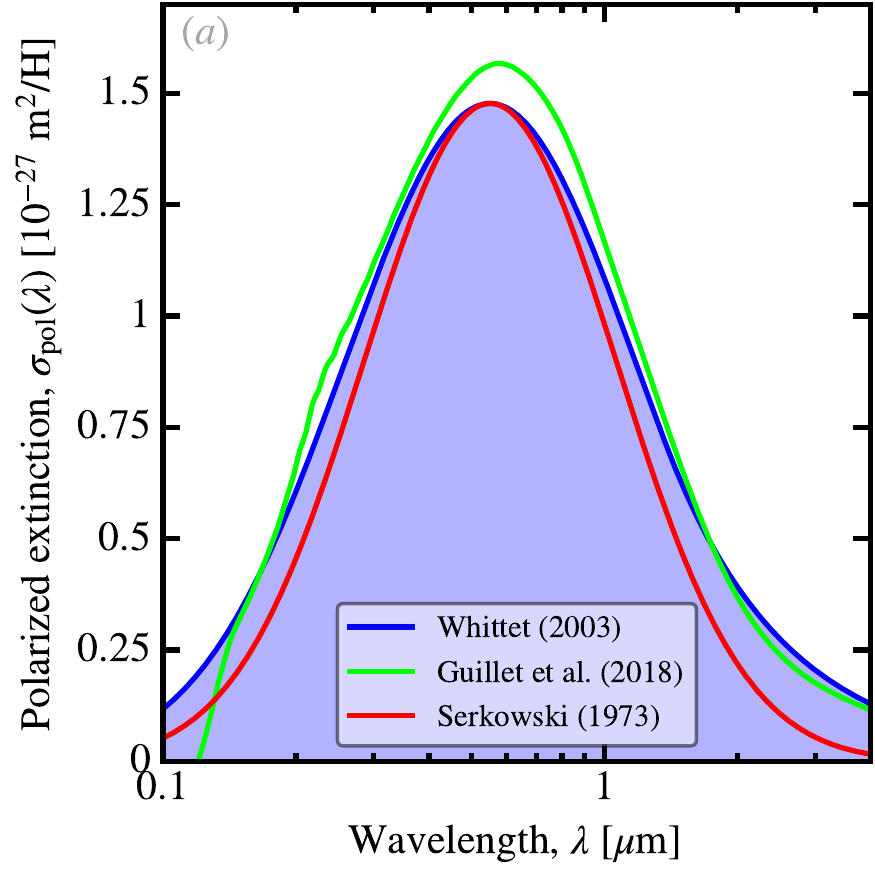} & 
    \includegraphics[width=0.48\textwidth]{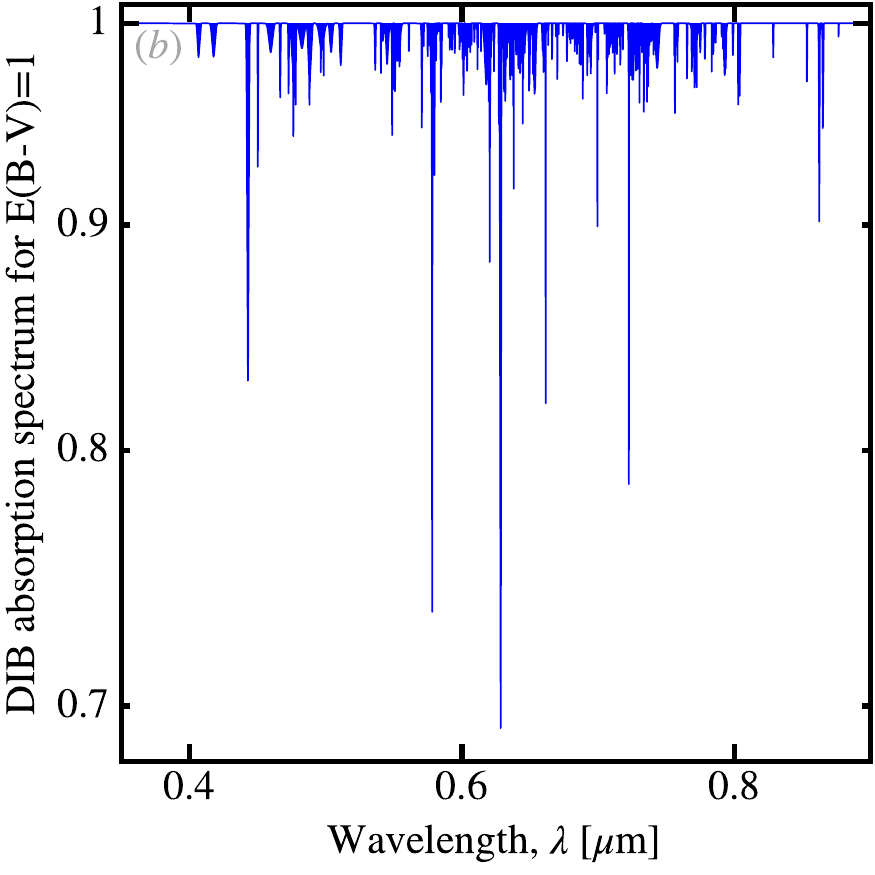} \\
  \end{tabular}
  \newcap{Polarized extinction and DIBs}%
         {Panel~\textit{(a)} shows the wavelength-dependent polarized 
          extinction.
          The plotted quantity is the ratio between the polarized optical depth 
          and the H column density, 
          $\sigma_\sms{pol}(\lambda)\equiv\tau_\sms{pol}(\lambda)/N_\sms{H}$
          The blue curve represents the synthetic, compromise fit of 
          \citet{whittet03}.
          The red curve is the original \citet{serkowski73} profile.
          Both have been normalized so that $p(V)/E(B-V)=0.13$ 
          \citep{hensley21}.
          The green curve shows the model E of \citet{guillet18}.
          Panel~\textit{(b)} shows the average absorption spectrum of \hDIB s
          from the study of \citet{jenniskens94}.
          It is for given for a typical $E(B-V)=1$.
          \CClicence}
  \label{fig:polaobs}
\end{figure}

The light from a background source seen through a cloud containing
 elongated grains, with their rotation axis aligned along the magnetic field, is partially polarized (\cf\ \refsec{sec:intropola}).
In the \hMW, the wavelength-dependent polarization fraction follows the empirical law of \citet{serkowski73}, shown in \refsubfig{fig:polaobs}{a}.
It runs from the near-\hUV\ (\hNUV) to the \hNIR, peaking around $\lambda\simeq0.55$~\tmic.
It is well reproduced by models with elongated grains \citep[\cf\ \refsubfig{fig:polaobs}{a} and][]{guillet18}.
The \expression{polarized extinction fraction}, $p(\lambda)$, is often quoted: $p(\lambda)\equiv C_\sms{pol}(\lambda)/C_\sms{ext}(\lambda)$, where $C_\sms{pol}$ and $C_\sms{ext}$ are the polarized and total cross-sections.
\takeaway{The interstellar polarized extinction peaks around 
          $\lambda_\sms{max}\simeq0.55$~\tmic, and its fraction is 
          $p(\lambda)/A(V)\lesssim3\,\%/$mag \citep{andersson15}.}

    \subsubsection{Diffuse Interstellar Bands}
    \label{sec:DIBs}

\hDIB s are ubiquitous absorption features in the $\simeq0.4-2$~\tmic\ range (\cf\ \refsubfig{fig:polaobs}{b}).
They are too broad to originate in atoms or simple molecules.
They have to come from large molecules and/or small grains.
Over 500 of them have been detected in the \hISM\ \citep{fan19}.
They are empirically associated with dust, as their strength correlates with $E({B-V})$ at low values, but they disappear in denser sightlines \citep[\eg][]{lan15}.
To first order, \hDIB s correlate with each other, but there are some notable 
differences, suggesting that they have different carriers \citep{herbig95}.
For instance, the so-called C$_2$ \hDIB s \citep{thorburn03} appear to be found 
preferentially in diffuse molecular clouds.
They remain largely unidentified, although four of them have been attributed to 
the ionized buckminsterfullerene, C$_{60}^+$, a football-shaped carbon molecule \citep{campbell15,walker15}.
The \hMIR\ transitions of this molecule, as well as C$_{70}$, had been detected in the \hISM, a few years before \citep{cami10}.

  \subsection{Emission}

As we have discussed in \refsec{sec:heatcool}, dust emits thermally in the \hIR.
This thermal emission is also partially polarized.
We will see in this section that there are also non-thermal emission components.

    \subsubsection{Infrared Continuum and Features}
    \label{sec:IRobs}

\begin{figure}[htbp]
  \includegraphics[width=\textwidth]{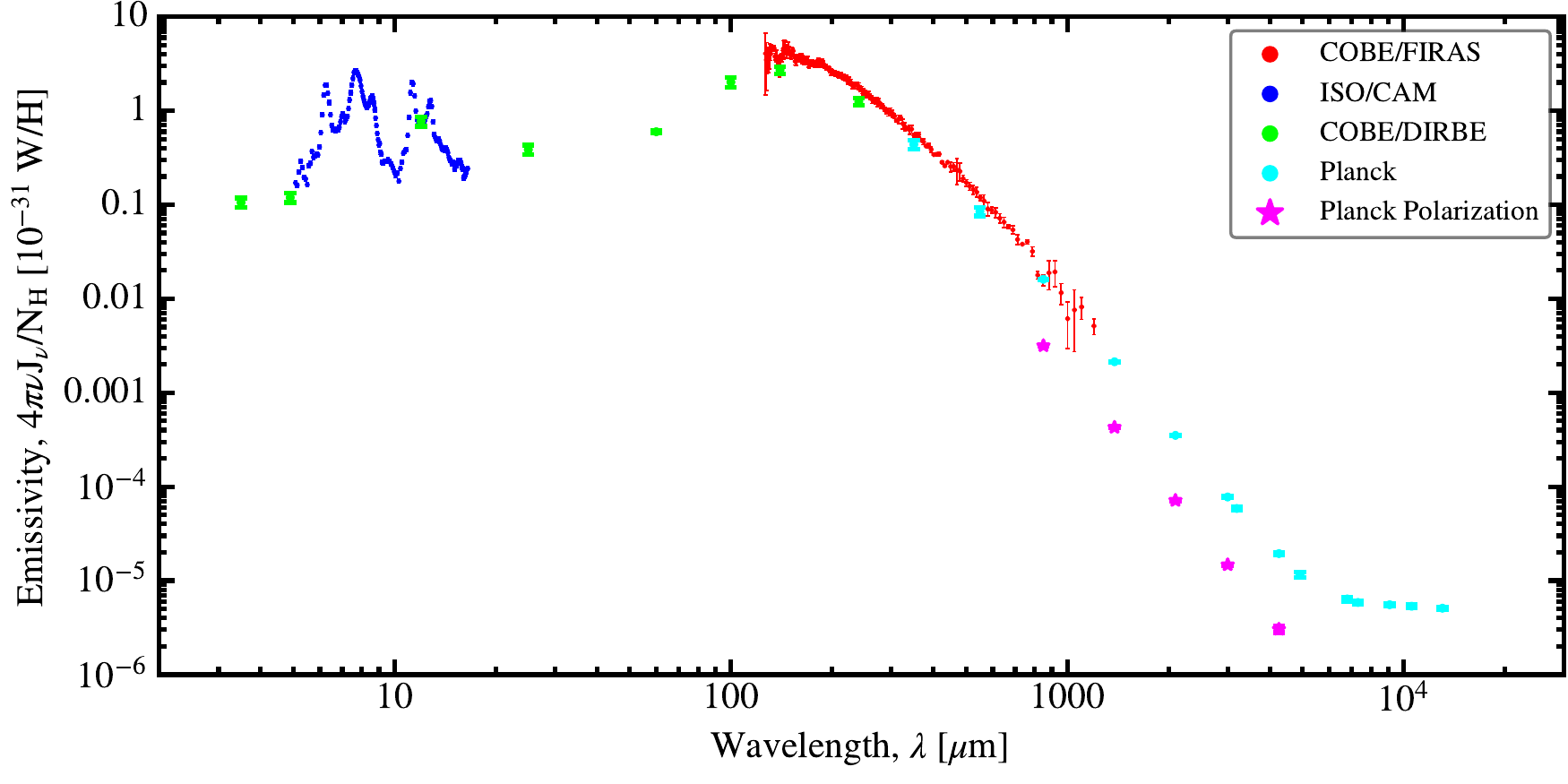}
  \newcap{Galactic diffuse ISM SED}%
         {These observations are the typical constraints on the emission of
          dust models.
          The \hDIRBE\ data are from \citet{dwek97};
          the ISOCAM spectrum is from \citet{flagey06};
          the \hFIRAS\ spectrum has been reprocessed by \citet{compiegne11}; and
          the \hplanck\ data are from 
          \citet{planck-collaboration14b,planck-collaboration15d}.
          The polarized emission is from \citet{planck-collaboration15d}:
          it is the linearly polarized intensity \refeqp{eq:pol}, 
          $4\pi\nu P_\nu/N_\sms{H}$.
          \CClicence}
  \label{fig:obsIR}
\end{figure}

\paragraph{Observations of the diffuse ISM.}
\reffig{fig:obsIR} represents the \hNIR-to-cm \hSED\ of the diffuse Galactic \hISM.
Those are the observations used to constrain the dust models we will discuss in \refsec{sec:dustmodels}.
The challenge of building such a data set is ensuring that these fluxes correspond to the emission of the most diffuse regions of the \hMW, characterized by its H column density ($N_\sms{H}\simeq10^{24}$~m$^{-2}$).
The disk of the \hMW\ contains the densest regions (\cf\ \refsubfig{fig:allsky}{b}).
It is also important to ensure avoiding denser regions, as grain properties evolve, probably due to the accretion of mantles \citep[\eg][]{ysard15}.
These observations therefore focus at high Galactic latitude, $b$, and low $N_\sms{H}$.
For instance, \citet{compiegne11} used data at $|b|>6^\circ$ and $N_\sms{H}<5.5\E{24}$~m$^{-2}$.
\citet{hensley21} gives a more a complete discussion about the homogenization of the different datasets.
At these emission levels, there are several contaminations that need to be subtracted.
\begin{description}
  \item[The zodiacal foreground] is the \hMIR\ emission from the interplanetary 
    dust in the Solar system disk.
  \item[The \hCIB] that we have already discussed in \reffig{fig:ISRF} is the 
    accumulated emission of background galaxies.
    Its \hSED\ is very similar to the emission of the \hMW, and thus difficult 
    to subtract.
  \item[The \hCMB] is the mm emission shown in \reffig{fig:ISRF}.
\end{description}

\paragraph{MIR features.}
The average \hMIR\ spectrum of the diffuse Galactic \hISM\ is represented in \reffig{fig:obsIR} (in blue).
This particular spectrum corresponds to a smaller patch of the sky, and is scaled on the \hDIRBE\ 12~\tmic\ photometry \citep{flagey06,compiegne11}.
There are indeed no \hMIR\ spectroscopic all sky surveys.
The \hMIR\ constraints of dust models prior to \citet{compiegne11} were only the \hDIRBE\ broadbands.
This difference in \hMIR\ coverage has consequences on the derived abundances and profiles of the aromatic feature carriers that we will discuss on \refsec{sec:dustmodels}.
The profiles and relative intensities of the main aromatic features can alternatively be constrained by a combination of laboratory data and the emission of nearby gas-rich galaxies \citep[\eg][]{draine07,hensley21}.

    \subsubsection{Polarized Emission}
    \label{sec:polaIR}

We have seen in \refsec{sec:intropola} that elongated grains emit polarized \hIR\ radiation.
Although the polarized submm emission of the \hISM\ had been measured from various balloon-borne observatories \citep{benoit04,bennett13}, the \hplanck\ satellite provided the first all sky survey in several bands \citep{planck-collaboration15d}.
These observations point toward one major result: large \hISM\ grains have homogeneous properties.
In other words, the \hIR\ emission can not originate in the mixing of several heterogeneous grain populations.
Small grains have a negligible polarization effect.
The models of \citet{guillet18}, which account both for total intensity and polarization, indeed provide the best fit for a single population of large composite astrosilicates with a-C mantles.
In parallel, \citet{planck-collaboration20} showed that the polarized \hSED\ was consistent with a single \hMBB\ with $\beta\simeq1.5$ and $T\simeq20$~K.
\takeaway{The maximum polarization fraction at 850~\tmic\ is $\simeq20\,\%$ \citep{planck-collaboration20b}.}

    \subsubsection{Non-Thermal Emission}
    \label{sec:AMEobs}

\paragraph{Spinning Grains.}
The \hAME\ is a centimeter continuum excess that can not be accounted for by the extrapolation of dust models, free-free, synchrotron and molecular line emission (\reffig{fig:dustobs}).
It was first detected in the \hMW\ \citep{kogut96b}.
\citet{draine98} promptly proposed that it was arising from the dipole emission 
of fastly rotating ultrasmall grains.
The candidate carriers were thought to be \hPAH s.
The \expression{Wilkinson Microwave Anisotropy Probe}
(\hWMAP; $\lambda\simeq3.2-13$~mm; 2001-2010) and \hplanck\ data of the Galaxy were successfully fitted with spinning dust models, including \hPAH s \citep{miville-deschenes08,ysard10,planck-collaboration11e}.
The cm \hSED\ in \reffig{fig:obsIR} is dominated by spinning grain emission.
In the \hMW, the \hAME\ correlates with all tracers of dust emission \citep[\eg][]{hensley16}.
However, the \hAME\ intensity increases with the \hISRF\ intensity, while \hPAH s are
destroyed in high \hISRF s.
\citet{hensley16} thus proposed that the carriers of the \hAME\ could be 
nano-silicates, rather than \hPAH s.
Refining the modeling of the \hMIR\ \hSED, \citet{bell19} showed that \hAME\ correlates better with the emission from charged \hPAH s, in the Galactic region $\lambda$-Orionis.
This will be discussed in more details in \refsec{sec:AME}.

\paragraph{Photoluminescence.}
We have seen in \refsec{sec:epiph} that the \hERE\ excess emission was thought to originate in the photoluminescence of dust grains.
In reflection nebulae, \hERE\ appears to be excited by \hFUV\ photons \citep[$11\;\textnormal{eV}\lesssim h\nu\lesssim13.6\;\textnormal{eV}$; \eg][]{lai17}.
It disappears if the exciting star has an effective temperature $T_\sms{eff}\lesssim10^4$~K.
The conversion efficiency, that is the rate of photoluminescent photons per absorbed \hUV\ photon, seems to be around $\simeq1\,\%$.
\hERE\ being seen in reflection nebulae, it is expected to be a general property of interstellar grains.
There is however a debate about the detection of \hERE\ toward cirrus clouds and its conversion efficiency \citep[\cf\ the discussion in][]{hensley21}.
\hERE\ is observed in C-rich \hPN e (containing predominantly carbonaceous grains) and not in O-rich \hPN e \citep[containing predominantly silicates grains;][]{witt04}.
The carriers should thus be carbon grains, such as \hPAH s.

  \subsection{Elemental Abundances in Grains}
  \label{sec:depletions}

The logarithmic abundance of an element E, relative to H, is often noted:
\begin{equation}
  \epsilon(E)\equiv12+\log\left(\frac{N_\sms{E}}{N_\sms{H}}\right),
\end{equation}
$N_\sms{E}$ being its column density.
The number abundance ratio can also be noted E/H instead of $N_\sms{E}/N_\sms{H}$, when it is not directly derived from the measure of a column density.
An element in the \hISM\ belongs either to the gas or to the dust phase.
If we know the total or \expression{reference} abundance of an element E in the \hISM, we can thus infer its abundance locked in dust grains, by measuring its abundance in the gas phase.
This difference is the \expression{depletion}.
The \expression{logarithmic depletion} of an element E is defined as \citep[][]{jenkins09}:
\begin{equation}
  \delta(E)\equiv\epsilon(E_\sms{gas})-\epsilon(E_\sms{ref}),
  \label{eq:depletion}
\end{equation}
The observable $\delta(E)$ is a measure of the ratio between the abundance of an element E observed in the gas phase to its total assumed abundance.
The abundance of element E, locked in grains, is thus:
\begin{equation}
  \left(\frac{E_\sms{dust}}{H}\right) 
  = \left(\frac{E_\sms{ref}}{H}\right)\left(1-10^{\delta(E)}\right).
  \label{eq:invdepletion}
\end{equation}
Note that, in \refeq{eq:invdepletion}, we do not differentiate the origin of H, as H is predominantly in the gas phase: $H_\sms{ref}\simeq H_\sms{gas}\gg H_\sms{dust}$.

    \subsubsection{Measuring ISM Abundances}
    \label{sec:ISMabund}

\begin{figure}[htbp]
  \includegraphics[width=\textwidth]{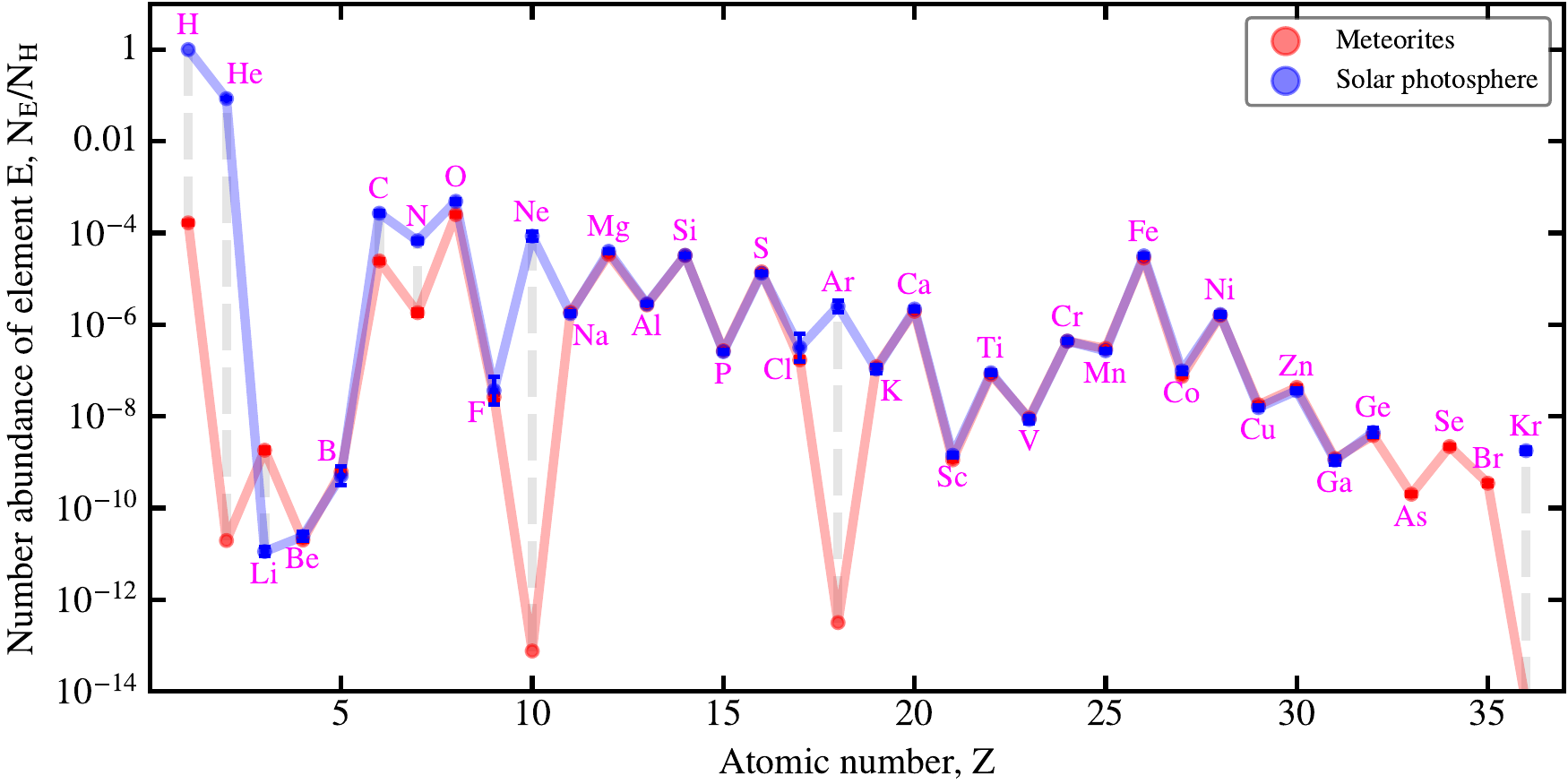}
  \newcap{Solar abundances}%
         {The two lines show the elemental abundances relative to H of the Solar
          system, as a function of the atomic number, from \citet{asplund09}.
          The red line and circles correspond to meteorites, and
          the blue line and circles correspond to the Solar photosphere.
          \CClicence}
  \label{fig:abundsun}
\end{figure}

\paragraph{Solar abundances.}
The abundance of elements and their isotopes are the most accurately known in the Solar system \citep[\eg][for a review]{asplund09}.
Those are thus used as a reference in the \hISM.
The abundances of the protosolar nebula, at the time the Sun formed, 4.56~Gyr ago, can be determined the two following ways.
\begin{description}
  \item[Meteorites,] analyzed with mass spectroscopy, provide the most precise
    abundances.
    The most primitive meteorites are the carbonaceous (CI) chondrites.
    The issue with meteorites is that the most volatile elements (\ie\ the 
    lightest ones and the noble gases) have been depleted due to 
    high-temperature processes within the Solar nebula 
    \citep[\eg][]{hellmann20}.
  \item[Solar photosphere] absorption spectroscopy is less precise, as it 
    requires some modeling.
    It however provides more reliable abundances of the volatile elements.
\end{description}
These abundances are compared in \reffig{fig:abundsun}.
We see that both tracers are in very good agreement, except for the volatile elements.
It is common to define the mass fractions of H, He, and elements heavier than He ($M_\sms{ISM}$ being the total \hISM\ mass):
\begin{equation}
  X\equiv\frac{M_\sms{H}}{M_\sms{ISM}}, \;\;\;\;\;\; 
  Y\equiv\frac{M_\sms{He}}{M_\sms{ISM}},\;\;\;\;\;\; 
  Z\equiv\frac{M_\sms{>He}}{M_\sms{ISM}},\;\;\;\;\;\;\mbox{with}\;\;\; X+Y+Z=1.
  \label{eq:abund}
\end{equation}
In the literature, the ratio $Z$ is unanimously called \expression{metallicity}.
Some even call the elements heavier than He, \expression{metals}, which is even worse, knowing what we have learned in \refsec{sec:metalvsdieletric}.
This is one of the worst choices of terminology in the whole history of sciences.
It is however difficult to avoid using the term \expression{metallicity}.
We will thus reluctantly use it in the rest of this manuscript.

\paragraph{Present-day Solar abundances.}
The abundances displayed in \reffig{fig:abundsun} are present-day photos\-pheric values.
They are however not perfectly representative of the present-day abundances of the Solar neighborhood \hISM.
To go from the former to the latter, a factor $+0.03$~dex has to be added to the heavy element abundances of \reffig{fig:abundsun} to account for diffusion in the Sun \citep{turcotte02}.
This provides protosolar abundances.
Present-day abundances can then be inferred by modeling the chemical evolution of the \hMW\ during the last 4.56~Gyr \citep[\eg][]{chiappini03,bedell18}.
This leads to correcting each element with a different factor, up to $\simeq0.2$~dex \citep[\cf\ \eg][for the correction of the major dust constituents]{hensley21}.
The present-day Solar photospheric abundances are \citep{asplund09}:
\begin{equation}
  X_\odot = 0.7381, \;\;\;\;\;\; Y_\odot = 0.2485,\;\;\;\;\;\; Z_\odot = 0.0134.
  \label{eq:sun}
\end{equation}
To put it in words, three quarters of the gas mass is made of H, one quarter is made of He, and only $1.3\,\%$ is made of heavy elements, in the \hMW.
Besides H and He, the most abundant species in the \hISM\ are O and C ($M_\sms{O}/M_\sms{ISM}\simeq8.0\E{-3}$ and $M_\sms{C}/M_\sms{ISM}\simeq2.8\E{-3}$).
These abundances can be used as references in \refeq{eq:depletion}.
Alternatively, B stars or young F and G stars can provide a more direct estimate of the abundances in nowadays \hISM. 
These abundances are however more difficult to estimate accurately.

    \subsubsection{Depletions}  
    \label{sec:depletionstrength}

\begin{figure}[htbp]
  \begin{tabular}{cc}
    \includegraphics[width=0.48\textwidth]{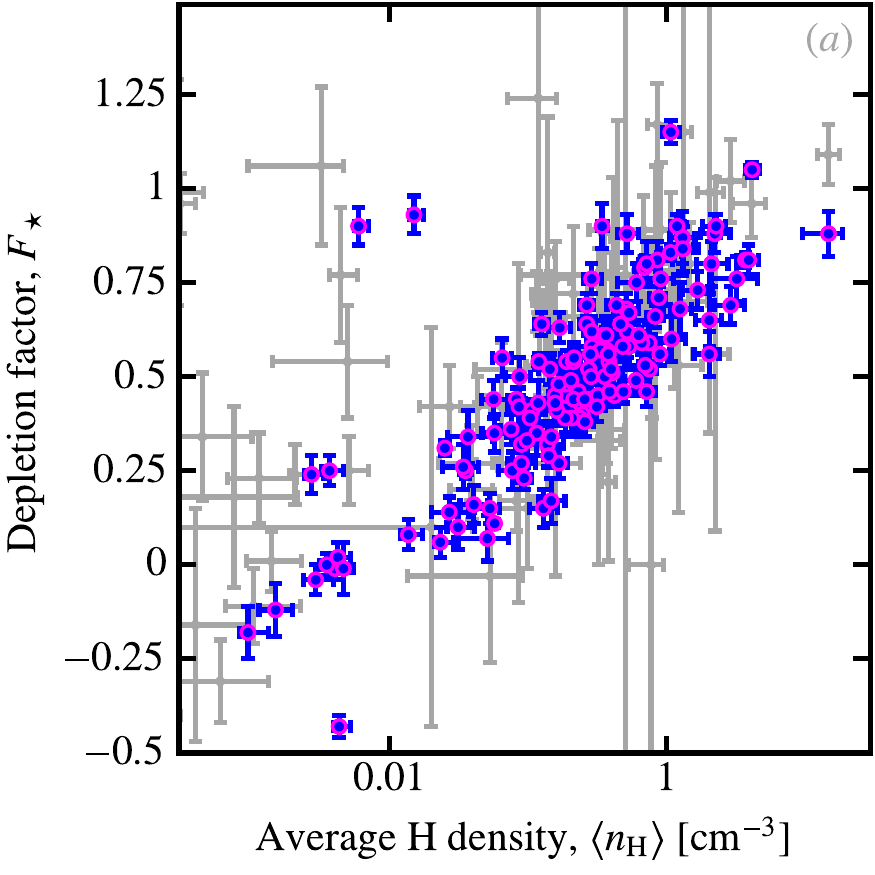} &
    \includegraphics[width=0.48\textwidth]{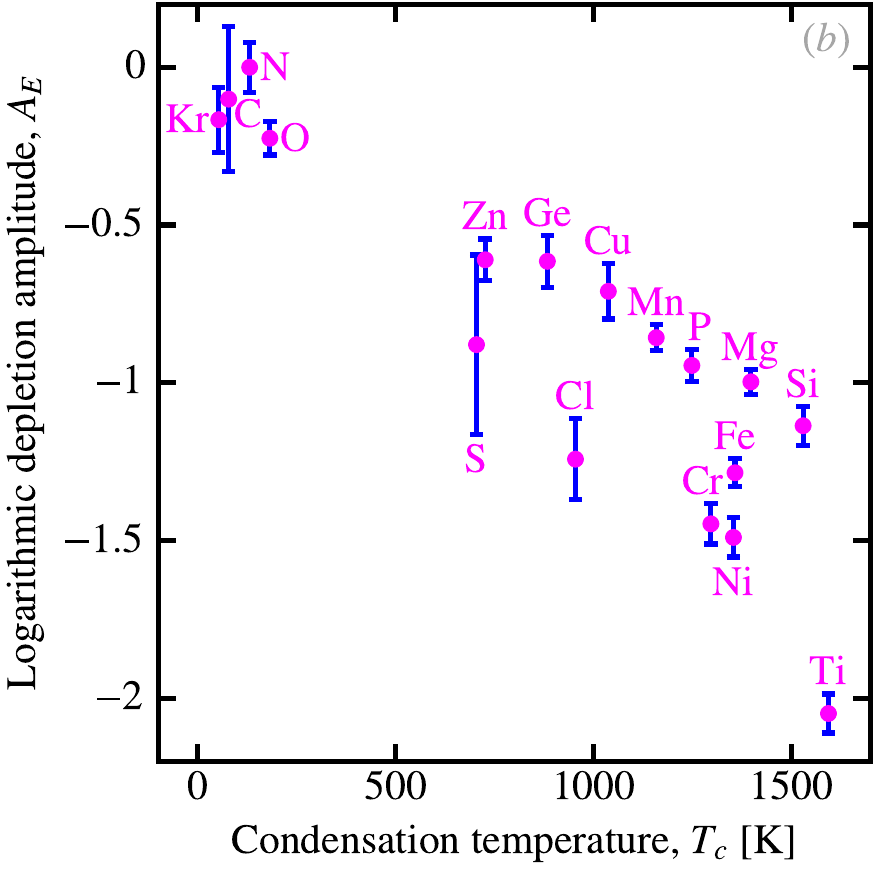} \\
  \end{tabular}
  \newcap{Depletion variations within the MW}%
         {Panel~\textit{(a)}: depletion factor \refeqp{eq:Fstar} as a function 
          of the average density,
          $\langle n_\sms{H}\rangle\equiv(N_\sms{\hi}+N_\sms{\hmol})/d$,
          where $d$ is the distance to the star \citep{jenkins09}.
          The points represented in grey are those for which the uncertainty on
          $F_\star$ is greater than 0.07.
          Panel~\textit{(b)}: depletion amplitude (from \refeqnp{eq:Fstar}) as a
          function of condensation temperature \citep{lodders03,jenkins09}.
          \CClicence}
  \label{fig:depletions}
\end{figure}

\paragraph{The depletion strength.}
The abundances in the gas phase are most reliably measured by absorption spectroscopy toward stars.
Gas atoms in the neutral \hISM\ are essentially in their ground state.
Most of the corresponding transitions are in the \hUV\ ($\lambda=0.0912-0.3$~\tmic).
\citet{jenkins09} compiled and homogenized the abundances of 17 elements measured along 243 sightlines, throughout the literature, to propose a unified representation of the depletions in the \hMW.
\citet{jenkins09} showed that the logarithmic depletions of each element are all linearly related, and controlled by a single parameter, $F_\star$, called the \expression{depletion strength factor}:
\begin{equation}
  \delta(E)\simeq A_E\times F_\star+B_E.
  \label{eq:Fstar}
\end{equation}
The factors $A_E$ and $B_E$ are empirically determined for each element.
The depletion factor accounts for the fact that depletions are different along different sightlines.
They however vary according to \refeq{eq:Fstar}.
This effect is due to dust growth in the \hISM.
It is supported by the good correlation between $F_\star$ and the average density of the \hISM, demonstrated on \refsubfig{fig:depletions}{a}.
When the density of the \hISM\ increases, the collisional rate of a grain with heavy elements increases.
A fraction of these elements stick on the grain surface and grow mantles.
\begin{itemize}
  \item When $F_\star\simeq0$, we are in the most diffuse \hISM, the depletion
    is $\delta(E)\simeq B_\sms{E}$. 
  \item When $F_\star\simeq1$, such as toward $\zeta$~Oph, we are sampling the 
    dense \hISM.
    The amplitude of the depletion, $A_\sms{E}$, reflects the composition of the 
    grain mantles.
\end{itemize}

\begin{figure}[htbp]
  \begin{tabular}{cc}
    \includegraphics[width=0.48\textwidth]{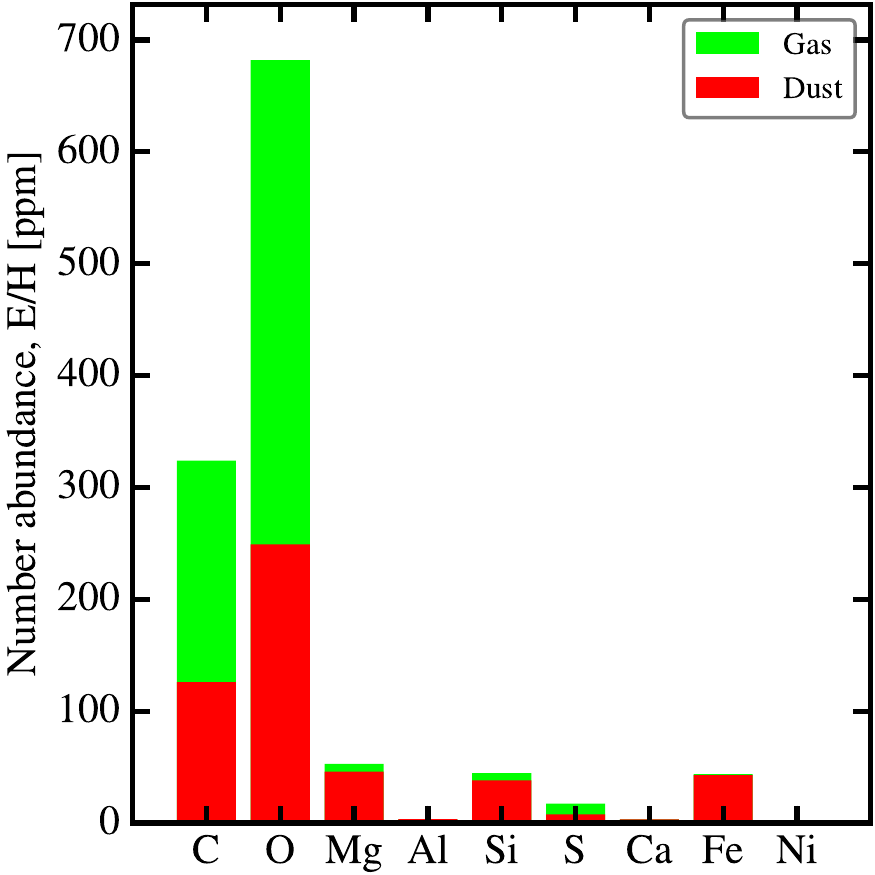} & 
    \includegraphics[width=0.48\textwidth]{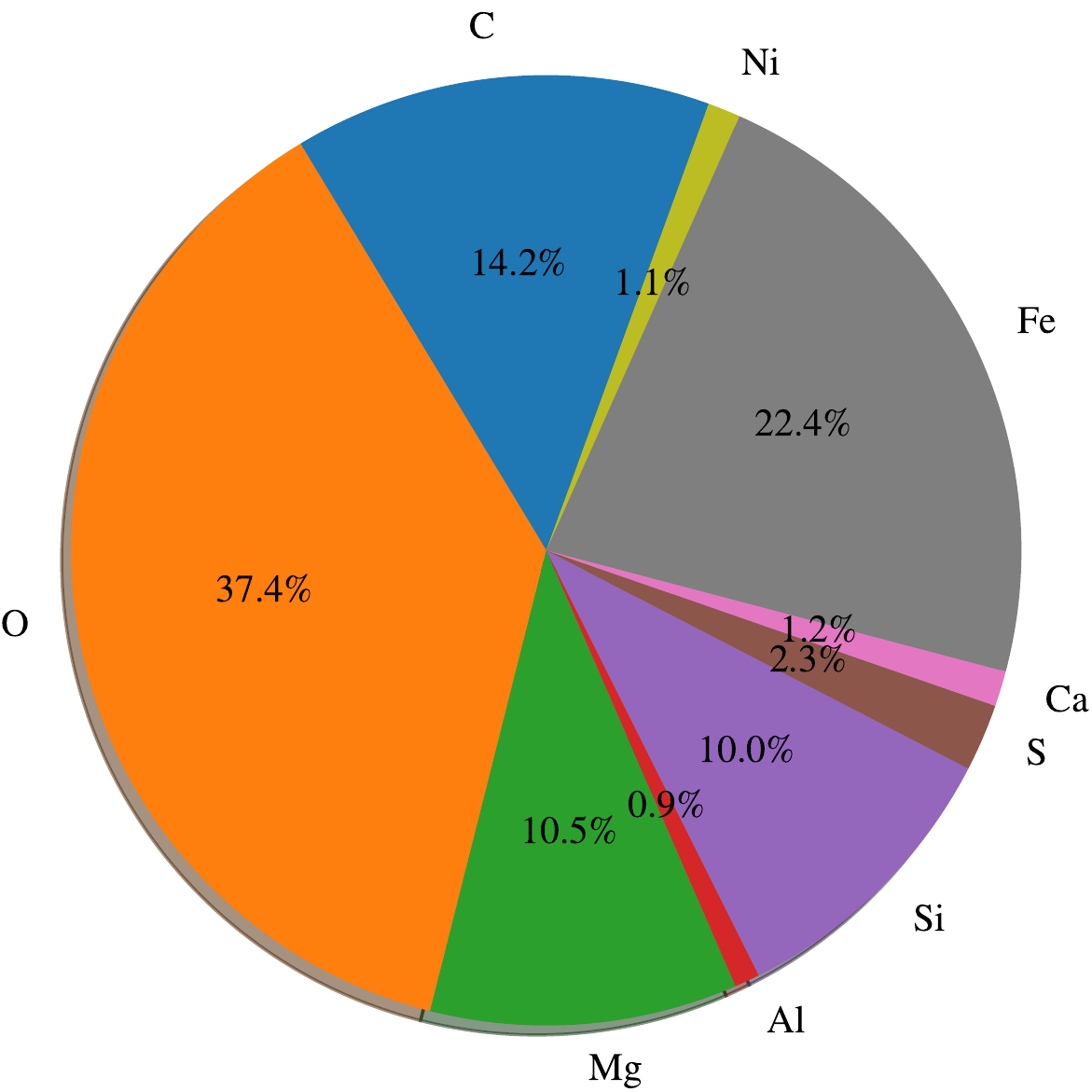} \\
    \textit{(a)} Number abundance, relative to H &
    \textit{(b)} Grain composition, in mass \\
  \end{tabular}
  \newcap{MW dust composition inferred from depletions}%
         {Panel~\textit{(a)}: number abundances of the main depleted elements, 
          relative to H, in \expression{part per million} (ppm),
          into dust (red) and gas (green).
          These values correspond to the \hMW, for $F_\star=0.5$ 
          \citep[Table~2 of][]{hensley21}.
          The top of each histogram represent the total \hISM\ abundance.
          Panel~\textit{(b)}: mass fraction of the different elements locked in
          grains, in the \hMW, for $F_\star=0.5$.
          \CClicence}
  \label{fig:depletions_comp}
\end{figure}

\paragraph{Volatile and refractory elements.}
Not all the most abundant elements in the \hISM\ enter the dust composition.
Some elements such as N or the noble gases are not significantly depleted.
\refsubfig{fig:depletions}{b} shows a general relation between the depletion amplitude and the condensation temperature of the most abundant heavy elements.
The most depleted elements are those which have a high condensation temperature.
For that reason, elements are often classified in the two following categories.
\begin{description}
  \item[Volatile elements] are the elements with low condensation 
    temperatures (C, N, O, noble gases).
    A moderate temperature is sufficient to remove them from the grains.
    These elements thus exist mainly in the gas phase.
  \item[Refractory elements] are the elements with high condensation 
    temperatures (those are essentially the metals). 
    They can be present in grains up to high temperatures.
    Their abundance in the gas phase therefore exhibits large variations as 
    a function of environment.
\end{description}
We note that, although C and O are two of the main dust constituents, these elements are classified as volatile.
These elements are indeed mainly in the gas phase, as their depletion is moderate (\cf\ \refsubfig{fig:depletions_comp}{a}). 
However, this modest depletion is sufficient to account for a large fraction of the dust mass.

\paragraph{Inferred dust composition.}
Since the individual depletion of each element can be inferred, it provides the unique prospect of constraining the average composition of dust grains.
This composition changes with density, as mantles grow.
Following \citet{hensley21}, we quote depletions for $F_\star=0.5$ as they correspond to $\langle n_\sms{H}\rangle\simeq0.3$~cm$^{-3}$, which is appropriate for the diffuse \hISM.
From this vantage point, the \hdustiness\ of the diffuse Galactic \hISM\ is:
\begin{equation}
  Z_\sms{dust}\equiv\frac{M_\sms{dust}}{M_\sms{gas}}\simeq\frac{1}{126\pm20}
    \simeq0.0079\pm0.0012.
\end{equation}
The dust-to-metal mass ratio is thus:
\begin{equation}
   DM\equiv\frac{Z_\sms{dust}}{Z}\simeq\frac{1}{2\pm0.26}\simeq0.592\pm0.093.
\end{equation}
The results of \citet{jenkins09} indicate that the \hdustiness\ is $\simeq2.7$ times higher at $F_\star=1$ than at $F_\star=0$.
The number and mass abundance in grains is represented in \reffig{fig:depletions_comp}.
The carbonaceous-to-silicate mass ratio is:
\begin{equation}
  \frac{M_\sms{C-dust}}{M_\sms{Sil.}} \simeq 0.177\pm0.085.
\end{equation}
Finally, we can have an idea of what the stoichiometry of silicates should be:
\begin{equation}
 \textnormal{SiO}_{6.6\pm2.5}\textnormal{Mg}_{1.21\pm0.16}\textnormal{Fe}_{1.13\pm0.14}.
\end{equation}
We note it results in a higher Si:O ratio than in olivine (1:4) and pyroxene (1:3) (\cf\ \refsec{sec:dustanalog}).
It is currently difficult to understand where all the depleted oxygen is, even if it also forms various oxides, such as Fe$_2$O$_3$, Al$_2$O$_3$, \etc

  \subsection{Direct Measures}
  \label{sec:direct}

Direct characterization of interstellar grains is possible in a few particular situations:
\begin{inlinelist}
  \item presolar grain inclusions in meteorites;
  \item interstellar grains entering the heliosphere; or 
  \item study of dust analogs in the laboratory.
\end{inlinelist}

    \subsubsection{Meteorite Inclusions}
    \label{sec:meteorites}

\begin{figure}[htbp]
  \begin{tabular}{cc}
    \includegraphics[width=0.48\textwidth]{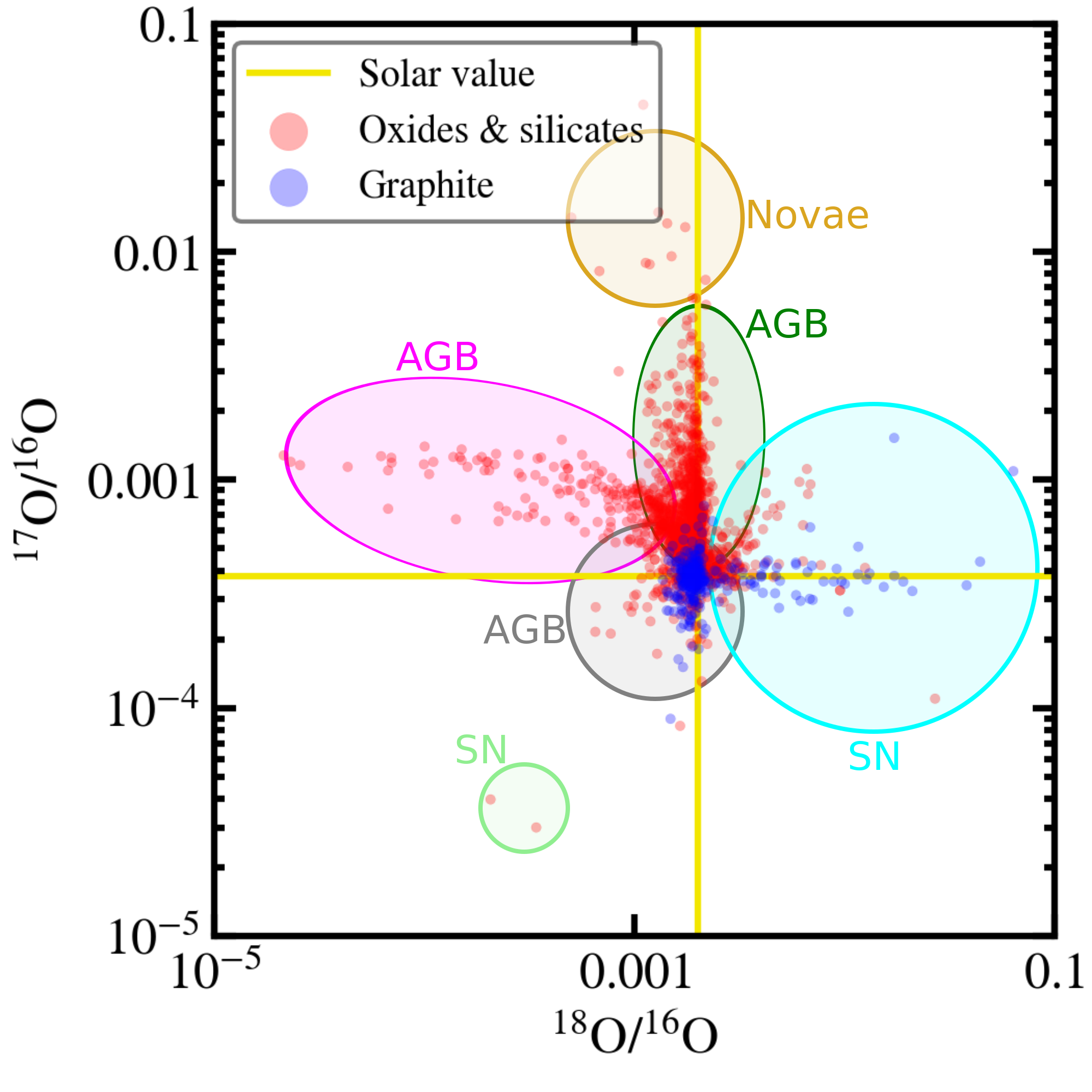} & 
    \includegraphics[width=0.48\textwidth]{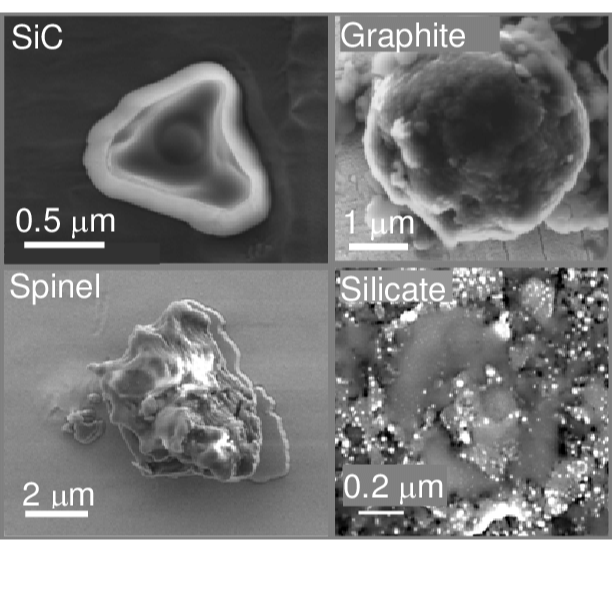} \\
    \textit{(a)} Isotopic abundances & \textit{(b)} Presolar grains \\
  \end{tabular}
  \newcap{Presolar grains in meteorites}%
   {Panel~\textit{(a)} shows the oxygen isotopic ratios of meteoritic 
    presolar grains, from the
    \href{https://presolar.physics.wustl.edu/presolar-grain-database/}{Presolar 
    Grain Database of Washington University} \citep{hynes09,stephan20}.
    Panel~\textit{(b)} shows pictures of presolar grains from primitive 
    meteorites \citep{hoppe10}. 
    The SiC grain is from a \hSN, the graphite from an \hAGB\ star or a \hSN, 
    and the spinel and silicate grains are from \hAGB\ stars.
    \uline{Credit:}
    \begin{inlinelistalph}
      \item \cclicence;
      \item courtesy of the 
        \href{https://www.mpg.de/153030/chemistry}{Max Planck Institute for Chemistry}, 
        with permission from Peter \familyname{Hoppe}.
    \end{inlinelistalph}}
  \label{fig:meteorites}
\end{figure}

\paragraph{Grain identification.}
Primitive meteorites contain presolar grains, that is grains that formed in the \hISM\ before being incorporated in the early Solar nebula \citep[\eg][]{hoppe00}.
They are believed to have remained relatively unaltered since the formation of the Solar system.
They can be identified by their isotopic anomalies (\cf\ \refsubfig{fig:meteorites}{a}).
Carbonaceous chondrites that we have mentioned in \refsec{sec:ISMabund} are of particular interest \citep[\eg][]{nittler19}.
Interstellar grains identified in meteorites can have one of the following compositions (\cf\ \refsubfig{fig:meteorites}{b}): 
\begin{itemize}
  \item Silicon carbide (SiC);
  \item Graphite (C sp$^2$);
  \item Silicates (SiO$_{3-4}$);
  \item Nanodiamonds (C sp$^3$);
  \item Silicon nitride (Si$_3$N$_4$);
  \item Corundum (Al$_2$O$_3$);
  \item Spinel (MgAl$_2$O$_4$); and 
  \item Titanium oxide (TiO$_2$).
\end{itemize}
The size of these grains ranges from a few tenths of nanometers to a few microns.
Their isotopic ratios are consistent with condensation in the ejecta of \hSN e or \expression{Asymptotic Giant Branch} (\hAGB) stars (\cf\ \refsubfig{fig:meteorites}{a}).

\paragraph{Limitations.}
Overall, the current analysis of presolar grains in meteorites suffers from several biases.
The search for presolar grains in meteorites uses chemical treatments dissolving the silicate matrix \citep{draine03c}.
It is the likely reason why:
\begin{inlinelist}
  \item most grains are crystalline stardust,
  \item why so few silicate grains are found, and
  \item why the smallest grains are not detected.
\end{inlinelist}

    \subsubsection{Interplanetary Dust}

\paragraph{ISD flux and cometary dust.}
We have seen in \refsec{sec:graincollect} that several spacecrafts have collected \expression{Interplanetary Dust Particles} (\hIDP) \textit{in situ}.
Among these \hIDP s, several grains have been shown to be of interstellar origins, because of the direction and speed of their flow.
Cometary dust also provides important clues, as comets formed during the early epoch of the Solar system.
They should contain pristine material.
A class of \hIDP s called \expression{Glass with Embedded Metals and Sulfides} \citep[\hGEMS;][]{bradley94,keller08}, are presolar.
They have sizes ranging from $\simeq0.1$ to 0.5~\tmic.

\begin{figure}[htbp]
  \includegraphics[width=\textwidth]{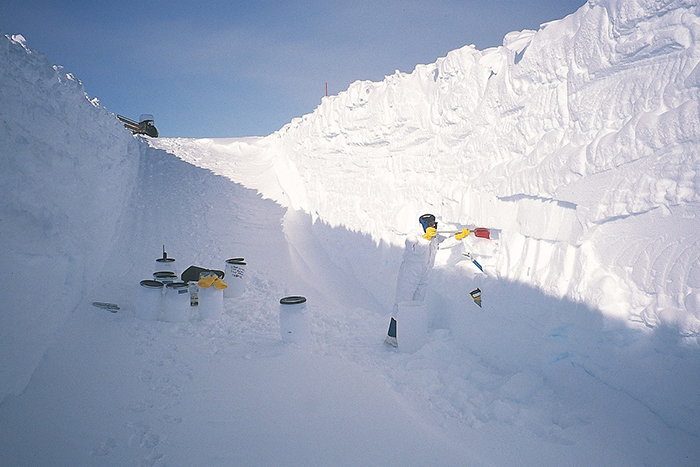}
  \newcap{Micrometeorite collection in Antarctica}%
         {Collecting micrometeorites in the central Antarctic regions, at Dome 
          C, in 2002. 
          \uline{Credit:} \href{https://www.cnrs.fr/en/more-5000-tons-extraterrestrial-dust-fall-earth-each-year}{Jean Duprat, Cécile Engrand, courtesy of CNRS Photothèque.}}
  \label{antartica}
\end{figure}
\paragraph{Micrometeorites.}
In addition to grain collection in space, \hIDP s entering the atmosphere become micrometeorites.
These can be collected on Earth and analyzed in the laboratory.
Antarctica is particularly interesting to that purpose, because of the absence of pollution and the possibility to sample the snow \citep[\cf\ \reffig{fig:meteorites};][for a review]{rojas21}.

    \subsubsection{Laboratory Measurements}
    \label{sec:lab}

Dust analogs, that is solids we think are making up \hISD, can be extensively studied in the laboratory \citep[\eg][for a review]{henning10b}.
We can distinguish at least two general types of experiments:
\begin{inlinelist}
  \item spectroscopic characterization; and
  \item reactivity and processing.
\end{inlinelist}
\begin{figure}[htbp]
  \includegraphics[width=\textwidth]{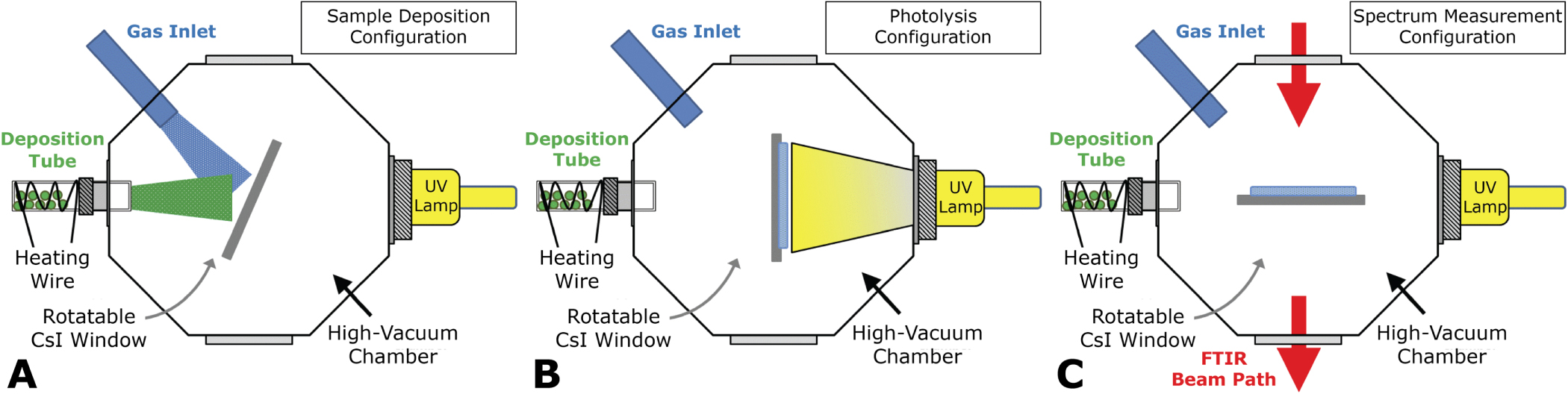}
  \newcap{NASA Ames PAH experiment}{A typical setup for a matrix-isolation 
    experiment: (A) sample deposition configuration, (B) \hUV\ photolysis 
    configuration, and (C) configuration for collecting the \hIR\ spectrum.
    \uline{Credit:} \citet{mattioda20}, with permission from Andy \familyname{Mattioda}.}
  \label{fig:experiment}
\end{figure}

\begin{figure}[htbp]
  \includegraphics[width=\textwidth]{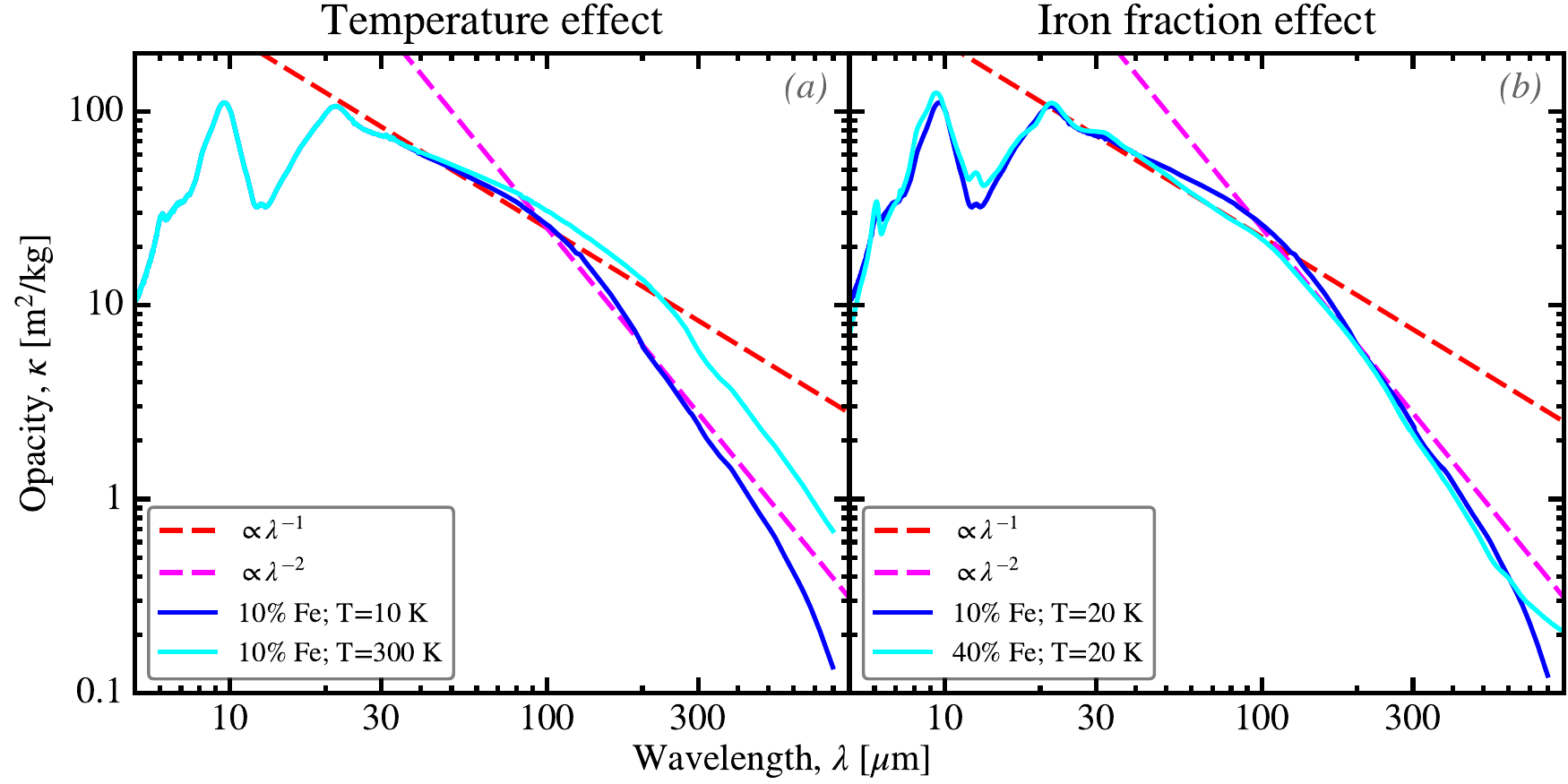}
  \newcap{Laboratory measurement of silicate opacities}%
         {These data are from the amorphous silicate samples of 
          \citet{demyk17b}.
          Among other parameters, this study samples the effects of:
          \begin{inlinelistalph} 
            \item temperature; and 
            \item iron fraction.
          \end{inlinelistalph}
          \CClicence}
  \label{fig:labsil}
\end{figure}

\paragraph{Spectroscopic characterization.}
Two general steps are required to perform such measures:
\begin{inlinelist}
  \item synthesizing the target compound;
  \item measuring its optical properties, usually in a rather narrow spectral 
    regime.
\end{inlinelist}
The details of these steps depend a lot on the nature of the compound,
 and on the spectral range explored.
\reffig{fig:experiment} shows an example of a particular experimental device to measure \hPAH\ properties in the \hIR, at \href{https://www.astrochemistry.org/pahdb/experimental/3.00/default/view}{NASA Ames}.
Different groups across the world specialize in such measures on \hPAH s \citep[\eg][]{useli-bacchitta10,bauschlicher18}, carbon grains \citep[\eg][]{mennella98,dartois16}, silicates \citep[\eg][]{dorschner95,demyk17b} and ices \citep[\eg][]{white09}, among others.
\reffig{fig:labsil} shows some of the results of the silicate study of \citet{demyk17b,demyk17a}.

\paragraph{Grain reactivity and evolution.}
Other experiments tackle the reactivity on grain surface \citep[\eg\ water formation;][]{dulieu10}.
Grain evolution in the \hISM\ is also studied.
For instance, the photoproduction of \hHAC\ \citep{dartois05}, the ion absorption on carbon grains \citep{mennella03}, the processing under high energy \citep[to mimic cosmic rays, \eg][]{dartois13}.
Some laboratory samples can even be exposed to space conditions  \citep{kebukawa19}, onboard the \expression{International Space Station} (\hISS).

\section{State-of-the-Art Dust Models}
\label{sec:dustmodels}

A dust model is defined by the abundance and size distribution of several grain components, characterized by their composition (\hPAH, graphite, silicates, \etc).
We now review how the Galactic observables we have presented in \refsec{sec:dustobs} are used to constrain modern dust models.
These models are therefore specific to the Galactic diffuse \hISM.
When using such a model to interpret other observations, we can vary the intensity and spectral shape of the \hISRF, to account for local variations.
In principle, we can also vary the abundance of each component, and some parameters of the size distribution to fit observations of other systems.
A dust model is a parametric framework that we can use to interpret any dust observable. 
There are however some limitations that we will discuss in \refchap{chap:dustprop}.

  \subsection{Composition and Size Distributions of Different
                 Models}

There has been a large number of dust models in the past.
We discuss here only some of the most recent ones \citep{zubko04,draine07,compiegne11,siebenmorgen14,jones17,guillet18}.

    \subsubsection{Diversity in Composition}
    \label{sec:modelcompo}

\paragraph{Inherent degeneracies of dust models.}
Different models make different choices in terms of composition.
This is because, even with all the constraints we have listed in \refsec{sec:dustobs}, there are still numerous degeneracies. 
Several dust mixtures can fit the same observables.
This has been best demonstrated by \citet[][hereafter \citetalias{zubko04}]{zubko04}.
\citetalias{zubko04} fitted the \hUV-to-\hNIR\ extinction, \hIR\ emission and elemental depletions with different compositions, including: \hPAH s, graphite, different types of amorphous carbons, silicates, and composite grains.
They also varied the reference abundances used to estimate elemental depletions.
In the end, they showed 15 different dust mixtures providing satisfying fits to the Galactic diffuse \hISM\ observables.
\begin{table}[!htbp]
  \centering
  \setlength\arrayrulewidth{2pt}
  \arrayrulecolor{white}
  \begin{tabularx}{\linewidth}{|>{}r%
                                |>{\columncolor{coltabcell}}X%
                                |>{\columncolor{coltabcell}}X%
                                |>{\columncolor{coltabcell}}X%
                                |>{\columncolor{coltabcell}}X%
                                |>{\columncolor{coltabcell}}X%
                                |>{\columncolor{coltabcell}}X|}
    \hline
      & \cellcolor{coltabsep}
        \rotatebox[origin=c]{90}{~~\citet[][BARE-GR-S]{zubko04}~~}
      & \cellcolor{coltabsep}
        \rotatebox[origin=c]{90}{\citet{draine07}}
      & \cellcolor{coltabsep}
        \rotatebox[origin=c]{90}{\citet{compiegne11}}
      & \cellcolor{coltabsep}
        \rotatebox[origin=c]{90}{\citet{siebenmorgen14}}
      & \cellcolor{coltabsep}
        \rotatebox[origin=c]{90}{\citet{jones17}}
      & \cellcolor{coltabsep}
        \rotatebox[origin=c]{90}{\citet{guillet18}} \\
    \hline
      \rowcolor{coltabhead}
      \multicolumn{7}{|c|}{Observational constraints accounted for} \\
    \hline
      \cellcolor{coltabsep} \hUV-to-\hNIR\ extinction
        & $\checkmark$
        & $\checkmark$
        & $\checkmark$
        & $\checkmark$
        & $\checkmark$
        & $\checkmark$ \\
      \cellcolor{coltabsep} Polarized extinction
        &
        & 
        & 
        & $\checkmark$
        & 
        & $\checkmark$ \\
      \cellcolor{coltabsep} \hMIR\ extinction
        & 
        & 
        & $\checkmark$
        & 
        & $\checkmark$
        & $\checkmark$ \\
      \cellcolor{coltabsep} Albedo
        &
        &
        & $\checkmark$
        & 
        & $\checkmark$
        & $\checkmark$ \\
      \cellcolor{coltabsep} \hNIR-to-mm emission
        & $\checkmark$
        & $\checkmark$
        & $\checkmark$
        & $\checkmark$
        & $\checkmark$
        & $\checkmark$ \\
      \cellcolor{coltabsep} Polarized emission
        & 
        & 
        & 
        & 
        & 
        & $\checkmark$ \\
      \cellcolor{coltabsep} Elemental depletions
        & $\checkmark$
        & 
        & $\checkmark$
        & $\checkmark$
        & $\checkmark$
        & $\checkmark$ \\
    \hline
      \rowcolor{coltabhead}
      \multicolumn{7}{|c|}{Composition of the dust mixture} \\
    \hline
      \cellcolor{coltabsep} \hPAH s
        & $\checkmark$
        & $\checkmark$
        & $\checkmark$
        & $\checkmark$
        & 
        & \\
      \cellcolor{coltabsep} Small \hHAC\ grains
        & 
        & 
        & $\checkmark$
        & $\checkmark$
        & $\checkmark$
        & $\checkmark$ \\
      \cellcolor{coltabsep} Large \hHAC\ grains
        & 
        & 
        & $\checkmark$
        & $\checkmark$
        & $\checkmark$
        & $\checkmark$ \\
      \cellcolor{coltabsep} Small graphite grains
        & $\checkmark$
        & $\checkmark$
        & 
        & $\checkmark$
        & 
        & \\
      \cellcolor{coltabsep} Large graphite grains
        & $\checkmark$
        & $\checkmark$
        & 
        & $\checkmark$
        & 
        & \\
      \cellcolor{coltabsep} Small silicate grains
        & $\checkmark$
        & 
        & 
        & $\checkmark$
        & 
        & \\
      \cellcolor{coltabsep} Large silicate grains
        & $\checkmark$
        & $\checkmark$
        & $\checkmark$
        & $\checkmark$
        & $\checkmark$
        & $\checkmark$ \\
      \cellcolor{coltabsep} Grain mantles
        & 
        & 
        & 
        & 
        & $\checkmark$
        & $\checkmark$ \\
      \cellcolor{coltabsep} Grain inclusions
        & 
        & 
        & 
        & 
        & $\checkmark$
        & $\checkmark$ \\
    \hline
  \end{tabularx}
  \newcap{Comparison between different dust models}%
         {The goal of this table is to illustrate the diversity of 
          observational constraints and the possible choices of dust mixtures.
          In the first part of the table, some checkmarks are questionable.
          A given model may indeed not actually use a particular constraint, 
          but end up being consistent with it, whereas another one may use it 
          but provide an imperfect fit.
          In the second part, the difference between ``small'' and ``large'' 
          grains is around a radius of $a\simeq10$~nm corresponding to the
          typical transition radius for stochastically heated grains (\cf\
          \refsec{sec:stochastheat}).}
  \label{tab:dustmodels}
\end{table}

\paragraph{Common compositional choices.}
A dust model accounting for at least the \hUV-to-\hMIR\ extinction and the \hIR\ emission must have the following features.
\begin{description}
  \item[PAHs or small a-C(:H)] are necessary to account for the aromatic 
    features.
    Among the models we discuss here, only \citetalias{jones17} also accounts
    for the 3.4~\tmic\ aliphatic feature.
    In addition, \hPAH s or small \hHAC\ account for a large fraction of the
    2175~$\r{A}$ extinction bump.
  \item[Silicate grains] are necessary to account for the 9.7 and 18~\tmic\
    silicate features.
    In addition, even if depletions are not actually fitted, they indicate that
    about 2/3 of dust mass must reside in some form of silicate grains (\cf\ 
    \refsec{sec:depletions}).
  \item[Large carbon grains] are necessary to account for the bulk of the \hFIR\
    emission with a reasonable \hdustiness.
    Large, uncoated silicate grains are indeed not emissive enough to explain
    the \hFIR\ \hSED\ without requiring more heavy elements locked up in grains 
    than what is available in the \hISM.
    Graphite and, even more, \hHAC\ will increase the overall emissivity of the
    large grain mixture to the desirable level, using the second most abundant 
    dust specie available.
\end{description}
In addition to these choices, grain mantles and/or inclusions can also enter the composition.
Some have been explored by \citetalias{zubko04}.
These are an essential part of the \citetalias{jones17} model, which is designed as an evolution model (\cf\ \refsec{sec:themis}).
The mantle thickness is indeed one of the parameters quantifying grain evolution through the \hISM.
Another important parameter is the shape of the grains.
Elongated grains are necessary to account for the polarization in extinction and emission (\cf\ \refsec{sec:intropola}).
\citet{siebenmorgen14} designed a model accounting for the polarized extinction.
The model of \citet{guillet18} is currently the only one also accounting for the polarized emission measured by \hplanck.
\reftab{tab:dustmodels} summarizes the differences between the most recent dust models.

    \subsubsection{Difference in Size Distributions}
    \label{sec:model_sizedist}

\paragraph{Origin of the Size Distribution.}
The size distribution of interstellar grains is a complex balance between the formation and destruction processes that we will discuss in \refchap{chap:dustevol}.
Two of these processes explain quite naturally two widely-used functional forms.
\begin{description}
  \item[Collisional fragmentation] of an initial distribution of large grains 
    leads to a power-law size distribution with an index close to $\simeq-3.5$, 
    similar to the \citetalias{mathis77} size distribution (\refeqnp{eq:MRN}; 
    $f(a)\propto a^{-3.5}$).
    This result was demonstrated for asteroids by \citet[][with a different 
    index, in his case]{hellyer70}.
    \citet{dorschner82} explained the interstellar size distribution as a 
    result of collisions in the circumstellar envelopes where grains are 
    produced.
    A simplified demonstration of this process is given in Chap.~7 of
    \citet{kruegel03}.
  \item[Turbulent grain growth] results in a log-normal dust size distribution
    \citep{mattsson20}.
\end{description}
\begin{figure}[htbp]
  \includegraphics[width=\textwidth]{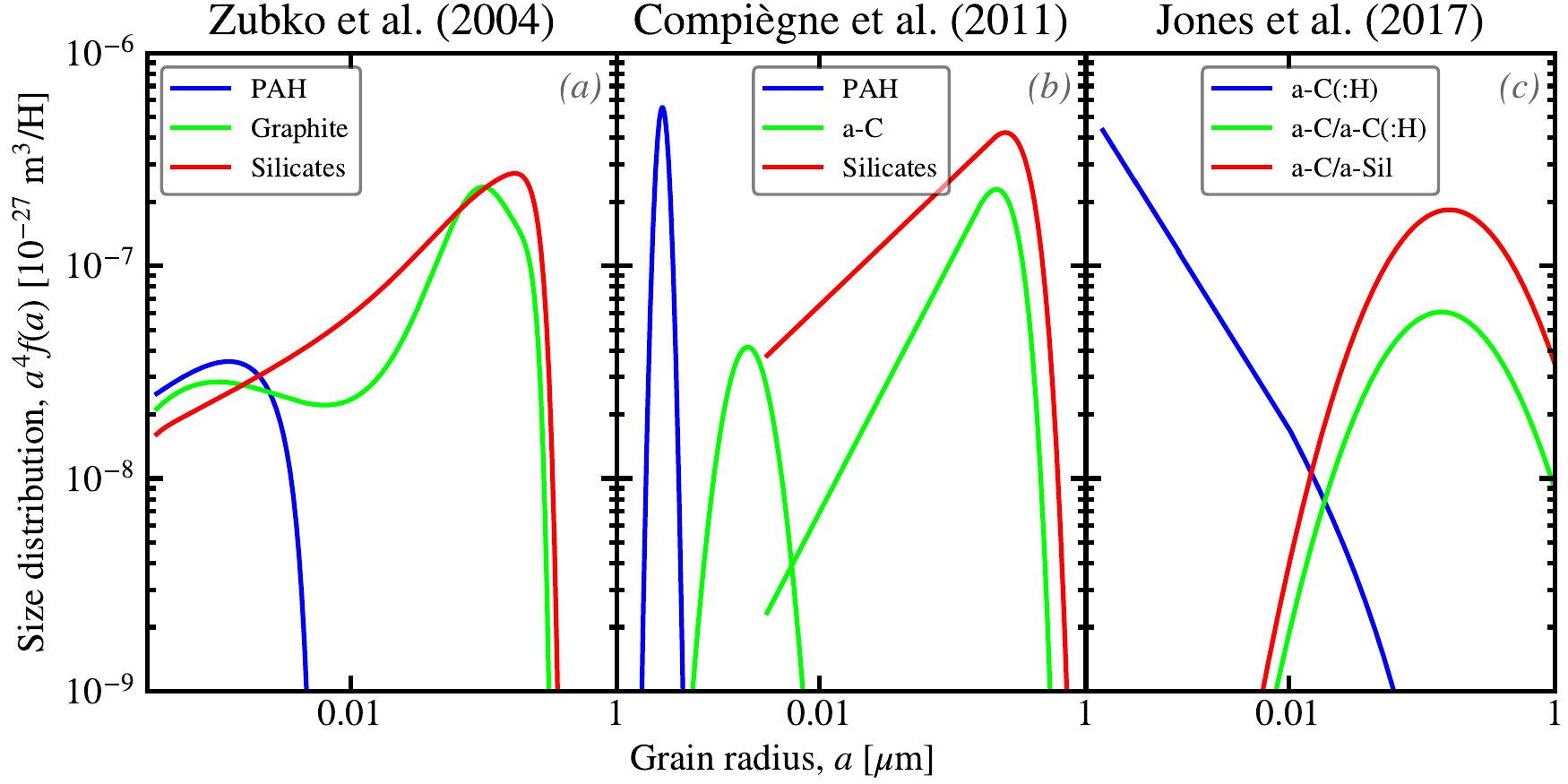}
  \newcap{Size distribution of several dust models}%
         {This figure shows the size distribution of the 
          \citet[][BARE-GR-S]{zubko04},
          \citet{compiegne11} and \citetalias{jones17} models.
          \CClicence}
  \label{fig:sizedist}
\end{figure}

\paragraph{Comparison Between Different Models.}
\reffig{fig:sizedist} compares the size distributions of three of the models we are discussing in this section.
The displayed size distributions, $a\times f(a)$, are multiplied by $a^3$ so that they are mass-weighted.
Although they manage to fit the same observables, these size distributions are quite different.
\citetalias{zubko04} adopt complex functional forms (\refsubfig{fig:sizedist}{a}).
\citet{compiegne11} use log-normal size distributions for \hPAH s and small a-C, and power-law with an exponential cut-off for large grains (\refsubfig{fig:sizedist}{b}).
It is the opposite for \citetalias{jones17}, which uses log-normals for large grains and a power-law with an exponential cut-off for small \hHAC\  (\refsubfig{fig:sizedist}{c}).
Despite these differences, we notice the common features that we have listed in \refsec{sec:modelcompo}.
\begin{itemize}
  \item In the three panels of \reffig{fig:sizedist}, the aromatic feature 
    carrying grains (in blue) are the smallest carbon grains.
    Their size distribution needs to peak around 
    $3\;\r{A}\lesssim a\lesssim50\;\r{A}$, for these grains to fluctuate to 
    high enough temperatures.
    Current observables are not accurate enough to distinguish differences 
    in the shape of $f(a)$ in this range.
  \item Intermediate-size grains ($5\;\textnormal{nm}\lesssim 
    a\lesssim20\;\textnormal{nm}$) are necessary 
    to account for the \hMIR\ continuum.
    \citetalias{zubko04} use both graphite and silicate grains, resulting in the
    presence of the 9.7 and 18~\tmic\ features in emission.
    \citet{compiegne11} use a separate a-C log-normal component (in green).
    For the \citetalias{jones17} model, the \hMIR\ continuum is accounted for 
    by the tail of the small \hHAC\ (in blue).
  \item The large grain distributions (in blue and red) all peak around 
    $a\simeq0.1$~\tmic.
    They need to drop sharply above this value, as larger grains tend to have 
    lower equilibrium temperatures (\cf\ \refsec{sec:Teq}) that would broaden
    the \hFIR\ emission peak of the \hSED.
\end{itemize}

  \subsection{The Model Properties}
  \label{sec:modelprop}

Each model computes the panchromatic opacity, albedo and emissivity of its grain mixture.
There are slight differences between different models, because they use different data sets and because the coverage of these data sets is not complete.
The properties of the dust mixture of a model are simply the properties of its individual grains, integrated over the size distribution.
For a given function $X(a)$, we note:
\begin{equation}
  \langle X\rangle_a \equiv \int_{a_-}^{a_+}X(a)f(a)\ddiff a.
\end{equation}
The general properties defined in \refsec{sec:calcQabs} and \refsec{sec:stochastheat} 
can therefore be generalized as:
\begin{eqnarray}
  \langle m_\sms{dust}\rangle_a
    & = & \left\langle\frac{4}{3}\pi a^3\rho\right\rangle_a 
  \label{eq:SD1} \\
  \kappa_\sms{abs/sca}(\lambda)
    & = & \frac{\displaystyle%
                \left\langle\pi a^2Q_\sms{abs/sca}(a,\lambda)\right\rangle_a}%
               {\displaystyle\left\langle m_\sms{dust}\right\rangle_a}  \\
  g(\lambda) 
    & = & \frac{\displaystyle%
                \left\langle 
                  g(a,\lambda)\pi a^2Q_\sms{sca}(a,\lambda)\right\rangle_a}%
               {\displaystyle%
                \left\langle\pi a^2Q_\sms{sca}(a,\lambda)\right\rangle_a} \\
  \epsilon_\nu(\lambda)
    & = & 
    \frac{\displaystyle%
          \left\langle4\pi Q_\sms{abs}(a,\lambda)
            \pi\int_0^\infty \frac{\dd P(T,a)}{\dd T}B_\nu(\lambda,T)\ddiff T\right\rangle_a}%
         {\displaystyle\left\langle m_\sms{dust}\right\rangle_a}.
  \label{eq:SD2}
\end{eqnarray}

    \subsubsection{Extinction and Emission}

\begin{figure}[htbp]
  \begin{tabular}{cc}
    \includegraphics[width=0.48\textwidth]{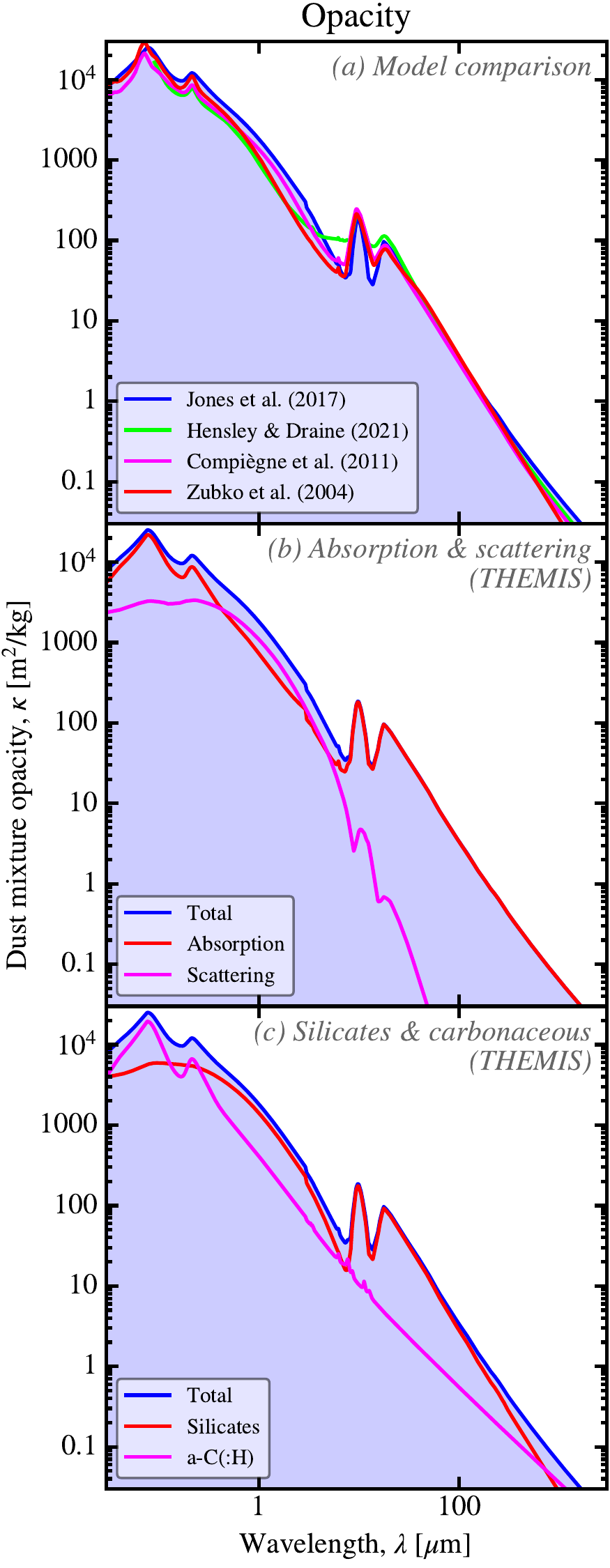} &
    \includegraphics[width=0.48\textwidth]{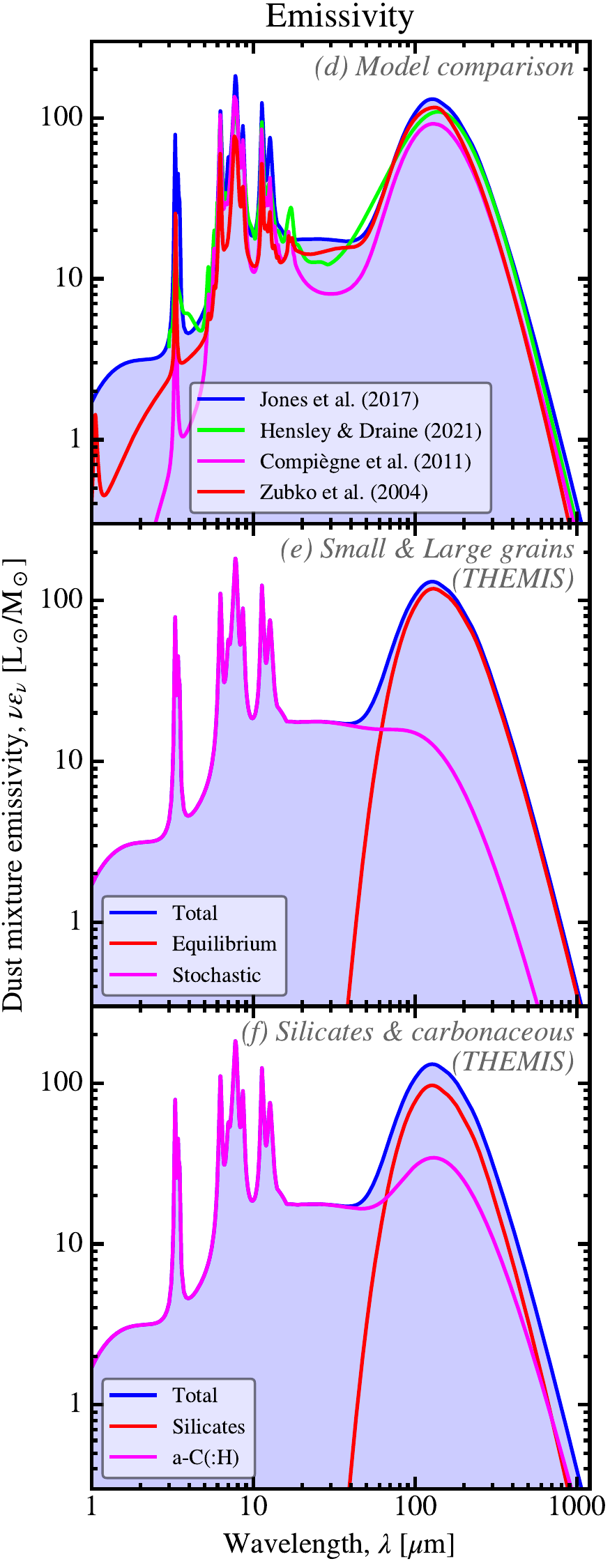} \\
  \end{tabular}
  \newcap{Model opacity and emissivity}%
         {Panel~\textit{(a)} compares the total opacities of: the BARE-GR-S mixture of 
          \citet{zubko04}; the 
          \citet{compiegne11} model; the \citetalias{jones17} model 
          \citep{jones17} and the synthetic observations of \citet{hensley21}.
          Panel~\textit{(b)} shows the decomposition of the opacity into 
          absorption and scattering (for \citetalias{jones17}).
          Panel~\textit{(c)} shows the decomposition of the opacity into
          silicates and carbonaceous grains (for \citetalias{jones17}).
          Panel~\textit{(d)} compares the emissivity of the different dust 
          models.
          Panel~\textit{(e)} shows the decomposition of the emissivity into 
          large and small grains (for \citetalias{jones17}).
          Panel~\textit{(f)} shows the decomposition of the emissivity into
          silicates and carbonaceous grains (for \citetalias{jones17}).
          \CClicence}
  \label{fig:kappaSED}
\end{figure}

\paragraph{The opacity.}
\refsubfig{fig:kappaSED}{a} compares the panchromatic opacity of different models.
At first order, the four models are in good agreement.
\begin{itemize}
  \item The discrepancies in the \hUV-to-\hNIR\ range are due to the different 
    compositions.
    For instance, the \citetalias{jones17} model tends to have more opaque 
    material.
    The observations of the Galaxy being expressed per H atom, this model
    therefore requires a slightly lower \hdustiness.
  \item The \citet{hensley21} model has a much flatter \hMIR\ continuum, to 
    account for Galactic-center-type extinction curves (\cf\ 
    \refsec{sec:extinctionMIR}).
  \item At wavelengths longer than 100~\tmic, the \citetalias{jones17} and 
    \citet{hensley21} models have a flatter opacity than the other ones.
    \hplanck\ constraints are indeed higher in this regime than the 
    extrapolation of the \hCOBE/\hFIRAS\ spectrum.
\end{itemize}
The bottom left panels of \reffig{fig:kappaSED} show the decomposition of the opacity of the \citetalias{jones17} model.
\begin{description}
  \item[Scattering and absorption] are shown in 
    \refsubfig{fig:kappaSED}{b}.     
    We can see that scattering is dominant only in the \hNIR\ range.
    This scattering component mainly originates in large grains (\cf\ 
    \refsec{sec:calcQabs}).
  \item[Carbonaceous and silicates] are shown in \refsubfig{fig:kappaSED}{c}.
    We can see that carbon grains dominate the \hUV\ and submm opacity.
    In the particular case we have displayed (the \citetalias{jones17} model), 
    silicate grains are coated with \hHAC\ mantles.
    The carbon component of a model made of bare grains would be sensibly 
    higher.
\end{description}

\paragraph{The SED.}
\refsubfig{fig:kappaSED}{d} compares the \hSED\ of the same four models as previously.
The shapes of these \hSED s are relatively similar.
\begin{itemize}
  \item The differences in the peak of the \hFIR\ \hSED\ is due to the 
    difference in composition, mirroring the difference in opacity previously 
    discussed.
  \item The difference in the level of the \hMIR\ continuum is due to the 
    paucity of observational constraints in this regime.
  \item The difference in the level of the aromatic feature emission is due to:
    \begin{inlinelist}
      \item the different \hdustiness; and
      \item the different sets of \hMIR\ constraints \citepalias[\cf\ 
        Sect.~3.1.1 of][for a discussion of this discrepancy]{galliano21}.
    \end{inlinelist}
\end{itemize}
The bottom right panels of \reffig{fig:kappaSED} show the decomposition of the \citetalias{jones17} model in sizes and composition.
\begin{description}
  \item[Small and large grains] are shown in \refsubfig{fig:kappaSED}{e}.
    We can see that the small grains are responsible from the \hMIR\ emission, 
    as they are stochastically heated (\cf\ \refsec{sec:stochastheat}).
  \item[Carbonaceous and silicates] are shown in \refsubfig{fig:kappaSED}{f}.
    Carbon grains are responsible for the entire \hMIR\ emission and for a 
    small fraction of the \hFIR\ peak.
    Amorphous-carbon-coated-silicates are responsible for most of the \hFIR\ 
    emission peak.
\end{description}

  \subsubsection{The Fitted Constraints}
  \label{sec:themis}

We now demonstrate the fit of the observational constraints by one the models, \citetalias{jones17}.
We start by presenting this model in more depth.

\paragraph{The THEMIS model.}
It is a laboratory-data-based model.
As we have previously discussed, it uses two populations of grains:
\begin{inlinelist}
  \item \hHAC\ grains with the optical properties of 
    \citet{jones12a,jones12b,jones12c}; and
  \item a-Silicates with Fe and FeS inclusions and \hHAC\ mantles, whose 
    optical properties have been computed by \citet{kohler15}.
\end{inlinelist}
The largest \hHAC\ are coated with a-C.
A first version was presented by \citet{jones13} and updated by \citet{jones17}.
At the time this manuscript is being written, a new version is in preparation including the laboratory optical properties of silicates measured by \citet{demyk17b,demyk17a}.
It is an evolution model.
The hydrogenation of \hHAC, their size distribution, as well as the mantle thickness of the large grains are parameters evolving with the \hISRF\ intensity and the density of the \hISM.

\begin{figure}[!htbp]
  \begin{tabular}{cc}
    \includegraphics[width=0.685\textwidth]{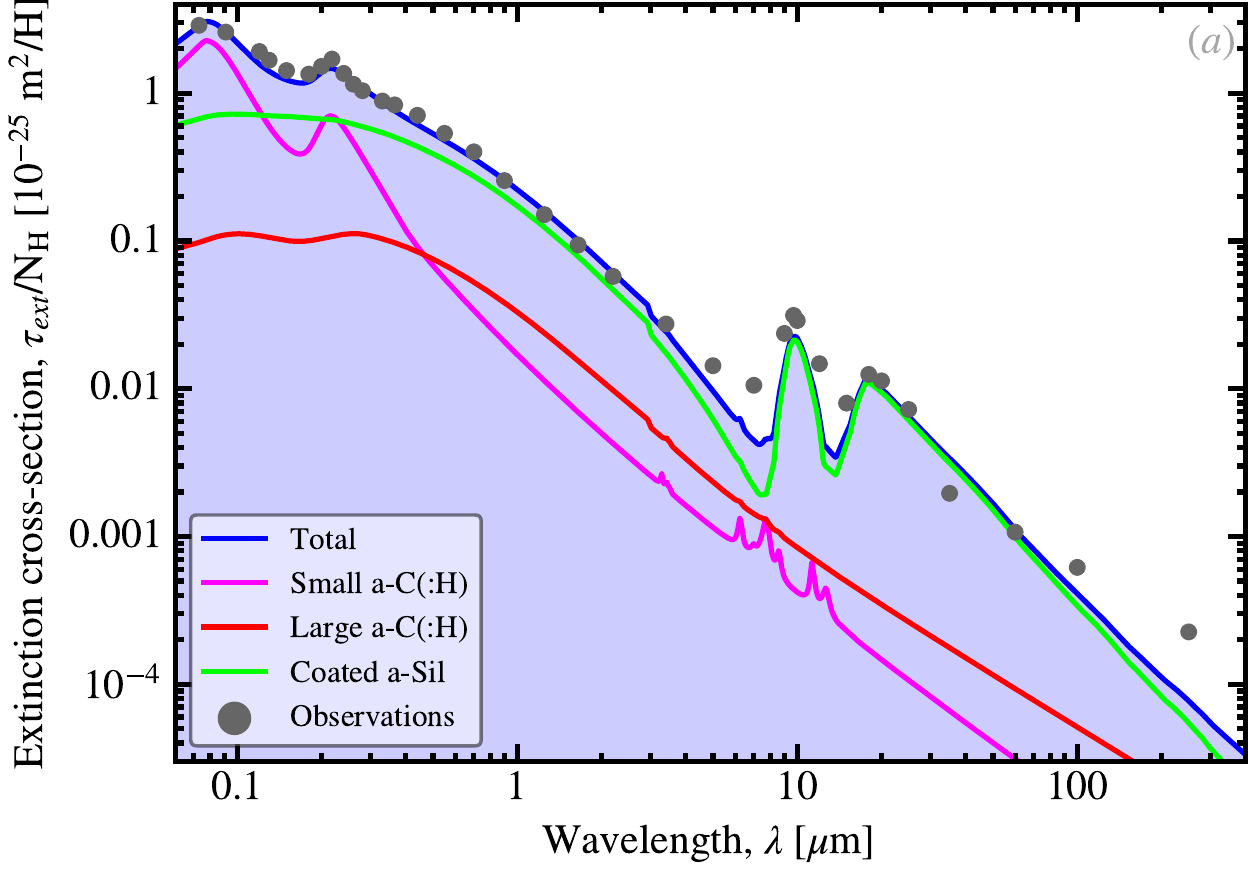} &
    \includegraphics[width=0.275\textwidth]{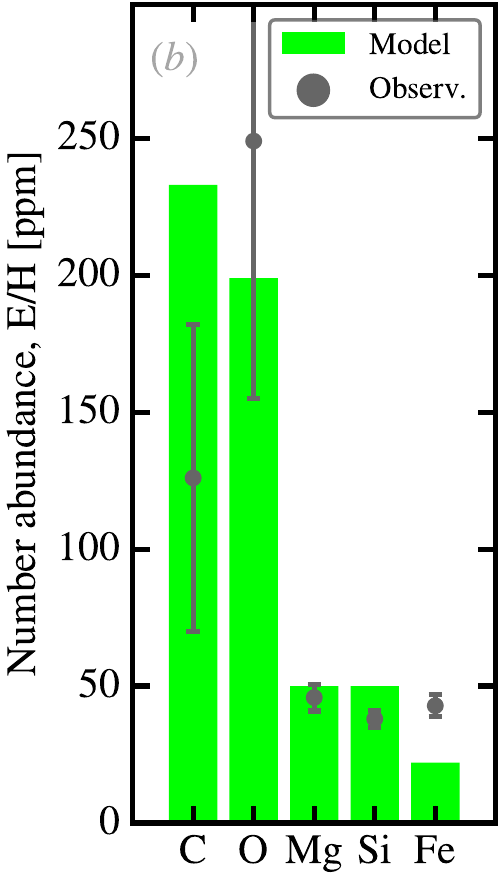} \\
  \end{tabular}
  \begin{tabular}{cc}
    \includegraphics[width=0.48\textwidth]{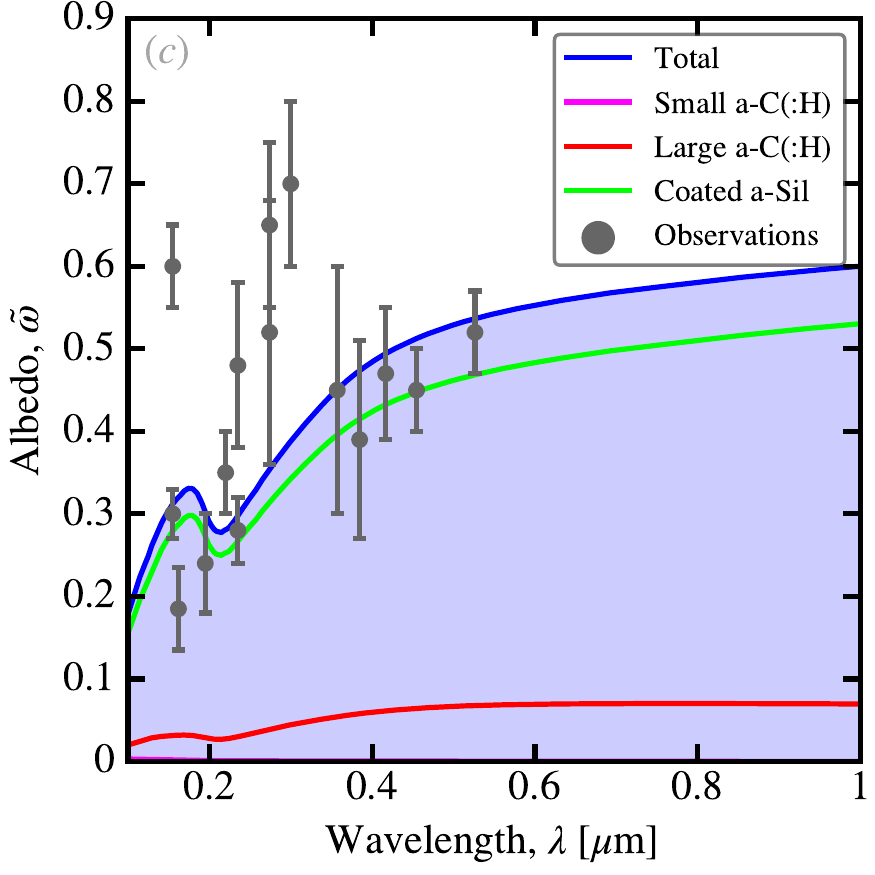} & 
    \includegraphics[width=0.48\textwidth]{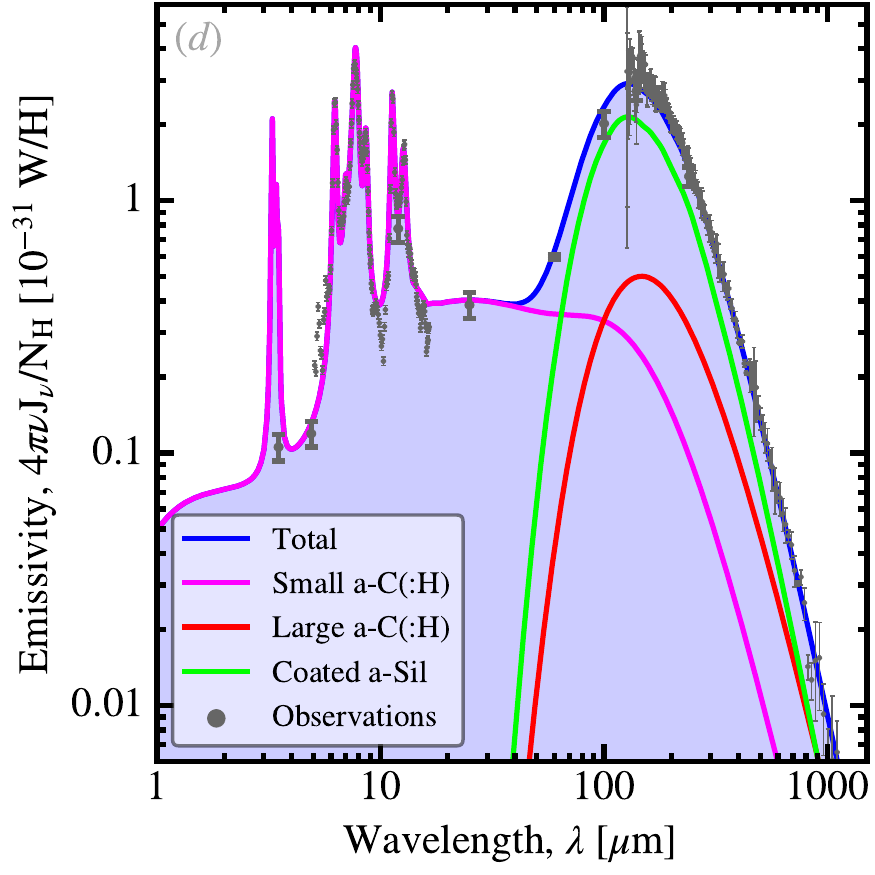} \\
  \end{tabular}
  \newcap{\citetalias{jones17} fit of the Galactic constraints}%
         {Panel~\textit{(a)}: fit of the diffuse Galactic \hISM\ extinction 
          curve with the THEMIS model by \citet{jones13}.
          The observations (black dots) are from \citet{mathis90} and are not
          provided with uncertainties.
          Panel~\textit{(b)}: albedo fit with the \citet{jones13} model.
          The observations are from: 
          \citet{lillie76,morgan76,morgan80,chlewicki88,hurwitz91,witt97}.
          Panel~\textit{(c)}: emissivity fit with the \citet{jones13} model.
          The observations are those of \reffig{fig:obsIR}.
          \CClicence}
  \label{fig:THEMIS}
\end{figure}

\paragraph{Discussion of the fit.}
\reffig{fig:THEMIS} shows the fit of the diffuse Galactic \hISM\ constraints by the \citetalias{jones17} model.
\begin{description}
  \item[The extinction curve] is well fitted except in the 10~\tmic\ range
    (\refsubfig{fig:THEMIS}{a}).
    This region is the most problematic because of:
    \begin{inlinelist}
      \item the uncertainty about the profiles of the astrophysical silicate 
        mixture features, which is common to every model; and
      \item the uncertainty about the shape of the continuum in this range 
        (\cf\ \refsec{sec:extinctionMIR}).
    \end{inlinelist}
    The synthetic observed extinction curve used to constrain this model
    is provided by \citet{mathis90} without error bars.
  \item[The elemental depletions] are relatively well fitted except for Fe
    (\refsubfig{fig:THEMIS}{b}).
    This is a common problem of contemporary dust models (\cf\ 
    \refsec{sec:depletions}).
  \item[The albedo] is relatively well fitted (\refsubfig{fig:THEMIS}{c}). 
    The problem is that the observational constraints themselves are rather 
    scattered.
    Some constraints in the \hUV\ range are inconsistent.
    This is because albedo measurements come from a diversity of regions
    (\hDGL, reflection nebulae; \cf\ \refsec{sec:extinctionUV}), which are 
    difficult to homogenize.
    In addition, the asymmetry parameter derived from these observations is 
    rather high.
    The albedo is thus measured at the tail of the scattering phase function, 
    which adds another layer of uncertainties.
  \item[The emissivity] is well fitted (\refsubfig{fig:THEMIS}{d}).
    There are no problem with this component.
    This is important because this model is used to analyze the \hIR\ emission 
    of galaxies and Galactic regions.
\end{description}

  \subsection{Some Useful Quantities}
  \label{sec:themis_quant}

We finish this chapter by listing a few quantities and formulae, useful to make simple estimates and approximations.
Unless otherwise noted, these quantities are computed using the \citetalias{jones17} dust model, and might slightly differ if another model is considered.

    \subsubsection{Grain Sizes, Areas and Masses}
    \label{sec:themis_sizedist}
   
\paragraph{For the MRN size distribution.}
The grain surface is important for chemical reactions and for the photoelectric effect.
For a \citetalias{mathis77} size distribution \refeqp{eq:MRN}, the average grain surface is:
\begin{equation}
  \langle S_\sms{dust}\rangle_a
    = \pi\int_{a_-}^{a_+} \underbrace{f(a)}_{\propto a^{-3.5}} 
    a^2\ddiff a \propto\frac{1}{\sqrt{a_+}}-\frac{1}{\sqrt{a_-}}
    \simeq \frac{1}{\sqrt{a_-}}.
\end{equation}
where $a$ is the grain radius, $f(a)$, the size distribution from \refeq{eq:MRN}, and $a_-$ and $a_+$, the minimum and maximum sizes ($a_+\gg a_-$).
\takeaway{The grain surface is thus dominated by small grains.}
The average grain volume is:
\begin{equation}
  \langle V_\sms{dust}\rangle_a
    = \frac{4\pi}{3}\int_{a_-}^{a_+} \underbrace{f(a)}_{\propto a^{-3.5}} 
    a^3\ddiff a \propto \sqrt{a_+}-\sqrt{a_-}\simeq \sqrt{a_+},
\end{equation}
For a given grain species, the volume is proportional to the mass, thus the average grain mass is $\langle m_\sms{dust}\rangle_a\propto\sqrt{a_+}$, too.
\takeaway{The grain mass is dominated by large grains.}

\paragraph{The case of the THEMIS model.}
The size distribution of the \citetalias{jones17} model (\refsubfig{fig:sizedist}{c}) can be split into three components: 
\begin{inlinelist}
  \item small \hHAC;
  \item big \hHAC; and
  \item silicates.
\end{inlinelist}
\reftab{tab:sizedistmoments} gives the first moments of the size distribution of these three components as well as of the total.
The first line indicates that most grains are small grains.
This is also reflected in the last column of each line: the value of each parameter is very close to its value for small \hHAC.
\begin{table}[htbp]
  \centering
  \setlength\arrayrulewidth{2pt}
  \arrayrulecolor{white}
  \begin{tabularx}{\linewidth}{|>{}X%
                                |>{\columncolor{coltabcell}}r%
                                |>{\columncolor{coltabcell}}r%
                                |>{\columncolor{coltabcell}}r%
                                |>{\columncolor{coltabcell}}r|}
    \hline
      & \cellcolor{coltabhead}\textbf{Small \hHAC}
      & \cellcolor{coltabhead}\textbf{Large \hHAC}
      & \cellcolor{coltabhead}\textbf{a-Silicates}
      & \cellcolor{coltabhead}\textbf{Total} \\
    \hline
      \cellcolor{coltabhead} Grain number fraction, $\langle1\rangle_a$
      & $10^9$ \hppb & 36 \hppb & 80 \hppb & $10^9$ \hppb \\
    \hline
      \cellcolor{coltabhead} Average radius, $\langle a\rangle_a$
      & 0.54 nm & 12 nm & 13 nm & 0.54 nm \\
    \hline
      \cellcolor{coltabhead} Average area, $\pi\langle a^2\rangle_a$
      & 1.02 nm$^2$ & 1140 nm$^2$ & 1510 nm$^2$ & 1.03 nm$^2$ \\     
    \hline
      \cellcolor{coltabhead} Average mass, $4/3\pi\rho\langle a^3\rangle_a$
      & 1040 amu & $1.2\E{8}$ amu & $2.5\E{8}$ amu & 2890 amu \\     
    \hline
      \cellcolor{coltabhead} Mass fraction
      & $23\,\%$ & $8.1\,\%$ & $69\,\%$ & $100\,\%$ \\     
    \hline
  \end{tabularx}
  \newcap{Moments of the \citetalias{jones17} size distribution}%
         {The number fractions are expressed in \expression{part per billion} 
          (\hppb) and the masses in \expression{atomic mass unit} (amu; 
          \reftab{tab:constants}).}
  \label{tab:sizedistmoments} 
\end{table}                                                                    

\paragraph{Mass fraction of small grains.}
The mass fraction of aromatic feature emitting grains \citepalias[\ie\ \hHAC\  smaller than $a\simeq1.5$~nm; \cf\ Fig.~1 of][]{galliano21} is $q_\sms{AF}\simeq 17\,\%$.
Other models, using \hPAH s instead of small \hHAC, use different values, because \hPAH s have more aromatic bonds per C atom than \hHAC.
A smaller \hPAH\ mass is thus required to account for the same aromatic band strength.
The mass fraction of \hPAH s is $q_\sms{PAH}\simeq4.6\,\%$ for the \citet{zubko04} and \citet{draine07} models, and $q_\sms{PAH}\simeq7.7\,\%$ for the \citet{compiegne11} model.
The difference between the two latter values is due to the different sets of \hMIR\ constraints they use \citepalias[\cf\ \refsec{sec:IRobs}; see also Sect.~3.1.1 of][for a discussion]{galliano21}.
For the \citetalias{jones17} model, the mass fraction of the grains responsible for the \hMIR\ continuum (\ie\ small \hHAC\ with radii $a\gtrsim1.5$~nm) is $q_\sms{MIRcont}\simeq6\,\%$.

\paragraph{Dustiness and other ratios.}
\reftab{tab:massthemis} gives various number and mass ratios for the \citetalias{jones17} model.
The \hdustiness\ and the dust-to-H mass ratios are equivalent quantities, there is just a factor $(1-Y_\odot-Z_\odot)$ difference.
The third line tells us that there are about 2 dust grains per million H atoms in the \hISM.
The last line indicates that about $40\,\%$ of the mass of heavy elements in the diffuse Galactic \hISM\ is locked-up in dust grains.
\begin{table}[htbp]
  \centering
  \setlength\arrayrulewidth{2pt}
  \arrayrulecolor{white}
  \begin{tabularx}{\linewidth}{|>{}X%
                               |>{\columncolor{coltabcell}}r%
                               |>{\columncolor{coltabcell}}r%
                               |>{\columncolor{coltabcell}}r%
                               |>{\columncolor{coltabcell}}r|}
    \hline
      & \cellcolor{coltabhead}\textbf{Small \hHAC}
      & \cellcolor{coltabhead}\textbf{Large \hHAC}
      & \cellcolor{coltabhead}\textbf{a-Silicates} 
      & \cellcolor{coltabhead}\textbf{Total} \\
    \hline
      \cellcolor{coltabhead} Dustiness, 
      $Z_\sms{dust}\equiv M_\sms{dust}/M_\sms{gas}$
      & $1/800$ & $1/2260$ & $1/270$ & $1/183$ \\
      \cellcolor{coltabhead}
      & $\simeq1.3\E{-3}$ & $\simeq4.4\E{-4}$ & $\simeq3.8\E{-3}$ 
      & $\simeq5.5\E{-3}$ \\      
    \hline
      \cellcolor{coltabhead} Dust-to-H mass ratio, 
      & $1/600$ & $1/1700$ & $1/200$ & $1/138$ \\
      \cellcolor{coltabhead}$Y_\sms{dust}\equiv M_\sms{dust}/M_\sms{H}$
      & $\simeq1.7\E{-3}$ & $\simeq5.9\E{-4}$ & $\simeq5.0\E{-3}$ 
      & $\simeq7.3\E{-3}$ \\
    \hline
      \cellcolor{coltabhead} Dust-to-H number ratio, $N_\sms{dust}/N_\sms{H}$
      & 5300 \hppb & 0.05 \hppb & 0.02 \hppb & 1905 \hppb \\
    \hline
      \cellcolor{coltabhead} Dust-to-metal mass ratio, $Z_\sms{dust}/Z$
      & $1/11\simeq9\,\%$ & $1/30\simeq3\,\%$ & $28\,\%$ & $41\,\%$ \\
    \hline
  \end{tabularx}
  \newcap{Dustiness and other ratios for the \citetalias{jones17} model}%
         {For the second to fourth columns, the dust mass ($M_\sms{dust}$) or 
          number ($N_\sms{dust}$) are those of the sole component.
          Therefore, the sum of the second to fourth columns is equal to the 
          fifth column.}
  \label{tab:massthemis}
\end{table}

    \subsubsection{Opacity and Emissivity}

\begin{figure}[htbp]
  \includegraphics[width=\textwidth]{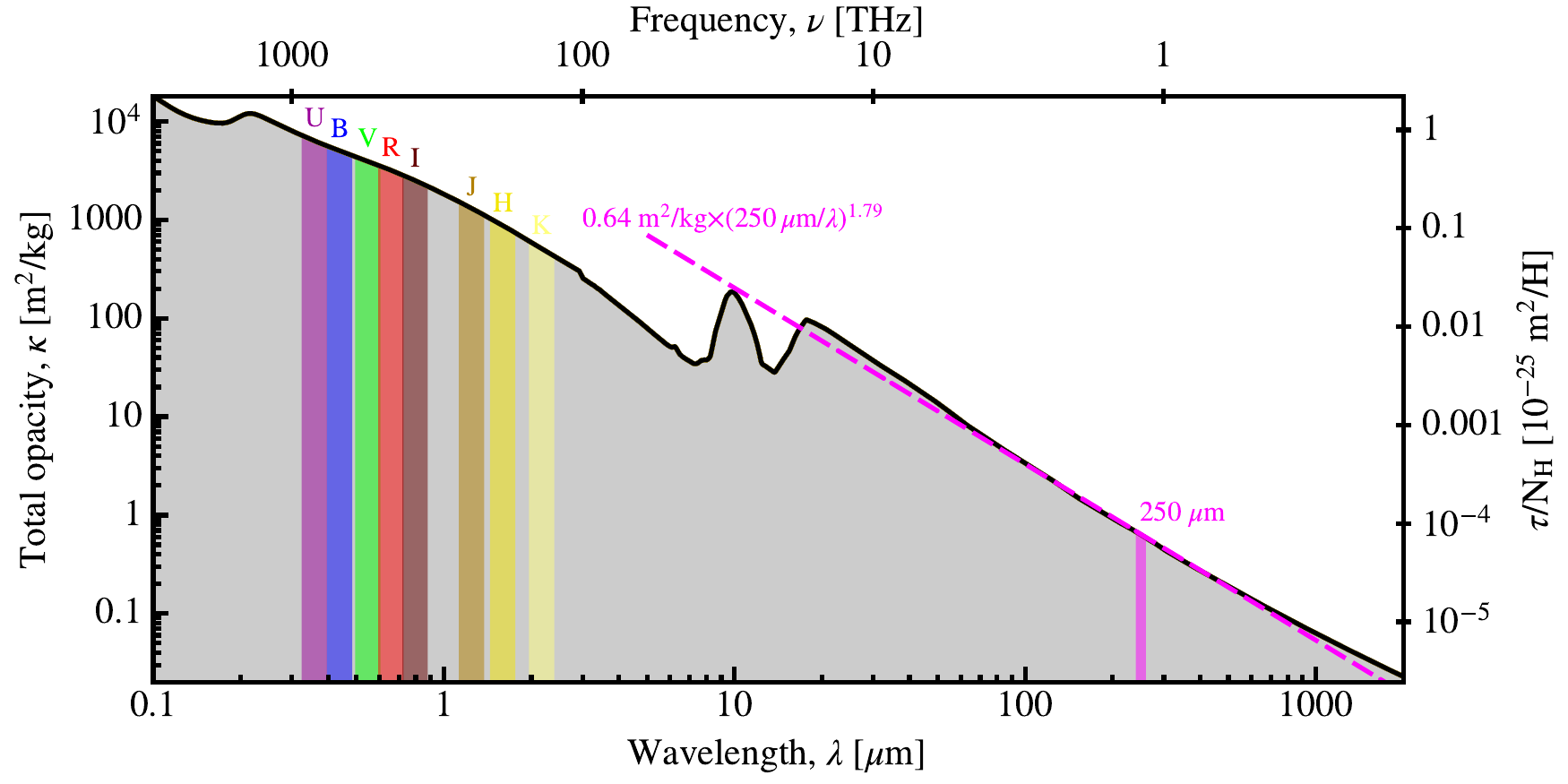}
  \newcap{IR approximation of the opacity}%
         {This figure shows the extinction opacity of the 
          \citetalias{jones17} model.
          We have highlighted the different photometric bands we quote in  
          \reftab{tab:opacity}.
          We also show the power-law approximation of the \hIR\ opacity (in 
          magenta; \refeqnp{eq:themis_kappa_proxy}).
          \CClicence}
  \label{fig:themis_kappa_proxy}
\end{figure}

\paragraph{Optical properties.}
\reftab{tab:opacity} gives the opacity, $\kappa$, $\tau/N_\sms{H}$, and the albedo, $\tilde\omega$, at the central wavelengths of the photometric filters displayed in \reffig{fig:themis_kappa_proxy}.
The opacity, $\kappa$ is expressed per mass of dust, whereas $\tau/N_\sms{H}$ is expressed per H atom in the gas phase.
The two quantities are related by:
\begin{equation}
  \frac{\tau(\lambda)}{N_\sms{H}}
    \equiv\kappa(\lambda)\times Z_\sms{dust}
    \times\frac{m_\sms{H}}{1-Y_\odot-Z_\odot},
\end{equation}
where $m_\sms{H}$ is the mass an H atom (\reftab{tab:constants}).
\reffig{fig:themis_kappa_proxy} displays a useful approximation, valid for $20\;\emic\lesssim\lambda\lesssim1$~mm:
\begin{equation}
  \kappa(\lambda)\simeq 0.64\;\textnormal{m}^2/\textnormal{kg}\times
    \left(\frac{250\;\emic}{\lambda}\right)^{1.79}.
  \label{eq:themis_kappa_proxy}
\end{equation}
\begin{table}[htbp]
  \centering
  \setlength\arrayrulewidth{2pt}
  \arrayrulecolor{white}
  \begin{tabularx}{\linewidth}{|>{}X|>{}l%
                               |>{\columncolor{coltabcell}}r%
                               |>{\columncolor{coltabcell}}r%
                               |>{\columncolor{coltabcell}}r%
                               |>{\columncolor{coltabcell}}r|}
    \hline
      & 
      & \cellcolor{coltabhead}\textbf{Small \hHAC}
      & \cellcolor{coltabhead}\textbf{Large \hHAC}
      & \cellcolor{coltabhead}\textbf{a-Silicates}
      & \cellcolor{coltabhead}\textbf{Total} \\      
    \hline
      \cellcolor{coltabhead}U band
      & \cellcolor{coltabhead}$\kappa$ 
      & 5934 m$^2$/kg & 9980 m$^2$/kg & 6107 m$^2$/kg & 6381 m$^2$/kg 
      \\
      \cellcolor{coltabhead}(0.36 \tmic)
      & \cellcolor{coltabhead}$\tau/N_\sms{H}$ 
      & $1.7\E{-26}$ m$^2$/H & $9.8\E{-27}$ m$^2$/H 
      & $5.1\E{-26}$ m$^2$/H & $7.8\E{-26}$ m$^2$/H 
      \\
      \cellcolor{coltabhead}
      & \cellcolor{coltabhead}$\tilde{\omega}$ 
      & $3.4\E{-3}$ & 0.43 & 0.60 & 0.45 
      \\
      \cellcolor{coltabhead} & \cellcolor{coltabhead}$g$
      & 0.19 & 0.57 & 0.56 & 0.56 \\
    \hline
      \cellcolor{coltabhead}B band
      & \cellcolor{coltabhead}$\kappa$ 
      & 3245 m$^2$/kg & 8549 m$^2$/kg & 5180 m$^2$/kg & 5008 m$^2$/kg 
      \\
      \cellcolor{coltabhead}(0.44 \tmic)
      & \cellcolor{coltabhead}$\tau/N_\sms{H}$ 
      & $9.1\E{-27}$ m$^2$/H & $8.4\E{-27}$ m$^2$/H 
      & $4.3\E{-26}$ m$^2$/H & $6.1\E{-26}$ m$^2$/H 
      \\
      \cellcolor{coltabhead}
      & \cellcolor{coltabhead}$\tilde{\omega}$ 
      & $3.4\E{-3}$ & 0.46 & 0.62 & 0.51 
      \\
      \cellcolor{coltabhead} & \cellcolor{coltabhead}$g$
      & 0.18 & 0.55 & 0.54 & 0.54 \\
    \hline
      \cellcolor{coltabhead}V band
      & \cellcolor{coltabhead}$\kappa$ 
      & 2023 m$^2$/kg & 7027 m$^2$/kg & 4272 m$^2$/kg & 3979 m$^2$/kg 
      \\
      \cellcolor{coltabhead}(0.55 \tmic)
      & \cellcolor{coltabhead}$\tau/N_\sms{H}$ 
      & $5.7\E{-27}$ m$^2$/H & $6.9\E{-27}$ m$^2$/H 
      & $3.6\E{-26}$ m$^2$/H & $4.8\E{-26}$ m$^2$/H 
      \\
      \cellcolor{coltabhead}
      & \cellcolor{coltabhead}$\tilde{\omega}$ 
      & $2.8\E{-3}$ & 0.48 & 0.64 & 0.54 
      \\
      \cellcolor{coltabhead} & \cellcolor{coltabhead}$g$
      & 0.16 & 0.54 & 0.53 & 0.53 \\
    \hline
      \cellcolor{coltabhead}R band
      & \cellcolor{coltabhead}$\kappa$ 
      & 1395 m$^2$/kg & 5789 m$^2$/kg & 3543 m$^2$/kg & 3231 m$^2$/kg 
      \\
      \cellcolor{coltabhead}(0.66 \tmic)
      & \cellcolor{coltabhead}$\tau/N_\sms{H}$ 
      & $3.9\E{-27}$ m$^2$/H & $5.7\E{-27}$ m$^2$/H 
      & $3.0\E{-26}$ m$^2$/H & $3.9\E{-26}$ m$^2$/H 
      \\
      \cellcolor{coltabhead}
      & \cellcolor{coltabhead}$\tilde{\omega}$ 
      & $2.3\E{-3}$ & 0.48 & 0.65 & 0.56 
      \\
      \cellcolor{coltabhead} & \cellcolor{coltabhead}$g$
      & 0.14 & 0.53 & 0.51 & 0.52 \\
    \hline
      \cellcolor{coltabhead}I band
      & \cellcolor{coltabhead}$\kappa$ 
      & 920 m$^2$/kg & 4487 m$^2$/kg & 2762 m$^2$/kg & 2479 m$^2$/kg 
      \\
      \cellcolor{coltabhead}(0.80 \tmic)
      & \cellcolor{coltabhead}$\tau/N_\sms{H}$ 
      & $2.6\E{-27}$ m$^2$/H & $4.4\E{-27}$ m$^2$/H 
      & $2.3\E{-26}$ m$^2$/H & $3.0\E{-26}$ m$^2$/H 
      \\
      \cellcolor{coltabhead}
      & \cellcolor{coltabhead}$\tilde{\omega}$ 
      & $1.7\E{-3}$ & 0.48 & 0.66 & 0.58 
      \\
      \cellcolor{coltabhead} & \cellcolor{coltabhead}$g$
      & 0.12 & 0.52 & 0.50 & 0.50 \\
    \hline
      \cellcolor{coltabhead}J band
      & \cellcolor{coltabhead}$\kappa$ 
      & 398 m$^2$/kg & 2380 m$^2$/kg & 1453 m$^2$/kg & 1286 m$^2$/kg 
      \\
      \cellcolor{coltabhead}(1.25 \tmic)
      & \cellcolor{coltabhead}$\tau/N_\sms{H}$ 
      & $1.1\E{-27}$ m$^2$/H & $2.4\E{-27}$ m$^2$/H 
      & $1.2\E{-26}$ m$^2$/H & $1.6\E{-26}$ m$^2$/H 
      \\
      \cellcolor{coltabhead}
      & \cellcolor{coltabhead}$\tilde{\omega}$ 
      & $7.5\E{-4}$ & 0.45 & 0.70 & 0.61 
      \\
      \cellcolor{coltabhead} & \cellcolor{coltabhead}$g$
      & 0.07 & 0.49 & 0.46 & 0.47 \\
    \hline
      \cellcolor{coltabhead}H band
      & \cellcolor{coltabhead}$\kappa$ 
      & 260 m$^2$/kg & 1636 m$^2$/kg & 982 m$^2$/kg & 869 m$^2$/kg 
      \\
      \cellcolor{coltabhead}(1.60 \tmic)
      & \cellcolor{coltabhead}$\tau/N_\sms{H}$ 
      & $7.3\E{-28}$ m$^2$/H & $1.6\E{-27}$ m$^2$/H 
      & $8.2\E{-27}$ m$^2$/H & $1.1\E{-26}$ m$^2$/H 
      \\
      \cellcolor{coltabhead}
      & \cellcolor{coltabhead}$\tilde{\omega}$ 
      & $4.6\E{-4}$ & 0.42 & 0.71 & 0.61 
      \\
      \cellcolor{coltabhead} & \cellcolor{coltabhead}$g$
      & 0.05 & 0.48 & 0.44 & 0.45 \\
    \hline
      \cellcolor{coltabhead}K band
      & \cellcolor{coltabhead}$\kappa$ 
      & 156 m$^2$/kg & 1005 m$^2$/kg & 573 m$^2$/kg & 512 m$^2$/kg 
      \\
      \cellcolor{coltabhead}(2.18 \tmic)
      & \cellcolor{coltabhead}$\tau/N_\sms{H}$ 
      & $4.4\E{-28}$ m$^2$/H & $9.9\E{-28}$ m$^2$/H 
      & $4.8\E{-27}$ m$^2$/H & $6.2\E{-27}$ m$^2$/H 
      \\
      \cellcolor{coltabhead}
      & \cellcolor{coltabhead}$\tilde{\omega}$ 
      & $2.5\E{-4}$ & 0.37 & 0.68 & 0.58 
      \\
      \cellcolor{coltabhead} & \cellcolor{coltabhead}$g$
      & 0.03 & 0.45 & 0.42 & 0.43 \\
  \end{tabularx}  
  \newcap{Optical properties of the \citetalias{jones17} model}%
         {This table gives several optical properties at each of the center of
          the photometric bands displayed in \reffig{fig:themis_kappa_proxy}.
          The opacity, $\kappa$, is the cross-section per mass of the dust 
          component, whereas
          $\tau/N_\sms{H}$ is the cross-section per H atom in the gas phase.
          The albedo is $\tilde{\omega}\equiv\kappa_\sms{sca}/\kappa$ 
          \refeqp{eq:albedo}, and $g$ is the asymmetry parameter 
          \refeqp{eq:costheta}.}
  \label{tab:opacity}
\end{table}

\begin{figure}[htbp]
  \includegraphics[width=\textwidth]{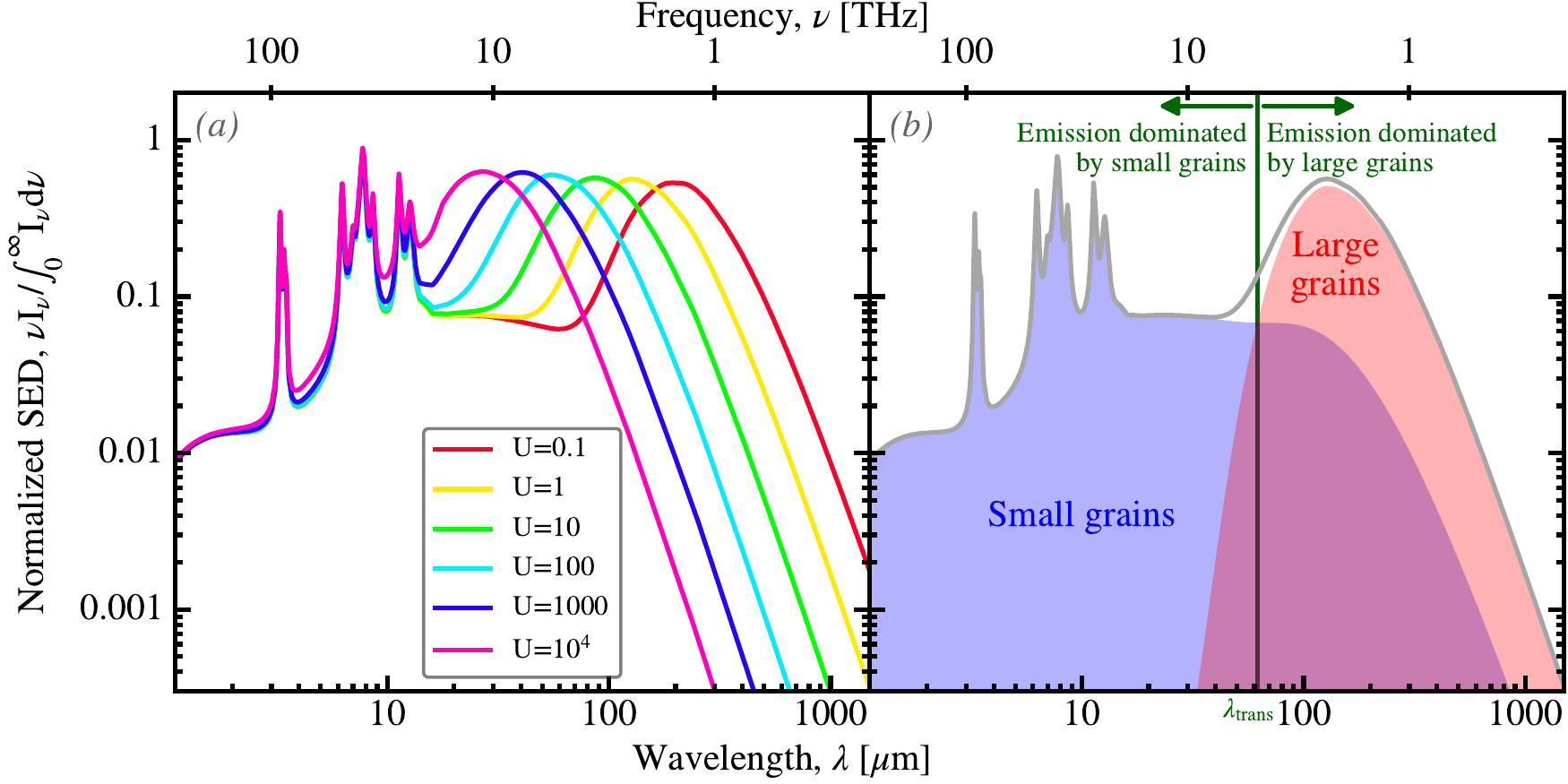}
  \newcap{Effect of U on the SED}%
         {Panel~\textit{(a)} shows the \hSED\ of the \citetalias{jones17} model 
          for different \hISRF\ intensities, $U$.
          Panel~\textit{(b)} shows the intensity of the \citetalias{jones17}
          model for $U=1$  (\refsec{sec:Teq}).
          We have decomposed the emission into large (a-Silicates and big 
          \hHAC) and small (small \hHAC) grains.
          We show the transition wavelength between the two heating regimes,
          $\lambda_\sms{trans}(U)$.
          In both panels, the $y$-axis is the intensity normalized by its 
          bolometric integral.
          Each displayed \hSED\ therefore emits the same total intensity.
          \CClicence}
  \label{fig:themis_emiss_proxy}
\end{figure}

\paragraph{Heating regimes.}
It is important to understand which regime is dominated by large grains at equilibrium with the \hISRF\ (\refsec{sec:Teq}), and which one is dominated by small, stochastically heated grains (\refsec{sec:stochastheat}).
\refsubfig{fig:themis_emiss_proxy}{a} shows the variation of the \hSED\ as a function of the \hISRF\ intensity, $U$ (\refsec{sec:Teq}).
We can see that when the intensity increases, the emission by large grains shifts to shorter wavelengths, as their equilibrium temperature increases.
On the contrary, the emission by small, out-of-equilibrium grains stays constant, as these grains are heated by single photon events.
Only their total intensity increases, which is hidden in \refsubfig{fig:themis_emiss_proxy}{a} by the normalization of the intensity.
We can estimate the transition wavelength, $\lambda_\sms{trans}(U)$, as the wavelength where the intensity of the small and large grains are equal.
This is demonstrated in \refsubfig{fig:themis_emiss_proxy}{b}.
The values of $\lambda_\sms{trans}(U)$ for the grid of $U$ displayed in \refsubfig{fig:themis_emiss_proxy}{a} is given in \reftab{tab:wtrans}.
\begin{table}[htbp]
  \centering
  \setlength\arrayrulewidth{2pt}
  \arrayrulecolor{white}
  \begin{tabularx}{\linewidth}{|>{}X%
                               |>{\columncolor{coltabcell}}X%
                               |>{\columncolor{coltabcell}}X%
                               |>{\columncolor{coltabcell}}X%
                               |>{\columncolor{coltabcell}}X%
                               |>{\columncolor{coltabcell}}X%
                               |>{\columncolor{coltabcell}}X|}
    \hline
      & \cellcolor{coltabhead}\textbf{U=0.1} 
      & \cellcolor{coltabhead}\textbf{U=1} 
      & \cellcolor{coltabhead}\textbf{U=10} 
      & \cellcolor{coltabhead}\textbf{U=100} 
      & \cellcolor{coltabhead}\textbf{U=1000} 
      & \cellcolor{coltabhead}\textbf{U=10}$^{\bm{4}}$ \\
    \hline
      \cellcolor{coltabhead} $\lambda_\sms{trans}(U)$ 
      & 88~\tmic & 62~\tmic & 43~\tmic & 31~\tmic & 22~\tmic & 17~\tmic \\
    \hline
  \end{tabularx}
  \newcap{Transition wavelengths between small and large grain emission}%
         {}
  \label{tab:wtrans}
\end{table}

\paragraph{Emissivity.}
\reftab{tab:themis_emiss} gives the emissivity of the \citetalias{jones17} model.
The emissivity is proportional to $U$.
We give only the value for $U=1$.
We quote the following two values.
\begin{description}
  \item[The emissivity,] \textit{per se}, integrated over the wavelengths, is 
    the emitted power per unit dust mass:
    \begin{equation}
      \epsilon\equiv\int_0^\infty\epsilon_\nu(\nu)\ddiff\nu.
    \end{equation}
  \item[The emitted power per H atom,] also integrated over the wavelengths, is 
    expressed per H atom in the gas phase:
    \begin{equation}
      \frac{4\pi I}{N_\sms{H}}
        = \int_0^\infty \frac{4\pi I_\nu(\nu)}{N_\sms{H}}\ddiff\nu.
    \end{equation}
\end{description}
These two quantities are related by:
\begin{equation}
  \frac{4\pi I}{N_\sms{H}}=\epsilon\times Z_\sms{dust}
           \times\frac{m_\sms{H}}{1-Y_\odot-Z_\odot}.
\end{equation}
\begin{table}[htbp]
  \centering
  \setlength\arrayrulewidth{2pt}
  \arrayrulecolor{white}
  \begin{tabularx}{\linewidth}{|>{}X%
                               |>{\columncolor{coltabcell}}r%
                               |>{\columncolor{coltabcell}}r%
                               |>{\columncolor{coltabcell}}r%
                               |>{\columncolor{coltabcell}}r|}
    \hline
      & \cellcolor{coltabhead}\textbf{Small \hHAC}
      & \cellcolor{coltabhead}\textbf{Large \hHAC}
      & \cellcolor{coltabhead}\textbf{a-Silicates}
      & \cellcolor{coltabhead}\textbf{Total} \\
    \hline
      \cellcolor{coltabhead}$\epsilon$
      & $98\times U\;\eLsun/\eMsun$
      & $25\times U\;\eLsun/\eMsun$
      & $97\times U\;\eLsun/\eMsun$
      & $221\times U\;\eLsun/\eMsun$ \\
    \hline
      \cellcolor{coltabhead}$4\pi I/N_\sms{H}$
      & $2.3\E{-31}\times U\:$W/H
      & $6.0\E{-32}\times U\:$W/H
      & $2.3\E{-31}\times U\:$W/H
      & $5.2\E{-31}\times U\:$W/H \\
    \hline
  \end{tabularx}
  \newcap{Emissivity of the \citetalias{jones17} model}%
         {These emissivities are proportional to $U$.}
  \label{tab:themis_emiss}
\end{table}


\stopcontents
\newchapter{The Grain Properties of Nearby Galaxies}
\markboth{\chaptername\ \thechapter.\ Grain Properties}{}
\label{chap:dustprop}
\citesmart{Without data, you're just another person with an opinion.}%
          {(Attributed to W.~Edwards \textsc{Deming})}
\minitoc

\noindent
In this chapter, we review how the models presented in \refsec{sec:dustmodels} are used to derive the grain properties of nearby galaxies.
The term \citengl{dust properties} is vague.
In the literature, it often indistinctively encompasses the three following categories \citep*{galliano18}.
\begin{description}
  \item[The dust mixture constitution] is characterized by:
    \begin{itemize}
      \item the chemical composition of the bulk material and its stoichiometry
        (\refsec{sec:dustanalog});
      \item the structure of the grains (crystalline, amorphous, porous, 
        aggregated, \etc; \refsec{sec:QabsnonMie});
      \item the presence of heterogeneous inclusions (\refsec{sec:QabsnonMie});
      \item the presence of organic and/or icy mantles 
        (\refsec{sec:QabsnonMie});
      \item the shape of the grains (\refsec{sec:intropola});
      \item their size distribution (\refsec{sec:modelprop});
      \item their abundance relative to the gas (\refsec{sec:depletions}).
    \end{itemize}
  \item[The dust physical conditions] are the state a grain enters, when 
    exposed to a particular environment:
    \begin{itemize}
      \item thermal excitation due to
        \begin{inlinelist}
          \item radiative heating (equilibrium or stochastic; 
            \refsecs{sec:Teq}{sec:stochastheat}); and/or
          \item collisional heating in a hot plasma (\refsec{sec:coll});
        \end{inlinelist}
      \item grain charging by exchange of electrons with the gas  
        (\refsec{sec:PE});
      \item alignment of elongated grains with the magnetic field 
        (\refsec{sec:intropola});
      \item grain rotation (\refsec{sec:intropola}; \refsec{sec:AMEobs}).
    \end{itemize}
  \item[The dust observables] arise from a grain mixture experiencing a 
    particular set of physical conditions.
    \begin{description}
      \item[Emission] of partially polarized components (\refsec{sec:polaIR}):
        \begin{inlinelist}
          \item a thermal continuum (\hIR\ to mm; \refsec{sec:IRobs});
          \item molecular and solid-state features (\hMIR; \refsec{sec:IRobs});
          \item a possible microwave emission (cm; \refsec{sec:AMEobs});
          \item a possible luminescence (visible; \refsec{sec:AMEobs}).
        \end{inlinelist}
      \item[Absorption] of the light from a background source by    
        (\refsec{sec:extinction}):
        \begin{inlinelist}
          \item a continuum (X-ray to \hMIR; \refsec{sec:extinctionUV});
          \item atomic, molecular and solid-state features 
            (X-rays, \hUV\ and \hMIR; \refsec{sec:extinctionMIR}), 
            including \hDIB s (visible; \refsec{sec:DIBs})
            and ices (\hMIR; \refsec{sec:extinctionMIR}).
        \end{inlinelist}
        The escaping light can be partially polarized as a result 
        (\refsec{sec:polavis}).
      \item[Scattering] of the light from a bright source in our direction, and 
        its polarization (X-rays to \hNIR; \refsec{sec:extinction}).
      \item[Depletion] patterns seen through the gas-phase elemental 
        abundances (\refsec{sec:depletions}).
    \end{description}
\end{description}

\section{Spectral Energy Distribution Modeling}
\label{sec:SEDmodel}

\hSED\ modeling is one of the main methods to empirically derive the dust properties of a region or a galaxy \citep[\cf\ \eg][for a review]{galliano18}.
The inherent complexity of astrophysical sources requires to account for the diversity of physical conditions within the studied region.
The treatment of radiative transfer, even in an extremely approximated fashion, is thus necessary.

  \subsection{Radiative Transfer}
  \label{sec:RT}

Radiative transfer is the method solving the propagation of multiple rays of light, emitted by one or several sources, through a macroscopic heterogeneous medium.
It accounts for the scattering, absorption and emission, at each point and along each direction, in the studied region.

    \subsubsection{Definition of the Main Radiative Transfer 
                       Quantities}

Radiative transfer deals with several quantities that are often improperly defined or mixed together in the literature: intensity, flux, emissivity, brightness, \etc\
We have already seen some of these quantities in \refsec{sec:heatcool}.
We now define them and explicit their differences \citep[\cf\ Chap.~1 of][for a complete review]{rybicky79}.
In what follows, we assume stationary systems.
The time variable, $t$, is used only to denote constant rates.

\begin{figure}[htbp]
  \includegraphics[width=\textwidth]{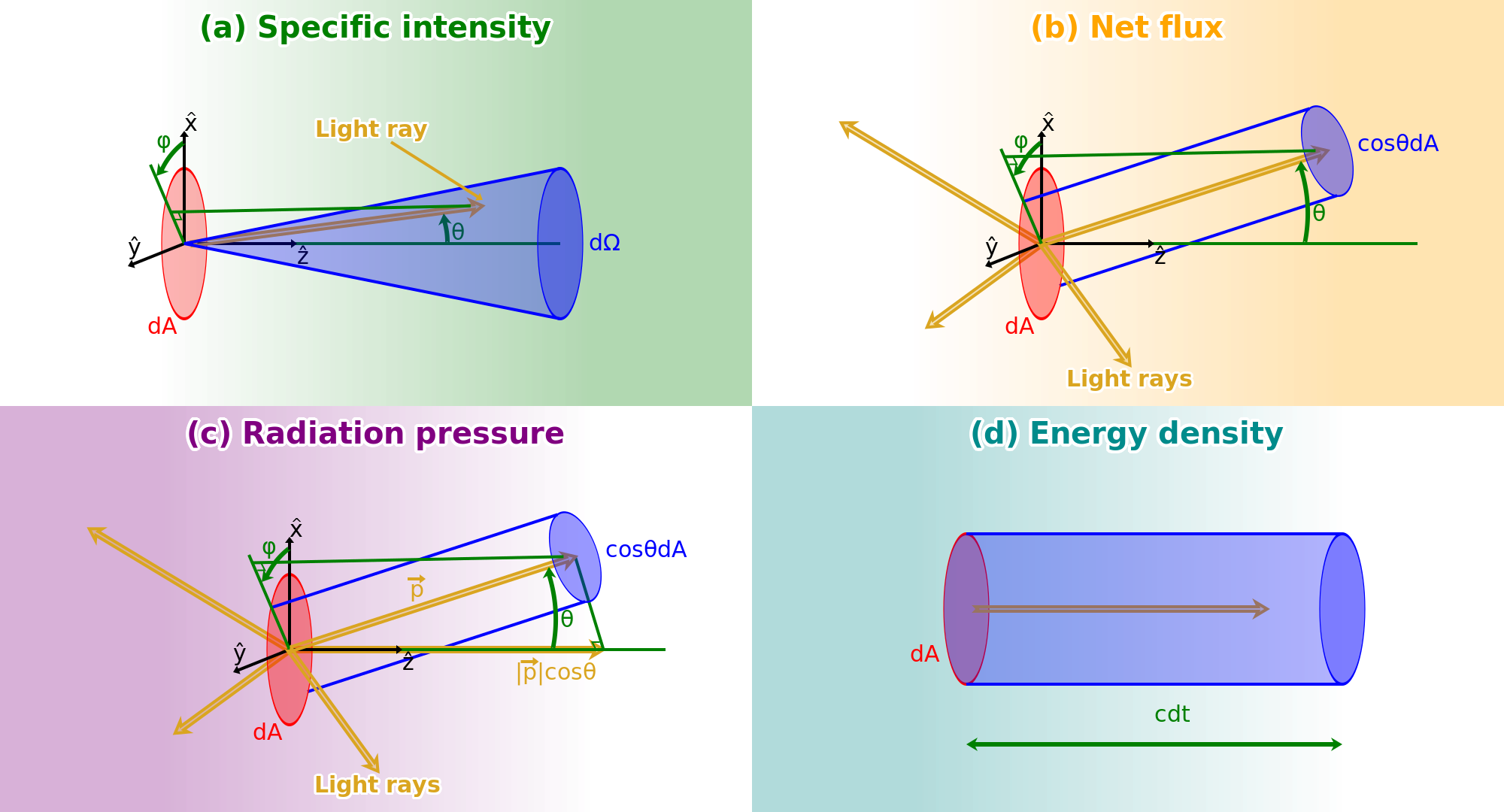}
  \newcap{Moments of the specific intensity}%
         {Panel~\textit{(a)} represents the specific intensity \refeqp{eq:Inu}.
          It is the monochromatic power of light rays per area, $\dd A$, and
          within the solid angle $\dd\Omega$.
          Panel~\textit{(b)} demonstrates the calculation of the flux 
          \refeqp{eq:Fnu}.
          We have represented several rays, with different directions and 
          intensities.
          The effective area perpendicular to rays going through $\dd A$, but 
          that are inclined at an angle $\theta$, is only $\cos\theta\ddiff A$.
          Panel~\textit{(c)} represents the radiation pressure \refeqp{eq:pnu}.
          It is similar to panel~\textit{(b)}, except that we have shown 
          the component of the momentum vector, $\vect{p}$, perpendicular to 
          the surface, $|\vect{p}|\cos\theta$.
          Only this component contributes to the pressure on $\dd A$.
          Panel~\textit{(d)} represents the energy density \refeqp{eq:unu}.
          Between times $t$ and $t+\dd t$, photons going through $\dd A$ are 
          encompassed within the cylinder of volume $c\dd t\dd A$.
          \CClicence}
  \label{fig:Inu}
\end{figure}

\paragraph{The moments of the specific intensity.}
The primary radiative transfer quantity is the \expression{specific intensity} or \expression{brightness} (\cf\ \refsubfig{fig:Inu}{a}):
\begin{equation}
  I_\nu(\nu,\vect{r},\theta,\phi)\equiv\frac{\dd E}{\dd t\dd A\dd\Omega\dd\nu}.
  \label{eq:Inu}
\end{equation}
The specific intensity is the electromagnetic energy, $E$, per unit time, $t$, area, $A$, solid angle, $\Omega$, and frequency, $\nu$.
It therefore quantifies the infinitesimal power carried by a monochromatic light ray.
This quantity depends on the position in the region,$\vect{r}$, and on the direction of propagation, $(\theta,\phi)$.
We adopt the spherical coordinate conventions used in \expression{physics} (\reffig{fig:Inu}; \cf\ \refapp{sec:integrals}): 
\begin{itemize}
  \item the \expression{polar angle}, $0\le\theta<\pi$, is the angle with the 
    $z$ axis;
  \item the \expression{azimuthal angle}, $0\le\phi<2\pi$, is the rotation angle
    in the $(x,y)$ plane;
  \item the solid angle element is 
    $\ddiff\Omega\equiv\dd\cos\theta\dd\phi=\sin\theta\dd\theta\dd\phi$, with:
    \begin{equation}
      \iint_\sms{sphere}\ddiff\Omega
      =\int_0^{2\pi}\ddiff\phi\int_{-1}^1\ddiff\cos\theta=4\pi.
    \end{equation}
\end{itemize}
The first moments of the specific intensity, relative to its angular distribution, are physically meaningful.
\begin{description}
  \item[The mean intensity] is the zeroth order moment of 
    \refeq{eq:Inu}:
    \begin{equation}
      J_\nu(\nu,\vect{r}) \equiv \frac{1}{4\pi}\iint_\sms{sphere} 
        I_\nu(\nu,\vect{r},\theta,\phi)\ddiff\Omega.
      \label{eq:Jnu}
    \end{equation}
    It is the specific intensity averaged over all directions.
    It is often used to quantify the \hISRF, $4\pi J_\nu$ accounting for rays 
    coming from all directions.
    In the \expression{isotropic} case, we have
    $I_\nu(\nu,\vect{r},\theta,\phi)=J_\nu(\nu,\vect{r})$, 
    $\forall(\theta,\phi)$.
  \item[The net flux] (\cf\ \refsubfig{fig:Inu}{b}) is the first order moment 
    of \refeq{eq:Inu}:
    \begin{equation}
      F_\nu(\nu,\vect{r}) \equiv \iint_\sms{sphere} 
        I_\nu(\nu,\vect{r},\theta,\phi)\cos\theta\ddiff\Omega.
      \label{eq:Fnu}
    \end{equation}
    It represents the monochromatic power per unit area passing through a    
    surface element perpendicular to $\hat{z}$.
    The $\cos\theta$ factor is there to account for the reduction of the 
    density of rays that are not perpendicular to the surface.
    In the isotropic case, 
    $F_\nu=J_\nu2\pi\int_{-1}^1\cos\theta\ddiff\cos\theta=0$, because there is 
    the same amount of flux passing through the area in both directions.
  \item[The radiation pressure] (\cf\ \refsubfig{fig:Inu}{c}) is the second 
    order moment of \refeq{eq:Inu}:
    \begin{equation}
      p_\nu(\nu,\vect{r}) \equiv \frac{1}{c}\iint_\sms{sphere} 
       I_\nu(\nu,\vect{r},\theta,\phi)\cos^2\theta\ddiff\Omega
      \label{eq:pnu}
    \end{equation}
    It is the \expression{momentum flux} carried by the photons, as $p=h\nu/c$ 
    is the momentum of a single photon.
    The $\cos^2\theta$ term comes from two sources:
    \begin{inlinelist}
      \item one $\cos\theta$ comes from the reduced fraction of inclined rays,
        similar to the flux;
      \item the second $\cos\theta$ factor comes from the fact the pressure is
        the momentum vector component that is perpendicular to the surface.
    \end{inlinelist}
\end{description}

\paragraph{Specific energy density.}
The radiative energy within a volume element at a given time is the \expression{specific energy density}:
\begin{equation}
  u_\nu(\nu,\vect{r},\theta,\phi)\equiv \frac{\dd E}{\dd V\dd\Omega\dd\nu}
    = \frac{\dd E}{c\dd t\dd A\dd\Omega\dd\nu}.
  \label{eq:unu}
\end{equation}
The second equality comes from the fact that $\dd V=c\dd t\dd A$, $\dd V$ being a volume element (\cf\ \refsubfig{fig:Inu}{d}).
The specific intensity is a power per unit area, whereas the energy density is an energy per unit volume.
Both quantities are linked, combining \refeq{eq:Inu} and \refeq{eq:unu}:
\begin{equation}
  u_\nu(\nu,\vect{r},\theta,\phi) = \frac{I_\nu(\nu,\vect{r},\theta,\phi)}{c}.
\end{equation}
If we integrate \refeq{eq:unu} over all directions, we get, using \refeq{eq:Jnu}:
\begin{equation}
  U_\nu(\nu,\vect{r}) 
    = \iint_\sms{sphere}u_\nu(\nu,\vect{r},\theta,\phi)\ddiff\Omega 
    = \frac{4\pi}{c}J_\nu(\nu,\vect{r}).
  \label{eq:Unu}
\end{equation}

\paragraph{Emission coefficient and emissivity.}
The monochromatic \expression{emission coefficient} is the power radiated in a given direction, per unit volume and frequency:
\begin{equation}
  j_\nu(\nu,\vect{r},\theta,\phi)
   \equiv\frac{\dd E_\sms{em}}{\dd t\dd V\dd\Omega\dd\nu}.
  \label{eq:jnu}
\end{equation}
We have also seen, in \refsec{sec:heatcool}, the \expression{emissivity}:
\begin{equation}
  \epsilon_\nu(\nu,\vect{r},\theta,\phi)\equiv
    4\pi\frac{\dd E_\sms{em}}{\dd t\dd m\dd\Omega\dd\nu}
    = 4\pi\frac{\dd E_\sms{em}}{\dd t\rho\dd V\dd\Omega\dd\nu},
  \label{eq:epsnu}
\end{equation}
where $\rho$ is the mass density of the \hISM, and $\dd m=\rho\dd V$, its mass element.
The factor $4\pi$, in \refeq{eq:epsnu}, makes $\epsilon_\nu$ the solid angle fraction of monochromatic power emitted in a given direction, per unit mass.
In \refsec{sec:heatcool}, we were considering the volume and mass of a grain, whereas, here, we are considering the volume and mass of the \hISM.
Combining \refeq{eq:jnu} and \refeq{eq:epsnu}, we get:
\begin{equation}
  j_\nu(\nu,\vect{r},\theta,\phi) 
    = \frac{\rho(\vect{r})}{4\pi}\epsilon_\nu(\nu,\vect{r},\theta,\phi).
\end{equation}

\paragraph{Extinction coefficient and opacity.}
The amount of specific intensity absorbed and scattered along an infinitesimal path length, $\dd l$, in the direction $(\theta,\phi)$, is the \expression{extinction coefficient}, $\alpha\ge0$, defined such that:
\begin{equation}
  \frac{\dd I_\nu(\nu,\vect{r},\theta,\phi)}{\dd l}
    =-\alpha(\nu,\vect{r})\,I_\nu(\nu,\vect{r},\theta,\phi).
  \label{eq:alpha}
\end{equation}
Similarly to the convention we have adopted for $\kappa$ in \refsec{sec:heatcool}, we pose $\alpha\equiv\alpha_\sms{ext}=\alpha_\sms{abs}+\alpha_\sms{sca}$ to distinguish absorption and scattering..
\refeq{eq:alpha} can be expressed with microscopic quantities, assuming the absorbers and scatterers have a cross-section $C_\sms{ext}(\nu,\vect{r})$ and a density $n(\vect{r})$:
\begin{equation}
   \alpha(\nu,\vect{r}) = n(\vect{r})\,C_\sms{ext}(\nu,\vect{r}).
\end{equation}
If the composition of the \hISM\ is homogeneous, then $C_\sms{ext}(\nu,\vect{r})=C_\sms{ext}(\nu)$.
The \expression{opacity}, that we have seen in \refsec{sec:heatcool}, is related to $\alpha$ by:
\begin{equation}
  \alpha(\nu,\vect{r}) = \rho(\vect{r})\,\kappa(\nu,\vect{r}).
  \label{eq:alphakappa}
\end{equation}

\paragraph{Mean free path.}
The \expression{mean free path} of a photon with frequency, $\nu$, at position $\vect{r}$ can be defined as:
\begin{equation}
  l_\sms{mean}(\nu,\vect{r})\equiv\frac{1}{\alpha(\nu,\vect{r})}
  =\frac{1}{\rho(\vect{r})\kappa(\nu,\vect{r})}.
  \label{eq:freepath}
\end{equation}
It is the average length a photon will be able to travel before being absorbed or scattered.
We can make the same remark as for the emissivity:
in \refsec{sec:heatcool}, we were considering the cross-section of dust particles per mass of grain, whereas \refeq{eq:alphakappa} gives the cross-section of the whole \hISM\ per mass of \hISM.
\reftab{tab:freepath} gives typical values of $l_\sms{mean}$ in a homogeneous medium, assuming the dust constitution of the \citetalias{jones17} model (\cf\ \refsec{sec:themis}).
\takeaway{In the diffuse \hISM\ (\hWNM; $n_\sms{H}\simeq0.3\;\textnormal{cm}^{-3}$; \reftab{tab:ISMism}), the mean free path of a photon in the visible range is of the order of a kiloparsec.}
\begin{table}[htbp]
  \centering
  \setlength\arrayrulewidth{2pt}
  \arrayrulecolor{white}
  \begin{tabularx}{\linewidth}{|>{}X%
                               |>{\columncolor{coltabcell}}r%
                               |>{\columncolor{coltabcell}}r%
                               |>{\columncolor{coltabcell}}r%
                               |>{\columncolor{coltabcell}}r%
                               |>{\columncolor{coltabcell}}r|}
    \hline
      & \cellcolor{coltabhead}\textbf{\hHIM}
      & \cellcolor{coltabhead}\textbf{\hWNM}
      & \cellcolor{coltabhead}\textbf{\hCNM}
      & \multicolumn{2}{c|}{\cellcolor{coltabhead}\textbf{Molecular clouds}}
      \\
      & \cellcolor{coltabhead} $n_\sms{H}=0.003\;\textnormal{cm}^{-3}$
      & \cellcolor{coltabhead} $n_\sms{H}=0.3\;\textnormal{cm}^{-3}$
      & \cellcolor{coltabhead} $n_\sms{H}=30\;\textnormal{cm}^{-3}$
      & \cellcolor{coltabhead} $n_\sms{H}=10^4\;\textnormal{cm}^{-3}$
      & \cellcolor{coltabhead} $n_\sms{H}=10^6\;\textnormal{cm}^{-3}$ 
      \\
    \hline
      \cellcolor{coltabhead}$l_\sms{mean}(U)$ 
      & 139 kpc & 1.39 kpc & 13.9 pc & 0.0417 pc & 86.1 a.u.
      \\
    \hline
      \cellcolor{coltabhead}$l_\sms{mean}(B)$ 
      & 177 kpc & 1.77 kpc & 17.7 pc & 0.0532 pc & 110 a.u.
      \\
    \hline
      \cellcolor{coltabhead}$l_\sms{mean}(V)$ 
      & 223 kpc & 2.23 kpc & 22.3 pc & 0.0669 pc & 138 a.u.
      \\
    \hline
      \cellcolor{coltabhead}$l_\sms{mean}(R)$ 
      & 275 kpc & 2.75 kpc & 27.5 pc & 0.0824 pc & 170 a.u.
      \\
    \hline
      \cellcolor{coltabhead}$l_\sms{mean}(I)$ 
      & 358 pc & 3.58 kpc & 35.8 pc & 0.107 pc & 222 a.u.
      \\
    \hline
      \cellcolor{coltabhead}$l_\sms{mean}(J)$ 
      & 691 kpc & 6.91 kpc & 69.1 pc & 0.207 pc & 427 a.u.
      \\
    \hline
      \cellcolor{coltabhead}$l_\sms{mean}(H)$ 
      & 1021 kpc & 10.2 kpc & 102 pc & 0.306 pc & 632 a.u.
      \\
    \hline
      \cellcolor{coltabhead}$l_\sms{mean}(K)$ 
      & 1734 kpc & 17.3 kpc & 173 pc & 0.52 pc & 1073 a.u.
      \\
    \hline
  \end{tabularx}
  \newcap{Mean free path as a function of wavelength and density}%
         {These quantities were computed by rewriting \refeq{eq:freepath} as
          $1/l_\sms{mean}(\lambda)
          =n_\sms{H}m_\sms{H}Y_\sms{dust}\kappa(\lambda)$, taking
          $Y_\sms{dust}\equiv M_\sms{dust}/M_\sms{H}$ from \reftab{tab:massthemis} 
          and $\kappa$ values from \reftab{tab:opacity}.
          We quote $l_\sms{mean}$ at the same photometric bands as in 
          \reftab{tab:opacity}.
          These values correspond to Solar metallicity, $Z=Z_\odot$.
          At first order, for metallicities $Z\gtrsim0.2\;Z_\odot$, we can 
          assume linearity: $l_\sms{mean}(Z)\simeq l_\sms{mean}(Z_\odot)\times 
          Z_\odot/Z$ (\cf\ \refchap{chap:dustevol}). 
          The different \hISM\ phases quoted here (\hHIM, \hWNM, \hCNM) will be defined
          in detail in \refsec{sec:ISMism}.}
  \label{tab:freepath}
\end{table}

    \subsubsection{The Radiative Transfer Equation}
    \label{sec:eqRT}

\begin{figure}[htbp]
  \includegraphics[width=\textwidth]{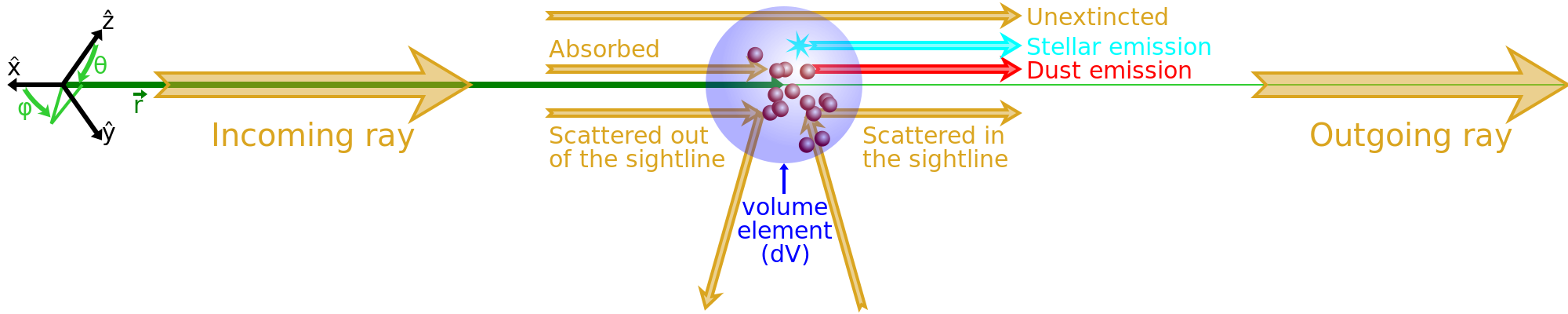}
  \newcap{The radiative transfer equation}%
         {This figure represents only the different processes contributing to
          the variation of the specific intensity at one location, $\vect{r}$, 
          in one direction, $(\theta,\phi)$.
          The blue sphere represents a volume element.
          The red spheres within represent dust grain and the cyan star 
          represents an actual star that would be present in the volume 
          element. 
          \CClicence}
  \label{fig:eqRT}
\end{figure}

The radiative transfer equation accounts for the variation of the specific intensity under the effects of absorption, scattering and emission \citep[\cf][for a review in the case of a dusty medium]{steinacker13}.
It is schematically represented in \reffig{fig:eqRT}.
This equation can be written:
\begin{equation}
  \begin{array}{rl}
  \displaystyle\frac{\dd I_\nu(\nu,\vect{r},\theta,\phi)}{\dd l} =
    & - \underbrace{\alpha_\sms{abs}(\nu,\vect{r})
                  I_\nu(\nu,\vect{r},\theta,\phi)}_\sms{absorption}
     - \underbrace{\alpha_\sms{sca}(\nu,\vect{r})
            I_\nu(\nu,\vect{r},\theta,\phi)}_\sms{scattering out of the sightline}
    \\
    & + \underbrace{\alpha_\sms{sca}(\nu,\vect{r})
         2\pi\int_{-1}^1\Phi(\cos\theta^\prime,\nu)
           I_\nu(\nu,\vect{r},\theta(\theta^\prime),\phi(\theta^\prime))
           \ddiff\cos\theta^\prime}_\sms{scattering in the sightline}
    \\
    & +\underbrace{j_\nu^\sms{dust}(\nu,\vect{r})}_\sms{dust emission}
      +\underbrace{j_\nu^\star(\nu,\vect{r})}_\sms{stellar emission}.
  \end{array}
  \label{eq:RTl}
\end{equation}
\begin{description}
  \item[Scattering in the sightline:] this term is the integral 
    of $\alpha_\sms{sca}I_\nu$ (\ie\ the scattered intensity) over the phase 
    function, $\Phi$ \refeqp{eq:phasefunction}.
    This expression depends on the relative angle, $\theta^\prime$, between the 
    incident rays and the scattered direction $(\theta,\phi)$.
    This is the term that makes this equation an 
    \expression{integro-differential equation}, coupling all the directions 
    together.
    This is why, numerical methods are required to solve \refeq{eq:RTl}.
    If we assume isotropic scattering (\ie\ $\langle\cos\theta\rangle=0$, 
    corresponding to the Rayleigh regime; \cf\ \refsec{sec:calcQabs}), this term
    simplifies and becomes: $\alpha_\sms{sca}(\nu,\vect{r})J_\nu(\nu,\vect{r})$.
  \item[Emission:]
    we have assumed that the dust and stellar emissions are both isotropic, 
    (\ie\ independent of $\theta$ and $\phi$).
    This is a reasonable assumption in the \hISM.
    If we also assume that there is only one grain species, and that this 
    species is at equilibrium with the radiation field, we can explicit:
    $j_\nu^\sms{dust}(\nu)=\alpha(\nu)B_\nu(\nu,T_\sms{eq})$.
    This simplification still contains the problem that to determine the
    equilibrium temperature, $T_\sms{eq}$, we need to integrate 
    $\alpha_\sms{abs}I_\nu$ over all directions, and all frequencies.
\end{description}
\refeq{eq:RTl} can be rewritten, using the optical depth, $\tau$, defined such that $\dd\tau(\nu)=\alpha(\nu,l)\ddiff l$, or:
\begin{equation}
  \tau(\nu,l) = \int_0^l\rho(l^\prime)\kappa(\nu,l^\prime)\ddiff l^\prime.
  \label{eq:opticaldepth}
\end{equation}
\refeq{eq:freepath} implies that $\tau(l_\sms{mean})=1$.
At a given wavelength, a medium is:
\begin{inlinelist}
  \item \expression{optically thin} or \expression{transparent}, if $\tau<\ll1$;
  and
  \item \expression{optically thick} or \expression{opaque}, if $\tau\gg1$.
\end{inlinelist}
Replacing $\dd l$ by $\dd\tau/\alpha$ in \refeq{eq:RTl}, we obtain:
\begin{equation}
  \frac{\dd I_\nu(\nu,\vect{r},\theta,\phi)}{\dd\tau}
  = - I_\nu(\nu,\vect{r},\theta,\phi) + S_\nu(\nu\vect{r},\theta,\phi),
  \label{eq:RTt}
\end{equation}
where the \expression{source function}, $S_\nu$, includes all the terms added
to the specific intensity:
\begin{equation}
  S_\nu(\nu,\vect{r},\theta,\phi) = 
    \tilde{\omega}(\nu,\vect{r})
         2\pi\int_{-1}^1\Phi(\cos\theta^\prime,\nu)
           I_\nu(\nu,\vect{r},\theta(\theta^\prime),\phi(\theta^\prime))
           \ddiff\cos\theta^\prime
     + \frac{j_\nu^\sms{dust}(\nu,\vect{r})+j_\nu^\star(\nu,\vect{r})}%
            {\alpha(\nu,\vect{r})}.
\end{equation}
We will discuss exact numerical solutions to this equation in \refsec{sec:MCRT}.
For now, we discuss trivial solutions, when some processes are assumed negligible.

\paragraph{Propagation in vacuum.}
Let's assume we have a star of radius, $R$, and surface temperature, $T_\star$, located at $\vect{r}=\vect{0}$.
The flux at $r=R$ has to be integrated only over the hemisphere where a surface element of the star emits: $F_\nu(R)=2\pi\int_0^1 B_\nu(T_\star)\cos\theta\ddiff\cos\theta=\pi B_\nu(T_\star)$.
If there is no \hISM\ around the star, \refeq{eq:RTl} simply becomes $\dd I_\nu/\dd l=0$.
The solution is thus $I_\nu=B_\nu(T_\star)$ along the directions coming from the star and 0 in all other directions.
At an arbitrary distance, $r$, from the star, the solid angle it occupies is $\Omega_\star=R^2/r^2$.
The flux is thus $F_\nu(r)=\pi B_\nu(T_\star)R^2/r^2$, which is the classic $1/r^2$ dilution of the flux.

\paragraph{Emission only.}
Let's assume we are observing, at submm wavelengths, a molecular cloud constituted of equilibrium grains at $T\simeq10$~K, with opacity $\kappa$.
At these wavelengths, the extinction is negligible.
\refeq{eq:RTl} is therefore simply $\dd I_\nu/\dd l=\rho\kappa B_\nu(T)$ within the cloud.
The solution is thus $I_\nu(l)=\kappa B_\nu(T)\int_{l_0}^l\rho(l)\ddiff l$, where $l_0$ is the position of the edge of the cloud along the direction of the sightline.
If the density is constant, we get $I_\nu(l)=\rho\kappa B_\nu(T)\times(l-l_0)$, which can be simplified as $I_\nu(l)=\tau(l)B_\nu(T)$.
This expression of the brightness is often used in radio-astronomy.

\paragraph{Absorption only.}
Let's assume we are observing a background star through a cold molecular cloud, in the \hMIR.
At these wavelengths, the albedo is close to 0 (\refsec{sec:calcQabs}).
If we make the assumption that the background star is much brighter than the thermal emission of the cloud, \refeq{eq:RTl} simply becomes $\dd I_\nu/\dd l=-\alpha I_\nu$.
The solution is therefore $I_\nu(l)=I_\nu^\star\exp\left[-\tau(l)\right]$ along the direction coming from the star and 0 in all other directions.

\begin{figure}[htbp]
  \includegraphics[width=\textwidth]{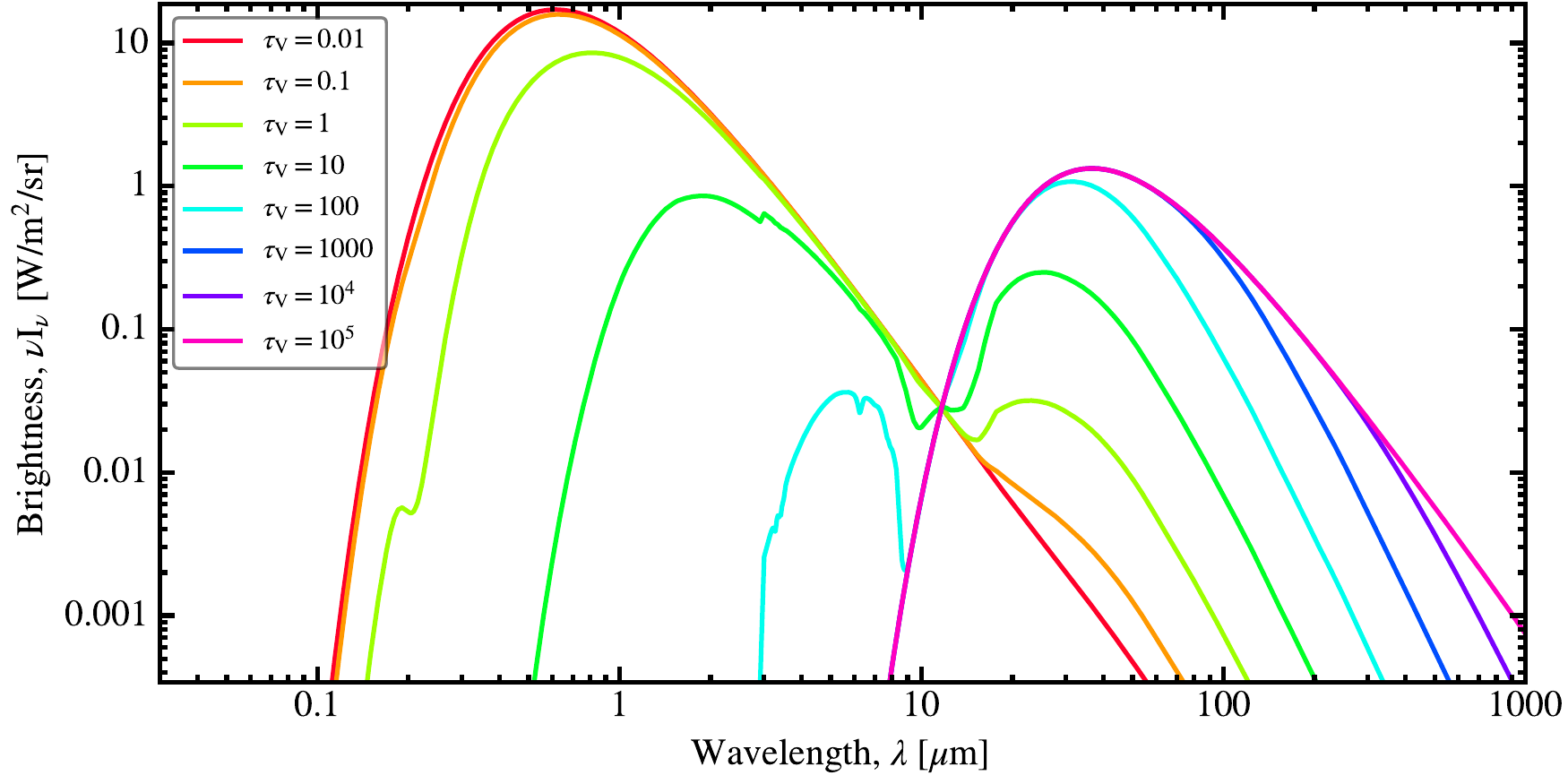}
  \newcap{Solution to the radiative transfer equation for an isothermal cloud, 
          without scattering}%
         {We have assumed that we are looking at a background star 
          ($T_\star=6000$~K), through an homogeneous cloud of grains at 
          equilibrium temperature, $T=100$~K, with the \citetalias{jones17} 
          opacity.
          We have diluted the stellar emission by a factor $10^{-6}$, which
          could correspond to a situation where the actual angular area of the 
          star is $10^{-6}$ times the beam of the telescope.
          We show the \hSED\ according to \refeq{eq:selfabs}, varying the 
          optical depth in the V band, $\tau_\sms{V}$.
          We have included very large $\tau_\sms{V}$, up to $10^5$, in order to
          demonstrate the asymptotic behavior of the dust emission tending 
          toward a perfect black body.
          In typical \hISM\ studies, it is however rare to find $\tau_\sms{V}$ 
          values higher than $\simeq100$.
          \CClicence}
  \label{fig:RTabsem}
\end{figure}

\paragraph{Emission and absorption.}
We can merge the two previous cases.
It could correspond to a hot molecular cloud, observed at \hMIR\ wavelengths.
Its thermal emission is absorbed by the cloud itself.
The solution of \refeq{eq:RTt}, in this case, is:
\begin{equation}
  I_\nu(\tau) = I_\nu^\star\exp\left(-\tau\right)
             + \int_0^\tau\exp\left(\tau^\prime-\tau\right)S_\nu(\tau^\prime)\ddiff\tau^\prime.
\end{equation}
If the cloud contains a single grain species at temperature $T$, the solution becomes:
\begin{equation}
  I_\nu(\tau) = B_\nu(T) + \left[I_\nu^\star-B_\nu(T)\right]\times\exp(-\tau).
\end{equation}
If we look in a direction away from the background star, we get the classical self-absorption formula:
\begin{equation}
  I_\nu(\tau) = B_\nu(T)\times\left[1-\exp(-\tau)\right].
  \label{eq:selfabs}
\end{equation}
This solution is displayed in \reffig{fig:RTabsem}.
\refeq{eq:selfabs} has the following two limit regimes.
\begin{description}
  \item[Optically thin:] if $\tau\ll1$, $I_\nu\simeq\tau B_\nu(T)$, which is the 
    \citengl{emission only} solution.
    The emission from the cloud is a grey body (\refsec{sec:heatcool}).
  \item[Optically thick:] if $\tau\gg1$, $I_\nu\simeq B_\nu(T)$, which means the 
    cloud is not anymore a \expression{grey} body, but a \expression{perfect} 
    one.
\end{description}

\begin{figure}[htbp]
  \begin{tabular}{cc}
    \includegraphics[width=0.75\textwidth]{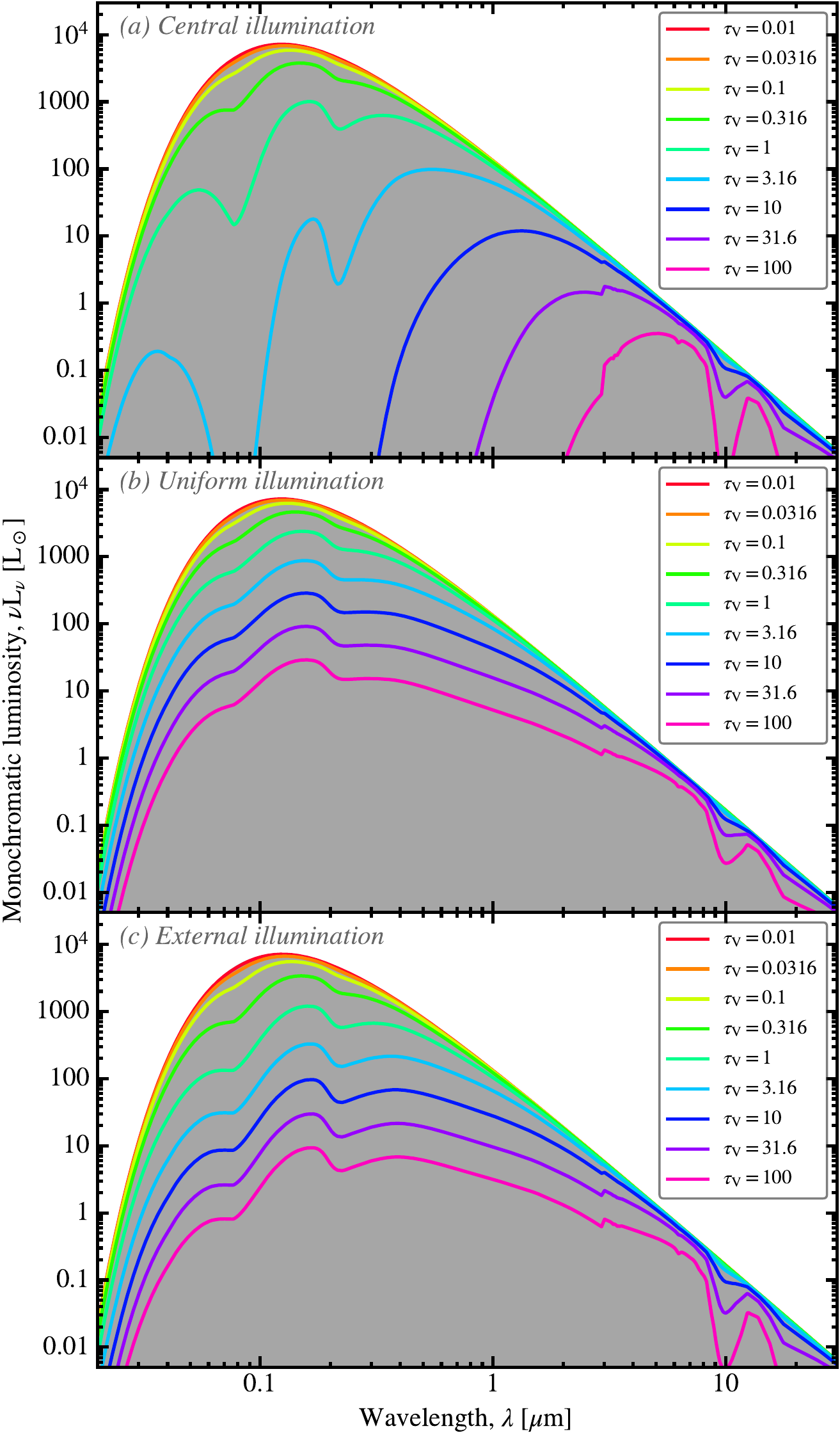} &
    \includegraphics[width=0.25\textwidth]{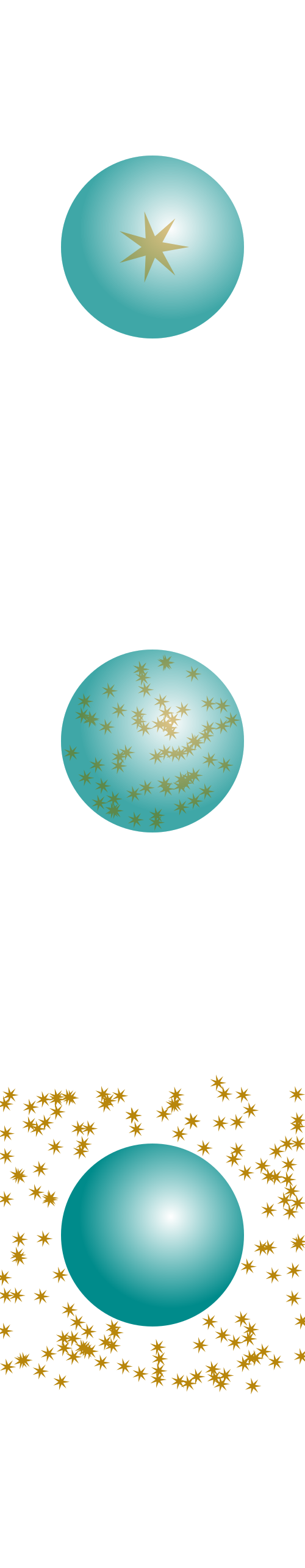} \\
  \end{tabular}
  \newcap{Escaping radiation from a spherical cloud}%
         {In the three panels, we show the escaping \hSED\ of a homogeneous 
          spherical cloud, made of \citetalias{jones17} grains.
          We vary the V-band optical depth, $\tau_\sms{V}$.
          The illuminating source is a star ($T_\star=3\E{4}$~K), with a 
          bolometric luminosity $L_\star=10^4\eLsun$.
          In panel~\textit{(a)}, we show the case of central illumination.
          All the power is in the central source.
          Panel~\textit{(b)} shows the case of uniform illumination.
          The total power of the sources within the cloud is $L_\star$.
          Panel~\textit{(c)} shows external illumination. 
          In this case, the flux at the surface of the cloud is 
          $L_\star/4\pi R_\sms{cl}^2$.
          \CClicence}
  \label{fig:eqRT_hom}
\end{figure}

\paragraph{Scattering and absorption with central illumination.}
Let's assume that we have a homogeneous spherical cloud, of radius $R_\sms{cl}$, and a central isotropically illuminating source, with monochromatic luminosity, $L_\nu^\star(\lambda)$.
If $L_\nu^\sms{esc}(\lambda)$ is the monochromatic luminosity escaping the cloud, the \expression{escape fraction} can be defined as:
\begin{equation}
  P_\nu^\sms{esc}(\lambda) 
    \equiv \frac{L_\nu^\sms{esc}(\lambda)}{L_\nu^\star(\lambda)}
    = \exp\left[-\tau_\sms{eff}(\lambda)\right].
  \label{eq:Pesc}
\end{equation}
The second equality defines $\tau_\sms{eff}$ as the \expression{effective optical depth} of the medium.
\begin{description}
  \item[Optically thick:] 
    In the case of pure isotropic scattering 
    ($g=\langle\cos\theta\rangle\simeq0$), it can be shown that the 
    \expression{net displacement} of a photon after $N_\sms{sca}$ interactions 
    is: $l_\sms{eff}=\sqrt{N_\sms{sca}}l_\sms{mean}$ \citep[\cf\ Chap.~1 
    of][]{rybicky79}.
    The probability a photon will be absorbed, at the end of a free path, is 
    $1-\tilde{\omega}$.
    The mean number of free paths can thus be estimated by
    $N_\sms{sca}(1-\tilde{\omega})=1$, or: $N_\sms{sca}=1/(1-\tilde{\omega})$.    
    The optical depth of the cloud is $\tau\simeq R_\sms{cl}/l_\sms{mean}$ 
    (\refeqnp{eq:freepath} and \refeqnp{eq:opticaldepth}).
    The same way, the effective optical depth can be written 
    $\tau_\sms{eff}\simeq R_\sms{cl}/l_\sms{eff}$.
    We thus get:
    \begin{equation}
      \tau_\sms{eff}\simeq\sqrt{1-\tilde{\omega}}\tau.
      \label{eq:taucenthick}
    \end{equation}
  \item[Optically thin:]
    in this case, there is a low probability of interaction.
    \refeq{eq:Pesc} tells us that $\tau_\sms{eff}$ accounts only for photons 
    that have been absorbed.
    We therefore simply have \citep[\cf\ Chap.~1 of][]{rybicky79}:  
    \begin{equation}
      \tau_\sms{eff}=(1-\tilde{\omega})\tau.
      \label{eq:taucenthin}
    \end{equation}
    This formula simply subtracts among the few photons that may have 
    interacted with the grains those which have been scattered.
    Contrary to \refeq{eq:taucenthick}, it is valid for any value of $g$.
\end{description}
\citet[][hereafter \citetalias{varosi99}]{varosi99} have proposed an empirical approximation to interpolate the two regimes of \refeq{eq:taucenthick} and \refeq{eq:taucenthin}, resulting in the following escape fraction:
\begin{equation}
  P_\sms{esc}^\sms{cen}(\lambda) 
    \simeq
    \exp\left[-(1-\tilde{\omega}(\lambda))^{\chi(\lambda)}\tau(\lambda)\right],
  \label{eq:Pesccen}
\end{equation}
with:
\begin{equation} 
  \chi(\lambda)\equiv1-\frac{1}{2}
    \left[1-\exp\left(-\frac{\tau(\lambda)}{2}\right)\right]\sqrt{1-g(\lambda)}.
\end{equation}
They have benchmarked this approximation with a Monte-Carlo radiative transfer model (\cf\ \refsec{sec:MCRT}).
Both are in very good agreement in most of the astrophysically relevant parameter space.
This solution is displayed in \refsubfig{fig:eqRT_hom}{a}.

\paragraph{Scattering and absorption with uniform illumination.}
\citetalias{varosi99} have also proposed an approximation in the case where the internal illumination of the cloud is uniform.
They start from the escape probability of a homogeneous sphere with uniform illumination, without scattering, given by \citet{osterbrock89}:
\begin{equation}
  P_\sms{esc}^\sms{nosca}=\frac{3}{4\tau}\left[1-\frac{1}{2\tau^2}
    +\left(\frac{1}{\tau}+\frac{1}{2\tau^2}\right)
    \exp\left(-2\tau\right)\right].
\end{equation}
A demonstration of this formula is given in Appendix C of \citetalias{varosi99}.
\citetalias{varosi99} find, with a recursive argument, that the following expression is in relatively good agreement with a Monte-Carlo radiative transfer model:
\begin{equation}
  P_\sms{esc}^\sms{uni}(\lambda)\simeq
   \frac{P_\sms{esc}^\sms{nosca}(\lambda)}{1-\tilde{\omega}(\lambda)
             \left[1-P_\sms{esc}^\sms{nosca}(\lambda)\right]}.
  \label{eq:Pescuni}
\end{equation}
\citetalias{varosi99} show that \refeq{eq:Pescuni} perfectly agrees with the exact solution for a particular value of $g(\tau)$.
The largest discrepancies, of about $20\,\%$, are found at high optical depth, for $g$ close to 0.
\refeq{eq:Pescuni} is demonstrated in \refsubfig{fig:eqRT_hom}{b}.
We can see that, at a given $\tau_\sms{V}$, this geometry has an overall larger escape fraction than central illumination.

\paragraph{Scattering and absorption with external illumination.}
There are also expressions in the case of a homogeneous spherical cloud, externally illuminated by an isotropic radiation field.
\citetalias{varosi99} derive the following approximation for the \expression{absorbed fraction}:
\begin{equation}
  P_\sms{abs}^\sms{ext}(\lambda)\equiv1-P_\sms{esc}^\sms{ext}(\lambda)
    \simeq \frac{4\tau(\lambda)\left[1-\tilde{\omega}(\lambda)\right]}{3}
      P_\sms{esc}^\sms{uni}(\lambda).
  \label{eq:Pabsext}
\end{equation}
The absorbed fraction is indeed more relevant in the case of external illumination, as it would be difficult to observationally separate the escaping radiation from the cloud and the ambient \hISRF.
On the contrary the absorbed fraction is meaningful if one wants to evaluate the heating of the cloud.
This approximation is demonstrated in \refsubfig{fig:eqRT_hom}{c}.

\seppar
The three formulae for spherical clouds, given in \refeq{eq:Pesccen}, \refeq{eq:Pescuni} and \refeq{eq:Pabsext}, do not allow us to model the internal heating of the cloud.
These expressions are indeed global escape and absorbed fractions, but they do not account for the gradient of illumination within the cloud that would lead to a gradient of heating rate.

    \subsubsection{Approximations for Clumpy Media}
    \label{sec:clumpy}

The \hISM\ is a highly heterogeneous medium, with contrast densities of several orders of magnitude.
A useful approximation is the \expression{clumpy medium}, composed of:
\begin{inlinelist}
  \item a diffuse, uniform \expression{InterClump Medium} (\hICM), 
    characterized by its density, $n_\sms{ICM}$; and
  \item dense, spherical clumps, with density, $n_\sms{C}$, 
    radius, $r_\sms{C}$, and volume filling factor, $f_\sms{V}$.
\end{inlinelist}

\paragraph{Effective optical depth of a clumpy medium.}
Let's assume that we are looking at a background star through a cloud, and that the albedo of the grains is negligible.
Along a given sightline, we have seen in \refsec{sec:eqRT} that the brightness in the direction of the star is $I_\nu=I_\nu^\star\times\exp\left(-\tau\right)$.
If the clumpy structure of the cloud is unresolved, the brightness measured in the telescope beam can be written as the sum of $N$ sightlines, some passing through the \hICM, others through the clumps:
\begin{equation}
  I_\nu^\sms{clumpy} = \frac{I_\nu^\star}{N}\sum_{i=1}^N\exp\left(-\tau_i\right).
  \label{eq:Inuclump}
\end{equation} 
This is the general expression.
In the case of an homogeneous medium, \refeq{eq:Inuclump} simplifies: $I_\nu=I_\nu^\star\exp(-\tau_\sms{hom})$, where $\tau_\sms{hom}$ is the optical depth of the homogeneous medium.
In order to have the same dust mass and opacity as in the clumpy medium, we need to have:
\begin{equation}
  \tau_\sms{hom}\equiv\frac{1}{N}\sum_{i=1}^N\tau_i.
\end{equation}
In the homogeneous medium, $\rho\times L$ is indeed the mass surface density of a cloud of depth $L$.
If there is a statistical distribution of clumps in the beam, it must be identical in the clumpy medium.
Thus, the brightness of the homogeneous medium is:
\begin{equation}
  I_\nu^\sms{hom}=\exp\left(-\frac{1}{N}\sum_{i=1}^N\tau_i\right)
               =\left[\prod_{i=1}^N\exp(-\tau_i)\right]^{1/N}.
\end{equation}
Invoking the \expression{arithmetic-mean/geometric-mean inequality}\footnote{Stating that $\sum_{i=1}^N a_i/N\ge\prod_{i=1}^Na_i^{1/N}$, provided that $a_i\ge0$, $\forall i$.} \citep[\eg\ page 456 of][]{cauchy1821}, we conclude that: $I_\nu^\sms{clumpy}\ge I_\nu^\sms{hom}$.
An important consequence of this result is that, from an observational point of view, we can miss a large mass of dust hidden in clumps. 
\citetalias{varosi99} discuss this result in more detail.
\takeaway{The effective optical depth of a clumpy medium is always lower than that of a homogeneous medium with the same dust constitution and dust mass.}

\paragraph{The mega-grains approximation.}
\citet{neufeld91} proposed a simple approach to explain the leakage of Ly-$\alpha$ photons by galaxies.
He treated dusty gas clumps, in an empty \hICM, as large grains with their own albedo and asymmetry parameter.
This idea was then further developed by \citet{hobson93}, in cases where the \hICM\ is non empty.
They named it the \expression{mega-grains approximation}, as clumps are treated as grains, although they have macroscopic sizes.
They applied this approach to a clumpy infinite slab, externally illuminated on one side, and compared the results to a Monte-Carlo radiative transfer model.
\citetalias{varosi99} then refined some of the expressions of \citet{hobson93} and applied them to the case of a spherical clumpy cloud, with the three types of illuminations we have discussed in \refsec{sec:eqRT}: 
\begin{inlinelist}
  \item central;
  \item uniform; and 
  \item external.
\end{inlinelist}
\citetalias{varosi99} systematically benchmarked their results with a Monte-Carlo radiative transfer code.
We briefly review their results in the rest of this section.

\begin{figure}[htbp]
  \begin{tabular}{cc}
    \includegraphics[width=0.75\textwidth]{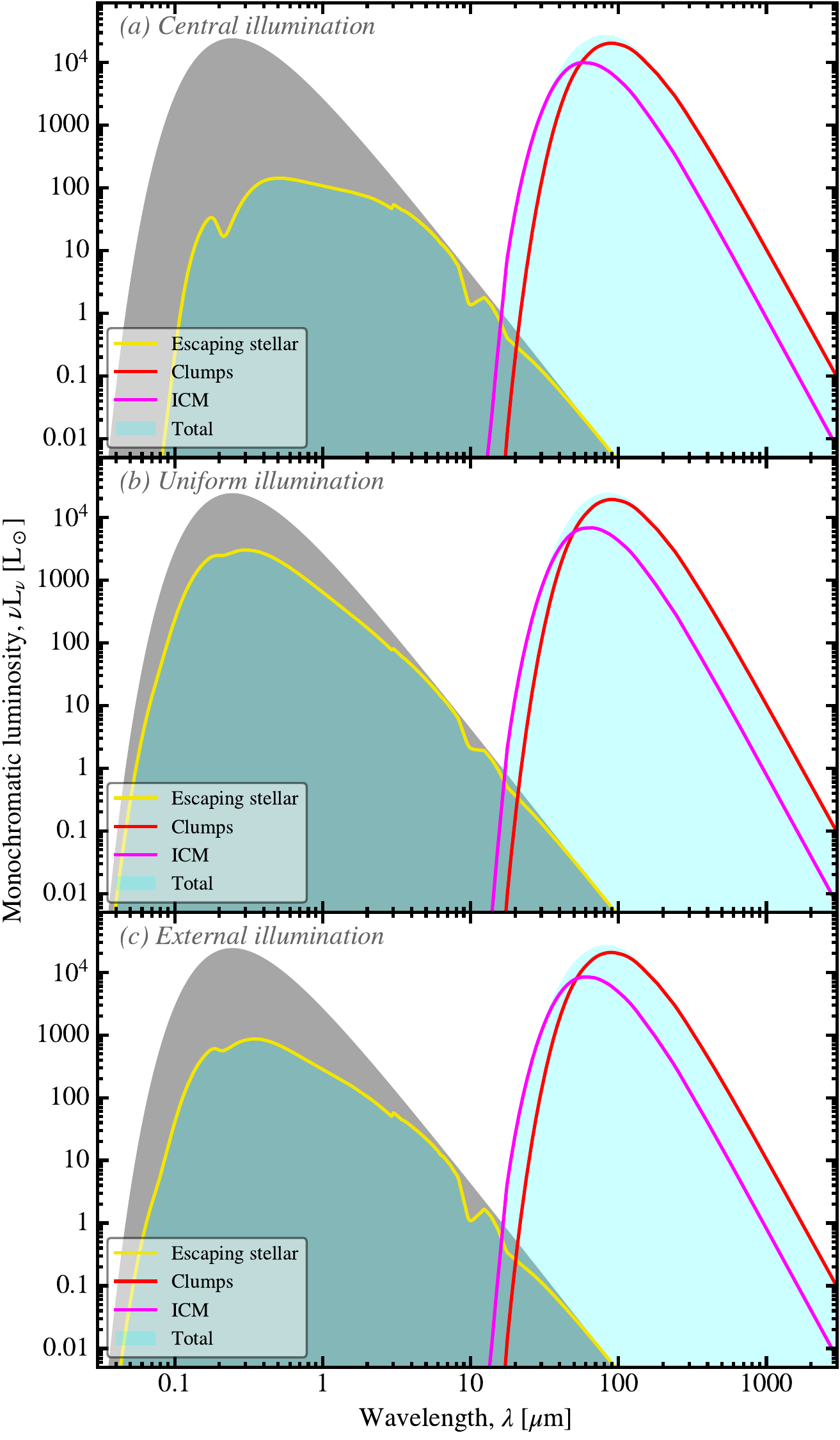} &
    \includegraphics[width=0.25\textwidth]{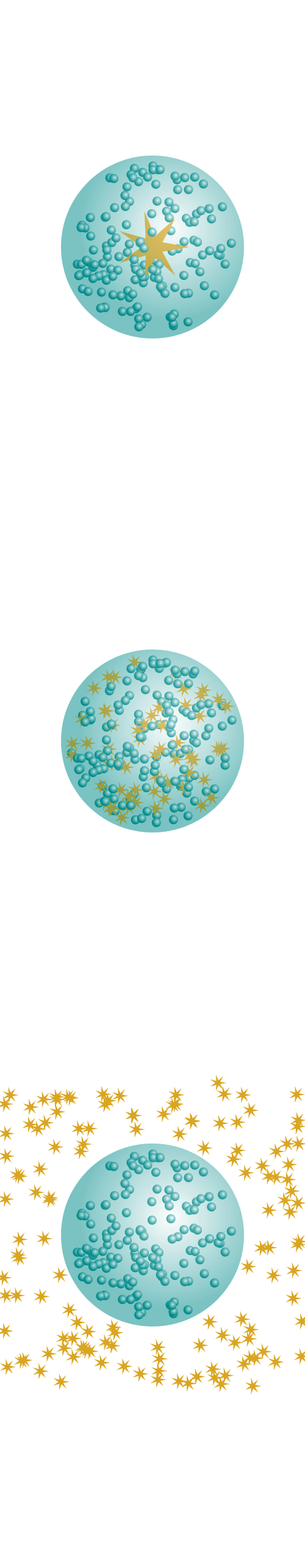} \\
  \end{tabular}
  \newcap{Escaping SED from clumpy spherical clouds}%
         {The three panels are the \hSED\ of a $R_\sms{cl}=1$~pc cloud, 
          containing $r_\sms{C}=0.05$~pc clumps, computed with the
          mega-grains approximation.
          The density is $n_\sms{ICM}=1000$~cm$^{-3}$ in the \hICM\ and 
          $n_\sms{C}=10^5$~cm$^{-3}$ in the clumps (volume filling factor, 
          $f_\sms{V}=20\,\%$).
          The grains have \citetalias{jones17} optical properties and are 
          assumed to be at thermal equilibrium.
          We have displayed:
          \begin{inlinelist}
            \item the intrinsic stellar luminosity
              ($T_\star=1.5\E{4}$~K; $L_\star=3.3\E{4}\eLsun$), in grey;
            \item the escaping stellar radiation, in yellow;
            \item the clump luminosity, in red;
            \item the \hICM\ luminosity, in magenta;
            \item the total escaping \hSED, in cyan.
          \end{inlinelist}
          \CClicence}
  \label{fig:eqRT_clumpy}
\end{figure}

\paragraph{Escape fractions for a clumpy medium.}
\citetalias{varosi99} derived a series of expressions, based on \refeq{eq:Pesccen}, \refeq{eq:Pescuni} and \refeq{eq:Pabsext}, but replacing grain properties by effective mega-grains properties.
These analytical approximations are all summarized in Sect.~5 of \citetalias{varosi99}.
We demonstrate these analytical expressions for the three types of illuminations in \reffig{fig:eqRT_clumpy}.
Overall, they provide a good agreement with Monte-Carlo radiative transfer calculations.
They are also very easy to compute.
The weakest point concerns the treatment of the grain heating.
The mega-grains formalism allows us to separate the absorbed fractions in the clumps and in the \hICM.
It thus provides different heating rates in the two phases.
In \reffig{fig:eqRT_clumpy}, we can clearly see that the clump emission (red) is significantly colder than that of the \hICM\ (magenta).
It however does not allow us to estimate the gradient of radiation field within the \hICM\ and within clumps.
This is the most dramatic in the case of central illumination.
In order to obtain a more realistic \hSED\ for this particular case, \citetalias{varosi99} used an \textit{ad hoc} prescription, assuming a power-law distribution of equilibrium grain temperatures controlled by several tuning parameters depending on the grain type.

    \subsubsection{Rigorous Solutions}
    \label{sec:MCRT}

The radiative transfer equation \refeqp{eq:RTl} can be solved numerically.
There are two main classes of methods \citep[][for a review]{steinacker13}.
\begin{description}
  \item[Monte-Carlo Radiative Transfer] (\hMCRT) consists in simulating the 
    random walk of photons, from their sources (stars, \hAGN s or thermal 
    emission from dust grains) to the outside of the region, through their 
    multiple scatterings.
  \item[Ray-tracing] numerically solves the radiative transfer 
    equation on a discretized grid along multiple sightlines.
    It is more difficult to implement than \hMCRT, and can be 
    computationally more intensive.
    It however allows the user a better assessment of the numerical errors.
\end{description}
\hMCRT\ is by far the most popular method.
In this section, we briefly review its principle and apply it to an example.

\paragraph{Setting the model.}
To solve the radiative transfer equation \refeqp{eq:RTl}, we need to specify the following physical ingredients.
\begin{description}
  \item[A 3D spatial grid of dust density] has to be defined.
    Different coordinate systems can be chosen.
    In case there are steep density gradients (\eg\ clumps), one has to make 
    sure the transitions are finely enough sampled.
  \item[A 3D distribution of primary emitters,] as well as their \hSED\ 
    needs to be defined.
    In this section, we will consider only stars, but \hAGN s can be treated 
    the same way.
    It is possible to account both for:
    \begin{inlinelist}
      \item discrete emission, such as an individual star or a cluster; and 
      \item diffuse emission, such as unresolved stellar distribution.
    \end{inlinelist}
  \item[The dust properties] need to include, at least: the opacity, the albedo
    and the asymmetry parameter.
    It is possible to account for a mixture of dust grains, with different 
    compositions and sizes.
    The size-distribution-integrated quantities of \refeqs{eq:SD1}{eq:SD2} are 
    used for the transfer of stellar photons.
    It is however necessary to treat the thermal emission of individual 
    species, especially if they are stochastically heated.
\end{description}

\begin{figure}[htbp]
  \includegraphics[width=\textwidth]{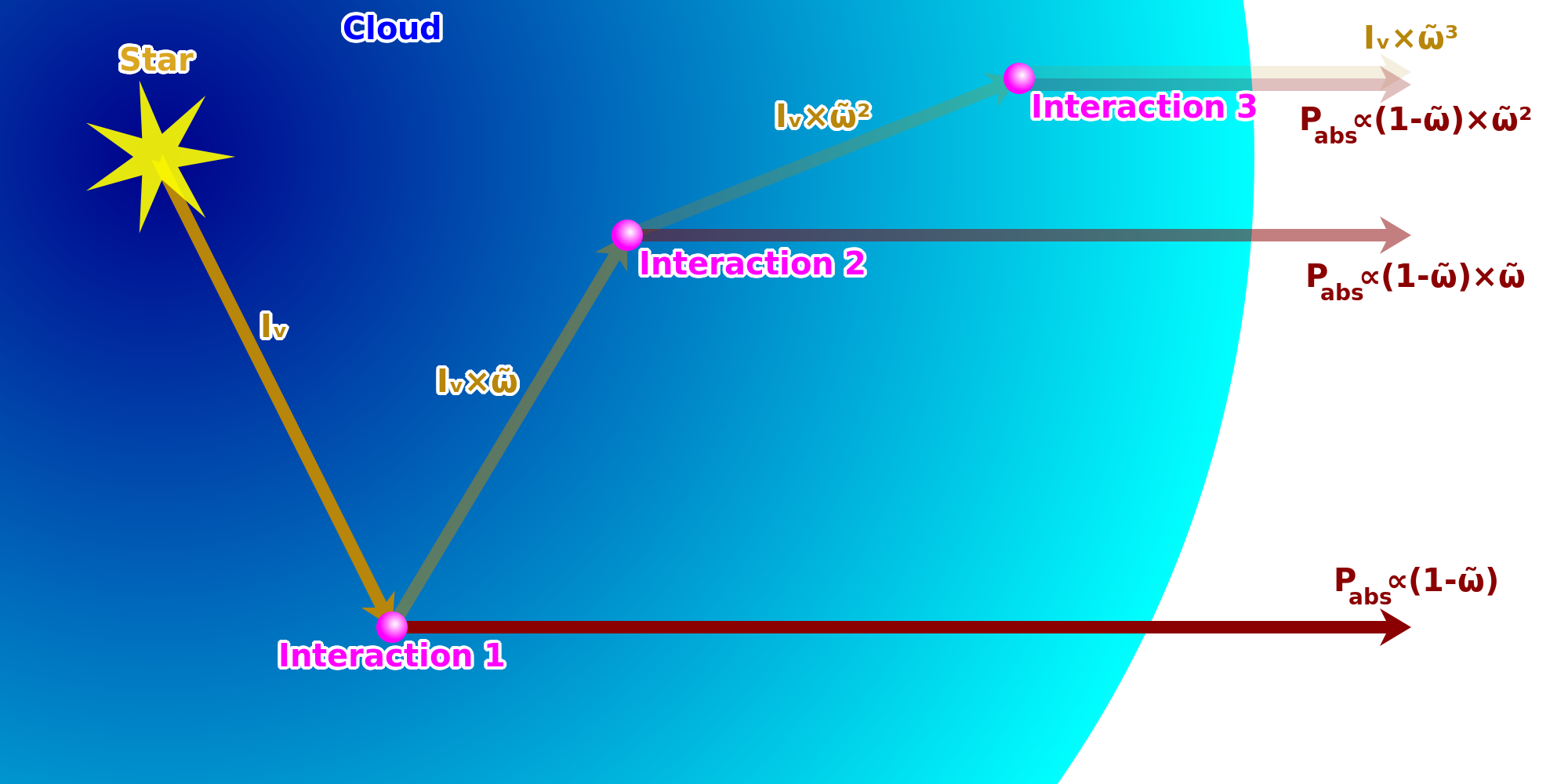}
  \newcap{Principle of Monte-Carlo radiative transfer}%
         {The blue area represents a dust cloud.
          We follow the journey of a single photon emitted by a star into the
          cloud and scattered at different locations.
          A single photon will be either scattered or absorbed.
          However, for numerical efficiency, we consider both solutions, 
          weighted by their probability (scattering: $\tilde{\omega}$; 
          absorption: $1-\tilde{\omega}$).
          The dark red arrows correspond to emitted \hIR\ photons.
          In principle, these photons can also be scattered and absorbed by the 
          cloud.
          The arrows exiting the cloud to the right represent what would be 
          measured by an observer.
          \CClicence}
  \label{fig:MCRT}
\end{figure}

\paragraph{The principle of Monte-Carlo radiative transfer.}
To compute a \hMCRT\ model, we need to draw a large number of photons (typically $\simeq10^6$ per wavelength bin), and execute the following steps.
The procedure is schematically represented on \reffig{fig:MCRT}.
\begin{enumerate}
  \item At each wavelength, photons are randomly drawn from primary 
    emitters, proportionally to their specific intensity.
    The emission angle, $(\theta,\phi)$, is randomly chosen, if the emitters 
    are isotropic, which is the case for stars.
  \item The interaction probability of a photon with a dust grain is then drawn
    from $1-\exp(-\tau)$, along the original direction of the photon.
    On interaction, one accounts for the probabilities of:
    \begin{inlinelist}
      \item absorption (proportional to $1-\tilde{\omega}$); and 
      \item scattering (proportional to $\tilde{\omega}$).
    \end{inlinelist}
    We could randomly choose between absorption and scattering, and thus 
    terminate the path of the photons $1-\tilde{\omega}$ of the times.
    However, forced scattering, the way it is shown in \reffig{fig:MCRT},
    allows us to track the absorption and scattering probabilities at each 
    interaction, in a numerically more efficient way.
    The scattering angle is randomly drawn from the scattering phase function
    \refeqp{eq:HG41}, setting the new direction of the photon.
  \item We iterate this process as long as needed, until the photon exits the
    nebula.
    The average number of interactions increases with the effective optical 
    depth of the nebula.
    At each interaction, $(1-\tilde{\omega})\times I_\nu$ is absorbed by the 
    grain, and $\tilde{\omega}\times I_\nu$ is scattered.
    After $N$ interactions, the scattered intensity is 
    $\tilde{\omega}^N\times I_\nu$ and the absorbed power is proportional to 
    $(1-\tilde{\omega})\times\tilde{\omega}^{N-1}$.
    For a typical $\tilde{\omega}\simeq0.5$, after three scatterings, the 
    intensity is reduced by a factor $\simeq0.13$.
  \item Once all the photons at all wavelengths have been drawn, the thermal
    emission of all dust species within each cell can be computed.
    \hIR\ photons are then drawn and scattered through the nebula, the same way
    as stellar photons.
    The overall opacity and albedo are usually much lower in the \hIR.
    The computation of \hIR\ radiative transfer is thus usually much faster.
    In the most embedded regions, the \hIR\ radiation absorbed by dust grains
    can be significant.
    One therefore has to recompute the \hIR\ transfer a few times, until an 
    energy balance is reached.
\end{enumerate}
These steps constitute the most basic implementation of \hMCRT.
Numerous optimizations have however been proposed in the last fifty years \citep[\eg][]{witt77a,witt77b,witt77c,witt77d,yusef-zadeh84,whitney92,wood97a,wood97,varosi99,baes01,gordon01,misselt01,steinacker02,steinacker06,wood08,baes11,camps15,siebenmorgen15,natale15,juvela19}.
Improvements and optimizations include:
\begin{inlinelist}
  \item massive parallelization, and the use of GPU;
  \item the production of synthetic photometric images;
  \item the treatment of polarization by scattering.
\end{inlinelist}

\begin{figure}[htbp]
  \includegraphics[width=\textwidth]{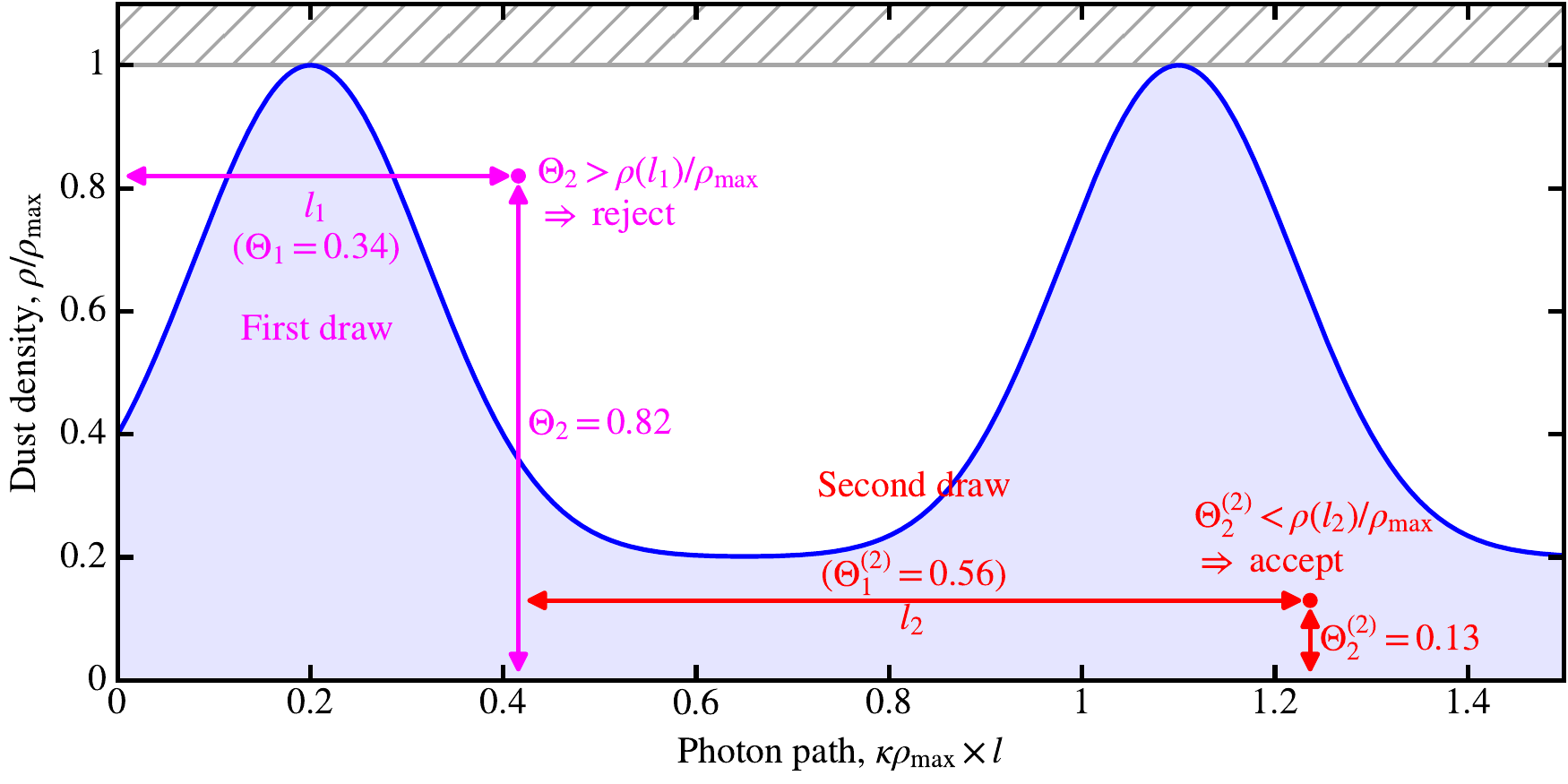}
  \newcap{Drawing photons in a Monte-Carlo radiative transfer model}%
         {This figure shows the dust density of a medium (in blue; normalized 
          to its peak value, $\rho_\sms{max}$), as a function of the path length 
          of the photon, $l$, across the nebula.
          The $x$-axis is multiplied by $\kappa\rho_\sms{max}$ so that it 
          corresponds to the optical depth the medium would have if the density 
          was everywhere $\rho_\sms{max}$.
          Two random draws are represented.
          In magenta, we show a first draw, bringing the photon to 
          $l_1=-\ln(1-\Theta_1)/\kappa\rho_\sms{max}$.
          The interaction at $l_1$ is rejected because 
          $\Theta_2>\rho(l_1)/\rho_\sms{max}$.
          A second similar draw brings the photon at $l=l_1+l_2$, where the 
          interaction is accepted.
          After this draw, one needs to draw the scattering angle.
          \CClicence}
  \label{fig:drawMCRT}
\end{figure}

\begin{figure}[htbp]
  \begin{tabular}{cc}
    \includegraphics[width=0.48\textwidth]{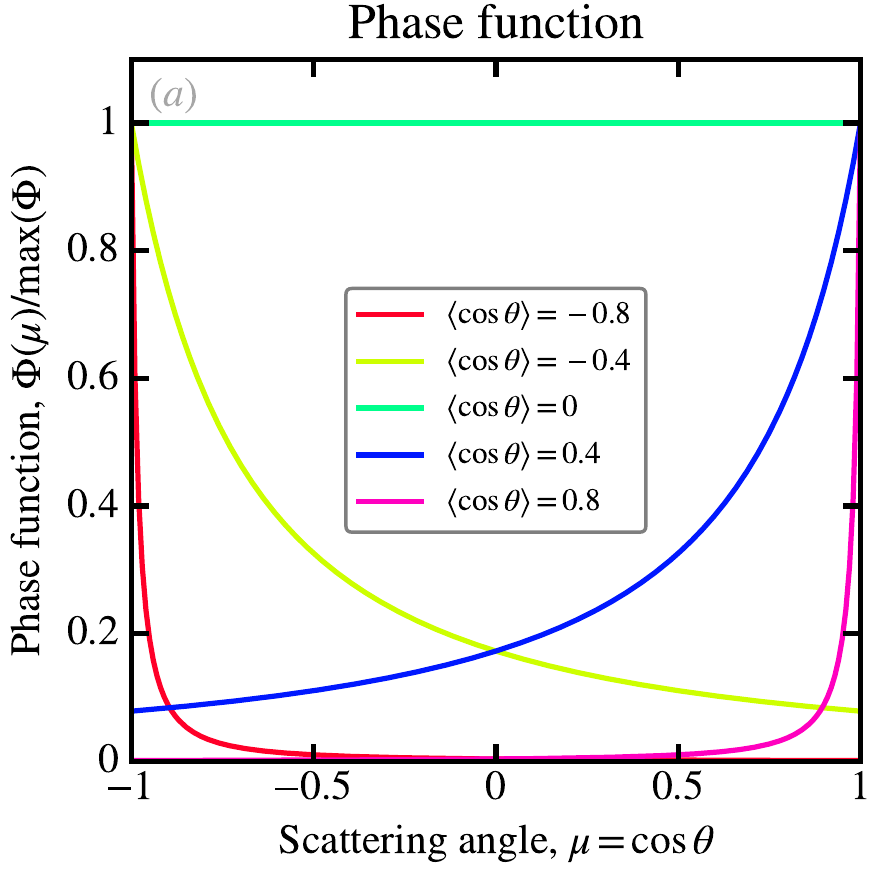} &
    \includegraphics[width=0.48\textwidth]{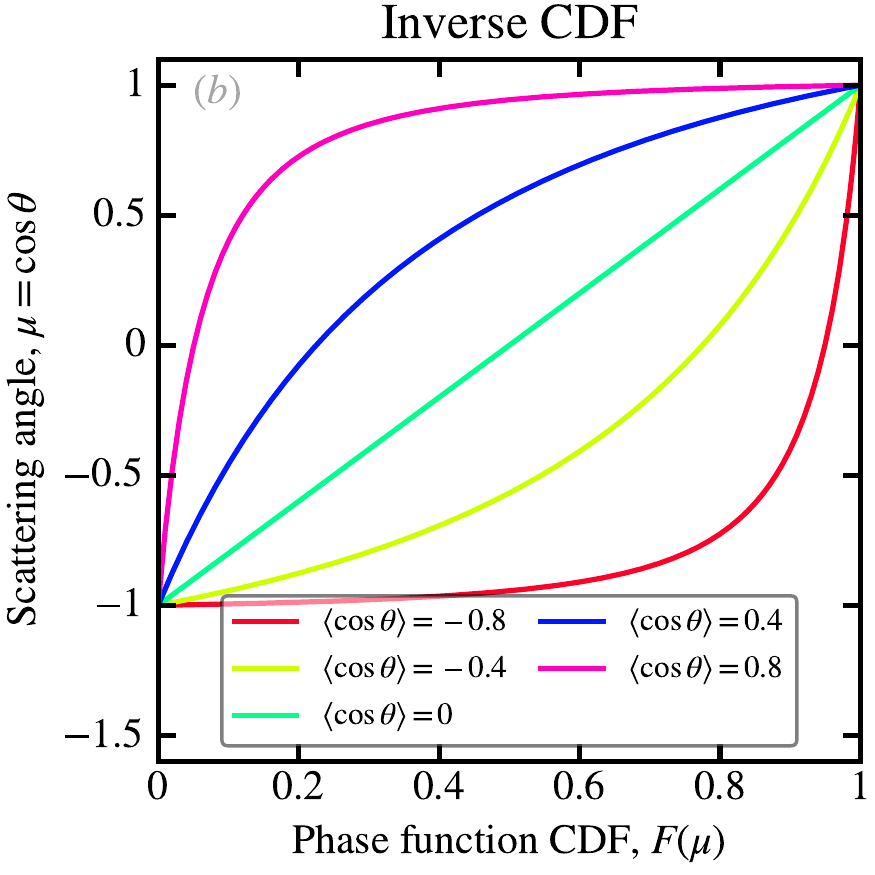} \\
  \end{tabular}
  \newcap{Drawing scattering angles in a Monte-Carlo radiative transfer model}%
         {Panel~\textit{(a)} shows the \citet{henyey41} scattering phase 
          function \refeqp{eq:HG41}, for different values of the asymmetry 
          parameter, $g=\langle\cos\theta\rangle$.
          Panel~\textit{(b)} shows the inverse \hCDF\ 
          \refeqp{eq:invHG41}, for different values of $g$.
          \CClicence}
  \label{fig:drawMCRT_angles}
\end{figure}

\paragraph{Numerical method to randomly draw photons.}
To simulate the random walk of a photon, there are two sets of random variables to draw:
\begin{inlinelist}
  \item random interaction events; and
  \item scattering angles.
\end{inlinelist}
\begin{description}
  \item[Drawing interaction events] along the path of a photon consists in 
    drawing a path length from the following \expression{Probability Density 
    Function} (\hPDF):
    \begin{equation}
      p(l) = 1- \exp\left(-\int_{0}^l\kappa\rho(l^\prime)l^\prime
             \ddiff l^\prime\right).
    \end{equation}
    For simplicity, $l$ corresponds to the path length along the direction of 
    the photon, starting at $l=0$.
    Drawing a variable from this distribution can be achieved the following way 
    \citep[\eg][]{varosi99}.
    It uses the \expression{rejection method} (\cf\ 
    \refapp{sec:random_rejection}).
    These steps are represented on \reffig{fig:drawMCRT}.
    \begin{enumerate}
      \item First draw a uniform random variable between 0 and 1, $\Theta_1$.
      \item Set $l_0=-\ln(1-\Theta_1)/\kappa\rho_\sms{max}$, where 
        $\rho_\sms{max}$ is the maximum of the density along the path.
      \item Draw a second uniform random variable between 0 and 1, $\Theta_2$.
      \item If $\Theta_2\le \rho(l_0)/\rho_\sms{max}$, then the interaction is 
        accepted, and a new scattering angle can be drawn.
        On the contrary, if $\Theta_2>\rho(l_0)/\rho_\sms{max}$, the interaction
        is rejected. 
        One then needs to go back to the first step, starting from $l_0$, this 
        time.
    \end{enumerate}
  \item[Scattering angles] are drawn from the scattering phase function 
    (\refeqnp{eq:HG41}; \refsubfig{fig:drawMCRT_angles}{a}).
    If we place a spherical coordinate reference frame at the position of 
    interaction with its $z$-axis aligned with the incoming direction of the 
    photon, then the two spherical angle, $(\theta,\phi)$, that will determine
    the new direction of the photon are drawn the following way.
    \begin{enumerate} 
      \item The new polar angle, $\theta$, is drawn by inverting the 
        \expression{Cumulative Distribution Function} (\hCDF) (\cf\ 
        \refapp{sec:random_CDF}).
        In case we use the \citet{henyey41} phase function, the inverse of the 
        \hCDF\ can be analytically derived (\cf\ 
        \refsubfig{fig:drawMCRT_angles}{b}):
        \begin{equation}
          F^{-1}(\Theta_3) 
           = \frac{1}{2g}\left[1+g^2-\left(\frac{1-g^2}{1-g+2g\Theta_3}
             \right)^2 \right],
          \label{eq:invHG41}
        \end{equation}
        where $\Theta_3$ is a uniform random variable between 0 and 1.
      \item The new azimuthal angle, $\phi$, is simply drawn from a uniform 
        distribution between 0 and $2\pi$, because there is a symmetry of 
        revolution around the direction of the photon.
    \end{enumerate}
\end{description}

\begin{figure}[htbp]
  \begin{tabular}{cc}
    \includegraphics[width=0.48\textwidth]{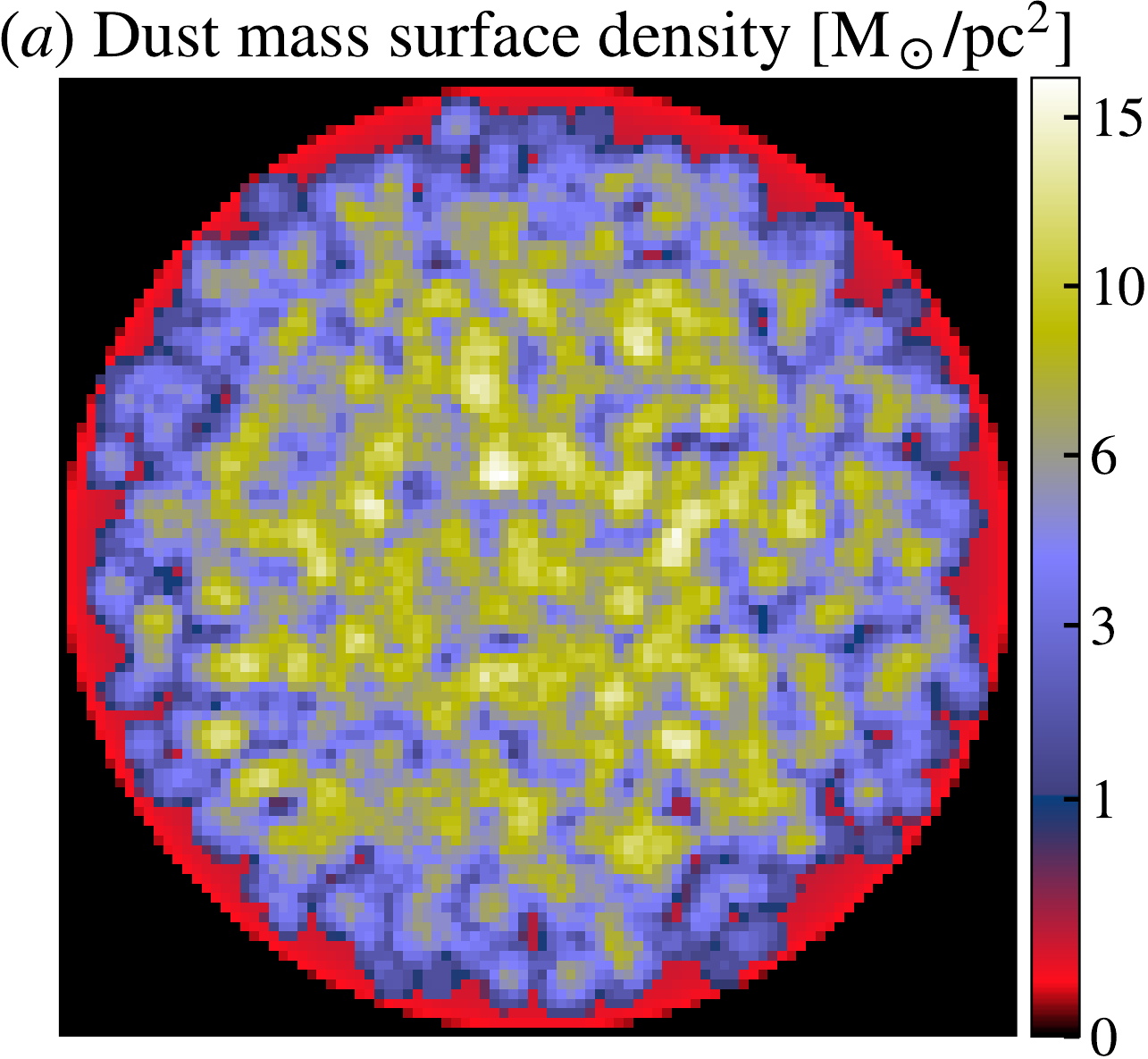} &
    \includegraphics[width=0.48\textwidth]{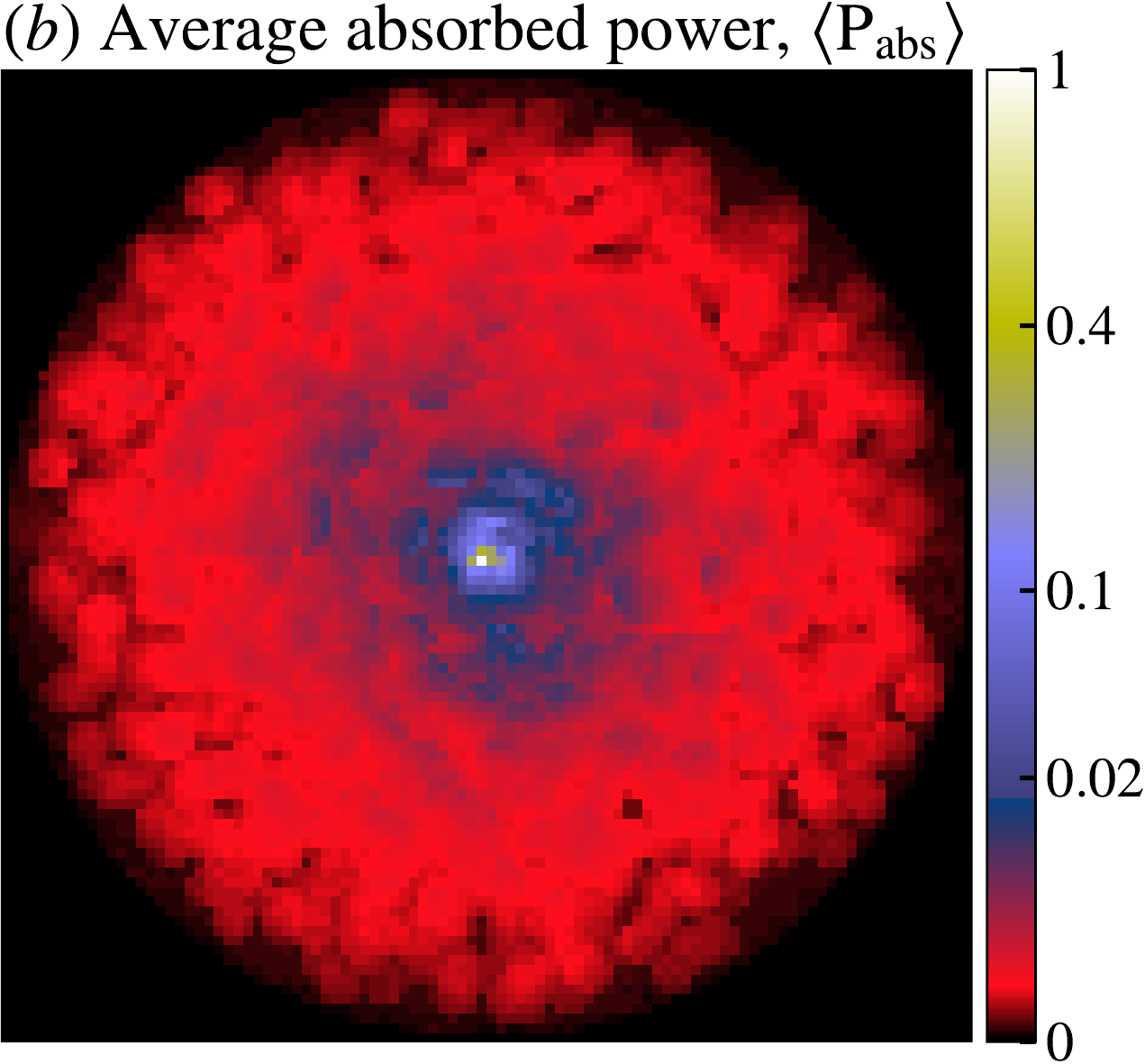} \\
  \end{tabular}
  \newcap{Spatial distributions of the clumpy radiative transfer model}%
         {Panel~\textit{(a)} shows the projected dust mass surface density 
          of the cloud.
          Panel~\textit{(b)} shows the average projected absorbed fraction,
          $\langle P_\sms{abs}\rangle$.
          The star is located at the center of the cloud.
          Notice that most of the power is absorbed in the central region.
          \CClicence}
  \label{fig:MCRTim}
\end{figure}

\begin{figure}[htbp]
  \begin{tabular}{ccc}
    \includegraphics[width=0.31\textwidth]{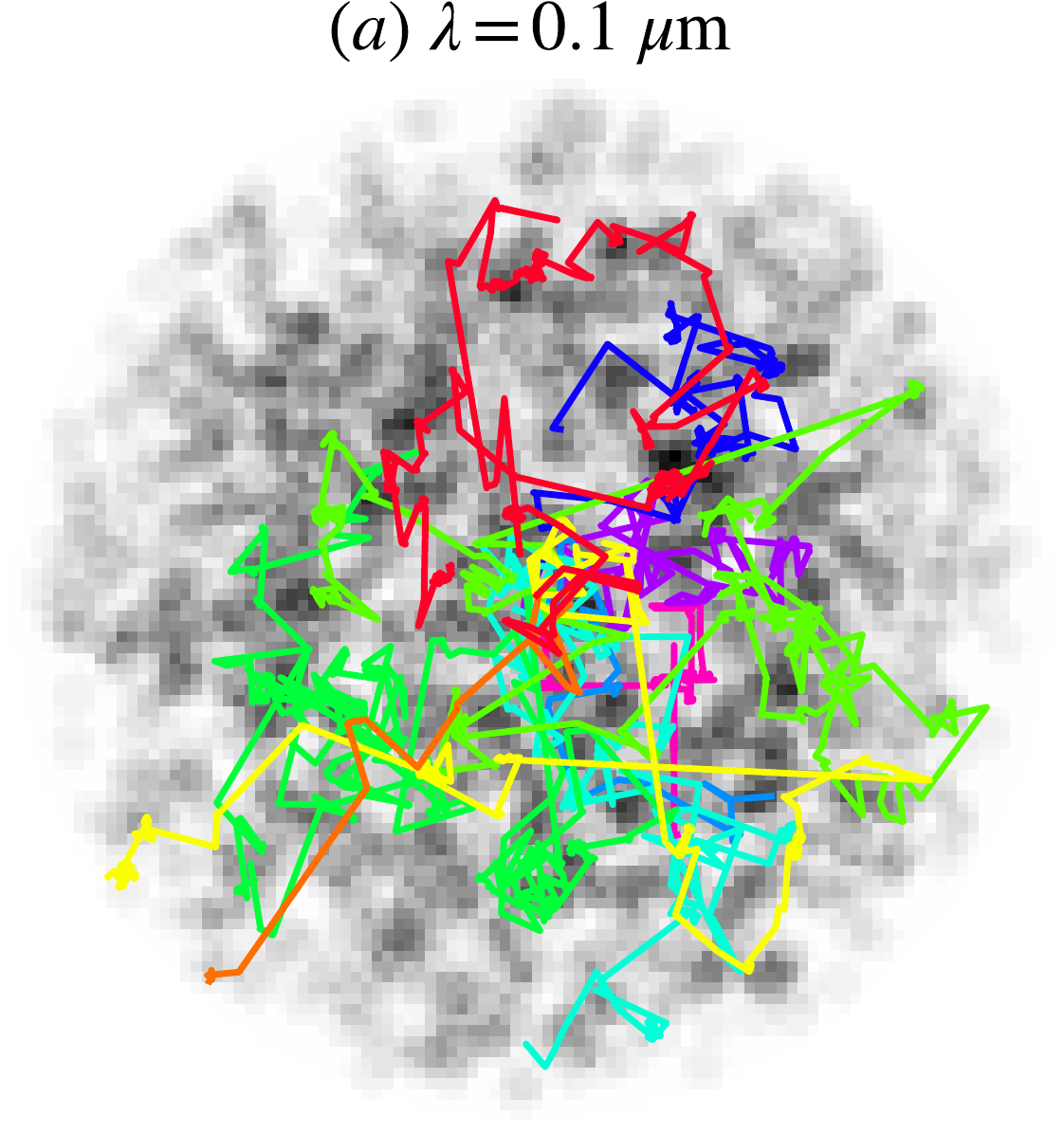} &
    \includegraphics[width=0.31\textwidth]{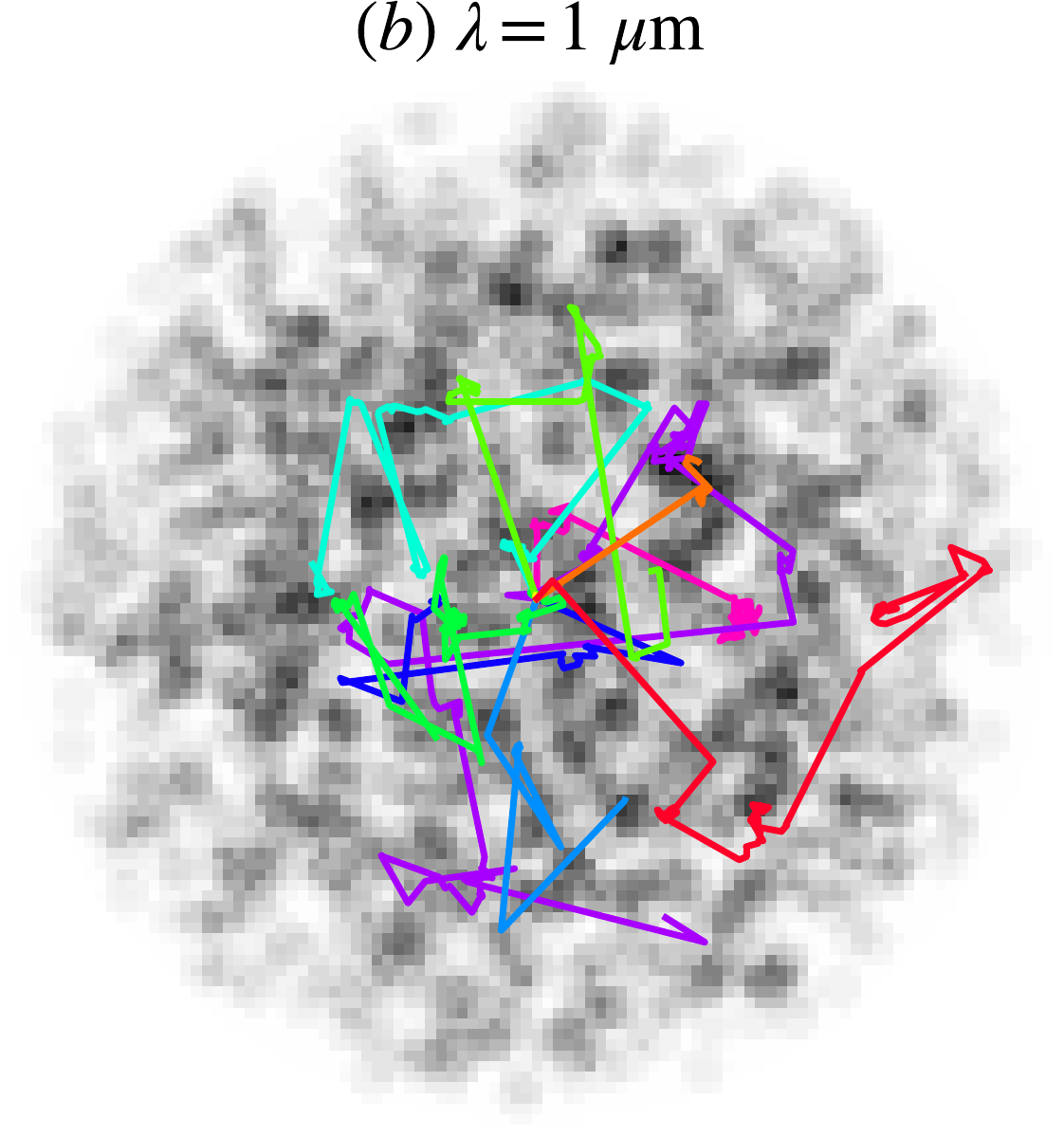} &
    \includegraphics[width=0.31\textwidth]{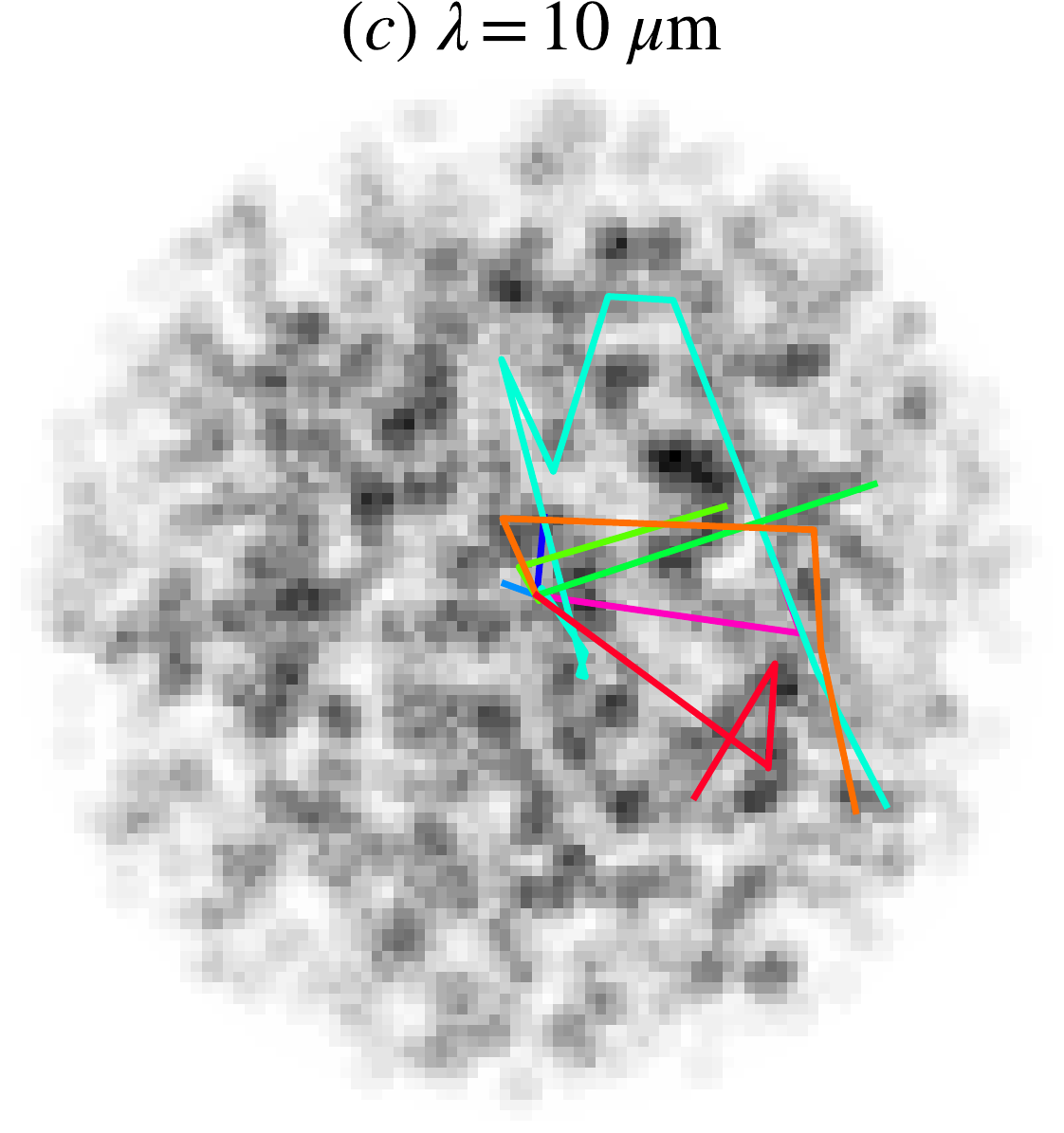} \\
  \end{tabular}
  \newcap{Random photon path within a clumpy medium}%
         {The three grey-scale images represent the projected column density of 
          \refsubfig{fig:MCRTim}{a}.
          We have overlaid the random path of 10 photons before they exit the
          cloud, at three different wavelengths.
          These draws are those computed by the \hMCRT\ model.
          For a given wavelength, the different colors represent different 
          photons.
          These figures demonstrate the fact that, at short wavelength, the 
          mean free-path is shorter, resulting in a larger number of
          scatterings.
          \CClicence}
  \label{fig:MCRT_demo}
\end{figure}

\begin{figure}[htbp]
  \includegraphics[width=\textwidth]{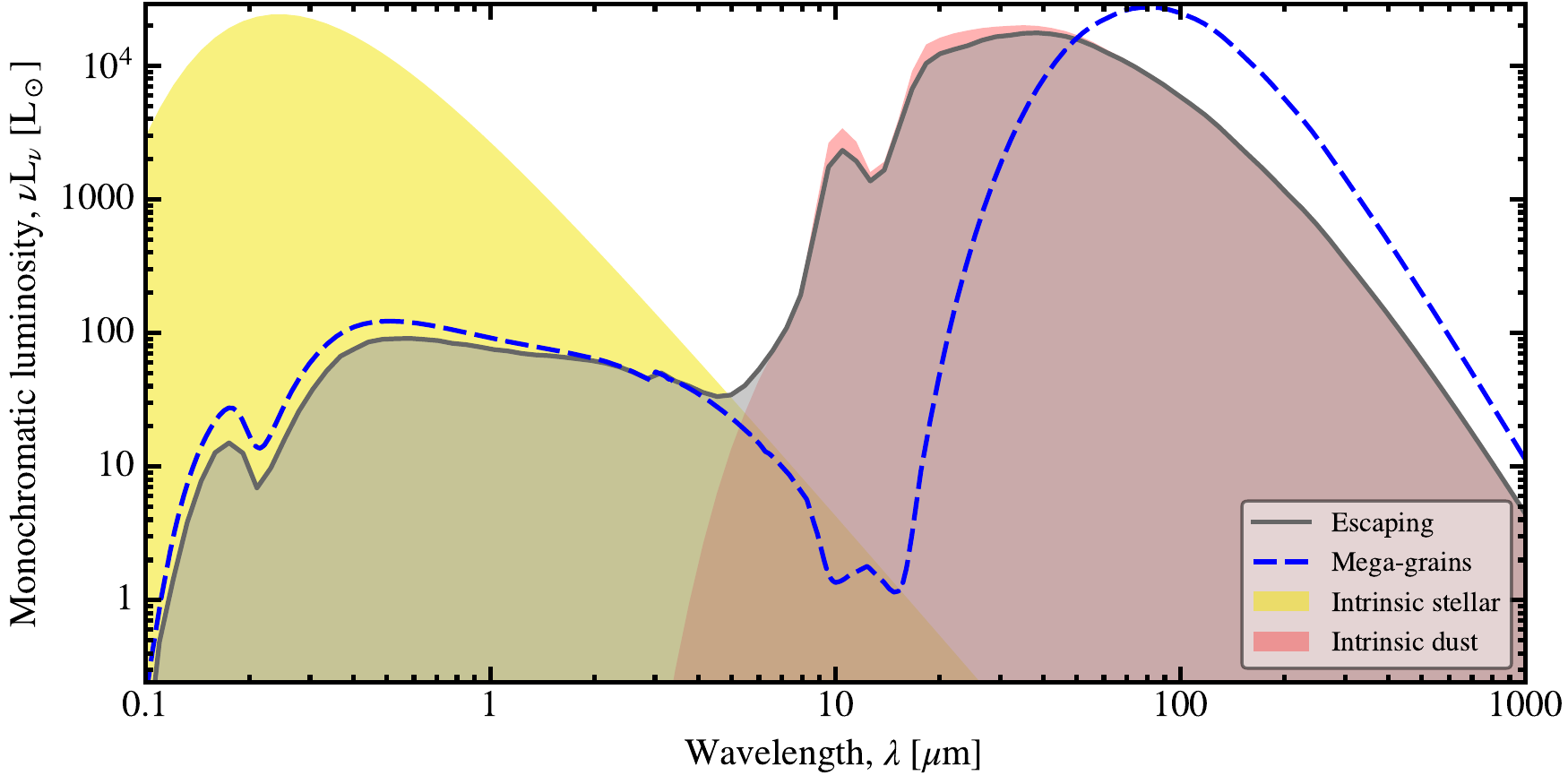}
  \newcap{Total SED from the clumpy radiative transfer model}%
         {The yellow curve represents the intrinsic \hSED\ of the central star.
          The red curve is the intrinsic emission (non self-extincted) of the
          dust emission of the whole cloud.
          The grey curve shows the total escaping \hSED\ of the cloud.
          We have overlaid the mega-grains approximation (\refsec{sec:clumpy}), 
          in dashed-blue, for comparison.
          \CClicence}
  \label{fig:MCRTSED}
\end{figure}

\paragraph{Example of a clumpy medium.}
For this manuscript, we have developed a \hMCRT\ model, following the procedure previously described.
We have applied it to a spherical clumpy cloud, similar to those discussed in \refsec{sec:clumpy}.
The radius is $R_\sms{s}=1$~pc, with an \hICM\ density, $n_\sms{ICM}=1000$~cm$^{-3}$. The $r_\sms{C}=0.05$~pc clumps have a density, $n_\sms{C}=10^5$~cm$^{-3}$, with a filling factor, $f_\sms{V}=20\,\%$.
We assume \citetalias{jones17} grain constitution, and assume all grains are at thermal equilibrium.
This cloud is centrally illuminated by a $T_\star=1.5\E{4}$~K star, with $L_\star=3.3\E{4}\eLsun$.
\begin{description}
  \item[The projected map] of the dust mass surface density is shown in \refsubfig{fig:MCRTim}{a}.
The average absorbed fraction, $\langle P_\sms{abs}\rangle$ is shown in \refsubfig{fig:MCRTim}{b}.
The latter quantity is the projected average of the absorbed fraction of photons passing through the cell.
This fraction is highly concentrated in the central region, as most of the power is absorbed by clumps close to the star.
Those clumps are hot, whereas clumps on the outskirt are colder, because they are essentially heated by fewer photons that have been multiply scattered.
\reffig{fig:MCRT_demo} represents a few photon paths at three different wavelengths.
  \item[The total SED] of this cloud is shown in \reffig{fig:MCRTSED}.
We notice that a fraction of the dust emission is self-absorbed.
This is mainly the 10 and 18~\tmic\ silicate emission of the hottest central clumps being absorbed by the outer ones.
We have also compared our model to the mega-grains approximation (\refsec{sec:clumpy}; in dashed blue).
We see this approximation is quite good in the \hUV-to-\hNIR\ range.
It is however discrepant in the \hIR.
This is because there is a very strong gradient of heating conditions, seen in \refsubfig{fig:MCRTim}{b}, whereas the mega-grains approximation accounts only for the total absorbed fraction in the clumps and the \hICM\footnote{\citetalias{varosi99} solved this issue by assuming a power-law distribution of temperatures that we have discussed in \refsec{sec:clumpy}.
This is however an \textit{ad hoc} solution that needs to be calibrated for each dust species, by using an actual \hMCRT\ model.
Our goal being to demonstrate what a \hMCRT\ model brings, we chose to compare it only to the \hSED\ that the mega-grains approximation allows us to derive, by itself.}.
\end{description}
Despite its intensive numerical requirements, \hMCRT\ is thus the most flexible and accurate way to compute the \hSED\ of an interstellar cloud.
It is however important to make sure that the spatial resolution of the density grid is fine enough to resolve the shortest mean free path of photons.
Otherwise, we would be smearing out a potential sub-grid temperature gradient.
For the present model, we have 0.01~pc cells for $l_\sms{mean}(U)\simeq0.005$~pc.
\takeaway{To accurately compute a radiative transfer model, it is necessary 
          to resolve scale-lengths of the order of the mean free path of \hUV\ 
          photons.}

  \subsection{Approximate Treatments of the Mixing of Physical
                  Conditions}
  \label{sec:mixT}

Ideally, each time we study a Galactic region or a galaxy, we should solve the radiative transfer equation \refeqp{eq:RTl}.
This is however, most of the time, impossible, because of the lack of constraints on the actual \hthreeD\ structure of the region.
Even if we have a collection of high-angular resolution multiwavelength images of our object, the matter and stellar distributions along the sightline have to be inferred.
This inference is possible when the large-scale geometry of the object is quite regular, for instance: a protostellar disk and its jet, or a galactic disk and its bulge, \etc\
We will discuss the \hMCRT\ modeling of disk galaxies in \refsec{sec:MCRTgal}.
Otherwise, most often, we need to adopt empirical approaches that allows us to constrain the dust properties, despite our uncertainty of the spatial structure of the region.

    \subsubsection{The Historical Model: the MBB}
    \label{sec:MBB}

\begin{figure}[htbp]
  \includegraphics[width=\textwidth]{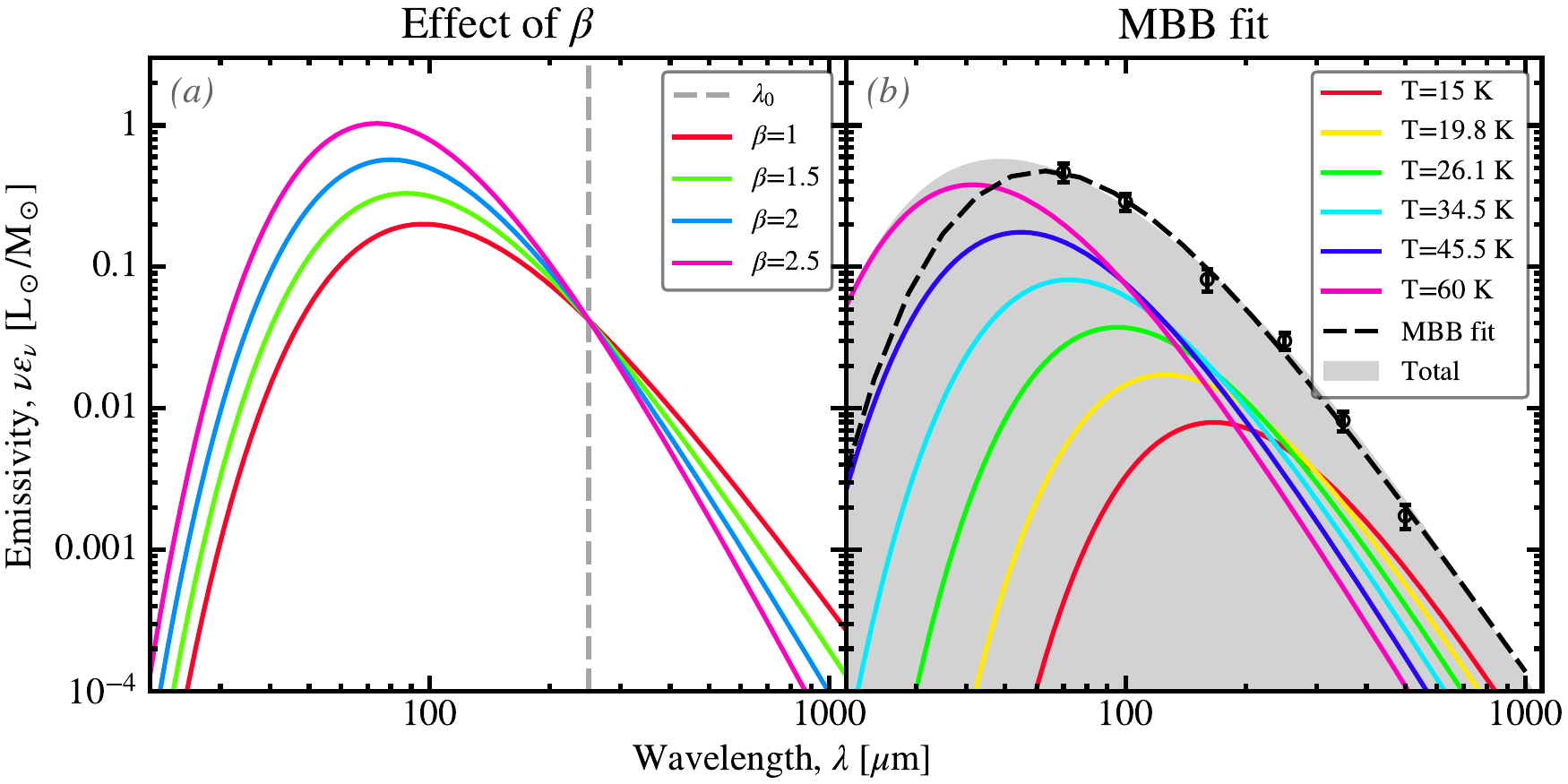}
  \newcap{MBB fitting}%
         {Panel~\textit{(a)} demonstrate the effect of the emissivity index, 
          $\beta$ \refeqp{eq:kappaMBB}, on the \hSED\ of a \hMBB.
          The temperature is $T=30$~K, with 
          $\kappa(\lambda_0=250\emic)=0.64$~m$^2$/kg
          \refeqp{eq:themis_kappa_proxy}.
          Panel~\textit{(b)} illustrates the limitation of the isothermal
          assumption.
          The black error bars are synthetic observations (noise has been 
          added).
          They are sampling the grey-filled curve, which is the integral of
          \hMBB s (color curves) times a power-law distribution of temperatures
          ($T_\sms{min}=15$~K, $T_\sms{max}=60$~K, and index 4).
          These \hMBB s have \citetalias{jones17} optical properties
          ($\beta=1.79$; \refeqnp{eq:themis_kappa_proxy}).
          Yet the fitted value (black dashed curve) is significantly lower:
          $\beta=1.12\pm0.30$.
          The fitted temperature falls in the middle of the range, toward 
          the high end: $T=45\pm8$~K.
          \CClicence}
  \label{fig:demoMBB}
\end{figure}

The \hMBB\ (\cf\ \refsec{sec:kirchhoff}) is historically the most widely-used dust \hSED\ model. 
It is controlled by the three following parameters (\refeqnp{eq:BBQ} and \refeqnp{eq:kappaMBB}).
\begin{enumerate}
  \item The dust mass, $M_\sms{dust}$, is a scaling parameter.
  \item The equilibrium temperature, $T$, controls the emission peak 
    wavelength.
  \item The \expression{emissivity index}, $\beta$ \refeqp{eq:kappaMBB}, 
    controls the long-wavelength slope of the \hSED.
    This is demonstrated in \refsubfig{fig:demoMBB}{a}.
\end{enumerate}
Its physical assumptions are simplistic:
\begin{inlinelist}
  \item the \hIR\ emission is optically thin;
  \item the dust is made of a single species of grains at thermal 
    equilibrium with the \hISRF; and
  \item the opacity is a power-law.
\end{inlinelist}
By inferring both $T$ and $\beta$, this model is designed to constrain both the dust excitation and its optical properties.
This model was popularized by \citet{hildebrand83}, in the \hIRAS\ days (\cf\ \refsec{sec:chronology}).
At the time, it was well adapted, being a simple, but still physical model, with only three parameters to fit four broadbands (the four \hIRAS\ bands at 12, 25, 60 and 100~\tmic).
It has however several limitations that are often disregarded in the literature.
\begin{description}
  \item[The mixing of physical conditions] is not accounted for by the \hMBB.
    It means that the \hSED\ fit of a complex region, containing a 
    gradient of temperatures, is biased \citep[\eg][]{juvela12,galliano18}.
    This is demonstrated in \refsubfig{fig:demoMBB}{b}.
    The grey-filled curve is the \hSED\ coming from a power-law distribution
    of temperatures ($T=15$ to 60~K; the color curves).
    The intrinsic emissivity index of the grains making up this region is 
    $\beta=1.79$.
    Yet, fitting such a \hSED\ with a single \hMBB\ leads to a compromise 
    temperature ($T=45\pm8$~K) and a systematically lower emissivity index, 
    $\beta=1.12\pm0.30$.
    This is because a gradient of temperature tends to flatten the \hFIR\ slope,
    similarly to a lower $\beta$ value.
    \takeaway{The emissivity index derived from a single \hMBB\ fit is always
              lower than its true, intrinsic value.}
  \item[Stochastic heating] is formally equivalent to a gradient of temperatures
    (\cf\ \refsec{sec:stochastheat}).
    Stochastically heated grains dominate the emission at \hMIR\ 
    wavelengths (\cf\ \reffig{fig:themis_emiss_proxy}).
    A single \hMBB\ is thus biased at short wavelengths by small grains, the
    same way it is biased at long wavelengths by cold dust.
    Solving this issue by fitting a linear combination of two or three \hMBB s
    can palliate this problem, but is usually not sufficient.
    In addition, fitting an out-of-equilibrium emission with equilibrium grains
    is physically incorrect.
    It renders the interpretation of the parameters of the hot \hMBB\ 
    unreliable.
  \item[Realistic opacities are more complex] than a power-law.
    Laboratory data show that emissivity indices of actual materials are 
    wavelength-dependent quantities, $\beta(\lambda)$ (\cf\ \eg\ 
    \reffig{fig:labsil}).
    Additionally, the somehow arbitrary choice of the two other parameters in 
    \refeq{eq:kappaMBB} --~the reference wavelength, $\lambda_0$, and the
    scaling of the opacity at this wavelength, $\kappa_0$~-- have dramatic 
    consequences on the derived dust mass.
    It is thus important to limit the potential range of variations of these
    parameters.
    The Kramers-Kronig relations (\refsec{sec:KK}) impose that $\beta\ge1$,
    and most compounds studied in the laboratory have $\beta<2.5$
    (\cf\ \eg\ the
    \href{https://www.astro.uni-jena.de/Laboratory/Database/databases.html}%
    {Jena  database}).
    We recommend calibrating $\kappa_0$ on laboratory data or on 
    well-constrained dust models (such as \refeqnp{eq:themis_kappa_proxy}).
    Contrary to what \citet{hildebrand83} recommended, it is probably better 
    to choose a reference wavelength, $\lambda_0$, around the peak of the 
    \hFIR\ \hSED\ (between 100 and 300~\tmic), rather than in the submm.
    This way, variations of $\beta$ will impact only moderately the derived 
    dust mass.
    \takeaway{A \hMBB\ fit infers parameters whose physical meaning is 
              difficult to assess.}
  \item[The noise-induced degeneracy] between $T$ and $\beta$ is well-documented
    \citep[\eg][]{shetty09,kelly12,galliano18a}.
    A false negative correlation arises when a series of \hMBB\ fits are 
    performed with standard least-squares, maximum likelihood, or 
    non-hierarchical Bayesian methods.
    This degeneracy prevents to explore the potential correlation of these 
    parameters that laboratory data and solid-state models suggest 
    \citep[\eg][]{mennella98,meny07}.
    Luckily, hierarchical Bayesian methods are efficient at removing false, 
    noise-induced correlations \citep[\eg][]{kelly12,galliano18a}. 
    \refchap{chap:method} is entirely devoted to fitting techniques.
\end{description}

  \subsubsection{A Phenomenological, Composite Approach}
  \label{sec:dale}

\begin{figure}[htp]
  \begin{tabular}{cc}
    \includegraphics[width=0.48\textwidth]{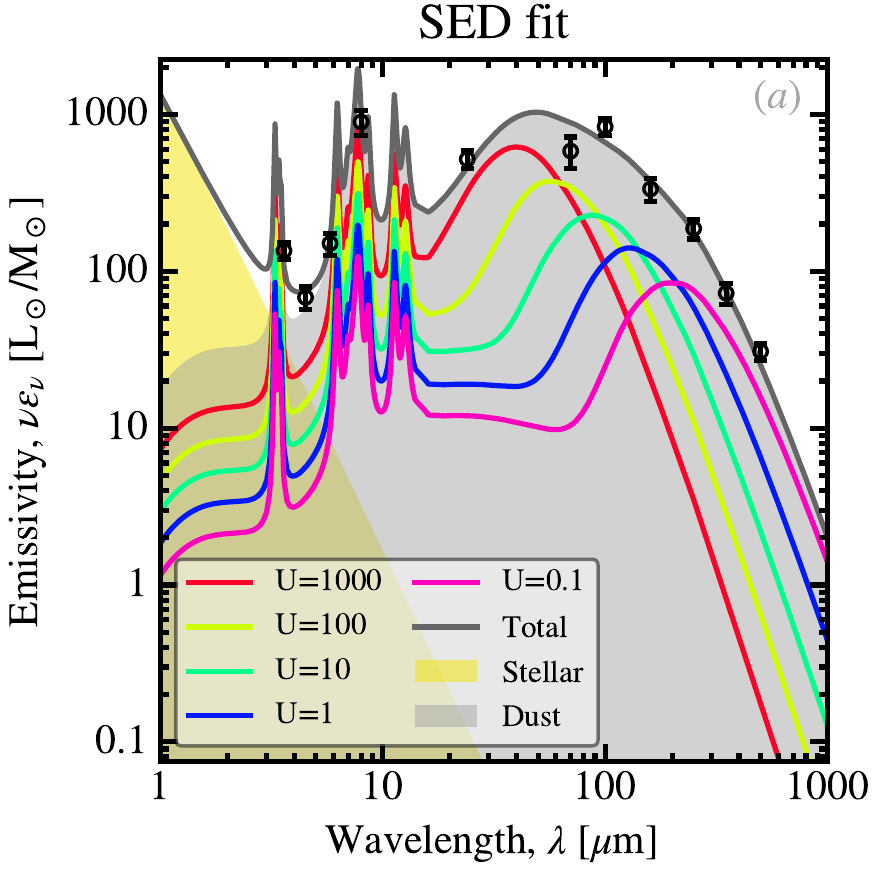} &
    \includegraphics[width=0.48\textwidth]{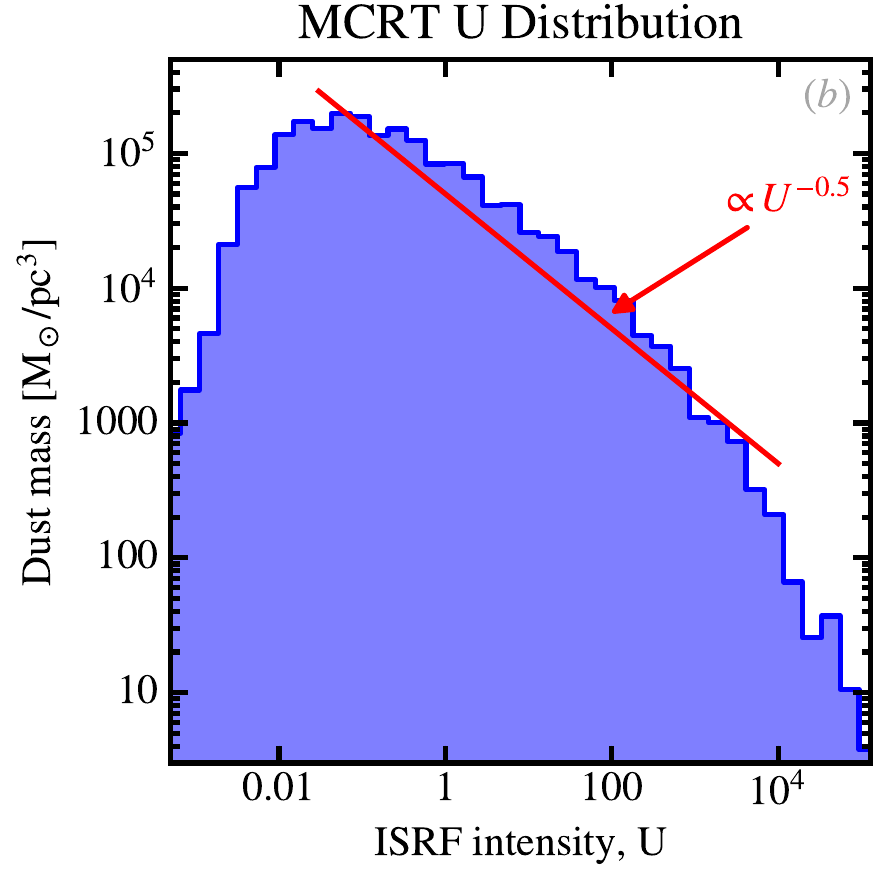} \\
  \end{tabular}
  \newcap{Phenomenological mixing of physical conditions}%
         {Panel~\textit{(a)} shows an example of \hSED\ fit using 
          the composite approach \refeqp{eq:dale}.
          The black error bars represent synthetic observations (noise has been 
          added). 
          These observations have been fitted with the total model represented 
          by the dark grey line.
          This total model is the sum of a stellar (in yellow) and a dust (in 
          light grey) components.
          The dust component is the integral of dust models illuminated by 
          different $U$ \refeqp{eq:dale}.
          The rainbow curves represent several bins of \hISRF.
          The sum of these rainbow curves is the light-grey-filled curve.
          The blue curve in panel~\textit{(b)} shows the distribution of dust 
          mass per bin of \hISRF\ intensity, $U$ (\refsec{sec:Teq}), for each 
          cell in the \hMCRT\ simulation presented in \refsec{sec:MCRT}.
          The value of $U$ has been derived from the grain equilibrium 
          temperature, using \refeq{eq:U2T}.
          The red line shows a power-law approximation of this distribution.
          With the parametrization of \refeq{eq:dale}, it would correspond to:
          $U_-=0.03$, $\Delta U=10^4$ and $\alpha=0.5+1=1.5$ (the +1 comes from 
          taking the derivative of $M_\sms{dust}(U)$ to get \refeqnp{eq:dale}).
          \CClicence}
  \label{fig:powerU}
\end{figure}

Dust models provide useful frameworks to model \hSED s (\refsec{sec:dustmodels}).
Without the possibility to compute the radiative transfer, we are however facing the problem of the mixing of physical conditions.
A prescription, proposed by \citet{dale01}, has proven to be a powerful solution to this issue.
It consists in assuming that the dust mass is distributed in regions with different starlight intensities, $U$, following a power-law:
\begin{equation}
  \def\arraystretch{2}
  \frac{1}{M_\sms{dust}}\frac{\dd M_\sms{dust}}{\dd U} = U^{-\alpha}\times\left\{
  \begin{array}{ll}
    \displaystyle
    \frac{1-\alpha}{(U_-+\Delta U)^{1-\alpha}-U_-^{1-\alpha}}
    & \mbox{ if } \alpha\ne1 \\
    \displaystyle
    \frac{1}{\ln(U_-+\Delta U)-\ln U_-}
    & \mbox{ if } \alpha = 1
  \end{array}
  \right\}
  \mbox{ for } U_-\le U\le U_-+\Delta U.
  \label{eq:dale}
\end{equation}
The idea is that the shape of the observed \hSED\ is used to constrain this distribution of \hISRF s, assuming a dust mixture constitution.
By lack of a better term, we call this approach the \expression{composite} model.
The free parameters are:
\begin{itemize}
  \item the total dust mass, $M_\sms{dust}$, acting as a scaling parameter;
  \item the power-law index, $\alpha$;
  \item the minimum starlight intensity, $U_-$;
  \item the width of the starlight intensity distribution, $\Delta U$.
\end{itemize}
It thus provides a way to account for the potential complexity of the region without having to model the radiative transfer.
The model \hSED\ is then simply:
\begin{equation}
  L_\nu^\sms{dust}(\lambda) 
    = \int_{U_-}^{U_-+\Delta U}\epsilon_\nu(\lambda,U)\frac{\dd 
      M_\sms{dust}}{\dd U}\ddiff U,
\end{equation}
where $\epsilon_\nu(\lambda,U)$ is the monochromatic emissivity of the dust model, exposed to a single starlight intensity, $U$ (\reffig{fig:kappaSED}).
\refsubfig{fig:powerU}{a} shows an example of a \hSED\ fit, using \refeq{eq:dale}.
We have a added a free-scaling black body, at $T_\star=30,000$~K, to account for the stellar continuum that may be contaminating the \hMIR\ photometric bands.
The composite approach is flexible enough to be usable in a diversity of environments.
\citet{dale01} lists several simple geometries for which \refeq{eq:dale} is actually the solution.
It is also adapted to more complex \hISM\ topologies.
For instance, \refsubfig{fig:powerU}{b} shows the dust mass distribution as a function of $U$, for each cell in the \hMCRT\ simulation of \refsec{sec:MCRT}.
Despite the complex, clumpy structure of this cloud, it can be reasonably well-approximated by a power-law (shown in red).

\paragraph{The average starlight intensity.}
The parameters of \refeq{eq:dale} do not have a very clear physical meaning.
Besides, they are often degenerate: the uncertainties on the three parameters $U_-$, $\Delta U$ and $\alpha$ are strongly correlated.
It can be more efficient to quote the average of the distribution: 
\begin{equation}
  \def\arraystretch{2}
  \langle U\rangle\equiv\frac{1}{M_\sms{dust}}
  \int_{U_-}^{U_-+\Delta U} U\frac{\dd M_\sms{dust}}{\dd U}\ddiff U
  = 
  \left\{
  \begin{array}{ll}
    \displaystyle
    \frac{1-\alpha}{2-\alpha}\times
    \frac{\left(U_-+\Delta U\right)^{2-\alpha}-U_-^{2-\alpha}}
         {\left(U_-+\Delta U\right)^{1-\alpha}-U_-^{1-\alpha}}
    & \mbox{ if } \alpha\neq1 \;\&\; \alpha\neq2 \\
    \displaystyle
    \frac{\Delta U}{\ln\left(U_-+\Delta U\right)-\ln U_-}
    & \mbox{ if } \alpha = 1 \\
    \displaystyle
    \frac{\ln\left(U_-+\Delta U\right)-\ln U_-}
         {U_-^{-1}-\left(U_-+\Delta U\right)^{-1}}
    & \mbox{ if } \alpha = 2. \\
  \end{array}
  \right.
  \label{eq:Uav}
\end{equation}
This parameter quantifies the \expression{average starlight intensity}, heating the bulk of the dust mass.
It is the equivalent of the equilibrium temperature of a \hMBB, as it controls the peak emission wavelength, except that it accounts for the mixing of physical conditions and the stochastic heating of small grains.
The \expression{Total InfraRed} (\hTIR) luminosity, $L_\sms{TIR}$, can be expressed:
\begin{equation}
  L_\sms{TIR}\equiv\int_{3\emic}^{1000\emic}L_\lambda(\lambda)\ddiff\lambda
  \simeq\int_0^\infty L_\nu^\sms{dust}(\nu)\ddiff\nu 
  = \epsilon\times M_\sms{dust}\times\langle U\rangle,
  \label{eq:TIR}
\end{equation}
where the constant $\epsilon\equiv\int\epsilon_\nu\ddiff\nu$ is the bolometric emissivity of the dust model.
For the \citetalias{jones17} mixture, heated by the \citet{mathis83} \hISRF, it is $\epsilon=221\eLsun/\Msun/U$ (\reftab{tab:themis_emiss}).

\begin{figure}[htbp]
  \begin{tabular}{cc}
    \includegraphics[width=0.48\textwidth]{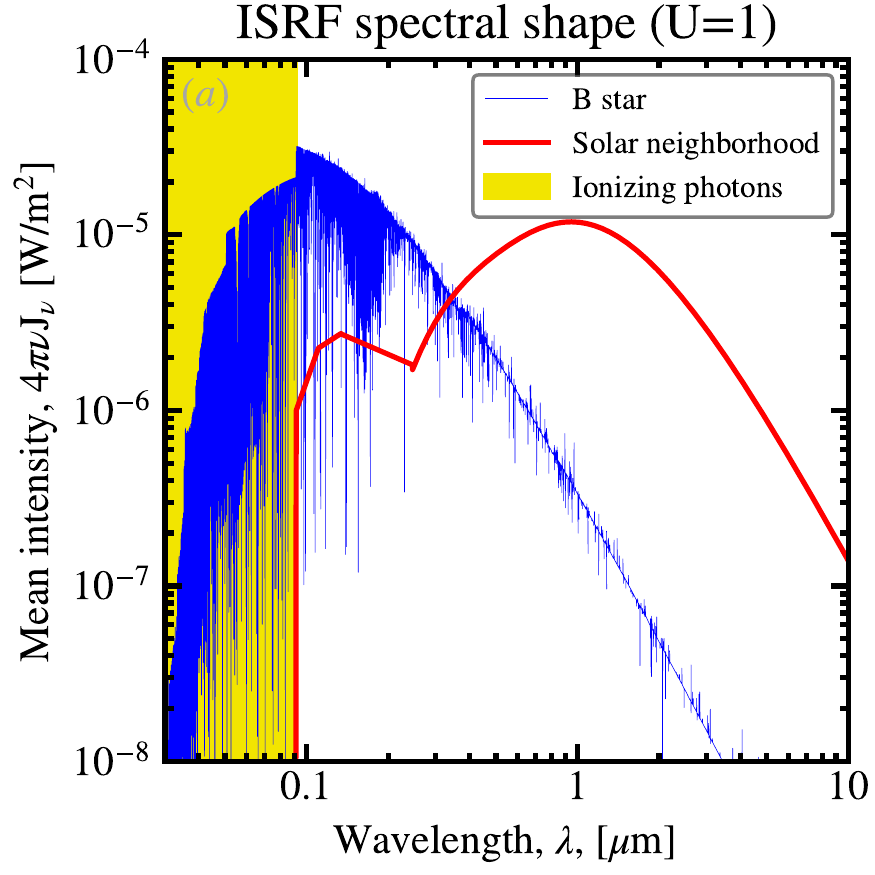} &
    \includegraphics[width=0.48\textwidth]{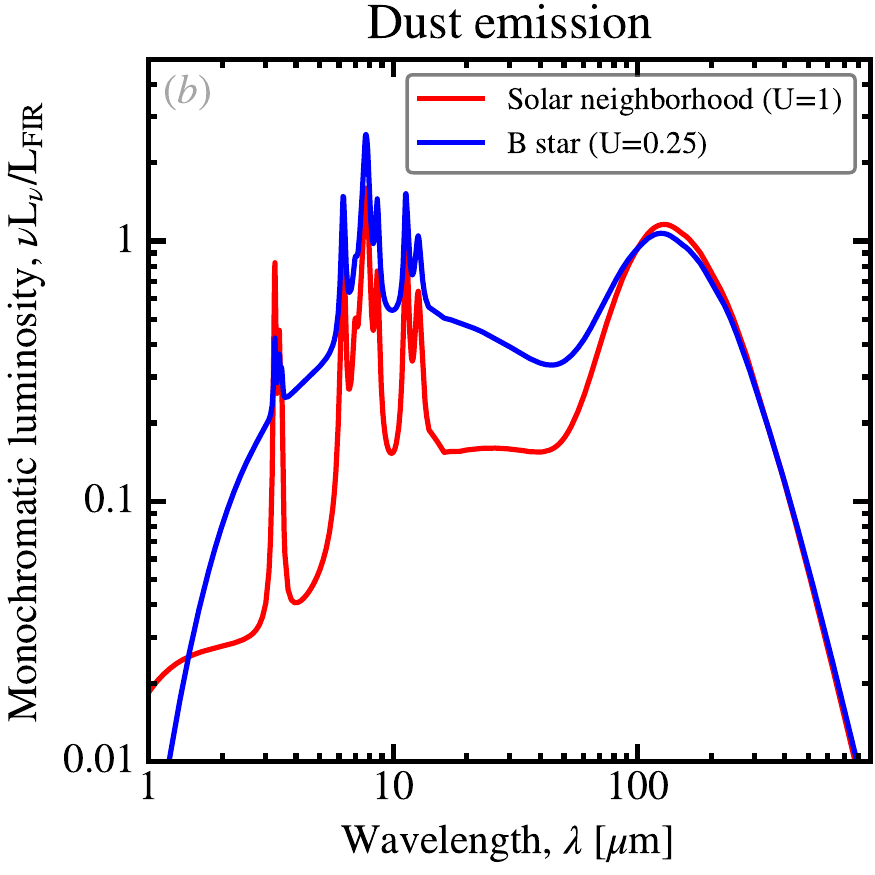} \\
  \end{tabular}
  \newcap{Effect of the ISRF hardness on the SED}%
         {Panel~\textit{(a)} represents different \hISRF s:
          \begin{inlinelist}
            \item the Solar neighborhood \citep[red;][]{mathis83}; and
            \item a B star \citep[$T_\star=30\,000$~K; blue;][]{lanz07}.
          \end{inlinelist}
          We have represented the spectral range of ionizing photons in yellow.
          Both \hISRF s are normalized so that $U=1$ \refeqp{eq:U}.
          Panel~\textit{(b)} shows the \hSED s of the \citetalias{jones17} 
          model, heated by the two \hISRF s.
          We show the \hSED\ for different values of $U$, in order to obtain 
          the same \hFIR\ \hSED, as this parameter is inferred by the fitter.
          The \hSED s are normalized by the \hFIR\ luminosity, $L_\sms{FIR}$, 
          between $\lambda=60$ and 200~\tmic.
          The B star heating leads to an increase in the $\lambda=6-9\emic$ 
          emission by a factor $\simeq1.8$, compared to the Solar neighborhood.
          \CClicence}
  \label{fig:ISRFhardness}
\end{figure}

\paragraph{Constraining the dust properties.}
The composite approach of \refeq{eq:dale} allows us to constrain parameters that are not extremely sensitive to radiative transfer effects \citep[\ie\ to variations of the local intensity and spectral shape of the \hISRF; \cf\ \eg][for a discussion]{galliano18,galliano21}.
\begin{description}
  \item[The dust mass,] $M_\sms{dust}$, is dominated by large grains (\cf\ 
    \reftab{tab:sizedistmoments}).
    The heating of these large grains is sensitive only to the power they 
    absorb, and not to the spectral shape of the \hISRF, contrary to small 
    grains (\cf\ \refsec{sec:stochastheat}).
    The assumption of a constant \hISRF\ shape therefore does not significantly 
    bias $M_\sms{dust}$.
  \item[The average starlight intensity,] $\langle U\rangle$ \refeqp{eq:Uav},
    is empirically constrained by the shape of the \hFIR\ emission, dominated
    by large grains.
    This quantity is thus not particularly biased, similarly to $M_\sms{dust}$.
    In addition, \refeq{eq:TIR} tells us that 
    $L_\sms{TIR}\propto M_\sms{dust}\langle U\rangle$.
    Yet, $L_\sms{TIR}$ is a weakly model-dependent quantity, as it is simply the 
    integral of the \hSED\ model passing through the observed fluxes.
    $L_\sms{TIR}$ and $M_\sms{dust}$ being reliable, $\langle U\rangle$ has to be.
  \item[The mass fraction of PAH,] $q_\sms{PAH}$ (or the mass fraction of 
    small \hHAC, $q_\sms{AF}$, for the \citetalias{jones17} model),
    discussed in \refsec{sec:themis_sizedist}, 
    controls the strength of the \hMIR\ aromatic features (\cf\ 
    \refsec{sec:IRobs}).
    These features, being carried by small, stochastically heated grains, are
    sensitive to the spectral shape of the \hISRF\
    (\cf\ \refsubfig{fig:themis_emiss_proxy}{a}).
    This is demonstrated in \reffig{fig:ISRFhardness}.
    We have considered two extreme \hISRF s (\refsubfig{fig:ISRFhardness}{a}).
    The bias on the aromatic feature emission is at most a factor $\simeq1.8$ 
    (\refsubfig{fig:ISRFhardness}{b}).
\end{description}
\takeaway{\refeq{eq:dale} provides acceptable estimates of $M_\sms{dust}$, 
          $\langle U\rangle$ and $q_\sms{PAH}$ (or $q_\sms{AF}$).}

\begin{figure}[htbp]
  \includegraphics[width=\textwidth]{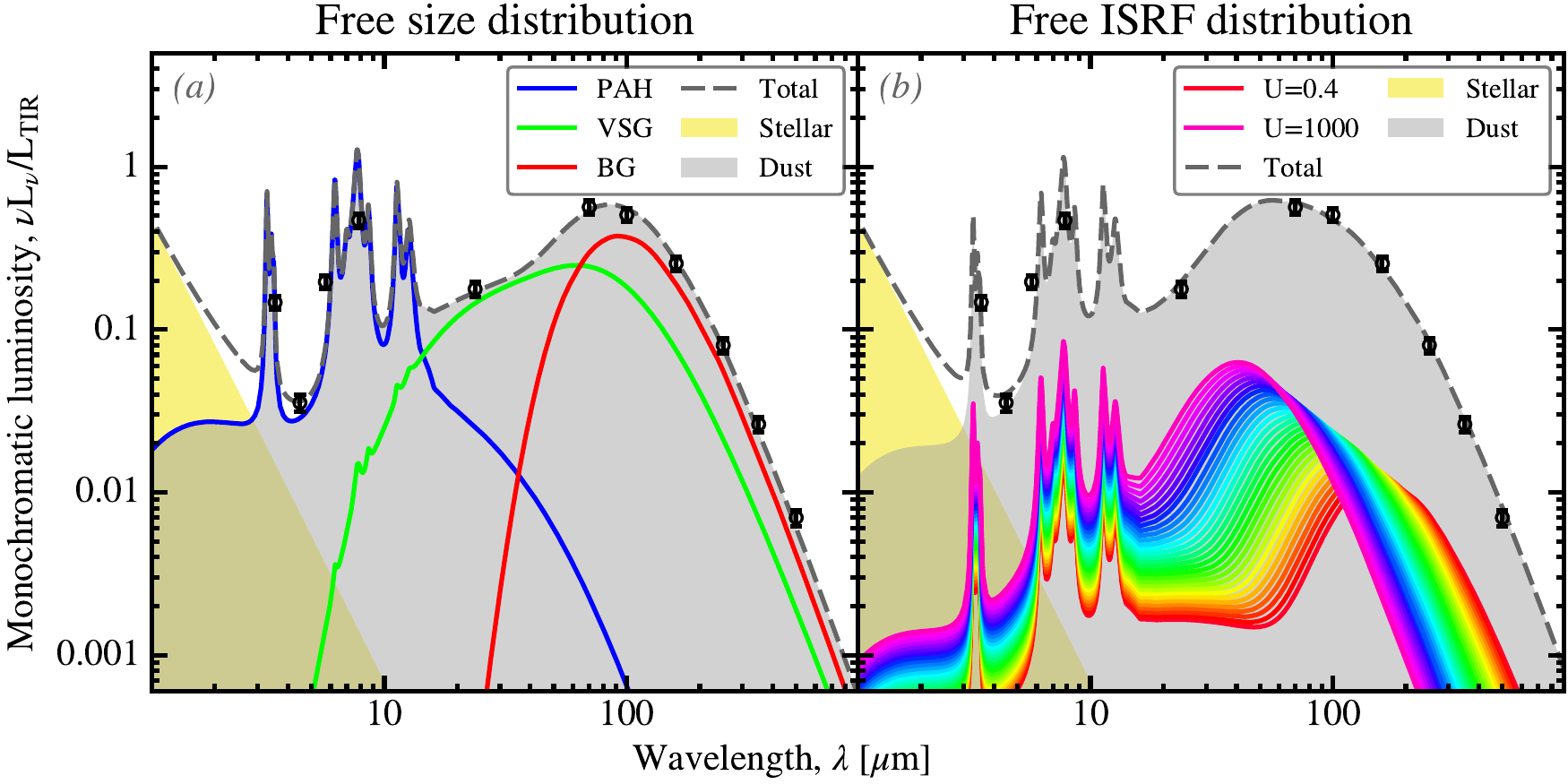}
  \newcap{Degeneracy of the grain size and ISRF distributions}%
         {The two panels show the same synthetic observations (black error bars)
          fitted by two different models \citep{galliano18}.
          Panel~\textit{(a)} represents a fit by playing on the proportions of
          grains of different sizes: \hPAH s, \hVSG s and \hBG s.
          Panel~\textit{(b)} represents a fit using \refeq{eq:dale}.
          In both cases, we have used the \citetalias{jones17} dust mixture.
          \CClicence}
  \label{fig:degeneracy_sizedist}
\end{figure}

\paragraph{Limitations of the composite approach.}
Using \refeq{eq:dale} to model typical broadband \hSED s of Galactic regions and nearby galaxies is an efficient method.
However, as any approximation, it has some limitations.
\begin{description}
  \item[Variation of the grain constitution] within the observed region is 
    expected, as dust evolves with \hISRF\ intensity and \hISM\ density
    (\cf\ \refchap{chap:dustevol}, which is devoted to dust evolution).
    The composite approach accounts for the fact that there are variations of 
    the physical conditions within the region.
    However, assuming that the dust constitution is homogeneous biases the 
    derived properties.
    The most problematic aspect is certainly the variation of the overall \hFIR\
    opacity.
    If there is significant mantle accretion in dense regions, we might be
    mixing together regions with different $\kappa$ (\cf\ 
    \reffig{fig:aggregates}).
    This is unavoidable and this has to be pondered when discussing the
    modeling results.
  \item[The degeneracy] between the grain size and \hISRF\ distributions
    prevents constraining the former.
    This is because a \hMIR\ excess due to an enhancement of the small grain 
    emission looks similar to the presence of hot equilibrium grains (such as 
    compact \hii\ regions)
    \reffig{fig:degeneracy_sizedist} demonstrates this degeneracy.
    It shows the fit of the same synthetic observations, either varying the 
    fraction of small grains, or the mixing of \hISRF s.
    Additional constraints, on the geometry of the region and its radiation 
    field, are necessary to attempt breaking this degeneracy (\cf\ 
    \refsec{sec:sizedist}).
\end{description}

\begin{figure}[htbp]
  \begin{tabular}{cc}
    \includegraphics[width=0.48\textwidth]{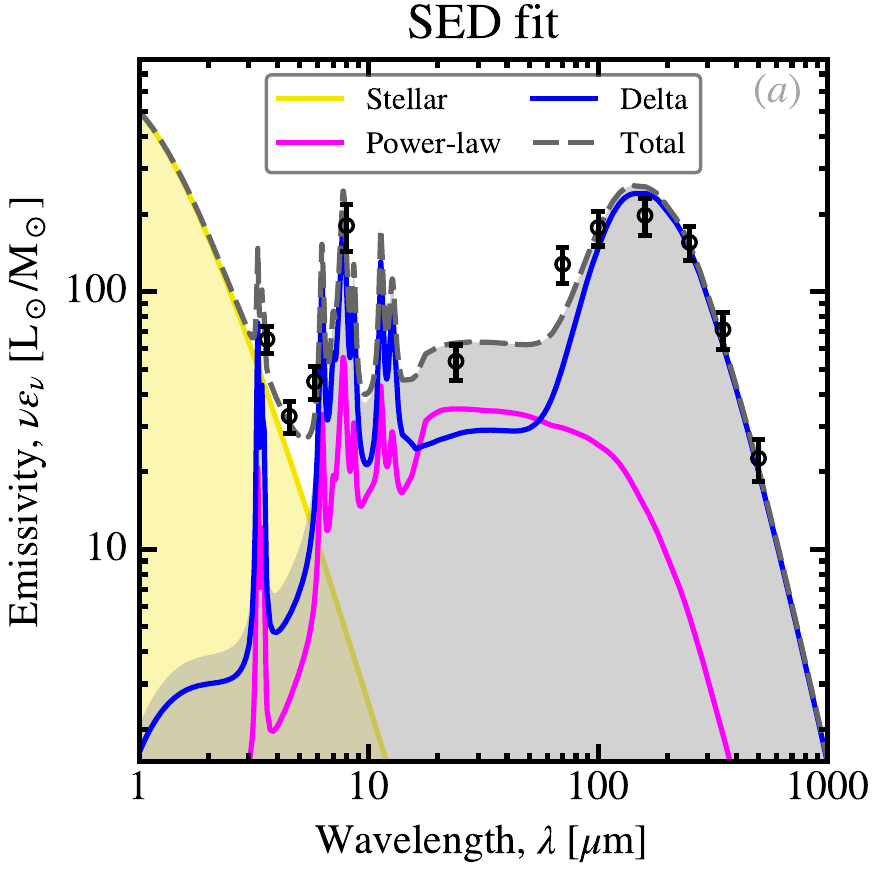} &
    \includegraphics[width=0.48\textwidth]{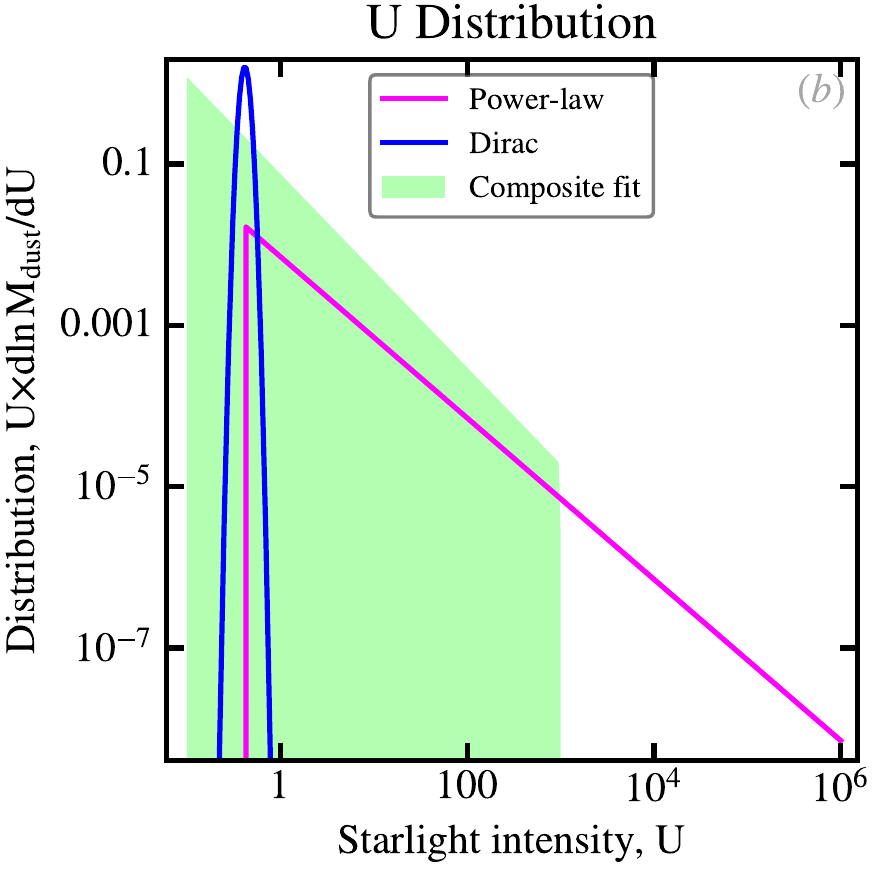} \\
  \end{tabular}
  \newcap{SED fit with the \citet{draine07} starlight intensity distribution}%
         {The black error bars in panel~\textit{(a)} are synthetic observations.
          They are fitted using \refeq{eq:DL07}:
          \begin{inlinelist}
            \item the power-law component in magenta;
            \item the Dirac component in blue; and
            \item an additional stellar continuum in yellow.
          \end{inlinelist}
          The corresponding starlight intensity distribution is shown in 
          panel~\textit{(b)}, with the same color code.
          We have overlaid, in green, a composite model fit to the same 
          synthetic observations.
          \CClicence}
  \label{fig:DL07}
\end{figure}

\begin{figure}[htbp]
  \includegraphics[width=\textwidth]{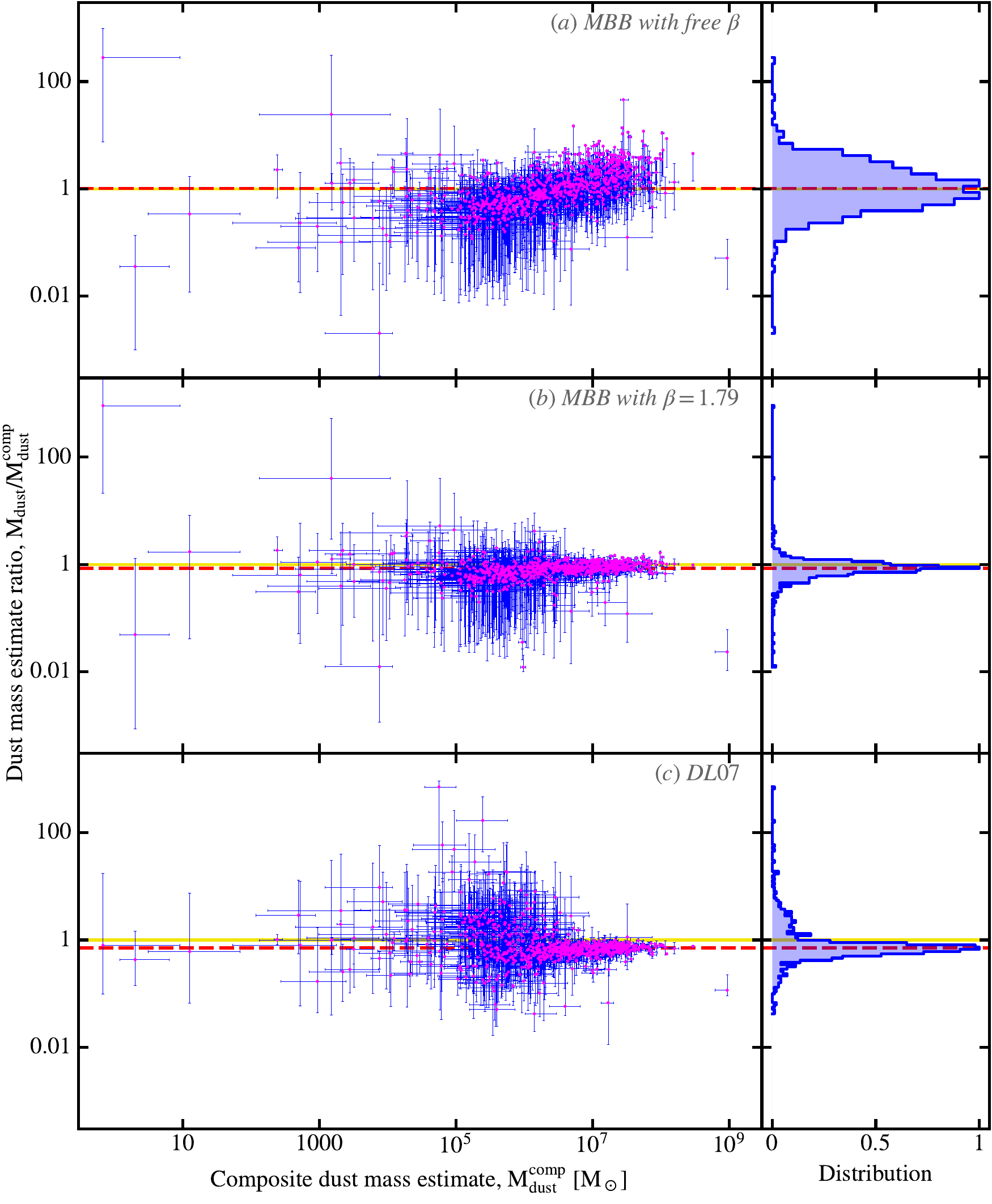}
  \newcap{Comparison of different SED models}%
         {The data are from the SED modeling of the DustPedia and DGS nearby 
          galaxies by \citet[][$\simeq800$ galaxies of all 
          types]{galliano21}.
          The $x$-axis of the left panels shows the dust mass derived by 
          fitting the \hSED\ of each galaxy using the composite approach 
          \refeqp{eq:dale}.
          The $y$-axis shows the ratio of the $x$-axis and the dust mass 
          estimate using one of the three alternate models:
          \begin{inlinelist}
            \item \hMBB; 
            \item \hMBB\ with fixed $\beta=1.79$;
            \item \citet{draine07} distribution.
          \end{inlinelist}
          Each point with an error bar corresponds to one particular galaxy.
          The yellow line shows the 1:1 ratio, and the dashed red line, the 
          median of the sample.
          The right panels show the histogram of the distribution.
          \CClicence}
  \label{fig:compare_fit}
\end{figure}

\paragraph{An alternative distribution.}
The starlight distribution of \refeq{eq:dale} is not the only possible one.
\citet{draine07} proposed the following:
\begin{equation}
 \left(\frac{1}{M_\sms{dust}}\frac{\dd M_\sms{dust}}{\dd U}\right)_\sms{DL07}
  = \gamma\times\left( \frac{1}{M_\sms{dust}}
    \frac{\dd M_\sms{dust}}{\dd U}\right)_\sms{\refeq{eq:dale}}
    + (1-\gamma)\times\delta(U-U_-),
  \label{eq:DL07}
\end{equation}
which simply is the linear combination of the power-law distribution of \refeq{eq:dale}, with a Dirac distribution centered at $U=U_-$, fixing $\alpha=2$ and $U_-+\Delta U=10^6$.
The power-law distribution is supposed to account for star-forming regions, with a large gradient of starlight intensity, and the Dirac represents the diffuse \hISM, uniformly illuminated.
The parameter $\gamma$ controls the relative weight of these two components.
This distribution was elaborated when \hspitz\ data were being analyzed.
There was no coverage beyond $\lambda=160\emic$.
There was thus no constraint on the Rayleigh-Jeans regime of the \hSED.
The role of the Dirac component was to fit the \hSED\ peak, avoiding the dust mass to explode by lack of constraint on the cold dust distribution.
This parametrization however became problematic when \hhersc\ data arrived.
The \hFIR-submm slope of the observed \hSED\ can indeed not be fitted as well with this model \refeqp{eq:DL07} as with the composite approach \refeqp{eq:dale}. 
This is because the model's slope is the intrinsic slope of the grain mixture.
This is demonstrated in \refsubfig{fig:DL07}{a}.
In \refsubfig{fig:DL07}{b}, we see that the starlight intensity distribution fit can not go down as low as the composite model.
The Dirac component fits the \hFIR\ peak with a compromise $U_-$.
This is reminiscent of the discussion we had about \hMBB\ fits, in \refsec{sec:MBB}.
This was demonstrated by \citetalias{galliano21}, who compared different approaches by fitting the \hSED s of about 800 galaxies of the \hDustPedia\ project \citep{davies17} and \expression{Dwarf Galaxy Sample} \citep[\hDGS;][]{madden13}.
\reffig{fig:compare_fit} compares the composite approach, as a reference, to the following models \citepalias[see the complete discussion in][]{galliano21}.
\begin{description}
  \item[A MBB with $\beta$ free] infers dust masses that are, on average over 
    the whole sample, consistent with the composite approach 
    (\refsubfig{fig:compare_fit}{a}).
    However, we see there is a bias as a function of the mass of the galaxy.
    The \hMBB\ approach tends to overestimate the mass of late-type galaxies 
    (mostly the high-mass objects), and underestimates the mass of early-type
    galaxies.
    This is because the \hSED\ of early-type galaxies is fitted with a 
    smaller $\beta$ (thus a lower temperature, and a lower dust mass).
  \item[A MBB with $\beta$ fixed] tends to underestimate the dust mass on 
    average.
    In \refsubfig{fig:compare_fit}{b}, the median of the ratio is $\simeq0.85$.
    This is the effect we have discussed in \refsec{sec:MBB}: the mixing of 
    physical conditions is fitted by a compromise temperature, underestimating 
    the coldest dust.
  \item[The \citet{draine07} distribution] \refeqp{eq:DL07} results in a 
    similar bias as the \hMBB.
    For the distribution in \refsubfig{fig:compare_fit}{c}, the median of the 
    ratio is $\simeq0.71$, even lower than in the case of the \hMBB.
    The reason is the same: the uniformly illuminated component dominating the
    \hFIR-submm \hSED\ can not account for the distribution of cold dust.
\end{description}

    \subsubsection{Panchromatic Empirical SED Models}
    \label{sec:panchromatic}

Several codes in the literature model the \expression{panchromatic}\footnote{Literally for \expression{all wavelengths}, from the X-rays to the radio.} \hSED\ of galaxies, with a simplified treatment of the radiative transfer \citep[\eg][]{silva98,charlot00,galliano08a,da-cunha08,boquien19,fioc19}.
They include the emission by stellar populations, in addition to dust \hSED s.

\begin{figure}[htbp]
  \includegraphics[width=\textwidth]{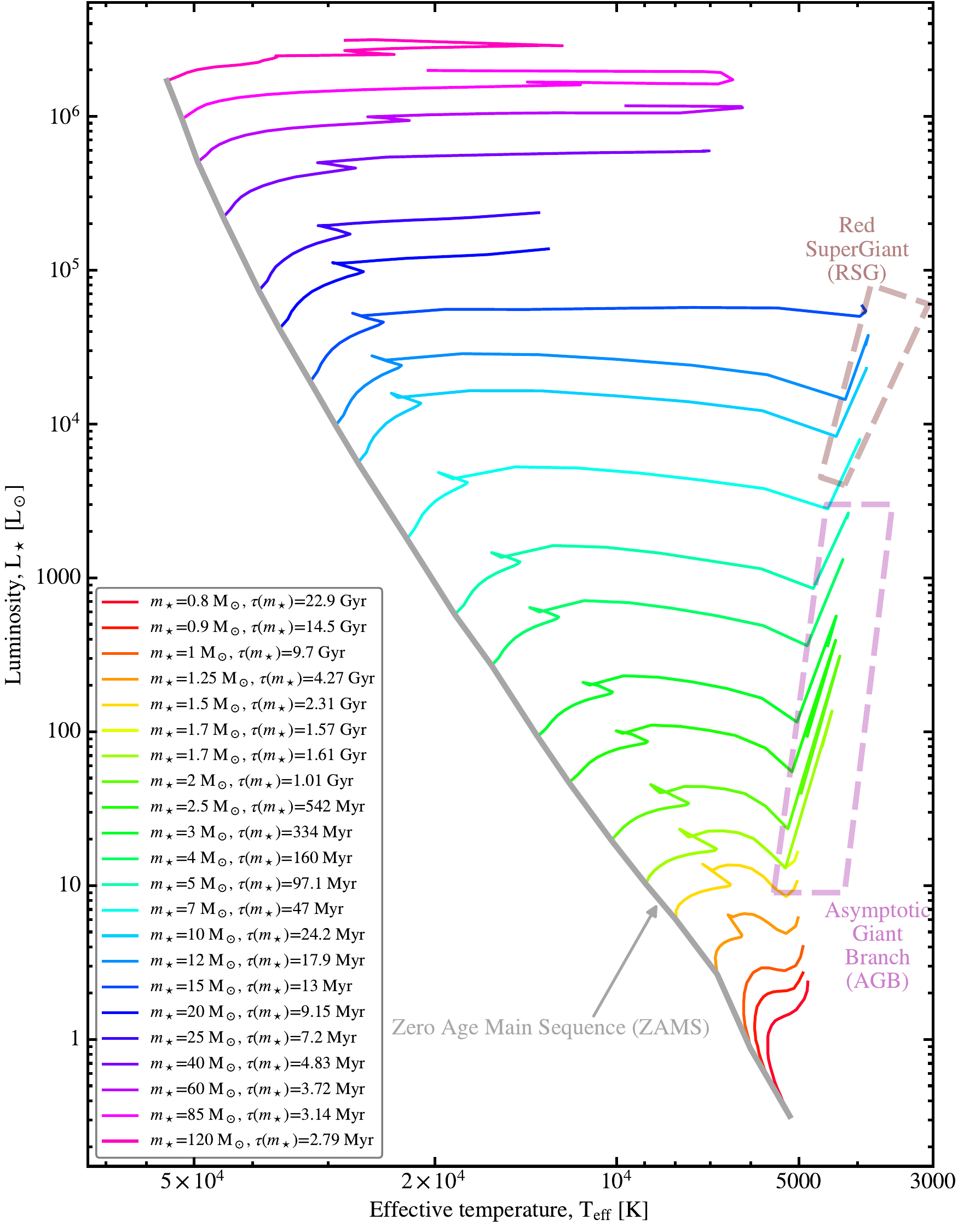}
  \newcap{Stellar isochrones}%
         {This figure shows theoretical tracks of individual star evolution
          from the model of \citet{schaller92}, at initial metallicity, 
          $Z=0.008$.
          This representation is known as a \expression{Hertzsprung-Russell 
          diagram}.
          Stars of a given mass, $m_\star$ (a given color), start from the 
          \expression{Zero Age Main Sequence} (\hZAMS; grey) and evolve along 
          their \expression{Main Sequence} (\hMS) track until their death, 
          after a time $\tau(m_\star)$.
          We have highlighted the late evolutionary stages: 
          \begin{inlinelist}
            \item red supergiants for massive stars; and 
            \item \hAGB\ for \expression{Low- and Intermedate-Mass Stars} 
              (\hLIMS).
          \end{inlinelist}
          \CClicence}
  \label{fig:stellar_isochrones}
\end{figure}

\begin{figure}[htbp]
  \includegraphics[width=\textwidth]{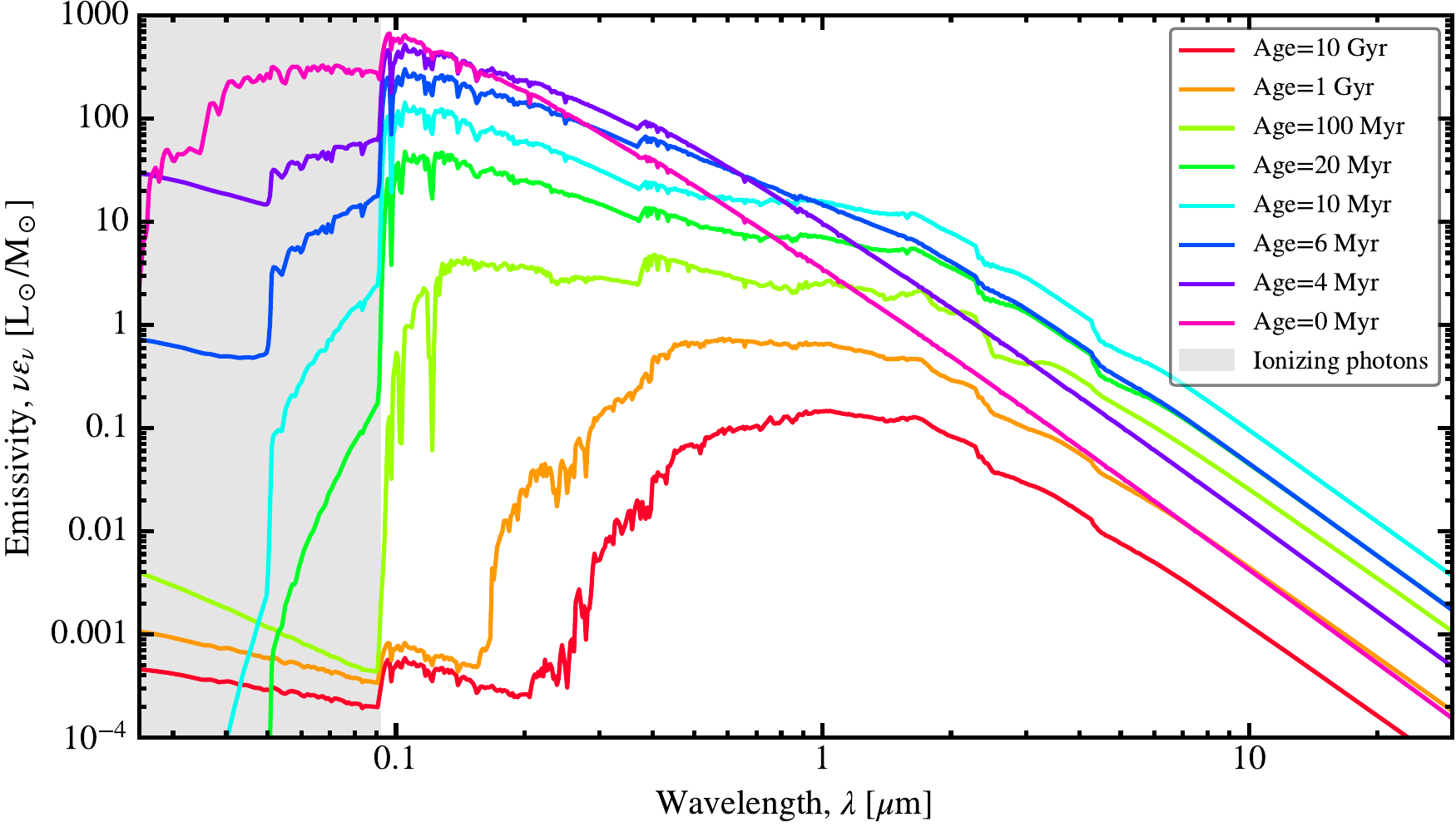}
  \newcap{Evolution of stellar SEDs as a function of time}%
         {The different curves show the evolution, as a function of time, of the
          \hSED\ of a stellar population produced by an instantaneous burst,
          at Solar metallicity.
          We have assumed the \citet{salpeter55} \hIMF.
          These models were generated using the code of \citet{fioc97}.
          Notice, in particular, the fast decrease of ionizing photons with
          time.
          \CClicence}
  \label{fig:stellar_SED}
\end{figure}

\paragraph{Stellar SEDs.}
Stars of different masses have different spectra and lifetimes.
This is shown in \reffig{fig:stellar_isochrones}, in the form of a \expression{Hertzsprung-Russell diagram}.
Massive stars have:
\begin{inlinelist}
  \item the highest effective temperatures, their \hSED s peaking in the \hUV;
  \item the highest luminosities; and
  \item the shortest lifetimes, of only a few million years.
\end{inlinelist}
This is the opposite for low-mass stars: their \hSED\ peaks in the \hNIR, and they live longer than several hundred million years.
These characteristics have a profound impact on the variation of stellar \hSED s with time.
The intrinsic emission of a stellar population can be simulated using \expression{evolutionary synthesis} \citep[\eg][]{fioc97}.
This approach follows, at each time step, the formation and evolution of stars with different masses, $m_\star$.
When these stars are born, all populations contribute to the \hSED.
It is shown as the magenta curve in \reffig{fig:stellar_SED}.
This initial \hSED\ is dominated by massive stars and peaks in the \hUV.
Then, as the most massive stars, which also have the shortest lifetime, die, their contribution to the \hSED\ is suppressed.
Consequently, when these stars disappear, the \hUV-side of the \hSED\ decreases.
After several hundred million years, the \hSED\ peaks in the \hNIR, as it originates only in low-mass stars (orange and red curves in \reffig{fig:stellar_SED}).
There is also a drastic evolution of the emissivity as a function of time, as
low-mass stars are significantly less luminous (\cf\ \reffig{fig:stellar_isochrones}).
To compute such synthetic spectra, the following quantities need to be defined.
\begin{itemize}
  \item The \expression{Star Formation History} (\hSFH) expresses
    the variation as a function of time of the \expression{Star Formation 
    Rate} (\hSFR), $\psi(t)=\dd m_\star/\dd t$.
    Several forms are used in the literature: instantaneous burst, 
    exponential, delayed, \etc\ \citep[\eg][]{lee10}.
  \item The \expression{Initial Mass Function} (\hIMF) gives the 
    \hPDF\ to form a star of mass $m_\star$.
    The historical \hIMF\ of \citet{salpeter55} is less commonly used nowadays.
    For instance, an alternate \hIMF\ has been proposed by \citet{chabrier03}.
\end{itemize}
We will more extensively discuss these quantities in \refsec{sec:cosmicdustevol}, when modeling the chemical evolution of galaxies.
We have summarized in \reftab{tab:stars} the main properties of the different stellar classes.
\begin{table}[htbp]
  \centering
  \setlength\arrayrulewidth{2pt}
  \arrayrulecolor{white}
  \begin{tabularx}{\linewidth}{|>{\columncolor{coltabhead}}X%
                                |>{\columncolor{coltabcell}}r%
                                |>{\columncolor{coltabcell}}r%
                                |>{\columncolor{coltabcell}}r%
                                |>{\columncolor{coltabcell}}r|}
    \hline
      \rowcolor{coltabhead}
      \cellcolor{white} 
      & \textbf{Effective temperature, T$\bm{_\sms{eff}}$}
      & \textbf{Initial mass, $\bm{m_\star}$} 
      & \textbf{Initial luminosity, L$\bm{_\star}$} 
      & \textbf{Lifetime, $\bm{\tau(m_\star)}$} \\
    \hline
      O & $\ge30\,000$ K & $\ge16\eMsun$ & $\ge30\,000\eLsun$ & $\le12$~Myr \\
      B & $10\,000-30\,000$ K & $2.1-16\eMsun$ & $25-30\,000\eLsun$ 
        & 12~Myr$-$1~Gyr \\
      A & $7\,500-10\,000$ K & $1.4-2.1\eMsun$ & $5-25\eLsun$ & $1-3$~Gyr \\
      F & $6\,000-7\,500$ K & $1.04-1.4\eMsun$ & $1.5-5\eLsun$ & $3-10$~Gyr \\
      G & $5\,200-6\,000$ K & $0.8-1.04\eMsun$ & $0.6-1.5\eLsun$ & $10-25$~Gyr\\
      K & $3\,700-5\,200$ K & $0.45-0.8\eMsun$ & $0.08-0.6\eLsun$ & \ldots \\
      M & $2\,400-3\,700$ K & $0.08-0.45\eMsun$ & $\le0.08\eLsun$ & \ldots \\
    \hline
  \end{tabularx}
  \newcap{Basic properties of the main stellar classes}{}
  \label{tab:stars}
\end{table}

\begin{figure}[htbp]
  \includegraphics[width=\textwidth]{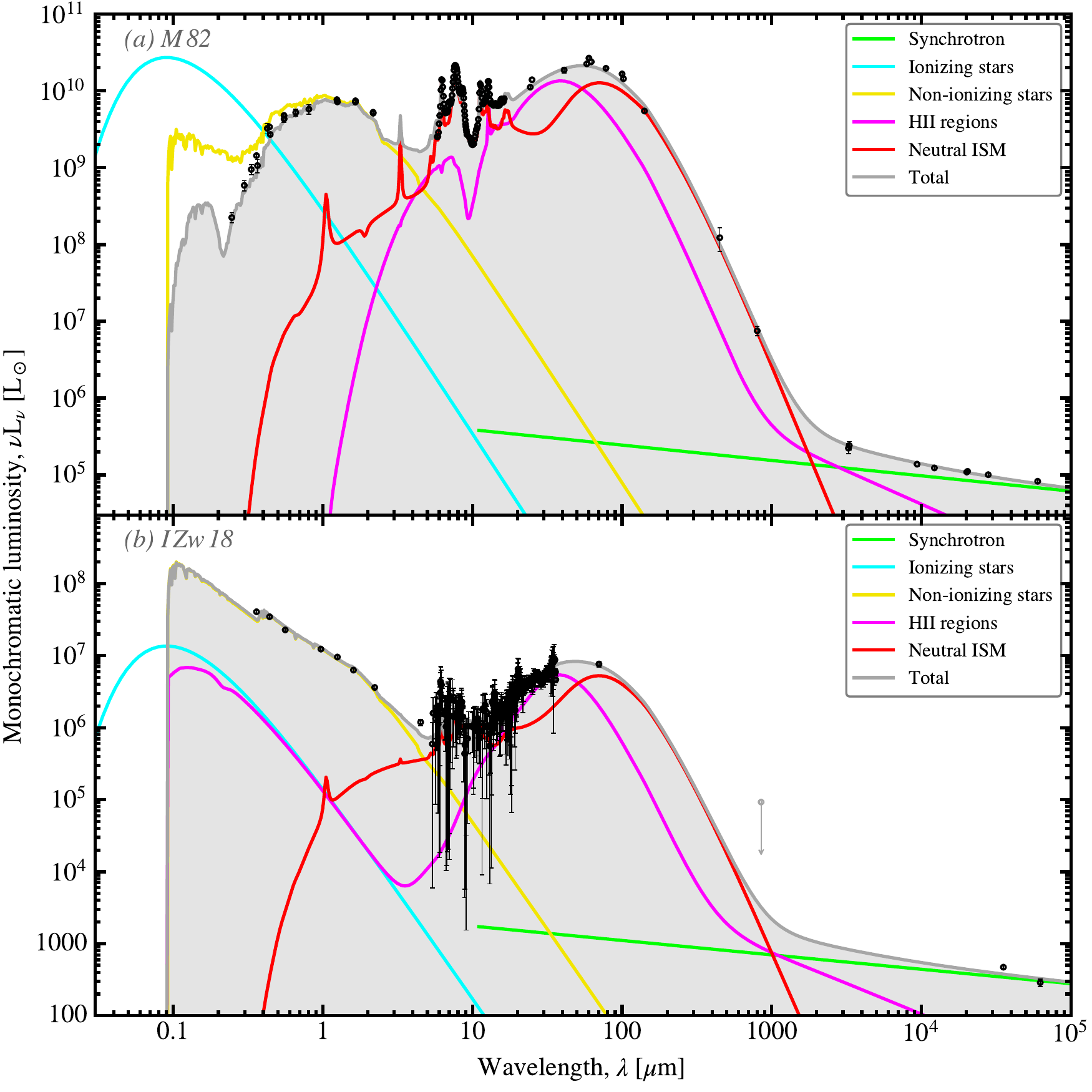}
  \newcap{Multiphase SEDs of galaxies}%
         {This figure shows the \hSED\ of two galaxies, modeled by 
          \citet{galliano08a}: the prototypical starburst, \M{82}, and the 
          lowest metallicity galaxy, \izw.
          The black error bars represent the observed fluxes.
          The total \hSED\ model is shown in grey.
          The stars are split in two populations:
          \begin{inlinelist}
            \item young, ionizing stars that are in \hii\ regions; and
            \item older, non-ionizing stars that are in the rest of the galaxy.
          \end{inlinelist}
          We show the intrinsic emission of both stellar components.
          The dust emission originates in:
          \begin{inlinelist}
            \item \hii\ regions, heated by ionizing stars; and
            \item the neutral \hISM, heated by non-ionizing stars and by the
              escaping radiation from \hii\ regions.
          \end{inlinelist}
          \CClicence}
  \label{fig:sed_G08}
\end{figure}

\paragraph{Putting everything together.}
Empirical panchromatic \hSED\ models usually include the following physical ingredients.
\begin{description}
  \item[One or several stellar populations] are modeled.
    Their escaping radiation is fitted to the \hUV-to-\hNIR\ \hSED, allowing 
    the user to constrain: the age, the \hSFH\ timescale, the \hSFR\ and the 
    total stellar mass.
    The \expression{attenuation}\footnote{Attenuation is not equivalent to 
    extinction. 
    The extinction is the sum of scattering and absorption along the sightline 
    toward a point source, whereas the attenuation is a global account of the 
    reddening of an ensemble of stars, potentially mixed with the dust.
    The extinction is directly linked to dust properties and is independent of 
    geometry. 
    The attenuation is a synthetic quantity depending both on the dust 
    properties and on the \hISM\ topology \citep[\eg][for a recent 
    discussion]{buat19}.} 
    is accounted for, by assuming simple topologies, such as those discussed in 
    \citet{calzetti94}.
    The intrinsic stellar power, $L_\star$, is derived from the 
    \expression{energy balance}: $L_\star=L_\sms{UV-NIR}+L_\sms{TIR}$, where 
    $L_\sms{UV-NIR}$ is the escaping \hUV-to-\hNIR\ power.
    This equation simply states that the intrinsic power emitted by stars is 
    either escaping the galaxy in the \hUV-to-\hNIR\ range, or has been 
    absorbed by dust.
    It implicitly assumes that the emission from the galaxy is isotropic, which
    is not necessarily the case for disk galaxies.
  \item[Dust emission] is modeled, either by solving the radiative transfer 
    equation \refeqp{eq:RTl} in a simple geometry, or by adopting a distribution
    of starlight intensities (\cf\ \refsec{sec:dale}).
    The dust properties are not always consistent with those used to account
    for attenuation.
  \item[Nebular emission,] in the form of lines or continuum, can be included.
    In case of star-forming galaxies, it is dominated by \hii\ regions
    \citep[\eg][]{charlot01}.
    Some models also include the gaseous emission from \hAGN s.
    These tracers are used to refine some of the diagnostics.
\end{description}
We illustrate this approach with the model of \citet{galliano08a}, applied to two galaxies, in \reffig{fig:sed_G08}.
This particular model separates the emission from \hii\ regions, which are powered by massive ionizing stars (cyan).
The magenta curve shows the escaping radiation from \hii\ regions.
It includes the dust emission and the free-free continuum.
The escaping radiation from \hii\ regions, as well as non-ionizing field stars (yellow), heat the neutral \hISM\ (red).
The degeneracy between both dust components was solved by using radio observations to constrain the level of the free-free emission.
The gas density in \hii\ regions, impacting the equilibrium temperature of large grains, is constrained by the \hMIR\ continuum.
The remaining emission is thus assumed to come from the neutral \hISM.
We will discuss in more detail the results of this model, in \refsec{sec:cosmicdustevol}.

\paragraph{Limitations of empirical panchromatic models.}
The approach we have just described has several advantages.
In a single fit, it provides estimates of the \hSFR, the age of the stellar populations, the stellar mass, and the dust properties.
Its major limitation however resides in the sensitivity of its results to the assumed \hISM\ topology.
The \hISM\ indeed has a fractal structure, with several orders of magnitude of contrast density \citep[\eg][]{combes00}.
The optical depth from the model thus probably underestimates the total dust column density (\cf\ \refsec{sec:clumpy}).
In addition, the extinction and emission properties of these models are usually not consistent.
The modeling of the microscopic grain constitution and of their macroscopic spatial distribution can differ from one side of the electromagnetic spectrum to the other.
For instance, assuming a \citet{calzetti94} attenuation law and taking a mean radiation field to account for dust heating is virtually equivalent to decoupling the extinction and emission.
Finally, the consistency brought by modeling all multi-wavelength tracers at once, which is \textit{a priori} positive, leads to propagating the systematic uncertainties, due to arbitrary choices of \hISM\ topologies, to parameters that could have been considered independent of these effects, if they had been modeled separately ($M_\sms{dust}$, $\langle U\rangle$, $q_\sms{PAH}$; \refsec{sec:dale}).

    \subsubsection{The Matryoshka Effect}

The main limitation of dust studies lies in the near impossibility to constrain at the same time the microscopic dust properties and their sub-pixel macroscopic spatial distribution.
All the approaches we have discussed in this chapter face this problem.
It can be illustrated with what \citet{galliano18} called the \expression{matryoshka effect}.
This empirical effect originates in the impact of the spatial resolution of the observations on the constrained parameters.

\begin{figure}[htbp]
  \begin{tabular}{*{10}{c}}
    \includegraphics[width=0.0733\textwidth]{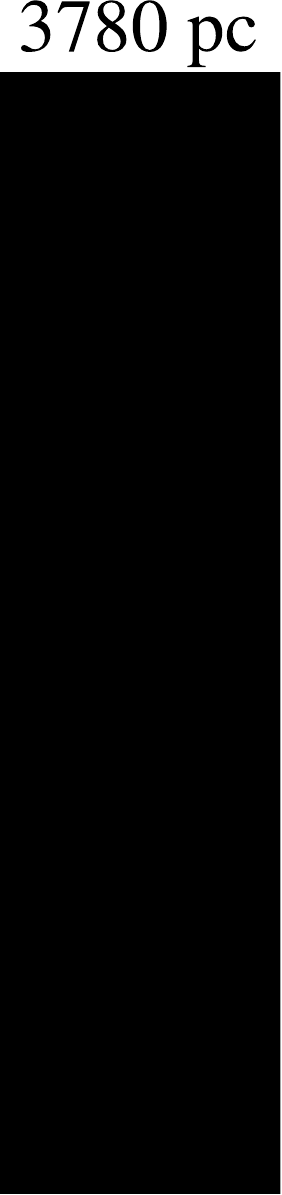} &
    \includegraphics[width=0.0733\textwidth]{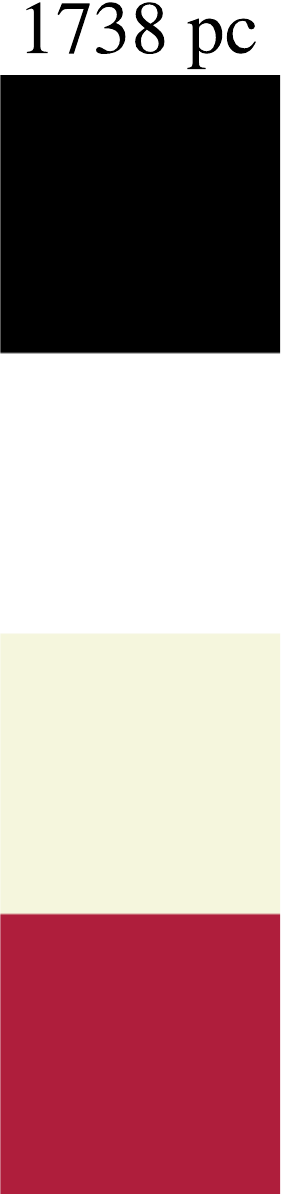} &
    \includegraphics[width=0.0733\textwidth]{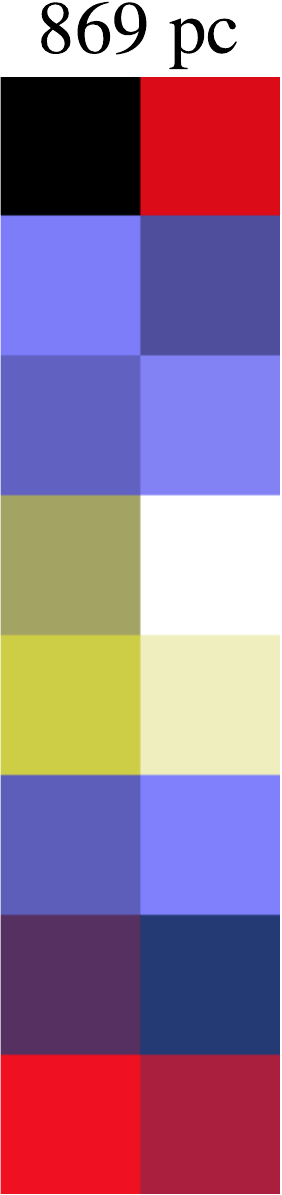} &
    \includegraphics[width=0.0733\textwidth]{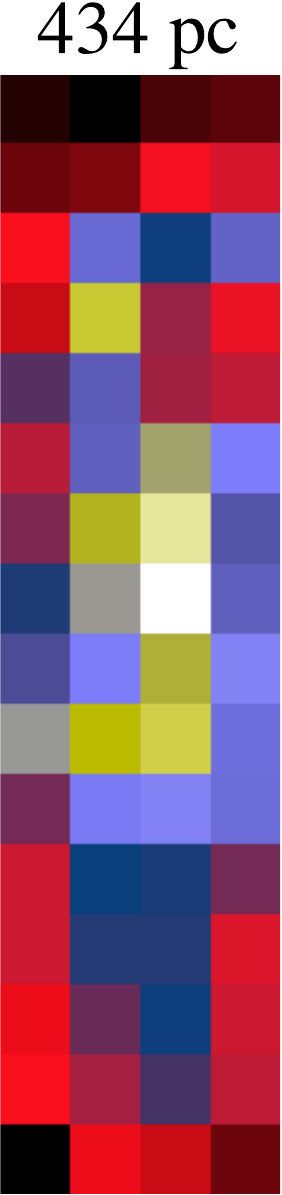} &
    \includegraphics[width=0.0804\textwidth]{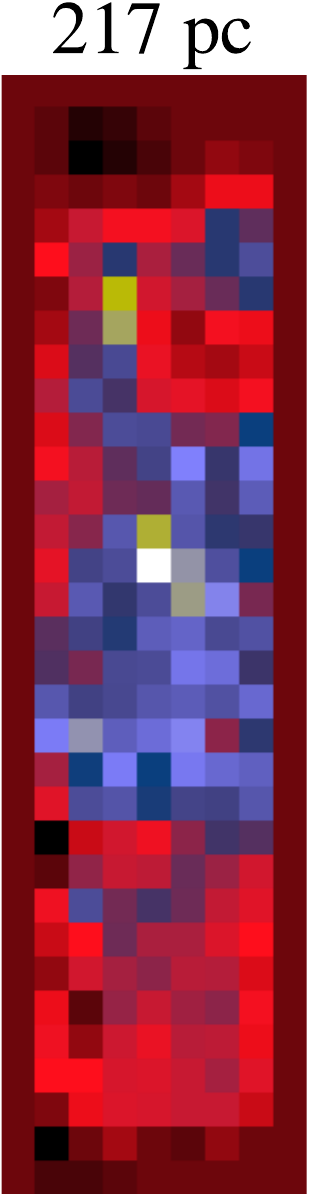} &
    \includegraphics[width=0.0804\textwidth]{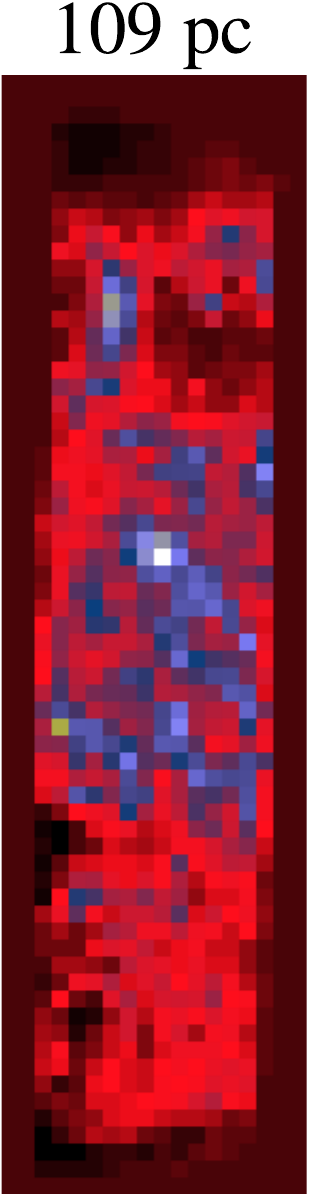} &
    \includegraphics[width=0.0804\textwidth]{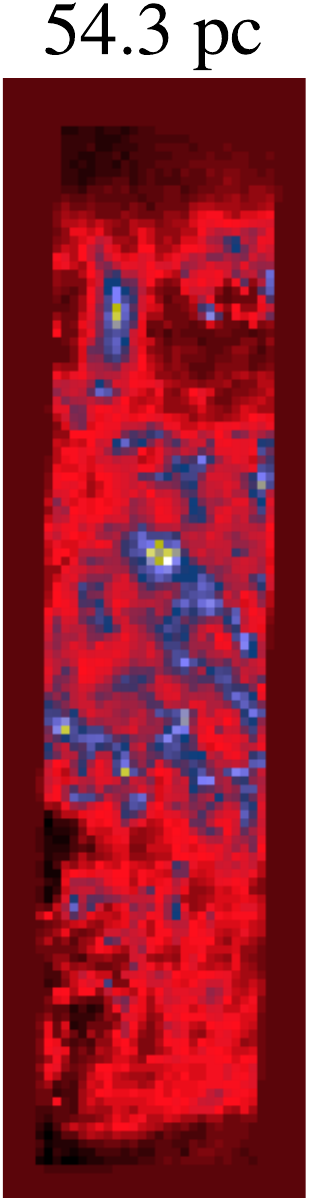} &
    \includegraphics[width=0.0801\textwidth]{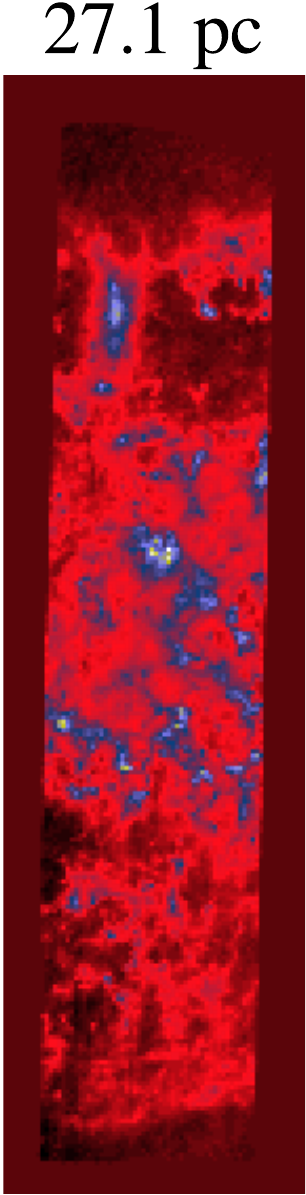} &
    \includegraphics[width=0.0790\textwidth]{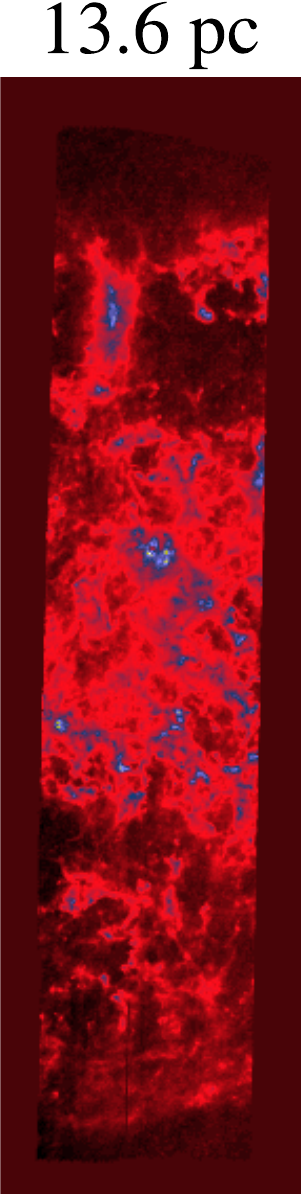} &
    \includegraphics[width=0.0803\textwidth]{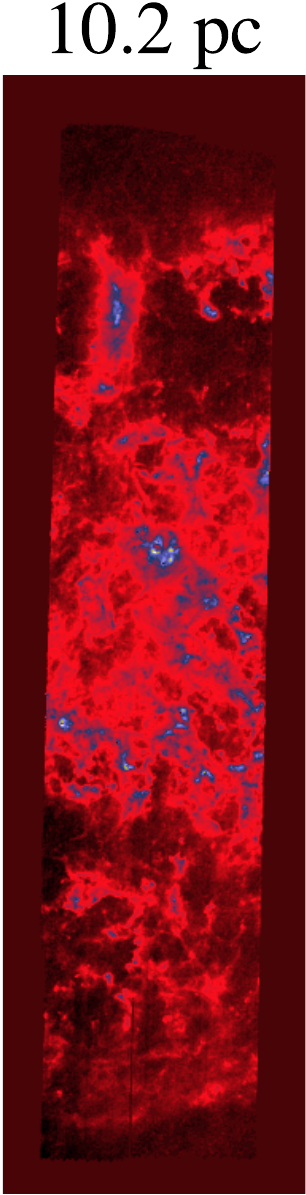} \\
  \end{tabular}
  \includegraphics[width=\textwidth]{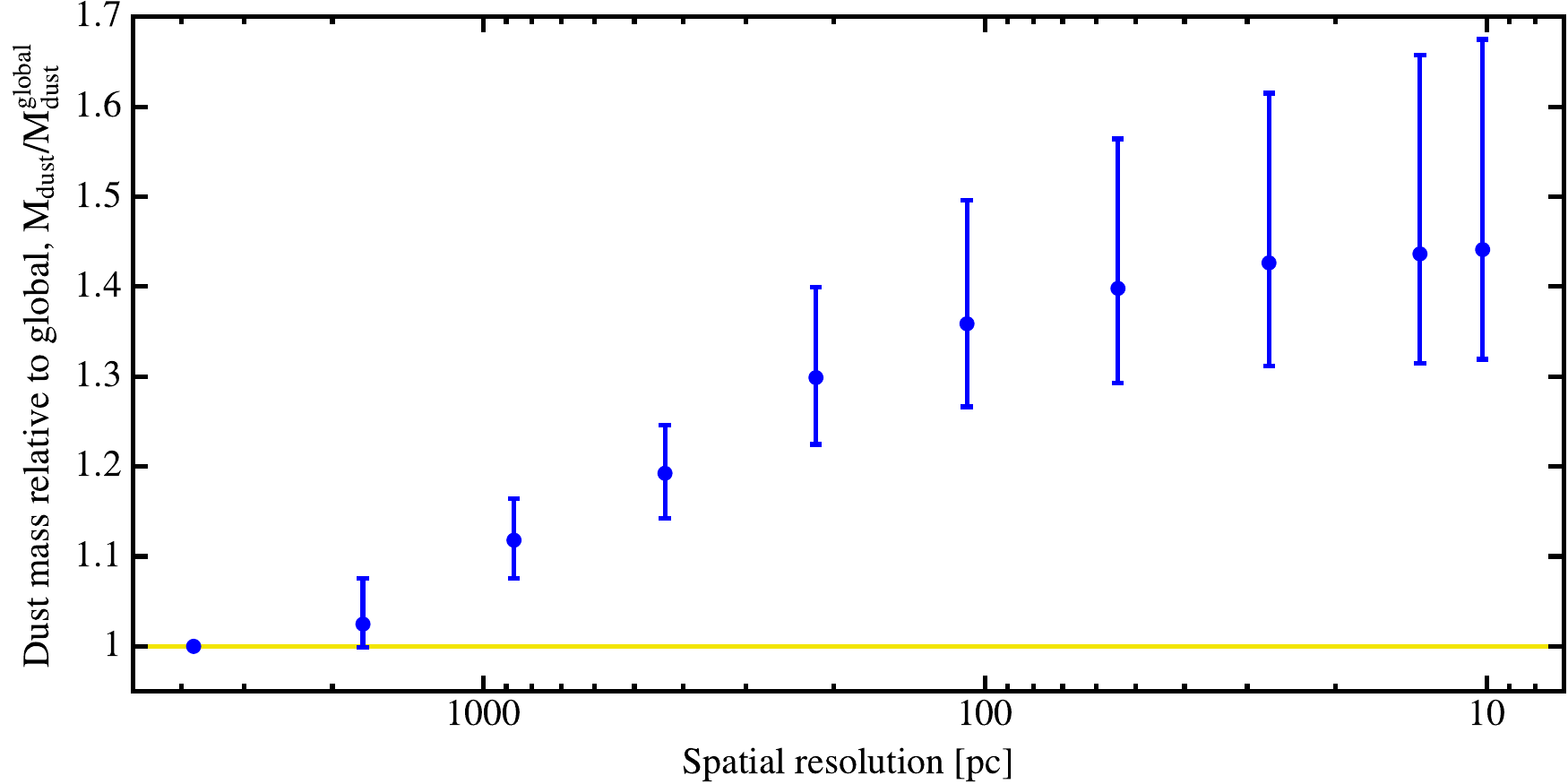}
  \newcap{Matryoshka effect demonstrated on the LMC}%
         {On top are nine \SPIREiii\ images of a strip covering about one 
          fourth of the \hLMC, observed by \citet{meixner10}.
          These images are rebinned at different spatial resolutions (indicated 
          on the top).
          The bottom panel shows the dust mass derived by modeling the \hSED\
          of each pixel of these images and summing them to derive the total 
          mass in the strip \citep{galliano11}.
          It is normalized by the mass obtained when modeling the global
          (\ie\ one pixel) \hSED\ of the strip, $M_\sms{dust}^\sms{global}$.
          \CClicence}
  \label{fig:matryoshka}
\end{figure}

\paragraph{Demonstration on the LMC.}
\reffig{fig:matryoshka} demonstrates the effect with the modeling of the dust mass in a strip covering one fourth of the \hLMC, by \citet{galliano11}.
The different images on top show the spatial resolution that is used.
The first image is one single large pixel.
The second one is divided in four pixels, and so on.
The curve in the bottom panel of \reffig{fig:matryoshka} shows the dust mass in the strip derived by summing all the pixels, at each resolution.
We see that this mass increases with spatial resolution, until reaching a plateau around $\simeq10$~pc.
For this particular region, the discrepancy with the global mass is about $\simeq50\,\%$.
\citet{galliano11} interpreted this effect by noticing that the cold dust, which accounts for most of the mass and, at the same time, is the least luminous, is diluted into the warm dust emission when we sum all the regions together.
With spatial resolution however, we can better separate the bright and cold regions.
It is thus reasonable to assume that the most accurate estimate of the dust mass is the one obtained at the highest resolution.
This assumption is confirmed by the fact that the length-scale at the growth curve plateau ($\simeq10$~pc) corresponds roughly to the mean free-path of a U-band photon at a density $n_\sms{H}\simeq100$~cm$^{-3}$ \citep[\reftab{tab:freepath}; the \hLMC\ has a half-Solar metallicity;][]{pagel03}.
This is the typical density of the \hCNM\ and diffuse \hmol\ phase (\cf\ \reftab{tab:ISMism}).
It is possible that, if we could increase the resolution, we would see another plateau around $n_\sms{H}\simeq10^4$~cm$^{-3}$ ($\simeq0.1$~pc), corresponding to dense \hmol\ clouds.

\paragraph{Generalization.}
This effect has been independently confirmed by \citet{galametz12}, \citet{roman-duval14} and \citet{aniano20}, although it is less important if the maximum spatial resolution is not as high as ours.
\takeaway{The dust mass derived at high spatial resolution is always larger than its global estimate.}

  \subsection{Application to Nearby Galaxies}

We now review the application of \hSED\ models to observations of nearby galaxies, aimed at constraining the grain properties in different environments.
We illustrate the different aspects using our own projects and collaborations.

    \subsubsection{The Different Types of Galaxies}
    \label{sec:galdesc}

There is a wide diversity of galaxy types.
Several systems of classification have been developed, through the years.
In particular, the Hubble-de Vaucouleurs system, although outdated, is still used nowadays.

\begin{figure}[htbp]
  \includegraphics[width=\textwidth]{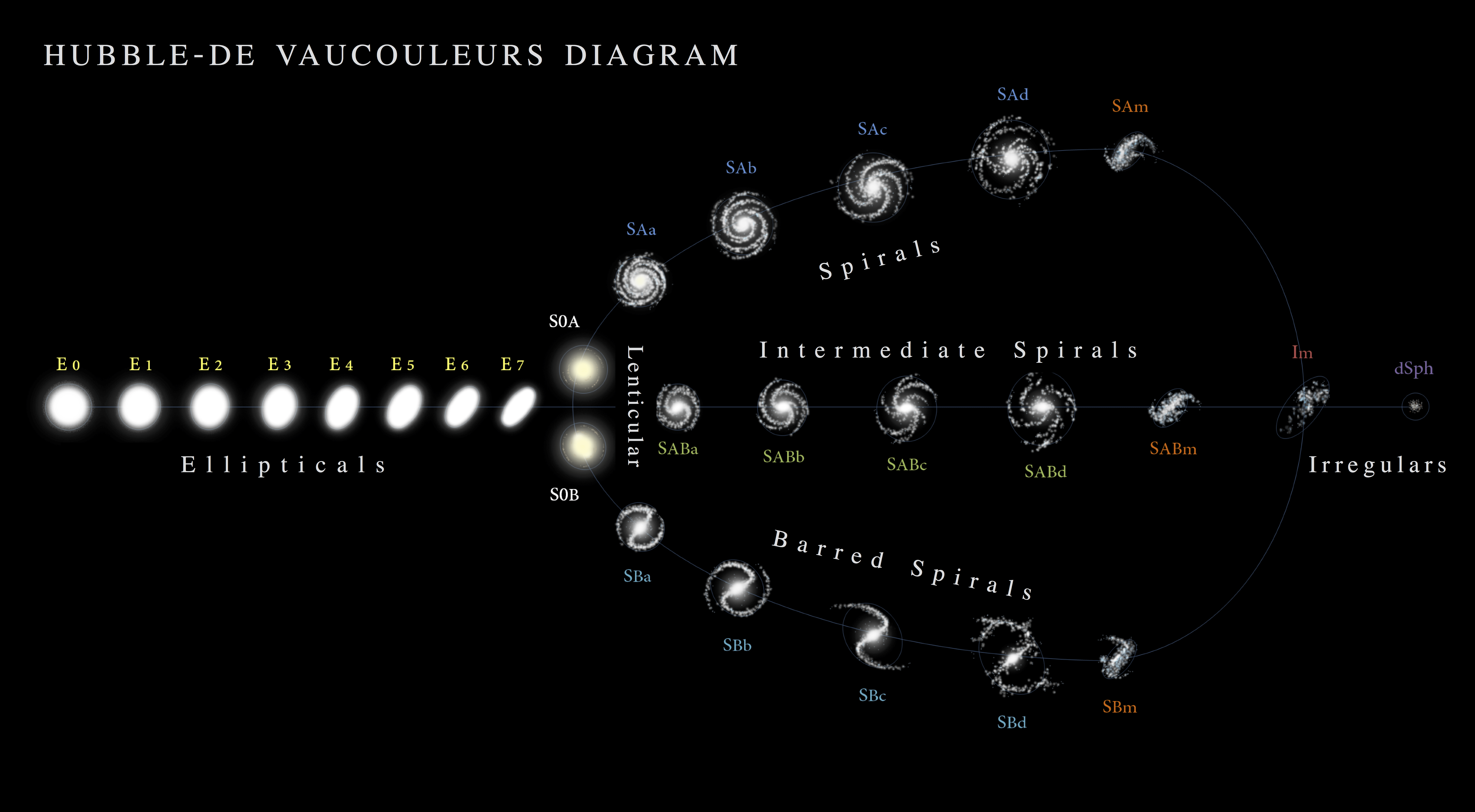}
  \newcap{Hubble-de-Vaucouleurs galaxy morphology diagram}%
         {\uline{Credit:} A.\ Ciccolella / M.\ De Leo, \href{https://commons.wikimedia.org/wiki/File:Hubble_-_de_Vaucouleurs_Galaxy_Morphology_Diagram.png}{Wikipedia}, licensed under
                \href{https://creativecommons.org/licenses/by-sa/3.0/deed.en}%
                     {CC BY-SA 3.0}.}
  \label{fig:hubble}
\end{figure}

\paragraph{The outdated galaxy morphological classification.}
The most famous observational system of morphological classification is the \expression{Hubble tuning fork} or \expression{Hubble-de Vaucouleurs diagram}, represented in \reffig{fig:hubble}.
It was originally developed by \citet{hubble36}, and refined by \citet{de-vaucouleurs59}.
It is based on the morphological characteristics of galaxies in the visible range: presence and thickness of spiral arms, bars, rings, \etc\
There are essentially three classes of galaxies (left, center and right parts of \reffig{fig:hubble}):
\begin{inlinelist}
  \item ellipticals, also called \expression{Early-Type Galaxies} (\hETG);
  \item spirals, also called \expression{Late-Type Galaxies} (\hLTG); and
  \item irregulars and dwarf spheroidals.
\end{inlinelist}
There are sub-categories with abstruse notations (SAa, E2, \etc) that would be a waste of time to describe here.
Overall, this is a mid-XX$^\sms{th}$-century empirical classification, based on stellar dynamics, that does not take into account the \hIR\ information (especially the \hSF\ activity), nor the radio properties (presence of an \hAGN).
Recently, with the \hDustPedia\ collaboration \citep{davies17}, we explored the sensitivity of dust properties to galaxy types \citep[\eg][]{davies19,bianchi18,nersesian19}.
We did not find any clear systematic differences between adjacent sub-categories in \reffig{fig:hubble}, and we found a large scatter within each class.
There are overall trends between the three main categories, because they are linked to the gas fraction, metallicity and stellar populations.
We will discuss those throughout this manuscript.
The terminology \citengl{late/early-type} was introduced to denote the evolution of galaxies along the sequence.
We now know that the sequence is reversed: early-type galaxies are the oldest objects, and late-types, the youngest ones.
In addition, the most numerous galaxies in the local Universe, dwarf galaxies are under-represented in this diagram.
They are part of the irregulars, which is a category by default.
Finally, this Hubble-de Vaucouleurs classification was based on the local Universe, while galaxies at high redshift can exhibit different morphologies, such as clumpy chains and tadpoles \citep[\eg][]{elmegreen15}.
We have developed an alternate, non-parametric classification, taking into account \hIR\ morphology \citep{baes20}.
It emphasizes the clumpy nature of the dust distribution in local galaxies.
\takeaway{Stellar morphology is not particularly relevant to \hISM\ studies.}

\begin{figure}[htbp]
  \begin{tabular}{ccc}
    \textbf{\textit{(a)} Irregular, dwarf} 
    & \textbf{\textit{(b)} Spiral} & \textbf{\textit{(c)} Elliptical} \\
    (\izw) & (\M{33}) & (Centaurus$\,$A) \\
    \includegraphics[width=0.2685\textwidth]{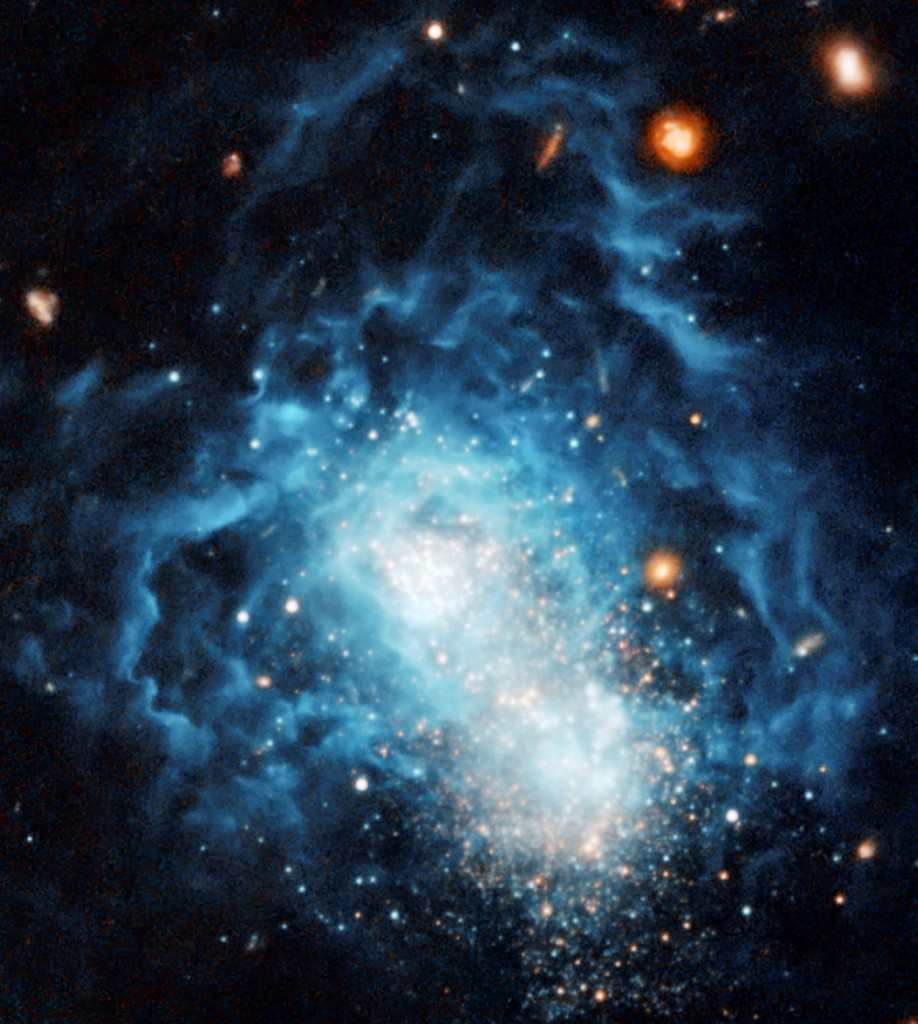} &
    \includegraphics[width=0.3695\textwidth]{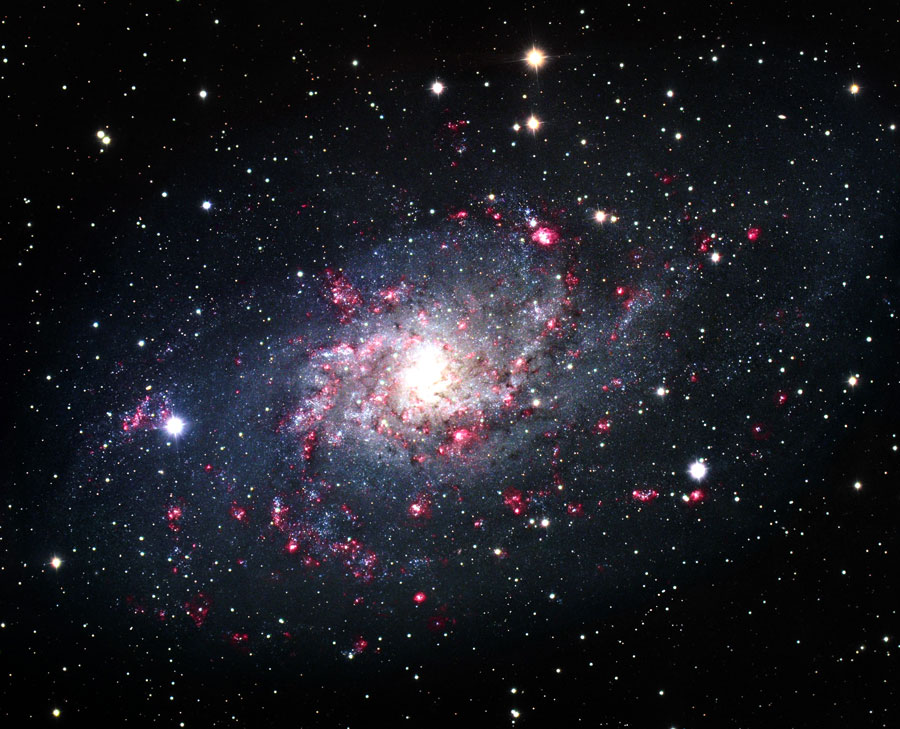} &
    \includegraphics[width=0.307\textwidth]{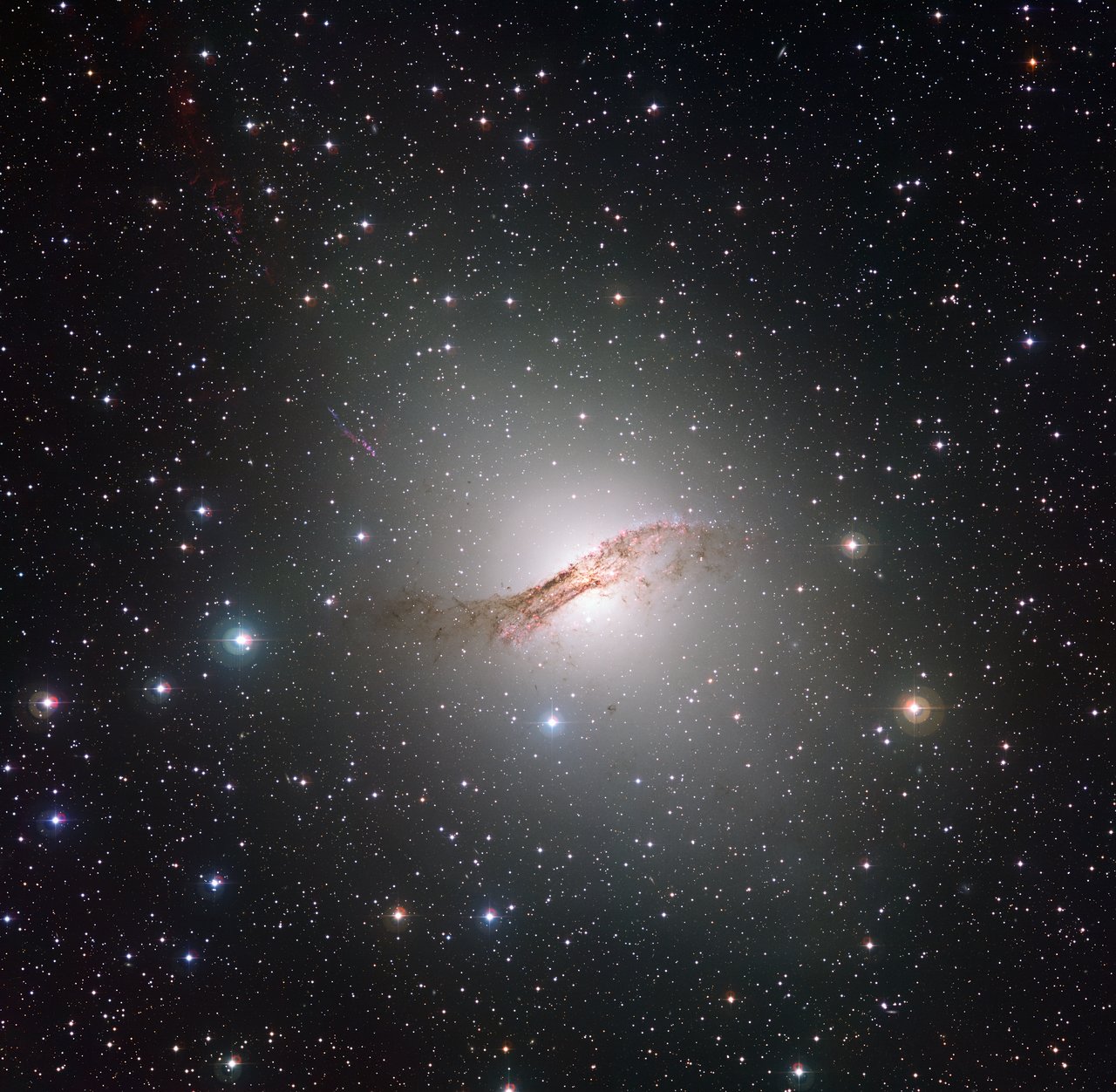} \\
    Low metallicity ($\simeq1/35\eZsun$) 
    & Solar metallicity ($\simeq1/2\eZsun$) 
    & High metallicity ($\simeq3\eZsun$) \\
  \end{tabular}
  \newcap{Visible-range image of three nearby galaxies}%
         {Panel~\textit{(a)} shows an image of the extremely low-metallicity,
          dwarf galaxy, \izw, obtained with the instrument ACS onboard the 
          \hHST\ \citep{aloisi07}.
          Panel~\textit{(b)} shows the spiral galaxy, \M{33}.
          Star-forming regions, traced by the \haline\ line, are shown in red.
          Panel~\textit{(c)} shows the elliptical galaxy, Centaurus~A.
          It has a warped dust lane and an \hAGN.
          \uline{Credit:} 
          \begin{inlinelistalph}
            \item \citet{cignoni09}, licensed under
                  \href{https://creativecommons.org/licenses/by-sa/3.0/deed.en}%
                       {CC BY-SA 3.0};
            \item  
              \href{https://apod.nasa.gov/apod/ap191003.html}%
                   {Image Data: Subaru Telescope (NAOJ), Hubble Space Telescope -- Image 
          Processing: Robert Gendler -- Additional Data: BYU, Robert Gendler, 
          Johannes Schedler, Adam Block -- Copyright: Robert Gendler, Subaru 
          Telescope, NAOJ}, with permission from Robert \familyname{Gendler};
            \item 
              \href{https://www.eso.org/public/images/eso1221a/}{courtesy of ESO}, 
              licensed under 
              \href{https://creativecommons.org/licenses/by/4.0/}{CC BY 4.0}.
          \end{inlinelistalph}}
  \label{fig:galim}
\end{figure}

\paragraph{Galactic properties that matter to ISM studies.}
A few global quantities, such as the metallicity, the \hSFR\ or the gas fraction are more relevant to assess the general properties of the \hISM\ of a galaxy.
The three main categories of the Hubble-de Vaucouleurs diagram can be characterized the following way.
\begin{description}
  \item[Irregular galaxies] (\eg\ \refsubfig{fig:galim}{a}) can contain large 
    amounts of atomic gas that typically extend to twice their Holmberg 
    radius\footnote{The \expression{Holmberg radius} is the radius within which 
    the B-band surface brightness of the galaxy is 26.5 magnitudes per 
    squared arcsecond.} \citep[\eg][]{huchtmeier81}.
    \begin{description}
      \item[At Solar metallicity,] irregulars are rich in dust.
        Their visible and \hMIR\ scale-lengths are very similar 
        \citep{hunter06}. 
        The dust emission in irregular galaxies is clumpy and warm, with hot 
        dust and \hUIB\ emissions mostly observed towards bright \hii\ regions. 
        This suggests that massive stars are a major source of heating in these
        environments \citep[\eg][]{hunter06}.
      \item[At low metallicity,] irregulars are dwarf galaxies.
        \hDustiness\ increases with metallicity 
        \citep[\cf\ \refsec{sec:cosmicdustevol};][]{remy-ruyer14,galliano21}.
        The \hISM\ in these objects is less dusty and thus, more transparent.
        Similarly to irregular galaxies, massive stars are a major source of 
        heating in these objects \citep[\eg][]{walter07}, and they are 
        permeated by \hSN-triggered shock waves \citep[\eg][]{oey96,izotov07}.
        Finally, these galaxies exhibit large \hi\ envelopes.
        The \hIR\ emitting region can correspond to only $\simeq20-30\,\%$ of 
        the total mass of the system \citep[\eg][]{walter07}.
        A particularly important type of dwarf galaxies are \expression{Blue 
        Compact Dwarf Galaxies} (\hBCD).
        These galaxies are actively star forming.
        As their name indicates, they have blue optical colors, because they 
        contain several \hSSC s, rich in young massive stars, and they are 
        weakly attenuated by dust.
    \end{description}
  \item[Late-type galaxies] (\eg\ \refsubfig{fig:galim}{b}) have a roughly 
    Solar metallicity and an \hISM\ accounting for $\simeq10-30\,\%$ of their 
    baryonic mass.
    \begin{description}
      \item[Scale-lengths] of disk galaxies are of the order of a few kpc (up to
        $\simeq10$~kpc).
        Their diameter tends to be systematically smaller in the \hMIR\ than in 
        the visible \citep{malhotra96}, whereas
        the arm/inter-arm contrast is larger in the \hMIR\ than in 
        the visible \citep{vogler05}. 
        This is also seen in \COio, \haline\ or radio continuum
        \citep{sauvage96,walsh02,vogler05}.
        This is because these different tracers are indicators of star 
        formation activity, which is enhanced along the spiral arms.
        In the \hIR, the disk scale-length increases with wavelength 
        \citep{hippelein03}, and is larger in the \hFIR\ than in the visible
        \citep{tuffs96,alton98,haas98,davies99,trewhella00,fritz12,casasola17}.
        This \hFIR\ colour gradient observed in the disk suggests that part of 
        the \hFIR\ emission arises from grains heated by the radially 
        decreasing diffuse \hISRF, the extended parts of the disk being cold
        \citep{block94,stevens05,hinz12}.
        \hFIR\ scale-lengths do not vary strongly as a function of galaxy type 
        and are on average $\simeq10\,\%$ larger than the stellar scale-lengths 
        \citep[\eg][]{munoz-mateos09,hunt15}.
      \item[Scale heights] of disk galaxies are typically of the order of a few 
        hundred parsecs.
        Outside our Milky Way, edge-on galaxies are ideal objects to constrain 
        this parameter \citep[\eg][]{xilouris99}. 
        Radiative transfer codes are robust tools to derive such structural 
        parameters (\cf~\refsec{sec:MCRTgal}).
    \end{description}
  \item[Early-type galaxies] (\eg\ \refsubfig{fig:galim}{c}) possess very 
    little dust: the average dust-to-stellar mass ratio is $\simeq50$ times 
    lower than that of spiral galaxies \citep{smith12,galliano21}. 
    Dust lanes are, however, commonly detected in elliptical galaxies 
    \citep[\eg][]{sadler85,gomez10}. 
    \citet{jura87} for instance found that half of nearby ellipticals are 
    detected at \hIRAS\ wavelengths.
    \citet{smith12} found that elliptical galaxies detected at 250~\tmic\ 
    tend to have higher X-ray luminosities.
    This was further explored by \citetalias{galliano21}.
    We will come back to this point in \refsec{sec:XETG}.
\end{description}

    \subsubsection{Large-Scale Dust Distribution in Galaxies}
    \label{sec:MCRTgal}

The dust spatial distribution, that is the value of various dust parameters in different regions or pixels of a galaxy, can be determined in the \hMW\ and nearby galaxies.
This determination however requires good quality, homogenized multi-wavelength images of the studied objects.
This is therefore significantly more complex than modeling the \hSED\ of a point source.

\begin{figure}[htbp]
  \includegraphics[width=\textwidth]{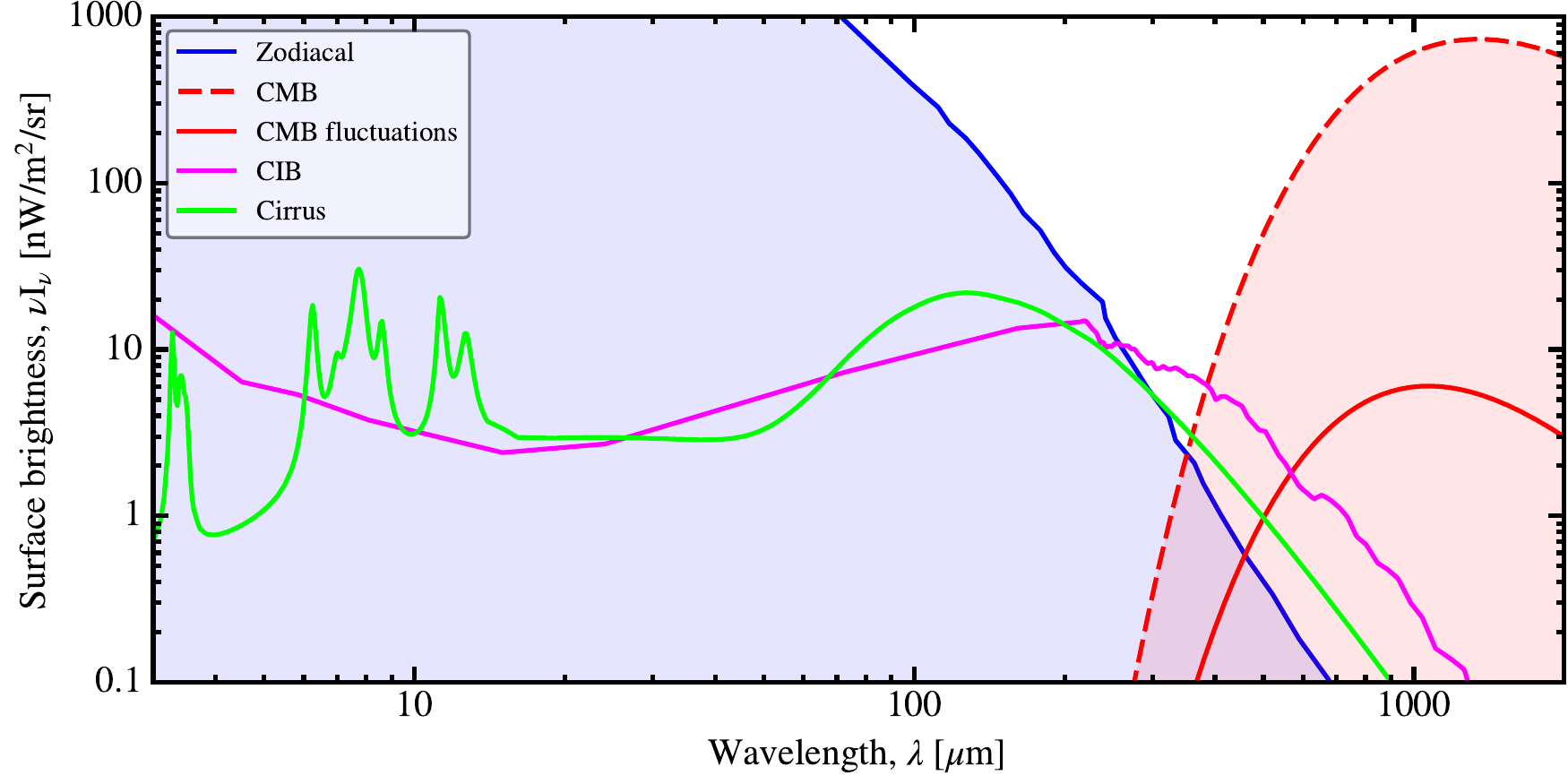}
  \newcap{Principal sources of contaminations encountered when modeling SEDs}%
         {The zodiacal spectrum (blue) has been computed using the model of 
          \citet{fixsen02}.
          The \hCMB\ spectrum is a $T=2.73$~K black body (red dashed line).
          The solid red line corresponds to the level of \hCMB\ fluctuations.
          The \hCIB\ (green) comes from the compilation of \citet{dole06}.
          The Galactic cirrus emission is the \citetalias{jones17} model scaled
          by a column density $N_\sms{H}=10^{24}$~H/m$^2$.
          \CClicence}
  \label{fig:contaminations}
\end{figure}

\paragraph{Homogeneous multi-wavelength data sets.}
To accurately model \hSED s of galaxies, the observed fluxes must originate, at each wavelength, in the same region, and must trace only the emission we are modeling (\ie\ the dust emission, the escaping stellar emission, \etc).
The following artifacts can be encountered.
\begin{description}
  \item[Contaminations] in the telescope beam can have several origins. 
    Most of them are displayed in \reffig{fig:contaminations}.
    \begin{description}
      \item[Foreground zodiacal emission] originates in dust grains from the 
        Solar system \citep[\eg][]{fixsen02,reach03,rowan-robinson13,kruger19}.
        Its intensity depends greatly on the angle above the planetary disk.
        It is prominent in the \hMIR\ (\cf\ \reffig{fig:contaminations}).
        The zodiacal emission is quite homogeneous at the typical angular scale
        of nearby galaxies (a few degrees or less).
        It can thus be efficiently subtracted, using the emission outside the
        target.
        Otherwise, several models compute its synthetic spectrum 
        \citep[\eg][]{kruger19}.
      \item[Foreground Galactic emission] originates in dust grains from 
        cirrus clouds within the \hMW.
        Its spectrum does not vary significantly with pointing direction, as the
        Galactic \hISRF\ is relatively homogeneous.
        However, its total intensity scales with the column density of the
        \hISM\ between the observer and the target galaxy.
        The H column density is the lowest at high Galactic latitude 
        ($N_\sms{H}\simeq10^{24}$~H/m$^2$).
        The fractal structure of interstellar clouds \citep[\eg][]{elmegreen96} 
        results in a high degree of structure at small angular scales.
        In other words, it is difficult to subtract this emission using 
        off-target areas.
        This contamination can not always be subtracted.
        Surface brightness being independent of distance, the emission from the
        diffuse \hISM\ of nearby galaxies is similar to that of the \hMW, 
        making the study of the former particularly challenging.
      \item[Background galaxy emission] constitutes the \hCIB\ 
        \citep[\eg][]{dole06}.
        Its \hSED\ looks similar to the diffuse Galactic \hISM\ (\cf\ 
        \reffig{fig:contaminations}).
        Spatially, it is granular, as it is the sum of numerous point sources.
        It is difficult to accurately subtract.
        It is thus another component that makes studying the diffuse \hISM\
        of nearby galaxies challenging.
      \item[Cosmological background emission] is a $T=2.73$~K black body 
        \citep[\cf\ \refsec{sec:Teq};][]{mather94}.
        It contaminates essentially the mm regime.
        Its emission is globally isotropic with some fluctuations.
        The amplitude of these fluctuations is shown as a solid red line in
        \reffig{fig:contaminations}.
        These fluctuations are a bit more difficult to subtract.
        However, in general this source of contamination is not the most 
        challenging to remove.
      \item[Non-dusty contamination,] such as gas-phase line emission or the 
        radio continuum need to be subtracted.
        For instance, in the submm regime, the CO lines and the free-free 
        continuum can account for $10-20\,\%$ of the emission around 
        $\lambda\simeq1$~mm \citep[\cf\ \reffig{fig:dustobs} and
        \eg][]{galliano03,galliano05}.
        Additional independent constraints are necessary, such as 
        spectroscopic observations of the contaminating lines, and
        long-wavelength radio continuum fluxes that probe the free-free
        and synchrotron emission.
    \end{description}
  \item[Differences in angular resolution] are due to variations of the beam 
    size across the observed \hSED.
    If not corrected, this can have dramatic consequences, especially if the
    studied region has large gradients of emission on scales of the order of the
    largest beam size.
    It is however easy to correct.
    One needs to identify, among the instruments used, which one has the largest
    beam, $\Omega_\sms{max}$.
    One then simply needs to degrade the angular resolution of all the other 
    wavelengths, to the resolution $\Omega_\sms{max}$.
    This degradation is performed by convolving the images by a kernel, which
    is the $\Omega_\sms{max}$ beam deconvolved by the beam at the nominal 
    wavelength.
    Such kernels are provided, for instance, by \citet{aniano11} for a wide 
    variety of instruments.
  \item[Differences in field of view] come from the fact that the different 
    instruments do not necessarily have the same orientation on the sky and the 
    same pixel size.
    To model a spatially-resolved \hSED, it is thus necessary to reproject every
    image on a single grid.
    Numerous methods are available to perform this reprojection 
    \citep[\eg][]{bertin02}.
    It can become problematic only when there are missing areas, such as 
    incomplete maps or masked regions (\eg\ because of saturation).
    The linear pixel size of the final grid can be as low as $\simeq1/3$ of the 
    largest beam size (Nyquist sampling).
  \item[Uncertainties] must be taken into account as rigorously as possible.
    Ideally, we should not only determine the uncertainties of each pixel at 
    each wavelength, but also their correlations.
    This can be simplified by separating the two major sources of uncertainties:
    noise and calibration effects.
    The noise comes from the instrument.
    It is usually provided by the data reduction pipeline.
    It thus needs to be propagated through the various sources of homogenization
    we have discussed.
    In particular, resolution degradation increases the median signal-to-noise 
    ratio, as it has an averaging effect on the noise spatial distribution.
    It also creates correlations between pixels.
    To our knowledge, the best method to propagate noise uncertainties, which is 
    also the simplest to implement, is by way of Monte-Carlo simulations
    \citep[\eg][]{galliano11}.
    The principle is the following.
    \begin{enumerate}
      \item We generate a large number, $N\simeq100$, of images with flux:
        \begin{equation}
        F_\nu^{(i)}(x,y,\lambda) 
          =F_\nu(x,y,\lambda)+\delta(i,x,y,\lambda)\times\sigma_\nu(x,y,\lambda)
        \end{equation}
        ($i=1,N$), where $F_\nu$ and $\sigma_\nu$ are the flux and uncertainty
        coming from the data reduction pipeline.
        The random variable, $\delta$, is independent, normal with mean 0 and
        standard deviation 1.
      \item We then perform the different homogenization steps (contamination 
        subtraction, degradation, reprojection) on each random samples.
      \item We now have, for each pixel and each wavelength of the final 
        homogenized maps, a distribution of $N$ values.
        The standard deviation of this distribution gives the uncertainty and
        we can compute correlation coefficients between different pixels or 
        wavelengths.
    \end{enumerate}
    Calibration uncertainties can be computed afterwards, as they are 
    proportional to the flux.
    These uncertainties are fully correlated between pixels and partially  
    correlated between wavelengths \citepalias[\eg][]{galliano21}.
\end{description}
This technical data preparation can be tedious, but it is crucial as dust model results directly rely on it.
A significant effort has been put into providing homogenized databases of nearby galaxies.
Among them, the most important surveys are the following.
\begin{itemize}
  \item \hSAGE\ \citep{meixner06,meixner13} were a series of surveys of 
    the Magellanic clouds with \hspitz\ and \hhersc.
  \item The \hDGS\ \citep{madden13,remy-ruyer13} was a survey of nearby dwarf 
    galaxies with \hhersc, for which we also added \hspitz\ and ancillary data 
    \citep{bendo12,remy-ruyer15}.
  \item SINGS/KINGFISH \citep{kennicutt03,kennicutt11} was a series of \hspitz\
    and \hhersc\ surveys of nearby galaxies.
  \item \hDustPedia\ \citep{davies17,clark18} was a European collaboration 
    project, which built a homogenized sample of all available data for 
    $\simeq800$ nearby galaxies.
\end{itemize}

\paragraph{Properties of individual galaxies.}
Numerous studies have presented the \hSED\ modeling of nearby galaxies, and their derived dust properties, either globally or spatially-resolved.
We have participated to several such projects
\citep[\eg][]{whaley09,galametz09,boselli10,galametz10,ohalloran10,eales10,cortese10,sauvage10,bendo10,roussel10,gordon10,davies10,boquien10,skibba12,de-looze12,galametz13,ciesla14,gordon14,galametz16,bianchi18,nersesian19}.
Reviewing them would be unwieldy, here.
In general, these studies provide dust parameters (mass, starlight intensity, \hPAH\ mass fraction, \etc) of different objects, which can be used in combination with other tracers to refine our understanding of the studied source.
They also provide scaling relations and calibrations of various diagnostics such as \hSFR\ tracers.
These results can also be used to train machine-learning models that could predict the \hSED\ of a poorly-observed galaxy \citep[\eg][]{dobbels20}.

\paragraph{Identifying dust heating sources.}
A particular question, that has been tackled by several studies, is the identification of the sources responsible for dust heating within galaxies.
In the \hMW, \hthreeD\ reconstruction of the \hISM\ distribution showed that the heating by young, O/B stars (\reftab{tab:stars}) was prominent in molecular regions, whereas the atomic phase was mainly heated by lower-mass stars \citep[\eg][]{sodroski97,paladini07}.
In nearby galaxies, this depends on the \hSF\ activity of the galaxy.
For instance, we showed that \hPAH s were essentially heated by field stars in the quiescent galaxy \ngc{2403}.
These molecules are however heated by the escaping radiation from \hii\ regions in the more actively star-forming object, \M{83} \citep{jones15}.
More generally, with the \hDustPedia\ sample, we found that dust in \hETG s was mainly heated by old stars \citep{nersesian19}.
It is only when considering more gas-rich galaxies that the contribution of young stars becomes more important.
It can account for up to $\simeq60\,\%$ of the dust luminosity in extreme late-type galaxies \citep[Sm--Irr, \reffig{fig:hubble};][]{nersesian19}.
These different heating sources have an impact on the global escape fraction (\ie\ the fraction of stellar radiation leaving the galaxy unattenuated).
Massive stars being embedded in molecular cocoons, they have a lower escape fraction than \expression{Low- and Intermediate-Mass Stars} (\hLIMS) which occupy lower density regions.
In the \hDustPedia\ sample, we showed that the escape fraction was on average $\simeq81\,\%$, with mild variations across galaxy types \citep{bianchi18}.
It is slightly lower in \hLTG s ($\simeq75\,\%$).
We emphasize that this nearby galaxy sample lacks the deeply enshrouded star formation of \hLIRG s and \hULIRG s, where the global escape fraction can drop down to $\simeq1\,\%$ \citep[\eg][]{clements96}.
The question of the dust heating contribution can now be tackled with more accuracy using \hthreeD\ \hMCRT\ models.

\begin{figure}[htbp]
  \includegraphics[width=\textwidth]{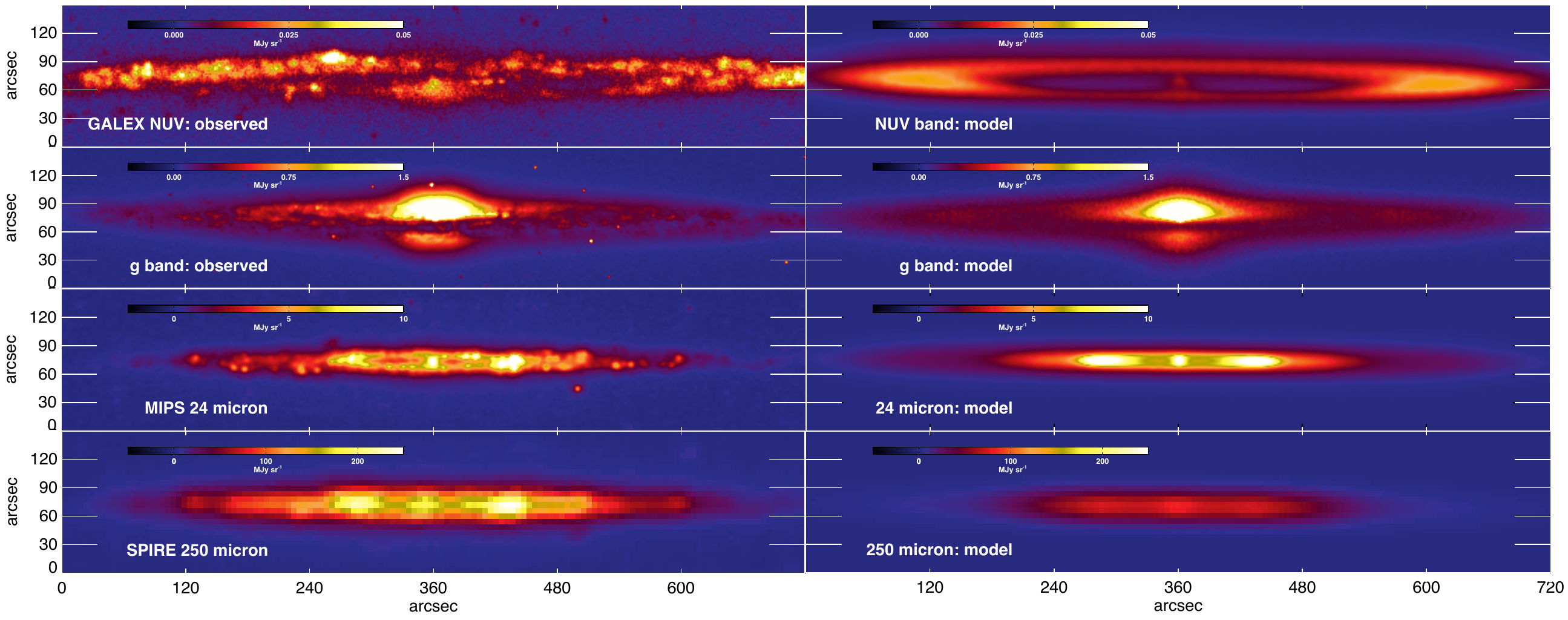}
  \newcap{Radiative transfer modeling of \ngc{4565}}%
         {The observations (left column) are compared to the modeled flux 
          distribution (right column).
          \uline{Credit:} \citet{de-looze12}.}
  \label{fig:RT}
\end{figure}

\paragraph{Large-scale radiative transfer models of galaxies.}
Applying a \hthreeD\ \hMCRT\ model to reproduce the spatial flux distribution of galaxies, in all wavebands, is not straightforward.
Indeed, the observations provide only 2D projected constraints.
This is why most studies favor edge-on galaxies, as the images of such objects provide constraints on both the radial and azimuthal distributions, assuming axisymmetry (\reffig{fig:RT}).
Several studies have modeled the effect of extinction on the optical data of 
disk galaxies using such codes \citep[\eg][]{xilouris99,alton04,bianchi07}.
They were able to answer the recurring question about the optical thickness of 
disk galaxies \citep{disney89}.
In particular, \citet{xilouris99} found that the face-on optical depth of 
typical spiral galaxies is less than one, in all optical bands.
Concerning dust heating, recent progress has been made, especially by the \hDustPedia\ collaboration, using the \hMCRT\ code SKIRT \citep{baes15}.
\begin{itemize}
  \item In \M{31}, \citet{viaene17} showed that $90\,\%$ of the dust could be
    heated by the evolved stellar populations.
  \item \citet{nersesian20a} showed that dust heating by young stars 
    accounts on average for $\simeq60\,\%$ in four face-on barred spirals.
  \item \citet{viaene20} modeled the strong \hAGN\ \ngc{1068}.
    We showed that $\simeq80\,\%$ of the heating is coming from young stars,
    and only a few percents from the central engine.
    However, dust heating by \hAGN\ represent about $\simeq90\,\%$ within 
    the central $\simeq100$~pc.
  \item \citet{nersesian20b} studied \M{51}, showing that globally 
    $\simeq71\,\%$ of the heating was coming from young stars.
    Surprisingly, we also found that \ngc{5195}, the companion of 
    \M{51} was responsible for $\simeq6\,\%$ of this heating.
\end{itemize}
A model such as SKIRT can also be used to model the radiative transfer in simulations of galaxies \citep[\eg][]{trcka20}.
Finally, these models account for the energy balance between the escaping
\hUV-visible light and the re-emitted \hIR-submm radiation.
Interestingly enough, several studies report a deficit of modeled
\hFIR\ emission by a factor $\simeq3-4$, compared to the observations
\citep{alton00,alton04,dasyra05,de-looze12,de-looze12b}.
This discrepancy is thought to result from a lack of detail in modeling the geometry. 
In particular, the presence of young stars, deeply embedded in molecular clouds, at sub-grid resolutions, could compensate for this deficit without significantly altering the extinction \citep[\cf\ \eg][]{baes10}.

    \subsubsection{Constraining the Grain Opacity}
    \label{sec:kappaLMC}

Dust masses derived from \hSED\ fits directly depend on the assumed grain opacity.
Using the \hMBB\ param\-etrization \refeqp{eq:kappaMBB}, both the scaling, $\kappa_0$, and the emissivity index, $\beta$, are important.
There are particular situations, where we can reverse the process and constrain these two parameters:
\begin{enumerate}
  \item if we are observing a region that we can assume uniformly illuminated, 
    then we can infer $\beta$;
  \item if we have an independent constraint on the dust mass, then we can 
    infer $\kappa_0$.
\end{enumerate}

\paragraph{Studies of the emissivity index.}
There are numerous publications presenting \hMBB\ fits of nearby galaxies.
However, as discussed in \refsec{sec:MBB}, the derived emissivity index, $\beta$, is degenerate with temperature mixing.
The best constraints on the intrinsic $\beta$ are obtained in the submm regime,
where only massive amounts of very cold dust ($T\lesssim 10$~K) could bias the 
value.
\reftab{tab:MBBfit} lists \expression{effective emissivity indices}, $\beta_\sms{eff}$, for several objects, obtained with \hplanck, with constraints up to $850\emic$.
It appears that all the values are lower than 2, and that low-metallicity systems have a lower $\beta_\sms{eff}$ than higher metallicity galaxies.
\citet{boselli12}, studying a volume-limited sample with \hhersc\ (up to 
$500\emic$), also found an average $\beta_\sms{eff}\simeq1.5$, and hinted that 
low-metallicity objects tend to have $\beta_\sms{eff}<1.5$.
In \M{33}, $\beta_\sms{eff}$ derived from \hersc\ observations is around 2 in the center and decreases down to 1.3 in the outer parts \citep{tabatabaei14}.
On the other hand, the outer regions of \M{31} exhibit a steeper slope  ($\beta_\sms{eff}\simeq2.3$) than in its center \citep{draine14}.
This contradictory behaviour does not appear to originate in fit biases, as 
both increasing and decreasing trends of $\beta_\sms{eff}$ with radius are found in the sample of \citet{hunt15}.
\begin{table}[htbp]
  \centering
  \setlength\arrayrulewidth{2pt}
  \arrayrulecolor{white}
  \begin{tabularx}{\linewidth}{|>{}X%
                                |>{\columncolor{coltabsep}}r%
                                |>{\columncolor{coltabsep}}r%
                                |>{\columncolor{coltabsep}}r%
                                |>{\columncolor{coltabsep}}r|}
    \hline
      \rowcolor{coltabhead}\cellcolor{white}
      & \textbf{Milky Way} & \textbf{\M{31}} & \textbf{LMC} & \textbf{SMC} \\
    \hline
      \cellcolor{coltabhead} Temperature & $19.7\pm1.4$~K 
        & $18.2\pm1.0$~K & $21.0\pm1.9$~K & $22.3\pm2.3$~K \\
    \hline
      \cellcolor{coltabhead} $\beta_\sms{eff}$ & $1.62\pm0.10$ 
        & $1.62\pm0.11$ & $1.48\pm0.25$ & $1.21\pm0.27$ \\
    \hline
      \cellcolor{coltabhead} Reference
        & \hyperlinkcite{planck-collaboration14c}%
                   {\planck~\citeyearpar{planck-collaboration14c}}
        & \hyperlinkcite{planck-collaboration15c}%
                   {\planck~\citeyearpar{planck-collaboration15c}}
        & \hyperlinkcite{planck-collaboration11}%
                   {\planck~\citeyearpar{planck-collaboration11}}
        & \hyperlinkcite{planck-collaboration11}%
                   {\planck~\citeyearpar{planck-collaboration11}} \\
    \hline
  \end{tabularx}
  \newcap{Free emissivity index MBB fits of nearby galaxies 
          by the \planck\ collaboration}{}
  \label{tab:MBBfit}
\end{table}

\begin{figure}[htbp]
  \includegraphics[width=\textwidth]{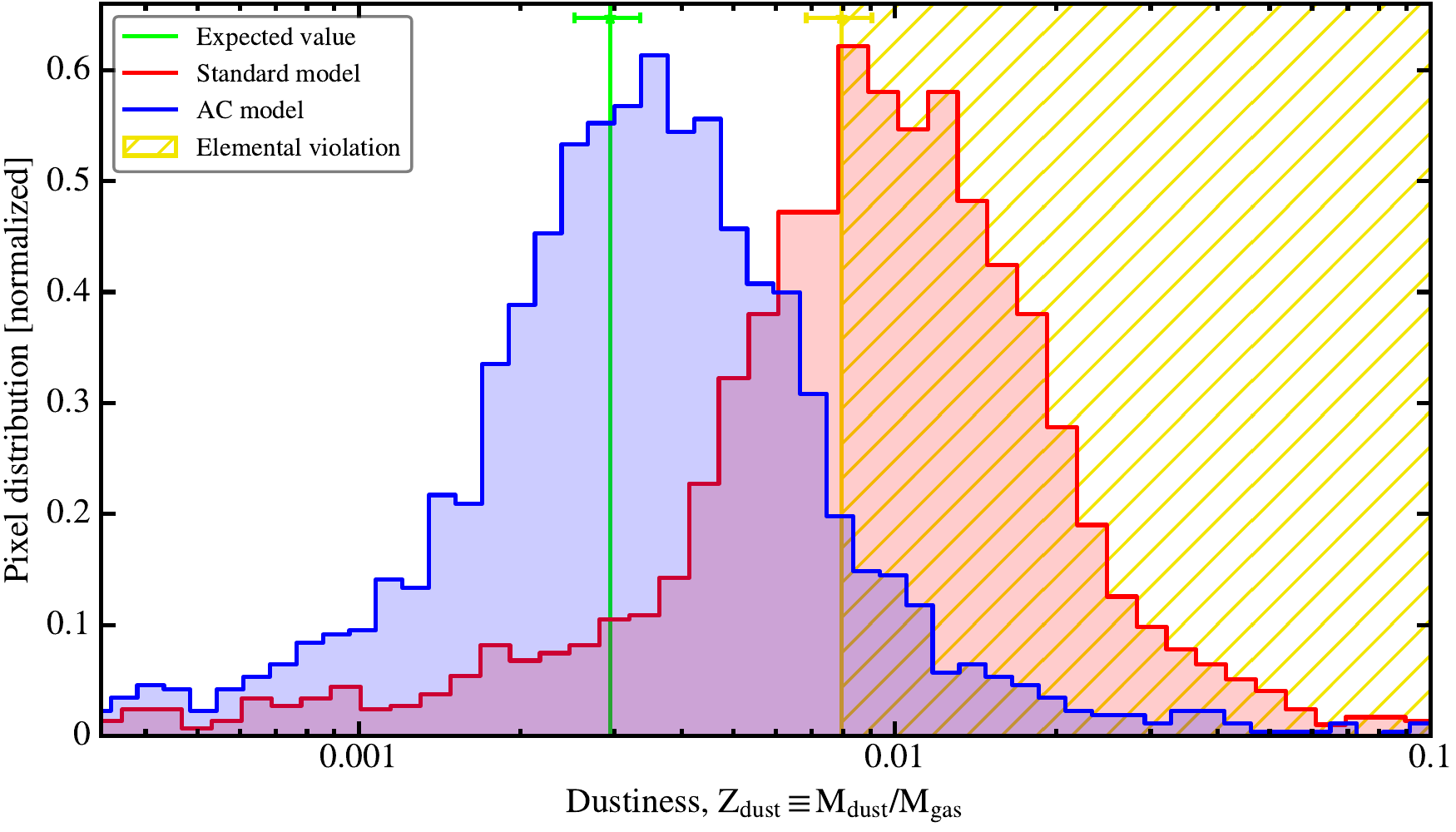}
  \newcap{Dust mass discrepancy in the LMC}%
         {The histograms are the pixel distribution of the \hdustiness\ in the 
          \hLMC\ strip of \citet{galliano11}.
          We show two models:
          \begin{inlinelist}
            \item the \citengl{standard model} (red), based on the 
              \citet{draine07} optical properties;
            \item the \citengl{AC model} (blue), replacing graphite by 
              amorphous carbon to boost the emissivity.
          \end{inlinelist}
          Most pixels of the standard model are in the hatched area, thus 
          violating the constraints put by heavy element abundances.
          We also show the uncertainties on the limit \hdustiness\ (yellow error
          bar) and on the expected one (green error bar).
          \CClicence}
  \label{fig:strip_LMC_G2D}
\end{figure}

\paragraph{Grain opacity in the LMC.}
An important result, that has often been misunderstood, concerns the grain opacity in the \hLMC.
\citet{galliano11}, modeling a strip covering one fourth of the \hLMC\ (\cf\ \reffig{fig:matryoshka}), with the composite approach (\refsec{sec:dale}), found that the \hdustiness\ distribution in this galaxy was most of the time larger than the maximum value it could in principle reach.
This maximum value is set by the elemental abundances.
The fraction of elements locked-up in grains can indeed not be larger than the amount available in the \hISM.
The metallicity of the \hLMC\ is $Z_\sms{LMC}\simeq1/2\eZsun$ \citep{pagel03}.
The maximum \hdustiness\ of the \hLMC\ is thus $Z_\sms{dust}^\sms{max}\simeq Z_\sms{LMC}\simeq0.007$.
The dust mixture that was used is an update of the \citet[][BARE-GR-S]{zubko04}.
It is essentially based on the \citet{draine07} optical properties, a pre-\hhersc\ model.
This discrepancy is shown in \reffig{fig:strip_LMC_G2D} (the red histogram).
The only explanation is that this grain mixture is not emissive enough to account for the observed \hFIR-submm emission.
We thus proposed an alternate dust model, simply replacing graphite by amorphous carbons \citep[the ACAR sample of][]{zubko96}, without altering the total carbon fraction.
This simple modification boosts the emissivity by a factor\footnote{This factor is not precise, as the slope of the opacity is also changed. The difference in emissivity is thus not a sole scaling.} $\simeq2-3$.
With this new model, most of the \hdustiness\ distribution is centered around its expected value, and is clear from the forbidden range (yellow).
It was called the \expression{AC model} (blue histogram in \reffig{fig:strip_LMC_G2D}).
The tail of the distribution in the forbidden zone originates in cold regions, where the uncertainty is large.
The conclusion of this modeling was that \hLMC\ grains must be a factor $\simeq2-3$ more emissive than the \citet{draine07} model.
We actually presented a preliminary version of this result during \hhersc's science demonstration phase \citep{meixner10}.

\paragraph{Confirmation in other systems.}
The conclusion of the \citet{galliano11} study was that grains in the \hLMC\ had to be more emissive than in the \hMW, because the \citet{draine07} model was at the time consistent with the \hMW.
This last statement however happened to be inexact.
\citet{planck-collaboration16} modeled the all sky dust emission, using also the \citet{draine07} model.
The $A(V)$ estimated along the sightlines of $\gtrsim200\,000$ \expression{Quasi-Stellar Objects} (\hQSO) was systematically lower than their dust-emission-derived $A(V)$.
Their comparison of emission and extinction thus indicates that the Galactic 
opacity is in fact also a factor of $\simeq2$ higher than previously assumed.
In addition, in \M{31}, \citet{dalcanton15} derived a high spatial resolution 
map of $A(V)$.
As in the Galaxy, the emission-derived $A(V)$ map \citep{draine14} was found to be a factor of $\simeq2.5$ higher.
We emphasize that, although each of these studies found evidence of local variations of the emissivity as a function of the density (\cf~\refsec{sec:mantles}), the overall opacity seems to be scaled up compared 
to \citet{draine07}.
In other words, in all the environments where enough data is available to constrain $\kappa$, it is found a factor of $\simeq2-3$ higher than the original \citet{draine07} properties.
Dust models therefore need to use an opacity consistent with these constraints.
This is the case of the \citetalias{jones17} model.
Its \hFIR-submm opacity is very close to our AC model \citep[\cf\ Fig.~4 of][]{galliano18}.
\takeaway{It is reasonable to adopt the \citetalias{jones17} grain opacity 
          (\cf\ \refsec{sec:themis_quant}), when modeling galaxies.}

\paragraph{The opacity in nearby galaxies.}
The \hDustPedia\ collaboration conducted several studies aimed at constraining the grain opacity in nearby galaxies.
First, \citet{bianchi19} studied the actual emissivity of 204 late-type galaxies, that is the ratio of \hIR\ emission to H column density.
We found an emissivity $\epsilon_\nu(250\emic)\simeq0.82\pm0.07$~MJy/sr/(1020 H/cm$^2$), consistent with the \hMW, except for the hottest sources.
These estimates were derived using global fluxes, integrated over the whole galaxy.
In parallel, \citet{clark19} modeled in details the two face-on galaxies, \M{74} and \M{83}.
We could map the grain opacity.
This was done by converting metallicity maps into oxygen depletion maps, and comparing those to the dust mass.
The derived opacities were quoted at $\lambda=500\emic$: $\kappa(500\emic)\simeq0.11-0.25$~m$^2$/kg in \M{74}, and $\kappa(500\emic)\simeq0.15-0.80$~m$^2$/kg in \M{83}. 
These values are consistent with the \hhersc-\hplanck-revised opacities of the \citetalias{jones17} model ($\kappa(500\emic)\simeq0.19$~m$^2$/kg; \cf\ \reffig{fig:themis_kappa_proxy}).

    \subsubsection{Constraining the Size Distribution}
    \label{sec:sizedist}

We have seen in \refsec{sec:dale} that there is a degeneracy between the grain size and starlight intensity distributions.
This degeneracy arises from the fact that it is observationally difficult to distinguish the \hMIR\ emission of a hot region from the \hMIR\ emission of small grains.
In the early 2000s, we did not know it was impossible, so we did it \citep{galliano03,galliano05}.
We modeled the \hSED\ of the following four \hBCD s: \ngc{1569} ($Z\simeq1/4\eZsun$), \iizw\  ($Z\simeq1/6\eZsun$), \hen\  ($Z\simeq1\eZsun$)and \ngc{1140}  ($Z\simeq1/3\eZsun$), to infer their size distribution.
We interpreted these results in light of shock processing.

\begin{figure}[htbp]
  \includegraphics[width=\textwidth]{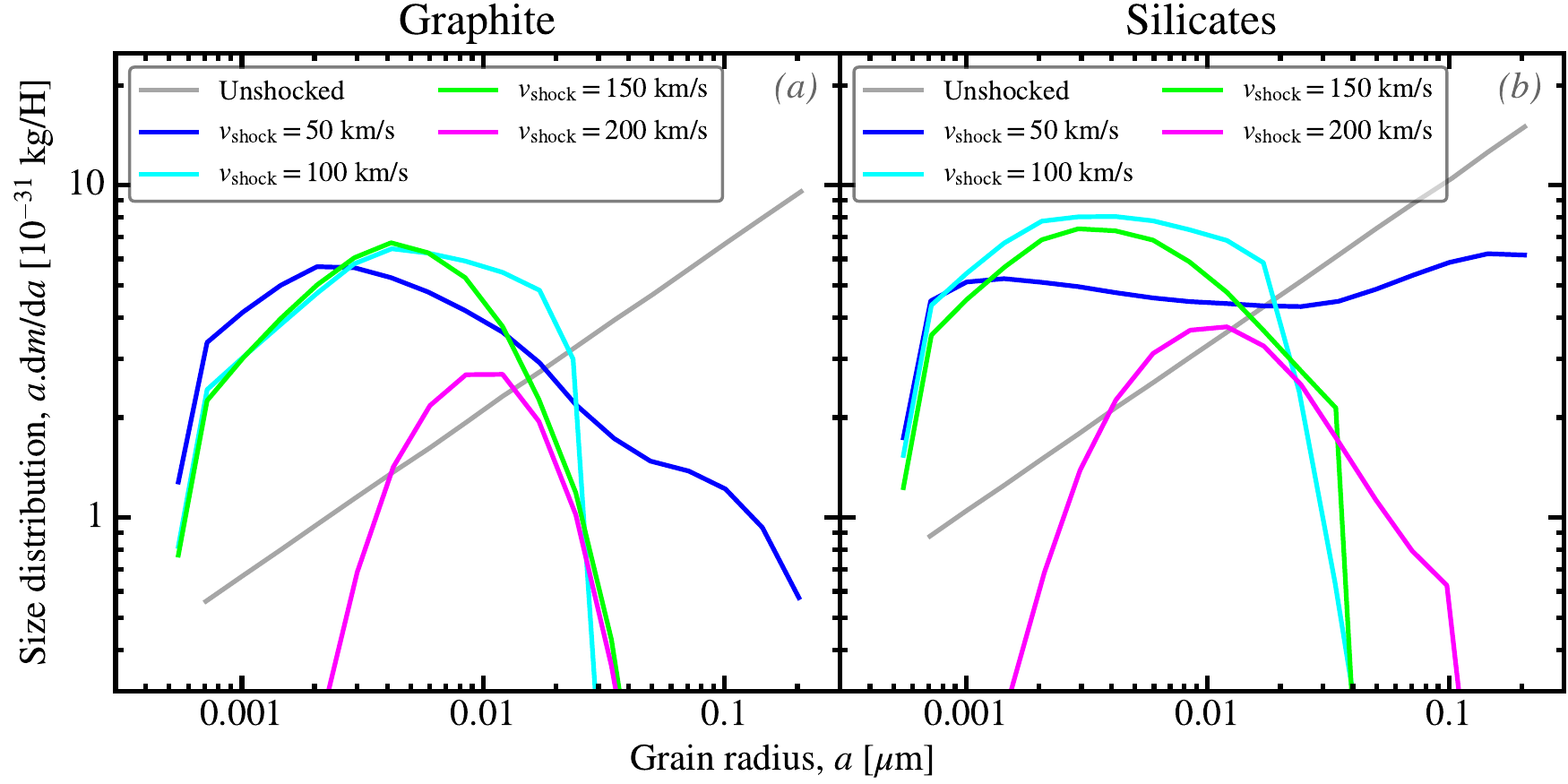}
  \newcap{Effect of a blast wave on the grain size distribution}%
         {In both panels, we demonstrate the processing of a 
          \citetalias{mathis77} size distribution (grey) by shock waves of 
          different velocities, $v_\sms{shock}$.
          Panel~\textit{(a)} shows graphite, and panel~\textit{(b)}, 
          silicates, both from the \citet{jones96} model.
          \CClicence}
  \label{fig:sizedist_shock}
\end{figure}

\paragraph{Grain processing by shock waves.}
Shock waves from \hSN e process dust grains, while sweeping the \hISM.
\begin{itemize}
  \item At high velocity, $v_\sms{shock}\gtrsim1\,000$~km/s, they vaporize most 
    of the dust \citep[\eg][]{dwek98,slavin15}.
  \item At intermediate velocity, 
    $50\;\textnormal{km/s}\lesssim v_\sms{shock}\lesssim200\;\textnormal{km/s}$, 
    shattering and fragmentation become dominant, altering the size 
    distribution \citep[\eg][]{jones96,bocchio14}.
\end{itemize}
\reffig{fig:sizedist_shock} shows the model of \citet{jones96} for graphite and silicates.
In both panels, the grey curve is the initial, \citetalias{mathis77} size distribution.
The color curves show the size distribution obtained after the mixture has been swept by a shock of velocity, $v_\sms{shock}$.
The main effect of the blast wave is to fragment and shatter grains, turning large grains into smaller grains.
The qualitative effect is similar for both compositions.
This is best seen at low velocity (blue curves).
Large grains are depleted and there is an excess of small grains.
At higher velocity, the distribution tends toward a log-normal centered around $a\simeq10$~nm (magenta curve). 
Vaporization also leads to a net loss of dust mass.
This model has since then been refined by \citet{bocchio14}, who applied it to \citetalias{jones17} grains.
We will discuss more extensively dust processing by \hSN\ blast waves in \refchap{chap:dustevol}.

\paragraph{The modeling strategy.}
\citet{galliano03,galliano05} modeled the UV-to-mm global \hSED\ of the four \hBCD s, using the \citetalias{desert90} dust model, the stellar evolutionary synthesis code \ncode{PÉGASE} \citep[][\reffig{fig:stellar_SED}]{fioc97}, and the photoionization code, \ncode{CLOUDY} \citep{ferland98}.
The modeling scheme was the following.
\begin{enumerate}
  \item We assume that the \hISM\ is concentrated in a spherical shell.
    At the center of this shell are the stellar populations.
  \item We model the escaping \hUV-visible \hSED\ with two \ncode{PÉGASE} 
    stellar populations, a young and an old one.
  \item We further constrain the \hUV-visible fit by selecting intrinsic stellar
    spectra that have the appropriate hardness.
    The hardness of the \hISRF\ is constrained by matching the observed
    \hISO\ \neiiiline/\neiiline\ and \sivline/\neiiiline\ ratios, using 
    \ncode{CLOUDY}.
  \item The intrinsic \hISRF\ is then used to heat the \citetalias{desert90}
    dust mixture.
    We fit the dust emission to the \hIR-to-mm observations, by varying the size
    distribution.
  \item We iterate this process a few times, as the dust size distribution
    impacts the extinction curve, which is circularly used to constrain the 
    stellar populations.
\end{enumerate}
This model is self-consistent \textit{per se}.
It however avoids the degeneracy between size and \hISRF\ distributions by assuming a simple geometry (the shell).
The dust is thus uniformly illuminated in this model, which is an unrealistic assumption for a star-forming galaxy.
It is therefore likely that some of the emission we have attributed to small grains originates in hot regions.
The results \citet{galliano03,galliano05} obtained, that we will discuss in the following paragraphs, are nevertheless qualitatively consistent with the properties we expect in these environments.
It is possible that only a fraction of the \hMIR\ emission originates in compact \hii\ regions, not excluding an overabundance of small grains.

\begin{figure}[!htbp]
  \includegraphics[width=\textwidth]{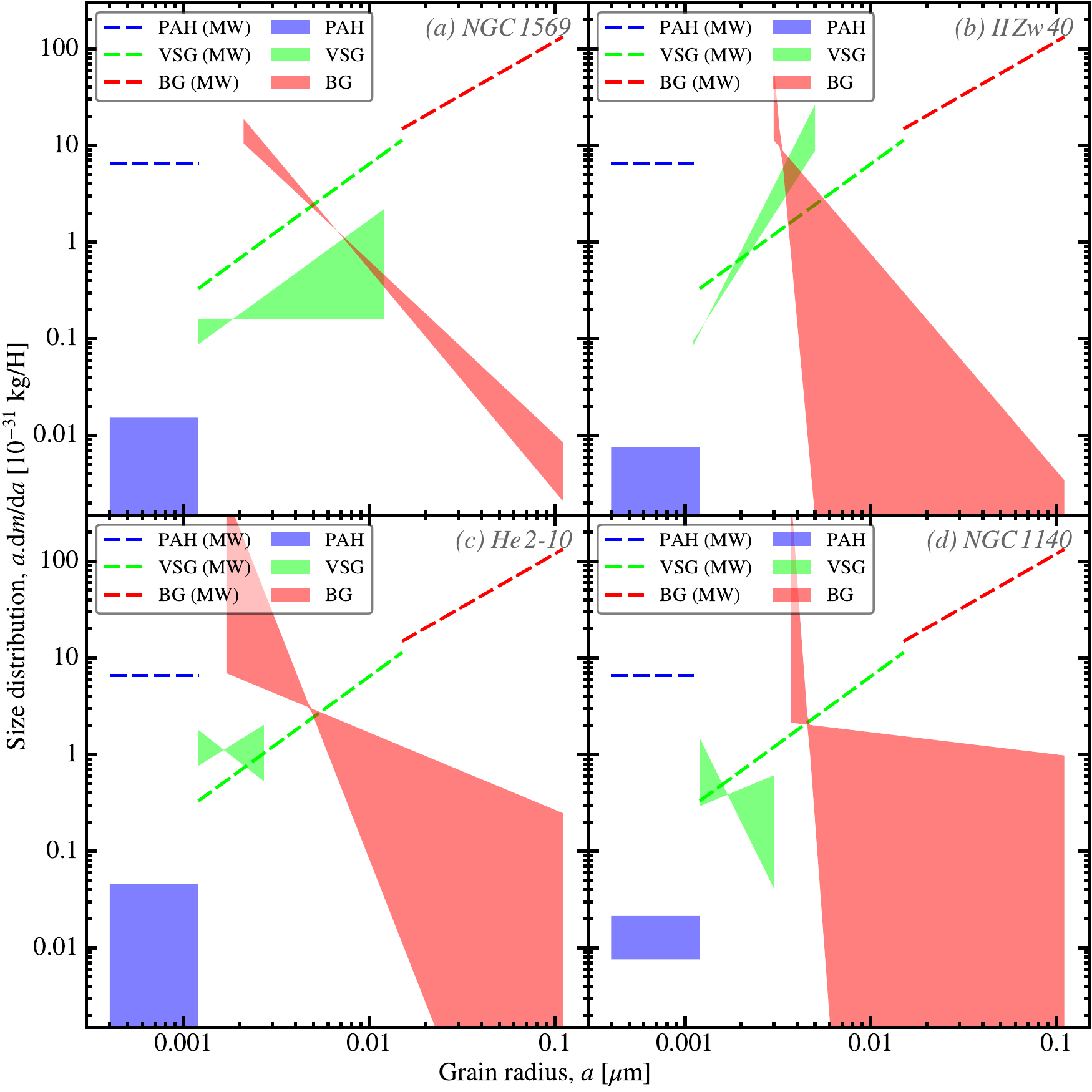}
  \newcap{Grain size distribution in four dwarf galaxies}%
         {In each panel, we show the \hMW\ grain size distribution of the
          \citetalias{desert90} model, as dashed lines.
          Those are the three components presented in \refsec{sec:histomodel}:
          \hPAH, \hVSG\ and \hBG.
          The filled curves show the likely range of size distribution inferred
          by the \hSED\ fits of \citet{galliano03,galliano05}.
          \CClicence}
  \label{fig:dilz_sizedist}
\end{figure}

\paragraph{The grain size distribution in four dwarf galaxies.}
The inferred size distribution, in the four \hBCD s, are shown in \reffig{fig:dilz_sizedist}.
We display the three components of the \citetalias{desert90} model: \hPAH s, \hVSG s and \hBG s (\cf\ \refsec{sec:histomodel}).
The most striking features, common to the four objects, are the following.
\begin{itemize}
  \item \hPAH s are under-abundant, compared to the \hMW.
    This was one of the first attempt at constraining the \hPAH\ abundance in 
    low-metallicity environments.
    Nowadays, there are overwhelming evidence of a general trend between 
    the aromatic feature strength and metallicity.
    We will come back to this point in \refsec{sec:PAHevol}.
  \item The global grain size distribution is dominated by small grains with
    radii of a few nanometers.
    These size distributions are thus qualitatively consistent with 
    shock-processed grains.
    This is encouraging, as we know that these environments are permeated by
    numerous \hSN\ blast waves, coming from their young stellar populations 
    \citep[\eg][]{oey96,izotov07}.
\end{itemize}
Two studies used a similar approach, assuming uniform illumination, and fitting the \hSED\ varying the size distribution of the \citetalias{desert90} model, in \ngc{1569} \citep{lisenfeld02} and in the \hLMC\ \citep{paradis09}.
They also concluded to an overabundance of small grains.

\begin{figure}[htbp]
  \includegraphics[width=\textwidth]{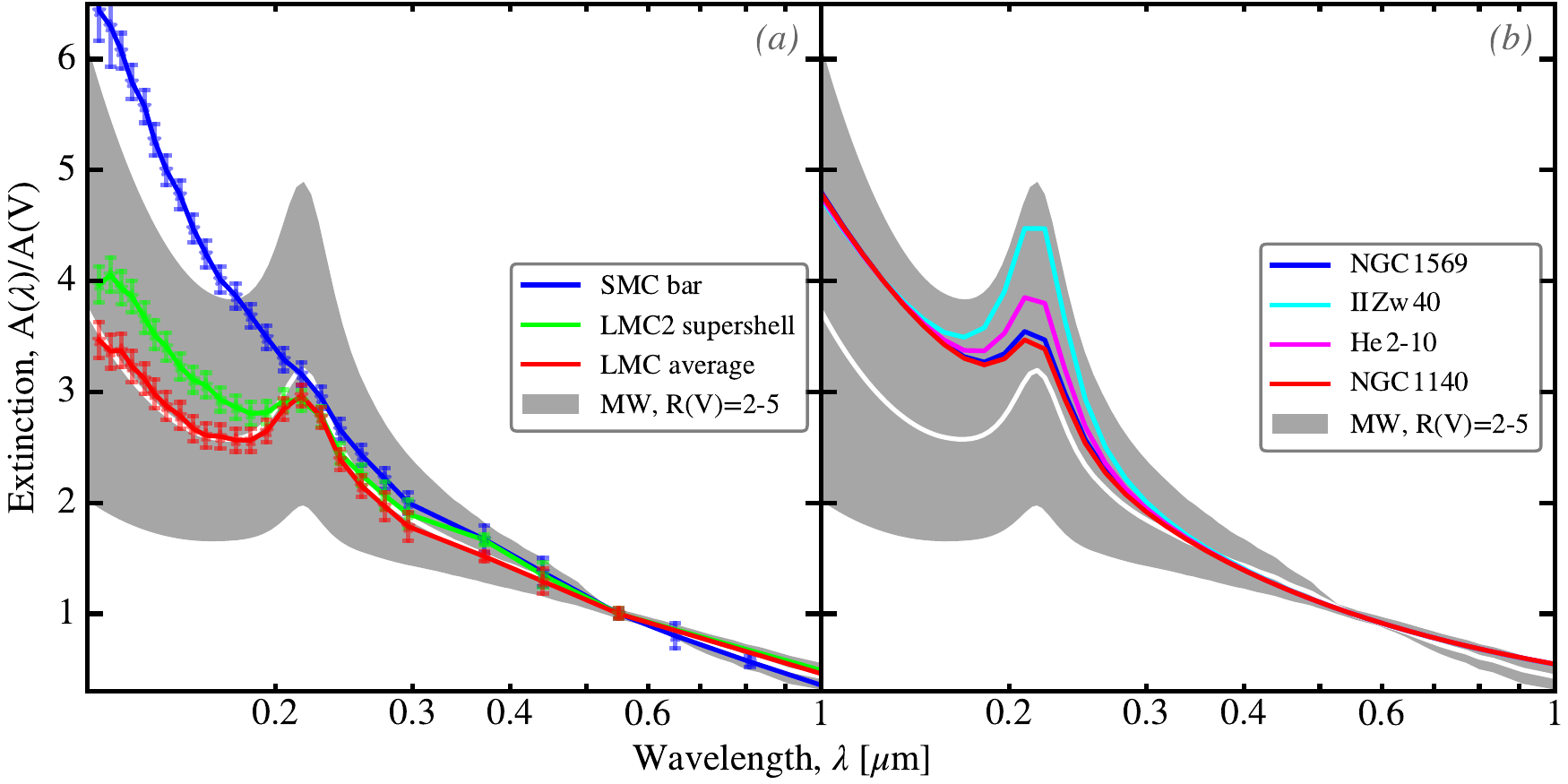}
  \newcap{Extinction curves of low-metallicity environments}%
         {In both panels, we show as a reference the range of Galactic 
          extinction curves, with $R(V)=2-5$, from the \citet{fitzpatrick19} 
          sample (\cf\ \reffig{fig:extinction}), in grey.
          The white curve is the average Galactic extinction, with $R(V)=3.1$.
          In panel~\textit{(a)}, we overlay several extinction curves within the
          Magellanic clouds, from the \citet{gordon03} sample: 
          \begin{inlinelist}
            \item the average of the \hLMC\ (red); 
            \item the \hLMC 2 supershell (green), near the massive star-forming 
              region \xxxdor; and
            \item the \hSMC\ bar (blue).
          \end{inlinelist}
          In panel~\textit{(b)}, we show the extinction of the four \hBCD s
          modeled by \citet{galliano03,galliano05}.
          \CClicence}
  \label{fig:dilz_extinction}
\end{figure}

\paragraph{Consequence on the extinction curves.}
We have briefly mentioned in \refsec{sec:histodustcont} that the extinction curves in the Magellanic clouds were systematically different from the \hMW.
\refsubfig{fig:dilz_extinction}{a} compares the extinction towards different sightlines in the \hLMC\ and \hSMC, to the range of extinction curves in the \hMW.
We see that the \hLMC\ ($Z\simeq1/2\eZsun$) is on average (red curve) similar to the \hMW.
However, toward the massive star-forming region \xxxdor\ (\hLMC2 supershell; green curve), it is steeper, with a weaker 2175~$\r{A}$ bump.
When we go to the \hSMC\ ($Z\simeq1/5\eZsun$; blue curve), the difference is more pronounced: the extinction curve has a very steep \hUV-rise and lacks the 2175~$\r{A}$ bump.
The origin of these variations are still debated.
Our four \hBCD s brought an interesting perspective on this open question.
The extinction curves derived from the size distributions of \reffig{fig:dilz_sizedist} are shown in \refsubfig{fig:dilz_extinction}{b}.
We can see that they are systematically steeper than the average \hMW\ ($R(V)=3.1$).
They also have a 2175~$\r{A}$ that is either weaker (\ngc{1569}, \hen\ and \ngc{1140}) or similar (\iizw) to the \hMW\footnote{We note that the strength of the bump is controlled by the carbon-to-silicate grain ratio, a parameter that is poorly constrained by the \hSED\ fit.}.
In other words, these extinction curves lie between the \hLMC\ and \hSMC, consistent with the metallicity range of these \hBCD s.
The modeling of \citet{galliano03,galliano05} therefore provides a coherent view of the grain properties in these environments, where shock waves have an instrumental role in shaping the grain sizes.

\section{Studies Focussing on Specific Spectral Domains}

\refsec{sec:SEDmodel} was devoted to modeling the whole \hIR\ \hSED, which is necessary to estimate the total dust content of galaxies.
There are however several other important properties that can be self-consistently studied by focussing on a particular wavelength range.

  \subsection{Scrutinizing Mid-IR Spectra}

Mid-IR spectra have been extensively observed since the first light of \hISO. 
\hspitz\ and \hAKARI\ have extended our knowledge of this spectral range and the \hJWST\ will likely revolutionize it.

    \subsubsection{The Aromatic Feature Spectrum}
    \label{sec:PAH}

Until now, we have discussed \hPAH s from a general point of view, and how to estimate their mass fraction.
We now focus on the information that the analysis of their detailed \hMIR\ emission spectrum can bring.
In what follows, we interchangeably use the terms \hUIB s, aromatic features and \hPAH\ bands.
The only case where these terms are not equivalent is when discussing the \hUIB s around 3~\tmic, coming from a mixture of aromatic and aliphatic features, that can not be solely attributed to \hPAH s.

\begin{figure}[!htbp]
  \includegraphics[width=\textwidth]{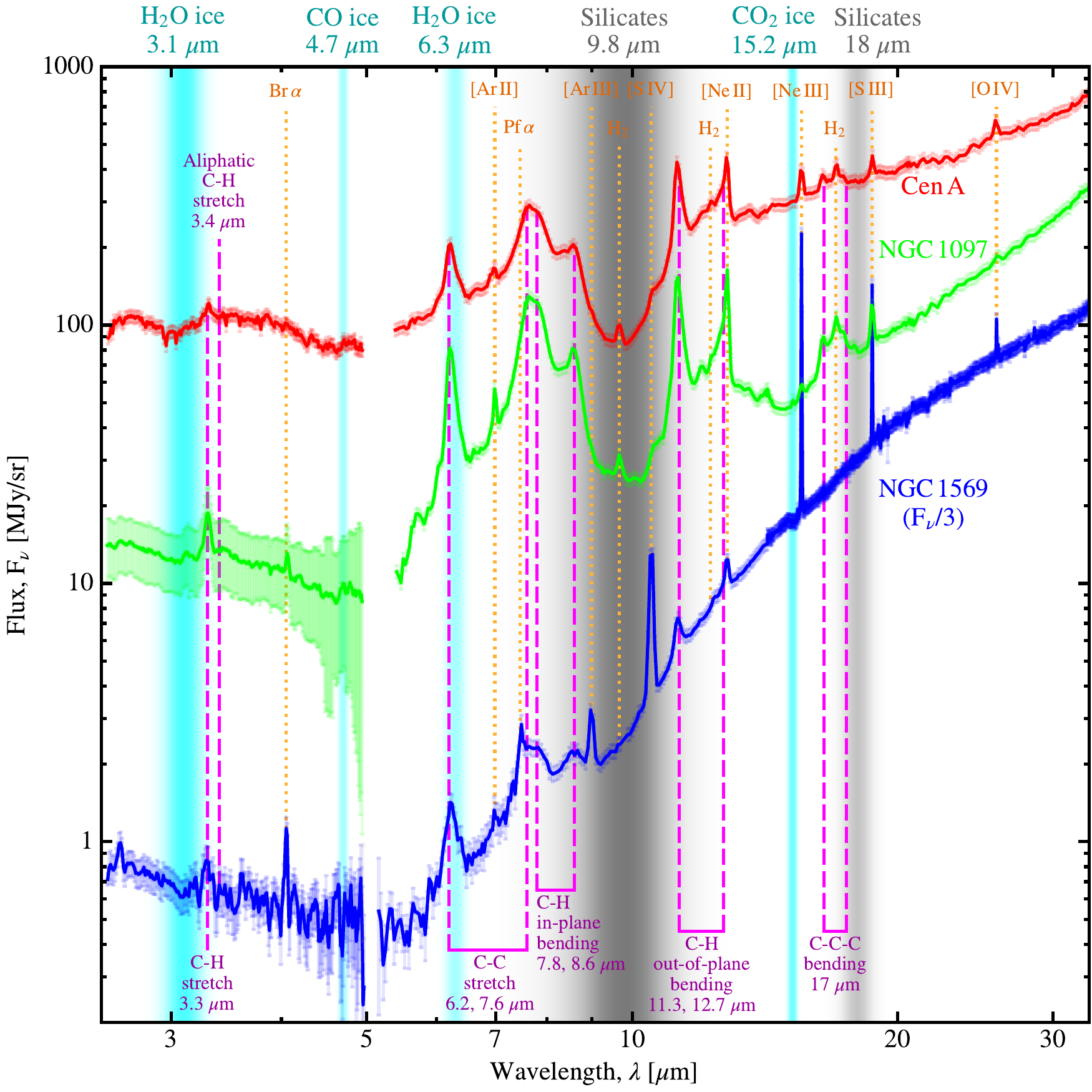}
  \newcap{MIR spectra of galaxies}%
         {We show the \hAKARI\ and \hspitz\ spectra of three galaxies:
          \begin{inlinelist}
            \item Centaurus$\,$A (red) is an \hETG\ with a powerful \hAGN;
            \item \ngc{1097} (green) is a \hLTG\ with a low-luminosity \hAGN; 
              and
            \item \ngc{1569} (blue) is \hBCD\ that we have scaled down by a 
              factor of 3.
          \end{inlinelist}
          These spectra were prepared for \citet{galliano18}.
          We have indicated most of the relevant features.
          Silicate absorption bands are shown in grey, and 
          ice absorption bands in cyan (\cf\ \refsec{sec:extinctionMIR}).
          Gas-phase emission lines are indicated in orange, and carbon-dust 
          emission features, in magenta.
          \CClicence}
  \label{fig:specMIR}
\end{figure}

\paragraph{MIR spectra of galaxies.}
\reffig{fig:specMIR} illustrates the diversity of \hMIR\ spectra encountered in different environments.
It shows three extreme galaxies from the \citeprep{hu21b} sample.
\begin{description}
  \item[Gas-rich, Solar-metallicity galaxies,] such as \ngc{1097} (green 
    spectrum in \reffig{fig:specMIR}), have bright aromatic features.
    These features are emitted by both their diffuse \hISM\ and their \hPDR s.
    The level and steepness of their \hMIR\ continuum, longward 10~\tmic, 
    depend on their \hSFR.
    Galaxies with a low \hSF-activity tend to have a flatter continuum, as the
    emission from hot equilibrium grains in \hii\ regions is lower.
  \item[Low-metallicity galaxies,] such as \ngc{1569} (blue spectrum in 
    \reffig{fig:specMIR}), have an integrated spectrum very similar to an \hii\
    region \citep[\eg][]{peeters02,martin-hernandez02}.
    They have weak or undetected aromatic features, which we will extensively 
    discuss in \refsec{sec:PAHevol}.
    Their strong ionic lines result from the combination of their young 
    stellar population and lower dust screening.
    For the same reason, their \hMIR\ continuum is significantly steeper than
    Solar-metallicity objects having the same \expression{specific Star 
    Formation Rate} ($\hsSFR\equiv\hSFR/M_\star$).
    In addition to small, stochastically-heated grains, the \hMIR\ continuum of
    \hBCD s originates partly in large grains at thermal equilibrium with the 
    radiation field in \hii\ regions.
    The latter grains are indeed hot enough to significantly emit at shorter 
    wavelengths than Solar-metallicity objects.
  \item[AGNs,] such as Centaurus$\,$A (red spectrum in \reffig{fig:specMIR}),
    have a rather flat \hNIR-to-\hMIR\ continuum 
    \citep[\eg][]{laurent00,spoon07}.
    This continuum originates in the strong gradient of temperature due to the 
    central illumination by the \hAGN\ (this is the same qualitative effect as 
    in our \hMCRT\ simulation; \refsec{sec:MCRT}).
    In particular, temperatures of the underlying continuum can reach higher 
    values in the central region, than normal galaxies.
    It explains why the \hNIR\ continuum does not drop to zero.
    These central hot regions produce a bright \hMIR\ continuum, attenuated by
    the entire disk, resulting in deep silicate absorption bands.
\end{description}

\begin{figure}[htbp]
  \includegraphics[width=\textwidth]{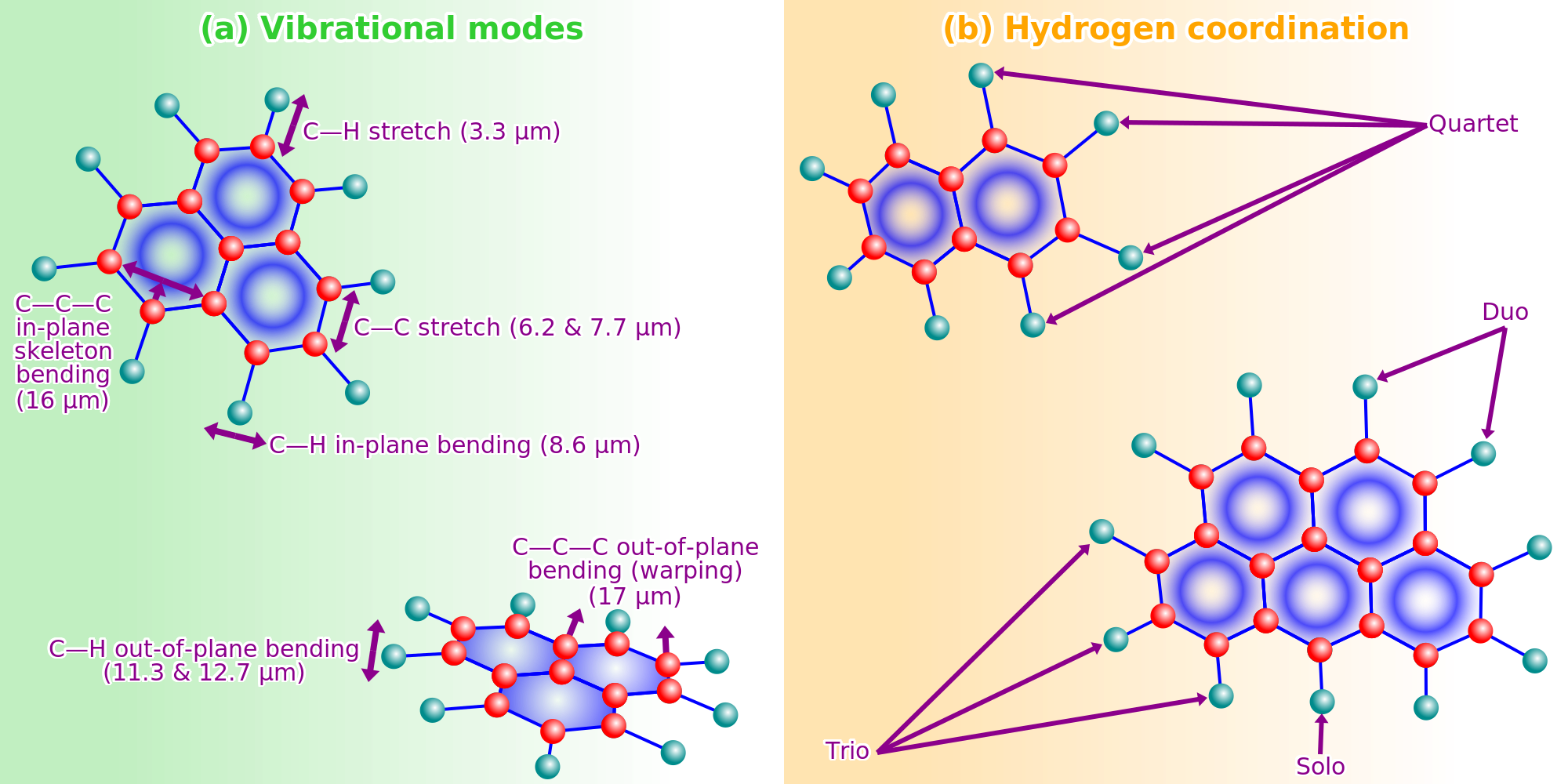}
  \newcap{Vibrational modes of PAHs}%
         {Panel~\textit{(a)} represents the main in-plane and out-of-plane
          vibrational modes.
          Panel~\textit{(b)} gives examples of solo, duo, trio and quartet
          H sites.
          In both panels, red spheres represent C atoms, and cyan spheres, H 
          atoms.
          \CClicence}
  \label{fig:modesPAH}
\end{figure}

\begin{figure}[htbp]
  \includegraphics[width=\textwidth]{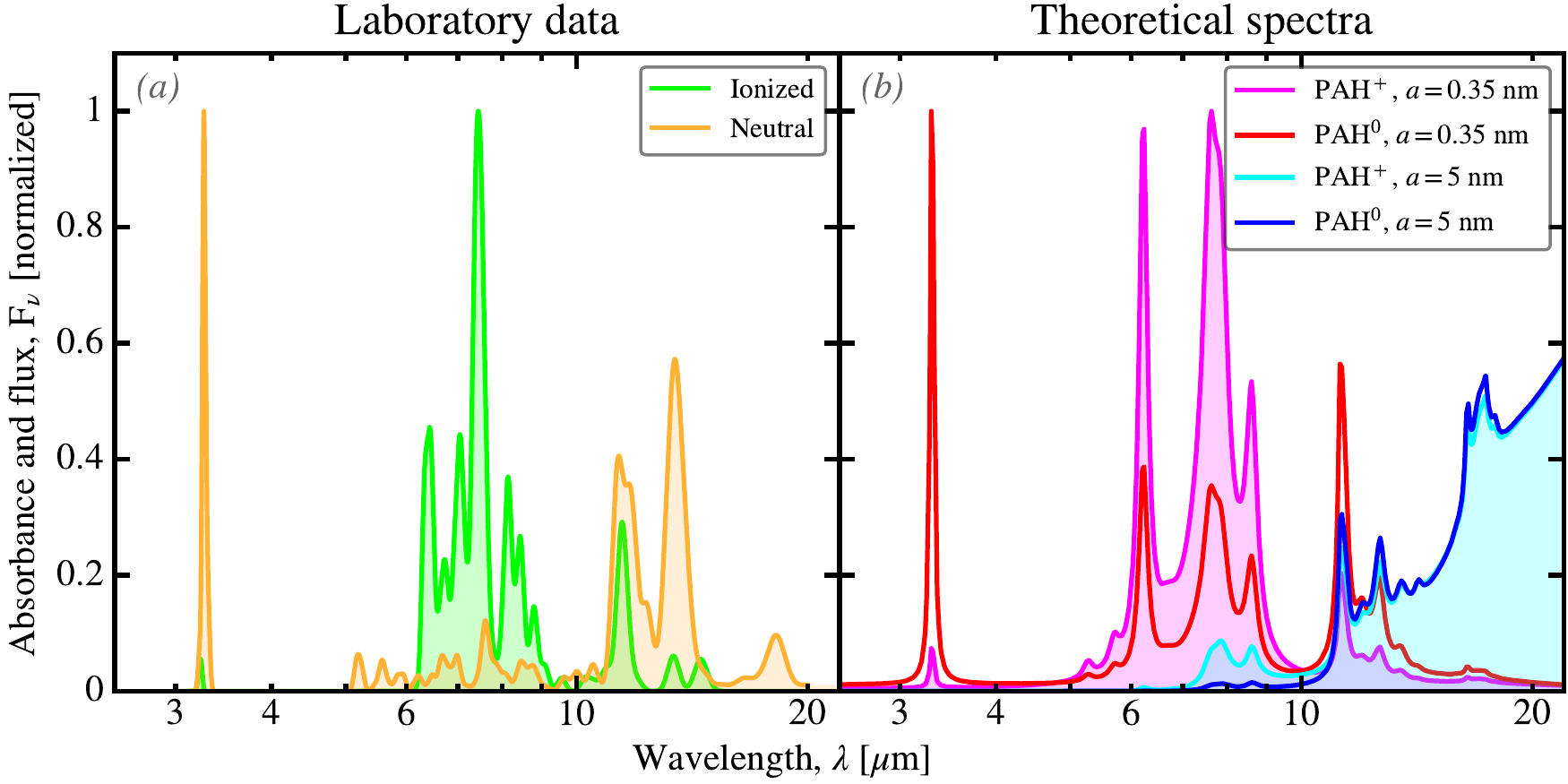}
  \newcap{Laboratory and theoretical PAH spectra}%
         {Panel~\textit{(a)} shows the absorption coefficient of neutral and 
          cationic \hPAH s measured in the laboratory by \citet{allamandola99}.
          Panel~\textit{(b)} shows theoretical emission spectra of neutral and 
          cationic \hPAH s, with radii $a=0.35$~nm and $a=5$~nm.
          We have used the \citet{draine07} optical properties and computed the
          stochastic heating for the Solar neighborhood \hISRF\ ($U=1$).
          \CClicence}
  \label{fig:ionPAH_lab}
\end{figure}

\paragraph{Laboratory and theoretical PAH physics.}
Although we do not know the exact composition of the interstellar carbon grain mixture responsible for the aromatic feature emission, the brightest bands have been attributed to the main vibrational modes of \hPAH s.
There are still some debates about the origin of the weakest features \citep[\cf\ \eg][]{allamandola99,verstraete01,tielens08,boersma10,jones13}.
In \reffig{fig:specMIR}, we have labeled the different bands with a given mode.
These modes are schematically represented in \refsubfig{fig:modesPAH}{a}.
\begin{description}
  \item[The charge] of the molecules is one of the most important parameters
    controlling the ratio between the C--H and C--C bands.
    \refsubfig{fig:ionPAH_lab}{a} shows the laboratory data of
    \citet{allamandola99}.
    The two spectra are the sum of several neutral and cationic molecules.
    It is clear that C--C and C--C--C bands (6.2, 7.7 and 17~\tmic\ complexes) 
    are predominantly carried by ionized \hPAH s, whereas C--H bands (3.3, 11.3 
    and 12.7~\tmic\ complexes) are carried essentially by neutral \hPAH s.
  \item[Dehydrogenation] has a similar effect to ionization. 
    However, for \hPAH s with more than $\simeq25$ C atoms (\ie\ the bulk of 
    interstellar \hPAH s), hydrogenation 
    through reactions with abundant atomic H is more important than H loss
    through unimolecular dissociation \citep[\cf\ \eg][]{hony01}.
    Thus, dehydrogenation does not have a detectable effect on the \hUIB\ 
    spectrum.
  \item[The molecular structure] is another factor.
    C--H \expression{Out-Of-Plane} (\hOOP) bending modes have different 
    frequencies, depending on the number of H atom per aromatic cycle (\cf\ 
    \refsubfig{fig:modesPAH}{b}).
    The 11.3~\tmic\ band corresponds to a solo H, found on straight molecular 
    edges, whereas the 12.7~\tmic\ one corresponds to a trio, found on corners 
    of the molecules.
    The solo-to-trio intensity ratio, $I_{11.3}/I_{12.7}$, is thus an
    indicator of \hPAH\ compactness ($I_\lambda$ being the integrated intensity of
    the feature centered at $\lambda\emic$).
  \item[The size] of the \hPAH s affects the relative intensity of the different
    bands.
    This is demonstrated in \refsubfig{fig:ionPAH_lab}{b}.
    Small \hPAH s (magenta and red spectra) fluctuate up to temperatures 
    higher than large ones (blue and cyan spectra).
    Short-wavelength bands are therefore more pumped in small \hPAH s, whereas
    large \hPAH s emit predominantly long-wavelength features.
  \item[ISRF hardness] has an effect similar to the size, as a higher mean 
    photon energy causes the grain to fluctuate up to higher temperatures
    \citep[\refsec{sec:PAHband} and \eg][]{galliano08b}.
    We have however seen, in \reffig{fig:ISRFhardness}, that this effect was 
    probably less than a factor $\simeq2$, in astrophysically relevant cases.
\end{description}
\takeaway{The charge and size of the \hPAH s are the main parameters 
          controlling their emission spectrum.}

    \subsubsection{Spectral Decomposition Methods}
    \label{sec:decompMIR}

To study variations of the aromatic feature spectrum with environmental conditions, one needs to measure the intensity of each band.
This task is not as simple as it appears.
There are indeed several challenges.
\begin{itemize}
  \item Aromatic bands are broad features with large wings.
    They are usually modeled with Lorentz or Drude profiles 
    \citep[\eg][]{smith07,galliano08b,lai20}.
    Their intensity drops quite slowly, away from their peak.
    Consequently, deblending adjacent features and distinguishing features from
    the continuum is difficult.
  \item The exact number and spectral shape of the different aromatic features
    is not precisely known \citep[\eg][]{verstraete96,boulanger98}.
    High-spectral-resolution, high-signal-to-noise-ratio spectra allow us
    to identify up to $\simeq40$ individual bands, some being very weak.
    The assumption about the number and spectral properties of these bands
    therefore adds another layer of uncertainties.
  \item Absorption by silicates and ices introduce a degeneracy between 
    $A(V)$ and the band strength.
    In particular, unless the 9.8~\tmic\ band is deep enough to be 
    unambiguously constrained (such as in Cen~A; \reffig{fig:specMIR}), 
    the $8-12\emic$ region can equivalently be modeled with:
    \begin{inlinelist}
      \item 8.6~\tmic\ and 11.3~\tmic\ bands and no extinction; or 
      \item weaker 8.6~\tmic\ and 11.3~\tmic\ bands, a higher underlying 
        continuum and some extinction.
    \end{inlinelist}
\end{itemize}
\hMIR\ spectral decomposition methods are therefore an essential tool to properly study \hPAH\ features.

\begin{figure}[htbp!]
  \begin{tabular}{cc}
    \includegraphics[width=0.69\textwidth]{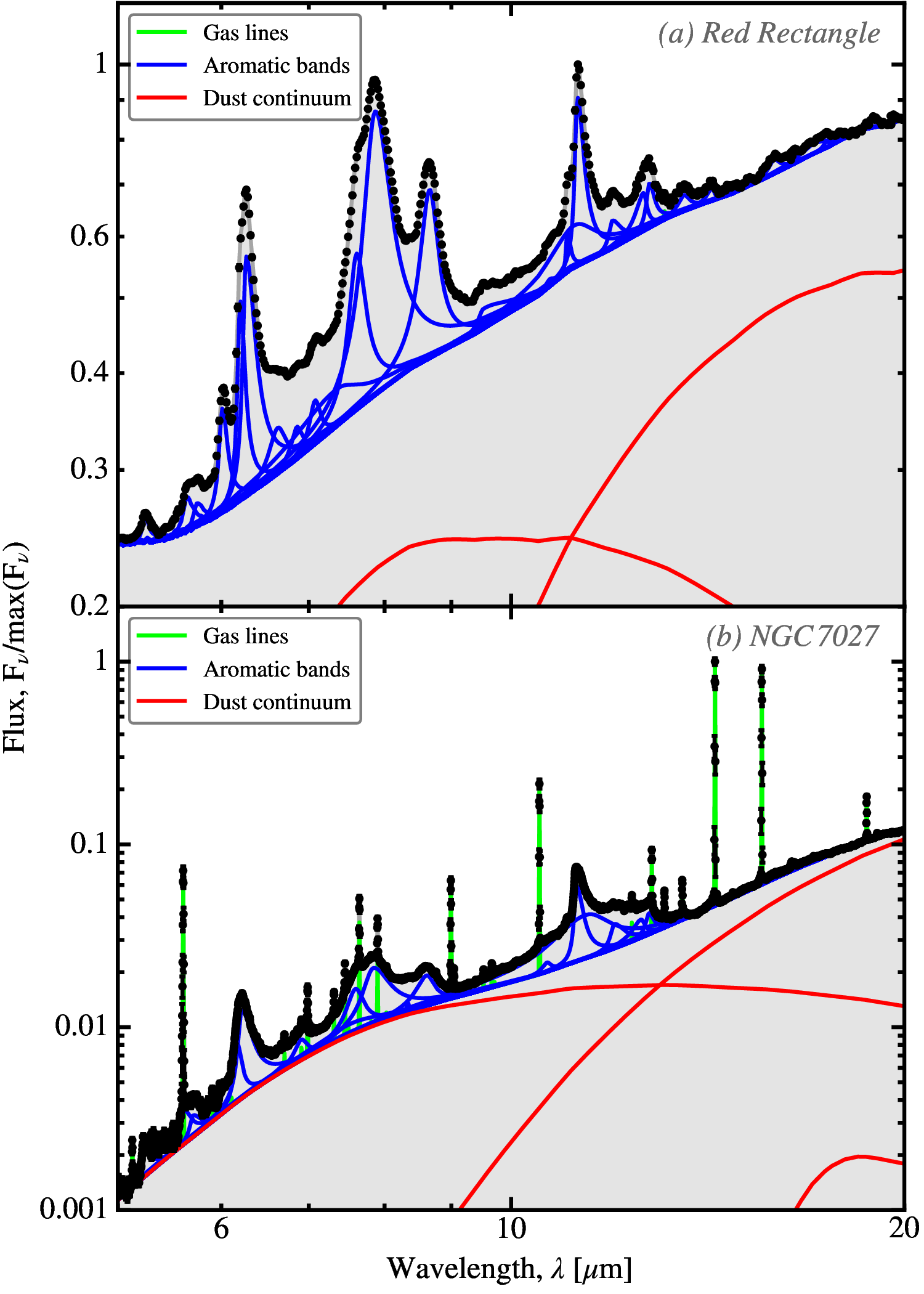} 
    & 
    \includegraphics[width=0.275\textwidth]{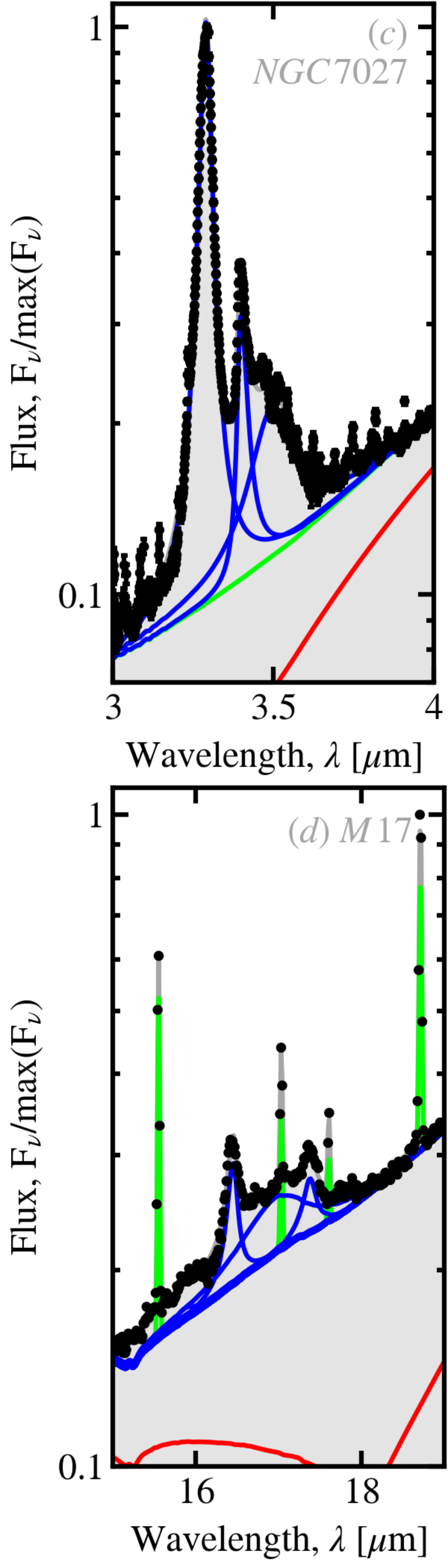} \\
  \end{tabular}
  \newcap{Empirical calibration of UIB profiles}%
         {In each panel, the black dots with error bars (barely visible)
          represent the observations.
          These are \hISO/SWS spectra except for \M{17}, which is a \hspitz/IRS
          spectrum.
          The fitted model is the sum of the different individual components: 
          \begin{inlinelist}
            \item grey bodies for the dust continuum;
            \item Gaussian profiles for gas lines; and
            \item split-Lorentz profiles \refeqp{eq:splitlorentz} for \hUIB s.
          \end{inlinelist}
          \CClicence}
  \label{fig:caliband}
\end{figure}

\paragraph{Calibrating feature properties.}
Since the properties of the different \hUIB s are not \textit{a priori} known, we need to empirically infer them.
In our work, we parametrize the band spectral profile with a \expression{split-Lorentz function} \citep{hu21}:
\begin{equation}
  F_\nu = I\times\left\{
  \begin{array}{ll}
    \displaystyle\frac{2}{\pi}\frac{\Delta\nu_s^2}{\Delta\nu_s+\Delta\nu_l}
      \frac{1}{(\nu-\nu_0)^2+\Delta\nu_s^2} & \mbox{ for } \nu\ge\nu_0 \\
    \displaystyle\frac{2}{\pi}\frac{\Delta\nu_l^2}{\Delta\nu_s+\Delta\nu_l}
      \frac{1}{(\nu-\nu_0)^2+\Delta\nu_l^2} & \mbox{ for } \nu<\nu_0, \\
  \end{array}
  \right. 
  \label{eq:splitlorentz}
\end{equation}
where $\nu_0$ is the central frequency of the feature, and $\Delta\nu_s$ and $\Delta\nu_l$ are its widths on the short- and long-wavelength sides.
Having an asymmetric feature is indeed necessary to accurately fit good quality spectra.
This asymmetry may originate in the anharmonicity of the transition responsible for the band, or may be due to unresolved blended features.
The parameters characterizing each individual features, $\nu_0$, $\Delta\nu_s$ and $\Delta\nu_l$, could be derived from each fit.
However, most of them would be quite uncertain, using an average \hspitz\ spectrum.
For that reason, we have calibrated these parameters (\ie\ inferred their reference value), using high-resolution, high-signal-to-noise-ratio spectra of Galactic regions.
This calibration is demonstrated in \reffig{fig:caliband}.
\begin{enumerate}
  \item We have first fitted the \hISO/SWS spectrum of the reflection nebula, 
    the \expression{Red Rectangle}, in order to calibrate the narrow bands
    (\cf\ \refsubfig{fig:caliband}{a}).
    The very wide bands, also called \expression{plateaus}, are not very 
    prominent in this region.
  \item We then fix the narrow band parameters and infer the 
    $\lambda\simeq8\emic$ and $\lambda\simeq12\emic$ plateau parameters
    by fitting the \hISO/SWS spectrum of the planetary nebula, \ngc{7027}
    (\cf\ \refsubfig{fig:caliband}{b}).
    The $\lambda\simeq17\emic$ complex is too weak to be calibrated using this 
    spectrum.
  \item The $\lambda\simeq17\emic$ complex is calibrated by fitting the \hIRS\
    spectrum of the \hPDR, \M{17} (\cf\ \refsubfig{fig:caliband}{c}).
  \item The aromatic and aliphatic bands in the $\lambda\simeq3\emic$ region
    are calibrated by fitting the $\lambda=3-5\emic$ spectrum of \ngc{7027}
    (\cf\ \refsubfig{fig:caliband}{d}).
    The continuum in this range is indeed clean and there are no other blended
    bands.
\end{enumerate}
\reftab{tab:caliband} gives the resulting band parameters.
With these parameters fixed, we can fit even low-signal-to-noise-ratio spectra varying only the intensity of each band (parameter $I$ in \refeqnp{eq:splitlorentz}).
\begin{table}[htbp]
  \centering
  \setlength\arrayrulewidth{2pt}
  \arrayrulecolor{white}
  \begin{tabularx}{\linewidth}{|>{\columncolor{coltabcell}}X%
                                |>{\columncolor{coltabcell}}X%
                                |>{\columncolor{coltabcell}}X%
                                |>{\columncolor{coltabcell}}X|}
    \hline
      \rowcolor{coltabhead}
      $\bm{\lambda_0}$ & $\bm{\Delta\lambda_s}$ 
      & $\bm{\Delta\lambda_l}$ & \textbf{Type} \\
    \hline
      \rowcolor{coltabsep}
      3.291 \tmic & 0.020 \tmic & 0.019 \tmic & Main \\
    \hline
      \rowcolor{coltabsep}
      3.399 \tmic & 0.011 \tmic & 0.024 \tmic & Main \\
    \hline
      3.499 \tmic & 0.077 \tmic & 0.071 \tmic & Small \\
    \hline
      5.239 \tmic & 0.025 \tmic & 0.058 \tmic & Small \\
    \hline
      5.644 \tmic & 0.040 \tmic & 0.080 \tmic & Small \\
    \hline
      5.749 \tmic & 0.040 \tmic & 0.080 \tmic & Small \\
    \hline
      6.011 \tmic & 0.040 \tmic & 0.067 \tmic & Small \\
    \hline
      \rowcolor{coltabsep}
      6.203 \tmic & 0.031 \tmic & 0.060 \tmic & Main \\
    \hline
      \rowcolor{coltabsep}
      6.267 \tmic & 0.037 \tmic & 0.116 \tmic & Main \\
    \hline
      6.627 \tmic & 0.120 \tmic & 0.120 \tmic & Small \\
    \hline
      6.855 \tmic & 0.080 \tmic & 0.080 \tmic & Small \\
    \hline
      7.079 \tmic & 0.080 \tmic & 0.080 \tmic & Small \\
    \hline
      \rowcolor{coltabsep}
      7.600 \tmic & 0.480 \tmic & 0.502 \tmic & Plateau \\
    \hline
      \rowcolor{coltabsep}
      7.617 \tmic & 0.119 \tmic & 0.145 \tmic & Main \\
    \hline
      \rowcolor{coltabsep}
      7.870 \tmic & 0.170 \tmic & 0.245 \tmic & Main \\
    \hline
      8.362 \tmic & 0.016 \tmic & 0.016 \tmic & Small \\
    \hline
      \rowcolor{coltabsep}
      8.620 \tmic & 0.183 \tmic & 0.133 \tmic & Main \\
    \hline
      9.525 \tmic & 0.107 \tmic & 0.600 \tmic & Small \\
    \hline
      10.707 \tmic & 0.100 \tmic & 0.100 \tmic & Small \\
    \hline
      11.038 \tmic & 0.027 \tmic & 0.073 \tmic & Small \\
    \hline
      \rowcolor{coltabsep}
      11.238 \tmic & 0.053 \tmic & 0.153 \tmic & Main \\
    \hline
      \rowcolor{coltabsep}
      11.400 \tmic & 0.720 \tmic & 0.637 \tmic & Plateau \\
    \hline
      11.796 \tmic & 0.021 \tmic & 0.021 \tmic & Small \\
    \hline
      11.950 \tmic & 0.080 \tmic & 0.222 \tmic & Small \\
    \hline
      \rowcolor{coltabsep}
      12.627 \tmic & 0.200 \tmic & 0.095 \tmic & Main \\
    \hline
      \rowcolor{coltabsep}
      12.761 \tmic & 0.081 \tmic & 0.140 \tmic & Main \\
    \hline
      13.559 \tmic & 0.160 \tmic & 0.161 \tmic & Small \\
    \hline
      14.257 \tmic & 0.152 \tmic & 0.059 \tmic & Small \\
    \hline
      15.893 \tmic & 0.178 \tmic & 0.200 \tmic & Small \\
    \hline
      16.483 \tmic & 0.100 \tmic & 0.059 \tmic & Small \\
    \hline
      \rowcolor{coltabsep}
      17.083 \tmic & 0.496 \tmic & 0.562 \tmic & Plateau \\
    \hline
      17.428 \tmic & 0.100 \tmic & 0.100 \tmic & Small \\
    \hline
      17.771 \tmic & 0.031 \tmic & 0.075 \tmic & Small \\
    \hline
      18.925 \tmic & 0.037 \tmic & 0.116 \tmic & Small \\
    \hline
  \end{tabularx}
  \newcap{UIB profile parameters}%
         {These are the parameters of \refeq{eq:splitlorentz}.
          We have converted the frequencies in wavelengths: 
          $\nu_0\equiv c/\lambda_0$, 
          $\Delta\nu_s\equiv c/(\lambda_0-\Delta\lambda_s)-c/\lambda_0$,
          $\Delta\nu_l\equiv c/\lambda_0-c/(\lambda_0+\Delta\lambda_l)$.
          These are the parameters used for the work of \citetprep{hu21}.}
  \label{tab:caliband}
\end{table}

\begin{figure}[htbp]
  \includegraphics[width=\textwidth]{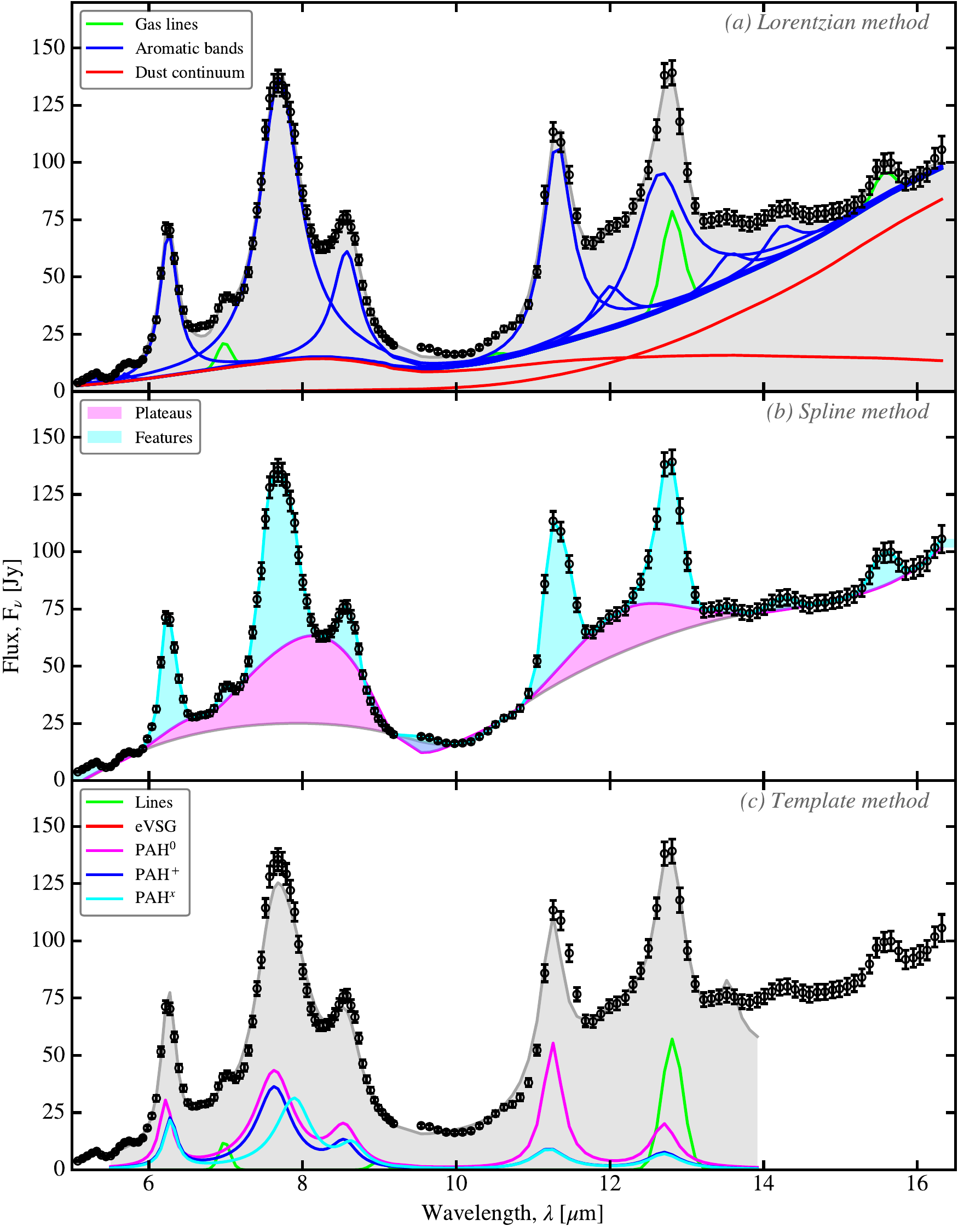}
  \newcap{MIR spectral fitting methods}%
         {Demonstration of the three \hMIR\ fitting methods, described in this 
          manuscript, on the total \hISO/CAM spectrum of the starburst galaxy,
          \M{82} \citep{galliano08b}.
          The observations (the black error bars) are identical in each panel.
          The models of panels~\textit{(a)} and \textit{(c)} are least-squares 
          fits, whereas panel~\textit{(b)} is a spline interpolation.
          Panels~\textit{(a)} and \textit{(b)} are from \citet{galliano08b}.
          The fit of panel~\textit{(c)} has been performed only up to 
          $\lambda=14\emic$, using \ncode{PAHtat} \citep{pilleri12}.
          In addition to the four components of \reffig{fig:PAHtat}, 
          a continuum and the main gas lines are simultaneously fitted.
          \CClicence}
  \label{fig:fit_MIRspec}
\end{figure}

\paragraph{Fitting every feature at once.}
The total \hMIR\ spectrum contains the emission from different physical processes that have to be separated in order to accurately measure \hUIB\ intensities (\cf\ \reffig{fig:specMIR}).
Several models have tackled this problem since the \hISO\ days \citep[\eg][]{verstraete96,boulanger98,laurent00,madden06,smith07,galliano08b,mori12,lai20}.
\begin{description}
  \item[Individual UIBs] can be fitted with Lorentz profiles   
    \citep[\eg][]{galliano08b}, Drude profiles \citep[\eg][]{smith07} or
    split-Lorentz profiles (\cf\ \reftab{tab:caliband} and 
    \reffig{fig:caliband}; \eg\ \citeprep{hu21}).
  \item[The dust continuum] can be fitted with a linear combination of grey 
    bodies.
    The opacity of these grey bodies can be a simple \hMBB\ power-law  
    \citep[\eg][]{smith07} or the opacity of realistic materials (\eg\ 
    \citeprep{hu21}).
    This dust continuum likely originates in a combination of small, 
    stochastically-heated grains, and large, hot equilibrium grains.
    The fitted parameters (mass and temperature) of each grey body therefore 
    have to be considered as nuisance variables.
    They can not be interpreted physically, because of the uncertainty of their
    heating mechanism.
    They are employed only to decompose the continuum and \hUIB\ emissions.
  \item[Gas lines] can be fitted with Gaussian profiles. 
    \reftab{tab:MIRlines} lists the most prominent \hMIR\ gas lines and their 
    central wavelengths.
    The width of the profile is determined by the spectral resolution of the 
    instrument, except at very high spectral resolution
    ($R\equiv\lambda/\Delta\lambda\gtrsim30\,000$), where the true width of the
    line can be resolved.
  \item[The stellar continuum] only contributes at short wavelength, and can
    be modeled with a Rayleigh-Jeans law.
  \item[The extinction,] which is essentially absorption at these wavelengths,
    can be modeled with a general extinction curve (\cf\ \eg\ 
    \reffig{fig:kappaSED}), assuming a simple geometry (\cf\ \eg\ 
    \refsec{sec:eqRT}).
    The most prominent dust features are the two silicate bands at 9.8 and 
    18~\tmic\ (\cf\ \reffig{fig:specMIR}).
    Ice absorption can also be non negligible.
    They can be taken into account the same way, as icy band profiles are 
    well-constrained (\cf\ \refsubfig{fig:MIRext}{b}).
    The optical depth ratio between different ice species and the silicates 
    however varies from one source to the other \citep[\eg][]{yamagishi15}.
    The optical depth therefore has to be independently estimated for each
    species (\ie\ one needs to derive $\tau_\sms{dust}$, $\tau_\sms{\hmol O}$, 
    $\tau_\sms{CO}$, \etc).
\end{description}
\refsubfig{fig:fit_MIRspec}{a} demonstrates such a fitting method on the total spectrum of \M{82} \citep{galliano08b}.
It is labeled as the \expression{Lorentzian method}, as the \hUIB s are modeled with Lorentz profiles.
This earlier model did not use all the bands given in \reftab{tab:caliband}, that we are now using.
\begin{table}[htbp]
  \centering
  \setlength\arrayrulewidth{2pt}
  \arrayrulecolor{white}
  \begin{tabularx}{\linewidth}{|>{\columncolor{coltabcell}\hsize=\hsize}X
                                |>{\columncolor{coltabcell}\hsize=\hsize}X
                                |>{\columncolor{coltabcell}\hsize=\hsize}X
                                |>{\columncolor{coltabcell}\hsize=\hsize}X|}
    \hline
      \rowcolor{coltabhead}
      \textbf{Central wavelength} & \textbf{Species} & \textbf{Transition} 
        & \textbf{Type} \\
    \hline
      4.052 \tmic & \hi & Brackett $\alpha$ & recombination \\
    \hline
      \rowcolor{coltabsep}
      5.511 \tmic & \hmol & 0--0 S(7) & ro-vibrational \\
    \hline
      5.908 \tmic & \hi & Humphreys $\gamma$ & recombination \\
    \hline
      \rowcolor{coltabsep}
      6.109 \tmic & \hmol & 0--0 S(6) & ro-vibrational \\
    \hline
      \rowcolor{coltabsep}
      6.910 \tmic & \hmol & 0--0 S(5) & ro-vibrational \\
    \hline
      6.985 \tmic & \arii & 2P$^{3/2}$--2P$^{1/2}$ & forbidden \\
    \hline
      7.460 \tmic & \hi & Pfund $\alpha$ & recombination \\
    \hline
      7.502 \tmic & \hi & Humphreys $\beta$ & recombination \\
    \hline
      \rowcolor{coltabsep}
      8.025 \tmic & \hmol & 0--0 S(4) & ro-vibrational \\
    \hline
      8.991 \tmic & \ariii & 3P$^2$--3P$^1$ & forbidden \\
    \hline 
      \rowcolor{coltabsep}
      9.665 \tmic & \hmol & 0--0 S(3) & ro-vibrational \\
    \hline 
      10.511 \tmic & \siv & 2P$^{3/2}$--2P$^{1/2}$ & forbidden \\
    \hline
      \rowcolor{coltabsep}
      12.279 \tmic & \hmol & 0--0 S(2) & ro-vibrational \\
    \hline
      12.369 \tmic & \hi & Humphreys $\alpha$ & recombination \\
    \hline
      12.814 \tmic & \neii & 2P$^{3/2}$--2P$^{1/2}$ & forbidden \\
    \hline
      15.555 \tmic & \neiii & 3P$^2$--3P$^1$ & forbidden \\
    \hline
      \rowcolor{coltabsep}
      17.035 \tmic & \hmol & 0--0 S(1) & ro-vibrational \\
    \hline
      18.713 \tmic & \siii & 3P$^2$--3P$^1$ & forbidden \\
    \hline
      21.829 \tmic & \ariii & 3P$^1$-3P$^0$ & forbidden \\
    \hline
      25.890 \tmic & \oiv & 2P$^{3/2}$--2P$^{1/2}$ & forbidden \\
    \hline
      \rowcolor{coltabsep}
      28.219 \tmic & \hmol & 0--0 S(0) & ro-vibrational \\
    \hline
      33.481 \tmic & \siii & 3P$^1$--3P$^0$ & forbidden \\
    \hline
      34.815 \tmic & \siII & 2P$^{3/2}$--2P$^{1/2}$ & forbidden \\
    \hline
      35.349 \tmic & \feii & 6D$^{5/2}$--6D$^{7/2}$ & forbidden \\
    \hline 
      36.014 \tmic & \neiii & 3P$^1$--3P$^0$ & forbidden \\
    \hline
  \end{tabularx}
  \newcap{Most prominent MIR gas lines}{}
  \label{tab:MIRlines}
\end{table}

\begin{figure}[htbp]
  \includegraphics[width=\textwidth]{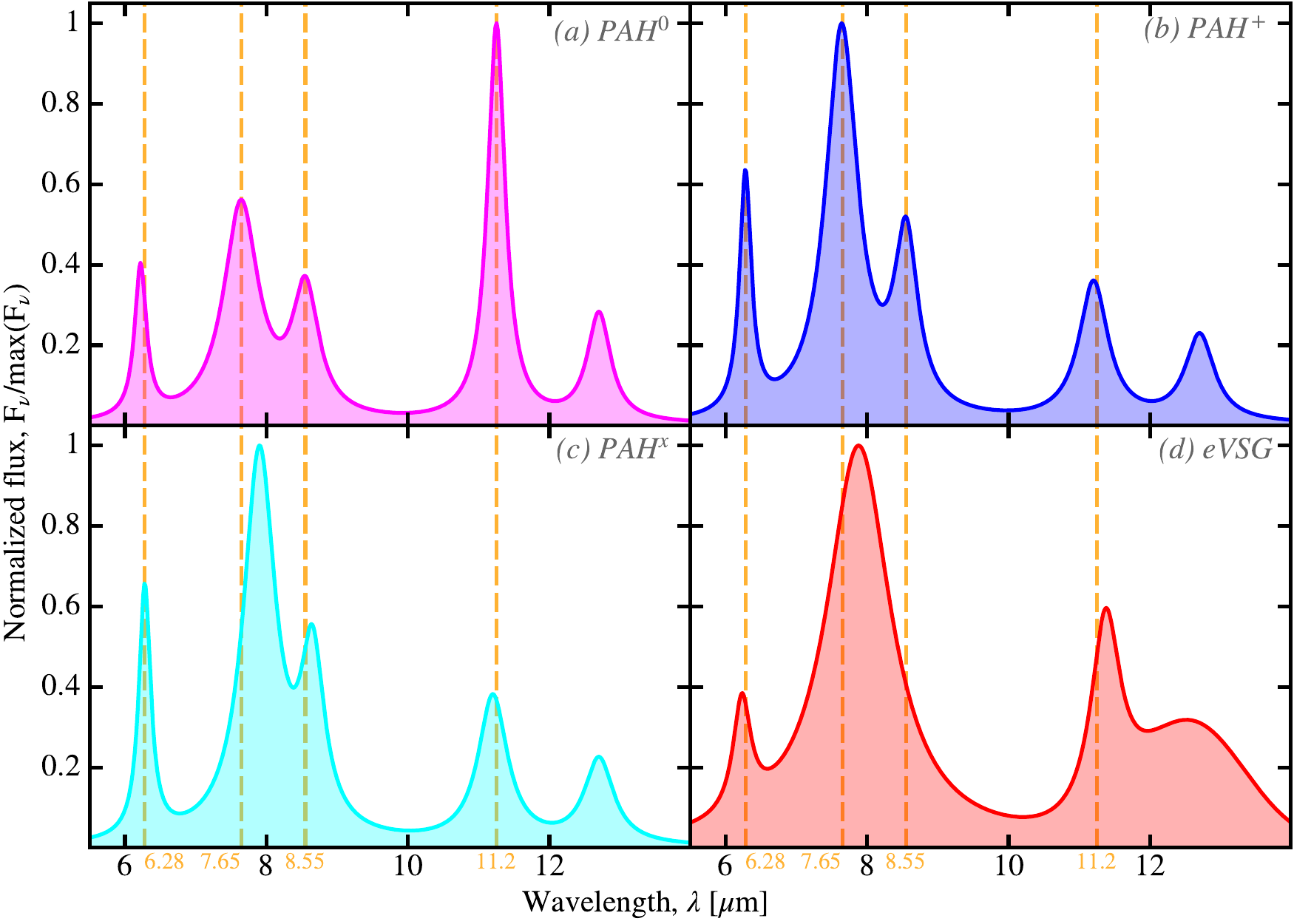}
  \newcap{PAH and small grain templates}%
         {We show the four components of \ncode{PAHtat} \citep{pilleri12}.
          Vertical yellow dashed lines indicate the same wavelengths in the four
          panels, helping visualizing the band shifts of some components.
          \CClicence}
  \label{fig:PAHtat}
\end{figure}

\paragraph{Alternative methods.}
Several other \hMIR\ spectral fitting methods have been dicussed in the literature.
The two following ones are worth mentioning.
\begin{description}
  \item[The spline method] consists in interpolating the spectrum under the main
    band complexes, using spline functions 
    \citep[\eg][]{vermeij02,galliano08b}.
    It is demonstrated in \refsubfig{fig:fit_MIRspec}{b}.
    A first spline interpolation defines the continuum and a second one defines 
    the band plateaus.
    The areas between the two continua (magenta filled) represent the plateaus
    in \refsubfig{fig:fit_MIRspec}{b}.
    The bands are then simply the intensity above the plateaus (cyan filled).
    The advantages of this method are: 
    \begin{inlinelist}
      \item it is simple to implement;
      \item it is very fast to run; and
      \item it does not suffer from the degeneracies between blended features, 
        or between band wings and the continuum.
    \end{inlinelist}
    It has however several limitations: 
    \begin{inlinelist}
      \item the choice of the spline anchor points is arbitrary, which  
        results in systematic differences with other methods;
      \item at medium-spectral resolution (typical of \hISO/CAM or \hspitz/IRS 
        SL-LL modes) it is impossible to deblend features and lines, such as
        the 12.7~\tmic\ \hUIB\ and the \neiiline\ line; and
      \item this method does not provide meaningful uncertainty estimates.
    \end{inlinelist}
    \citet{galliano08b} performed a systematic comparison of the Lorentzian and
    spline methods, and found that, although band intensities were different, 
    the trends between band intensities or band ratios were consistent with 
    the two methods.
  \item[The template method] consists in fitting a small number of synthetic
    spectra characteristics of different regions or different species.
    We have demonstrated this method in \refsubfig{fig:fit_MIRspec}{c} using
    \ncode{PAHtat} \citep{pilleri12}.
    This method uses four main components.
    These components were extracted from a sample of Galactic \hPDR s and
    \hPN e, using blind-signal separation methods \citep{berne07,joblin08}.
    In that sense, they are empirical, synthetic small grain spectra.
    These components are the following.
    \begin{description}
      \item[Neutral PAHs] (PAH$^0$) are shown in \refsubfig{fig:PAHtat}{a}.
        Their brightest C--C band is centered at $\lambda\simeq7.65\emic$.
      \item[Ionized PAHs] (PAH$^+$) are shown in \refsubfig{fig:PAHtat}{b}.
        They have similar band centers as PAH$^0$, but different intensity
        ratios, as we have previously seen (\cf\ \reffig{fig:ionPAH_lab}).
        Both PAH$^0$ and PAH$^+$ have similar spectral characteristics as
        class $\mathcal{A}$ spectra of \citet{peeters02b}, found in \hii\ 
        regions, and thought to be \citengl{processed} \hPAH s.
      \item[Large ionized PAHs] (PAH$^x$), shown in \refsubfig{fig:PAHtat}{c}, 
        have a peculiar redshifted C--C band at $\lambda\simeq7.9\emic$,
        observed in \hPN e by \citet{joblin08}.
        They are assumed to be $\simeq100$~C atom grains.
        Those are reminiscent of the class $\mathcal{B}$ spectra of 
        \citet{peeters02b}.
      \item[Evaporating VSGs] (\heVSG s), shown in \refsubfig{fig:PAHtat}{d},
        are thought to be PAH clusters of $\simeq500$~C atoms.
        Their spectra have the characteristics of class $\mathcal{C}$ of 
        \citet{peeters02b} seen in post-\hAGB\ stars, with a broad 7.9~\tmic\
        band and no 8.6~\tmic\ feature.
        \citet{peeters02b} conjectured these could be \citengl{freshly-formed}
        carbon grains.
        The carrier of the broad 7.9~\tmic\ band could be destroyed by the 
        ionizing radiation, during its journey in the \hISM\ 
        \citep[\eg][]{joblin08}.
        In this scenario, \hPAH s are formed by the destruction of \hVSG s at
        the \hUV-illuminated edges of molecular clouds.
        Class $\mathcal{C}$ is not well studied. 
        Some novae show spectra similar to Class $\mathcal{C}$, and could be 
        due to nitrogen impurities in hydrocarbon compounds \citep{endo21}.
    \end{description}
\end{description}
The advantages of this method are that:
\begin{inlinelist}
  \item fits are not extremely degenerate, even at low signal-to-noise ratios,
    because of the small number of free parameters, compared to the Lorentzian
    method;
  \item the four different classes are physically meaningful, providing a clear
    interpretation of the results.
\end{inlinelist}
However, its lack of flexibility prevents accurate fits, that could overlook new information present in the observations.

    \subsubsection{PAH Band Ratio Studies}
    \label{sec:PAHband}

Applying a spectral decomposition method to a set of \hMIR\ spectra allows us to study the variations of several band ratios that contain physical 
information about the small carbon grain properties.

\begin{figure}[htbp]
  \includegraphics[width=\textwidth]{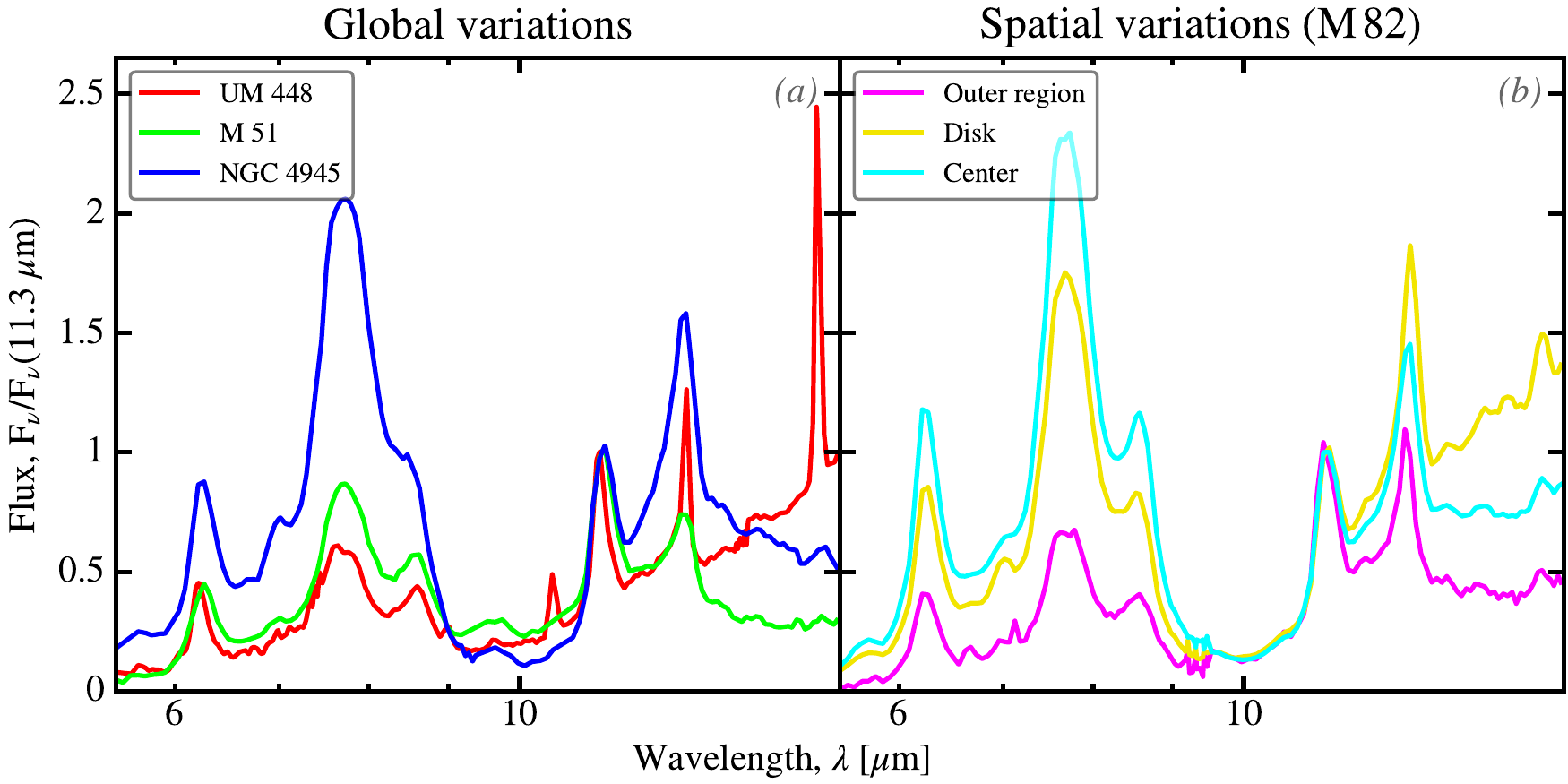}
  \newcap{Diversity of MIR spectra among and within galaxies}%
         {Panel~\textit{(a)} shows the integrated spectra of three galaxies:
          \begin{inlinelist}
            \item \um{448}, a \hBCD;
            \item \M{51}, a prototypical \hLTG; and
            \item \ngc{4945}, a \hLIRG.
          \end{inlinelist}
          Panel~\textit{(b)} shows the spectra of three different regions within
          the starbursting irregular galaxy, \M{82}.
          In both panels, the monochromatic flux is normalized to its value at
          11.3~\tmic, in order to demonstrate band ratio variations.
          These are \hISO\ spectra \citep{galliano08b}.
          \CClicence}
  \label{fig:ionPAH_obs}
\end{figure}

\begin{figure}[htbp]
  \includegraphics[width=\textwidth]{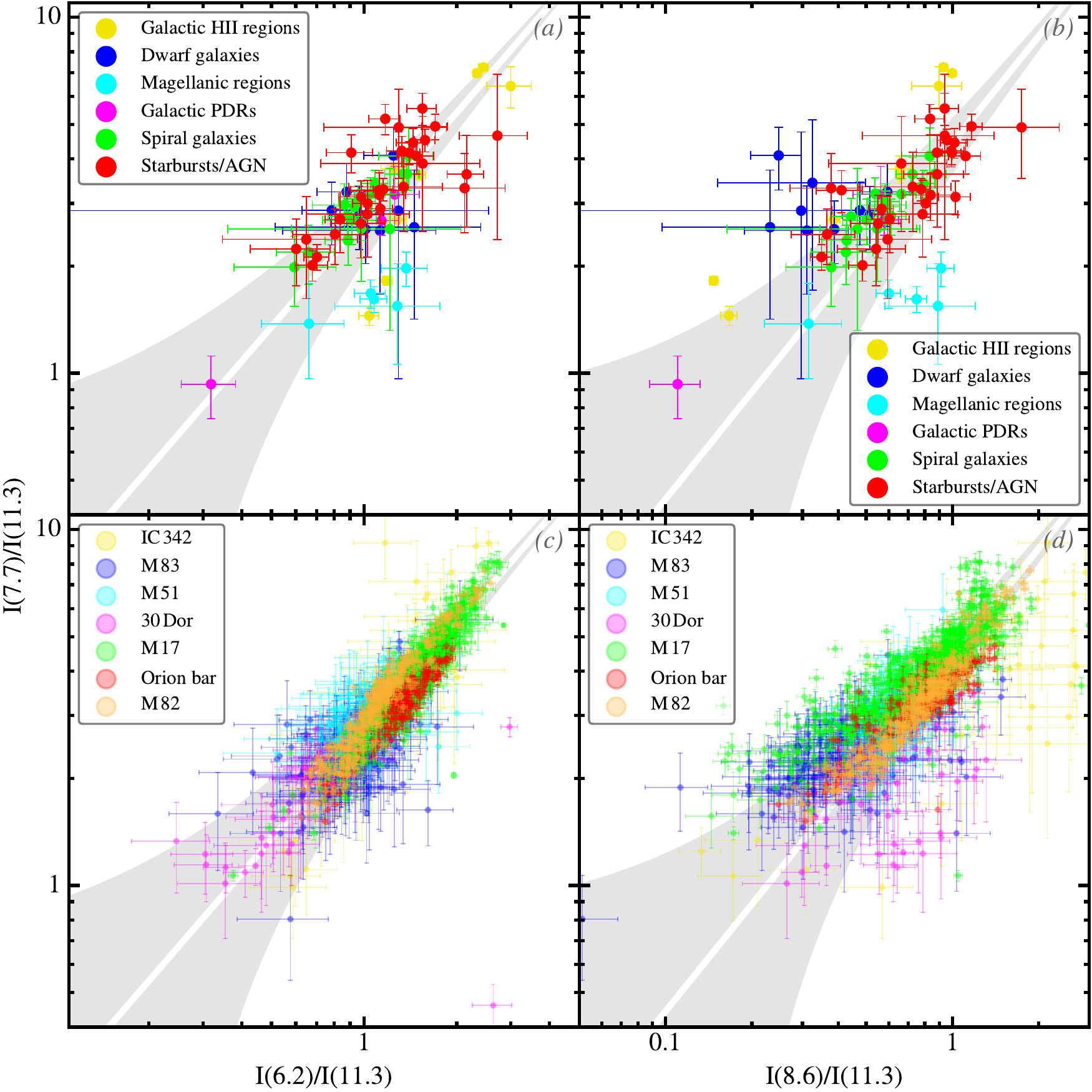}
  \newcap{PAH band ratio correlations inside and among galaxies}%
         {The top two panels show the values of the ratios derived from 
          integrated spectra of galaxies, as well as Galactic and Magellanic 
          regions.
          The bottom two panels show pixel distribution of the same ratios, 
          within a variety of objects.
          These data are a combination of \hISO/CAM and \hspitz/IRS spectra,
          fitted using the Lorentzian method \citep{galliano08b}.
          The grey-filled curves represent the $\pm1\sigma$ linear fit to the 
          integrated source correlations \citep[\cf\ Table~2 of][]{galliano08b}.
          The grey curves are identical in panels~\textit{(a)} and \textit{(c)}
          and in panels~\textit{(b)} and \textit{(d)}, to guide the eye when 
          comparing integrated and resolved correlations.
          \CClicence}
  \label{fig:corrband}
\end{figure}

\paragraph{Observed band ratios in galaxies.}
\reffig{fig:ionPAH_obs} demonstrates the diversity of spectra among galaxies (panel~\textit{a}) or within one (panel~\textit{b}).
This figure emphasizes the differences in terms of aromatic band intensity.
All these spectra are normalized by the 11.3-\tmic\ feature intensity.
Yet, they exhibit large variations of their 6-to-9-\tmic\ features.
This can be more precisely quantified by studying the correlation between specific band ratios, such as in \reffig{fig:corrband} \citep{galliano08b}.
The quantity $I(\lambda)$ is simply the intensity of the feature centered at $\lambda\emic$.
The bottom two panels show the results for integrated galaxies and Galactic regions, whereas the bottom two panels show the results for a few spatially-resolved sources.
Overall the trends are similar for both types.
It means that this is a multiscale relation, valid at sub-pc scales (within the Orion bar or \M{17}) and $\simeq10$~kpc scales (among integrated galaxies).
These three displayed band ratios span about an order of magnitude, and are linearly correlated with each other.
It implies that the 6.2, 7.7 and 8.6~\tmic\ features are tied together, while the 11.3~\tmic\ can vary independently.
This is what was illustrated in \reffig{fig:ionPAH_obs}.
The only parameter that can explain such a variation is the charge of the \hPAH s (\cf\ \reffig{fig:ionPAH_lab}).
\takeaway{The variation of the \hPAH\ charge can explain most of the \hUIB\ variations observed in the nearby Universe.}

\begin{figure}[htbp]
  \includegraphics[width=\textwidth]{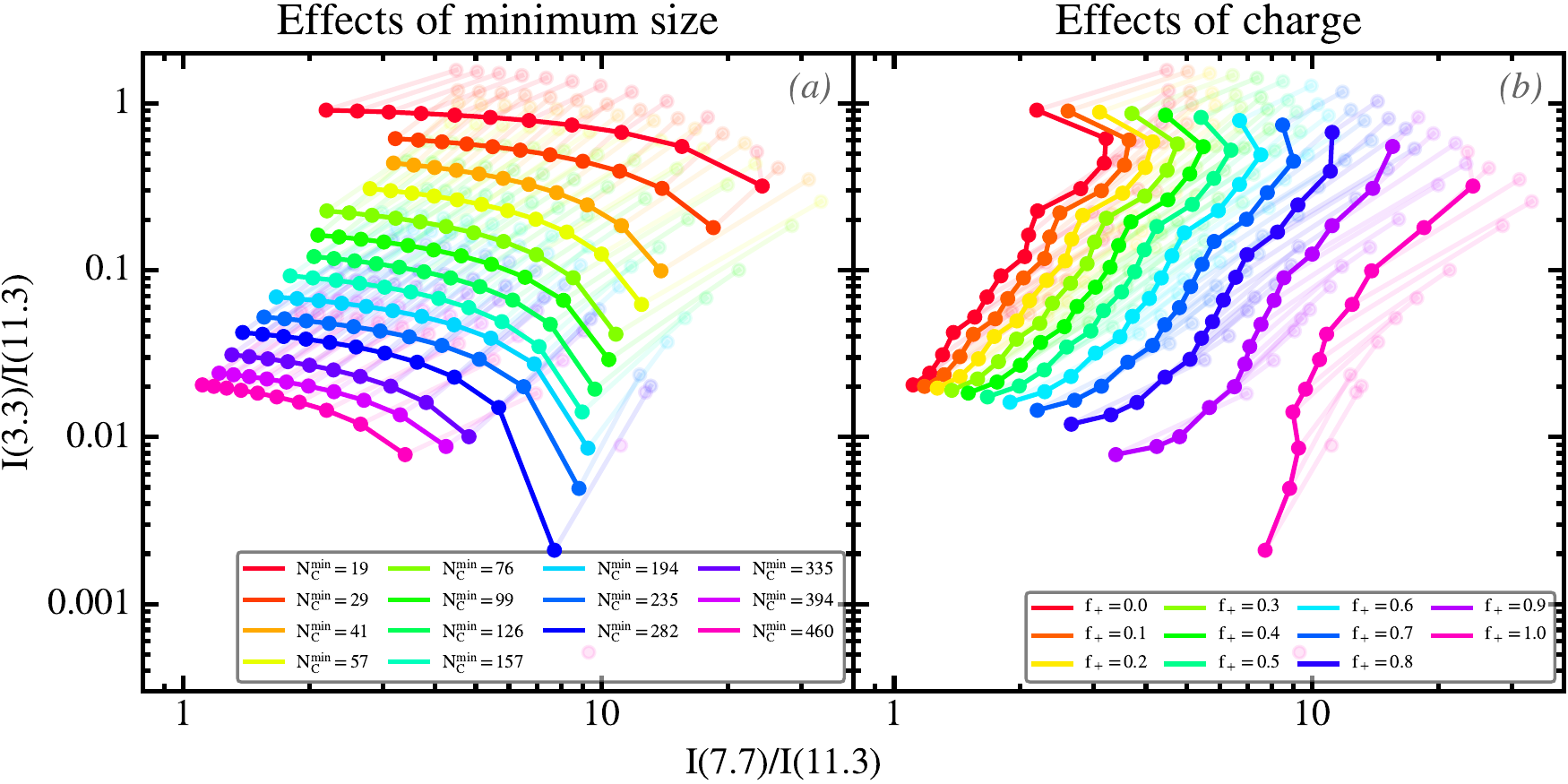}
  \newcap{Theoretical MIR band ratio variations}%
         {Both panels show the same grid points, but are highlighted 
          differently.
          Each circle is the theoretical band ratio estimated from the 
          stochastic emission spectrum of a PAH mixture having the 
          \citet{draine07} optical properties and the 
          \citet[][BARE-GR-S]{zubko04} size distribution.
          The solid color symbol are mixtures heated with the 
          Solar neighborhood \hISRF\ (\reffig{fig:ISRF}) with $U=1$, whereas 
          the pastel symbols are heated by a black body at $T=30\,000$~K with
          $U=10^4$.
          In panel~\textit{(a)}, we highlight the effect of changing the    
          minimum size cut-off, in number of C atoms, $N_\sms{C}^\sms{min}$.
          In panel~\textit{(b)}, we show the effect of the charge fraction, 
          $f_+$.
          \CClicence}
  \label{fig:bandMIR_grid}
\end{figure}

\paragraph{Effects of ionization and size.}
Although ionization is the main driver of the \hUIB\ relative variations in galaxies, other effects can play a role in specific environments.
The most important one of these secondary effects is the \hPAH\ size distribution.
\reffig{fig:bandMIR_grid} shows numerical simulations of several key \hUIB\ ratios \citep{hu21}.
We have varied:
\begin{inlinelist}
  \item the minimum \hPAH\ size expressed in number of C atoms, 
    $N_\sms{C}^\sms{min}$, highlighted in \refsubfig{fig:bandMIR_grid}{a};
  \item the \hPAH\ charge fraction, $f_+$, highlighted in 
    \refsubfig{fig:bandMIR_grid}{b};
  \item the ISRF intensity and hardness.
\end{inlinelist}
To illustrate the last effect, we have computed the model grid for the Solar neighborhood \hISRF\ \citep{mathis83}.
This values are the bright grid points.
We have also computed the grid for a hot star spectrum, with $U=10^4$ (the faint grid points).
The effect of the \hISRF\ is non negligible, but it is less drastic than that of the charge and size.
Most astrophysically relevant \hISRF s will be intermediate between the two extreme cases we have displayed.
\reffig{fig:bandMIR_grid} illustrates the following points \citep[see also][for a more complete calculation using \expression{Density Functional Theory}]{rigopoulou21}.
\begin{description}
  \item[$I(3.3)/I(11.3)$] is mainly a size tracer. 
    Both 3.3 and 11.3~\tmic\ bands (C--H modes) are indeed primarily emitted by 
    neutral \hPAH s.
    Charge thus does not significantly impact this ratio.
    The 3.3~\tmic\ feature is mainly carried by the smallest \hPAH s, whereas
    the 11.3~\tmic\ feature is carried by intermediate sizes.
  \item[$I(7.7)/I(11.3)$,] on the opposite, is essentially a charge tracer.
    It is still sensitive to the size, because of the relatively large 
    difference in wavelength of both features.
    The degeneracy due to this additional dependence can be broken, using
    $I(3.3)/I(11.3)$.
\end{description}
Such band ratio diagrams have been used to demonstrate systematic variations of the \hPAH\ size distribution in various environments.
We studied the spatial variations of $I(3.3)/I(11.3)$ in \ngc{1097} (\reffig{fig:specMIR}) and showed it was systematically lower in the central region, close to the \hAGN\ \citep{wu18a}.
The most likely explanation is that the hard radiation field from the central engine is selectively destroying the smallest \hPAH s.
This effect was also shown by \citet{smith07} and \citet{sales10}, on global scales.

\paragraph{The particular case of low-metallicity environments.}
In low-metallicity systems, band ratio variations can be more difficult to probe, as the band equivalent widths are lower, and thus more uncertain (\cf\ \refsec{sec:PAH2VSG}).
In the \hLMC, \citet{mori12} found different trends in neutral and ionized sightlines.
Toward the latter, there are evidences that \hPAH s have a lower charge (as a consequence of the higher recombination rate) and are on average larger (due to the destruction of the smallest \hPAH s).
In contrast, in the \hSMC, \citet{sandstrom12} found very weak $I(6-9)/I(11.3)$ ratios and weak 8.6 and 17~\tmic\ bands, implying small weakly ionized \hPAH s.
This last point is consistent with the trend of $I(17)/I(11.3)$ with $12+\log(\textnormal{O/H})$ found by \citet{smith07}.
This was also noted by \citet{galliano08b}, who found that low-metallicity systems tends to lie on average toward the lower left corner of \reffig{fig:corrband}, whereas the upper left corner is essentially populated by Solar-metallicity sources.
However, \citet{hunt10} argued that \hBCD s exhibit a deficit of small \hPAH s.
If there is a smooth variation of \hPAH\ size distribution with metallicity, these results are in contradiction.
\citet{sandstrom12} noted that these \hBCD s are more extreme environments than the \hSMC, and that photodestruction could dominate the \hPAH\ processing (\cf\ \refsec{sec:PAHevol}).
We note that the solution to this apparent controversy might alternatively 
reside in the difference in studied spatial scales.
In the Magellanic Clouds, \hspitz\ spectroscopy gives a spatial resolution of a few parsecs, compared to a few hundred in nearby \hBCD s.
The fact is that the \hLMC\ and \hSMC\ exhibit strong spatial variations of 
their \hUIB\ spectrum.
\citet{whelan13} showed a diversity of \hMIR\ spectral properties in the \hSMC.
In this study, we demonstrated that the \hPAH\ emission in a region like N$\,$66 is dominated by its diffuse component, and not by its bright clumps, where \hPAH s are destroyed.
At the other extreme, the molecular cloud SMC-B1\#1 shows unusually high \hUIB\ equivalent widths \citep{reach00}.
Also, the $I(11.2)/I(12.7)$ ratio indicates that \hPAH s are more compact in \xxxdor\ and more irregular outside \citep{vermeij02}.
All these elements suggest that there is a complex balance of processes shaping 
the \hMIR\ spectra throughout low-metallicity environments.

\begin{figure}[htbp]
  \begin{tabular}{cc}
    \includegraphics[width=0.48\textwidth]{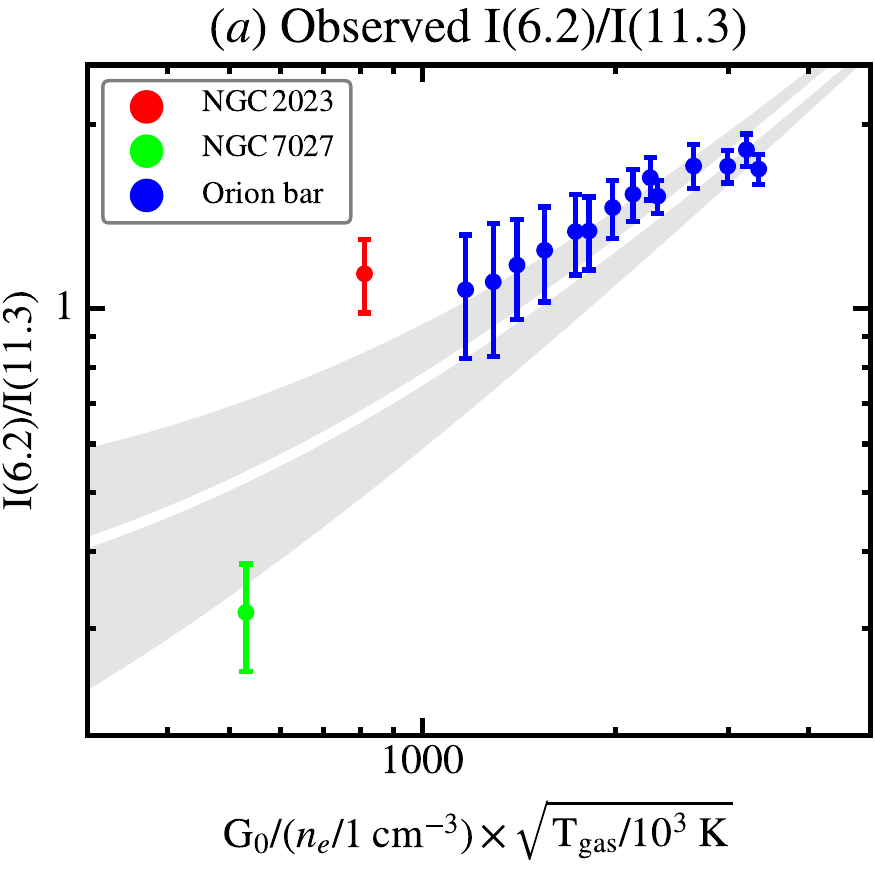} &
    \includegraphics[width=0.48\textwidth]{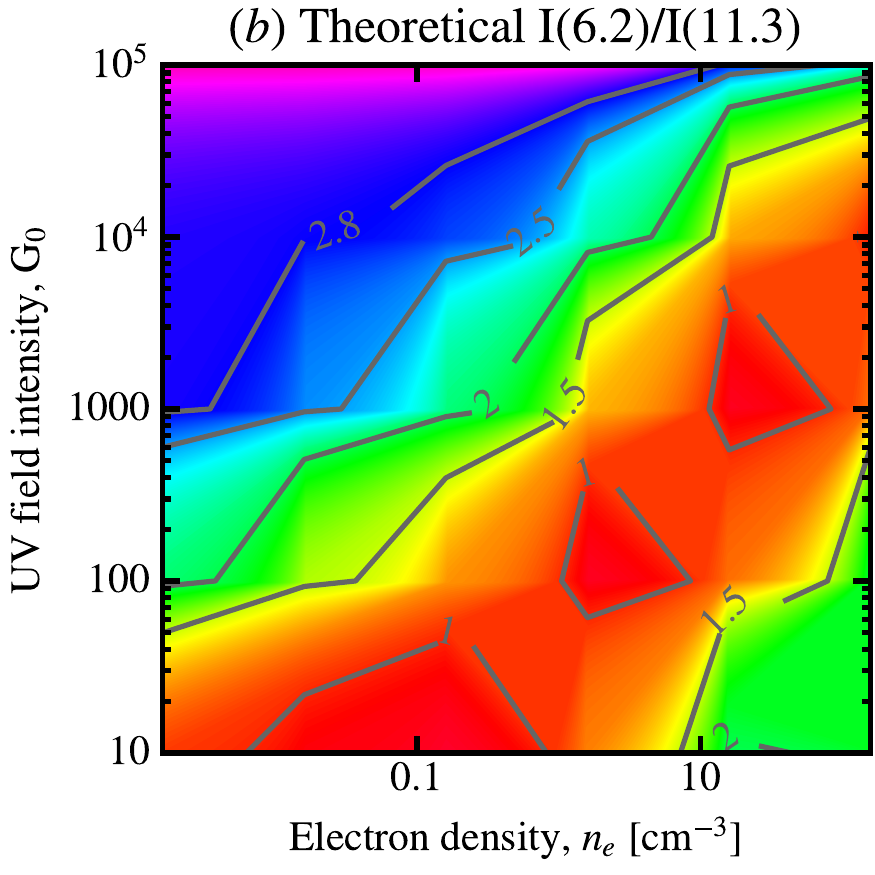} \\
  \end{tabular}
  \newcap{Calibration of PAH ratio diagnostics}%
         {Panel~\textit{(a)} shows the empirical correlation between the 
          observed 6.2-to-11.3~\tmic\ ratio and the modeled parameter,
          $\gamma$ \refeqp{eq:gamma}, in three galactic \hPDR s 
          \citep{galliano08b}.
          Panel~\textit{(b)} shows the theoretical variation of the band ratio
          as a function of $G_0$ and $n_e$ \citep{galliano09}. 
          This parameter grid was computed implementing stochastic heating in 
          the \hPDR\ model of \citet{kaufman06}, using a fixed \hPAH\ size 
          distribution \citep[the BARE-GR-S of][]{zubko04}.
          \CClicence}
  \label{fig:PAHcalib_G0ne}
\end{figure}

\begin{figure}[htbp]
  \includegraphics[width=\textwidth]{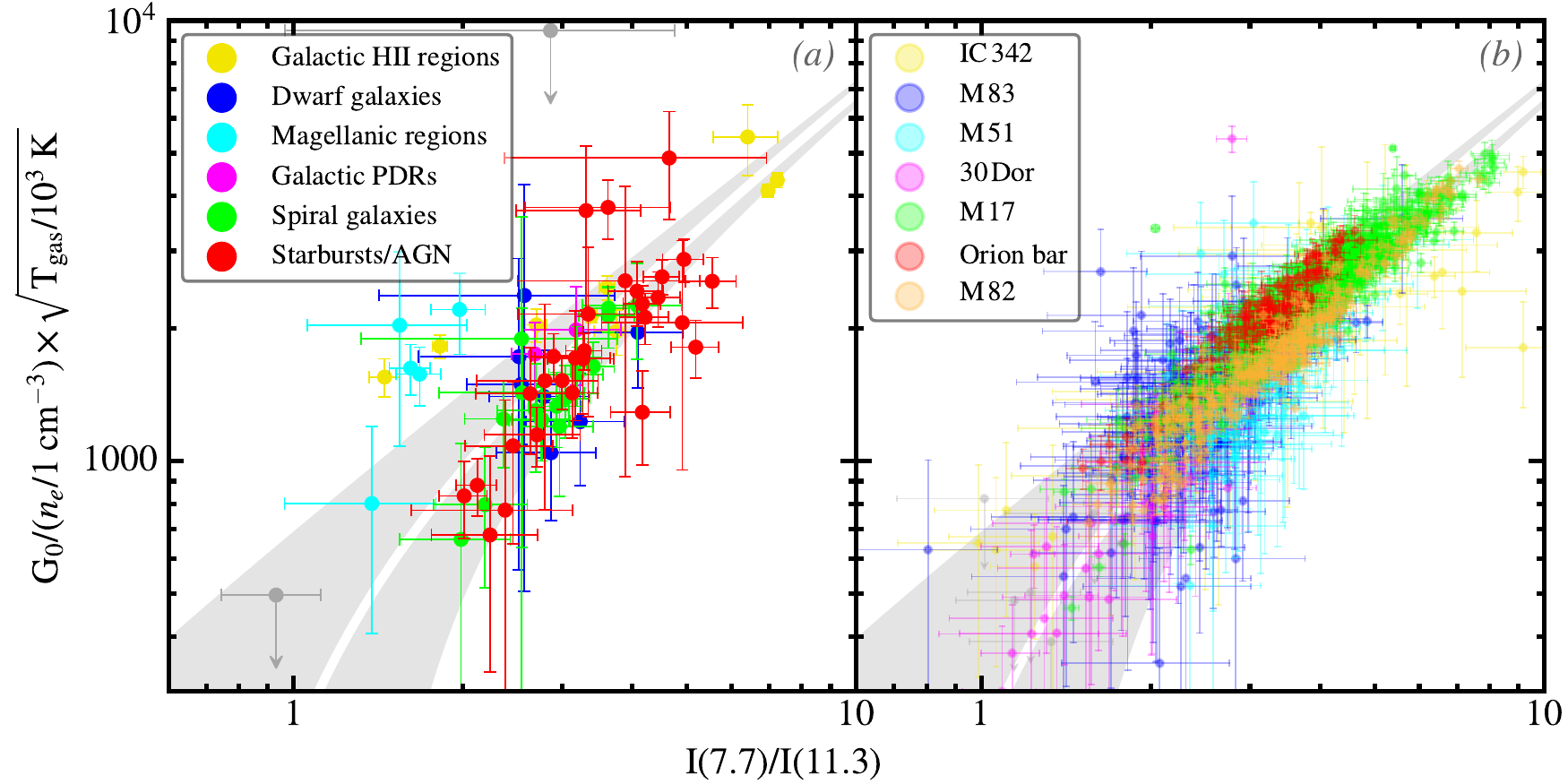}
  \newcap{PAH band ratio as diagnostics of the physical conditions}%
         {These are the data shown in \reffig{fig:corrband} where the 
          $I(6.2)/I(11.3)$ ratio as been converted to $G_0/n_e\times\sqrt{T}$ 
          using \refeq{eq:PAHgamma} \citep[Eq.~(5) of][]{galliano08b}.
          \CClicence}
  \label{fig:PAH2G0n}
\end{figure}

\paragraph{UIBs as diagnostics of the physical conditions.}
The fact that ionization dominates the \hUIB\ variation in galaxies opens the possibility to use specific observed band ratios to quantify the physical conditions.
The charge of an ensemble of molecules is indeed the balance between:
\begin{inlinelist}
  \item the ionizing photon rate; and 
  \item the electronic recombination rate.
\end{inlinelist}
The first quantity is usually quantified by the variable $G_0$, defined as the integral of the \hISRF\ in the \hFUV\ \citep[\eg][]{hollenbach97}:
\begin{equation}
  G_0\equiv\frac{\displaystyle\int_{6\,\textnormal{eV}}^{13.6\,\textnormal{eV}} 
                 I_E(E)\ddiff E}{1.6\E{-6}\;\textnormal{W/m}^2}.
  \label{eq:G0}
\end{equation}
The recombination rate is roughly proportional to $n_{e}/\sqrt{T_\sms{gas}}$ \citep[$n_e$ being the electron density, and $T_\sms{gas}$, the gas temperature; \eg][]{de-jong77}.
The ratio of these two rates, often called the \expression{photoionization parameter}, therefore quantifies this equilibrium \citep[\eg\ Chap.~5 of][]{tielens05}:
\begin{equation}
  \gamma\equiv\frac{G_0}{n_e}\sqrt{T_\sms{gas}}.
  \label{eq:gamma}
\end{equation}
The electron density can be related to the total H density by considering that most electrons in the neutral gas come from the photoionization of C.
We thus have $n_e\simeq(C/H)n_\sms{H}\simeq2\E{-4}n_\sms{H}$ (\cf\ \refsec{sec:ISMabund}).
\citet{galliano08b} measured the $I(6.2)/I(11.3)$ ratio in Galactic regions where $G_0$, $n_\sms{H}$ and $T_\sms{gas}$ had been reliably estimated (\refsubfig{fig:PAHcalib_G0ne}{a}).
It allowed us to propose an empirical relation between $\gamma$ and $I(6.2)/I(11.3)$:
\begin{equation}
  \frac{I(6.2)}{I(11.3)}\simeq
  \frac{G_0/(n_e/1\;\textnormal{cm}^{-3})\sqrt{T_\sms{gas}/10^3\;\textnormal{K}}}%
       {1990}+0.26\pm0.16.
  \label{eq:PAHgamma}
\end{equation}
In other words, measuring $I(6.2)/I(11.3)$ provides an estimate of $\gamma$.
With such a relation, the diagrams of \reffig{fig:corrband} can now be turned into the diagnostics of \reffig{fig:PAH2G0n}.
This relation has, since then, been refined by several studies, especially \citet{boersma16}.
Although unique, the diagnostics of \refeq{eq:PAHgamma} has however the following limitations.
\begin{itemize}
  \item The average value of a quantity such as $\gamma$ is problematic to
    interpret.
    If the observation includes a mix of regions with different densities 
    and \hISRF\ intensities, the weight of these different regions in 
    $\langle\gamma\rangle$ will be non-trivial.
  \item As we have seen in \refsec{sec:decompMIR}, there are systematic 
    discrepancies between different fitting methods.
    The calibration of \refeq{eq:PAHgamma} is therefore specific to the details
    of the fitting method used.
  \item The $I(6.2)/I(11.3)$ has a relatively narrow dynamical range.
    It varies at most by one order of magnitude in our sample.
    The relation is unclear outside this range.
\end{itemize}
Finally, the calibration of \refeq{eq:PAHgamma} can also be estimated theoretically.
\refsubfig{fig:PAHcalib_G0ne}{b} shows the theoretical $I(6.2)/I(11.3)$ ratio as a function of $G_0$ and $n_e$ \citep{galliano09}.
It has been derived by computing the stochastic emission of \hPAH s within the \hPDR\ model of \citet{kaufman06}.
Such model indeed computes the charge balance of \hPAH s at each point within the cloud, where $G_0$ and $n_e$ are known.
One grid point value corresponds to a whole cloud.

\paragraph{Other properties.}
We end this section by briefly reviewing other properties that can be studied with observations of \hUIB s.
\begin{description}
  \item[PAH compactness] can be probed by studying the relative variations of 
    C--H \hOOP\ bending modes (\cf\ \refsubfig{fig:modesPAH}{a}).
    These bending modes depend on the number of H atoms per aromatic cycle 
    (\cf\ \refsubfig{fig:modesPAH}{b}).
    We have seen in \refsec{sec:PAH} that, in particular, the solo-to-trio 
    intensity ratio, $I_{11.3}/I_{12.7}$, is an indicator of \hPAH\ compactness.
    This ratio was scrutinized in Galactic regions (evolved stars, \hii\  
    regions, reflection nebulae and \hYSO s) by \citet{hony01}.
    Their results are consistent with the scenario where large ($\simeq100-150$ 
    C atoms), compact \hPAH s are formed in winds of evolved stars, and degraded
    into smaller, irregular molecules in the \hISM.
  \item[Small a-C(:H) hydrogenation] can be studied with the $I(3.4)/I(3.3)$ 
    ratio.
    There is a debate whether aromatic features are carried by \hPAH s or 
    small \hHAC.
    The 3.4~\tmic\ aliphatic feature however can not be carried by pure 
    \hPAH s, it must come from either \hHAC\ grains \citep{jones13} or 
    \expression{Hydrogenated \hPAH s} (HPAH) with one of several aliphatic 
    groups \citep[\eg\ Fig.~1 of][]{marciniak21}.
    The $I(3.4)/I(3.3)$ aliphatic-to-aromatic ratio shows regional 
    variations in the \hISM, as the result of structural changes in the 
    hydrocarbons through \hUV\ processing \citep[\eg][]{jones13}.
    \citet{mori14} showed that the $I(3.4)/I(3.3)$ decreases with the 
    ionization of \hPAH s, in Galactic \hii\ regions.
    \citet{yamagishi12} detected the 3.4~\tmic\ feature in the superwind of 
    \M{82}.
    They found that the $I(3.4)/I(3.3)$ ratio increases with distance from 
    the center.
    They interpreted this trend as the production of small \hHAC, by 
    shattering of larger grains in this harsh halo.
    Similarly, \citet{kondo12} found a higher $I(3.4)/I(3.3)$ ratio in 
    the nuclear bar of \ngc{1097}, suggesting that the gas flow towards the   
    center could lead to the formation of small \hHAC\ by shattering.
    We note that, alternatively, the $I(3.4)/I(3.3)$ ratio can increase 
    with the accretion of \hHAC\ mantles in denser regions \citep{jones13}.
    This feature can also be seen in extinction, in \hAGN s 
    \citep[\eg][]{mason07}.
    \citetprep{hu21} modeled the spatially resolved \hAKARI\ and \hspitz\ 
    spectra in \M{82}.
    In this study, we found a negative correlation between the $I(3.4)/I(3.3)$ 
    and \sivline/\neiiline\ ratios.
    The latter is a \hISRF\ hardness indicator.
    This result thus demonstrates the dehydrogenation of \hHAC\ grains by hard
    \hISRF s.
  \item[SFR indicators] are one of the most sought after astrophysical 
    diagnostics.
    It happens that \hUIB s can be used to that purpose.
    \citet{peeters04} showed that the 6.2~\tmic\ feature intensity correlates 
    well with \hSFR, making it a reliable estimator.
    Alternatively, \citet{shipley16} have calibrated a \hSFR\ estimator based 
    on the integrated power of the 6.2, 7.7 and 11.3~\tmic\ features.
    The reason of this correlation is the same as for the \hTIR-\hSFR\ 
    correlation \citep[\eg][for a review]{kennicutt12}.
    At first order, the \hUIB\ strength is indeed correlated with \hTIR.
    \citet{peeters04} however note that \hUIB s are biased towards B stars.
    In addition, we note that such a tracer will be underestimating the \hSFR\
    at low metallicity, because of the variation of the relative \hUIB\ 
    strength, as we will see in \refsec{sec:PAH2VSG}.
\end{description}

    \subsubsection{Variations of the Aromatic Feature Strength}
    \label{sec:PAH2VSG}

The evolution of the shape of the \hUIB\ spectrum, probed by studying band ratio variations, is not the only diagnostics of the small carbon grain properties.
The overall aromatic feature strength, relative to the continuum (\ie\ to the emission of the rest of the dust populations) shows drastic variations across environments (\cf\ \reffig{fig:specMIR}).
These variations trace the evolution of the mass fraction of their carriers~-- \hPAH\ or small \hHAC.

\begin{figure}[htbp]
  \includegraphics[width=\textwidth]{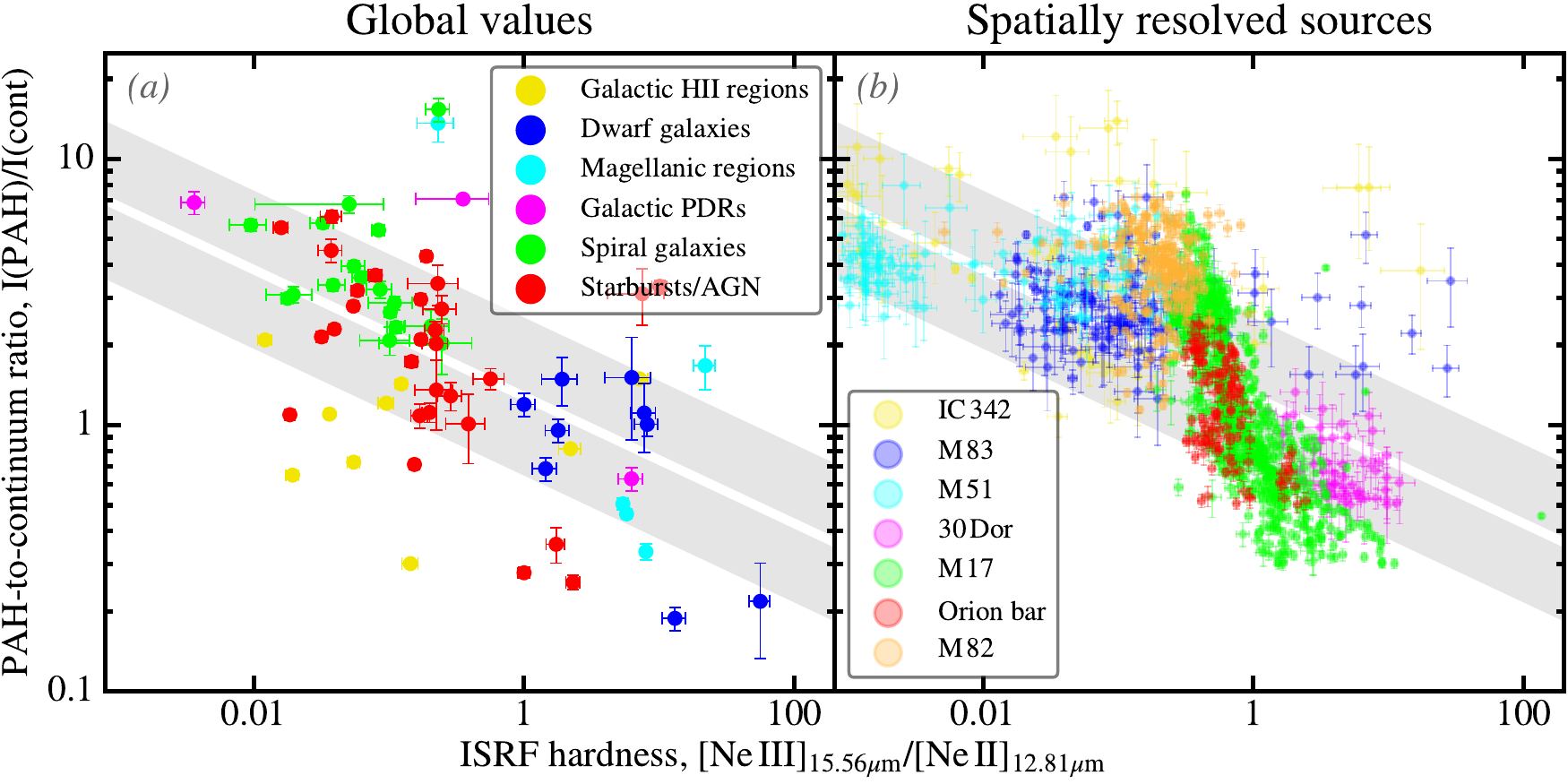}
  \newcap{Effect of ISRF hardness on PAH strength}%
         {These results are from the spectral decomposition of \citet{madden06}
          and \citet{galliano08b}.
          Panel~\textit{(a)} shows integrated sources, whereas 
          panel~\textit{(b)} shows pixel-by-pixel distributions.
          The white line with the grey band shows the affine fit to the global 
          values, $\pm1\sigma$, in log-log space.
          It is the same in both panels.
          \CClicence}
  \label{fig:PAH2ISRF}
\end{figure}

\paragraph{Effect of ISRF hardness.}
\hPAH\ and small \hHAC\ are known to be sensitive to hard an intense radiation fields.
They tend to evaporate near massive stars, and can be assumed to be fully depleted in \hii\ regions \citep[\eg][]{cesarsky96b,galametz13,galliano18}.
This effect can be quantified by studying the variation of the aromatic feature strength with a tracer of the \hISRF\ hardness.
\begin{enumerate}
  \item The aromatic feature strength can be traced with I(PAH)/I(cont), the 
    \hPAH-to-\hMIR-continuum ratio, where I(PAH) is the 
    sum of the intensities of every aromatic feature, and I(cont) is the 
    integrated intensity of the continuum, over an arbitrary wavelength range 
    \citep[10-16~\tmic\ in the case of][]{madden06}.
  \item The hardness of the \hISRF\ can conveniently be traced by the 
    \neiiiline/\neiiline\ ratio, as both of these \hMIR\ lines are usually 
    bright.
    \citet{madden06} demonstrated, using a photoionization model, that 
    \neiiiline/\neiiline\ is around unity when the \hISM\ is heated by a young 
    star cluster, and drops rapidly after a few million years, as massive stars 
    die.
\end{enumerate}
Such a trend is shown on \reffig{fig:PAH2ISRF}, with the \citet{madden06} results\footnote{The results of \citet{madden06} consisted in a first spectral decomposition of \hISO/CAM spectra, that we have refined in \citet{galliano08b}, and added \spitz/IRS spectra to the sample.}.
It clearly indicates that \hPAH s are less abundant in regions permeated with a hard \hISRF, at all spatial scales down to $\simeq0.1$~pc.
Note however that both tracers do not come from the same physical region:
\begin{inlinelist}
  \item PAH being destroyed in \hii\ regions are tracing the neutral gas;
  \item \neiiiline\ and \neiiline\ obviously come from the ionized gas.
\end{inlinelist}
The correlation therefore reflects the mixing of phases within the beam.

\begin{figure}[htbp]
  \begin{tabular}{cc}
    \includegraphics[width=0.48\textwidth]{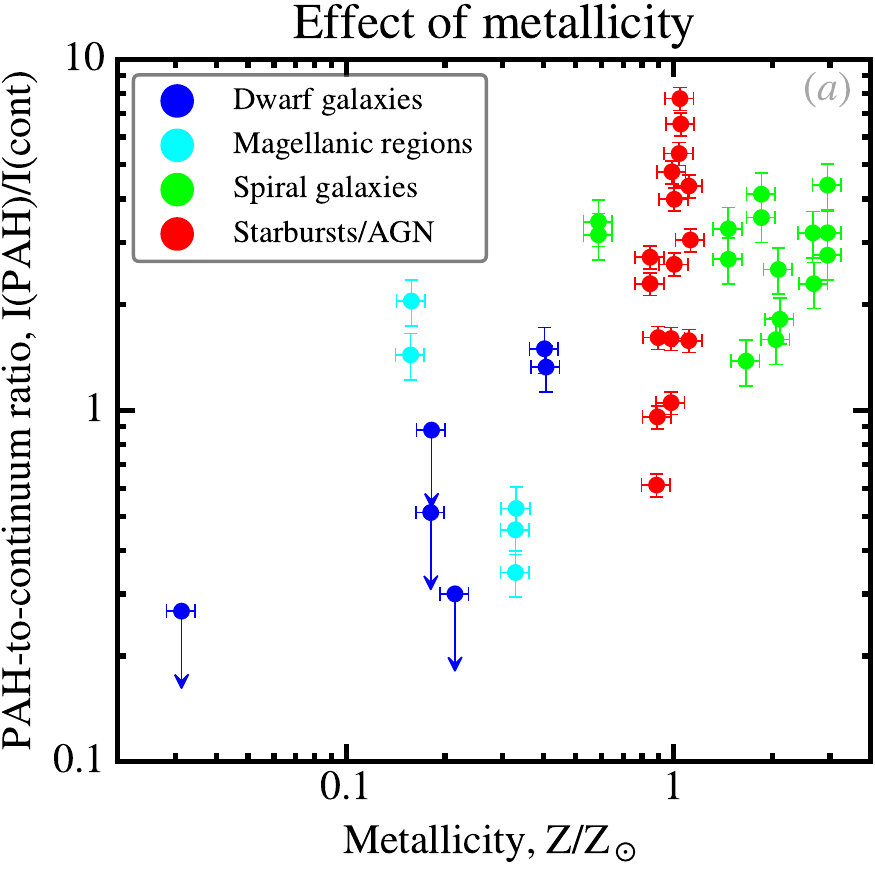} &
    \includegraphics[width=0.48\textwidth]{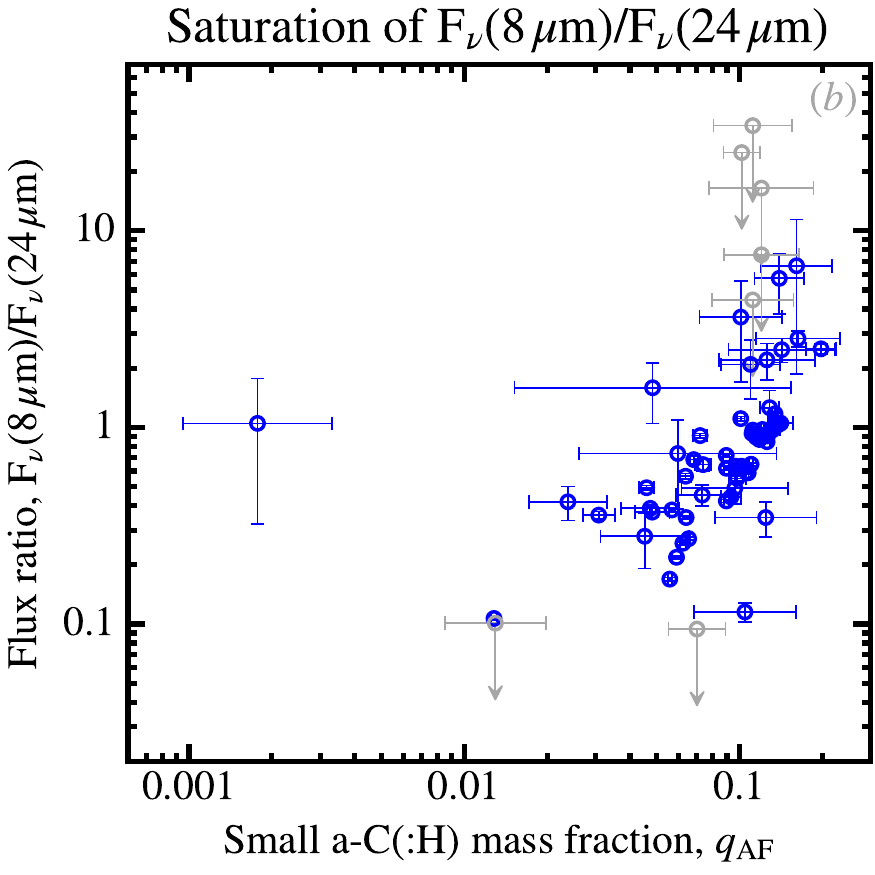} \\
  \end{tabular}
  \newcap{Effect of metallicity on the PAH strength}%
         {Panel~\textit{(a)} shows the trend of PAH relative strength with 
          metallicity, by \citet{madden06}.
          This was the first spectroscopy-based demonstration of this relation.
          Panel~\textit{(b)} illustrates the bias of the 
          $F_\nu(8\emic)/F_\nu(24\emic)$ ratio as a tracer of PAH strength.
          It shows the flux ratio as a function of the modeled fraction of
          small \hHAC\ (the grains carrying the aromatic features), $q_\sms{AF}$,
          in the \hDustPedia\ sample \citepalias{galliano21}.
          Only galaxies with both \IRACiv\ and \MIPSi\ fluxes are 
          displayed.
          \CClicence}
  \label{fig:PAH2Z}
\end{figure}

\paragraph{Effect of metallicity.}
The effects of \hISRF\ and metallicity are often degenerate, as at low metallicity:
\begin{inlinelist}
  \item stars of a given mass have a systematically higher effective 
    temperature, because of line blanketing effects;
  \item the \hISM\ is less opaque, because of the lower \hdustiness, and thus 
    more permeated by \hUV\ radiation.
\end{inlinelist}
The highest \neiiiline/\neiiline\ ratios are found in \hBCD s, as well as the lowest I(PAH)/I(cont).
\refsubfig{fig:PAH2Z}{a} shows the evolution of I(PAH)/I(cont) as a function of metallicity \citep{madden06}.
There is a paucity of \hPAH s in low-metallicity environments.
The questions is whether this is the result of their increased destruction, or if they have not been produced.
We will come back to this question, when discussing dust evolution in \refsec{sec:PAHevol}.
\takeaway{\hPAH s are under-abundant in low-metallicity environments.}

\paragraph{The absence of metallicity threshold.}
The relation of \citet{madden06} was the first spectrosco\-py-based demonstration of the effect.
Shortly before, \citet{engelbracht05} showed the broadband correlation of $F_\nu(8\emic)/F_\nu(24\emic)$ as a function of metallicity.
They showed both quantities were clearly correlated.
They however argued there were essentially two populations, below and above $12+\log(\textnormal{O/H})\simeq8$.
\citet{galliano08a} showed that this was a bias due to the saturation of \IRACiv\ as a \hPAH\ tracer at low metallicity.
When the aromatic feature strength becomes indeed low, $F_\nu(8\emic)/F_\nu(24\emic)$ is not anymore a measure of I(PAH)/I(cont), but is a measure of the temperature of the continuum.
This is illustrated on \refsubfig{fig:PAH2Z}{b}.
We show that when the actual mass fraction of small \hHAC\ drops below $\simeq10\,\%$, $F_\nu(8\emic)/F_\nu(24\emic)$ becomes insensitive to its value.

  \subsection{Long-Wavelength Properties}

At long wavelengths, in the submm-to-cm range, dust emission does not always behave as the extrapolation from the Rayleigh-Jeans law ($F_\nu\propto\nu^{2+\beta}$).
Two peculiar phenomena have been widely discussed in the literature:
\begin{inlinelist}
  \item the \expression{submillimeter excess}; and
  \item the \expression{Anomalous Microwave Emission} (\hAME).
\end{inlinelist}

    \subsubsection{The Elusive Submillimeter Excess}
    \label{sec:submmex}

\begin{figure}[htbp]
  \includegraphics[width=\textwidth]{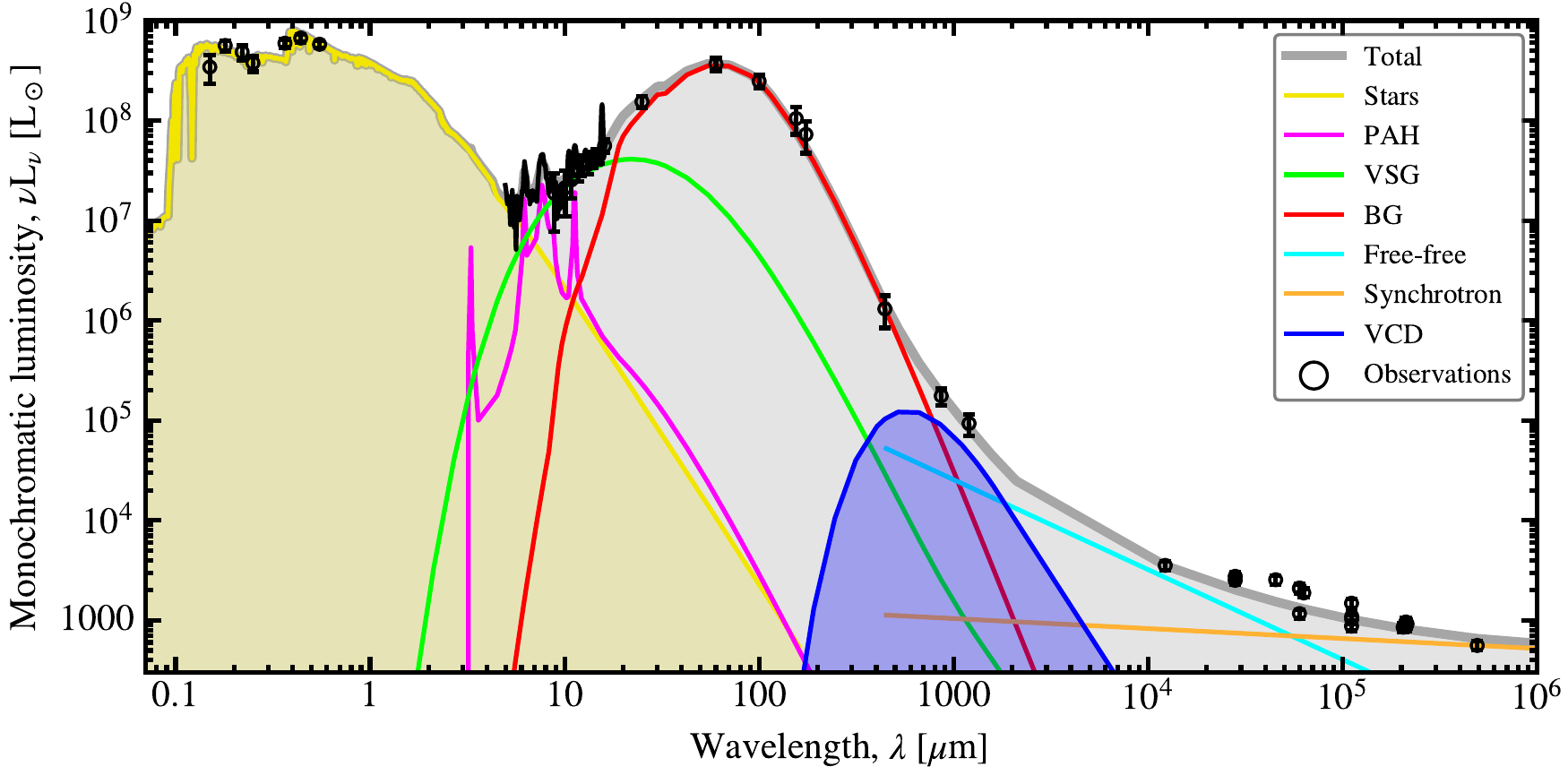}
  \newcap{Submillimeter excess in \ngc{1569}}%
         {This is the best fit model of \citet{galliano03}, discussed in 
          \refsec{sec:sizedist}.
          We focus here on the blue component, which is a $T=5$~K, $\beta=1$
          \hMBB\ accounting for $\simeq80\,\%$ of the dust mass
          of the galaxy.
          This component has been fitted to account for the submm excess.
          \CClicence}
  \label{fig:submmex1569}
\end{figure}

An excess emission above the modeled dust continuum is often observed, longward $\lambda\simeq500\emic$.
The most significant reports of this \citext{submm excess} can not be accounted for by: 
\begin{inlinelist}
  \item thermal dust emission;
  \item free-free and synchrotron continua; and 
  \item molecular line emission 
\end{inlinelist}
\citep[\cf\ \reffig{fig:submmex1569};][]{galliano03}.
\begin{enumerate}
  \item The first occurrence of such an excess was unveiled by \citet{reach95}, 
    studying the \hCOBE\ observations of the Milky Way.
    Their \hIR--submm \hSED\ could be fitted with a \hMBB\ ($\beta=2$; 
    \cf~\refsec{sec:MBB}), and an additional $4-7$~K component.
  \item A few years later, \citet{lisenfeld02} and \citet{galliano03} found a 
    statistically significant excess in the dwarf galaxy \ngc{1569}, at 
    850~\tmic\ and 1.3~mm.
    It is shown in \reffig{fig:dustobs}.
  \item Several subsequent studies confirmed the presence of an excess in other 
    late-type galaxies 
    \citep[\eg][]{dumke04,bendo06a,zhu09,galametz09,galametz11},
    including the global \hSED s of the Magellanic clouds 
    \citep{israel10,bot10}.
  \item \hhersc\ and \hplanck\ opened the way to more detailed 
    characterizations.
\end{enumerate}
Studying this excess is important, as:
\begin{inlinelist}
  \item it could bias dust mass estimates;
  \item it potentially contains untapped physical information about the \hISM.
\end{inlinelist}

\paragraph{Possible explanations.}
The origin of the submm excess is currently debated.
The following explanations have been proposed.
These different scenarios are not exclusive.
\begin{description}
  \item[Very cold dust] (\hVCD), that is equilibrium grains colder than 
    $T\simeq10$~K, can be used to fit the excess.
    The emission of such a grain population indeed peaks in the submm range.
    Cold dust being weakly emissive, this scenario however leads to massive 
    amounts of grains     
    \citep[$40-80\,\%$ of the total dust mass; \eg][]{galliano03,galliano05}.
    \citet{galliano03} showed that \hVCD\ would be realistic only if this 
    component was distributed in a few number of dense, parsec-size clumps.
    The existence of such cold dust indeed requires it to be shielded from all
    \hUV-visible light.
    Using the spatially resolved observations of the 500~\tmic\ excess in the 
    \hLMC, \citet{galliano11} concluded that this explanation is unrealistic, as
    it would require at least one clump in each of the $\simeq90\,000$ pixels 
    of this study.
  \item[Temperature-dependent emissivity.]
    Laboratory measures on cosmic grain analogs exhibit an increase of the
    \hFIR-submm opacity as a function of temperature 
    \citep[\eg][]{mennella98,demyk17b,demyk17a}.
    \citet{meny07} designed the so-called \expression{Disordered Charge 
    Dis\-tri\-bution/Two-Level System} (\hDCD/\hTLS) model, predicting an 
    increase of $\kappa(\lambda_0)$ and a decrease of $\beta$ with the 
    temperature of amorphous grains.
    It reproduces the submm excess observed in the Galaxy by \citet{paradis12} 
    and in the \hLMC\ by 
    \citet[][coupled with spinning grains; \cf~\refsec{sec:AME}]{bot10}.
    However, this model can not account for the submm excess in the \hSMC\ 
    \citep{bot10}.
  \item[Magnetic grains.]
    \citet{draine12} showed that the submm excess of the \hSMC\ could be 
    attributed to magnetic nanoparticles (Fe, Fe$_3$O$_4$, 
    $\gamma$-Fe$_2$O$_3$\footnote{$\gamma$-Fe$_2$O$_3$ is the notation to 
    design ferromagnetic Fe$_2$O$_3$, called \expression{maghemite}.
    It must be distinguished from its non-magnetic form, noted 
    $\alpha$-Fe$_2$O$_3$, called \expression{hematite}.
    Both have distinct crystalline structures: cubic lattice for maghemite; 
    trigonal crystals for hematite \citep[\eg][]{lecznar77}.}). 
    Thermal fluctuations in the magnetization of these grains can produce 
    strong magnetic dipole emission, since ferromagnetic materials are known to 
    have large opacities at microwave frequencies. 
    This hypothesis seems to be consistent with the observed elemental 
    abundances of the \hSMC\ and could also be responsible for the excess 
    detected in other environments.
    These magnetic nanoparticles could be free-flying or inclusions in larger
    grains.
\end{description}

\begin{figure}[htbp]
  \includegraphics[width=\textwidth]{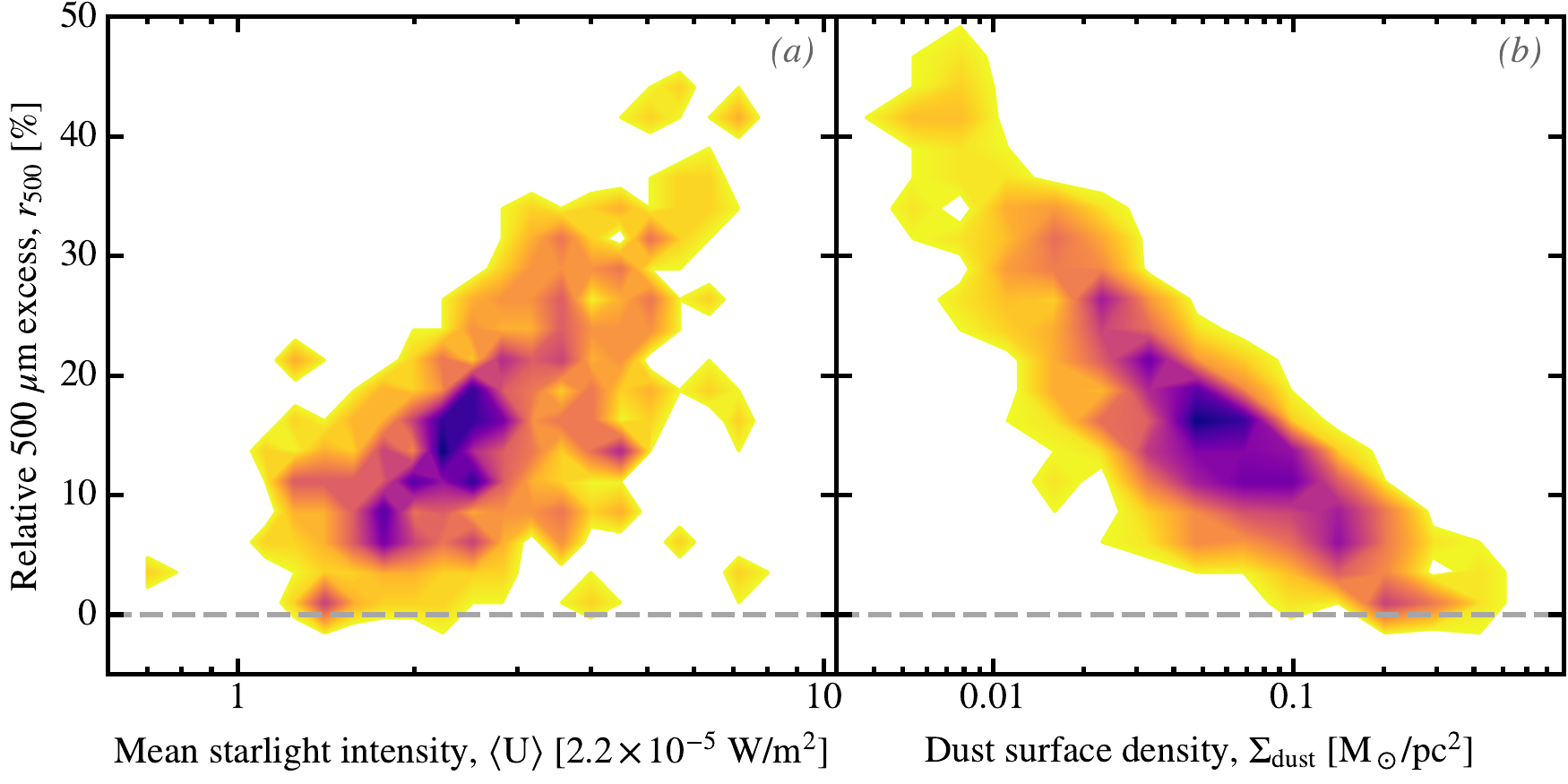}
  \newcap{Spatially-resolved submm excess in the LMC}%
         {In both panels, the color density represents the number of pixels
          per bin of parameters in the star-forming region N$\,$44 of the \hLMC\
          \citep{galliano11}.
          The \SPIREiii\ excess is shown as a function of:
          \begin{inlinelistalph}
            \item mean starlight intensity \refeqp{eq:Uav}; and
            \item dust mass surface density.
          \end{inlinelistalph}
          \CClicence}
  \label{fig:r500}
\end{figure}

\paragraph{Empirical properties of the excess.}
Since the origin of the submm excess is still unknown, most studies focus on characterizing its observed properties.
\begin{description}
  \item[Low-metallicity systems] are the environments where the excess appears 
    the be the most prominent.
    This is the reason why it has been detected in most nearby \hBCD s 
    \citep[\eg][]{galliano03,dumke04,galliano05,galametz09,bot10,galametz11}.
    \citet{galliano11} have shown, using spatially-resolved observations of the 
    \hLMC, that the \SPIREiii\ excess, $r_{500}$, varies up to $\simeq40\,\%$ in 
    certain regions.
    It is correlated with the mean starlight intensity, $\langle U\rangle$, and 
    anticorrelated with the dust surface density, as shown in \reffig{fig:r500}.
    The correlation with $\langle U\rangle$ could be a confirmation of the 
    \hDCD/\hTLS\ effect, as $\langle U\rangle$ directly controls the temperature
    of the grains.
    The correlation with surface density is however significantly better than 
    with \hISRF, a trend that the \hDCD/\hTLS\ model alone fails to explain.
  \item[In the Milky Way,] \citet{paradis12} showed that the \SPIREiii\ excess 
    becomes significant in the peripheral regions ($>35^{\circ}$), as well as 
    towards some \hii\ regions.
    This is qualitatively consistent with the trend we found  in the \hLMC, as 
    these peripheral regions are also the most diffuse ones.
    The relative amplitude of the excess can rise up to $\simeq20\,\%$.
  \item[In other galaxies,] the excess is more difficult to detect 
    \citep[\eg][]{remy-ruyer15,dale17}.
     When resolved in non-barred spirals, the submm excess is primarily 
     detected in the disk outskirts, thus at low-surface density 
     \citep[\eg][]{hunt15}.
\end{description}
\takeaway{The submm excess is more prominent in low-metallicity environments,
          and in diffuse regions.}

\paragraph{Reality of the phenomenon.}
The reality of the submm excess has been questioned for the two following reasons.
We try to address these criticisms in order to support its likeliness.
\begin{description}
  \item[The excess is model-dependent.]
    Different dust opacities lead to different amplitudes of the excess.
    Some over-parametrized models have found excesses virtually everywhere.
    For instance, the \expression{Broken-Emissivity Modified Black Body} 
    (\hBEMBB) is a \hMBB\ with two free-varying emissivity indices, 
    $\beta_1$ and $\beta_2$, below and above $\lambda=\lambda_b$ 
    \citep{gordon14}.
    With such a model, any excess can be fitted, but the derived $\beta_1$, 
    $\beta_2$ and $\lambda_b$ do not necessarily correspond to existing solids.
    This model is also extremely degenerate, as shown by \citet{galliano18a}.
    A \hBEMBB\ fit of the diffuse Galactic \hISM\ \hSED\ (\cf\ 
    \refsubfig{fig:THEMIS}{d}) by \citet{gordon14} gives a 
    $r_{500}\simeq48\pm11\,\%$ excess starting at $\lambda_b\simeq300\pm30\emic$.
    Yet, this is the \hSED\ that everyone uses to calibrate dust models. 
    It does not exhibit an actual excess.
    It illustrates that probing the submm excess with models which are not 
    based on realistic optical properties is a non-sense.
    It must be studied with a model as physical as possible.
  \item[The technically-challenging nature of submm-mm observations] 
    is also questioning the reality of the excess:
    \begin{inlinelist}
      \item observations in this regime are difficult from the ground (\cf\ 
        \refsec{sec:atmosphere});
      \item the calibration of these observations is often uncertain;
      \item data reduction methods have problems dealing with the diffuse 
        emission, which is where the excess appears to be the most prominent.
    \end{inlinelist}
    In particular, \hplanck\ data have lead to revise the calibration of 
    the \hCOBE/\hFIRAS\ \hFIR-submm spectrum of the diffuse Galactic \hISM\ 
    \citep{planck-collaboration14a}.
    With this new calibration, the excess of \citet{reach95}, discussed 
    earlier, is not significant anymore.
    There are however several sound indications that these concerns are not 
    enough to doubt the reality of the phenomenon.
    \begin{itemize}
      \item The \hSED\ shape of low-metallicity systems is well characterized 
        in this regime.
        It has been observed at different wavelengths, with different 
        instruments.
        It is still present with the latest \hhersc\ calibration \citep{dale17}.
      \item Reports of a deficit are very rare.
      \item In the Magellanic clouds, \citet{planck-collaboration11} showed 
        that, while the submm excess in the integrated \hSED\ of the \hLMC\ was 
        consistent with \hCMB\ fluctuations, the \hSMC\ excess was 
        significantly above this level.
    \end{itemize}
\end{description}

    \subsubsection{The Anomalous Microwave Emission}
    \label{sec:AME}
    
As we have seen in \refsec{sec:AMEobs}, the \hAME\ is a centimeter continuum excess that can not be accounted for by:
\begin{inlinelist}
  \item the extrapolation of dust thermal emission; 
  \item molecular line emission; and 
  \item free-free and synchrotron continua (\reffig{fig:dustobs}).
\end{inlinelist}
Its \hSED\ looks like a bump peaking around $\lambda=1$~cm (\cf\ \reffig{fig:dustobs}).
It is commonly attributed to the dipole emission of fastly rotating, small dust grains.
The faster grains rotate, the shorter the frequency of the emission peak is.

\paragraph{The AME in extragalactic environments.}
In nearby galaxies, the first unambiguous detection of an \hAME\ has been obtained in an outer region of \ngc{6946} \citep{murphy10,scaife10}.
Follow up observations showed evidence for \hAME\ in eight regions of this galaxy \citep{hensley15}.
This study showed that the spectral shape of this \hAME\ is consistent with spinning dust, but with a stronger \hAME-to-\hPAH-surface-density ratio, hinting that other grains could be the carriers.
Overall, the \hAME\ fraction is highly variable, in nearby galaxies.
\citet{peel11} put upper limits on the \hAME\ in \M{82}, \ngc{253} and \ngc{4945}.
These upper limits suggest that \hAME/100~\tmic\ is lower than in the Milky Way, in these objects.
In \M{31}, \citet{planck-collaboration15c} report a $2.3\sigma$ \citext{measurement} of the \hAME, consistent with the Galactic properties.
Finally, \citet{bot10}, fitting the \hNIR-to-radio \hSED\ of the \hLMC\ and \hSMC, tentatively explained the submm/mm excess with the help of spinning dust, in combination with a modified submm dust emissivity (\cf\ \refsec{sec:submmex}).
They concluded that, if spinning grains are responsible for this excess, their emission must peak at 139~GHz (\hLMC) and 160~GHz (\hSMC), implying large \hISRF\ intensities and densities.
\citet{draine12} argued that such fastly rotating grains would need a \hPDR\ phase with a total luminosity more than two orders of magnitude brighter than the \hSMC.

\paragraph{Controversy about the carriers of the AME.}
Although \hPAH s have been considered the most likely carriers of the \hAME, \citet{hensley17} argued that nanosilicates could be a viable alternative.
They showed that nanosilicates can indeed account for the \hAME, without violating the other observational constraints (depletions, emission, extinction; \cf\ \refsec{sec:dustobs}).
This claim relies on the earlier findings of \citet{hensley16}, showing that \hAME\ correlates better with dust radiance, $\mathcal{R}$\footnote{The radiance is the spectral integral of the specific intensity: $\mathcal{R}\equiv\int I_\nu\ddiff\nu$.}, in the \hMW.
\citet{hensley16} also found some fluctuations in the \hPAH\ emission relative to the \hAME\ intensity, traced by \WISEiii.
The fact is that there is no observational evidence of nanosilicates in the \hISM.
In particular, these grains would emit the 9.8 and 18~\tmic\ features, as they would be stochastically heated.
These bands would eventually remain diluted in the aromatic feature emission, provided the abundance of nanosilicates is low enough.
We can however note that in \hii\ regions, where \hPAH s are destroyed, we can see the underlying continuum, and silicate features in emission are rarely present \citep[\eg][]{peeters02}.
They can be seen only in the most compact \hii\ regions, and their 9.8-to-18~\tmic\ ratio indicates that they originate in large equilibrium grains.
A weak correlation does not always indicate an absence of causality.
This issue might reside in the fact that \citet{hensley16} used the \WISEiii\ band as a tracer of \hPAH\ intensity, whereas this broad photometric band also encompasses a significant fraction of the underlying continuum, and is biased towards neutral \hPAH s.
We have tried to address this controversy by refining the constraints on the \hPAH\ emission.

\begin{figure}[!htbp]
  \includegraphics[width=\textwidth]{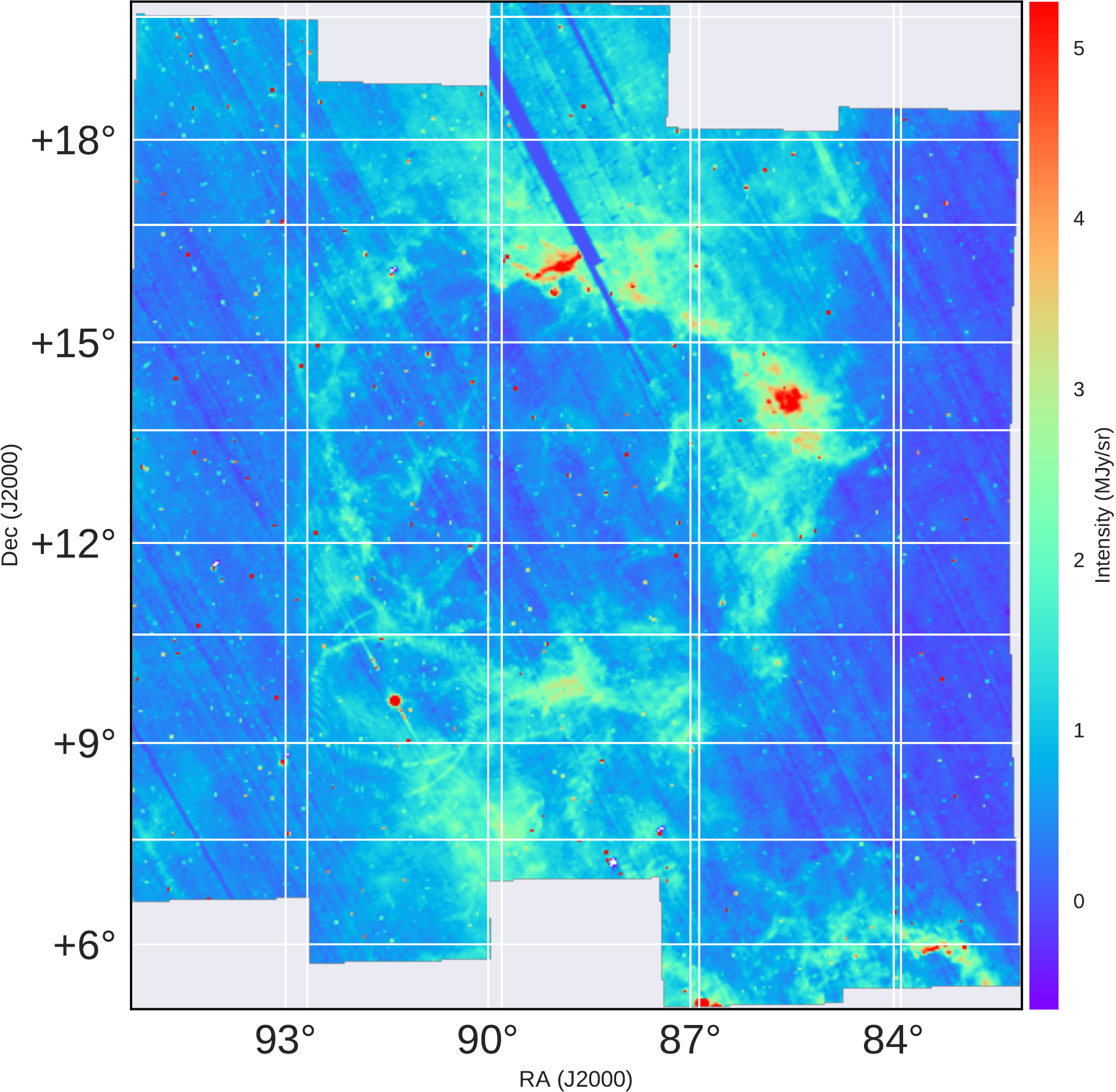}
  \newcap{AKARI 9~\tmic\ map of $\lambda$-Orionis}%
         {This is a mosaic of broadband photometric images of \hPAH s within 
          this Galactic molecular ring.
          \uline{Credit:} \citet{bell19}.}
  \label{fig:lori}
\end{figure}

\begin{figure}[htbp]
  \begin{tabular}{cc}
    \includegraphics[width=0.48\textwidth]{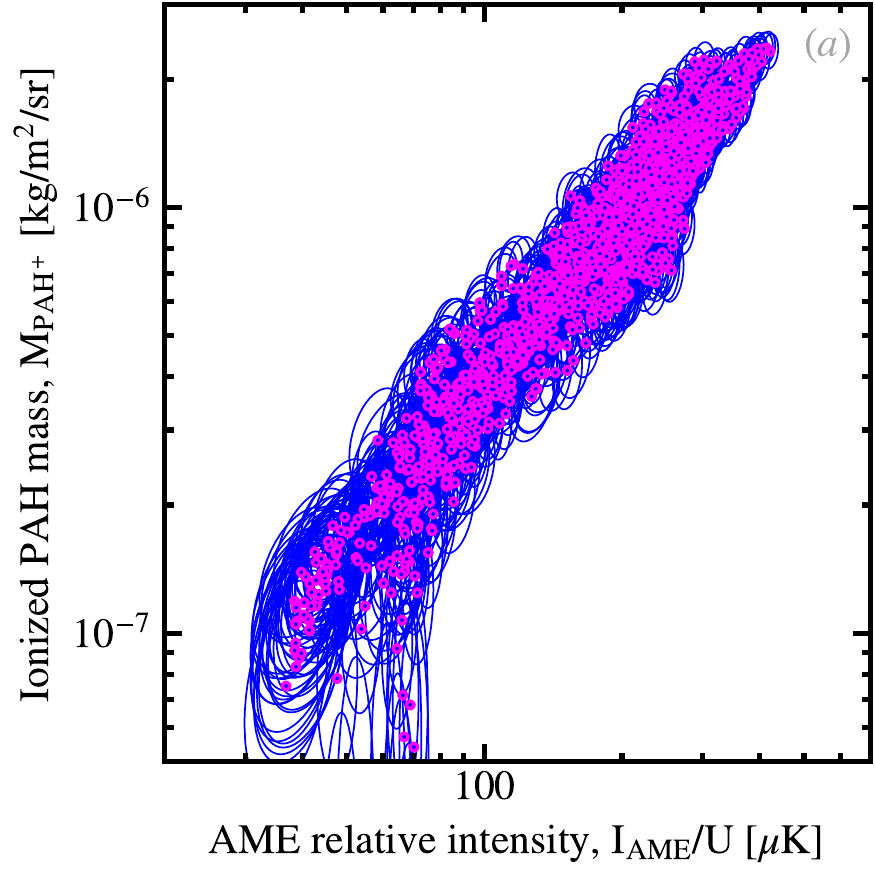} &
    \includegraphics[width=0.48\textwidth]{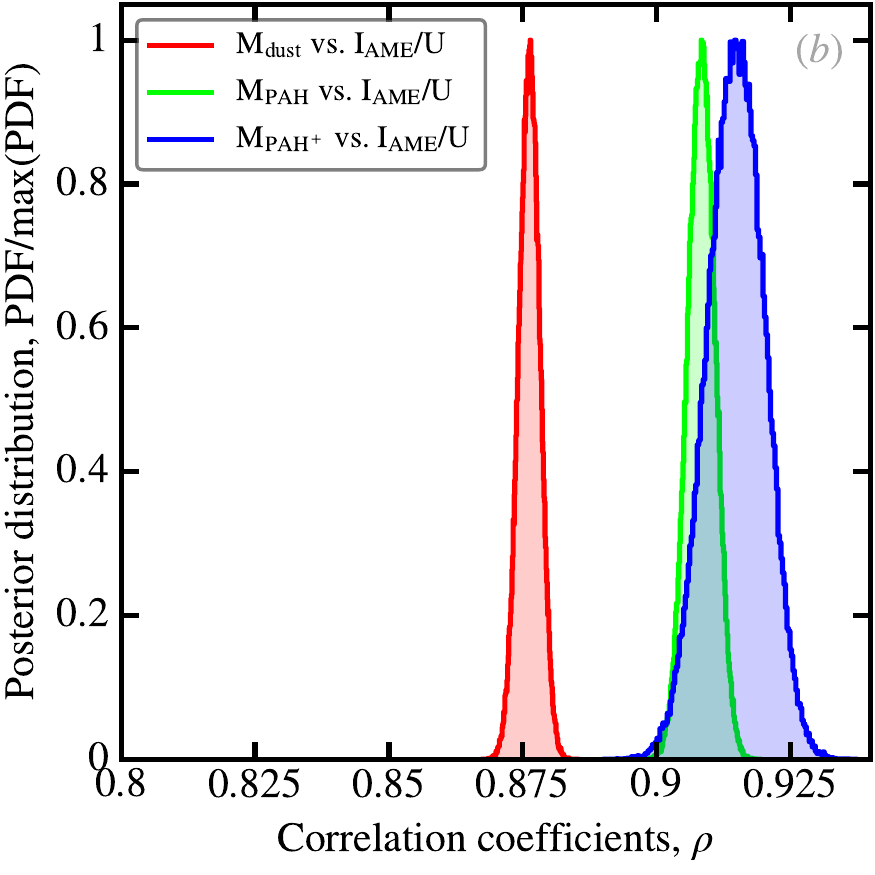} \\
  \end{tabular}
  \newcap{AME correlation with ionized PAHs in $\lambda$-Orionis}%
         {Panel~\textit{(a)} shows the pixel correlation between 
          $I_\sms{AME}/U$ and $M_\sms{PAH+}$, derived from the \hSED\ modeling
          of \citep{bell19}.
          Each point and its uncertainty ellipse represent the posterior \hPDF\
          of a pixel.
          Panel~\textit{(b)} shows the posterior \hPDF s of the correlation 
          coefficients between two sets of parameters (such as        
          panel~\textit{a}).
          The correlation between $I_\sms{AME}/U$ and $M_\sms{PAH+}$ is the best
          of our study, and it is significantly better than with $M_\sms{dust}$.
          \CClicence}
  \label{fig:AME}
\end{figure}

\paragraph{Correlation with charged PAHs.}
\citet{bell19} focussed on the 10$^\circ$-wide Galactic molecular ring surrounding the O-type star, $\lambda$-Orionis (\cf\ \reffig{fig:lori}).
We chose this region, because the \hplanck\ data indicate there is a large gradient of \hAME\ intensity.
We fitted the spatially-resolved \hSED, at $0.25^\circ$ angular sampling, using the \hAKARI\ 9~\tmic\ and \hIRAS\ 12~\tmic\ bands to constrain the \hPAH\ abundance, and longer wavelength bands for the rest of the emission.
We used the dust \hSED\ code \citetalias{galliano18a} \citep[\expression{HiERarchical Bayesian Inference for dust Emission;}][\cf\ \refsec{sec:HerBIE}]{galliano18a}.
We were able to address the controversy, thanks to the combination of:
\begin{inlinelist}
  \item rigorous \hSED\ fitting, allowing us to account for all the available
    information, not only a few broadbands;
  \item better observational constraint on the \hPAH\ emission (9 and 12~\tmic);
    and
  \item focussing on a clean region rather than the whole sky.
\end{inlinelist}
Our results are shown in \reffig{fig:AME}.
We have found very good pixel-to-pixel correlations between the \hAME\ intensity, derived by \citet{planck-collaboration16a}, and the dust and \hPAH s surface densities from \hSED\ fitting (\eg\ \cf\ \refsubfig{fig:AME}{a}).
Our analysis however show that, if the dust mass per pixel is very well correlated with the intensity of \hAME\ per $U$, $I_\sms{AME}/U$ ($\rho\simeq0.88$), the correlation is better with the mass of \hPAH s, and even better with the mass of ionized \hPAH s ($\rho\simeq0.92$; \cf\ \refsubfig{fig:AME}{b}).
Our Bayesian results also show that there is a 0 probability that the total dust could correlate better with \hAME\ than with \hPAH s.
Our impression is that the study of \citet{hensley16} may have too quickly interpreted a poor correlation as an absence of causality.
The scatter in \WISEiii\ was likely not the result of an intrinsic scatter of the \hPAH/\hAME\ correlation, but rather a combination of observational artefacts:
\begin{inlinelist}
  \item noise;
  \item contribution of the continuum and ionic lines.
\end{inlinelist}
Our \hSED\ model and our better \hMIR\ coverage allowed us to more accurately recover the true \hPAH\ column density.
\takeaway{\hAME\ correlates better with charged \hPAH s.}

\section{Dust in Relation with the Gaseous and Stellar 
         Contents}

We end this chapter with a short introduction to \expression{ISMology}, since the knowledge of the \hISM\ as a whole is crucial to understanding \hISD.
We discuss a few of our results in this domain and illustrate the way dust can be used to better understand the gas.
The video lectures and accompanying slides of the \href{https://ismgalaxies2021.sciencesconf.org/resource/page/id/8}{2021 \citengl{ISM of galaxies} summer school} (35 hours of lecture in total), that we have organized, can provide a good introduction to this subject.
Otherwise, the books of \citet{spitzer78}, \citet{tielens05} and \citet{draine11b} are references.

  \subsection{The Phases of the ISM}
  \label{sec:ISMism}

The \hISM\ is intrinsically a \expression{multi-phase} medium, with large contrast densities and differences in temperatures.
The order of magnitude of its average density is $n_\sms{gas}\simeq1$~cm$^{-3}$.
We list the physical characteristics of the main phases in \reftab{tab:ISMism}.
We will discuss each phase individually in the rest of this section.
\begin{table}[htbp]
  \centering
  \setlength\arrayrulewidth{2pt}
  \arrayrulecolor{white}
  \begin{tabularx}{\linewidth}{|>{\columncolor{coltabhead}}X%
                                |>{\columncolor{coltabcell}}X%
                                |>{\columncolor{coltabcell}}X%
                                |>{\columncolor{coltabcell}}X|}
    \hline
      \rowcolor{coltabhead}
      \cellcolor{white} & \textbf{Density} & \textbf{Temperature} 
      & \textbf{Volume filling factor} \\
    \hline
      \hHIM  & $\simeq3\E{-3}$ cm$^{-3}$ & $\simeq10^6$ K & $\simeq50\,\%$ \\
    \hline
      \hii\ regions & $\simeq1-10^5$ cm$^{-3}$ & $\simeq10^4$ K 
                                                             &$\lesssim1\,\%$\\
    \hline
      \hWIM & $\simeq0.1$ cm$^{-3}$ & $\simeq10^4$ K & $\simeq25\,\%$ \\
    \hline
      \hWNM & $\simeq0.3$ cm$^{-3}$ & $\simeq10^4$ K & $\simeq30\,\%$ \\
    \hline
      \hCNM & $\simeq30$ cm$^{-3}$ & $\simeq100$ K & $\simeq1\,\%$ \\
    \hline
      Diffuse \hmol & $\simeq100$ cm$^{-3}$ & $\simeq50$ K & $\simeq0.1\,\%$ \\
    \hline
      Dense \hmol & $\simeq10^3-10^6$ cm$^{-3}$ & $\simeq10$ K & 
        $\simeq0.01\,\%$ \\
    \hline
  \end{tabularx}
  \newcap{Phases of the ISM}%
         {Adapted from \citet{tielens05} and \citet{draine11b}.
          The sum of the filling factors is slightly larger than $100\,\%$,
          because these estimates are rough.}
  \label{tab:ISMism}
\end{table}

\begin{figure}[!htbp]
  \includegraphics[width=\textwidth]{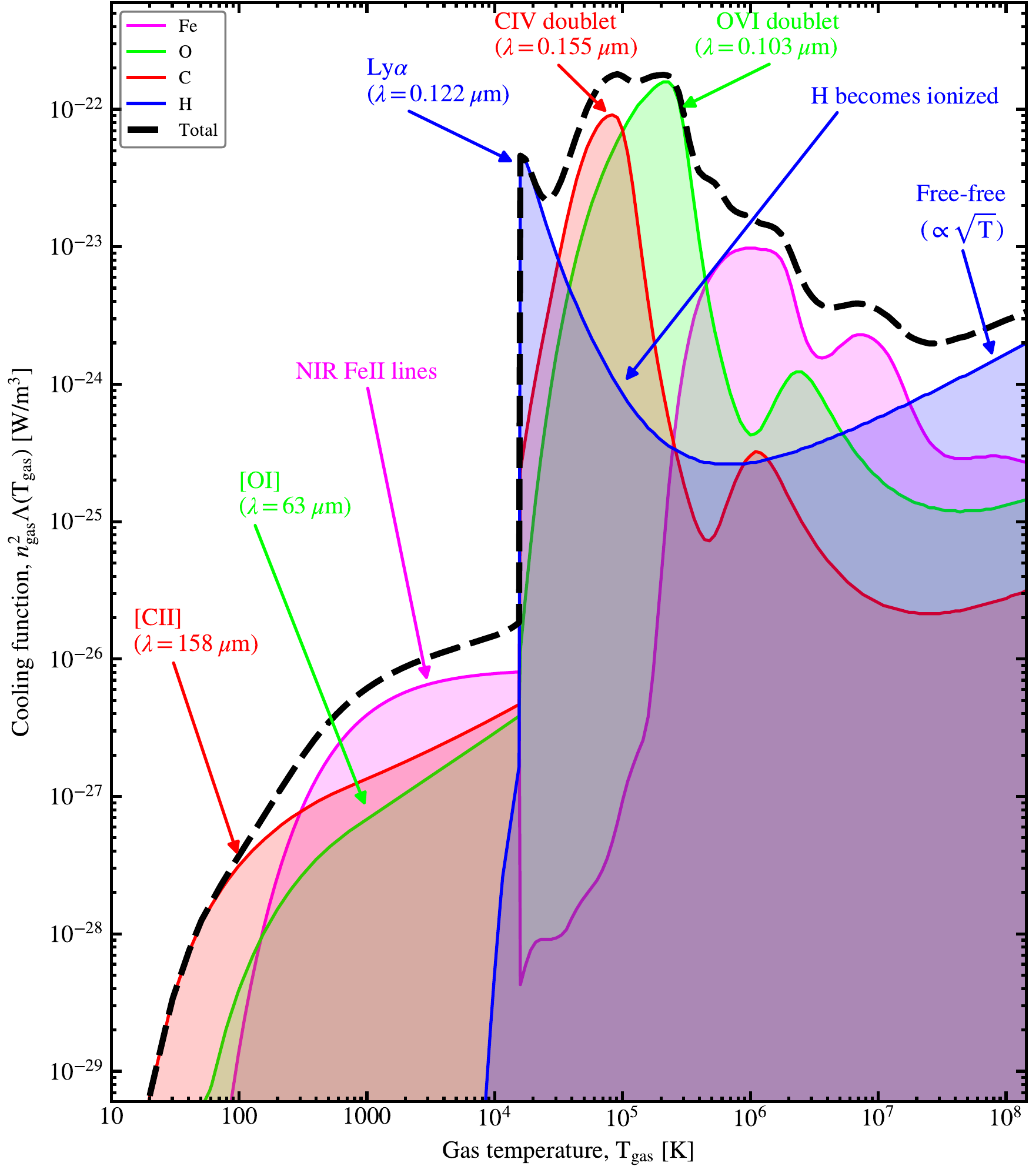}
  \newcap{Cooling function of the ISM at Solar metallicity}%
         {The data below $T_\sms{gas}=10^4$~K is from \citet{dalgarno72}, 
          and above, from \citet{schure09}.
          We have highlighted the main cooling elements.
          In the neutral regime, we have assumed an electron fraction of 
          $x\equiv n_e/n_\sms{H}=10^{-4}$.
          \CClicence}
  \label{fig:cooling}
\end{figure}

\paragraph{The cooling function.}
The way the gas cools across phases has a decisive impact on the multiphase structure of the \hISM.
It is possible to estimate its cooling rate as a function of temperature, or, in other words, how the thermal energy of the gas is converted into radiation.
This quantity is called the \expression{cooling function}.
It is represented on \reffig{fig:cooling}.
We have highlighted the dominant processes in the different temperature regimes.
\begin{itemize}
  \item At low-temperatures, in neutral atomic regions, \ciiline\ is the main 
    coolant.
  \item Around $T_\sms{gas}\simeq10^4$~K, there is drastic increase of cooling 
    efficiency, thanks to the \lyaline\ line of H.
  \item Above $T_\sms{gas}\gtrsim10^4$~K, the gas becomes rapidly ionized.
    The cooling is then channelled through various ions.
  \item At $T_\sms{gas}\gtrsim10^7$~K, in coronal phases, the gas cools mainly   
    \textit{via} free-free emission.
\end{itemize}
The exact shape of this cooling function can vary sensibly.
It depends on:
\begin{inlinelist}
  \item metallicity, as this parameter impacts the relative abundances of the 
    different species; and
  \item the radiation field, which impacts the ionization state of the gas.
\end{inlinelist}

    \subsubsection{The Neutral Atomic Gas}
    \label{sec:PE}

The neutral atomic gas is the most abundant phase in the \hMW: it accounts for $\simeq60\,\%$ of the total \hISM\ mass, and $\simeq8\,\%$ of the total baryonic mass (stars and \hISM).
It fills up about $\simeq30\,\%$ of the \hMW\ volume.
The neutral gas is far from thermal equilibrium, but it is roughly in pressure equilibrium, with:
\begin{equation}
  \frac{P_\sms{gas}}{k}=n_\sms{gas}\times T_\sms{gas}\simeq3000\;
  \textnormal{K/cm}^3.
  \label{eq:Pgas}
\end{equation}

\begin{figure}[htbp]
  \includegraphics[width=\textwidth]{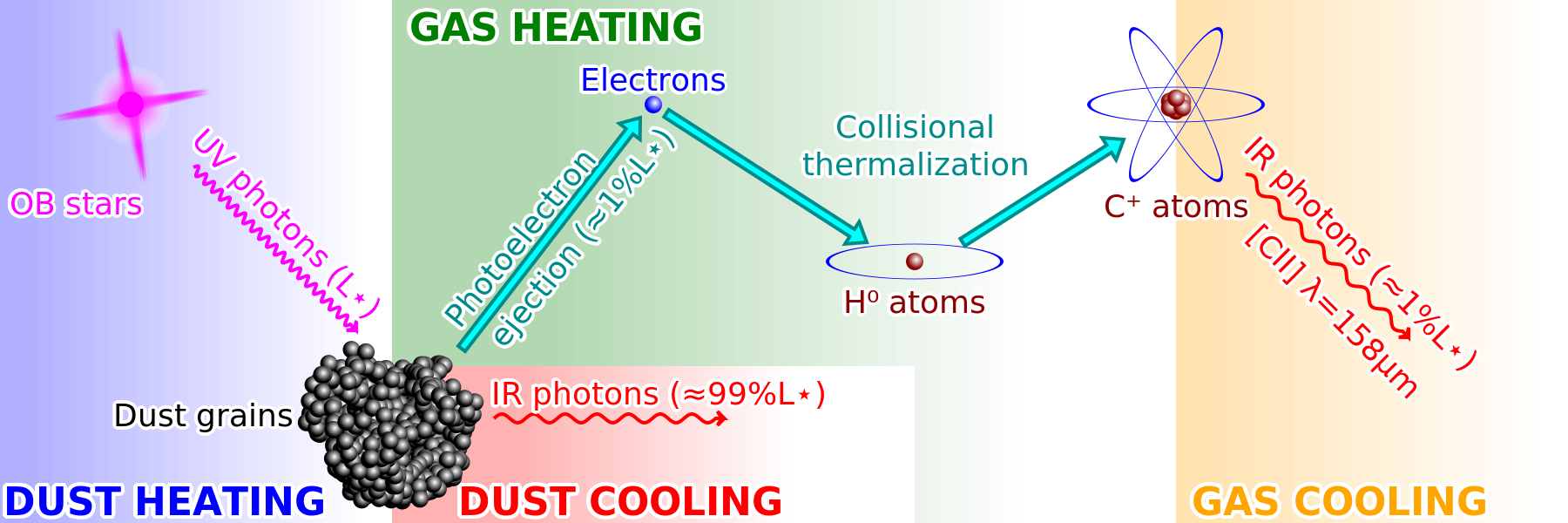}
  \newcap{Photoelectric heating in PDRs}%
         {We have implicitly assumed that the stellar power, $L_\star$, is 
          also the total power absorbed by dust, 
          as \hPDR s are mostly optically thick at \hUV\ wavelengths.
          \CClicence}
  \label{fig:photoel}
\end{figure}

\paragraph{The photoelectric heating.}
In neutral regions, the direct heating of the gas by absorption of stellar \hUV\ photons is not efficient, because only a small fraction of these photons can be absorbed through the different available atomic lines.
In these regions, the heating of the gas is indirect.
Dust grains absorb much more efficiently \hUV\ photons, with their spectrally continuous cross-section.
The absorption of an energetic photon (of a few eV) can lead to the ejection of an electron, \textit{via} the photoelectric effect.
This electron will then collide with the gas and heat it.
This is the \expression{photoelectric heating} of the gas \citep{de-jong77,draine78,bakes94,weingartner01b,kimura16}.
We have schematically represented this process in \reffig{fig:photoel}.
The overall efficiency of this process is related to the integrated surface of dust grains.
According to \reftab{tab:sizedistmoments}, this surface is dominated by small grains.
The smallest grains, especially \hPAH s, are therefore responsible for most of this heating.
\citet{wolfire95} give an expression for the photoelectric heating rate:
\begin{equation}
  n_\sms{gas}\Gamma\simeq\epsilon_\sms{PE}(\gamma,T_\sms{gas})\times   
    G_0\times\frac{n_\sms{gas}}{1\;\textnormal{cm}^{-3}}
    \times10^{-25}\;\textnormal{W/m}^{-3},
  \label{eq:PE}
\end{equation}
where $\epsilon_\sms{PE}$ is the efficiency of conversion of \hFUV\ energy into gas heating (it is a few percents), and $G_0$ and $\gamma$ have been defined in \refeqs{eq:G0}{eq:gamma}.

\begin{figure}[htbp]
  \includegraphics[width=\textwidth]{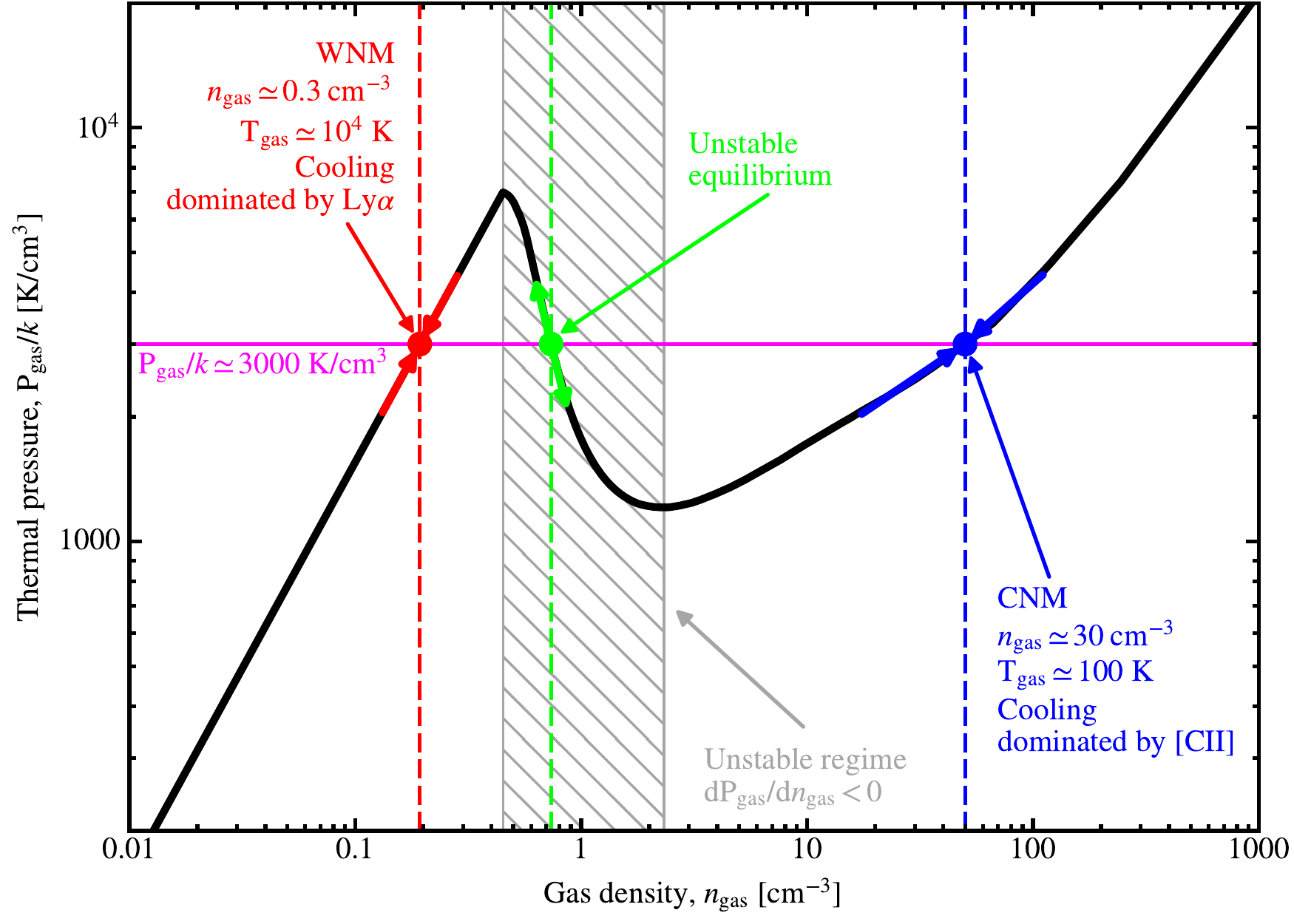}
  \newcap{Neutral ISM phase diagram}%
         {The black line is \refeq{eq:CNMWNM}, with the cooling function of 
          \reffig{fig:cooling}.
          The horizontal magenta line is the typical pressure of the \hISM\
          \refeqp{eq:Pgas}.
          There are three pressure equilibrium solutions (the three dots), but 
          one (the green dot) lies in the unstable regime (grey hatched).
          The arrows along each curve, near the dots, show the direction the gas
          would move if there was a perturbation around the equilibrium.
          \CClicence}
  \label{fig:phase_diagram}
\end{figure}

\paragraph{The two stable neutral atomic phases.}
For simplicity, let's assume the heating of the neutral \hISM\ is assured only by photoelectric heating.
Let's also use \refeq{eq:PE} with a fixed value of the photoelectric heating efficiency, $\epsilon_\sms{PE}=4.9\,\%$, and $G_0=1.7$.
The equilibrium is obtained when cooling and heating are balanced:
\begin{equation}
  n_\sms{gas}\Gamma=n_\sms{gas}^2\Lambda(T),
\end{equation}
which becomes, using \refeq{eq:Pgas}:
\begin{equation}
  P_\sms{gas}/k = \frac{\Gamma\times T}{\Lambda(T)}.
  \label{eq:CNMWNM}
\end{equation}
This is the black line we have represented in \reffig{fig:phase_diagram}.
There are three pressure equilibrium positions (the three dots), at the value of \refeq{eq:Pgas}.
We have hatched in grey the regime corresponding to unstable solutions.
In this regime, the pressure indeed decreases with density.
Thus, a small pressure increase, above the green dot, will decrease the density. It will thus make the gas expand and its temperature increase, moving further away from the green dot.
On the opposite, a small decrease of the pressure, below the green dot, will increase the gas density, and make its temperature decrease, at the same time.
The gas will thus collapse and move further away from the green dot.
The two only stable solutions correspond to the two main neutral atomic phases of the \hISM.
\begin{description}
  \item[The Warm Neutral Medium] (\hWNM) is a diffuse phase with density of
    the order of $n_\sms{gas}\simeq0.3\;\textnormal{cm}^{-3}$ and temperature
    $T_\sms{gas}\simeq10^4$~K. 
    It is heated essentially by the photoelectric effect, and also by cosmic 
    rays.
    It cools \text{via} \hUV-optical emission lines, essentially \lyaline.
  \item[The Cold Neutral Medium] (\hCNM) is a denser phase with density of the
    order of $n_\sms{gas}\simeq30\;\textnormal{cm}^{-3}$ and temperature
    $T_\sms{gas}\simeq100$~K.
    It is heated essentially by the photoelectric effect, and also by cosmic 
    rays.
    It cools essentially \textit{via} fine-structure lines, \ciiline\ and also
    \oiline.
\end{description}
Both of these phases are observed with \hiline, and \hUV-optical absorption lines.
\takeaway{There are two distinct neutral atomic phases in pressure equilibrium:
          the \hWNM\ and the \hCNM.}

    \subsubsection{The Ionized Gas}

The ionized gas accounts for a moderate mass fraction of the \hISM\ in the \hMW, but fills up a large volume. 
It is $\simeq23\,\%$ of the \hISM\ mass and $\simeq3\,\%$ of the total baryonic mass (stars and \hISM).

\paragraph{The Hot Ionized Medium (HIM).}
The \hHIM\ is a coronal gas ($n_\sms{gas}\simeq3\E{-3}\;\textnormal{cm}^{-3}$, $T_\sms{gas}\simeq10^6$~K; \reftab{tab:ISMism}) filling up $\simeq50\,\%$ of the \hMW\ volume.
It is at pressure equilibrium with the \hWNM\ and \hCNM.
\citet{mckee77} showed that this phase was heated by \hSN\ shockwaves.
It is composed of overlapping \hSN\ bubbles, and is sometimes referred to as the \expression{intercloud medium}.
This gas, which pervades everywhere, can be found significantly above the Galactic disk, contrary to the other phase, which are contained to the disk.
It is observed through \hFUV\ absorption lines of ions, and diffuse soft X-ray and synchrotron emissions.

\paragraph{\hii\ regions.}
\hii\ regions are short-lived dense ionized regions around OB star associations.
The gas is ionized by photons from the massive stars.
These regions are not in equilibrium, they are expanding.
\citet{stromgren39} has estimated the radius of an homogeneous sphere of ionized gas, around a star of H-ionizing photon rate, $Q_0$.
Such a sphere is called a \expression{Strömgren sphere}.
Its radius, $R_\sms{s}$, can be estimated by balancing photoionization and recombination:
\begin{equation}
  \underbrace{Q_0}_\sms{total ionization rate}
  = \underbrace{\frac{4\pi}{3}R_\sms{s}^3n_p}_\sms{number of protons}\times
        \underbrace{n_e\alpha_\sms{B}}_\sms{recombination rate},
  \label{eq:Rs0}
\end{equation}
where $\alpha_\sms{B}$ is the \expression{case B recombination rate}.
The product $n_e\alpha_\sms{B}$ is the electronic recombination rate to any H level $n\ge2$:
\begin{equation}
  \alpha_\sms{B}(T_e)\simeq 2.6\E{-13}T_e^{-3/4} \textnormal{cm}^3/\textnormal{s}.
\end{equation}
Recombination down to $n=1$ will indeed produce an ionizing photon that will be absorbed by another H atom.
Rearranging \refeq{eq:Rs0}, Strömgren's radius is thus:
\begin{equation}
  R_\sms{s} = \left(\frac{3Q_0}{4\pi n_\sms{gas}^2\alpha_\sms{B}}\right)^{1/3}.
  \label{eq:Rs}
\end{equation}
For an O5 star, with $n_\sms{gas}=10^3\;\textnormal{cm}^{-3}$, $R_\sms{s}\simeq1$~pc.
This estimate can be refined, accounting for He and other atomic species, as well as dust screening \citep[\eg][for an extensive lecture]{osterbrock89}.

\paragraph{The Warm Ionized Medium (WIM).}
The \hWIM\ is a diffuse phase ($n_\sms{gas}\simeq0.1\;\textnormal{cm}^{-3}$, $T_\sms{gas}\simeq10^4$~K; \reftab{tab:ISMism}) filling up about $\simeq25\,\%$ of the \hMW\ volume.
It is roughly in pressure equilibrium with the other thermal phases, although it can be found expanding in some regions.
This gas is photoionized by photons from OB star associations, escaping from \hii\ regions.
The electrons ejected by this photoionization provide the main heating source.
It is observed through optical lines, essentially \haline, as well as some fine-structure lines and free-free emission.

    \subsubsection{The Molecular Gas}
    \label{sec:H2}

The molecular gas constitutes only a moderate fraction of the \hISM\ mass and a small fraction of its volume.
It is however crucial, as it is where stars are formed.
In the \hMW, $\simeq17\,\%$ of the \hISM\ mass is molecular, which corresponds to $\simeq2\,\%$ of the total baryonic mass (stars and \hISM).

\begin{figure}[htbp]
  \includegraphics[width=\textwidth]{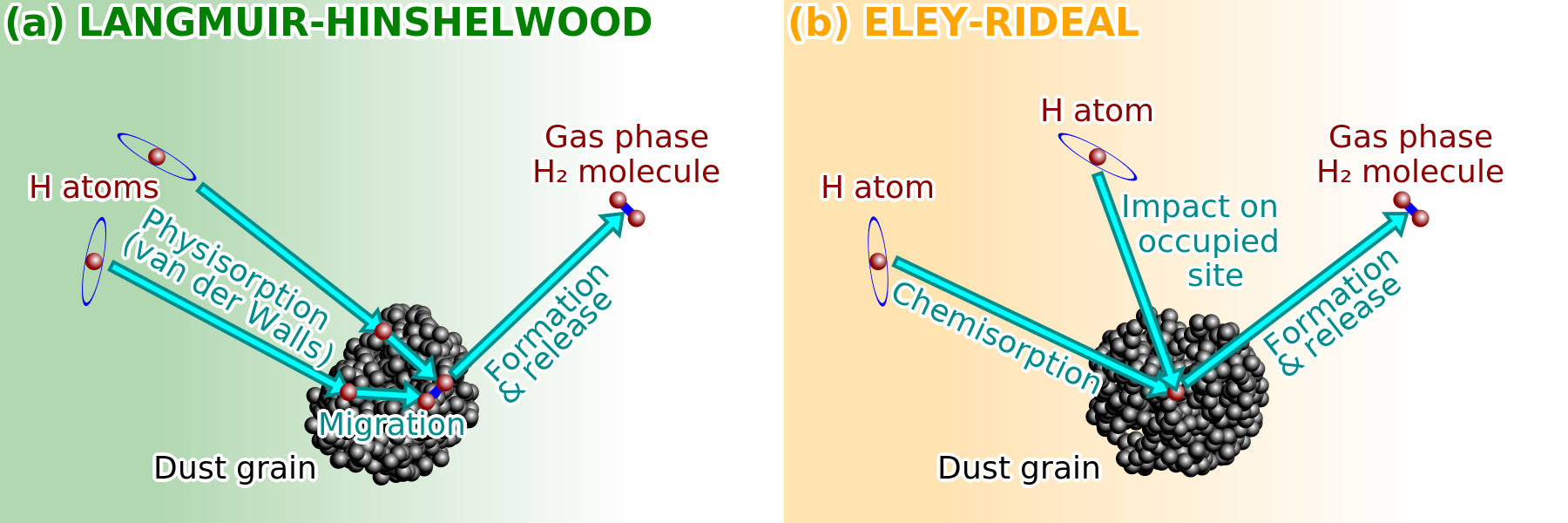}
  \newcap{\hmol\ formation on grain surface}%
         {Panel~\textit{(a)} illustrates the \expression{Langmuir-Hinshelwood
          mechanism}, which is driven by physisorption.
          H atoms are first captured by the grains and migrate to form \hmol.
          The energy released by the formation liberates the molecule.
          Panel~\textit{(b)} demonstrates the \expression{Eley-Rideal 
          mechanism}, which is driven by chemisorption.
          In this case, an H atom is absorbed at a site where another H atom is
          already present.
          See \citet{bron14b} for more details.
          \CClicence}
  \label{fig:formH2}
\end{figure}

\paragraph{Molecular hydrogen formation.}
The formation of the most abundant molecule of the Universe, \hmol, is inefficient in the gas phase.
This is because the freshly formed molecule needs to evacuate 4.5~eV to remain stable.
Yet, due to its symmetry, \hmol\ does not have rotational transitions that would allow it to radiate at these energy levels.
\hmol\ formation thus takes place on grain surfaces \citep[\eg][for a review]{bron14}.
We have represented the two main processes on \reffig{fig:formH2}: the so-called \expression{Langmuir-Hinshelwood} (physisorption) and \expression{Eley-Rideal} (chemisorption) mechanisms.
In both cases, the grain serves as a catalyst and helps evacuate the formation energy when the newly formed molecule is released in the gas phase.
Similarly to the photoelectric heating, this process happening on grain surfaces, it takes place primarily on small grains.
Temperature fluctuations therefore play an important role in its efficiency \citep[\eg][]{le-bourlot12}.

\paragraph{The diffuse molecular gas.}
Molecular gas can be observed at moderate densities ($n_\sms{gas}\simeq100\;\textnormal{cm}^{-3}$; \reftab{tab:ISMism}).
This is often associated with the \hCNM\ with large enough column densities to allow \hmol\ to be self-shielded (\ie\ its \hUV\ electronic lines are optically thick).
It is also heated by photoelectric effect and cosmic rays.
It cools primarily \textit{via} \ciiline\ and can be observed through \hUV\ absorption lines.

\begin{figure}[htbp]
  \includegraphics[width=\textwidth]{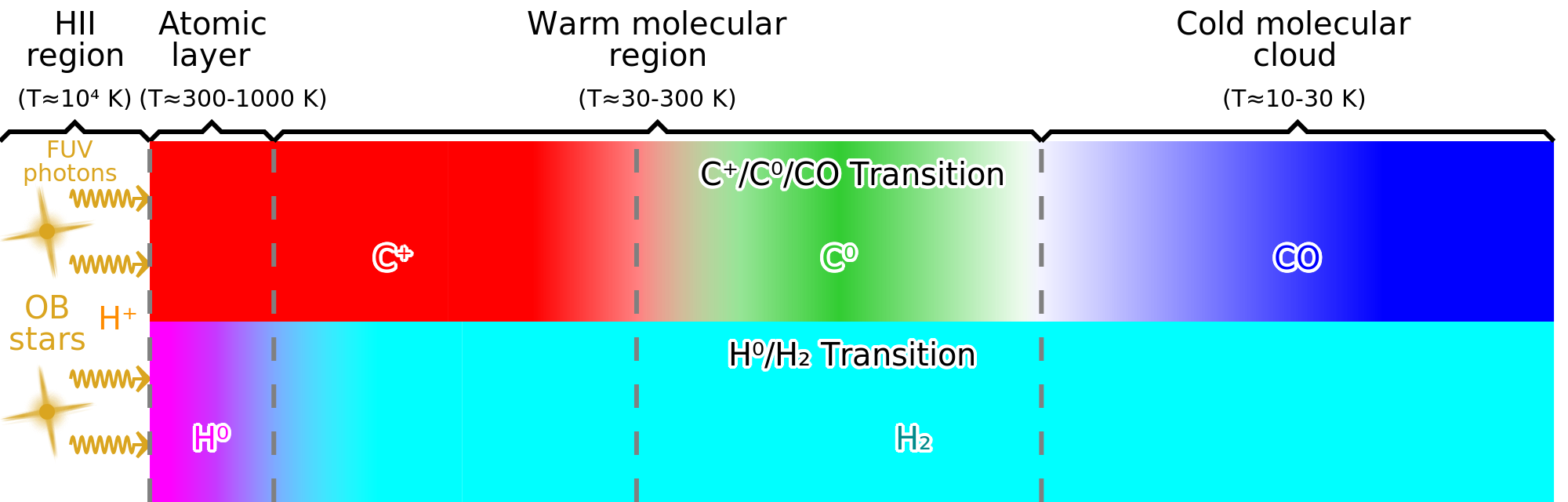}
  \includegraphics[width=\textwidth]{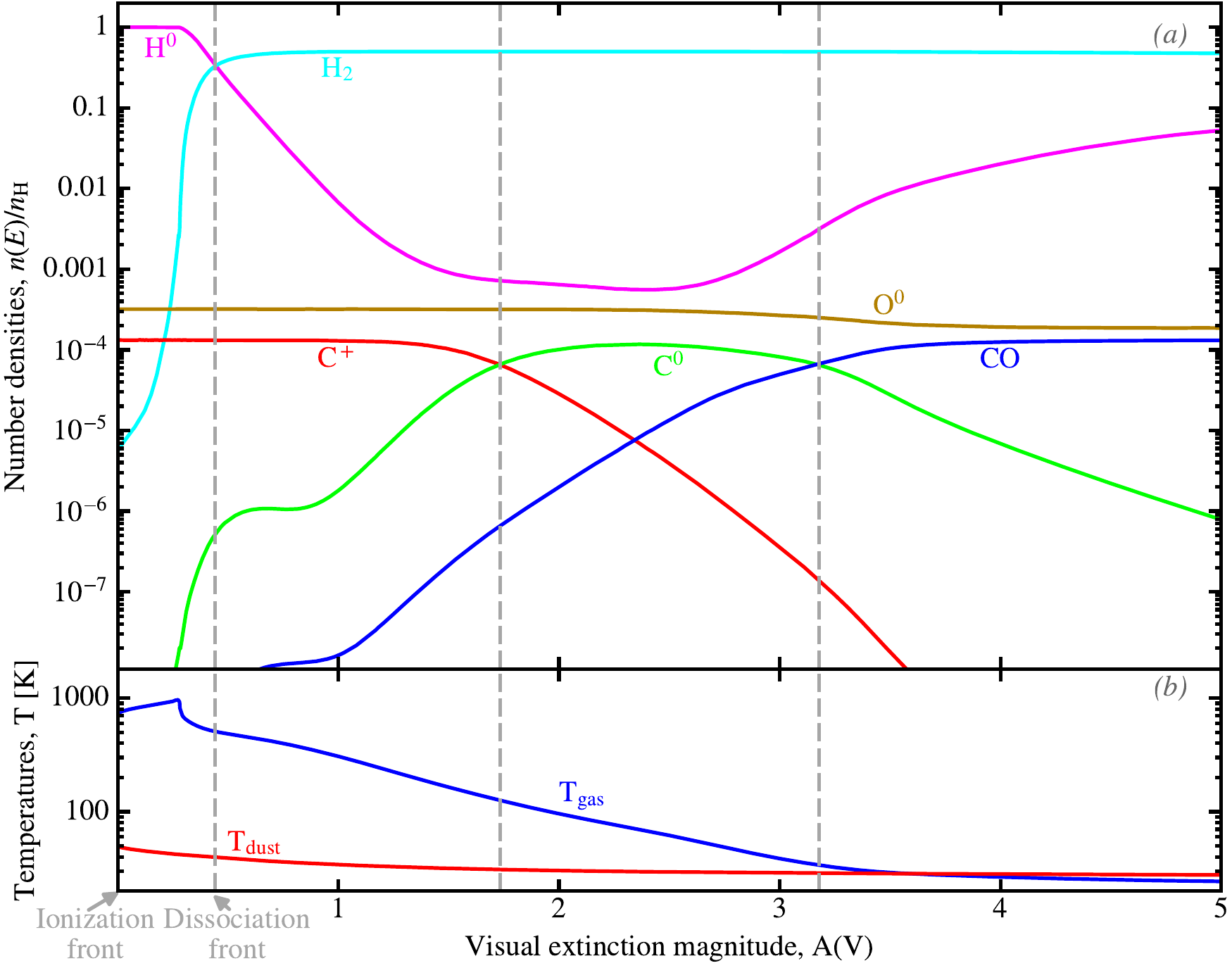}
  \newcap{Structure of a PDR}%
          {The top drawing schematically represents the structure of a \hPDR.
           The ionizing stars are illuminating the cloud from the left.
           In the \hii\ region, H is fully ionized.
           We enter the \hPDR\ at the ionization front.
           H is essentially neutral and atomic in this layer.
           We then pass the dissociation front around $A(V)\simeq0.5-1$, where
           H is essentially molecular (\hmol).
           Around $A(V)\simeq1.5-2$, C becomes neutral, and around 
           $A(V)\simeq3$, CO becomes the dominant carbon-bearing species.
           We are at this stage in the cold molecular cloud.
           The two bottom panels show the results of an isobaric \hPDR\ model 
           for $P_\sms{gas}=10^7\;\textnormal{K/cm}^{3}$ and $G_0=10^5$ 
           \citep[using the \ncode{Meudon PDR code};][]{le-petit06}.
           Panel~\textit{(a)} shows the evolution as a function of A(V) of the 
           densities of the most important species.
           These densities are normalized by the total H density, 
           $n_\sms{H}\equiv n(H^+)+n(H^0)+2n(H_2)$.
           Panel~\textit{(b)} shows the evolution of the gas and dust 
           temperatures.
           The exact densities, A(V) and temperatures are specific to the 
           particular model we have run, but they give a rough idea of the 
           typical values of these parameters.
           \CClicence}
  \label{fig:PDR}
\end{figure}

\paragraph{Photodissociation regions.}
Same as \hii\ regions, \hPDR s are a phase defined by the presence of massive stars in their vicinity.
They are not a stable phase of the \hISM, but they are very important since they are the interface between the \hii\ region and the molecular cloud \citep[\eg][for a review]{hollenbach97}.
Because of their high density and their high $G_0$, they absorb a significant fraction of the \hFUV\ radiation by OB stars and reradiate it in the \hIR, as dust thermal emission and fine structure lines.
Since they are the regions where molecules are being dissociated by \hFUV\ photons and subsequently recombine, they are the place of many important chemical reactions.
\reffig{fig:PDR} shows the structure of a typical \hPDR.
We have performed an isobaric run with the \expression{Meudon PDR code} \citep{le-petit06}, for $P_\sms{gas}=10^7\;\textnormal{K/cm}^{3}$ and $G_0=10^5$.
We show the variation of the abundances of the main species, and we represent in the upper part the \hi/\hmol\ and \cii/\ci/CO transitions.
This figure demonstrates that \hmol, being efficiently self-shielded, exists at lower A(V), whereas CO appears deeper into the cloud.

\paragraph{Dense molecular clouds.}
Dense molecular clouds ($n_\sms{gas}\simeq10^3-10^6\;\textnormal{cm}^{-3}$; \reftab{tab:ISMism}) contain most of the molecular gas, concentrated in a small volume.
These molecular clouds are gravitationally bound. 
Their structure is clumpy and filamentary.
The gas motions are controlled by turbulence.
They can be arranged in \expression{molecular complexes} of sizes up to $\simeq100$~pc \citep[\eg][]{solomon87}.
The densest cores are collapsing and will lead to star formation.
One of the most challenging issue of ISMology is the difficulty to measure the mass of molecular clouds.
As we have mentioned earlier, \hmol\ is a symmetric molecule.
It thus does not have any dipole moment allowing rotational transitions.
Its first transitions are its rovibrational levels (\hmolooo, \hmolooi\ and so on) that need temperatures of a few hundred K to be pumped.
Cold molecular clouds are thus primarily traced by the second most abundant molecule, CO, which is asymmetric.
CO rotational lines, \COio\ and \COiitoi, are the most commonly observed transitions.
The conversion of an observed \COio\ line intensity, $I_\sms{CO}$, to a \hmol\ column density requires the knowledge of a \expression{CO-to-\hmol\ conversion factor}, $X_\sms{CO}$, such that \citep[\eg][]{bolatto13}:
\begin{equation}
  N(\textnormal{\hmol}) \simeq X_\sms{CO}\times I_\sms{CO}.
  \label{eq:XCO}
\end{equation}
The $X(CO)$ factor has been calibrated on Galactic molecular clouds: $X(CO)\simeq2\E{20}\;\textnormal{cm}^{-2}(\textnormal{K.km/s})^{-1}$.
This value is however metallicity dependent, as we will see in \refsec{sec:darkgas}.

  \subsection{Dust as a Diagnostic Tool}

We have already discussed the potential of dust as a tracer of the physical conditions in the \hISM, especially in \refsec{sec:PAHband}.
We review here a few examples where dust tracers were used to refine the results of studies dedicated to star formation or gas physics.

    \subsubsection{Dust to Study Star Formation}

\paragraph{Star formation rates.}
Dust-related \hSFR\ tracers rely on the fact that young stars are extremely luminous and are  enshrouded with dust. 
If the clouds are optically thick and if their covering factor is unity, the OB star luminosity is: $L_\sms{OB}\simeq L_\sms{TIR}$.
Contrary to a common misconception, this is independent of dust properties.
Assuming a typical \hIMF, burst age and metallicity, $L_\sms{OB}$ can be 
converted to: $\textnormal{SFR}\simeq 10^{-10}\times L_\sms{TIR}/L_\odot$ \citep[\eg][]{kennicutt98b}.
The contribution of old stars to $L_\sms{TIR}$ is negligible for high enough \hSFR s.
Alternatively, monochromatic \hSFR\ indicators have been proposed.
\citet{calzetti07} and \citet{li10} found that the 24 and 70~\tmic\ monochromatic luminosities were good local \hSFR\ indicators (on spatial scales of $\simeq 0.5-1$~kpc): $\textnormal{SFR}\simeq 2611\times [L_\nu(24\emic)/(L_\odot/\textnormal{Hz})]^{0.885}$ and $\textnormal{SFR}\simeq 1547\times L_\nu(70\emic)/(L_\odot/\textnormal{Hz})$.
We have discussed the use of aromatic features as \hSFR\ tracers in \refsec{sec:PAHband}.
Finally, several composite indicators have been calibrated \citep{hao11}.
By combining \hFUV\ or \haline\ measurements with the 24~\tmic\ or \hTIR\ indicators, they account for the fact that star-forming regions are not completely opaque.

\paragraph{Resolving star formation.}
\citet{hony15} performed a comparison of different \hSFR\ estimators with the actual stellar content of the star-forming region N$\,$66, in the \hSMC.
In this study, we derived the stellar surface density, $\Sigma_\star$, based on individual star counts from \hHST\ photometry.
When compared to the dust mass surface density, $\Sigma_\sms{dust}$, derived from \hSED\ modeling, we found a significant scatter, at $\simeq6$~pc linear resolution.
The \hSFR s derived from $\Sigma_\sms{dust}$ or \haline\ underestimate the more reliable, $\Sigma_\star$-derived \hSFR, by up to a factor of $\simeq10$.
This is likely due to ionizing photons escaping the region\footnote{We saw a similar discrepancy in the center of \M{83}, where the \niiline\ line was significantly more extended than other \hSFR\ tracers \citep{wu15}.}.
Finally, converting our $\Sigma_\sms{dust}$ map to a gas mass surface density map, $\Sigma_\sms{gas}$, we found that the pixels of our region are lying above the \expression{Schmidt-Kennicutt relation}\footnote{The Schmitt-Kennicutt relation is the empirical correlation between $\Sigma_\sms{SFR}$ and $\Sigma_\sms{gas}$ for a wide diversity of galaxies \citep{schmidt59,kennicutt98}.}.
This might be due to low density gas, inefficient at forming stars, that is included in global Schmidt-Kennicutt relations, but absent when zooming on star-forming regions.

    \subsubsection{Photodissociation Regions}

\begin{figure}[htbp]
  \begin{tabular}{cc}
    \bfseries\textit{(a)} Three color image of \xxxdor &
    \bfseries\textit{(b)} $A(V)$ estimates \\
    \includegraphics[width=0.435\textwidth]{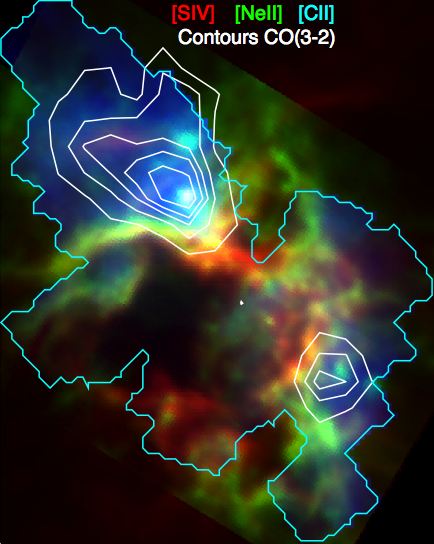} &
    \includegraphics[width=0.545\textwidth]{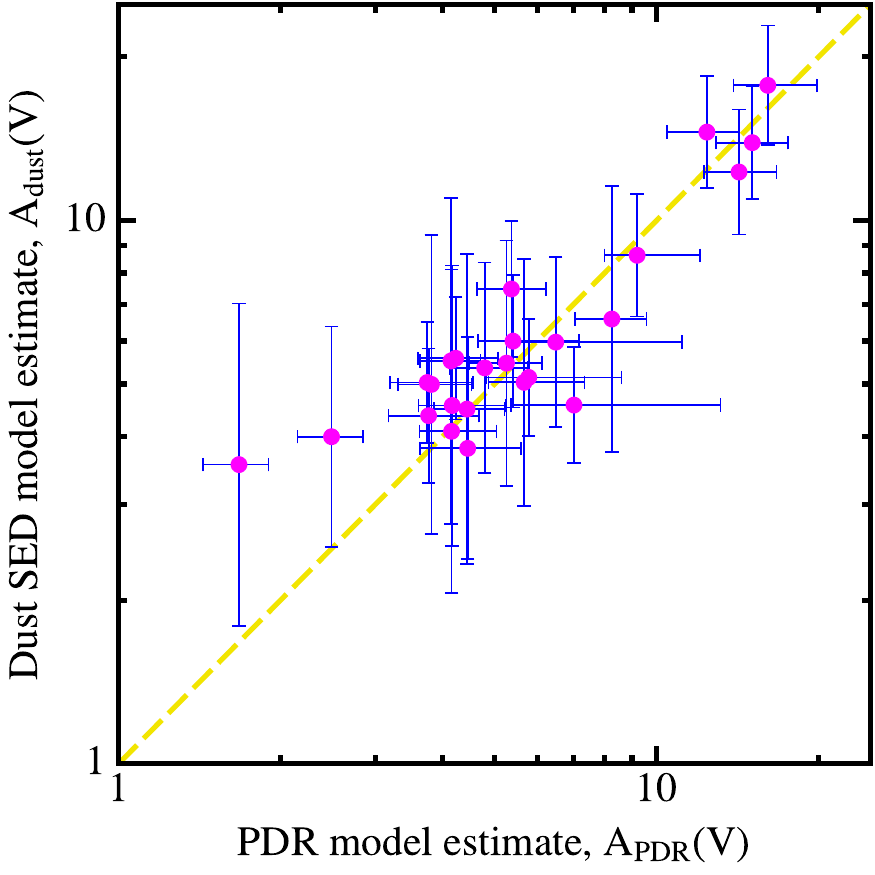} \\
  \end{tabular}
  \newcap{Comparison of visual extinctions in \xxxdor}%
         {Panel~\textit{(a)} shows a map of \xxxdor, seen through:
          \begin{inlinelist}
            \item \sivline\ \citep[red; heavily-ionized gas;][]{indebetouw09};
            \item \neiiline\ \citep[green; moderately ionized
              gas;][]{indebetouw09};
            \item \ciiline\ \citep[blue; neutral gas;][]{chevance16}; and
            \item \COiiitoii\ \citep[white contours; molecular 
              gas;][]{minamidani11}.
          \end{inlinelist}
          At the center of the image lies the \hSSC, R$\,$136.
          Panel~\textit{(b)} shows the comparison between the estimates of the 
          visual extinction magnitudes, $A(V)$ \refeqp{eq:Alambda}, in 
          different regions of panel~\textit{(a)}.
          The $x$-axis shows the $A(V)$ value derived from the \hPDR\ modeling
          of the available gas lines, whereas the $y$-axis shows the value of
          $A(V)$ inferred from dust \hSED\ modeling.
          \uline{Credit:} 
          \begin{inlinelistalph}
            \item \citet{chevance16};
            \item \cclicence.
          \end{inlinelistalph}}
  \label{fig:30Dor_Av}
\end{figure}

\paragraph{PDR Properties.}
We have participated to numerous studies aiming at constraining the \hPDR\ properties in low-metallicity environments \citep[][]{cormier10,cormier12,lebouteiller12,cormier15,chevance16,lebouteiller17,wu18b,cormier19,lebouteiller19}.
The common point of these studies is their use of numerous fine structure lines observed by \hhersc\ and \hspitz, as well as the dust emission, to constrain the main parameters of a \hPDR\ model ($G_0$, $n_\sms{gas}$, filling factor).
The challenge lies in the multiple degeneracies, due to the fact that a given line can trace several phases.
For instance, \ciiline\ comes from the ionized gas, the neutral gas and from an important fraction of molecular clouds (\reffig{fig:PDR}).
Such a degeneracy can be solved by using additional lines to constrain the properties of these different phases.
Dust tracers are also useful, either as a constraint or as a self-consistency test.
As an example, \citet{chevance16} modeled the gas properties in \xxxdor\ and derived the typical depth of \hPDR s, in terms of visual extinction magnitude, $A_\sms{PDR}(V)$.
\reffig{fig:30Dor_Av} shows the comparison of $A_\sms{PDR}(V)$ to the value of this parameter, derived from \hSED\ modeling, $A_\sms{dust}(V)$.
We can see that both quantities are in good agreement, validating the \hPDR\ results.

\paragraph{Photoelectric heating.}
Assuming that \ciiline\ is the main gas coolant, the \expression{photoelectric efficiency}, that we already discussed in \refsec{sec:PE}, $\epsilon_\sms{PE}$, can be approximated by the gas-to-dust cooling ratio: $\epsilon_\sms{PE}\simeq L_\sms{\cii}/L_\sms{TIR}$.
Detailed studies usually add other lines to the gas cooling rate, like \oiline, to have a more complete estimate \citep[\eg][]{lebouteiller12,cormier15,lebouteiller19}.
Overall, \citet{smith17} found that $0.1\,\%\lesssim \epsilon_\sms{PE}\lesssim 1\,\%$, with an average of $\langle\epsilon_\sms{PE}\rangle\simeq0.5\,\%$, in a sample of 54 nearby galaxies.
It appears that $\epsilon_\sms{PE}$ is lower when the dust temperature is higher \citep{rubin09,croxall12}.
This is not likely the result of the destruction of the \hUIB\ carriers, as their intensity correlates the best with the \ciiline\ emission \citep[\eg][]{helou01}.
It is rather conjectured to be due to the saturation of grain charging 
in \hUV-bright regions.
The shape of the \hISRF\ also has a consequence on the accuracy with which $L_\sms{TIR}$ represents the true \hUV, photoelectric-efficient flux.
Indeed, \citet{kapala17} showed that the variations of $\epsilon_\sms{PE}$ in \M{31} could be explained by the contribution of old stars to $L_\sms{TIR}$.
Finally, one of the most puzzling features is that $\epsilon_\sms{PE}$ is higher at low metallicity \citep{poglitsch95,madden97,cormier15,smith17,madden20}, while the \hUIB\ strength drops in these systems (\refsec{sec:PAHevol}).
This is currently poorly understood.
However, in the extreme case of \izw\ ($Z\simeq1/35\,Z_\odot$), where no \hUIB\ is detected \citep[\eg][]{wu06} and the photoelectric heating is estimated to be negligible, the gas-cooling-to-\hTIR\ ratio is still $\simeq1\,\%$ \citep{lebouteiller17}.
In this instance, we have shown the gas could be heated by X-rays.

    \subsubsection{The Molecular Gas and its Dark Layer}
    \label{sec:darkgas}

\begin{figure}[htbp]
  \includegraphics[width=\textwidth]{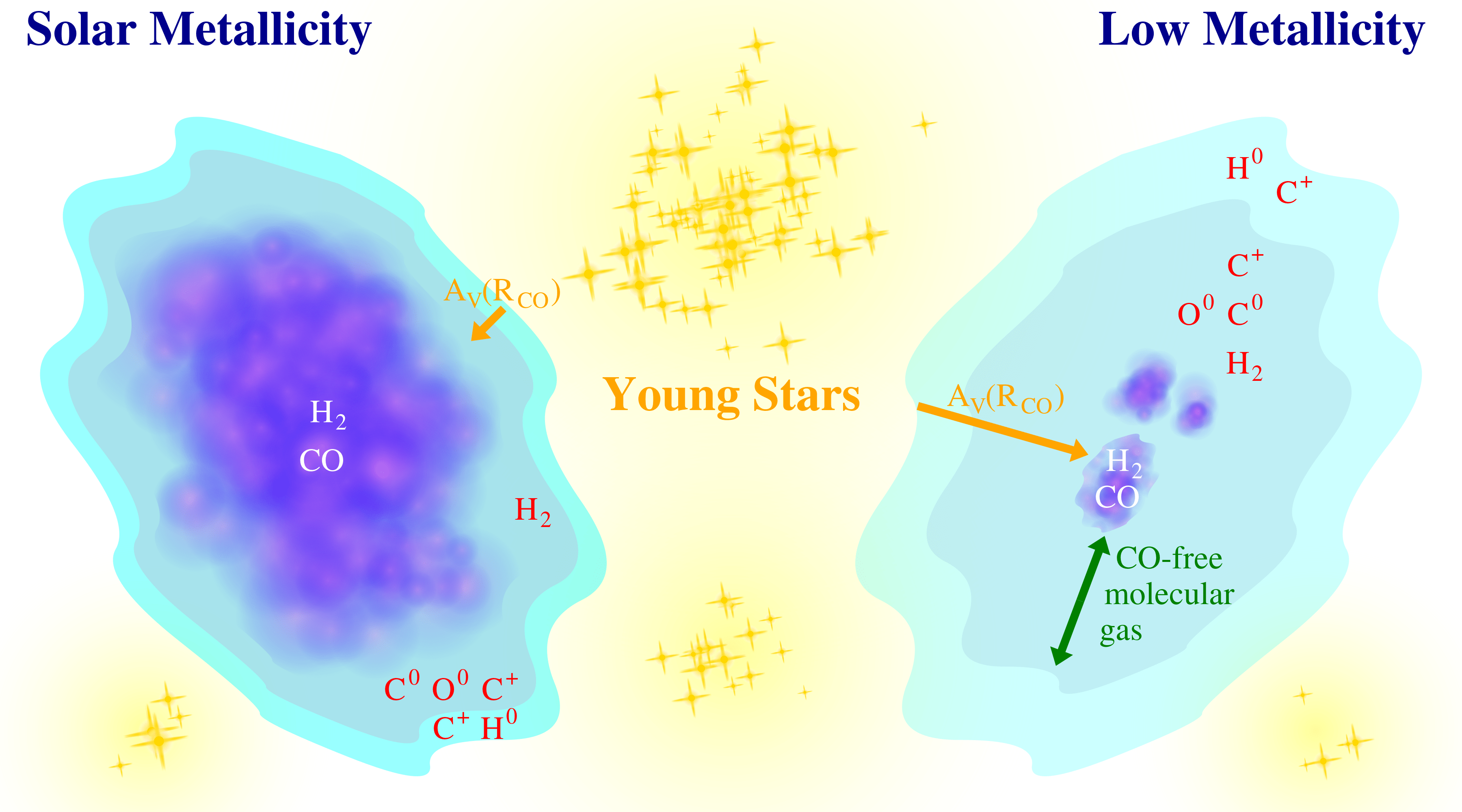}
  \newcap{Metallicity effect on the CO-dark gas}%
         {We have represented two molecular clouds photodissociated by nearby 
          OB star associations.
          The left cloud has a Solar metallicity.
          Its CO core is efficiently shielded by dust.
          This is not the case for the right cloud, which has a low metallicity.
          In this cloud, the lower grain abundance causes a lower dust
          screening.
          Consequently, CO is photodissociated deeper into the cloud, whereas
          \hmol\ remains self-shielded.
          \uline{Credit:} \citet{madden20}.}
  \label{fig:darkCO}
\end{figure}

\paragraph{The dark gas.}
We have discussed in \refsec{sec:H2} that the photodissociation of \hmol\ and CO at different depths into molecular clouds leads to biases in gas mass estimates.
\hmol\ is indeed self-shielded.
It therefore exists at column densities roughly independent of metallicity.
On the contrary, CO, which is significantly less abundant, is not self-shielded (\ie\ its electronic lines remain optically thin at large column densities).
Consequently, CO needs to be shielded by dust to survive.
It thus exists only deeper into molecular clouds.
The \hmol\ gas that exists in regions where CO is photodissociated is often referred to as the \expression{CO-dark molecular gas}.
Other tracers can be used to constrain this dark gas: 
\begin{inlinelist}
  \item dust emission \citep[\eg][]{israel97,leroy11};
  \item \ciiline\ \citep[\eg][]{madden97}; 
  \item and $\gamma$-rays \citep[\eg][]{grenier05}.
\end{inlinelist}
Using $I_\sms{CO}$ to derive $N(\textnormal{\hmol})$ with \refeq{eq:XCO} can therefore be biased if the dark gas fraction is significantly larger than in the \hMW, where $X(CO)$ has been calibrated.
This is what happens in low-metallicity systems, where the \hdustiness\ is lower, because of the lower abundance of heavy elements (\cf\ \refsec{sec:G2DvsZ}).
This is represented in \reffig{fig:darkCO}.
We see that in the low-metallicity cloud (on the right), CO cores are much smaller, because of the increased photodissociation of this molecule, compared to the left cloud (Solar metallicity).
It results that the mass of CO-dark \hmol\ is significantly larger at low metallicity.
We have investigated this \hISM\ component in the \hLMC, using dust mass surface density, concluding it could account between $\simeq10\,\%$ and $\simeq100\,\%$ of the total molecular gas mass \citep{galliano11}.
Recently, we studied dark gas in the star-forming region N$\,$11 of the \hLMC, modeling the full set of \hIR\ emission lines \citep{lebouteiller19}.
We showed that most of the molecular gas in this region is CO-dark and that \ciiline\ traces mostly this component.
We extended this analysis to a sample of nearby dwarf galaxies \citep{madden20}.
We found that $\simeq70-100\,\%$ of the molecular gas mass is not traced by \COio.
\takeaway{The molecular gas content of low-metallicity systems is dominated by dark gas.}

\begin{figure}[htbp]
  \begin{tabular}{cc}
    \bfseries\textit{(a)} Pressure/ISRF relation &
    \bfseries\textit{(b)} SFR map of \M{83} \\
    \includegraphics[width=0.435\textwidth]{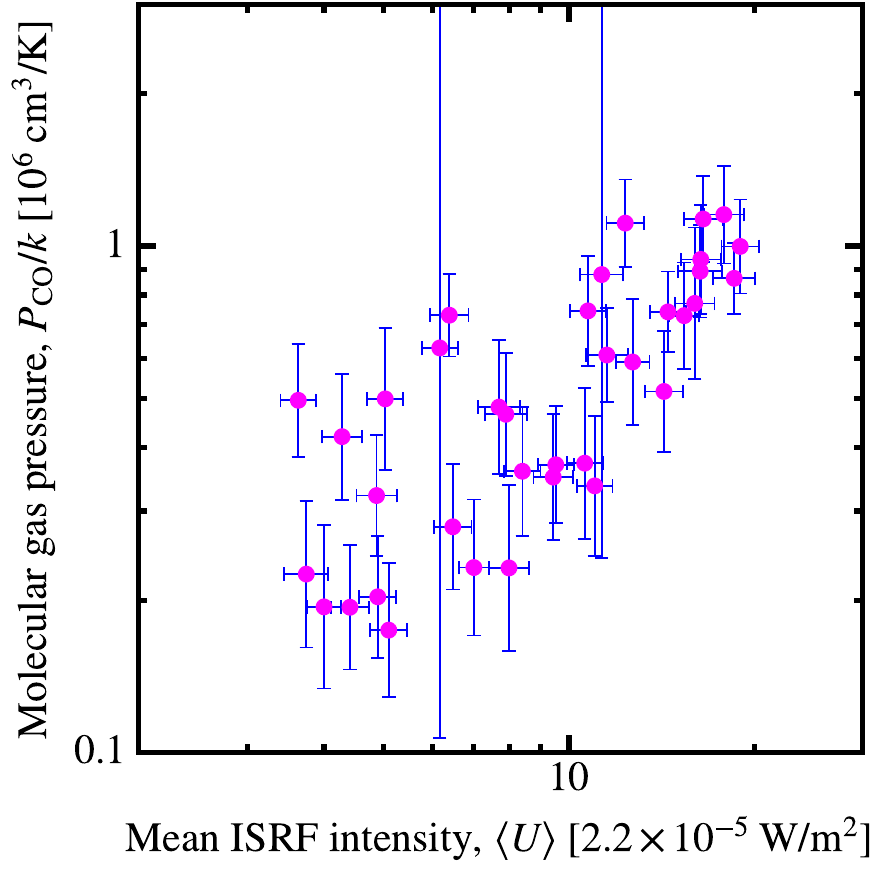} &
    \includegraphics[width=0.545\textwidth]{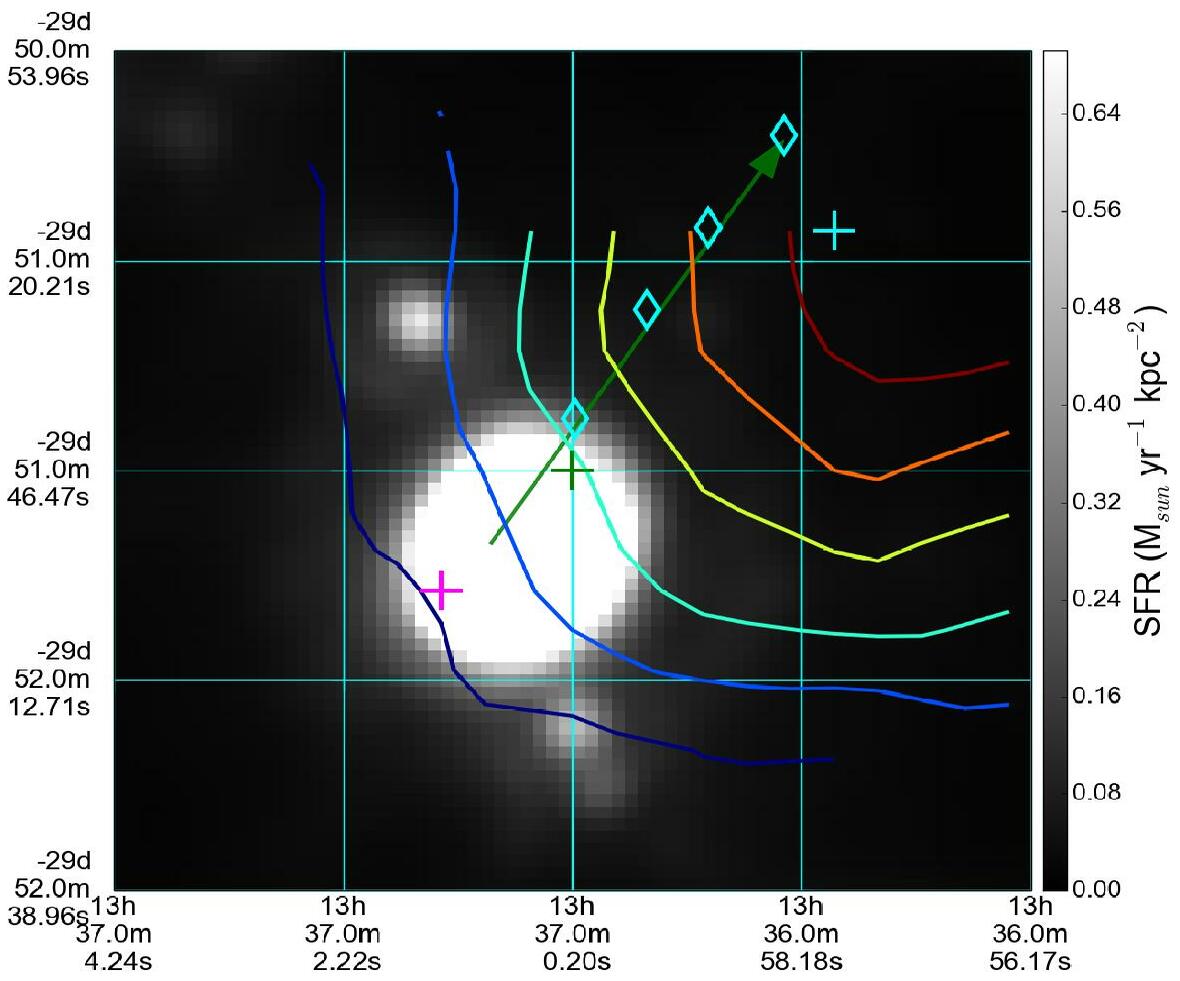} \\
  \end{tabular}
  \newcap{Molecular gas pressure in the center of \M{83}}%
         {Panel~\textit{(a)} displays the relation between the average 
          starlight intensity derived from \hSED\ modeling, and the molecular 
          gas pressure, inferred from CO \hSLED\ modeling, in different regions
          in the center of \M{83} \citep{wu15}.
          Panel~\textit{(b)} shows the \hSFR\ map of the central region of 
          \M{83}.
          The color contours represent the molecular gas pressure, from
          $P_\sms{CO}=4\E{5}\;\textnormal{K/cm}^3$ (blue) to 
          $P_\sms{CO}=2.4\E{6}\;\textnormal{K/cm}^3$ (red).
          The four diamonds indicates the position of the four radio sources
          reported by \citet{maddox06}.
          \uline{Credit:} 
          \begin{inlinelistalph}
            \item \cclicence; 
            \item \citet{wu15}.
          \end{inlinelistalph}}
  \label{fig:m83}
\end{figure}

\paragraph{Pressure and radiation field.}
The pressure in molecular clouds can be significantly larger than in the pressure equilibrium phases of the \hISM: $P_\sms{gas}\simeq6\E{7}\;\textnormal{K/cm}^{-3}$ in the Orion bar \citep{goicoechea16}; compared to $P_\sms{gas}\simeq3000\;\textnormal{K/cm}^{-3}$ in the \hHIM, \hWIM\ and \hCNM\ \refeqp{eq:Pgas}.
We have studied the physical conditions of the molecular gas in the central region of the starbursting galaxy, \M{83} \citep{wu15}.
We used the CO \expression{Spectral Line Energy Distribution} (\hSLED) observed by \hhersc\ to estimate its column density and pressure, $N(CO)$ and $P_\sms{CO}$, in different regions.
We have also performed \hSED\ modeling to estimate the mean starlight intensity heating the grains, $\langle U\rangle$ \refeqp{eq:Uav}.
This allowed us to show that both quantities are correlated (\cf\ \refsubfig{fig:m83}{a}).
We also noted that the pressure gradient was oriented along a chain of radio sources, corresponding to a radio jet (\refsubfig{fig:m83}{b}).
We derived a similar correlation between the \hISRF\ strength and the gas thermal pressure in the Carina nebula \citep{wu18b}.
Such a correlation was also found by \citet{joblin18} in the Orion bar.
They argue that the photoevaporation of the \hPDR\ can explain this relation.


\newchapter{Modeling Cosmic Dust Evolution}
\markboth{\chaptername\ \thechapter.\ Dust Evolution}{}
\label{chap:dustevol}
\citesmart{This is not a new result --~\citet{draine79} reached the conclusion that grain destruction was rapid and that regrowth of dust in the ISM was required to explain the observed depletions.
The numbers basically haven't changed appreciably since then; the argument has been reiterated a number of times (...). Nevertheless, some authors continued to hold the view that the solids in the interstellar medium were primarily formed in stars.}{\citep[Bruce T.\ \familyname{Draine};][]{draine09}}
\minitoc

\begin{figure}[htbp]
  \includegraphics[width=\textwidth]{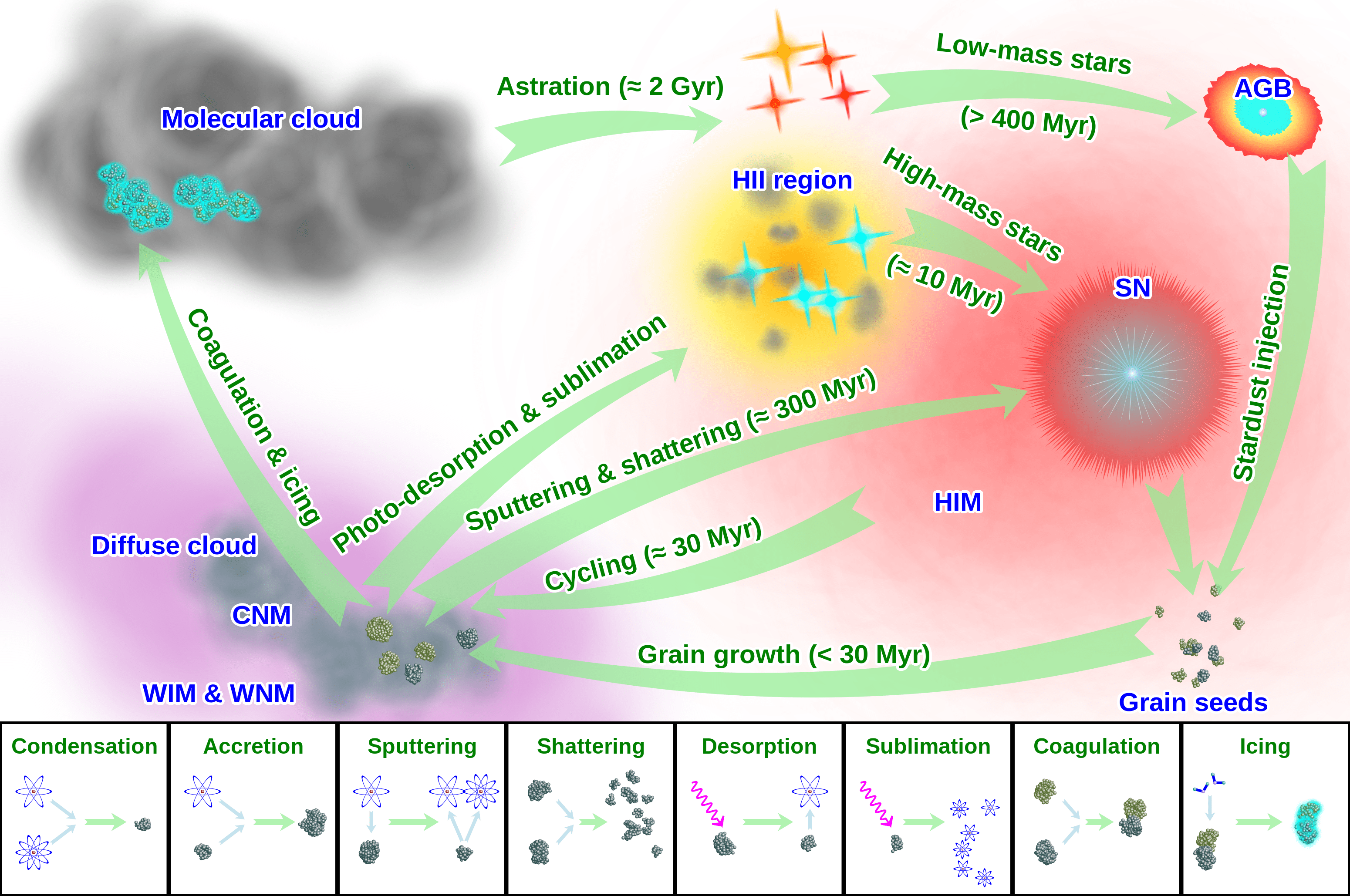}
  \newcap{The interstellar dust lifecycle}%
         {This is the schematic representation of dust evolution through the 
          \hISM.
          The upper part illustrates the different sources, sinks and 
          environments where grains are processed.
          The timescales are indicative and will be discussed in the rest of 
          this chapter.
          The eight small bottom panels focus on the microscopic processes.
          \CClicence}
  \label{fig:dustevol}
\end{figure}

\noindent
This chapter focusses on the study of dust evolution in all interstellar environments, at all spatial scales.
Dust evolution is the variation of the constitution of a grain mixture with time, under the effects of its environment.
The timescales of evolution being significantly longer than the career of a scientist, we usually study spatial variations of the dust content in a region, or the variations among a sample of galaxies.
These different observations are then compared, being considered as snapshots at different evolutionary stages.
The environmental parameters that are commonly used to quantify dust evolution are:
\begin{inlinelist}
  \item the \hISM\ density and the \hISRF\ intensity and hardness, for 
    spatially-resolved studies;
  \item the metallicity and star formation rate, for global galactic studies.
\end{inlinelist}
The main processes responsible for dust evolution are represented on \reffig{fig:dustevol}.
\begin{description}
  \item[Grain Formation]
    is the dust mass build-up by:
    \begin{itemize}
      \item grain condensation in the ejecta of core-collapse \hSN e and 
        \hAGB\ stars (\ie\ making grains from scratch by condensing atoms);
      \item grain (re-)formation in the \hCNM\ and molecular clouds, by 
        accretion of atoms and molecules onto grains:
        \begin{inlinelist}
          \item grain growth;
          \item mantle accretion; and 
          \item ice formation.
        \end{inlinelist}
    \end{itemize}
  \item[Grain Processing]
    is the alteration of the grain constitution in the \hISM\ by:
    \begin{itemize}
      \item shattering and fragmentation by grain-grain collisions in 
        low-velocity shocks (modification of the size distribution);
      \item structural modifications by high energy photons or cosmic ray
        impacts;
      \item grain-grain coagulation in cold regions.
    \end{itemize}
  \item[Grain Destruction]
    is the full or partial removal of the elements constituting the grains by:
    \begin{itemize}
      \item erosion and evaporation by thermal or kinetic sputtering (gas-grain 
        collision in a hot gas or a shock);
      \item photodesorption of atoms and molecules;
      \item thermal evaporation;
      \item astration (incorporation into stars).
    \end{itemize}
\end{description}

\section{Stellar Evolution}

Stars have a crucial impact on \hISD: 
\begin{inlinelist}
  \item they synthesize the heavy elements that constitute dust grains
    (\reffig{fig:abundsun});
  \item they also directly produce dust seeds in their ejecta;
  \item the shock waves of \hSN e erode and vaporize the grains;
  \item the radiative and mechanical feedback of massive stars carve the \hISM\
    and process the grains.
\end{inlinelist}

  \subsection{The Fate of Stars of Different Masses}

A star can be conceptualized as a sphere of gas in hydrostatic equilibrium, where the gravity is counterbalanced by the thermal pressure sustained by nuclear reactions in its core \citep[\eg][for an introduction]{deglinnocenti16}.
The energy produced in the core is carried out through radiation, convection or conduction.
The initial mass of a star, and in a lesser extent its initial metallicity, determine its future evolution.

    \subsubsection{Nucleosynthesis}
    \label{sec:nucleosynthesis}

The nuclear reactions in stellar interiors, on top of being the fuel of stars, lead to the production of heavy elements.
A fraction of these freshly synthesized elements are injected back into the \hISM, during the final stages of stellar evolution.

\begin{figure}[htbp]
  \includegraphics[width=\textwidth]{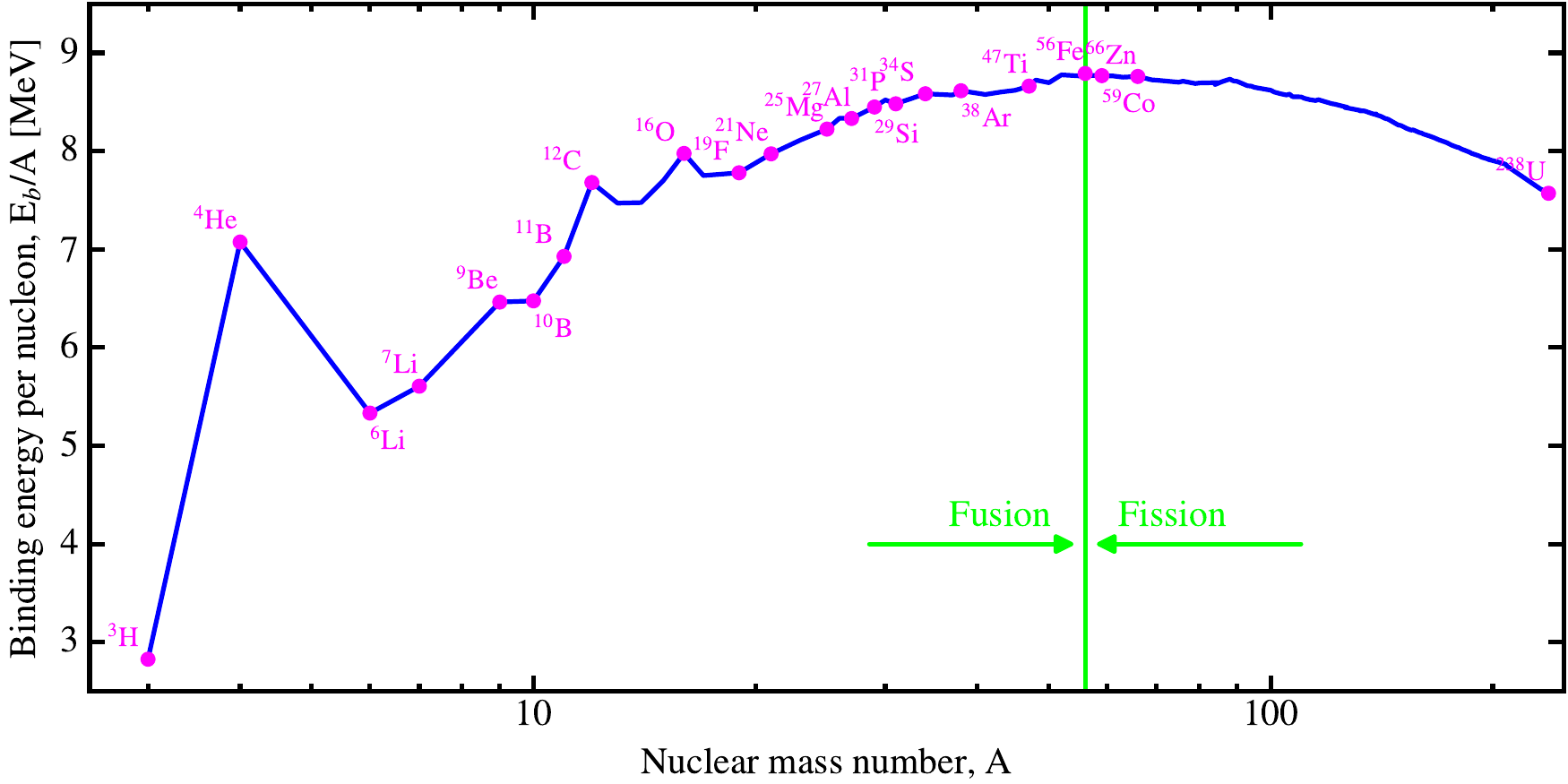}
  \newcap{Average nuclear binding energies per nucleon}%
         {We have displayed the experimental data compiled by 
          \citet{ghahramany12}.
          The most stable elements are around $^{56}$Fe.
          Below, fusion is exothermic, above, it is fission.
          \CClicence}
  \label{fig:nuclear_binding}
\end{figure}

\paragraph{Nuclear binding energies.}
A fundamental quantity to determine the efficiency of nuclear reactions to sustain the thermal pressure within a star is the nuclear binding energy of an element of mass A \citep[number of nucleons; \cf\ \eg\ Chaps.~1-2 of][for a review]{pagel97}.
This quantity is represented in \reffig{fig:nuclear_binding} for the most relevant nuclei.
To have an exothermic reaction, that will be able to counterbalance gravity, one needs to synthesize elements of higher binding energies.
The curve of \reffig{fig:nuclear_binding} reaches a maximum around $^{56}$Fe.
\begin{description}
  \item[Fusion] of elements lighter than $^{56}$Fe is exothermic.
    Since the initial composition of a star is $\simeq3/4$ H and $\simeq1/4$ He,
    stellar nucleosynthesis takes this way.
  \item[Fission] of elements heavier than $^{56}$Fe is exothermic.
    This is the process implemented in nuclear reactors to generate electricity 
    (through heat).
\end{description}

\paragraph{Primordial nucleosynthesis.}
Before the first stars appeared, $^2$H and $^4$He, as well as elements up to $^7$Li, were synthesized during the first 15 minutes after the big bang \citep[\eg][]{pagel97,calura04,johnson19}.
The temperature was at this time around $T\simeq10^9$~K.
This primordial nucleosynthesis was brief, as the Universe was expanding and cooling.
It is estimated that after 20 minutes, the temperature was too low to synthesize new elements.
The \expression{primordial abundances} refer to the elements produced during these first minutes (\cf\ \refeqnp{eq:abund}):
\begin{equation}
    X_\sms{primordial}\simeq0.76,\;\;\;\;\;\;
    Y_\sms{primordial}\simeq0.24,\;\;\;\;\;\;
    Z_\sms{primordial}\simeq0.00.
\end{equation}

\paragraph{Stellar nucleosynthesis.}
Once the temperature at the center of a collapsing protostar becomes high enough ($T\simeq10^7$~K), \expression{thermonuclear} reactions\footnote{In thermonuclear reactions, the high temperature gives nuclei enough kinetic energy to overcome their Coulomb barrier, and allows them to fuse with each other.} are initiated.
Several chains and cycles of reactions occur in stars, at different stages.
The most important ones are the following \citep[\eg][]{filippone86,pagel97,silva-aguirre18}.
\begin{description}
  \item[Proton fusion,] also called \expression{p-p chain}, is a series of 
    nuclear reactions converting 4$^1$H into $^4$He.
    This reaction chain has three branches cycling through various light 
    elements (D, Li, B, Be).
    It is the dominant process in stellar interiors with 
    $T\lesssim2\E{7}$~K, that is for stars with mass $m_\star\lesssim1\eMsun$. 
    \footnote{The Sun's core is at $T\simeq1.5\E{7}$~K.}
  \item[CNO cycle] is another series of nuclear reactions converting 4$^1$H 
    into $^4$He.
    Contrary to the p-p chain, this cycle requires pre-existing C, N or O (\ie\
    it requires a non-zero-metallicity star).
    This cycle can be broken into:
    \begin{inlinelist}
      \item a \expression{CN cycle}, starting with $^{12}$C and 4$^1$H, ending 
        with $^{12}$C and $^4$He; and
      \item a \expression{NO cycle}, starting with $^{15}$N and 3$^1$H, ending 
        with $^{14}$N and $^4$He.
    \end{inlinelist}
    This cycle is more efficient than the p-p chain for $T\gtrsim2\E{7}$~K, that
    is for stars with $m_\star\gtrsim1\eMsun$.
    In practice, both happen simultaneously, but with different efficiencies.
  \item[The triple $\alpha$ process] 
    is a series of nuclear reactions converting 3$^4$He into $^{12}$C (the 
    $^4$He nucleus is indeed called the \expression{$\alpha$ particle}).
    It also produces $^{16}$O and $^{20}$Ne as byproducts.
    This process starts when the star has converted $\simeq10\,\%$ of its H into
    He.
    It requires temperatures of $T\simeq10^8$~K.
  \item[Heavier element fusion] happens essentially in massive stars 
    ($m_\star\gtrsim8\eMsun$), when the temperature of the core reaches 
    $T\simeq10^9$~K.
    Several successive phases are then possible: C burning, Ne burning and O 
    burning, producing up to $^{28}$Si.
    The last series of reactions are the $\alpha$ ladder, which produce 
    elements up to Fe and Ni.
\end{description}
\takeaway{\expression{H burning}, which encompasses both the p-p chain and the 
          CNO cycle, represents the longest phase in the lifetime of a star,
          whereas \expression{He burning} lasts only $\simeq10\,\%$ 
          of its existence.}

    \subsubsection{Brief Outline of Stellar Evolution}
    \label{sec:stellarevol}

\begin{figure}[htbp]
  \includegraphics[width=\textwidth]{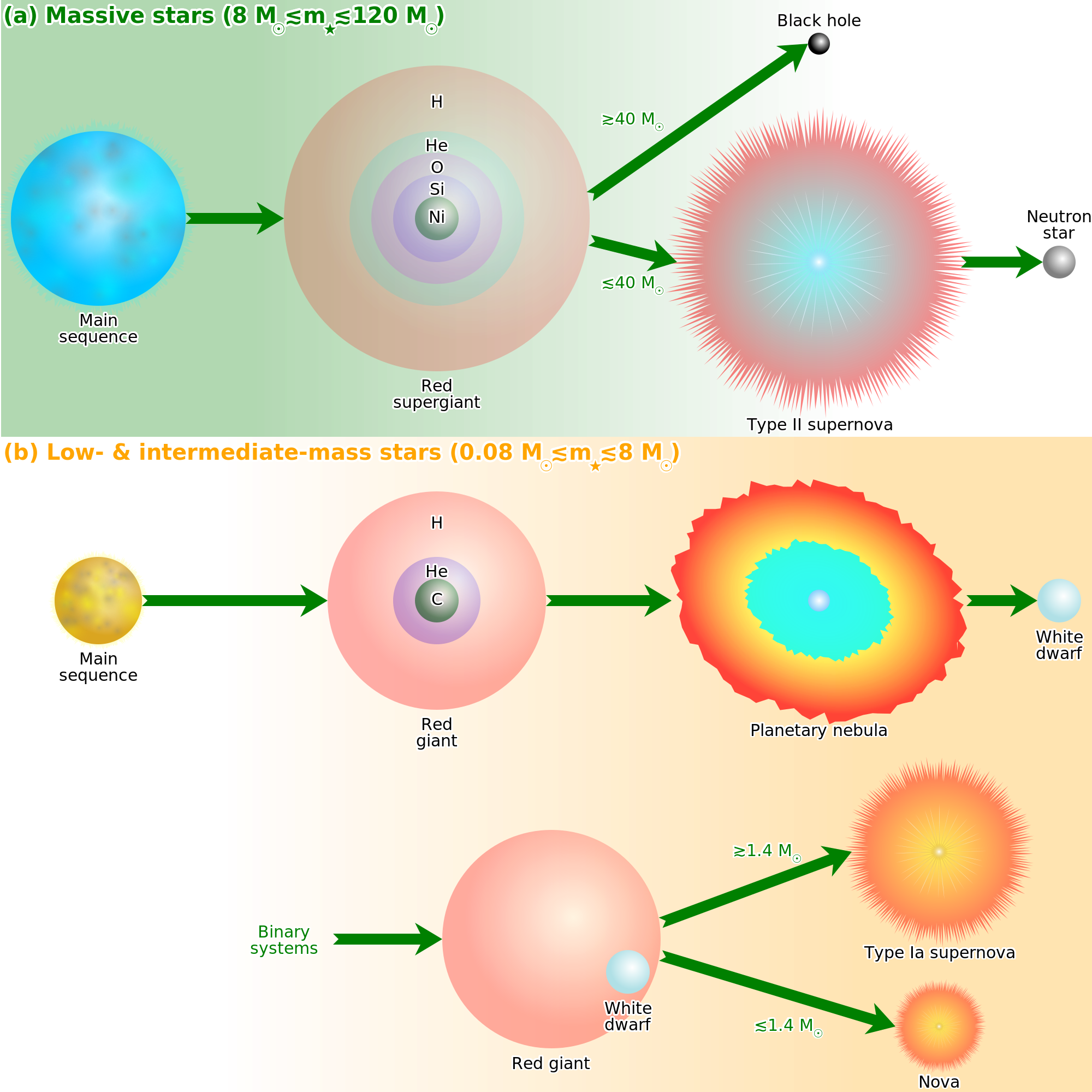}
  \newcap{Schematic representation of stellar evolution}%
         {\CClicence}
  \label{fig:stellarevol}
\end{figure}

Stars are born from the collapse of molecular clouds into protostars \citep[\eg][for a review]{motte18}.
Protostars accrete matter until their winds and radiation pressure stops this process, leading to a \expression{pre-main sequence star} (pre-\hMS).
Pre-\hMS\ stars exhibit violent winds and bipolar jets, clearing away the remaining molecular cocoon they were born in.
They contract until the temperature in their core is high enough to initiate H fusion ($T\simeq10^7$~K).
Below $m_\star\le0.08\eMsun$, we get a \expression{brown dwarf}, which is a compact object not massive enough to sustain nuclear reactions.
\reffig{fig:stellarevol} schematically represents the different stages of evolution of low- and high-mass stars.

\paragraph{The Main Sequence.}
Once nuclear reactions ignite, stars are on the \expression{Zero-Age Main Sequence} (\hZAMS; grey line in \reffig{fig:stellar_isochrones}).
We have already briefly discussed stellar evolution in \refsec{sec:panchromatic}.
The different types of stars, their mass, luminosities and lifetimes are given in \reffig{fig:stellar_isochrones} and in \reftab{tab:stars}.
\begin{enumerate}
  \item As long as stars are in their H burning phase (\cf\ 
    \refsec{sec:nucleosynthesis}) they are \expression{Main Sequence} (\hMS) 
    stars.
    They move slowly along their tracks in \reffig{fig:stellar_isochrones}.
  \item Once the core has exhausted its H, it contracts by lack of fuel.
    This contraction increases the temperature, allowing the He burning to 
    start (\cf\ \refsec{sec:nucleosynthesis}).
    Due to the increase of central temperature, the outer layers expand.
    The star is now a \expression{red giant}.
  \item This process repeats, with He burning (triple $\alpha$ process;  
    \refsec{sec:nucleosynthesis}).
\end{enumerate}

\paragraph{The late stages of massive stars.} 
Massive stars ($8\eMsun\le M_\star<120\eMsun$) are the hottest and most luminous ones (\cf\ \reftab{tab:stars}).
They are short-lived ($\tau(m_\star)\lesssim30$~Myr; \reffig{fig:stellar_isochrones}).
\begin{enumerate}
  \item The process of burning heavier elements is repeated beyond C, resulting 
    in an \citengl{onion} structure (\cf\ \refsubfig{fig:stellarevol}{a}).
    The combustion of each element is exponentially faster.
    These stars are, at this point, \expression{red supergiants} (\cf\ 
    \reffig{fig:stellar_isochrones}).
  \item Once the core is made of Fe, the star can not anymore produce energy by
    nuclear fusion.
    It therefore collapses \citep[\eg][]{heger03}.
    \begin{description}
      \item[If $m_\star\lesssim40\eMsun$,] the collapse is halted by the 
        \expression{degeneracy pressure}\footnote{The degeneracy pressure is 
        due to the fact that fermions can not occupy the same state. In very 
        dense environments, this leads to a pressure: electron degeneracy 
        pressure in white dwarfs; neutron degeneracy pressure in neutron 
        stars.} of neutrons.
        The outer layers of the stars then explode as a 
        \expression{type~II supernova} (\hSNII) or \expression{core-collapse 
        supernova}, leaving a \expression{Neutron Star} (\hNS) in the center.
      \item[If $m_\star\gtrsim40\eMsun$,] the degeneracy pressure of the neutron 
        core is not sufficient to sustain the collapse.
        The remnant is not anymore a \hNS, but a \expression{Black Hole} (\hBH).
        This is also the approximate mass range where the star leaves a remnant 
        without exploding as a \hSN, ending as a \textit{collapsar} 
        \citep{heger03}.
        These two phenomena (ending as a collapsar and leaving a \hBH) are not
        necessarily concomitant.
        The exact masses above which these two phenomema occur are not
        accurately known and depend on other parameters, such as stellar 
        rotation.
        For simplicity, we have represented both phenomena on the same branch 
        in \reffig{fig:stellarevol}.
    \end{description}
\end{enumerate}

\paragraph{The late stages of LIMS.}
\expression{Low- and Intermediate-Mass Stars} (\hLIMS; $0.08\eMsun\le M_\star<8\eMsun$) are less luminous than massive stars, but they are the most numerous.
Their lower gravity allow them to burn their elements slower than massive stars, and therefore to live longer (several Gyrs, on average; \reffig{fig:stellar_isochrones}).
\begin{enumerate}
  \item Their mass does not allow them to start C burning.
    \hLIMS\ enter the \expression{Asymptotic Giant Branch} 
    \citep[\hAGB; \cf\ 
    \reffig{fig:stellar_isochrones}; \eg][for reviews]{van-winckel03,herwig03}.
    They are larger and more luminous than red giants and are thermally pulsing.
  \item The contraction of the core is halted by electron degeneracy pressure.
    The thermal pulses lead the outer shell to expand progressively, creating a 
    \expression{Planetary Nebula} (\hPN), leaving a \expression{White Dwarf} 
    (\hWD) in the center.
  \item The maximum mass a white dwarf can reach is the \expression{Chandrasekar
    mass}, $m_\sms{Chandra}\simeq1.4\eMsun$.
    If a white dwarf of mass $m_\sms{WD}$ happens to be in a binary system with 
    another red giant, it will accrete some of its mass.
    \begin{description}
      \item[If $m_\sms{WD}\gtrsim m_\sms{Chandra}$,] 
        the excess of mass above the Chandrasekar limit re-ignites the 
        thermonuclear reactions.
        It follows a \expression{type Ia SN} (\hSNIa), which disrupts the 
        binary system.
      \item[If $m_\sms{WD}\lesssim m_\sms{Chandra}$,] thermonuclear reactions are 
       ignited at the surface of the white dwarf.
       It ensues a \expression{nova}, that does not disrupt the binary system.
    \end{description}
\end{enumerate}

    \subsubsection{Parametrizing Star Formation}
    \label{sec:SFH}

Star formation is a complex process involving stars of different masses being formed at different times.
At the scale of a star-forming region or an entire galaxy, \hSF\ can be described statistically.

\begin{figure}[htbp]
  \begin{tabular}{cc}
    \includegraphics[width=0.48\textwidth]{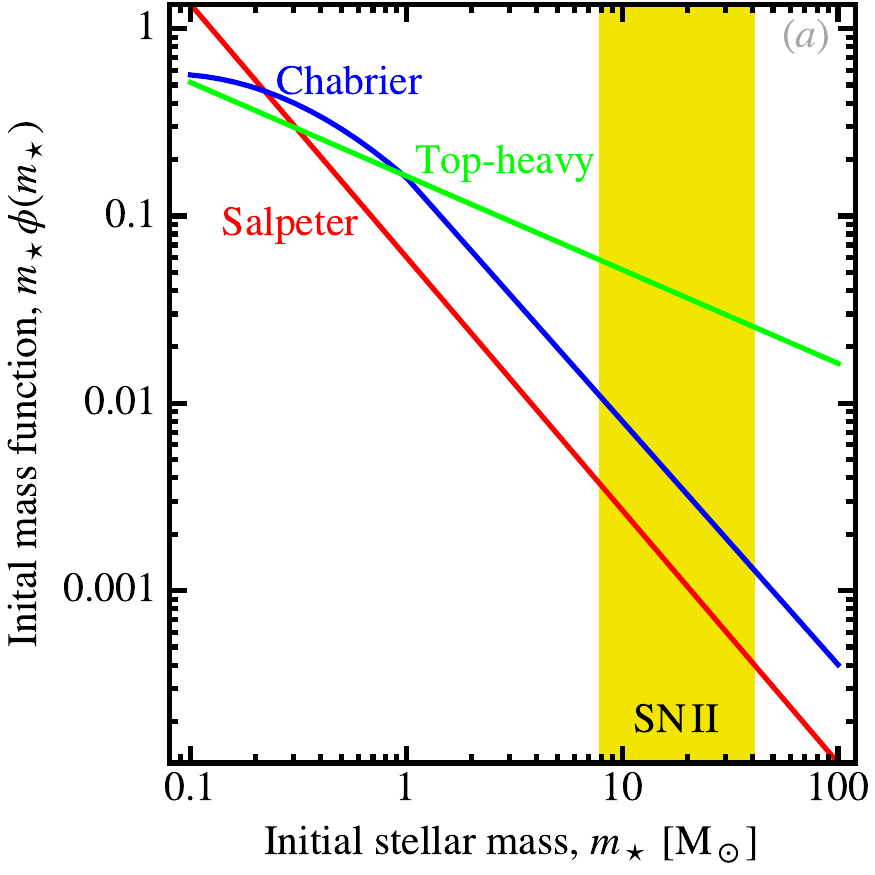} &
    \includegraphics[width=0.48\textwidth]{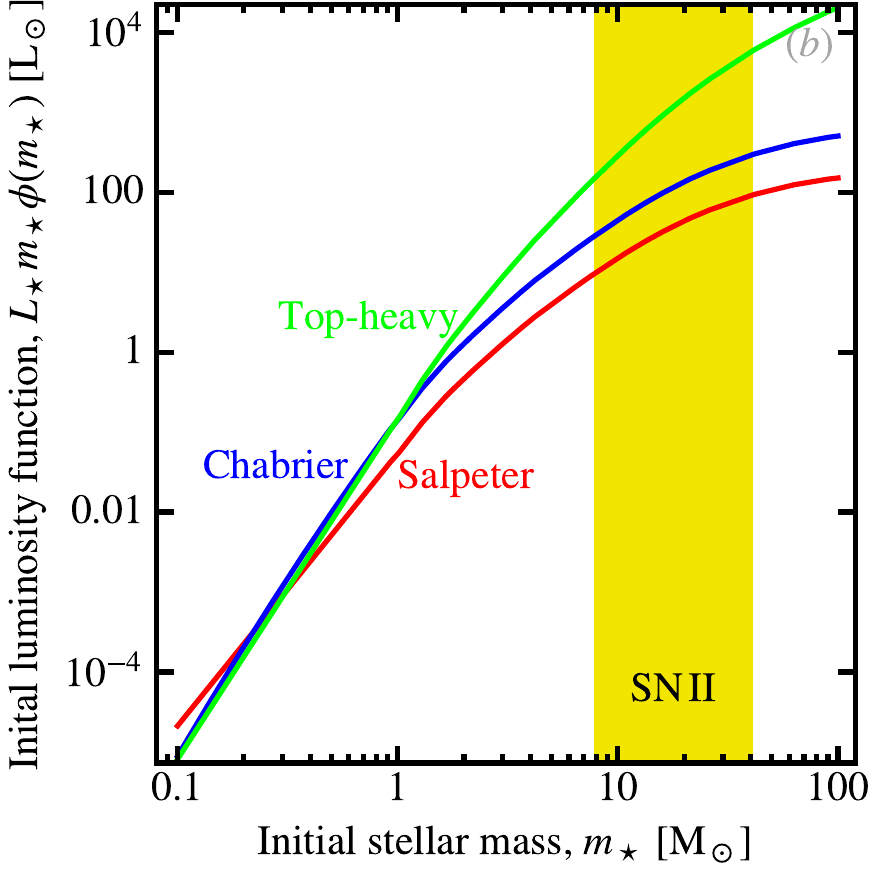} \\
  \end{tabular}
  \newcap{Initial mass functions}%
         {We compare the three \hIMF s discussed in the text: 
          \begin{inlinelist}
            \item \citet{salpeter55} \refeqp{eq:salpeter};
            \item \citet{chabrier03} \refeqp{eq:chabrier}; and 
            \item Top-heavy \refeqp{eq:topheavy}.
          \end{inlinelist}
          Panel~\textit{(a)} shows the number distribution of stars,
          $m_\star\phi(m_\star)$.
          Panel~\textit{(b)} shows the luminosity distribution, $L_\star 
          m_\star\phi(m_\star)$, where $L_\star$ comes from 
          \reffig{fig:stellar_isochrones}.
          We have indicated in yellow the range $8\le m_\star<40\eMsun$, 
          corresponding to \hSNII.
          \CClicence}
  \label{fig:IMF}
\end{figure}

\paragraph{Initial mass functions.}
\expression{Initial Mass Functions} (\hIMF) express the number distribution of stars of mass $m_\star$ born at a given time: $\phi(m_\star)\equiv\dd N_\star/\dd m_\star$.
\hIMF s are usually expressed in $\Msun^{-1}$, and are normalized as\footnote{Careful though, some authors quote \hIMF s, normalized as $\int m_\star\phi(m_\star)\ddiff m_\star=1$.}:
\begin{equation}
  \int_{m_-}^{m_+}\phi(m_\star)\ddiff m_\star = 1,
\end{equation}
where $m_-=0.1\eMsun$ and $m_+=100\eMsun$ are the lower and upper masses.
The average stellar mass is defined as:
\begin{equation}
  \langle m_\star\rangle\equiv\int_{m_-}^{m_+}m_\star\phi(m_\star)\ddiff m_\star. 
  \label{eq:mstar}
\end{equation}
The fraction of stars ending their life as a core-collapse \hSN\ is:
\begin{equation}
  f_\sms{SN}\equiv\int_{m^\sms{SN}_-}^{m^\sms{SN}_+}\phi(m_\star)\ddiff m_\star, 
  \label{eq:fSN}
\end{equation}
with $m_-^\sms{SN}=8\eMsun$ and $m_+^\sms{SN}=40\eMsun$ \citep[\eg][]{heger03}.
Several \hIMF s have been proposed in the literature \citep[see also][]{kroupa01}.
\begin{description}
  \item[The Salpeter IMF] \citep{salpeter55} was the first one proposed.
    It is defined as:
    \begin{equation}
      \phi_\sms{Salp}(m_\star) 
        \equiv \frac{(1-\alpha)\times m_\star^{-\alpha}}{m_+^{1-\alpha}-m_-^{1-\alpha}},
      \;\;\;\mbox{ with }\;\;\;\alpha=2.35
      \label{eq:salpeter}
    \end{equation}
    where the lower and upper masses, $m_-$ and $m_+$, are usually taken as 
    $m_-=0.1\eMsun$ and $m_+=100\eMsun$, although the original 
    \citet{salpeter55} study constrained the index of the power-law only up to 
    $m_+=10\eMsun$.
  \item[The Chabrier IMF] \citep{chabrier03} for individual stars is defined, 
    within the same mass range, as:
    \begin{equation}
      \phi_\sms{Chab}(m_\star) \equiv\left\{\begin{array}{ll}
        0.16033\times m_\star^{-2.3} & \mbox{ for } m_\star>1\eMsun \\
        \displaystyle
        \frac{0.5718}{m_\star}\exp\left(-\frac{(\log_{10}(m_\star)
           +1.1023729 )^2}{0.9522}\right)
        & \mbox { for } m_\star\le1\eMsun.
      \end{array}\right.
      \label{eq:chabrier}
    \end{equation}
  \item[The top-heavy IMF] \citep[\eg][]{dwek07} is often invoked at 
    high-redshift.
    It is defined as:
    \begin{equation}
      \phi_\sms{Top}(m_\star) 
        \equiv \frac{(1-\alpha)\times m_\star^{-\alpha}}{m_+^{1-\alpha}-m_-^{1-\alpha}},
      \;\;\;\mbox{ with }\;\;\;\alpha=1.50
      \label{eq:topheavy}
    \end{equation}
\end{description}
These \hIMF s are compared in \reffig{fig:IMF}.
Some of their properties are listed in \reftab{tab:IMF}.
The \hIMF\ is thought to be a universal property of interstellar media.
The different \hIMF s of \reftab{tab:IMF} have consequences on the stellar properties.
The current consensus is that the Chabrier \hIMF\ might be more appropriate than Salpeter's, at least at low redshift.
\begin{table}[htbp]
  \centering
  \setlength\arrayrulewidth{2pt}
  \arrayrulecolor{white}
  \begin{tabularx}{\linewidth}{|>{\columncolor{coltabhead}}X%
                                |>{\columncolor{coltabcell}}r%
                                |>{\columncolor{coltabcell}}r%
                                |>{\columncolor{coltabcell}}r|}
    \hline
      \rowcolor{coltabhead}
      \cellcolor{white} & \textbf{Salpeter} & \textbf{Chabrier} 
      & \textbf{Top-Heavy} \\
    \hline
      Average mass, $\langle m_\star\rangle$
      & $0.351\eMsun$ & $0.673\eMsun$ & $3.16\eMsun$ \\
    \hline
      \hSNII\ fraction, $f_\sms{SN}$
      & $0.239\,\%$ & $0.724\,\%$ & $6.38\,\%$ \\
    \hline
      Mass fraction of massive stars
      & $13.9\,\%$ & $22.6\,\%$ & $74.1\,\%$ \\
    \hline
      \hLIMS\ luminosity ($t_\star=0$, $Z=0.008$)
      & $4.81\eLsun/\Msun$ & $13.9\eLsun/\Msun$ & $55.3\eLsun/\Msun$ \\
    \hline
      Massive star luminosity ($t_\star=0$, $Z=0.008$)
      & $180\eLsun/\Msun$ & $579\eLsun/\Msun$ & $1.38\E{4}\eLsun/\Msun$ \\
    \hline
      Total luminosity ($t_\star=0$, $Z=0.008$)
      & $185\eLsun/\Msun$ & $593\eLsun/\Msun$ & $1.38\E{4}\eLsun/\Msun$ \\
    \hline
  \end{tabularx}
  \newcap{IMF properties}%
         {These properties are integrated over three different \hIMF s:
          Salpeter \refeqp{eq:salpeter}; Chabrier \refeqp{eq:chabrier}; and
          Top-heavy \refeqp{eq:topheavy}.
          For the last three lines, we used the \hZAMS\ stellar luminosities 
          of \reffig{fig:stellar_isochrones} (initial metallicity $Z=0.008$).}
  \label{tab:IMF}
\end{table}

\begin{figure}[htbp]
  \includegraphics[width=\textwidth]{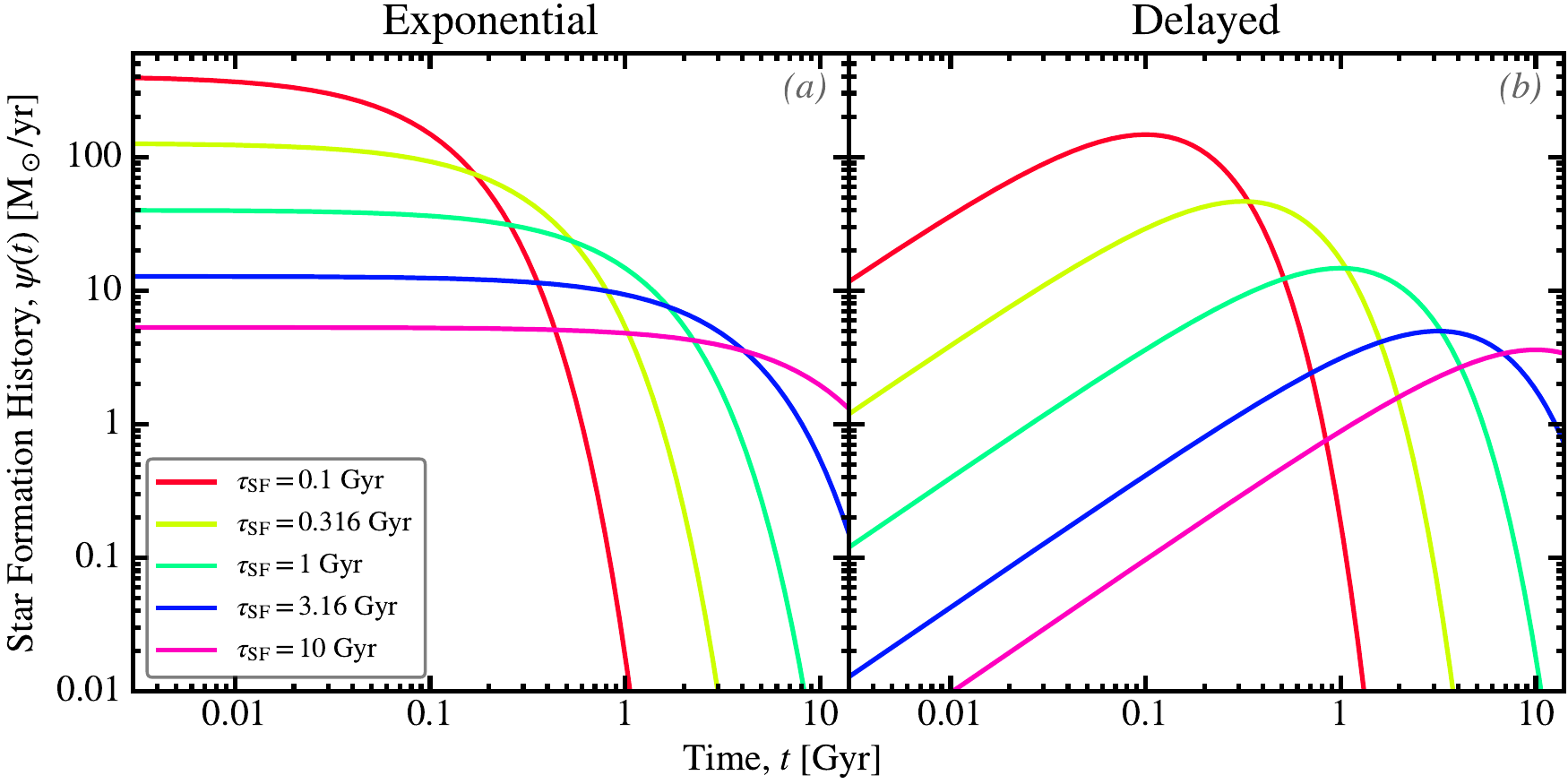}
  \newcap{Parametric star formation histories}%
         {Panel~\textit{(a)} represents the exponential \hSFH\ 
          \refeqp{eq:SFHexp}, for different values of $\tau_\sms{SF}$.
          Panel~\textit{(b)} represents the delayed \hSFH\ \refeqp{eq:SFHdel},
          for the same grid of $\tau_\sms{SF}$.
          In both panels, the \hSFH s are normalized so that the integrated mass
          of stars formed in $\Delta t=14$~Gyr (about the age of the Universe) 
          is $M_\star=4\E{10}\eMsun$ (roughly the \hMW\ stellar mass).
          \CClicence}
  \label{fig:SFH}
\end{figure}

\paragraph{Parametric star formation histories.}
We have already briefly discussed \expression{Star Formation Histories} (\hSFH) in \refsec{sec:panchromatic}.
The \hSFH\ quantifies the \hSFR, $\psi(t)$, as a function of time, $t$, of a star-forming region or galaxy.
Several parametric forms are commonly used in the literature.
As we will see in \refsec{sec:cosmicdustevol}, their parameters can be inferred by fitting a set of observations.
\begin{description}
  \item[The exponential SFH] (\cf\ \refsubfig{fig:SFH}{a}) is parametrized by a 
    timescale, $\tau_\sms{SF}$, and a \hSFR\ at $t=0$, $\psi_0$:
    \begin{equation}
      \psi_\sms{exp}(t)\equiv\psi_0\exp\left(-\frac{t}{\tau_\sms{SF}}\right).
      \label{eq:SFHexp}
    \end{equation}
  \item[The delayed SFH] \citep[\cf\ \refsubfig{fig:SFH}{b};][]{lee10} has the 
    same number of parameters as the exponential SFH, but its \hSFR\ peak is 
    \expression{delayed} at $t=\tau_\sms{SF}$:
    \begin{equation}
      \psi_\sms{del}(t)\equiv\psi_0\frac{t}{\tau_\sms{SF}}
        \exp\left(-\frac{t}{\tau_\sms{SF}}\right).
      \label{eq:SFHdel}
    \end{equation}
\end{description}
It is possible to combine several \hSFH s to account for the complex history of a galaxy.
A useful quantity, deriving from the \hSFH, is the \expression{stellar birth rate}, which is the average number of stars born per unit time:
\begin{equation}
  B(t)\equiv\frac{\psi(t)}{\langle m_\star\rangle}.
  \label{eq:birth}
\end{equation}
The rate of \hSNII, $R_\sms{SN}(t)$, can be approximated from this quantity:
\begin{equation}
  R_\sms{SN}(t)\equiv\int_{m_-^\sms{SN}}^{m_+^\sms{SN}}B\left(t-\tau(m_\star)\right)
    \times\phi(m_\star)\ddiff m_\star
   \simeq B(t) f_\sms{SN},
  \label{eq:RSN}
\end{equation}
where $\tau(m_\star)$ is the lifetime of a star of mass $m_\star$ (\cf\ \reffig{fig:stellar_isochrones}).
The approximation, in the second part of \refeq{eq:RSN}, comes from the fact that the lifetime of massive stars ($\tau(m_\star)\lesssim30$~Myr) is usually much smaller than the \hSF\ timescale: $\tau(m_\star)\ll\tau_\sms{SF}$.

  \subsection{Elemental and Dust Yields}

Stars, in their late stages, return to the \hISM\ a fraction of the heavy elements they have synthesized.
Stellar ejecta are:
\begin{inlinelist}
  \item stellar winds; 
  \item planetary nebulae;
  \item novae;
  \item \hSNIa; and
  \item \hSNII.
\end{inlinelist}
In addition, the temperature in these ejecta can be low enough (\cf\ \refsubfig{fig:depletions}{b}) to condense grains, that we refer to as \expression{stardust}.

    \subsubsection{Injection of Heavy Elements in the ISM}

\begin{figure}[htbp]
  \includegraphics[width=\textwidth]{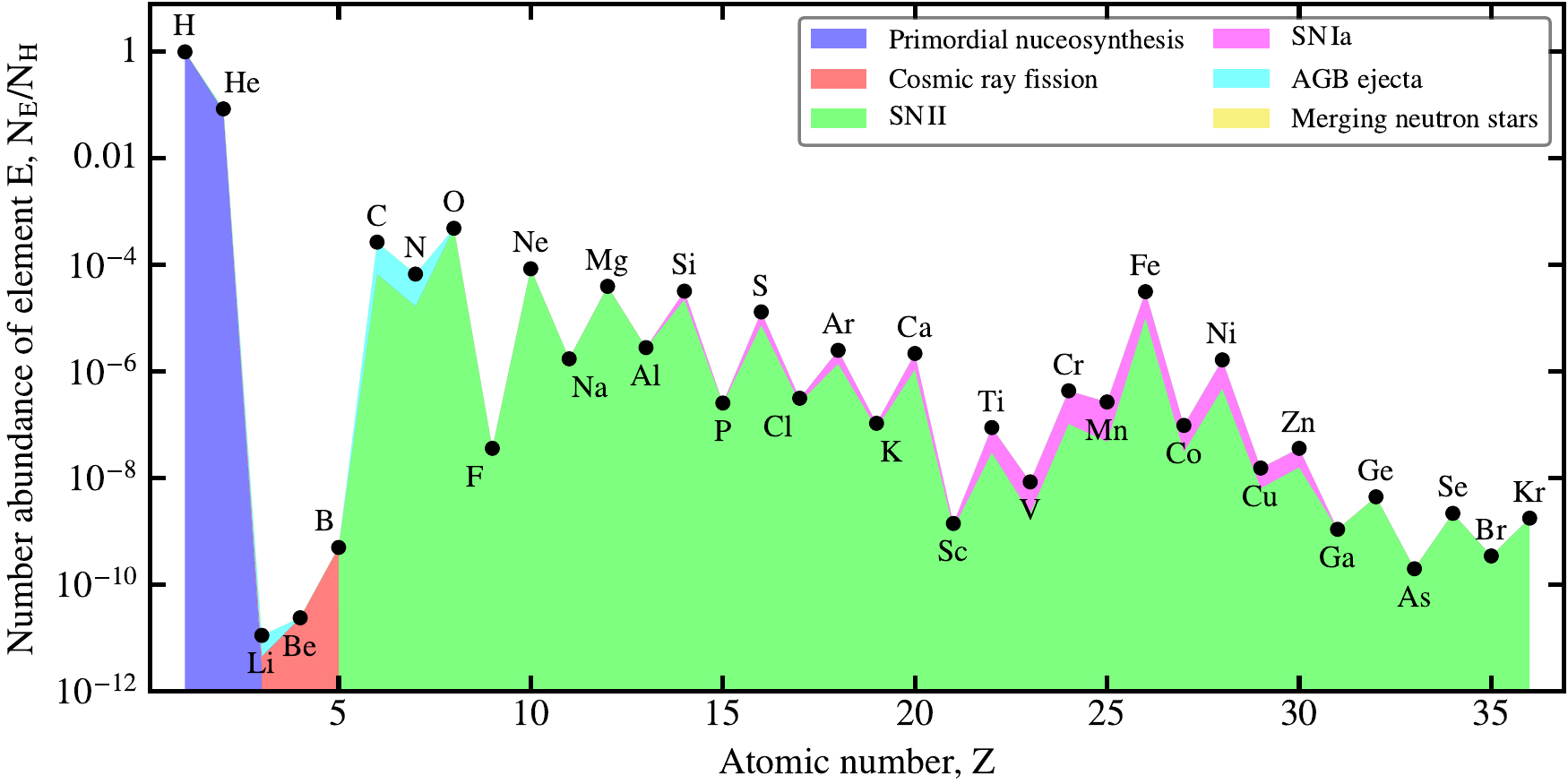}
  \newcap{Nucleosynthesis origin of the main elements}%
         {This curve represents the Solar abundances of \citet{asplund09} that
          we have shown in \reffig{fig:abundsun}.
          We have indicated the fraction of the different sources of  
          nucleosynthesis where these elements originate from 
          \citep[Table S1 of][]{johnson19}.
          Since we have stopped at $Z=36$, none of the elements shown are coming
          from merging neutron stars.
          \CClicence}
  \label{fig:nucleosynthesis}
\end{figure}

\reffig{fig:nucleosynthesis} gives the approximate fraction of each element produced in different environments, for the Solar neighborhood \citep[\eg][]{johnson19}.
These proportions depend on the past \hSFH\ of the system we are considering.
\begin{description}
  \item[Primordial nucleosynthesis] (\cf\ \refsec{sec:nucleosynthesis}) is 
    responsible for the production of most of the light elements.
  \item[\snii] account for most of the \hISD-relevant elements, 
    except C.
  \item[AGB ejecta] (\ie\ \hLIMS\ winds and \hPN e) account for a significant
    fraction of C and N, and heavier elements we have not displayed here.
  \item[\snia] are responsible for synthesizing a significant 
    fraction of the metals around Fe.
  \item[Other processes,] such as cosmic ray fission and merging neutron stars,
    are not very relevant to \hISD.
\end{description}

\begin{figure}[htbp]
  \begin{tabular}{cc}
    \includegraphics[width=0.48\textwidth]{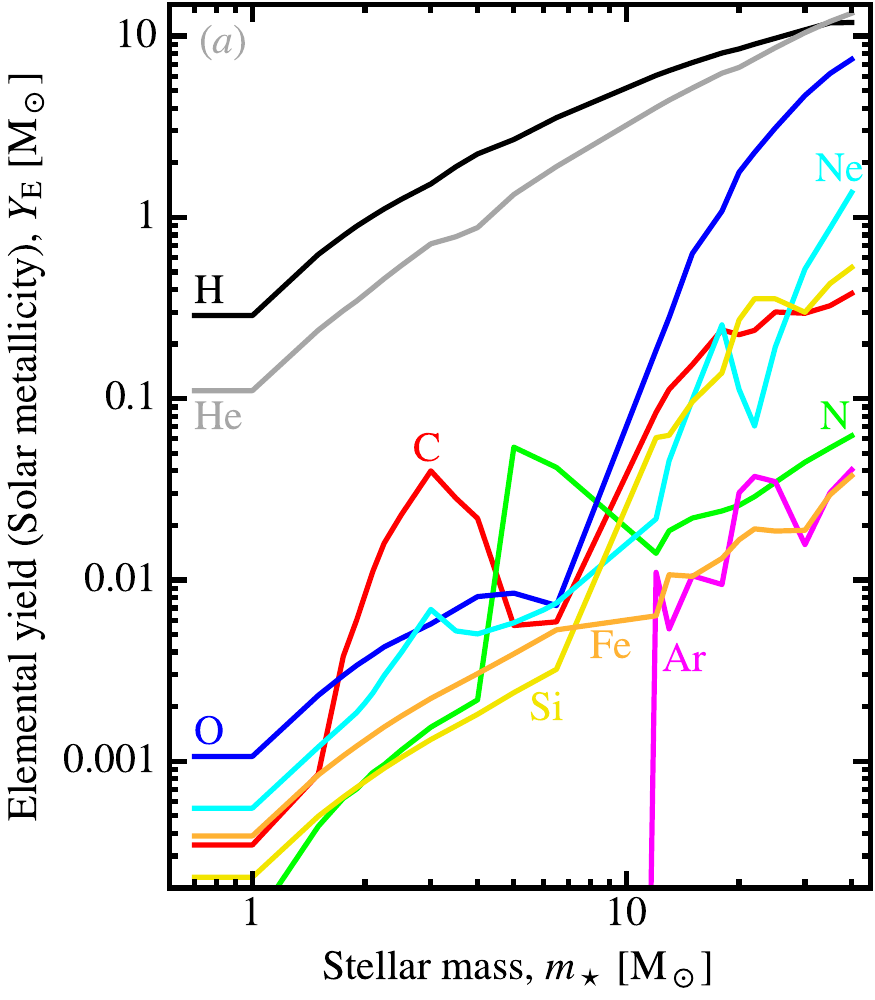} &
    \includegraphics[width=0.48\textwidth]{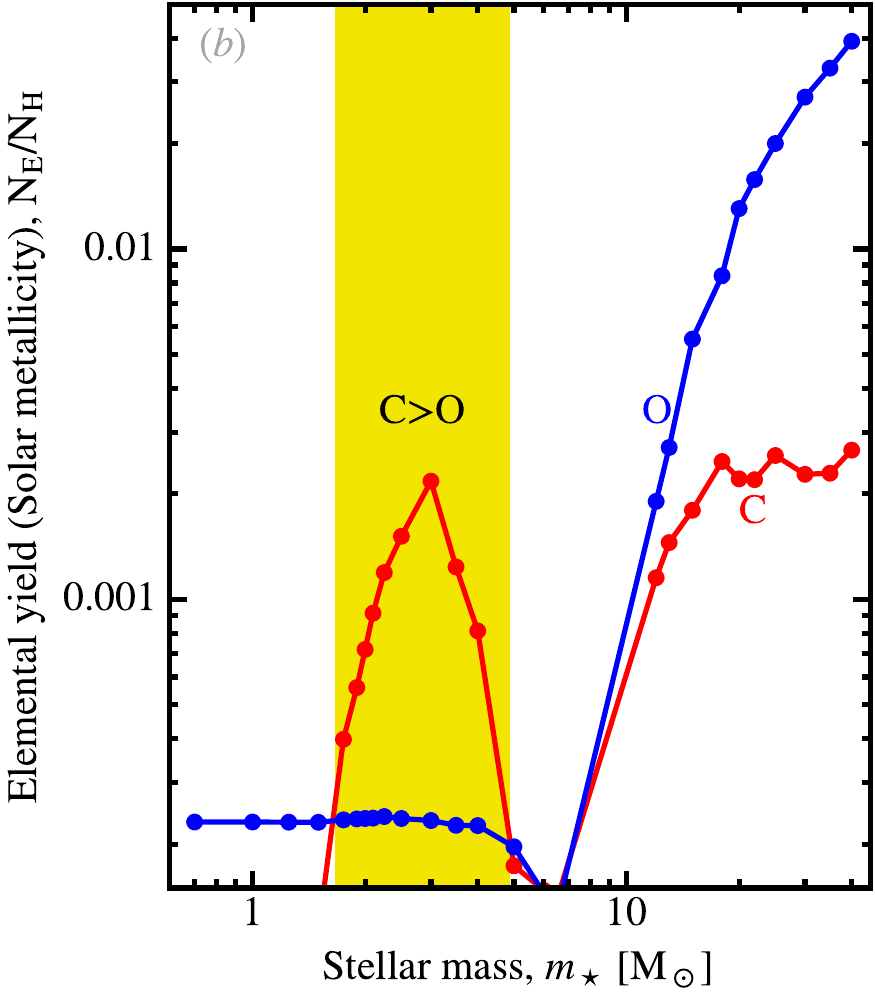} \\
  \end{tabular}
  \newcap{Solar metallicity elemental stellar yields}%
         {Panel~\textit{(a)} shows the mass of the most abundant elements 
          ejected by stars of different masses.
          Panel~\textit{(b)} focusses on the stellar mass range where the 
          number abundance of C exceeds that of O (in yellow).
          These data are the yields used by the chemical evolution model of
          \citet{galliano08a}.
          They are an homogenization of the yields of 
          \citet{karakas03b,karakas03a}, for \hLIMS, and of \citet{woosley95}, 
          for high-mass stars.
          \CClicence}
  \label{fig:stellar_yields}
\end{figure}

\paragraph{Stellar elemental yields.}
A stellar yield, $Y_\sms{E}(m_\star)$, is the mass of an element E injected into the \hISM\ by a star of mass $m_\star$, at the end of its lifetime.
These yields can be constrained observationally, but they are essentially determined theoretically \citep[\eg][for a review]{karakas14}.
\refsubfig{fig:stellar_yields}{a} shows the yields of the most important elements.
An important quantity determining the type of dust grains that will form in the ejecta is the C/O ratio.
Indeed, when the temperature cools down enough, C and O tend to combine to form CO molecules.
The excess atom will thus be the only one left to form stardust.
Therefore, stellar ejecta with $C<O$ will form primarily O-rich grains (silicates and oxides), whereas stellar ejecta with $C>O$ will form mainly carbon grains and SiC.
\refsubfig{fig:stellar_yields}{b} compares the number abundances of C and O ejected by stars of different masses.
We can see that carbon grains originate mainly in \hLIMS\ around $m_\star\simeq3\eMsun$.
\takeaway{\hSNII\ are responsible for most O-rich stardust, while \hLIMS\ 
          produce most C-rich grain seeds.}

\paragraph{Metallicity estimates.}
The heavy elements injected by the successive stellar populations increase the metallicity of the \hISM, Z.
In that sense, the metallicity is an indicator of \citengl{the fraction of baryonic matter that has been converted into heavier elements by means of stellar nucleosynthesis} \citep{kunth00}, such that:
\begin{equation}
  \left\{\begin{array}{rcl}
    X_\sms{primordial} & \simeq & 0.76 \\
    Y_\sms{primordial} & \simeq & 0.24 \\
    Z_\sms{primordial} & \simeq & 0.00 \\
  \end{array}\right.
  \longrightarrow
  \left\{\begin{array}{rcl}
    X_\odot & \simeq & 0.74 \\
    Y_\odot & \simeq & 0.25 \\
    Z_\odot & \simeq & 0.01. \\
  \end{array}\right.  
\end{equation}
The measure of metallicity is not completely straightforward and is vigorously debated \citep[\eg][for a review]{kewley19}.
In external galaxies, it is usually estimated by modeling observations of nebular optical lines, coming from \hii\ regions.
Several methods, allowing an observer to convert a few line ratios into a metallicity estimate, have been proposed.
These methods have been calibrated on particular \hii\ regions, modeling the photoionization, making unavoidable assumptions about the stellar populations and the topology of the gas.
We have systematically compared several calibrations over the \hDustPedia\ sample \citep{de-vis19}.
We have favored the ``S'' calibration from \citet{pilyugin16}, as it is the most reliable down to low metallicities.

    \subsubsection{Production of Stardust}
    \label{sec:stardust}

At the scale of a galaxy, the two most important sources of stardust are:
\begin{inlinelist}
  \item \hAGB\ stars, encompassing both \hLIMS\ winds and \hPN e; and
  \item \hSNII.
\end{inlinelist}

\paragraph{AGB stars.}
Most of the dust production in \hLIMS\ is believed to occur during the \expression{Thermally-Pulsing Asymptotic Giant Branch} (\hTPAGB) phase \citep{gail09}.
In addition, \hLIMS\ with $m_\star\lesssim1\eMsun$ do not condense grains \citep[\eg][]{ferrarotti06}.
Theoretical models concur that only a fraction of the available heavy elements will go into stardust \citep{morgan03,ventura12}: 
\begin{equation}
  \delta_\sms{LIMS}\equiv\frac{m^\sms{ej}_\sms{stardust}}{m^\sms{ej}_\sms{Z}}
    \simeq10-40\,\%,
  \label{eq:LIMS}
\end{equation}
$m^\sms{ej}_\sms{stardust}$ and $m^\sms{ej}_\sms{Z}$  being the ejected mass of stardust and heavy elements.
Observations and modeling of the circumstellar envelopes of post-\hAGB\ stars 
are consistent with these values \citep[\eg][]{ladjal10}.

\paragraph{\snii.}
There is solid evidence that grains form in \expression{SuperNova Remnants} (\hSNR), as the ejected gas cools down.
Theoretical estimates of the net dust yield of a single \hSNII\ range in the literature from $Y_\sms{SN}\simeq10^{-3}$ to $Y_\sms{SN}\simeq1\eMsun/\textnormal{SN}$ \citep[\eg][]{todini01,ercolano07,bianchi07b,bocchio16b,marassi19}.
From an observational point of view, measuring the dust mass produced \textit{in situ} by a single \hSNII\ is quite difficult, as it implies disentangling the freshly-formed dust from the surrounding \hISM.
It also carries the usual uncertainty about dust optical properties.
A decade ago, the largest dust yield ever measured was $Y_\sms{SN}\simeq0.02\eMsun$ \citep[in SN$\,$2003gd;][]{sugerman06}.
The \hhersc\ space telescope has been instrumental in estimating the cold mass of \hSNR s.
The yields of the three most well-studied \hSNR s are now an order of magnitude higher:
\begin{description}
  \item[Cassiopeia A:] $Y_\sms{SN}\simeq0.04-1.1\eMsun$
    \citep{barlow10,arendt14,de-looze17,bevan17,priestley19};
  \item[The Crab nebula:] $Y_\sms{SN}\simeq0.03-0.23\eMsun$ 
    \citep{gomez12,temim13,de-looze19};
  \item[SN~1987A:] $Y_\sms{SN}\simeq0.45-0.8\eMsun$ \citep{dwek15,matsuura15}.
\end{description}
Most of the controversy however lies in the fact that, while large amounts could form in \hSNII\ ejecta \citep[\eg][]{matsuura15,temim17}, a large fraction of freshly formed grains would not survive the reverse shock \citep[$v\simeq1000$~km/s; \eg][]{nozawa06,micelotta16,kirchschlager19}.
In all the cases we have listed above, the newly-formed grains have indeed not yet experienced the reverse shock \citep{bocchio16b}.
The net yield is thus expected to be significantly lower.
Even if $\simeq10-20\,\%$ of the dust condensed in an \hSNII\ ejecta survives its reverse shock \citep[\eg][]{nozawa06,micelotta16,bocchio16b}, we have to also consider the fact that massive stars are born in clusters.
The freshly-formed dust injected by a particular \hSNII, having survived the reverse shock, will thus be exposed to the forward shock waves of nearby \hSN e \citep[\eg][]{martinez-gonzalez18}. 
Overall, \hSNII\ dust yields are largely uncertain.
We will extensively discuss their empirical constraint, from a statistical point of view, in \refsec{sec:cosmicdustevol}.
We will show that we can infer the average dust yield per \hSNII, $\langle Y_\sms{SN}\rangle$.
The corresponding timescale is then simply:
\begin{equation}
  \frac{1}{\tau_\sms{SN-cond}(t)}\equiv
    \frac{\langle Y_\sms{SN}\rangle}{M_\sms{dust}(t)}R_\sms{SN}(t).
  \label{eq:SNcond}
\end{equation}

\paragraph{Indirect evidence.}
The best constraints on the fraction of \hISD\ which is stardust might be indirect.
The clear correlation between the depletion factor, $F_\star$, and the average density of the \hISM, $\langle n_\sms{H}\rangle$, that we have discussed in \refsubfig{fig:depletions}{a}, has been shown to require rapid destruction and reformation into the \hISM\ \citep[\eg][]{draine79,tielens98,draine09}.
The rest of the grains needs to form in the \hISM.
\citet[][\citetalias{draine09}]{draine09} gives a series of additional arguments concluding that, in the \hMW, stardust has to be less than $10\,\%$ of \hISD.
\begin{enumerate}
  \item A first argument given by \citetalias{draine09} is based on the typical 
    lifetime of \hISD.
    \begin{enumerate}
      \item From the literature, the stardust injection rate is roughly
        $\dot{M}_\sms{stardust}\simeq5\E{-3}\eMsun/\textnormal{yr}$.
      \item Noting that the typical lifetime of a dust grain in the \hMW\ is 
        $\tau_\sms{ISD}\simeq3\E{8}$~yr \citep[\cf\ 
        \refsec{sec:shockdest};][]{jones96}, the present stardust mass in the 
        \hISM\ should be:
        $M_\sms{stardust}\simeq\dot{M}_\sms{stardust}\times\tau_\sms{ISD}
        \simeq1.5\E{6}\eMsun$.
      \item The \hISD\ mass in the \hMW\ is roughly 
        $M_\sms{ISD}\simeq M_\sms{ISM}/183\simeq2.7\E{7}\eMsun$ 
        (\cf\ \reftab{tab:massthemis}).
    \end{enumerate}     
    Stardust is thus only $M_\sms{stardust}/M_\sms{ISD}\simeq5\,\%$ of the total 
    \hISD.
  \item A second argument given by \citetalias{draine09} is based on the study
    of \hIDP s in meteorites (\cf\ \refsec{sec:direct}).
    \begin{enumerate}
      \item Stardust silicate grains in meteorites, identified with their 
        isotopic anomalies (\cf\ \reffig{fig:meteorites}), appear to be 
        $\simeq20\,\%$ crystalline.
      \item Silicates in the \hISM\ are less than $\lesssim2\,\%$ crystalline 
        \citep[\cf\ \refsec{sec:extinctionMIR};][]{kemper04}.
    \end{enumerate}
    Therefore, the fraction of stardust is only less than 
    $\lesssim2/20\simeq10\,\%$ of \hISD.
    Also, the fact that \hISD\ is mainly amorphous, whereas circumstellar grains
    are essentially crystalline, is another argument in favor of rapid 
    destruction and reformation in the \hISM.
\end{enumerate}
\takeaway{In the \hMW, stardust represents only a few percents of the \hISD\ content.}

\section{Dust Evolution Processes in the ISM}
\label{sec:dustevolISM}

Most of the important dust evolution processes occur in the \hISM.
These effects can be studied by looking at spatial variations of the dust properties in a region.

  \subsection{Grain Formation and Transformation}

\expression{Grain formation} is the transfer of elements from the gas phase to the dust phase, therefore increasing the \hdustiness.

    \subsubsection{Evidence of Grain Growth and Coagulation in the 
                        ISM}
    \label{sec:mantles}

We have just discussed stardust (\cf\ \refsec{sec:stardust}) which is thought to produce grain seeds onto which mantle can grow.
We now focus on the dominant process in Solar metallicity systems: the accretion of gas phase atoms and molecules.
Grain-grain coagulation does not result in grain formation \textit{per se}, as it does not affect the \hdustiness.
It however follows grain growth and has similar effects on the \hFIR\ opacity.

\paragraph{The evidence brought by depletions.}
As we have discussed in \refsec{sec:stardust}, the clearest evidence of grain growth in the \hISM\ is provided by the good correlation between the depletion factor and the average density of the \hISM\ (\cf\ \refsubfig{fig:depletions}{a}).
It implies that atoms and molecules from the gas phase are progressively building up grain mantles, when going into denser regions.
This observed behaviour is also consistent with the progressive de-mantling and disaggregation of cloud-formed, mantled and coagulated grains injected into the low density \hISM, following cloud disruption. 
It is perhaps not unreasonable to hypothesise that dust growth in the \hISM\ occurs on short timescales during cloud collapse rather than by dust growth in the quiescent diffuse \hISM. 
In this alternative interpretation, the arrow of time is in the opposite sense and requires rapid dust growth, through accretion and coagulation, in dense molecular regions and slow de-mantling and disaggregation in the diffuse \hISM\ \citep[\eg][]{jones09}. 
Given that astronomical observations provide only single-time snapshots, it will seemingly be difficult to determine the direction of the time-arrow of dust evolution.

\paragraph{FIR opacity variations.}
We have seen in \reffig{fig:aggregates} that the growth of mantles has an impact on the \hFIR\ opacity \citep[\cf][]{kohler14,kohler15}.
Yet, there is clear evidence of \hFIR\ opacity variations in the \hMW.
The main factor seems to be the density of the medium.
For instance, both \citet{stepnik03} and \citet{roy13} found that the \hFIR\
dust cross-section per H atom increases by a factor of $\simeq3$ from the 
diffuse \hISM\ to the molecular cloud they targeted.
\citet{stepnik03} noticed that this opacity variation is accompanied by the 
disappearance of the small grain emission.
They concluded that grain coagulation could explain these variations.
In the diffuse \hISM, \citet{ysard15} showed that the variation of emissivity, 
including the $\beta-T$ relation (\cf~\refsec{sec:MBB}), could be explained by 
slight variations of the mantle thickness of the \citetalias{jones17} model.
For that reason, the \citetalias{jones17} model aims at describing the evolution of grain mantles as a function of density and \hISRF, as we have seen in \reffig{fig:aggregates}:
\begin{inlinelist}
  \item in the diffuse \hISM, the grains are supposed to have a thin 
    a-C mantle, largely dehydrogenated (aromatic) by \hUV\ photons;
  \item in denser regions, the mantle thickness is hypothesized to increase and
    to become more hydrogenated (aliphatic), because of the progressive 
    shielding of stellar photons;
  \item in molecular clouds, grains are thought to be coagulated and iced.
\end{inlinelist}
The \citetalias{jones17} model predicts a factor of $\simeq2.5$ dust mass increase in going from \hMW\ diffuse to dense clouds.
This would correspond to a factor up to $\simeq7$ in terms of dust emissivity per H atom \citep{kohler15}.

    \subsubsection{Studies of the Magellanic Clouds}

In nearby galaxies, studies of the local grain processing are difficult to conduct, as the emissivity variations are smoothed out by the mixing of dense and diffuse regions.
Even when potential evolutionary trends are observed, their interpretation is often degenerate with other factors.
The Magellanic clouds are the most obvious systems where this type of study can be attempted.
The insights provided by depletion studies (\cf~\refsec{sec:depletions}) show that there are clear variations of the fraction of heavy elements locked-up in dust, and these variations correlate with the density \citep{tchernyshyov15,jenkins17}.
Since the coagulation and the accretion of mantles lead to an increase of \hFIR\ emissivity \citep[\cf\ \reffig{fig:aggregates};][]{kohler15}, we should expect emissivity variations in the Magellanic clouds.
Indeed, \citet{roman-duval17} studied the trends of gas surface density (derived from \hi\ and CO) as a function of dust surface density (derived from the \hIR\ emission), in these galaxies.
They found that the observed \hdustiness\ of the \hLMC\ increases smoothly by a factor of $\simeq 3$ from the diffuse to the dense regions.
In the \hSMC, the same variation occurs, with a factor of $\simeq7$.
They argue that optically thick \hi\ and CO-free \hmol\ gas (\cf~\refsec{sec:darkgas}) can not explain these trends, and that grain growth is thus the most likely explanation.

\begin{figure}[htbp]
  \includegraphics[width=\textwidth]{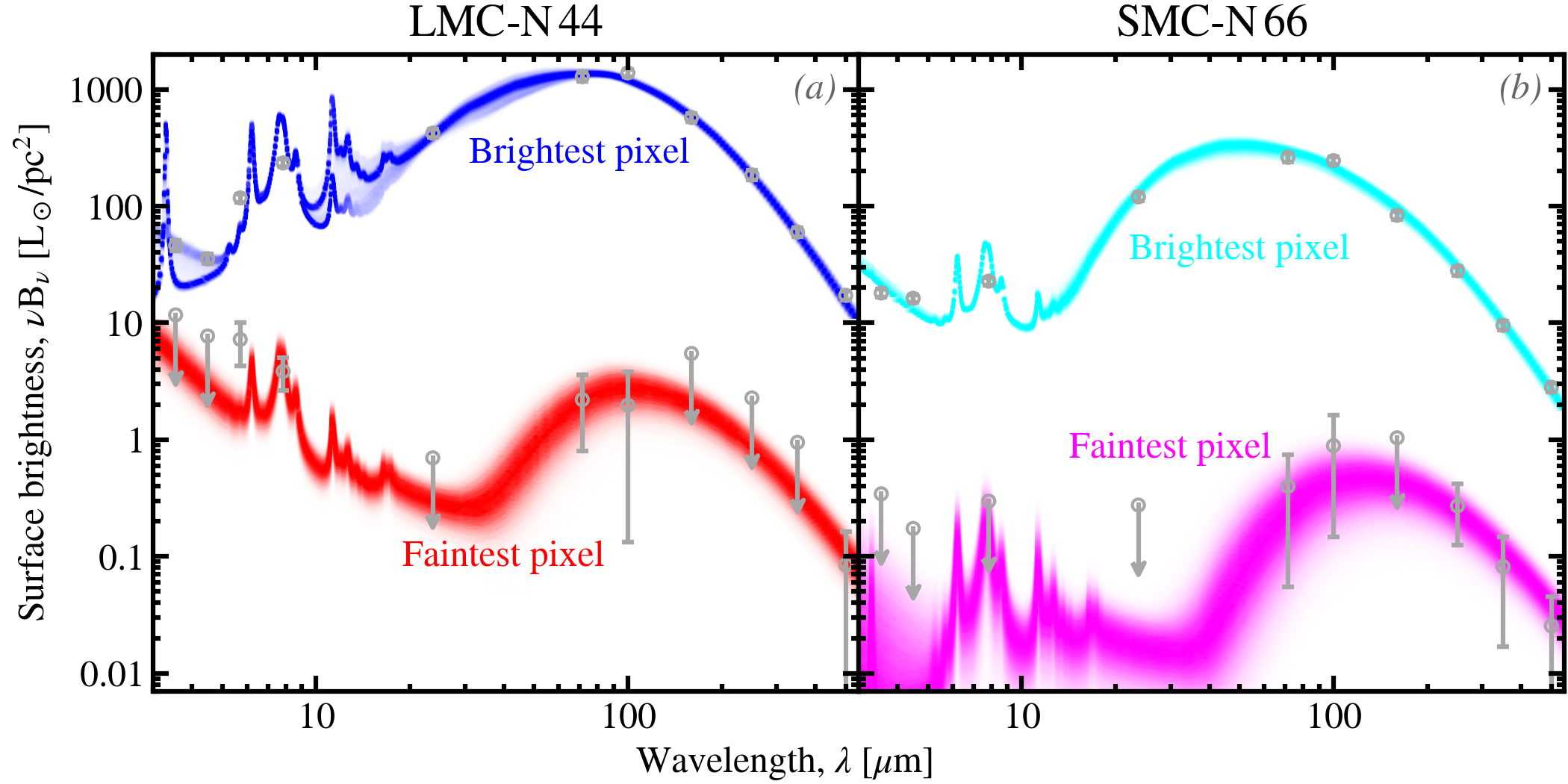}
  \newcap{Spatially-resolved SED fits in N$\,$44 and N$\,$66}%
         {In both panels, we show the brightest and the faintest pixels of our
          entire map.
          The grey error bars are the observations, and the colored densities
          are the Bayesian posterior distributions of \hSED\ models 
          \citep{galliano17}.
          \CClicence}
  \label{fig:sed_N44N66}
\end{figure}

\paragraph{Spatially-resolved SED fitting of LMC-N$\,$44 and 
                SMC-N$\,$66.}
We have conducted a similar study, focussing on two massive star-forming regions, rather than the whole galaxies\footnote{As a reminder, the metallicities of the Magellanic clouds are: $Z_\sms{LMC}\simeq Z_\odot/2$ and $Z_\sms{SMC}\simeq Z_\odot/5$ \citep{pagel03}.}:
\begin{inlinelist}
  \item N$\,$44 in the \hLMC; and 
  \item N$\,$66 in the \hSMC
\end{inlinelist}
\citep{galliano17}.
Our maps were 200~pc wide regions, with a spatial resolution of $\simeq15$~pc.
The \hMIR-to-submm data were coming from the \hSAGE\ surveys \citep[\hspitz\ and \hhersc\ data;][]{meixner06,meixner13}.
We used the hierarchical Bayesian \hSED\ model, \citetalias{galliano18a}, with the AC dust composition of \citet[][\cf\ \refsec{sec:kappaLMC}]{galliano11}.
The goal was to perform a spatially-resolved modeling of the dust properties, in a region with a strong gradient of physical conditions, in order to probe dust processing, as a function of density, \hISRF\ and metallicity.
The wide range of physical conditions can be estimated by looking at the range of \hSED s shown in \reffig{fig:sed_N44N66}.
In both panels, the faintest pixels show a rather cold \hSED, peaking around $\lambda\simeq100\emic$, whereas the brightest pixels peak around $\lambda\simeq60\emic$, with a very broad \hFIR\ bump, indicating a wide range of \hISRF\ intensities, typical of compact \hSF\ regions (\cf\ \refsec{sec:dale}).

\begin{figure}[htbp]
  \includegraphics[width=\textwidth]{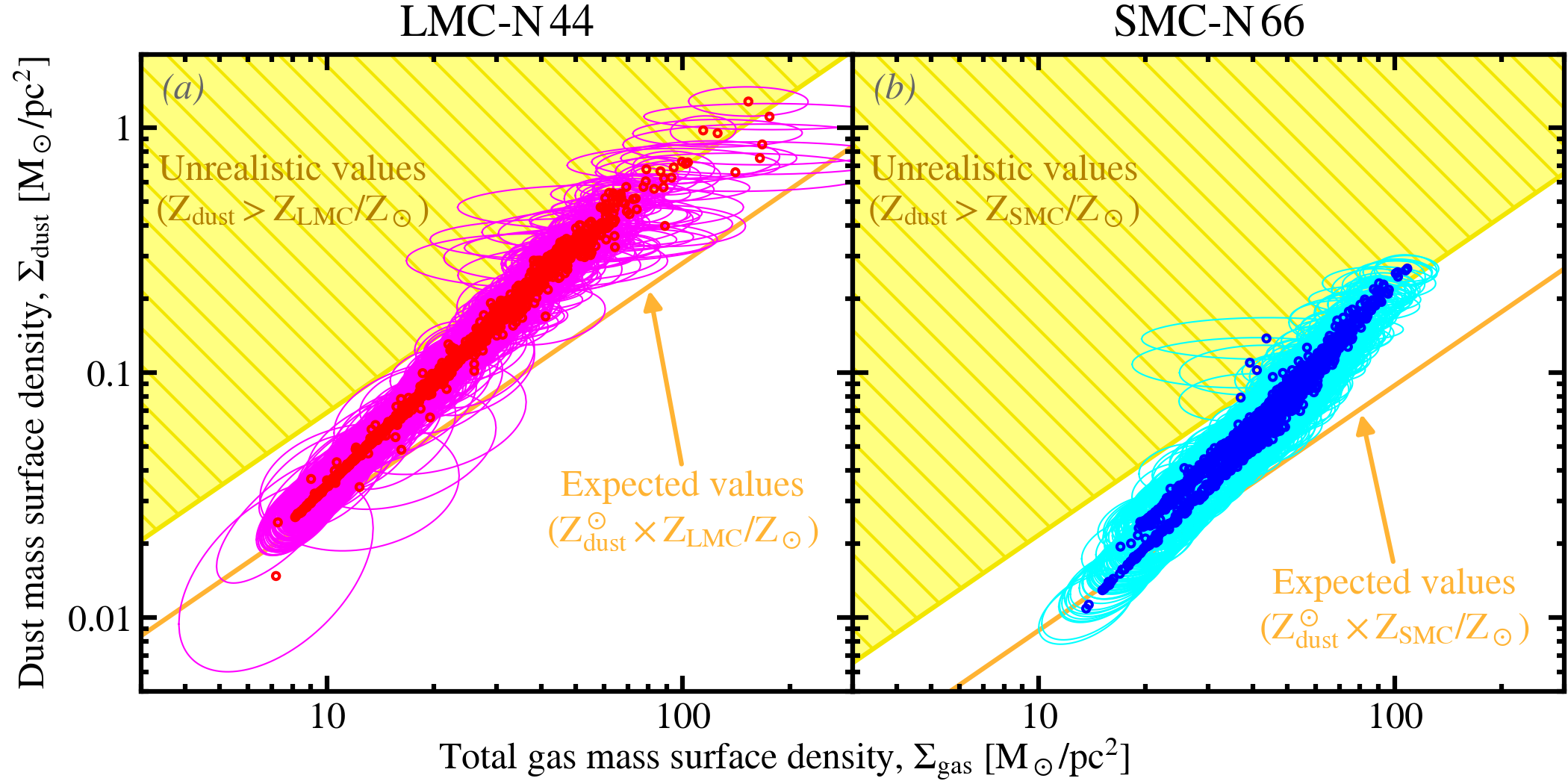}
  \newcap{Dust-to-gas mass surface density relation in N$\,$44 and N$\,$66}%
         {In both panels, we show the trends of dust mass surface density as a
          a function of the total gas mass (atomic and molecular) surface 
          density derived from the \hSED\ fitting of \reffig{fig:sed_N44N66}
          \citep{galliano17}.
          Each point, with its uncertainty ellipse, corresponds to $\simeq15$~pc
          pixel.
          The orange line corresponds to the \hdustiness\ of the \hMW\ (\cf\ 
          \reftab{tab:massthemis}) scaled by the metallicity of the region.
          The yellow hatched area corresponds to a \hdustiness\ larger than the
          metallicity, that is an unrealistic value requiring more heavy
          elements being locked up in grains than what is available in the 
          \hISM.
          \CClicence}
  \label{fig:correl_N44N66}
\end{figure}

\paragraph{Derived dust-gas relations.}
We have compared the derived dust and total gas column densities.
The latter was estimated from the \hiline\ and \COio\ measurements \citep{meixner06,gordon11,meixner13}.
The results are displayed in \reffig{fig:correl_N44N66}.
The orange line represents the Galactic \hdustiness\ (\cf\ \reftab{tab:massthemis}) scaled by the metallicity, therefore representing the Galactic dust-to-metal mass ratio.
This line corresponds to the values we would expect if the dust constitution was close to the diffuse \hISM\ of the \hMW\ and was not evolving with density.
The most diffuse pixels in both regions are consistent with this value.
The hatched yellow area corresponds to a \hdustiness\ larger than the metallicity, that is requiring more heavy elements in dust than what is available in the \hISM.
Overall, the trends of \reffig{fig:correl_N44N66} indicate a non-linear dust-to-gas relation, with a variation of the observed \hdustiness\ by a factor of $\simeq3$, similar to the studies we have reviewed at the beginning of this section.
The high density pixels lie in the forbidden zone.
The possible causes are the following.
\begin{description}
  \item[Grain growth:] a part of this trend is likely the result of the 
    evolution of the \expression{true} \hdustiness, due to mantle accretion in 
    denser regions.
    The yellow hatched area can however be considered as a hard upper limit, as 
    in practice, not all heavy elements are refractory: even in the densest 
    molecular clouds, there are gas phase CO, HCN, \etc\
    This sole factor is therefore probably not sufficient to explain the full 
    extent of the \expression{observed} \hdustiness\ variation.
  \item[Emissivity increase:] 
    grain growth and coagulation are accompanied by an increase of \hFIR\ 
    emissivity \citep[\cf\ \reffig{fig:aggregates};][]{kohler15}.
    This effect is naturally expected and would amplify the increase of the
    \expression{observed} \hdustiness\ with density, as the constant emissivity
    assumption of our \hSED\ model would lead us to overestimate the dust mass
    of dense regions.
  \item[Contribution of dark gas:] we have not accounted for CO-dark gas (\cf\ 
    \refsec{sec:darkgas}).
    This component can potentially bias the molecular gas mass estimate in    
    translucent regions by up to a factor $\simeq100$ \citep{madden20}.
    It could also explain a part of the trend, as it would result in an
    underestimate of the total gas mass.
    This effect should however be significative at intermediate column 
    densities, and decrease toward the densest regions, where CO would dominate.
    It might thus not be the main cause of the non-linearity of our trends.
\end{description}
\takeaway{There is multiple evidence of dust evolution as a function of 
          density, consistent with grain growth and coagulation, and the 
          consequent increase of emissivity.}

    \subsubsection{Quantifying Grain Growth}
    \label{sec:graingrowth}

We now discuss the way grain growth can be approximately quantified.
The following relations are rather uncertain, because of the lack of constraint on grain structure and composition.
They however provide a framework to study grain growth efficiency.

\paragraph{Accretion timescale.}
Timescales for grains to accrete atoms are widely discussed in the literature \citep[\eg][]{dwek98,edmunds01,draine09,hirashita11,zhukovska16,priestley21}.
First, the collision rate of an atom E of mass $m_\sms{E}$, with a grain of radius $a$ is:
\begin{equation}
  \frac{1}{\tau_\sms{coll}(a,E)}\equiv\underbrace{\pi a^2}_\sms{grain 
    cross-section} 
    \times
    \underbrace{n_\sms{E}}_\sms{gas density of E}
    \times
    \underbrace{\sqrt{\frac{8kT_\sms{gas}}{\pi m_\sms{E}}}}_\sms{Maxwellian 
      velocity of E}.
\end{equation}
In this equation, we have implicitly neglected Coulomb interaction (\ie\ we have assumed that the grain and the atom are both neutral, which is a reasonable assumption in the \hCNM).
Second, the growth rate of a grain of mass $m_\sms{d}(a)$, due to accretion following these collisions, can be written:
\begin{equation}
  \left(\frac{\dd m_\sms{d}}{\dd t}\right)_\sms{acc}(a,E)
  = \underbrace{\mathcal{S}}_\sms{sticking probability}\times
    \underbrace{\frac{m_\sms{E}}{f_\sms{E}}}_\sms{gained mass}\times
    \underbrace{\frac{1}{\tau_\sms{coll}(a,E)}}_\sms{rate},
\end{equation}
where $0\le\mathcal{S}\lesssim1$ is the \expression{sticking coefficient}, that is the probability the atom will be bound with the grain after the collision.
The factor $f_\sms{E}$ is the mass fraction of element E within the grain.
We choose E as a \expression{key element} \citep{zhukovska08}, that is the element in the grain make-up that will have the longest collision time.
\begin{description}
  \item[For silicates,] the key element is Si, with $f_\sms{Si}\simeq0.16$ for 
    olivine and $f_\sms{Si}\simeq0.24$ for pyroxene (assuming Fe:Mg=1:1).
    In other words, for each collision with Si, there are more collisions with 
    O, Fe and Mg.
    Therefore, the dust mass gained between two Si collisions is the mass of a 
    full crystal unit (SiO$_4$MgFe for olivine SiO$_3$Mg$_{0.5}$Fe$_{0.5}$ for 
    pyroxene; \cf\ \refsec{sec:silicates}).
  \item[For carbon grains,] the mass is essentially C, as H is negligible.
    We thus have $f_\sms{C}\simeq1$.
\end{description}
Finally, it is convenient to express this quantity as an \expression{accretion timescale}, $\tau_\sms{acc}(a)$:
\begin{equation}
  \frac{1}{\tau_\sms{acc}(a)}\equiv
  \frac{1}{m_\sms{d}(a)}\left(\frac{\dd m_\sms{d}}{\dd t}\right)_\sms{acc}(a,E)
    = \mathcal{S}\frac{\sqrt{m_\sms{E}}}{f_\sms{E}}\frac{3n_\sms{E}}{2a\rho}
      \sqrt{\frac{2kT_\sms{gas}}{\pi}},
  \label{eq:grow}
\end{equation}
where we have simply developed $m_\sms{d}(a)=4/3\pi a^3\rho$ in the second equality, $\rho$ being the mass density of the grain.
The density of the element E can be written as a function of the total H density, assuming its abundance scales with metallicity:
\begin{equation}
  n_\sms{E}\simeq\left(\frac{Z}{Z_\odot}\right)\left(\frac{E}{H}\right)_\odot 
    n_\sms{H}.
\end{equation}
We therefore see that the grain growth timescale roughly obeys the following proportionality (assuming $\mathcal{S}=1$ and olivine composition of silicates):
\begin{equation}
  \tau_\sms{acc}(a) \simeq \frac{100\;\textnormal{cm}^{-3}}{n_\sms{H}}
   \times \frac{Z_\odot}{Z} \times 
    \sqrt{\frac{100\;\textnormal{K}}{T_\sms{gas}}}\times 
    \frac{a}{100\;\textnormal{nm}}
    \times\left\{
    \begin{array}{ll}
      57\;\textnormal{Myr} & \mbox{for silicates} \\
      41\;\textnormal{Myr} & \mbox{for carbon grains.}
    \end{array}
    \right.
  \label{eq:proxgrow}
\end{equation}
As said above, these estimates are uncertain.
We especially have no idea of the sticking probability, $\mathcal{S}$.
\refeq{eq:proxgrow} however provides a description of the sensitivity of grain growth to density, size and metallicity.
It is also indicative of the lower limit of these timescales.

\begin{figure}[htbp]
  \includegraphics[width=\textwidth]{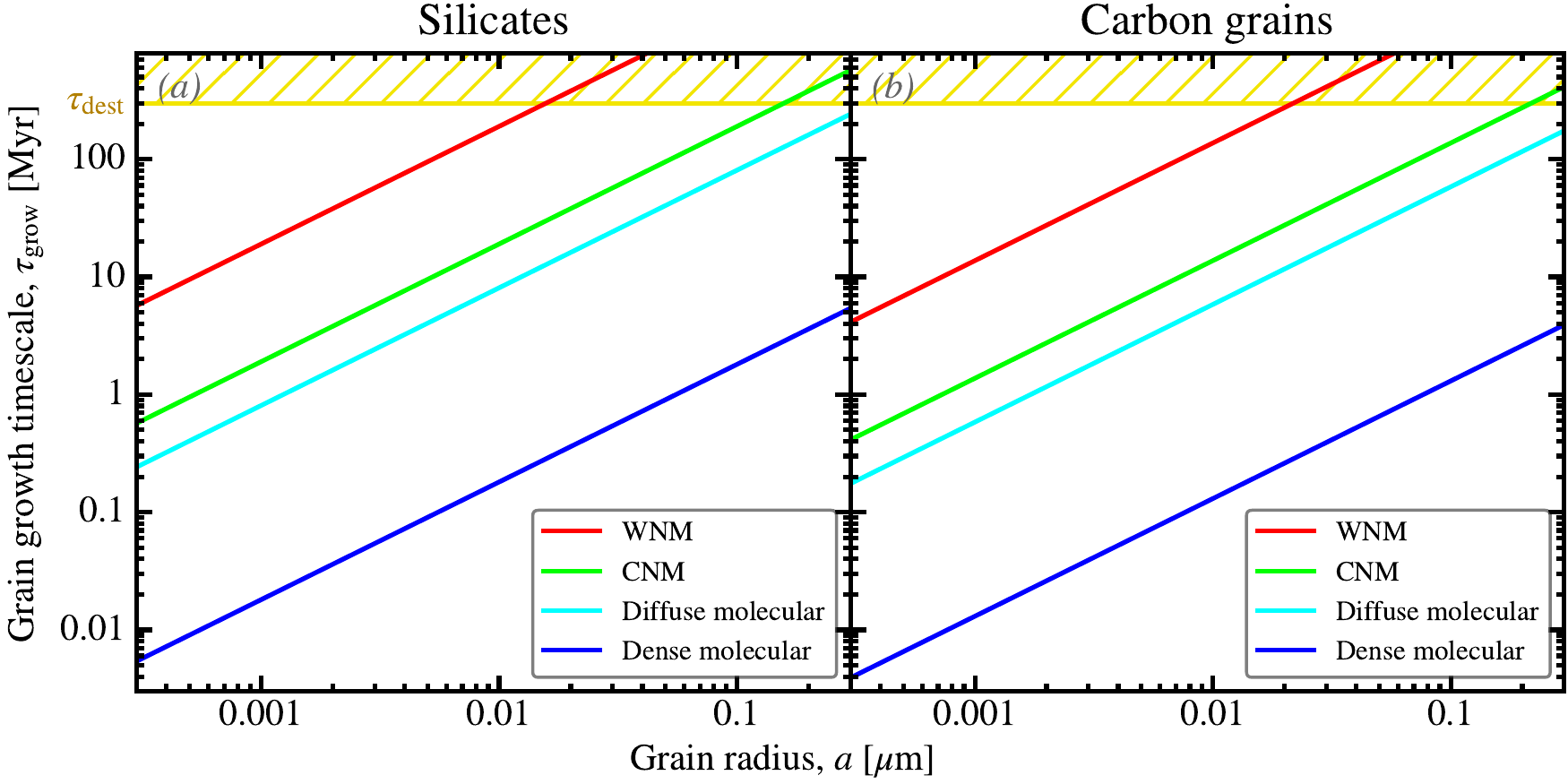}
  \newcap{Grain growth timescales}%
         {Both panels represent \refeq{eq:proxgrow} for several of the phases
          in \reftab{tab:ISMism} (taking $n_\sms{H}=10^4$~cm$^{-3}$ for the dense 
          molecular phase).
          We have assumed $\mathcal{S}=1$ for both silicates and carbon grains.
          We have highlighted timescales longer than the typical shock 
          destruction timescale, $\tau_\sms{SN-dest}\simeq300$~Myr 
          (\cf\ \refsec{sec:shockdest}), in hatched yellow.
          \CClicence}
  \label{fig:growtime}
\end{figure}

\paragraph{Grain growth in different ISM phases.}
\reffig{fig:growtime} displays \refeq{eq:proxgrow} for carbon and silicate grains in the most relevant \hISM\ phases.
Timescales longer than the typical destruction timescales by \hSNII\ blast waves are irrelevant.
That is the reason why this range is hatched in yellow in \reffig{fig:growtime}.
\begin{description}
  \item[In the WNM,] only the smallest grains ($a\lesssim10$~nm) could grow.
    The \hWNM\ is thus not very suitable for grain growth.
  \item[In the CNM,] all grain sizes can grow in less than 300~Myr.
    The \hCNM\ is thus a phase where dust growth could happen.
  \item[In diffuse molecular clouds,] growth timescales are roughly similar to 
    the \hCNM.
    The same conclusion therefore applies.
  \item[In dense molecular clouds,] all relevant interstellar grain sizes can 
    grow in less than $\simeq10$~Myr, which is also the typical lifetime of 
    these clouds, in star-forming regions.
\end{description}
These timescales are consistent with the picture painted by the variation of elemental depletions across phases (\cf\ \refsubfig{fig:depletions}{a}).
To estimate a global growth timescale, let's consider the radius corresponding to the average mass of the \citetalias{jones17} model, in \reftab{tab:sizedistmoments}:
\begin{equation}
  \sqrt[3]{\langle a^3\rangle_a}\simeq
  \left\{
  \begin{array}{ll}
    31\;\textnormal{nm} & \mbox{for silicates} \\
    28\;\textnormal{nm} & \mbox{for large a-C(:H).}
  \end{array}
  \right.
  \label{eq:meansizethemis}
\end{equation}
With these sizes, a typical accretion time in the \hCNM\ would be $\tau_\sms{acc}\simeq58$~Myr for silicates, and $\tau_\sms{acc}\simeq38$~Myr for large \hHAC.
\takeaway{Grains can possibly grow in the \hCNM, on timescales of 
          $\gtrsim30-60$~Myr, and faster in molecular clouds.}

\paragraph{Relation to global parameters.}
As we will see in \refsec{sec:cosmicdustevol}, it is convenient to relate the grain growth timescale to global galaxy parameters.
\citet{mattsson12a} proposed a relation based on the following assumptions.
\begin{enumerate}
  \item Most of grain growth happens in molecular clouds.
    The mass surface density of these molecular clouds, $\Sigma_\sms{mol}$, is 
    proportional to the \hSFR\ 
    surface density, $\Sigma_\sms{SFR}$ \citep[\eg][]{kennicutt98}.
    Grain growth rate is thus proportional to $\Sigma_\sms{SFR}$:
    \begin{equation}
      \frac{\Sigma_\sms{gas}}{\tau_\sms{grow}}\propto\Sigma_\sms{SFR}.
    \end{equation}
  \item The grain growth rate is also proportional to the fraction of available 
    heavy elements in the gas.
    It implies that:
    \begin{equation}
      \frac{\Sigma_\sms{gas}}{\tau_\sms{grow}}\propto\Sigma_\sms{SFR}\times 
       \left(1-\frac{Z_\sms{dust}}{Z}\right)Z,
      \label{eq:graingrowth0}
    \end{equation}
    where $Z_\sms{dust}$ is the \hdustiness, and $Z_\sms{dust}/Z$, the 
    dust-to-metal mass ratio (\cf\ \refsec{sec:depletionstrength}).
    By subtracting $Z_\sms{dust}/Z$, we account for the fact that the fraction 
    of heavy elements already locked up in grains does not contribute to grain 
    growth.
  \item The other parameters in \refeq{eq:proxgrow} are assumed to not vary 
    significantly.
    These parameters are the mean grain size, the mean gas velocity and mean
    density of molecular clouds.
\end{enumerate}
The grain growth rate proposed by \citet{mattsson12a} can thus be parametrized as a function of global galactic quantities and a phenomenological, dimensionless parameter, $\epsilon_\sms{grow}$, containing all our uncertainties.
The goal is to empirically infer $\epsilon_\sms{grow}$, as we will see in \refsec{sec:cosmicdustevol}.
\refeq{eq:graingrowth0} thus becomes:
\begin{equation}
  \frac{1}{\tau_\sms{grow}(t)}
    = \epsilon_\sms{grow}\frac{\psi(t)}{M_\sms{gas}(t)}(Z(t)-Z_\sms{dust}(t)),
  \label{eq:eps_grow}
\end{equation}
where we have replaced the ratio of surface densities by the ratio of the quantities, and have explicited the temporal dependencies.
In the case of the \hMW\ ($\psi\simeq1.3\eMsun$/yr; $M_\sms{gas}\simeq7\E{9}\eMsun$), a grain growth timescale of $\tau_\sms{grow}\simeq60$~Myr \refeqp{eq:proxgrow} corresponds to $\epsilon_\sms{grow}\simeq10^4$.

  \subsection{Grain Destruction}

We now discuss grain destruction, that is the return of heavy elements from the grains to the gas phase.
Note that fragmentation and shattering by shock waves (at $v\lesssim200$~km/s), that we have discussed in \refsec{sec:shockdest}, simply rearrange the size distribution without destroying the dust.
Shocks however have a pulverization effect, accompanying the other processes, that are difficult to differentiate from an observational point of view.

    \subsubsection{Photodestruction of Small Grains}
    \label{sec:PAHevol}

Due to thermal spikes, small grains have a certain probability that one of their atom will be ejected.
This is a runaway process leading to the complete sublimation of the dust grain.

\begin{figure}[htbp]
  \includegraphics[width=\textwidth]{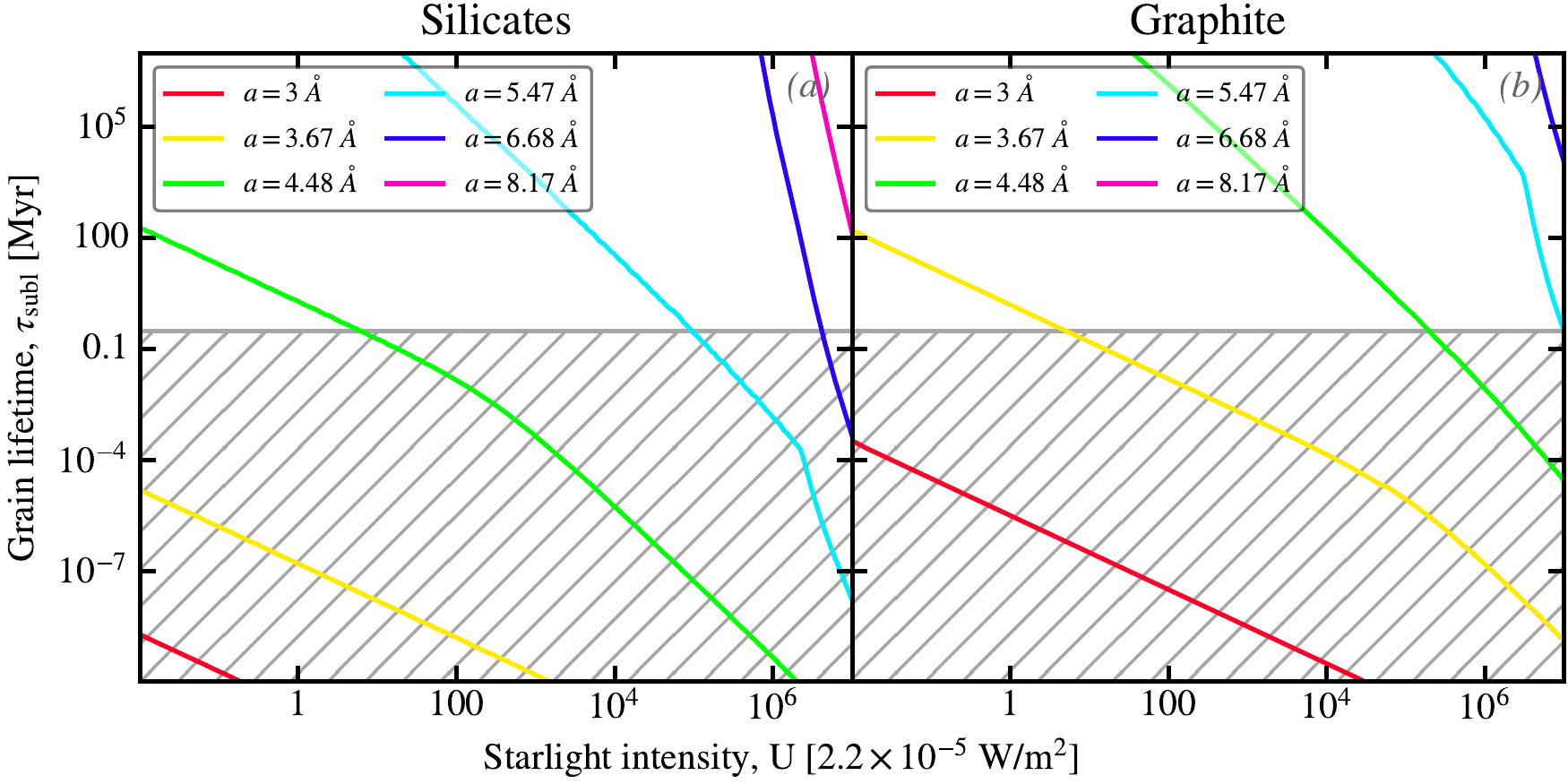}
  \newcap{Lifetimes of small grains in a radiation field}%
         {In both panels, we plot the mean sublimation times from 
          \refeq{eq:sublimation}, for different grain radii (color lines), as
          a function of the starlight intensity of the \citet{mathis83} \hISRF:
          \begin{inlinelistalph}
            \item for the silicates of \citet{weingartner01}; and 
            \item for the graphite of \citet{laor93}.
          \end{inlinelistalph}
          \CClicence}
  \label{fig:sublimation}
\end{figure}

\paragraph{Photodesorption and sublimation.}
Following the formalism of \citet{guhathakurta89}, we consider a cluster $X_N$ containing $N$ atoms of $X$ ($X$ can be C, Fe, Si, O, \etc).
The ejection of an atom from the grain is balanced by the return of an atom from the gas phase.
The rate of the reaction $X_N + X \leftrightarrows X_{N+1}$ is $R_NA_N$, where the total grain surface is $A_N=4\sigma_N\simeq4\pi a^2$.
\citet{guhathakurta89} write the sublimation rate as:
\begin{equation}
  \frac{\dd N}{\dd t} = - R_N(T)\times S_N(T) \times A_N,
\end{equation}
and provide the following rates for graphite and silicate:
\begin{eqnarray}
  R_{N+1}^\sms{gra}(T) & \simeq & 4.6\E{33} \left(\frac{\alpha_N}{0.1}\right)
  \exp\left(-\frac{B_\sms{gra}}{kT}\right) \;\textnormal{m}^{-2}\,\textnormal{s}^{-1}, \\
  R_{N+1}^\sms{sil}(T) & \simeq & 4.9\E{34} \left(\frac{\alpha_N}{0.1}\right)
  \exp\left(-\frac{B_\sms{sil}}{kT}\right) \;\textnormal{m}^{-2}\,\textnormal{s}^{-1},
\end{eqnarray}
with the binding energy per atom:
\begin{eqnarray}
  B_\sms{gra}/k & = & 81\,200-20\,000N^{-1/3}\;\textnormal{K}, \\
  B_\sms{sil}/k & = & 68\,100-20\,000N^{-1/3}\;\textnormal{K}.
\end{eqnarray}
The sticking coefficients, $\alpha_N$, is unknown and is arbitrarily chosen by the authors to be $\alpha_N\simeq0.1$.
Assuming the surface free energy is about $2\E{4}$~K, the term $-20\,000N^{-1/3}$ accounts for the surface tension, making it easier to release an atom when the grain is smaller.
Finally, the \expression{suppression factor}, $S_N(T)<1$, accounts for the suppression of the thermal fluctuations in a thermally isolated particle.
This factor is:
\begin{equation}
  S_N(T) = \left(\frac{1+\gamma}{\gamma}\right)^b
  \frac{\Gamma(\gamma f+1)\Gamma(\gamma f+f-b)}
       {\Gamma(\gamma f-b+1)\Gamma(\gamma f +f)},
\end{equation}
where $f=3N-6$ is the number of vibrational degrees of freedom (\cf\ \refsec{sec:heatcap}).
In addition, the mean number of quanta per degree of freedom is $\gamma=H(T)/(\hbar \omega_0 f)$, and the number of quanta necessary to release a particle is $b=B/(\hbar\omega_0)$.
\citet{guhathakurta89} take $\hbar\omega_0=0.75\Theta$, where $\Theta$ is the Debye temperature (taking $\Theta_\sms{gra}=420\;\textnormal{K}$ and $\Theta_\sms{sil}=470\;\textnormal{K}$; \cf\ \refsec{sec:debye}).
The mean lifetime is then integrated over the temperature distribution:
\begin{equation}
  \frac{1}{\tau_\sms{subl}} = \int\frac{\dd P}{\dd T} R_N(T) S_N(T) A_N\ddiff T.
  \label{eq:sublimation}
\end{equation}
\citet{guhathakurta89} assume that a grain does not survive if it has a lifetime $\lesssim10^{13}\;\textnormal{s}\simeq0.3$~Myr.
\reffig{fig:sublimation} displays these lifetimes for silicates and graphite bathed in the \citet{mathis83} \hISRF.
Although the exact numbers are to be taken with caution, we can conclude the following.
\begin{enumerate}
  \item In the diffuse \hISM\ of the \hMW\ ($U=1$), silicates larger than 
    $a\simeq4.5\;\r{A}$ and graphite $a\simeq3.5\;\r{A}$ can survive.
  \item When $a$ increases, the lifetimes become exponentially longer, meaning
    that for grains larger than $a\gtrsim6\;\r{A}$, the survival of these 
    grains will not be very sensitive to the assumptions in 
    \refeq{eq:sublimation}.
\end{enumerate}
The hardness of the \hISRF, that we have not represented here, will however increase the minimum size a grain needs to have in order to survive.
The vicinity of OB associations will thus be environments where the smallest grains can be photodestroyed.

\begin{figure}[htbp]
  \begin{tabular}{cc}
    \includegraphics[width=0.48\textwidth]{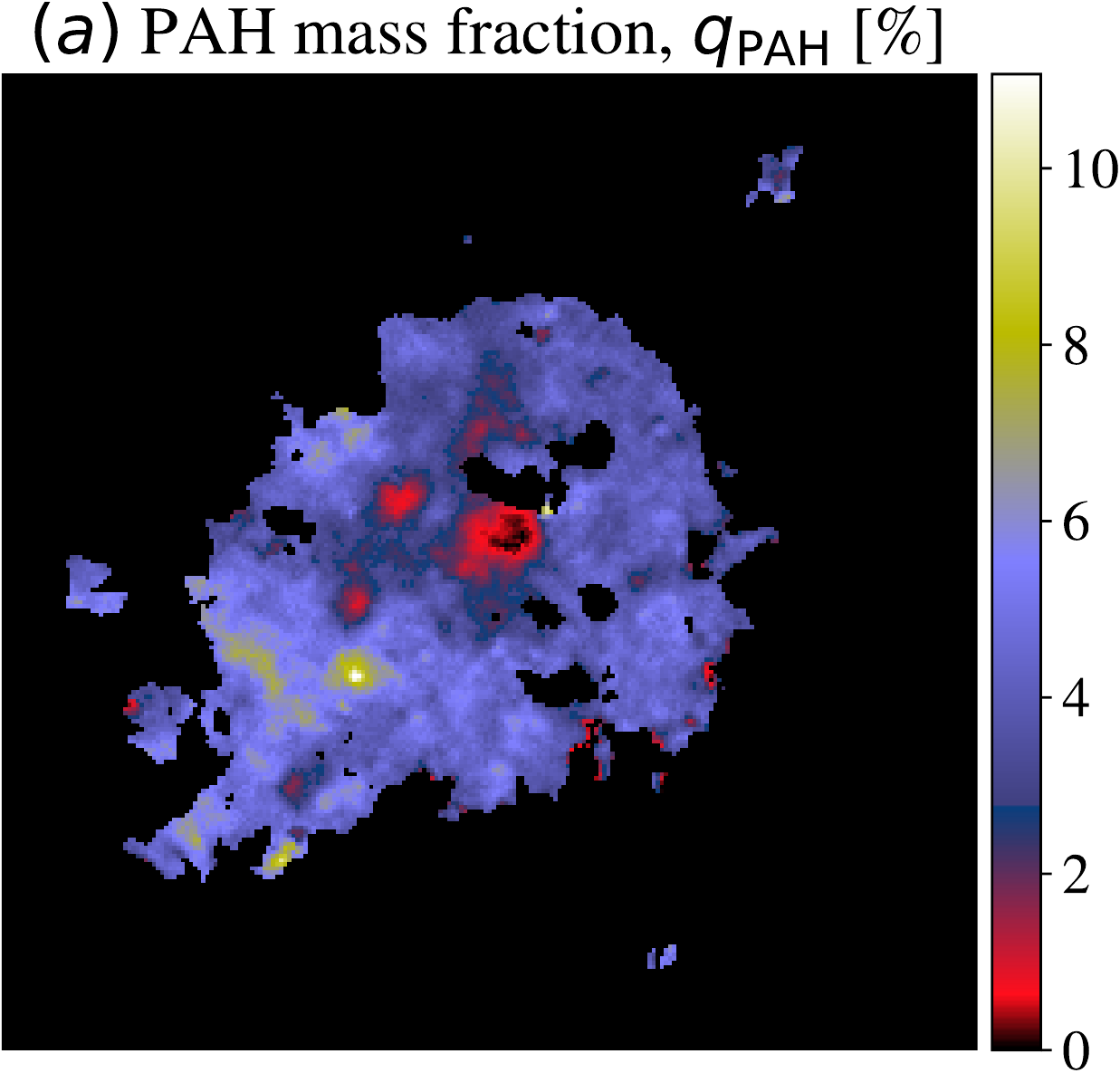} &
    \includegraphics[width=0.48\textwidth]{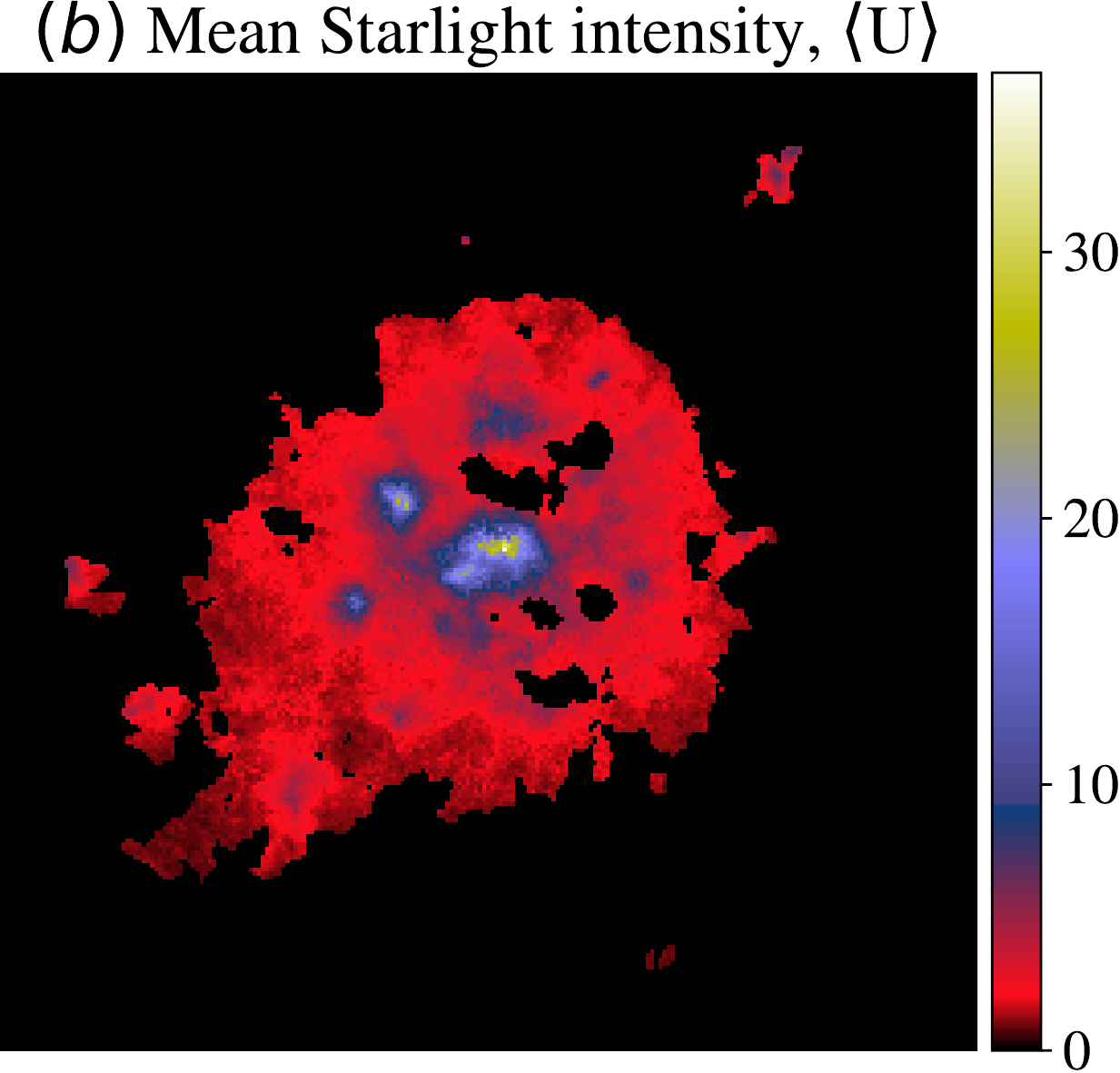} \\
  \end{tabular}
  \newcap{Carving out of PAHs by UV photons in N$\,$11}%
         {Both maps have identical fields of view, centered on the star cluster
          N$\,$11B \citep[LH10;][]{lucke70}, and are 400~pc wide.
          The quantities displayed are the results from the spatially-resolved
          \hSED\ modeling, with the composite approach \citep[\cf\ 
          \refsec{sec:dale};][]{galametz16}: 
          \begin{inlinelistalph}
            \item the \hPAH\ mass fraction, $q_\sms{PAH}$; and
            \item the mean starlight intensity, $\langle U\rangle$, in units
              of $2.2\E{-5}$~W/m$^2$.
          \end{inlinelistalph}
          The black areas have been masked.
          \CClicence}
  \label{fig:N11}
\end{figure}

\paragraph{Evidence in resolved regions.}
This last point is observationally verified in countless regions.
It can be conveniently witnessed, as the smallest grains are the carriers of the \hMIR\ continuum, which is well separated from the rest of the emission (\cf\ \refsubfig{fig:themis_emiss_proxy}{b}).
In addition, small carbon grains carry the series of aromatic features (\cf\ \refsec{sec:PAH}).
The disappearance of these features in regions of enhanced \hISRF\ is very likely the sign of the destruction of these grains by hard \hUV\ photons.
This is, for instance, evident in one of our studies of the massive star-forming region, N$\,$11, in the \hLMC\ \citep{galametz16}.
This region contains several blobs, with embedded star clusters.
The maps of the \hPAH\ mass fraction, $q_\sms{PAH}$ (\cf\ \refsec{sec:dale}), is shown in \refsubfig{fig:N11}{a}.
It has been derived by modeling the spatially-resolved \hSED\ of \hspitz\ and \hhersc\ images.
Comparing this image to the mean starlight intensity in \refsubfig{fig:N11}{b}, we see that \hPAH s are strongly depleted in the blobs where $\langle U\rangle$ is enhanced.
The photodestruction is evident.
In the case of this massive region, a bright star cluster such as N$\,$11B can clear \hPAH s out over a region of typically $\simeq50$~pc.
\takeaway{Aromatic features are severely depleted around star-forming regions.}

    \subsubsection{Thermal Sputtering}
    \label{sec:XETG}

Another destruction mechanism is grain erosion and vaporization by collisions with energetic ions, either in coronal plasmas (\expression{thermal sputtering}; $T\gtrsim10^6$~K), or in shock waves (\expression{kinetic sputtering}; $v\gtrsim100$~km/s).
There is an abundant literature on the subject \citep[\eg][see also the review by \citeauthor{dwek92}, \citeyear{dwek92}]{draine79,dwek80,tielens94,jones96,jones04,nozawa06,micelotta10b,bocchio12,bocchio14,hu19}.
We start by discussing thermal sputtering in this section, and will review kinetic sputtering in \refsec{sec:shockdest}.

\begin{figure}[htbp]
  \includegraphics[width=\textwidth]{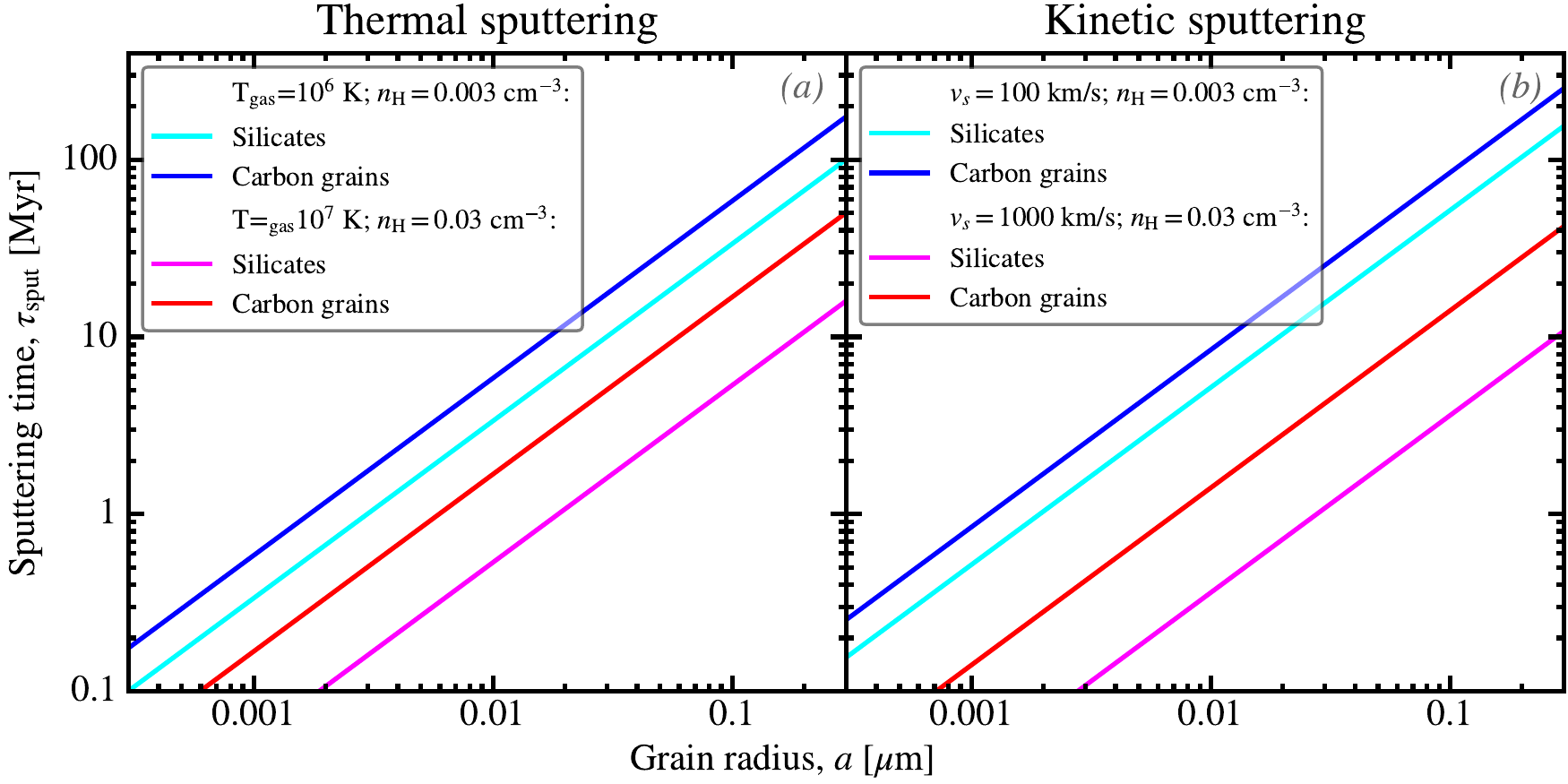}
  \newcap{Thermal and kinetic sputtering times of silicates and carbon grains}%
         {Panel~\textit{(a)} show the mean lifetimes of grains in a hot gas,
          computed from \refeq{eq:thsput}, using the sputtering yields of 
          \citet{nozawa06} fitted by \citet{hu19}.
          Panel~\textit{(b)} show the mean lifetimes of grains in a shock of 
          velocity $v_s$, computed from \refeq{eq:kinsput}, using the same 
          sputtering yields.
          \CClicence}
  \label{fig:sputtering}
\end{figure}

\paragraph{Sputtering times.}
The evolution of a grain of radius, $a$, and mass, $m_\sms{d}(a)=4/3\pi a^3\rho$, subjected to sputtering in a gas of density $n_\sms{H}$, can be expressed \citep[\eg][]{hu19}:
\begin{equation}
  \frac{\dd m_\sms{d}(a)}{\dd t}=3m_\sms{d}(a)\frac{1}{a}\frac{\dd a}{\dd t}
  = 3\frac{m_\sms{d}(a)}{a}n_\sms{H}Y_\sms{sput}(T_\sms{gas},v_\sms{s}),
\end{equation}
where we have hidden all the microphysics into the \expression{sputtering yield}, $Y_\sms{sput}(T_\sms{gas},v_s)\equiv\dd a/\dd t/n_\sms{H}$.
This quantity depends on:
\begin{inlinelist}
  \item the gas temperature, $T_\sms{gas}$, in case of thermal sputtering; or
  \item the shock velocity, $v_s$, in case of kinetic sputtering.
\end{inlinelist}
A detailed derivation of $Y_\sms{sput}$ can be found in \citet[][Sect.~5]{nozawa06}.
For our simple discussion, we will adopt their yields, for silicate and carbon grains, fitted by \citet{hu19}.
In the thermal case, the \expression{sputtering rate} can be expressed as:
\begin{equation}
  \frac{1}{\tau_\sms{sput}^\sms{th}(a,T_\sms{gas},n_\sms{H})}
    \equiv\frac{1}{m_\sms{d}}\frac{\dd m_\sms{d}}{\dd t}
    = \frac{3n_\sms{H}}{a}Y_\sms{sput}(T_\sms{gas}).
  \label{eq:thsput}
\end{equation}
\refsubfig{fig:sputtering}{a} shows the lifetimes of grains in a coronal plasma.
With \refeq{eq:thsput}, in the \hHIM\ (\cf\ \reftab{tab:ISMism}), typical grains \refeqp{eq:meansizethemis} have lifetimes of:
\begin{inlinelist}
  \item $\tau_\sms{sput}^\sms{th}\simeq11$~Myr, for silicates; and
  \item $\tau_\sms{sput}^\sms{th}\simeq16$~Myr, for carbon grains.
\end{inlinelist}
In the case of \hSNII\ blastwaves, grains stay in post-shock conditions for only $\simeq10^4$~yr.
Dust destruction by thermal sputtering is thus not the dominant process in the shocked \hISM.
\takeaway{Grains have short lifetimes in coronal plasmas.}

\begin{figure}[htbp]
  \begin{tabular}{cc}
    \includegraphics[width=0.48\textwidth]{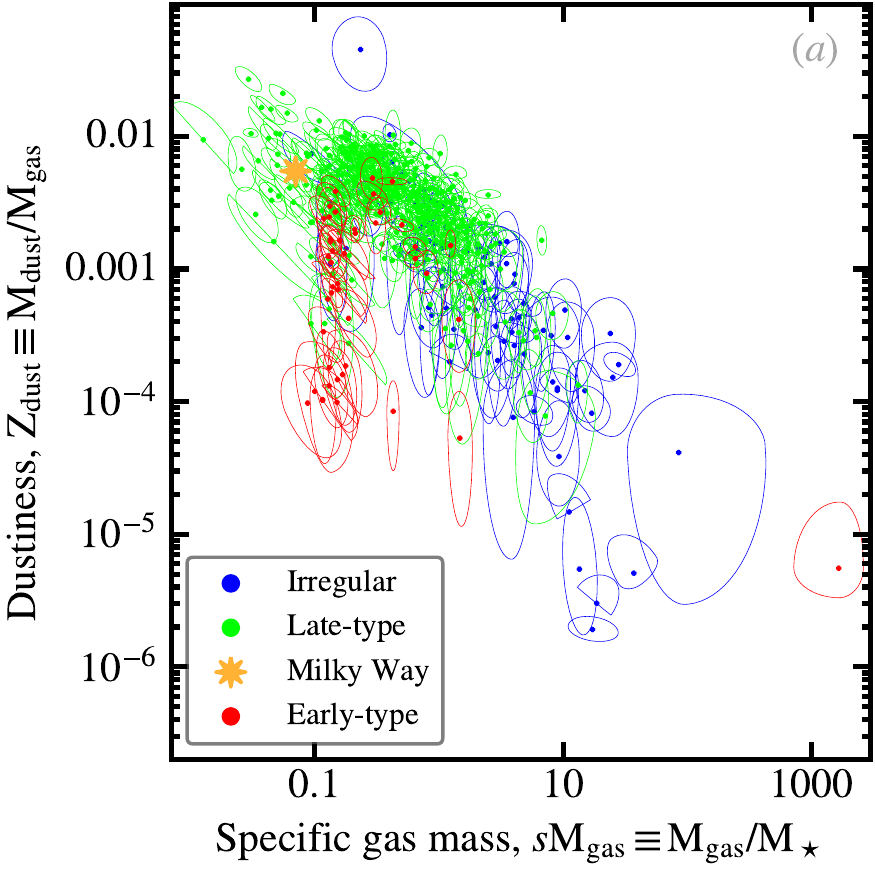} &
    \includegraphics[width=0.48\textwidth]{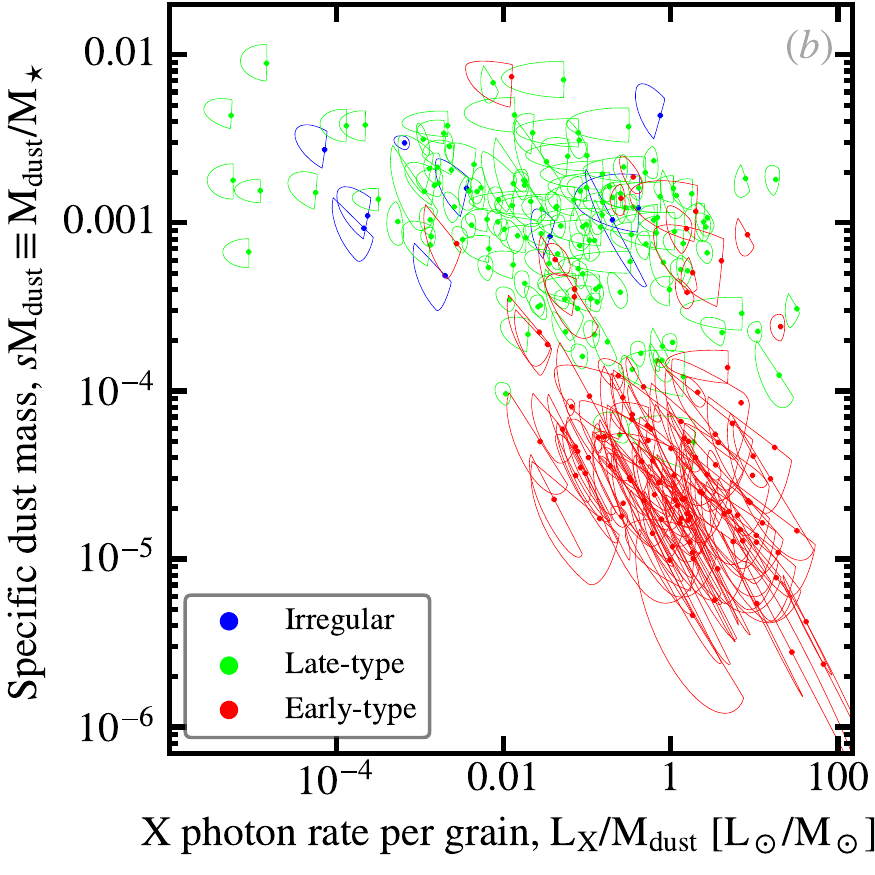} \\
  \end{tabular}
  \newcap{Evidence of thermal sputtering in elliptical galaxies}%
         {Both panels show the results of the \hSED\ fitting of 800 nearby 
          galaxies \citepalias{galliano21}, that we have started to discuss in 
          \refsec{sec:dale}.
          We have represented the measurements as \expression{Skewed 
          Uncertainty Ellipses} \citepalias[\hSUE; Appendix~F of][]{galliano21}.
          Each \hSUE\ corresponds to one whole galaxy.
          Galaxies are color coded according to their Hubble stage, $T$:
          \begin{inlinelist}
            \item \hETG s ($T\le0$) in red;
            \item \hLTG s ($0<T<9$) in green; and
            \item Irregulars ($T\ge9$) in blue.
          \end{inlinelist}
          Panel~\textit{(a)} shows how the \hdustiness\ scales with the 
          specific gas mass.
          We see that most \hETG s (red) are distributed along a vertical
          branch, below the main trend.
          Panel~\textit{(b)} shows how the specific dust mass varies with 
          X-ray luminosity, $L_\sms{X}$, over dust mass, $M_\sms{dust}$.
          \hETG s occupy the lower right corner of this relation.
          \CClicence}
  \label{fig:XETG}
\end{figure}

\paragraph{Early-type galaxies.}
We have seen in \refsec{sec:galdesc} that \hETG s tend to be characterized by a diffuse X-ray emission, originating in a permeating coronal gas.
This \hHIM\ is likely filling most of their \hISM.
This has consequences on the dust properties.
This can be seen in \refsubfig{fig:XETG}{a}, looking at a classic scaling relation between the \hdustiness\ and the \expression{specific gas mass}\footnote{Given an extensive quantity, $Q$, it is common, in extragalactic astronomy, to define the corresponding intensive \expression{specific} quantity, $sQ\equiv Q/M_\star$, by dividing by the stellar mass.}, $sM_\sms{gas}\equiv M_\sms{gas}/M_\star$.
Most \hETG s appear to be distributed on a vertical branch, below the main trend.
They appear to be depleted in dust, at a given specific gas mass.
Investigating the contribution of the X-ray emitting coronal gas, we have displayed the specific dust mass, as a function of the X-ray-luminosity-to-dust-mass ratio, $L_\sms{X}/M_\sms{dust}$, in \refsubfig{fig:XETG}{b} \citepalias{galliano21}.
The $L_\sms{X}/M_\sms{dust}$ ratio quantifies the X-ray photon rate per dust grain.
We see that ellipticals occupy the lower right corner of this relation: they have a high photon rate per dust grain and a low specific dust mass.
We have just shown that grains in a hot gas have a short lifetime \refeqp{eq:thsput}.
The correlation of \refsubfig{fig:XETG}{b} is thus likely the result of enhanced thermal sputtering in \hETG s.

    \subsubsection{Destruction by SN Blast Waves}
    \label{sec:shockdest}

We now focus on the effect of kinetic sputtering.
This process leads to erosion and vaporization of grains in \hSNII\ blast waves.
As we will see, this happens to be the major dust grain destruction mechanism.
The kinetic sputtering rate is similar to the thermal case \refeqp{eq:thsput}, except that the sputtering yield now depends on the shock velocity, $v_s$ (\cf\ \refsubfig{fig:sputtering}{b}):
\begin{equation}
  \frac{1}{\tau_\sms{sput}^\sms{kin}(a,v_s,n_\sms{H})}
    = \frac{3n_\sms{H}}{a}Y_\sms{sput}(v_s).
  \label{eq:kinsput}
\end{equation}
In addition, grain shattering in grain-grain collisions is an important dust destruction mechanism in \hSNII\ blast waves \citep[\eg][]{kirchschlager21}.

\paragraph{Evidence in resolved regions.}
Although the efficiency of the process is debated, the reality of dust destruction by \hSNII\ shock waves is rather consensual.
This process can even be observed in spatially-resolved \hSNR s.
In particular, \hISO\ and \hspitz\ \hMIR\ spectra of pre-shock and post-shock matter show systematic differences in, for instance: 
\begin{inlinelist}
  \item 3C$\,$391 \citep{reach02};
  \item SN$\,$1987A \citep{dwek08,arendt16}; and
  \item Puppis$\,$A \citep{arendt10b}.
\end{inlinelist}
The post-shock \hISM\ exhibits:
\begin{itemize}
  \item the disappearance of the aromatic features;
  \item the disappearance of the dust continuum.
\end{itemize}

\paragraph{Global model prescription.}
Similarly to what we did for grain growth \refeqp{eq:eps_grow}, it is convenient to express the dust destruction rate as a function of global galactic quantities.
Such a formula was proposed by \citet{dwek80}:
\begin{equation}
  \frac{1}{\tau_\sms{SN-dest}(t)}
    = \frac{m_\sms{gas}^\sms{dest}}{M_\sms{gas}(t)}R_\sms{SN}(t),
  \label{eq:SNdest}
\end{equation}
where $R_\sms{SN}(t)$ is the \hSNII\ rate \refeqp{eq:RSN}, and $m_\sms{gas}^\sms{dest}$ is an empirical parameter quantifying the destruction efficiency.
The latter represents the gas mass swept by a single \hSNII\ blast wave, within which all grains are destroyed.
\refeq{eq:SNdest} can be understood the following way.
\begin{enumerate}
  \item A single \hSNII\ destroys a mass
    $Z_\sms{dust}\times m_\sms{gas}^\sms{dest}$ of dust.
  \item Knowing the \hSNII\ rate, the dust mass destroyed per unit time is 
    therefore:
    \begin{equation}
      \left(\frac{\dd M_\sms{dust}}{\dd t}\right)_\sms{dest}
      = Z_\sms{dust}m_\sms{gas}^\sms{dest}\times R_\sms{SN}(t).
    \end{equation}
  \item The dust destruction rate is then simply 
    $1/\tau_\sms{SN-dest}
      =\left(\dd M_\sms{dust}/\dd t\right)_\sms{dest}/M_\sms{dust}$,
    which gives \refeq{eq:SNdest}.
\end{enumerate}

\paragraph{The destruction efficiency.}
The dust destruction efficiency, quantified by the parameter $m_\sms{gas}^\sms{dest}$ in \refeq{eq:SNdest}, ranges in the literature between $\simeq100\eMsun$ and $\simeq1000\eMsun$.
It can be roughly estimated with the following arguments \citep[][slightly adapting his numbers]{draine09}:
\begin{enumerate}
  \item In a \hISM\ of density $n_\sms{H}\simeq1$~cm$^{-3}$, a \hSNII, with 
    energy $E_0\simeq10^{43}$~J ($10^{51}$~erg), produces a blast wave that stays 
    in the Sedov-Taylor phase (adiabatic expansion), until reaching a velocity 
    $v_s\simeq200$~km/s \citep[Eq.~39.22 of][]{draine11b}.
  \item Grains are primarily destroyed in the Sedov-Taylor phase.
    At the end of this phase, the radius of our blast wave is 
    $R_\sms{Sedov}\simeq24$~pc \citep[Eq.~39.21 of][]{draine11b}, corresponding
    to a total gas mass of $M_\sms{Sedov}\simeq1900\eMsun$, which is a rough 
   estimate of our efficiency parameter, $m_\sms{gas}^\sms{dest}$.
  \item At solar metallicity (using the \hdustiness\ of 
    \reftab{tab:massthemis}), this corresponds to 
    $M_\sms{dust}^\sms{dest}\simeq10\eMsun$.
\end{enumerate}
This estimate corresponds to a dust lifetime of $\tau_\sms{SN-dest}\simeq370$~Myr, in the \hMW.
This is a value close to what is found by  more detailed, theoretical studies \citep[\eg][]{jones96}.
Note however that a recent re-estimate, using hydrodynamical simulations, and accounting for the role of dust mantles found:
\begin{inlinelist}
  \item a shorter lifetime for carbon grains; but 
  \item a significantly longer lifetime ($\tau_\sms{SN-dest}\simeq2-3$~Gyr) for 
    silicates \citep{slavin15}.
\end{inlinelist}
We will give our own take on this timescale in \refsec{sec:dustime}.
\takeaway{A single \hSNII\ blast wave can destroy up to $\lesssim10\eMsun$ of 
          dust, at Solar metallicity, resulting in a dust lifetime of 
          $\tau_\sms{SN-dest}\gtrsim400$~Myr.}

\section{Cosmic Dust Evolution}
\label{sec:cosmicdustevol}

\begin{figure}[htbp]
  \includegraphics[width=\textwidth]{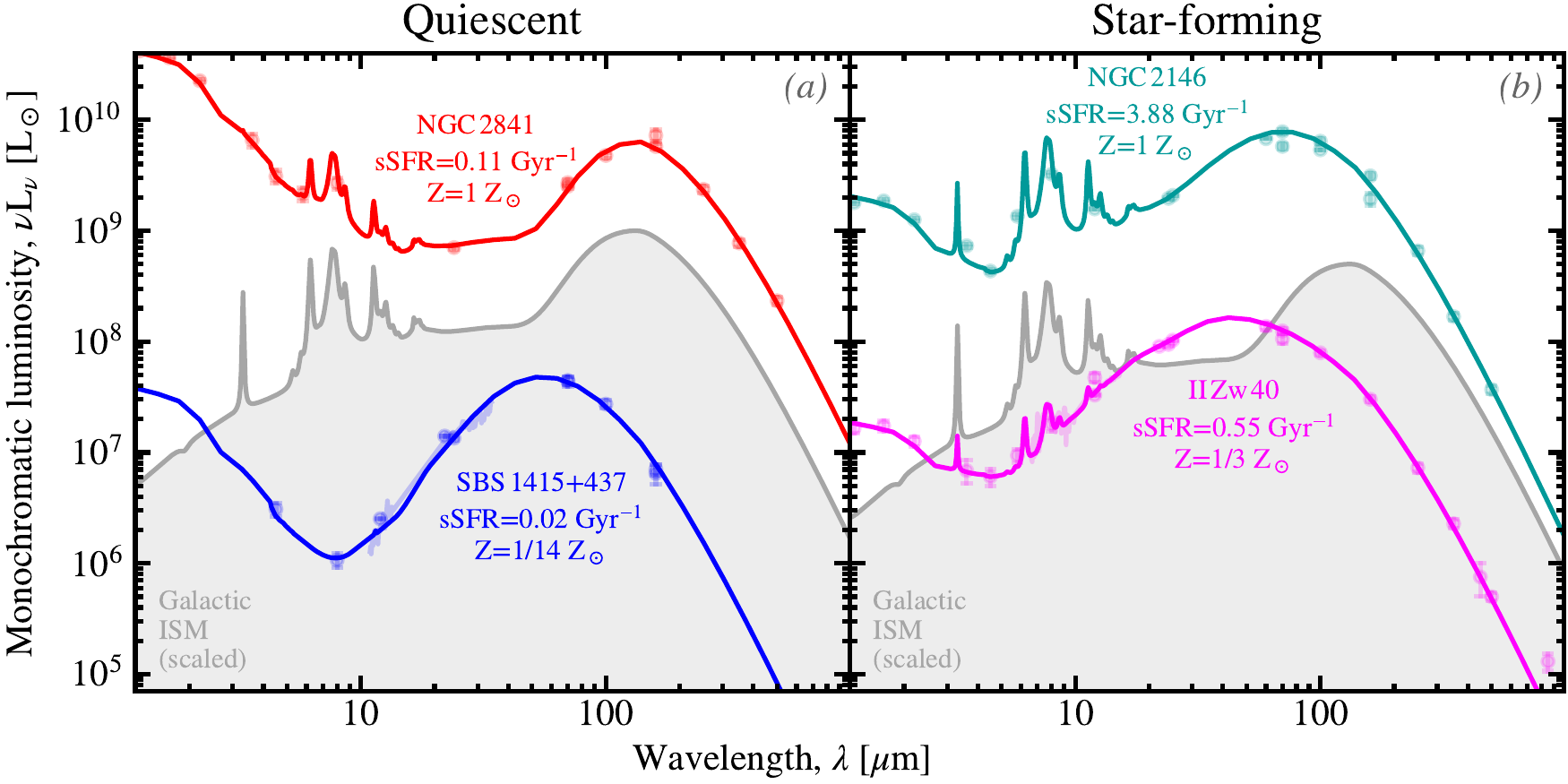}
  \newcap{Effects of dust evolution on the SEDs of galaxies.}%
         {Each panel displays the observations and the \hSED\ model of two
          nearby galaxies \citep{remy-ruyer15}, on top of the \hSED\ of the 
          diffuse Galactic \hISM\ (in grey).
          The red curve in panel~\textit{(a)} shows a quiescent 
          Solar-metallicity galaxy. 
          Apart from the stellar continuum, it is identical to the diffuse
          \hISM.
          In contrast, the blue curve represents a low-metallicity quiescent
          system.
          Its dust properties are notably different: 
          \begin{inlinelist}
            \item weak or absent \hUIB s; 
            \item overall hotter dust (\hFIR\ peaks at shorter wavelengths);
            \item a somehow broader \hFIR\ spectrum, resulting from a 
              distribution of starlight intensities and/or an overabundance of 
              small grains (\cf\ \refsec{sec:dale}).
          \end{inlinelist}
          This \hSED\ is qualitatively similar to the \hSED\ of a compact 
          \hii\ region \citep[\eg][]{peeters02}.
          The red curve in panel~\textit{(b)} shows a Solar-metallicity
          galaxy with a sustained star formation activity.
          Compared to its quiescent counterpart, it has a much hotter and 
          broader \hFIR\ emission, originating at least partly in bright
          \hPDR s.
          The starbursting low-metallicity galaxy (blue curve in 
          panel~\textit{b}) has the same features as its quiescent 
          counterpart, with a broader \hFIR\ emission.
          \CClicence}
  \label{fig:selectSEDs}
\end{figure}

\expression{Cosmic dust evolution} is the modeling of dust evolution from a global point of view, at the scale of a galaxy, over cosmic times\footnote{We assume that the age of the Universe is $t_\sms{present}\simeq14$~Gyr \citep[\eg][]{hogg99}.}.
At galaxy-wide scales, most dust evolution processes can be linked to star formation:
\begin{inlinelist}
  \item formation of molecular clouds and their subsequent evaporation;
  \item stellar ejecta;
  \item \hSN\ shock waves;
  \item \hUV\ and high-energy radiation.
\end{inlinelist}
The characteristic timescale of these processes is relatively short
(of the order of the lifetime of massive stars; $\tau\lesssim10$~Myr) and their 
effect is usually localized around star-forming regions.
For these reasons, the \hsSFR\ is an indicator of sustained dust processing.
However, the dust lifecycle is a hysteresis.
There is a longer term evolution, resulting from the progressive elemental 
enrichment of the \hISM, which becomes evident on timescales of $\simeq1$~Gyr.
This evolutionary process can be traced by the metallicity.
\reffig{fig:selectSEDs} illustrates these two timescales by comparing the \hSED s of a few galaxies.

  \subsection{Constraining the Dust Build-Up in Galaxies}
  \label{sec:G2DvsZ}

We start by focussing on the build-up of the total dust mass.
We will discuss the evolution of small carbon grains in \refsec{sec:PAHvsZ}.
For terminological consistency of the rest of the discussion, let's define the following metallicity regimes:
\begin{description}
  \item[Very low metallicity:] $Z\lesssim0.2\times Z_\odot$;
  \item[Low metallicity:] 
    $0.2\times Z_\odot\lesssim Z\lesssim0.45\times Z_\odot$;
  \item[Normal metallicity:] $Z\gtrsim0.45\times Z_\odot$.
\end{description}
We will see in \refsec{sec:dustime} that these ranges correspond to dust evolution regimes of nearby galaxies.

    \subsubsection{Cosmic Dust Evolution Models}
    \label{sec:dustevolmodel}

The first model accounting for the evolution of the gas content of galaxies and its cycle with star formation was presented by \citet{schmidt59}.
The Eqs.~7-9 of \citet{schmidt59} are the basic equations for the evolution of the gas mass as a result of the successive waves of star formation.
Subsequent studies included the heavy element enrichment of the gas, therefore accounting for the chemical evolution of galaxies \citep[\eg][for an early review]{audouze76}.
\citet{dwek80} then initiated the first cosmic dust evolution model, by including grain processing in the gas enrichment modeling.
\citet{dwek98} modeled the radial trends in the \hMW, accounting for the individual elemental yields by stars of different initial masses.
Such models have since then been refined \citep[\eg][]{morgan03,dwek07,zhukovska08,galliano08a,hirashita11,asano13,rowlands14,zhukovska14,feldmann15,de-looze20,galliano21}.
These models are nowadays used to account for subgrid physics in numerical simulations of galaxy evolution \citep[\eg][]{hou17,aoyama20}.

\paragraph{Physical ingredients and assumptions.}
The different cosmic dust evolution models we have just cited above all have differences.
They however have a common set of physical ingredients and assumptions.
In what follows, we describe the model of \citetalias{galliano21}, which is well representative of the diversity found in the literature.
\begin{description}
  \item[Star-formation regulation:] 
    the evolution is driven by the \hSFH\ of the galaxy (\cf\ 
    \refsec{sec:SFH}), considered as a box, 
    where the mixing of freshly injected elements and grains is assumed
    instantaneous.
    This box can be closed or include the effects of \expression{infall} and 
    \expression{outflow}.
    The infall and outflow rates are usually assumed to be proportional to the
    \hSFR:
    \begin{eqnarray}
      R_\sms{in}(t) & \equiv & \delta_\sms{in}\times\psi(t) \label{eq:Rin} \\
      R_\sms{out}(t) & \equiv & \delta_\sms{out}\times\psi(t). \label{eq:Rout} 
    \end{eqnarray}
  \item[Stellar evolution and ejecta:]
    in each time interval, $[t_0,t_0+\Delta t]$, a mass $\psi(t_0)\Delta t$ 
    of stars is formed.
    The fraction of stars of different initial masses, $m_\star$, is given by 
    the particular \hIMF\ we have assumed (usually Salpeter or Chabrier; \cf\ 
    \refsec{sec:SFH}).
    These stars have different lifetimes, $\tau(m_\star)$ (\cf\ 
    \reffig{fig:stellar_isochrones}).
    They return to the \hISM\ a fraction of their gas, freshly formed heavy 
    elements and dust grain seeds, after a time $t_0+\tau(m_\star)$.
    This is called the \expression{delayed injection} process.
    After this time $t_0+\tau(m_\star)$, a fraction of the initial stellar mass
    is locked in a remnant 
    \citep[white dwarf, neutron star or black hole; \eg][]{ferreras00}.
  \item[Dust evolution:]
    grain sources and sinks are estimated with the formulae we have discussed
    earlier in this chapter:
    \begin{description}
      \item[Stardust condensation] is accounted for by assuming a mean fraction
        of ejected dust by \hLIMS\ and \hSNII:
        \begin{inlinelist}
          \item for \hLIMS, we have used $\delta_\sms{LIMS}=15\,\%$ 
            \refeqp{eq:LIMS};
          \item for \hSNII, we use \refeq{eq:SNcond}.
        \end{inlinelist}
      \item[Grain growth] is accounted for by \refeq{eq:eps_grow}.
      \item[SN blast wave destruction] is accounted for by \refeq{eq:SNdest}.
      \item[Astration] is simply the fraction of dust consumed by \hSF, at a 
        rate $Z_\sms{dust}(t)\times\psi(t)$.
    \end{description}
\end{description}

\paragraph{The ejected masses.}
The dust evolution differential equations we will discuss below depend on the gas, heavy element and dust masses ejected by stars, at time $t$:
\begin{eqnarray}
  e(t) & = & \int_{\min[m_\star(t-\tau(m_\star))]}^{m_+}\left[m_\star-r(m_\star)\right]\times 
             B(t-\tau(m_\star))\times\phi(m_\star)\ddiff m_\star \label{eq:ejg} \\
  e_\sms{Z}(t) 
    & = & \int_{\min[m_\star(t-\tau(m_\star))]}^{m_+}Y_\sms{Z}(m_\star)\times
    B(t-\tau(m_\star))\times\phi(m_\star)\ddiff m_\star \label{eq:ejZ} \\
  e_\sms{dust}(t)
    & = & \int_{\min[m_\star(t-\tau(m_\star))]}^{m_-^\sms{SN}}
          Y_\sms{Z}(m_\star)\delta_\sms{LIMS}\times
    B(t-\tau(m_\star))\times\phi(m_\star)\ddiff m_\star \nonumber\\
    & + & \int_{\min\left[m_\star(t-\tau(m_\star)),m_-^\sms{SN}\right]}^{m_+^\sms{SN}}
      \langle Y_\sms{SN}\rangle\times
      B(t-\tau(m_\star))\times\phi(m_\star)\ddiff m_\star. \label{eq:ejd}
\end{eqnarray}
These three equations are essentially the integral of the products of three terms: $\int f(m_\star)\times B(t-\tau(m_\star))\times\phi(m_\star)\ddiff m_\star$.
\begin{enumerate}
  \item The term $f(m_\star)$ is the mass of gas, heavy elements and 
    dust ejected by a star of initial mass, $m_\star$.
    For the gas mass, we have subtracted the remnant mass, $r(m_\star)$.
    The term $Y_\sms{Z}(m_\star)$ is the total heavy element yield 
    (\cf\ \reffig{fig:stellar_yields}).
  \item The term $B(t-\tau(m_\star))$ is the stellar birth rate, $\tau(m_\star)$ 
    ago.
    It gives the number of stars of initial mass $m_\star$, dying at time $t$.
  \item The third term is the \hIMF, giving the number of stars per mass bin.
    It is the number density we integrate over.
  \item In \refeqs{eq:ejg}{eq:ejd}, the lower bound of the integrals, 
    $m_\star(\tau=t-\tau(m_\star))$, is the mass of stars having a lifetime 
    $t-\tau(m_\star)$.
    We thus take the minimum mass of stars dying at time $t$ 
    (assuming star formation has started at $t=0$).
    Low-mass stars ($m_\star\lesssim0.9\eMsun$) are irrelevant for chemical 
    evolution, as their lifetime is longer than the age of the Universe.
\end{enumerate}

\paragraph{The equations of evolution.}
The physical ingredients and assumptions we have discussed earlier in this section translate into four coupled differential equations describing the temporal evolution of the stellar, gas, heavy element and dust masses, $M_\star$, $M_\sms{gas}$, $M_\sms{Z}$ and $M_\sms{dust}$:
\begin{align}
  \frac{\dd M_\star}{\dd t} 
    & = & \psi(t) & - e(t) & & & &
  \label{eq:evolstar} \\
  \frac{\dd M_\sms{gas}}{\dd t}
    & = & -\psi(t) & + e(t) & + R_\sms{in}(t) & - R_\sms{out}(t) & &
  \label{eq:evolgas} \\
  \frac{\dd M_\sms{Z}}{\dd t}
    & = & -Z(t)\psi(t) & + e_\sms{Z}(t) & + 0\times R_\sms{in}(t)
          & - Z(t)R_\sms{out}(t) 
          & + \frac{M_\sms{dust}}{\tau_\sms{SN-dest}(t)}
          & - \frac{M_\sms{dust}}{\tau_\sms{grow}(t)}
  \label{eq:evolmet} \\
  \frac{\dd M_\sms{dust}}{\dd t}
    & = & - \underbrace{Z_\sms{dust}(t)\psi(t)}_\sms{astration} 
        & + \underbrace{e_\sms{dust}(t)}_\sms{ejecta} 
        & + \underbrace{0\times R_\sms{in}(t)}_\sms{infall}
        & - \underbrace{Z_\sms{dust}(t)R_\sms{out}(t)}_\sms{outflow}
        & - \underbrace{\frac{M_\sms{dust}}{\tau_\sms{SN-dest}(t)}}_\sms{SN 
                                                                   destruction}
        & + \underbrace{\frac{M_\sms{dust}}{\tau_\sms{grow}(t)}}_\sms{grain 
                                                                   growth}.
  \label{eq:evoldust}
\end{align}
\refeqs{eq:evolstar}{eq:evoldust} simply express the time derivative of the mass, on the left-hand side, and the sum of the different individual rates on the right-hand side, some positive, some negative.
We can note a few points.
\begin{enumerate}
  \item We have neglected metallicity variations in the total gas mass 
    \refeqp{eq:evolgas}, as it represents at most $\simeq2\,\%$.
  \item We have assumed that the infalling gas was free of heavy elements and 
    dust.
    It is therefore 0 in \refeqs{eq:evolmet}{eq:evoldust}.
  \item On the contrary, the outflowing gas is carrying away heavy elements and
    dust.
    This is the reason why we loose these quantities at rates $Z.R_\sms{out}$ and
    $Z_\sms{dust}.R_\sms{out}$, respectively.
  \item Dust destruction by \hSNII\ removes mass to the dust content, at a rate
    $M_\sms{dust}/\tau_\sms{SN-dest}$, but returns it as gas-phase heavy elements.
  \item This is the opposite for grain growth which removes mass to gas-phase 
    heavy elements, at a rate $M_\sms{grow}/\tau_\sms{grow}$, and puts it into 
    grains.
\end{enumerate}

\begin{figure}[!htbp]
  \includegraphics[width=\textwidth]{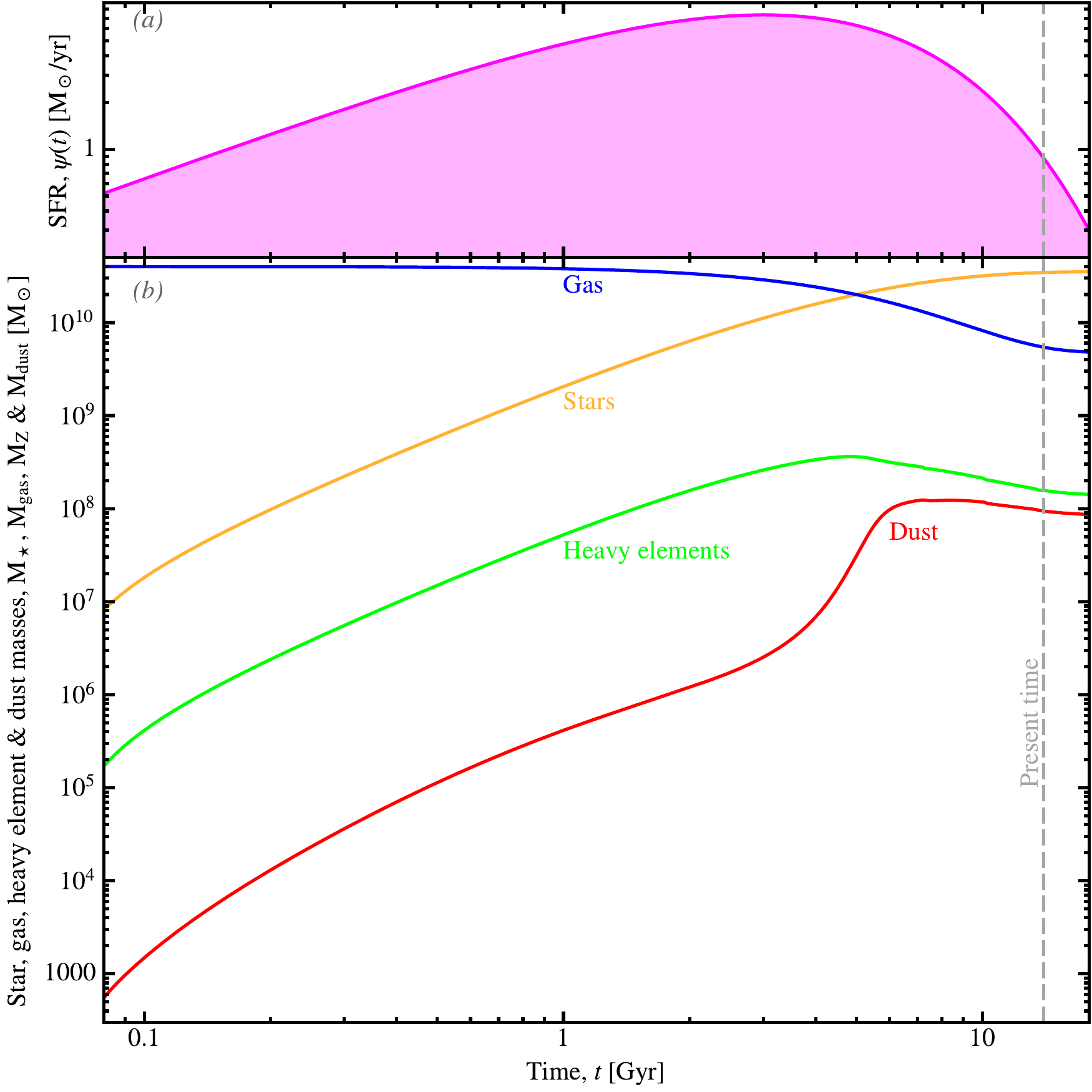}
  \newcap{Dust evolution tracks for a MW-like galaxy}%
         {Panel~\textit{(a)} displays the adopted \hSFH.
          Panel~\textit{(b)} shows the evolution as a function of time of the
          four quantities of \refeqs{eq:evolstar}{eq:evoldust}.
          These quantities approximately reach the \hMW\ values at 
          $t=t_\sms{present}$.
          The parameters are those of 
          \refeqs{eq:dustevolpar1}{eq:dustevolpar2}.
          \CClicence}
  \label{fig:dustevolMW}
\end{figure}

\paragraph{Dust evolution tracks.}
We now present a set of solutions to \refeqs{eq:evolstar}{eq:evoldust}.
For simplicity, we adopt the Chabrier \hIMF\ \refeqp{eq:chabrier} and the delayed \hSFH\ \refeqp{eq:SFHdel}.
The total baryonic, initial mass of the galaxy is assumed to be $M_\sms{ini}=4\E{10}\eMsun$.
\reffig{fig:dustevolMW} shows the time evolution of the main quantities, for a \hMW-like galaxy.
We have adopted the following parameters:
\begin{eqnarray}
  \tau_\sms{SFH}  = & 3\;\textnormal{Gyr} & \mbox{\refeqp{eq:SFHdel}}  
    \label{eq:dustevolpar1} \\
  \psi_0         = & 20\eMsun/\textnormal{yr} & \mbox{\refeqp{eq:SFHdel}} \\
  \delta_\sms{in} = & 0.05               & \mbox{\refeqp{eq:Rin}} \\
  \delta_\sms{out} = & 0.05              & \mbox{\refeqp{eq:Rout}} \\
  \langle Y_\sms{SN}\rangle = & 0.007\eMsun/\textnormal{SN} 
                      & \mbox{\refeqp{eq:SNcond}} \\
  \epsilon_\sms{grow} = & 1000 & \mbox{\refeqp{eq:eps_grow}} \\
  m_\sms{gas}^\sms{dest} = & 1000\eMsun/\textnormal{SN} & 
    \mbox{\refeqp{eq:SNdest}}. \label{eq:dustevolpar2}
\end{eqnarray}
\refsubfig{fig:dustevolMW}{a} shows the \hSFH\ we have adopted.
The time evolution of the individual quantities are represented in \refsubfig{fig:dustevolMW}{b}.
We can note the following points.
\begin{enumerate}
  \item The total baryonic content is conserved: $M_\sms{ini}=M_\star+M_\sms{gas}$.
    The stellar mass is increasing with time, while the gas gets depleted.
    In this particular simulation, they cross over around 
    $t_\sms{cross}\simeq5$~Gyr.
  \item The mass of heavy elements follows the stellar mass at 
    $t\lesssim t_\sms{cross}$, two orders of magnitude lower, as it is controlled
    by the accumulative stellar enrichment.
    Above $t\gtrsim t_\sms{cross}$, the net mass of heavy elements decreases due 
    to astration, as the \hISM\ becomes more tenuous.
  \item The dust mass follows the trend of heavy elements two orders of 
    magnitude lower, for $t\lesssim3$~Gyr, as it is dominated by dust production
    by \hSNII.
    Around $t\simeq3$~Gyr, the metallicity is high enough to render grain growth
    efficient.
    This is what \citet{asano13} have conceptualized as the 
    \expression{critical metallicity}.
    This is an important quantity, that depends on the \hSFH.
    It roughly delineates the two regimes dominated by \hSNII\ production and 
    grain growth in the \hISM.
    In the particular case of \reffig{fig:dustevolMW}, it is 
    $Z_\sms{crit}\simeq Z_\odot/3$.
    Above $t\gtrsim t_\sms{cross}$, dust production is dominated by grain 
    growth, and $M_\sms{dust}/M_\sms{Z}\simeq0.5$, which is 
    roughly the Galactic dust-to-metal mass ratio (\cf\ 
    \reftab{tab:massthemis}).
\end{enumerate}

\begin{figure}[!htbp]
  \includegraphics[width=\textwidth]{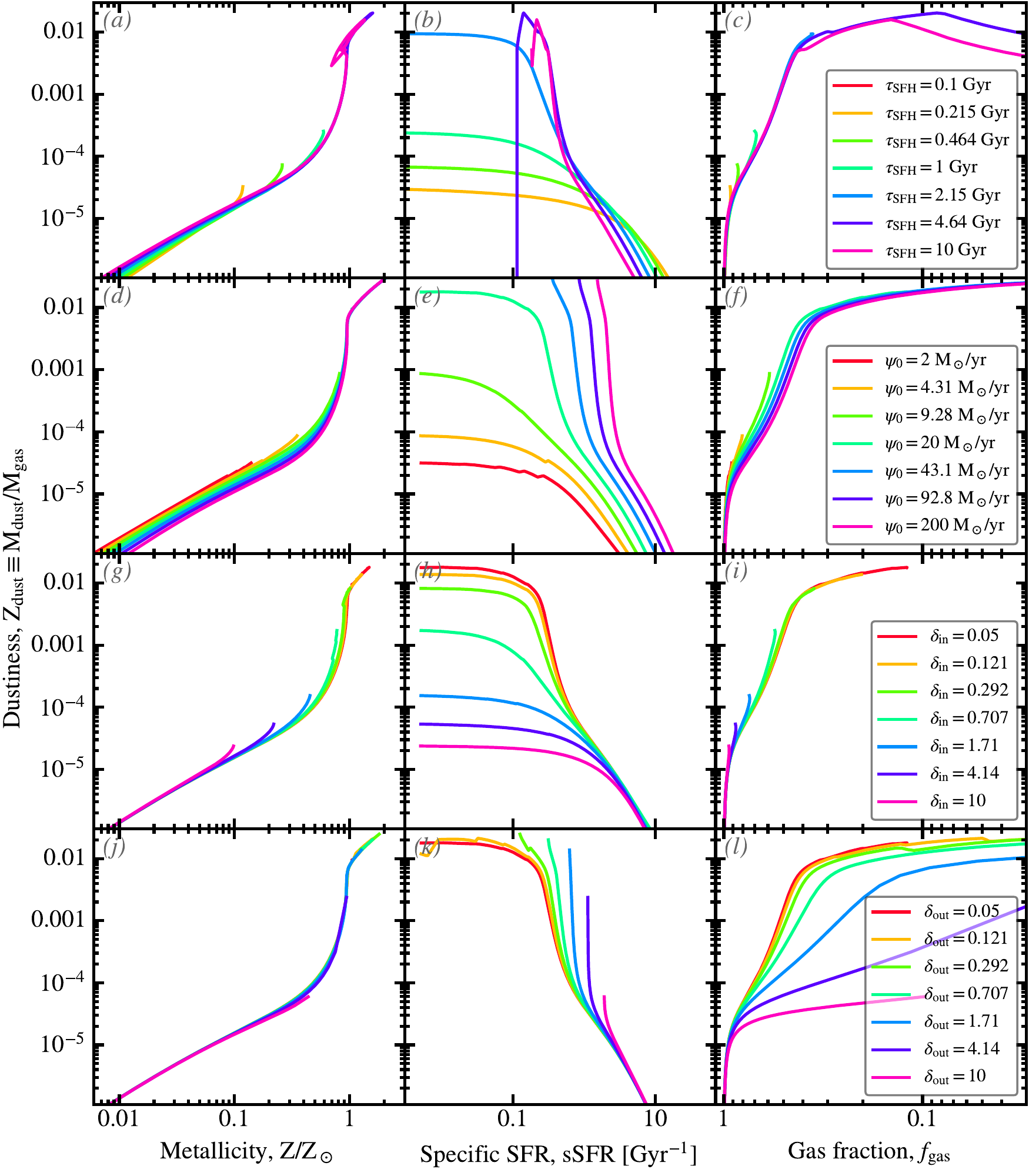}
  \newcap{Effects of SFH-related parameters on dust evolution}%
         {Each row of panels represents the evolution of the \hdustiness\ as a
          function of metallicity, \hsSFR\ and gas fraction.
          In each row, we vary a particular parameter, whose values are given 
          in the right panel.
          The other parameters are kept to their values in 
          \refeqs{eq:dustevolpar1}{eq:dustevolpar2}.
          \CClicence}
  \label{fig:dustevolSFH}
\end{figure}

\begin{figure}[!htbp]
  \includegraphics[width=\textwidth]{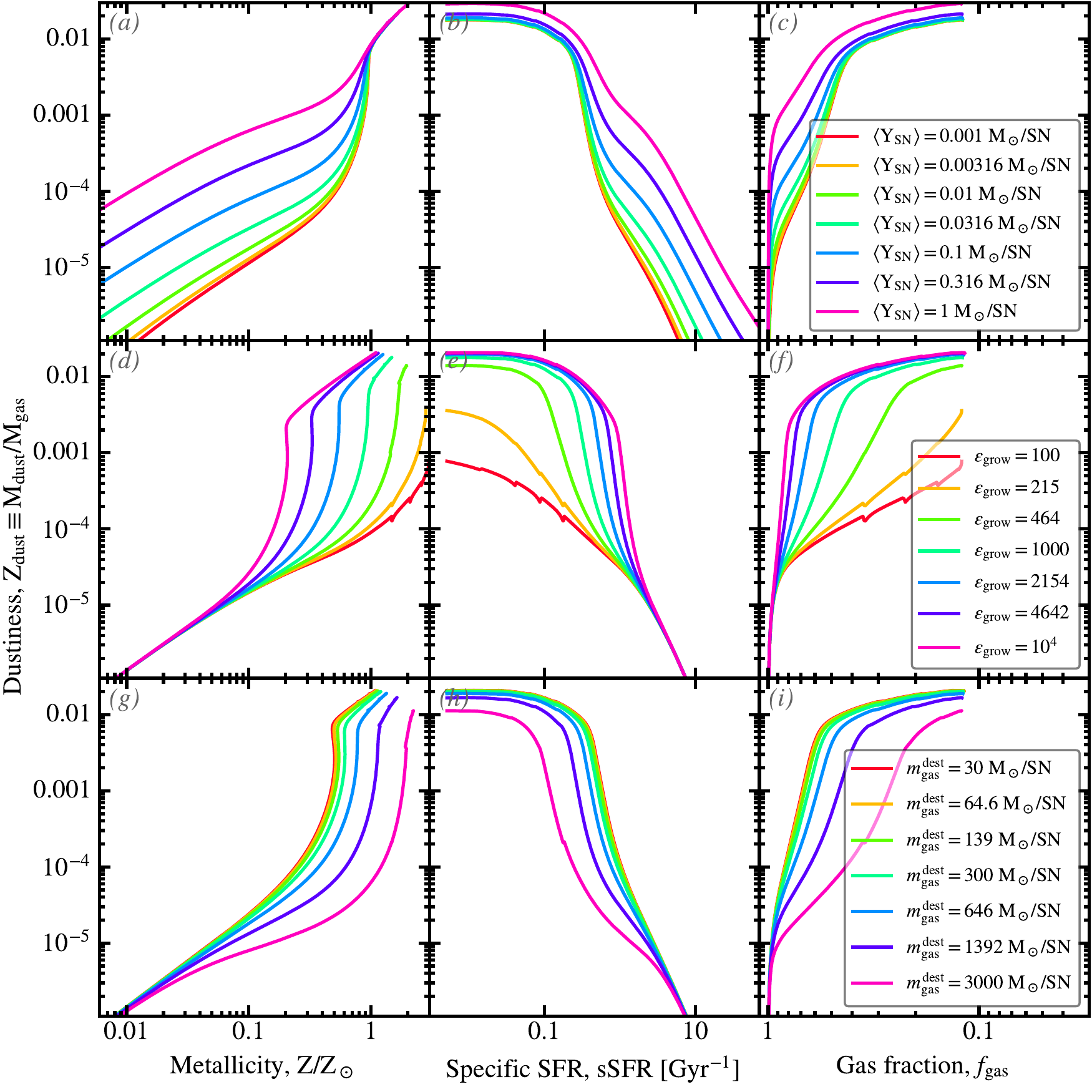}
  \newcap{Effects of tuning parameters on dust evolution}%
         {This figure is similar to \reffig{fig:dustevolSFH}, except that we 
          have varied the dust evolution tuning parameters.
          \CClicence}
  \label{fig:dustevoldust}
\end{figure}

\paragraph{Effects of the individual parameters.}
We end this section by demonstrating the effects of the seven parameters of \refeqs{eq:dustevolpar1}{eq:dustevolpar2} on dust evolution.
We vary each parameter, one by one, keeping the other ones to their values in \refeqs{eq:dustevolpar1}{eq:dustevolpar2}.
We represent the \hdustiness, as a function of:
\begin{inlinelist}
  \item metallicity;
  \item \hsSFR; and
  \item gas fraction, $f_\sms{gas}\equiv M_\sms{gas}/(M_\star+M_\sms{gas})$.
\end{inlinelist}
For completeness, we have represented the four SFH-related parameters in \reffig{fig:dustevolSFH}.
Our center of interest is however the three dust evolution tuning parameters, represented in \reffig{fig:dustevoldust}.
\begin{description}
  \item[The \snii\ dust yield,] $\langle Y_\sms{SN}\rangle$ \refeqp{eq:SNcond},
    has essentially a scaling effect on the \hdustiness, below the critical 
    metallicity, and has no effect above (\refsubfig{fig:dustevoldust}{a-c}).
    This is the reason why constraining this parameter requires 
    very-low-metallicity objects.
    In the very-low-metallicity regime, dust and heavy elements directly come 
    from stellar ejecta. 
    Both quantities are thus roughly proportional, using our simple 
    prescriptions.
    When $\langle Y_\sms{SN}\rangle\lesssim0.003\eMsun$/SN, this parameter is 
    so low, that stardust injection by \hLIMS\ becomes dominant, below the 
    critical metallicity.
  \item[The grain growth efficiency,] $\epsilon_\sms{grow}$ \refeqp{eq:eps_grow},
    has also a scaling effect on the \hdustiness, but above the critical 
    metallicity (\refsubfig{fig:dustevoldust}{d-f}).
    It has no effect at very low metallicity, because 
    $1/\tau_\sms{grow}\propto Z$ (\refeqnp{eq:eps_grow} with $Z\gg Z_\sms{dust}$).
    The tracks for the lowest values of $\epsilon_\sms{grow}$ (in red on 
    \reffig{fig:dustevoldust}) exhibit a quasi-linear \hdustiness-trend with 
    metallicity, over the whole range\footnote{The sawtooth features in 
    \refsubfig{fig:dustevoldust}{d}-\hyperref[fig:dustevoldust]{f}, for the two 
    lowest values of $\epsilon_\sms{grow}$ (red and orange), are numerical
    artefacts due to the fact that grain growth is so low, that \hSNII\ can
    clear dust faster than our time resolution.}, because grain growth never
    reaches the efficiency of stardust condensation, in these extreme cases.
    The timescale ratio between \hSNII\ production and \hISM\ grain growth is
    (using \refeqnp{eq:dustiness}, \refeqnp{eq:birth}, \refeqnp{eq:RSN}, 
    \refeqnp{eq:SNcond} and \refeqnp{eq:eps_grow}):
    \begin{equation}
      \frac{\tau_\sms{SN-cond}}{\tau_\sms{grow}} 
       = \frac{\epsilon_\sms{grow}\langle m_\star\rangle}%
              {f_\sms{SN}\langle Y_\sms{SN}\rangle}
         Z_\sms{dust}(Z-Z_\sms{dust}).
      \label{eq:taucondtaugrow}
    \end{equation}
    Since we usually have $Z\gtrsim2\times Z_\sms{dust}$, we can write the 
    rough proportionality: 
    $\tau_\sms{SN-cond}/\tau_\sms{grow}\propto Z\times Z_\sms{dust}$, with 
    $Z_\sms{dust}$ being a steep function of $Z$.
    This is the reason why the relative efficiency of the two processes is so
    strongly metallicity dependent.
  \item[The \snii\ destruction efficiency,] $m_\sms{gas}^\sms{dest}$ 
    \refeqp{eq:SNdest}, has the effect of decreasing the \hdustiness, at 
    low to normal metallicity (\refsubfig{fig:dustevoldust}{g-i}).
    It however has no noticeable effect at very low metallicity.
    This is because the average dust mass destroyed by a single \hSNII\ blast 
    wave is $m_\sms{dust}^\sms{dest}=Z_\sms{dust}\times m_\sms{gas}^\sms{dest}$ and 
    $Z_\sms{dust}\ll 1$.
\end{description}

    \subsubsection{Empirical Inference of Dust Evolution Timescales}
    \label{sec:dustime}

We now discuss some empirical estimates of the three main grain evolution parameters ($\langle Y_\sms{SN}\rangle$, $\epsilon_\sms{grow}$ and $m_\sms{gas}^\sms{dest}$), derived from fitting dust scaling relations.
This is not a new topic.
Several studies have attempted to tackle this issue
\citep[\eg][]{lisenfeld98,morgan03,galliano08a,mattsson12b,remy-ruyer14,de-vis17,nanni20,de-looze20}.
We have recently published such a study \citepalias{galliano21}.
We will try to demonstrate the progress it has brought to the field.

\begin{figure}[!htbp]
  \includegraphics[width=\textwidth]{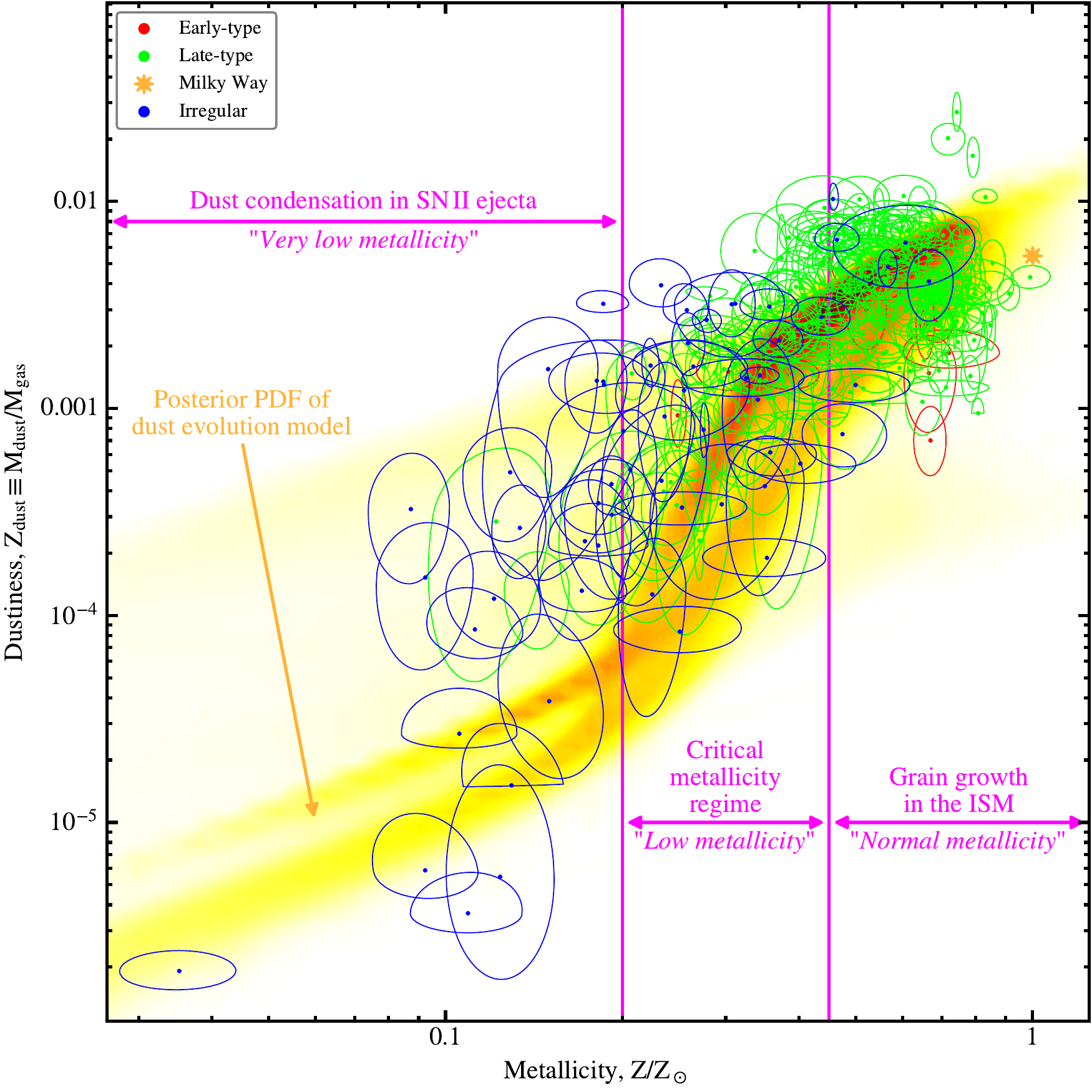}
  \newcap{Dustiness-metallicity relation fitted with a dust evolution model}%
         {The \hSUE s are the results of the \hSED\ fitting of the 
          \hDustPedia/\hDGS\ sample.
          Each \hSUE\ represents one galaxy, color-coded according to its type
          (\cf\ \reffig{fig:XETG}).
          We have overlaid the posterior probability distribution of the dust
          evolution model of \refeqs{eq:evolstar}{eq:evoldust} as a 
          yellow-orange density.
          This fit was used by \citetalias{galliano21} to infer the values of 
          the three dust evolution tuning parameters.
          This is only a sub-sample of our 800 galaxies, as not all of them
          had reliable metallicity measurements.
          \CClicence}
  \label{fig:dustevol_fit}
\end{figure}

\paragraph{Fitting dust scaling relations.}
We have used the \hSED\ modeling results of the \hDustPedia/\hDGS\ sample we have already discussed in \reffig{fig:compare_fit} and \refsec{sec:XETG} \citepalias{galliano21}.
These results were obtained using the \expression{composite approach} (\refsec{sec:dale}), with the \citetalias{jones17} grain properties, and a hierarchical Bayesian model (\refsec{sec:HerBIE}).
These results are an estimate of:
\begin{inlinelist}
  \item the dust mass, $M_\sms{dust}$, for all galaxies \citepalias{galliano21};
  \item the stellar mass, $M_\star$ and \hSFR\ for most of them 
    \citep{remy-ruyer15,nersesian19};
  \item the metallicity, $Z$, for about half the sample 
    \citep{madden13,de-vis19}.
\end{inlinelist}
The dust evolution model of \refsec{sec:dustevolmodel} has then been fitted to our estimates of $M_\star$, $M_\sms{gas}$, $M_\sms{dust}$, $Z$ and \hSFR.
We have adopted a hierarchical Bayesian approach (\cf\ \refsec{sec:HerBIE}), varying the following set of parameters.
\begin{description}
  \item[The SFH-related parameters,] $\tau_\sms{SFH}$ \refeqp{eq:SFHdel}, 
    $\psi_0$ \refeqp{eq:SFHdel}, $\delta_\sms{in}$ \refeqp{eq:Rin} and 
    $\delta_\sms{out}$ \refeqp{eq:Rout} are varied individually for each galaxy.
    In other words, we have assumed that each galaxy has a particular, 
    independent \hSFH.
  \item[The dust evolution tuning parameters,] 
    $\langle Y_\sms{SN}\rangle$ \refeqp{eq:SNcond}, $\epsilon_\sms{grow}$ 
    \refeqp{eq:eps_grow} and $m_\sms{gas}^\sms{dest}$ \refeqp{eq:SNdest}
    were varied, assuming they were common to every galaxy.
    In other words, we have assumed that the efficiencies of the dust evolution 
    processes were universal.
\end{description}
\reffig{fig:dustevol_fit} shows the fitted \hdustiness-metallicity relation, from our study \citepalias{galliano21}.
We have represented the estimated observed quantities (\hSUE s) on top of the posterior \hPDF\ of our dust evolution model.
This figure clearly shows the physical origin of the three regimes we had arbitrarily defined at the beginning of \refsec{sec:G2DvsZ}.
\begin{enumerate}
  \item At very low metallicity ($Z\lesssim0.2\times Z_\odot$), grain 
    production is dominated by condensation in \hSNII\ ejecta.
    We see a roughly-linear \hdustiness\ trend with metallicity, with a rather 
    low dust-to-metal mass ratio
    ($Z_\sms{dust}/Z\simeq10^{-4}$):
    \begin{equation}
      \frac{Z_\sms{dust}}{Z_\sms{dust}^\odot}\simeq10^{-4}\times\frac{Z}{Z_\odot}
      \;\;\;\;\;\;\mbox{for}\;\;\;\;\;\;Z\lesssim0.2\times Z_\odot.
      \label{eq:Zdvlz}
    \end{equation}
  \item At low metallicity 
    ($0.2\times Z_\odot\lesssim Z\lesssim0.45\times Z_\odot$), 
    we are in what we have called the critical metallicity regime.
    The \hdustiness\ rises sharply with metallicity, as grain growth in the 
    \hISM\ kicks in.
    The critical metallicity regime of individual galaxies is usually narrower 
    (\cf\ \reffigs{fig:dustevolSFH}{fig:dustevoldust}).
    It is broadened here, because the \hPDF\ is the superimposition of all the 
    galaxies, having different \hSFH s.
  \item At normal metallicity ($Z\gtrsim0.45\times Z_\odot$), we have another 
    roughly-linear \hdustiness\ trend with metallicity, sustained by grain 
    growth in the \hISM, with a Galactic dust-to-metal mass ratio 
    ($Z_\sms{dust}/Z\simeq0.5$):
    \begin{equation}
      \frac{Z_\sms{dust}}{Z_\sms{dust}^\odot}\simeq\frac{Z}{Z_\odot}.
      \;\;\;\;\;\;\mbox{for}\;\;\;\;\;\;Z\gtrsim0.45\times Z_\odot.
      \label{eq:Zdnz}      
    \end{equation}
\end{enumerate}

\begin{figure}[!htbp]
  \includegraphics[width=\textwidth]{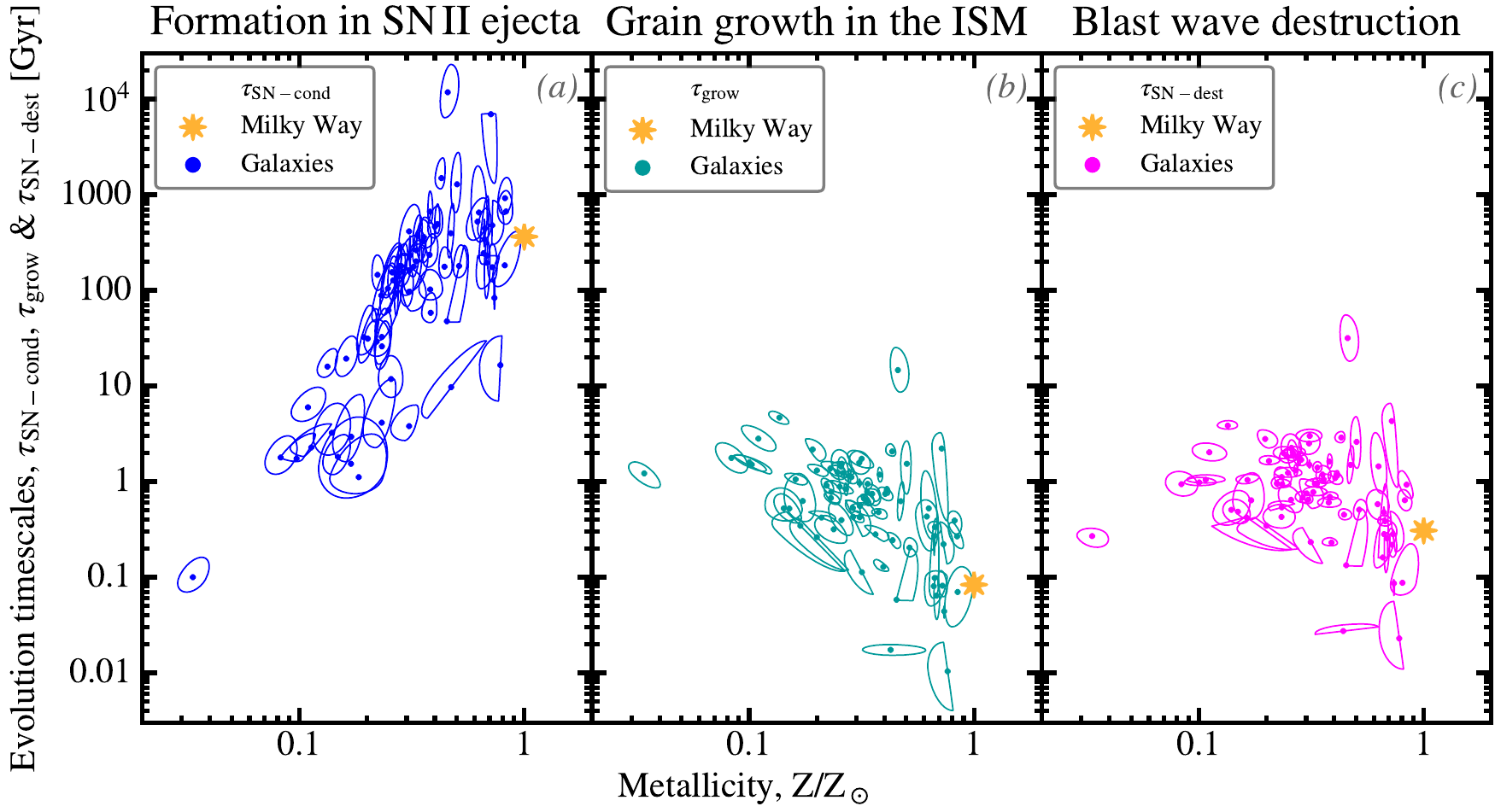}
  \newcap{Empirical estimates of dust evolution timescales as a function of  
          metallicity}%
         {The three panels represent the dust evolution timescales inferred
          from the fit of \reffig{fig:dustevol_fit} \citepalias{galliano21}:
          \begin{inlinelistalph}
            \item $\tau_\sms{SN-cond}$ \refeqp{eq:SNcond};
            \item $\tau_\sms{grow}$ \refeqp{eq:eps_grow}; and
            \item $\tau_\sms{SN-dest}$ \refeqp{eq:SNdest}. 
          \end{inlinelistalph}
          Each \hSUE\ corresponds to one galaxy.
          \CClicence}
  \label{fig:dustevol_timescales}
\end{figure}

\paragraph{Evolutionary timescales as a function of metallicity.}
The dust evolution fitting of \reffig{fig:dustevol_fit} allowed us to infer the values of the three tuning parameters.
Accounting for possible systematic biases, we concluded the following \citepalias{galliano21}:
\begin{equation}
  \langle Y_\sms{SN}\rangle\lesssim0.03\eMsun/\textnormal{SN},
  \;\;\;\;\;\;\;\;\;
  \epsilon_\sms{grow}\gtrsim3000,
  \;\;\;\;\;\;\;\;\;
  m_\sms{gas}^\sms{dest}\gtrsim1200\eMsun/\textnormal{SN}.
  \label{eq:tuning}
\end{equation}
These efficiencies can be translated into timescales of the three dust evolution processes, in each galaxy.
We have represented these timescales as a function of metallicity, in \reffig{fig:dustevol_timescales}.
The derived timescales for the \hMW\ are represented as a yellow star, although they were not used in the fit.
We note the following points
\begin{enumerate}
  \item The timescale for dust condensation in \hSNII\ ejecta rises very 
    abruptly with metallicity (\cf\ \refsubfig{fig:dustevol_timescales}{a}).
    It is realistic (\ie\ shorter than the age of the Universe) only for 
    very-low- and low-metallicity systems.
    At normal metallicity, another process needs to be invoked.
    \refeq{eq:taucondtaugrow} can be approximated by (assuming our limits in 
    \refeqnp{eq:tuning} are close to the true values):
    \begin{equation}
      \frac{\tau_\sms{SN-cond}}{\tau_\sms{grow}}\simeq1000\times
        \frac{Z}{Z_\odot}\times\frac{Z_\sms{dust}}{Z_\sms{dust}^\odot}.
    \end{equation}
    \begin{description}
      \item[At low very metallicity] ($Z\lesssim0.2\times Z_\odot$), 
        \refeq{eq:Zdvlz} gives 
        $\tau_\sms{SN-cond}/\tau_\sms{grow}\simeq0.1\times(Z/Z_\odot)^2\ll1$,
        which is another way to show that grain growth is inefficient in this
        regime.
      \item[At normal metallicity] ($Z\gtrsim0.45\times Z_\odot$), 
        \refeq{eq:Zdnz} gives 
        $\tau_\sms{SN-cond}/\tau_\sms{grow}\simeq1000\times(Z/Z_\odot)^2\gg1$, 
        which is another way to show that grain growth is now dominant.
    \end{description}
  \item The grain growth and blast wave destruction timescales 
    (\cf\ \refsubfig{fig:dustevol_timescales}{b-c}) have rather 
    similar features, because their ratio is (using \refeqnp{eq:birth}, 
    \refeqnp{eq:RSN}, \refeqnp{eq:SNdest} and \refeqnp{eq:eps_grow}):
    \begin{equation}
      \frac{\tau_\sms{SN-dest}}{\tau_\sms{grow}}
       = \frac{\epsilon_\sms{grow}\langle m_\star\rangle}%
              {f_\sms{SN}m_\sms{gas}^\sms{dest}}(Z-Z_\sms{dust})
       \simeq 4\times\frac{Z}{Z_\odot}.
    \end{equation}
    The two processes thus balance each other around 
    $Z_\sms{crit}\simeq Z_\odot/4$.
    This is where our value of the critical metallicity comes from.
    We note that, for the \hMW, we find $\tau_\sms{grow}^\sms{MW}\simeq80$~Myr 
    and $\tau_\sms{SN-dest}^\sms{MW}\simeq300$~Myr, close to the values we had 
    expected in \refsec{sec:graingrowth} and \refsec{sec:shockdest}.
    Yet, we did not put any prior on the Galactic values.
    This is a indication in favor of the consistency of our analysis.
  \item The ratio of the timescales for dust condensation in \hSNII\ ejecta and
    destruction by \hSNII\ blast waves is (using \refeqnp{eq:SNcond} and 
    \refeqnp{eq:SNdest}):
    \begin{equation}
      \frac{\tau_\sms{SN-cond}}{\tau_\sms{SN-dest}}
      = \frac{m_\sms{gas}^\sms{dest}}{\langle Y_\sms{SN}\rangle}Z_\sms{dust}
      \simeq500\times\frac{Z_\sms{dust}}{Z_\sms{dust}^\odot}.
    \end{equation}
    \begin{description}
      \item[At very low metallicity] ($Z\lesssim0.2\times Z_\odot$),
        \refeq{eq:Zdvlz} gives 
        $\tau_\sms{SN-cond}/\tau_\sms{SN-dest}\simeq0.5\times(Z/Z_\odot)\ll1$,
        showing that \hSNII\ are net dust producers.
      \item[At normal metallicity] ($Z\gtrsim0.45\times Z_\odot$), 
        \refeq{eq:Zdnz} gives 
        $\tau_\sms{SN-cond}/\tau_\sms{SN-dest}\simeq500\times(Z/Z_\odot)\gg1$,
        showing that \hSNII\ are net dust destroyers.
    \end{description}
\end{enumerate}
\takeaway{\hSNII\ are net dust destroyers, except at very low metallicity.}

\paragraph{Methodological remarks.}
The study this section relies on \citepalias{galliano21} was the first rigorous empirical determination of the dust evolution tuning parameters in \refeq{eq:tuning}, using a wide enough metallicity range to unambiguously constrain these quantities.
We emphasize two important points.
\begin{description}
  \item[Fitting dust evolution models is instrumental.]
    Numerous studies, trying to tackle the issues of cosmic dust 
    evolution, simply overlaid dust evolution tracks (such as those in 
    \reffigs{fig:dustevolSFH}{fig:dustevoldust}), on top of dust scaling
    relations (such as those in \refsubfig{fig:XETG}{a} and 
    \reffig{fig:dustevol_fit}).
    The issue with this approach is that two quantities of a given galaxy could 
    be fitted with two different \hSFH s, at different ages.
    This is obviously inconsistent, but can give the appearance a good 
    agreement with the data.
    We have avoided this pitfall with our hierarchical Bayesian approach.
    It allows us to avoid mutually inconsistent explanations of different trends
    and correlations.
    Overall, performing a rigorous fit does not help getting better solutions, 
    but it definitely helps avoiding bad ones.
  \item[Low-metallicity systems are crucial.]
    As we have shown in \reffig{fig:dustevoldust}, the effect of dust 
    condensation in \hSNII\ ejecta can only be probed at very low metallicity.
    It is therefore necessary to have a good enough sampling of this regime.
    Without a good coverage at very low metallicity, the solution will 
    consequently be degenerate.
    It will be impossible to disentangle the contributions of grain growth
    and stardust production.
    The relevance of dwarf galaxies here is not necessarily that they can be 
    considered as analogs of primordial distant galaxies, but that they sample 
    a particular, key dust production regime.
\end{description}

\paragraph{The controversy about stardust.}
We have opened this chapter with a citation of \citetalias{draine09}, about the belief that \hISD\ could be mainly stardust.
It is unfortunate that this discussion can sometimes turn into an ideological debate, in the literature.
\begin{description}
  \item[In the nearby Universe,] for instance, \citet{de-looze20} recently 
    tried to show that \hSNII\ could be net dust producers at normal 
    metallicity.
    They used a rather similar approach to ours.
    The only difference is that they did not have very-low-metallicity 
    constraints.
    Their results were therefore clearly degenerate, but they forced their 
    interpretation in favor of stardust.
    We will come back to these issues, from an epistemological point of 
    view, in \refchap{chap:method}.
  \item[In the distant Universe,] this debate is also vigorous, as dusty 
    galaxies are found at high redshifts 
    \citep[$z\gtrsim6$; \eg][]{dwek07,valiante09,dwek14,laporte17}.
    At this time, we are only $\simeq400$~Myr after the reionization.
    We thus need a rapid source of dust, and stardust is seriously considered.
    However, our results confirm that grain growth can happen on timescales
    shorter than $\lesssim100$~Myr, provided that the \hISM\ has been enriched
    by a first generation of stars, up to the critical metallicity.
    This therefore provides a simple solution to this conundrum.
  \item[In between,] measurements in absorption by \expression{Damped    
    Lyman-Alpha systems} (\hDLA), along the sightline towards a \hQSO\ or a
    \expression{Gamma-Ray Burst} (\hGRB), can be used to estimate the 
    metallicity and depletion of some elements in these systems, by comparing 
    volatile and refractory abundances \citep[\eg][]{de-cia16}.
    These data produce a quasi-linear \hdustiness-metallicity trend, much
    flatter than \reffig{fig:dustevol_fit} 
    \citepalias[\eg\ Fig.~10 of][]{galliano21}.
    If this trend is correct, it is consistent with stardust production at all
    metallicities.
    It is however difficult to understand the discrepancy between these systems
    and nearby galaxies.
    \citetalias{galliano21} conjectured it was possible that these estimates 
    could be biased due to the dilution of heavy element absorption lines in 
    near-pristine clouds, along the sightline, within the same velocity bin.
\end{description}
To try to rise above a mere ideological debate, we should not lose sight of the big picture, as the truth is the whole.
\reftab{tab:dustorigin} summarizes the observational evidence in favor of one scenario and the other.
\begin{table}[htbp]
  \centering
  \setlength\arrayrulewidth{2pt}
  \arrayrulecolor{white}
  \begin{tabularx}{\linewidth}{|>{\columncolor{coltabhead}}X
                                |>{\columncolor{coltabcell}}c
                                |>{\columncolor{coltabcell}}c|}
    \hline
      \rowcolor{coltabhead}
      \cellcolor{white} & \textbf{Stardust} & \textbf{ISM} \\
      \rowcolor{coltabhead}
      \cellcolor{white} & \textbf{origin} & \textbf{origin} \\
    \hline
      Elemental depletions 
      (\cf\ \refsec{sec:depletions} \&\ \refsec{sec:mantles})
      & & $\checkmark$ \\
    \hline
      Nearby galaxy \hdustiness-metallicity trend 
      (\cf\ \refsec{sec:dustime})
      & & $\checkmark$ \\  
    \hline
      Individual \hSNR s
      (before the reverse shock; \cf\ \refsec{sec:stardust})
      & $(\checkmark)$ & \\  
    \hline
      Individual \hSNR s
      (accounting for the reverse shock; \cf\ \refsec{sec:stardust})
      & & $\checkmark$ \\  
    \hline
      Isotopic ratios of \hIDP s in meteorites
      (\cf\ \refsec{sec:stardust})
      & & $\checkmark$ \\  
    \hline
      Distant dusty galaxies
      (\cf\ \refsec{sec:dustime})
      & $(\checkmark)$ & $(\checkmark)$ \\  
    \hline
      \hDLA\ \hdustiness-metallicity trend 
      (\cf\ \refsec{sec:dustime})
      & $(\checkmark)$ & \\  
    \hline
      \hISD\ is mainly amorphous, while \hCSD\ is crystalline 
      (\cf\ \refsec{sec:stardust})
      & & $(\checkmark)$ \\
    \hline
      Emissivity variation as a function of \hISM\ density 
      (\cf\ \refsec{sec:mantles})
      & & $(\checkmark)$ \\
    \hline
  \end{tabularx}
  \newcap{Summary of the observational evidence about interstellar dust origin 
          at normal metallicity}%
         {Check marks between parenthesis indicate uncertain evidence.}
  \label{tab:dustorigin}
\end{table}

\paragraph{Limitations of our approach.}
Although our approach was successful in providing a unique, rigorous estimate of the dust evolution tuning parameters, and in deriving timescales as a function metallicity, it has several limitations.
\begin{description}
  \item[Systematic uncertainties of the data]
    could bias the observed \hdustiness\ of our galaxy, displayed in 
    \reffig{fig:dustevol_fit}.
    We have estimated the different potential biases
    on both the dust and gas mass estimates \citepalias[Sect.~4.1.3
    of][]{galliano21}.
    At normal metallicity, our measurements are consistent with the \hMW.
    At very low metallicity, we could suffer from:
    \begin{inlinelist}
      \item the possible overestimate of the gas mass because of the extended 
        gas halo of dwarf galaxies (\refsec{sec:galdesc});
      \item the systematic variation of the grain opacity with metallicity
        (\refsec{sec:mantles});
      \item the potential overabundance of small grains at very low metallicity
        (\refsec{sec:sizedist});
        and
      \item the possible presence of unaccounted for \hVCD\ 
        (\refsec{sec:submmex}).
    \end{inlinelist}
    \citetalias{galliano21} concluded that these biases can not, in total, be 
    larger than a factor of $\simeq4$, which is consequent, but not sufficient 
    to produce a linear  \hdustiness-metallicity relation, that would be 
    consistent with \hSNII\ dust production at all epochs.
  \item[The universality of the tuning parameters]
    is a questionable assumption.
    As we have shown in \refsec{sec:dustevolISM}, all these parameters hide
    information about, in particular:
    \begin{inlinelist}
      \item the typical grain size distribution;
      \item sticking coefficients, which are dependent on grain structure and
        composition; 
      \item the topology of the \hISM;
      \item stellar evolution.
    \end{inlinelist}
    Exploring these variations with environment is however premature.
    We need independent estimates of the different factors we have just listed.
  \item[The simplicity of the dust evolution model] 
    can cause discrepancies when trying to account for the observed \hSFR s.
    In particular, our model failed at reproducing the trend of s$M_\sms{dust}$ 
    with \hsSFR\ \citepalias[\cf\ Sect.~5.2.3 of][]{galliano21}.
    The likely explanation is that our parametric \hSFH\ is too simple.
    This could be solved by adding another \hSF\ component, to account for a 
    potential recent burst.
\end{description}

  \subsection{Evolution of the Aromatic Feature Carriers}
  \label{sec:PAHvsZ}

We close this chapter with a discussion about the trend followed by the grains carrying the aromatic feature emission.
We have already discussed this point in \refsec{sec:PAH2VSG}, from a spectroscopic point of view.
We now give a more general point of view, based on \hSED\ modeling, and discuss the different scenarios.
We remind the reader that aromatic features can be emitted by \hPAH s\footnote{Alternative acronym: \expression{Poor-people's Amorphous Hydrocarbon}...} or small \hHAC s.
This is a debated modeling choice (\cf\ \refsec{sec:dustmodels}).
In the present section, we will assume that small \hHAC s are the carriers.
Their mass fraction is $q_\sms{AF}$ (\cf\ \refsec{sec:themis_sizedist}).
Small \hHAC\ and \hPAH s emit similar aromatic feature strengths if $q_\sms{PAH}\simeq0.45\times q_\sms{AF}$ \citepalias{galliano21}.

    \subsubsection{The Different Evolution Scenarios}
    \label{sec:PAHscenarios}

Aromatic features are significantly weaker in low-metallicity systems, compared to normal galaxies (\cf\ \refsec{sec:PAH2VSG}).
This fact could indicate an increasing formation efficiency as a function of $Z$.
However, low-metallicity environments have also their \hISM\ bathed with a hard, permeating \hISRF\ (\cf\ \refsec{sec:darkgas}).
Knowing that small \hHAC s are massively destroyed by such an \hISRF\ (\cf\ \refsec{sec:PAHevol}), this trend could result from the increased suppression of aromatic features at low metallicity.
Several scenarios have been proposed to explain these trends.

\paragraph{Enhanced destruction at low metallicity.}
\citet{madden06} proposed that small \hHAC s are more efficiently destroyed at low metallicity.
This was supported by the relation in \reffig{fig:PAH2ISRF}, between the strength of the aromatic features and the \neiiiline/\neiiline\ tracing the hardness of the \hISRF.
The fact that dwarf galaxies have in general harder, more intense \hISRF\ is linked to the following facts.
\begin{itemize}
  \item Low-metallicity stars, at a given initial mass, tend to be hotter, 
    because of line-blanketing effects, therefore emitting a harder \hISRF\
    \citep[\eg][]{martins02}.
  \item The opacity of the \hISM\ is lower at low metallicity, because of the 
    low \hdustiness, allowing photons to travel farther away from ionizing 
    stars \citep[\eg][]{madden20}.
  \item Most observational samples are biased, because they tend to select
    starburst or post-starburst dwarf galaxies, quiescent dwarf galaxies being
    challenging to observe at \hIR\ wavelengths.
    We will address this selection effect in \refsec{sec:discussPAH}.
\end{itemize}
Another mechanism, proposed by \citet{ohalloran06}, is the destruction of aromatic feature carriers by \hSNII\ blast waves.
However, this scenario is less satisfactory, as blast waves tend to destroy all dust species \citep[\eg][]{reach02}.
They therefore do not constitute a consistent explanation for the selective destruction of small \hHAC s.

\paragraph{Inhibited formation efficiency at low metallicity.}
Several scenarios based on metallicity-dependent production mechanisms, in stellar ejecta or in the \hISM, have been proposed.
\begin{description}
  \item[The delayed injection of C,] by \hLIMS\ in their post-\hAGB\ phase, was 
    proposed by \citet{dwek05}.
    \citet{galliano08a} conducted a quantitative comparison, using a dust 
    evolution model comparable to \refeqs{eq:evolstar}{eq:evoldust}.
    We showed that it provided a consistent account of the observed trend of 
    $q_\sms{AF}$ with metallicity.
    We also took into account the small \hHAC\ photodestruction in \hii\ 
    regions, in the \hSED\ modeling, and estimated it was not sufficient to 
    produce the trend.
    One major problem of this scenario is however that the volatility of 
    small \hHAC s, that we see spatially, requires a mechanism to reform them 
    in the \hISM.
  \item[Shattering of large C grains] leads to the formation of small 
    \hHAC s.
    \citet{seok14} have implemented this process in a cosmic dust evolution
    model and showed it could reproduce the $q_\sms{AF}$-$Z$ trend.
    The range of \hSFH s required to cover the whole $q_\sms{AF}-Z$ trend is
    however wider than that needed to reproduce the \hdustiness-metallicity
    trend of the same sample (\cf\ our discussion in \refsec{sec:dustime} about 
    the importance of fitting dust evolution models).
  \item[Formation in molecular clouds] is another interesting scenario, as the 
    molecular mass fraction is known to rise with metallicity 
    \citep[\eg][]{schruba12}.
    \citet{greenberg00} proposed that aromatic feature carriers could form on 
    grain surfaces in molecular clouds and be photoprocessed in the diffuse 
    \hISM.
    \citet{sandstrom10} and \citet{chastenet19} showed that the spatial 
    distribution of $q_\sms{AF}$ is consistent with this scenario in the 
    Magellanic clouds: $q_\sms{AF}$ is higher in molecular clouds.
\end{description}

    \subsubsection{The Observed Trends}
    \label{sec:discussPAH}

\begin{figure}[!htbp]
  \includegraphics[width=\textwidth]{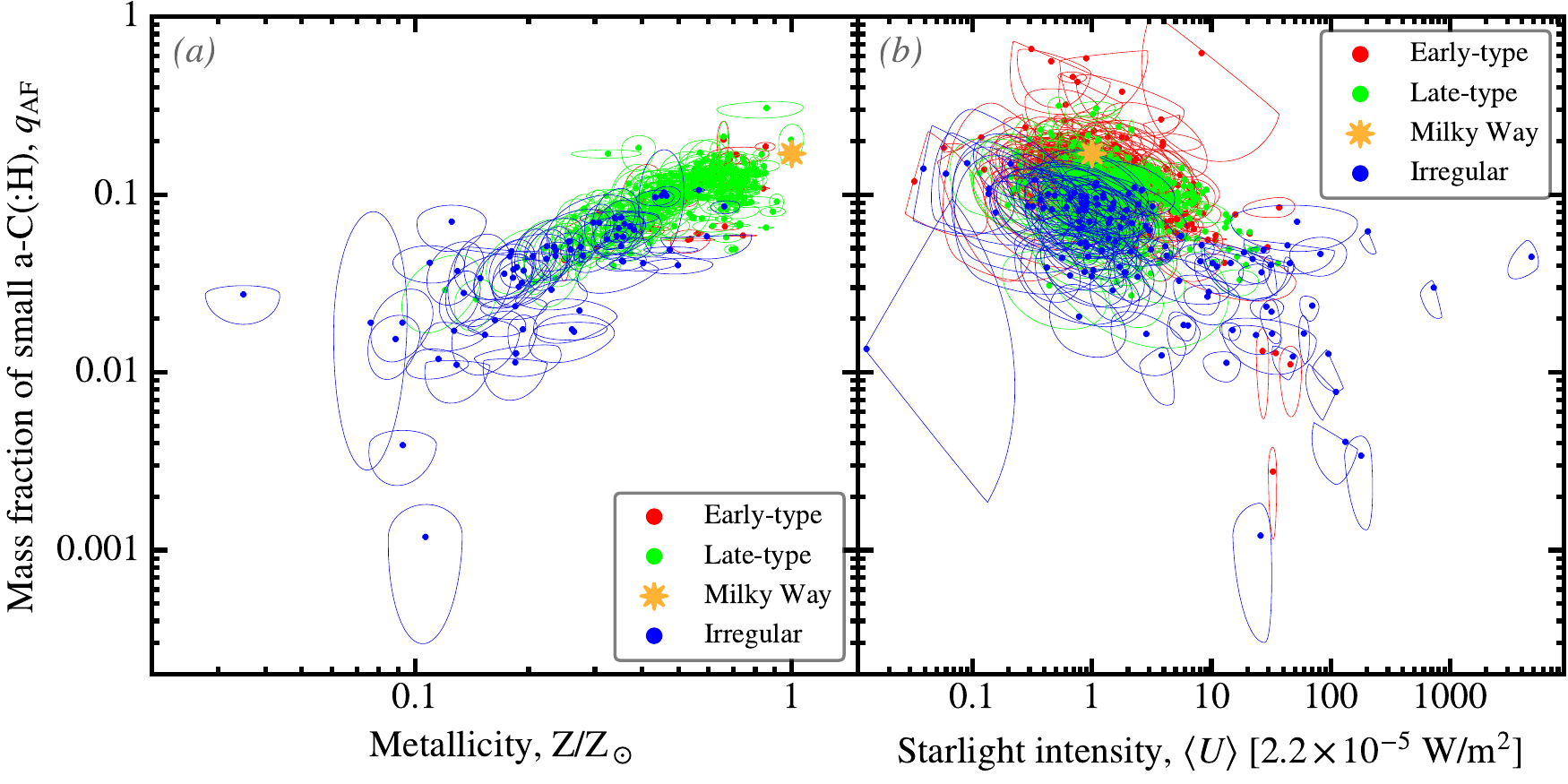}
  \newcap{Evolution of the mass fraction of small a-C(:H) grains 
          with metallicity and starlight intensity}%
         {In both panels, we show the mass fraction of 
          aromatic-feature-emitting grains, $q_\sms{AF}$ (\cf\ 
          \refsec{sec:themis_sizedist}) derived from the \hSED\ fit of 
          \citetalias{galliano21}.
          This quantity is displayed as a function of metallicity and 
          mean starlight intensity, $\langle U\rangle$ \refeqp{eq:Uav}.
          Each \hSUE\ corresponds to one galaxy, color-coded according to its
          type (\cf\ \reffig{fig:XETG}).
          \CClicence}
  \label{fig:dustevol_qAF}
\end{figure}

\citetalias{galliano21} have derived $q_\sms{AF}$ in each galaxy of the \hDustPedia/\hDGS\ sample, that we have already amply discussed earlier in this chapter.
\reffig{fig:dustevol_qAF} shows the evolution of this quantity with:
\begin{inlinelistalph}
  \item the metallicity, $Z$; and
  \item the mean starlight intensity, $\langle U\rangle$ \refeqp{eq:Uav}.
\end{inlinelistalph}

\paragraph{A better correlation with metallicity.}
\refsubfig{fig:dustevol_qAF}{a} shows a clear linear rising trend of $q_\sms{AF}$ with metallicity \citepalias[Eq.~9 of][]{galliano21}, and a decreasing trend of $q_\sms{AF}$, with $\langle U\rangle$ quantifying the intensity of the \hISRF.
Both correlations could be explained by any of the scenarios discussed in \refsec{sec:PAHscenarios}.
We have however found that the correlation with metallicity is significantly better \citepalias[\cf\ detailed discussion in Sect.~4.2.2 of][]{galliano21}.
This result is worth noting, especially since several studies focussing on a narrower metallicity range concluded the opposite \citep[\eg][]{gordon08,wu11}. It probably relies on the fact the metallicities we have adopted in this study \citep{de-vis19} correspond to well-sampled galaxy averages, while in the past a single metallicity, often central, was available and may have not been representative of the entire galaxy. 
This result suggests that photodestruction, although real at the scale of star-forming regions, might not be the dominant mechanism at galaxy-wide scales and that one needs to invoke one of the inhibited formation processes discussed in \refsec{sec:PAHscenarios}.
\takeaway{At global scales, the mass fraction of small \hHAC\ seems to correlate
          better with metallicity than with the \hISRF.}

\paragraph{The global point of view.}
Overall, the $q_\sms{AF}-Z$ trend might have several origins.
We think we can be confident about the following facts.
\begin{description}
  \item[Small a-C(:H) photodestruction is real] and it is enhanced at low 
    metallicity.
    It is however difficult to firmly establish if it is sufficient or not to
    explain the $q_\sms{AF}-Z$ trend.
  \item[The C/O ratio varies as a function of metallicity] 
    \citep[\eg][]{pagel03}.
    For instance, in the \hSMC, 
    $\textnormal{(C/O)}\simeq1/4\times\textnormal{(C/O)}_\odot$.
    It means that if the \hHAC-to-metal mass ratio is Galactic, the abundance
    of small \hHAC s will be at most 1/4 Galactic.
  \item[In terms of filling factors of a multiphase ISM,] we can 
    assume that small \hHAC s are absent of the \hHIM, \hWIM\ and \hii\ regions,
    and are present in the other phases.
    Yet, the \hISM\ of low-metallicity systems appears to be permeated with
    ionized gas, and their molecular cloud filling factor is lower.
\end{description}
What makes this question difficult to tackle is the diversity of spatial scales needed to properly balance the different processes.
Ideally, we would indeed need to account for the following.
\begin{itemize}
  \item A large fraction of the aromatic feature power comes from the   
    \hUV-illuminated edge of molecular clouds (sub-pc-scales).
    This is the region where small \hHAC\ will have the higher emissivity, and 
    it is at the edge of the region where they are massively destroyed.
  \item Most of the mass of small \hHAC\ is in the diffuse, weakly-illuminated 
    phases, filling an important volume of the galaxy (100 kpc-scales).
\end{itemize}
To know the origin of the $q_\sms{AF}-Z$ trend, we would need to reliably estimate the contribution to the integrated emission of both of these components, resolving sub-pc scales, in order to account for the enhanced emissivity in \hPDR s.
This is out of reach of current facilities.
\begin{figure}[htbp]
  \includegraphics[width=\textwidth]{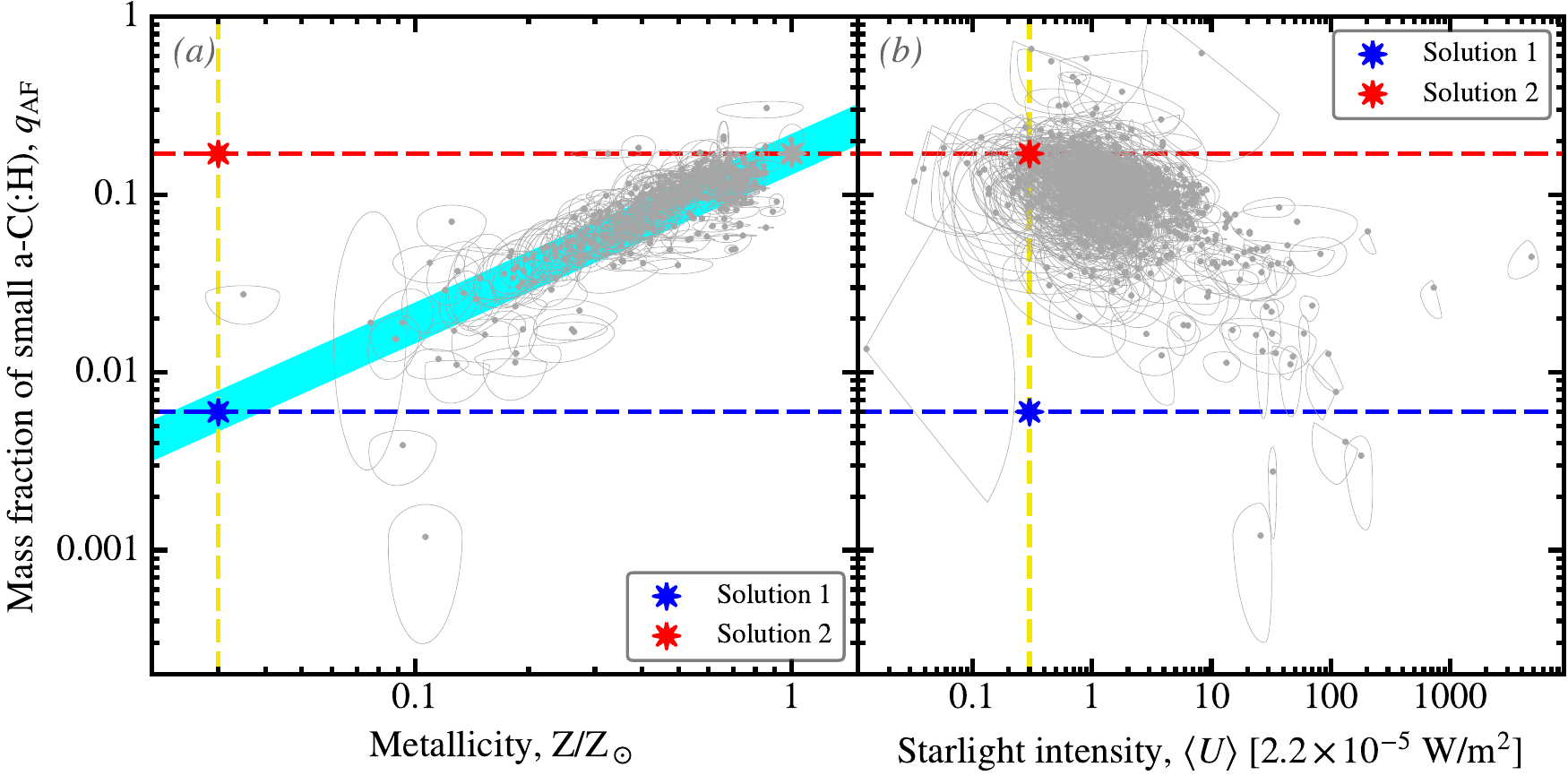}
  \newcap{The potential of quiescent very-low-metallicity galaxies to 
          understand the origin of small a-C(:H) grains}%
         {The data in both panels are identical to \reffig{fig:dustevol_qAF}.
          We have simply added the hypothetical observations of a quiescent
          very-low-metallicity galaxy (solutions 1 and 2).
          The cyan stripe in panel~\textit{(a)} is the analytical fit of
          the trend given in Eq.~(10) of \citetalias{galliano21}.
          \CClicence}
  \label{fig:qAF_LSB}
\end{figure}

\paragraph{The potential of quiescent very-low-metallicity galaxies.}
An alternative would be to observe quiescent very-low-metallicity systems. 
We know such a population of galaxies exist \citep[\eg][]{lara-lopez13}.
Let's assume that we can estimate $q_\sms{AF}$ for a galaxy with $Z\simeq0.03\eZsun$ and $\langle U\rangle\simeq0.3$.
We have represented the two possible solutions on \reffig{fig:qAF_LSB}.
We see that:
\begin{enumerate}
  \item if we find $q_\sms{AF}\simeq0.006$, it will be consistent with the 
    $Z$ trend (\refsubfig{fig:qAF_LSB}{a}), but not with the 
    $\langle U\rangle$ trend (\refsubfig{fig:qAF_LSB}{b}); or
  \item if we find $q_\sms{AF}\simeq0.17$, it will be consistent with the 
    $\langle U\rangle$ trend (\refsubfig{fig:qAF_LSB}{b}), but not with the 
    $Z$ trend (\refsubfig{fig:qAF_LSB}{a}).
\end{enumerate}
Such observations would require a sensitive \hMIR-to-\hFIR\ observatory, such as what \hSPICA\ \citep{van-der-tak18} could have been.


\newchapter{Methodological Effort and Epistemological Reflection}
\markboth{\chaptername\ \thechapter.\ Methodology}{}
\label{chap:method}
\citesmart{(...) the Bayesian method is easier to apply and yields the same or 
better results.
(...) the orthodox results are satisfactory only when they agree closely (or exactly) with the Bayesian results. No contrary example has yet been produced.
(...) We conclude that orthodox claims of superiority are totally unjustified;
today, the original statistical methods of Bayes and Laplace stand in a position of proven superiority in actual performance, that places them beyond the reach of mere ideological or philosophical attacks. It is continued teaching and use of orthodox methods that is in need of justification and defense.}{\citep[Edwin T.\ \familyname{Jaynes};][]{jaynes76}}
\minitoc

\noindent
We have stressed earlier that \hISD\ studies were essentially empirical, because of the complexity of their subject.
Most of the time, they consist in interpreting data with formulae and models.
Yet, comparing observations to models is a very wide methodological topic.
It also has \expression{epistemological}\footnote{Epistemology is the philosophical study of the nature, origin and limits of human knowledge. By extension, it is the philosophy of science.} consequences.
All the knowledge we derive about \hISD\ depends on the way it was inferred.
The question of the methods we use, and the way we articulate different results to build a comprehensive picture of the \hISM, is thus of utmost importance.

\section{Understanding the Opposition between 
          Bayesians and Frequentists}
\label{sec:BvF}

Historically, two competitive visions of the way empirical data should be quantitatively compared to models have emerged, the 
\begin{inlinelist}
  \item Bayesian, and
  \item frequentist methods.
\end{inlinelist}
We personally follow the Bayesian method and will try to give arguments in favor of its superiority.
An efficient way to present the Bayesian approach is to compare it to its alternative, and to show how both methods differ.
There is a large literature on the subject.
The book of \citet{gelman04} is a reference to learn Bayesian concepts and techniques.
The posthumous book of \citet{jaynes03} is more theoretical, but very enlightening.
Otherwise, several reviews have been sources of inspiration for what follows \citep{jaynes76,loredo90,lindley01,bayarri04,wagenmakers08,hogg10b,lyons13,vanderplas14}.
A good introduction to frequentist methods can be found in the books of \citet{barlow89} and \citet{bevington03}.

  \subsection{Two Conceptions of Probability and Uncertainty}

Bayesian and frequentist methods differ by:
\begin{inlinelist}
  \item the meaning they attribute to probabilities; and 
  \item the quantities they consider random.
\end{inlinelist}
Their radically different approaches and the various bifurcations the two methods take to address a given problem originate from these sole conceptions.

    \subsubsection{The Concept of Conditional Probability}

The concept of \expression{conditional probability} is central to what follows.
As we will see, Bayesian and frequentist approaches differ on this aspect.

\paragraph{The meaning of conditional probabilities.}
If $A$ and $B$ are two logical propositions, the conditional probability, noted \tpcond{A}{B}, is the probability of $A$ being true, knowing $B$ is true.
To give an astronomical example, let's assume that:
\begin{itemize}
  \item $A$ is the probability per unit time to observe a \hSNIa, when pointing 
    a telescope at a random star; and
  \item $B$ is the probability to observe a binary system, when pointing a 
    telescope at a random star.
\end{itemize}
Since most stars are \hLIMS, and that they have a lifetime $\langle\tau\rangle\simeq10$~Gyr, we can estimate the probability to observe a \hSNIa, knowing we are observing a binary system:
\begin{equation}
  \pcond{\textnormal{\snia}}{\textnormal{binary}}
    \simeq0.1\;\textnormal{Gyr}^{-1}.
\end{equation}
On the contrary the probability to observe a binary system, knowing we are observing a \hSNIa\ is:
\begin{equation}
  \pcond{\textnormal{binary}}{\textnormal{\snia}}=1,
\end{equation}
because \hSNIa\ happen only in binary systems.
We see that $\pcond{A}{B}\ne\pcond{B}{A}$. 
In our example, the two quantities do not even have the same units.

\paragraph{All probabilities are conditional.}
In practice, all probabilities are conditional.
In the previous example, we have implicitly assumed that our \expression{possibility space}, that is the ensemble of cases we can expect out of the experiment we are conducting, contained all the events where we are actually observing a star, when we are pointing our telescope at its coordinates.
However, what if there is suddenly a cloud in front of the telescope?
We would then need to account for these extra possibilities, which is equivalent to adding conditions.
For instance, if we are conducting the same experiment from a ground-based telescope in London, during winter, we will get:
\begin{equation}
  \pcond{\textnormal{\snia}}{\textnormal{binary}\wedge\textnormal{London}
                \wedge\textnormal{winter}}\ll0.1\;\textnormal{Gyr}^{-1},
\end{equation}
where the symbol $\wedge$ denotes the logical \citengl{and}.
\takeaway{All probabilities are conditional and the possible conditions are 
          limited by our own knowledge.}

\begin{figure}[htbp]
  \includegraphics[width=\textwidth]{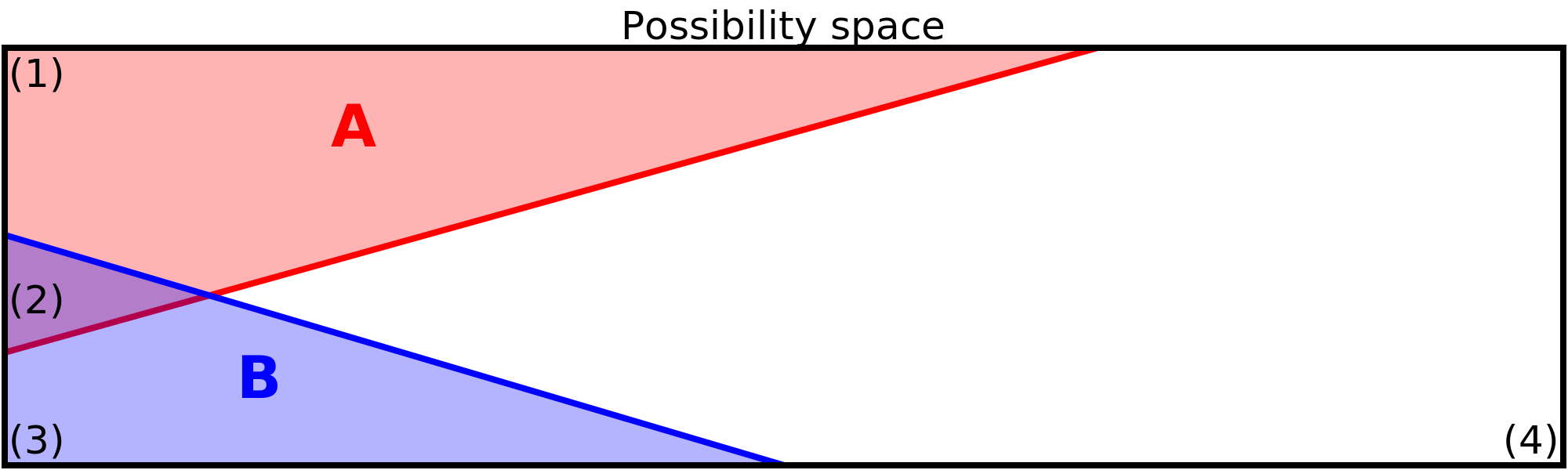}
  \newcap{Venn diagram to demonstrate Bayes' rule}{\CClicence}
  \label{fig:venn}
\end{figure}

\paragraph{Bayes' rule.}
Thomas \textsc{Bayes} derived, in the middle of the XVIII$^\textnormal{th}$ century, a formula to reverse the event and the condition, in conditional probabilities.
\reffig{fig:venn} shows a \expression{Venn diagram}, that is a graphic representation of a possibility space.
We have shown an arbitrary event A, in red, and another event B, in blue.
The intersection of A and B, numbered (2), can be written $A\wedge B$.
The probability of A, knowing B, is the probability of A, when B is considered as the new possibility space.
In other words:
\begin{equation}
  \pcond{A}{B}=\frac{(2)}{(2)+(3)}=\frac{\proba{A\wedge B}}{\proba{B}}.
\end{equation}
We therefore have: $\proba{A\wedge B}=\pcond{A}{B}\proba{B}$.
By symmetry of A and B, we have $\proba{A\wedge B}=\proba{B\wedge A}$, thus $\pcond{A}{B}\proba{B}=\pcond{B}{A}\proba{A}$, which gives \expression{Bayes' rule}:
\begin{equation}
  \pcond{A}{B}=\frac{\pcond{B}{A}\proba{A}}{\proba{B}}.
  \label{eq:Bayes}
\end{equation}

    \subsubsection{The Bayesian and Frequentist Assumptions}
    \label{sec:BvFassumptions}

We review here the assumptions of both methods, in a general, abstract way.
We will illustrate this presentation with concrete examples in \refsec{sec:BvFexample}.
Let's consider we are trying to estimate a list of $n$ physical parameters, $\vecprob{x}=(x_i)_{i=1,\ldots,n}$, using a set of $m$ observations, $\vecprob{d}=(d_j)_{j=1,\ldots,m}$.
Let's also assume that we have a model, $f$, predicting the values of the observables, $\vecprob{d}_\sms{mod}$, for any value of $\vecprob{x}$: $\vecprob{d}_\sms{mod}=f\left(\vecprob{x}\right)$.

\paragraph{The Bayesian viewpoint.}
The Bayesian approach considers that there is an objective truth, but that our \expression{knowledge} is only partial and subjective.
Bayesians thus assume the following.
\begin{description}
  \item[Probabilities] quantify the plausibility of a proposition, when we only 
    have incomplete knowledge.
    With this assumption, probabilities can be assigned to parameters and 
    hypotheses.
    More precisely, the true value of a parameter is considered fixed and an 
    hypothesis is either true or false, but our knowledge of the value of this
    parameter, or of the plausibility of this hypothesis, can be described by
    random variables.
    Probabilities are therefore used to quantify our partial, subjective 
    knowledge.
  \item[The method] then consists in sampling the probability distribution of 
    the parameters, conditional on the data, using Bayes' rule:
    \begin{equation}
      \underbrace{\pcond{\vecprob{x}}{\vecprob{d}}}_\sms{posterior}
      \propto
      \underbrace{\pcond{\vecprob{d}}{\vecprob{x}}}_\sms{likelihood}
      \times
      \underbrace{\proba{\vecprob{x}}}_\sms{prior}.
      \label{eq:posterior}
    \end{equation}
    Compared to \refeq{eq:Bayes}, \refeq{eq:posterior} misses the denominator, 
    \tproba{\vecprob{d}}.
    This is because this distribution is independent of our variables, that are 
    the physical parameters.
    If we were to explicit it, it would be:
    \begin{equation}
      \proba{\vecprob{d}}
      =\int\pcond{\vecprob{d}}{\vecprob{x}}\proba{\vecprob{x}}
        \ddiff^n\vecprob{x},
    \end{equation}
    which is simply the normalization factor of 
    \tpcond{\vecprob{x}}{\vecprob{d}}.
    This factor can thus be estimated by numerically normalizing the posterior, 
    hence the proportionality we have used in \refeq{eq:posterior}.
    The three remaining terms are the following.
    \begin{description}
      \item[The posterior distribution,] $\pcond{\vecprob{x}}{\vecprob{d}}$, 
        is what we are interested in.
        It is literally the \hPDF\ of the physical parameters (what we want),
        knowing the data (what we have).
      \item[The likelihood,] $\pcond{\vecprob{d}}{\vecprob{x}}$, is the 
        probability of the data, for a fixed value of the parameters.
        It can be computed using our model, $f\left(\vecprob{x}\right)$.
        For instance, assuming that our observations are affected by 
        uncorrelated, Gaussian noise, with standard deviations 
        $\vecprob{\sigma}=(\sigma_j)_{j=1,\ldots,m}$, we can write:
        \begin{equation}
          \pcond{\vecprob{d}}{\vecprob{x}}
          = \frac{1}{\left(2\pi\right)^{m/2}\prod_{j=1}^m\sigma_j}
            \exp\left(-\sum_{j=1}^m
            \frac{\left(f_j(\vecprob{x})-d_j\right)^2}{2\sigma_j^2}\right).
          \label{eq:LH}
        \end{equation}
      \item[The prior distribution,] \tproba{\vecprob{x}}, is a unique feature 
        of the Bayesian approach.
        It literally quantifies all our prior knowledge about the values of 
        $\vecprob{x}$.
        We will give concrete examples of what that could mean in the following 
        sections.
    \end{description}
  \item[The result] of the Bayesian approach is the posterior, as it contains
    all the information we want on the parameters, informed by the observations 
    and our prior knowledge.
    We can then decide to synthesize this information by, for instance:
    \begin{itemize}
      \item quoting moments of the posterior: means, 
        $\left\langle x_i\middle|\vecprob{d}\right\rangle$, 
        standard deviation,
        $\sigma\left(x_i\middle|\vecprob{d}\right)$, \etc;
      \item quoting their correlation coefficient:
        $\rho\left(x_i,x_j\middle|\vecprob{d}\right)$, \etc;
      \item testing hypotheses: \tPcond{x_i>\textnormal{const}}{\vec{d}}, 
        \etc;
      \item and so on.
    \end{itemize}
\end{description}

\paragraph{The frequentist viewpoint.}
The frequentist approach also considers that there is an objective truth, but it differs with the Bayesian viewpoint by rejecting its subjectivity.
Frequentists thus assume the following.
\begin{description}
  \item[Probabilities] are the limit to infinity of the 
    \expression{occurrence frequency} of an event, in a sequence of repeated 
    experiments, under identical conditions\footnote{It is often considered as 
    the scientific definition of probabilities, while we will show later that 
    the Bayesian definition has more practical applications.}.
    This repeated event can be, for instance, the measure of a quantity.
    The rejection of the use of probabilities as a quantification of knowledge
    forbids frequentists to consider parameters or hypotheses as random 
    variables.
    In other words, physical quantities have a single, true value, and 
    hypotheses are either true or false.
    Only the data, which are tainted with uncertainties, can be considered as 
    random variables.
  \item[The method] then consists in using the likelihood, 
    \tpcond{\vecprob{d}}{\vecprob{x}}, to perform several tests, the most
    well-known being the \expression{maximum likelihood}.
    We can see here that frequentists consider the probability distribution of 
    the data, given the physical parameters.
    The data are thus considered as \expression{variables}, and the physical
    parameters, fixed.
  \item[The results] consist in describing what values of the physical 
    parameters we would find if we were to repeat the experiment in the same 
    conditions.
    Frequentists then compute the uncertainties on their estimate of the 
    parameters by simulating new data that could have been obtained in the same 
    conditions.
    We thus end-up with a distribution of parameter values, that is not a 
    probability distribution.
    For instance, hypothesis testing can not be performed the Bayesian way, 
    because we dot not have a conditional probability of the parameter.
    We will see in \refsec{sec:phacking} that we need to resort to the infamous 
    significance tests, instead.
\end{description}
\takeaway{Bayesians do not tamper with the data, 
          whereas frequentists account for hypothetical data that have not 
          actually been obtained.}

  \subsection{Comparison of the Two Approaches on Simple 
                  Cases}
  \label{sec:BvFexample}

We now compare the two approaches on a series of simple examples, in order to demonstrate in which situations the two approaches may differ.

    \subsubsection{Simple Case: When the Two Approaches Agree}
    \label{sec:BvFsimple}

\begin{figure}[htbp]
  \begin{tabular}{cc}
    \includegraphics[width=0.48\textwidth]{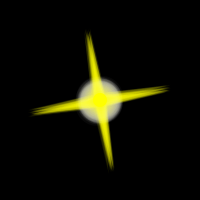} &
    \includegraphics[width=0.48\textwidth]{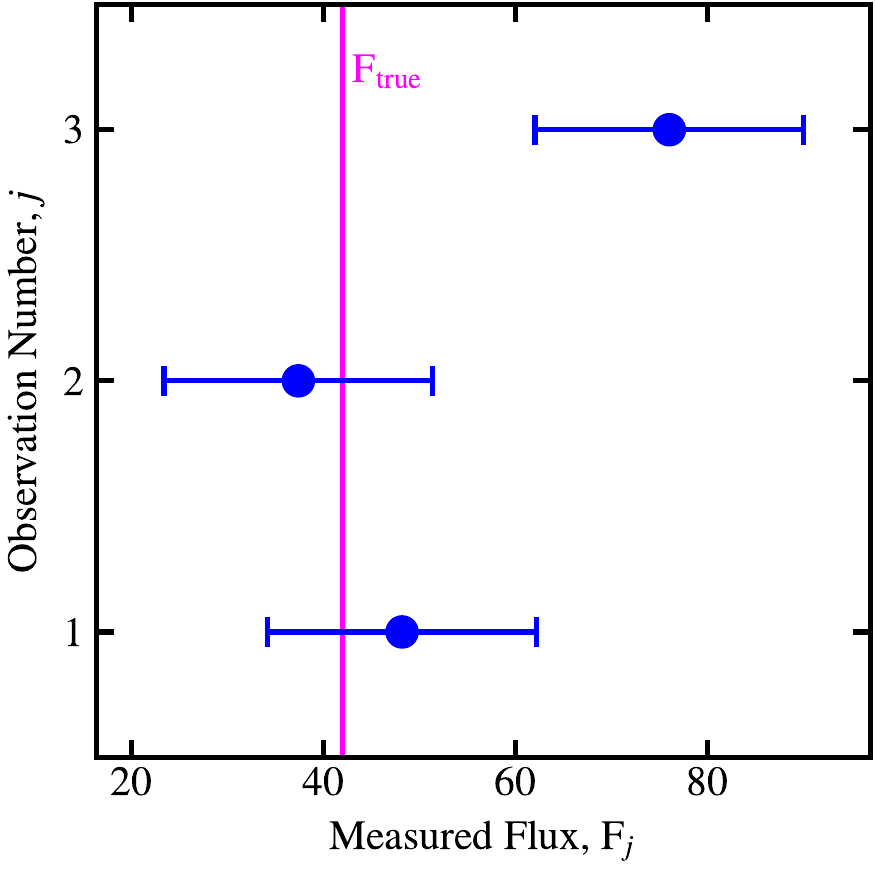} \\
  \end{tabular}
  \newcap{Simulation of the measure of a stellar flux to compare Bayesian and 
          frequentist methods}%
         {The blue dots with error bars represent the successive measures and 
          their uncertainties.
          The magenta line shows the true value.
          The units are arbitrary.
          \CClicence}
  \label{fig:BvFsimple}
\end{figure}

Let's assume we are trying to estimate the flux of a star, $F_\star$, in a given photometric filter, with the following assumptions.
\begin{itemize}
  \item The true flux of the star is $F_\sms{true}=42$ (in arbitrary units).
  \item We have performed $m=3$ repeated measures, $d_j=F_j$ $(j=1,\ldots,m)$.
  \item Each flux has Gaussian \expression{independent, identically 
    distributed} (\hiid) uncertainties: $\sigma_j=\sigma_\sms{F}=14$ 
    $(\forall j=1,\ldots,m)$.
  \item The physical quantity we want to estimate is simply the value of the 
    flux, $F_\star$.
    This corresponds thus to the simplest case, where we have only $n=1$ 
    parameter and the model is the identity: $F_\sms{mod}=f(x)=x$.
\end{itemize}
This is represented in \reffig{fig:BvFsimple}.
There is an analytic solution to this simple case:
\begin{equation}
  F_\star\simeq
    \frac{\sum_{j=1}^m F_j}{m}\pm\frac{\sigma_\sms{F}}{\sqrt{m}}=53.9\pm8.1.
  \label{eq:BvFsimplesol}
\end{equation}

\begin{figure}[htbp]
  \begin{tabular}{cc}
    \includegraphics[width=0.48\textwidth]{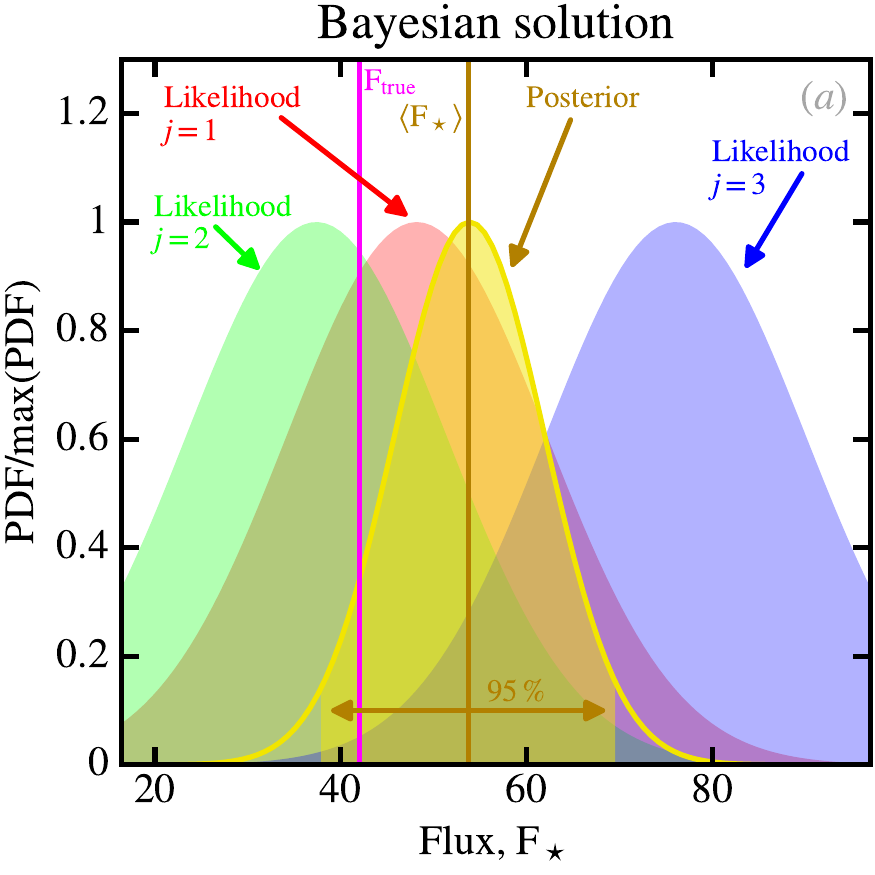} &
    \includegraphics[width=0.48\textwidth]{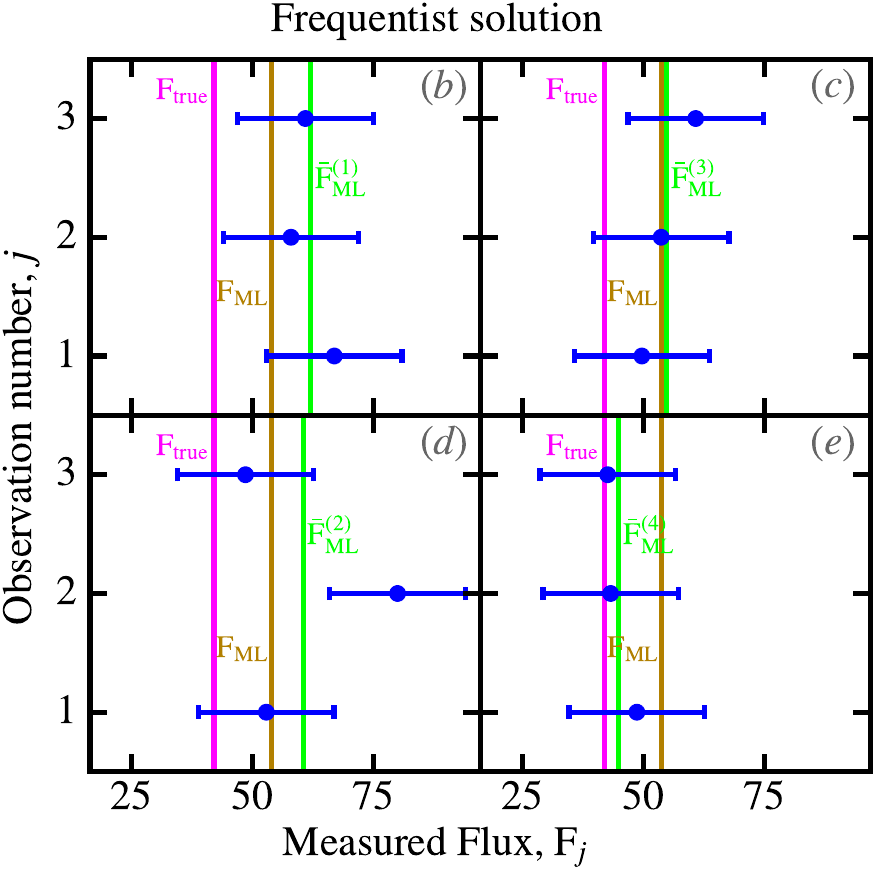} \\
  \end{tabular}
  \newcap{Bayesian and frequentist solutions to the problem of 
          \reffig{fig:BvFsimple}}%
         {Panel~\textit{(a)} shows the likelihood of the individual measures,
          in red, green and blue.
          The posterior, which is the product of the three likelihoods, is shown
          in yellow.
          We have filled its area corresponding to the $95\,\%$ credible range.
          The posterior average, $\langle F_\star\rangle$ is shown in dark 
          yellow.
          Panels~\textit{(b-e)} shows random reproduction of the measures.
          The blue dots with error bars are randomly drawn around the maximum 
          likelihood, $F_\sms{ML}$, using a bootstrapping method.
          The estimated value of $F_\star$ resulting from these draws, 
          $\bar{F}_\sms{ML}^{(k)}$, is shown in green.
          In all panels, the true value, $F_\sms{true}$, is shown in magenta.
          \CClicence}
  \label{fig:BvFsimplecomp}
\end{figure}

\paragraph{The Bayesian solution.}
The Bayesian solution is obtained by sampling the posterior distribution in \refeq{eq:posterior}: \tpcond{F_\star}{F_1,F_2,F_3}.
\begin{description}
  \item[The likelihood term] is exactly \refeq{eq:LH}, in our case, as we have 
    Gaussian \hiid\ uncertainties.
  \item[The choice of the prior] is the arbitrary part of the Bayesian model.
    For this first example, let's assume that we have no idea what the flux 
    should be.
    We thus take a flat, \expression{uninformative prior}, 
    $\proba{F_\star}\propto1$.
    This is a subjective choice that has the following consequences.
    \begin{enumerate}
      \item This particular choice is called an \expression{improper prior},
        meaning it can not be normalized, as: 
        $\int_{-\infty}^\infty\proba{F_\star}\ddiff F_\star=\infty$.
        In practice, we thus need to choose lower and upper bounds, 
        $[F_-,F_+]$, beyond which the prior is 0.
        Since we are measuring a positive quantity\footnote{Our variable is the 
        true flux. It is positive. Measured fluxes can occasionally be 
        negative because of noise fluctuations.}, we can take $F_-=0$ as the 
        lower bound.
        The upper bound could be taken as several times the flux of a 
        120~\tMsun\ star at the distance of the source.
        We thus have:
        \begin{equation}
          \proba{F_\star} = \left\{
          \begin{array}{ll}
            \displaystyle\frac{1}{F_+-F_-} & \mbox{for } F_-\le F_\star\le F_+ \\
            0 & \mbox{otherwise.}
          \end{array}\right.
        \end{equation}
      \item This prior is also subjective, as it depends on the choice of the
        variable.
        If we decide to study $\ln F_\star$, instead of $F_\star$, it  
        will lead to a different result.
        We will address this issue in \refsec{sec:Bayesfactor}.
        We can summarize it the following way.
        \begin{itemize}
          \item If the choice of the prior matters in the 
            final solution, it means that the weight of evidence brought by the 
            data is weak.
            It is therefore natural that the way we decide to quantify our prior
            knowledge is important.
            This is an aspect of the Bayesian method we need to embrace.
          \item On the opposite, if the weight of evidence brought by the data 
            is large, the choice of the prior will not matter significantly.
            In other words, if the width of the likelihood is much narrower 
            than a dex, the difference between multiplying by \tproba{F_\star}
            or \tproba{\ln F_\star} will be negligible.
        \end{itemize}
    \end{enumerate}
\end{description}
The solution is represented in \refsubfig{fig:BvFsimplecomp}{a}.
We have sampled the posterior using a \expression{Markov Chain Monte-Carlo} method \citep[\hMCMC; using the code \ncode{emcee} by][]{foreman-mackey13}.
We will come back to \hMCMC\ methods in \refsec{sec:mcmc}.
The estimated value is $F_\star\simeq53.9\pm8.1$ ($1\sigma$ uncertainty).
This is exactly the analytical solution in \refeq{eq:BvFsimplesol}.
The $95\,\%$ \expression{credible range}, that is the range centered on $\langle F_\star\rangle$ containing a $95\,\%$ probability, is $\CR{F_\star}=[37.8,69.8]$.

\paragraph{The frequentist solution.}
There are different ways to approach this problem in the frequentist tradition.
The most common solution would be to use a \expression{Maximum-Likelihood Estimation} (\hMLE) method.
\begin{description}
  \item[Maximum-likelihood value.]
    There are numerical tools to compute the \hMLE\ of complex models.
    We have used a \expression{Levenberg-Marquardt algorithm} 
    \citep[\eg][]{markwardt09}.
    The \hMLE\ value of $F_\star$ we derive is $F_\sms{ML}=53.9$, which is in 
    agreement with the analytical solution in \refeq{eq:BvFsimplesol}.
    It however does not provide uncertainties.
  \item[Uncertainty estimates.]
    A common solution to estimate the uncertainties on $F_\sms{ML}$ would be to
    perform \expression{bootstrapping}.
    Following the frequentist conception of probabilities, we randomly draw 
    new observations around the \hMLE\ value, a large number of times.
    We thus obtain a set of synthetic repeated measures, assuming the population
    distribution has $F_\sms{ML}$ for mean.
    For each new set, we derive a new \hMLE\ value, $\bar{F}_\sms{ML}^{(k)}$ 
    $(k=1,\ldots,3000)$.
    The first four draws are represented in \refsubfig{fig:BvFsimplecomp}{b-e}.
    The standard deviation of this sample is $8.1$, in agreement with 
    \refeq{eq:BvFsimplesol}.
    We can also compute the \expression{$95\,\%$ confidence interval}, from this
    sample: $\CI{F_\star}=[37.8,69.8]$.
\end{description}

\paragraph{A few remarks.}
We can see that both methods give the same exact result, which is also consistent with the analytical solution \refeqp{eq:BvFsimplesol}.
This is because the assumptions were simple enough to make the two approaches equivalent:
\begin{enumerate}
  \item by assuming a flat prior, we removed the effect of this Bayesian 
    peculiarity;
  \item the symmetry and the \hiid\ nature of the noise made the sampling
    of the likelihood as a function of the parameters, or as a function of the 
    data, identical.
\end{enumerate}
Note also our subtle choice of terminology: 
\begin{inlinelist}
  \item we talk about \expression{credible range} in the Bayesian case, as 
    this term designates the quantification of our beliefs; while
  \item we talk about \expression{confidence interval} in the frequentist 
    case, as it concerns our degree of confidence in the results, if the 
    experiment was repeated a large number of times, assuming the population
    mean is the maximum likelihood.
\end{inlinelist}

    \subsubsection{ Benefits of Using an Informative Prior}
    \label{sec:BvFprior}

A first way to find differences between the Bayesian and frequentist approaches is to explore the effect of the prior.
To that purpose, let's keep the same experiment as in \refsec{sec:BvFsimple}, but let's assume now that the star we are observing belongs to a cluster, and we know its distance.

\begin{figure}[htbp]
  \begin{tabular}{cc}
    \includegraphics[width=0.48\textwidth]{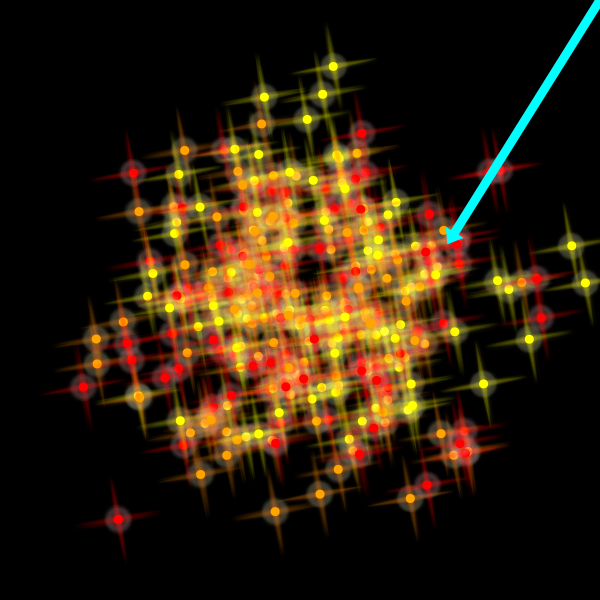} &
    \includegraphics[width=0.48\textwidth]{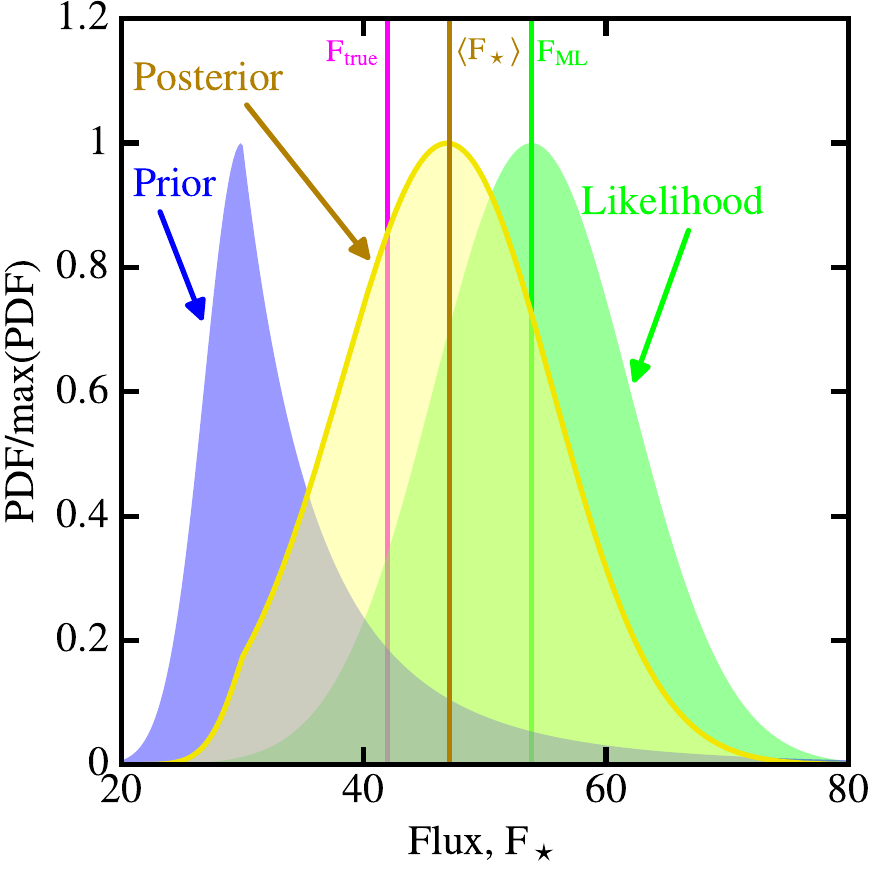} \\
  \end{tabular}
  \newcap{The benefits of using an informative prior}%
         {The green curve shows the total likelihood. 
          It is identical to the yellow curve in \reffig{fig:BvFsimplecomp}.
          The prior, which is taken as the luminosity function of the cluster,
          is represented in blue.
          The posterior, in yellow, is the product of these two distributions.
          \CClicence}
  \label{fig:BvF_prior}
\end{figure}

\paragraph{The Bayesian improvement.}
Contrary to \refsec{sec:BvFsimple}, where we had to guess a very broad, flat prior, we can now refine this knowledge, based on the expected luminosity function, scaled at the known distance of the cluster.
This is represented on \reffig{fig:BvF_prior}.
The posterior distribution (yellow) is now the product of the likelihood (green) and prior (blue).
The frequentist solution has not changed, as it can not account for this kind of information.
We can see that the maximum \textit{a posteriori} is now closer to the true value than the maximum likelihood.
This can be understood the following way.
\begin{description}
  \item[The true flux] has been drawn from the luminosity function, because we 
    have randomly targetted a star in this cluster.
    This is the definition of the luminosity function, which we happen to have
    chosen as the prior.
    This is often noted $F_\sms{true}\sim\proba{F_\star}$, the $\sim$ symbol 
    meaning \citengl{distributed as}.
  \item[The observed flux] has then been drawn from a normal law centered on  
    $F_\sms{true}$ with variance $\sigma_\sms{F}^2$, that can be written:
    $F_{j}|F_\sms{true}\sim\mathcal{N}(F_\sms{true},\sigma_\sms{F}^2)$.
    This is equivalent to saying that the observed flux has been drawn from 
    the posterior $\proba{F_\star}
    \times\mathcal{N}(F_\sms{true},\sigma_\sms{F}^2)$.
    Sampling this distribution is thus the best choice we can make, considering
    the information we have.
    This is why we get an advantage over the frequentist result.
\end{description}
If we perform several such measures, there will be some Bayesian solutions that will get corrected farther away from the true flux.
This is a consequence of stochasticity.
On \reffig{fig:BvF_prior}, keeping our value of $F_\sms{true}$, this will be the case if the noise fluctuation, $-\delta F$ is negative, that is if the observed flux is lower than $F_\sms{true}$ ($F_j=F_\sms{true}-\delta F<F_\sms{true}$).
However, this deviation will be less important than the correction we would benefit from if the same fluctuation was positive, because the prior would be higher: $\proba{F_\sms{true}-\delta F}>\proba{F_\sms{true}+\delta F}$ (\cf\ \reffig{fig:BvF_prior}).
The prior would therefore correct less the likelihood on the left side, in this particular case.
Thus, on average, taking into account this informative prior will improve the results.

\paragraph{Accumulation of data.}
In this case and the previous one (\refsec{sec:BvFsimple}), we had three observations of the same flux.
The posterior distribution we sampled was:
\begin{equation}
  \pcond{F_\star}{F_1,F_2,F_3}\propto\proba{F_\star}\times\pcond{F_1}{F_\star}
    \times\pcond{F_2}{F_\star}\times\pcond{F_3}{F_\star}.
  \label{eq:prior1}
\end{equation}
This is because we considered the three measures as part of the same experiment.
However, we could have chosen to analyze the data as they were coming.
After the first flux, we would have inferred:
\begin{equation}
  \pcond{F_\star}{F_1}\propto\proba{F_\star}\times\pcond{F_1}{F_\star}.
\end{equation}
This posterior would have been wider (\ie\ more uncertain), as we would have had only one data point.
Note that such an inference would have not been possible with a frequentist method, as we would have had one parameter for one constraint (\ie\ zero degree of freedom).
What is interesting to note is that the analysis of the second measure, can be seen as taking into account the first measure in the prior:
\begin{equation}
  \pcond{F_\star}{F_1,F_2}\propto
    \proba{F_\star}\pcond{F_1}{F_\star}\times\pcond{F_2}{F_\star}
    = 
    \underbrace{\pcond{F_\star}{F_1}}_\sms{new prior}
    \times\pcond{F_2}{F_\star},
\end{equation}
and so on.
For the third measure, the new prior would be \tpcond{F_\star}{F_1,F_2}:
\begin{equation}
  \pcond{F_\star}{F_1,F_2,F_3}
    \propto\underbrace{\pcond{F_\star}{F_1,F_2}}_\sms{new prior}
    \times\pcond{F_3}{F_\star}.
  \label{eq:prior2}
\end{equation}
which is formally equivalent to \refeq{eq:prior1}, but is a different way of looking at the prior.
Notice that the original prior, \tproba{F_\star}, appears only once in the product.
The more we accumulate data, the less important it becomes.
\takeaway{The Bayesian approach is an optimal framework to account for the accumulation of knowledge.}

    \subsubsection{Case Where the Two Approaches Differ: Non 
                        Gaussianity and Few Data}
    \label{sec:BvFasym}

The other reason why the two approaches might differ is because Bayesians sample \tpcond{\vec{x}}{\vec{d}}, whereas frequentists produce a series of tests based on \tpcond{\vec{d}}{\vec{x}}.
The difference becomes evident when we consider non-Gaussian errors with small data sets.

\begin{figure}[htbp]
  \includegraphics[width=\textwidth]{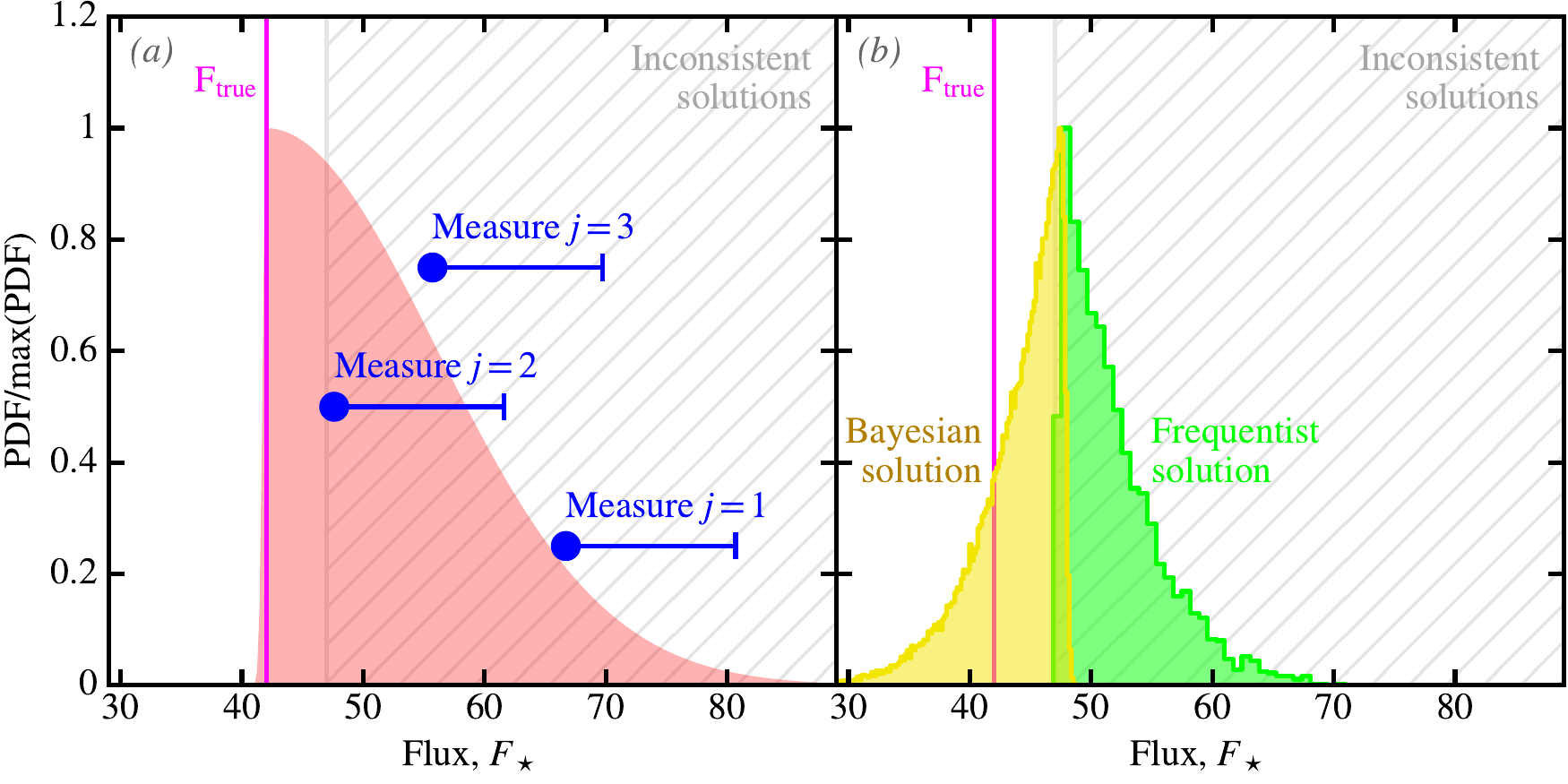}
  \newcap{Flux measures with a non-linear detector}%
         {Panel~\textit{(a)} shows the noise distribution, in red, and the 
          three observations in blue.
          Panel~\textit{(b)} displays the Bayesian posterior in yellow and the 
          frequentist bootstrapping in green.
          In both panels, the true flux is shown in magenta and the zone of
          inconsistent solutions is hatched in grey.
          \CClicence}
  \label{fig:BvF_asym}
\end{figure}

\paragraph{Flux with a non-linear detector.}
Let's assume that we are measuring again the flux from the same star ($F_\star=42$), with $m=3$ repetitions, but that our detector is now highly non-linear.
This non-linearity translates into a heavily-skewed split-normal noise distribution (\cf\ \refsec{sec:statistics}):
\begin{equation}
  p(F_j|F_\star) = 
  \left\{
  \begin{array}{ll}
    \displaystyle\frac{1}{\sqrt{2\pi}\lambda}
      \exp\left(-\frac{(F_j-F_\star)^2}{2\lambda^2}\right)
    & \mbox{if } F_j\le F_\star \\
    \displaystyle\frac{1}{\sqrt{2\pi}\lambda\tau}
      \exp\left(-\frac{(F_j-F_\star)^2}{2\lambda^2\tau^2}\right)
    & \mbox{if } F_j> F_\star. \\
  \end{array}
  \right.
  \label{eq:asymLH}
\end{equation}
with $\lambda=0.3$ and $\tau=50$.
This noise distribution is the red curve in \refsubfig{fig:BvF_asym}{a}.
In practice, this could for example be a very accurate detector suffering from transient effects.
The measured value would be systematically higher than the true flux\footnote{The distribution in \refsubfig{fig:BvF_asym}{a} has a very narrow tail on the lower side of $F_\sms{true}$.
It can thus in principle be lower, but this is such a low probability event that, for the clarity of the discussion, we will assume this is unlikely.}.
We have simulated three such measures in \refsubfig{fig:BvF_asym}{a}, in blue.
This problem was adapted from example 5 of \citet{jaynes76}.

\paragraph{The solutions.}
Knowing that the measured flux is always greater than or equal to the true flux, it is obvious that the solution should be lower than the lowest measured flux: $F_\star\le\min_jF_j=47.6$, in our particular case.
This flux range, corresponding to inconsistent values, has been hatched in grey, in \reffig{fig:BvF_asym}.
\begin{description}
  \item[The Bayesian solution] is obtained the same way as before, by sampling
    the posterior.
    We again assume a flat prior, and take the likelihood in \refeq{eq:asymLH}.
    The posterior is shown in yellow, in \refsubfig{fig:BvF_asym}{b}.
    It has zero probability in the range that we qualified as 
    \citengl{inconsistent}, and the true value falls in a high probability 
    domain.
    The mean and standard deviation of the posterior give us 
    $F_\star\simeq43.8\pm3.6$, with $\CR{F_\star}=[34.6,47.9]$.
  \item[The frequentist solution] is obtained the same way as before.
    The maximum likelihood and the whole bootstrapping sample however falls in 
    the \citengl{inconsistent} domain.
    We get $F_\star\simeq51.9\pm4.0$, with $\CI{F_\star}=[47.4,62.5]$.
    We see here that the frequentist solutions fails at inferring the true flux.
    On top of that, it gives only \citengl{inconsistent} solutions.
    In this particular case, this is because of the asymmetry of the noise, 
    which breaks the symmetry between \tpcond{F_j}{F_\star} and 
    \tpcond{F_\star}{F_j} that we had in \refsec{sec:BvFsimple}.
    This can be seen in \refsubfig{fig:BvF_asym}{b}.
    The frequentist solution is the mirror symmetric of the Bayesian posterior 
    for that reason.
\end{description}
When $m$ increases, the frequentist solution gets closer and closer to the true flux.
However, a bootstrapping analysis will reject the true flux in $100\,\%$ of the cases.

\paragraph{The reason of the frequentist failure.}
The failure of the frequentist approach is a direct consequence of its conception of probability \citep[\cf\ \eg][for a more detailed discussion and more examples]{vanderplas14}.
The frequentist method actually succeeds in returning the result it pretends to give: predicting a confidence interval where the solution would fall 95\,\%\ of the time, if we repeated the same procedure a large number of times.
This is however not equivalent to giving the credible range where the true value of the parameter has a 95\,\%\ probability to be (the Bayesian solution).
With the frequentist method, we have no guarantee that the \expression{true} flux will be in the confidence interval, only the \expression{solution}.
We can see that the main issue with the frequentist approach is that it is difficult to interpret, even in a simple problem such as that of \reffig{fig:BvF_asym}.
\citengl{Bayesians address the question everyone is interested in by using assumptions no-one believes, while Frequentists use impeccable logic to deal with an issue of no interest to anyone} \citep{lyons13}.
In the previous citation, the \citengl{assumption no-one believes} is the subjective choice of the prior, and the \citengl{issue of no interest to anyone} is the convoluted way frequentists formulate a problem, to avoid assigning probabilities to parameters.
\takeaway{Frequentist results can be inconsistent in numerous practical 
         applications, and they never perform better than Bayesian methods.}

  \subsection{Numerical Methods to Solve Bayesian Problems}
  \label{sec:mcmc}

Bayesian problems are convenient to formulate as they consist in laying down all the data, the model, the noise sources and the nuisance variables to build a posterior, using Bayes' rule.
The Bayesian results are also convenient to interpret as they all consist in using the posterior, which gives the true probability of the parameters.
However, in between, estimating the average, standard deviation, correlation coefficients of parameters, or testing hypotheses can be challenging, especially if there are a lot of parameters or if the model is complex.
Fortunately, several numerical methods have been introduced to make these tasks simpler.
Most of these methods are based on \expression{Markov Chain Monte-Carlo} (\hMCMC\footnote{Numerous authors publish articles claiming to have solved a problem using \citengl{MCMC methods}. This is not the best terminology to our mind, especially knowing that MCMCs can be used to sample any distribution, not only a Bayesian posterior. These authors should state instead to have solved a problem in a Bayesian way (what), using a MCMC numerical method (how). The same way, we tell our students to say that they \citengl{modeled the photoionization}, rather than they \citengl{used Cloudy}.}), which are a class of algorithms for sampling \hPDF s.

    \subsubsection{Sampling the Posterior Distribution}

\paragraph{Markov Chains Monte-Carlo.}
A \hMCMC\ draws samples from the posterior. 
In other words, it generates a \expression{chain} of $N$ values of the parameters, $\vec{x}_k$ ($k=1,\ldots,N$).
These parameter values are not uniformly distributed in the parameter space, but their number density is proportional to the posterior \hPDF.
Consequently, there are more points where the probability is high, and almost none where the probability is low.
It has several advantages.
\begin{itemize}
  \item Contrary to grid-based sampling or standard Monte-Carlo techniques, 
    we spend most of our computing time estimating our model (which can be 
    costly) where it matters, and not much time where it does not.
  \item From the \hMCMC, it is very simple and easy to estimate moments of a 
    given parameter, marginalizing over the other ones.
    For instance, if we have two parameters, $x$ and $y$, the average of $x$,
    marginalizing over $y$ would be:
    \begin{equation}
      \langle x\rangle\equiv\iint x.p(x,y)\ddiff x\ddiff y
      \simeq \frac{1}{N}\sum_{k=1}^{N}x_k,
    \end{equation}
    where the second equality is simply the average of the sample.
    We could do the same for the standard deviation, or the correlation 
    coefficient.
  \item It also makes hypothesis testing very easy.
    For instance, if we want to know the probability that $x>y$, we literally
    compute the fraction of points in the \hMCMC, where $x_k>y_k$.
\end{itemize}
All these operations would have been much more expensive, in terms of computing time, if we had to numerically solve the integral.
In particular, computing the normalization of the whole posterior would have been costly.
From a technical point of view, a \hMCMC\ is a random series of values of a parameter where the value at step $k$ depends only the value at step $k-1$.
We briefly discuss below the two most used algorithms.
A good presentation of these methods can be found in the book of \citet{gelman04} or in the \expression{Numerical recipes} \citep{press07}.

\begin{figure}[htbp!]
  \includegraphics[width=\textwidth]{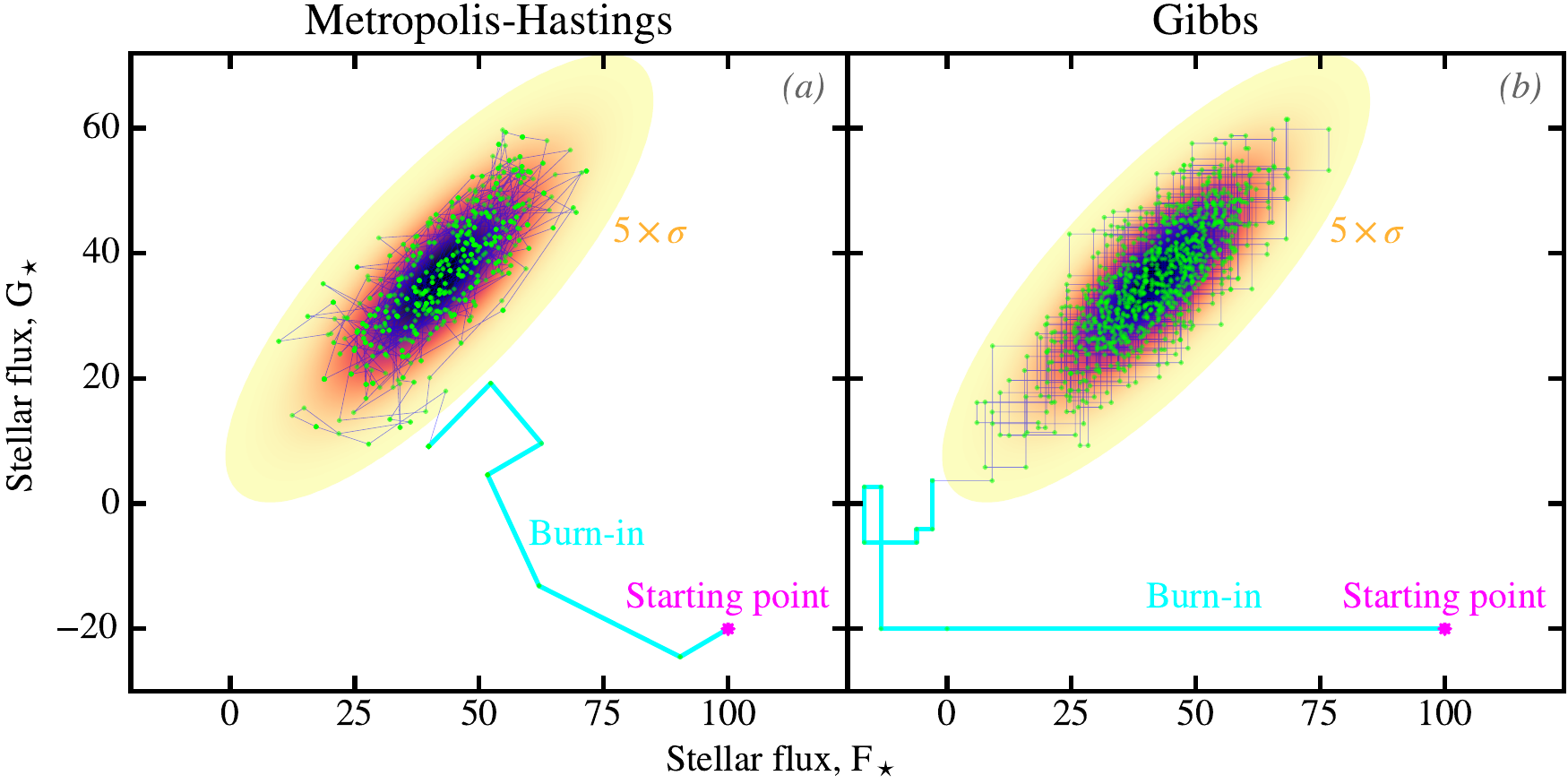} \\
  \includegraphics[width=\textwidth]{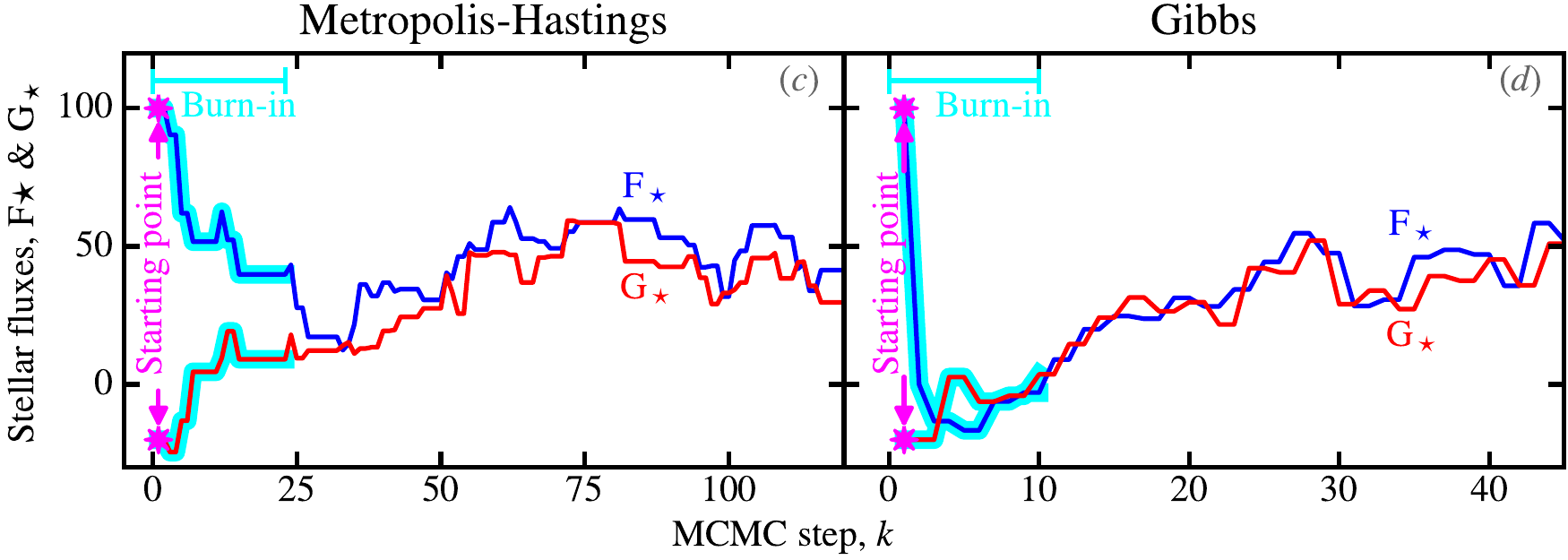} \\
  \includegraphics[width=\textwidth]{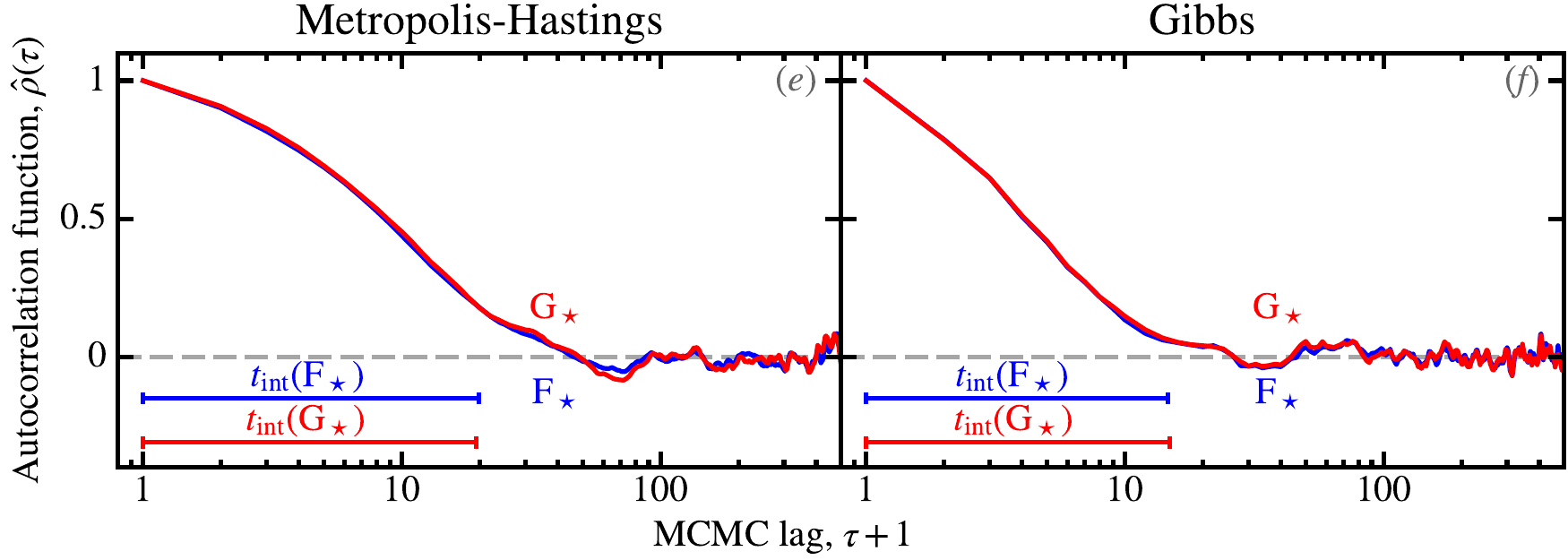} 
  \newcap{Markov Chain Monte-Carlo algorithms}%
         {Panels~\textit{(a)} and \textit{(b)} show contours, up to $5\sigma$, 
          of the posterior distribution of \refeq{eq:distsamp}.
          In both panels, the blue line with the green dots represent the 
          \hMCMC, for the first 1000 steps, starting from the magenta star.
          The burn-in phase is highlighted in cyan.
          Panels~\textit{(c)} and \textit{(d)} represent the \hMCMC s of the two
          parameters, at the start of the chain.
          We have highlighted the burn-in phase in cyan.
          Panels~\textit{(e)} and \textit{(f)} represent the \hACF s of both
          parameters.
          We have also quoted the integrated autocorrelation times, 
          $t_\sms{int}$.
          As indicated, left panels demonstrate the Metropolis-Hastings 
          algorithm, while right panels show Gibbs sampling.
          \CClicence}
  \label{fig:mcmc}
\end{figure}

\paragraph{The Metropolis-Hastings algorithm.}
To illustrate this method and the next one, let's consider again the measure of the flux of our star, with the difference, this time, that we would be observing it through two different photometric filters.
\begin{itemize}
  \item Let's call $F_\star$ and $G_\star$ the fluxes in these two bands, with
    uncertainties $\sigma_\sms{F}=12$ and $\sigma_\sms{G}=10$.
  \item Let's assume that the uncertainties in the filters are correlated, with 
    a correlation coefficient, $\rho=0.8$, because, for instance, of the way 
    the calibration was performed.
  \item Let's assume that we make only $m=1$ measure in each band, with 
    observed fluxes $F_1=42$ and $G_1=36$.
    The posterior, assuming a flat prior, is simply a bivariate normal 
    distribution centered on the observed flux, with covariance matrix:
    \begin{equation}
      V=\left(
      \begin{array}{cc}
        \sigma_\sms{F}^2 & \rho\sigma_\sms{F}\sigma_\sms{G} \\
        \rho\sigma_\sms{F}\sigma_\sms{G} & \sigma_\sms{G}^2 \\
      \end{array}
      \right).
    \end{equation}
    The posterior is thus, noting $\vec{x}=(F_\star,G_\star)$ and 
    $\vec{d}=(F_1,G_1)$:
    \begin{equation}
      \pcond{\vec{x}}{\vec{d}} \propto 
        \exp\left(-\frac{1}{2}(\vec{x}-\vec{d})^TV^{-1}(\vec{x}-\vec{d})\right).
      \label{eq:distsamp}
    \end{equation}
    Contours of this distribution are represented in \refsubfig{fig:mcmc}{a-b}.
\end{itemize}
The algorithm proposed by \citet{metropolis53} and generalized by \citet{hastings70} is the most popular method to sample any \hPDF.
This is a \expression{rejection method}, similar to what we have discussed for \hMCRT s, in \refsec{sec:MCRT} (\cf\ also \refapp{sec:random_rejection}).
\begin{description}
  \item[A proposal distribution,] \tpcond{\vec{x}_k}{\vec{x}_{k-1}}, first 
    needs to be chosen.
    The choice of this distribution is instrumental in the sampling efficiency:
    \begin{inlinelist}
      \item if it is too wide, a lot of draws will be rejected;
      \item if it is too narrow, the sampling steps are going to be small, and 
        more iterations are going to be necessary to sample the posterior.
    \end{inlinelist}
    For our present example, we choose a bivariate normal distribution,  
    centered on $\vec{x}_{k-1}$, whose width is the noise of our data, 
    $s_\sms{F}=\sigma_\sms{F}$ and $s_\sms{G}=\sigma_\sms{G}$:
    \begin{equation}
      \pcond{\vec{x}_k}{\vec{x}_{k-1}}\propto
      \exp\left[-\frac{1}{2}\left(\frac{F_{\star,k}-F_{\star,k-1}}{s_\sms{F}}\right)^2
                -\frac{1}{2}\left(\frac{G_{\star,k}-G_{\star,k-1}}{s_\sms{G}}\right)^2
          \right]
      \label{eq:MHprop}
    \end{equation}
  \item[The method] then consists in the following steps, iterated $N$ times.
    \begin{enumerate}
      \item At each iteration, $k$, we draw a new set of parameters, 
        $\vec{x}_k$, from the proposal distribution: 
        $\vec{x}_k\sim\pcond{\vec{x}_k}{\vec{x}_{k-1}}$.
      \item With this new value, we compute the \expression{acceptance 
        probability}, defined as:
        \begin{equation}
          \alpha_k = \min\left(1,
            \frac{\pcond{\vec{x}_k}{\vec{d}}\pcond{\vec{x}_{k-1}}{\vec{x}_k}}%
                 {\pcond{\vec{x}_{k-1}}{\vec{d}}\pcond{\vec{x}_k}{\vec{x}_{k-1}}}
            \right).
          \label{eq:MHaccept}
        \end{equation}
        The $\min$ function is here to make sure we get a result between 0 and 
        1.
        In our case, we have chosen a symmetric proposal distribution.
        \refeq{eq:MHprop} thus implies that:
        \begin{equation} 
        \frac{\pcond{\vec{x}_{k-1}}{\vec{x}_k}}{\pcond{\vec{x}_k}{\vec{x}_{k-1}}}=1
        \;\;\;\Rightarrow\;\;\;
        \alpha_k=\min\left(1,\frac{\pcond{\vec{x}_k}{\vec{d}}}%
                                  {\pcond{\vec{x}_{k-1}}{\vec{d}}}\right).
        \end{equation}
        We just need to estimate our posterior at one position (assuming we 
        saved the value of \tpcond{\vec{x}_{k-1}}{\vec{d}}, after the previous 
        iteration).
        In addition, since we need only the ratio of two points in the 
        posterior, we do not need to normalize it.
        This is the reason why this algorithm is so efficient.
      \item To update $\vec{x}_k$, we draw a random variable, $\theta_k$, 
        uniformly distributed between 0 and 1.
        \begin{itemize}
          \item If $\theta_k\le\alpha_k$, we accept the new value, $\vec{x}_k$.
          \item If $\theta_k>\alpha_k$, we reject the new value and keep the 
            old one, $\vec{x}_k=\vec{x}_{k-1}$.
        \end{itemize}
    \end{enumerate}
    The initial value of the chain has to be a best guess.
\end{description}
The sampling of \refeq{eq:distsamp} with the Metropolis-Hastings method is shown in \refsubfig{fig:mcmc}{a}.

\paragraph{Gibbs sampling.}
The number of parameters, $n$, determines the dimension of the posterior.
The higher this number is, the smaller the \expression{support of the function} is (\ie\ the region where the probability is non negligible).
Metropolis-Hastings methods therefore will have a high rejection rate, if $n\gg1$, requiring longer chains to ensure convergence.
\expression{Gibbs sampling} \citep{geman84} provides an alternative \hMCMC\ method, where all draws are accepted.
Its drawback is that it requires normalizing, at each iteration, the \expression{full conditional distribution}:
\begin{equation}
  \pcond{x_i}{\vec{x}_-,\vec{d}}\equiv
  \pcond{x_i}{x_1,\ldots,x_{i-1},x_{i+1},\ldots,x_n,\vec{d}}
  \label{eq:fullcond}
\end{equation}
that is the posterior fixing all parameters except one.
This is only a one dimensional \hPDF, though, much less computing-intensive than the whole posterior.
The method consists, at each iteration $k$, to cycle through the different parameters, and draw a new value from \refeq{eq:fullcond}:
\begin{equation}
  x_{i,k}\sim
  \pcond{x_{i,k}}{\underbrace{x_{1,k},\ldots,x_{i-1,k}}_\sms{already updated}  
                ,\underbrace{x_{i+1,k-1},\ldots,x_{n,k-1}}_\sms{not yet updated},
                 \vec{d}}.
\end{equation}
Since this distribution has an arbitrary form, random drawing can be achieved numerically using the \expression{\hCDF\ inversion method} (\refapp{sec:random_CDF}).
\refsubfig{fig:mcmc}{b} represents the Gibbs sampling of \refeq{eq:distsamp}.
The squared pattern comes from the fact that we alternatively sample each parameter, keeping the other one fixed.

    \subsubsection{Post-Processing MCMCs}
    \label{sec:postmcmc}

\begin{figure}[htbp]
  \begin{tabular}{cc}
    \includegraphics[width=0.48\textwidth]{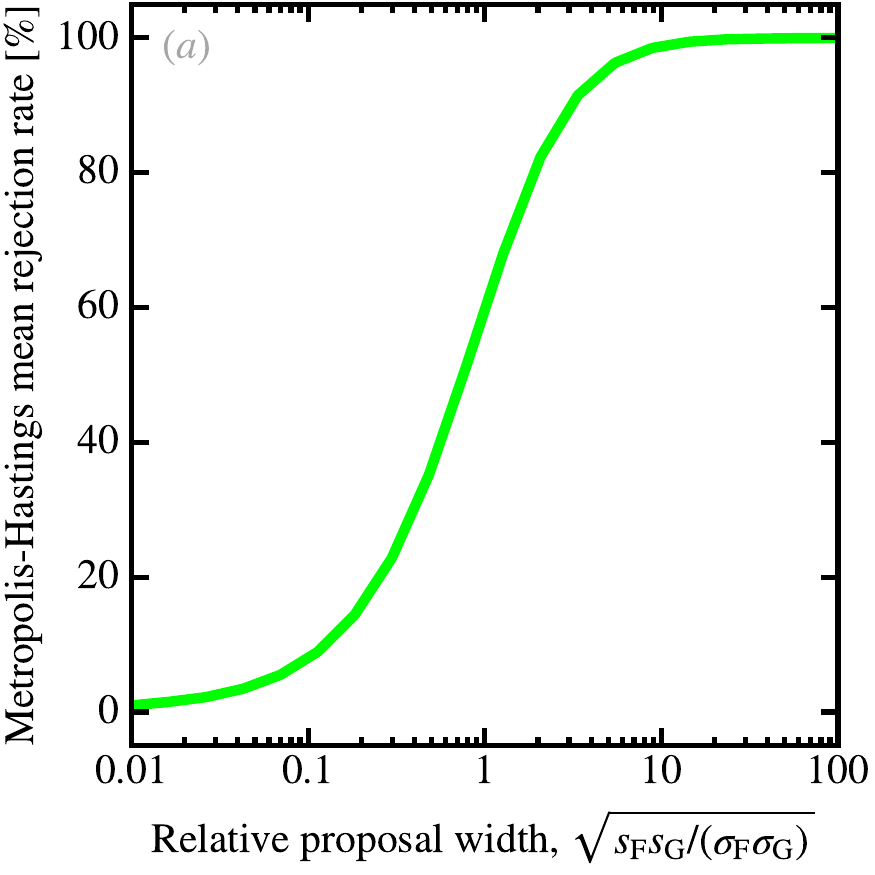} &
    \includegraphics[width=0.48\textwidth]{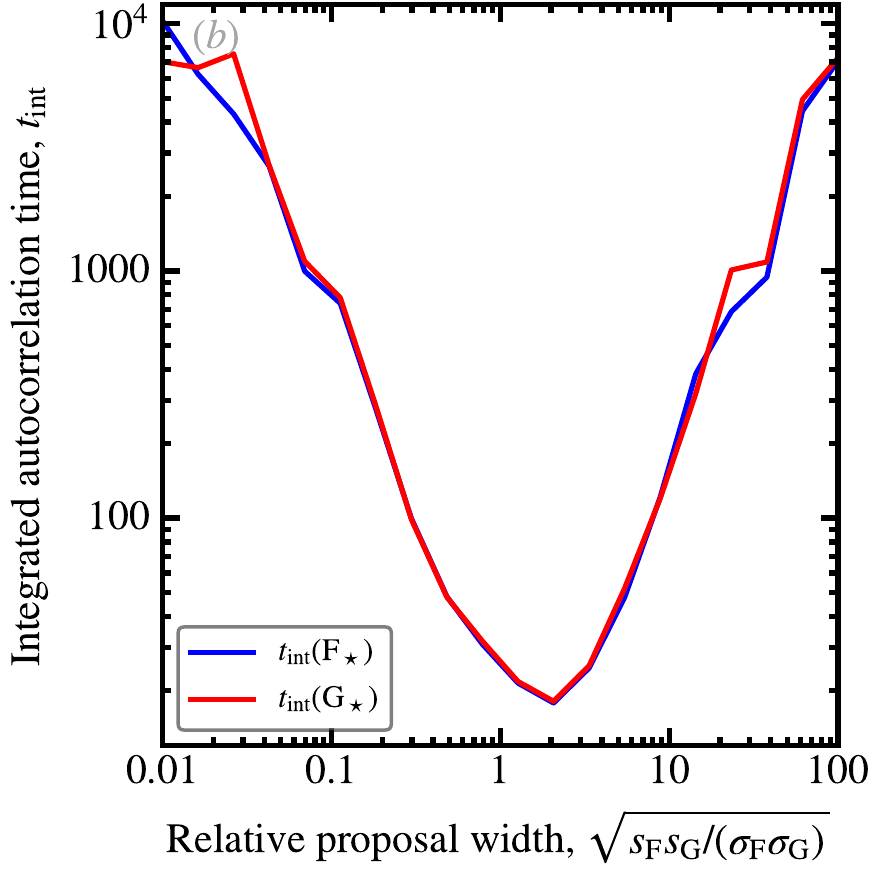} \\
  \end{tabular}
  \newcap{Importance of the choice of the Metropolis-Hastings proposal 
          distribution}%
         {Panel~\textit{(a)} represents the Metropolis-Hastings mean rejection 
          rate, varying the width of the proposal distribution 
          \refeqp{eq:MHprop}, when sampling \refeq{eq:distsamp}.
          Panel~\textit{(b)} represents the corresponding integrated 
          autocorrelation time for the two parameters.
          \CClicence}
  \label{fig:MHrejection}
\end{figure}

\paragraph{Assessing convergence.}
One of the most crucial questions, when using a \hMCMC\ method, is how long a chain do we need to run.
To answer that question, we need to estimate if the \hMCMC\ has converged toward the stationary posterior.
Concretely, it means that we want to make sure the sampling of the posterior is homogeneous, and that the moments and hypothesis testing we will perform will not be biased, because some areas of the parameter space have only been partially explored.
The reason why the sampling of the parameter space might be incomplete is linked to the two following factors.
\begin{description}
  \item[The burn-in] refers to the first drawn values, before the \hMCMC\ could 
    find the support of the posterior.
    This can be seen in \refsubfig{fig:mcmc}{a-b} (highlighted in cyan).
    The arbitrary starting value (magenta star) is outside the $5\sigma$ contour
    of the \hPDF.
    The \hMCMC\ thus walks a few steps before finding the probable region.
    This burn-in phase is also highlighted in cyan, in 
    \refsubfig{fig:mcmc}{c-d}.
    This burn-in phase could actually be significantly longer, for several 
    non-exclusive reasons:
    \begin{inlinelist}
      \item a larger number of parameters;
      \item a more degenerate model, with several local maxima;
      \item a less lucky choice of initial conditions; or
      \item a very well-constrained model, resulting in a very small support 
        over the whole parameter space.
    \end{inlinelist}
    There is no universal method to identify burn-in, it needs to 
    be investigated carefully, most of the time.
    Running several \hMCMC s, starting from initial conditions distributed 
    over the whole parameter space, is usually efficient.
  \item[The autocorrelation] of the \hMCMC\ results from the fact that the 
    parameter value at step $k+1$ depends on step $k$.
    If several successive iterations stay in the same region of the posterior, 
    this will create a portion of correlated values.
    The \expression{AutoCorrelation Function} \citep[\hACF; \eg][]{sokal96} is 
    an essential tool to determine the correlation length of a \hMCMC.
    The \hACF, $\hat{\rho}$, of a given parameter, depends on the 
    \expression{lag}, $\tau$, that is the number of steps between two arbitrary 
    iterations:
    \begin{equation}
      \hat{\rho}(\tau)\equiv
      \frac{N}{N-\tau}
      \frac{\displaystyle\sum_{k=1}^{N-\tau}(x_k-\langle x\rangle)
            (x_{k+\tau}-\langle x\rangle)}%
           {\displaystyle\sum_{k=1}^{N}(x_k-\langle x\rangle)^2}.
    \end{equation}
    This is the correlation coefficient of the parameter with itself, 
    shifted by $\tau$ steps.
    The \hACF s of our example are displayed in \refsubfig{fig:mcmc}{e-f}.
    We can see that the \hACF\ starts at 1, for $\tau=0$.
    It then drops over a few steps and oscillates around 0.
    The typical lag after which the \hACF\ has dropped to 0, corresponds to the 
    average number of steps necessary to draw independent values.
    This typical lag can be quantified, by the \expression{integrated 
    autocorrelation time}, $t_\sms{int}$\footnote{This quantity is problematic 
    to compute. \citet{sokal96} and \citet{foreman-mackey13} discuss an 
    algorithm to evaluate it numerically.}:
    \begin{equation}
      t_\sms{int}\equiv1+2\sum_{i=\tau}^{N}\hat{\rho}(\tau).
    \end{equation}
    It is represented in \refsubfig{fig:mcmc}{e-f}.
    It corresponds roughly to the average number of steps needed to go from one 
    end of the posterior to the other.
    Different parameters of a given \hMCMC\ can in principle have very 
    different $t_\sms{int}$ \citep[\eg][]{galliano18a}.
    To make sure that our posterior is properly sampled, we thus need to let
    our \hMCMC\ run a large number of steps, times $t_\sms{int}$, after burn-in.
    The \expression{effective sample size}, 
    $N_\sms{eff}\equiv N/t_\sms{int}$, quantifies the effective number
    of steps that can be considered independent.
    We need $N_\sms{eff}\gg1$.
\end{description}
With the Metropolis-Hastings algorithm, the integrated autocorrelation time will depend heavily on the choice of the proposal distribution.
We have explored the effect of the width of this distribution on $t_\sms{int}$.
In \refeq{eq:MHprop}, instead of taking $s_\sms{F}=\sigma_\sms{F}$ and $s_\sms{G}=\sigma_\sms{G}$, we have varied this parameter.
\refsubfig{fig:MHrejection}{a} represents the mean rejection rate as a function of $\sqrt{s_\sms{F}s_\sms{G}/(\sigma_\sms{F}\sigma_\sms{G})}$.
\begin{itemize}
  \item When $\sqrt{s_\sms{F}s_\sms{G}/(\sigma_\sms{F}\sigma_\sms{G})}\ll1$, 
    the proposal is much narrower than the posterior.
    The proposed steps are thus very small, and a lot of them are necessary to 
    cross the posterior.
    This is why $t_\sms{int}$ is very large in this case (\cf\ 
    \refsubfig{fig:MHrejection}{b}).
  \item When $\sqrt{s_\sms{F}s_\sms{G}/(\sigma_\sms{F}\sigma_\sms{G})}\gg1$, the 
    proposal distribution is much larger than the posterior.
    Most proposals therefore falls outside the support of the posterior.
    They are therefore rejected close to 100\,\% of the times.
    This is why $t_\sms{int}$ is also very large in this case.
\end{itemize}
\refsubfig{fig:MHrejection}{b} shows that the only range where $t_\sms{int}$ is reasonable is when the width of the proposal distribution is comparable to the width of the posterior.

\begin{figure}[!htbp]
  \includegraphics[width=\textwidth]{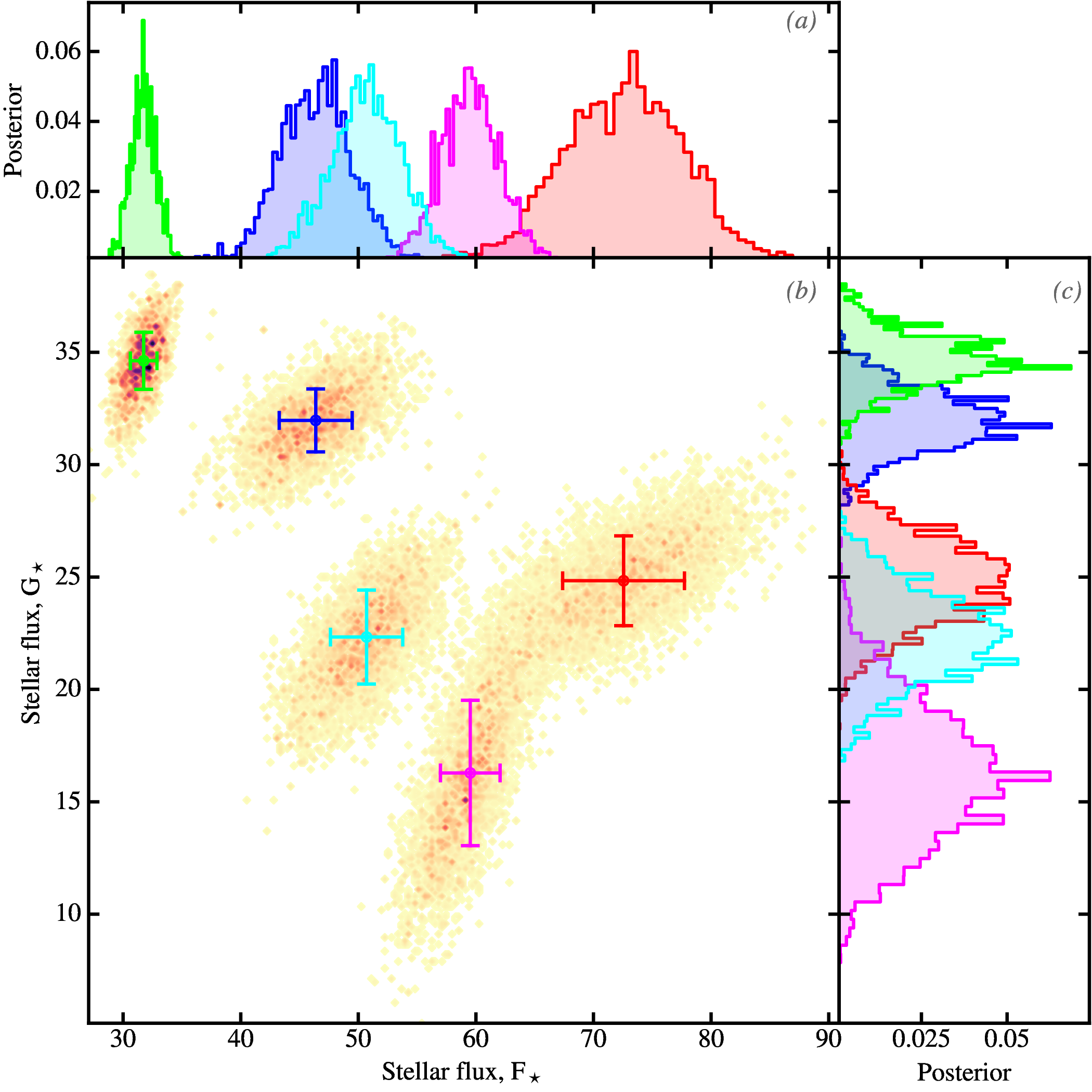}
  \newcap{Post-processing of the MCMC of a sample of sources}%
         {In panel~\textit{(b)}, the contours  represent the posterior of five 
          stars, observed through the two photometric bands (fluxes $F_\star$ 
          and $G_\star$).
          The margin plots represent the marginalized distribution of the 
          posterior of each individual star.
          The error bars in panel~\textit{(b)} are plotted from the mean and
          standard deviations of the posterior.
          \CClicence}
  \label{fig:mcmcpost}
\end{figure}

\begin{figure}[htbp]
  \includegraphics[width=\textwidth]{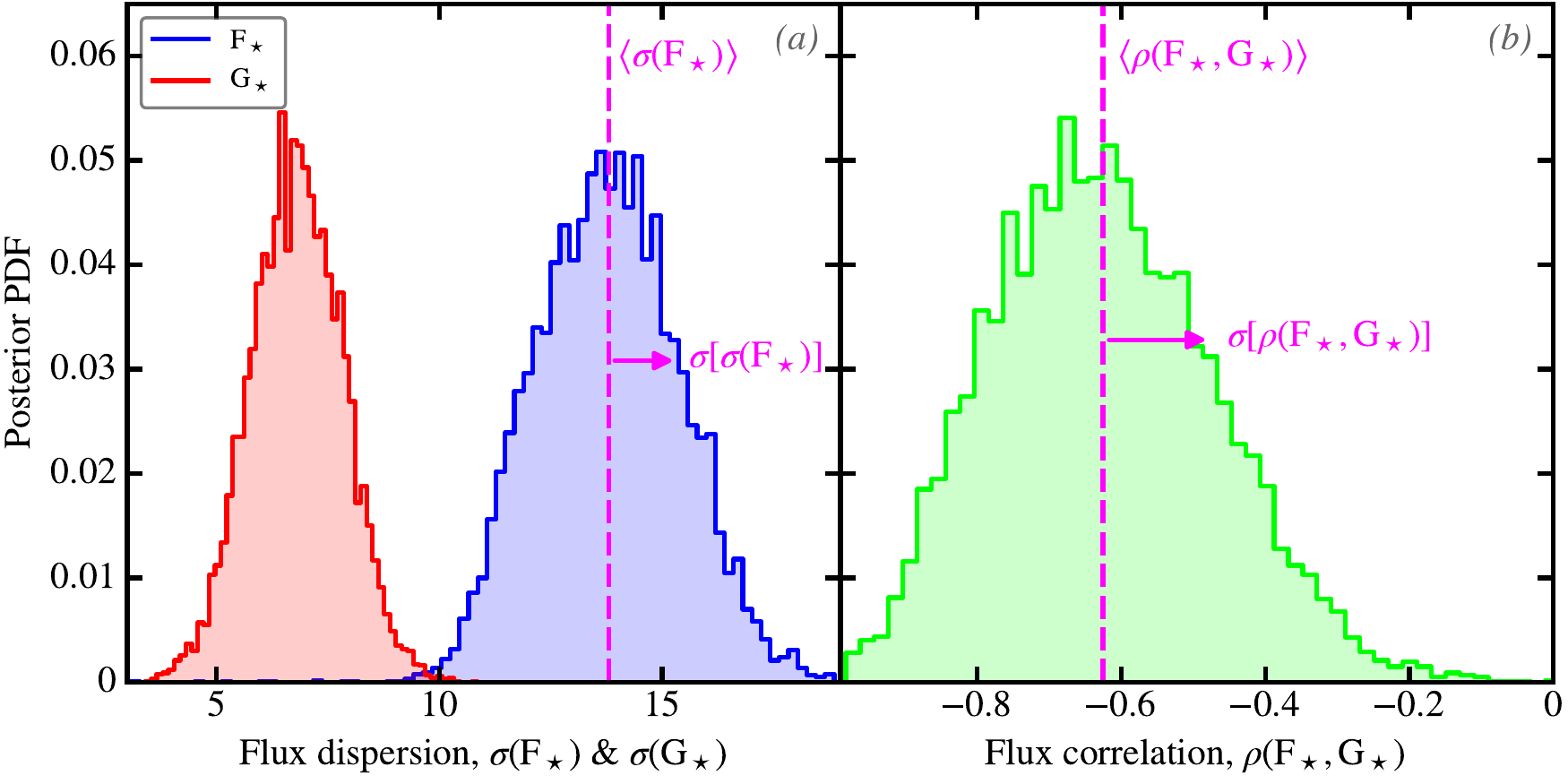}
  \newcap{MCMC statistics of a sample of sources}%
         {Panel~\textit{(a)} represents the posterior of the standard deviation
          of the sample in \reffig{fig:mcmcpost}.
          It represents the \hPDF\ of the dispersion of the sample, not the 
          width of individual \hPDF s.
          Similarly, panel~\textit{(b)} represents the correlation 
          coefficient of the flux distribution.
          \CClicence}
  \label{fig:mcmcpoststat}
\end{figure}

\paragraph{Parameter inference.}
Numerous quantities can be inferred from a \hMCMC.
We have previously seen that the average, uncertainties, and various tests can be computed using the posterior of the parameters of a source.
This becomes even more powerful when we are analyzing a sample of sources.
To illustrate this, let's assume we are now observing $N_\star=5$ stars, through the same photometric bands as before.
\refsubfig{fig:mcmcpost}{b} shows the posterior \hPDF\ of the two parameters of the five stars.
It is important to distinguish the following two types of distributions.
\begin{description}
  \item[The posterior of individual stars] are represented in 
    \reffig{fig:mcmcpost}.
    The error bars in panel~\textit{(b)} correspond to the mean and 
    standard deviation of the marginal distributions in panels~\textit{(a)} and
    \textit{(c)}.
    They represent the uncertainty on the measured fluxes of each 
    individual star.
    They are the moments of a given parameter, over the whole \hMCMC.
    This is what we have focussed on, until now.
  \item[The statistic distribution across the sample] is represented in 
    \reffig{fig:mcmcpoststat}.
    In panel~\textit{(a)}, we have shown the distribution of the standard 
    deviation of the sample, at each step $k$ in the \hMCMC:
    \begin{equation}
      \sigma(F_\star)_k\equiv\frac{1}{N_\star-1}\sum_{i=1}^{N_\star}
        \left(F_{i,k}-\frac{1}{N_\star}\sum_{j=1}^{N_\star}F_{j,k}\right)^2,
    \end{equation}
    where $F_{i,k}$ is the observed flux of the star $i$, at the \hMCMC\ 
    iteration $k$.
    It is how we can quantify the dispersion of the sample.
    We can thus quote the sample dispersion as 
    (\cf\ \refsubfig{fig:mcmcpoststat}{a}):
    \begin{equation}
      \sigma(F_\star)\simeq
        \langle\sigma(F_\star)\rangle\pm\sigma\left[\sigma(F_\star)\right].
    \end{equation}
    In our example, we have: $\sigma(F_\star)\simeq13.8\pm1.6$ and 
    $\sigma(G_\star)\simeq6.8\pm1.0$.
    We can see that these values correspond roughly to the intrinsic scatter 
    between individual stars, in \refsubfig{fig:mcmcpost}{b}, but they are 
    larger than the uncertainty on the flux of individual stars.
    We can do the same for the correlation coefficient, as shown in 
    \refsubfig{fig:mcmcpoststat}{b}: $\rho(F_\star,G_\star)\simeq-0.63\pm0.15$.
    Notice that it is negative, because the correlation between stellar 
    \expression{fluxes}, in \refsubfig{fig:mcmcpost}{b}, points toward the 
    lower right corner.
    However, the correlation between the \expression{uncertainties} on $F_\star$ 
    and $G_\star$ is positive: the individual ellipses point in the other 
    direction, toward the upper right corner.
    We have deliberately simulated data with these two opposite correlations to 
    stress the difference between the individual likelihood properties and those
    of the ensemble.
    Finally, we could have done the same type of estimate for the mean of the 
    sample: 
    $\langle F_\star\rangle\simeq52.2\pm1.5$ and 
    $\langle G_\star\rangle\simeq26.0\pm1.0$.
\end{description}

\begin{figure}[!htbp]
  \includegraphics[width=\textwidth]{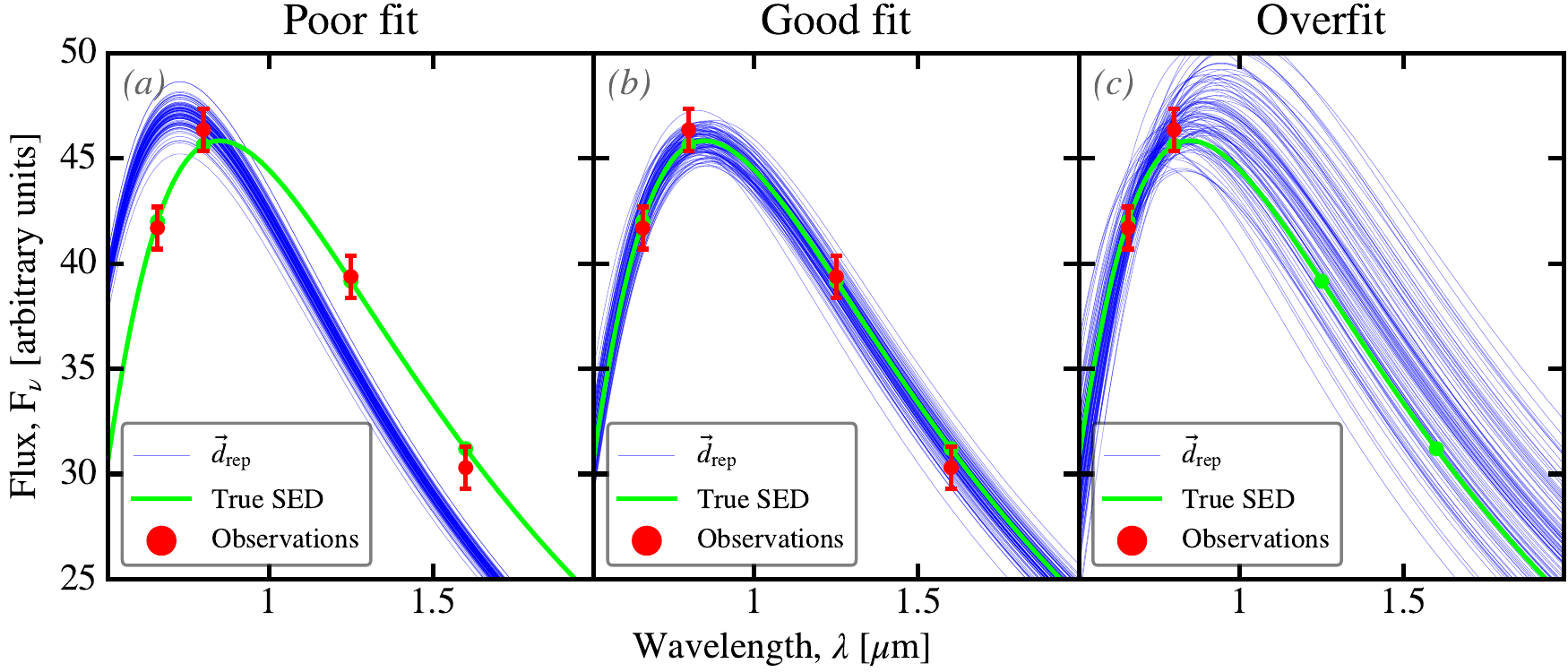} \\
  \includegraphics[width=\textwidth]{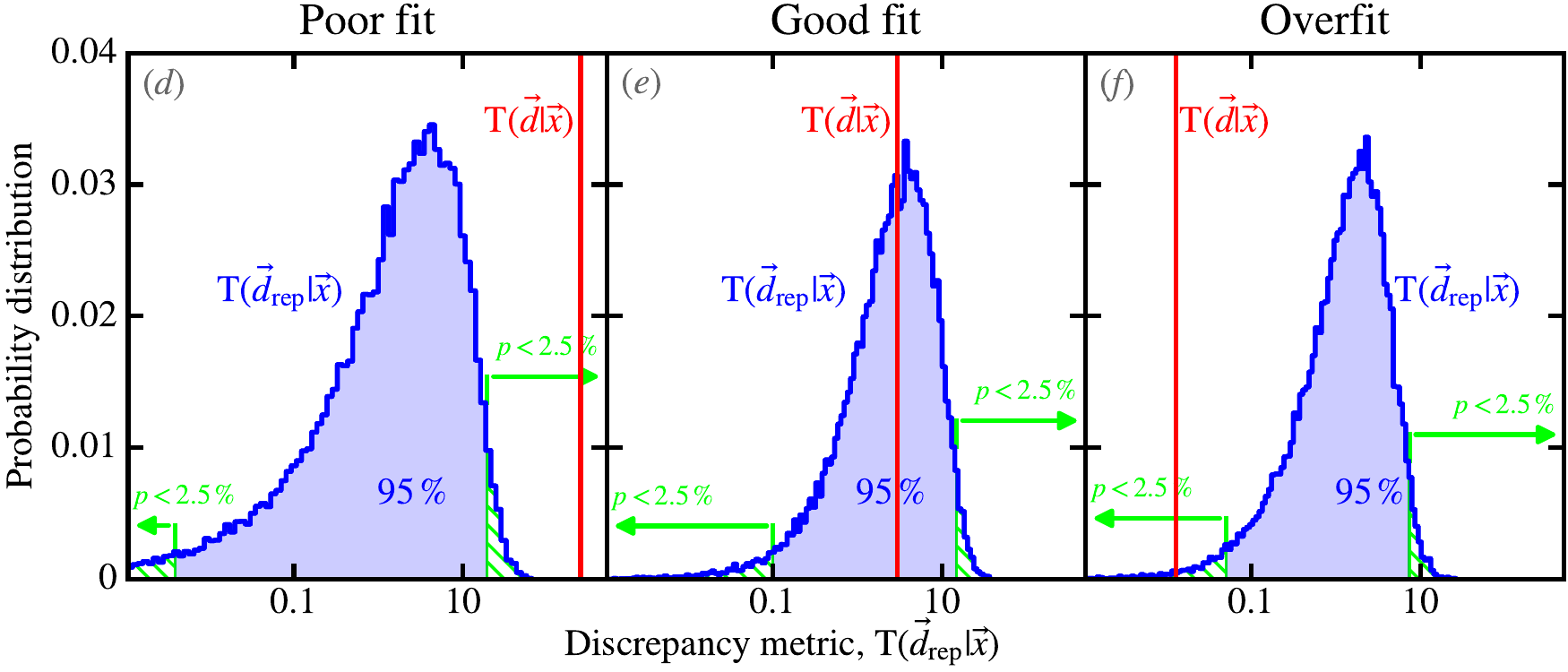} 
  \newcap{Demonstration of the use of posterior predictive \textit{p}-values}%
         {The top three panels show the fit of an observed stellar \hSED\
          (red error bars) with a blackbody.
          The true model ($T_\star=6000$~K) is shown in green.
          The blue lines are a subsample of the inferred model (at different
          steps in the \hMCMC).
          In panel~\textit{(a)}, we have fixed the temperature to 
          $T_\star=7000$~K, different from its true value.
          We thus vary only $\Omega_\star$, resulting in a poor fit.
          In panel~\textit{(b)}, we let both $T_\star$ and $\Omega_\star$
          vary, in order to get a good fit.
          In panel~\textit{(c)}, we fit only two fluxes, with our two 
          parameters.
          We are thus overfitting the data.
          The three bottom panels show the distribution of the discrepancy 
          metric for the replicated data set (in blue), corresponding to the
          upper panels.
          We compare it to $T\left(\vec{d}\middle|\vec{x}\right)$, in red.
          The low probability ranges are hatched in green.  
          This example has been generated using \ncode{emcee} 
          \citep{foreman-mackey13}.
          \CClicence}
  \label{fig:ppp}
\end{figure}

\paragraph{Quantifying the goodness of a fit.}
It is important, in any kind of model fitting, to be able to assess the quality of the fit.
In the frequentist approach, this is done with the chi-squared test, which is limited in its assumptions to normal \hiid\ noise, without nuisance parameters.
In the Bayesian approach, the same type of test can be done, accounting for the full complexity of the model (non-Gaussian errors, correlations, nuisance parameters, priors).
This test is usually achieved by computing \expression{posterior predictive $p$-values} \citep[\hppp; \eg\ Chap.~6 of][]{gelman04}.
To illustrate how \hppp s work, let's consider now that we are observing the same star as before, through four bands (R, I, J, H) and are performing a blackbody Bayesian fit to this \hSED, varying the temperature, $T_\star$, and the dilution factor, $\Omega_\star$.
This is represented in \refsubfig{fig:ppp}{b}.
The principle is the following.
\begin{enumerate}
  \item We generate a set of \expression{replicated data}, $\vec{d}_\sms{rep}$, 
    from our posterior:
    \begin{equation}
      \pcond{\vec{d}_\sms{rep}}{\vec{d}}\equiv
        \int\pcond{\vec{d}_\sms{rep}}{\vec{x}}
            \pcond{\vec{x}}{\vec{d}}\ddiff\vec{x}.
    \end{equation}
    If we sampled our posterior with a \hMCMC, this integral can simply be 
    computed by evaluating our model (the blackbody, in the present case), for
    values of our drawn parameters:
    $\vec{d}_\sms{rep}=\left\{f(\vec{x}_k)\right\}_{k=1}^\sms{N}$.
  \item We evaluate the mean and standard deviation of this replicated data set,
    $\left\langle\vec{d}_\sms{rep}\middle|\vec{x}\right\rangle$ and 
    $\sigma\left(\vec{d}_\sms{rep}\middle|\vec{x}\right)$.
    These quantities are the average and the dispersion of the predicted flux
    in the different bands.
    They will serve as position and scale references, when comparing model and
    observations.
  \item We compute a \expression{discrepancy metric}, 
    $T\left(\vec{d}\middle|\vec{x}\right)$.
    Several choices are possible, but the most common is to adopt a chi-squared
    equivalent:
    \begin{equation}
      T\left(\vec{d}\middle|\vec{x}\right)\equiv
      \sum_{j=1}^m\frac{\left[d_j
               -\left\langle d_j\middle|\vec{x}\right\rangle\right]^2}%
                      {\sigma\left(d_j\middle|\vec{x}\right)^2}.
      \label{eq:discppp}
    \end{equation}
    We compute this quantity both for the replicated set, 
    $T\left(\vec{d}_\sms{rep}\middle|\vec{x}\right)$, which is the blue 
    distribution in \refsubfig{fig:ppp}{e}, and for the observations, 
    $T\left(\vec{d}\middle|\vec{x}\right)$, which is the red line in 
    \refsubfig{fig:ppp}{e} (it is a single value).
    To be clear, only the data term in \refeq{eq:discppp} changes between 
    $T\left(\vec{d}_\sms{rep}\middle|\vec{x}\right)$ and 
    $T\left(\vec{d}\middle|\vec{x}\right)$.
    The quantities $\left\langle d_j\middle|\vec{x}\right\rangle$ and 
    $\sigma\left(d_j\middle|\vec{x}\right)$ are identical in both cases.
  \item The quality of the test is assessed by comparing both quantities.
    To that purpose, we compute the following probability:
    \begin{equation}
      p_\sms{B}\equiv
      \pcond{T\left(\vec{d}_\sms{rep}\middle|\vec{x}\right)\ge
             T\left(\vec{d}\middle|\vec{x}\right)}{\vec{d}}.
    \end{equation}
    If the difference between the replicated set and the data is solely due to
    statistical fluctuations, we should have on average $p_\sms{B}\simeq50\,\%$.
    The fit is considered bad, at the $95\,\%$ level, if $p_\sms{B}<2.5\,\%$ or
    $p_\sms{B}>97.5\,\%$.
\end{enumerate}
We have illustrated this test in \reffig{fig:ppp}, varying the number of parameters and observational constraints, in order to explore the different possible cases.
\begin{description}
  \item[A good fit] is shown in \refsubfig{fig:ppp}{b}.
    We have varied both $T_\star$ and $\Omega_\star$ to fit the 4 fluxes.
    \refsubfig{fig:ppp}{e} shows that $T\left(\vec{d}\middle|\vec{x}\right)$
    falls in the high probability range of 
    $T\left(\vec{d}_\sms{rep}\middle|\vec{x}\right)$.
    In other words, the average deviation of the replicated data, relative to 
    the reference we have chosen, 
    $\left\langle\vec{d}_\sms{rep}\middle|\vec{x}\right\rangle$, is comparable 
    to the deviation of the actual data relative to the same reference.
    The observations could thus have likely been drawn from our posterior.
  \item[A poor fit] is shown in \refsubfig{fig:ppp}{a}.
    We have intentionally fixed the temperature of the fit at $T_\star=7000$~K,
    while the true value is $T_\star=6000$~K.
    We see that $p_\sms{B}\simeq1$, in \refsubfig{fig:ppp}{d}.
    The difference between the observations and the model, can thus not be 
    explained by the scatter of the model.
    It is the sign of a bad fit.
    In our case, it is because the model is bad (wrong choice of fixed 
    temperature).
  \item[An overfit] is shown in \refsubfig{fig:ppp}{c}.
    This time we fit only two fluxes.
    With a chi-squared fit, we would have 0 degrees of freedom.
    We see that $p_\sms{B}\simeq0$, in \refsubfig{fig:ppp}{f}.
    The average model therefore gets too close to the observations.
    The \hppp\ tells us this is very unlikely.
    It is however not an issue in terms of derived parameters.
    We can see that the true model (green) is well among the sampled model 
    (blue), and the inferred parameters are consistent with their true values.
\end{description}
There is no issue with fitting even fewer constraints than parameters, with a Bayesian approach.
The consequence is that the posterior is going to be very wide along the dimensions corresponding to the poorly constrained parameters.
However, the results will be consistent, and the derived probabilities will be meaningful.
\takeaway{Contrary to the frequentist approach, we can fit Bayesian models with more parameters than data points.}

  \subsection{Decision Making and Limitations of the 
                  Frequentist Approach}

We now resume our comparison of the Bayesian and frequentist approaches, started in \refsec{sec:BvFexample}.
We focus more on the interpretation of the results and synthesize the advantages and inconveniences of both sides.

    \subsubsection{Hypothesis Testing}
    \label{sec:Bayesfactor}
    \label{sec:phacking}

Until now, we have seen how to estimate parameters and their uncertainties.
It is however sometimes necessary to be able to make decisions, that is to choose an outcome or its alternative, based on the observational evidence.
Hypothesis testing consists in assessing the likeliness of a \expression{null hypothesis}, noted $H_0$.
The \expression{alternative hypothesis} is usually noted $H_1$, and satisfies the logical equation $H_1=\neg H_0$, where the $\neg$ symbol is the logical negation.
The priors necessarily obey $\proba{H_0}+\proba{H_1}=1$.
To illustrate this process let's go back to our first example, in \refsec{sec:BvFexample}.
We are observing a star with true flux $F_\sms{true}$, $m$ times, with an uncertainty $\sigma_\sms{F}$ on each individual flux measurements, $F_i$.
This time, we want to know if $F_\sms{true}\le F_\sms{test}$, for a given $F_\sms{test}$.

\begin{figure}[htbp]
  \begin{tabular}{cc}
    \includegraphics[width=0.48\textwidth]{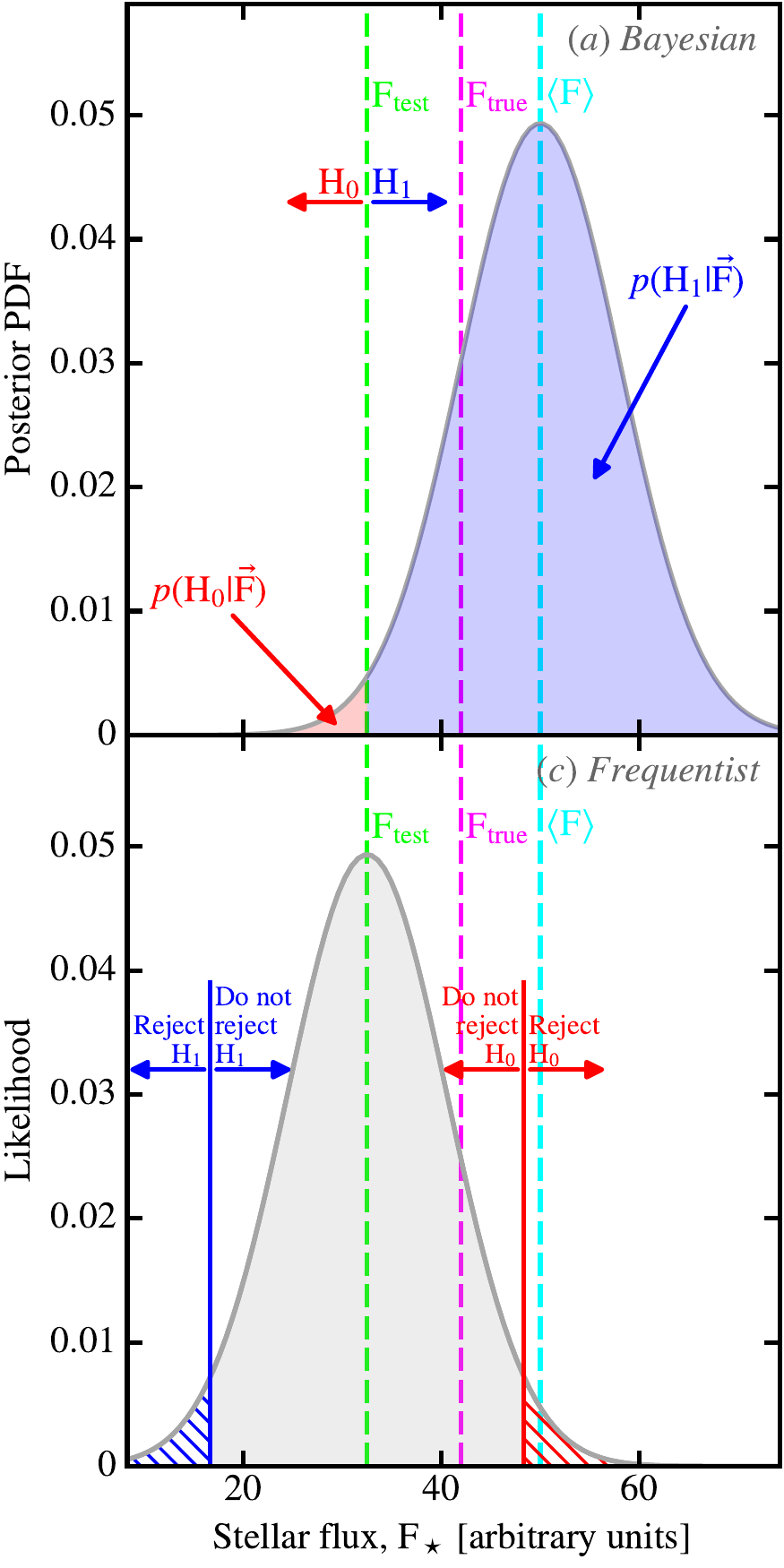} &
    \includegraphics[width=0.48\textwidth]{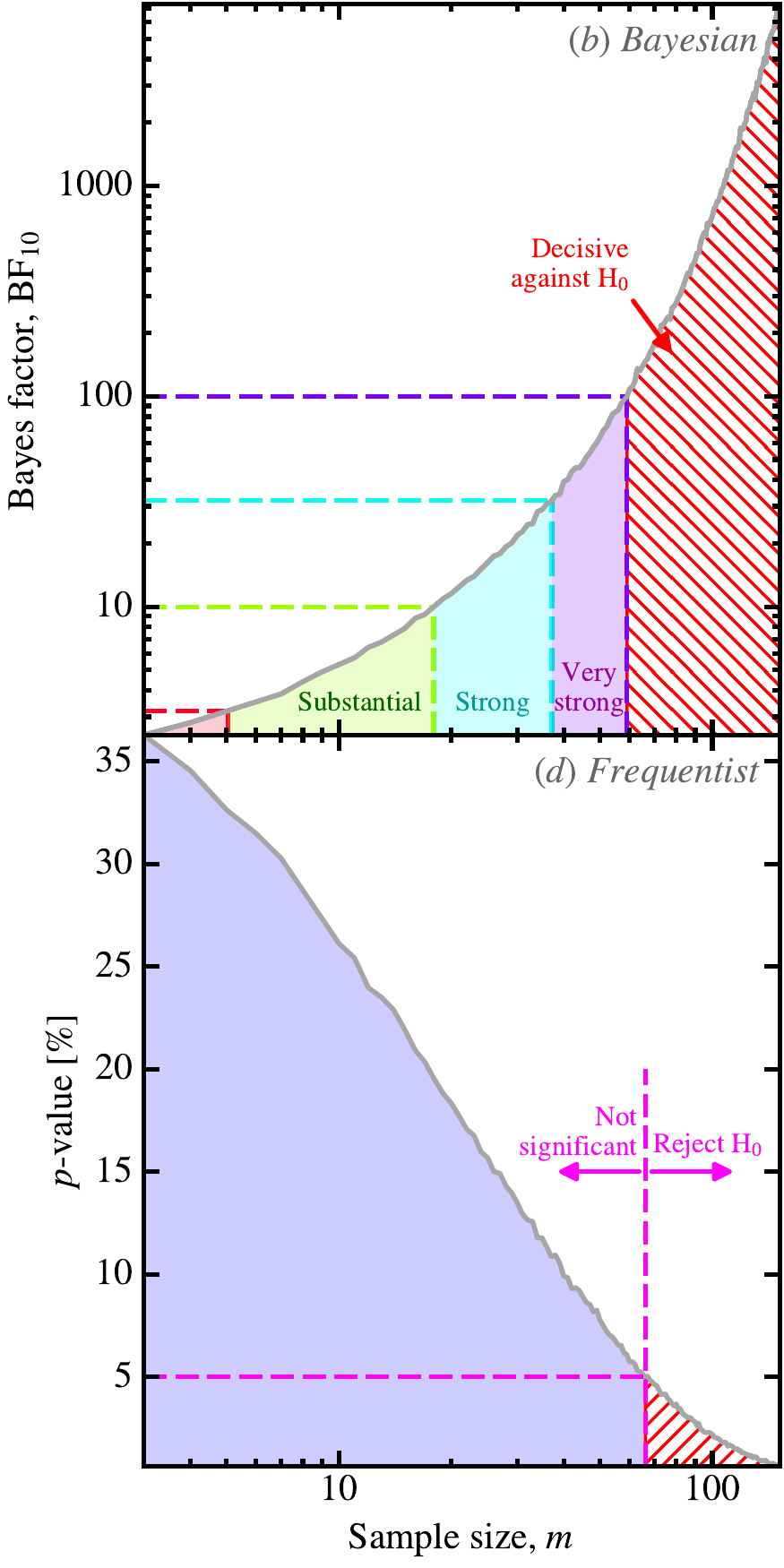} \\
  \end{tabular}
  \newcap{Bayesian and frequentist hypothesis testing}%
         {Panel~\textit{(a)} represents the calculation of the posterior odds
          \refeqp{eq:Bfac}.
          Notice the distribution is centered on the average of the observed 
          fluxes, $\langle F\rangle$, and we consider $F_\sms{test}$ as a
          variable.
          We want to assess the null hypothesis,
          $H_0=(F_\sms{true}\le F_\sms{test})$.
          In the particular example we have plotted, $H_0$ is false.
          The Bayes factor is simply the ratio of the blue and red areas.
          Panel~\textit{(c)} represents \hNHST\ for the same problem.
          Notice, this time, the distribution is centered on $F_\sms{test}$, 
          which is considered fixed, and the observations, $\langle F\rangle$, 
          are considered variable.
          We have plotted the two-tailed $p$-value decisions.
          The alternative to rejection is not acceptance, but the absence of 
          significance.
          The two right panels illustrate the variation of the Bayes factors
          and $p$-values, as a function of the sample size, $m$.
          In these particular examples, we have chosen a fixed $F_\sms{test}=38$
          and $F_\sms{true}=42$, such that $H_0$ is false.
          The uncertainty is $\sigma_\sms{F}=14$.
          The more we have data, the more we can constrain the solution.
          We can see that both Bayesian and frequentist methods conclude the
          right solution ($H_0$ is false), around the same sample size 
          ($m\simeq60-70$ in our case).
          Below this value, both methods are inconclusive.
          The Bayesian credibility scale is however more continuous than the 
          frequentist \hNHST, though.
          The latter tells us our data are useless below $m\simeq70$.
          To avoid stochastic fluctuations, that would make the figure less 
          easy to read, we have randomly drawn 10\,000 samples for each value of
          $m$ and we show the averages.
          \CClicence}
  \label{fig:hyptest}
\end{figure}

\paragraph{Bayesian hypothesis testing.}
Bayesian hypothesis testing consists in computing the \expression{posterior odds} of the two complementary hypotheses:
\begin{equation}
  \underbrace{\frac{\pcond{H_1}{\vec{d}}}{\pcond{H_0}{\vec{d}}}}_\sms{posterior
                                                                     odds}
  =
  \underbrace{\frac{\pcond{\vec{d}}{H_1}}{\pcond{\vec{d}}{H_0}}}_\sms{Bayes 
                                                                     factor}  
  \times
  \underbrace{\frac{\proba{H_1}}{\proba{H_0}}}_\sms{prior odds}.
  \label{eq:Bfac}
\end{equation}
The posterior odds is the ratio of the posterior probabilities of the two hypotheses. 
It is literally the odds we would use for gambling (\eg\ a posterior odd of 3 corresponds to a 3:1 odd in favor of $H_1$).
The important term in \refeq{eq:Bfac} is the \expression{Bayes factor}, usually noted $BF_{10}\equiv\pcond{\vec{d}}{H_1}/\pcond{\vec{d}}{H_0}$.
It quantifies the weight of evidence, brought by the data, \expression{against} the null hypothesis.
It tells us how much our observations changed the odds we had against $H_0$, prior to collecting the data.
\reftab{tab:Bfac} gives a qualitative scale to decide upon Bayes factors.
We see that it is a continuous credibility range going from rejection to confidence.
The posterior of our present example, assuming a wide flat prior, is:
\begin{equation}
  \pcond{F_\star}{F_1,\ldots,F_m} = \frac{1}{\sqrt{2\pi}\sigma_\sms{F}/\sqrt{m}}
   \exp\left(-\frac{1}{2}\frac{(F_\star-\langle 
     F\rangle)^2}{\sigma_\sms{F}^2/m}\right),
\end{equation}
where $\langle F\rangle=\sum_{i=1}^m F_i/m$.
The posterior probability of $H_0=(F_\sms{true}\le F_\sms{test})$ is then simply:
\begin{equation}
  \pcond{H_0}{F_1,\ldots,F_m} 
    = \int_{-\infty}^{F_\sms{test}}\pcond{F_\star}{F_1,\ldots,F_m}\ddiff F_\star
    = \frac{1}{2}\left[1+
      \erf\left(\frac{1}{\sqrt{2}}\frac{F_\sms{test}-\langle 
        F\rangle}{\sigma_\sms{F}/\sqrt{m}}\right)\right].
  \label{eq:postodd}
\end{equation}
It is represented in \refsubfig{fig:hyptest}{a}.
This \hPDF\ is centered in $\langle F\rangle$, since, as usual in the Bayesian approach, it is conditional on the data.
\refsubfig{fig:hyptest}{a} shows the complementary posteriors of $H_0$ (red) and $H_1$ (blue), which are the incomplete integrals of the \hPDF.
When we vary $F_\sms{test}$, the ratio of the two posteriors, $BF_{10}$, changes.
Assuming we have chosen a very wide, flat prior, such that $\proba{H_1}/\proba{H_0}\simeq1$, the Bayes factor becomes:
\begin{equation}
  BF_{10}\simeq\frac{1}{\pcond{H_0}{F_1,\ldots,F_m}}-1.
  \label{eq:BF10}
\end{equation}
\refsubfig{fig:hyptest}{b} represents the evolution of the Bayes factor as a function of the sample size, $m$.
In this particular simulation, $H_0$ is false.
We see that, when $m$ increases, we accumulate evidence against $H_0$, going through the different levels of \reftab{tab:Bfac}.
The evidence is decisive around $m\simeq60$, here.
\begin{table}[htbp]
  \setlength\arrayrulewidth{2pt}
  \arrayrulecolor{white}
  \centering 
  \begin{tabularx}{0.7\linewidth}{|>{\columncolor{coltabcell}}l%
                                   |>{\columncolor{coltabcell}}X|}
    \hline
      \rowcolor{coltabhead}
      \textbf{Bayes factor, BF$\bm{_{10}}$} 
      & \textbf{Strength of evidence against H$_0$} \\
    \hline
      1--3.2 & Barely worth mentioning \\
    \hline
      3.2--10 & Substantial \\
    \hline
      10--32 & Strong \\
    \hline
      32--100 & Very strong \\
    \hline
      >100 & Decisive \\
    \hline
  \end{tabularx}
  \newcap{Jeffreys strength of evidence scale}%
         {This scale translates the value of a Bayes factor in a qualitative 
          decision \citep{jeffreys39}.
          Below one, we consider $BF_{01}$ instead, and discuss the evidence in
          favor of $H_0$.}
  \label{tab:Bfac}  
\end{table}

\paragraph{Frequentist hypothesis testing.}
Instead of computing the credibility of $H_1$ against $H_0$, the frequentist approach relies on the potential rejection of $H_0$.
It is called \expression{Null Hypothesis Significance Test} \citep[\hNHST; \eg][]{ortega17}.
It consists in testing if the observation average, $\langle F\rangle$, could have confidently been drawn out of a population centered on $F_\sms{test}$. 
In that sense, it is the opposite of the Bayesian case.
This is demonstrated in \refsubfig{fig:hyptest}{c}.
We see that the distribution is centered on $F_\sms{test}$, which is assumed fixed, and the observations, $\langle F\rangle$, are assumed variable.
Since frequentists can not assign a probability to $F_\sms{test}$, they estimate a $p$-value, that is the probability of observing the data at hands, assuming the null hypothesis is true.
It is based on the following statistics: 
\begin{equation}
  t_m\equiv\frac{\langle F\rangle-F_\sms{test}}{\sigma_\sms{F}/\sqrt{m}}.
\end{equation}
The $p$-value of this statistics is then simply:
\begin{equation}
  p_0 = \frac{1}{2}\left[1+\erf\left(\frac{t_m}{\sqrt{2}}\right)\right].
\end{equation}
Notice it is identical to \refeq{eq:postodd}, but different from the Bayes factor \refeqp{eq:BF10}.
It however does not mean the same thing.
A significance test at $p_0=0.05$ does not tell us that the probability that the null hypothesis is $5\,\%$.
It means that the null hypothesis will be rejected $5\,\%$ of the time\footnote{The acceptance of a wrong hypothesis (false positive) is called \citengl{type I error}, whereas the rejection of a correct hypothesis (false negative) is called \citengl{type II error}.}.
\hNHST\ decision making is represented in \refsubfig{fig:hyptest}{c}.
It shows one of the most common misconceptions about \hNHST: the absence of rejection of $H_0$ does not mean that we can accept it.
It just mean that the results are not significant.
Accepting $H_0$ requires rejecting $H_1$, and vice versa.
\refsubfig{fig:hyptest}{d} shows the effect of sample size on the $p$-value, using the same example as we have discussed for the Bayesian case.
The difference is that the data are not significant until $m\simeq70$.

\paragraph{The Jeffreys-Lindley's paradox.}
Although the method and the interpretations are different, Bayesian and frequentist tests give consistent results, in numerous applications.
There are however particular cases, where both approaches are radically inconsistent.
This ascertainment was first noted by \citet{jeffreys39} and popularized by \citet{lindley57}.
\citet{lindley57} demonstrated the discrepancy on an experiment similar to the example we have been discussing in this section, with the difference that a \expression{point} null hypothesis is tested: $H_0=(F_\sms{true}=F_\sms{test})$.
\citet{lindley57} shows that there are particular cases, where the posterior probability of $H_0$ is $1-\alpha$, and $H_0$ is rejected at the $\alpha$ level, at the same time.
This \citengl{statistical paradox}, known as the \expression{Jeffreys-Lindley's paradox} has triggered a vigorous debate, that is still open nowadays \citep[\eg][]{robert14}.
The consensus about the paradox is that \expression{there is no paradox}.
The discrepancy simply arises from the fact that both approaches answer different questions, as we have been illustrating at several occasions in this chapter, and that these different interpretations can sometimes be inconsistent.

\paragraph{The recent controversy about frequentist significance tests.}
\hNHST\ has recently been at the center of an important controversy across all empirical sciences.
We have already discussed several of the issues with frequentist significance tests.
Let's summarize them here \citep[\eg][]{ortega17}.
\begin{enumerate}
  \item Frequentist tests are conditional on model parameters and thus consider
    data that have not actually been observed.
    The general frequentist approach is difficult to grasp, even for advanced
    statisticians.
    It can easily lead to false interpretations.
  \item \hNHST\ is prone to overestimates and can state effects even if none 
    exist.
    If we repeat an experiment a sufficient number of times, we will always
    end up rejecting $H_0$.
    This potentially leads to a large number of false positives.
  \item There is a variety of statistics that one can test, and they are not 
    all going to give the same result.
    In addition, the significance level is subjective and there are no clear
    rules how to choose the $p$-value ($p=0.05$, $p=0.01$, \etc).
    There is therefore some subjectivity in the frequentist approach.
    It is not in the prior, it is in the significance assessment.
\end{enumerate}
\expression{Data dredging} or \expression{$p$-hacking} has come into the spotlight during the last twenty years, although it was known before \citep[\eg][]{smith02,simmons11,head15}.
It points out that numerous scientific studies could be wrong, and several discoveries could have been false positives.
This is particularly important in psychology, medical trials, \etc, but could affect any field using $p$-values.
In 2016, the \expression{American Statistical association} published a \citengl{Statement on statistical significance and $p$-values} \citep{wasserstein16}, saying that: \citengl{widespread use of 'statistical significance' (generally interpreted as 'p<0.05') as a license for making a claim of a scientific finding (or implied truth) leads to considerable distortion of the scientific process}.
They suggested \citengl{moving toward a 'post p<0.05' era}.
While some recommendations have been proposed to use $p$-values in a more controlled way \citep[\eg][]{simmons11}, by deciding the sample size and significance level before starting the experiment, some researchers have suggested abolishing \hNHST\ \citep[\eg][]{loftus96,anderson00}.
Several journals have stated that they will no longer publish articles reporting $p$-values (\eg\ \expression{Basic \&\ Applied Social Psychology}, in 2015, and \expression{Political Analysis}, in 2018).
\takeaway{Frequentist $p$-values are to be used with caution.}

    \subsubsection{Pros and Cons of the two Approaches}

We finish this section by summarizing the advantages and inconveniences of both approaches.
This comparison is synthesized in \reftab{tab:BvF}.
\newcommand{\PRO}{\textcolor{blue}{\fbox{PRO}}}
\newcommand{\CON}{\textcolor{magenta}{\fbox{CON}}}
\begin{table}[htbp]
  \centering
  \setlength\arrayrulewidth{2pt}
  \arrayrulecolor{white}
  \begin{tabularx}{\linewidth}{|>{\columncolor{coltabcell}}X%
                                |>{\columncolor{coltabcell}}X|}
    \hline
      \rowcolor{coltabhead}
      \textbf{Bayesian approach} & \textbf{Frequentist approach} \\
    \hline
      \CON\ choice of prior is subjective & 
      \PRO\ likelihood is not subjective \\
    \hline
      \PRO\ can account for non-Gaussian errors, nuisance parameters, complex 
      models \&\ prior information & 
      \CON\ very limited in terms of the type of noise, the
      complexity of the model \&\ can not deal with nuisance parameters \\
    \hline
      \PRO\ the posterior makes sense (conditional on the data) \&\ is easy to 
        interpret &
      \CON\ samples non-observed data, arbitrary choice of estimator
        \&\ $p$-value \\
    \hline
      \PRO\ probabilistic logic \implic\ continuum between skepticism \&\ 
        confidence &
      \CON\ boolean logic \implic\ a proposition is either true or false,
        which leads to false positives \\
    \hline
      \PRO\ based on a master equation (Bayes' rule) \implic\ easier to learn 
        \&\ teach &
      \CON\ difficult to learn \&\ teach (collection of \textit{ad hoc} cooking 
        recipes)\\
    \hline
      \CON\ heavy computation & \PRO\ fast computation \\
    \hline
      \PRO\ works well with small samples \&\ heterogeneous data sets &
      \CON\ does not work well with small samples, can not mix samples \&\ 
        require fixing the sample size and significance level before 
        experimenting \\
    \hline
      \PRO\ holistic \&\ flexible: can account for all data \&\ theories &
      \CON\ strict: can account only for data related to a particular 
        experiment \\
    \hline
      \PRO\ conservative & \CON\ can give ridiculous answers \\
    \hline
  \end{tabularx}
  \newcap{Pros and cons of the Bayesian and frequentist methods}{}
  \label{tab:BvF}
\end{table}

\paragraph{Hypotheses and information that can be taken into account.}
The two methods diverge on what information can be included in the analysis.
\begin{description}
  \item[Bayesian] models can account for a maximum amount of information:
    \begin{itemize}
      \item all types of noise and uncertainties, whether non-Gaussian, 
        partially or fully correlated (\cf\ \refsec{sec:BvFasym});
      \item nuisance parameters, that are parameters necessary to estimate the 
        model, but whose values are not relevant (we will give an example in
        \refsec{sec:nuisance});
      \item complex, non-linear models with more parameters than observational 
        constraints (\cf\ \refsec{sec:postmcmc});
      \item any kind of prior information (\cf\ \refsec{sec:BvFprior}).
    \end{itemize}
  \item[Frequentist] analysis is limited in that way:
    \begin{itemize}
      \item complex, correlated noise is difficult to include in the likelihood;
      \item nuisance parameters can not be included and we can not have more
        parameters than observations;
      \item complex models, with a lot of degeneracies fail frequentist 
        approaches, such as maximum likelihood methods;
      \item no prior information can be included.
    \end{itemize}
\end{description}
In favor of the frequentist approach, we can note that the likelihood is perfectly objective, whereas the choice of the prior is subjective.
The subtlety is however that this choice is \expression{subjective}, as it depends on the knowledge we believe we have prior to the observation, but it is not \expression{arbitrary}, as a prior can be rationally constructed.
In addition, when the strength of evidence is large, the prior becomes unimportant.
The prior is important only when the data are very noisy or unconvincing.
In that sense, the prior does not induce a \expression{bias of confirmation}.

\paragraph{Analysis and interpretation.}
As we have seen throughout \refsec{sec:BvF}, the point of view of the two approaches is very different.
\begin{description}
  \item[Inference] is performed on the posterior, in the Bayesian approach.
    It gives the probability of the parameters, knowing the data.
    It makes sense and is easy to interpret.
    On the contrary, frequentists sample data that have not actually been 
    observed and base their results on arbitrary choices of statistics and 
    estimators.
    The results are difficult to interpret, as they consist in describing what
    could be the observations, for a given set of model parameters (\cf\ 
    \refsec{sec:phacking}).
  \item[The underlying logic] 
    is probabilistic, in the Bayesian approach.
    There is a continuum between skepticism and confidence that makes any data
    worth taking into account.
    Bayes factors quantify the strength of evidence brought by these data
    (\cf\ \refsec{sec:Bayesfactor}).
    On the contrary, frequentist logic is \expression{Platonic}, a proposition
    is either true or false.
    With real-life uncertainties and stochasticity, this leads to false 
    positives.
    By refusing to assign probabilities to parameters and hypotheses, 
    frequentists rely on $p$-values, while those are only one particular tool
    of the Bayesian analysis (\eg\ \refsec{sec:postmcmc}).
    Credibility is progressive whereas significance is dichotomic.
  \item[Learning and teaching]
    of the Bayesian method is considerably easier, because it is based on a 
    master equation, Bayes' rule.
    All Bayesian problems start with \refeq{eq:Bayes}, which is developed to 
    account for all the details we are modeling.
    On the contrary, frequentists methods are a collection of \textit{ad hoc} 
    cooking recipes, whose derivation is often obscure 
    \citep[\eg\ R.~Fisher's book, \citengl{Statistical methods for research 
    workers};][]{fisher25}.
  \item[Computation] 
   of Bayesian problems is intensive, as sampling the posterior is challenging
   (\cf\ \refsec{sec:mcmc}).
   One of the advantages of frequentist methods is that they are usually fast.
   Even a Bayesian can use them, for instance, to find good \hMCMC\ 
   starting points \citep[\eg][Sect.\ 4.2.1]{galliano18a} or compute quick 
   estimators \citep[\eg][Appendix F.2]{galliano21}.
\end{description}

\paragraph{Overall applicability.}
In practice, choosing one approach over the other depends on the situation.
There are however a lot of arguments in favor of the Bayesian point of view.
\begin{description}
  \item[Sample size and data collection] are one of the major issues with 
    frequentist methods.
    We have seen that \hNHST\ was problematic in that aspect, and that the 
    stopping condition of an experiment could bias its significance.
    It is recommended to use large samples, and decide of the size and 
    significance level, before conducting the measures.
    Consequently, contrary to Bayesians, frequentists can not
    \begin{inlinelist}
      \item analyze partial data sets, as the stopping point could be 
        instrumental in forcing one outcome over another (concept of 
        $p$-hacking discussed in \refsec{sec:phacking}), or
      \item combine heterogeneous data sets, as the limiting frequency would 
         not have any sense.
    \end{inlinelist}
  \item[Consistency] of the results is also an issue with the frequentist 
    approach, as it can only account for data related to a particular
    experiment.
    On the contrary, the Bayesian approach is more flexible.
    We will even discuss in \refsec{sec:prior} that it is potentially holistic.
    It is at the same time more conservative, as the prior tends to prevents 
    aberrant results, while we have seen that frequentist methods can give 
    ridiculous answers (\cf\ \refsec{sec:BvFasym}).
\end{description}
\takeaway{For all these reasons, the Bayesian approach is more well-suited for 
          most problems encountered in empirical sciences.}

\section{Bayesianism, an Alternative to Popper's Scientific 
          Method}

The Bayesian and frequentist approaches lead to radically different epistemological points of view, that have important consequences on the way we study \hISD.
We start by briefly brushing the history of the competition between these two systems.
We then discuss their consequences on the scientific method.

  \subsection{Bayes Formula Throughout History}

The History of the introduction of probability in sciences and the subsequent competition between Bayesians and frequentists is epic.
The book of \citet{mcgrayne11} gives an invaluable overview of this controversy, that started two centuries ago.

    \subsubsection{The Origins}

\begin{figure}[htbp]
  \begin{tabular}{ccc}
    \includegraphics[width=0.31\textwidth]{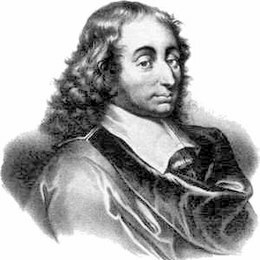} &
    \includegraphics[width=0.31\textwidth]{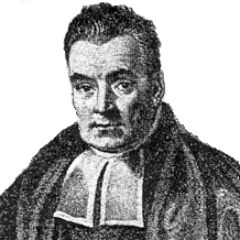} &
    \includegraphics[width=0.31\textwidth]{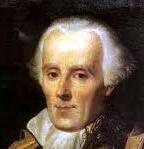} \\
    \textit{(a)} Blaise \familyname{Pascal} &
    \textit{(b)} Thomas \familyname{Bayes} &
    \textit{(c)} Pierre-Simon \familyname{Laplace} \\
    (1623--1662) & (1702--1761) & (1749--1827) \\
  \end{tabular}
  \newcap{The probability pioneers}%
         {\uline{Credit:} 
          \begin{inlinelistalph}
            \item \href{https://commons.wikimedia.org/wiki/File:Blaise_pascal.jpg}%
              {Wikipedia}, public domain;
            \item \href{https://commons.wikimedia.org/wiki/File:Thomas_Bayes.gif}%
              {Wikipedia}, public domain;
            \item \href{https://commons.wikimedia.org/wiki/File:Pierre-Simon,_marquis_de_Laplace_(1745-1827)_-_Gu\%C3\%A9rin.jpg}{Wikipedia}, public domain.
          \end{inlinelistalph}}
  \label{fig:peopleBayes}
\end{figure}

\paragraph{The emergence of the concept of probability.}
In antique societies, randomness was interpreted as the language of the gods.
\citet{hacking06} argues that the notion of probability emerged around 1660, in western Europe.
Before this date, \citengl{probable} meant only \citengl{worthy of approbation} or \citengl{approved by authority}.
In a few years, during the Renaissance, there was a transition of the meaning of \citengl{probable} from \citengl{commonly approved} to \citengl{probed by evidence}, what Gaston \familyname{Bachelard} would have called an \expression{epistemological break}.
The time was ready for the idea. 
The \expression{Thirty Years' War} (1618--1648), which had caused several millions of deaths throughout western Europe, had just ended.
It consolidated the division into Catholic and Lutheran states of a continent that had been religiously homogeneous for almost a thousand years.
\citengl{Probabilism is a token of the loss of certainty that characterizes the Renaissance, and of the readiness, indeed eagerness, of various powers to find a substitute for the older canons of knowledge. 
Indeed the word 'probabilism' has been used as a name for the doctrine that certainty is impossible, so that probabilities must be relied on} \citep[][page 25]{hacking06}.
The first book discussing the concept of probability, applied to games of fortune, was published in 1657 by Christiaan \familyname{Huygens} \citep{huygens1657}.
It is however Blaise \familyname{Pascal} (\cf\ \refsubfig{fig:peopleBayes}{a}) who is considered the pioneer in the use of probability as a quantification of beliefs.
His \expression{wager}\footnote{Pascal's wager states that it is rational to act as if God existed.
If God indeed exists, we will be rewarded, which is a big win.
If He doesn't, we will have only renounced to some material pleasures, which is not a dramatic loss.} is known as the first example of decision theory \citep{pascal1670}.

\paragraph{The discovery of Bayes.}
Thomas \familyname{Bayes} (\cf\ \refsubfig{fig:peopleBayes}{b}) was an XVIII$^\sms{th}$ English Presbyterian minister.
Coming from a nonconformist family, he had read the work of Isaac \familyname{Newton}, David \familyname{Hume} and Abraham \familyname{de Moivre} \citep{mcgrayne11}.
His interest in game theory led him to imagine a thought experiment.
\begin{description}
  \item[His thought experiment] was developed between 1746 and 1749.
    He was trying to infer the position of a ball on a pool table behind him, 
    that he could not see.
    His idea was to be able to start from a guess and refine it using some 
    information.
    \begin{enumerate}
      \item His assistant would throw on the table a first ball, 
        whose position is to be inferred.
      \item His assistant would then throw a second ball and tell him if it 
        landed on the left or the right of the first one.
      \item This procedure would be repeated until Bayes could infer the 
        quadrant were the first ball is.
    \end{enumerate}
    He derived \refeq{eq:Bayes} to solve this problem.
  \item[The essay] presenting his formula \citep{bayes1763} was published after 
    his death by Richard \familyname{Price}.
    Bayes defined the probability of an event as \citengl{the ratio between the 
    value at which an expectation depending on the happening of the event ought 
    to be computed, and the value of the thing expected upon its happening}.
    Richard \familyname{Price} added that his formula provides the 
    \expression{probability of causes} and can thus be applied to prove the 
    existence of God.
\end{description}

\paragraph{The contribution of Laplace.}
Pierre-Simon \familyname{Laplace} (\cf\ \refsubfig{fig:peopleBayes}{c}) was the son of a small estate owner, in Normandie.
His father pushed him towards a religious career, that led him to study theology.
He however quit at age 21 and moved to Paris, where he met the mathematician Jean \familyname{Le Rond d'Alembert}, who helped him to get a teaching position.
Laplace then had a successful scientific and political career \citep[\cf][for a complete biography]{hahn05}.
Among his many other scientific contributions, Laplace is the true pioneer in the development of statistics using Bayes' rule.
Some authors even argue that we should call the approach presented in \refsec{sec:BvF} \citengl{Bayesian-Laplacian} rather than simply \citengl{Bayesian}.
After having read the memoir of Abraham \familyname{de Moivre}, he indeed understood that probabilities could be used to quantify experimental uncertainties.
His 1774 memoir on \citengl{the probability of causes by events} \citep{laplace1774} contains the first practical application of Bayes' rule.
His \expression{rule of succession}, giving the probability of an event knowing how many times it happened previously, was applied to give the probability that the Sun will rise again.
Laplace rediscovered Bayes' rule.
He was only introduced to Bayes' essay in 1781, when Richard \familyname{Price} came to Paris.
Laplace had a Bayesian conception of probabilities: \citengl{in substance, probability theory is only common sense reduced to calculation; it makes appreciate with accuracy what just minds can feel by some sort of instinct, without realizing it} \citep{laplace1812}.

    \subsubsection{The Frequentist Winter}
    \label{sec:frequentistwinter}

\begin{figure}[htbp]
  \begin{tabular}{ccc}
    \includegraphics[width=0.31\textwidth]{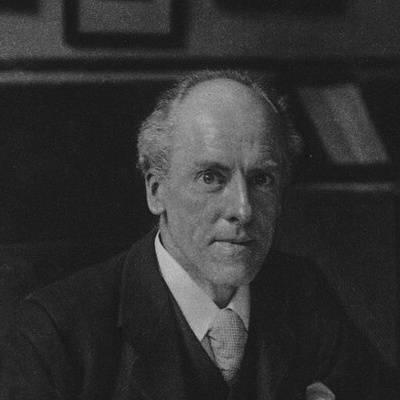} &
    \includegraphics[width=0.31\textwidth]{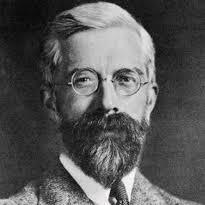} &
    \includegraphics[width=0.31\textwidth]{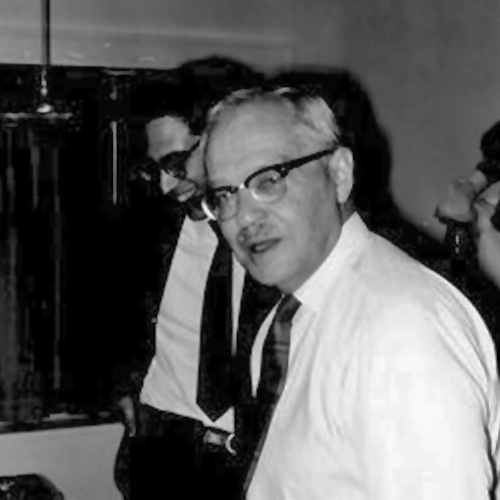} \\
    \textit{(a)} Karl \familyname{Pearson} &
    \textit{(b)} Ronald \familyname{Fisher} &
    \textit{(c)} Jerzy \familyname{Neyman} \\
    (1857--1936) & (1890--1962) & (1894--1981) \\
  \end{tabular}
  \newcap{The frequentist promoters}%
         {\uline{Credit:} 
          \begin{inlinelistalph}
            \item \href{https://commons.wikimedia.org/wiki/File:Karl_Pearson.jpg}%
              {Wikipedia}, public domain;
            \item \href{https://www.bibmath.net/bios/index.php?action=affiche&quoi=fisher}%
              {Bibmath}, public domain;
            \item \href{https://fr.wikipedia.org/wiki/Jerzy_Neyman\#/media/Fichier:Jerzy_Neyman2.jpg}{Wikipedia}, licensed under 
             \href{https://creativecommons.org/licenses/by-sa/2.0/}{CC BY-SA 2.0 DE}.
          \end{inlinelistalph}}
  \label{fig:peopleFreq}
\end{figure}

\paragraph{The rejection of Laplace's work.}
The frequentist movement was initiated by British economist John Stuart \familyname{Mill}, only ten years after the death of Laplace.
There were several reasons for this reaction \citep{loredo90,mcgrayne11}.
\begin{itemize}
  \item The idea that probability should represent a \expression{degree of 
    plausibility} seemed too vague to provide the foundation for a mathematical 
    theory.
    People also realized there was no clear way to assign priors.
  \item The computation of Bayesian solutions by hand was crippling.
  \item Laplace was despised in England for his support to Napoléon.
\end{itemize}
Mill's disdain for the Bayesian approach was unhinged:
\citengl{a very slight improvement in the data, by better observations, or by taking into fuller consideration the special circumstances of the case, is of more use than the most elaborate application of the calculus of probabilities founded on the data in their previous state of inferiority. 
The neglect of this obvious reflection has given rise to misapplications of the calculus of probabilities which have made it the real opprobrium of mathematics.} \citep{mill1843}.
The early anti-Bayesian movement was led by English statistician Karl \familyname{Pearson} (\cf\ \refsubfig{fig:peopleFreq}{a}).
Pearson developed:
\begin{inlinelist}
  \item the chi-squared test;
  \item the standard-deviation;
  \item the correlation coefficient;
  \item the $p$-value;
  \item the \expression{Principal Component Analysis} (\hPCA).
\end{inlinelist}
His book, \citengl{The Grammar of Science} \citep{pearson1892}, was very influential, in particular to the young Albert \familyname{Einstein}.
Despite these great contributions, Pearson was a social Darwinist and a eugenicist.

\paragraph{The golden age of frequentism (1920-1930).}
Ronald \familyname{Fisher} (\cf\ \refsubfig{fig:peopleFreq}{b}) followed the way opened by Pearson.
He is the most famous representative of the frequentist movement.
He developed:
\begin{inlinelist}
  \item the maximum likelihood;
  \item \hNHST;
  \item the F-distribution and the F-test.
\end{inlinelist}
His 1925 book, \citengl{Statistical Methods for Research Workers} \citep{fisher25}, was widely used in academia and industry.
Despite the criticism we can address to the frequentist approach, Fisher's contributions gave guidelines to rigorously interpret experimental data, that brought consistency to science.
Fisher, who was like Pearson a eugenicist, was also paid as a consultant by the \citengl{Tobacco Manufacturer's Standing Committee}.
He spoke publicly against a 1950 study showing that tobacco causes lung cancer, by resorting to \citengl{correlation does not imply causation} \citep{fisher57}.
Besides Pearson and Fisher, Jerzy \familyname{Neyman} (\cf\ \refsubfig{fig:peopleFreq}{c}) was also a prominent figure of frequentism at this time.
These scientists, also known for their irascibility, made sure that nobody revived the methods of Laplace.
\citet{mcgrayne11} estimates that this golden era culminated in the 1920s-1930s.

\begin{figure}[htbp]
  \begin{tabular}{ccc}
    \includegraphics[width=0.31\textwidth]{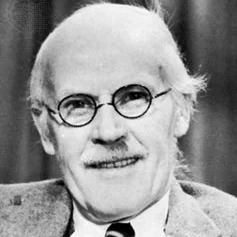} &
    \includegraphics[width=0.31\textwidth]{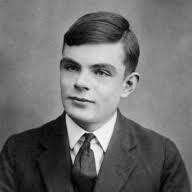} &
    \includegraphics[width=0.31\textwidth]{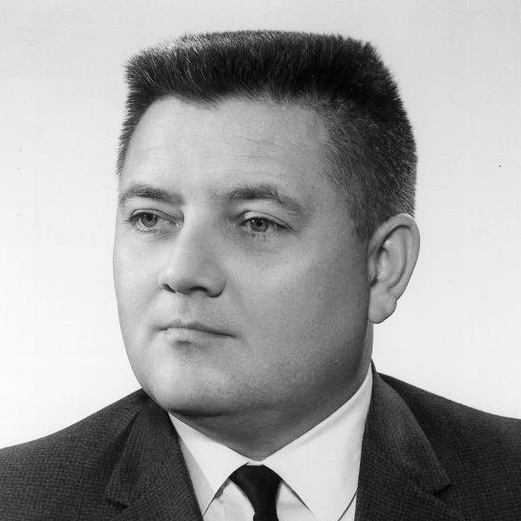} \\
    \textit{(a)} Harold \familyname{Jeffreys} &
    \textit{(b)} Alan \familyname{Turing} &
    \textit{(c)} Edwin Thompson \familyname{Jaynes} \\
    (1891--1989) & (1912--1954) & (1922--1998) \\
  \end{tabular}
  \newcap{The Bayesian resistance}%
         {\uline{Credit:} 
          \begin{inlinelistalph}
            \item \href{https://www.britannica.com/biography/Harold-Jeffreys}%
                       {Encyclopedia Britannica}, used for non-commercial, educational 
              purpose;
            \item \href{https://commons.wikimedia.org/wiki/File:Alan_Turing_Aged_16.jpg}%
              {Wikipedia}, public domain;
            \item \href{https://commons.wikimedia.org/wiki/File:ETJaynes1.jpg}{Wikipedia},
              public domain.
          \end{inlinelistalph}}
  \label{fig:peopleResist}
\end{figure}

\paragraph{The Bayesian resistance.}
Several prominent scientists, who were not intimidated by Fisher and his colleagues, perpetuated the Bayesian approach \citep{mcgrayne11}.
Among them, we can cite the following two.
\begin{description}
  \item[Harold \familyname{Jeffreys}] (\cf\ \refsubfig{fig:peopleResist}{a})
    was a British geophysicist and mathematician.
    In 1926, he performed a Bayesian analysis of earthquake records and 
    inferred that the Earth had a liquid core.
    This discovery could not have been possible with the frequentist approach,
    as the data were very scarce.
    Jeffreys initiated the Bayesian revival and was an early critic of \hNHST.
    His book, \citengl{Theory of probability} \citep{jeffreys39}, popularized 
    the use of Bayes factors, as we have seen in \refsec{sec:Bayesfactor}.
  \item[Alan \familyname{Turing}] (\cf\ \refsubfig{fig:peopleResist}{b})
    was the founder of theoretical computer science and artificial intelligence.
    He also brilliantly put into practice Bayesian techniques during World
    War II.
    He secretly worked at Bletchley Park, near London, to decode the 
    communications between German U-boats, that were using the cryptographic 
    \expression{Enigma} machine.
    Turing built a mechanical computer, called \citengl{The Bomb}, which he used
    to test combinations.
    He used Bayesian priors to reduce the number of combinations, looking for
    frequent German words and meteorological terms.
    He even developed a unit quantifying the weight of evidence (Bayes factor), 
    named the \citengl{ban}, after the city of Banbury where punch cards were 
    printed.
\end{description}

    \subsubsection{The Bayesian Renaissance}

\paragraph{After World War II.}
The first computers were built during the war.
Bayesian techniques were now becoming feasible.
Their use rapidly increased in the early 1950s.
\citet{mcgrayne11} note a few of them.
\begin{description}
  \item[Competitive businesses,] that usually favor pragmatic solutions over 
    ideology, turned naturally to the Bayesian approach.
    Arthur \familyname{Bailey} was a pioneer in applying Bayes' rule to estimate
    insurance premiums.
    This was the only way to calculate the probability of catastrophic events, 
    that had never happened before.
    This is the type of situations were frequentist methods are unusable.
    Howard \familyname{Raiffa} and Robert \familyname{Schlaifer} taught Bayes'
    rule for business and decision-making.
  \item[Nuclear security] is another example of a discipline requiring to 
    estimate the probability of rare, unprecedented events.
    Frequentist studies concluded that nuclear plant incidents were unlikely, 
    but would be catastrophic if they happened.
    In the 1970s, Norman \familyname{Rasmussen} estimated, in a Bayesian way,
    that it was the opposite: they were likely, but not necessarily 
    catastrophic.
    The Three Mile Island incident (1979) proved him right.
    In another area, Bayesian search algorithms were used to find lost nuclear 
    bombs and Russian nuclear submarines.
  \item[Epidemiological] studies, using Bayesian techniques, were pioneered by 
    Jerome \familyname{Cornfield}, who ridiculed Fisher's attempt at 
    minimizing the link between smoking and lung cancer.
  \item[In academia,] Dennis \familyname{Lindley} and Jimmie 
    \familyname{Savage} were actively promoting Bayesian methods and showing 
    the limitations of frequentist techniques, claiming that \citengl{Fisher is 
    making Bayesian omelet without breaking Bayesian eggs}.
\end{description}

\paragraph{The great numerical leap forward.}
In the 1970s, the increasing power of computers opened new horizons to the Bayesian approach.
The Metropolis-Hastings algorithm \citep[\cf\ \refsec{sec:mcmc};][]{hastings70} provided a fast, easy-to-implement method, which rendered Bayesian techniques more attractive.
Gibbs sampling \citep[\cf\ \refsec{sec:mcmc};][]{geman84}, which can be used to solve complex problems, put Bayes' rule into the spotlight.
We can note the following achievements.
\begin{description}
  \item[The human genome] was decoded using Bayesian methods \citep{beaumont04}.
  \item[In astrophysics,] Bayesian techniques were first applied to analyze the
    neutrino flux from SN$\,$1987$\,$A \citep{loredo89}.
\end{description}
At the same time, Edwin \familyname{Jaynes} (\cf\ \refsubfig{fig:peopleResist}{c}) was working at solidifying the mathematical foundations of the Bayesian approach.
It culminated in his posthumous book, \citengl{Probability Theory: The Logic of Science} \citep{jaynes03}.

\paragraph{Bayesian techniques, nowadays.}
Looking at the contemporary literature, it appears that Bayesians have won over frequentists\footnote{There are still a few frequentist trolls roaming Wikipedia's mathematical pages.}.
In a lot of cases, this is however only a fashion trend due to the fact that the word \citengl{Bayesian} became hip in the 2010s, in astrophysics.
There are already misuses of Bayesian methods.
This is unavoidable.
This might result from the fact that there is still a generation of math teachers and the majority of statistical textbooks ignoring the Bayesian approach.
The difference with $p$-hacking is that Bayesian-hacking is easier to spot, because interpreting posterior distributions is less ambiguous than \hNHST.
The supremacy of Bayesian techniques is ultimately demonstrated by the success of \expression{Machine-Learning} (\hML).
\hML\ has Bayesian foundations.
It is a collection of probabilistic methods.
Using \hML\ can be seen as performing posterior inference, based on the evidence gathered during the training of the neural network.

  \subsection{Bayesian and Popperian Epistemologies}

The following epistemological considerations have been expressed in several texts \citep[\eg][]{good75,loredo90,jaynes03,hoang20}.
The book of \citet{jaynes03} is probably the most rigorous on the subject, while the book of \citet{hoang20} provides an accessible overview.

    \subsubsection{The Epistemological Debate at the Beginning of 
                        the XX$^\sms{th}$ Century}

Epistemology treats several aspects of the development of scientific theories, from their imagination to their validation.
We will not discuss here how scientists can come up with ground-breaking ideas.
We will only focus on how a scientific theory can be experimentally validated.

\begin{figure}[htbp]
  \begin{tabular}{ccc}
    \includegraphics[width=0.31\textwidth]{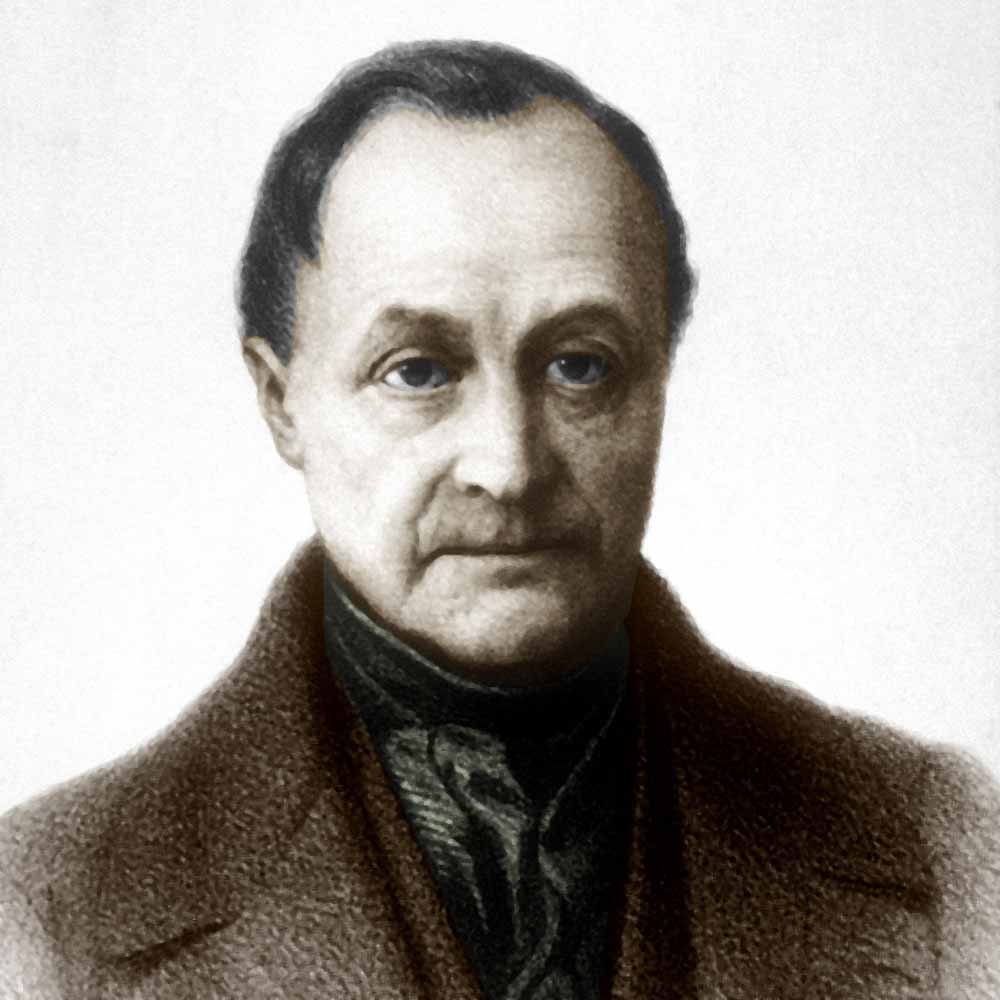} &
    \includegraphics[width=0.31\textwidth]{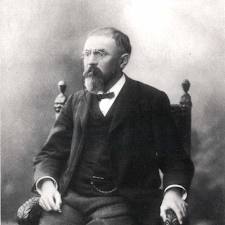} &
    \includegraphics[width=0.31\textwidth]{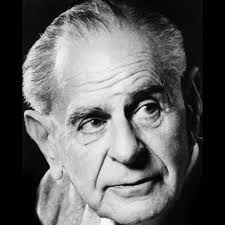} \\
    \textit{(a}) Auguste \familyname{Comte} &
    \textit{(b)} Henri \familyname{Poincaré} &
    \textit{(c)} Karl \familyname{Popper} \\
    (1798--1857) & (1854--1912) & (1902--1994) \\
  \end{tabular}
  \newcap{Three figures of modern epistemology}%
         {\uline{Credit:} 
          \begin{inlinelistalph}
            \item \href{https://commons.wikimedia.org/wiki/File:Auguste_Comte.jpg}%
                       {Wikipedia}, public domain;
            \item \href{https://www.estrepublicain.fr/education/2011/06/16/henri-poincare-savant-universel}%
              {Est Republicain}, public domain;
            \item \href{https://commons.wikimedia.org/wiki/File:Karl_Popper.jpg}%
                       {Wikipedia}, not licensed.
          \end{inlinelistalph}}
  \label{fig:peopleEpist}
\end{figure}

\paragraph{Scientific positivism.}
The epistemological point of view of Auguste \familyname{Comte} (\cf\ \refsubfig{fig:peopleEpist}{a}) had a considerable influence on the XIX$^\sms{th}$ century epistemology, until the beginning of the XX$^\sms{th}$ century.
Comte was a French philosopher and sociologist, who developed a complex classification of sciences and theorized their role in society.
What is interesting for the rest of our discussion is that he was aiming at \expression{demarcating} sciences from theology and metaphysics.
He proposed that we need to renounce to understand the absolute \expression{causes} (why), to focus on the mathematical \expression{laws} of nature (how).
In that sense, his system, called \expression{positivism}, is not an empiricism \citep[\eg][]{grange02}.
Comte stressed that a theoretical framework is always necessary to interpret together different experimental facts. 
His approach is a reconciliation of empiricism and idealism, where both viewpoints are necessary to make scientific discoveries.
Positivism is not scientism, either.
Comte's view was, in substance, that science provides a knowledge that is rigorous and certain, but at the same time only partial and relative.

\paragraph{Conventionalism and verificationism.}
At the beginning of the XX$^\sms{th}$ century, two complementary epistemological points of view were debated.
\begin{description}
  \item[Conventionalism] 
    was represented, in physics and mathematics, by Henri \familyname{Poincaré} 
    (\cf\ \refsubfig{fig:peopleEpist}{b}).
    This philosophy considers that human intuitions about the physical world are
    possibly flawed.
    Some of our scientific principles are only \expression{conventions}.
    Poincaré, having worked on Lobachev\-skian geometries, was taking Euclidean 
    geometry as an example \citep[\eg][]{bland11}.
    These conventions should thus be chosen so that they agree with the physical
    reality.
  \item[Verificationism]
    is another doctrine deriving from scientific positivism     
    \citep[\eg][]{okasha01}.
    It postulates that a proposition has a cognitive meaning 
    only if it can be \expression{verified} by experience.
\end{description}

    \subsubsection{Popper's Logic of Scientific Discovery}

We now review the epistemology developed by Karl \familyname{Popper} (\cf\ \refsubfig{fig:peopleEpist}{c}), which is the center of our discussion.
Popper was an Austrian-born British philosopher who had a significant impact on the modern scientific method\footnote{If journalists are reading these lines, we stress that the \expression{peer-review process} has nothing to do with the scientific method. It is just a convenient editorial procedure that filters poorly thought-out studies.}.
His concepts of falsifiability and reproducibility are still considered as the standards of the scientific method, nowadays.
His major book, \citengl{The Logic of Scientific Discovery} \citep{popper59}, was originally published in German in 1934, and rewritten in English in 1959.
Before starting, it is important to make the distinction between the following two terms.
\begin{description}
  \item[Deduction] is inferring the truth of a specific case from a general 
    rule.
    For instance, knowing that \hSNIa\ occur only in binary systems, if we
    observe a \hSNIa, we can \expression{deduce} that it originates from a
    binary system.
  \item[Induction] is inferring a general conclusion based on individual cases.
    For instance, if we observe a few \hSNR s with massive amounts of 
    freshly-produced dust, we can \expression{induce} that \hSN e produce 
    massive amounts of dust.
\end{description}

\paragraph{The criticism of induction.}
Popper's reflection focusses on the methods of empirical sciences.
The foundation of his theory is the rejection of inductive logic, as he deems that it does not provide a suitable \expression{criterion of demarcation}, that is a criterion to distinguish empirical sciences from mathematics, logic and metaphysics.
According to him, the conventionalist and verificationist approaches are not rigorous enough.
Popper argues that only a deductivist approach provides a reliable empirical method: \citengl{hypotheses can only be empirically tested and only after they have been advanced} \citep[][Sect.\ I.1]{popper59}.
Deductivism is however not sufficient in Popper's mind.

\paragraph{Falsifiability.}
Popper criticizes conventionalists who \citengl{evade falsification by using ad hoc modifications of the theory}.
For that reason, verifiability is not enough.
Empirical theories must be \expression{falsifiable}, that is they must predict experimental facts that, if empirically refuted, will prove them wrong.
His system, which was afterward called \expression{falsifiabilism}, is the combination of:
\begin{inlinelist}
  \item deductivism; and
  \item \textit{modus tollens}.  
\end{inlinelist}
The \expression{modus tollens} is the following logical proposition:
\begin{equation}
  \left((T\Rightarrow D)\wedge\neg D\right)\Rightarrow \neg T.
\end{equation}
Put in words, it can be interpreted as: if a theory T predicts an observational fact D, and this fact D happens to be wrong, then we can \expression{deduce} that the theory T is wrong.
This principle is to be strictly applied: \citengl{one must not save from falsification a theory if it has failed} \citep[][Sect.\ II.4]{popper59}.
Let's take a pseudo-science example to illustrate Popper's point.
Let's assume that a \citengl{ghost expert} pretends a ghost inhabits a given haunted house.
The verificationist approach would consist in saying that, to determine if there is really a ghost, we need to go there and see if it shows up.
Popper would argue that, if the ghost did not appear, our expert would claim that it was because it was intimidated or it felt our skepticism.
Falsifiabilism would dictate to set experimental conditions beforehand by agreeing with the expert: if the ghost does not appear in visible light, in this house, at midnight, on a particular day, then we will deduce that this ghost theory is wrong.
The ghost expert would probably not agree with such strict requirements.
Popper would thus conclude that ghostology is not an empirical science.

\paragraph{Reproducibility.}
A difficulty of the empirical method is relating perceptual experiences to concepts.
Popper argues that the objectivity of scientific statements lies in the fact that they can be \expression{inter-subjectively} tested.
In other words, if several people, with their own subjectivity, can perform the same empirical tests, they will rationally come to the same conclusion.
This requires \expression{reproducibility}.
Only repeatable experiments can be tested by anyone.
Reproducibility is also instrumental in avoiding coincidences.

\paragraph{Parsimony (Ockham's razor).}
A fundamental requirement of scientific theories is that they should be the simplest possible.
Unnecessarily complex theories should be eliminated.
This is the \expression{principle of parsimony}.
Popper is aware of that and includes it in his system.
This is however not the most convincing point of his epistemology.
His idea is that a simple theory is a theory that has a high \expression{degree of falsifiability}, which he calls \citengl{empirical content} \citep[][Sect.\ II.5]{popper59}.
In other words, according to him, the simplest theories are those that have the highest prior improbability, whereas complex theories tend to have special conditions that help them evade falsification.

\paragraph{Popper's epistemology is frequentist.}
It is obvious that Popper's system has a frequentist frame of mind.
It was indeed conceived at the golden age of frequentism (\cf\ \refsec{sec:frequentistwinter}).
\begin{description}
  \item[The requirement of falsifiability] is reminiscent of \hNHST\ (\cf\ 
    \refsec{sec:phacking}).
    Frequentists only accept a hypothesis by rejecting (\ie\ falsifying) its 
    alternative.
    Rigorous \hNHST\ requires setting up the detailed experimental procedure,
    the stop condition and the significance of the outcome beforehand.
    This is exactly what Popper requires to make sure the conditions of 
    falsifiability are not tampered with.
    This makes it impossible to test several theories with a single experiment, 
    contrary to the Bayesian approach.
  \item[Platonic logic] is at the center of Popper's epistemology.
    The last third of his book \citep{popper59} is actually devoted to 
    probabilities.
    He favors their \expression{objective} interpretation, in terms of 
    frequency.
    This is because he does not even conceive the possibility to assign a 
    probability to a theory.
    This is also the weakness of the frequentist approach (\cf\ 
    \refsec{sec:BvF}).
  \item[Repeatability] is the necessary condition to satisfy the assumption that
    experimental uncertainties are the limiting frequency of the result's
    fluctuations.
    This also makes it impossible to account for sparse or unique constraints.
  \item[Accumulation of knowledge] is impossible in this approach, as each 
    individual experiment must be considered independently.
    It is impossible to account for prior knowledge.
    The Popperian approach, if it was actually applied by scientists, would even
    lead to a decrease of knowledge. 
    Indeed, if we indefinitely try to falsify a theory, we will end up by 
    rejecting it, just by luck\footnote{This could be called 
    \citengl{P(opper)-hacking}.}.
\end{description}

    \subsubsection{Verifiability, Holisticity and Parsimony: the 
                        Bayesian Alternative}

We now discuss how the Bayesian approach provides an alternative to Popper's epistemology.
\citet{jaynes03} demonstrates that probabilities, in the Bayesian sense, could be the foundation of a rigorous scientific method.
\citet{hoang20} even argues that Bayes' rule is the optimal way to account for experimental data.

\paragraph{Falsifiability and the limits of Platonic logic.}
The refusal of Popper and frequentists to adopt probabilistic logic is the reason why their decision upon experimental evidence is so convoluted.
The application of Platonic logic to the physical reality indeed presents some issues.
One of the most famous aporias is \citengl{Hempel's paradox} \citep{hempel45}.
It states the following.
\begin{description}
  \item[Hempel's paradox:]
    if a logical \expression{proposition} is true, its 
    \expression{contraposition} is necessarily true:
    \begin{equation}
      \underbrace{(A \Rightarrow B)}_\sms{proposition} 
        \Leftrightarrow
        \underbrace{(\neg B \Rightarrow \neg A)}_\sms{contraposition}.
    \end{equation}
    The example taken by \citet{hempel45}, is \citengl{all ravens are black}
    ($\textnormal{raven}\Rightarrow\textnormal{black}$).
    The contraposition is \citengl{anything that is not black is not a raven} 
    ($\neg\textnormal{black}\Rightarrow\neg\textnormal{raven}$).
    From an experimental point of view, if we want to \expression{corroborate} 
    that all ravens are black, we can either:
    \begin{inlinelist}
      \item find black ravens (\ie\ verifying the proposition); or 
      \item find anything that is neither black nor a raven, such as a red 
        apple (\ie\ verifying the contraposition).
    \end{inlinelist}
    The second solution is obviously useless in practice.
    To take an astrophysical example, finding a quiescent \hi\ cloud would be 
    considered as a confirmation that star formation occurs only in \hmol\ 
    clouds.
    This is one of the reasons why Popper requires falsifiability.
  \item[A Bayesian solution] to the paradox was proposed by \citet[][who was 
    Turing's collaborator at Bletchley Park]{good60}.
    First, in probabilistic logic:
    \begin{equation}
      (A \Rightarrow B) \Leftrightarrow \pcond{B}{A}=1.
    \end{equation}
    The difference is that, with probabilities, we can deal with uncertainty, 
    that is $0<\pcond{B}{A}<1$.
    Second, Bayes factors quantify the strength of evidence (\cf\ 
    \refsec{sec:Bayesfactor}), and it is different in the case of a black raven 
    or a red apple.
    \cite{good60} shows that the strength of evidence is negligible in the case
    of a red apple.
    The Bayesian solution is thus the most sensible one.
    Bayes factors are therefore the tool needed to avoid requiring 
    falsifiability.
    We can adopt a verificationist approach and discuss if our data brought 
    significant evidence.
\end{description}
\takeaway{Bayesianism does not require falsifiability.
          Bayes factors provide a way to quantify the strength of evidence
          brought by any data set.}

\begin{figure}[htbp]
  \begin{tabular}{cc}
    \includegraphics[width=0.48\textwidth]{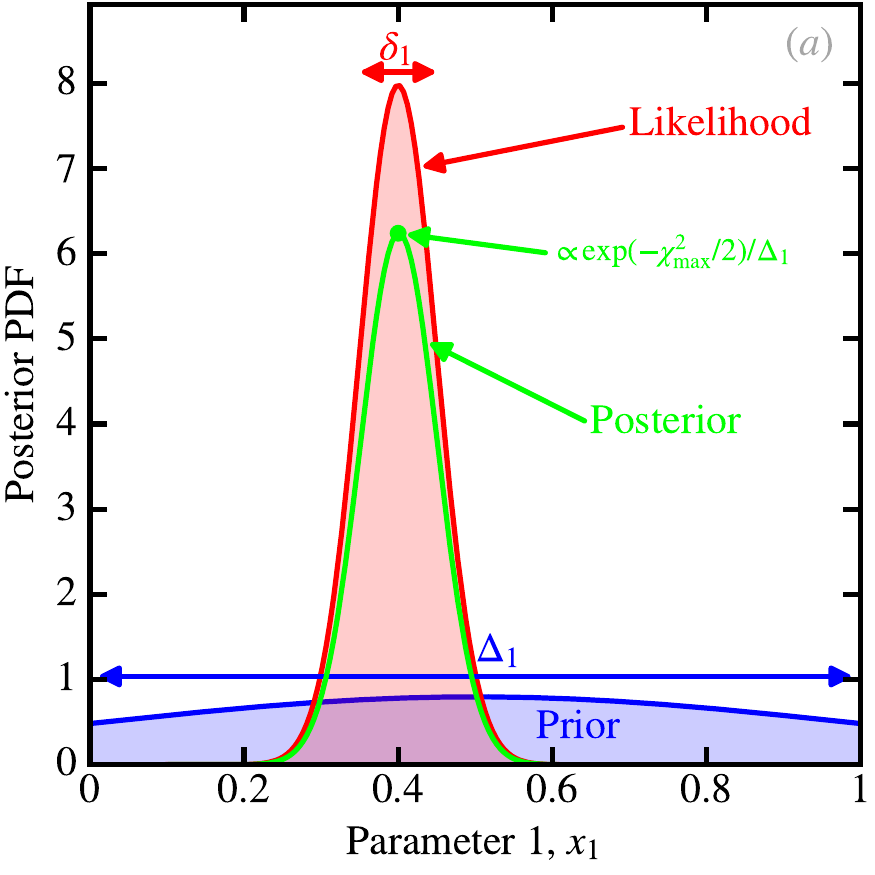} &
    \includegraphics[width=0.48\textwidth]{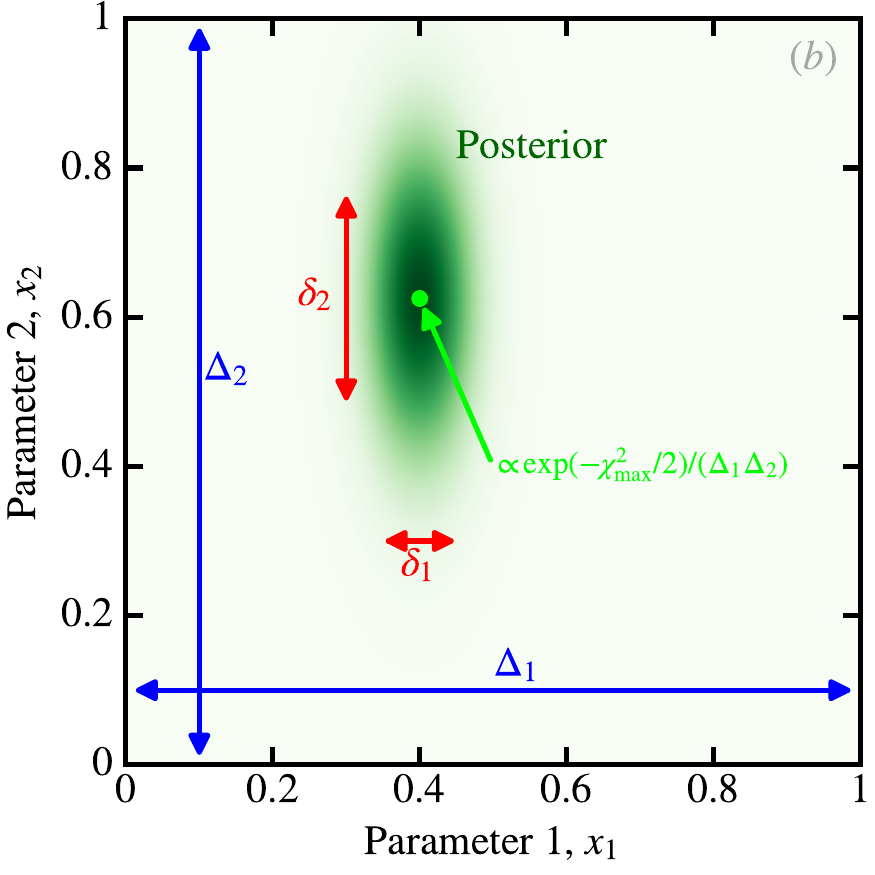} \\
  \end{tabular}
  \newcap{Bayes factors and parsimony}%
         {Panel~\textit{(a)} represents the likelihood of a one-parameter model,
          in red. 
          Its width is $\delta_1$.
          The prior is shown in blue and is wider (width $\Delta_1$).
          The posterior, in green, is the product of the two.
          Its peak, the green dot, is roughly proportional to 
          $\exp(-\chi^2_\sms{max}/2)/\Delta_1$ \refeqp{eq:parsimony2}.
          Panel~\textit{(b)} represents the same model, adding an extra model
          parameter, $x_2$.
          In the case we have represented, the parameter does not
          improve the fit, so that the chi-squared is similar.
          The maximum \textit{a posteriori} is the green dot.
          It is roughly proportional to 
          $\exp(-\chi^2_\sms{max}/2)/(\Delta_1\Delta_2)$ \refeqp{eq:parsimony3}.
          \CClicence}
  \label{fig:parsimony}
\end{figure}

\paragraph{Parsimony is hard-coded in Bayes' rule.}
The principle of parsimony is directly implied by the use of Bayes factors.
To illustrate this point, let's assume we are fitting a two-parameter model to a data set, $\vec{d}$, and we want to know if we can fix the second parameter (model $M_1$) or let it free (model $M_2$).
The Bayes factor \refeqp{eq:Bfac} is simply:
\begin{equation}
  BF_{21}=\frac{\pcond{\vec{d}}{M_2}}{\pcond{\vec{d}}{M_1}}=
    \frac{\displaystyle\iint\pcond{\vec{d}}{x_1,x_2}\pcond{x_1,x_2}{M_2}
          \ddiff x_1\ddiff x_2}%
         {\displaystyle\int\pcond{\vec{d}}{x_1}\pcond{x_1}{M_1}\ddiff x_1}.
  \label{eq:parsimony0}
\end{equation}
Let's assume that we have a Gaussian model.
We can write the product of the prior and the likelihood, in this case:
\begin{equation}
  \underbrace{\pcond{\vec{d}}{x_1}}_\sms{likelihood}
  \underbrace{\pcond{x_1}{M_1}}_\sms{prior}
    = \underbrace{\frac{1}{(\sqrt{2\pi}\sigma)^m}
                          \exp\left(-\frac{\chi^2}{2}\right)}_\sms{likelihood}
      \underbrace{\frac{1}{\Delta_1}}_\sms{prior},
  \label{eq:parsimony1}
\end{equation}
where we have assumed that the prior was flat over $\Delta x_1$ and that we had $m$ observations with uncertainty $\sigma$.
Notice that the product in \refeq{eq:parsimony1} is proportional but not equal to the posterior.
We indeed have not divided it by \tproba{\vec{d}}, as this is the quantity we want to determine.
If we approximate the posterior by a rectangle, we obtain the following rough expression, using the notations in \refsubfig{fig:parsimony}{a}:
\begin{equation}
  \pcond{\vec{d}}{x_1}\pcond{x_1}{M_1}
    \simeq\frac{\delta_1}{\Delta_1}\exp\left(-\frac{\chi^2_\sms{max}}{2}\right).
  \label{eq:parsimony2}
\end{equation}
Let's assume that adding parameter $x_2$ does not improve the fit.
The $\chi^2_\sms{max}$ thus stays the same.
This is represented in \refsubfig{fig:parsimony}{b}.
With the same assumptions as in \refeq{eq:parsimony2}, we obtain:
\begin{equation}
  \pcond{\vec{d}}{x_1,x_2}\pcond{x_1,x_2}{M_2}
    \simeq
    \frac{\delta_1\delta_2}{\Delta_1\Delta_2}\exp\left(-\frac{\chi^2_\sms{max}}{2}
         \right).
  \label{eq:parsimony3}
\end{equation}
\refeq{eq:parsimony0} thus becomes:
\begin{equation}
  BF_{21}\simeq\frac{\delta_2}{\Delta_2}\ll 1.
\end{equation}
It thus tell us that model $M_1$ is more credible.
The penalty of adding the extra parameter is $\delta_2/\Delta_2$.
More generally, we can encounter the three following situations.
\begin{enumerate}
  \item If the fit is indeed better, we will have:
    $BF_{21}\simeq\delta_2/\Delta_2\exp[-(\chi_2^2-\chi_1^2)/2]$.
    Model $M_2$ will be more credible only if the increased chi-squared 
    compensates the penalty.
  \item If the fit is not better, but parameter $x_2$ is not constrained, we
    will have $\delta_2\simeq\Delta_2$, and $BF_{21}\simeq1$.
    Both models will be equivalent because we will not have changed our prior
    knowledge about $x_2$.
  \item If the fit is not better, but parameter $x_2$ has a smaller support than
    its prior, model $M_2$ will be less credible than $M_1$.
    This is the example of \reffig{fig:parsimony}.
\end{enumerate}

\paragraph{Comparison of the Bayesian and Popperian methods.}
We can now compare both approaches.
The following arguments are summarized in \reftab{tab:bayesianism}.
\begin{description}
  \item[Theory corroboration] requires falsifiability, in the Popperian 
    approach.
    Experiments need to be reproducible and the detailed procedure needs to be 
    defined before starting acquiring data.
    The Bayesian approach is more flexible because falsifiability is not 
    required and heterogeneous data sets can be accounted for.
    For instance, the first detection of gravitational waves \citep{ligo16} and
    the first direct image of a black hole \citep{eventhorizon19}, combined, 
    bring a large weight of evidence in favor of general relativity.
    On the contrary, a strict Popperian would argue that:
    \begin{inlinelist}
      \item both experiments are independent and should not be mixed together; 
      and
      \item these experiments are not falsifiable, as an absence of detection
        could have been blamed on their complexity.
    \end{inlinelist}
    In that sense, the Popperian approach, if it was actually put in practice,
    would be extremely wasteful and would slow down the progress of science.
    Fortunately, the majority of scientists do not apply the Popperian method, 
    most of them unconsciously.
    Our impression is that most scientists apply the Bayesian principles, at 
    least qualitatively.
    This is probably because our own brain is Bayesian \citep[\eg][]{meyniel17}.
  \item[Reproducibility] is a requirement of the Popperian 
    approach\footnote{Nowadays, scientists call \citengl{reproducibility} the 
    action of providing the data and the codes a publication was prepared with.
    This is a good practice, but this is not Popper's reproducibility.
    It should rather be qualified as \citengl{open source}.}, whereas
    Bayesians can account for unique, unreproducible data, such as earthquakes,
    \hSN e, \hGRB s, \etc\
    The Bayesian approach will benefit from reproducible experiments, as they 
    will increase the strength of evidence, but it is not a requirement.
  \item[Accumulation of knowledge] is natural in Bayesianism, as the prior is
    there to account for what any previous experience has taught us.
    This is what we demonstrated in \refsec{sec:BvFprior}:
    \begin{equation}
      \underbrace{\proba{\vec{x}}}_\sms{new prior}
      = \underbrace{1}_\sms{initial prior}\times
        \underbrace{\pcond{\vec{d}_1}{\vec{x}}\times\ldots
               \times\pcond{\vec{d}_N}{\vec{x}}}_\sms{accumulated previous data}.
    \end{equation}
  \item[Logical decision] is Platonic for Popperians.
    They see theories either right or wrong, until proven otherwise.
    The probabilistic logic of Bayesians is more progressive, because it uses a
    continuous credibility scale and accounts for all previous data.
    It makes it more conservative.
    Yet, Epistemological breaks are still possible if new data bring a large 
    strength of evidence in favor of a new theory.
  \item[Scientist or lawyer?]
    The Popperian approach forces scientists to design experiments to test a
    single theory, whereas Bayesians can naturally compare several theories and
    use all available data.
    The latter favors a less doctrinal attitude, less ideological, that should 
    be promoted in science.
    On the contrary, Popperians have an attitude closer to that of lawyers, as
    they are forced to defend one particular case, by attacking (falsifying) 
    the alternatives\footnote{Concerning the topics discussed in 
    \refsec{sec:cosmicdustevol}, we have seen this attitude in a small group of 
    people trying to prove \hISM\ dust is stardust, constantly ignoring the big 
    picture summarized in \reftab{tab:dustorigin}.}.
\end{description}
\begin{table}[htbp]
  \centering
  \setlength\arrayrulewidth{2pt}
  \arrayrulecolor{white}
  \begin{tabularx}{\linewidth}{|>{\columncolor{coltabhead}}l%
                                |>{\columncolor{coltabcell}}X%
                                |>{\columncolor{coltabcell}}X|}
    \hline
      \rowcolor{coltabhead}\cellcolor{white} & \textbf{Bayesian}
        & \textbf{Popperian} \\
    \hline
      Corroboration & Verifiability \&\ strength of evidence 
        & Requires falsifiability \\
    \hline
      Logic & Probabilistic: continuity between skepticism \&\ confidence
        & Platonic: theories are either true or false at a given time \\
    \hline
      Repeatability & Can account for unique data
        & Requires reproducibility \\
    \hline
      Experimental data & Can account for small, heterogeneous data sets 
        & Experimental settings need to be defined beforehand \\
    \hline
      External data & Holistic approach
      & No possibility to account for any data outside of the experiment \\
    \hline
      Parsimony &
      Bayes factors eliminate unnecessary complex theories
        & The most falsifiable theories are preferred \\
    \hline
      Knowledge growth & Prior accounts for past knowledge 
        & Each experiment is independent \\
    \hline
      Attitude & Universal approach: can test any theory
        & Partial approach: only one theory can be tested \\
    \hline
      Application & Most scientists are unconsciously pragmatic Bayesians
        & Strict Popperians are rare \&\ probably not very successful \\
    \hline
  \end{tabularx}
  \newcap{Comparison of Bayesian and Popperian epistemologies}%
         {}
  \label{tab:bayesianism}
\end{table}

\section{Relevance for Interstellar Dust Studies}
\label{sec:HB4ISD}

We now demonstrate what the Bayesian methods, which constitute a true epistemological approach, can bring to \hISD\ studies.
In particular, we advocate that hierarchical Bayesian models are an even better application of this approach.
We illustrate this point with our own codes.

  \subsection{The Particularities of Interstellar Dust Studies}
  \label{sec:particularities}

We have discussed throughout this manuscript the difficulty to constrain dust properties using a variety of observational constraints.
For instance, we have seen that there is a degeneracy between small and hot equilibrium grains in the \hMIR\ (\cf\ \refsec{sec:dale}), or between the effects of the size and charge of small \hHAC\ (\cf\ \refsec{sec:PAHband}).
In a sense, we are facing what mathematicians call an \expression{ill-posed problem}, with the difference that we do not have the luxury of rewriting our equations, because they are determined by the observables.
We detail these issues below.

    \subsubsection{Complexity of the Physics}

\paragraph{Degeneracy between microscopic and macroscopic properties.}
When we observe a \ciiline\ line, we know it comes from a C$^+$ atom, and we can characterize very precisely the physical nature of this atom.
On the contrary, if we observe thermal grain emission, we know it can come from a vast diversity of solids, with different sizes, shapes, structures, and composition.
Even if we observe a feature, such as the 9.8~\tmic\ silicate band or an aromatic feature, we still have a lot of uncertainty about the physical nature of its carrier.
This is the fundamental difference between \hISM\ gas and dust physics.
The complexity of gas modeling comes from the difficulty to determine the variation of the environmental conditions within the telescope beam.
This difficulty thus comes from our uncertainty about the \expression{macroscopic} distribution of \hISM\ matter and of the energy sources (stars, \hAGN s, \etc) in galaxies.
We also have the same issue with dust.
When studying \hISD, we therefore constantly face uncertainties about both the \expression{microscopic} and \expression{macroscopic} properties.
Assuming we have a model that accounts for variations of both the dust constitution and the spatial distribution of grains relative to the stars, the Bayesian approach is the only way to consistently explore the credible regions of the parameter space, especially if several quantities are degenerate.
We will give some examples in \refsec{sec:HerBIE}.

\paragraph{Heterogeneity of the empirical constraints.}
A dust model, such as those discussed in \refsec{sec:dustmodels}, has been constrained from a variety of observables (\cf\ \refsec{sec:dustobs}):
\begin{inlinelist}
  \item from different physical processes, over the whole electromagnetic 
    spectrum;
  \item originating from different regions in the \hMW;
  \item with prior assumptions coming from studies of laboratory analogues and
    meteorites.
\end{inlinelist}
When we use such a model to interpret a set of observations, we should in principle account for all the uncertainties that went into using these constraints:
\begin{itemize}
  \item the uncertainties on the observations used to design the model
    (depletions, extinction curves, \etc);
  \item the uncertainties on the laboratory data (opacity, density, \etc);
  \item the prior probability of the different assumptions.
\end{itemize}
Obviously, only the Bayesian approach can account for these, especially knowing that these different uncertainties will likely be non-Gaussian and partially correlated.
This is however an ambitious task, and it has been done only approximately \citep[\eg\ Sect.\ 4.1.3 of][]{galliano21}.
This is a direction that future dust studies should take.

    \subsubsection{Entanglement of the Observations}

\paragraph{The observables are weakly informative.}
Another issue with \hISD\ studies is that the observables, taken individually, bring a low weight of evidence.
A single broadband flux is virtually useless, but a few fluxes, strategically distributed over the \hFIR\ \hSED, can unlock the dust mass and starlight intensity (\cf\ \refsec{sec:dale}).
Yet, these different fluxes come from different observation campaigns, with different instruments.
If we add that the partially-correlated calibration uncertainties of the different instruments often dominate the error budget, we understand that the Bayesian approach is the only one that can rigorously succeed in this type of analysis.
We will discuss the treatment of calibration uncertainties in \refsec{sec:calib}.

\paragraph{Contaminations are challenging to subtract.}
The different sources of foreground and background contaminations, that we have discussed in \refsec{sec:MCRTgal}, will become more and more problematic with the increasing sensitivity of detectors.
Indeed, probing the diffuse \hISM\ of galaxies requires to observe surface brightnesses similar to the \hMW\ foreground.
In addition, the \hCIB\ is even brighter at submm wavelengths (\cf\ \reffig{fig:contaminations}).
These two contaminations, the \hMW\ and the \hCIB, have very similar \hSED s and a complex, diffuse spatial structure.
Accurately separating these different layers therefore requires probabilistic methods, using redundancy on large-scales \citep[\eg][]{planck-collaboration16a}.
It requires modeling every component at once.
Bayesian and \hML\ methods are the most obvious solutions.

  \subsection{The Principles of Hierarchical Bayesian Inference}

We now discuss the formalism of hierarchical Bayesian inference, applied to \hSED\ modeling.
This method has been presented by \citet{kelly12} and \citet{galliano18a}.

    \subsubsection{Non-Hierarchical Bayesian Formalism for SED 
                        Modeling}

\paragraph{Posterior of a single source.}
Let's assume that we are modeling a single observed \hSED\ (\eg\ one pixel or one galaxy), sampled in $m$ broadband filters, that we have converted to monochromatic luminosities: $L_\nu^\sms{obs}(\lambda_j)$ ($j=1,\ldots,m$).
Let's assume that these observations are affected by normal \hiid\ noise, with standard-deviation $\sigma_\nu^\sms{noise}(\lambda_j)$.
If we have a \hSED\ model, depending on a set of parameters $\vec{x}$, such that the predicted monochromatic luminosities in the observed bands is $L_\nu^\sms{mod}(\lambda_j,\vec{x})$, we can write that:
\begin{equation}
  L_\nu^\sms{obs}(\lambda_j) = L_\nu^\sms{mod}(\lambda_j,\vec{x}) 
    +\epsilon(\lambda_j)\sigma_\nu^\sms{noise}(\lambda_j),
  \label{eq:BBsingle}
\end{equation}
where $\epsilon(\lambda_j)\stackrel{\sms{iid}}{\sim}\mathcal{N}(0,1)$.
In other words, our observations are the model fluxes plus some random fluctuations distributed with the properties of the noise.
The distribution of the parameters, $\vec{x}$, is what we are looking for.
We can rearrange \refeq{eq:BBsingle} to isolate the random variable:
\begin{equation}
  \epsilon(\lambda_j,\vec{x}) = 
    \frac{L_\nu^\sms{obs}(\lambda_j)-L_\nu^\sms{mod}(\lambda_j,\vec{x})}%
         {\sigma_\nu^\sms{noise}(\lambda_j)}.
  \label{eq:BBeps}
\end{equation}
Since we have assumed \hiid\ noise, the likelihood of the model is the product of the likelihoods of each individual broadbands:
\begin{equation}
  \pcond{\vec{L}_\nu^\sms{obs}}{\vec{x}}
    =\prod_{j=1}^m\proba{\epsilon(\lambda_j,\vec{x})},
   \label{eq:LHBB}
\end{equation}
were $\vec{L}_\nu^\sms{obs}\equiv\{L_\nu^\sms{obs}(\lambda_j)\}_{j=1,\ldots,m}$.
If we assume a flat prior, the posterior is \refeqp{eq:posterior}:
\begin{equation}
  \pcond{\vec{x}}{\vec{L}_\nu^\sms{obs}}
  \propto \prod_{j=1}^m\proba{\epsilon(\lambda_j,\vec{x})}.
  \label{eq:postBB}
\end{equation}
The difference is that:
\begin{inlinelist}
  \item in \refeq{eq:LHBB}, the parameters, $\vec{x}$, are assumed fixed, the 
    different \tproba{\epsilon(\lambda_j,\vec{x})} are thus independent;
    whereas
  \item in \refeq{eq:postBB}, the observations, $\vec{L}_\nu^\sms{obs}$, are 
    assumed fixed, the different terms in the product are now correlated, 
    because each \tproba{\epsilon(\lambda_j,\vec{x})} depends on all the 
    parameters.
\end{inlinelist}

\paragraph{Modeling several sources together.}
If we now model $n$ sources with observed luminosities, $\vec{L}_\nu^{\sms{obs},i}$ ($i=1,\ldots,n$), to infer a set of parameters, $\vec{x}_i$, the posterior of the source sample will be:
\begin{equation}
  \pcond{\vec{x}_1,\ldots,\vec{x}_n}%
        {\vec{L}_\nu^{\sms{obs},1},\ldots,\vec{L}_\nu^{\sms{obs},n}}
  \propto
  \prod_{i=1}^n\prod_{j=1}^mp(\epsilon_i(\lambda_j,\vec{x}_i))
  =\prod_{i=1}^n\pcond{\vec{x}_i}{\vec{L}_\nu^{\sms{obs},i}}.
  \label{eq:postBBall}
\end{equation}
Notice that, in the second equality, the different \tpcond{\vec{x}_i}{\vec{L}_\nu^{\sms{obs},i}} are independent, as each one depends on a distinct set of parameters, $\vec{x}_i$.
The sampling of the whole posterior distribution will thus be rigorously equivalent to sampling each individual \hSED, one by one.
\takeaway{With a non-hierarchical Bayesian approach, the sources in a sample
          are independently modeled.}

    \subsubsection{The Introduction of Nuisance Variables}
    \label{sec:calib}

Nuisance variables are parameters we need to estimate to properly compare our model to our observations.
The particular value of these variables is however not physically meaningful, and we end up marginalizing the posterior over them.
The Bayesian framework is particularly well-suited for the treatment of nuisance parameters.

\paragraph{Calibration uncertainties.}
Calibration errors originate from the uncertainty on the conversion of detector readings to astrophysical flux (typically ADU/s to Jy/pixel).
Detectors are calibrated by observing a set of \expression{calibrators}, that are bright sources with well-known fluxes.
The uncertainties in the observations of these calibrators and on the true flux of the calibrators translate into a \expression{calibration uncertainty}.
\begin{description}
  \item[Correlation between sources:]
    the offset resulting from this uncertainty will be the same for every 
    observations made with a given detector.
    For instance, if an instrument's calibration factor is $5\,\%$ higher than 
    what it should be\footnote{In this example, $5\,\%$ is not the calibration 
    \expression{uncertainty}, but the \expression{error} made because of the 
    calibration uncertainty.}, all high signal-to-noise measures made with this 
    instrument will report a flux higher by $5\,\%$ than its true value.
    The calibration uncertainty, for a given broadband filter, will thus be
    perfectly correlated between all our sources.
  \item[Partial correlation between wavelengths:]
    instruments are often cross-calibrated and use similar calibrators.
    The correlation procedure will thus induce a partial correlation between
    different broadband fluxes.
    An example of this type of correlation is discussed in Appendix~A of 
    \citet{galliano21}.
\end{description}

\paragraph{Introduction into the posterior.}
To account for calibration uncertainties, we can rewrite \refeq{eq:BBeps} as:
\begin{equation}
  \epsilon(\lambda_j,\vec{x},\delta_j) = 
    \frac{L_\nu^\sms{obs}(\lambda_j)
          -L_\nu^\sms{mod}(\lambda_j,\vec{x})\times(1+\delta_j)}%
         {\sigma_\nu^\sms{noise}(\lambda_j)}.
\end{equation}
We have now multiplied the model by $(1+\delta_j)$, where $\delta_j\sim\mathcal{N}(0,\mathbb{V}_\sms{cal})$ is a random variable following a centered multivariate normal law\footnote{We could have taken a different distribution, such as a Student's $t$ \citep[\eg\ Eq.\ 31 of][]{galliano18a}.} with covariance matrix, $\mathbb{V}_\sms{cal}$.
This random variable, which represents a correction to the calibration factor, is multiplicative: it scales the flux up and down.
$\mathbb{V}_\sms{cal}$ contains all the partial correlations between wavelengths \citep[\cf\ Appendix A of][]{galliano21}.
The important point to notice is that the $\delta_j$ do not depend on the individual object (index $i$), they are unique for the whole source sample.
The posterior of \refeq{eq:postBBall} now becomes:
\begin{equation}
  \pcond{\vec{x}_1,\ldots,\vec{x}_n,\vec{\delta}}%
        {\vec{L}_\nu^{\sms{obs},1},\ldots,\vec{L}_\nu^{\sms{obs},n}}
  \propto
  \proba{\vec{\delta}}\times
  \prod_{i=1}^n\underbrace{\pcond{\vec{x}_i}{\vec{L}_\nu^{\sms{obs},i},\vec{\delta}}%
     }_{\displaystyle\prod_{j=1}^m\proba{\epsilon_i(\lambda_j,\vec{x}_i,\delta_j)}},
  \label{eq:postBBcal}
\end{equation}
where $\proba{\vec{\delta}}=\mathcal{N}(0,\mathbb{V}_\sms{cal})$ is the prior on $\delta$.
We can make the following remarks.
\begin{enumerate}
  \item The prior on the calibration \expression{errors}, $\vec{\delta}$, is 
    the calibration \expression{uncertainty}, \tproba{\vec{\delta}}, quoted by 
    the different instrument teams (all the information is included in 
    $\mathbb{V}_\sms{cal}$).
    It means that, by sampling \refeq{eq:postBBcal}, we will infer values of 
    $\vec{\delta}$ that are potentially more accurate than the calibration 
    coefficients provided by the instrument teams.
    In practice, this is however not the case, because:
    \begin{inlinelist}
      \item the calibration sources are usually the brightest and the most
        well-constrained, it is unlikely to reach the same accuracy observing
        galaxies;
      \item observations of galaxies suffer from contaminations adding
        several biases that are difficult to take into account; and
      \item models used to interpret observations of galaxies are not as 
        accurate as models of typical calibrators, such as stars or Uranus.
    \end{inlinelist}
  \item By sampling \refeq{eq:postBBcal}, we are inferring a single value of 
    the $\delta_j$ factors.
    This is because the calibration has been done only once, and stays the same.
    The randomness it introduces is a single draw.
    It is not a reproducible event.
    The calibration uncertainty can not be considered as the limiting frequency
    of  a repeated procedure.
    This is why frequentists could not treat these uncertainties.
  \item The calibration factors in \refeq{eq:postBBcal} link the posteriors of 
    the individual sources together.
    We have noted that they were independent in \refeq{eq:postBBall}.
    This is not the case anymore, as they all depend on $\vec{\delta}$.
    Thus:
    \begin{inlinelist}
      \item inferring $\vec{\delta}$ is made possible by simultaneously 
        sampling several sources, sharing the same calibration coefficients;
      \item the presence of $\vec{\delta}$, at the same time, improves the fit
         of individual \hSED s.
    \end{inlinelist}
\end{enumerate}
The posterior of the parameters is, in the end, the marginalization over $\vec{\delta}$ of \refeq{eq:postBBcal}:
\begin{equation}
  \pcond{\vec{x}_1,\ldots,\vec{x}_n}%
        {\vec{L}_\nu^\sms{obs,1},\ldots,\vec{L}_\nu^\sms{obs,n}}
  = \int\pcond{\vec{x}_1,\ldots,\vec{x}_n,\vec{\delta}}%
        {\vec{L}_\nu^\sms{obs,1},\ldots,\vec{L}_\nu^\sms{obs,n}}
    \ddiff^m\vec{\delta}.
\end{equation}
\takeaway{Calibration errors can be rigorously taken into account as nuisance
          parameters.}

    \subsubsection{The Role of the Hyperparameters}

\paragraph{Accounting for the evidence brought by each source.}
In \refsec{sec:BvFprior}, we have stressed that, when performing a sequential series of measure, we can use the previous posterior as the new prior. 
Yet, the posterior in \refeq{eq:postBBcal} does not allow us to do so, as:
\begin{inlinelist}
  \item the posteriors of individual sources all depend on $\vec{\delta}$,
    they must therefore be sampled at once; and
  \item the parameters, $\vec{x}_i$, are not identical,
    we are not repeating the same measure several times as in 
    \refeq{eq:prior2}, we are observing different sources. 
\end{inlinelist}
Accounting for the accumulation of evidence is thus not as straightforward as in \refeq{eq:prior2}.
There is however a way to use an informative prior, consistently constrained by the sample.
It is the \expression{Hierarchical Bayesian} (\hHB) approach.
To solve the conundrum that we have just exposed, we can make the following assumptions.
\begin{enumerate}
  \item We can assume that the parameters of the different sources 
    are drawn from a \expression{common distribution}, but this distribution is 
    unknown.
    For instance, the dust masses in the pixels of a galaxy span only a few 
    dexes.
    They are not arbitrarily distributed.
    Their distribution results from the complex physics at play: dynamics, 
    star formation, dust evolution, \etc
  \item We can reasonably approximate this common distribution with a 
    particular functional form, such as a multivariate Gaussian or a Student's 
    $t$.
    Such a distribution is parametrized by its average, $\vec{\mu}$, and its 
    covariance matrix, $\mathbb{V}$.
    These parameters are called \expression{hyperparameters}, because they 
    control the distribution of physical parameters.
    This is why this approach is called \expression{hierarchical}. 
    There are two layers of modeling:
    \begin{inlinelist}
      \item the common distribution of parameters, controlled by a set of 
        hyperparameters;
      \item the \hSED\ model controlled by as many sets of parameters as there 
        are sources.
    \end{inlinelist}
    The average $\vec{\mu}$ will therefore represent the mean of each \hSED\
    model parameters ($M_\sms{dust}$, $\langle U\rangle$, $q_\sms{AF}$, \etc; 
    \cf\ \refsec{sec:dale}) and, $\mathbb{V}$, their intrinsic scatter and 
    correlations (such as the correlation between $\langle U\rangle$ and 
    $q_\sms{AF}$; \cf\ \refsec{sec:PAHvsZ}).
  \item This common distribution, controlled by hyperparameters, is treated as
    the prior of our \hSED\ model parameters.
    We can then infer the values of the hyperparameters when sampling the whole
    posterior and marginalize over them in the end.
\end{enumerate}

\paragraph{The hierarchical posterior.}
With the \hHB\ approach, the full posterior of our source sample is:
\begin{equation}
  \pcond{\vec{x}_1,\ldots,\vec{x}_n,\vec{\delta},\vec{\mu},\mathbb{V}}%
        {\vec{L}_\nu^\sms{obs,1},\ldots,\vec{L}_\nu^\sms{obs,n}}
  \propto
  \prod_{i=1}^n
    \underbrace{\pcond{\vec{L}_\nu^\sms{obs,i}}%
                      {\vec{x}_i,\vec{\delta}}}_\sms{source likelihoods}
    \times\underbrace{\pcond{\vec{x}_i}{\vec{\mu},\mathbb{V}}}_\sms{parameter prior}
    \times\underbrace{\proba{\vec{\mu}}\proba{\mathbb{V}}}_\sms{hyperparameter prior}
    \times\underbrace{\proba{\vec{\delta}}}_\sms{calibration prior}.
  \label{eq:postHB}
\end{equation}
\begin{description}
  \item[The hyperprior:]
    compared to \refeq{eq:postBBcal}, we have introduced a new term:
    $\pcond{\vec{x}_i}{\vec{\mu},\mathbb{V}}\times\proba{\vec{\mu}}
    \proba{\mathbb{V}}$.
    This is the hierarchical prior.
    The term \tpcond{\vec{x}_i}{\vec{\mu},\mathbb{V}} is what we have 
    previously called the \expression{common distribution} of parameters.
    It is the actual prior on the parameters, and it is parametrized by the 
    hyperparameters.
    The other terms, $\proba{\vec{\mu}}\proba{\mathbb{V}}$ are the necessary 
    priors on $\vec{\mu}$ and $\mathbb{V}$, that we can assume rather flat 
    \citep[\cf\ Sect.\ 3.2.4 of][for more details]{galliano18a}.
    The elements of $\vec{\mu}$ are drawn one by one, using Gibbs sampling 
    (\cf\ \refsec{sec:mcmc}).
    Regarding the elements of $\mathbb{V}$, we independently draw each 
    standard-deviation and correlation coefficient 
    \citep[\cf\ Sect.\ 3.2.4 of][]{galliano18a}.
  \item[The parameter space] corresponding to \refeq{eq:postHB} has 
    dimension (noting $q$ the number of \hSED\ model parameters):
    \begin{equation}
       N_\sms{dim} = \underbrace{n\times q}_\sms{SED model parameters}
                  + \underbrace{q}_\sms{elements of $\vec{\mu}$}
                  + \underbrace{q}_\sms{diagonal of $\mathbb{V}$}
                  + \underbrace{q\times(q-1)/2}_\sms{number of correlations}
                  + \underbrace{m}_\sms{calibration errors}.
    \end{equation}
    For the \expression{composite} model ($q=7$; \cf\ \refsec{sec:dale}), 
    constrained by $m=10$ wavelengths, and for $n=1000$ sources or pixels, we 
    would have to sample a  $N_\sms{dim}\simeq7000$ dimension parameter space.
\end{description}

  \subsection{Hierarchical Bayesian Models for ISD Studies}
  \label{sec:HerBIE}

We now present practical illustrations of \hSED\ modeling with the \hHB\ approach.
The first \hHB\ dust \hSED\ model was presented by \citet{kelly12}.
It was restrained to single \hMBB\ fits.
\citet{veneziani13} then presented a \hHB\ model that could be applied to a combination of \hMBB s.
The \expression{HiERarchical Bayesian Inference for dust Emission} code  \citep[\citetalias{galliano18a};][]{galliano18a}, was the first \hHB\ model, and to this day the only one to our knowledge, to properly account for full dust models, with:
\begin{inlinelist}
  \item realistic optical properties;
  \item complex size distributions;
  \item rigorous stochastic heating;
  \item mixing of physical conditions;
  \item photometric filter and color corrections; and
  \item partially-correlated calibration errors.
\end{inlinelist}
The following examples have been computed with \citetalias{galliano18a}.

    \subsubsection{Efficiency and Comparison to Other Methods}
    \label{sec:nuisance}

To demonstrate the efficiency of the \hHB\ method and the fact that it performs better than its alternatives, we rely on the simulations presented by \citet{galliano18a}.
These simulations are simply obtained by randomly drawing \hSED\ model parameters from a multivariate distribution, for a sample of sources.
This distribution is designed to mimic what we observe in typical star-forming galaxies.
The \hSED\ model used is the \expression{composite} approach (\cf\ \refsec{sec:dale}), except when we discuss \hMBB s.
Each set of parameters result in an observed \hSED, that we integrate into the four \hIRAC, the three \hPACS\ and the three \hSPIRE\ bands.
We add noise and calibration errors.
This way we can test fitting methods and assess their efficiency by comparing the inferred and the true values.

\paragraph{Close look at a fit.}
\refsubfig{fig:demoHB_comp}{a-b} shows the posterior \hSED\ \hPDF\ of the faintest and brightest pixels in a simulation, fitted in a \hHB\ fashion.
\begin{description}
  \item[For the faintest pixel] (\refsubfig{fig:demoHB_comp}{a}), we can see 
    that the \hPDF\ is wider, because it is less constrained.
    Half of the observations (green) are indeed only upper limits.
    The \hSED\ looks however realistic and matches very well its truth (in 
    red).
    This is one of the advantages of the \hHB\ approach: when a source is
    poorly constrained, its posterior is dominated by the prior, which has
    been informed by the whole sample.
    Thus, despite few information on this particular source, we obtain realistic
    parameters and \hSED, because the rest of the sample is providing 
    information about the typical shape of an \hSED\ in that luminosity range.
  \item[For the brightest pixel] (\refsubfig{fig:demoHB_comp}{b}), the \hSED\ 
    is much tighter.
    There are however spectral domains where the uncertainty can increase.
    For instance, notice that there is a wider spread around 3~\tmic\ and 
    around 11~\tmic.
    This is because the charge of the \hPAH s gives the model a degree of 
    freedom in this range (\cf\ \refsec{sec:PAHband}).
    Yet, the model is poorly constrained.
    The true \hSED\ lies in the tail of the distribution.
    It is still consistent, though.
  \item[The calibration errors] are shown in \refsubfig{fig:demoHB_comp}{c}.
    The red dots represent the biases we have introduced into the synthetic 
    observations.
    These biases are the same for each pixel in the simulation.
    These errors are treated as nuisance parameters in \refeq{eq:postHB}.
    The blue error bars show the inference of these biases.
    We see that they are most of the time consistent within $1\sigma$.
    We also see that they are most of the time consistent with zero.
    This is what we discussed in \refsec{sec:calib}: typical galaxy observations
    are not accurate enough to refine the calibration of the instruments.
    The only outlier is the \SPIREii\ point.
    It is however less than $3\sigma$ away from zero.
  \item[The posterior] distribution of these two pixels is shown in 
    \refsubfig{fig:demoHB_post}{b}.
    We have represented the \hPDF\ of two parameters, $M_\sms{dust}$ and 
    $\langle U\rangle$, marginalizing over the rest.
    We see that the faintest pixel has a larger uncertainty, and that both 
    posteriors are close to their true values (red dots).
    If we now compare these results to the same exact simulation fitted in a
    standard, non-hierarchical Bayesian way (\refsubfig{fig:demoHB_post}{a}), 
    we notice that the \hPDF\ of the brightest pixel is very much the same, but 
    the \hPDF\ of the faintest pixel is now much wider, covering a large 
    fraction of the parameter space.
    This is because, as we have noted earlier, sources are individually fitted
    with a standard Bayesian method, they thus do not benefit from the 
    information provided by the rest of the sample, through the prior.
\end{description}
\takeaway{In a \hHB\ model, the least-constrained sources are more corrected by 
          the prior than the brightest ones.}
\begin{figure}[htbp]
  \begin{tabular}{cc}
    \includegraphics[width=0.64\textwidth]{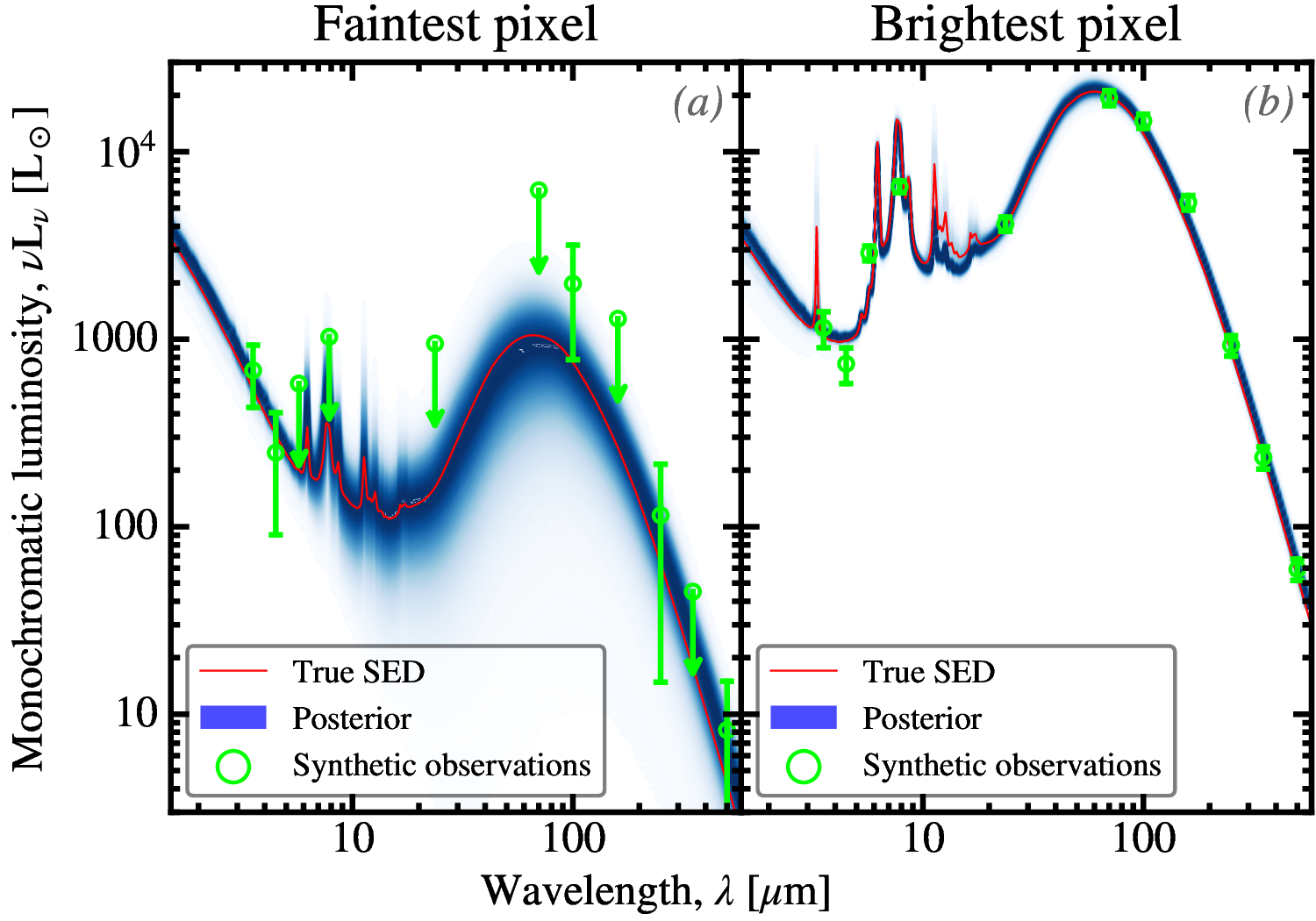} &
    \includegraphics[width=0.32\textwidth]{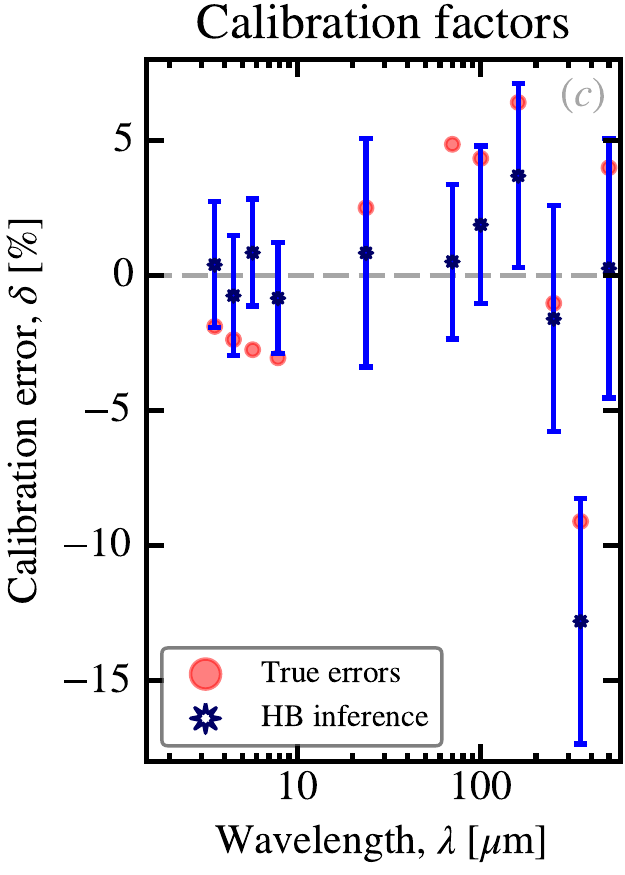} \\
  \end{tabular}
  \newcap{Example of hierarchical Bayesian SED fits}%
         {The blue contours in panels \textit{(a)} and \textit{(b)} represent 
          the posterior \hSED\ of the faintest and brightest pixels in a 
          simulation presented by \citet{galliano18a}.
          This simulation reproduces typical conditions in star-forming 
          galaxies.
          The synthetic observations, including noise and calibration errors, 
          are in green.
          The true model is shown in red, for reference.
          Panel \textit{(c)} shows the simulated calibration errors in red, and
          their inferred posterior values, in blue.
          \CClicence}
  \label{fig:demoHB_SED}
\end{figure}
\begin{figure}[htbp]
  \includegraphics[width=\textwidth]{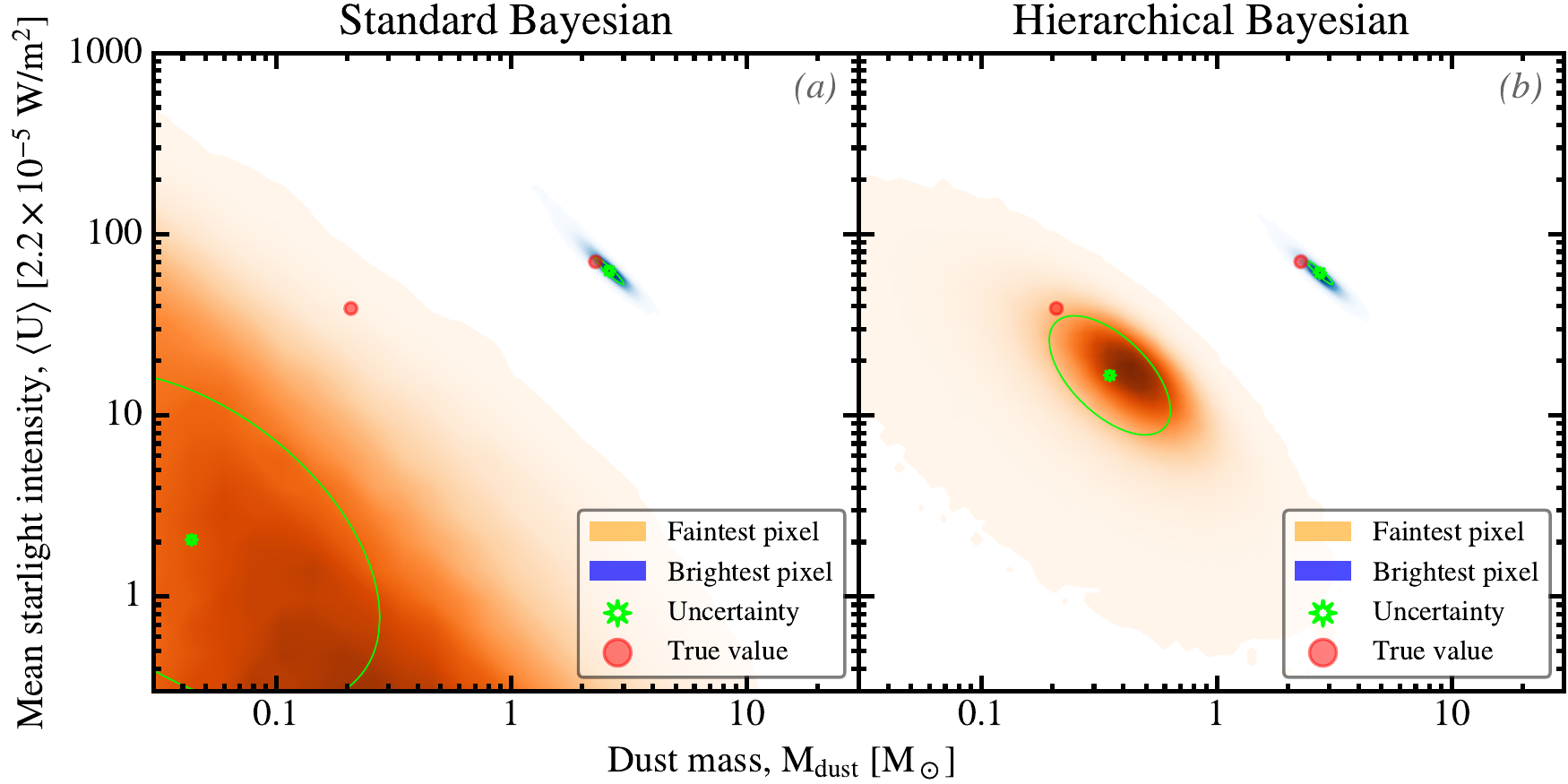}
  \newcap{Posterior distributions with standard and hierarchical Bayesian 
          methods}%
         {In each panel, we show the marginalized posterior of the dust mass,
          $M_\sms{dust}$, and mean starlight intensity, $\langle U\rangle$ 
          \refeqp{eq:Uav}, for the two pixels in \reffig{fig:demoHB_SED}.
          Panel \textit{(a)} corresponds to the case of a non-hierarchical 
          Bayesian fit, and panel \textit{(b)}, to a full hierarchical fit.
          In both panels, the orange contours represent the faintest pixel and 
          the blue contours, the brightest one.
          The uncertainty ellipses corresponding to these posteriors are the
          green ellipses.
          The true values are the red dots.
          \CClicence}
  \label{fig:demoHB_post}
\end{figure}

\paragraph{Comparison between different approaches.}
We have been discussing the two extreme pixels of our simulation.
Let's now look at the whole source sample and compare several methods.
\reffig{fig:demoHB_comp} shows the same parameter space as in \reffig{fig:demoHB_post}, but for all sources, with different methods.
\begin{description}
  \item[The least-squares] method, which is frequentist, is shown in green, in 
    \refsubfig{fig:demoHB_comp}{a}.
    We note the following points.
    \begin{enumerate}
      \item The inferred parameters have large individual uncertainties (\ie\ 
        big ellipses).
      \item The overall sample is quite scattered, covering three orders of 
        magnitude, while the true values are within a dex.
      \item There is a false negative correlation between $M_\sms{dust}$ and 
        $\langle U\rangle$.
    \end{enumerate}
    This false correlation is typical of frequentist methods, but not exclusive.
    It is the equivalent of the $\beta-T$ degeneracy we have already discussed 
    in \refsec{sec:MBB}.
    Because of the way the model is parametrized, if we slightly overestimate 
    the dust mass, we will indeed need to compensate by decreasing $\langle 
    U\rangle$, to account for the same observed fluxes, and vice versa.
    This false correlation is thus induced by the noise.
  \item[The non-hierarchical Bayesian] method, in 
    \refsubfig{fig:demoHB_comp}{b}, provides a more accurate fit of the 
    brightest sources (high $M_\sms{dust}$ and high $\langle U\rangle$).
    The faintest sources are however quite scattered.
    There is still a false correlation, due to the same reasons as for the 
    least-squares, but it is less significant.
  \item[The hierarchical Bayesian] method, in \refsubfig{fig:demoHB_comp}{c},
    on the contrary, provides an unbiased statistical account of the sample.
    We note the following points.
    \begin{enumerate}
      \item The uncertainty of individual sources (blue ellipses) is moderate.
        It is never larger than the scatter of the true sample.
      \item The inferred mean and scatter of the sample properties are very
        close to their true values.
        Their comparison is given in \reftab{tab:refgrid}.
      \item There is no false correlation. 
        The inferred correlation coefficient is consistent with zero, its true 
        value.
    \end{enumerate}
    From a general point view, a \hHB\ method is efficient at removing the 
    scatter between sources that is due to the noise.
    In addition, the inferred uncertainties on the parameter of a source are 
    never larger than the intrinsic scatter of the sample, because this scatter 
    is also the width of the prior.
    For instance, in \refsubfig{fig:demoHB_comp}{c}, the lowest signal-to-noise
    sources (lower left side of the distribution) have uncertainties (blue 
    ellipses) similar to the scatter of the true values (red points), because
    the width of the prior matches closely this distribution.
    On the opposite, the uncertainties of high signal-to-noise sources (upper 
    right side) are much smaller, they are thus not significantly affected by 
    the prior.
\end{description}
\takeaway{\hHB\ methods are efficient at recovering the true, intrinsic scatter
          of parameters and their correlations.}
\begin{figure}[htbp]
  \includegraphics[width=\textwidth]{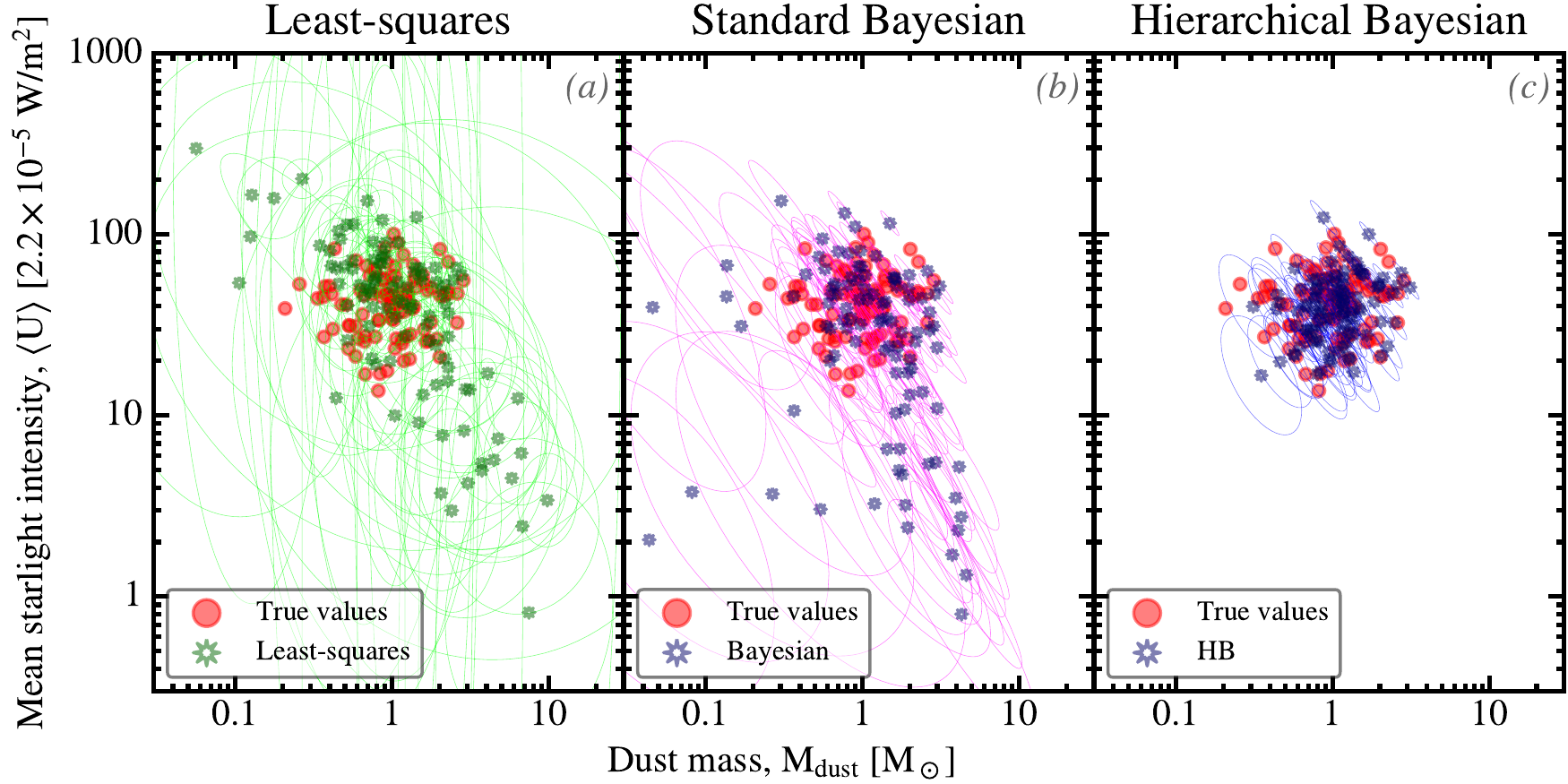}
  \newcap{Comparison of least-squares, standard Bayesian and HB methods}%
         {Each panel shows the full simulation of \reffig{fig:demoHB_comp},
          fitted with three different methods. 
          The true values are the red dots.
          They are identical in the three panels.
          We have 100 sources.
          We represent the same parameter space as in \reffig{fig:demoHB_post}.
          The ellipses represent the posterior of:
          \begin{inlinelistalph}
            \item a least-squares fit;
            \item the non-hierarchical Bayesian fit of
              \refsubfig{fig:demoHB_post}{a}; and
            \item the hierarchical Bayesian fit of 
              \refsubfig{fig:demoHB_post}{c}.
          \end{inlinelistalph}
          \CClicence}
  \label{fig:demoHB_comp}
\end{figure}
\begin{table}[htbp]
  \setlength\arrayrulewidth{2pt}
  \arrayrulecolor{white}
  \centering
  \begin{tabularx}{0.7\linewidth}{|>{\columncolor{coltabhead}}X%
                                   |>{\columncolor{coltabcell}}X%
                                   |>{\columncolor{coltabcell}}X|}
    \hline
      \rowcolor{coltabhead}\cellcolor{white} & \textbf{HB} & \textbf{True} \\
    \hline
      $\langle \ln M_\sms{dust}\rangle$ & $0.053\pm0.136$ & 0 \\
    \hline
      $\sigma(\ln M_\sms{dust})$ & $0.477\pm0.065$ & 0.5 \\
    \hline
      $\langle \ln\langle U\rangle\rangle$ & $3.64\pm0.29$ & 3.742 \\
    \hline
      $\sigma(\ln\langle U\rangle)$ & $0.65\pm0.20$ & 0.4 \\
    \hline
      $\rho(\ln M_\sms{dust},\ln\langle U\rangle)$ & $-0.088\pm0.144$ & 0 \\
    \hline
  \end{tabularx}
  \newcap{Inferred statistical properties of the HB model in 
          \refsubfig{fig:demoHB_comp}{c}}%
         {These quantities are the inferred moments of the source distribution.}
  \label{tab:refgrid}
\end{table}

\begin{figure}[htbp]
  \includegraphics[width=\textwidth]{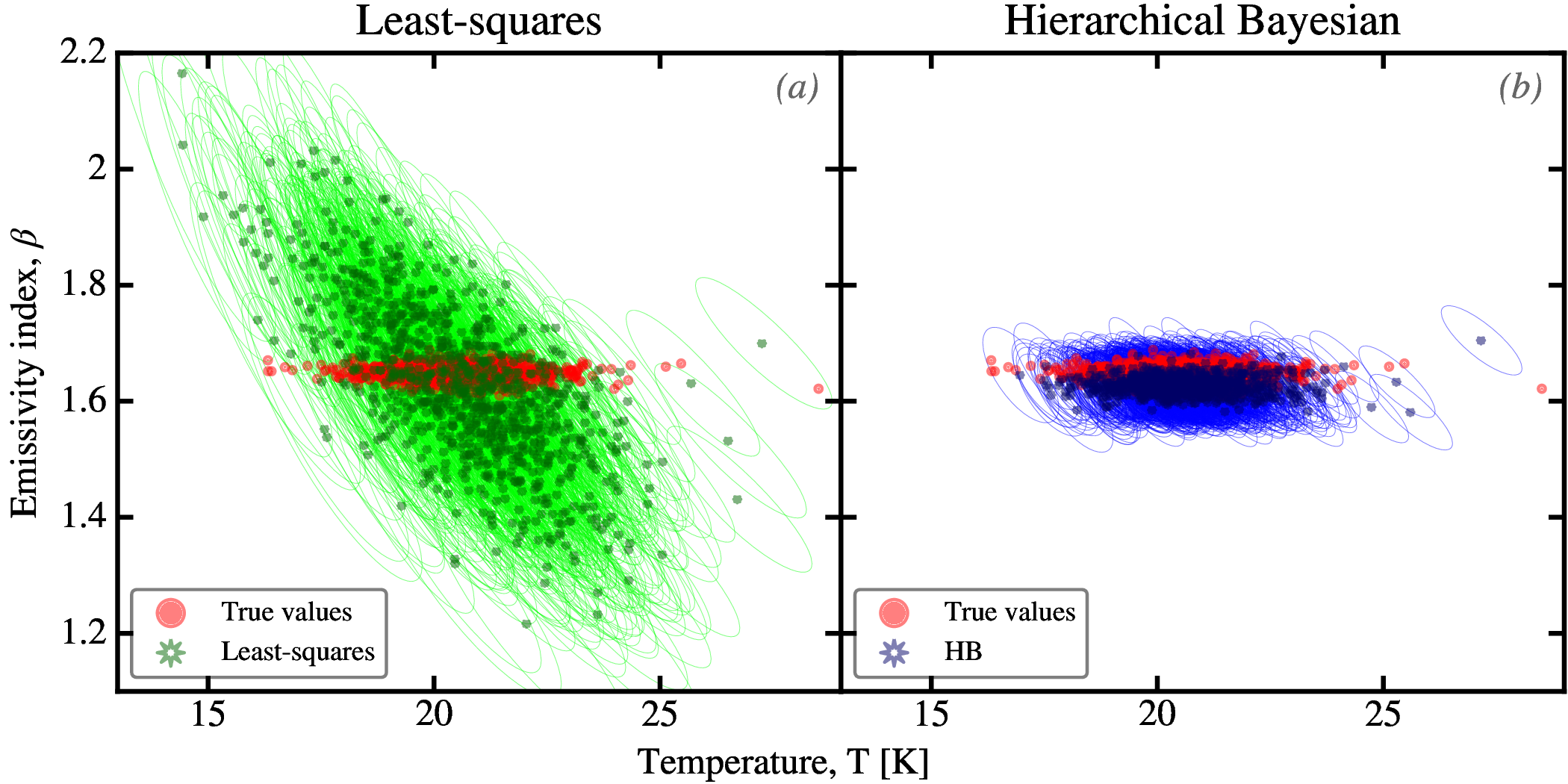}
  \newcap{Solving the emissivity-index-temperature degeneracy of MBBs 
          with a HB model}%
         {The red dots in both panels represent a \hMBB\ simulation of 1000
          sources in the $T-\beta$ plane \citep{galliano18a}.
          Noise and calibration uncertainties have been added to the synthetic 
          \hSED s corresponding to these values.
          They have been fitted with:
          \begin{inlinelistalph}
            \item a least-squares method, in green; and
            \item a \hHB\ method, in blue.
          \end{inlinelistalph}
          \CClicence}
  \label{fig:demoHB_MBB}
\end{figure}

\paragraph{The emissivity-index-temperature degeneracy of MBBs.}
We have just seen that \hHB\ methods are efficient at removing false correlations between inferred properties.
We emphasize that \hHB\ methods do not systematically erase correlations if there is a true one between the parameters \citep[\cf\ the tests performed in Sect.\ 5.1 of][with intrinsic positive and negative correlations]{galliano18a}.
This potential can obviously be applied to the infamous $\beta-T$ correlation discussed in \refsec{sec:MBB}.
This false correlation has been amply discussed by \citet{shetty09}.
\citet{kelly12} showed, for the first time, that \hHB\ methods could be used to solve the degeneracy.
\reffig{fig:demoHB_MBB} shows the results of \citet{galliano18a} on that matter.
We can see that the false $\beta-T$ negative correlation obtained with a least-squares fit (\refsubfig{fig:demoHB_MBB}{a}) is completely eliminated with a \hHB\ method (\refsubfig{fig:demoHB_MBB}{b}).

    \subsubsection{The Role of the Prior}
    \label{sec:prior}

We now further develop and illustrate the instrumental role of the hierarchical prior.

\begin{figure}[htbp]
  \includegraphics[width=\textwidth]{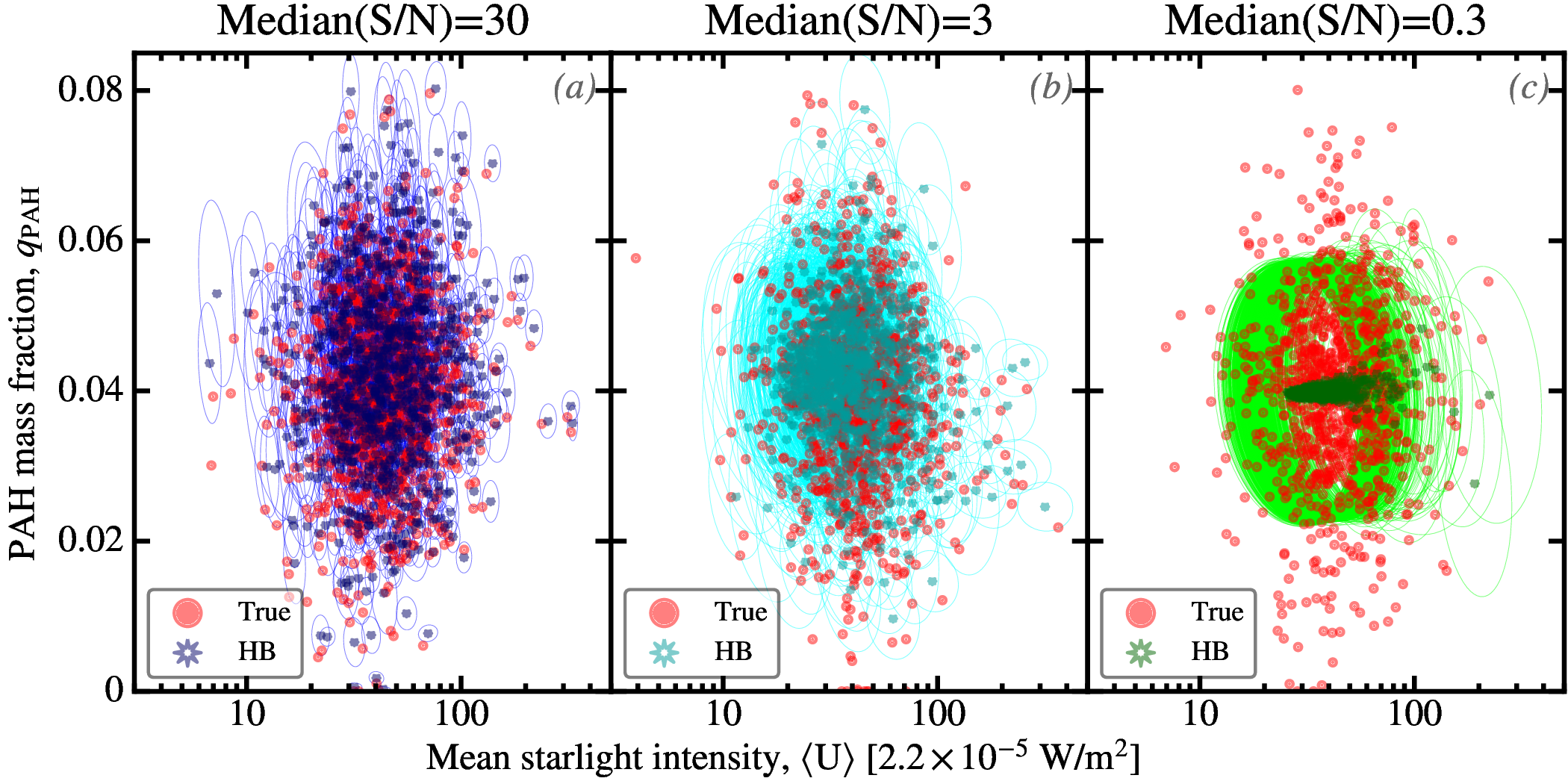}
  \newcap{Demonstration of the effect of the prior in a HB model}%
         {The red dots in the different panels represent three simulations of 
          1000 sources, with similar physical properties, similar calibration
          uncertainties, but different signal-to-noise ratios (S/N).
          Each simulation has been fitted with our \hHB\ code 
          \citep{galliano18a}.
          We represent the marginalized posterior of the PAH mass fraction,
          $q_\sms{PAH}$, and mean starlight intensity, $\langle U\rangle$ 
          (\cf\ \refsec{sec:dale}).
          \CClicence}
  \label{fig:demoHB_prior}
\end{figure}

\paragraph{Linking the different sources.}
We have noted in \refeq{eq:postHB} that the hierarchical prior was breaking the independency between the different sources in the sample, encountered in the non-hierarchical Bayesian case.
This is because the properties of the prior (the hyperparameters) are inferred from the source distribution, and the properties of the individual sources are affected, in return, by this prior.
This is illustrated in \reffig{fig:demoHB_prior}.
This figure shows the \hHB\ fits of three simulations, varying the median signal-to-noise ratio of the sample.
We have represented a different parameter space, this time.
\begin{description}
  \item[At high signal-to-noise] (\refsubfig{fig:demoHB_prior}{a}), we see that 
    the uncertainty of individual sources is significantly smaller than the 
    scatter of their properties.
    The prior thus does not play an important role.
    This is a case where a non-hierarchical Bayesian fit would give very 
    similar results.
  \item[At intermediate signal-to-noise] (\refsubfig{fig:demoHB_prior}{b}), the
    brightest sources (on the right side) still have small uncertainties.
    However, most of the points now have uncertainties comparable to the sample
    scatter, because the posterior of each individual source starts to be
    dominated by the prior.
    We note that the maximum \textit{a posteriori} (cyan stars) starts to
    cluster in the center.
  \item[At low signal-to-noise] (\refsubfig{fig:demoHB_prior}{c}), the inferred
    values of $q_\sms{PAH}$ (green stars) are now almost similar for every 
    source.
    We are in the case where the uncertainty on each individual source is so 
    large that we can not recover their individual values.
    We can however give their most likely value, based on the distribution of 
    the sample.
    This type of result has to be interpreted in a Bayesian way, to be 
    consistent (\ie\ performing tests on the \hMCMC).
    The fact that the inferred values of $q_\sms{PAH}$ all collapsed on a single 
    point does not mean we would deduce that the points all have the same value.
    If we were performing some tests, we would realize that they are 
    uncorrelated: if we were randomly drawing parameter values from the 
    posterior, they would be scattered with a distribution similar to the true 
    values.
    In a sense, we obtain here a result similar to stacking the sources, but in 
    a smarter way, as some parameters are better constrained than others.
    For instance, in \refsubfig{fig:demoHB_prior}{c}, we see that we have only 
    access to $\langle q_\sms{PAH}\rangle$ ($y$-axis), but we resolve the 
    $\langle U\rangle$ of individual sources ($x$-axis).
\end{description}
\takeaway{\hHB\ methods are useful when the parameter uncertainty of
          some sources is comparable to or larger than the scatter of this 
          property over the whole sample.}

\begin{figure}[htbp]
  \includegraphics[width=\textwidth]{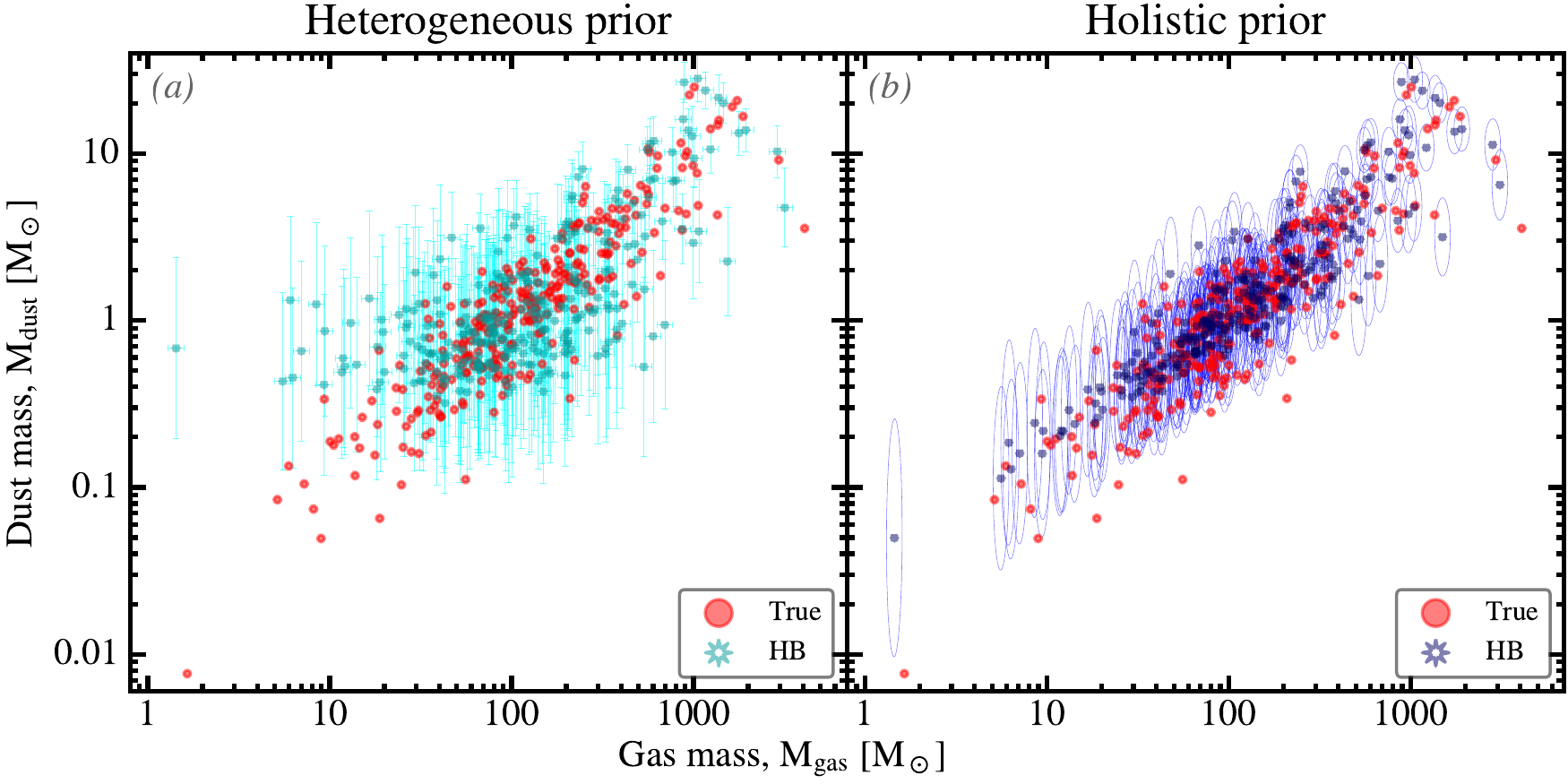}
  \newcap{The holistic approach: inclusion of external parameters into the 
          prior}%
         {Both panels represent the same simulation (red dots) of 300 sources,
          with physical properties typical of star-forming galaxies, including
          noise and calibration errors \citep{galliano18a}.
          We have added a parameter that is external to our \hSED\ model, the
          gas mass, $M_\sms{gas}$.
          In panel \textit{(a)}, we show the \hHB\ inference of the dust mass
          as a function of the synthetic observations of the gas mass (cyan 
          error bars).
          In panel \textit{(b)}, we show the \hHB\ inference of the parameters
          when they are both in the prior (blue ellipses).
          \CClicence}
  \label{fig:demoHB_extra}
\end{figure}

\paragraph{Holistic prior.}
From a statistical point of view, it is always preferable to treat all the variables we are interested in as if they were drawn from the same multivariate distribution, however complex it might be.
\citet{stein56} showed that the usual estimator of the mean ($\sum_i X_i/N$) of a multivariate normal variable is inadmissible (for more than two variables), that is we could always find a more accurate one.
In other words, if we were interested in analyzing together several variables, not necessarily correlated, such as the dust and stellar masses, it would always be more suitable to use estimators that combine all of them.
This is known as \expression{Stein's paradox}.
Although \citet{stein56}'s approach was frequentist, this is a general conclusion.
From a Bayesian point of view, it means that we should put all the variables we are interested in analyzing, even if they are not \hSED\ model parameters, in the prior.
In addition, if these external variables happen to be correlated with some \hSED\ model parameters, they will help refining their estimates.
This is illustrated in \reffig{fig:demoHB_extra}.
We show in both panels a simulation of the correlation between the dust mass, which is a \hSED\ model parameter, and the gas, which is not.
When performing a regular \hHB\ fit, and plotting the correlation as a function of $M_\sms{gas}$, we obtain the correlation in \refsubfig{fig:demoHB_extra}{a}.
We see that the agreement with the true values breaks off at low mass (also the lowest signal-to-noise).
If we now include $M_\sms{gas}$ in the prior\footnote{From a technical point of view, \expression{including} an external parameter in the prior, such as $M_\sms{gas}$, can be seen as adding an identity model component: $M_\sms{gas}=f(M_\sms{gas})$.
Concretely, it means that, at each iteration, we sample $M_\sms{gas}$ from its uncertainty distribution, and the distribution of $M_\sms{gas}$ and its potential correlations with the other parameters inform the prior.}, we obtain the correlation in \refsubfig{fig:demoHB_extra}{b}.
It provides a much better agreement with the true values.
This is because adding $M_\sms{gas}$ in the prior brought some extra information.
The information provided by a non-dusty parameter helped refine the dust \hSED\ fit.
For instance, imagine that you have no constraints on the dust mass of a source, but you know its gas mass.
You could infer its dust mass by taking the mean \hdustiness\ of the rest of the sample.
This \expression{holistic} prior does that, in a smarter way, as it accounts for all the correlations in a statistically consistent way.
\takeaway{The \hHB\ approach allows an optimal, holistic treatment of all the 
          quantities of interest, even if they are not related to the dust.}

    \subsubsection{Other Developments}
    \label{sec:otherHB}

\hSED\ modeling is far from the only possible application of \hHB\ methods.
We briefly discuss below two other models we have developed.

\paragraph{Cosmic dust evolution.}
The dust evolution model we have discussed in \refsec{sec:cosmicdustevol} has been fitted to galaxies by \citet{galliano21}, in a \hHB\ way.
To be precise, we have used the output of \citetalias{galliano18a}, $M_\sms{dust}$, $M_\sms{gas}$, $M_\star$, SFR and metallicity, as observables.
We then have modeled the \hSFH-related parameters in a \hHB\ way, and have assumed that the dust efficiencies were common to all galaxies (\ie\ we inferred one single value for the whole sample).
These \expression{common} parameters were not in the hierarchical prior, because their value is the same for all galaxies.
They were however sampled with the other parameters, in a consistent way.
We were successful at recovering dust evolution timescales consistent with the \hMW\ at Solar metallicity (\cf\ \refsec{sec:dustime}).
One important improvement would be to treat everything within the same \hHB\ model: 
\begin{inlinelist}
  \item the \hSED; 
  \item the stellar and gas parameters; and 
  \item the dust evolution.
\end{inlinelist}
This is something we plan to achieve in the near future.

\paragraph{MIR spectral decomposition.}
The type of \hMIR\ spectral decomposition that we have discussed in \refsec{sec:decompMIR} could also benefit from the \hHB\ approach.
Although the model is mostly linear, there are a lot of degeneracies between the uncertainties of adjacent blended bands, such as the 7.6 and 7.8~\tmic\ features.
In addition, plateaus and weak features are usually poorly constrained and their intensity can considerably bias the fit at low signal-to-noise.
\citetprep{hu21} have developed such a \hHB\ \hMIR\ spectral decomposition tool.
Its efficiency has been assessed on simulated data, and it is now being applied to \M{82}.
This model will be valuable to analyze the spectra from the \hJWST.


\newchapter{Conclusion and Prospective}
\markboth{\chaptername\ \thechapter.\ Prospective}{}
\label{chap:prosp}
\citesmart{We look at the present through a rear view mirror. We march 
           backwards into the future.}{\citep[Marshall 
                                            \familyname{McLuhan};][]{mcluhan67}}
\minitoc

\section{What Have We Learned About ISD in the Past 
          Decade?}
\label{sec:learned}

We have seen in \refsec{sec:particularities} that \hISD\ studies were particular because of the inherent complexity of the dust make-up and of the low weight of evidence provided by individual observables.
This sometimes leads colleagues from other fields to think that our subject is messy and stalling.
If we however integrate over a large number of studies, we can delineate some clear breakthroughs.
I try the exercise of listing them here\footnote{I have, with many others, contributed to the first 5 items of \refsec{sec:conclprop} and to the first 4 items of \refsec{sec:conclevol}.}.
This is of course a subjective account of the progress, biased by my own interests.
The observations acquired by \hhersc\ and \hplanck\ played an instrumental role in these.

  \subsection{About Dust Properties}
  \label{sec:conclprop}

\paragraph{Grain opacity.}
\hhersc\ and \hplanck\ have brought invaluable information about the \hFIR\ and submm optical properties of dust grains, in the \hMW\ and nearby galaxies (\cf\ \refsec{sec:kappaLMC}).
Before that, the commonly-used opacities were a factor of $\simeq2-3$ lower, biasing the dust masses one would estimate.
We still do not know the exact constitution of interstellar grains and its evolution, but we know relatively well the zero level \hFIR-submm opacity they should have in the diffuse \hISM\ of the \hMW\ and a few nearby galaxies.

\paragraph{Scaling relations.}
The large number of galaxies observed by \hhersc\ and the effort to build homogeneous ancillary data samples have provided consistent estimates of the dust, stellar and gaseous properties for a large number of objects, with spatial resolution in numerous cases (\cf\ \refsecs{sec:dustevolISM}{sec:cosmicdustevol}).
These relations provide snapshots of galaxy evolution at different stages.
They are now well sampled over most of the parameter space (metallicity, gas fraction and specific star formation rate).
They are the main benchmarks dust evolution models must account for.
The most important ones and the information they convey are the following.
\begin{itemize}
  \item The \hdustiness-metallicity relation exhibits two distinct regimes 
    of dust production:
    \begin{inlinelist}
      \item at very low metallicity, dominated by stardust; and 
      \item at high metallicity, dominated by grain growth in the \hISM.
    \end{inlinelist}
  \item The \hdustiness-gas-fraction relation links the gas depletion 
    timescale and the evolution probed by the \hdustiness-metallicity relation.
  \item Scaling relations linking the dust content to the diffuse X-ray emission
    are promising tools to constrain grain sputtering timescales.
\end{itemize}

\paragraph{Dust properties of low-metallicity systems.}
I have emphasized in \refsec{sec:cosmicdustevol} that low-metallicity systems (\ie\ dwarf galaxies) were crucial to constrain dust evolution models, because they sample a different grain production regime than higher metallicity objects.
\hhersc\ has provided us with the first \hFIR\ \hSED s of extremely low-metallicity galaxies that were decisive in understanding the early stages of dust evolution.

\paragraph{The Submillimeter excess.}
Twenty years after its discovery the submm excess is still a mysterious epiphenomenon (\cf\ \refsec{sec:submmex}).
In the last years:
\begin{inlinelist}
  \item several studies have hinted that it could be more prominent in diffuse 
    regions of galaxies; and
  \item a new physical process that could explain its origin has been proposed 
    (magnetic grains).
\end{inlinelist}

\paragraph{The AME.}
The \hAME\ has, for the first time, been detected in extragalactic systems (\cf\ \refsec{sec:AME}).
To our humble opinion, the debate about its origin (\hPAH s or nanosilicates) is closed, in favor of \hPAH s.

\paragraph{Stoichiometry and grain structure.}
Important progress has been made constraining the grain structure and stoichiometry using X-ray edge absorption.
The results are sometimes difficult to understand, such as the high crystalline fraction discussed in \refsec{sec:Xrays}.
The technics are however promising and will revolutionize our understanding of the dust constitution, when \hATHENA\ will be observing.

\paragraph{Dust models.}
A few dust models have been published in the last decade.
The \citetalias{jones17} model (\cf\ \refsec{sec:themis}) is, in our opinion, the most innovative for the following reasons.
\begin{itemize}
  \item It is designed as an evolution model.
    Some fundamental properties of the constitution of the grain mixture
    (\hHAC\ hydrogenation, size distribution, mantle thickness) can be 
    varied to account for different observables.
    This way, we have a few physically-grounded parameters to empirically 
    explore the effects of dust evolution.
    It is an important progress over classic, static models.
  \item It is laboratory-data based.
  \item The model can be adapted to account for polarization constraints.
  \item It is the only recent model accounting for the 3.4~\tmic\ aliphatic 
    feature.
  \item It is one of the only models to be consistent with the revised 
    \hFIR-submm opacity that I have mentioned earlier.
\end{itemize}

\paragraph{Polarization.}
Whole-sky submm polarized emission maps have been produced by \hplanck, at several wavelengths (\cf\ \refsec{sec:polaIR}).
On top of permitting unprecedented studies of the magnetic field, they provide evidence that the bulk of the large grain emission is homogeneous in size and composition.

\paragraph{Laboratory data.}
Long-wavelength (\hFIR-submm) measures of a diversity of dust analogues have been produced (\cf\ \refsec{sec:lab}).
They provide the necessary data that, when consistently included in dust models, will help us to:
\begin{inlinelist}
  \item better constrain the evolution of grain mantles; and 
  \item characterize more precisely the submm excess.
\end{inlinelist}

  \subsection{About Dust Evolution}
  \label{sec:conclevol}

\paragraph{Dust sources.}
Our understanding of the dust production mechanisms has considerably progressed, thanks to \hhersc\ (\cf\ \refsecs{sec:dustevolISM}{sec:cosmicdustevol}).
\begin{description}
  \item[The modeling of scaling relations,] in particular the 
    \hdustiness-metallicity relation, provides clear evidence that:
    \begin{inlinelist}
      \item dust growth in the \hISM\ is the prominent grain production regime 
        around Solar metallicity; and 
      \item condensation in \hSNII\ ejecta dominates at very low metallicity.
    \end{inlinelist}
  \item[Observations of individual SNRs] have shown that large amounts of
    dust could be produced short\-ly after \hSNII\ explosions.
    It suggests that a large fraction of these freshly-formed grains must be 
    destroyed by the reverse shock.
\end{description}

\paragraph{Evolution of the Aromatic Feature Carriers.}
Before the 2010s, \hISO\ and \hspitz\ were crucial to understand the variations of the strength of the \hUIB s (\cf\ \refsec{sec:PAH}).
In the 2010s, the detailed modeling of the \hFIR\ \hSED\ permitted by \hhersc, allowed us to understand more finely how the \expression{abundance} of the grains carrying these features evolves.
To our mind, the important points are that:
\begin{inlinelist}
  \item their mass fraction is better correlated with metallicity than with the
    strength of the \hISRF; and
  \item they are spatially associated with molecular clouds.
\end{inlinelist}

\paragraph{Thermal sputtering.}
\hETG s, due to the paucity of their \hISM, had been poorly studied before \hhersc.
We have now been able to characterize their dust content.
It appears that these environments exhibit a dust deficit due to grain destruction in their permeating coronal gas.
They are thus potentially interesting laboratories to constrain sputtering timescales (\cf\ \refsec{sec:XETG}).

\paragraph{Emissivity Variations in the ISM.}
The good coverage and sensitivity of \hhersc\ and \hplanck\ allowed us to better characterize the way the \hFIR-submm emissivity evolves from the diffuse \hISM\ to dense regions, in the \hMW\ and the Magellanic clouds (\cf\ \refsec{sec:dustevolISM}).
It is now clear that the increase of emissivity with \hISM\ density, resulting from mantle accretion and grain coagulation, is a universal process.

\paragraph{Distant objects.}
The dust content of numerous galaxies at very high redshifts ($z>6$) has been constrained, thanks to \hALMA.
It appears that dust-rich objects existed only a few 100~Myrs after reionization, requiring fast grain build-up.
In our opinion, this can be explained with rapid dust growth in the \hISM\ (\cf\ \refsec{sec:cosmicdustevol}).

\section{What Are the Open Questions for the Next 
          Decade?}
\label{sec:questions}

I now try to delineate a few open questions that should occupy us during the next decade.
The list below is as subjective as \refsec{sec:learned}.
This is not everything we \expression{should} do, but rather everything we \expression{will be able} to do, knowing the available observing facilities.

  \subsection{Extragalactic Dust}
  \label{sec:questextra}

\paragraph{Studies of diffuse dust in nearby galaxies.}
The diffuse \hISM\ of the \hMW\ is the only medium used to constrain dust models, because it provides simultaneous information about extinction, emission and elemental depletions (\cf\ \refsec{sec:dustmodels}).
The combination of these different constraints is crucial to solve the degeneracies between emissivity and size distribution.
This is why dust models are calibrated on these data, and why we do not yet have reliable dust models for the \hSMC, for instance.
Such observations are however available, in a fragmented way, in external systems such as the Magellanic clouds and \M{31}.
An effort should be made to produce an homogeneous data set of similar constraints in a few external galaxies and to build dust models using it.
This implies several challenges.
\begin{itemize}
  \item The extraction of the emission from the diffuse \hISM\ is required to
    make sure we have homogeneous physical conditions and are not biased by the
    variation of the \hFIR\ opacity.
    The large beam of \hplanck\ renders this task difficult in external 
    galaxies, but can be compensated by using other observations (\eg\ \hhersc\ 
    and ground-based submm data).
  \item Additional observations are probably needed to cover the missing 
    information.
    Facilities such as the \expression{Multi Unit Spectroscopic Explorer} 
    \citep[\hMUSE;][for abundances]{bacon10} and \hALMA\ (for the submm 
    continuum) provide exceptional data that have not yet been utilized to 
    their full potential in our field.
\end{itemize}
Having different extragalactic dust models would allow us to:
\begin{inlinelist}
  \item understand how the diffuse dust properties vary as a function of
    metallicity; and
  \item have a more robust way to study external galaxies, by using models that
    take into account the effect of cosmic dust evolution on the grain mixture.
\end{inlinelist}

\paragraph{Quiescent low-metallicity galaxies.}
I have emphasized several times the crucial role low-metal\-licity systems play in constraining dust evolution.
These objects are however faint and we usually observe those which are actively star-forming.
We therefore suffer from a selection effect that hides the nature of low-metallicity \expression{quiescent} systems.
I have given an example of what such systems could bring to our understanding of the evolution of \hMIR\ features in \reffig{fig:qAF_LSB}.
The question is in which quadrant of this figure they will fall?
Obtaining \hJWST\ spectra of these objects would probably be a game changer.

\paragraph{Circumgalactic dust.}
Grains present in the immediate vicinity of galaxies, either in the infalling or outflowing gas, are currently poorly known.
Yet, infall and outflow might be an important mechanism regulating the \hdustiness\ of galaxies.
\hJWST\ and \hALMA\ observations might be able to characterize circumgalactic grain properties beyond the nearby Universe, because of their resolving power.
For local objects, \hNIKA\ is currently acquiring mm maps of infalls and outflows in nearby galaxies.

  \subsection{Dust Evolution Modeling}

\paragraph{Local evolution modeling.}
I have emphasized that a dust model such as \citetalias{jones17} provides a unique framework to model \hSED s, taking into account dust evolution.
Its current limitation is however that we lack a quantitative link between the evolution parameters (\hHAC\ hydrogenation, size distribution and mantle thickness) and the environmental conditions (density and \hISRF\ intensity).
This can be achieved by empirically calibrating the tuning parameters of the evolution processes discussed in \refsec{sec:dustevolISM}.
The goal would be to have a reliable dust model predicting the constitution of the grain mixture as a function of $n_\sms{gas}$ and $G_0$.

\paragraph{Cosmic dust evolution models.}
The empirical modeling of cosmic dust evolution, that was the center of \refsec{sec:cosmicdustevol}, calls for several improvements that could greatly change our understanding of the matter.
\begin{enumerate}
  \item If we want to be able to precisely constrain the grain growth and 
    \hSNII\ blast-wave destruction timescales, we need the most accurate 
    possible stellar elemental and dust yields.
    This is an effort asked to the circumstellar community, both modelers and 
    observers.
  \item We need to adopt a more consistent approach between the different 
    physical elements involved (dust, gas and stellar emissions and evolutions; 
    \cf\ \refsec{sec:otherHB}).
  \item When modeling \hSED s at scales of several tens of parsecs or larger, 
    we need to properly account for the mixing of physical conditions.
    This is currently done phenomenologically (\cf\ \refsec{sec:dale}).
    Ideally, we should move toward fitting models accounting for the statistical
    distribution of dust and stars, with a wide range of topologies, accounting 
    for:
    \begin{inlinelist}
      \item dust evolution in the \hISM, as a function of $n_\sms{gas}$ and 
        $G_0$, as I have mentioned earlier; and
      \item radiative transfer.
    \end{inlinelist}
\end{enumerate}

  \subsection{Dusty Epiphenomena}

\paragraph{Long-wavelength properties.}
The current submm-to-cm ground-based observatories (such as \hNIKA, \hALMA, \etc) open windows to progress on our understanding of the submm excess (\cf\ \refsec{sec:submmex}) and of the \hAME\ (\cf\ \refsec{sec:AME}).
By combining these new observations with archival \hspitz, \hAKARI\ and \hhersc\ data, we should be able to systematically test the different possible scenarios.
In addition, this analysis should be performed with dust models including state-of-the-art submm-mm laboratory opacities of interstellar grain analogues, in order to provide a reliable baseline.

\paragraph{DIBs.}
\hDIB s have been extensively observed with \hGaia.
Since we do not know their nature, unbiased exploration of how their properties vary with all available \hISM\ tracers should be performed, in a big data way.

  \subsection{The Need for a New FIR Observatory}

I have illustrated all along this manuscript what modeling the \hIR\ \hSED\ of various regions can bring.
In particular, the \hFIR\ regime is crucial to properly constrain the peak of the large equilibrium grain emission.
This is the only way we can quantify the total dust mass and its excitation conditions.
These quantities are necessary to interpret any other observables, such as the strength of the aromatic features, the submm excess, \etc\
We currently have good archival data for most nearby galaxies.
We however lack:
\begin{inlinelist}
  \item continuous \hMIR-to-\hFIR\ spectroscopy to better constrain \hSED\ 
    models and study the various solid-state features in emission and 
    absorption; and
  \item deep observations of quiescent low-metallicity systems (\cf\ 
    \refsec{sec:questextra}).
\end{inlinelist}
After the cancellation of \hSPICA, our community should regroup around a new project.

  \subsection{The Public Image of Interstellar Dust}

On a \expression{public relation} viewpoint, we should think about the way our field is represented.
\begin{description}
  \item[Among colleagues,] there is still a distinction between \expression{the 
    \hISM} (\ie\ the \hMW) and the \hISM\ of other galaxies.
    This hierarchy between \expression{intragalactic} and 
    \expression{extragalactic} \hISM\ is becoming less and less justified.
    In addition, \expression{interstellar media} (extragalactic \hISM) provide 
    unique constraints on \hISM\ physics, as I have illustrated along this 
    manuscript.
    The plural is justified, as they are characterized by different heating and
    cooling mechanisms, different grain formation processes, \etc\
    We should motivate the new generation of astronomers to consider 
    ISMology as a field where, depending on the studied physical process, we can
    use a galaxy or a Galactic region as a laboratory.
  \item[To the outside world,] \citengl{dust} physics is not very appealing.
    What we do is important, but we need a better name.
    The \href{https://www.epa.gov/sites/default/files/2014-03/documents/ffrrofactsheet_emergingcontaminant_nanomaterials_jan2014_final.pdf}{US environmental protection agency} 
    defines \expression{nanoparticles} as having sizes roughly between 1 and 
    100~nm.
    This is very close to the range of sizes of interstellar grains (\cf\ 
    \refsec{sec:themis_sizedist}).
    We could thus call our object of study \expression{Cosmic NanoParticles} 
    (\hCNP), when talking to the outside world, and keep talking about 
    \expression{dust} between us.
\end{description}

\section{Current Future Projects}

I finish by listing the future projects I am currently considering working on.
These are motivated by the challenges I have listed in \refsec{sec:questions}.

  \subsection{The \textit{Modelosaur} Approach}

It is more and more becoming a necessity to consistently model the different processes (dust, gas and stellar physics), in a large Bayesian framework.
I plan to progressively include more physical processes in \citetalias{galliano18a} to account for a wider diversity of observables:
\begin{enumerate}
  \item the modeling of stellar evolution;
  \item the consistency between \hSED\ modeling, and dust and chemical 
    evolution;
  \item the account of local dust evolution and of the systematic uncertainties 
    of the model \citepalias[depending on the developers of][]{jones17};
  \item radiative transfer grids of different dust-star topologies;
  \item including the constraints from photoionization and photodissociation 
   lines.
\end{enumerate}
A first step will be the project ICED (IAS-CEA Evolution of Dust) that I have put together.
It will consist in modeling the spatial distribution of the dust properties in nearby galaxies, to constrain local grain evolution.
This will be done with the data from \hDustPedia, as well as from the \hNIKA\ guaranteed time project IMEGIN.

  \subsection{Out-of-the-Box Idea Bin}

\begin{itemize}
  \item One way to solve the degeneracy between size distribution and emissivity
    is to observe a given dust mixture illuminated by a time-varying \hISRF.
    This way, we can see the same grains exposed to two different fields, 
    at two different times.
    This is possible with the light echo of \hSN e \citep[\eg][]{arendt16}, but
    very limited.
    Nearby Cepheids provide periodic variable sources, that modern 
    observatories give us access to.
    I am currently working on a feasibility study of the observations of their
    echo with the \hJWST\ and \hALMA.
  \item I currently have an intern working on the feasibility of using 
    \expression{Natural Language Processing} (\hNLP) methods, in collaboration 
    with \href{https://iris.ai/}{IRIS.AI}, to tackle the origin of \hDIB s.
    \hNLP\ is a machine-learning method that, when trained on a large corpus of
    scientific articles, can be used to find relations between concepts, 
    formulate new research directions, \etc\
    It is aimed at dealing with information overload.
    The idea of the project is to see if we can automatically find clues in the 
    chemistry and material science literature that could
    be relevant to \hDIB s.
  \item I am planning to use machine-learning to accelerate the interpolation 
    of large model grids in \citetalias{galliano18a}.
    Machine-learning, trained on hydrodynamical simulations, could also be 
    useful to simulate realistic dust-star topologies (accounting for 
    clustering, \etc), for which we would solve the radiative transfer.
    These grids would be \hSED\ model building blocks.
\end{itemize}


\appendix\stopcontents\startcontents

\newchapter{List of Acronyms}
\citesmart{Education is what is left after you have forgotten all you have 
           learned.}{(Forgotten author)}
\minitoc

\section{General Acronyms}

\setlongtables
\begin{longtable}{|l|l|}
  \hline
    \textbf{Acronym} & \textbf{Expression} \\
  \hline
\endhead
    3D      & \expression{3-Dimensional} \\
    ACF     & \expression{AutoCorrelation Function} \\
    AGB     & \expression{Asymptotic Giant Branch} stars \\
    AGN     & \expression{Active Galactic Nucleus} \\
    AME     & \expression{Anomalous Microwave Emission} \\
    BCD     & \expression{Blue Compact Dwarf Galaxies} \\
    BEMBB   & \expression{Broken-Emissivity Modified Black Body} \\
    BG      & \expression{Big Grain} \\
    BH      & \expression{Black Hole} \\
    CCD     & \expression{Charge-Coupled Device} \\
    CDF     & \expression{Cumulative Distribution Function} \\
    CGS     & \expression{Centimetre-Gram-Second} \\
    CIB     & \expression{Cosmic Infrared Background} \\
    CMB     & \expression{Cosmic Microwave Background} \\
    CNP     & \expression{Cosmic NanoParticles} \\
    CNM     & \expression{Cold Neutral Medium} \\
    DCD     & \expression{Disordered Charge Distribution} \\
    DDA     & \expression{Discrete Dipole Approximation} \\
    DGL     & \expression{Diffuse Galactic Light} \\
    DGS     & \expression{Dwarf Galaxy Sample} \\
    DIB     & \expression{Diffuse Interstellar Bands} \\
    DLA     & \expression{Damped Lyman-Alpha systems} \\
    EMT     & \expression{Effective Medium Theory} \\
    ERE     & \expression{Extended Red Emission} \\
    ETG     & \expression{Early-Type Galaxy} \\
    eVSG    & \expression{evaporating Very Small Grains} \\
    FIR     & \expression{Far-InfraRed} \\
    FUV     & \expression{Far-UltraViolet} \\
    FWHM    & \expression{Full Width at Half Maximum} \\
    GEMS    & \expression{Glass with Embedded Metals and Sulfides} \\
    GRB     & \expression{Gamma-Ray Burst} \\
    HB      & \expression{Hierarchical Bayesian} \\
    HDR     & \expression{Habilitation à Diriger des Recherches} \\
    HIM     & \expression{Hot Ionized Medium} \\
    ICM     & \expression{InterClump Medium} \\
    IDP     & \expression{Interplanetary Dust Particles} \\
    iid     & \expression{independent, identically distributed} \\
    IMF     & \expression{Initial Mass Function} \\
    IR      & \expression{InfraRed} \\
    ISD     & \expression{InterStellar Dust} \\
    ISM     & \expression{InterStellar Medium} \\
    ISRF    & \expression{InterStellar Radiation Field} \\
    ISS     & \expression{International Space Station} \\
    LIRG    & \expression{Luminous InfraRed Galaxies} \\
    LMC     & \expression{Large Magellanic Cloud} \\
    LIMS    & \expression{Low- and Intermediate-Mass Stars} \\
    LTG     & \expression{Late-Type Galaxy} \\
    MBB     & \expression{Modified Black Body} \\
    MCRT    & \expression{Monte-Carlo Radiative Transfer} \\
    MCMC    & \expression{Markov Chain Monte-Carlo} \\
    MIR     & \expression{Mid-InfraRed} \\
    MKS     & \expression{Meter-Kilogram-Second} \\
    MKSA    & \expression{Meter-Kilogram-Second-Ampere} \\
    ML      & \expression{Machine-Learning} \\
    MLE     & \expression{Maximum-Likelihood Estimation} \\
    MS      & \expression{Main Sequence} \\
    MW      & \expression{Milky Way} \\
    NHST    & \expression{Null Hypothesis Significance Test} \\
    NIR     & \expression{Near-InfraRed} \\
    NLP     & \expression{Natural Language Processing} \\
    NS      & \expression{Neutron Star} \\
    NUV     & \expression{Near-UltraViolet} \\
    OOP     & \expression{Out-Of-Plane} \\
    PAH     & \expression{Polycyclic Aromatic Hydrocarbon} \\
    PCA     & \expression{Principal Component Analysis} \\
    PDF     & \expression{Probability Density Function} \\
    PDR     & \expression{PhotoDissociation Regions} \\
    PN      & \expression{Planetary Nebula} \\
    ppb     & \expression{part per billion} \\
    ppp     & \expression{posterior predictive $p$-value} \\
    QSO     & \expression{Quasi-Stellar Object} \\
    RAT     & \expression{Radiative Alignment Torques} \\
    SSC     & \expression{Super Star Cluster} \\
    SED     & \expression{Spectral Energy Distribution} \\
    SF      & \expression{Star Formation} \\
    SFH     & \expression{Star Formation History} \\
    SFR     & \expression{Star Formation Rate} \\
    SI      & \expression{International System of units} \\
    SLED    & \expression{Spectral Line Energy Distribution} \\
    SMC     & \expression{Small Magellanic Cloud} \\
    SN      & \expression{SuperNova} \\
    \snia   & \expression{Type Ia SuperNova} \\
    \snii   & \expression{Type II SuperNova} \\
    SNR     & \expression{SuperNova Remnant} \\
    sSFR    & \expression{specific Star Formation Rate} \\
    SUE     & \expression{Skewed Uncertainty Ellipse} \\
    TIR     & \expression{Total InfraRed} \\
    TLS     & \expression{Two-Level System} \\
    TPAGB   & \expression{Thermally-Pulsing Asymptotic Giant Branch} \\
    TMA     & \expression{Too Many Acronyms} \\
    UIB     & \expression{Unidentified Infrared Bands} \\
    ULIRG   & \expression{UltraLuminous InfraRed Galaxies} \\
    UV      & \expression{UltraViolet} \\
    VCD     & \expression{Very Cold Dust} \\
    VSG     & \expression{Very Small Grain} \\
    YSO     & \expression{Young Stellar Object} \\
    WD      & \expression{White Dwarf} \\
    WIM     & \expression{Warm Ionized Medium} \\
    WNM     & \expression{Warm Neutral Medium} \\
    WR      & \expression{Wolf Rayet} star \\
    ZAMS    & \expression{Zero-Age Main Sequence} \\
  \hline
  \newcap{List of acronyms used throughout the manuscript}{}
  \label{tab:acronym}
\end{longtable}

\section{Telescope and Instrument Acronyms}

\setlongtables
\begin{longtable}{|l|l|}
  \hline
    \textbf{Acronym} & \textbf{Telescope or Instrument} \\
  \hline
\endhead
  ALMA    & \expression{Atacama Large Millimeter/submillimeter Array} \\
  APEX    & \expression{Atacama Pathfinder Experiment} \\
  ATHENA  & \expression{Advanced Telescope for High ENergy Astrophysics} \\
  BLAST   & \expression{Balloon-borne Large Aperture Submillimeter Telescope} \\
  COBE    & \expression{COsmic Background Explorer} \\
  CSO     & \expression{Caltech Submilleter Observatory} \\
  DIRBE   & \expression{Diffuse Infrared Background Experiment} \\
  DMR     & \expression{Differential Microwave Radiometer} \\
  FIRAS   & \expression{Far-InfraRed Absolute Spectrophotometer} \\
  FIS     & \expression{Far-Infrared Surveyor} \\
  FUSE    & \expression{Far Ultraviolet Spectroscopic Explorer} \\
  FTS     & \expression{Fourier Transform Spectrometer} \\
  HFI     & \expression{High Frequency Instrument} \\
  HIFI    & \expression{Heterodyne Instrument for the Far-Infrared} \\
  HST     & \expression{Hubble Space Telescope} \\
  IRAC    & \expression{InfraRed Array Camera} \\
  IRAM    & \expression{Institut de RadioAstronomie Millimétrique} \\
  IRAS    & \expression{InfraRed Astronomical Satellite} \\
  IRC     & \expression{InfraRed Camera} \\
  IRS     & \expression{InfraRed Spectrograph} \\
  IRTF    & \expression{InfraRed Telescope Facility} \\
  ISO     & \expression{Infrared Space Observatory} \\
  IUE     & \expression{International Ultraviolet Explorer} \\
  JCMT    & \expression{James Clerk Maxwell Telescope} \\
  JWST    & \expression{James Webb Space Telescope} \\
  KAO     & \expression{Kuiper Airborne Observatory} \\
  LFI     & \expression{Low Frequency Instrument} \\
  MIPS    & \expression{Multiband Imaging Photometer for Spitzer} \\
  MIRI    & \expression{Mid-InfraRed Instrument} \\
  MUSE    & \expression{Multi Unit Spectroscopic Explorer} \\
  NCT     & \expression{Nuclear Compton Telescope} \\
  NIKA2   & \expression{New IRAM Kids Arrays} \\
  NIRcam  & \expression{Near-InfraRed Camera} \\
  NIRISS  & \expression{Near-InfraRed Imager and Slitless Spectrograph} \\
  NIRspec & \expression{Near-InfraRed Spectrograph} \\
  OAO     & \expression{Orbiting Astronomical Observatory} \\
  PACS    & \expression{Photodetector Array Camera and Spectrometer} \\
  PILOT   & \expression{Polarized Instrument for the Long-wavelength  
                        Observation of the Tenuous ISM} \\
  PRONAOS & \expression{PROjet National pour l'Observation Submillimétrique} \\
  SOFIA   & \expression{Stratospheric Observatory for Infrared Astronomy} \\
  SPICA   & \expression{SPace Infrared telescope for Cosmology and 
                        Astrophysics} \\
  SPIRE   & \expression{Spectral and Photometric Imaging REceiver} \\
  UKIRT   & \expression{United Kingdom InfraRed Telescope} \\
  VLT     & \expression{Very Large Telescope} \\
  WIRO    & \expression{Wyoming InfraRed Observatory} \\
  WISE    & \expression{Wide-field Infrared survey Explorer} \\
  WMAP    & \expression{Wilkinson Microwave Anisotropy Probe} \\
  \hline
  \newcap{List of instrumental acronyms used throughout the manuscript}{}
  \label{tab:telescope}
\end{longtable}

\section{Model and Project Acronyms}

\setlongtables
\begin{longtable}{|l|l|}
  \hline
    \textbf{Acronym} & \textbf{Model Name} \\
  \hline
\endhead
    DGS     & \expression{Dwarf Galaxy Survey} \\
    DustPedia& \expression{A definitive study of dust in the local Universe} \\
    HerBIE  & \expression{HiERarchical Bayesian Inference for dust Emission} \\
    HERITAGE& \expression{HERschel Inventory of The Agents of Galaxy Evolution}
            \\
    SAGE    & \expression{Surveying the Agents of Galaxy Evolution} \\
    THEMIS  & \expression{The Heterogeneous Evolution Model for Interstellar 
                          Solids} \\
  \hline
  \newcap{List of model acronyms used throughout the manuscript}{}
  \label{tab:modelnames}
\end{longtable}

\section{Denomination of the Main Spectral Windows}

\begin{figure}[htbp]
  \includegraphics[width=\textwidth]{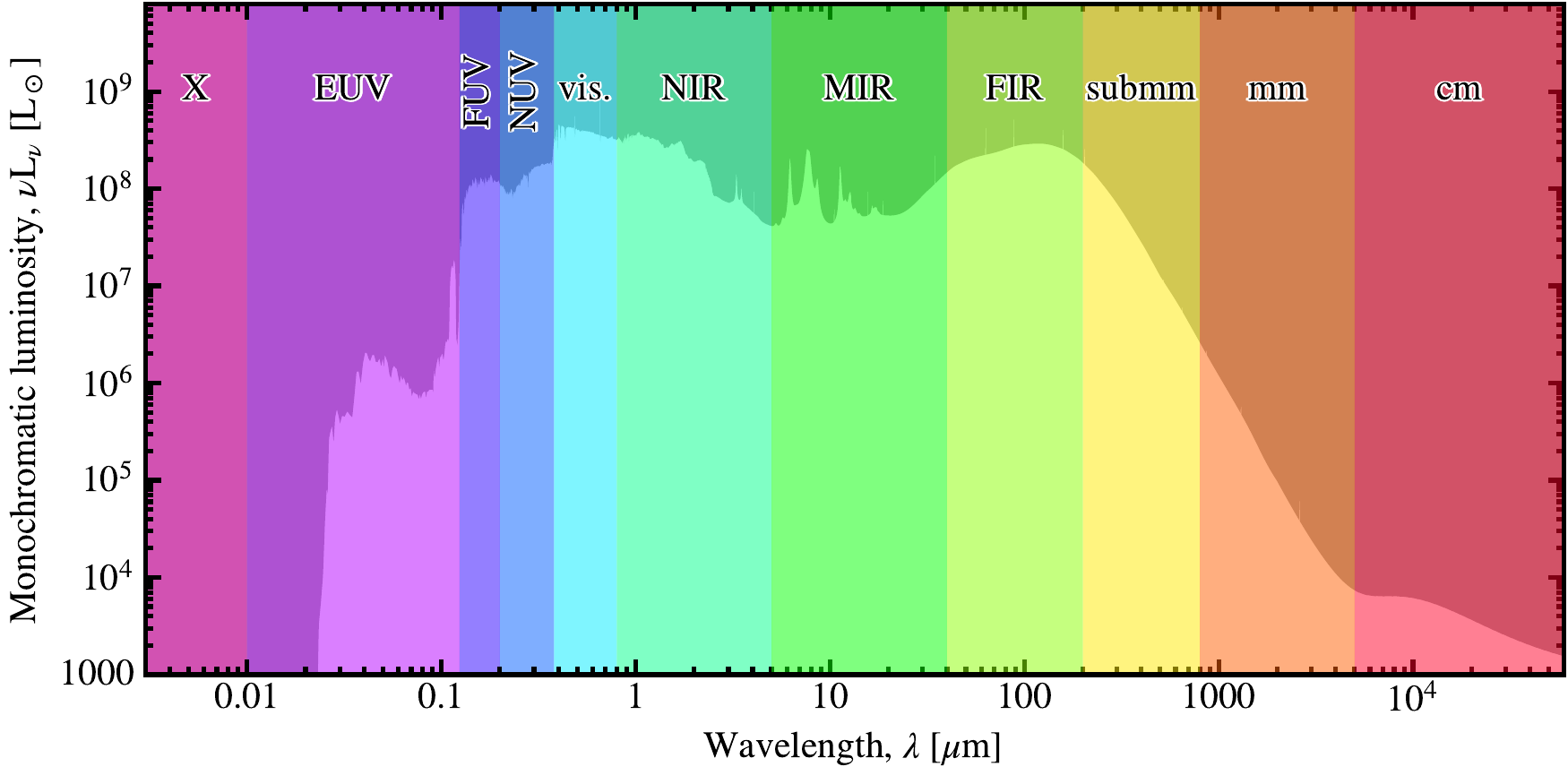}
  \newcap{Spectral domains represented over the SED of a nearby galaxy}%
         {See \reftab{tab:spectralrange} for the acronyms.
          \CClicence}
  \label{fig:spectral_domains}
\end{figure}

\reftab{tab:spectralrange} gives the acronyms and spectral ranges of the most important electromagnetic domains.
For each domain, we give the interval in photon wavelength, frequency and energy.
There are slight variations of these intervals across the literature.
The physical phenomena listed in the right column are indicative of the typical most dominant emission at the scale of a galaxy.
\reffig{fig:spectral_domains} shows where these different domains fall on the \hSED\ of a nearby galaxy.
\begin{table}[htbp]
  \centering
  \setlength\arrayrulewidth{2pt}
  \arrayrulecolor{white}
  \begin{tabularx}{\linewidth}%
    {|>{\columncolor{coltabcell}}l%
     |>{\columncolor{coltabcell}}l%
     |>{\columncolor{coltabcell}}r%
     |>{\columncolor{coltabcell}}r%
     |>{\columncolor{coltabcell}}X|}
    \hline
      \rowcolor{coltabhead}
      \textbf{Abbreviation}
      & \textbf{Name}
      & \multicolumn{2}{c|}{\textbf{Spectral range}}
      & \textbf{Predominant physical origin} \\
      \rowcolor{coltabhead}
      & & \textbf{Start} & \textbf{End} & \\     
    \hline
      \rowcolor{coltabsep} \multicolumn{5}{|c|}{\textsc{High Energies}} \\
    \hline
      \cellcolor{coltabhead}
      $\gamma$ & $\gamma$ rays & \ldots & 0.01 nm
        & Cosmic rays \\
      \cellcolor{coltabhead}
      & & \ldots & 30 EHz & \\
      \cellcolor{coltabhead}
      & & \ldots & 0.12 GeV & \\
    \hline
      \cellcolor{coltabhead}
      X & X rays & 0.01 nm & 10 nm
        & Accretion disks \&\ coronal plasmas \\
      \cellcolor{coltabhead}
      & & 30 EHz & 0.03 EHz & \\
      \cellcolor{coltabhead}
      & & 120 keV & 0.12 keV & \\
    \hline
      \rowcolor{coltabsep} \multicolumn{5}{|c|}{\textsc{UV-Visible}} \\
    \hline
      \cellcolor{coltabhead}
      EUV & Extreme-UV & 10 nm & 124 nm
        & Massive stars \\
      \cellcolor{coltabhead}
      & & 30 PHz & 2.4 PHz & \\
      \cellcolor{coltabhead}
      & & 120 eV & 10 eV & \\
    \hline
      \cellcolor{coltabhead}
      FUV & Far-UltraViolet & 124 nm & 200 nm
        & Massive stars \\
      \cellcolor{coltabhead}
      & & 2.4 PHz & 1.5 PHz & \\
      \cellcolor{coltabhead}
      & & 10 eV   & 6.2 eV & \\
    \hline
      \cellcolor{coltabhead}
      NUV & Near-UV & 200 nm & 380 nm
        & Massive stars \\
      \cellcolor{coltabhead}
      & & 1.5 PHz & 0.8 PHz & \\
      \cellcolor{coltabhead}
      & & 6.2 eV  & 3.3 eV & \\
    \hline
      \cellcolor{coltabhead}
      vis. & Visible & 0.38 \tmic & 0.8 \tmic 
        & Intermediate \&\ low-mass stars \\
      \cellcolor{coltabhead}
      & & 0.8 PHz & 0.4 PHz & \\
      \cellcolor{coltabhead}
      & & 3.3 eV & 1.5 eV & \\
    \hline
      \rowcolor{coltabsep} \multicolumn{5}{|c|}{\textsc{Cold Universe}} \\
    \hline
      \cellcolor{coltabhead}
      NIR & Near-InfraRed & 0.8 \tmic & 5\tmic 
        & Circumstellar material \\
      \cellcolor{coltabhead}
      & & 400 THz & 60 THz & \\
      \cellcolor{coltabhead}
      & & 1.5 eV & 0.25 eV & \\
    \hline
      \cellcolor{coltabhead}
      MIR & Mid-InfraRed & 5 \tmic & 40 \tmic
        & Aromatic features \&\ hot dust \\
      \cellcolor{coltabhead}
      & & 60 THz & 7.5 THz & \\
      \cellcolor{coltabhead}
      & & 250 meV & 30 meV & \\
    \hline
      \cellcolor{coltabhead}
      FIR & Far-InfraRed & 40 \tmic & 200 \tmic
        & Large ISM grains \\
      \cellcolor{coltabhead}
      & & 7.5 THz & 1.5 THz & \\
      \cellcolor{coltabhead}
      & & 30 meV & 6 meV & \\
    \hline
      \cellcolor{coltabhead}
      submm & Submillimeter & 200 \tmic & 800 \tmic
        & Cold dust \\
      \cellcolor{coltabhead}
      & & 1.5 THz & 0.4 THz & \\
      \cellcolor{coltabhead}
      & & 6 meV & 1.5 meV & \\
    \hline
      \rowcolor{coltabsep} \multicolumn{5}{|c|}{\textsc{Radio/Microwave}} \\
    \hline
      \cellcolor{coltabhead}
      mm & Millimeter & 0.8 mm & 5 mm
        & Cold dust \&\ free-free \\
      \cellcolor{coltabhead}
      & & 400 GHz & 60 GHz & \\
      \cellcolor{coltabhead}
      & & 1.5 meV & 0.25 meV & \\
    \hline
      \cellcolor{coltabhead}
      cm & Centimeter & 0.5 cm & 6 cm
        & Free-free \&\ synchrotron \\
      \cellcolor{coltabhead}
      & & 60 GHz & 5 GHz & \\
      \cellcolor{coltabhead}
      & & 250 $\mu$eV & 20 $\mu$eV & \\
    \hline
  \end{tabularx}
  \newcap{Denomination of the main spectral windows}%
         {}
  \label{tab:spectralrange}
\end{table}

\newchapter{Astronomers and Units}
\citesmart{I hope all Americans will do everything in their power to introduce the French metrical system. (...) I look upon our English system as a wickedly, brain-destroying system of bondage under which we suffer. The reason why we continue to use it, is the imaginary difficulty of making a change.}{\citep[William \familyname{Thomson}, Lord Kelvin;][]{thomson89}}
\minitoc

\section{Brief History of Unit Systems}

\paragraph{The necessity to unify disparate measures.}
During Antiquity and the Middle Age, measures of weight, length and duration were varying from one place to another.
The first government in History to try and homogenize measures was under king Henri III, in England.
The \href{https://mjp.univ-perp.fr/constit/uk1297c.htm}{1297 version} of the \expression{Magna Carta} (originally signed in 1215), expressed the will to define standards for measuring weights and distances. \citengl{Article 25. One measure of Wine shall be through our Realm, and one measure of Ale, and one measure of Corn, that is to say, the Quarter of London; and one breadth of dyed Cloth, Russets, and Haberjects, that is to say, two Yards within the lists. And it shall be of Weights as it is of Measures.}
The rest of the world kept using different yards and pounds, for several centuries.

\paragraph{Measuring the Earth.}
In France, the \expression{Académie royale des sciences} was funded in 1666 by Jean-Baptis\-te \familyname{Colbert}, under king Louis XIV, influenced by his secretary for sciences and arts, Charles \familyname{Perrault} \citep{debardat19}.
In 1667, the royal observatory was created and astronomers were tasked with providing more accurate maps of the realm.
Clergyman Jean \familyname{Picard} performed a series of measures by triangulation, with a unique measuring board, and estimated the size of the Earth (\href{https://gallica.bnf.fr/ark:/12148/btv1b7300361b/pdf
}{Mesure de la Terre, 1671}).
During the XVIII$^\sms{th}$ century, several scientific expeditions in Latin America or the North Pole (by La Condamine, Maupertuis, \etal) refined the measurement of the size of the Earth and confirmed its flattening around the poles.

\paragraph{The introduction of the metric system.}
It is only during the French revolution (1789--1799) that the metric system was introduced.
It was consistent with its time.
It was aimed at erasing differences in an abstract way, with new standards independent of the old human references (foot, inch, \etc).
One of its important features was that it was a decimal system, simplifying calculations.
It can be traced back to the months before the revolution, in 1789.
The \expression{cahiers de doléances} (register of grievances) expressed the wish to have unified measures throughout the realm.
In 1790, Charles-Maurice \familyname{de Talleyrand-Périgord}, a bishop elected at the recent national assembly, submitted a memoir to adopt a new system of weights and measures, contributed by Marie-Jean-Antoine \familyname{Critat de Condorcet} and Joseph-Jérôme \familyname{Lefrançois de Lalande} \citep{debardat19}.
In 1792, Jean-Baptiste \familyname{Delambre} and Pierre-François \familyname{Méchain} were charged with measuring the length of the meridian between Dunkerque and Barcelona \citep{alder15}.
The definition of the meter was then $1/10\,000\,000$ of the distance between the North Pole and the equator.
The kilogram was defined as the mass of one cubic decimeter of water.

\paragraph{Difficulty of adoption.}
The metric system was not adopted right away, even in France.
It was mocked by Napoléon, although Laplace promoted the advantage of its decimal system to him.
For a while, we kept \citengl{mesure usuelles}, which were a standardization of imperial units.
In Germany, around 1830, Carl-Friedrich \familyname{Gauss} formalized the metric unit system in physics, and proposed to add the seconds to meters and kilograms, leading to the \expression{\hCGS\ system} (Centimetre-Gram-Second).
During the 1851 World's fair in London, France promoted the metric system to the world.
It led to the \citengl{Treaty of the Metre}, signed in Paris in 1875, by seventeen countries adopting the metric system.
Great Britain, the Netherlands and Portugal were opposed. 
England, in particular, felt that adopting the French system would be a political defeat.
A diplomatic solution was proposed in adopting the Greenwich meridian, during the 1883 Geodetic Congress: \citengl{The Conference hopes that, if the whole world is agreed upon the unification of longitudes and hours in accepting the Greenwich meridian as the point of departure, Great Britain will find in this fact an additional motive to take on her side new steps in favour of the unification of weights and measures, by joining the Metrical Convention of May 20, 1875} \citep{geodetic1883}.
The Greenwich meridian was finally adopted during the 1884 International Meridian Conference, but England did not adopt the metric system\ldots

\paragraph{The international system.}
Since 1960, the \expression{International System of units} (\hSI) is the \expression{Meter-Kilogram-Second-Ampere system} (\hMKSA\ or \hMKS).
Astronomers are one of the last communities to use \hCGS\ units (and the Gaussian system for electrodynamics; \reftab{tab:EM}).
The continued use of a mix between imperial units and the \hCGS\ system is counterproductive.
The most dramatic example is the crash of the 1999 \expression{Mars Climate Observer} probe, because of a conversion mistake between imperial and metric units (\href{https://www.popularmechanics.com/space/moon-mars/news/a28632/the-dumb-mistake-that-doomed-a-mars-probe-in-1999/}{Popular Mechanics, 2017}).
\reftab{tab:unit} gives the correspondence between the \hMKS\ and \hCGS\ systems, and \reftabs{tab:constants}{tab:astroconst} list the fundamental constants in both systems.
\reftab{tab:EM} compares the \hMKSA\ and Gaussian systems for electrodynamics.

\begin{table}[htbp]
  \centering
  \setlength\arrayrulewidth{2pt}
  \arrayrulecolor{white}
  \begin{tabularx}{\linewidth}%
    {|>{\columncolor{coltabcell}}X%
     |>{\columncolor{coltabcell}}r%
      >{\columncolor{coltabcell}}l%
     |>{\columncolor{coltabcell}}r%
      >{\columncolor{coltabcell}}l|}
    \hline
      \rowcolor{coltabhead}\textbf{Quantity}
        & \multicolumn{2}{l|}{\textbf{International units (\hMKS)}} 
        & \multicolumn{2}{l|}{\textbf{Astronomer's units (\hCGS)}} \\
    \hline
      \rowcolor{coltabsep}\multicolumn{5}{|c|}{\textsc{General}} \\
    \hline
      \cellcolor{coltabhead}
      Length & 1 & m (meter) & $10^2$ & cm (centimeter) \\
    \hline
      \cellcolor{coltabhead}
      Force & 1 & N (Newton) & $10^5$ & dyn (dynes) \\
    \hline
      \cellcolor{coltabhead}
      Energy & 1 & J (joule) & $10^7$ & erg \\
    \hline
      \cellcolor{coltabhead}
      Power & 1 & W (watt) & $10^7$ & erg/s \\
    \hline
      \cellcolor{coltabhead}
      Flux (1 Jansky) & $10^{-26}$ & W/m$^2$/Hz & $10^{-23}$ & erg/s/cm$^2$/Hz \\
    \hline
      \rowcolor{coltabsep}\multicolumn{5}{|c|}{\textsc{Electromagnetism}} \\
    \hline
      \cellcolor{coltabhead}
      Charge & 1 & C (Coulomb) & $2.997\,924\,58\E{9}$
      & esu (electrostatic unit) \\
    \hline
      \cellcolor{coltabhead}
      Current & 1 & A (Ampere) & $2.997\,924\,58\E{9}$ & esu/s \\
    \hline
      \cellcolor{coltabhead}
      Electric potential & 1 & V (volt) & $1/299.792\,458$ & statV (statvolt) \\
    \hline
      \cellcolor{coltabhead}
      Electric field & 1 & V/m & $1/299\,79.245\,8$ & statV/cm \\
    \hline
      \cellcolor{coltabhead}
      Magnetic field & 1 & T (Tesla) & $10^4$ & G (Gauss) \\
    \hline
      \cellcolor{coltabhead}
      Magnetic flux & 1 & Wb (Weber) & $10^8$ & G.cm$^2$ \\ 
    \hline
      \cellcolor{coltabhead}
      Auxiliary field H & 1 & A/m & $4\pi\E{-3}$ & Oe (Oersted) \\
    \hline
      \rowcolor{coltabsep}\multicolumn{5}{|c|}{\textsc{Angular distance}} \\
    \hline
      \cellcolor{coltabhead}
      1 arcsec & $4.848\E{-6}$ & rad & & \\
    \hline
      \cellcolor{coltabhead}
      1 arcsec at 1 Mpc & 4.848 & pc & & \\
    \hline
  \end{tabularx}
  \newcap{Unit conversion}{}
  \label{tab:unit}
\end{table}

\begin{table}[htbp]
  \centering
  \setlength\arrayrulewidth{2pt}
  \arrayrulecolor{white}
  \begin{tabularx}{\linewidth}%
    {|>{\columncolor{coltabcell}}X%
     |>{\columncolor{coltabcell}}r%
      >{\columncolor{coltabcell}}l%
     |>{\columncolor{coltabcell}}r%
      >{\columncolor{coltabcell}}l|}
    \hline
      \rowcolor{coltabhead}\textbf{Quantity}
        & \multicolumn{2}{l|}{\textbf{International units (MKS)}} 
        & \multicolumn{2}{l|}{\textbf{Astronomer's units (CGS)}} \\
    \hline
      \rowcolor{coltabsep}\multicolumn{5}{|c|}{\textsc{Universal}} \\
    \hline
    \cellcolor{coltabhead}
    Light speed, $c$ & $2.999\,792\,458\E{8}$ & m/s
      & $2.999\,792\,458\E{10}$ & cm/s \\
    \hline
    \cellcolor{coltabhead}
    Newton constant, $\mathcal{G}$ & $6.674\,28\E{-11}$ & m$^3$/kg/s$^2$
      & $6.674\,28\E{-8}$ & cm$^3$/g/s$^2$ \\
    \hline
    \cellcolor{coltabhead}
    Planck constant, $h$ & $6.626\,068\,96\E{-34}$ & J.s
      & $6.626\,068\,96\E{-27}$ & erg.s \\
    \cellcolor{coltabhead}
    $\hbar\equiv h/2\pi$ & $1.054\,571\,628\E{-34}$ & J.s
      & $1.054\,571\,628\E{-27}$ & erg.s \\
    \hline
    \cellcolor{coltabhead}
    Magnetic constant, &&&&\\ 
    \cellcolor{coltabhead}
    $\mu_0\equiv4\pi\E{-7}$ & $1.256\,637\,061\,4\E{-6}$ & N/A$^2$ & & \\
    \hline
    \cellcolor{coltabhead}
    Electric constant, & & & & \\
    \cellcolor{coltabhead}
    $\epsilon_0\equiv1/\mu_0c^2$ & $8.854\,187\,817\E{-12}$ & F/m & & \\
    \hline
    \cellcolor{coltabhead}
    Elementary charge, $e$ & $1.602\,176\,487\E{-19}$ & C
      & $4.806\,529\,5\E{-10}$ & esu \\
    \hline
      \rowcolor{coltabsep}\multicolumn{5}{|c|}{\textsc{Atomic}} \\
    \hline
    \cellcolor{coltabhead}
    Electron mass, $m_e$ & $9.109\,382\,15\E{-31}$ & kg
      & $9.109\,382\,15\E{-28}$ & g \\
    \cellcolor{coltabhead}
    $m_ec^2$ & $8.187\,104\,38\E{-14}$ & J & $8.187\,104\,38\E{-7}$ & erg \\
    \cellcolor{coltabhead}
    $m_ec^2/e$ & $0.510\,998\,910$ & MeV & & \\
    \hline
    \cellcolor{coltabhead}
    Proton mass, $m_p$ & $1.672\,621\,637\E{-27}$ & kg
      & $1.672\,621\,637\E{-24}$ & g \\
    \cellcolor{coltabhead}
    $m_pc^2$ & $1.503\,277\,359\E{-10}$ & J & $1.503\,277\,359\E{-3}$ & erg \\
    \cellcolor{coltabhead}
    $m_pc^2/e$ & $0.938\,272\,013$ & GeV & & \\
    \hline
    \cellcolor{coltabhead}
    Rydberg, &&&&\\
    \cellcolor{coltabhead}
    $R_\infty\equiv\alpha^2m_ec/2h$ & $10\,973\,731.568\,527$ & m$^{-1}$
      & $109\,737.315\,685\,27$ & cm$^{-1}$ \\
    \cellcolor{coltabhead}
    $R_\infty c$ & $3.289\,841\,960\,361\E{15}$ & Hz & & \\
    \cellcolor{coltabhead}
    $R_\infty hc$ & $2.179\,871\,97\E{-18}$ & J & $2.179\,871\,97\E{-11}$ 
      & erg \\
    \cellcolor{coltabhead}
    $R_\infty hc/e$ & $13.605\,691\,93$ & eV & & \\
    \hline
      \rowcolor{coltabsep}\multicolumn{5}{|c|}{\textsc{Macroscopic}} \\
    \hline
    \cellcolor{coltabhead}
    Boltzmann constant, $k$ & $1.380\,650\,4\E{-23}$ & J/K
      & $1.380\,650\,4\E{-16}$ & erg/K \\
    \cellcolor{coltabhead}
    $k/e$
      & $8.617\,343\E{-5}$ & eV/K & & \\
    \cellcolor{coltabhead}
    $k/eh$ & $69.503\,56\E{10}$ & Hz/K & & \\
    \hline
    \cellcolor{coltabhead}
    Atomic mass unit, & & & & \\
    \cellcolor{coltabhead}
    $m_u\equiv m({\rm ^{12}C})/12$ & $1.660\,538\,782\E{-27}$ & kg
      & $1.660\,538\,782\E{-24}$ & g \\
    \hline
    \cellcolor{coltabhead}
    Avogadro number, $\mathcal{N}_A$ & $6.022\,141\,79\E{23}$ & mol$^{-1}$ & & \\
    \hline
    \cellcolor{coltabhead}
    Molar gas constant, &&&&\\
    \cellcolor{coltabhead}
    $R\equiv k\mathcal{N}_A$ & $8.314\,472$ & J/mol/K
      & $8.314\,472\E{7}$ & erg/mol/K \\
  \end{tabularx}
  \newcap{Fundamental constants}{}
  \label{tab:constants}
\end{table}

\begin{table}[htbp]
  \centering
  \setlength\arrayrulewidth{2pt}
  \arrayrulecolor{white}
  \begin{tabularx}{\linewidth}%
    {|>{\columncolor{coltabcell}}X%
     |>{\columncolor{coltabcell}}r%
      >{\columncolor{coltabcell}}l%
     |>{\columncolor{coltabcell}}r%
      >{\columncolor{coltabcell}}l|}
    \hline
      \rowcolor{coltabhead}\textbf{Quantity}
        & \multicolumn{2}{l|}{\textbf{International units (MKS)}} 
        & \multicolumn{2}{l|}{\textbf{Astronomer's units (CGS)}} \\
    \hline
      \rowcolor{coltabsep}\multicolumn{5}{|c|}{\textsc{General}} \\
    \hline
    \cellcolor{coltabhead}
    Astronomical unit, ${\rm a.u.}\equiv\langle\odot-\oplus\rangle$
      & $1.495\,979\E{11}$ & m & $1.495\,979\E{13}$ & cm \\
    \hline
    \cellcolor{coltabhead}
    Parsec, $\rm pc\equiv1\;a.u./1''$
      & $3.085\,678\E{16}$ & m & $3.085\,678\E{18}$ & cm \\
    \hline
      \rowcolor{coltabsep}\multicolumn{5}{|c|}{\textsc{Solar system}} \\
    \hline
    \cellcolor{coltabhead}
    Solar radius, $R_\odot$ & $6.959\,9\E{8}$ & m & $6.959\,9\E{10}$ & cm \\
    \hline
    \cellcolor{coltabhead}
    Solar mass, $M_\odot$ & $1.988\,9\E{30}$ & kg & $1.988\,9\E{33}$ & g \\
    \hline
    \cellcolor{coltabhead}
    Solar luminosity, $L_\odot$ & $3.846\E{26}$ & W & $3.846\E{33}$ & erg/s \\
    \hline
    \cellcolor{coltabhead}
    Earth radius, $R_\oplus$ & $6.378\,140\E{6}$ & m & $6.378\,140\E{8}$ & cm \\
    \hline
    \cellcolor{coltabhead}
    Earth mass, $M_\oplus$ & $5.974\E{24}$ & kg & $5.974\E{27}$ & g \\
    \hline
      \rowcolor{coltabsep}\multicolumn{5}{|c|}{\textsc{Galaxy}} \\
    \hline
    \cellcolor{coltabhead}
    Solar velocity around G.C., $\Theta_\circ$ & $220$ & km/s & & \\
    \hline
    \cellcolor{coltabhead}
    Distance sun-G.C., $R_\circ$ & $8.0$ & kpc & & \\
    \hline
    \cellcolor{coltabhead}
    Local disk density, $\rho_\sms{disk}$ & $3-12\E{-21}$ & kg/m$^3$
      & $3-12\E{-24}$ & g/cm$^3$ \\
    \cellcolor{coltabhead}
    $n_\sms{disk}$ & $1-5\E{6}$ & m$^{-3}$ & $1-5$ & cm$^{-3}$ \\
    \hline
    \cellcolor{coltabhead}
    Local halo density, $\rho_\sms{halo}$
      & $2-13\E{-22}$ & kg/m$^3$ & $2-13\E{-25}$ & g/cm$^3$ \\
    \cellcolor{coltabhead}
    $n_\sms{halo}$ & $10-60\E{4}$ & m$^{-3}$ & $0.1-0.6$ & cm$^{-3}$ \\
    \hline
      \rowcolor{coltabsep}\multicolumn{5}{|c|}{\textsc{Cosmology}} \\
    \hline
    \cellcolor{coltabhead}
    Hubble expansion rate, $H_0$ & $71$ & km/s/Mpc & & \\
    \hline
    \cellcolor{coltabhead}
    Critical density, $\rho_c\equiv3H_0^2/8\pi\mathcal{G}$
      & $1.399\,062\E{11}$ & $M_\odot$/Mpc$^3$ & $9.472\E{-30}$ & g/cm$^3$ \\
    \hline
    \cellcolor{coltabhead}
    Pressureless matter density, &&&&\\
    \cellcolor{coltabhead}
    $\Omega_M\equiv\rho_M/\rho_c$
      & $0.15\lesssim\Omega_M\lesssim0.45$ & & & \\
    \hline
    \cellcolor{coltabhead}
    Baryon density, $\Omega_B\equiv\rho_B/\rho_c$
      & $0.019\lesssim\Omega_B\lesssim0.046$ & & & \\
    \hline
    \cellcolor{coltabhead}
    Cosmological constant, &&&&\\
    \cellcolor{coltabhead}
    $\Omega_\Lambda\equiv\Lambda_c^2/3H_0^2$
      & $0.6\lesssim\Omega_\Lambda\lesssim0.8$ & & & \\
  \end{tabularx}
  \newcap{Astronomical constants}{}
  \label{tab:astroconst}
\end{table}

\begin{table}[htbp]
  \centering
  \setlength\arrayrulewidth{2pt}
  \arrayrulecolor{white}
  \begin{tabularx}{\linewidth}%
    {|>{\columncolor{coltabcell}}X%
     |>{\columncolor{coltabcell}}l%
     |>{\columncolor{coltabcell}}l}
    \hline
      \rowcolor{coltabhead}\textbf{Quantity}
        & \textbf{Rationalized MKSA} & \textbf{Gaussian units} \\
    \hline
      \cellcolor{coltabhead}
      & & \\
    Lorentz Force
      \cellcolor{coltabhead}
      & $\displaystyle
               \lvec{F}=q\left(\lvec{E}+\lvec{v}\wedge\lvec{B}\right)$
      & $\displaystyle
         \lvec{F}=q\left(\lvec{E}+\frac{\lvec{v}}{c}\wedge\lvec{B}\right)$ \\
      \cellcolor{coltabhead}
      & $\displaystyle
               \frac{\dd\lvec{F}}{\dd V}=\rho\lvec{E}+\lvec{j}\wedge\lvec{B}$
      & $\displaystyle
               \frac{\dd\lvec{F}}{\dd V}
               =\rho\lvec{E}+\frac{\lvec{j}}{c}\wedge\lvec{B}$ \\   
      \cellcolor{coltabhead}
      & & \\
    \hline
      \cellcolor{coltabhead}
      & & \\
      \cellcolor{coltabhead}
    Dielectric Constant \&\ Permeability
      & $\displaystyle\epsilon_0=\frac{10^7}{4\pi c^2}\;\;\;\;\;\;\;
               \mu_0 = 4\pi 10^{-7}$
      & $\displaystyle\epsilon_0=1\;\;\;\;\;\;\;\mu_0=1$ \\
      \cellcolor{coltabhead}
      & & \\
    \hline
      \cellcolor{coltabhead}
      & & \\
      \cellcolor{coltabhead}
    Displacement \&\ Magnetic Field
      & $\displaystyle\lvec{D}=\epsilon\lvec{E}+\lvec{P}$
      & $\displaystyle\lvec{D}=\epsilon\lvec{E}+4\pi\lvec{P}$ \\
      \cellcolor{coltabhead}
      & $\displaystyle\lvec{H}=\frac{\lvec{B}}{\mu}-\lvec{M}$
      & $\displaystyle\lvec{H}=\frac{\lvec{B}}{\mu}-4\pi\lvec{M}$ \\
      \cellcolor{coltabhead}
      & & \\
    \hline
      \cellcolor{coltabhead}
      & & \\
      \cellcolor{coltabhead}
    Maxwell Equations
      & $\displaystyle \lvec{\nabla}.\lvec{D} = \rho$
      & $\displaystyle \lvec{\nabla}.\lvec{D} = 4\pi\rho$
      \\
      \cellcolor{coltabhead}
      & $\displaystyle \lvec{\nabla}\wedge\lvec{H}
                 =\lvec{j}+\frac{\partial\lvec{D}}{\partial t}$
      & $\displaystyle \lvec{\nabla}\wedge\lvec{H}
                 =\frac{4\pi}{c}\lvec{j}
                   +\frac{1}{c}\frac{\partial\lvec{D}}{\partial t}$
      \\
      \cellcolor{coltabhead}
      & $\displaystyle \lvec{\nabla}\wedge\lvec{E}
               +\frac{\partial\lvec{B}}{\partial t}=\lvec{0}$
      & $\displaystyle \lvec{\nabla}\wedge\lvec{E}
                +\frac{1}{c}\frac{\partial\lvec{B}}{\partial t}=\lvec{0}$
      \\
      \cellcolor{coltabhead}
      & $\displaystyle \lvec{\nabla}.\lvec{B} = 0$
      & $\displaystyle \lvec{\nabla}.\lvec{B} = 0$
      \\
      \cellcolor{coltabhead}
      & & \\
    \hline
      \cellcolor{coltabhead}
      & & \\
      \cellcolor{coltabhead}
    Poynting Vector
      & $\displaystyle
               \lvec{\mathcal{P}} = \lvec{E}\wedge\lvec{H}$
      & $\displaystyle
                \lvec{\mathcal{P}} = \frac{c}{4\pi}\lvec{E}\wedge\lvec{H}$ \\
      \cellcolor{coltabhead}
      & & \\
    \hline
      \cellcolor{coltabhead}
      & & \\
      \cellcolor{coltabhead}
    Electromagnetic Power
      & $\displaystyle
               P = \iint_S \lvec{\mathcal{P}}.\lvec{\dd S}$
      & $\displaystyle
               P = \iint_S \lvec{\mathcal{P}}.\lvec{\dd S}$ \\
      \cellcolor{coltabhead}
      & & \\
    \hline
      \cellcolor{coltabhead}
      & & \\
      \cellcolor{coltabhead}
    Energy Density
      & $\displaystyle
        U=\frac{1}{2}\left(\epsilon\lvec{E}^2+\frac{\lvec{B}^2}{\mu}\right)$
      & $\displaystyle
        U=\frac{1}{8\pi}\left(\epsilon\lvec{E}^2+\frac{\lvec{B}^2}{\mu}\right)$
      \\
      \cellcolor{coltabhead}
      & & \\
    \hline
  \end{tabularx}
  \newcap{Classical electrodynamics}{}
  \label{tab:EM}
\end{table}

\section{Working with Units}

My personal experience in working with units led me to the following advices.
\begin{description}
  \item[Adopt specific units for each problem,] so that the quantities one 
    have to deal with are close to unity (in orders of magnitude).
    This is particularly important to avoid numerical problems.
    In ISMology, the \tmic\ is a good wavelength unit, and the cm$^{-3}$ a good
    density unit, and the pc a good distance unit.
  \item[Deciding which quantity should be logarithmic] must be based on the
    way the uncertainty on this quantity has been estimated.
    The conversion between linear and logarithmic quantities indeed is not
    straightforward.
    Keeping the same $1\sigma$ range is a good practice, but it is not 
    rigorously equivalent:
    \begin{equation}
      \log X\pm \sigma_{\log X} \;\;\;\nLeftrightarrow\;\;\;
        X_{\displaystyle-(1-10^{-\sigma_{\log X}})X}^{\displaystyle+(10^{\sigma_{\log X}}-1)X}.
    \end{equation}
    The underlying probability law is different in both cases.
    It is preferable to choose a representation of the quantity so that its
    uncertainty is the closest to a normal law.
    If we are estimating the uncertainty on a parameter, the dynamical range is
    important.
    If a quantity varies by more than one order of magnitude, it is often a 
    good choice to treat the logarithm of this quantity in a Bayesian model.
    Concerning fluxes, the magnitude system is logarithmic, but it is not a 
    decimal system (it uses a cumbersome 2.5 factor), and it relies on 
    arbitrary zero-point fluxes (\cf\ \reftab{tab:magnitude}).
    The only useful formula concerning magnitudes is to get out of them:
    \begin{equation}
      F_\nu(\lambda_0) = F_{\nu,0}\times10^{-0.4m(\lambda_0)}.
    \end{equation}
  \item[Perform conversions through the \hSI,] and avoid \hCGS\ units, which are
    are already deprecated outside astronomy.
    \hCGS\ are indeed boomer units, so is the Gaussian unit system in 
    electrodynamics.
    In terms of computing, it is important to have a reliable conversion module 
    in the different programming languages that one uses, to avoid stupid 
    mistakes.
\end{description}

\begin{table}[htbp]
  \centering
  \setlength\arrayrulewidth{2pt}
  \arrayrulecolor{white}
  \begin{tabularx}{0.6\linewidth}%
    {|>{\columncolor{coltabhead}}X%
     |>{\columncolor{coltabcell}}l%
     |>{\columncolor{coltabcell}}l}
    \hline
      \rowcolor{coltabhead}\cellcolor{white} 
      & \textbf{Central wavelength, $\bm{\lambda_0}$} 
      & \textbf{Zero-point flux, F$\bm{_{\nu,0}}$} \\
    \hline
      U & 0.36 \tmic & 1884 Jy \\
    \hline
      B & 0.44 \tmic & 4646 Jy \\
    \hline
      V & 0.55 \tmic & 3953 Jy \\
    \hline
      R & 0.66 \tmic & 2875 Jy \\
    \hline
      I & 0.80 \tmic & 2241 Jy \\
    \hline
      J & 1.25 \tmic & 1602 Jy \\
    \hline
      H & 1.60 \tmic & 1010 Jy \\
    \hline
      K & 2.18 \tmic & 630 Jy \\
    \hline
      L & 3.45 \tmic & 278 Jy \\
    \hline
      M & 4.75 \tmic & 153 Jy \\
    \hline
      N & 10.6 \tmic & 36.3 Jy \\
    \hline
  \end{tabularx}
  \newcap{The magnitude system}{}
  \label{tab:magnitude}
\end{table}

\newchapter{Useful Formulae}
\citesmart{There is no royal road to science, and only those who do not dread 
           the fatiguing climb of its steep paths have a chance of gaining its 
           luminous summits.}{\citep[Karl \familyname{Marx};][]{marx1872}}
\minitoc

\noindent
This appendix contains various unrelated formulae that I find useful in my daily practice, but that I do not necessarily remember.
The problem is that gathering them here suppresses any incentive to memorize them, therefore making it essential to have them written here.

\section{3D Quantities and Volume Integrals}
\label{sec:integrals}

\subsection{Differential Operators}

\begin{figure}[htbp]
  \includegraphics[width=\textwidth]{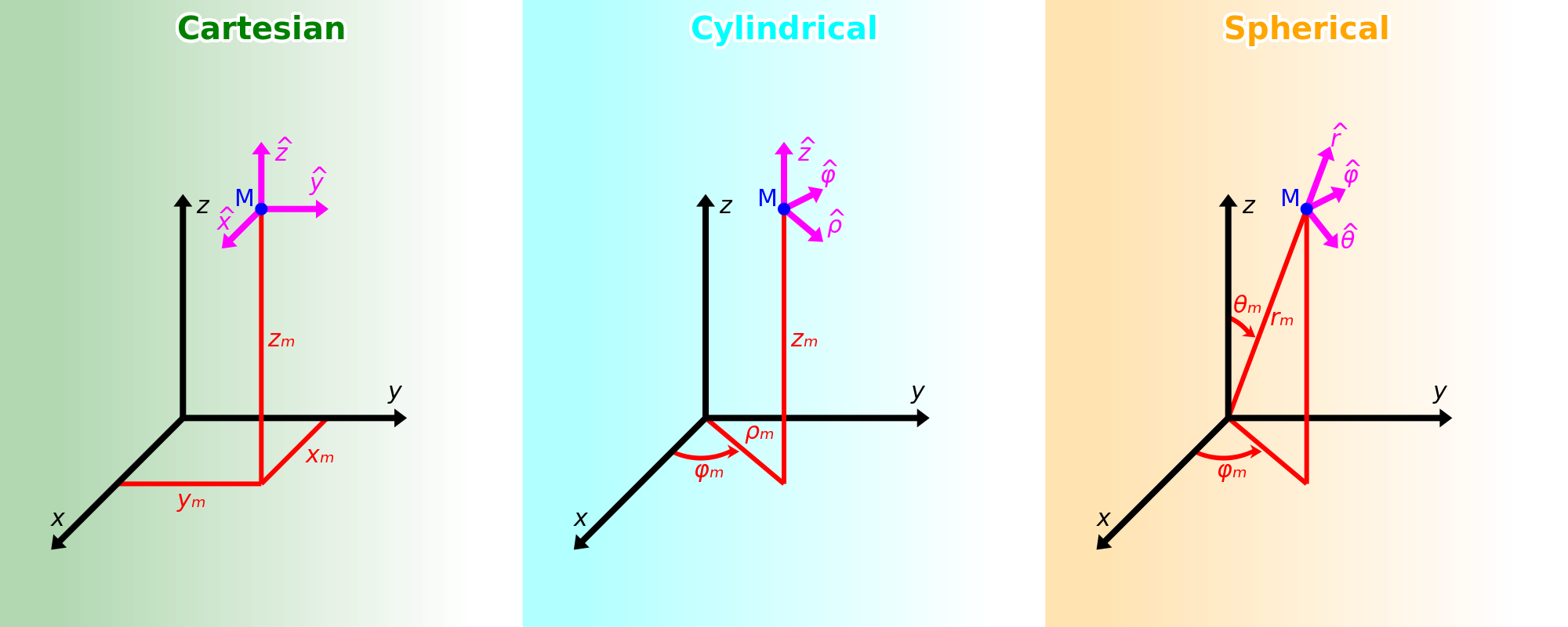}
  \begin{tabularx}{\linewidth}{XXX}
    $\dd V=\dd x.\dd y.\dd z$
    & $\dd V = \dd\rho.\dd\phi.\dd z$
    & $\dd V = \dd r.\dd\phi.\dd\cos\theta$ \\
  \end{tabularx}
  \newcap{Most common coordinate systems}{\CClicence}
  \label{fig:coords}
\end{figure}

The three most common coordinate systems are represented in \reffig{fig:coords}.
In what follows, we express the differential operators in these systems.

\subsubsection{Gradient}
\begin{description}
  \item[Cartesian:]
    $\lvec{\nabla}U=\displaystyle\frac{\partial U}{\partial x}\hat{x}+
                    \frac{\partial U}{\partial y}\hat{y}+
                    \frac{\partial U}{\partial z}\hat{z}$.
  \item[Cylindrical:]
    $\lvec{\nabla}U=\displaystyle\frac{\partial U}{\partial\rho}\hat{\rho}+
                    \frac{1}{\rho}\frac{\partial U}{\partial\phi}\hat{\phi}+
                    \frac{\partial U}{\partial z}\hat{z}$.
  \item[Spherical:]
    $\lvec{\nabla}U=\displaystyle\frac{\partial U}{\partial r}\hat{r}+
                    \frac{1}{r}\frac{\partial U}{\partial\theta}\hat{\theta}+
                    \frac{1}{r\sin\theta}
                             \frac{\partial U}{\partial\phi}\hat{\phi}$.
\end{description}

\subsubsection{Laplacian}
\begin{description}
  \item[Cartesian:]
    $\lvec{\nabla}^2 U = \displaystyle \frac{\partial^2 U}{\partial x^2} +
                \frac{\partial^2 U}{\partial y^2} +
                \frac{\partial^2 U}{\partial z^2}$.
  \item[Cylindrical:]
    $\lvec{\nabla}^2 U = \displaystyle 
                \frac{1}{\rho}\frac{\partial}{\partial\rho}
                  \left(\rho\frac{\partial U}{\partial\rho}\right) +
                \frac{1}{\rho^2}\frac{\partial^2 U}{\partial\phi^2} +
                \frac{\partial^2 U}{\partial z^2}$.
  \item[Spherical:]
    $\lvec{\nabla}^2 U = \displaystyle 
                \frac{1}{r}\frac{\partial^2}{\partial r^2}
                  \left(rU\right) +
                \frac{1}{r^2\sin\theta}\frac{\partial}{\partial\theta}
                  \left(\sin\theta\frac{\partial U}{\partial\theta}\right)
                  +
                \frac{1}{r^2\sin^2\theta}\frac{\partial^2 U}{\partial\phi^2}$.
\end{description}

\subsubsection{Divergence}
\begin{description}
  \item[Cartesian:]
    $\displaystyle\lvec{\nabla}.\lvec{A}=\frac{\partial A_x}{\partial x} +
     \frac{\partial A_y}{\partial y} + \frac{\partial A_z}{\partial z}$.
  \item[Cylindrical:]
    $\displaystyle\lvec{\nabla}.\lvec{A}=
     \frac{1}{\rho}\frac{\partial}{\partial\rho}\left(\rho A_\rho\right) +
     \frac{1}{\rho}\frac{\partial A_\phi}{\partial\phi} + 
     \frac{\partial A_z}{\partial z}$.
  \item[Spherical:]
    $\displaystyle\lvec{\nabla}.\lvec{A} =
     \frac{1}{r^2}\frac{\partial}{\partial r}\left(r^2 A_r\right) +
     \frac{1}{r\sin\theta}\frac{\partial}{\partial\theta}
       \left(\sin\theta A_\theta\right) + 
     \frac{1}{r\sin\theta}\frac{\partial A_\phi}{\partial\phi}$.
\end{description}

\subsubsection{Curl}
\begin{description}
  \item[Cartesian:]
    $\displaystyle\lvec{\nabla}\wedge\lvec{A} =
     \left(\frac{\partial A_z}{\partial y}-\frac{\partial A_y}{\partial z}
           \right)\hat{x} + 
     \left(\frac{\partial A_x}{\partial z}-\frac{\partial A_z}{\partial x}
           \right)\hat{y} +
     \left(\frac{\partial A_y}{\partial x}-\frac{\partial A_x}{\partial y}
           \right)\hat{z}$.
  \item[Cylindrical:]
    $\displaystyle\lvec{\nabla}\wedge\lvec{A} =
     \left(\frac{1}{\rho}\frac{\partial A_z}{\partial\phi}
           -\frac{\partial A_\phi}{\partial z}
           \right)\hat{\rho} + 
     \left(\frac{\partial A_\rho}{\partial z}-\frac{\partial A_z}{\partial \rho}
           \right)\hat{\phi} +
     \frac{1}{\rho}
       \left(\frac{\partial\rho A_\phi}{\partial\rho}
             -\frac{\partial A_\rho}{\partial\phi}
           \right)\hat{z}$.
  \item[Spherical:]
    $\displaystyle\lvec{\nabla}\wedge\lvec{A} =
     \frac{1}{r\sin\theta}\left(\frac{\partial\sin\theta A_\phi}{\partial\theta}
           -\frac{\partial A_\theta}{\partial\phi}\right)\hat{r} +
     \frac{1}{r\sin\theta}\left(\frac{\partial A_r}{\partial\phi}
           -\sin\theta\frac{\partial r A_\phi}{\partial r}
           \right)\hat{\theta} +
     \frac{1}{r}
       \left(\frac{\partial r A_\theta}{\partial r}
             -\frac{\partial A_r}{\partial\theta}
           \right)\hat{\phi}$.
\end{description}

\subsection{Vectorial Analysis}

\begin{eqnarray}
  \lvec{A}.\left(\lvec{B}\wedge\lvec{C}\right) & = & 
    \lvec{B}.\left(\lvec{C}\wedge\lvec{A}\right)
      = \lvec{C}.\left(\lvec{A}\wedge\lvec{B}\right) \\
  \lvec{A}\wedge\left(\lvec{B}\wedge\lvec{C}\right) 
     & = & \left(\lvec{A}.\lvec{C}\right)\lvec{B} - \left(\lvec{A}.\lvec{B}\right)\lvec{C} \\
  \left(\lvec{A}\wedge\lvec{B}\right).\left(\lvec{C}\wedge\lvec{D}\right)
     & = & \left(\lvec{A}.\lvec{C}\right)\left(\lvec{B}.\lvec{D}\right) 
         - \left(\lvec{A}.\lvec{D}\right)\left(\lvec{B}.\lvec{C}\right) \\
  \lvec{\nabla}\wedge\lvec{\nabla}\psi & = & \lvec{0} \\
  \lvec{\nabla}.\left(\lvec{\nabla}\wedge\lvec{A}\right) & = & 0 \\
  \lvec{\nabla}\wedge\left(\lvec{\nabla}\wedge\lvec{A}\right) 
      & = & \lvec{\nabla}\left(\lvec{\nabla}.\lvec{A}\right) - \lvec{\nabla}^2\lvec{A} \\
  \lvec{\nabla}.\left(\psi\lvec{A}\right) 
      & = & \lvec{A}.\lvec{\nabla}\psi + \psi\lvec{\nabla}.\lvec{A} \\
  \lvec{\nabla}\wedge\left(\psi\lvec{A}\right) 
      & = & \lvec{\nabla}\psi\wedge\lvec{A} + \psi\lvec{\nabla}\wedge\lvec{A} \\
  \lvec{\nabla}\left(\lvec{A}.\lvec{B}\right) 
      & = & \left(\lvec{A}.\lvec{\nabla}\right)\lvec{B} + \left(\lvec{B}.\lvec{\nabla}\right)\lvec{A}
        + \lvec{A}\wedge\left(\lvec{\nabla}\wedge\lvec{B}\right) 
        + \lvec{B}\wedge\left(\lvec{\nabla}\wedge\lvec{A}\right) \\
  \lvec{\nabla}.\left(\lvec{A}\wedge\lvec{B}\right) 
      & = & \lvec{B}.\left(\lvec{\nabla}\wedge\lvec{A}\right)-\lvec{A}.\left(\lvec{\nabla}
                                                         \wedge\lvec{B}\right) \\
  \lvec{\nabla}\wedge\left(\lvec{A}\wedge\lvec{B}\right)
      & = & \lvec{A}\left(\lvec{\nabla}.\lvec{B}\right) - \lvec{B}\left(\lvec{\nabla}.\lvec{A}\right)
        + \left(\lvec{B}.\lvec{\nabla}\right)\lvec{A} - \left(\lvec{A}.\lvec{\nabla}\right)\lvec{B}
\end{eqnarray}

\subsection{Integral Theorems}

\begin{eqnarray}
  \displaystyle\iiint_V\lvec{\nabla}.\lvec{A}\ddiff V 
    & = & \oiint \lvec{A}.\lvec{\dd S} \\
  \displaystyle \iiint_V \lvec{\nabla}\psi\ddiff V
    & = & \oiint \psi\,\lvec{\dd S} \\
  \displaystyle\iiint_V \lvec{\nabla}\wedge\lvec{A}\ddiff V
    & = & - \oiint \lvec{A}\wedge\lvec{\dd S} \\
  \displaystyle\iiint_V\left(\phi\lvec{\nabla}^2\psi
    +\lvec{\nabla}\phi.\lvec{\nabla}\psi\right)\ddiff V
    & = & \oiint\phi\lvec{\nabla}\psi.\lvec{\dd S} \\
  \displaystyle
    \iiint_V \left(\phi\lvec{\nabla}^2\psi-\psi\lvec{\nabla}^2\phi\right)\ddiff V
    & = & \oiint\left(\phi\lvec{\nabla}\psi
                     -\psi\lvec{\nabla}\phi\right).\lvec{\dd S} \\
  \displaystyle\iint_S\left(\lvec{\nabla}\wedge\lvec{A}\right).\lvec{\dd S}
    & = & \oint_C \lvec{A}.\lvec{\dd l} \\
  \displaystyle\iint_S\lvec{\nabla}\psi\wedge\lvec{\dd S}
    & = & - \oint_C \psi\,\lvec{\dd l}
\end{eqnarray}

\subsection{Dust Heating and Cooling: Two Ways of Slicing
               the Pis}

One of the most elementary equations for a grain is the relation between its absorption efficiency and the power it absorbs or emits, given in \refeq{eq:PabsBBQ} and \refeq{eq:PemBBQ}.
Yet, it often causes problems to newcomers, who, by trying to visualize the rays, get the numerical factor in the integrand ($4\pi^2a^2$) wrong.
I have seen several improper values: $16\pi^2a^2$, $4\pi a^2$, \etc\ 
The solution is of course to explicitly write the integral, which I do below.
I also provide two alternative visual solutions to count the factor the right way.

\begin{figure}[htbp]
  \includegraphics[width=\textwidth]{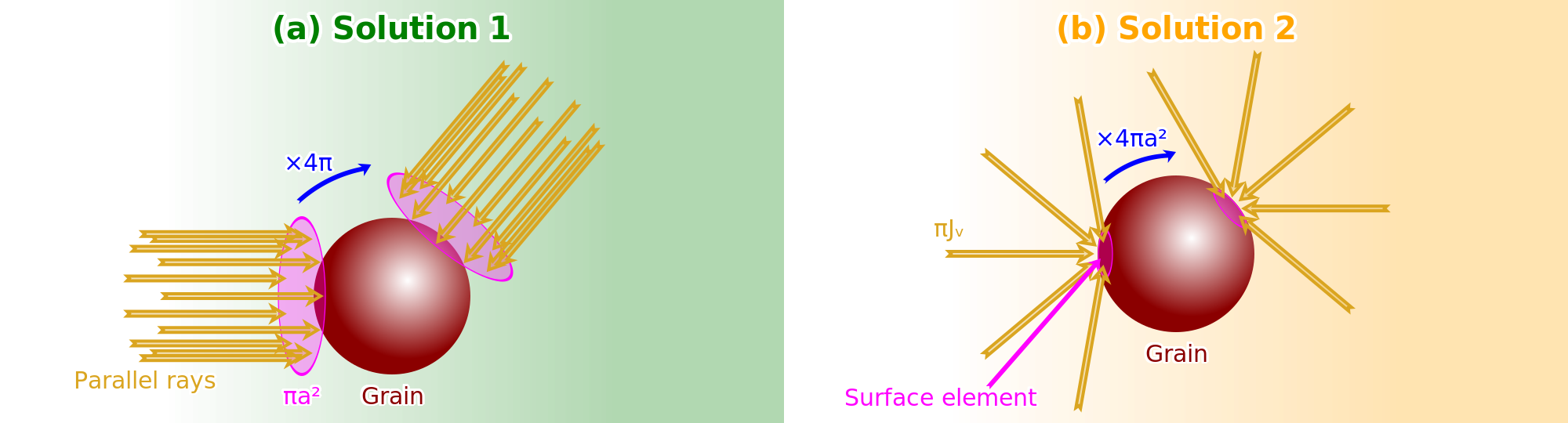}
  \newcap{Two ways of slicing the $\pi$s}%
         {This figure represents the two possible orders to integrate the power 
          absorbed by a grain.
          For the emitted power, the direction of the arrows is simply 
          reversed.
          \CClicence}
  \label{fig:cutpi}
\end{figure}
The power absorbed by a spherical grain of radius $a$, exposed to an isotropic \hISRF\ with mean intensity $J_\nu$, $P_\sms{abs}$, can expressed as:
\begin{equation}
  \dd P_\sms{abs}(a,\theta,\phi) 
    = \left(\int_0^\infty Q_\sms{abs}(a,\nu)J_\nu(\nu)\ddiff\nu\right)
      \ddiff A\ddiff\Omega,
\end{equation}
where $\dd A$ and $\dd\Omega=\dd\cos\theta\ddiff\phi$ are the grain surface and solid angle elements.
$\Omega$ indicates the direction of the incident rays.
The possible orders of the integrals over these two elements is illustrated in \reffig{fig:cutpi}.

  \subsubsection{Solution 1}

The case represented in \refsubfig{fig:cutpi}{a} corresponds to the case where we first integrate over $\ddiff A$.
Thus, parallel rays with a given direction $(\theta,\phi)$ intercept the grain on a surface $\pi a^2$:
\begin{equation}
  \dd P_\sms{abs}(a,\theta,\phi) 
    = \pi a^2\left(\int_0^\infty Q_\sms{abs}(a,\nu)J_\nu(\nu)\ddiff\nu\right)
      \ddiff\Omega.
\end{equation}
Then, we need to integrate over all the possible ray directions:
\begin{eqnarray}
  P_\sms{abs}(a) & = & 
    \pi a^2\left(\int_0^\infty Q_\sms{abs}(a,\nu)J_\nu(\nu)\ddiff\nu\right)
    \iint_\sms{sphere}\ddiff\Omega \\
  & = & 
    \int_0^\infty 4\pi^2 a^2Q_\sms{abs}(a,\nu)J_\nu(\nu)\ddiff\nu.
\end{eqnarray}

  \subsubsection{Solution 2}

The case represented in \refsubfig{fig:cutpi}{b} corresponds to the case where we first integrate the flux over the ray directions, on an incident surface:
\begin{eqnarray}
  \dd P_\sms{abs}(a) & = &
      \left(\int_0^\infty Q_\sms{abs}(a,\nu)J_\nu(\nu)\ddiff\nu\right)
      \int_0^1\cos\theta\ddiff\cos\theta\int_0^{2\pi}\ddiff\phi \\
  & = & 
    \pi\left(\int_0^\infty Q_\sms{abs}(a,\nu)J_\nu(\nu)\ddiff\nu\right).
\end{eqnarray}
There is only a $\pi$ factor here, as we integrate the flux on a surface element, somewhere on the grain.
The rest of the grain shields this surface from the radiation coming from the other hemisphere.
In addition, this integration is weighted by the inclination of the rays on the surface (this is the classical flux formula).
We thus have the flux received by a surface element of the grain, from all possible ray directions.
We now simply need to integrate over the whole grain surface: $4\pi a^2$:
\begin{equation}
  P_\sms{abs}(a) =
      \int_0^\infty 4\pi^2a^2Q_\sms{abs}(a,\nu)J_\nu(\nu)\ddiff\nu.
\end{equation}

\section{Statistics}

  \subsection{General Formulae}

    \subsubsection{Moments of a PDF}

\paragraph{Definitions.}
If we have a joint \hPDF, $p(x,y)$, of two variables, $X$ and $Y$, the first moments are the following.
\begin{description}
  \item[Normalization:] 
    $\displaystyle1=
     \iint_{-\infty}^\infty p(x,y)\ddiff x\ddiff y$.
  \item[Mean:] 
    $\displaystyle\langle X\rangle\equiv
     \iint_{-\infty}^\infty xp(x,y)\ddiff x\ddiff y$.
  \item[Variance:] 
    $\displaystyle V(X)\equiv\iint_{-\infty}^\infty
    \left[x-\langle X\rangle\right]^2p(x,y)\ddiff x\ddiff y$.
  \item[Skewness:] 
    $\displaystyle \gamma_1(X)\equiv\iint_{-\infty}^\infty
    \left[\frac{x-\langle X\rangle}{\sigma(X)}\right]^3p(x,y)\ddiff x\ddiff y$.
  \item[Covariance:] 
    $\displaystyle V(X,Y)\equiv\iint_{-\infty}^\infty
    \left[x-\langle X\rangle\right]
    \times\left[y-\langle Y\rangle\right]p(x,y)\ddiff x\ddiff y$.
  \item[Standard-deviation:] $\sigma(X)\equiv\sqrt{V(X)}$.
  \item[Correlation coefficient:] $\rho(X)\equiv V(X,Y)/\sigma(X)\sigma(Y)$.
\end{description}

\paragraph{Estimators.}
The most commonly used non-robust estimators are the following.
\begin{description}
  \item[Mean:] 
    $\displaystyle\langle X\rangle 
     \simeq \frac{1}{N}\sum_{i=1}^Nx_i
       \;\;\;\left(\pm\frac{\sigma}{\sqrt{N}}\right)$.
  \item[Standard-deviation:] 
   $\displaystyle
    \sigma(X)\simeq
      \sqrt{\frac{1}{N-1}\sum_{i=1}^N\left(x_i-\langle X\rangle\right)^2}
       \;\;\;\left(\pm\frac{\sigma}{\sqrt{2(N-1)}}\right)$.
  \item[Skewness:]
   $\displaystyle
    \gamma_1(X)\simeq
    \frac{N}{(N-1)(N-2)}
      \sum_{i=1}^N\left(\frac{x_i-\langle X\rangle}{\sigma(X)}\right)^3$.
  \item[Correlation coefficient:]
   $\displaystyle
    \rho(X,Y)\simeq
    \frac{1}{N-1}\sum_{i=1}^N
      \left(\frac{x_i-\langle X\rangle}{\sigma(X)}\right)
      \left(\frac{y_i-\langle Y\rangle}{\sigma(Y)}\right)$.
\end{description}

    \subsubsection{Marginalization}

\noindent
Marginalizing over a parameter, $\theta$:
\begin{equation}
  p(y) = \int_{-\infty}^\infty \proba{\theta}\pcond{y}{\theta}\ddiff\theta.
\end{equation}
Comparing two data sets, $y$ and $\tilde{y}$:
\begin{equation}
  \pcond{\tilde{y}}{y} = \int_{-\infty}^\infty 
    \pcond{\tilde{y}}{\theta}\pcond{\theta}{y}\ddiff\theta .
\end{equation}

    \subsubsection{Variable Change}

If we have two sets of random variables, $\vect{u}$ and $\vect{v}$, such that $\vect{v}=f(\vect{u})$, then the relation between their \hPDF\ is:
\begin{equation}
  p_v\left(\vect{v}\right) = |\mathbb{J}|\times p_u\left(f^{-1}\left(\vect{v}\right)\right),
\end{equation}
where the Jacobian of the transformation is:
\begin{equation}
  \mathbb{J}_{i,j}\equiv\frac{\partial u_i}{\partial v_j}.
\end{equation}

    \subsubsection{Combining Uncertainties}

If we have a set of random variables, $\vect{x}$, the uncertainty of an arbitrary function of these parameters, $f(\vect{x})$, is given by:
\begin{equation}
  \sigma^2\left(f(\vect{x})\right)
   = \left(\vect{\nabla}f(\vect{x})\right)^T\mathbb{V}\vect{\nabla}f(\vect{x}),
  \label{eq:deriverr}
\end{equation}
where $\mathbb{V}$ is the covariance matrix of the variable set.
In the 2D case, posing $\vect{x}=(a,b)$, we have:
\begin{eqnarray}
  \vect{\nabla}f=\left(
  \begin{array}{c}
    \displaystyle\frac{\partial f}{\partial a} \\ \\
    \displaystyle\frac{\partial f}{\partial b} 
  \end{array}
  \right)
& \mbox{\&} &
  \mathbb{V}=\left(
  \begin{array}{cc}
    \displaystyle\sigma_a^2 & \rho\sigma_a\sigma_b \\
    \displaystyle\rho\sigma_a\sigma_b & \sigma_b^2
  \end{array}
  \right),
\end{eqnarray}
and \refeq{eq:deriverr} gives the usual expression:
\begin{equation}
  \sigma^2_{f(\vect{x})} 
  = \left(\frac{\partial f}{\partial a}\right)^2\sigma_a^2 
      + \left(\frac{\partial f}{\partial b}\right)^2\sigma_b^2 
      + 2\left(\frac{\partial f}{\partial a}\right)
         \left(\frac{\partial f}{\partial b}\right)\rho\sigma_a\sigma_b.
  \label{eq:comberr2}
\end{equation}
Similarly, the covariance of two functions of the parameter set, $f(\vect{x})$ and $g(\vect{x})$, is:
\begin{equation}
  V(f(\vect{x}),g(\vect{x})) 
    = \left(\vect{\nabla}f(\vect{x})\right)^T\mathbb{V}
       \vect{\nabla}g(\vect{x}).
\end{equation}

\paragraph{Rant about systematics.}
There is a long-lasting laboratorian legend that \citengl{systematics must be non-quadratically summed}.
This is true in some cases and false in others.
Everything depends on what we are talking about.
\begin{enumerate}
  \item
    If we are measuring a flux, $F$, with noise, $\sigma_\sms{noise}$, and 
    calibration uncertainty (systematics), $\sigma_\sms{cal}$, the total 
    uncertainty on the flux will be, according to \refeq{eq:comberr2}: 
    $\sigma_\sms{tot}=\sqrt{\sigma_\sms{noise}^2+\sigma_\sms{cal}^2}$.
    This is because $\rho=0$.
    The fluctuations of the detector's signal at the time of the observation do 
    not have anything to do with the error the instrument's team made by 
    deriving the calibration factor.
  \item
    Now, if we are summing the flux in two pixels with same flux and noise 
    levels, we will get, using \refeq{eq:comberr2}: 
    $\sigma_\sms{tot}=\sqrt{2\sigma^2_\sms{noise}+4\sigma_\sms{cal}^2}$, because 
    the calibration factors of the pixels were correlated.
    The total calibration uncertainty is this time linearly summed: 
    $\sigma_\sms{cal tot}=2\sigma_\sms{cal}$, because $\rho=1$ (the error due to 
    the calibration uncertainty is the same for both pixels).
  \item If we are now summing two systematics, such as the calibration 
    and the background subtraction uncertainties, $\sigma_\sms{cal}$ and 
    $\sigma_\sms{back}$, we will sum them quadratically: 
    $\sigma_\sms{syst}=\sqrt{\sigma_\sms{cal}^2+\sigma_\sms{back}^2}$.
    This is because the error the instrument's team made deriving the 
    calibration factor is independent of the error we have made by selecting a
    region in one of the corners of our map, assuming it was free of galaxy 
    emission.
\end{enumerate}

  \subsection{Useful Probability Distributions}
  \label{sec:statistics}

    \subsubsection{Binomial Distribution}

Discrete probability distribution to get $r$ successes out of $n$ tries, each one having a probability $p$:
\begin{equation}
  P_\sms{binomial}(r|p,n) \equiv C_r^n p^r(1-p)^{n-r}
    = \frac{n!}{r!(n-r)!}p^r(1-p)^{n-r},
\end{equation}
with $\langle r\rangle=np$ and $\sigma(r)=\sqrt{np(1-p)}$.

    \subsubsection{Poisson Distribution}

Discrete probability distribution to get $r$ events per unit time knowing the mean expected number, $\lambda$, of such events per unit time:
\begin{equation}
  P_\sms{Poisson}(r|\lambda)\equiv\frac{e^{-\lambda}\lambda^r}{r!},
\end{equation}
with $\langle r\rangle=\lambda$ and $\sigma(r)=\sqrt{\lambda}$.
The superposition of two Poissonian events $(\lambda_a,\lambda_b)$ is also Poissonian with mean $\lambda=\lambda_a+\lambda_b$.
It is the limit of the binomial distribution to large numbers:
\begin{equation}
  P_\sms{Poisson}(r|\lambda) = \lim_{n\rightarrow\infty}
    P_\sms{binomial}\left(r\middle|\frac{\lambda}{n},n\right).
\end{equation}

    \subsubsection{Gaussian Distribution}

\begin{equation}
  P_\sms{Gauss}(x|\mu,\sigma)\equiv
    \frac{1}{\sqrt{2\pi}\sigma}\exp\left(-\frac{(x-\mu)^2}{2\sigma^2}\right),
\end{equation}
with $\langle x\rangle=\mu$, $\sigma(x)=\sigma$ and all superior moments equal to 0.
It is the limit of a Poisson distribution when $\lambda\gg1$: 
$\displaystyle P_\sms{Poisson}(r|\lambda)
  =\lim_{\lambda\gg1}P_\sms{Gauss}\left(r\middle|\lambda,\sqrt{\lambda}\right)$.

\paragraph{Multivariate form.}
A multivariate normal law of mean $\vect{\mu}$ and covariance matrix $\mathbb{V}$ is defined as:
\begin{equation}
  P_\sms{Gauss}(\vect{x}|\vect{\mu},\mathbb{V})\equiv
  \frac{1}{(2\pi)^{n/2}\sqrt{|\mathbb{V}|}}
    \exp\left(-\frac{1}{2}\left(\vect{x}-\vect{\mu}\right)^T\mathbb{V}^{-1}
              \left(\vect{x}-\vect{\mu}\right)\right).
\end{equation}

\paragraph{Error function.}
Noting $\Phi(x)$ the \hCDF\ of a reduced normal law, the \expression{error function} is defined such that:
\begin{equation}
  \Phi(x)=\frac{1}{2}\left(1+\erf\left(\frac{x}{\sqrt{2}}\right)\right).
\end{equation}
It is thus:
\begin{equation}
  \erf(x)\equiv\frac{2}{\sqrt{\pi}}\int_0^xe^{-t^2}\ddiff t
    = 2\Phi\left(x\sqrt{2}\right)-1.
\end{equation}

    \subsubsection{Student's \textit{t} Distribution}

It is defined as:
\begin{equation}
  P_\sms{Student}(x|f)
    \equiv\frac{1}{\sqrt{f\pi}}
    \frac{\displaystyle\Gamma\left(\frac{f+1}{2}\right)}%
         {\displaystyle\Gamma\left(\frac{f}{2}\right)}
    \left(1+\frac{x^2}{f}\right)^{-\frac{f+1}{2}},
\end{equation}
with $f>0$ being the \expression{degree of freedom}.
Its mean is 0 and its standard-deviation, for $f>2$, is $\sigma=\sqrt{f/(f-2)}$.

    \subsubsection{Split-Normal Distribution}

It is, to my mind, the most convenient asymmetric distribution:
\begin{equation}
  P_\sms{split-norm}(x|\mu,\lambda,\tau)
  \equiv\sqrt{\frac{2}{\pi}}\frac{1}{\lambda(1+\tau)}\times
  \left\{
  \begin{array}{ll}
    \displaystyle
    \exp\left(-\frac{1}{2}\left(\frac{x-\mu}{\lambda}\right)^2\right)
     & \mbox{ if } x\le \mu \\
    & \\
    \displaystyle
    \exp\left(-\frac{1}{2}\left(\frac{x-\mu}{\lambda\tau}\right)^2\right)
     & \mbox{ if } x> \mu. \\
  \end{array}\right.
\end{equation}
Posing:
\begin{equation}
  b=\frac{\pi-2}{\pi}(\tau-1)^2+\tau,
\end{equation}
the first moments are:
\begin{align}
  \langle X\rangle &= \mu+\sqrt{\frac{2}{\pi}}\lambda(\tau-1) \\
  \sigma(X) & = \sqrt{b}\lambda \\
  \gamma_1(X) & = b^{-3/2}\sqrt{\frac{2}{\pi}}(\tau-1)\times
    \left[\left(\frac{4}{\pi}-1\right)(\tau-1)^2+\tau\right].
\end{align}

    \subsubsection{Lorentzian Distribution}

\begin{equation}
  P_\sms{Lorentz}(x|\mu,\gamma)\equiv
  \frac{2}{\pi\gamma}
    \frac{1}{\displaystyle1+\left(\frac{x-\mu}{\gamma/2}\right)^2}.
\end{equation}
Its mean is $\mu$, its \hFWHM\ is $\gamma$, but its standard-deviation is not defined.

  \subsection{Drawing random variables from an arbitrary 
                distribution}
  \label{app:random}

    \subsubsection{The Rejection Method}
    \label{sec:random_rejection}

\begin{figure}[htbp]
  \begin{tabular}{cc}
    \includegraphics[width=0.48\textwidth]{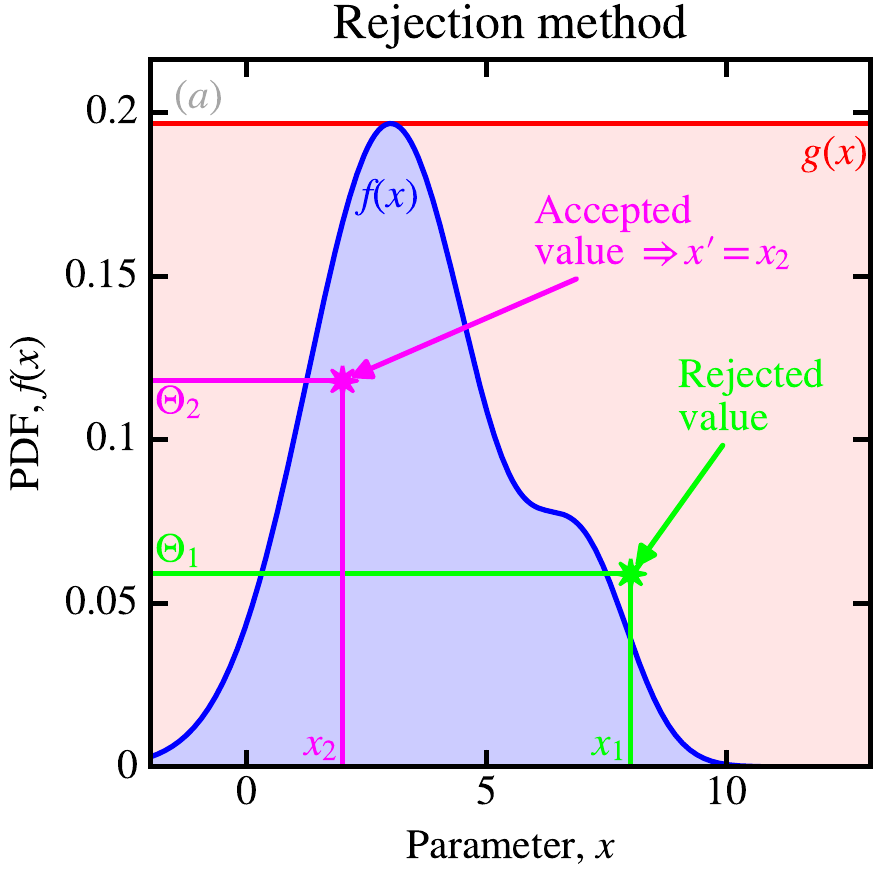} &
    \includegraphics[width=0.48\textwidth]{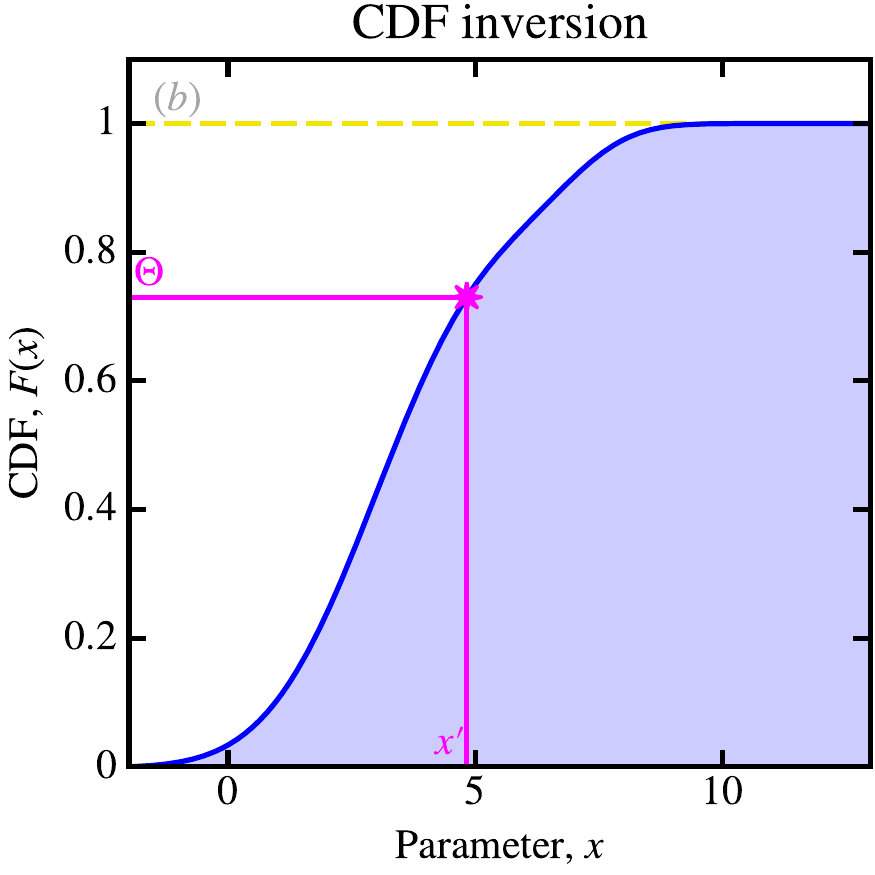} \\
  \end{tabular}
  \newcap{Methods for drawing random numbers from arbitrary distributions}%
         {Panel~\textit{(a)} represents the rejection method applied to the
          distribution in blue, with the proposal in red.
          We have represented a first rejected draw and a second accepted one.
          Panel~\textit{(b)} represents the CDF method applied to the 
          distribution in panel~\textit{(a)}.
          \CClicence}
  \label{fig:random_draw}
\end{figure}

The \expression{rejection method} is a widely used technique to draw a random variable, $x^\prime$, from an arbitrary \hPDF, $f(x)$.
It requires the ability to easily draw a random variable, $x_1$, from a \expression{proposal distribution}, $g(x)$, such that $g(x)\ge f(x)$ $\forall x$.
In case $f(x)$ is finite over $[x_\sms{min},x_\sms{max}]$, we can take:
\begin{equation}
  g(x) = \left\{
  \begin{array}{ll}
    \max{(f)} & \mbox{ for } x_\sms{min}\le x\le x_\sms{max} \\
    0 & \mbox{ elsewhere. } 
  \end{array}
  \right.
\end{equation}
The algorithm is the following.
\begin{enumerate}
  \item Draw a random variable $x_1$ from $g(x)$.
  \item Draw a uniform random variable between 0 and 1, $\Theta_1$.
  \item If $\Theta_1<f(x_1)/g(x_1)$, then $x^\prime=x_1$ is accepted. 
    Otherwise, if $\Theta_1\ge f(x_1)/g(x_1)$, this draw is rejected, and we 
    need to go back to the first step.
\end{enumerate}
The closer $g(x)$ is from $f(x)$, the lower the rejection rate will be, and the faster the method will be.
It is illustrated in \refsubfig{fig:random_draw}{a}.

    \subsubsection{Inverting the CDF}
    \label{sec:random_CDF}

A general Monte-Carlo technique to draw a random variable, $x^\prime$, from an arbitrary \hPDF, $f(x)$, consists in drawing a uniform random variable between 0 and 1, $\Theta$, and inverting the \expression{Cumulative Distribution Function} (\hCDF) of $f(x)$:
\begin{equation}
  F(x)\equiv\int_{-\infty}^x f(y)\ddiff y.
\end{equation}
The desired random variable is then simply:
\begin{equation}
  x^\prime=F^{-1}(\Theta).
\end{equation}
It is illustrated in \refsubfig{fig:random_draw}{b}.

\section{Trigonometry}

\begin{align}
  & \cos a = \cos b && \Leftrightarrow
    && a=b\;[2\pi]\;\;\;\vee\;\;\; a=-b \;[2\pi]
  \\
  & \sin a = \sin b && \Leftrightarrow
    && a=b\;[2\pi]\;\;\;\vee\;\;\; a=\pi-b \;[2\pi]
  \\
  & \tan a = \tan b && \Leftrightarrow
    && a=b\;[2\pi]
\end{align}

  \subsection{Transformations}

    \subsubsection{Rotations}

\begin{align}
  & \cos\left(\frac{\pi}{2}+x\right) = - \sin x &&
  \sin\left(\frac{\pi}{2}+x\right) = \cos x 
  \\
  & \cos\left(\frac{\pi}{2}-x\right) = \sin x &&
  \sin\left(\frac{\pi}{2}-x\right) = \cos x 
  \\
  & \cos(\pi-x) = -\cos x && \sin(\pi-x)=\sin x
  \\
  & \cos(\pi+x)=-\cos x && \sin(\pi+x)=-\sin x
  \\
  & \cos(-x)=\cos x && \sin(-x)=-\sin x
\end{align}

    \subsubsection{Relations Between Functions}

\begin{align}
  & \cos^2 x +\sin^2 x = 1 && 1+\tan^2x=\frac{1}{\cos^2x}
\end{align}
\begin{align}
  & \cos(2x)=\frac{1-\tan^2x}{1+\tan^2x} 
  && \sin(2x)=\frac{2\tan x}{1+\tan^2x}
  && \tan(2x)=\frac{2\tan x}{1-\tan^2x}
\end{align}

  \subsection{Addition}

    \subsubsection{Summing Angles}

\begin{align}
  & \cos(a-b) = \cos a .\cos b + \sin a .\sin b &&
  \cos(a+b) = \cos a .\cos b - \sin a .\sin b
  \\
  & \sin(a-b) = \sin a .\cos b - \cos a .\sin b &&
  \sin(a+b) = \sin a .\cos b + \cos a .\sin b
  \\
  & \tan(a-b) = \frac{\tan a - \tan b}{1 + \tan a .\tan b} &&
  \tan(a+b) = \frac{\tan a + \tan b}{1 - \tan a .\tan b}
\end{align}

    \subsubsection{Inverse Relations}

\begin{equation}
  \cos a .\cos b = \frac{1}{2}\left[\cos(a-b)+\cos(a+b)\right]
\end{equation}
\begin{equation}
  \cos a .\sin b = \frac{1}{2}\left[\sin(a+b)-\sin(a-b)\right]
\end{equation}
\begin{equation}
  \sin a .\sin b = \frac{1}{2}\left[\cos(a-b)-\cos(a+b)\right]
\end{equation}
\begin{align}
  & \cos a + \cos b 
    = 2\cos\left(\frac{a+b}{2}\right)\cos\left(\frac{a-b}{2}\right) &&
  \cos a - \cos b 
    = -2\sin\left(\frac{a+b}{2}\right)\sin\left(\frac{a-b}{2}\right)
  \\
  & \sin a + \sin b 
    = 2\sin\left(\frac{a+b}{2}\right)\cos\left(\frac{a-b}{2}\right) &&
  \sin a - \sin b 
    = 2\sin\left(\frac{a-b}{2}\right)\cos\left(\frac{a+b}{2}\right)
\end{align}

  \subsection{Linearization}

    \subsubsection{Squares and Cubes}

\begin{align}
  & \cos^2x=\frac{1+\cos(2x)}{2} &&
  \sin^2x=\frac{1-\cos(2x)}{2} &&
  \tan^2x=\frac{1-\cos(2x)}{1+\cos(2x)} 
  \\
  & \cos^3x=\frac{\cos(3x)+3\cos x}{4} &&
  \sin^3x=\frac{-\sin(3x)+3\sin x}{4} &&
  \tan^3x=\frac{-\sin(3x)+3\sin x}{\cos(3x)+3\cos x}
\end{align}

    \subsubsection{Inverse Relations}

\begin{equation}
  \cos(2x) \;\;\;=\;\;\;\ \cos^2x-\sin^2x \;\;\;=\;\;\; 2\cos^2x-1 
    \;\;\;=\;\;\; 1-2\sin^2x
\end{equation}
\begin{equation}
  \sin(2x) = 2\sin x .\cos x \;\;\;\;\;\;\;\;\;\;\;\;\;\;
  \tan(2x) = \frac{2\tan x}{1-\tan^2 x}
\end{equation}
\begin{equation}
  \cos(3x) = 4\cos^3x-3\cos x \;\;\;\;\;\;\;\; \sin(3x) = 3\sin x - 4\sin^3x 
    \;\;\;\;\;\;\;\; \tan(3x) = \frac{3\tan x-\tan^3x}{1-3\tan^2x}
\end{equation}


\backmatter
\chapter{Acknowledgements}
\fancyhead[LO,RE]{\textcolor{colhead}{\textsl{Acknowledgements}}}
\citesmart{Bien faire et laisser dire.}%
          {(devise d'\'Etienne Fran\c{c}ois \textsc{Sall\'e de Chou})}

\renewcommand{\thefigure}{D.1}
\begin{figure}[htbp]
  \includegraphics[width=\textwidth]{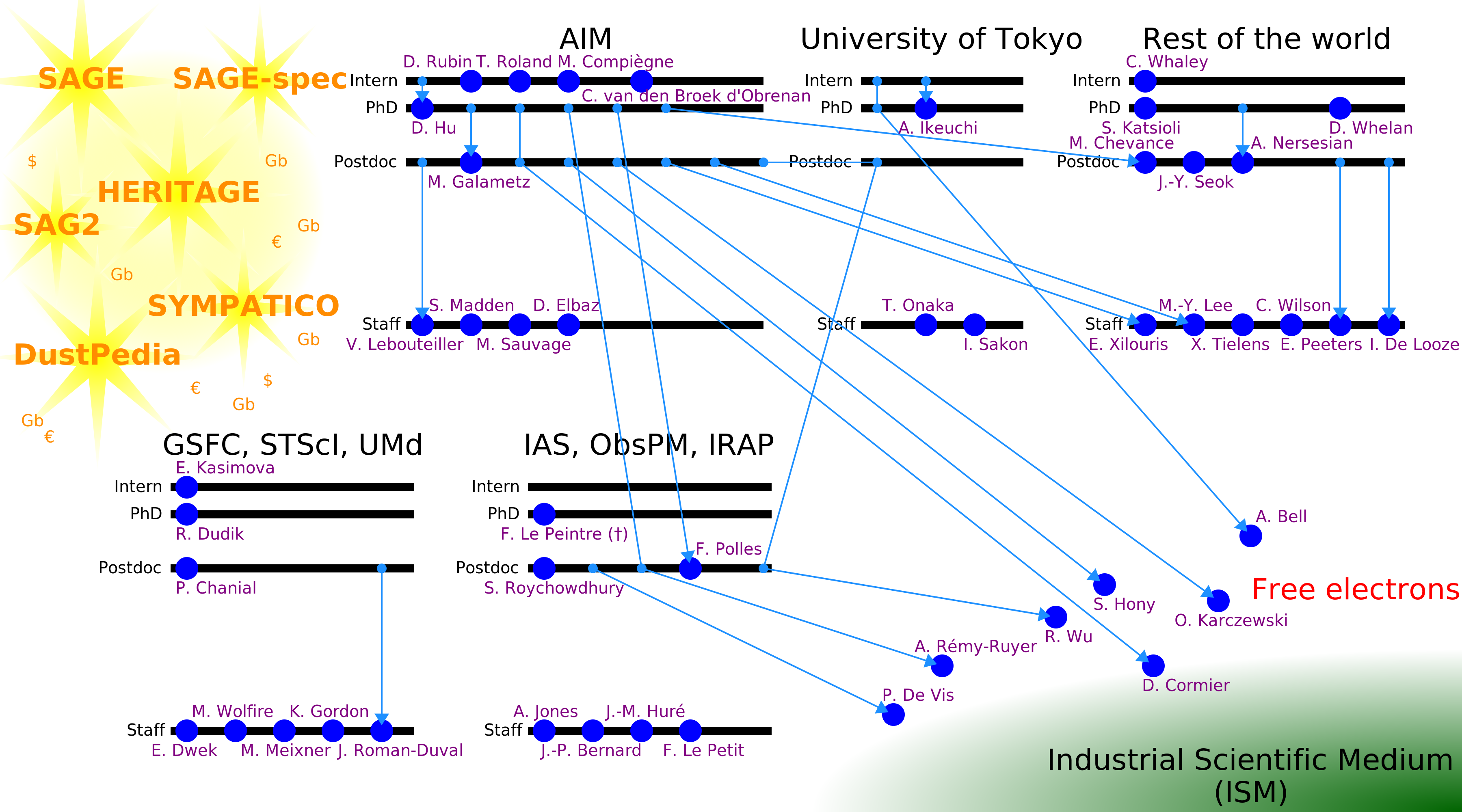}
  \newcap{Nerdy allegory of my collaboration network}%
         {Large collaborations are sources of funding and of Gigabytes of data.
          They subsequently trigger transitions between levels in laboratories.
          Collaborators usually become more bound to their academic 
          institution with time.
          We also notice some charge exchange and a few individuals becoming
          free electrons in the private sector.
          \CClicence}
  \label{fig:acknowledgements}
\end{figure}
\noindent
I am particularly grateful to the colleagues who have accepted to participate in my \hHDR\ committee and to read this whole manuscript: Véroni\-que \familyname{Buat}, Vassilis \familyname{Charmandaris}, Thomas \familyname{Henning}, Laurent \familyname{Verstraete}, Stéphane \familyname{Charlot} and François-Xavier \familyname{Désert}.
I thank them for their comments and the interesting discussions we had.
I also thank Hervé \familyname{Dole} for handling the administrative procedure of this accreditation.

I thank the people who willingly provided me with useful comments, clarifying the text and correcting some errors: Dangning \familyname{Hu}, Anthony \familyname{Jones}, Vianney \familyname{Lebouteiller}, Suzanne \familyname{Madden}, Marc-Antoi\-ne \familyname{Miville-Deschênes} and Takashi \familyname{Onaka}.
I also thank Christine \familyname{Joblin} for her insight on the 3.4~\tmic\ feature, and Matteo \familyname{Bugli} for a useful discussion about \hSN e and collapsars.
I thank Anthony \familyname{Jones} and Nathalie \familyname{Ysard} for providing me with their \hDDA\ results, plotted in \reffig{fig:aggregates}, and Chia-Yu \familyname{Hu} for his sputtering yield polynomial fits used to make \reffig{fig:sputtering}.
I am finally grateful to all the people who have granted me the permission to reproduce their figures in this manuscript: Lou \familyname{Allamandola}, Jean-Philippe \familyname{Bernard}, Robert \familyname{Gendler}, Peter \familyname{Hoppe}, Tom \familyname{Jarrett}, Andy \familyname{Mattioda} and Marc-Antoine \familyname{Miville-Deschênes}.
\takeaway{No dust grains were harmed in the making of this work.}

\clearpage
\fancyhead[LO,RE]{\textcolor{colhead}{\textsl{Bibliography}}}
\fancyhead[LE,RO]{}
\bibliographystyle{bib_notes}
\bibliography{references}

\clearpage
\pagestyle{empty}
\begin{mdframed}[linecolor=Prune,linewidth=1]
  \vspace{-.25cm}
  \selectlanguage{french}
  \paragraph*{Titre~:}
    Propriétés et évolution de la poussière interstellaire. 
    Le point de vue des galaxies proches.

  \begin{small}
    \vspace{-.25cm}
    \paragraph*{Mots clés~:} Milieu interstellaire -- Poussière -- Astronomie  
      infrarouge -- Évolution des galaxies -- Science des données.

    \vspace{-.5cm}
    \begin{multicols}{2}
      \paragraph*{Résumé~:} 
        La poussière interstellaire est un ingrédient physique clef des 
        galaxies, responsable de l'obscurcissement de la formation stellaire,
        de la régulation du chauffage et du refroidissement du gaz, et de la 
        croissance de la complexité chimique.
        Dans ce manuscrit, je donne une large revue des propriétés de la 
        poussière interstellaire et des techniques modernes utilisées pour 
        l'étudier.
        Je commence avec une introduction générale, présentant les principaux 
        concepts, en physique moléculaire et physique du solide, requis pour
        comprendre la littérature contemporaine sur le sujet.
        Je passe ensuite en revue les évidences empiriques que nous utilisons
        de nos jours pour contraindre les modèles à l'état de l'art.
        S'ensuit une longue discussion sur notre compréhension actuelle des 
        propriétés des grains dans les galaxies proches, avec un accent sur
        la modélisation des distributions spectrales d'énergie.
        Le chapitre suivant présente l'évolution des grains, à toutes les 
        échelles.
        Je passe en revue les différents processus microphysiques d'évolution 
        et la manière dont ils sont pris en compte dans les modèles d'évolution
        cosmique des grains.
        Je donne mon opinion sur l'origine de la poussière interstellaire dans
        les galaxies en fonction de la métallicité.
        Le dernier chapitre traite de méthodologie.
        J'y donne une introduction sur la méthode bayésienne et la compare aux
        techniques fréquentistes.
        J'y discute les conséquences épistémologiques des deux approches, et
        montre pourquoi le champ d'étude de la poussière interstellaire requiert
        un point de vue probabiliste.
        Je finis le manuscrit par un résumé des avancées majeures obtenues au 
        cours de la dernière décennie et je fais une prospective pour la 
        prochaine.
    \end{multicols}
  \end{small}
\end{mdframed}

\begin{mdframed}[linecolor=Prune,linewidth=1]
  \vspace{-.25cm}
  \selectlanguage{english}
  \paragraph*{Title:} A Nearby Galaxy Perspective on Interstellar Dust 
    Properties and their Evolution.

  \begin{small}
    \vspace{-.25cm}
    \paragraph*{Keywords:} Interstellar medium -- Dust -- Infrared astronomy 
      -- Galaxy evolution -- Data science.

    \vspace{-.5cm}
    \begin{multicols}{2}
      \paragraph*{Abstract:} 
      \abstractEN
    \end{multicols}
  \end{small}
\end{mdframed}

\fontfamily{fvs}\fontseries{m}\selectfont
\begin{tabular}{p{14cm}r}
  \multirow{3}{16cm}[+0mm]{{\color{Prune} Université Paris-Saclay, France}} & \\
\end{tabular}

\end{document}